\documentclass{aa}
\usepackage{graphicx}
\usepackage{txfonts}
\usepackage[colorlinks = true,
linkcolor = blue,
urlcolor  = blue,
citecolor = blue]{hyperref}
\begin{document}

\title{MEGADES: MEGARA galaxy disc evolution survey}

\subtitle{Data release I: Central fields}

\author{M. Chamorro-Cazorla\inst{1,2},
A. Gil de Paz\inst{1,2},
Á. Castillo-Morales\inst{1,2},
J. Gallego\inst{1,2},
E. Carrasco\inst{3},
J. Iglesias-P\'aramo\inst{4},
M. L. Garc\'ia-Vargas\inst{5},
S. Pascual\inst{1,2},
N. Cardiel\inst{1,2},
C. Catal\'an-Torrecilla\inst{1,2},
J. Zamorano\inst{1,2},
P. Sánchez-Blázquez\inst{1,2},
A. P\'erez-Calpena\inst{5},
P. G\'omez-\'Alvarez\inst{5}
\and
J. Jiménez-Vicente\inst{6,7}
}

\institute{\inst{1}Departamento de F\'isica de la Tierra y
Astrof\'isica, Fac. CC. Físicas, Universidad Complutense de Madrid, Plaza de las Ciencias, 1, Madrid, E-28040, Spain\\
\email{mchamorro@ucm.es}\\
\inst{2}Instituto de Física de Partículas y del Cosmos, IPARCOS-UCM, Fac. CC. Físicas, Universidad Complutense de Madrid, Plaza de las Ciencias 1, Madrid, E-28040, Spain\\
\inst{3}Instituto Nacional de Astrof\'isica, \'Optica
y Electr\'onica, Luis Enrique Erro No.1, C.P. 72840, Tonantzintla,
Puebla, Mexico\\
\inst{4}Instituto de Astrof\'isica de Andaluc\'ia-CSIC,  Glorieta de
la Astronom\'ia s/n, 18008, Granada, Spain\\ 
\inst{5}FRACTAL S.L.N.E.. C/ Tulip\'an 2, p13, 1A. E-28231, Las Rozas de Madrid, Spain\\
\inst{6}Departamento de Física Teórica y del Cosmos, Universidad de Granada, Campus de Fuentenueva, 18071, Granada, Spain\\
\inst{7}Instituto Carlos I de Física Teórica y Computacional, Universidad de Granada, 18071, Granada, Spain
}

% \institute{\inst{1}Departamento de Física de la Tierra y Astrofísica, Universidad Complutense de Madrid, E-28040 Madrid, Spain\\
%         \email{mchamorro@ucm.es}
%\and
%    University of Alexandria, Department of Geography, ...\\
%    \email{c.ptolemy@hipparch.uheaven.space}
%    \thanks{The university of heaven temporarily does not
%           accept e-mails}
%             }

\date{\today}

% \abstract{}{}{}{}{} 
% 5 {} token are mandatory

\abstract
{The main interest of the science team for the exploitation of the MEGARA instrument at the 10.4m Gran Telescopio Canarias (GTC hereafter) is devoted to the study of nearby galaxies. The focus lies on researching the history of star formation, and the chemical and kinematical properties of disc systems. We refer to this project as MEGADES: the MEGARA galaxy disc evolution survey. The initial goal of MEGADES is to provide a detailed study of the inner regions of nearby disc galaxies in terms of their spectrophotometric and chemical evolution, and to provide a dynamical characterisation by distinguishing the contribution of in situ and ex situ processes to the history of star formation and effective chemical enrichment of these regions. In addition, the dynamical analysis of these inner regions naturally includes the identification and characterisation of galactic winds that might be present in these regions. At a later stage, we will extend this study farther out in galactocentric distance. The first stage of this project encompasses the analysis of the central regions of 43 nearby galaxies observed with the MEGARA integral field unit for $\sim$ 114 hours, including both guaranteed time and open time observations. In this paper we provide a set of all the processed data products available to the community and early results from the analysis of these data regarding stellar continuum and ionised and neutral gas features.}


\authorrunning{M. Chamorro-Cazorla et al.}
\maketitle
%
%________________________________________________________________

\section{Introduction}

Whereas the fundamental principles of stellar evolution are reasonably well understood since the last century, the fundamental aspects of galaxy evolution continue to be the subject of fierce discussion within the astronomical community. This is due to the fact that the distance between galaxies, relative to their own size, is considerably smaller than in the case of stars, which means that the evolution of the vast majority of galaxies has proceeded under the influence of other galaxies, and did not occur in isolation, especially in early epochs when the universe was much denser than it is at present. Moreover, some processes that take place in galaxies and are considered to be internal secular processes, might be periodic, and can be triggered by the interaction of the galaxies with their environment, such as nuclear activity or intense star formation. However, even if these mechanisms are classified as secular (considering the definition of secularity given by \citealt{Kormendy_2004}), they leave an evident and clearly distinct imprint on the later evolution of these objects that helps to distinguish the evolution of the different galaxies. Some of the mechanisms capable of affecting the kinematic, chemical, and photometric properties of galaxies are major and minor mergers \citep{Toomre_1972}, bars, rings, density waves \citep{Lin_1964}, active galactic nuclei (AGN) feedback (\citealt{Croton_2006}, \citealt{Fabian_2012}), stellar diffusion, gas in-fall \citep{Sancisi_2008}, and, in the case of galaxies in clusters, also ram-pressure striping \citep{Gunn_1972} and galaxy harassment \citep{Moore_1996}. At the same time, galactic winds (GWs) constitute an important mechanism for redistributing dust and metals both in galaxies and towards the intergalactic medium (IGM) and have been invoked to reproduce the scale relations observed in galaxies, as well as to understand the apparent discrepancies between the theoretical and observed luminosity functions and to understand the evolution of galaxies (especially of high-redshift objects) through the green valley. The contribution of all these mechanisms is rather complex, therefore, determining the evolution of galaxies is in many aspects still a puzzle. 

In recent years, our knowledge about these questions has progressed significantly, mostly through studies carried out with integral field spectroscopy (IFS) instruments. These instruments are particularly well suited for understanding the role played by each of these mechanisms in the evolution of galaxies, as they provide two-dimensional spectroscopic information. This technique has already demonstrated its potential in the study of nearby galaxies such as SAURON\footnote{Spectroscopic Areal Unit for Research on Optical Nebulae} \citep{Zeeuw_2002}, ATLAS$^{\mathrm{3D}}$ \citep{Cappellari_2011}, CALIFA\footnote{The Calar Alto Legacy Integral Field Area survey} \citep{Sanchez_2012}, MaNGA\footnote{Mapping Nearby Galaxies at APO} \citep{Bundy_2015}, VENGA\footnote{The VIRUS-P Exploration of Nearby Galaxies} \citep{Blanc_2013} and SIGNALS\footnote{Star-formation, Ionized Gas and Nebular Abundances Legacy Survey} \citep{Rousseau_2019}. These surveys have enabled a great advance, especially in studies of galaxies at low redshift, because existing techniques did not allow performing diverse analyses like this before. 

At higher redshifts, we also count on surveys based on IFS observations such as SINS\footnote{Spectroscopic Imaging survey in the Near-infrared with SINFONI} \citep{Forster_2009}, MASSIVE\footnote{Mitchell spectrograph Assembly of Stars and Stuffwith Integral-field spectroscopy in the Visible} \citep{Contini_2012}, KMOS$^{\mathrm{3D,}}$\footnote{K-band Multi Object Spectrograph Survey} \citep{Wisnioski_2015}, and MAGIC\footnote{MUSE-gAlaxy Groups In Cosmos Survey} \citep{Mercier_2022}. The main difference between these studies and those based on observations made on nearby galaxies is that at high redshift, these surveys basically focus on the analysis of the brightest emission lines and do not study the stellar continuum because the sensitivity of the instruments prevents observing them correctly. Studies based on multi-band photometric observations that generate spectral energy distributions (SED), such as J-PAS\footnote{Javalambre Physics of the Accelerating Universe Astrophysical Survey} \citep{Benitez_2014} and SHARDS\footnote{Survey for High-z Absorption Red \& Dead Sources}  \citep{Perez_Gonzalez_2013}, represent the best alternative to these high-z surveys when the stellar continuum is to be studied. SED-based surveys have all the spatial information and are very powerful at high redshift in determining a large number of parameters (i.e. mean ages, redshifts, and stellar masses), but they lack the spectral resolution needed to measure individual lines or different kinematic components. However, we are now on the verge of an unprecedented revolution in the study of distant galaxies through the start of the JWST\footnote{James Webb Space Telescope} \citep{Gardner_2006} observations, in photometry with NIRcam\footnote{The JWST Near Infrared Camera} \citep{Rieke_2005} and in 2D spectroscopy with NIRSpec\footnote{The JWST Near Infrared Spectrograph} \citep{Bagnasco_2007}.

These surveys have also taught us that the amount of information that their data provide and the studies that can be carried out with them must rely on multi-wavelength information from multiple facilities, both ground-based and from space. This necessitates public sharing of the observations with the rest of the community, and with them, the legacy surveys, in order to be able to exploit them in the best possible way. 

In this paper, we present the first data release within MEGADES (MEGARA galaxy disc evolution survey), the legacy survey of the MEGARA instrument (\textit{Multi-Espectrógrafo  en  GTC de  Alta  Resolución  para  Astronomía}) (\citealt{armando_2018SPIE};  \citealt{carrasco_2018SPIE}). MEGARA is a high-resolution optical spectrograph installed on the 10.4\,m GTC telescope with two observing modes, the large compact bundle (LCB) mode, an IFU covering a field of view (FoV) of 12.5\,$\times$\,11.3\,arcsec$^{2}$ with a spatial resolution of 0.62 arcsec, and the multi-object spectroscopy (MOS) mode, which allows the simultaneous acquisition of 92 object spectra in an area of 3.5\,$\times$\,3.5\,arcmin$^{2}$ around the IFU. The observations can cover a spectral interval from 3,650 to 9,700\,\r{A} in different ranges with three possible resolutions, R=$\lambda$/FWHM $\sim$ 6,000, 12,000, and 20,000.

The unique characteristics of MEGARA and the large collecting area of GTC mean that the MEGADES survey is another step forward on our way to discovering the mechanisms that shape the evolution of galaxies through the unprecedented combination of depth, spatial, and especially spectral resolutions. MEGADES aims to improve our knowledge of the secular processes that galaxies undergo by studying the kinematic properties of stars and gas in all the phases in which the latter is present, determining the characteristics of stellar populations, analysing diagnostic diagrams of the emission line ratio and discovering the nature of the galaxy bulges in the sample. In this first paper, we present studies of the stellar and gas kinematics and spectral line flux measurements in the central regions of MEGADES galaxies. Together with this paper, we release the processed observations and the analysis products we obtained.

In $\S$\ref{sec: survey} we describe the scientific goals of the survey together with the sample of observed galaxies. $\S$\ref{sec: observations} describes the observations in detail. In $\S$\ref{sec: data_processing} we explain the process of data reduction and subsequent data refinements. We detail the analysis of the stellar kinematics and the spectral line fitting performed on the data in $\S$\ref{sec: analysis}. In $\S$\ref{sec: data_release} we detail all the data and products that are released together with this paper. Finally, we summarise in $\S$\ref{sec: summary} .
Throughout this paper, we assume a standard $\Lambda$CDM universe, whose cosmological parameters are H$_{0}$ = 70\,km\,s$^{-1}$\,Mpc$^{-1}$, $\Omega_{\Lambda}$ = 0.7, and $\mathrm{\Omega_{m}}$ = 0.3.

\section{Survey}
\label{sec: survey}

\subsection{MEGADES goals}

Since this is introductory paper to MEGADES that includes data from the central regions of the galaxies alone together with the first kinematic analyses of the stellar content and interstellar gas, the bulk of the MEGADES goals will be addressed in future papers. The long-term objective of MEGADES is to understand the impact of secular processes on disc evolution. Our primary goal is to test whether gas infall is one of the main mechanisms in the evolution of galaxies, and whether it drives the inside-out formation of discs. Therefore, we will test the model predictions for this scenario against our MEGADES observations by analysing the secular and external causes of the differences, that is, nuclear activity, stellar migration, minor mergers, intense star formation, and so on. We will use MEGARA high spatial and spectral resolution 2D spectroscopy to analyse the kinematic properties of the stellar component in order to detect different structures such as bars or inner discs in the central regions of MEGADES galaxies. We will also analyse line indices sensitive to age and metallicity \citep{Chamorro_Cazorla_2022}, and for the detected HII regions, the gas emission line spectrum, from which we will produce diagnostic diagrams and determine chemical abundances. Finally, we will study the presence and frequency of GWs in the MEGADES sample, both in its warm and cold phases. We will examine the kinematic properties and the shape of the H$\alpha$ emission line to distinguish different kinematic components to study the warm phase, and we will use the NaI D absorption doublet to analyse the cold phase.

With this first data release, we will be able to perform all these studies in the central regions of the galaxies in the sample and then place them in context with analyses of the outer emission disc when these observations become available. We will study the stellar populations that make up the bulges of the MEGADES galaxies together with their kinematical properties. With this information, we will be able to determine whether classical bulges dominate the galaxies in our sample or if there are more galaxies with pseudo-bulges \citep{Kormendy_2004}. We will also study the possible presence of outflows associated with nuclear starbursts and AGN, considering the existence of different kinematic components in the neutral or warm phase, and the differences that may exist between the two phases. In addition to all of this, we will be able to generate different diagnostic diagrams with the information contained in the different lines we will study.

\subsection{MEGADES sample}

\begin{figure*}[h]
        \centering
        \includegraphics[trim={7,1cm 0mm 8cm 1cm},clip, width=0.21\linewidth]{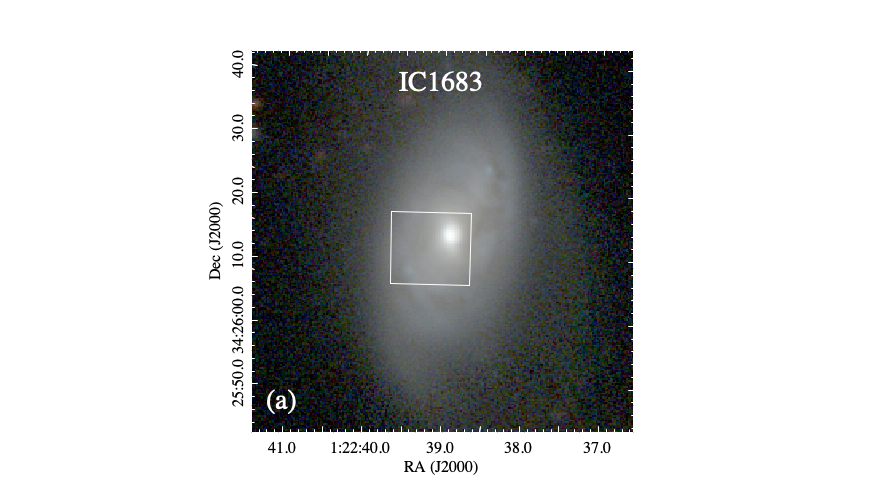}
        \includegraphics[trim={7,1cm 0mm 8cm 1cm},clip, width=0.21\linewidth]{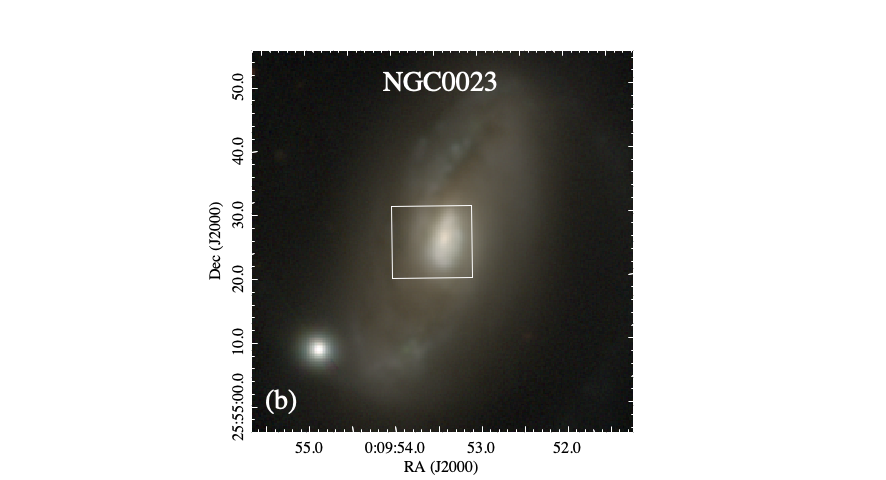}
        \includegraphics[trim={7,1cm 0mm 8cm 1cm},clip, width=0.21\linewidth]{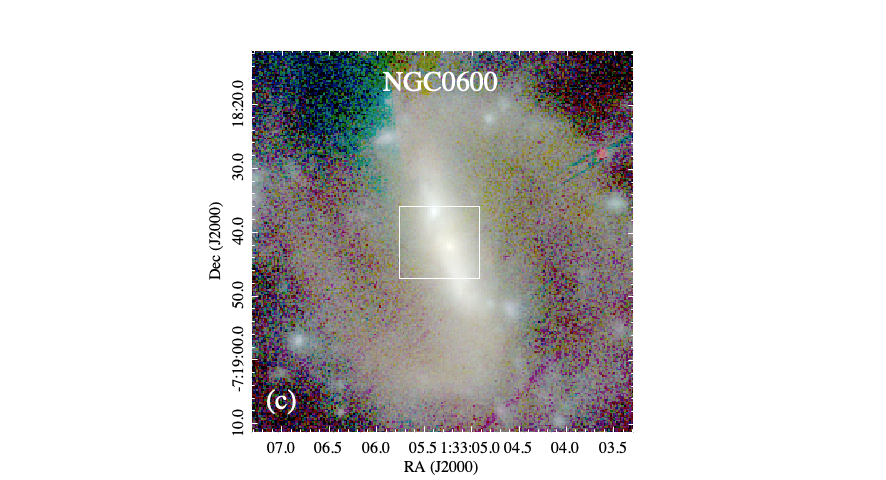}
        \includegraphics[trim={7,1cm 0mm 8cm 1cm},clip, width=0.21\linewidth]{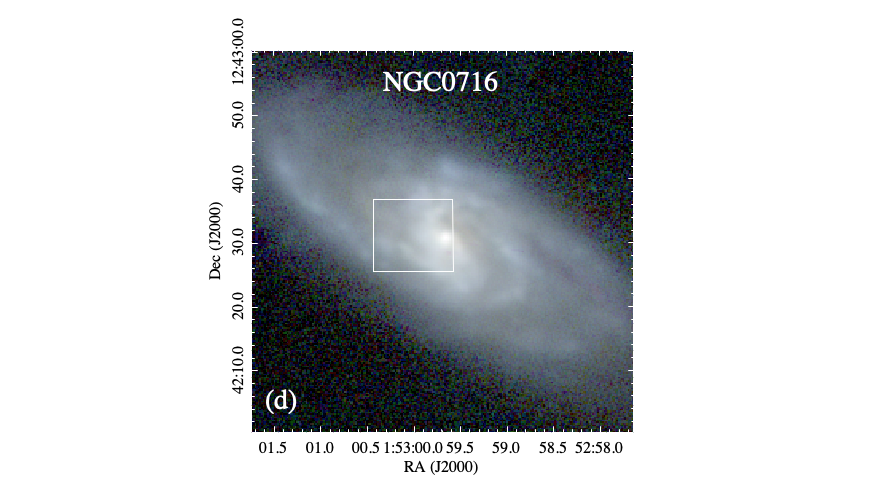}
        \includegraphics[trim={7,1cm 0mm 8cm 1cm},clip, width=0.21\linewidth]{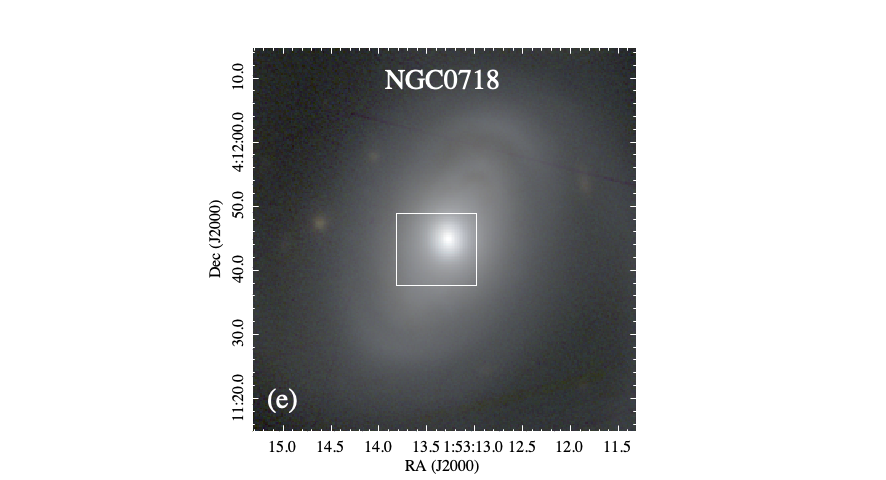}
        \includegraphics[trim={7,1cm 0mm 8cm 1cm},clip, width=0.21\linewidth]{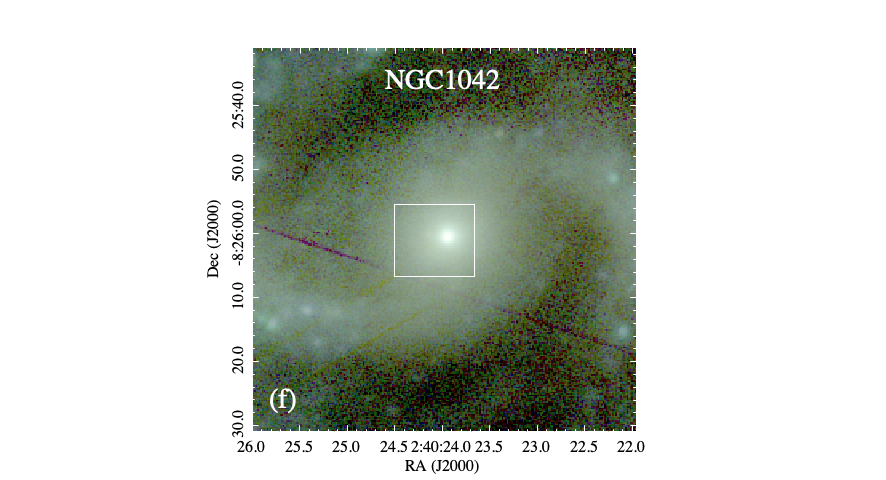}
        \includegraphics[trim={7,1cm 0mm 8cm 1cm},clip, width=0.21\linewidth]{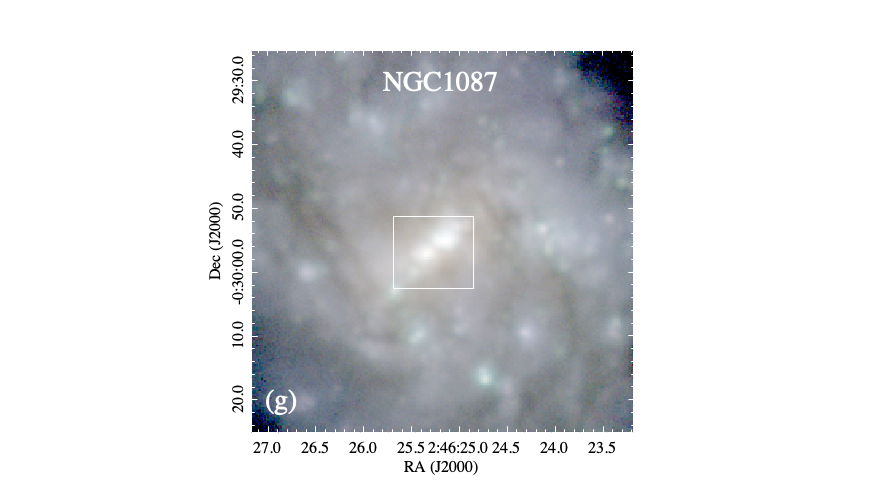}
        \includegraphics[trim={7,1cm 0mm 8cm 1cm},clip, width=0.21\linewidth]{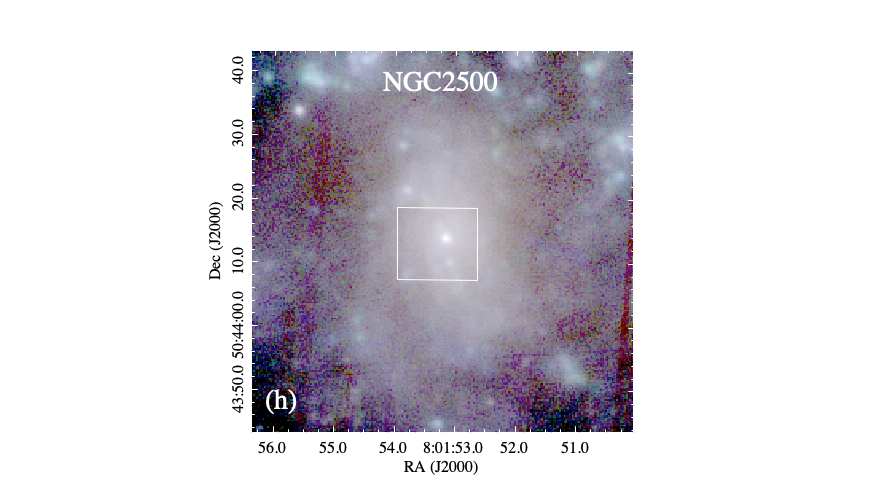}
        \caption{RGB images of some of the galaxies in the MEGADES sample from PanSTARRS observations (g, r, and i filters). The white box in the centre of each panel indicates the MEGARA IFU FoV. The images of all the galaxies in the  MEGADES sample can be found in the Appendix~\ref{appendix:additional_material}, Figure~\ref{fig:sample1}.}
        \label{fig:sample1_part}
\end{figure*}

The original MEGADES sample (see Table~\ref{table:galaxy_properties} and Figures~\ref{fig:sample1_part} and \ref{fig:sample1}) was extracted from the Spitzer survey of stellar structure in galaxies (S4G) sample \citep{Sheth_2010}). The S4G was designed as a volume- (d < 40 Mpc, |b| > 30$^{\circ}$), magnitude- (m$_{\mathrm{B,corr}}$ < 15.5 mag), and diameter-limited (D25 > 1\arcmin) survey. The use of the MEGARA IFU to carry out all observations makes a diameter-limited sample a reasonable idea, as has been the case for the CALIFA sample \citep{Sanchez_2012} or PHANGS-MUSE \citep{Emsellem_2022}.

In addition to the limitations of the S4G sample, we imposed further constraints to fit the sample to our scientific goals. We wished to avoid dwarf systems and elliptical galaxies because in most cases, they are not supported by rotation. We also removed galaxies with inclinations higher than 70$^{\circ}$ to be able to analyse the metallicity and have the possibility of deriving velocity ellipsoids.

Taking into account all the previous considerations, and because it is not practical to observe the entire S4G sample (2331 galaxies) because of the required telescope time, we must consider further selection criteria in our sample: We limited the diameter to a range in which on the lower side, we removed the smallest galaxies, and on the upper side, we limited the diameter to approximately the field of view of the MEGARA MOS (2.5\arcmin < D25 < 4\arcmin)\footnote{Note that as part of the outer-disc extension of MEGADES we plan to obtain MOS spectra of HII regions over the whole galaxies extension.}. We also limited the declination (Dec(J2000) > $-20^{\circ}$). Summarising all the above selection criteria, the galaxies in our sample are selected by distance d < 40 Mpc (z $\sim$ 0.0092), galactic latitude |b| > 30$^{\circ}$, declination Dec(J2000) > –20$^{\circ}$, apparent magnitude m$_{\mathrm{B,corr}}$ < 15.5 mag, apparent diameter 2.5\arcmin < D25 < 4\arcmin and inclination i < 70$^{\circ}$.

This paper includes a random subsample of 30 out of the 215 galaxies that meet the selection criteria of the MEGADES-S4G sample (see Table \ref{table:galaxy_properties}). However, to enrich our sample, we added 13 galaxies from the CALIFA sample (Calar Alto Legacy Integral Field Area Survey, marked with a dagger in Table \ref{table:galaxy_properties}) that show signs of containing interstellar NaI D and are candidates for hosting galactic winds in their neutral phase because their EW$_{\mathrm{ISM-NaI D}}$ > 1.5\,$\AA$. The CALIFA survey, with SDSS DR7\footnote{Sloan Digital Sky Survey Data Release 7} \citep{Abazajian_2009} as the parent sample, is limited in diameter (45\arcsec < D25 < 80\arcsec), redshift (0.005 < z < 0.03), galactic latitude above 20$^{\circ}$ , and declination above +7$^{\circ}$.

\begin{table*}
        \caption{Global properties of the galaxy sample.}              % title of Table
        \label{table:galaxy_properties}      % is used to refer this table in the text
        \centering                                      % used for centering table
        \resizebox{\textwidth}{!}{
        \begin{tabular}{l l l l l l l l l}          % centered columns (4 columns)
                \hline\hline                     % inserts two horizontal lines 
                \noalign{\smallskip}
                Name & Morphology$^{\mathrm{(1)}}$ & t$^{\mathrm{(2)}}$ & z$^{\mathrm{(1)}}$ & R.A.$^{\mathrm{(1)}}$ & Dec.$^{\mathrm{(1)}}$ & Scale$^{\mathrm{(3)}}$ & D$_{L}$ $^{\mathrm{(3)}}$ & Spectral class$^{\mathrm{(1)}}$ \\ %& $\log(M_{\star}/M_{\odot})$ & $m_{g}$ & PA (deg) & B/T \\    % table heading
                & & & & (J2000) & (J2000) & [kpc arcsec$^{-1}$]  & [Mpc] & \\
                \hline                                   % inserts single horizontal line
                \noalign{\smallskip}
                IC~1683$^{\dag}$ & S? & 2.7 & 0.01624 & 01$^\textup{h}$22$^\textup{m}$39$^\textup{s}$\hspace{-1mm}.000 & +34$^\textup{d}$26$^\textup{m}$13$^\textup{s}$\hspace{-1mm}.000 & 0.331 & 70.4 & - \\
                NGC~0023$^{\dag}$ & SB(s)a & 1.2 & 0.01523 & 00$^\textup{h}$09$^\textup{m}$53$^\textup{s}$\hspace{-1mm}.410 & +25$^\textup{d}$55$^\textup{m}$25$^\textup{s}$\hspace{-1mm}.600 & 0.310 & 66.0 & HII LIRG \\
                NGC~0600 & (R')SB(rs)d & 7.0 & 0.00614 & 01$^\textup{h}$33$^\textup{m}$05$^\textup{s}$\hspace{-1mm}.300 & -07$^\textup{d}$18$^\textup{m}$41$^\textup{s}$\hspace{-1mm}.100 & 0.127 & 26.4 & HII \\
                NGC~0716$^{\dag}$ & SBa: & 1.1 & 0.01521 & 01$^\textup{h}$52$^\textup{m}$59$^\textup{s}$\hspace{-1mm}.681 & +12$^\textup{d}$42$^\textup{m}$30$^\textup{s}$\hspace{-1mm}.490 & 0.310 & 65.9 & HII \\
                NGC~0718 & SAB(s)a & 1.0 & 0.00578 & 01$^\textup{h}$53$^\textup{m}$13$^\textup{s}$\hspace{-1mm}.300 & +04$^\textup{d}$11$^\textup{m}$45$^\textup{s}$\hspace{-1mm}.000 & 0.119 & 24.9 & - \\
                NGC~1042 & SAB(rs)cd & 6.0 & 0.00457 & 02$^\textup{h}$40$^\textup{m}$23$^\textup{s}$\hspace{-1mm}.967 & -08$^\textup{d}$26$^\textup{m}$00$^\textup{s}$\hspace{-1mm}.760 & 0.094 & 19.6 & NLAGN \\
                NGC~1087 & SAB(rs)c & 5.2 & 0.00508 & 02$^\textup{h}$46$^\textup{m}$25$^\textup{s}$\hspace{-1mm}.164 & -00$^\textup{d}$29$^\textup{m}$55$^\textup{s}$\hspace{-1mm}.140 & 0.105 & 21.8 & WR HII \\
                NGC~2500 & SB(rs)d & 7.0 & 0.00171 & 08$^\textup{h}$01$^\textup{m}$53$^\textup{s}$\hspace{-1mm}.210 & +50$^\textup{d}$44$^\textup{m}$13$^\textup{s}$\hspace{-1mm}.600 & 0.035 & 7.3 & HII \\
                NGC~2537 & SB(s)m pec & 8.8 & 0.00148 & 08$^\textup{h}$13$^\textup{m}$14$^\textup{s}$\hspace{-1mm}.640 & +45$^\textup{d}$59$^\textup{m}$23$^\textup{s}$\hspace{-1mm}.300 & 0.031 & 6.3 & HII \\
                NGC~2543$^{\dag}$ & SB(s)b & 3.0 & 0.00824 & 08$^\textup{h}$12$^\textup{m}$57$^\textup{s}$\hspace{-1mm}.922 & +36$^\textup{d}$15$^\textup{m}$16$^\textup{s}$\hspace{-1mm}.660 & 0.169 & 35.5 & - \\
                NGC~2552 & SA(s)m? & 9.0 & 0.00175 & 08$^\textup{h}$19$^\textup{m}$20$^\textup{s}$\hspace{-1mm}.532 & +50$^\textup{d}$00$^\textup{m}$34$^\textup{s}$\hspace{-1mm}.660 & 0.036 & 7.5 & - \\
                NGC~2967 & SA(s)c & 5.2 & 0.00626 & 09$^\textup{h}$42$^\textup{m}$03$^\textup{s}$\hspace{-1mm}.295 & +00$^\textup{d}$20$^\textup{m}$11$^\textup{s}$\hspace{-1mm}.180 & 0.169 & 26.9 & - \\
                NGC~3104 & IAB(s)m & 9.9 & 0.00201 & 10$^\textup{h}$03$^\textup{m}$57$^\textup{s}$\hspace{-1mm}.350 & +40$^\textup{d}$45$^\textup{m}$24$^\textup{s}$\hspace{-1mm}.900 & 0.042 & 8.6 & HII \\
                NGC~3485 & SB(r)b: & 3.2 & 0.00478 & 11$^\textup{h}$00$^\textup{m}$02$^\textup{s}$\hspace{-1mm}.380 & +14$^\textup{d}$50$^\textup{m}$29$^\textup{s}$\hspace{-1mm}.700 & 0.099 & 20.5 & AGN \\
                NGC~3507 & SB(s)b & 3.1 & 0.00327 & 11$^\textup{h}$03$^\textup{m}$25$^\textup{s}$\hspace{-1mm}.357 & +18$^\textup{d}$08$^\textup{m}$07$^\textup{s}$\hspace{-1mm}.620 & 0.068 & 14.0 & LINER \\
                NGC~3780 & SA(s)c: & 5.2 & 0.00798 & 11$^\textup{h}$39$^\textup{m}$22$^\textup{s}$\hspace{-1mm}.360 & +56$^\textup{d}$16$^\textup{m}$14$^\textup{s}$\hspace{-1mm}.400 & 0.164 & 34.4 & - \\
                NGC~3893 & SAB(rs)c: & 5.1 & 0.00323 & 11$^\textup{h}$48$^\textup{m}$38$^\textup{s}$\hspace{-1mm}.190 & +48$^\textup{d}$42$^\textup{m}$39$^\textup{s}$\hspace{-1mm}.000 & 0.067 & 13.9 & HII \\
                NGC~3982 & SAB(r)b: & 3.2 & 0.00371 & 11$^\textup{h}$56$^\textup{m}$28$^\textup{s}$\hspace{-1mm}.129 & +55$^\textup{d}$07$^\textup{m}$30$^\textup{s}$\hspace{-1mm}.860 & 0.077 & 15.9 & HII Sy2 \\
                NGC~3998* & SA0\^{}0(r)? & -2.2 & 0.00350 & 11$^\textup{h}$57$^\textup{m}$56$^\textup{s}$\hspace{-1mm}.133 & +55$^\textup{d}$27$^\textup{m}$12$^\textup{s}$\hspace{-1mm}.922 & 0.072 & 15.0 & Sy1 LINER \\
                NGC~4037 & SB(rs)b: & 3.3 & 0.00310 & 12$^\textup{h}$01$^\textup{m}$23$^\textup{s}$\hspace{-1mm}.670 & +13$^\textup{d}$24$^\textup{m}$03$^\textup{s}$\hspace{-1mm}.700 & 0.064 & 13.3 & - \\
                NGC~4041 & SA(rs)bc: & 4.0 & 0.00405 & 12$^\textup{h}$02$^\textup{m}$12$^\textup{s}$\hspace{-1mm}.202 & +62$^\textup{d}$08$^\textup{m}$14$^\textup{s}$\hspace{-1mm}.000 & 0.084 & 17.4 & HII \\
                NGC~4189 & SAB(rs)cd? & 5.9 & 0.00700 & 12$^\textup{h}$13$^\textup{m}$47$^\textup{s}$\hspace{-1mm}.270 & +13$^\textup{d}$25$^\textup{m}$29$^\textup{s}$\hspace{-1mm}.300 & 0.144 & 30.1 & HII \\
                NGC~4278* & E1-2 & -4.8 & 0.00207 & 12$^\textup{h}$20$^\textup{m}$06$^\textup{s}$\hspace{-1mm}.824 & +29$^\textup{d}$16$^\textup{m}$50$^\textup{s}$\hspace{-1mm}.722 & 0.043 & 8.9 & Sy1 LINER \\
                NGC~4593* & (R)SB(rs)b & 3.0 & 0.00831 & 12$^\textup{h}$39$^\textup{m}$39$^\textup{s}$\hspace{-1mm}.425 & -05$^\textup{d}$20$^\textup{m}$39$^\textup{s}$\hspace{-1mm}.340 & 0.171 & 35.8 & Sy1 \\
                NGC~4750* & (R)SA(rs)ab & 2.4 & 0.00540 & 12$^\textup{h}$50$^\textup{m}$07$^\textup{s}$\hspace{-1mm}.271 & +72$^\textup{d}$52$^\textup{m}$28$^\textup{s}$\hspace{-1mm}.720 & 0.111 & 23.2 & LINER \\
                NGC~5218$^{\dag}$ & SB(s)b? pec & 3.1 & 0.00949 & 13$^\textup{h}$32$^\textup{m}$10$^\textup{s}$\hspace{-1mm}.379 & +62$^\textup{d}$46$^\textup{m}$03$^\textup{s}$\hspace{-1mm}.910 & 0.195 & 40.9 & HII LINER \\
                NGC~5394$^{\dag}$ & SB(s)b pec & 3.1 & 0.01150 & 13$^\textup{h}$58$^\textup{m}$33$^\textup{s}$\hspace{-1mm}.652 & +37$^\textup{d}$27$^\textup{m}$12$^\textup{s}$\hspace{-1mm}.550 & 0.235 & 49.7 & HII LIRG\\
                NGC~5616$^{\dag}$ & SBc & 4.0 & 0.02810 & 14$^\textup{h}$24$^\textup{m}$20$^\textup{s}$\hspace{-1mm}.699 & +36$^\textup{d}$27$^\textup{m}$41$^\textup{s}$\hspace{-1mm}.120 & 0.564 & 122.9 & - \\
                NGC~5953$^{\dag}$ & SAa: pec & 0.1 & 0.00656 & 15$^\textup{h}$34$^\textup{m}$32$^\textup{s}$\hspace{-1mm}.383 & +15$^\textup{d}$11$^\textup{m}$37$^\textup{s}$\hspace{-1mm}.590 & 0.135 & 28.2 & Sy2 LINER \\
                NGC~5957* & (R')SAB(r)b & 2.9 & 0.00605 & 15$^\textup{h}$35$^\textup{m}$23$^\textup{s}$\hspace{-1mm}.214 & +12$^\textup{d}$02$^\textup{m}$51$^\textup{s}$\hspace{-1mm}.360 & 0.125 & 26.0 & LINER \\
                NGC~5963 & S pec & 4.1 & 0.00219 & 15$^\textup{h}$33$^\textup{m}$27$^\textup{s}$\hspace{-1mm}.860 & +56$^\textup{d}$33$^\textup{m}$34$^\textup{s}$\hspace{-1mm}.900 & 0.045 & 9.4 & - \\
                NGC~6027$^{\dag}$ & S0 pec & -1.5 & 0.01466 & 15$^\textup{h}$59$^\textup{m}$12$^\textup{s}$\hspace{-1mm}.537 & +20$^\textup{d}$45$^\textup{m}$48$^\textup{s}$\hspace{-1mm}.090 & 0.299 & 63.5 & HII \\
                NGC~6140 & SB(s)cd pec & 5.6 & 0.00303 & 16$^\textup{h}$20$^\textup{m}$58$^\textup{s}$\hspace{-1mm}.160 & +65$^\textup{d}$23$^\textup{m}$26$^\textup{s}$\hspace{-1mm}.000 & 0.063 & 13.0 & - \\
                NGC~6217 & (R)SB(rs)bc & 4.0 & 0.00454 & 16$^\textup{h}$32$^\textup{m}$39$^\textup{s}$\hspace{-1mm}.200 & +78$^\textup{d}$11$^\textup{m}$53$^\textup{s}$\hspace{-1mm}.400 & 0.094 & 19.5 & HII Sy2\\
                NGC~6339 & SBd & 6.3 & 0.00703 & 17$^\textup{h}$17$^\textup{m}$06$^\textup{s}$\hspace{-1mm}.500 & +40$^\textup{d}$50$^\textup{m}$41$^\textup{s}$\hspace{-1mm}.900 & 0.145 & 30.3 & - \\
                NGC~6412 & SA(s)c & 5.2 & 0.00438 & 17$^\textup{h}$29$^\textup{m}$37$^\textup{s}$\hspace{-1mm}.510 & +75$^\textup{d}$42$^\textup{m}$15$^\textup{s}$\hspace{-1mm}.900 & 0.090 & 18.8 & - \\
                NGC~7025$^{\dag}$ & Sa & 1.0 & 0.01657 & 21$^\textup{h}$07$^\textup{m}$47$^\textup{s}$\hspace{-1mm}.340 & +16$^\textup{d}$20$^\textup{m}$09$^\textup{s}$\hspace{-1mm}.100 & 0.350 & 74.6 & - \\
                NGC~7437 & SAB(rs)d & 6.7 & 0.00707 & 22$^\textup{h}$58$^\textup{m}$10$^\textup{s}$\hspace{-1mm}.060 & +14$^\textup{d}$18$^\textup{m}$30$^\textup{s}$\hspace{-1mm}.590 & 0.146 & 30.4 & - \\
                NGC~7479* & SB(s)c & 4.3 & 0.00792 & 23$^\textup{h}$04$^\textup{m}$56$^\textup{s}$\hspace{-1mm}.650 & +12$^\textup{d}$19$^\textup{m}$22$^\textup{s}$\hspace{-1mm}.400 & 0.163 & 34.1 & Sy2 LINER \\
                NGC~7591$^{\dag}$ & SBbc & 3.6 & 0.01654 & 23$^\textup{h}$18$^\textup{m}$16$^\textup{s}$\hspace{-1mm}.280 & +06$^\textup{d}$35$^\textup{m}$08$^\textup{s}$\hspace{-1mm}.900 & 0.337 & 71.7 & Sy LIRG LINER\\
                NGC~7738$^{\dag}$ & SB(rs)b & 3.0 & 0.02251 & 23$^\textup{h}$44$^\textup{m}$02$^\textup{s}$\hspace{-1mm}.059 & +00$^\textup{d}$30$^\textup{m}$59$^\textup{s}$\hspace{-1mm}.860 & 0.455 & 98.1 & - \\
                NGC~7787$^{\dag}$ & (R')SB(rs)0/a: & 0.4 & 0.02221 & 23$^\textup{h}$56$^\textup{m}$07$^\textup{s}$\hspace{-1mm}.823 & +00$^\textup{d}$32$^\textup{m}$58$^\textup{s}$\hspace{-1mm}.140 & 0.449 & 96.7 & - \\
                PGC~066559 & SB(s)dm pec & 8.1 & 0.00900 & 21$^\textup{h}$19$^\textup{m}$43$^\textup{s}$\hspace{-1mm}.040 & -07$^\textup{d}$33$^\textup{m}$12$^\textup{s}$\hspace{-1mm}.500 & 0.185 & 38.8 & - \\
                \hline       %inserts single line
        \end{tabular}}
        \parbox{185mm}{\footnotesize *Data products from these galaxies are under temporary embargo as they constitute the core of Hermosa-Muñoz et al. 2023 in preparation. \linebreak $^{\dag}$ CALIFA survey data. References. (1) NASA/IPAC extragalactic database. (2)   Morphological type code from Hyperleda. (3) \cite{Wright_2006}.}
\end{table*}

\section{MEGARA observations}
\label{sec: observations}
%\subsection{MEGARA}

All MEGADES observations (43 galaxies) of this first release, DR1, were taken with the MEGARA IFU at the GTC. The MEGARA IFU, or LCB, covers a field of view of 12.5 $\times$ 11.3\,arcsec$^{2}$ using 567 fibres in a hexagonal tessellation with a spaxel size of 0.62\,arcsec. In addition to these fibres, it also uses 56 fibres to simultaneously measure sky-background spectra with the observation of the scientific target in order to subtract them from the observations afterwards. These sky fibres are distributed in eight distinct regions in the outermost parts of the MOS field of view, located between 1.75 and 2\,arcmin from the IFU. The fundamental feature that makes MEGARA ideal for performing the analyses required in MEGADES is its spectral resolution. MEGARA has 18 different volume-phase holographic (VPH) transmission gratings that allow covering a range from 3650 to 9700\,\r{A} with low (LR), medium (MR), and high (HR) resolutions.

\begin{table}%[H]
        \caption{MEGARA VPH specifications.}              % title of Table
        \label{table:VPH_characteristics}      % is used to refer this table in the text
        \centering                                      % used for centering table
        \resizebox{0.48\textwidth}{!}{
        \begin{tabular}{c c c c c}          % centered columns (4 columns)
                \hline\hline                     % inserts two horizontal lines 
                \noalign{\smallskip}
                MEGARA VPH & Spectral Coverage  & R.L.D.* & $\mathrm{\Delta\lambda_{FWHM}}$ & R \\    % table heading
                & [$\AA$] & [$\AA$ px$^{-1}$] & [$\AA$] \\
                \hline                                   % inserts single horizontal line
                \noalign{\smallskip}
                LR-B & 4350.6 - 5250.8 & 0.23 & 0.79 & 6061 \\ % inserting body of the table
                LR-V & 5165.6 - 6176.2 & 0.27 & 0.94 & 6078 \\
                LR-R & 6158.3 - 7287.7 & 0.31 & 1.11 & 6100 \\
                \hline                                             %inserts single line
        \end{tabular}}
        \parbox{185mm}{\footnotesize *Reciprocal linear dispersion.}
\end{table}

\begin{figure}[h]
\center
%trim=left bottom right top
\includegraphics[trim={5mm 0 0 0}, clip, width=1\linewidth]{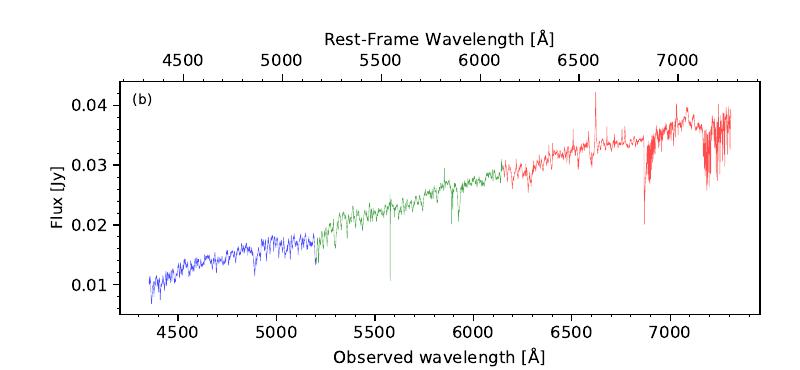}
\caption{Concatenated spectra of the galaxy NGC~0718 observed with the VPH LR-B, LR-V, and LR-R in blue, green, and red, respectively. The spectrum of each VPH is the result of coadding the 567 spectra measured simultaneously in one single exposure with the MEGARA IFU.}
\label{fig:MEGARA_spectrum}
\end{figure}

We observed the central regions of the MEGADES galaxies using three different VPHs, VPH480-LR (LR-B), VPH570-LR (LR-V), and VPH675-LR (LR-R), whose specifications can be found in Table ~\ref{table:VPH_characteristics}. This produces a combined spectrum covering
the spectral range from 4350 to 7288 Å (see Figure~\ref{fig:MEGARA_spectrum} for the combined MEGARA spectrum of NGC 0718). This range includes spectral features such as H$\beta$, $[\mathrm{O III}]\lambda5007$, NaI D, and H$\alpha$, as well as absorption features needed for the stellar populations analysis \citep{Chamorro_Cazorla_2022}.
For the 13 CALIFA galaxies, we only have LR-V and LR-R observations. They suffice for studying the possible presence of outflows in the neutral (NaI D) and the warm (H$\alpha$) components.

\begin{figure}[h]
\center
%trim=left bottom right top
\includegraphics[trim={0 0mm 0mm 10mm},clip,width=1\linewidth]{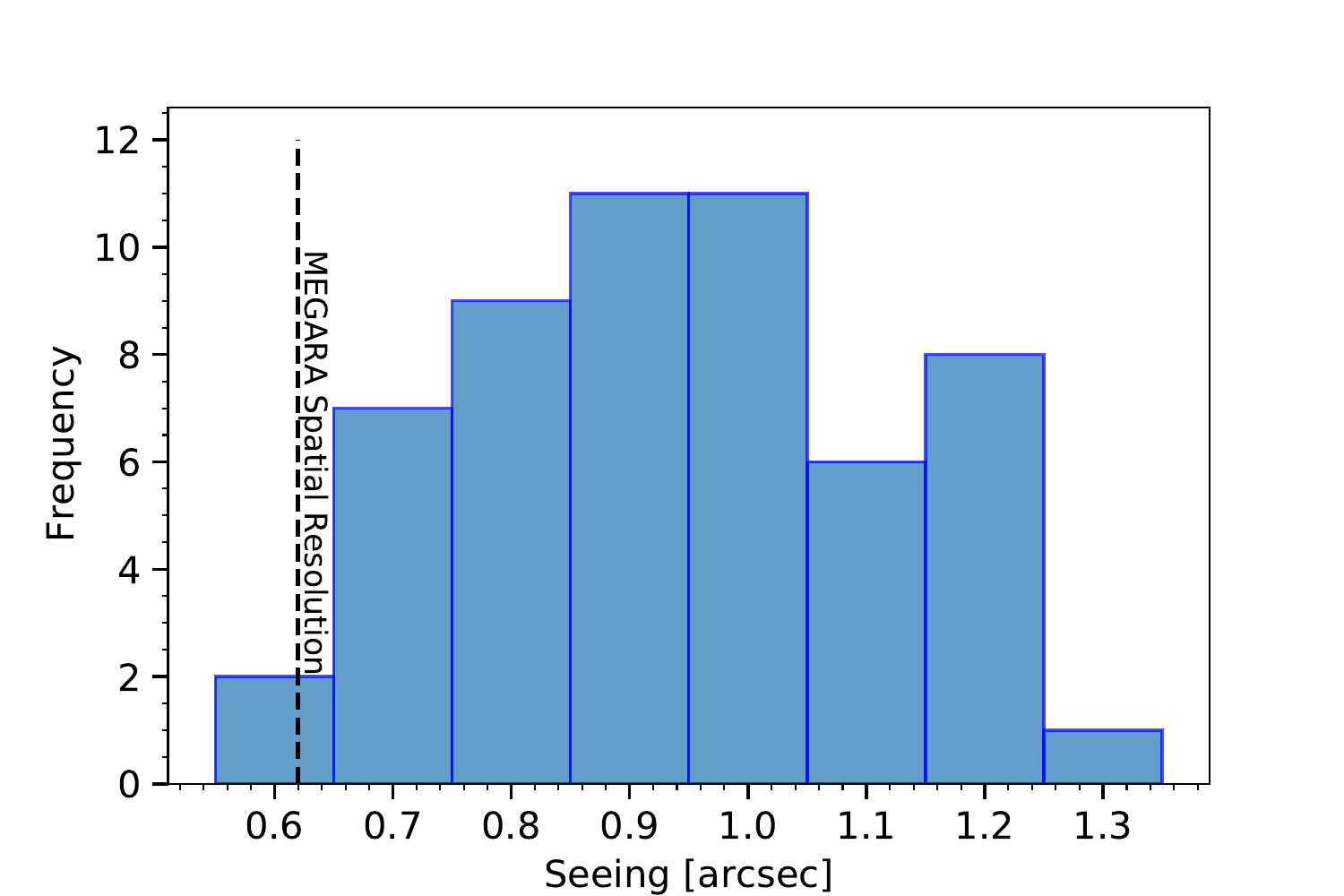}
\caption{Seeing values, provided by the monitors of observatory, and histogram for the nights of the MEGADES sample observations. The dashed black line indicates the size of the hexagonal spaxels that comprise the MEGARA IFU.}
\label{fig:seeing}
\end{figure}

Table~\ref{table:log} shows the observing dates, airmass, and exposure times for all the galaxies in the sample. Table~\ref{table:observing_nights} provides the observing conditions for each observing night with information
on the seeing and atmospheric transparency. Figure~\ref{fig:seeing} shows a histogram with the seeing values provided by the monitors of the observatory during the nights of MEGADES observations.

\section{Data processing}
\label{sec: data_processing}
\subsection{Basic reduction: MEGARA DRP}
\label{section:megara_drp}

We used the MEGARA data reduction pipeline (DRP) v0.12.0 \citep{pascual_2022} to process all the data provided by the telescope. This data reduction was performed according to the instructions provided in \cite{africa_castillo_morales_2020_3932063}\footnote{\href{https://doi.org/10.5281/zenodo.1974953}{DOI: 10.5281/zenodo.1974953}}.We have followed the same routines as we used to reduce the data for the galaxy NGC~7025 in \cite{Chamorro_Cazorla_2022}, considering the particularities of each individual observation.

We started by removing the level of bias present in the image. This bias is an electronic pedestal that is added to all images taken with MEGARA before the analogue-to-digital converter (ADC) to minimise errors associated with the conversion of voltages that would otherwise be very close to zero. For the purpose of removing this pedestal from the counts in the images, we used the bias calibration images and the task \textit{MegaraBiasImage}. These images are taken with zero exposure time, and the bias level is slightly different at the top and bottom of the image ($\sim$ 100 counts) because the MEGARA CCD (a CCD231-84 from E2V) is read through two diagonally opposed amplifiers. The use of two (instead of the four available) amplifiers minimises electronic cross-talk during the read-out process. The images were corrected from overscan and trimmed to match the physical size of the detector. Bad pixels are automatically masked in this step with the file \textit{master\_bpm.fits}, available in the pipeline. 

The trajectory of each one of our fibre spectra needs to be properly traced on the detector. To do this, we used the TraceMap calibration images obtained by illuminating the focal plane with a halogen lamp providing homogeneous illumination.  When performing this step, we were careful with the calibration images we used because the position of the fibres on the detector changes depending on the temperature of the telescope. It is recommended to use calibration images as close as possible to the observation of the scientific target to best minimise the temperature difference between them. The DRP tasks we used for this purpose were \textit{MegaraTraceMap} and \textit{MegaraModelMap}. The first task tracks the path of the light projected by the fibres onto the CCD, and the second task makes a more exhaustive model of it in order to perform the extraction, assuming each fibre has a Gaussian light profile across dispersion. In some cases, 11 observations taken with LR-B and 3 with LR-R, we had problems tracing the last fibre box (from fibre 603 to 623) because the instrument was not properly focused. In these cases, the spectra from these fibres were ignored. An example of the impact caused by doing this on our data is shown in Figure~\ref{fig:NGC4189_card_2}. In panels (t) and (v), continuum images taken with LR-B and LR-R, respectively, are shown, and all the spaxels corresponding to the out-of-focus fibres are shown in white. 

\begin{figure}[h]
\centering
%trim=left bottom right top
\includegraphics[trim={0 0mm 0mm 10mm},clip,width=1\linewidth]{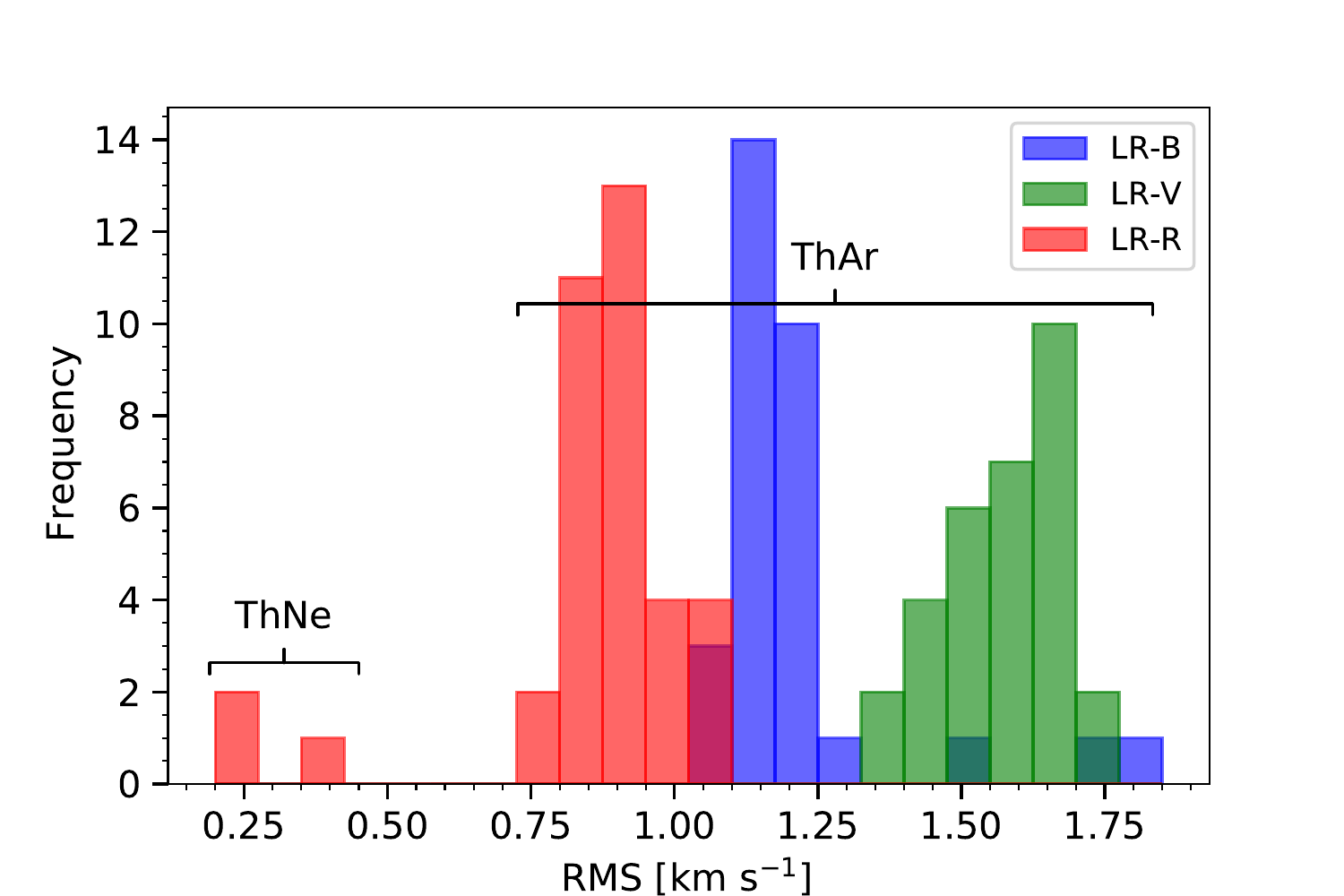}
\caption{Wavelength calibration RMS of all observations by VPH. Blue, green, and red histograms represent LR-B, LR-V, and LR-R observations, respectively.}
\label{fig:wavelength_rms}
\end{figure}

We then calibrated the wavelength. Here we used images that were acquired using ThAr or ThNe lamps, depending generally on the VPH we used for our observation. As a general rule, for LR-B, LR-V, and LR-R observations, we used ThAr lamps. For this task, it is necessary to confirm that the traces obtained in the previous tasks match the path followed by the projection of the fibres on the detector in the images of the arc lamps. If there is a small offset between the two, this can be corrected by indicating it in the recipe we use, \textit{MegaraArcCalibration}. In this paper, we only calibrated three observations with ThNe lamps: NGC~4189, NGC~2552, and NGC~3104 observed with LR-R, because these calibrations are even more precise, with RMS < 0.5\,km\,s$^{-1}$(see Figure~\ref{fig:wavelength_rms}). The median RMS values in $\mathrm{\AA}$ for LR-B, LR-V, and LR-R are 0.019, 0.030, and 0.020, respectively, and the value is 0.020 for the whole sample. We find that the RMS represents only 2.4\% (LR-B), 3.2\% (LR-V), and 1.8\% (LR-R) of the reciprocal linear dispersion of each VPH (see Table \ref{table:VPH_characteristics}). In Figure~\ref{fig:wavelength_rms} we show the average of the wavelength calibration RMS (in km s$^{-1}$) for each observation, separated by VPH, of the fittings after all observations are calibrated in wavelength.

MEGARA transmits the light from the focal plane of the telescope to the spectrograph via optical fibres. This means that each fibre transmits the light in a different way and with different wavelength dependences. The fibre-specific calibration of the variation in transmission as a function of wavelength is done with the \textit{MegaraFiberFlatImage} task using the same images as were used for the \textit{MegaraTraceMap} recipe. The median of the standard deviation values of the fiber-to-fiber sensitivity variation for each VPH in our sample exposures are 0.090, 0.086, and 0.096 for LR-B, LR-V, and LR-R, respectively.

The last correction to fully reduce our images is the flux calibration. For this purpose, we used spectra of standard stars available in the ESO spectrophotometric standards database (\citealt{Oke_1990}; \citealt{Hamuy_1992, Hamuy_1994}): BD+33 26 42, HD192281, HR153, HR718, HR1544, HR1996, HR3454, HR4468, HR4554, HR4963, HR5501, HR7596, HR7950, and HR8634. The exposures of standard stars were also examined to determine whether the previously computed traces were still valid for these observations.

\begin{figure}[h]
\centering
\includegraphics[trim={0 0mm 0mm 10mm},clip,width=1\linewidth]{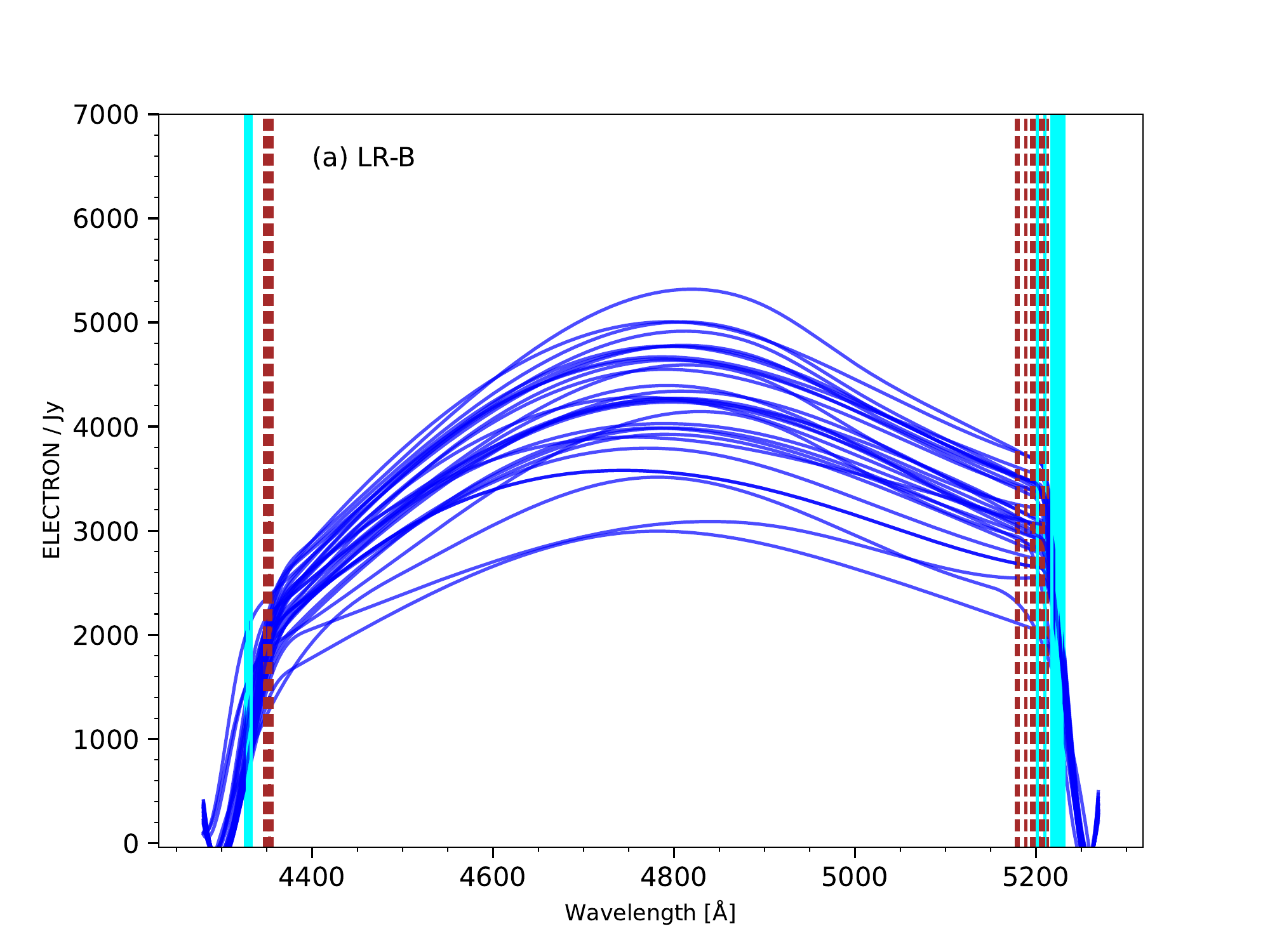}
\includegraphics[trim={0 0mm 0mm 10mm},clip,width=1\linewidth]{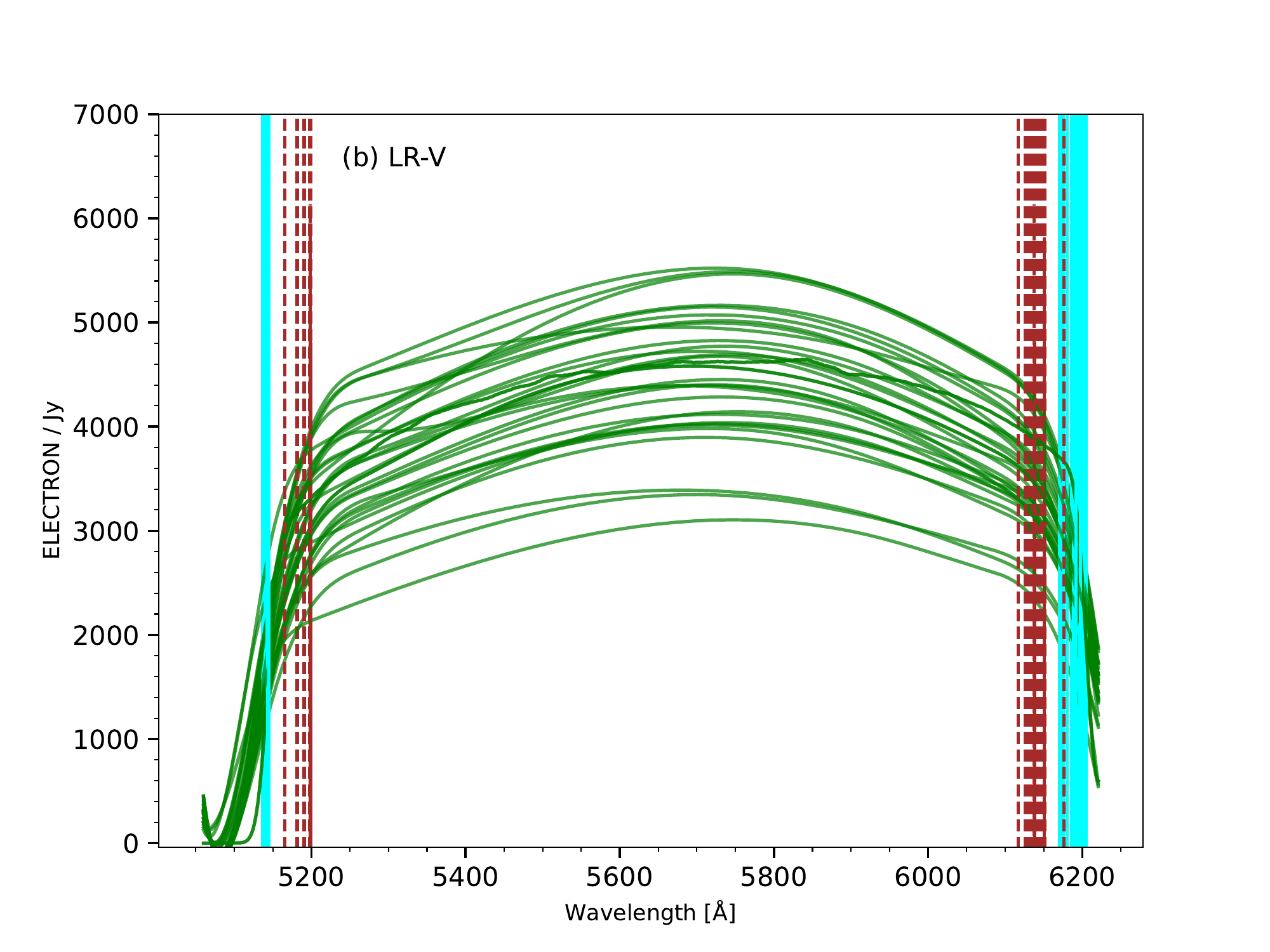}
\includegraphics[trim={0 0mm 0mm 10mm},clip,width=1\linewidth]{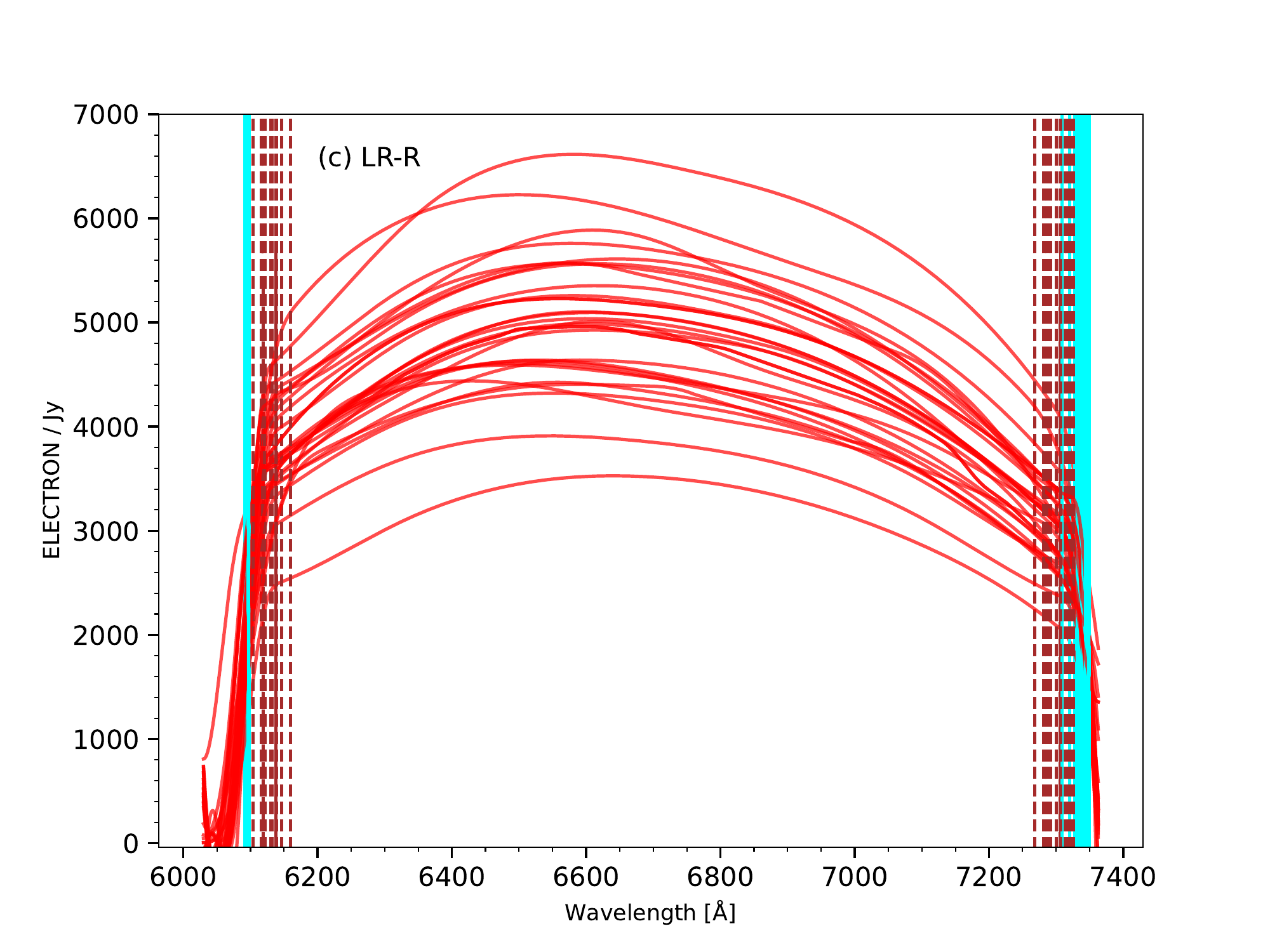}
\caption{Sensitivity curves for the flux calibration of all MEGADES observations. Panels (a), (b), and (c) show the sensitivity curves derived from all the standard stars observed with the LR-B, LR-V, and LR-R gratings, respectively. Cyan lines mark the spectral coverage after flux calibration in Angstroms for all fibres. Dashed brown lines indicate the spectral coverage after flux calibration in Angstroms, with a proper flux calibration.}
\label{fig:master_sensitivity}
\end{figure}

The task \textit{MegaraLcbAcquisition} determines the region of the MEGARA IFU field of view in which the standard star is located in order to know in which fibres the spectra to be considered are placed when they are compared with those of CALSPEC. To ensure that we do not miss flux when measuring in the image of the star, we computed the position of the centroid using the spectra of three concentric rings around the brightest spaxel (37 spaxels in total). In this step, an atmospheric extinction correction is also applied using a reference file provided by the GTC \citep{King_1985}. With this information, we ran the \textit{MegaraLcbStdStar} task, which produces the sensitivity curve needed to flux calibrate our data. This curve was saved in the \textit{master\_sensitivity.fits} file. Figure~\ref{fig:master_sensitivity} shows the sensitivity curves for the flux calibration obtained during the reduction of the whole sample. Absolute flux calibration is difficult to guarantee, but relative flux calibration is much more reliable. This is very important when measuring and comparing lines that have been observed with the same VPH. 

The file \textit{master\_sensitivity.fits} includes information in its headers that is very valuable when the MEGARA observations are analysed later. This information provides the spectral coverage after flux calibration in pixels and in Angstroms for at least one fibre (\textit{PIXLIMR1-2} and  \textit{WAVLIMR1-2}), common to all fibres (\textit{PIXLIMM1-2} and  \textit{WAVLIMM1-2}) and with a proper flux calibration (\textit{PIXLIMF1-2} and  \textit{WAVLIMF1-2}). The  \textit{WAVLIMM1-2} and  \textit{WAVELIMF1-2} ranges are shown in Figure~\ref{fig:master_sensitivity} using cyan and dashed brown lines, respectively.

This steps completes the calibration files that an observation of this kind requires. However, before applying all these corrections, we must verify two important details. The first detail is to confirm that the traces we used are correct for the scientific target image or to decide that we need to apply some offset on these traces, as in previous steps. The second detail is to determine whether any diffuse light in these images might spoil the observation because in some exposures, moonlight managed to sneak into the detector. Fortunately, this light only adds a few counts to the image and can easily be removed with one of the \textit{megaratools}, \textit{megaratools-diffuse\_light}. This tool makes a model of the diffuse light using information from the regions of the CCD that are not illuminated by the fibres. This model allows us to eliminate all traces of diffuse light from our observations. The observations affected by this diffuse light are identified in Table~\ref{table:log} with the dagger.

Finally, we applied all the corrections calculated in the data reduction process, including correction for diffuse light, if necessary. For this purpose, we ran the \textit{MegaraLcbImage} task with the science images. This task, in addition to applying all the corrections described in this section, also subtracts an average spectrum of the sky background measured simultaneously on the outermost 56 fibres of the MEGARA MOS. As final products, we have several images in row-stacked spectra (RSS) format. This format consists of a 2D image with 623 rows, one for each fibre (567 target and 56 sky background fibres) and 4300 columns of spectral information. The most relevant products we have at the end of the data reduction process with the MEGARA DRP are \textit{final\_rss.fits,} which is the reduced science image with sky subtraction applied, \textit{reduced\_rss.fits,} which is the reduced science image that still includes the sky background, and \textit{sky\_rss.fits, \textup{which}} contains the sky background information. The images we used for our analysis are the \textit{final\_rss.fits}.

There is one final caveat before we start the pertinent analysis.  If there is too much contrast between the brightest spaxel and the adjacent fibres in the \textit{final\_rss.fits}, there may be crosstalk in these fibres that the ModelMap could not account for. This usually happens when the galaxy is not well centred in the IFU. We must take this into account to avoid drawing incorrect conclusions when this effect is noticeable.

\subsection{Post-processing}

\subsubsection{Astrometry correction}

\begin{figure}[h]
\centering
\includegraphics[trim={0 0mm 0mm 10mm},clip,width=1\linewidth]{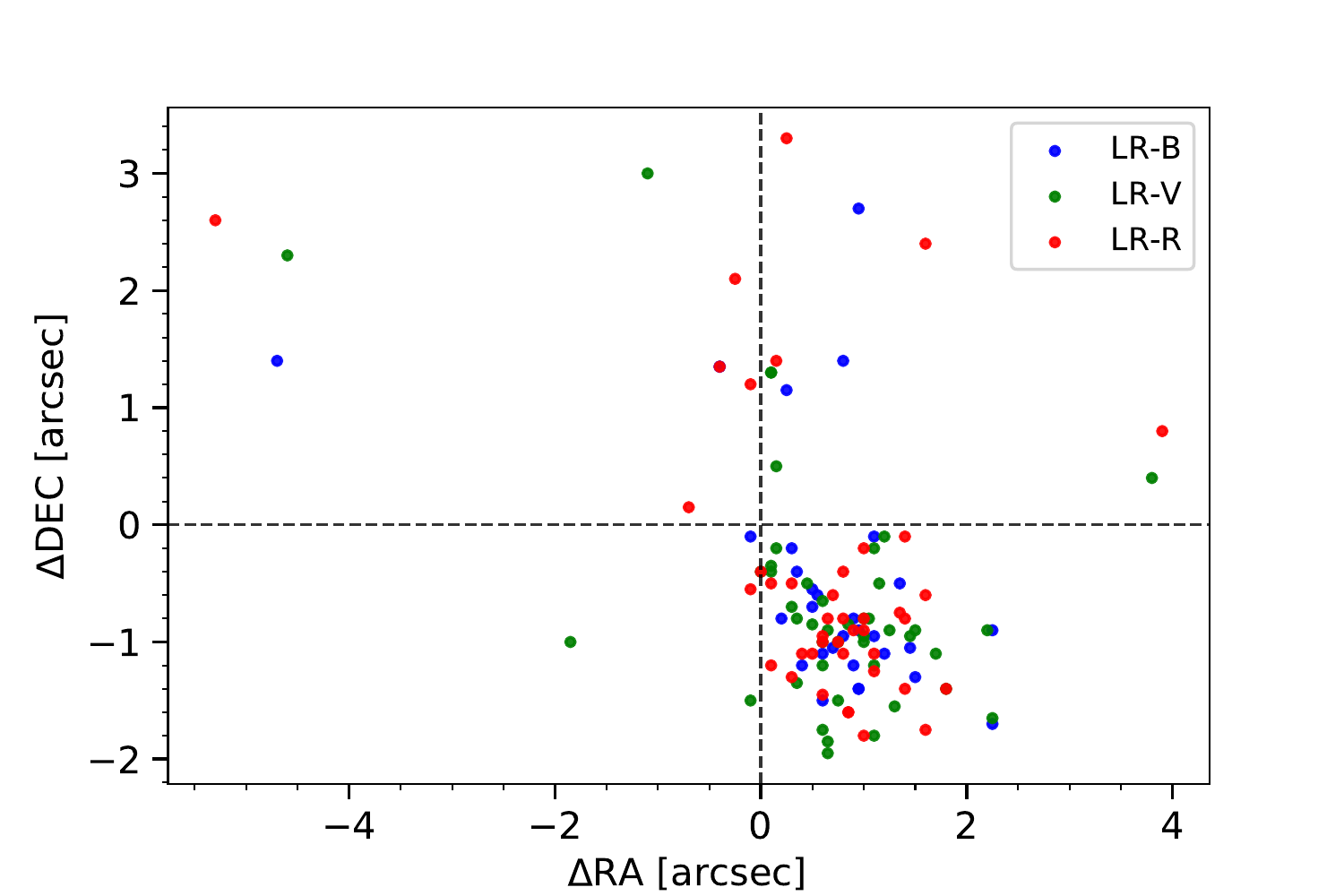}
\caption{Offsets applied to all MEGADES sample images for pointing correction. The blue, green, and red dots represent the offsets in the LR-B, LR-V, and LR-R observations, respectively.}
\label{fig:pointing_corrections}
\end{figure}

We may wish to compare our observations with observations from other studies, either photometric or spectroscopic. We are therefore interested in keeping the coordinates of our pointings as precise as possible. The GTC website claims that the pointing accuracy of the telescope is around 1 to 2\,arcsec. For this reason, we verified whether the coordinates that appear in the headers of our images correspond to the real coordinates of the objects. To do this, we compared our observations with the images available in the Pan-STARRS survey \citep{Chambers_2016} and measured the offsets between the peak of the brightest contours in the Pan-STARRS images and the brightest spaxel in the MEGARA observations. This procedure allowed us to ensure that our pointing is correct with an error $\leq$ 1 spaxel (0.62\,\arcsec). Figure \ref{fig:pointing_corrections} shows the offsets we applied to our observations, both in right ascension and declination, to match the Pan-STARRS pointing. In most cases, the deviation of the targets clearly follows the same trend.

The information of the new pointing has been included in the header of the \textit{final\_rss.fits} images by updating the following keywords:  \textit{RADEG},  \textit{DECDEG},  \textit{RA,} and  \textit{DEC} from extension 0 and  \textit{CRVAL1} and  \textit{CRVAL2} from extension 1 (FIBERS extension).  The FIBERS extension is the one used by the cube megaratool.

\subsubsection{Voronoi binning}

A systematic procedure was followed to analyse all the MEGADES data, using the final RSS as a departure point. First we separated the stellar signal from the interstellar contribution in our spectra. However, if our data are to be sufficiently robust to make this distinction, we must perform a Voronoi binning first. We used the Voronoi binning method by \cite{Cappellari_2003} to reach a signal-to-noise ratio (S/N) per angstrom of 10 at least. We calculated this signal-to-noise ratio in the continuum near the spectral lines of interest in each spectral range and took the redshift of each galaxy into account. For LR-B, we measured it between 4800\,$\mathrm{\AA}$ and 4850\,$\mathrm{\AA}$, for LR-V between 5920\,$\mathrm{\AA}$ and 5970\,$\mathrm{\AA,}$ and for LR-R between 6604\,$\mathrm{\AA}$ and 6654\,$\mathrm{\AA}$ (all rest-frame). We recall that each observation will have a different Voronoi binning independent from the rest of the observations because it only depends on the signal-to-noise ratio we measure in the spaxels of each exposure.

\begin{figure}[h]
\centering
\includegraphics[trim={0 0mm 0mm 10mm},clip,width=1\linewidth]{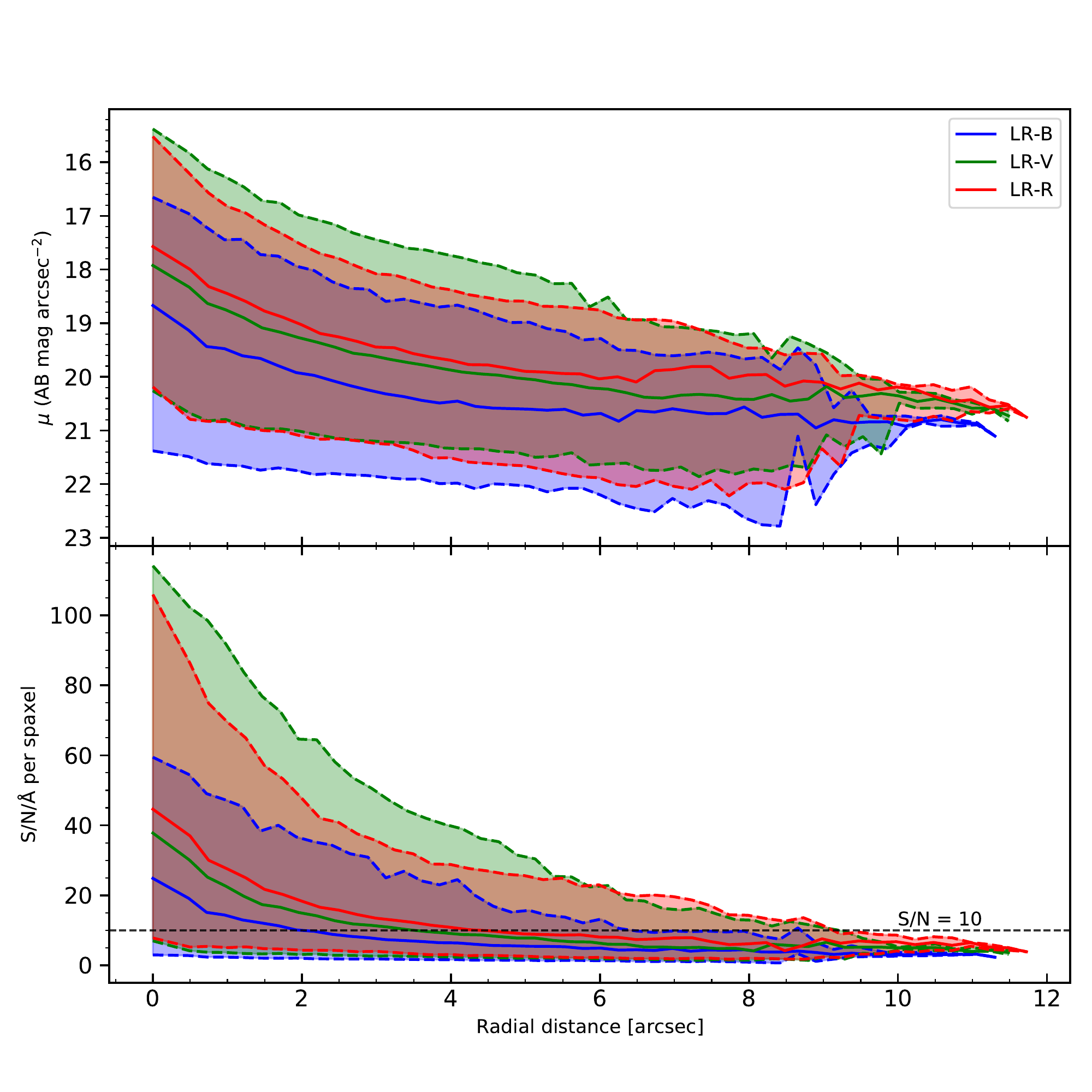}
\caption{Surface AB magnitude per spaxel and signal-to-noise ratio per Angstrom as a function of distance to the brightest spaxel for all spaxels in the MEGADES sample separated by VPH. Top panel: Surface AB magnitude per spaxel for all the spaxels in the sample as a function of the distance. In both cases, the solid lines represent the median S/N value for each distance and VPH, and the dashed lines encompass the 5th and 95th percentiles for each of them. Blue, green, and red correspond to the measurements for the LR-B, LR-V, and LR-R observations, respectively. Bottom panel: Signal-to-noise ratio per Angstrom of all the spaxels in the sample as a function of the distance. The dashed horizontal line identifies the S/N 10 value.}
\label{fig:sn_sample}
\end{figure}

Figure~\ref{fig:sn_sample} (bottom panel) shows the S/N as a function of the distance to the brightest spaxel in each observation for all spaxels in the MEGADES sample separated by VPH. The solid lines represent the median S/N value for each distance, and the dashed lines represent the corresponding 5th and 95th percentile bin. The S/N values decrease away from the centre of the galaxies, as expected. Furthermore, the small differences that may exist between the different VPHs become smaller with distance from the brightest spaxel. In the top panel of Figure~\ref{fig:sn_sample}, we show the surface AB magnitude per spaxel as a function of distance to the brightest spaxel, in a similar way as in the S/N figure. In this case, the behaviour of the curves is similar to that of the S/N because the noise level in the whole detector is rather homogeneous.

Out of the 65905 spaxels in our sample, 32493 ($\sim$ 49\%) have an S/N below 10. This means that all these spaxels are combined within a Voronoi region in their corresponding observation, and the rest of them are analysed individually. After performing the Voronoi binning of the data, we find that 91.6 \% of all Voronoi regions in the sample consist of 3 or fewer spaxels.

The median S/N values plotted in Figure~\ref{fig:sn_sample} and their intersection with the horizontal dashed line, which represents the S/N cut we made to create the Voronoi regions, show that we reach in single spaxels a distance of 2\,arcsec, 3.5\,arcsec, and 4\,arcsec for the LR-B, LR-V, and LR-R observations, respectively. The average sample distance is 34.8\,Mpc; therefore, this means that we can reach 0.33\,kpc (LR-B), 0.58\,kpc (LR-V), and 0.66\,kpc (LR-R) without binning.

\section{Analysis}
\label{sec: analysis}

For the analysis of the stellar continuum (and its subtraction for the line analysis), the same software as was used for the pilot study of the MEGADES sample performed in \citealt{Chamorro_Cazorla_2022} was employed; the penalized pixel-fitting (pPXF) by \citealt{Cappellari_2004} (see also \citealt{Cappellari_2017}).  pPXF  enables us to determine the kinematic characteristics
of the stellar component of galaxies and their stellar populations (which
will be discussed in future papers) in addition to enabling us to discern the features that have a stellar origin and those that do not. In section \ref{section: pPXF_fitting} we describe in detail how we applied this software to our data.

One of the products we obtain from the pPXF analysis are the residuals that remain after the best fit is subtracted from the original data. This means that all features found in these residual spectra, in emission and absorption, do not have a stellar origin. This information allows us to study the interstellar gas in the ionised and neutral phases present in these galaxies.

However, in contrast to the stellar information, the interstellar spectral lines are not analysed from a Voronoi binned region, but are analysed spaxel by spaxel. This means that we have to use the best-fitting spectra of the underlying stellar populations in Voronoi regions to subtract the stellar continuum from individual spaxels. The continuum level differs between the Voronoi regions and the induvidual spaxels that comprise these regions, therefore, we need to normalise each best-fitting Voronoi spectrum to the continuum level of the original spaxel in order to correctly subtract the stellar information. Moreover, since we have the continuum level information in the spectral range in which we have measured the signal, close to the lines of interest, we can add the same continuum level to the residuals as in the original data in order to measure the equivalent widths (EW) of the lines of interest in our spectra. In section \ref{section:emission_lines} we discuss in detail which lines we studied and the method we applied for this purpose.

\subsection{Stellar continuum fitting}
\label{section: pPXF_fitting}

\begin{figure*}[h]
\centering
\includegraphics[trim={0 0mm 5mm 5mm},clip, width=0.49\linewidth]{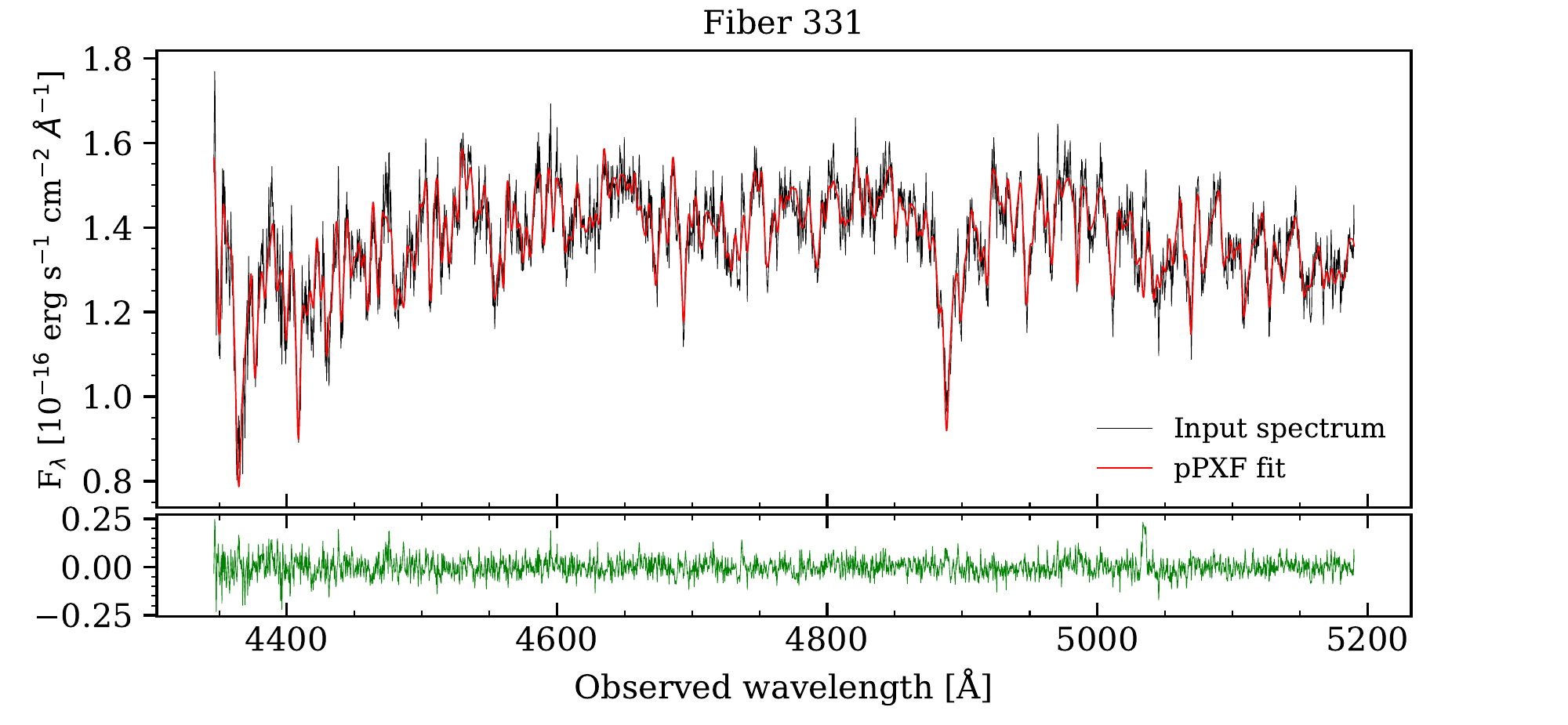}
\includegraphics[trim={0 0mm 5mm 5mm},clip, width=0.49\linewidth]{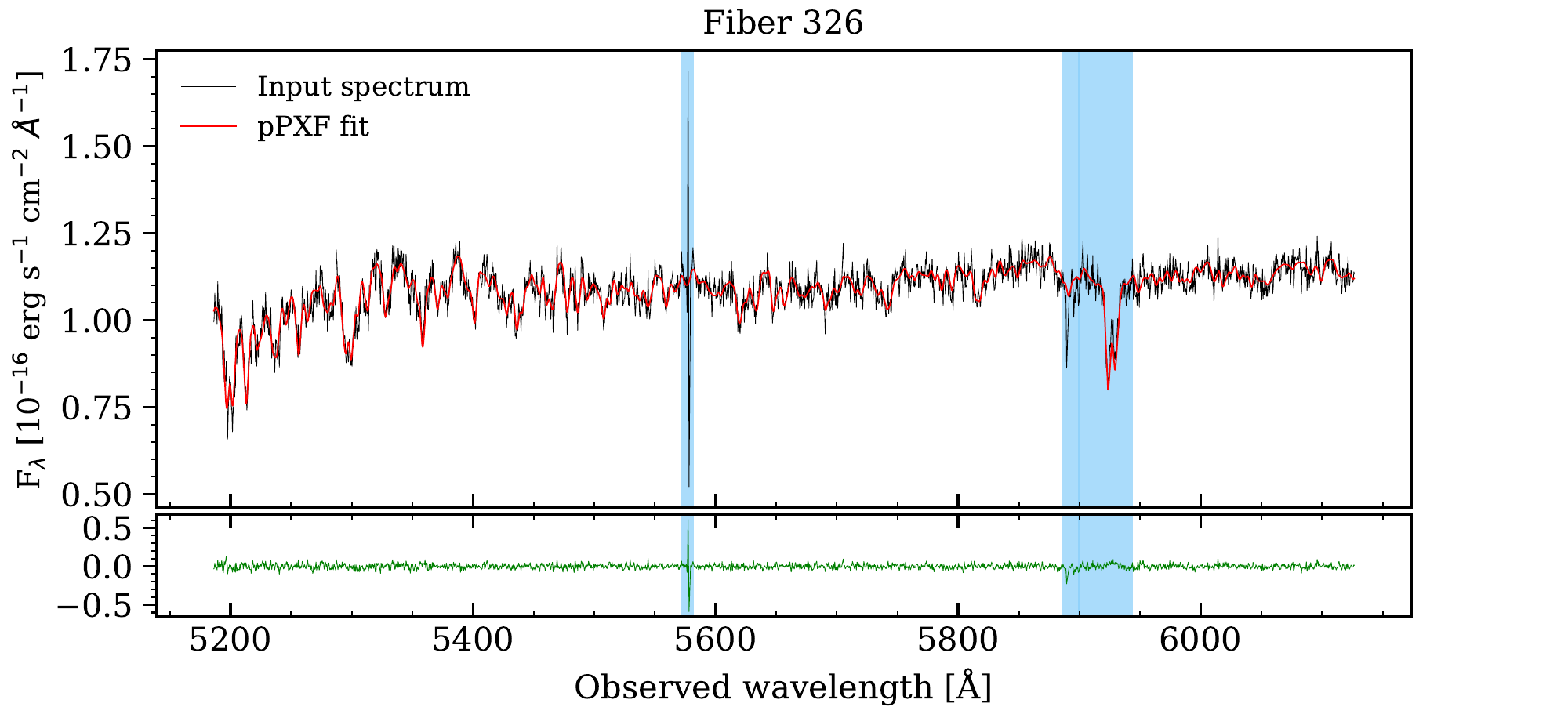}
\includegraphics[trim={0 0 5mm 5mm},clip, width=0.49\linewidth]{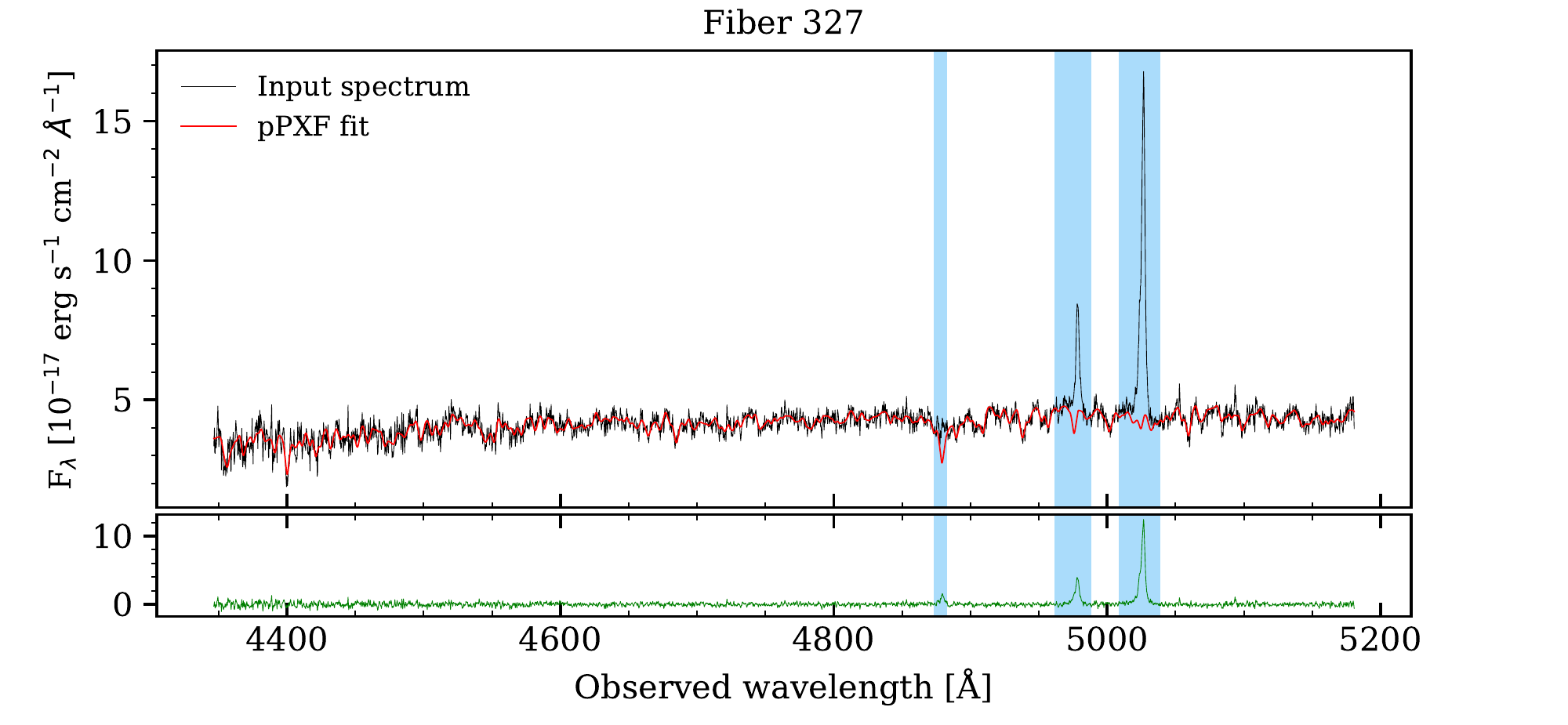}
\includegraphics[trim={0 0 5mm 5mm},clip, width=0.49\linewidth]{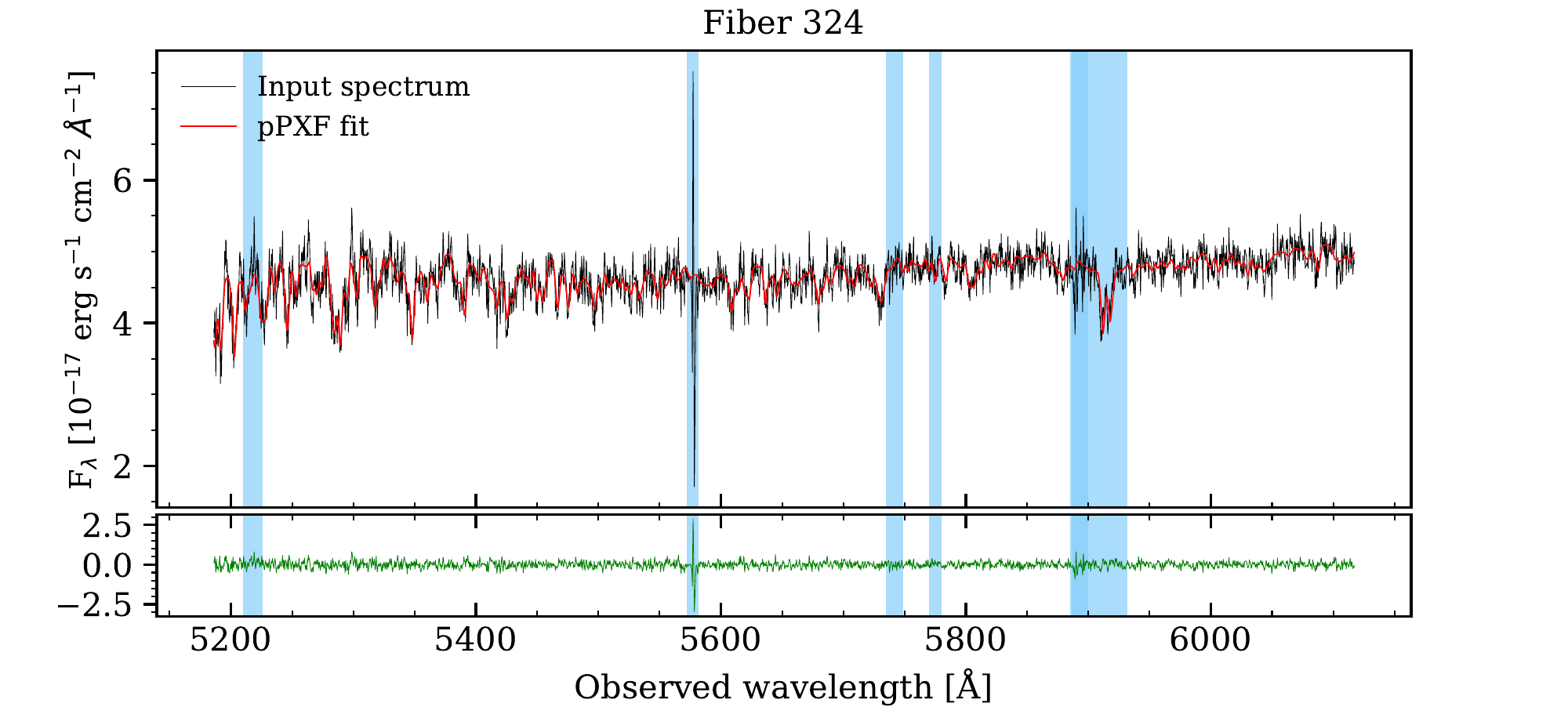}
\caption{Examples of stellar continuum fitting using pPXF for 2 galaxies of the sample. Top left panel: NGC~0718 spectrum in the LR-B spectral range. Top right panel: NGC~0718 spectrum in the LR-V spectral range. Bottom left panel: NGC~3982 spectrum in the LR-B spectral range. Bottom right panel: NGC~3982 spectrum in the LR-V spectral range. The black lines show the original spectrum observed with MEGARA for these spaxels. The red lines are the best fit performed by pPXF on the data. The green lines are the residuals resulting from subtracting the best model from the original data. The vertical blue regions are areas that were masked during the fitting.}
\label{fig:pPXF_fitting}
\end{figure*}

To extract all the information of the stellar continuum present in the spectra, we used, as mentioned above, the full spectral fitting analysis software pPXF. This programme is based on the analysis of the absorption lines in order to estimate the stellar populations present in the spectra. This means that it is essential to avoid the emission lines we may have in the spectra from affecting these fits. For this purpose, we masked the spectral regions in which this type of feature was present, and the regions in which the subtraction of the sky lines was not perfect. These masked regions may be different depending on the galaxy that is analysed, as not all of them have the same spectral features. These characteristics can even change from one spaxel to the next within the same observation. In the case of LR-V observations, it is also important to mask the wavelength range corresponding to the sodium doublet (NaI D) in addition to these areas because pPXF  distinguishes based on the remaining absorption lines the part of this absorption that comes from the stars and the part that comes from the interstellar medium. In Figure \ref{fig:pPXF_fitting} we show some examples of these fittings and the different masked regions depending on the analysed galaxy.

The process we followed to perform our analysis with pPXF is similar to the one followed in the study by \cite{Kacharov_2018}. We first analysed the properties of the stellar kinematics using additive and multiplicative polynomials of degree 10. After this step, we already know the velocity, velocity dispersion, skewness, and kurtosis in our data. This allows us to analyse the stellar populations by fixing their kinematic parameters. In this step, where we analyse the stellar populations, we set the degree of the additive polynomials to 0 and keep the degree of the multiplicative polynomials at 10. These latter fittings are used to separate the stellar signal from the interstellar component. Throughout all of our analyses, we use a first-order regularisation with a low factor (R = 5; see \citealt{Kacharov_2018}).

It is necessary to use stellar population synthesis models in order to be able to fit our spectra. In the case of this survey, we used the stellar population synthesis model predictions (SSP) by \cite{Vazdekis_2010} based on the MILES stellar library and Padova+00 isochrones (\citealt{SanchezBlazquez_2006} and \citealt{FalconBarroso_2011}). These models (350 in total) cover a very wide age range, from 0.063 Gyr to 17.78 Gyr, and have seven different levels of metallicity ([M/H]): $-2.32$, $-1.71$, $-1.31$, $-0.71$, $-0.40$, and 0.00 y +0.22. Its spectral coverage, ranging from 3525\,$\mathrm{\AA}$ to 7500\,$\mathrm{\AA}$, perfectly spans the spectral range of the MEGARA VPHs used in this survey. Although the FWHM of the MILES models (2.5\,\AA) is lower than that of our MEGARA observations (R$\sim$6000), the broadening of the lines along the line of sight in the central galaxy regions allows the use of these models. Of all the possible available initial mass functions (IMF), we used a unimodal IMF with a logarithmic slope of 1.3, that is, Salpeter \citep{Salpeter_1955}, and did not take any kind of $\alpha$-enhancement into account as they are not available in these models.

\subsection{Emission and absorption lines}
\label{section:emission_lines}

We followed different strategies for the analysis of the lines, depending on the VPH used for the observation. For the lines observed with LR-B, H$\beta,$ and $[\mathrm{O III}]\lambda5007$, as well as for the lines observed within LR-V, that is, NaI D, we employed the spectra obtained from the residuals of the stellar populations with a constant flux bias added to match the level of the original (reduced) spectra for EW measurement purposes. However, for the lines observed within LR-R, H$\alpha$, $[\mathrm{N II}]\lambda6584,$ and the two [SII] lines, $[\mathrm{S II}]\lambda6717$ and $[\mathrm{S II}]\lambda6731$, we used a different procedure because in the wavelength range covered by this VPH, not many spectral features are available to fit the stellar populations in a robust way. Therefore, we measured the lines in this spectral range directly on the reduced observations of the galaxy.

Only one of the four lines we measured in this setup can be affected by absorption features of variable intensity depending on the different underlying stellar populations present. This is H$\alpha$. To solve this problem, we combined the LR-V and LR-R observations so that we could take advantage of the information present in the LR-V spectra to estimate the H$\alpha$ component in absorption. We then corrected our flux and EW measurements for H$\alpha$ in emission in the following way: \\

${\rm EW}_{\rm total} = {\rm EW}_{\rm emission} + |{\rm EW}_{\rm absorption}|$\\

${\rm Flux}_{\rm total} = {\rm EW}_{\rm total} \times {\rm Flux}_{\rm continuum.}$\\

The pointing of the observations sometimes varies from one VPH to the next in the same galaxy, therefore the spaxel in which we measure the emission may be thought to have a different underlying stellar population as the spaxel in which we measure the absorption. However, since the offsets between observations are rather small (typically 1 spaxel) and the variations in the absorption of the H$\alpha$ line are not affected at these scales, we find that this correction is not affected by this effect, as will be shown in the MEGADES II paper.

All lines with a signal to noise ratio measured at the peak of the line higher than 3 were analysed individually using Gauss-Hermite models, except for NaI D, for which we used an anchored double-Gaussian model. We also fitted the continuum emission around the lines using the information available on both sides of the line. For this task, we made use of one of the tools developed for the analysis of MEGARA observations, the \textit{analyze\_rss} megaratool. The spectral ranges used for each line are listed in Table \ref{table:emission_lines_def}.

\begin{table}%[H]

\caption{Line-fitting window definitions in restframe.}              % title of Table
\label{table:emission_lines_def}      % is used to refer this table in the text
\centering                    % used for centering table
\resizebox{0.49\textwidth}{!}{
\begin{tabular}{c c c c}          % centered columns (4 columns)
\hline \hline
\noalign{\smallskip}
Ion & $\lambda_{0}$ & Line window & Continuum windows\\
& [\AA] & [\AA] & [\AA] \\
\hline
\noalign{\smallskip}
H$\beta$ & 4861.333 & 4848 - 4877 &  4828 - 4848  \& 4877 - 4892 \\
$[\mathrm{O III}]\lambda5007$ & 5006.843 & 4997 - 5017 & 4977 - 4997 \& 5017 - 5037 \\
NaI D & 5889.950 & 5883 - 5905 & 5850 - 5870 \& 5910 - 5930 \\
H$\alpha$ & 6562.819 & 6555 - 6573 & 6513 - 6533 \& 6598 - 6618 \\
$[\mathrm{N II}]\lambda6584$ & 6583.460 & 6575 - 6594 & 6513 - 6533 \& 6598 - 6618 \\
$[\mathrm{S II}]\lambda6717$ & 6716.440 & 6707 - 6724 & 6680 - 6700 \& 6751 - 6771 \\
$[\mathrm{S II}]\lambda6731$ & 6730.810 & 6724 - 6741 & 6680 - 6700 \& 6751 - 6771 \\
\hline                                             %inserts single line
\end{tabular}}
\end{table}

\section{Data release}
\label{sec: data_release}

MEGADES is a legacy survey, and therefore, we make all data public to the community in order to exploit its full potential. We deliver all our observations reduced and with corrected pointings, ready for science. In addition to the reduced observations, we also deliver the analyses of stellar kinematics performed with pPXF and the measurements of some spectral lines carried out with \textit{analyze\_rss}. The study of the stellar populations was performed on the observations with a Voronoi binning so that all analysed regions reach a signal-to-noise ratio of 10. On the other hand, the study of the spectral lines was performed spaxel by spaxel, without any kind of binning, in order to take full advantage of the spatial resolution of the instrument. In future papers, we will publish our analysis for the stellar populations and diagnostic diagrams using all measured lines.

Some of the data included in this data release have already demonstrated their scientific potential in the analysis of stellar kinematics \citep{Dullo_2019} and stellar populations \citep{Chamorro_Cazorla_2022}. Other MEGARA observations have also been successfully exploited to study spatially resolved outflows in cold and warm phases \citep{Catalan_Torrecilla_2020}. This type of studies was also carried out on AGN and LINER galaxies \citep{Cazzoli_2020, Cazzoli_2022}. Some other studies that have taken advantage of the MEGARA observational features include \citet{Mayya_2020} and \citet{Tendulkar_2021}.

\subsection{Data products for data release I}

Both the observations and the products we deliver with this paper are in RSS format. This format has already been explained in section~\ref{section:megara_drp}. For the case of the data products, each row in the two-dimensional data corresponds to a fibre of the MEGARA IFU, as in the case of the observations, but each column, instead of being a different wavelength, is a different property, depending on the analysis we performed. Information regarding which property is in which column of each product type (stellar kinematic analysis or spectral line analysis) is included within the headers of each file in extension 2. Tables  \ref{table:ppxf_products} and \ref{table:analyze_rss_products} show the information in each column of the files for the products of pPXF and line analysis, respectively.

All the file products are available on the MEGADES website (see $\S$\ref{sec: web}), and the list of files with the information they contain is as follows:

\begin{itemize}
\item $ [ \textit{OBJECT} ]\_[ \textit{VPH} ]\_\textit{final\_rss.fits}$: These are the final products of the MEGARA DRP. They are reduced RSS 2D images including the 623 fiber spectra for the LCB mode, all using a common flux calibration and wavelength solution with constant reciprocal dispersion for all fibers and 4300 resolution elements. Based on the averaged spectrum of all fibers to be used for sky subtraction (by default, all 56 sky fibers in the LCB), the DRP also performs the sky substraction correction to generate this product. There is one of these files per observation. \\

\item $ [ \textit{OBJECT} ]\_[ \textit{VPH} ]\_\textit{voronoi\_rss\_snr\_10.fits}$: These files contain the same information as the final\_rss.fits files with a Voronoi binning performed on them to achieve a signal to noise ratio of 10 at least. The format of this file is an RSS with the same dimensions as \textit{final\_rss.fits} (623 rows and 4300 columns). Spaxels belonging to the same Voronoi region share the same spectrum. There is one of these files per observation. \\

\item $ [ \textit{OBJECT} ]\_[ \textit{VPH} ]\_\textit{emission\_final\_rss.fits}$: These files contain the result of subtracting the best fit performed with pPXF on the Voronoi regions of the observations renormalised at the continuum level of each individual spaxel to the observed spectra of each individual spaxel. The format of this file is again an RSS with the same dimensions as \textit{final\_rss.fits}.  There is one of these files per observation. \\

\item $ [ \textit{OBJECT} ]\textit{\_LR-V\_stellar\_kin.fits}$: These files contain the stellar kinematic information for each galaxy. It is in RSS format with 623 rows, but this time, with only 12 columns instead of the 4300 columns of the previous files. In these 12 columns, we list the information given in Table~\ref{table:ppxf_products}, which is stored in extension 2 of these files (there is one per galaxy). Table~\ref{table:analyze_rss_products} shows the information contained in extension 2 of the header of these files with each of the characteristics described here and the channel to which they correspond.

\item $ [ \textit{OBJECT} ]\_[ \textit{VPH} ]\_[ \textit{SPECTRAL\_LINE} ]\textit{.fits}$: These files contain the information of the different spectral lines that we analysed in this paper. The format of these files is an RSS with 623 rows and 33 columns, one for each measured feature, and it has the same dimensions for all the lines we studied. The information available in each of the columns is provided in Table~\ref{table:analyze_rss_products}. For the specific nomenclature of the files containing the analysis of each line provided in this paper, depending on the line, replace $ [ \textit{VPH} ]\_[ \textit{SPECTRAL\_LINE} ]$ with \textit{LR-B\_Hbeta}, \textit{LR-B\_O5007}, \textit{LR-V\_NaD}, \textit{LR-R\_Halpha}, \textit{LR-R\_N6584}, \textit{LR-R\_S6717,} and \textit{LR-R\_S6731}. There is one of these files per analysed line and galaxy. Table~\ref{table:analyze_rss_products} shows the information contained in extension 2 of the header of these files with each of the characteristics described here and the file channel to which they correspond.

\end{itemize}

\begin{table}%[H]
        \caption{Information included in each channel of the products obtained with pPXF. This information is included in extension 2 of the header of the corresponding products.}              % title of Table
        \label{table:ppxf_products}    % is used to refer this table in the text
        \centering                                      % used for centering table
        \resizebox{0.49\textwidth}{!}{
        \begin{tabular}{c l}          % centered columns (4 columns)
                \hline\hline                     % inserts two horizontal lines 
                \noalign{\smallskip}
                Channel (CH) & Description \\   % table heading
                \hline                                   % inserts single horizontal line
                \noalign{\smallskip}
                01 & Velocity with heliocentric (if applied) in km/s \\
                02 & Error of CH 01 in km/s \\
                03 & Sigma in km/s corrected for instrumental sigma \\
                04 & Error of CH 03 in km/s \\
                05 & h3 \\
                06 & Error of CH 05 \\
                07 & h4 \\
                08 & Error of CH 06 \\
                09 & Velocity in km/s \\
                10 & Error of CH 09 in km/s \\
                11 & Sigma in km/s \\
                12 & Error of CH 11 in km/s \\
                \hline       %inserts single line
        \end{tabular}}
\end{table}

\begin{table*}[h]
        \caption{Information included in each channel of the products obtained with \textit{analyze\_rss.py}. This information is included in extension 2 of the header of the corresponding products.}              % title of Table
        \label{table:analyze_rss_products}    % is used to refer this table in the text
        \centering                                      % used for centering table
        \begin{tabular}{c l}          % centered columns (4 columns)
                \hline\hline                     % inserts two horizontal lines 
                \noalign{\smallskip}
                Channel (CH) & Description \\  % table heading
                \hline                                   % inserts single horizontal line
                \noalign{\smallskip}
                01 & Fitting method (0=gauss-hermite, 1=1gauss, 2=2gauss, 3=gaussian doublet) \\
                02 & Continuum level in cgs$^{\dagger}$ \\
                03 & RMS in cgs \\
                04 & S/N at the peak of the line \\
                05 & Flux from window\_data - window\_continuum in cgs \\
                06 & Flux from (window\_data - window\_continuum) / mean\_continuum (EW) in $\mathrm{\AA}$ \\
                07 & Flux from best-fitting function(s) in cgs \\
                08 & EW from best-fitting function(s) in $\mathrm{\AA}$ \\
                09 & Amplitude for methods 0 \& 1 \& 2 (first gaussian) \& 3 (first gaussian) in cgs \\
                10 & Central lambda (in $\mathrm{\AA}$) for methods 0 \& 1 \& 2 (first gaussian) \& 3 (first gaussian) \\
                11 & Sigma (in $\mathrm{\AA}$) for methods 0 \& 1 \& 2 (first gaussian) \& 3 (first gaussian) \\
                12 & h3 for method 0 \\
                13 & h4 for method 0 \\
                14 & Amplitude for method 2 (second gaussian) \& 3 (first gaussian) in cgs \\
                15 & Central lambda (in $\mathrm{\AA}$) for method 2 (second gaussian) \& 3 (second gaussian) \\
                16 & Sigma (in $\mathrm{\AA}$) for method 2 (second gaussian) \& 3 (second gaussian) \\
                17 & Velocity in km/s from CH 10 (1st gaussian) \\
                18 & Velocity with heliocentric correction in km/s from CH 10 (1st gaussian) \\
                19 & Sigma in km/s from CH 11 (1st gaussian) \\
                20 & Sigma in km/s from CH 11 corrected for instrumental sigma (1st gaussian) \\
                21 & Velocity in km/s from CH 15 (2nd gaussian) \\
                22 & Velocity with heliocentric correction in km/s from CH 15 (2nd gaussian) \\
                23 & Sigma in km/s from CH 16 (2nd gaussian) \\
                24 & Sigma in km/s from CH 16 corrected for instrumental sigma (2nd gaussian) \\
                25 & Flux from best-fitting 1st gaussian in cgs \\
                26 & Flux from best-fitting 2nd gaussian in cgs \\
                27 & Error of CH 05 (Flux from window\_data - window\_continuum) in cgs \\
                28 & Error of CH 06 ((Flux from window\_data - window\_continuum) / mean\_continuum) in $\mathrm{\AA}$ \\
                29 & Error of CH 07 (Flux from best-fitting function(s)) in cgs \\
                30 & Error of CH 08 (EW from best-fitting function(s)) in $\mathrm{\AA}$ \\
                31 & Error of CH 17 (Velocity in km/s from CH 10 (1st gaussian)) in km/s \\
                32 & Error of CH 21 (Velocity in km/s from CH 15 (2nd gaussian) in km/s \\
                33 & Best-fitting $\mathrm{\chi}^2$ in cgs \\
                \hline       %inserts single line
        \end{tabular}
        \parbox{185mm}{\footnotesize $^{\dagger}$erg\,s$^{-1}$\,cm$^{-2}$\,\AA$^{-1}$}
\end{table*}

\subsection{Maps}

In addition to the data release, this work also presents some of the results we obtained. In the appendix of this paper, we show some figures derived from the analyses described in section~\ref{sec: analysis}. A continuum image obtained using the LR-V VPH (5165\,\AA - 6150\,\AA) with the PanSTARRS r-band image isophotes overplotted in white is shwon in panel (a). Panel (b) shows the integrated spectrum of the different observations. We achieved this by first summing the sky-corrected spectra of all the IFU spaxels, and by combining the information in the different setups to cover the whole spectral range spanned by all the VPHs. We also present the results of the study of stellar kinematics in the central regions of the sample galaxies. In panels (c), (d), (e), and (f), we show the stellar velocity, velocity dispersion, skewness, and kurtosis maps, respectively, for each galaxy. All these results were computed on a Voronoi binning to reach a signal-to-noise ratio per $\AA$ of 10. The data corresponding to these maps are all included in this first data release. Panel (s) shows a continuum RGB image obtained from the MEGARA observations. To make these maps, we aligned the continuum images of LR-B, LR-V, and LR-R (shown in panels (t), (u), and (v), respectively) based on their corrected astrometry solution. In the case of observations with two out of the three observations, the LR-V contribution to the image was doubled.

For the galaxies in the S4G subsample, we also show the flux, EW, velocity, and velocity dispersion data for all the lines we studied in this paper, H$\beta$ (panels (g), (h), (i) and (j), respectively), $[\mathrm{O III}]\lambda5007$ (panels (k), (l), (m) and (n), respectively), H$\alpha$ (panels (w), (x), (y) and (z), respectively), $[\mathrm{N II}]\lambda6584$ (panels ($\alpha$), ($\beta$), ($\gamma$) and ($\delta$), respectively) and the two [SII] lines, $[\mathrm{S II}]\lambda6717$ (panels ($\epsilon$), ($\zeta$), ($\eta$) and ($\theta$), respectively) and $[\mathrm{S II}]\lambda6731$ (panels ($\iota$), ($\kappa$), ($\lambda$) and ($\mu$), respectively). For the NaI D line, we present the same results, except for the case of velocity dispersion (panels (o), (p), (q) and (r), respectively). Instead of this, we show the map of the ratio of the flux of the doublet lines. For the 13 CALIFA galaxies of the sample, we only show the analysis for NaI D, H$\alpha$, $[\mathrm{N II}]\lambda6584$, $[\mathrm{S II}]\lambda6717$ and $[\mathrm{S II}]\lambda6731$ lines because we have no observations in LR-B. The S/N of the spaxels plotted in these maps at the peak of the line is higher than 3, and the rest of them remain blank.

In the galaxies marked as embargoed in Table~\ref{table:galaxy_properties}, we masked out the central spaxel and two rings of spaxels around it (19 spaxels in total) due to the presence of AGNs, which makes the spectral features in the central regions more complex. In these areas, we would need more kinematic components to analyse these features. The studies related to these areas will be conducted in Hermosa-Muñoz et al. 2023 in preparation.

\subsection{Web interface}
\label{sec: web}

The MEGADES library database is a software tool to manage all the observations and data products obtained from the analyses performed on the entire MEGADES sample. The database has been developed in MySQL with all the data available to date and a web-based tool that allows handling them. The technologies involved in the development of this software included HTML, JavaScript and Java Servlets with a Tomcat server and a MySQL database.
Security reasons require a public user name and password to access the MEGADES database\footnote{\href{https://www.megades.es}{https://www.megades.es}}. This user name and password are "public" and "6BRLukU55E", respectively.

\section{Notes on some individual galaxies}
\label{sec: notes}

\textit{IC~1683}: The stellar component of this galaxy reveals a rotation pattern that appears to be misaligned with respect to the photometric semi-major axis of the galaxy. We detect ionised gas in most of the field. A star-forming ring is found to surround the position of the brightest region in the continuum. The compactness of this structure (3\,arcsec in diameter) possibly prevented its detection in molecular gas emission by \cite{Bolatto_2017}. The kinematics of this ionised gas component show a rather distorted rotation pattern in the central region that can be explained using two kinematic components: one component follows the galaxy general rotation field traced by the stellar kinematics, and the other component rotates with respect to the ring minor axis. We also detect extended neutral gas corresponding to the brightest areas in the stellar continuum.\\

\noindent{\textit{NGC~0023}: This galaxy shows stellar kinematics dominated by a rotation pattern in the NS direction with and a clear rotation-h3 anticorrelation. The line emission extends to the entire MEGARA IFU field of view in the shape of a circumnuclear star-forming ring that also coincides with a minimum in the width of the emission lines, with its east side partly obscured by an intervening dust lane (see Figure~\ref{fig:sample1}). The ionised-gas velocity field is more distorted than the stellar one. These distortions may be caused by the bar in its central region (\citealt{deVaculeurs_1976}; see however \citealt{Afanasev_1991}). This galaxy has a companion, NGC~0026, but \cite{Epinat_2008} found no signs of any interaction in their H$\alpha$ velocity maps. It shares several characteristics with IC~1683, including extended NaD absorption, although its velocity field is misaligned by as much as $\sim$40$^{\circ}$.} \\

\noindent{\textit{NGC~0718}: The early-type (SAB(s)a) spiral galaxy NGC~0718 shows stellar kinematics dominated by a rotation pattern in the NS direction (quite different from the RC3 photometric position angle of the disc) and a clear rotation-h3 anticorrelation. The line emission extends to the entire MEGARA IFU field of view, but its velocity field is more distorted than the stellar one. \cite{Laurikainen_2005} suggested the existence of a (secondary) nuclear bar extending up to 5\,arcsec, which could be responsible for its ionised-gas distorted kinematics as traced by the $[\mathrm{N II}]\lambda6584$ emission, since the H$\alpha$ emission is barely detectable above its corresponding stellar absorption. \cite{Diaz_Garcia_2020} identified the star formation in the inner regions of this barred galaxy as limited to the circumnuclear region and therefore having a passive bar.}\\

\noindent{\textit{NGC~1087}: The central region of this galaxy shows two bright regions in the continuum that do not resemble a bona fide bulge \citep{Eskridge_2002}, not even a boxy pseudo-bulge. This might be due in part to the  dust lane crossing in between them (see Figure~\ref{fig:sample1}). We measured a global stellar rotation pattern with isovelocity curves running parallel from N to S. The ionised gas shows a velocity pattern with a clear NE to SW gradient, similar to that derived from CO (see \cite{Leroy_2021}), but highly distorted and showing the largest velocity gradient along the position and orientation of the nuclear dust lane. The misalignment found between the stellar and ionised gas velocity fields was already reported by \cite{Martin_1997}. The irregular distribution of the ionised-gas velocity dispersion in the nuclear region of NGC~1087 makes this galaxy a good candidate for a multi-component analysis.} \\

\noindent{\textit{NGC~2537}:} In the Bear Paw galaxy, which is a blue compact dwarf (BCD) galaxy, most of its star formation is distributed in a ring with a diameter of 1.2\,kpc that is bright both in H$\alpha$ and CO \citep{GildePaz_2002} around a massive post-starburst ($\sim$30\,Myr) region \citep{GildePaz_2000} that corresponds to our MEGARA pointing. This results in a lack of bright line emission associated with the nuclear region, but faint widespread diffuse emission instead. Interestingly, the variations found in radial velocity within the field are about ($\pm$10\,km\,s$^{-1}$) of the expected differences in velocity between the galaxy main (underlying) disc and the molecular ring that formed as a consequence of the evolution of the massive nuclear starburst \citep{GildePaz_2002b}. \\

\noindent{\textit{NGC~2543}: The line emission extends over most of the MEGARA field of view, although we identify a ring-like feature showing the highest equivalent widths, which could indicate the presence of a young star-forming circumnuclear ring with a radius of $\sim$3\,arcsec (0.5\,kpc). The stellar and ionised-gas velocity fields are rather similar, although the stellar kinematics become noisy beyond the innermost 4-5\,arcsec. We also detected neutral gas absorption coinciding with the inner edge (with a radius of $\sim$2\,arcsec) of the ring showing high EW in emission that is also slightly misaligned with respect to the ionised component. While \cite{Tang_2008} detected a tidal bridge between this galaxy and PGC~80408 that produces a distortion in the H I disc of NGC~2543, it is unclear whether this interaction could have any effect on the properties of these central regions. The detection of a high-EW ring and the misaligned NaD nuclear absorption indicates a post-nuclear starburst scenario.}\\

\noindent{\textit{NGC~3893}: In this galaxy, we find a stellar velocity pattern from SE to NW with a minimum in the stellar velocity dispersion coinciding with the galaxy nuclear regions. Morphologically classified as an SAB(rs)c: galaxy with a S\'{e}rsic index for its bulge component as measured by \cite{Weinzirl_2009} of $n_{\mathrm{bulge}}$=2.05, close to the separation limit proposed by \cite{Fisher_2008}, we cannot conclude that this is secularly-formed pseudo-bulge, although its low velocity dispersion ($\leq$60\,km\,s$^{-1}$) points in this direction. We also find neutral gas absorption in the brightest areas in the continuum of the galaxy with a small rotation pattern within the detected spaxels. The most remarkable feature is the widespread ionised-gas emission, especially in the H$\beta$, H$\alpha$, $[\mathrm{N II}]\lambda6584,$ and $[\mathrm{S II}]\lambda6731$ lines, but not so much in $[\mathrm{O III}]\lambda5007$. The corresponding velocity field seems to be aligned with the stellar velocity field, but some distortions in all lines are detected.}\\

\noindent{\textit{NGC~3982}: This Seyfert 2 spiral galaxy has a cylindrical rotation pattern in its stellar component. We measured emission lines of ionised gas throughout the entire field of the MEGARA IFU and in all the lines presented in Figure~\ref{fig:NGC3982_card_1}. The velocity pattern of the ionised gas component is apparently formed by two distinct components. The rotation of one of them is aligned with the stellar component (PA$\sim$$-$165$^{\circ}$), and that of the other with a rotation pattern almost perpendicular to it. This could be caused by the 4-arcsecond-long bar in the bulge of this galaxy reported by \cite{Erwin_2005}. The nuclear region shows a maximum in velocity dispersion both in stars and the ionised gas ($\geq$70\,km\,s$^{-1}$).}\\

\noindent{\textit{NGC~3998}: This LINER galaxy shows a clear rotation pattern for its stellar component with a minor axis orientation of PA$\sim$45$^{\circ}$ and with the radial velocity clearly anti-correlated with the skewness (h3), as sign of a strong rotational support for the line-of-sight radial velocities measured. The ionised gas also shows a well-ordered rotation, although the minor axis PA is close to 0$^{\circ}$. Neutral gas detected in the form of NaI D absorption is also detected in most of the IFU field of view, with a velocity pattern aligned with that of the stellar component and very high equivalent widths of up to 2\,\AA. This galaxy lacks LR-R observations.}\\

\noindent{\textit{NGC~4041}: The faintness of the stellar continuum in this object does not allow us to properly measure stellar kinematics even after a Voronoi binning is applied. However, we detect a large amount of ionised gas in all of the spaxels of the field, with a rather cylindrical rotation aligned with the photometric semi-major axis of the galaxy and some small-scale (about one to two spaxels, or $\sim$100\,pc) distortions superimposed. For this object, our LR-V observations are significantly more shallow than for LR-B and LR-R.}\\

\noindent{\textit{NGC~4189}: In this barred spiral (SAB(rs)cd) galaxy, the rotation of the stellar component is very different from that seen in the ionised gas. Thus, while the line-of-sight stellar rotation is oriented from SW (redshifted) to the NE (blueshifted), $\sim$60$^{\circ}$ away from the bar major axis, the ionised-gas radial velocity field is well reproduced by the superposition of two rotation patterns, one aligned with the bar with a high amplitude, and the other, witha  lower amplitude, running roughly NS. The distorted velocity field and the kinematic and photometric misalignment in the H$\alpha$ component was observed by \cite{Chemin_2006}, although with our data, much more detail can be discerned, making this object (and our data) a prime target for the dynamical modelling of stars and gas in the presence of the non-axisymmetric potential of a stellar bar.}\\

\noindent{\textit{NGC~4278}: This LINER galaxy shows a clear anti-correlation between line-of-sight radial velocity and skewness (h3). We also measure ionised gas emission throughout the entire field of view, possibly associated with the halo component assigned to this object by \cite{Masegosa_2011}. The line emission we detected shows a radial velocity field with a rotation pattern slightly misaligned compared to the stellar one (PA$\sim$45$^{\circ}$ vs$.$ 30$^{\circ}$ for the corresponding receding semi-major axes, respectively). We also find a large amount of absorbing NaI D throughout the field. The radial velocities of this neutral gas are at odds with those measured from either the stars or the ionised gas. This peculiar velocity structure is compatible with the inflow reported in this object by \cite{Cazzoli_2018}. This galaxy lacks LR-R observations.}\\

\noindent{\textit{NGC~4593}: This Seyfert 1 galaxy has a very clear rotation pattern in its stellar component that is anti-correlated with h3 and a ring wide oriented along the major axis with low velocity dispersion and relatively high kurtosis (h4$\sim$+0.15). This region coincides with a lens in the continuum images from PanSTARRS (see Figure~\ref{fig:sample1}) that is outlined by high-EW line-emitting regions, some of which might be embedded in the dust lanes that according to the HST images, appear to spiral in towards the position of the AGN. The velocity pattern of the ionised gas is aligned with the velocity pattern of the stars in spite of some small-scale distorsions. We also detect neutral gas absorption over a large fraction of the circumnuclear ring.}\\

\noindent{\textit{NGC~4750}: The stellar component of this LINER galaxy shows a clear rotation that, again, it is clearly anti-correlated with h3. The analysis of the ionised-gas kinematics, as detected in all emission lines studied in this paper, shows a rotation pattern that is composed of a circumnuclear component that is clearly misaligned by almost 90$^{\circ}$ (clockwise) relative to the stellar velocity field and a more extended component whose receding side is at PA$\sim$0$^{\circ}$ (i.e. $\sim$45$^{\circ}$ counter-clockwise relative to the stars). The complex behaviour of the line emission found in NGC~4750 agrees with the finding of non-rotational motions in the narrow component by \cite{Cazzoli_2018}. We do not detect NaD in absorption, although this interstellar residual absorption has previously been reported by these same authors.}\\

\noindent{\textit{NGC~5218}: This interacting spiral galaxy shows a stellar rotation pattern misaligned with respect to its photometric semi-major axis. We detect gas in both its neutral and ionised phases over a large part of the field of view. The radial velocity and velocity dispersion maps in the case of the ionised gas component appear to be highly distorted, likely due to the interaction with its companion galaxy (NGC~5216) \citep{Elfhag_1996}, but also to the presence of a strong bar. NaD interstellar absorption is detected with equivalent widths as high as 8\,\AA\ in the nucleus, showing a velocity field roughly aligned with the stellar velocity field, but with two deep velocity minima (blueshifted by $\sim$100\,km\,s$^{-1}$ relative to the systemic velocity) found at both edges of the rotation axis.}\\

\noindent{\textit{NGC~5394}: This object shows morphological and kinematic features similar to those described for galaxy NGC~5218 object above. The reason in this case is proabably also that the galaxy is in close interaction with a (massive) companion,  NGC~5395 at 27\,kpc \citep{Epinat_2008} in the case of NGC~5394, that has led, among other things, to the development to two opposite stellar tidal tails (see Figure~\ref{fig:sample1}). Thus, NGC~5394 shows a regular stellar rotation pattern, but highly distorted ionised gas velocity and velocity dispersion maps. The NaD interstellar absorption is also very deep (with EW as high as 6\,\AA) and spatially extended. Its velocity field  appears to have a rotation pattern perpendicular to that of the stellar component.}\\

\noindent{\textit{NGC~5616}: This is a early edge-on spiral galaxy with a clear stellar rotation pattern along with a major axis and regions with stellar radial velocities that are close to the galaxy systemic velocity along with the minor axis. These correspond to regions located at larger galactocentric distances, with only a minor fraction of their motion taking place along the line of sight. On the east side of the galaxy centre, we identifya dust lane crossing the entire field of view  in
the PanSTARRS images and in our MEGARA false-colour images that is embedded by bright knots of star formation. The properties of this object as seen with MEGARA resemble those of UGC~10205 \citep{Catalan_Torrecilla_2020}, including the presence of some NaI D absorption that might be associated with the high gas column density and bright background continuum provided by the galaxy nucleus. This absorption shows a slight rotation pattern that is aligned with the stellar component.}\\

\noindent{\textit{NGC~5953}: This Seyfert 2 galaxy shows stellar kinematics that are dominated by a rotation pattern that is anti-correlated with h3. The line-of-sight kurtosis (h4) values show a positive correlation with the radial velocities. We were able to detect and measure NaI D in absorption over a large part of the field of view. Its velocity pattern is aligned perpendicularly to the stellar rotation, and it is highly distorted. We identify an emission ring around the brightest continuum region for the ionised-gas component (the continuum know located 2\,arcsec W of the nucleus appears to be a foreground star because it shows diffraction spikes in the HST images) that we relate to the star-forming ring with a radius of 4\,arcsec reported by \cite{Riffel_2006}. The innermost part of the ionised-gas radial velocity field shows a distortion that might be explained as due to an overimposed rotation pattern aligned NS of $\sim$2\,arcsec ($\sim$300\,pc) in length. According to \cite{Arp_1966}, this galaxy interacts with NGC~5954.}\\

\noindent{\textit{NGC~6027}: This lenticular galaxy shows a very clear rotation pattern in its stellar component that is anti-correlated with h3. We detect neutral gas in the form of blueshifted NaI D in absorption that appears to show an S-shaped morphology in radial velocity. The most striking feature of this object, which it is part of the Seyfert Sextet (\citealt{Arp_1966}; also known as HCG92), is ionised-gas emission in H$\alpha$ and $[\mathrm{N II}]\lambda6584$ that counter-rotates relative to the stars. This was first reported by \cite{Durbala_2008} and was interpreted as being a dwarf spiral that accretes during the assembly process of this compact group. The radial velocity gradient in the ionised gas seems to be steeper (and of a different sign) than that of the stars, possibly due to a larger contribution of the axisymmetric-drift correction to the latter.}\\

\noindent{\textit{NGC~6217}: The stellar component of this Seyfert 2 galaxy shows radial velocities that appear to be the superposition of two rotation patterns, both with low amplitudes, one aligned from E to W and the other with the receding side oriented towards a PA of $-$45$^{\circ}$. Both patterns are clearly visible when analysing the ionised-gas radial velocity field, with the latter being associated with a (young) circumnuclear star-forming ring of high equivalent width in emission and 3\,arcsec (300\,pc) in diameter. \cite{Hernandez_2005} reported another H$\alpha$ ring of about 43 arcsec in diameter. We also detect some neutral gas in the central part of the galaxy, especially in its NE region.}\\

\noindent{\textit{NGC~6339}: This barred spiral galaxy shows stellar kinematics with a rotation pattern that is perpendicular with respect to the photometric semi-major axis of the bar. Ionised-gas emission is found along the brightest continuum regions, with the brightest emission-line region located at the easternmost end of the field of view. The velocity distribution of the ionised gas is distorted and shows an S-shape along its central part that might be explained by two distinct kinematic components: one component that follows the stellar kinematics, and another component that is slightly deflected in the direction of the bright region to the east. This distortion might be caused by the effects of the bar or by star formation that is embedded in a dust lane that crosses the galaxy nuclear regions from N to S. This galaxy was classified as having both circumnuclear star formation and hosting star formation along the bar by \cite{Diaz_Garcia_2020}.}\\

\noindent{\textit{NGC~7025}: The stellar kinematics of this galaxy was studied in detail by \cite{Dullo_2019}. It shows a rotation pattern that is anti-correlated with h3. Its stellar kinematics along with its photometric properties led these authors to suggest that NGC~7025 hosts a pseudo-bulge. Using MEGARA data, \cite{Chamorro_Cazorla_2022} proposed that NGC~7025 suffered a minor merger $\sim$3.5-4.5\,Gyr ago that shaped its spatially resolved stellar populations. Here, we detect NaI D in absorption in the central regions of the galaxy and report a rotation pattern that follows the rotation of the stellar component. We also detect ionised gas, especially in the $[\mathrm{N II}]\lambda6584$ and $[\mathrm{O III}]\lambda5007$ lines, whose velocity fields also seem to follow the stellar velocity pattern, although with a highly distorted morphology that extends much farther away to the NE than to the SW relative to the galaxy nucleus. In the areas in which we detect H$\alpha$ emission, their velocities appear to be much higher than in the other (forbidden) emission lines.}\\

\noindent{\textit{NGC~7479}: This strongly barred Seyfert 2 galaxy \citep{Lumsden_2001} shows continuum emission in the nuclear regions that is severely affected by a dust lane that crosses from N to S and is visible even in the MEGARA false-colour images (see also Fig$.$\ref{fig:sample1}). This NS orientation also corresponds to the PA of the bar at larger spatial scales. The stellar velocity field shows a clear rotation pattern, with its receding side having a PA of $-$135$^{\circ}$. The ionised-gas emission is particularly bright in the emission lines located in the LR-R setup, possibly because of the high dust attenuations in the region. Its radial velocity maps indicate a complex behaviour, with a twist in the rotation axis relative to that of the stellar velocity field, and two regions (along with the major axis of the stellar radial velocity field) with high (gas) velocity dispersions ($\geq$100\,km\,s$^{-1}$).}\\

\noindent{\textit{NGC~7591}: This LINER galaxy shows a stellar velocity field with a clear rotation pattern. We detect ionised gas throughout the entire field of view of MEGARA. This galaxy was not observed in the LR-B setup. The velocity field of the ionised gas follows the stellar kinematics in almost all regions, although it is distorted with a twist in the central region that might be explained by the onset of several kinematic components along the line of sight. The high velocity dispersion values found ($\geq$100\,km\,s$^{-1}$), both in recombination and in the forbidden emission lines, also indicate the need of using multiple kinematic components in emission. We also detect neutral gas in the central regions of the galaxy, with a radial velocity field that is distinct from that of the stars or the ionised gas.}\\

\noindent{\textit{NGC~7738}: This is a strongly barred (SB(rs)b) spiral galaxy with its bar aligned in the NE to SW direction. Associated with this bar, a dust lane crosses the MEGARA field of view in the same direction, but slightly SE of the galaxy nucleus. The stellar kinematics of the galaxy show a rotation pattern that is anti-correlated with h3 and aligned with the photometric semi-major axis of the galaxy. We detect neutral gas in the form of NaI D in absorption in the brightest continuum regions. The velocity pattern of this intervening neutral gas approximately follows stellar rotation, although the absorption detected on top of the dust lane seems to be close to the galaxy systemic velocity. We also measure ionised gas emission coming from H$\alpha$ and $[\mathrm{N II}]\lambda6584$ lines (LR-B observations were not acquired for this object) whose velocity fields follow stellar rotation, although with a remarkably high velocity dispersion (almost 150\,km\,s$^{-1}$), which might be connected to the effects of the bar and the dust lane associated with it. The velocity field of the $[\mathrm{S II}]\lambda6717$ line is affected by atmospheric telluric absorptions.}\\

\noindent{\textit{NGC~7787}: In this early-type barred spiral galaxy ((R')SB(rs)0/a), the stellar component follows a rotation pattern that is anti-correlated with h3. We detect intervening neutral gas in the form of NaI D in absorption in the central regions of the galaxy. Its kinematics approximately follow the rotation pattern of the stars. We detect line emission arising from two distinct regions at either side of the nucleus, about 4\,arcsec apart (or 2\,kpc at the distance NGC~7787). These regions might be part of a (nuclear; see \citealt{Buta_1993}) ring of 1\,kpc in radius or, alternatively, indicate star formation at the extremes of a nuclear bar. The fact that the two regions appear closer in the emission line flux maps than in those showing the equivalent widths indicates that the continuum emission drops faster than the emission and that the emission is relatively extended in these regions. The velocity field of the $[\mathrm{S II}]\lambda6717$ line is affected by atmospheric telluric absorptions.}\\

\section{Summary}
\label{sec: summary}

The MEGARA galaxy disc evolution survey (MEGADES) is the scientific legacy project of MEGARA, the integral field and multi-object spectrograph installed on the 10.4 m GTC telescope.  With this project, we aim to deepen our understanding of the diverse secular processes that take place in galaxies and their influence on the evolution of disc galaxies (e.g. stellar migration, AGN feedback, and starbursts). For this purpose, we conduct several studies of the characteristics and kinematics of the stellar populations that form our galaxies, and we also study all possible aspects related to the ionised and neutral gas (through NaI D absorption) in the galaxies of the sample. Because this is a legacy project, we expect the scientific community to be able to take advantage of all the data and results we make public in this data release to ensure that this survey is as fruitful as possible and serves beyond the scientific objectives for which it was conceived.

We currently work with high spatial (0.62\,\arcsec) and spectral resolution (R $\sim$ 6000) observations made on the central regions of a total of 43 galaxies, 30 of which are a sub-sample of the S4G survey, and the remaining 13 were part of the CALIFA survey. The details of the sample are described in $\S$\ref{sec: survey}. All these observations have been made using the MEGARA IFU. For the galaxies belonging to the S4G sample, we have observations taken with three different VPHs, obtaining spectra in a spectral range between 4350.6\,$\AA$ and 7287.7\,$\AA$  at a resolution of R $\sim$ 6000. For the galaxies belonging to the CALIFA sample, we have observations taken with two VPHs, covering a spectral range from 5165.6\,$\AA$  to 7287.7\,$\AA$  at the same resolution. All the technical details of the instrument are presented in $\S$\ref{sec: observations}.

We also carefully detailed the whole process of data reduction and processing from observation to conversion into data ready for scientific use in $\S$\ref{sec: data_processing}. In this same section, we explained the post-processing performed on the reduced data to improve the quality of its pointings and correct them. We also performed a Voronoi binning on the observations to achieve a signal-to-noise ratio of 10 at least.

In addition to the observations ready for scientific use, in this paper we carried out the first studies on the data from the inner regions of the galaxies in the MEGADES sample. We performed an analysis of the stellar kinematics in the galaxies of the entire sample using the pPXF software based on observations made with the VPH LR-V (5165.6\,$\AA$ - 6176.2\,$\AA$). We present all maps of the kinematic components, that is, radial velocity, velocity dispersion, skewness, and kurtosis. We also carried out measurements on the emission lines of the ionised gas such as H$\beta$, $[\mathrm{O III}]\lambda5007$, H$\alpha$, $[\mathrm{N II}]\lambda6584$, $[\mathrm{S II}]\lambda6717,$ and $[\mathrm{S II}]\lambda6731$, and measurements on the neutral gas using the sodium doublet lines NaI D. We also present some of the features measured on these lines in maps. All these maps are gathered as individual galaxy cards in the appendix to facilitate their inspection.

All the observations together with the results of the analyses carried out in this paper are available to the community on the MEGADES website\footnote{\href{https://www.megades.es}{https://www.megades.es}}. To access the public data, one can log in with the username "public" and the password "6BRLukU55E".

\begin{acknowledgements}

This publication is based on data obtained with the MEGARA instrument at the Gran Telescopio
CANARIAS. MEGARA has been built by a Consortium led by the Universidad Complutense de Madrid (Spain)
and that also includes the Instituto de Astrofísica, Óptica y Electrónica (Mexico), Instituto de Astrofísica de
Andalucía (CSIC, Spain) and the Univesidad Politécnica de Madrid (Spain). MEGARA is funded by the
Consortium institutions and by GRANTECAN S.A.   

Based on observations made with the Gran Telescopio Canarias (GTC), installed in the Spanish Observatorio del Roque de los Muchachos of the Instituto de Astrof\'isica de Canarias, in the island of La Palma.

This work is based on data obtained with MEGARA instrument, funded by European Regional Development Funds (ERDF), through Programa Operativo Canarias FEDER 2014-2020.

This work has been supported by grant RTI2018-096188-B-I00 and AYA2016-75808-R  funded by program MCIN/AEI/10.13039/501100011033 of the Spanish Ministry of Science and Innovation.

This research has made use of the NASA/IPAC Extragalactic Database, which is funded by the National Aeronautics and Space Administration and operated by the California Institute of Technology.

%This research has made use of the SIMBAD database, operated at CDS, Strasbourg, France.

The Pan-STARRS1 Surveys (PS1) and the PS1 public science archive have been made possible through contributions by the Institute for Astronomy, the University of Hawaii, the Pan-STARRS Project Office, the Max-Planck Society and its participating institutes, the Max Planck Institute for Astronomy, Heidelberg and the Max Planck Institute for Extraterrestrial Physics, Garching, The Johns Hopkins University, Durham University, the University of Edinburgh, the Queen's University Belfast, the Harvard-Smithsonian Center for Astrophysics, the Las Cumbres Observatory Global Telescope Network Incorporated, the National Central University of Taiwan, the Space Telescope Science Institute, the National Aeronautics and Space Administration under Grant No. NNX08AR22G issued through the Planetary Science Division of the NASA Science Mission Directorate, the National Science Foundation Grant No. AST-1238877, the University of Maryland, Eotvos Lorand University (ELTE), the Los Alamos National Laboratory, and the Gordon and Betty Moore Foundation.
\end{acknowledgements}
\bibliographystyle{aa}
\bibliography{biblio}

\begin{thebibliography}{77}
\expandafter\ifx\csname natexlab\endcsname\relax\def\natexlab#1{#1}\fi

\bibitem[{{Abazajian} {et~al.}(2009){Abazajian}, {Adelman-McCarthy},
  {Ag{\"u}eros}, {Allam}, {Allende Prieto}, {An}, {Anderson}, {Anderson},
  {Annis}, {Bahcall}, {Bailer-Jones}, {Barentine}, {Bassett}, {Becker},
  {Beers}, {Bell}, {Belokurov}, {Berlind}, {Berman}, {Bernardi}, {Bickerton},
  {Bizyaev}, {Blakeslee}, {Blanton}, {Bochanski}, {Boroski}, {Brewington},
  {Brinchmann}, {Brinkmann}, {Brunner}, {Budav{\'a}ri}, {Carey}, {Carliles},
  {Carr}, {Castander}, {Cinabro}, {Connolly}, {Csabai}, {Cunha}, {Czarapata},
  {Davenport}, {de Haas}, {Dilday}, {Doi}, {Eisenstein}, {Evans}, {Evans},
  {Fan}, {Friedman}, {Frieman}, {Fukugita}, {G{\"a}nsicke}, {Gates},
  {Gillespie}, {Gilmore}, {Gonzalez}, {Gonzalez}, {Grebel}, {Gunn},
  {Gy{\"o}ry}, {Hall}, {Harding}, {Harris}, {Harvanek}, {Hawley}, {Hayes},
  {Heckman}, {Hendry}, {Hennessy}, {Hindsley}, {Hoblitt}, {Hogan}, {Hogg},
  {Holtzman}, {Hyde}, {Ichikawa}, {Ichikawa}, {Im}, {Ivezi{\'c}}, {Jester},
  {Jiang}, {Johnson}, {Jorgensen}, {Juri{\'c}}, {Kent}, {Kessler}, {Kleinman},
  {Knapp}, {Konishi}, {Kron}, {Krzesinski}, {Kuropatkin}, {Lampeitl},
  {Lebedeva}, {Lee}, {Lee}, {French Leger}, {L{\'e}pine}, {Li}, {Lima}, {Lin},
  {Long}, {Loomis}, {Loveday}, {Lupton}, {Magnier}, {Malanushenko},
  {Malanushenko}, {Mandelbaum}, {Margon}, {Marriner}, {Mart{\'\i}nez-Delgado},
  {Matsubara}, {McGehee}, {McKay}, {Meiksin}, {Morrison}, {Mullally}, {Munn},
  {Murphy}, {Nash}, {Nebot}, {Neilsen}, {Newberg}, {Newman}, {Nichol},
  {Nicinski}, {Nieto-Santisteban}, {Nitta}, {Okamura}, {Oravetz}, {Ostriker},
  {Owen}, {Padmanabhan}, {Pan}, {Park}, {Pauls}, {Peoples}, {Percival}, {Pier},
  {Pope}, {Pourbaix}, {Price}, {Purger}, {Quinn}, {Raddick}, {Re Fiorentin},
  {Richards}, {Richmond}, {Riess}, {Rix}, {Rockosi}, {Sako}, {Schlegel},
  {Schneider}, {Scholz}, {Schreiber}, {Schwope}, {Seljak}, {Sesar}, {Sheldon},
  {Shimasaku}, {Sibley}, {Simmons}, {Sivarani}, {Allyn Smith}, {Smith},
  {Smol{\v{c}}i{\'c}}, {Snedden}, {Stebbins}, {Steinmetz}, {Stoughton},
  {Strauss}, {SubbaRao}, {Suto}, {Szalay}, {Szapudi}, {Szkody}, {Tanaka},
  {Tegmark}, {Teodoro}, {Thakar}, {Tremonti}, {Tucker}, {Uomoto}, {Vanden
  Berk}, {Vandenberg}, {Vidrih}, {Vogeley}, {Voges}, {Vogt}, {Wadadekar},
  {Watters}, {Weinberg}, {West}, {White}, {Wilhite}, {Wonders}, {Yanny},
  {Yocum}, {York}, {Zehavi}, {Zibetti}, \& {Zucker}}]{Abazajian_2009}
{Abazajian}, K.~N., {Adelman-McCarthy}, J.~K., {Ag{\"u}eros}, M.~A., {et~al.}
  2009, \apjs, 182, 543

\bibitem[{{Afanasev} {et~al.}(1991){Afanasev}, {Zasov}, {Popravko}, \&
  {Silchenko}}]{Afanasev_1991}
{Afanasev}, V.~L., {Zasov}, A.~V., {Popravko}, G.~V., \& {Silchenko}, O.~K.
  1991, Soviet Astronomy Letters, 17, 325

\bibitem[{{Arp}(1966)}]{Arp_1966}
{Arp}, H. 1966, \apjs, 14, 1

\bibitem[{{Bagnasco} {et~al.}(2007){Bagnasco}, {Kolm}, {Ferruit}, {Honnen},
  {Koehler}, {Lemke}, {Maschmann}, {Melf}, {Noyer}, {Rumler}, {Salvignol},
  {Strada}, \& {Te Plate}}]{Bagnasco_2007}
{Bagnasco}, G., {Kolm}, M., {Ferruit}, P., {et~al.} 2007, in Society of
  Photo-Optical Instrumentation Engineers (SPIE) Conference Series, Vol. 6692,
  Cryogenic Optical Systems and Instruments XII, ed. J.~B. {Heaney} \& L.~G.
  {Burriesci}, 66920M

\bibitem[{{Benitez} {et~al.}(2014){Benitez}, {Dupke}, {Moles}, {Sodre},
  {Cenarro}, {Marin-Franch}, {Taylor}, {Cristobal}, {Fernandez-Soto}, {Mendes
  de Oliveira}, {Cepa-Nogue}, {Abramo}, {Alcaniz}, {Overzier},
  {Hernandez-Monteagudo}, {Alfaro}, {Kanaan}, {Carvano}, {Reis}, {Martinez
  Gonzalez}, {Ascaso}, {Ballesteros}, {Xavier}, {Varela}, {Ederoclite},
  {Vazquez Ramio}, {Broadhurst}, {Cypriano}, {Angulo}, {Diego}, {Zandivarez},
  {Diaz}, {Melchior}, {Umetsu}, {Spinelli}, {Zitrin}, {Coe}, {Yepes}, {Vielva},
  {Sahni}, {Marcos-Caballero}, {Shu Kitaura}, {Maroto}, {Masip}, {Tsujikawa},
  {Carneiro}, {Gonzalez Nuevo}, {Carvalho}, {Reboucas}, {Carvalho}, {Abdalla},
  {Bernui}, {Pigozzo}, {Ferreira}, {Chandrachani Devi}, {Bengaly}, {Campista},
  {Amorim}, {Asari}, {Bongiovanni}, {Bonoli}, {Bruzual}, {Cardiel}, {Cava},
  {Cid Fernandes}, {Coelho}, {Cortesi}, {Delgado}, {Diaz Garcia}, {Espinosa},
  {Galliano}, {Gonzalez-Serrano}, {Falcon-Barroso}, {Fritz}, {Fernandes},
  {Gorgas}, {Hoyos}, {Jimenez-Teja}, {Lopez-Aguerri}, {Lopez-San Juan},
  {Mateus}, {Molino}, {Novais}, {OMill}, {Oteo}, {Perez-Gonzalez}, {Poggianti},
  {Proctor}, {Ricciardelli}, {Sanchez-Blazquez}, {Storchi-Bergmann}, {Telles},
  {Schoennell}, {Trujillo}, {Vazdekis}, {Viironen}, {Daflon},
  {Aparicio-Villegas}, {Rocha}, {Ribeiro}, {Borges}, {Martins}, {Marcolino},
  {Martinez-Delgado}, {Perez-Torres}, {Siffert}, {Calvao}, {Sako}, {Kessler},
  {Alvarez-Candal}, {De Pra}, {Roig}, {Lazzaro}, {Gorosabel}, {Lopes de
  Oliveira}, {Lima-Neto}, {Irwin}, {Liu}, {Alvarez}, {Balmes}, {Chueca},
  {Costa-Duarte}, {da Costa}, {Dantas}, {Diaz}, {Fabregat}, {Ferrari},
  {Gavela}, {Gracia}, {Gruel}, {Gutierrez}, {Guzman}, {Hernandez-Fernandez},
  {Herranz}, {Hurtado-Gil}, {Jablonsky}, {Laporte}, {Le Tiran}, {Licandro},
  {Lima}, {Martin}, {Martinez}, {Montero}, {Penteado}, {Pereira}, {Peris},
  {Quilis}, {Sanchez-Portal}, {Soja}, {Solano}, {Torra}, \&
  {Valdivielso}}]{Benitez_2014}
{Benitez}, N., {Dupke}, R., {Moles}, M., {et~al.} 2014, arXiv e-prints,
  arXiv:1403.5237

\bibitem[{{Blanc} {et~al.}(2013){Blanc}, {Weinzirl}, {Song}, {Heiderman},
  {Gebhardt}, {Jogee}, {Evans}, {van den Bosch}, {Luo}, {Drory}, {Fabricius},
  {Fisher}, {Hao}, {Kaplan}, {Marinova}, {Vutisalchavakul}, \&
  {Yoachim}}]{Blanc_2013}
{Blanc}, G.~A., {Weinzirl}, T., {Song}, M., {et~al.} 2013, \aj, 145, 138

\bibitem[{{Bolatto} {et~al.}(2017){Bolatto}, {Wong}, {Utomo}, {Blitz}, {Vogel},
  {S{\'a}nchez}, {Barrera-Ballesteros}, {Cao}, {Colombo}, {Dannerbauer},
  {Garc{\'\i}a-Benito}, {Herrera-Camus}, {Husemann}, {Kalinova}, {Leroy},
  {Leung}, {Levy}, {Mast}, {Ostriker}, {Rosolowsky}, {Sandstrom}, {Teuben},
  {van de Ven}, \& {Walter}}]{Bolatto_2017}
{Bolatto}, A.~D., {Wong}, T., {Utomo}, D., {et~al.} 2017, \apj, 846, 159

\bibitem[{{Bundy} {et~al.}(2015){Bundy}, {Bershady}, {Law}, {Yan}, {Drory},
  {MacDonald}, {Wake}, {Cherinka}, {S{\'a}nchez-Gallego}, {Weijmans}, {Thomas},
  {Tremonti}, {Masters}, {Coccato}, {Diamond-Stanic}, {Arag{\'o}n-Salamanca},
  {Avila-Reese}, {Badenes}, {Falc{\'o}n-Barroso}, {Belfiore}, {Bizyaev},
  {Blanc}, {Bland-Hawthorn}, {Blanton}, {Brownstein}, {Byler}, {Cappellari},
  {Conroy}, {Dutton}, {Emsellem}, {Etherington}, {Frinchaboy}, {Fu}, {Gunn},
  {Harding}, {Johnston}, {Kauffmann}, {Kinemuchi}, {Klaene}, {Knapen},
  {Leauthaud}, {Li}, {Lin}, {Maiolino}, {Malanushenko}, {Malanushenko}, {Mao},
  {Maraston}, {McDermid}, {Merrifield}, {Nichol}, {Oravetz}, {Pan}, {Parejko},
  {Sanchez}, {Schlegel}, {Simmons}, {Steele}, {Steinmetz}, {Thanjavur},
  {Thompson}, {Tinker}, {van den Bosch}, {Westfall}, {Wilkinson}, {Wright},
  {Xiao}, \& {Zhang}}]{Bundy_2015}
{Bundy}, K., {Bershady}, M.~A., {Law}, D.~R., {et~al.} 2015, \apj, 798, 7

\bibitem[{{Buta} \& {Crocker}(1993)}]{Buta_1993}
{Buta}, R. \& {Crocker}, D.~A. 1993, \aj, 105, 1344

\bibitem[{{Cappellari}(2017)}]{Cappellari_2017}
{Cappellari}, M. 2017, \mnras, 466, 798

\bibitem[{{Cappellari} \& {Copin}(2003)}]{Cappellari_2003}
{Cappellari}, M. \& {Copin}, Y. 2003, \mnras, 342, 345

\bibitem[{{Cappellari} \& {Emsellem}(2004)}]{Cappellari_2004}
{Cappellari}, M. \& {Emsellem}, E. 2004, \pasp, 116, 138

\bibitem[{{Cappellari} {et~al.}(2011){Cappellari}, {Emsellem}, {Krajnovi{\'c}},
  {McDermid}, {Scott}, {Verdoes Kleijn}, {Young}, {Alatalo}, {Bacon}, {Blitz},
  {Bois}, {Bournaud}, {Bureau}, {Davies}, {Davis}, {de Zeeuw}, {Duc},
  {Khochfar}, {Kuntschner}, {Lablanche}, {Morganti}, {Naab}, {Oosterloo},
  {Sarzi}, {Serra}, \& {Weijmans}}]{Cappellari_2011}
{Cappellari}, M., {Emsellem}, E., {Krajnovi{\'c}}, D., {et~al.} 2011, \mnras,
  413, 813

\bibitem[{{Carrasco} {et~al.}(2018){Carrasco}, {Gil de Paz}, {Gallego},
  {Iglesias-P{\'a}ramo}, {Cedazo}, {Garc{\'\i}a Vargas}, {Arrillaga},
  {Avil{\'e}s}, {Bouquin}, {Carbajo}, {Cardiel}, {Carrera}, {Castillo Morales},
  {Castillo-Dom{\'\i}nguez}, {Esteban San Rom{\'a}n}, {Ferrusca},
  {G{\'o}mez-{\'A}lvarez}, {Izazaga-P{\'e}rez}, {Lefort}, {L{\'o}pez Orozco},
  {Maldonado}, {Mart{\'\i}nez Delgado}, {Morales Dur{\'a}n}, {M{\'u}jica},
  {Ortiz}, {P{\'a}ez}, {Pascual}, {P{\'e}rez-Calpena}, {Picazo},
  {S{\'a}nchez-Penim}, {S{\'a}nchez-Blanco}, {Tulloch}, {Vel{\'a}zquez},
  {V{\'\i}lchez}, {Zamorano}, {Aguerri}, {Barrado}, {Bertone}, {Cava},
  {Catal{\'a}n-Torrecilla}, {Cenarro}, {Ch{\'a}vez}, {Dullo}, {Eliche},
  {Garc{\'\i}a}, {Garc{\'\i}a-Rojas}, {Guichard}, {Gonz{\'a}lez-Delgado},
  {Guzm{\'a}n}, {Herrero}, {Hu{\'e}lamo}, {Hughes}, {Jim{\'e}nez-Vicente},
  {Kehrig}, {Marino}, {M{\'a}rquez}, {Masegosa}, {Mayya}, {M{\'e}ndez-Abreu},
  {Moll{\'a}}, {Mu{\~n}oz-Tu{\~n}{\'o}n}, {Peimbert}, {P{\'e}rez-Gonz{\'a}lez},
  {P{\'e}rez-Montero}, {Roca-F{\`a}brega}, {Rodr{\'\i}guez},
  {Rodr{\'\i}guez-Espinosa}, {Rodr{\'\i}guez-Merino},
  {Rodr{\'\i}guez-Mu{\~n}oz}, {Rosa-Gonz{\'a}lez}, {S{\'a}nchez-Almeida},
  {S{\'a}nchez Contreras}, {S{\'a}nchez-Bl{\'a}zquez}, {S{\'a}nchez},
  {Sarajedini}, {Silich}, {Sim{\'o}n-D{\'\i}az}, {Tenorio-Tagle}, {Terlevich},
  {Terlevich}, {Torres-Peimbert}, {Trujillo}, {Tsamis}, \&
  {Vega}}]{carrasco_2018SPIE}
{Carrasco}, E., {Gil de Paz}, A., {Gallego}, J., {et~al.} 2018, in Society of
  Photo-Optical Instrumentation Engineers (SPIE) Conference Series, Vol. 10702,
  Ground-based and Airborne Instrumentation for Astronomy VII, ed. C.~J.
  {Evans}, L.~{Simard}, \& H.~{Takami}, 1070216

\bibitem[{Castillo-Morales {et~al.}(2020)Castillo-Morales, Pascual, \&
  de~Paz}]{africa_castillo_morales_2020_3932063}
Castillo-Morales, A., Pascual, S., \& de~Paz, A.~G. 2020, MEGARA Data Reduction
  Cookbook

\bibitem[{{Catal{\'a}n-Torrecilla} {et~al.}(2020){Catal{\'a}n-Torrecilla},
  {Castillo-Morales}, {Gil de Paz}, {Gallego}, {Carrasco},
  {Iglesias-P{\'a}ramo}, {Cedazo}, {Chamorro-Cazorla}, {Pascual},
  {Garc{\'\i}a-Vargas}, {Cardiel}, {G{\'o}mez-Alvarez}, {P{\'e}rez-Calpena},
  {Mart{\'\i}nez-Delgado}, {Dullo}, {Coelho}, {Bruzual}, \&
  {Charlot}}]{Catalan_Torrecilla_2020}
{Catal{\'a}n-Torrecilla}, C., {Castillo-Morales}, {\'A}., {Gil de Paz}, A.,
  {et~al.} 2020, \apj, 890, 5

\bibitem[{{Cazzoli} {et~al.}(2020){Cazzoli}, {Gil de Paz}, {M{\'a}rquez},
  {Masegosa}, {Iglesias}, {Gallego}, {Carrasco}, {Cedazo},
  {Garc{\'\i}a-Vargas}, {Castillo-Morales}, {Pascual}, {Cardiel},
  {P{\'e}rez-Calpena}, {G{\'o}mez-Alvarez}, {Mart{\'\i}nez-Delgado}, \&
  {Hermosa-Mu{\~n}oz}}]{Cazzoli_2020}
{Cazzoli}, S., {Gil de Paz}, A., {M{\'a}rquez}, I., {et~al.} 2020, \mnras, 493,
  3656

\bibitem[{{Cazzoli} {et~al.}(2022){Cazzoli}, {Hermosa Mu{\~n}oz},
  {M{\'a}rquez}, {Masegosa}, {Castillo-Morales}, {Gil de Paz},
  {Hern{\'a}ndez-Garc{\'\i}a}, {La Franca}, \& {Ramos Almeida}}]{Cazzoli_2022}
{Cazzoli}, S., {Hermosa Mu{\~n}oz}, L., {M{\'a}rquez}, I., {et~al.} 2022, \aap,
  664, A135

\bibitem[{{Cazzoli} {et~al.}(2018){Cazzoli}, {M{\'a}rquez}, {Masegosa}, {del
  Olmo}, {Povi{\'c}}, {Gonz{\'a}lez-Mart{\'\i}n}, {Balmaverde},
  {Hern{\'a}ndez-Garc{\'\i}a}, \& {Garc{\'\i}a-Burillo}}]{Cazzoli_2018}
{Cazzoli}, S., {M{\'a}rquez}, I., {Masegosa}, J., {et~al.} 2018, \mnras, 480,
  1106

\bibitem[{Chambers {et~al.}(2016)Chambers, Magnier, Metcalfe, Flewelling,
  Huber, Waters, Denneau, Draper, Farrow, Finkbeiner, Holmberg, Koppenhoefer,
  Price, Rest, Saglia, Schlafly, Smartt, Sweeney, Wainscoat, Burgett, Chastel,
  Grav, Heasley, Hodapp, Jedicke, Kaiser, Kudritzki, Luppino, Lupton, Monet,
  Morgan, Onaka, Shiao, Stubbs, Tonry, White, Bañados, Bell, Bender, Bernard,
  Boegner, Boffi, Botticella, Calamida, Casertano, Chen, Chen, Cole, Deacon,
  Frenk, Fitzsimmons, Gezari, Gibbs, Goessl, Goggia, Gourgue, Goldman, Grant,
  Grebel, Hambly, Hasinger, Heavens, Heckman, Henderson, Henning, Holman, Hopp,
  Ip, Isani, Jackson, Keyes, Koekemoer, Kotak, Le, Liska, Long, Lucey, Liu,
  Martin, Masci, McLean, Mindel, Misra, Morganson, Murphy, Obaika, Narayan,
  Nieto-Santisteban, Norberg, Peacock, Pier, Postman, Primak, Rae, Rai, Riess,
  Riffeser, Rix, Röser, Russel, Rutz, Schilbach, Schultz, Scolnic, Strolger,
  Szalay, Seitz, Small, Smith, Soderblom, Taylor, Thomson, Taylor, Thakar,
  Thiel, Thilker, Unger, Urata, Valenti, Wagner, Walder, Walter, Watters,
  Werner, Wood-Vasey, \& Wyse}]{Chambers_2016}
Chambers, K.~C., Magnier, E.~A., Metcalfe, N., {et~al.} 2016, The Pan-STARRS1
  Surveys

\bibitem[{{Chamorro-Cazorla} {et~al.}(2022){Chamorro-Cazorla}, {Gil de Paz},
  {Castillo-Morales}, {Dullo}, {Gallego}, {Carrasco}, {Iglesias-P{\'a}ramo},
  {Cedazo}, {Garc{\'\i}a-Vargas}, {Pascual}, {Cardiel}, {P{\'e}rez-Calpena},
  {G{\'o}mez-{\'A}lvarez}, {Mart{\'\i}nez-Delgado}, \&
  {Catal{\'a}n-Torrecilla}}]{Chamorro_Cazorla_2022}
{Chamorro-Cazorla}, M., {Gil de Paz}, A., {Castillo-Morales}, A., {et~al.}
  2022, \aap, 657, A95

\bibitem[{{Chemin} {et~al.}(2006){Chemin}, {Balkowski}, {Cayatte}, {Carignan},
  {Amram}, {Garrido}, {Hernandez}, {Marcelin}, {Adami}, {Boselli}, \&
  {Boulesteix}}]{Chemin_2006}
{Chemin}, L., {Balkowski}, C., {Cayatte}, V., {et~al.} 2006, \mnras, 366, 812

\bibitem[{{Contini} {et~al.}(2012){Contini}, {Garilli}, {Le F{\`e}vre},
  {Kissler-Patig}, {Amram}, {Epinat}, {Moultaka}, {Paioro}, {Queyrel}, {Tasca},
  {Tresse}, {Vergani}, {L{\'o}pez-Sanjuan}, \& {Perez-Montero}}]{Contini_2012}
{Contini}, T., {Garilli}, B., {Le F{\`e}vre}, O., {et~al.} 2012, \aap, 539, A91

\bibitem[{{Croton} {et~al.}(2006){Croton}, {Springel}, {White}, {De Lucia},
  {Frenk}, {Gao}, {Jenkins}, {Kauffmann}, {Navarro}, \&
  {Yoshida}}]{Croton_2006}
{Croton}, D.~J., {Springel}, V., {White}, S. D.~M., {et~al.} 2006, \mnras, 365,
  11

\bibitem[{{de Vaucouleurs} {et~al.}(1976){de Vaucouleurs}, {de Vaucouleurs}, \&
  {Corwin}}]{deVaculeurs_1976}
{de Vaucouleurs}, G., {de Vaucouleurs}, A., \& {Corwin}, J.~R. 1976, Second
  reference catalogue of bright galaxies, 1976, 0

\bibitem[{{de Zeeuw} {et~al.}(2002){de Zeeuw}, {Bureau}, {Emsellem}, {Bacon},
  {Carollo}, {Copin}, {Davies}, {Kuntschner}, {Miller}, {Monnet}, {Peletier},
  \& {Verolme}}]{Zeeuw_2002}
{de Zeeuw}, P.~T., {Bureau}, M., {Emsellem}, E., {et~al.} 2002, \mnras, 329,
  513

\bibitem[{{D{\'\i}az-Garc{\'\i}a} {et~al.}(2020){D{\'\i}az-Garc{\'\i}a},
  {Moyano}, {Comer{\'o}n}, {Knapen}, {Salo}, \& {Bouquin}}]{Diaz_Garcia_2020}
{D{\'\i}az-Garc{\'\i}a}, S., {Moyano}, F.~D., {Comer{\'o}n}, S., {et~al.} 2020,
  \aap, 644, A38

\bibitem[{{Dullo} {et~al.}(2019){Dullo}, {Chamorro-Cazorla}, {Gil de Paz},
  {Castillo-Morales}, {Gallego}, {Carrasco}, {Iglesias-P{\'a}ramo}, {Cedazo},
  {Garc{\'\i}a-Vargas}, {Pascual}, {Cardiel}, {P{\'e}rez-Calpena},
  {G{\'o}mez-Alvarez}, {Mart{\'\i}nez-Delgado}, \&
  {Catal{\'a}n-Torrecilla}}]{Dullo_2019}
{Dullo}, B.~T., {Chamorro-Cazorla}, M., {Gil de Paz}, A., {et~al.} 2019, \apj,
  871, 9

\bibitem[{{Durbala} {et~al.}(2008){Durbala}, {del Olmo}, {Yun}, {Rosado},
  {Sulentic}, {Plana}, {Iovino}, {Perea}, {Verdes-Montenegro}, \&
  {Fuentes-Carrera}}]{Durbala_2008}
{Durbala}, A., {del Olmo}, A., {Yun}, M.~S., {et~al.} 2008, \aj, 135, 130

\bibitem[{{Elfhag} {et~al.}(1996){Elfhag}, {Booth}, {Hoeglund}, {Johansson}, \&
  {Sandqvist}}]{Elfhag_1996}
{Elfhag}, T., {Booth}, R.~S., {Hoeglund}, B., {Johansson}, L.~E.~B., \&
  {Sandqvist}, A. 1996, \aaps, 115, 439

\bibitem[{{Emsellem} {et~al.}(2022){Emsellem}, {Schinnerer}, {Santoro},
  {Belfiore}, {Pessa}, {McElroy}, {Blanc}, {Congiu}, {Groves}, {Ho}, {Kreckel},
  {Razza}, {Sanchez-Blazquez}, {Egorov}, {Faesi}, {Klessen}, {Leroy}, {Meidt},
  {Querejeta}, {Rosolowsky}, {Scheuermann}, {Anand}, {Barnes},
  {Be{\v{s}}li{\'c}}, {Bigiel}, {Boquien}, {Cao}, {Chevance}, {Dale},
  {Eibensteiner}, {Glover}, {Grasha}, {Henshaw}, {Hughes}, {Koch}, {Kruijssen},
  {Lee}, {Liu}, {Pan}, {Pety}, {Saito}, {Sandstrom}, {Schruba}, {Sun},
  {Thilker}, {Usero}, {Watkins}, \& {Williams}}]{Emsellem_2022}
{Emsellem}, E., {Schinnerer}, E., {Santoro}, F., {et~al.} 2022, \aap, 659, A191

\bibitem[{{Epinat} {et~al.}(2008){Epinat}, {Amram}, {Marcelin}, {Balkowski},
  {Daigle}, {Hernandez}, {Chemin}, {Carignan}, {Gach}, \&
  {Balard}}]{Epinat_2008}
{Epinat}, B., {Amram}, P., {Marcelin}, M., {et~al.} 2008, \mnras, 388, 500

\bibitem[{{Erwin}(2005)}]{Erwin_2005}
{Erwin}, P. 2005, \mnras, 364, 283

\bibitem[{{Eskridge} {et~al.}(2002){Eskridge}, {Frogel}, {Pogge}, {Quillen},
  {Berlind}, {Davies}, {DePoy}, {Gilbert}, {Houdashelt}, {Kuchinski},
  {Ram{\'\i}rez}, {Sellgren}, {Stutz}, {Terndrup}, \& {Tiede}}]{Eskridge_2002}
{Eskridge}, P.~B., {Frogel}, J.~A., {Pogge}, R.~W., {et~al.} 2002, \apjs, 143,
  73

\bibitem[{{Fabian}(2012)}]{Fabian_2012}
{Fabian}, A.~C. 2012, \araa, 50, 455

\bibitem[{{Falc{\'o}n-Barroso} {et~al.}(2011){Falc{\'o}n-Barroso},
  {S{\'a}nchez-Bl{\'a}zquez}, {Vazdekis}, {Ricciardelli}, {Cardiel}, {Cenarro},
  {Gorgas}, \& {Peletier}}]{FalconBarroso_2011}
{Falc{\'o}n-Barroso}, J., {S{\'a}nchez-Bl{\'a}zquez}, P., {Vazdekis}, A.,
  {et~al.} 2011, \aap, 532, A95

\bibitem[{{Fisher} \& {Drory}(2008)}]{Fisher_2008}
{Fisher}, D.~B. \& {Drory}, N. 2008, \aj, 136, 773

\bibitem[{{F{\"o}rster Schreiber} {et~al.}(2009){F{\"o}rster Schreiber},
  {Genzel}, {Bouch{\'e}}, {Cresci}, {Davies}, {Buschkamp}, {Shapiro},
  {Tacconi}, {Hicks}, {Genel}, {Shapley}, {Erb}, {Steidel}, {Lutz},
  {Eisenhauer}, {Gillessen}, {Sternberg}, {Renzini}, {Cimatti}, {Daddi},
  {Kurk}, {Lilly}, {Kong}, {Lehnert}, {Nesvadba}, {Verma}, {McCracken},
  {Arimoto}, {Mignoli}, \& {Onodera}}]{Forster_2009}
{F{\"o}rster Schreiber}, N.~M., {Genzel}, R., {Bouch{\'e}}, N., {et~al.} 2009,
  \apj, 706, 1364

\bibitem[{{Gardner} {et~al.}(2006){Gardner}, {Mather}, {Clampin}, {Doyon},
  {Greenhouse}, {Hammel}, {Hutchings}, {Jakobsen}, {Lilly}, {Long}, {Lunine},
  {McCaughrean}, {Mountain}, {Nella}, {Rieke}, {Rieke}, {Rix}, {Smith},
  {Sonneborn}, {Stiavelli}, {Stockman}, {Windhorst}, \&
  {Wright}}]{Gardner_2006}
{Gardner}, J.~P., {Mather}, J.~C., {Clampin}, M., {et~al.} 2006, \ssr, 123, 485

\bibitem[{{Gil de Paz} {et~al.}(2018){Gil de Paz}, {Carrasco}, {Gallego},
  {Iglesias-P{\'a}ramo}, {Cedazo}, {Garc{\'\i}a-Vargas}, {Arrillaga},
  {Avil{\'e}s}, {Bouquin}, {Carbajo}, {Cardiel}, {Carrera}, {Castillo-Morales},
  {Castillo-Dom{\'\i}nguez}, {Esteban San Rom{\'a}n}, {Ferrusca},
  {G{\'o}mez-{\'A}lvarez}, {Izazaga-P{\'e}rez}, {Lefort}, {L{\'o}pez-Orozco},
  {Maldonado}, {Mart{\'\i}nez-Delgado}, {Morales-Dur{\'a}n}, {Mujica},
  {P{\'a}ez}, {Pascual}, {P{\'e}rez-Calpena}, {Picazo}, {S{\'a}nchez-Penim},
  {S{\'a}nchez-Blanco}, {Tulloch}, {Vel{\'a}zquez}, {V{\'\i}lchez}, {Zamorano},
  {Aguerri}, {Barrado y Navascues}, {Berlanas}, {Bertone}, {Cava},
  {Catal{\'a}n-Torrecilla}, {Cenarro}, {Ch{\'a}vez}, {Dullo}, {Garc{\'\i}a},
  {Garc{\'\i}a-Rojas}, {Guichard}, {Gonz{\'a}lez-Delgado}, {Guzm{\'a}n},
  {Herrero}, {Hu{\'e}lamo}, {Hughes}, {Jim{\'e}nez-Vicente}, {Kehrig},
  {Marino}, {M{\'a}rquez}, {Masegosa}, {Mayya}, {M{\'e}ndez-Abreu},
  {Moll{\'a}}, {Mu{\~n}oz-Tu{\~n}{\'o}n}, {Peimbert}, {P{\'e}rez-Gonz{\'a}lez},
  {P{\'e}rez-Montero}, {Rodr{\'\i}guez}, {Rodr{\'\i}guez-Espinosa},
  {Rodr{\'\i}guez Merino}, {Rodr{\'\i}guez-Mu{\~n}oz}, {Rosa-Gonz{\'a}lez},
  {S{\'a}nchez-Almeida}, {S{\'a}nchez-Contreras}, {S{\'a}nchez-Bl{\'a}zquez},
  {S{\'a}nchez}, {Sarajedini}, {Silich}, {Sim{\'o}n-D{\'\i}az},
  {Tenorio-Tagle}, {Terlevich}, {Terlevich}, {Torres-Peimbert}, {Trujillo},
  {Tsamis}, \& {Vega}}]{armando_2018SPIE}
{Gil de Paz}, A., {Carrasco}, E., {Gallego}, J., {et~al.} 2018, in Society of
  Photo-Optical Instrumentation Engineers (SPIE) Conference Series, Vol. 10702,
  Ground-based and Airborne Instrumentation for Astronomy VII, ed. C.~J.
  {Evans}, L.~{Simard}, \& H.~{Takami}, 1070217

\bibitem[{{Gil de Paz} {et~al.}(2002{\natexlab{a}}){Gil de Paz}, {Silich},
  {Madore}, {S{\'a}nchez Contreras}, {Zamorano}, \& {Gallego}}]{GildePaz_2002}
{Gil de Paz}, A., {Silich}, S.~A., {Madore}, B.~F., {et~al.}
  2002{\natexlab{a}}, \apjl, 573, L101

\bibitem[{{Gil de Paz} {et~al.}(2002{\natexlab{b}}){Gil de Paz}, {Silich},
  {Madore}, {S{\'a}nchez Contreras}, {Zamorano}, \& {Gallego}}]{GildePaz_2002b}
{Gil de Paz}, A., {Silich}, S.~A., {Madore}, B.~F., {et~al.}
  2002{\natexlab{b}}, \apjl, 573, L101

\bibitem[{{Gil de Paz} {et~al.}(2000){Gil de Paz}, {Zamorano}, \&
  {Gallego}}]{GildePaz_2000}
{Gil de Paz}, A., {Zamorano}, J., \& {Gallego}, J. 2000, \aap, 361, 465

\bibitem[{{Gunn} \& {Gott}(1972)}]{Gunn_1972}
{Gunn}, J.~E. \& {Gott}, J.~Richard, I. 1972, \apj, 176, 1

\bibitem[{{Hamuy} {et~al.}(1994){Hamuy}, {Suntzeff}, {Heathcote}, {Walker},
  {Gigoux}, \& {Phillips}}]{Hamuy_1994}
{Hamuy}, M., {Suntzeff}, N.~B., {Heathcote}, S.~R., {et~al.} 1994, \pasp, 106,
  566

\bibitem[{{Hamuy} {et~al.}(1992){Hamuy}, {Walker}, {Suntzeff}, {Gigoux},
  {Heathcote}, \& {Phillips}}]{Hamuy_1992}
{Hamuy}, M., {Walker}, A.~R., {Suntzeff}, N.~B., {et~al.} 1992, \pasp, 104, 533

\bibitem[{{Hernandez} {et~al.}(2005){Hernandez}, {Carignan}, {Amram}, {Chemin},
  \& {Daigle}}]{Hernandez_2005}
{Hernandez}, O., {Carignan}, C., {Amram}, P., {Chemin}, L., \& {Daigle}, O.
  2005, \mnras, 360, 1201

\bibitem[{{Kacharov} {et~al.}(2018){Kacharov}, {Neumayer}, {Seth},
  {Cappellari}, {McDermid}, {Walcher}, \& {B{\"o}ker}}]{Kacharov_2018}
{Kacharov}, N., {Neumayer}, N., {Seth}, A.~C., {et~al.} 2018, \mnras, 480, 1973

\bibitem[{King(1985)}]{King_1985}
King, D. 1985

\bibitem[{{Kormendy} \& {Kennicutt}(2004)}]{Kormendy_2004}
{Kormendy}, J. \& {Kennicutt}, Robert~C., J. 2004, \araa, 42, 603

\bibitem[{{Laurikainen} {et~al.}(2005){Laurikainen}, {Salo}, \&
  {Buta}}]{Laurikainen_2005}
{Laurikainen}, E., {Salo}, H., \& {Buta}, R. 2005, \mnras, 362, 1319

\bibitem[{{Leroy} {et~al.}(2021){Leroy}, {Schinnerer}, {Hughes}, {Rosolowsky},
  {Pety}, {Schruba}, {Usero}, {Blanc}, {Chevance}, {Emsellem}, {Faesi},
  {Herrera}, {Liu}, {Meidt}, {Querejeta}, {Saito}, {Sandstrom}, {Sun},
  {Williams}, {Anand}, {Barnes}, {Behrens}, {Belfiore}, {Benincasa},
  {Be{\v{s}}li{\'c}}, {Bigiel}, {Bolatto}, {den Brok}, {Cao}, {Chandar},
  {Chastenet}, {Chiang}, {Congiu}, {Dale}, {Deger}, {Eibensteiner}, {Egorov},
  {Garc{\'\i}a-Rodr{\'\i}guez}, {Glover}, {Grasha}, {Henshaw}, {Ho}, {Kepley},
  {Kim}, {Klessen}, {Kreckel}, {Koch}, {Kruijssen}, {Larson}, {Lee}, {Lopez},
  {Machado}, {Mayker}, {McElroy}, {Murphy}, {Ostriker}, {Pan}, {Pessa},
  {Puschnig}, {Razza}, {S{\'a}nchez-Bl{\'a}zquez}, {Santoro}, {Sardone},
  {Scheuermann}, {Sliwa}, {Sormani}, {Stuber}, {Thilker}, {Turner}, {Utomo},
  {Watkins}, \& {Whitmore}}]{Leroy_2021}
{Leroy}, A.~K., {Schinnerer}, E., {Hughes}, A., {et~al.} 2021, \apjs, 257, 43

\bibitem[{{Lin} \& {Shu}(1964)}]{Lin_1964}
{Lin}, C.~C. \& {Shu}, F.~H. 1964, \apj, 140, 646

\bibitem[{{Lumsden} {et~al.}(2001){Lumsden}, {Heisler}, {Bailey}, {Hough}, \&
  {Young}}]{Lumsden_2001}
{Lumsden}, S.~L., {Heisler}, C.~A., {Bailey}, J.~A., {Hough}, J.~H., \&
  {Young}, S. 2001, \mnras, 327, 459

\bibitem[{{Martin} \& {Friedli}(1997)}]{Martin_1997}
{Martin}, P. \& {Friedli}, D. 1997, \aap, 326, 449

\bibitem[{{Masegosa} {et~al.}(2011){Masegosa}, {M{\'a}rquez}, {Ramirez}, \&
  {Gonz{\'a}lez-Mart{\'\i}n}}]{Masegosa_2011}
{Masegosa}, J., {M{\'a}rquez}, I., {Ramirez}, A., \&
  {Gonz{\'a}lez-Mart{\'\i}n}, O. 2011, \aap, 527, A23

\bibitem[{{Mayya} {et~al.}(2020){Mayya}, {Carrasco}, {G{\'o}mez-Gonz{\'a}lez},
  {Zaragoza-Cardiel}, {Gil de Paz}, {Ovando}, {S{\'a}nchez-Cruces},
  {Lomel{\'\i}-N{\'u}{\~n}ez}, {Rodr{\'\i}guez-Merino}, {Rosa-Gonz{\'a}lez},
  {Silich}, {Tenorio-Tagle}, {Bruzual}, {Charlot}, {Terlevich}, {Terlevich},
  {Vega}, {Gallego}, {Iglesias-P{\'a}ramo}, {Castillo-Morales},
  {Garc{\'\i}a-Vargas}, {G{\'o}mez-Alvarez}, {Pascual}, \&
  {P{\'e}rez-Calpena}}]{Mayya_2020}
{Mayya}, Y.~D., {Carrasco}, E., {G{\'o}mez-Gonz{\'a}lez}, V.~M.~A., {et~al.}
  2020, \mnras, 498, 1496

\bibitem[{{Mercier} {et~al.}(2022){Mercier}, {Epinat}, {Contini},
  {Abril-Melgarejo}, {Boogaard}, {Brinchmann}, {Finley}, {Krajnovi{\'c}},
  {Michel-Dansac}, {Ventou}, {Bouch{\'e}}, {Dumoulin}, \&
  {Pineda}}]{Mercier_2022}
{Mercier}, W., {Epinat}, B., {Contini}, T., {et~al.} 2022, \aap, 665, A54

\bibitem[{{Moore} {et~al.}(1996){Moore}, {Katz}, {Lake}, {Dressler}, \&
  {Oemler}}]{Moore_1996}
{Moore}, B., {Katz}, N., {Lake}, G., {Dressler}, A., \& {Oemler}, A. 1996,
  \nat, 379, 613

\bibitem[{{Oke}(1990)}]{Oke_1990}
{Oke}, J.~B. 1990, \aj, 99, 1621

\bibitem[{Pascual {et~al.}(2022)Pascual, Picazo, Cardiel, Castillo-Morales, \&
  de~Paz}]{pascual_2022}
Pascual, S., Picazo, P., Cardiel, N., Castillo-Morales, A., \& de~Paz, A.~G.
  2022, guaix-ucm/megaradrp: v0.12.0

\bibitem[{{P{\'e}rez-Gonz{\'a}lez} {et~al.}(2013){P{\'e}rez-Gonz{\'a}lez},
  {Cava}, {Barro}, {Villar}, {Cardiel}, {Ferreras}, {Rodr{\'\i}guez-Espinosa},
  {Alonso-Herrero}, {Balcells}, {Cenarro}, {Cepa}, {Charlot}, {Cimatti},
  {Conselice}, {Daddi}, {Donley}, {Elbaz}, {Espino}, {Gallego}, {Gobat},
  {Gonz{\'a}lez-Mart{\'\i}n}, {Guzm{\'a}n}, {Hern{\'a}n-Caballero},
  {Mu{\~n}oz-Tu{\~n}{\'o}n}, {Renzini}, {Rodr{\'\i}guez-Zaur{\'\i}n}, {Tresse},
  {Trujillo}, \& {Zamorano}}]{Perez_Gonzalez_2013}
{P{\'e}rez-Gonz{\'a}lez}, P.~G., {Cava}, A., {Barro}, G., {et~al.} 2013, \apj,
  762, 46

\bibitem[{{Rieke} {et~al.}(2005){Rieke}, {Kelly}, \& {Horner}}]{Rieke_2005}
{Rieke}, M.~J., {Kelly}, D., \& {Horner}, S. 2005, in Society of Photo-Optical
  Instrumentation Engineers (SPIE) Conference Series, Vol. 5904, Cryogenic
  Optical Systems and Instruments XI, ed. J.~B. {Heaney} \& L.~G. {Burriesci},
  1--8

\bibitem[{{Riffel} {et~al.}(2006){Riffel}, {Rodr{\'\i}guez-Ardila}, \&
  {Pastoriza}}]{Riffel_2006}
{Riffel}, R., {Rodr{\'\i}guez-Ardila}, A., \& {Pastoriza}, M.~G. 2006, \aap,
  457, 61

\bibitem[{{Rousseau-Nepton} {et~al.}(2019){Rousseau-Nepton}, {Martin},
  {Robert}, {Drissen}, {Amram}, {Prunet}, {Martin}, {Moumen}, {Adamo},
  {Alarie}, {Barmby}, {Boselli}, {Bresolin}, {Bureau}, {Chemin}, {Fernandes},
  {Combes}, {Crowder}, {Della Bruna}, {Duarte Puertas}, {Egusa}, {Epinat},
  {Ksoll}, {Girard}, {G{\'o}mez Llanos}, {Gouliermis}, {Grasha}, {Higgs},
  {Hlavacek-Larrondo}, {Ho}, {Iglesias-P{\'a}ramo}, {Joncas}, {Kam}, {Karera},
  {Kennicutt}, {Klessen}, {Lianou}, {Liu}, {Liu}, {de Amorim}, {Lyman},
  {Martel}, {Mazzilli-Ciraulo}, {McLeod}, {Melchior}, {Millan}, {Moll{\'a}},
  {Momose}, {Morisset}, {Pan}, {Pati}, {Pellerin}, {Pellegrini}, {P{\'e}rez},
  {Petric}, {Plana}, {Rahner}, {Ruiz Lara}, {S{\'a}nchez-Menguiano},
  {Spekkens}, {Stasi{\'n}ska}, {Takamiya}, {Vale Asari}, \&
  {V{\'\i}lchez}}]{Rousseau_2019}
{Rousseau-Nepton}, L., {Martin}, R.~P., {Robert}, C., {et~al.} 2019, \mnras,
  489, 5530

\bibitem[{{Salpeter}(1955)}]{Salpeter_1955}
{Salpeter}, E.~E. 1955, \apj, 121, 161

\bibitem[{{S{\'a}nchez} {et~al.}(2012){S{\'a}nchez}, {Kennicutt}, {Gil de Paz},
  {van de Ven}, {V{\'\i}lchez}, {Wisotzki}, {Walcher}, {Mast}, {Aguerri},
  {Albiol-P{\'e}rez}, {Alonso-Herrero}, {Alves}, {Bakos}, {Bart{\'a}kov{\'a}},
  {Bland-Hawthorn}, {Boselli}, {Bomans}, {Castillo-Morales}, {Cortijo-Ferrero},
  {de Lorenzo-C{\'a}ceres}, {Del Olmo}, {Dettmar}, {D{\'\i}az}, {Ellis},
  {Falc{\'o}n-Barroso}, {Flores}, {Gallazzi}, {Garc{\'\i}a-Lorenzo},
  {Gonz{\'a}lez Delgado}, {Gruel}, {Haines}, {Hao}, {Husemann},
  {Igl{\'e}sias-P{\'a}ramo}, {Jahnke}, {Johnson}, {Jungwiert}, {Kalinova},
  {Kehrig}, {Kupko}, {L{\'o}pez-S{\'a}nchez}, {Lyubenova}, {Marino},
  {M{\'a}rmol-Queralt{\'o}}, {M{\'a}rquez}, {Masegosa}, {Meidt},
  {Mendez-Abreu}, {Monreal-Ibero}, {Montijo}, {Mour{\~a}o}, {Palacios-Navarro},
  {Papaderos}, {Pasquali}, {Peletier}, {P{\'e}rez}, {P{\'e}rez}, {Quirrenbach},
  {Rela{\~n}o}, {Rosales-Ortega}, {Roth}, {Ruiz-Lara},
  {S{\'a}nchez-Bl{\'a}zquez}, {Sengupta}, {Singh}, {Stanishev}, {Trager},
  {Vazdekis}, {Viironen}, {Wild}, {Zibetti}, \& {Ziegler}}]{Sanchez_2012}
{S{\'a}nchez}, S.~F., {Kennicutt}, R.~C., {Gil de Paz}, A., {et~al.} 2012,
  \aap, 538, A8

\bibitem[{{S{\'a}nchez-Bl{\'a}zquez} {et~al.}(2006){S{\'a}nchez-Bl{\'a}zquez},
  {Peletier}, {Jim{\'e}nez-Vicente}, {Cardiel}, {Cenarro},
  {Falc{\'o}n-Barroso}, {Gorgas}, {Selam}, \&
  {Vazdekis}}]{SanchezBlazquez_2006}
{S{\'a}nchez-Bl{\'a}zquez}, P., {Peletier}, R.~F., {Jim{\'e}nez-Vicente}, J.,
  {et~al.} 2006, \mnras, 371, 703

\bibitem[{{Sancisi} {et~al.}(2008){Sancisi}, {Fraternali}, {Oosterloo}, \& {van
  der Hulst}}]{Sancisi_2008}
{Sancisi}, R., {Fraternali}, F., {Oosterloo}, T., \& {van der Hulst}, T. 2008,
  \aapr, 15, 189

\bibitem[{{Sheth} {et~al.}(2010){Sheth}, {Regan}, {Hinz}, {Gil de Paz},
  {Men{\'e}ndez-Delmestre}, {Mu{\~n}oz-Mateos}, {Seibert}, {Kim},
  {Laurikainen}, {Salo}, {Gadotti}, {Laine}, {Mizusawa}, {Armus},
  {Athanassoula}, {Bosma}, {Buta}, {Capak}, {Jarrett}, {Elmegreen},
  {Elmegreen}, {Knapen}, {Koda}, {Helou}, {Ho}, {Madore}, {Masters},
  {Mobasher}, {Ogle}, {Peng}, {Schinnerer}, {Surace}, {Zaritsky},
  {Comer{\'o}n}, {de Swardt}, {Meidt}, {Kasliwal}, \& {Aravena}}]{Sheth_2010}
{Sheth}, K., {Regan}, M., {Hinz}, J.~L., {et~al.} 2010, \pasp, 122, 1397

\bibitem[{{Tang} {et~al.}(2008){Tang}, {Kuo}, {Lim}, \& {Ho}}]{Tang_2008}
{Tang}, Y.-W., {Kuo}, C.-Y., {Lim}, J., \& {Ho}, P. T.~P. 2008, \apj, 679, 1094

\bibitem[{{Tendulkar} {et~al.}(2021){Tendulkar}, {Gil de Paz}, {Kirichenko},
  {Hessels}, {Bhardwaj}, {{\'A}vila}, {Bassa}, {Chawla}, {Fonseca}, {Kaspi},
  {Keimpema}, {Kirsten}, {Lazio}, {Marcote}, {Masui}, {Nimmo}, {Paragi},
  {Rahman}, {Pay{\'a}}, {Scholz}, \& {Stairs}}]{Tendulkar_2021}
{Tendulkar}, S.~P., {Gil de Paz}, A., {Kirichenko}, A.~Y., {et~al.} 2021,
  \apjl, 908, L12

\bibitem[{{Toomre} \& {Toomre}(1972)}]{Toomre_1972}
{Toomre}, A. \& {Toomre}, J. 1972, \apj, 178, 623

\bibitem[{{Vazdekis} {et~al.}(2010){Vazdekis}, {S{\'a}nchez-Bl{\'a}zquez},
  {Falc{\'o}n-Barroso}, {Cenarro}, {Beasley}, {Cardiel}, {Gorgas}, \&
  {Peletier}}]{Vazdekis_2010}
{Vazdekis}, A., {S{\'a}nchez-Bl{\'a}zquez}, P., {Falc{\'o}n-Barroso}, J.,
  {et~al.} 2010, \mnras, 404, 1639

\bibitem[{{Weinzirl} {et~al.}(2009){Weinzirl}, {Jogee}, {Khochfar}, {Burkert},
  \& {Kormendy}}]{Weinzirl_2009}
{Weinzirl}, T., {Jogee}, S., {Khochfar}, S., {Burkert}, A., \& {Kormendy}, J.
  2009, \apj, 696, 411

\bibitem[{{Wisnioski} {et~al.}(2015){Wisnioski}, {F{\"o}rster Schreiber},
  {Wuyts}, {Wuyts}, {Bandara}, {Wilman}, {Genzel}, {Bender}, {Davies},
  {Fossati}, {Lang}, {Mendel}, {Beifiori}, {Brammer}, {Chan}, {Fabricius},
  {Fudamoto}, {Kulkarni}, {Kurk}, {Lutz}, {Nelson}, {Momcheva}, {Rosario},
  {Saglia}, {Seitz}, {Tacconi}, \& {van Dokkum}}]{Wisnioski_2015}
{Wisnioski}, E., {F{\"o}rster Schreiber}, N.~M., {Wuyts}, S., {et~al.} 2015,
  \apj, 799, 209

\bibitem[{{Wright}(2006)}]{Wright_2006}
{Wright}, E.~L. 2006, \pasp, 118, 1711

\end{thebibliography}

%-------------------------------------------------------------------
%\clearpage
\onecolumn
\begin{appendix}
%\onecolumn
\section{Additional material}
\label{appendix:additional_material}
\begin{figure}[h!]
        \centering
        \includegraphics[trim={7,1cm 0mm 8cm 1cm},clip, width=0.21\linewidth]{Figuras/galaxias_fov_megara/fig1_IC1683.png}
        \includegraphics[trim={7,1cm 0mm 8cm 1cm},clip, width=0.21\linewidth]{Figuras/galaxias_fov_megara/fig1_NGC0023.png}
        \includegraphics[trim={7,1cm 0mm 8cm 1cm},clip, width=0.21\linewidth]{Figuras/galaxias_fov_megara/fig1_NGC0600.png}
        \includegraphics[trim={7,1cm 0mm 8cm 1cm},clip, width=0.21\linewidth]{Figuras/galaxias_fov_megara/fig1_NGC0716.png}
        \includegraphics[trim={7,1cm 0mm 8cm 1cm},clip, width=0.21\linewidth]{Figuras/galaxias_fov_megara/fig1_NGC0718.png}
        \includegraphics[trim={7,1cm 0mm 8cm 1cm},clip, width=0.21\linewidth]{Figuras/galaxias_fov_megara/fig1_NGC1042.png}
        \includegraphics[trim={7,1cm 0mm 8cm 1cm},clip, width=0.21\linewidth]{Figuras/galaxias_fov_megara/fig1_NGC1087.png}
        \includegraphics[trim={7,1cm 0mm 8cm 1cm},clip, width=0.21\linewidth]{Figuras/galaxias_fov_megara/fig1_NGC2500.png}
        \includegraphics[trim={7,1cm 0mm 8cm 1cm},clip, width=0.21\linewidth]{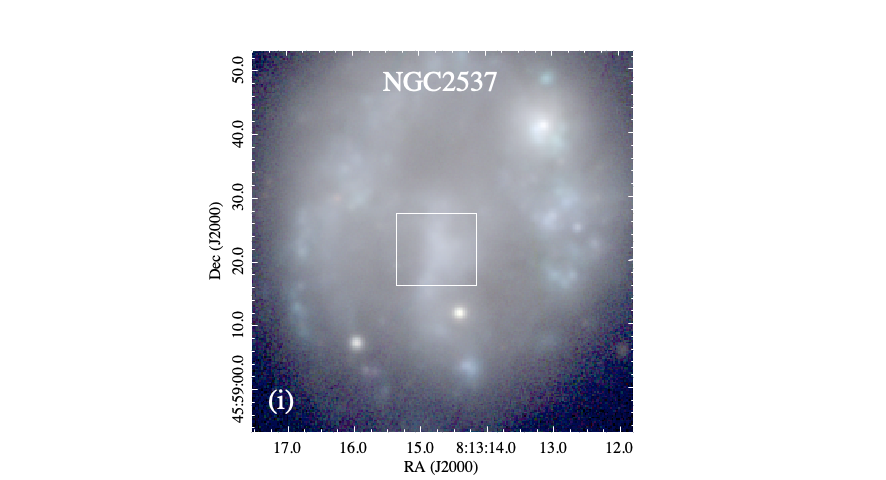}
        \includegraphics[trim={7,1cm 0mm 8cm 1cm},clip, width=0.21\linewidth]{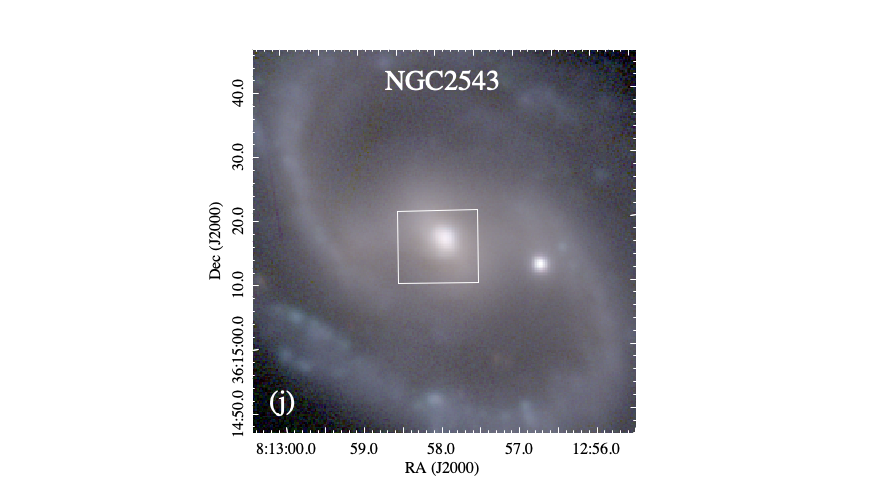}
        \includegraphics[trim={7,1cm 0mm 8cm 1cm},clip, width=0.21\linewidth]{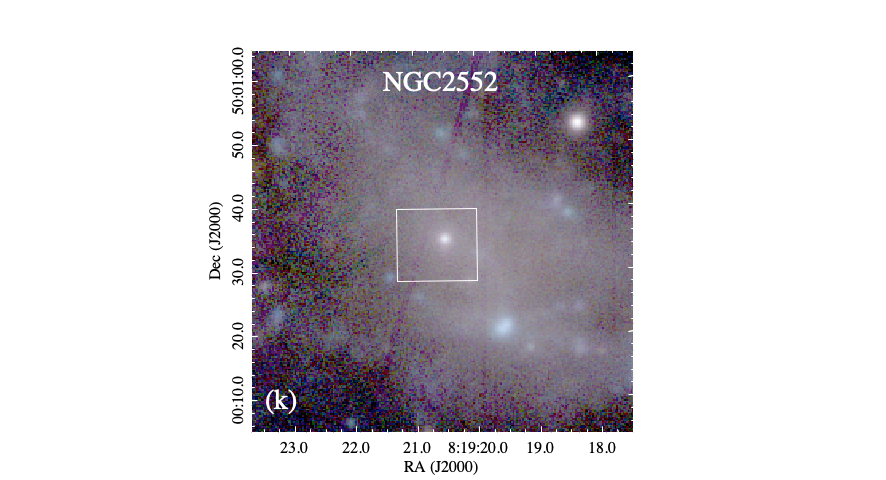}
        \includegraphics[trim={7,1cm 0mm 8cm 1cm},clip, width=0.21\linewidth]{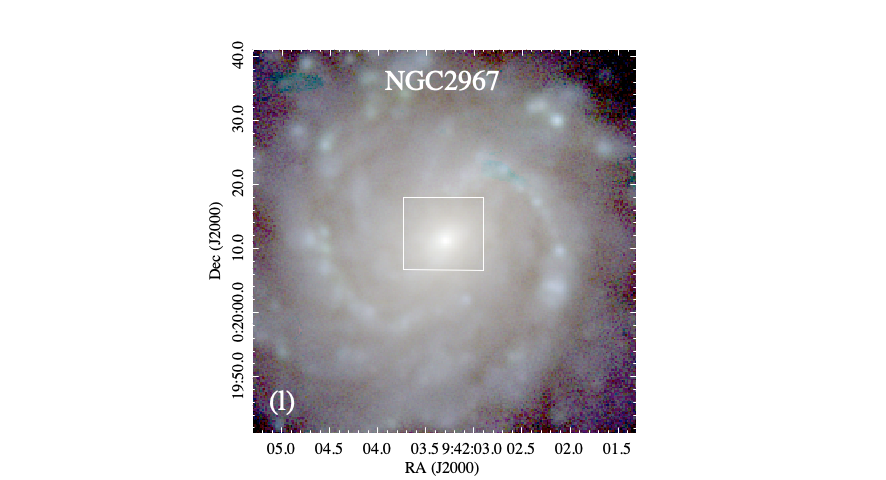}
        \includegraphics[trim={7,1cm 0mm 8cm 1cm},clip, width=0.21\linewidth]{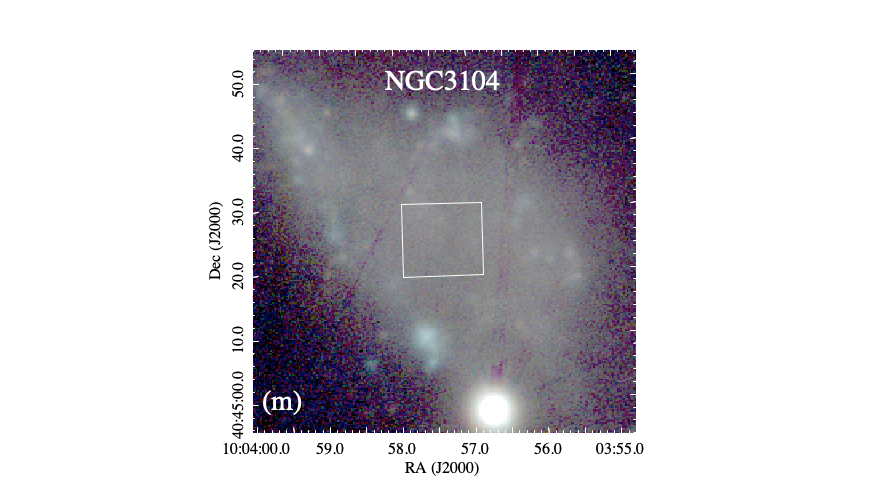}
        \includegraphics[trim={7,1cm 0mm 8cm 1cm},clip, width=0.21\linewidth]{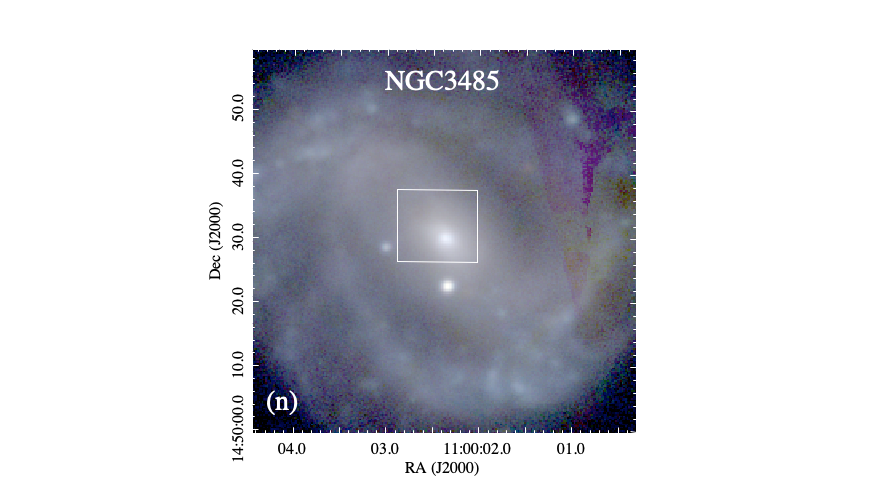}
        \includegraphics[trim={7,1cm 0mm 8cm 1cm},clip, width=0.21\linewidth]{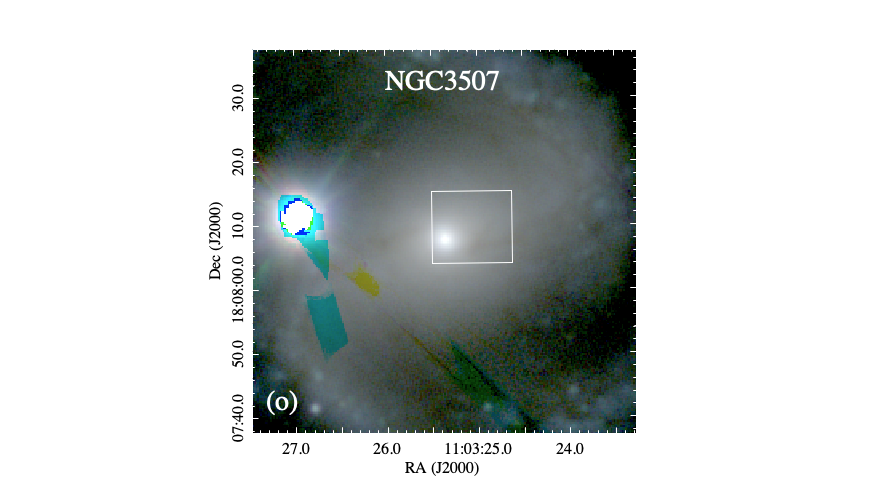}
        \includegraphics[trim={7,1cm 0mm 8cm 1cm},clip, width=0.21\linewidth]{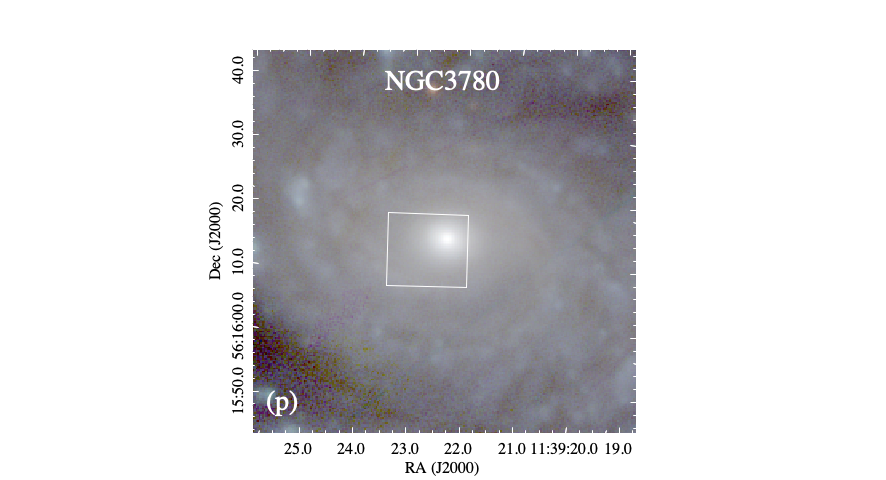}
        \includegraphics[trim={7,1cm 0mm 8cm 1cm},clip, width=0.21\linewidth]{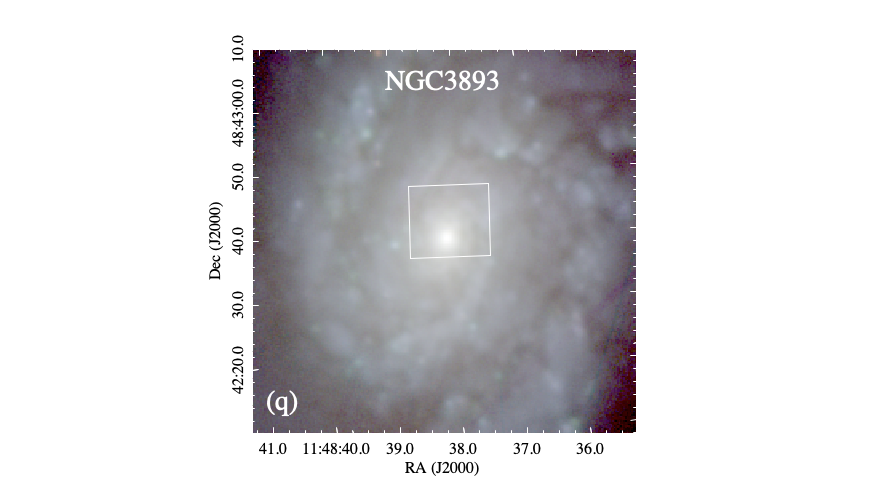}
        \includegraphics[trim={7,1cm 0mm 8cm 1cm},clip, width=0.21\linewidth]{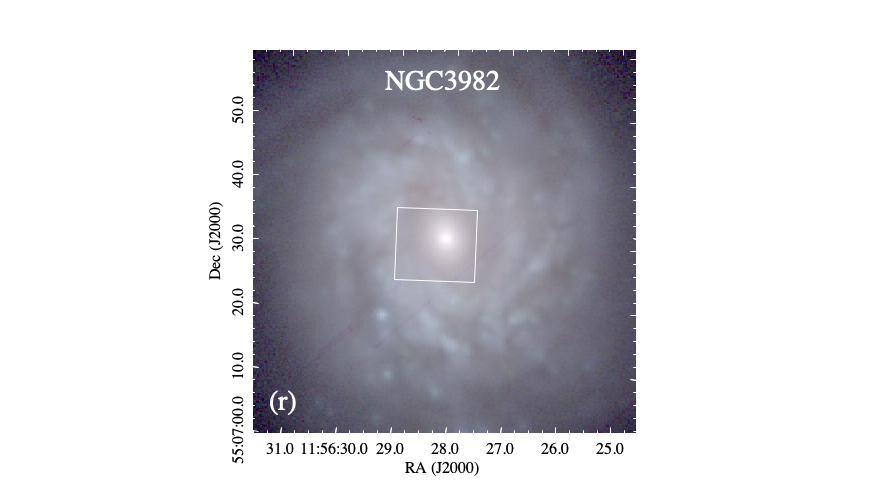}
        \includegraphics[trim={7,1cm 0mm 8cm 1cm},clip, width=0.21\linewidth]{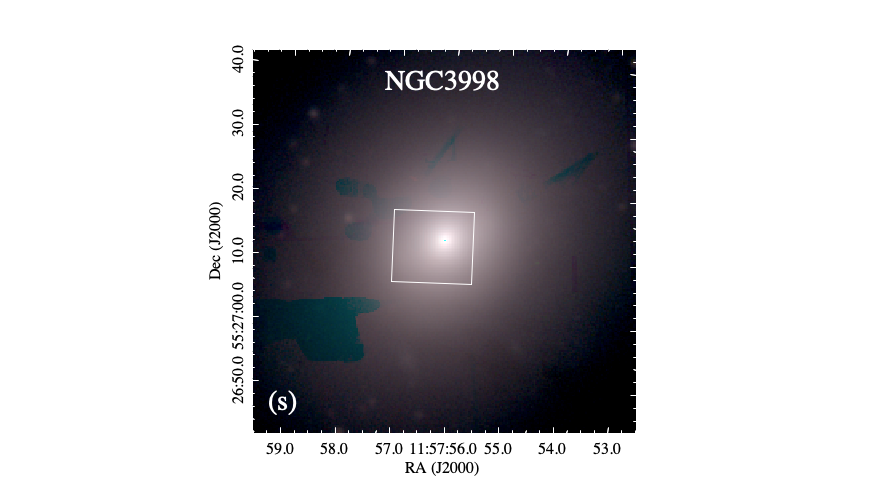}
        \includegraphics[trim={7,1cm 0mm 8cm 1cm},clip, width=0.21\linewidth]{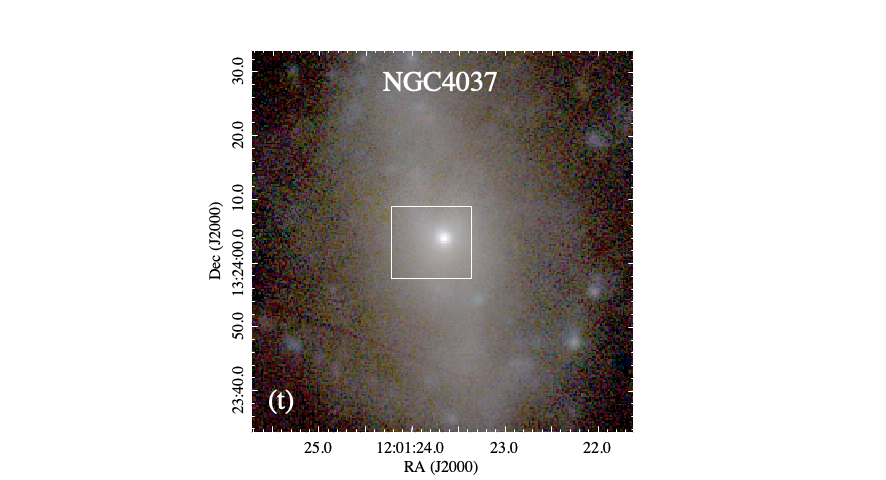}

        \caption{MEGADES sample: RGB images from PanSTARRS observations (g, r, and i filters). The white box in the centre of each panel indicates the MEGARA IFU FoV.}
        \label{fig:sample1}
\end{figure}

\setcounter{figure}{0}
%\newpage
\begin{figure}[ht!]
        \centering
        \includegraphics[trim={7,1cm 0mm 8cm 1cm},clip, width=0.21\linewidth]{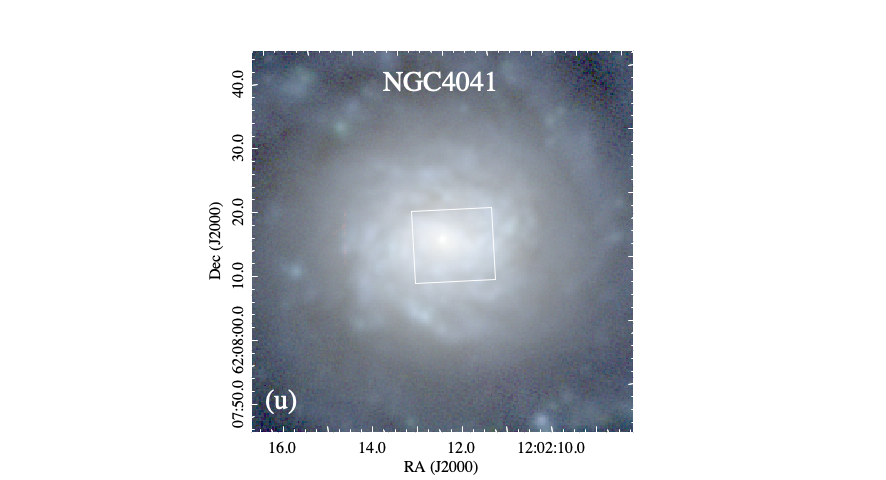}
        \includegraphics[trim={7,1cm 0mm 8cm 1cm},clip, width=0.21\linewidth]{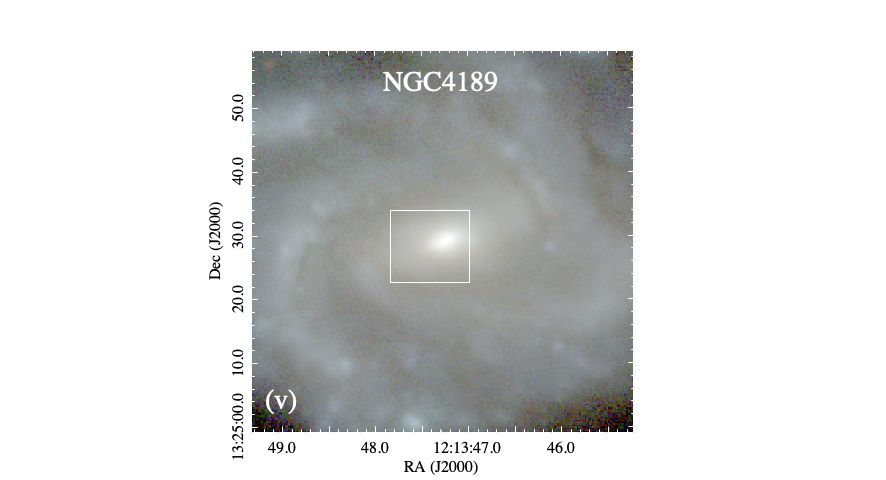}
        \includegraphics[trim={7,1cm 0mm 8cm 1cm},clip, width=0.21\linewidth]{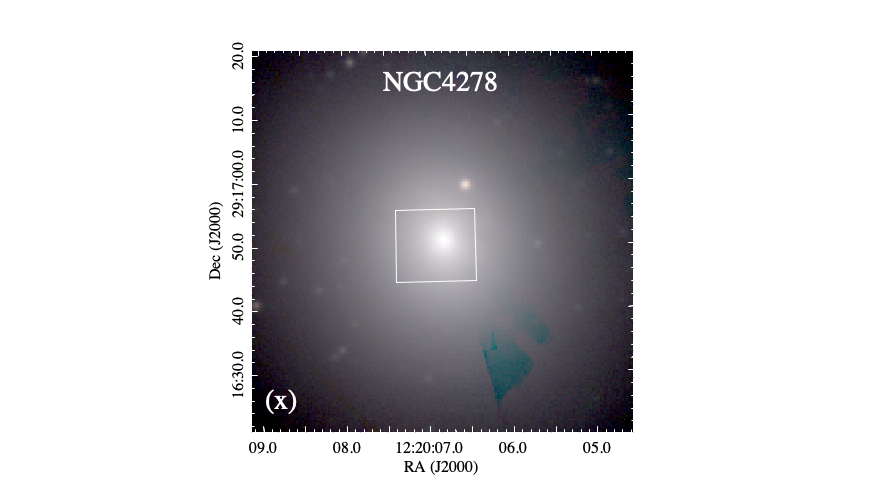}
        \includegraphics[trim={7,1cm 0mm 8cm 1cm},clip, width=0.21\linewidth]{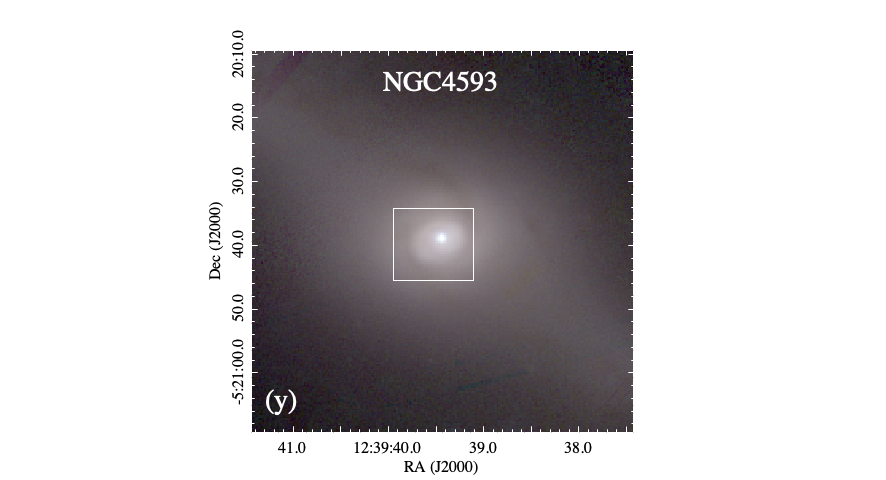}
        \includegraphics[trim={7,1cm 0mm 8cm 1cm},clip, width=0.21\linewidth]{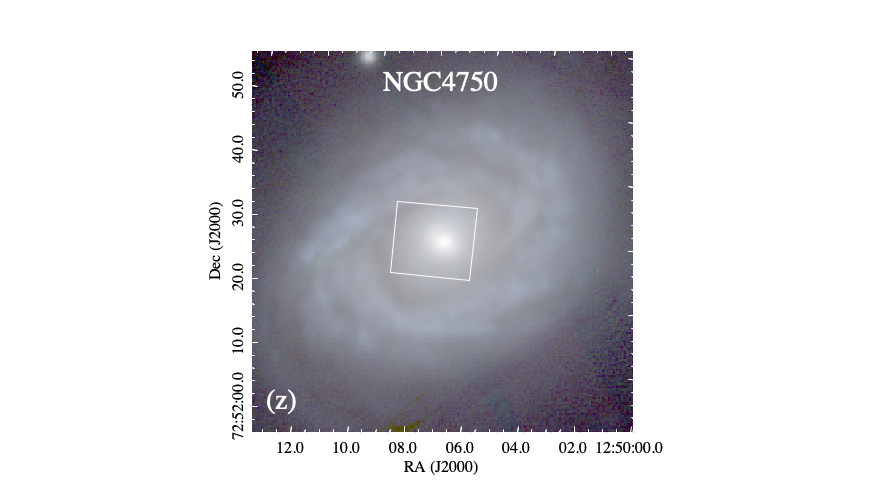}
        \includegraphics[trim={7,1cm 0mm 8cm 1cm},clip, width=0.21\linewidth]{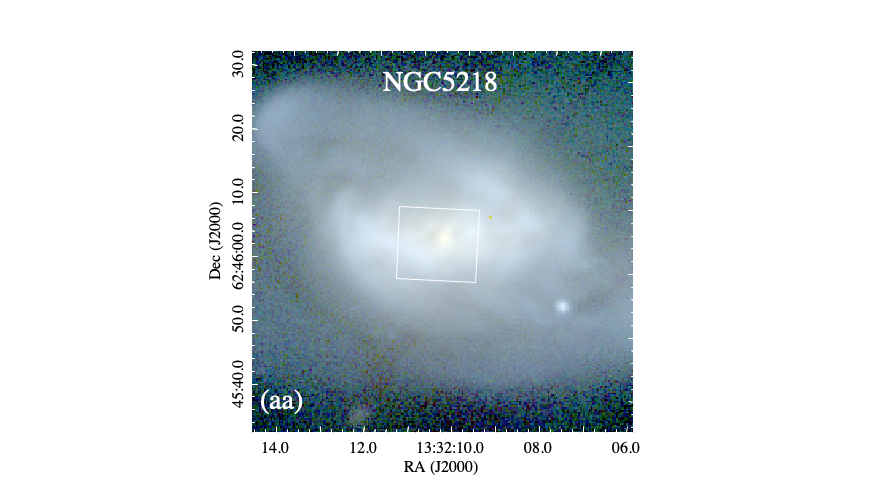}
        \includegraphics[trim={7,1cm 0mm 8cm 1cm},clip, width=0.21\linewidth]{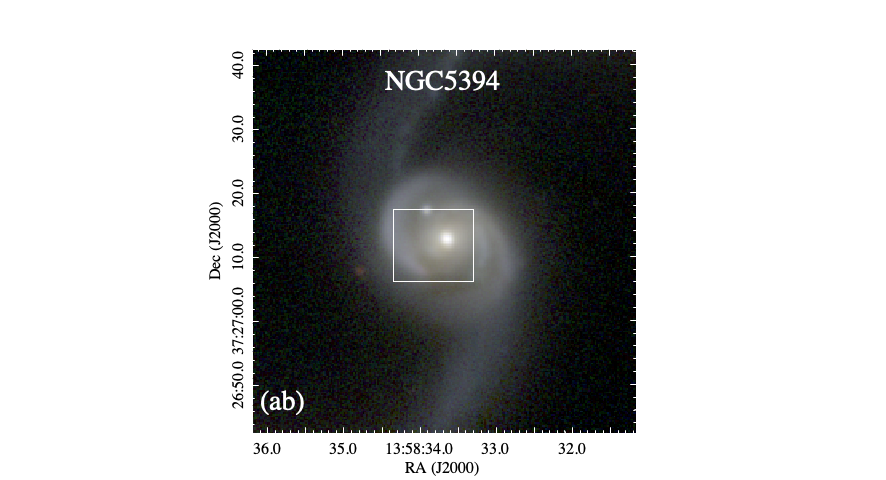}
        \includegraphics[trim={7,1cm 0mm 8cm 1cm},clip, width=0.21\linewidth]{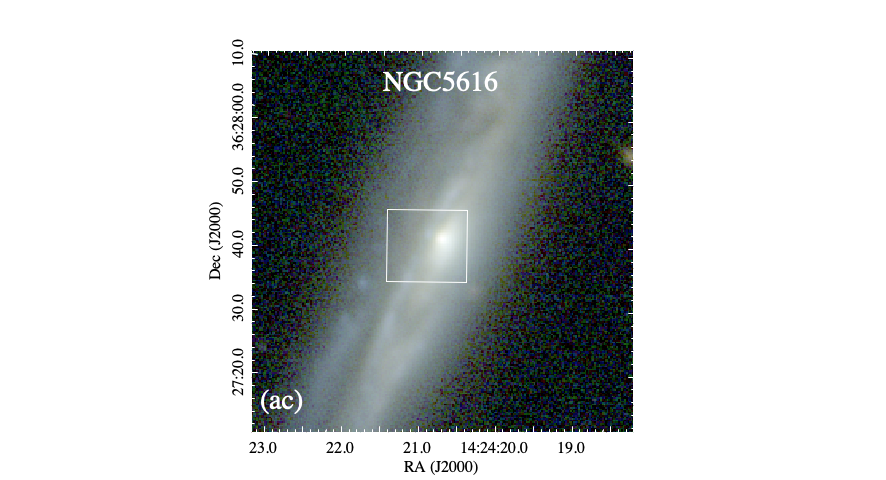}
        \includegraphics[trim={7,1cm 0mm 8cm 1cm},clip, width=0.21\linewidth]{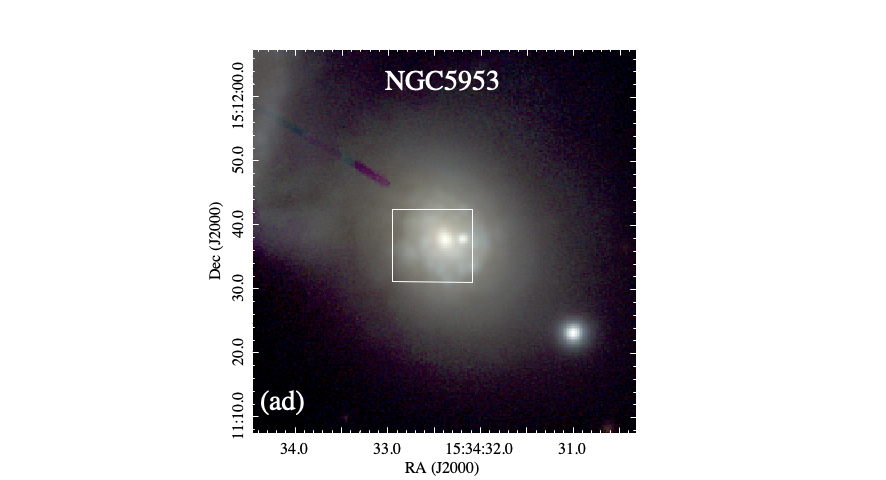}
        \includegraphics[trim={7,1cm 0mm 8cm 1cm},clip, width=0.21\linewidth]{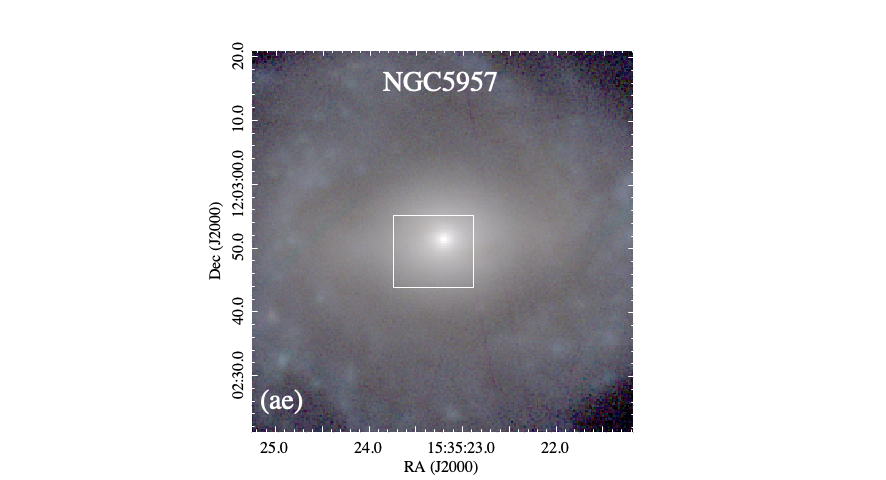}
        \includegraphics[trim={7,1cm 0mm 8cm 1cm},clip, width=0.21\linewidth]{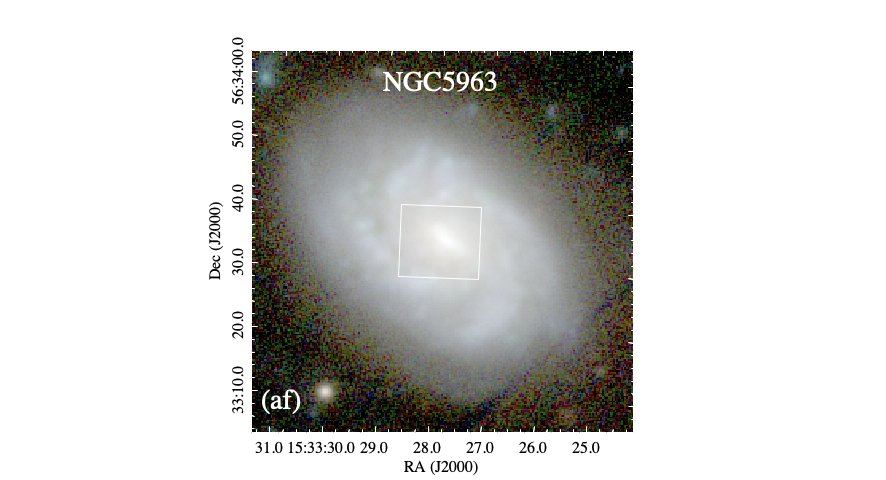}
        \includegraphics[trim={7,1cm 0mm 8cm 1cm},clip, width=0.21\linewidth]{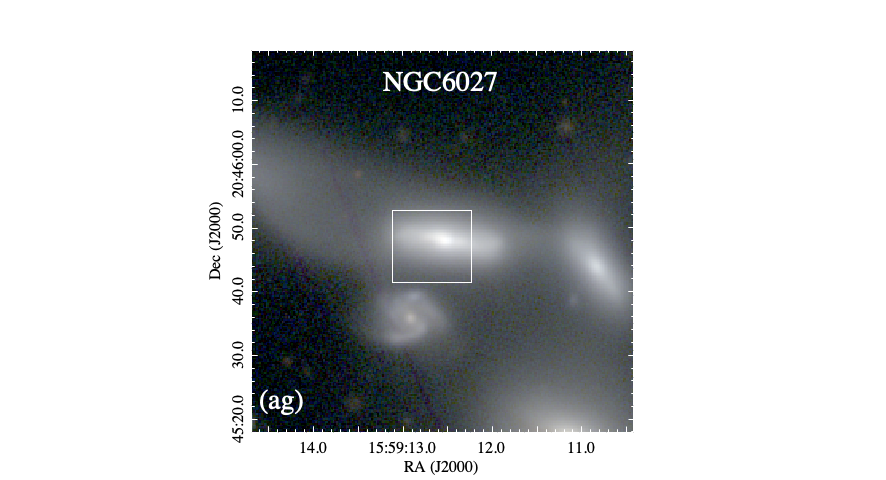}
        \includegraphics[trim={7,1cm 0mm 8cm 1cm},clip, width=0.21\linewidth]{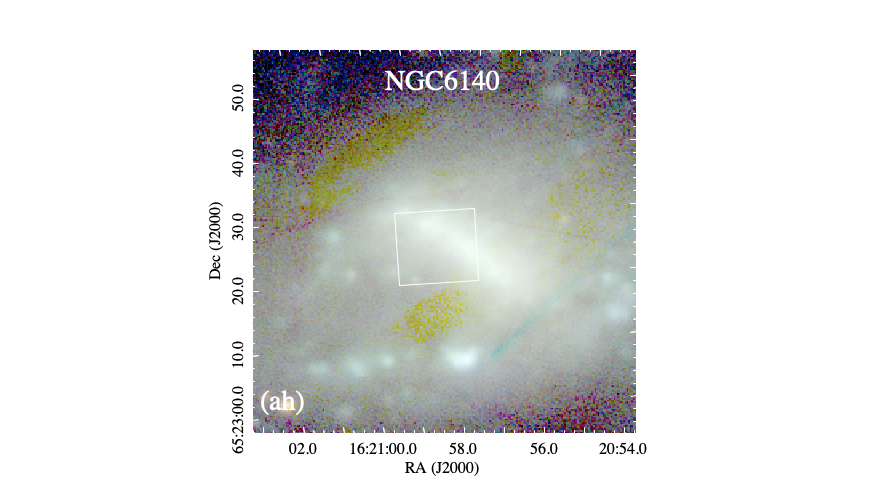}
        \includegraphics[trim={7,1cm 0mm 8cm 1cm},clip, width=0.21\linewidth]{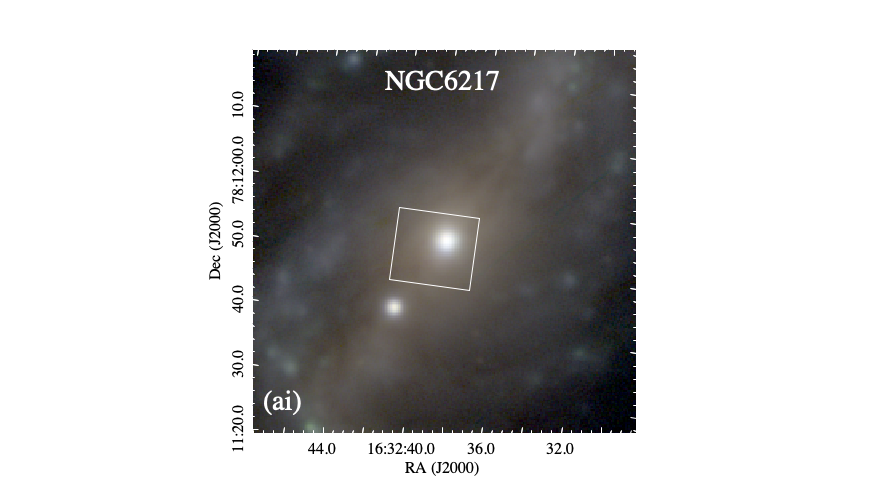}
        \includegraphics[trim={7,1cm 0mm 8cm 1cm},clip, width=0.21\linewidth]{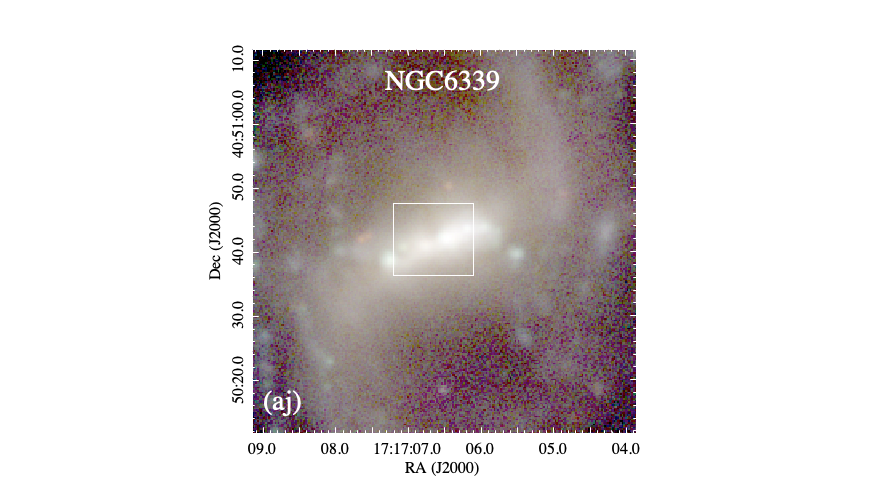}
        \includegraphics[trim={7,1cm 0mm 8cm 1cm},clip, width=0.21\linewidth]{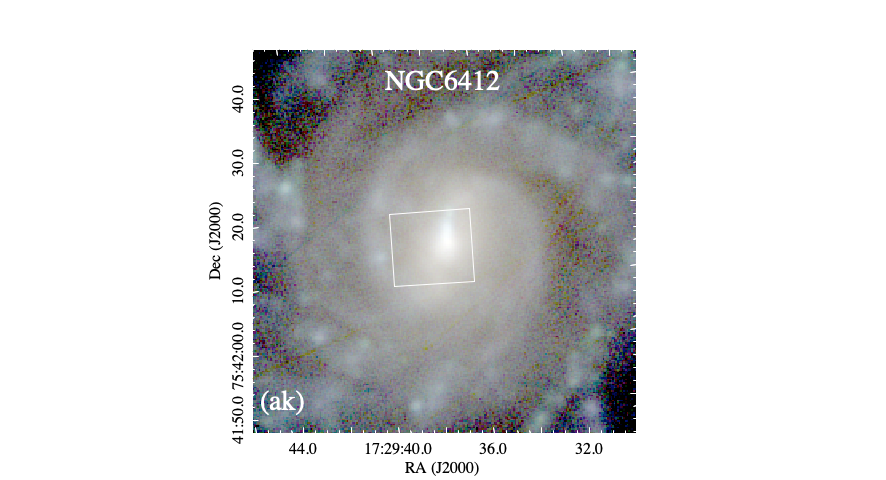}
        \includegraphics[trim={7,1cm 0mm 8cm 1cm},clip, width=0.21\linewidth]{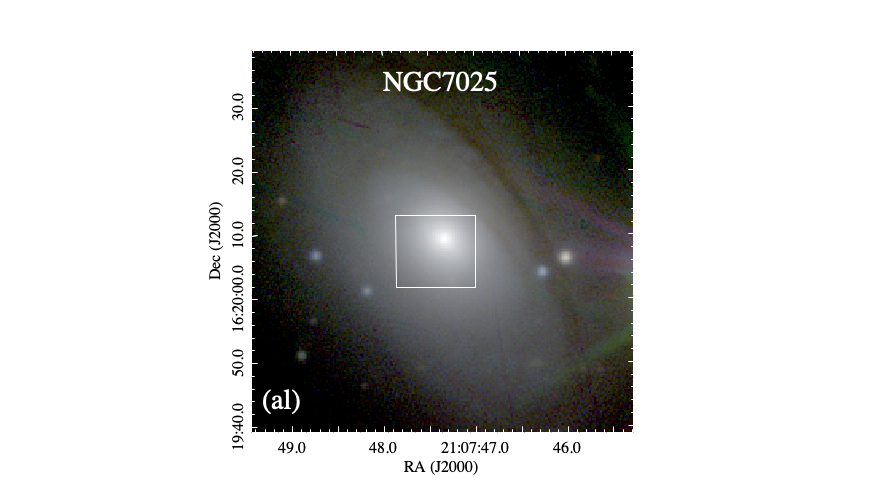}
        \includegraphics[trim={7,1cm 0mm 8cm 1cm},clip, width=0.21\linewidth]{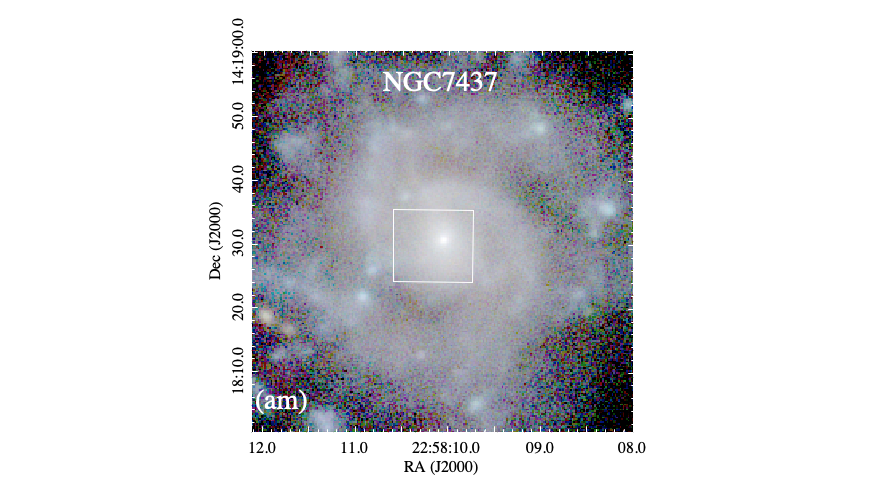}
        \includegraphics[trim={7,1cm 0mm 8cm 1cm},clip, width=0.21\linewidth]{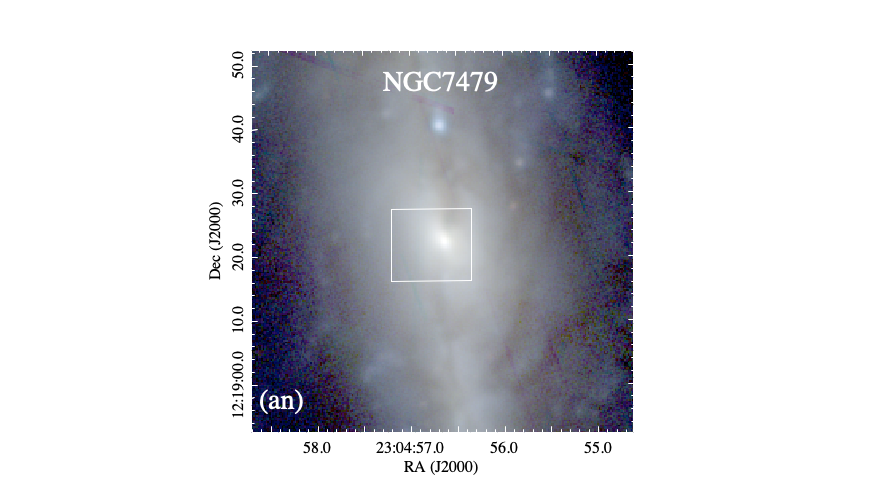}
        \includegraphics[trim={7,1cm 0mm 8cm 1cm},clip, width=0.21\linewidth]{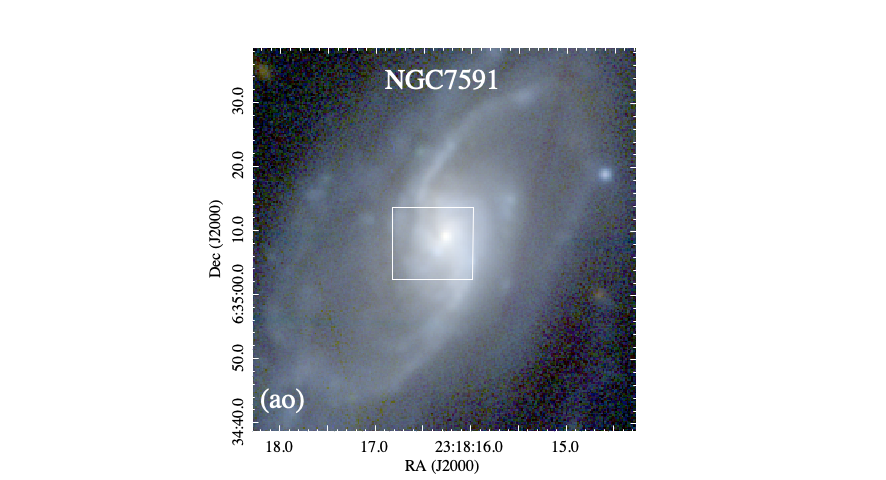}
        \includegraphics[trim={7,1cm 0mm 8cm 1cm},clip, width=0.21\linewidth]{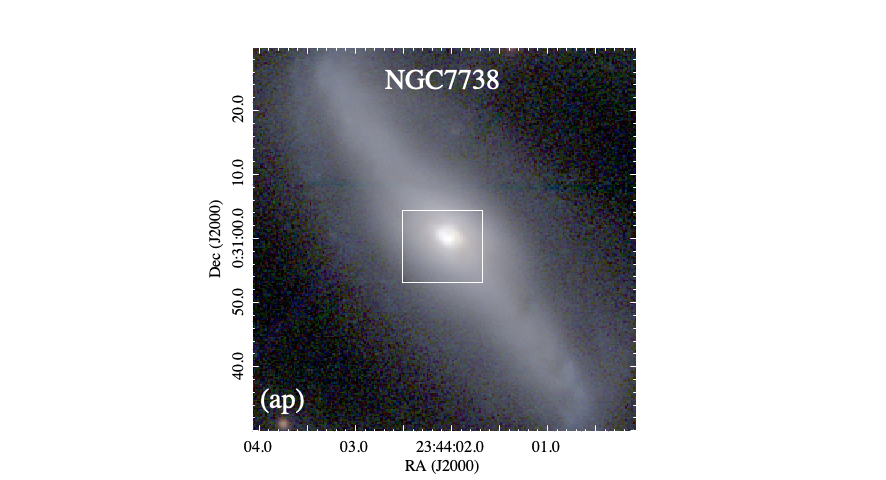}
        \includegraphics[trim={7,1cm 0mm 8cm 1cm},clip, width=0.21\linewidth]{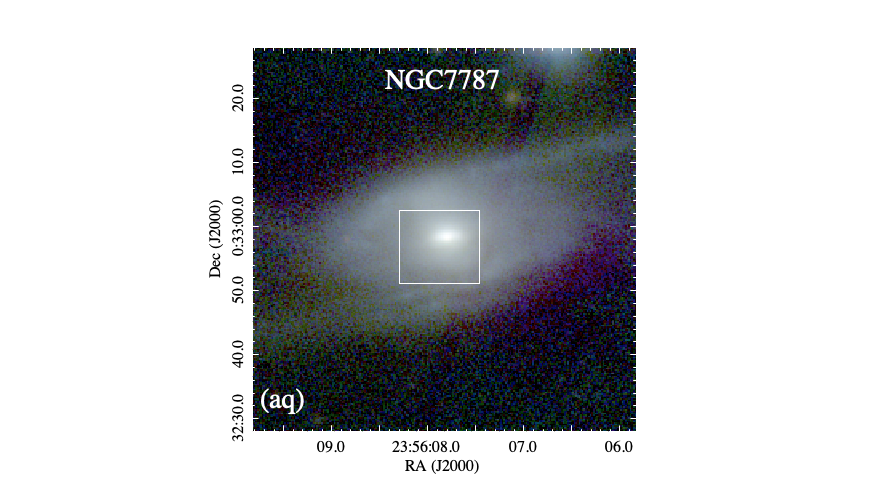}
        \includegraphics[trim={7,1cm 0mm 8cm 1cm},clip, width=0.21\linewidth]{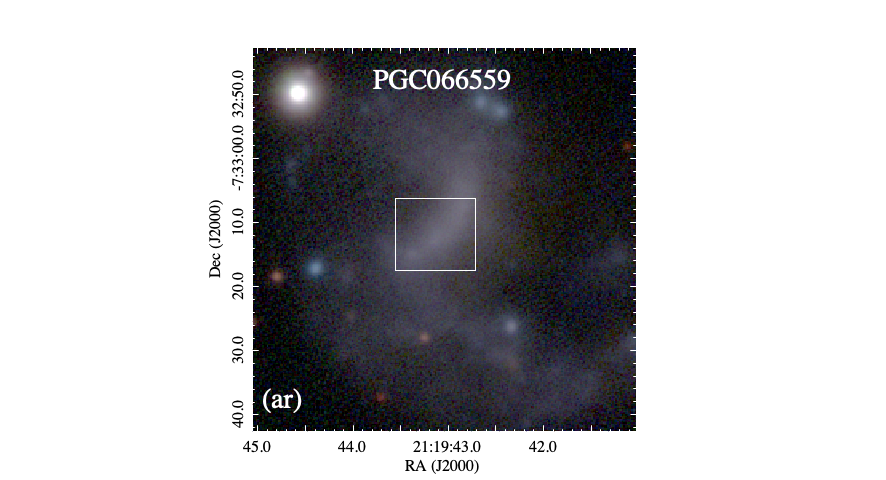}
        \hspace{3.9cm}
        \caption{(cont.) MEGADES sample: RGB images from PanSTARRS observations (g, r, and i filters). The white box in the centre of each panel indicates the MEGARA IFU FoV.}
\end{figure}

%\newpage
\begin{table}[h]
        \caption{Observing log.}              % title of Table
        \label{table:log}      % is used to refer this table in the text
        \centering                                      % used for centering table
        \resizebox{\textwidth}{!}{\begin{tabular}{l l c r l c r l c r}          % centered columns (4 columns)
                        \hline\hline                     % inserts two horizontal lines 
                        \noalign{\smallskip}
                        & \multicolumn{3}{c}{LR-B} & \multicolumn{3}{c}{LR-V} & \multicolumn{3}{c}{LR-R} \\
                        \hline 
                        \noalign{\smallskip}
                        Galaxy & Date & secz* & t$_\mathrm{exp}$ & Date & secz* & t$_\mathrm{exp}$ & Date & secz* & t$_\mathrm{exp}$ \\    % table heading
                        \hline                                   % inserts single horizontal line
                        \noalign{\smallskip}
                        IC~1683 & -- & -- & -- & 19 Sep. 2018 & 1.2 & 5$\times$720 & 19 Sep. 2018 & 1.01 & 5$\times$720 \\
                        NGC~0023 & -- & -- & -- & 22 Sep. 2019 & 1.03 & 3$\times$1200 & 22 Sep. 2019 & 1.14 & 3$\times$1200 \\
                        NGC~0600 & 27 Sep. 2019 & 1.24 & 3$\times$1200 & 27 Sep. 2019 & 1.34 & 3$\times$1200 & 27 Sep. 2019 & 1.48 & 3$\times$1200 \\
                        NGC~0716 & -- & -- & -- & 31 Dec. 2018 & 1.04 & 5$\times$720 & 31 Dec. 2018 & 1.06 & 5$\times$720 \\
                        NGC~0718 & 12 Oct. 2018$^{\dag}$ & 1.45 & 3$\times$1200 & 02 Oct. 2018$^{\dag}$ & 1.1 & 3$\times$1200 & 02 Oct. 2018$^{\dag}$ & 1.13 & 3$\times$1200 \\
                        NGC~1042 & 24 Oct. 2019 & 1.34 & 3$\times$1200 & 24 Oct. 2019 & 1.26 & 3$\times$1200 & 24 Oct. 2019 & 1.29 & 3$\times$1200 \\
                        NGC~1087 & 06 Nov. 2018$^{\dag}$ & 1.17 & 3$\times$1200 & 30 Nov. 2018$^{\dag}$ & 1.21 & 3$\times$1200 & 06 Nov. 2018 & 1.36 & 3$\times$1200 \\
                        NGC~2500 & 02 Dec. 2018$^{\dag}$ & 1.13 & 3$\times$1200 & 02 Dec. 2018$^{\dag}$ & 1.09 & 3$\times$600 & 02 Dec. 2018$^{\dag}$ & 1.08 & 3$\times$1200 \\
                        NGC~2537 & 08 May 2019 & 1.28 & 3$\times$1200 & 04 Dec. 2019 & 1.31 & 3$\times$1200 & 03 Dec. 2018$^{\dag}$ & 1.05 & 3$\times$1200 \\
                        NGC~2543 & -- & -- & -- & 30 Nov. 2018 & 1.06 & 5$\times$720 & 30 Nov. 2018 & 1.01 & 5$\times$720 \\
                        NGC~2552 & 01 Dec. 2019 & 1.15 & 3$\times$1200 & 01 Dec. 2019 & 1.08 & 3$\times$1200 & 01 Dec. 2019 & 1.07 & 3$\times$1200 \\
                        NGC~2967 & 30 Jan. 2019$^{\dag}$ & 1.26 & 3$\times$1200 & 30 Jan. 2019$^{\dag}$ & 1.15 & 3$\times$1200 & 30 Jan. 2019$^{\dag}$ & 1.15 & 3$\times$1200 \\
                        NGC~3104 & 02 Dec. 2019 & 1.11 & 3$\times$1200 & 02 Dec. 2019 & 1.26 & 3$\times$1200 & 02 Dec. 2019 & 1.33 & 3$\times$1200 \\
                        NGC~3485 & 02 Jan. 2019$^{\dag}$ & 1.12 & 3$\times$1200 & 02 Jan. 2019$^{\dag}$ & 1.12 & 3$\times$1200 & 11 Jan. 2019$^{\dag}$ & 1.12 & 3$\times$1200 \\
                        NGC~3507 & 03 Jan. 2019$^{\dag}$ & 1.02 & 3$\times$1200 & 03 Jan. 2019$^{\dag}$ & 1.04 & 3$\times$1200 & 03 Jan. 2019$^{\dag}$ & 1.1 & 3$\times$1200 \\
                        NGC~3780 & 31 Dec. 2019 & 1.29 & 3$\times$1200 & 31 Dec. 2019 & 1.18 & 3$\times$1200 & 01 Feb. 2019$^{\dag}$ & 1.24 & 3$\times$1200 \\
                        NGC~3893 & 12 Jan. 2019$^{\dag}$ & 1.06 & 3$\times$1200 & 11 Jan. 2019$^{\dag}$ & 1.06 & 3$\times$1200 & 11 Jan. 2019$^{\dag}$ & 1.08 & 3$\times$1200 \\
                        NGC~3982 & 20 Mar. 2021 & 1.14 & 3$\times$1200 & 19 Mar. 2021 & 1.21 & 3$\times$1200 & 19 Mar. 2021 & 1.25 & 3$\times$1200 \\
                        NGC~3998 & 15 Mar. 2021$^{\dag}$ & 1.24 & 3$\times$1200 & 15 Mar. 2021 & 1.4 & 3$\times$1200 & -- & -- & -- \\
                        NGC~4037 & 14 Mar. 2021$^{\dag}$ & 1.05 & 3$\times$1200 & 14 Mar. 2021 & 1.13 & 3$\times$1200 & 15 Mar. 2021 & 1.07 & 3$\times$1200 \\
                        NGC~4041 & 14 Jan. 2019$^{\dag}$ & 1.39 & 3$\times$1200 & 14 Jan. 2019$^{\dag}$ & 1.26 & 3$\times$1200 & 01 Feb. 2019$^{\dag}$ & 1.22 & 3$\times$1200 \\
                        NGC~4189 & 07 May 2019 & 1.07 & 3$\times$1200 & 08 May 2019 & 1.18 & 3$\times$1200 & 08 May 2019 & 1.43 & 3$\times$1200 \\
                        NGC~4278 & 08 May 2019 & 1.01 & 3$\times$1200 & 09 May 2019 & 1.07 & 3$\times$1200 & -- & -- & -- \\
                        NGC~4593 & 08 Feb. 2019 & 1.22 & 3$\times$1200 & 06 May 2019 & 1.32 & 3$\times$1200 & 08 Feb. 2019 & 1.25 & 3$\times$1200 \\
                        NGC~4750 & 09 Feb. 2019 & 1.4 & 3$\times$1200 & 07 May 2019 & 1.4 & 3$\times$1200 & 07 May 2019 & 1.55 & 3$\times$1200 \\
                        NGC~5218 & -- & -- & -- & 19 Mar. 2021 & 1.21 & 3$\times$1200 & 19 Mar. 2021 & 1.24 & 3$\times$1200 \\
                        NGC~5394 & -- & -- & -- & 09 May 2019 & 1.14 & 3$\times$1200 & 09 May 2019 & 1.44 & 3$\times$1200 \\
                        NGC~5616 & -- & -- & -- & 08 May 2019 & 1.49 & 3$\times$1200 & 02 Aug. 2019 & 1.45 & 3$\times$1200 \\
                        NGC~5953 & -- & -- & -- & 19 Mar. 2021 & 1.03 & 3$\times$1200 & 19 Mar. 2021 & 1.04 & 3$\times$1200 \\
                        NGC~5957 & 25 May 2019 & 1.07 & 3$\times$1200 & 29 May 2019 & 1.16 & 3$\times$1200 & 29 May 2019 & 1.67 & 3$\times$1200 \\
                        NGC~5963 & 04 May 2021 & 1.17 & 3$\times$1200 & 10 Jun. 2019 & 1.29 & 3$\times$1200 & 10 Jun. 2019 & 1.21 & 3$\times$1200 \\
                        NGC~6027 & -- & -- & -- & 06 May 2019 & 1.07 & 3$\times$1200 & 06 May 2019 & 1.21 & 3$\times$1200 \\
                        NGC~6140 & 27 May 2019$^{\dag}$ & 1.27 & 3$\times$1200 & 29 May 2019 & 1.26 & 3$\times$1200 & 30 May 2019 & 1.27 & 3$\times$1200\\
                        NGC~6217 & 20 Aug. 2018 & 1.6 & 3$\times$1200 & 20 Aug. 2018 & 1.7 & 3$\times$1200 & 21 Aug. 2018$^{\dag}$ & 1.84 & 3$\times$1200 \\
                        NGC~6339 & 10 Jul. 2019 & 1.02 & 3$\times$1200 & 11 Jul. 2019 & 1.07 & 3$\times$1200 & 11 Jul. 2019 & 1.31 & 3$\times$1200 \\
                        NGC~6412 & 30 May 2019 & 1.48 & 3$\times$1200 & 30 May 2019 & 1.53 & 3$\times$1200 & 11 Jun. 2019 & 1.6 & 3$\times$1200 \\
                        NGC~7025 & 01 Aug. 2017 & 1.09 & 3$\times$900 & 01 Aug. 2017 & 1.23 & 3$\times$900 & 01 Aug. 2017 & 1.37 & 3$\times$600 \\
                        NGC~7437 & 28 Jul. 2019 & 1.2 & 3$\times$1200 & 28 Jul. 2019 & 1.03 & 3$\times$1200 & 27 Dec. 2019 & 1.25 & 3$\times$1200 \\
                        NGC~7479 & 30 Jul. 2019 & 1.24 & 3$\times$1200 & 30 Jul. 2019 & 1.09 & 3$\times$1200 & 30 Jul. 2019 & 1.08 & 3$\times$1200 \\
                        NGC~7591 & -- & -- & -- & 29 Sep. 2018 & 1.48 & 5$\times$720 & 29 Sep. 2018 & 1.2 & 5$\times$720 \\
                        NGC~7738 & -- & -- & -- & 23 Sep. 2019 & 1.18 & 3$\times$1200 & 23 Oct. 2019 & 1.13 & 3$\times$1000 \\
                        NGC~7787 & -- & -- & -- & 31 Jul. 2019 & 1.2 & 3$\times$1200 & 31 Jul. 2019 & 1.13 & 3$\times$1200 \\
                        PGC~066559 & 11 Jul. 2019 & 1.25 & 3$\times$1200 & 11 Jul. 2019 & 1.26 & 3$\times$1200 & 01 Aug. 2019 & 1.24 & 3$\times$1200 \\
                        \hline                                             %inserts single line
        \end{tabular}}
        \parbox{185mm}{\footnotesize $^{\dag}$Observations affected by diffuse light. *Airmass measured at the beginning of the observation.}
\end{table}

\begin{table}[h]
        \caption{Observing nights.}              % title of Table
        \label{table:observing_nights}      % is used to refer this table in the text
        \centering                                      % used for centering table
        \begin{tabular}{c c c}          % centered columns (3 columns)
                \hline\hline                     % inserts two horizontal lines 
                \noalign{\smallskip}
                Date & Seeing ('') & Atm. Conditions \\    % table heading
                \hline                                   % inserts single horizontal line
                \noalign{\smallskip}
                01 Aug. 2017 & 0.7 & -- \\
                02 Aug. 2017 & 1.1  & -- \\
                20 Aug. 2018 & 0.6 - 0.7 & Clear \\
                21 Aug. 2018 & 0.8 & Clear \\
                19 Sep. 2018 & 0.9 & Clear \\
                29 Sep. 2018 & 0.8 & Clear \\
                02 Oct. 2018 & 0.7 & Clear \\
                12 Oct. 2018 & 0.9 & Clear \\
                06 Nov. 2018 & 1.0 & Clear \\
                30 Nov. 2018 & 0.8 & Clear \\
                02 Dec. 2018 & 1.0 - 1.1 & Clear \\
                03 Dec. 2018 & 0.8 & Clear \\
                31 Dec. 2018 & 0.8 - 1.0 & Clear \\
                02 Jan. 2019 & 0.9 & Clear \\
                03 Jan. 2019 & 0.8 & Clear \\
                11 Jan. 2019 & 0.9 & Clear \\
                12 Jan. 2019 & 0.9 & Clear \\
                14 Jan. 2019 & 1.1 - 1.2 & Clear \\
                30 Jan. 2019 & 1.2 & Clear \\
                01 Feb. 2019 & 1.2 & Clear \\
                08 Feb. 2019 & 0.7 & Photometric \\
                09 Feb. 2019 & 0.8 & Photometric \\
                06 May 2019 & 0.7 - 0.8 & Photometric \\
                07 May 2019 & 0.6 - 0.8 & Photometric \\
                08 May 2019 & 0.6 - 0.8 & Photometric  \\
                09 May 2019 & 0.5 - 0.6 & Photometric \\
                25 May 2019 & 0.8 & Clear \\
                27 May 2019 & 1.0 & Clear \\
                29 May 2019 & 1.0 - 1.1 & Clear \\
                30 May 2019 & 0.9 & Photometric \\
                10 Jun. 2019 & 1.1 - 1.2 & Clear \\
                11 Jun. 2019 & 1.0 & Clear \\
                10 Jul. 2019 & 1.3 & Clear \\
                11 Jul. 2019 & 1.0 - 1.2 & Clear \\
                28 Jul. 2019 & 1.0 & Clear \\
                30 Jul. 2019 & 1.0 & Photometric \\
                31 Jul. 2019 & 0.9 & Photometric \\
                01 Aug. 2019 & 1.0 & Photometric \\
                02 Aug. 2019 & 0.8 & Photometric \\
                22 Sep. 2019 & 0.6 & Clear \\
                23 Sep. 2019 & 1.1 & Clear \\
                27 Sep. 2019 & 0.9 - 1.0 & Photometric \\
                23 Oct. 2019 & 1.2 & Clear \\
                24 Oct. 2019 & 1.1 - 1.2 & Clear \\
                01 Dec. 2019 & 0.6 - 1.1 & Clear \\
                02 Dec. 2019 & 0.6 - 0.7 & Clear \\
                04 Dec. 2019 & 0.9 & Clear \\
                27 Dec. 2019 & 0.9 & Clear \\
                31 Dec. 2019 & 1.1 - 1.2 & Spectroscopic \\
                26 Jan. 2020 & 1.2 & Clear \\
                14 Mar. 2021 & 1.0 & Clear \\
                15 Mar. 2021 & 1.0 & Clear \\
                19 Mar. 2021 & 0.9 - 1.1 & Clear \\
                20 Mar. 2021 & 1.0 & Clear \\
                04 May 2021 & 1.1 & Clear \\
                \hline                                             %inserts single line
        \end{tabular}
\end{table}

%\vspace{5cm}
\clearpage
\section{Galaxy cards}

\begin{figure*}[h]
	\centering
	\includegraphics[clip, width=0.35\linewidth]{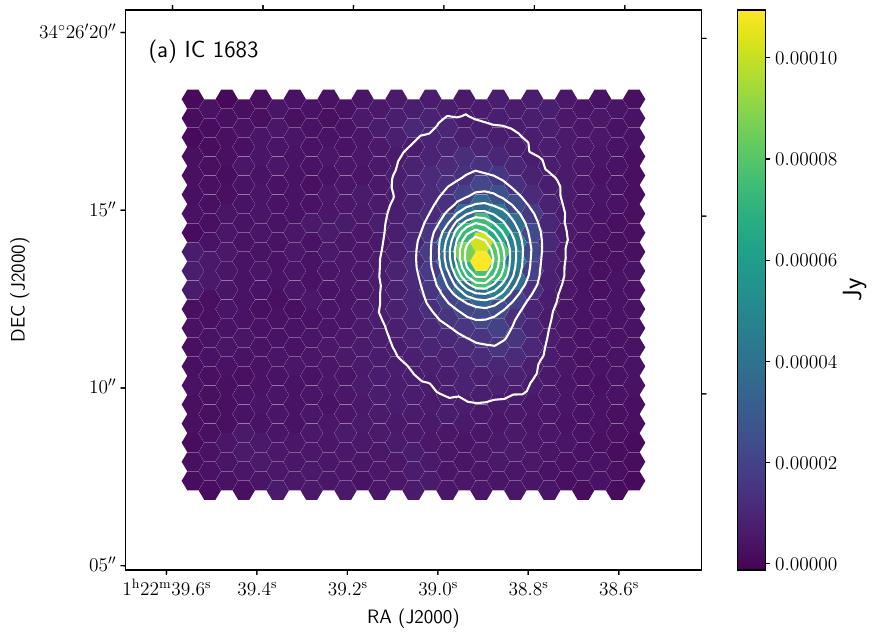}
	\includegraphics[clip, width=0.6\linewidth]{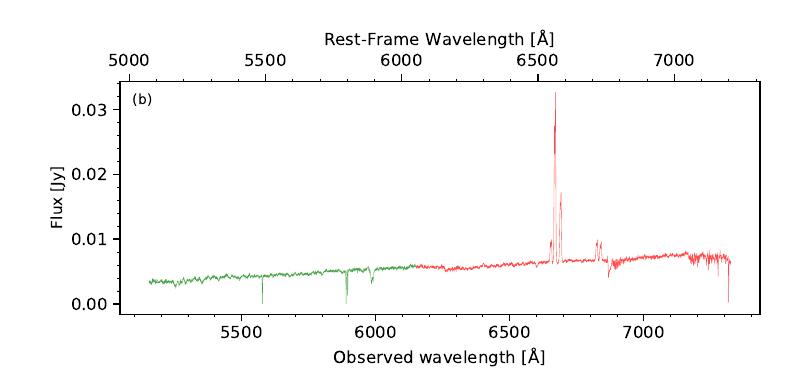}
	\includegraphics[clip, width=0.24\linewidth]{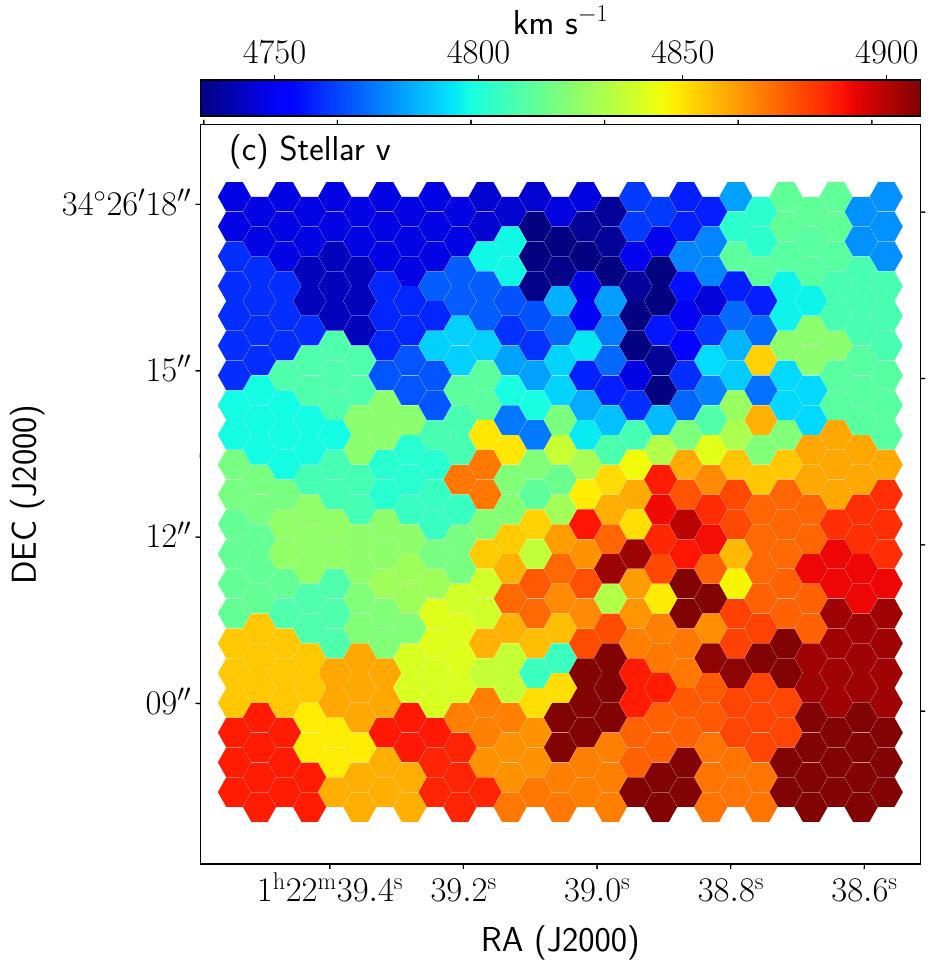}
	\includegraphics[clip, width=0.24\linewidth]{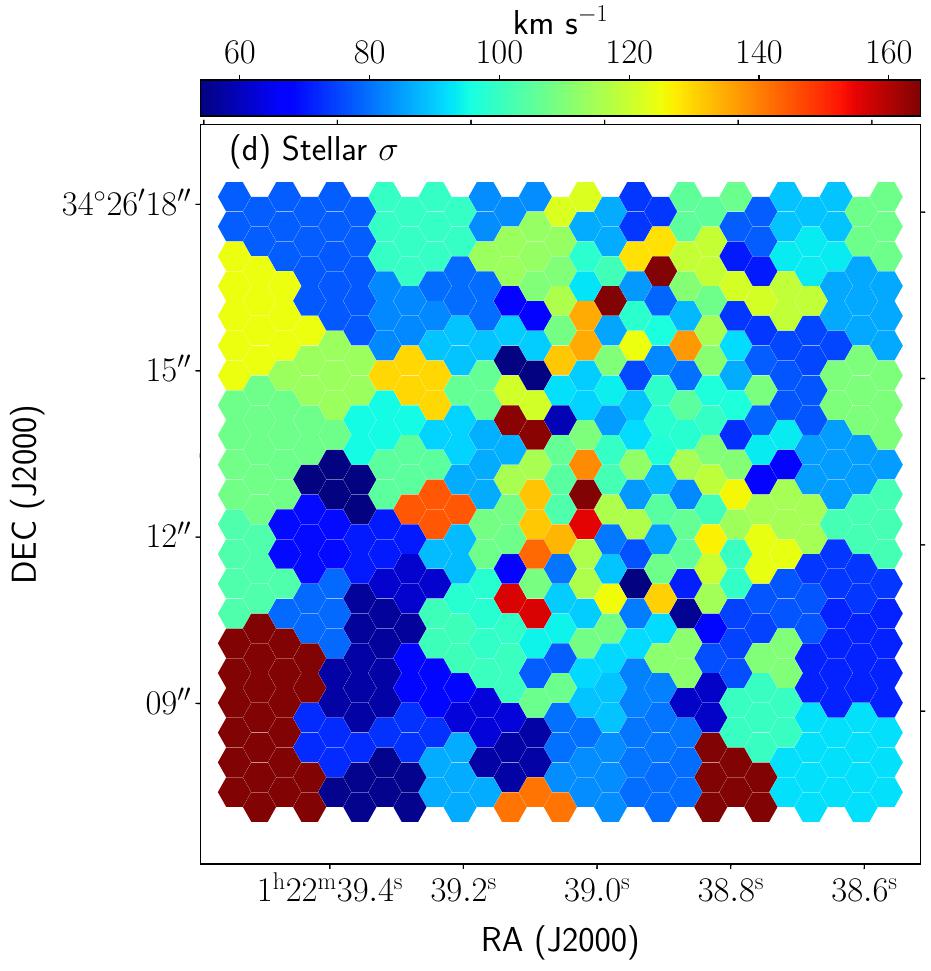}
	\includegraphics[clip, width=0.24\linewidth]{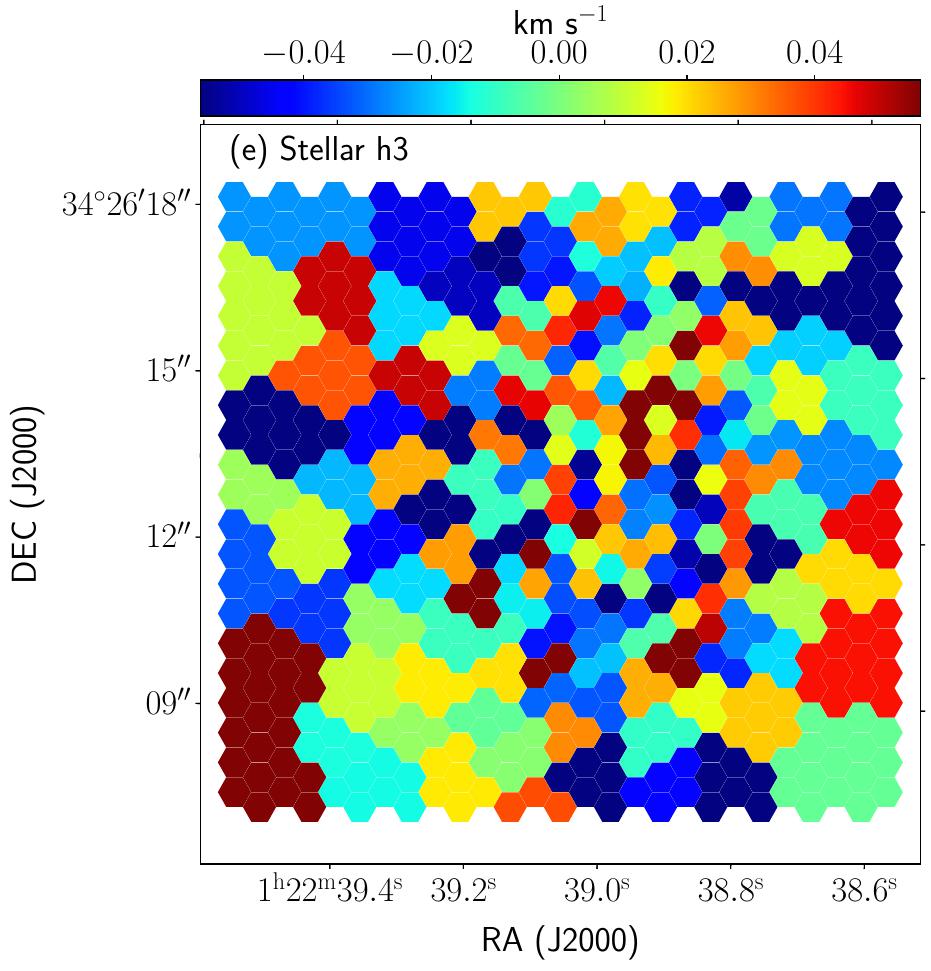}
	\includegraphics[clip, width=0.24\linewidth]{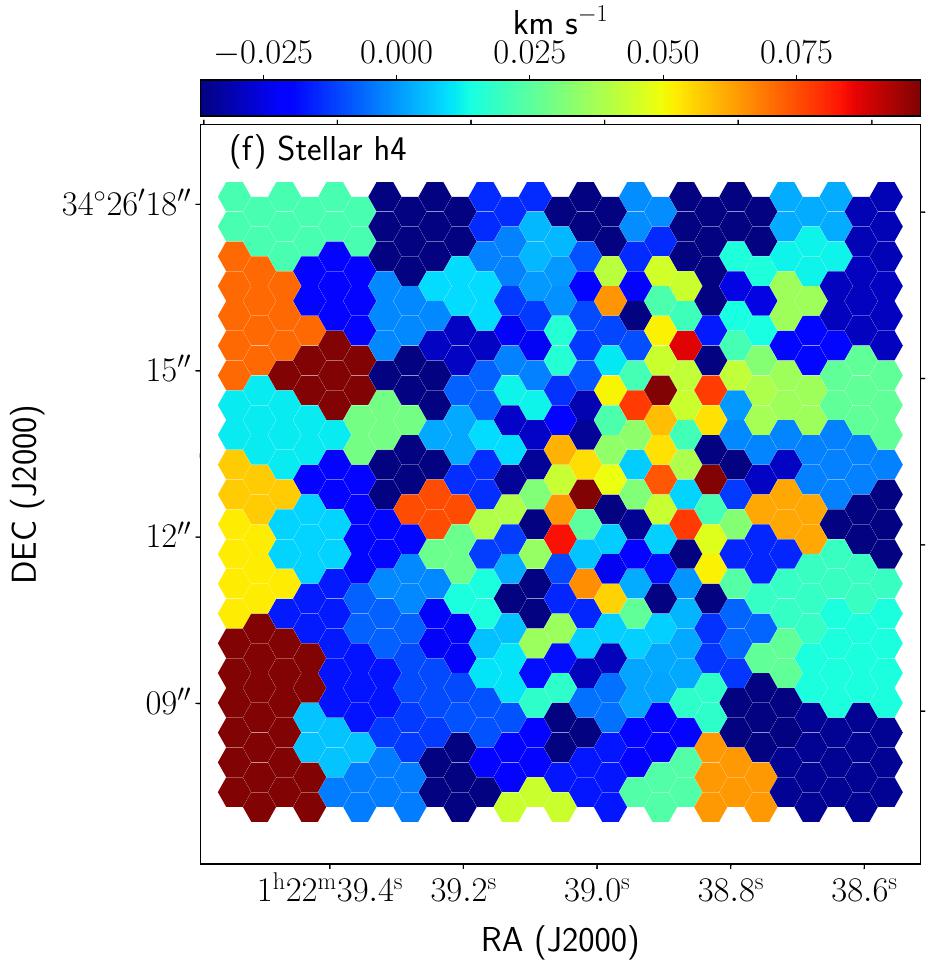}
	
	\vspace{7.8cm}
	
	\includegraphics[clip, width=0.24\linewidth]{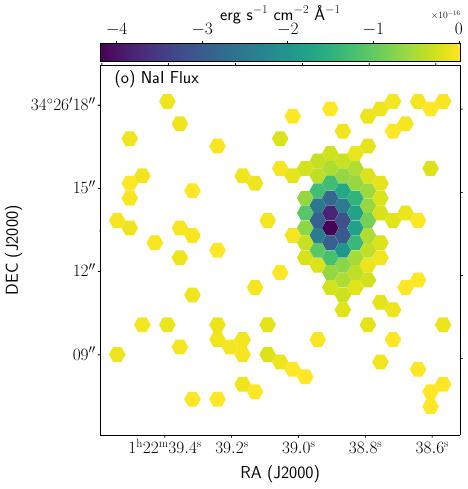}
	\includegraphics[clip, width=0.24\linewidth]{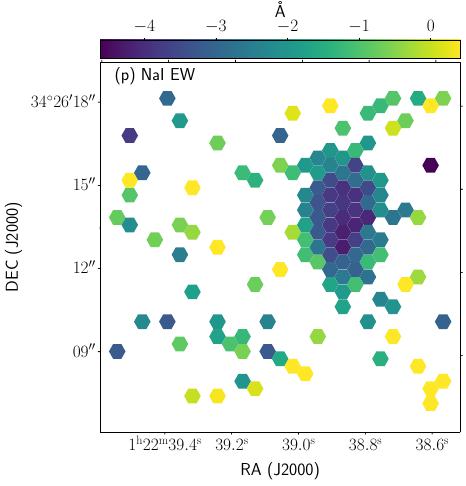}
	\includegraphics[clip, width=0.24\linewidth]{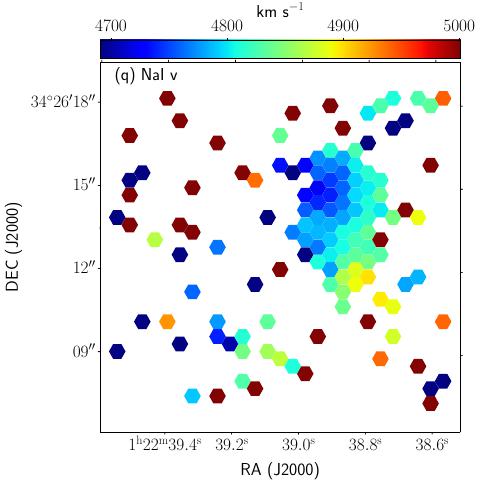}
	\includegraphics[clip, width=0.24\linewidth]{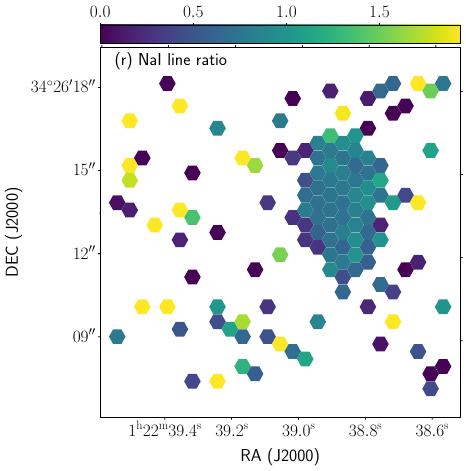}
	%\begin{figure*}
	%\vspace{-0cm}
	\caption{IC~1683 card.}
	\label{fig:IC1683_card_1}
\end{figure*}
\addtocounter{figure}{-1}
\begin{figure*}[h]
	\centering
	\includegraphics[clip, width=0.24\linewidth]{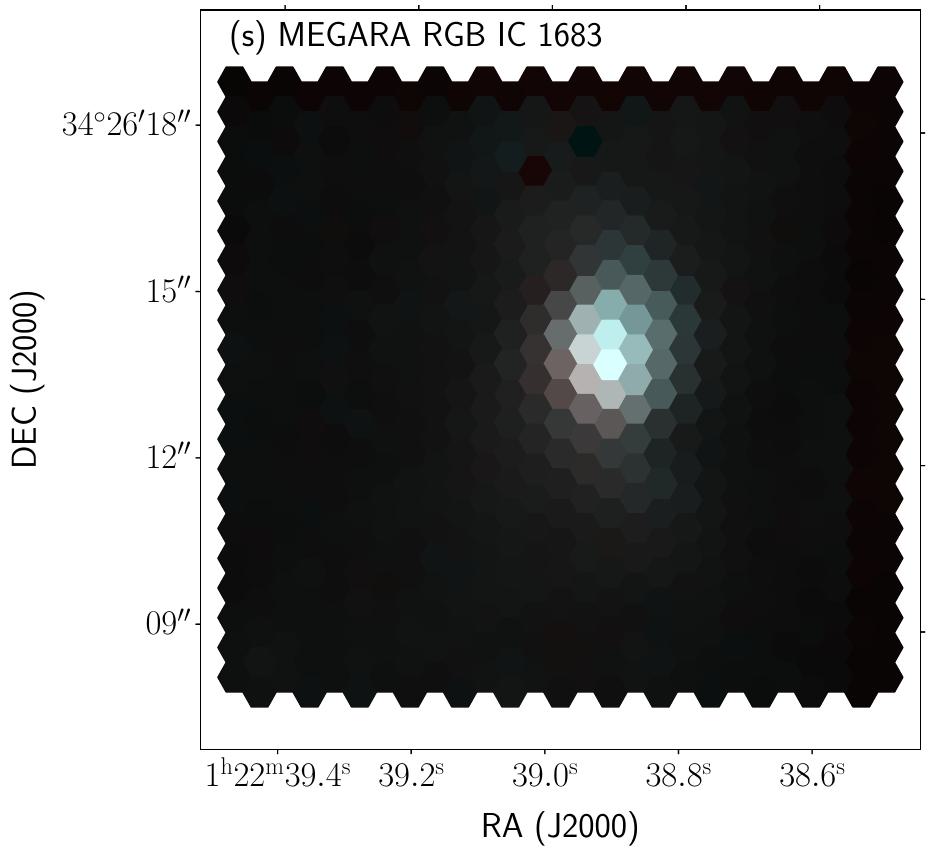}
	\hspace{4,4cm}
	\includegraphics[clip, width=0.24\linewidth]{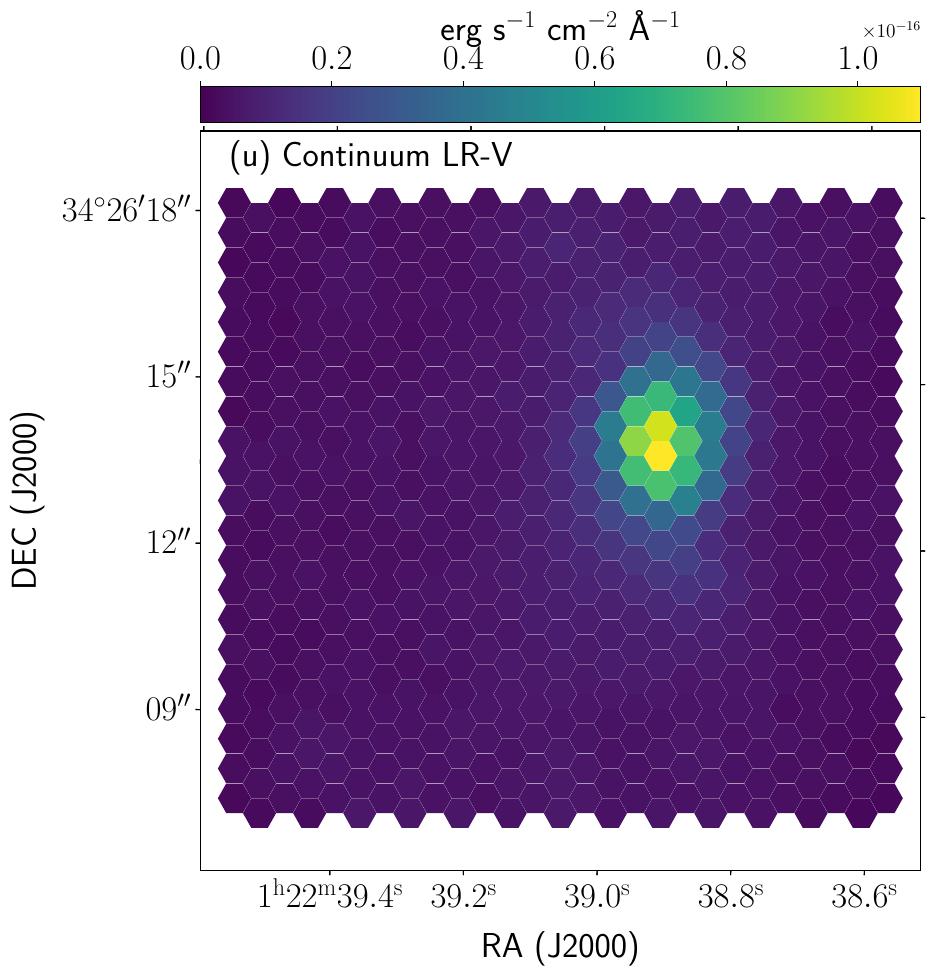}
	\includegraphics[clip, width=0.24\linewidth]{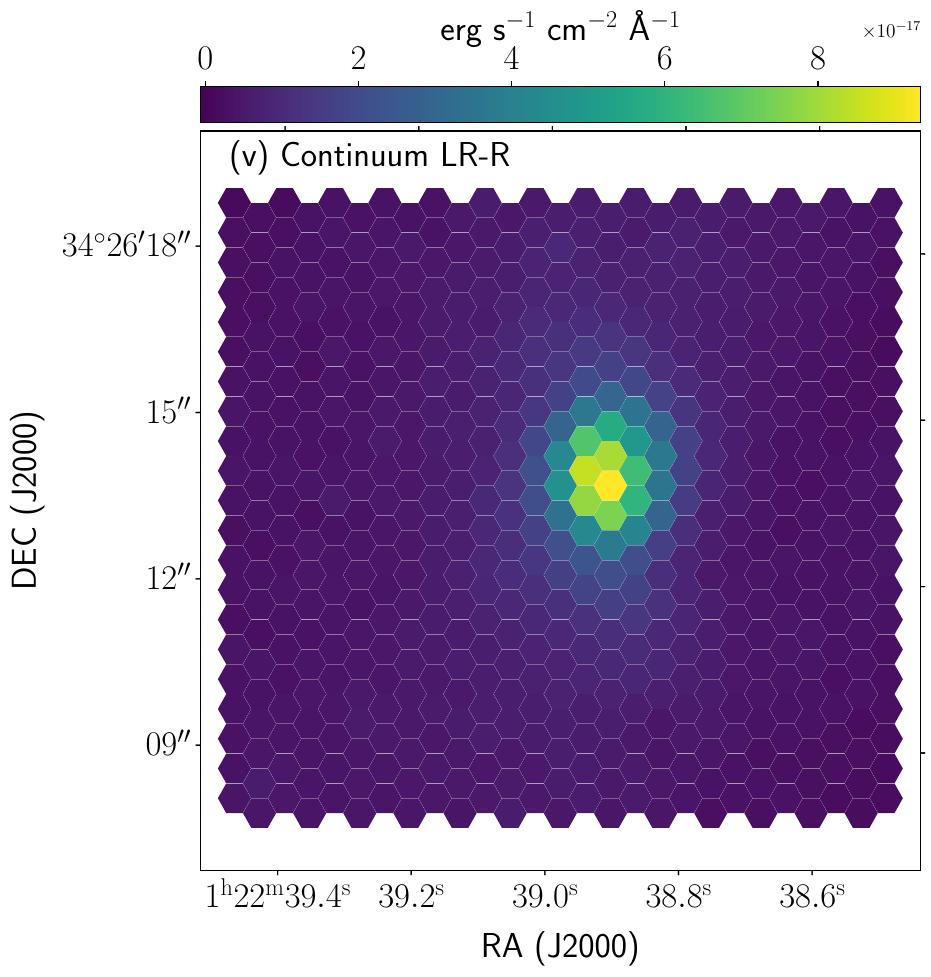}
	\includegraphics[clip, width=0.24\linewidth]{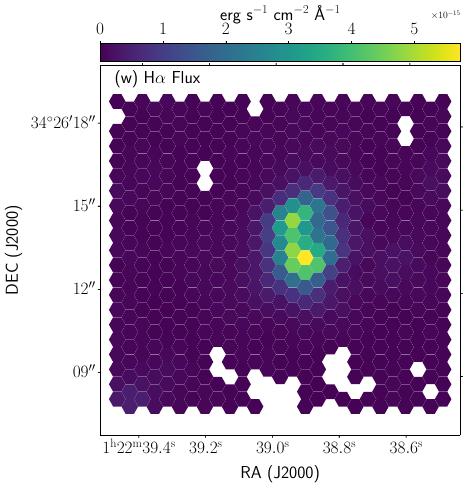}
	\includegraphics[clip, width=0.24\linewidth]{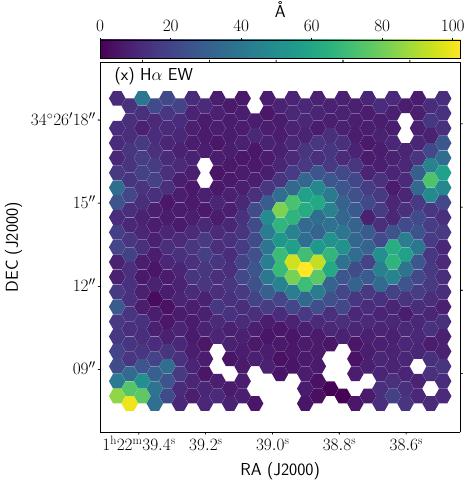}
	\includegraphics[clip, width=0.24\linewidth]{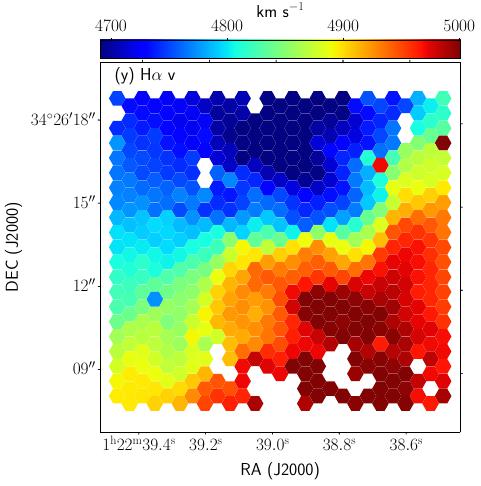}
	\includegraphics[clip, width=0.24\linewidth]{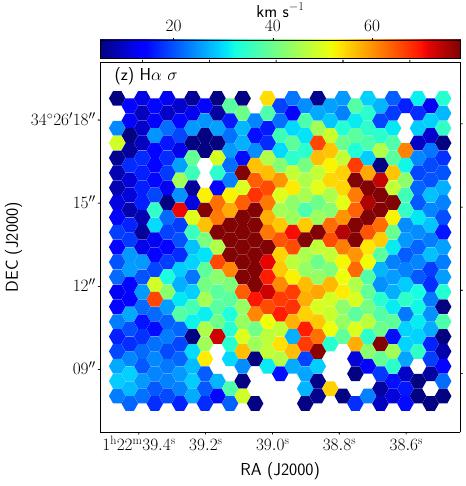}
	\includegraphics[clip, width=0.24\linewidth]{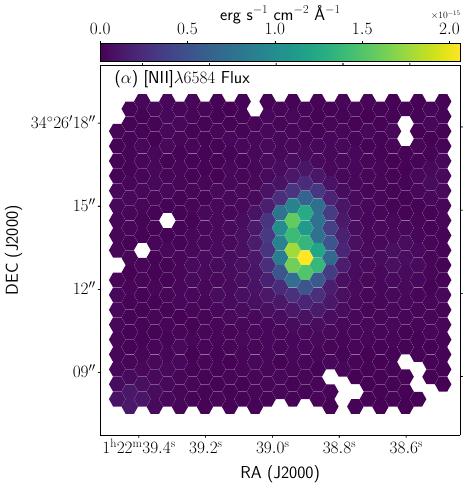}
	\includegraphics[clip, width=0.24\linewidth]{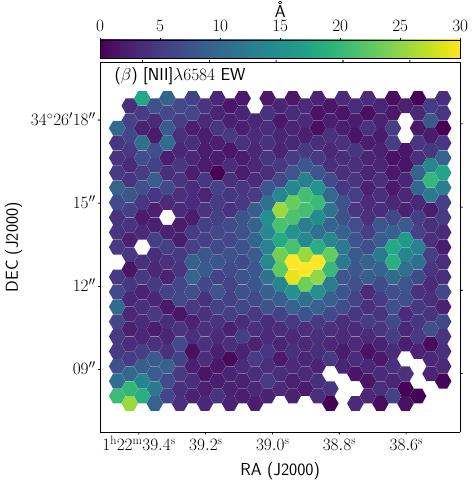}
	\includegraphics[clip, width=0.24\linewidth]{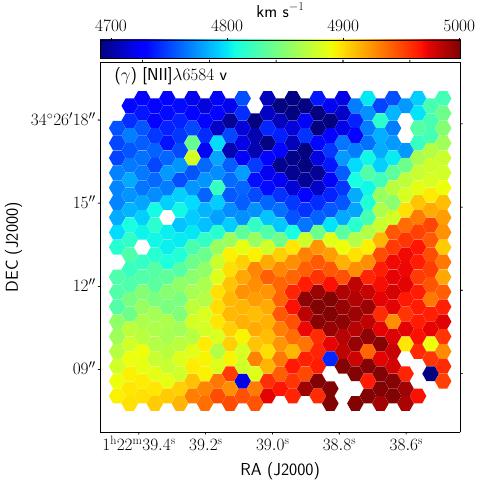}
	\includegraphics[clip, width=0.24\linewidth]{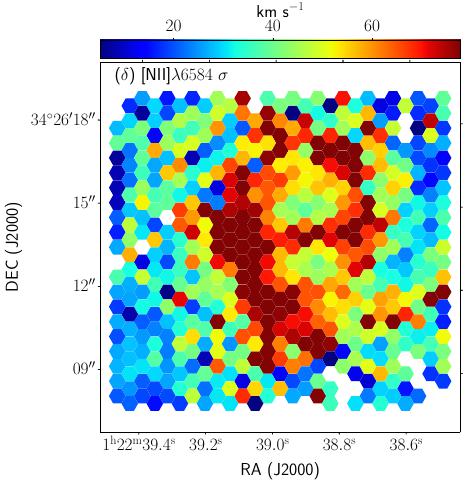}
	\includegraphics[clip, width=0.24\linewidth]{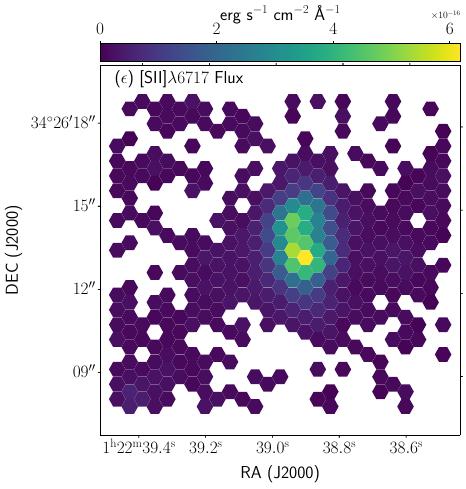}
	\includegraphics[clip, width=0.24\linewidth]{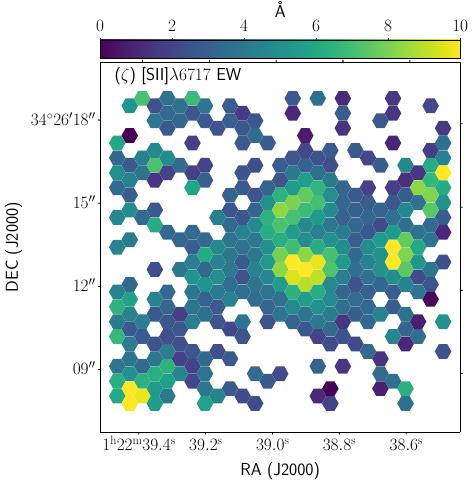}
	\includegraphics[clip, width=0.24\linewidth]{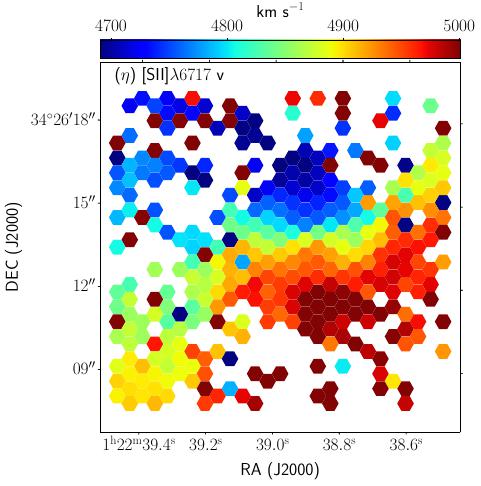}
	\includegraphics[clip, width=0.24\linewidth]{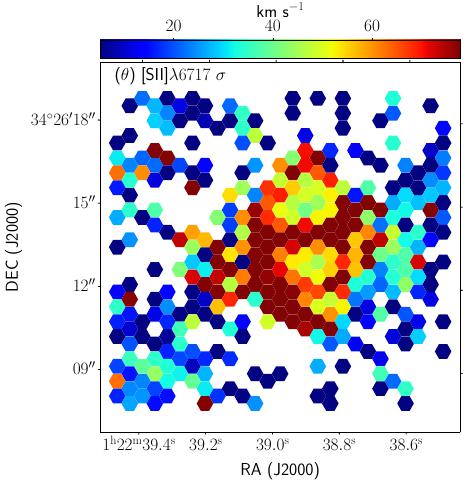}
	\includegraphics[clip, width=0.24\linewidth]{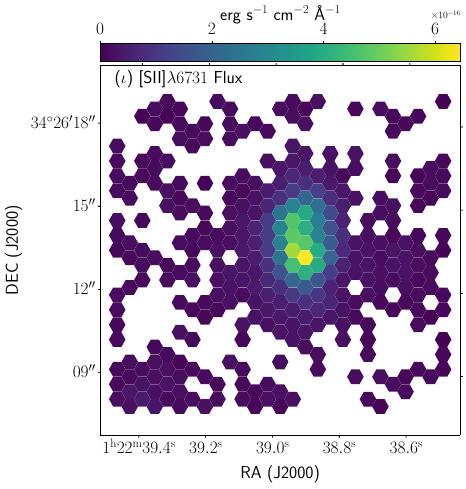}
	\includegraphics[clip, width=0.24\linewidth]{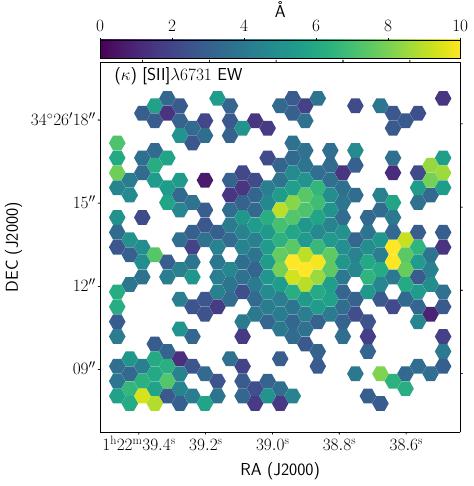}
	\includegraphics[clip, width=0.24\linewidth]{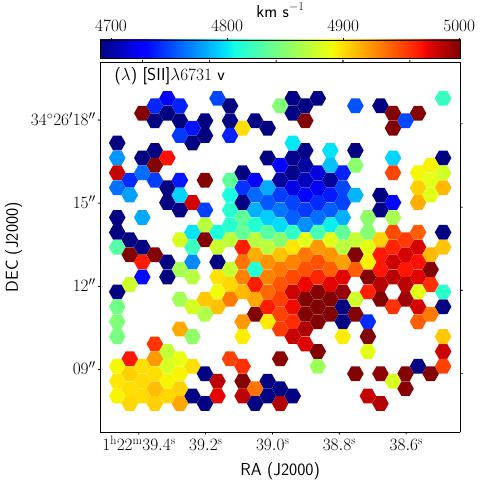}
	\includegraphics[clip, width=0.24\linewidth]{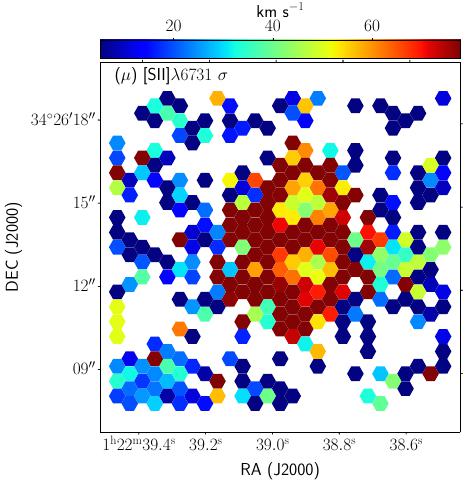}
	\caption{(cont.) IC~1683 card.}
	\label{fig:IC1683_card_2}
\end{figure*}

\begin{figure*}[h]
	\centering
	\includegraphics[clip, width=0.35\linewidth]{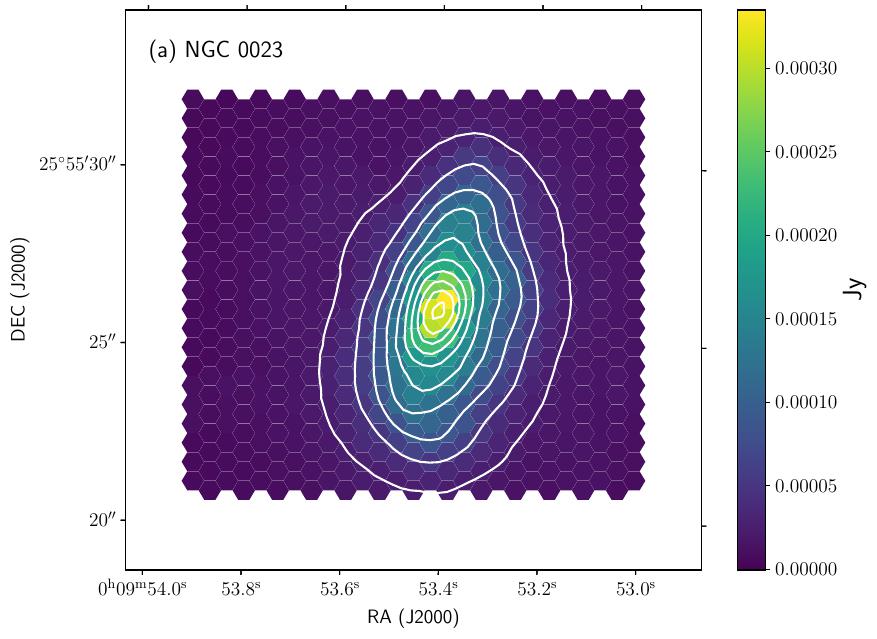}
	\includegraphics[clip, width=0.6\linewidth]{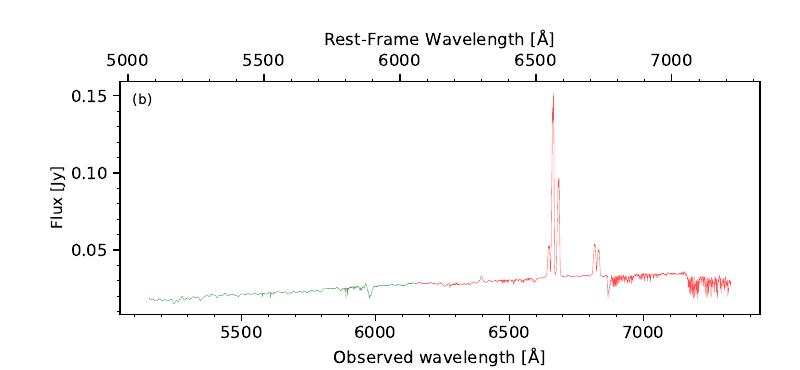}
	\includegraphics[clip, width=0.24\linewidth]{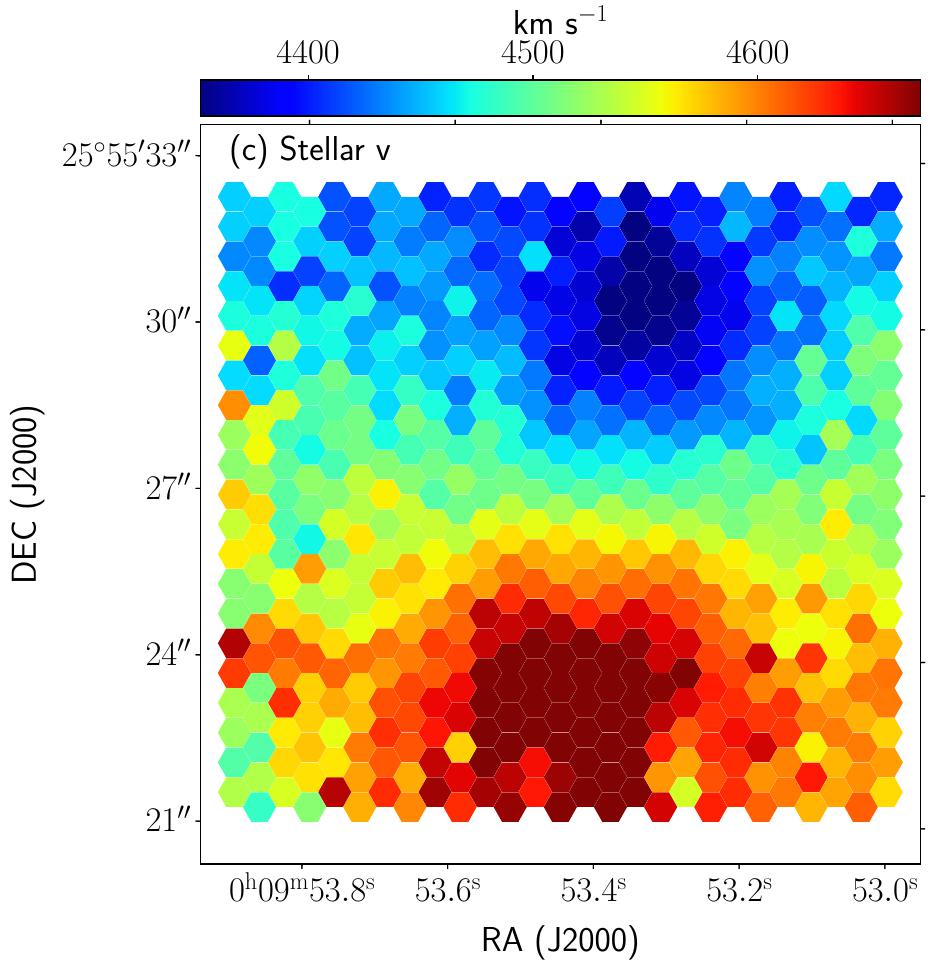}
	\includegraphics[clip, width=0.24\linewidth]{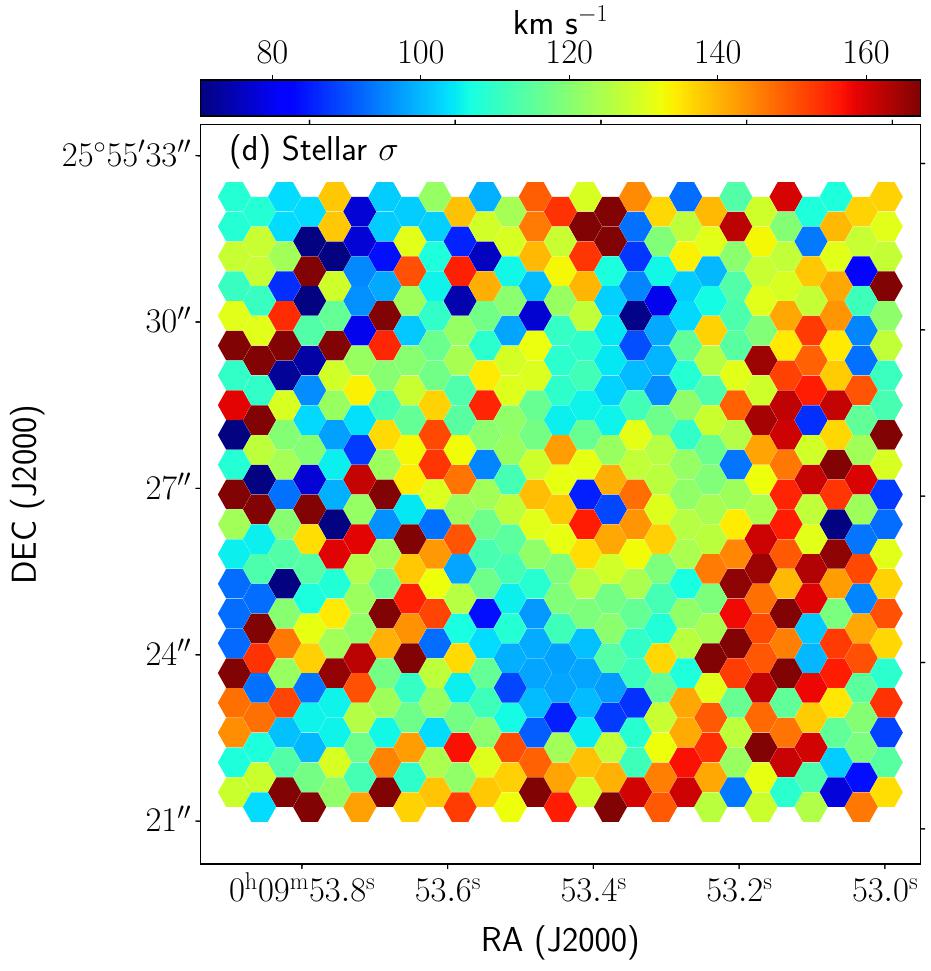}
	\includegraphics[clip, width=0.24\linewidth]{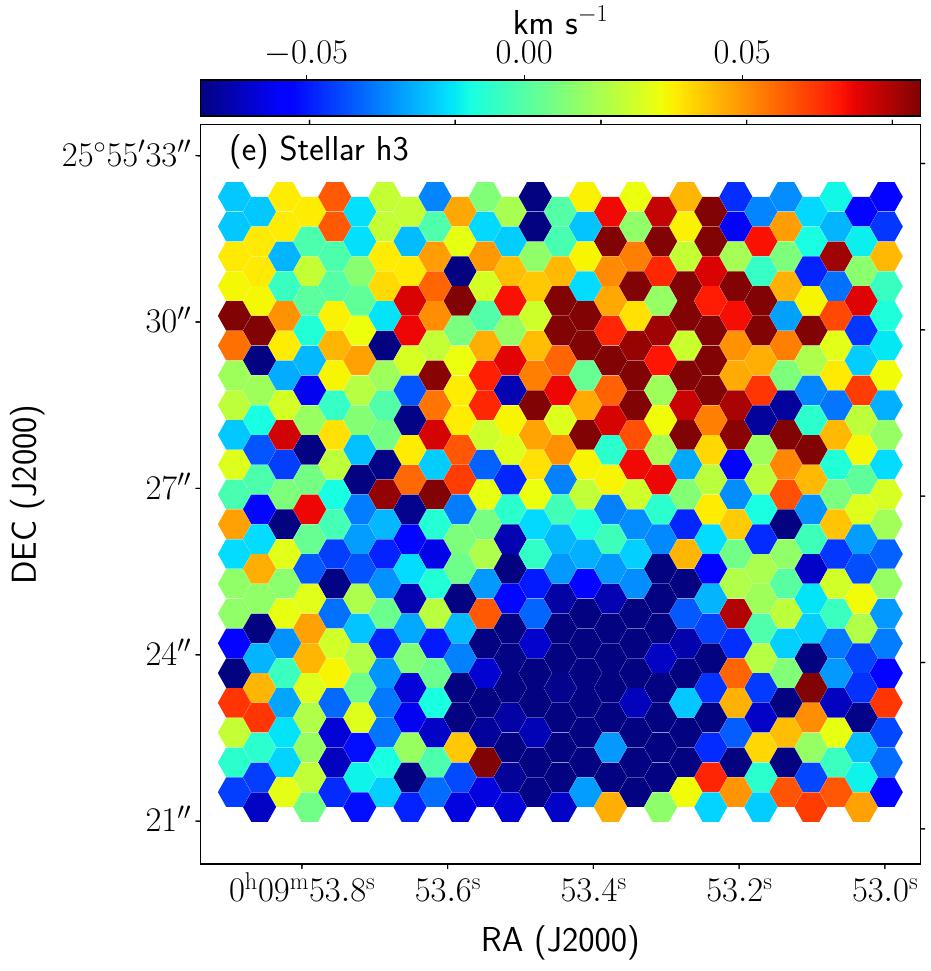}
	\includegraphics[clip, width=0.24\linewidth]{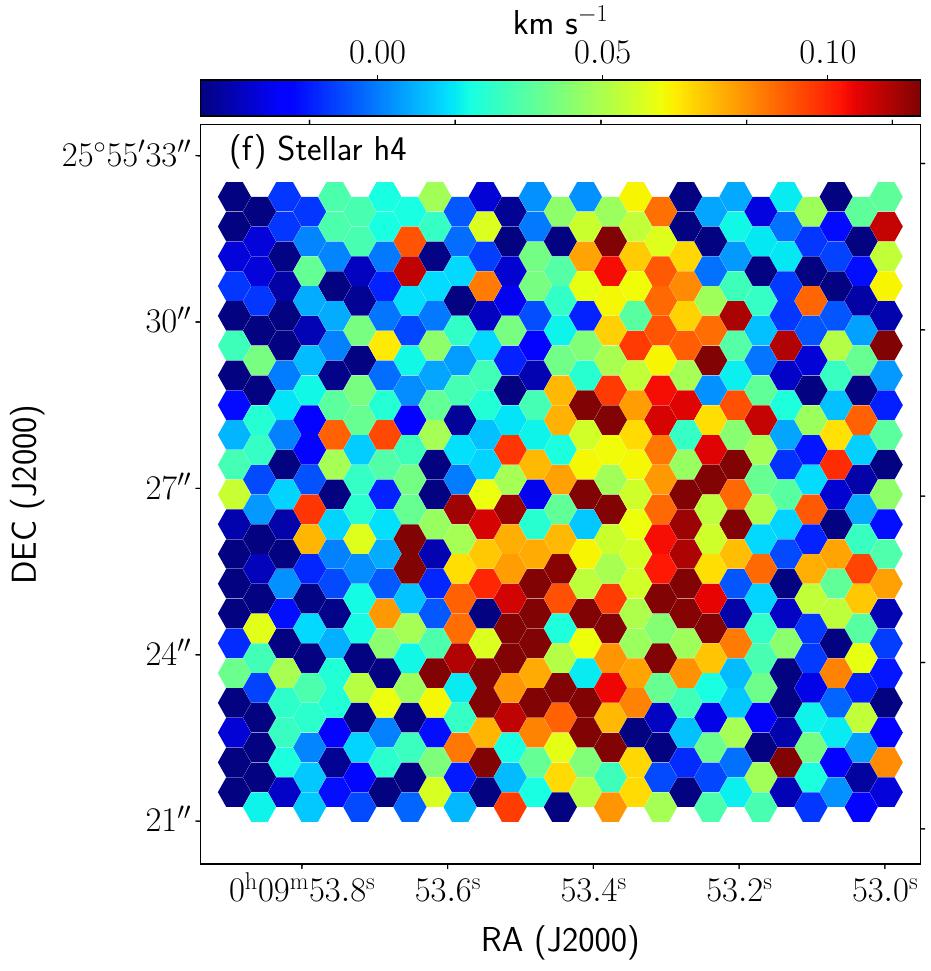}
	
	\vspace{8.8cm}
	
	\includegraphics[clip, width=0.24\linewidth]{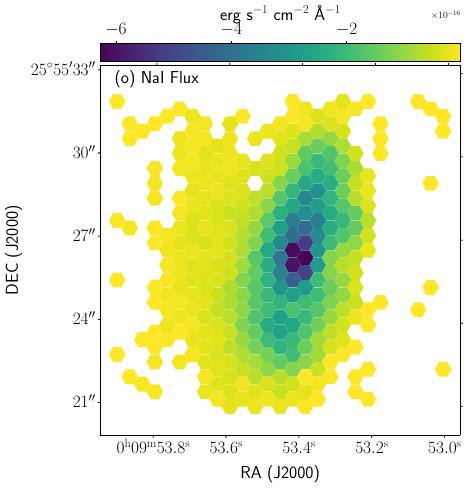}
	\includegraphics[clip, width=0.24\linewidth]{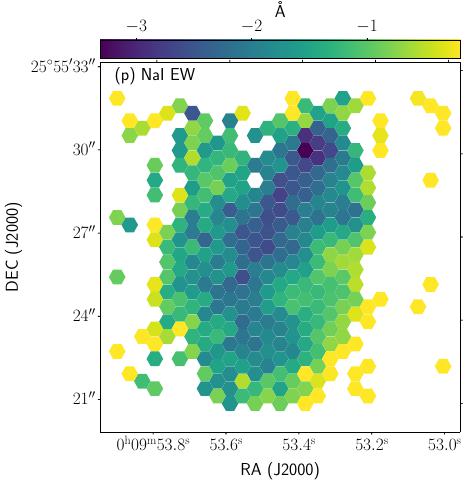}
	\includegraphics[clip, width=0.24\linewidth]{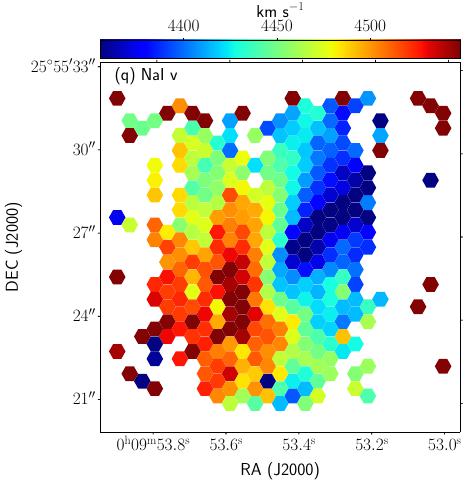}
	\includegraphics[clip, width=0.24\linewidth]{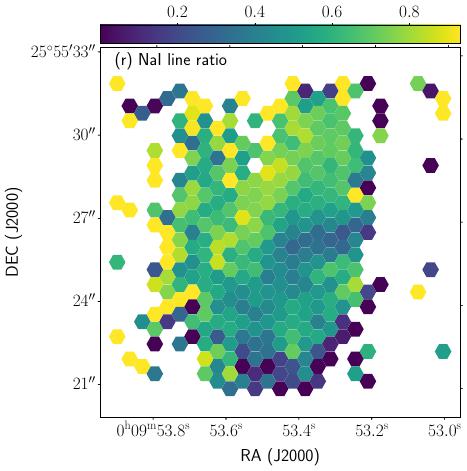}
	\caption{NGC~0023 card.}
	\label{fig:NGC0023_card_1}
\end{figure*}
\addtocounter{figure}{-1}
\begin{figure*}[h]
	\centering
	\includegraphics[clip, width=0.24\linewidth]{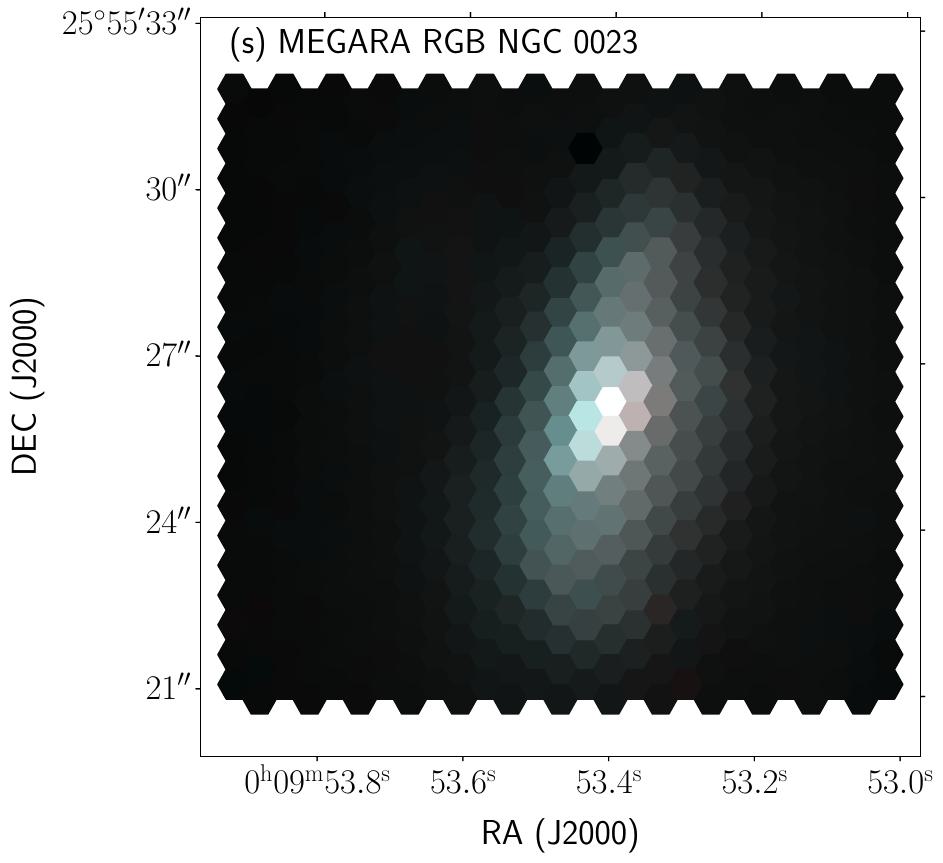}
	\hspace{4,4cm}
	\includegraphics[clip, width=0.24\linewidth]{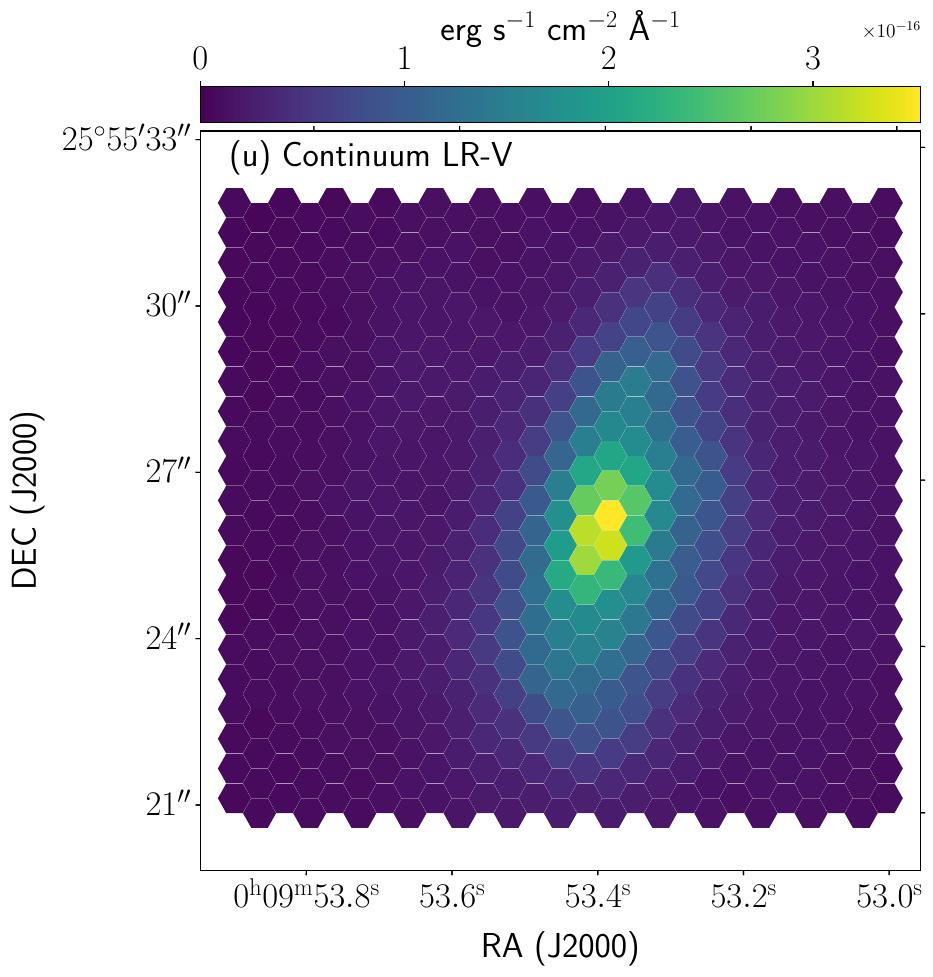}
	\includegraphics[clip, width=0.24\linewidth]{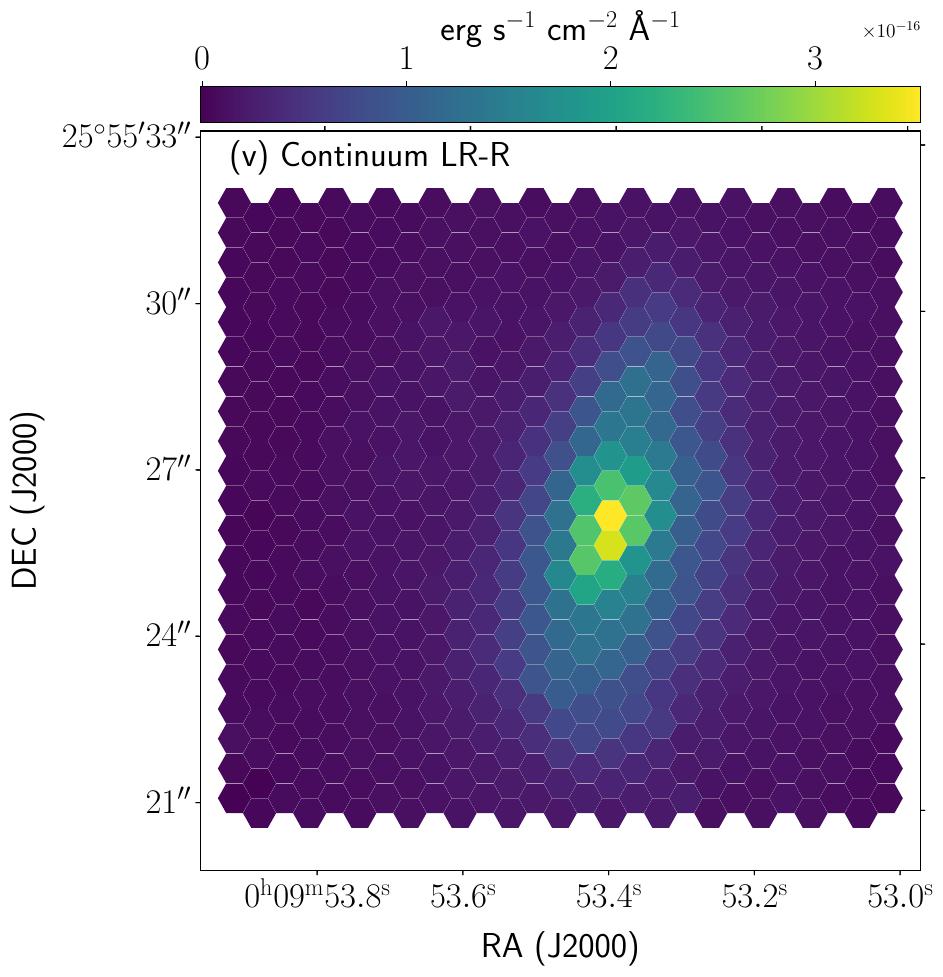}
	\includegraphics[clip, width=0.24\linewidth]{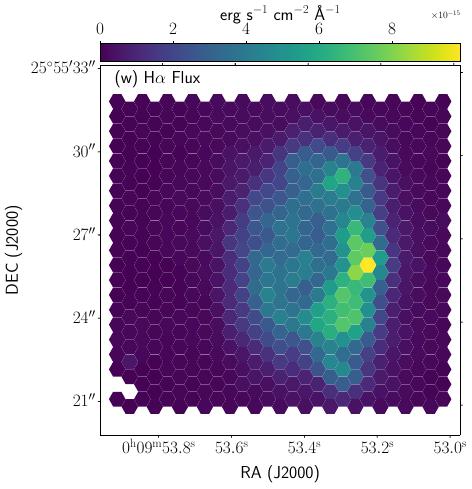}
	\includegraphics[clip, width=0.24\linewidth]{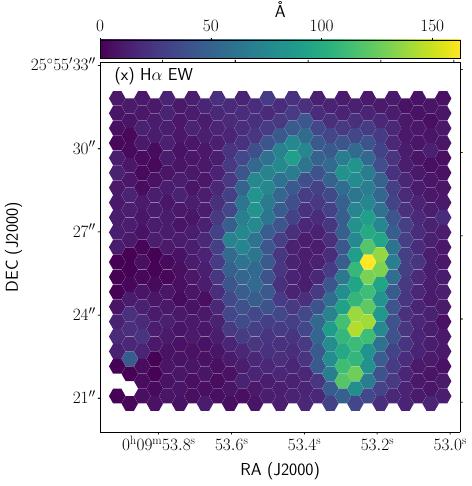}
	\includegraphics[clip, width=0.24\linewidth]{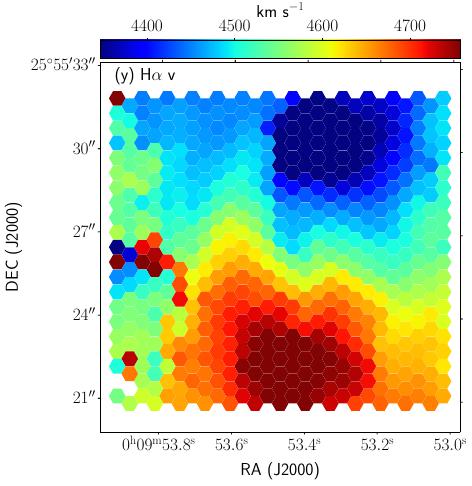}
	\includegraphics[clip, width=0.24\linewidth]{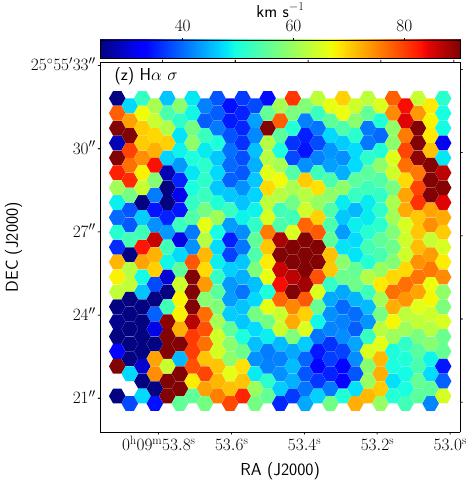}
	\includegraphics[clip, width=0.24\linewidth]{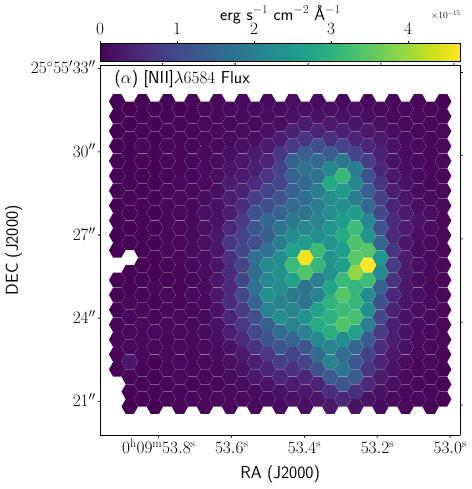}
	\includegraphics[clip, width=0.24\linewidth]{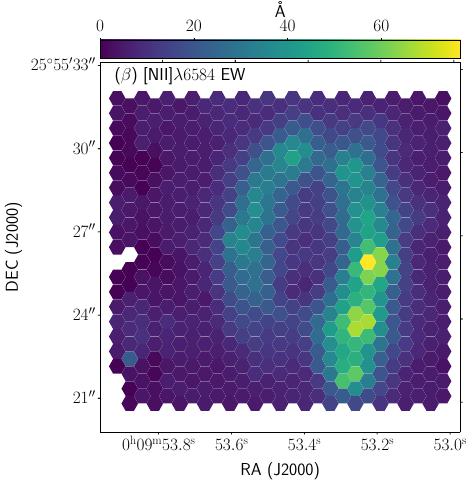}
	\includegraphics[clip, width=0.24\linewidth]{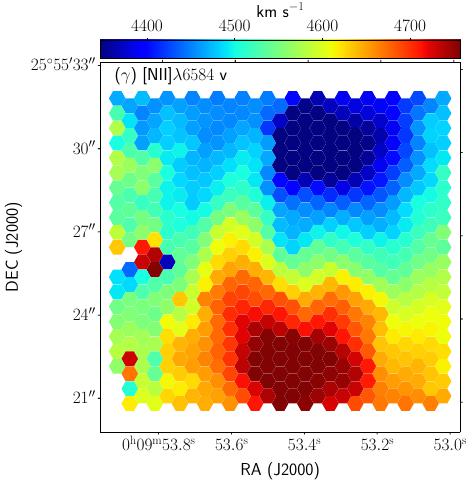}
	\includegraphics[clip, width=0.24\linewidth]{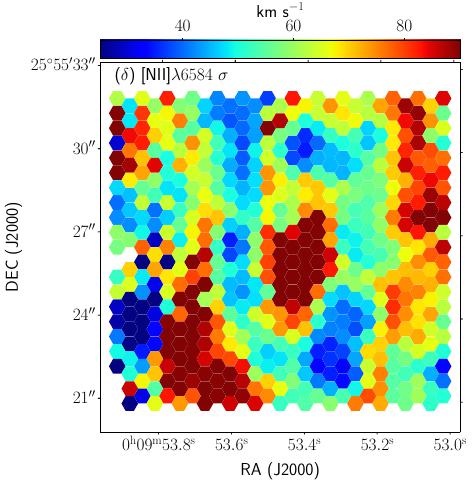}
	\includegraphics[clip, width=0.24\linewidth]{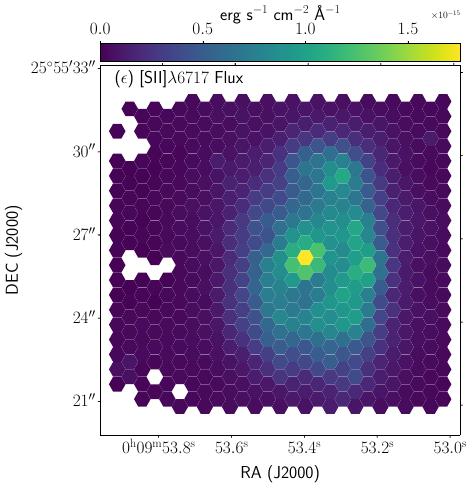}
	\includegraphics[clip, width=0.24\linewidth]{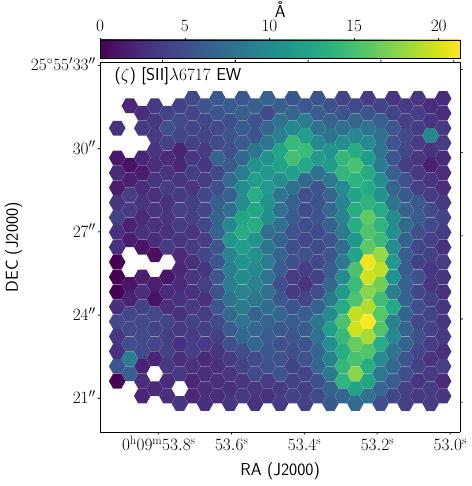}
	\includegraphics[clip, width=0.24\linewidth]{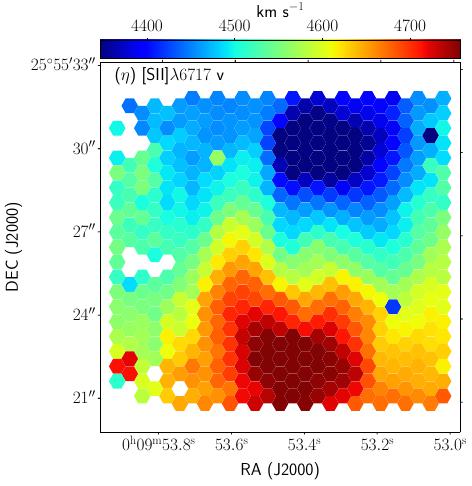}
	\includegraphics[clip, width=0.24\linewidth]{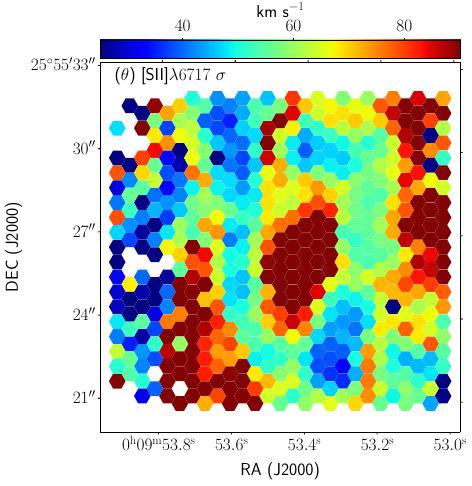}
	\includegraphics[clip, width=0.24\linewidth]{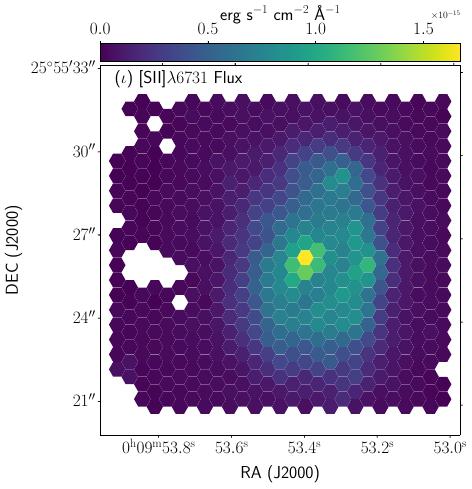}
	\includegraphics[clip, width=0.24\linewidth]{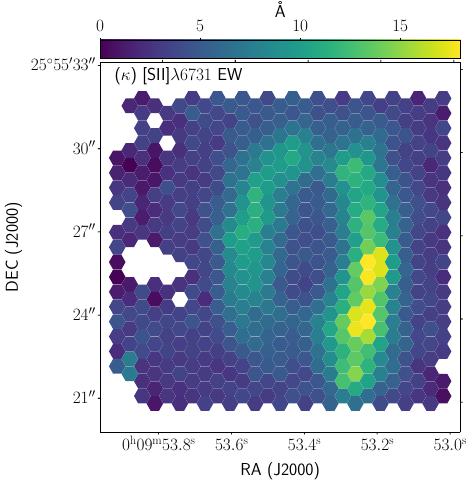}
	\includegraphics[clip, width=0.24\linewidth]{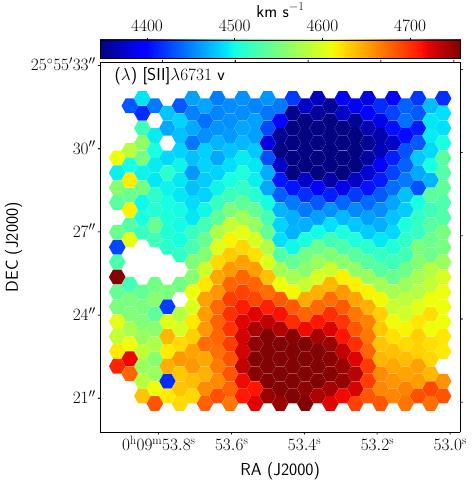}
	\includegraphics[clip, width=0.24\linewidth]{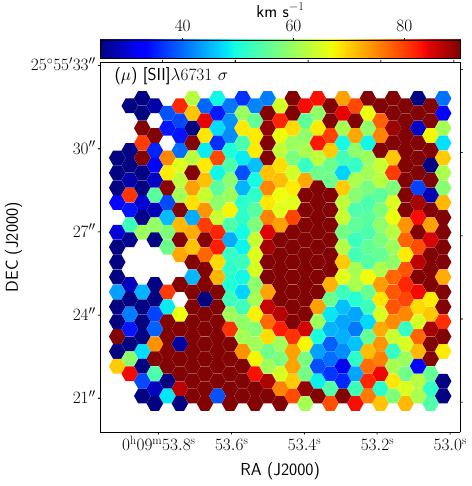}
	\caption{(cont.) NGC~0023 card.}
	\label{fig:NGC0023_card_2}
\end{figure*}

\begin{figure*}[h]
	\centering
	\includegraphics[clip, width=0.35\linewidth]{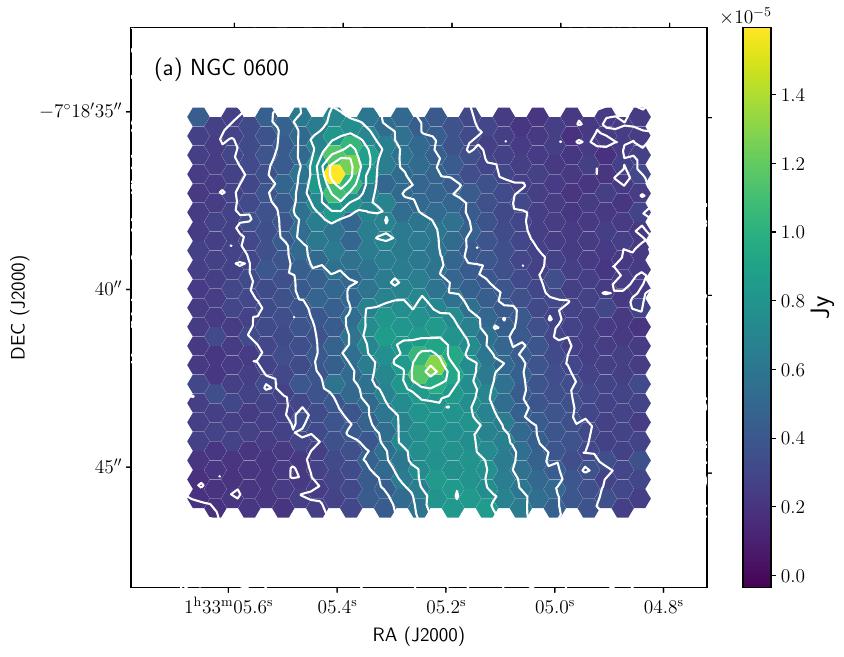}
	\includegraphics[clip, width=0.6\linewidth]{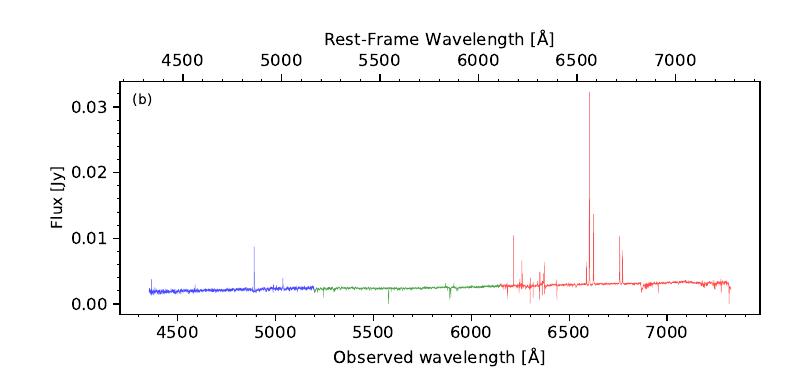}
	\includegraphics[clip, width=0.24\linewidth]{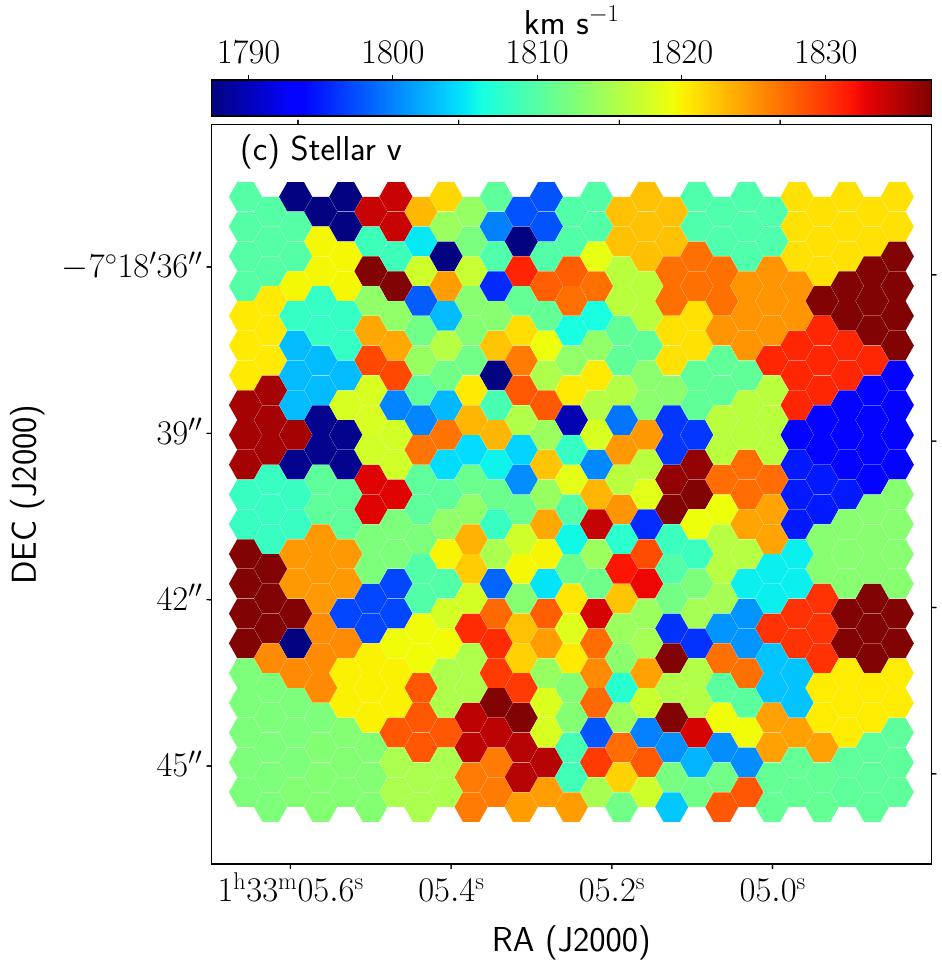}
	\includegraphics[clip, width=0.24\linewidth]{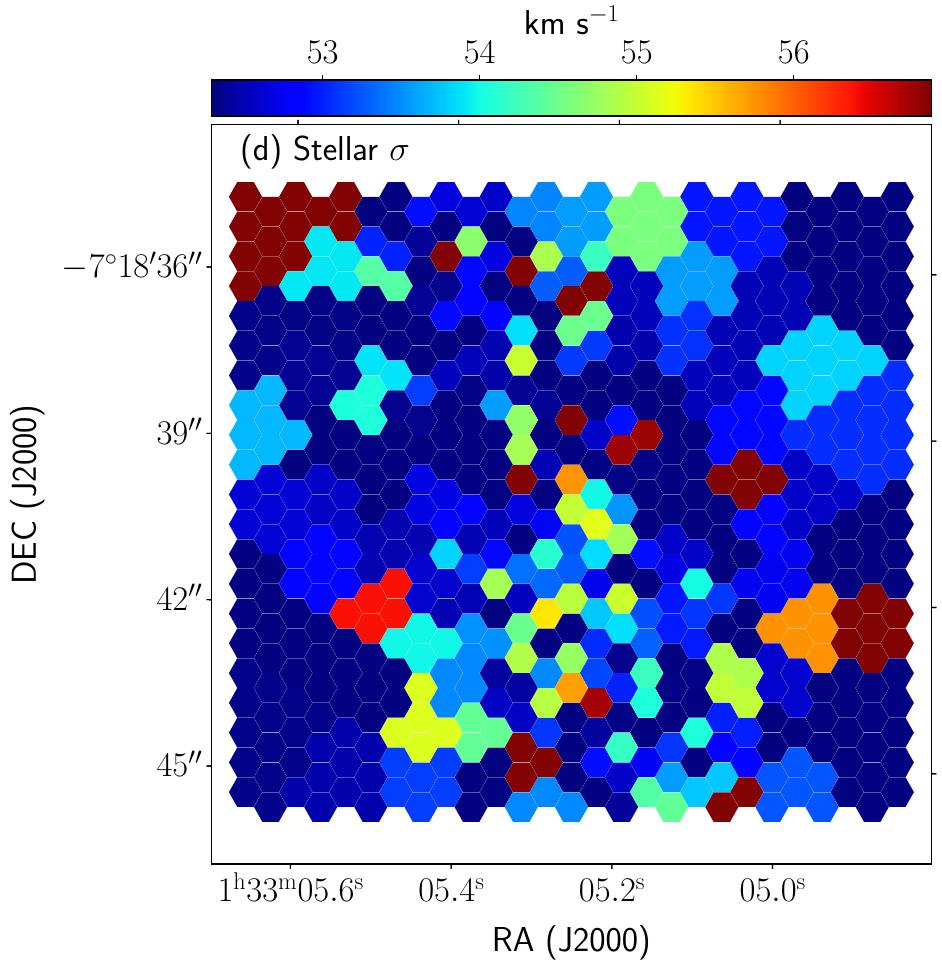}
	\includegraphics[clip, width=0.24\linewidth]{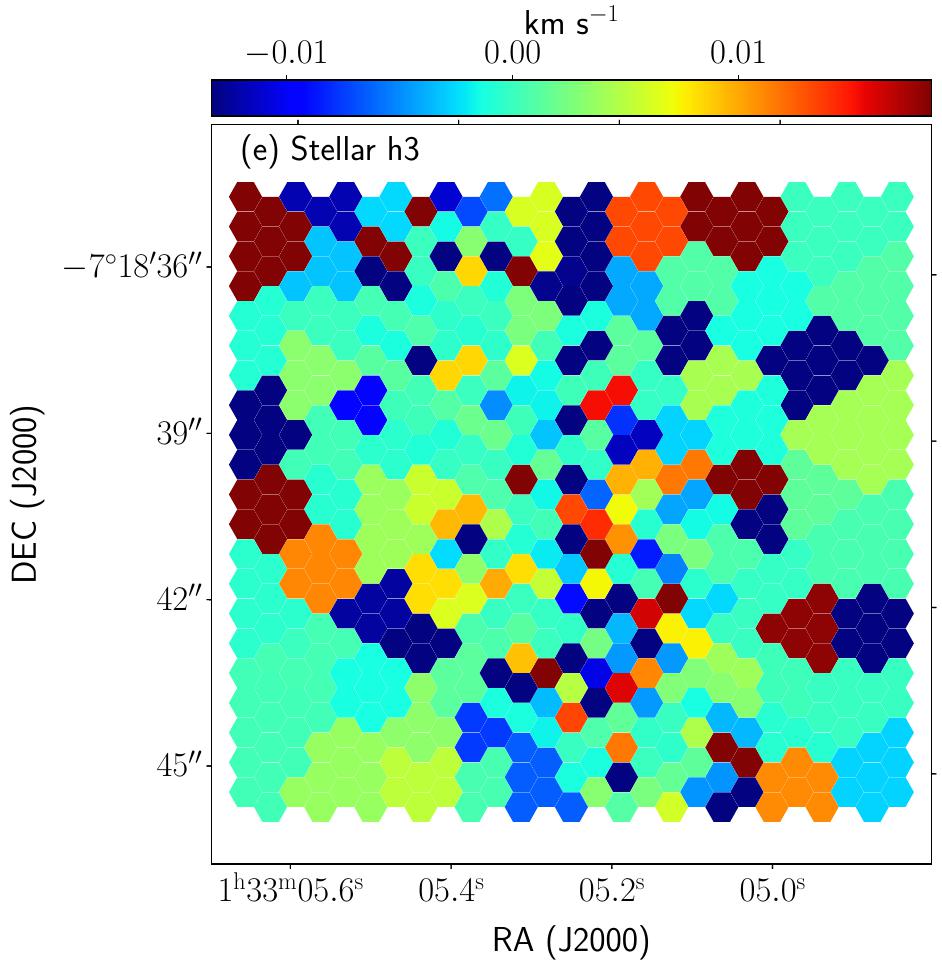}
	\includegraphics[clip, width=0.24\linewidth]{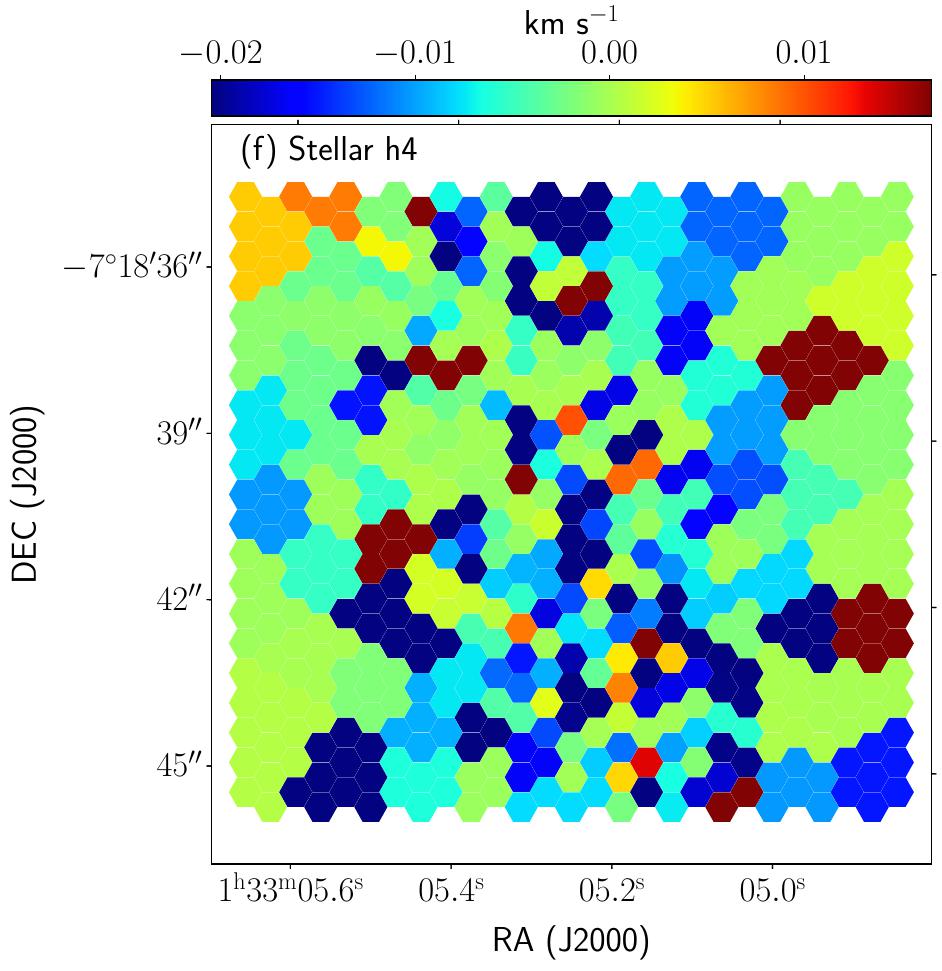}
	\includegraphics[clip, width=0.24\linewidth]{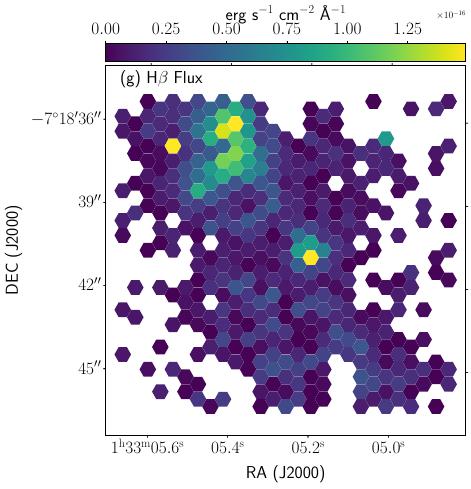}
	\includegraphics[clip, width=0.24\linewidth]{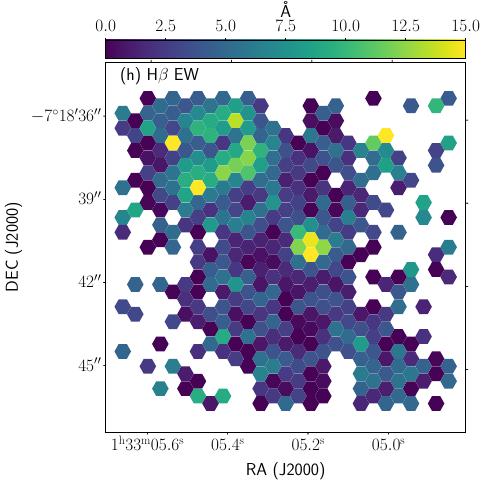}
	\includegraphics[clip, width=0.24\linewidth]{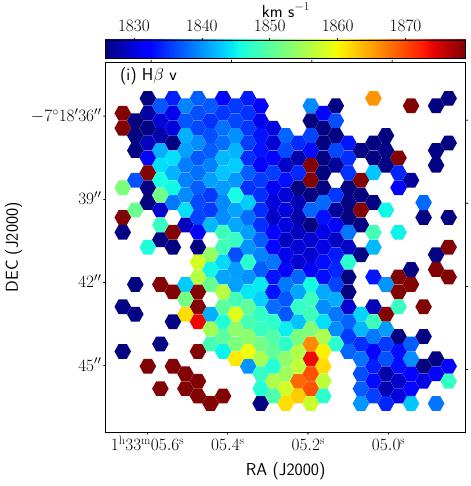}
	\includegraphics[clip, width=0.24\linewidth]{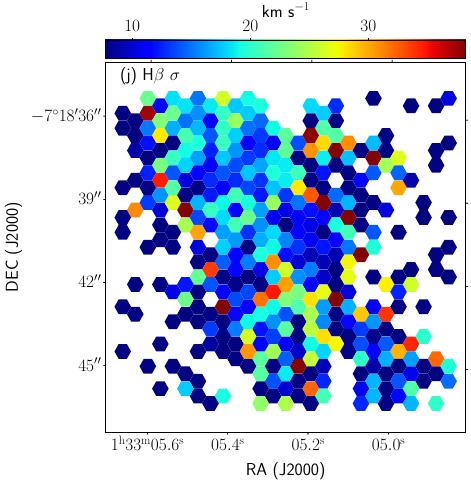}
	\includegraphics[clip, width=0.24\linewidth]{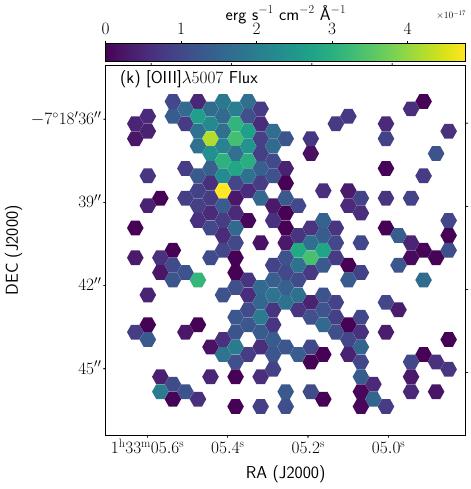}
	\includegraphics[clip, width=0.24\linewidth]{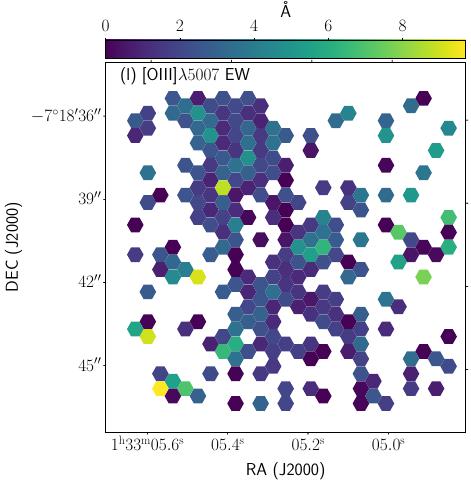}
	\includegraphics[clip, width=0.24\linewidth]{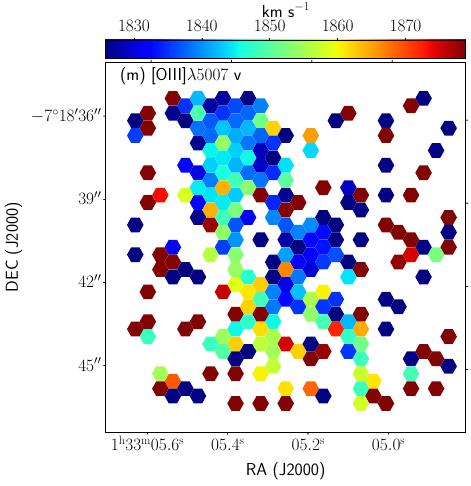}
	\includegraphics[clip, width=0.24\linewidth]{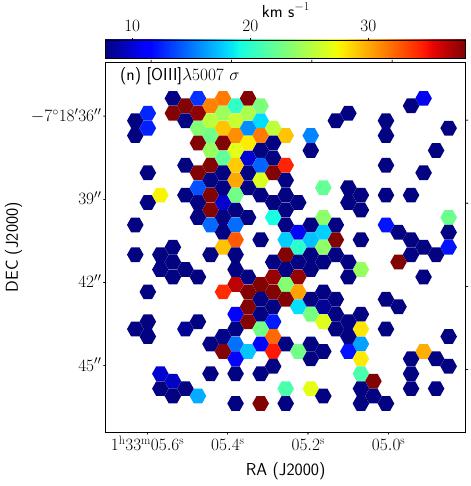}
	\vspace{5cm}
	\caption{NGC~0600 card.}
	\label{fig:NGC0600_card_1}
\end{figure*}
\addtocounter{figure}{-1}
\begin{figure*}[h]
	\centering
	\includegraphics[clip, width=0.24\linewidth]{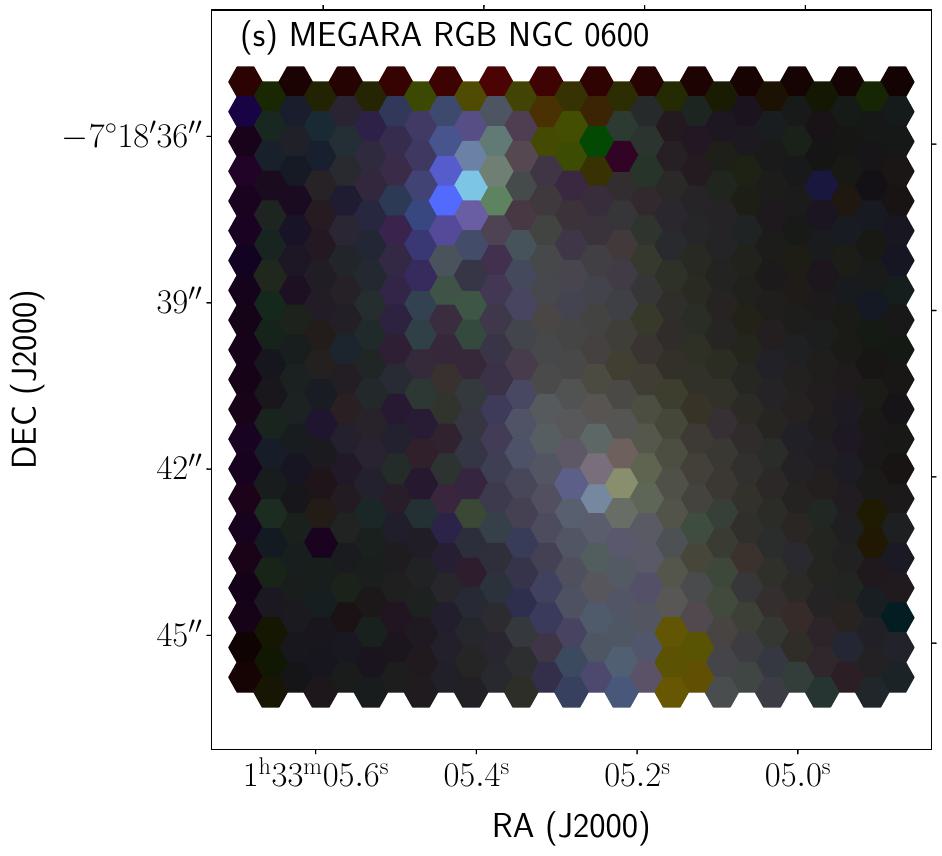}
	\includegraphics[clip, width=0.24\linewidth]{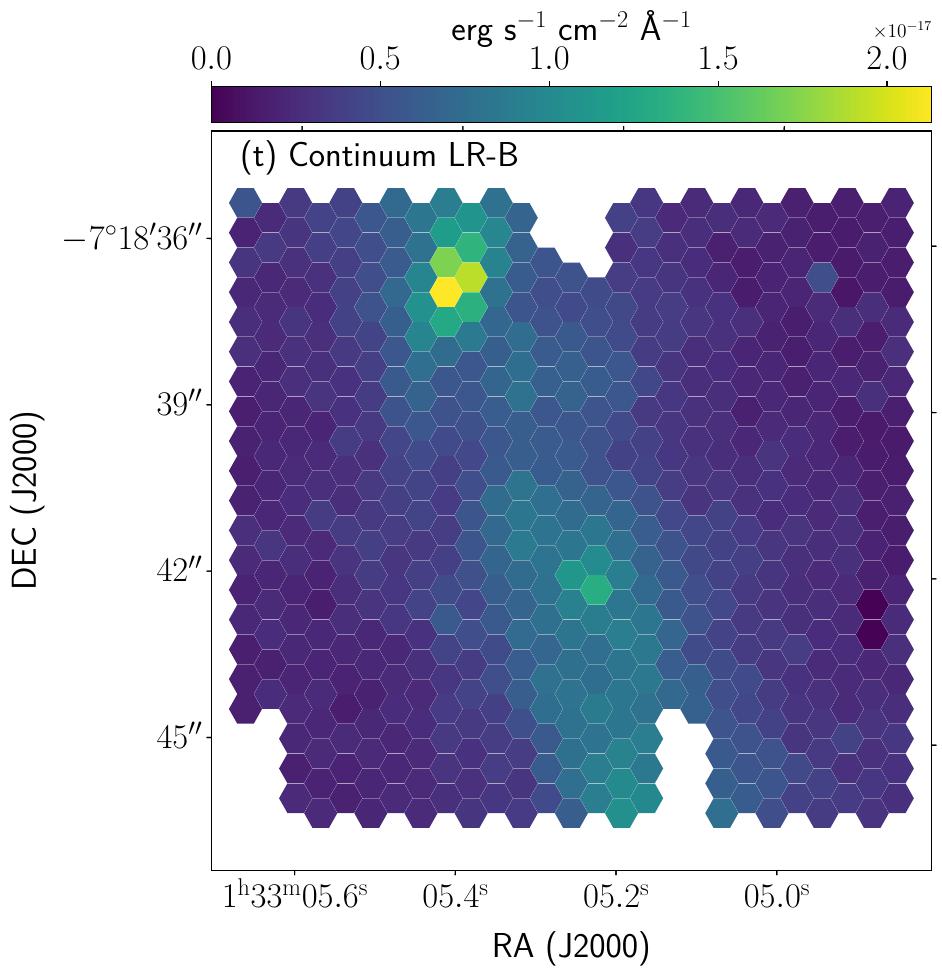}
	\includegraphics[clip, width=0.24\linewidth]{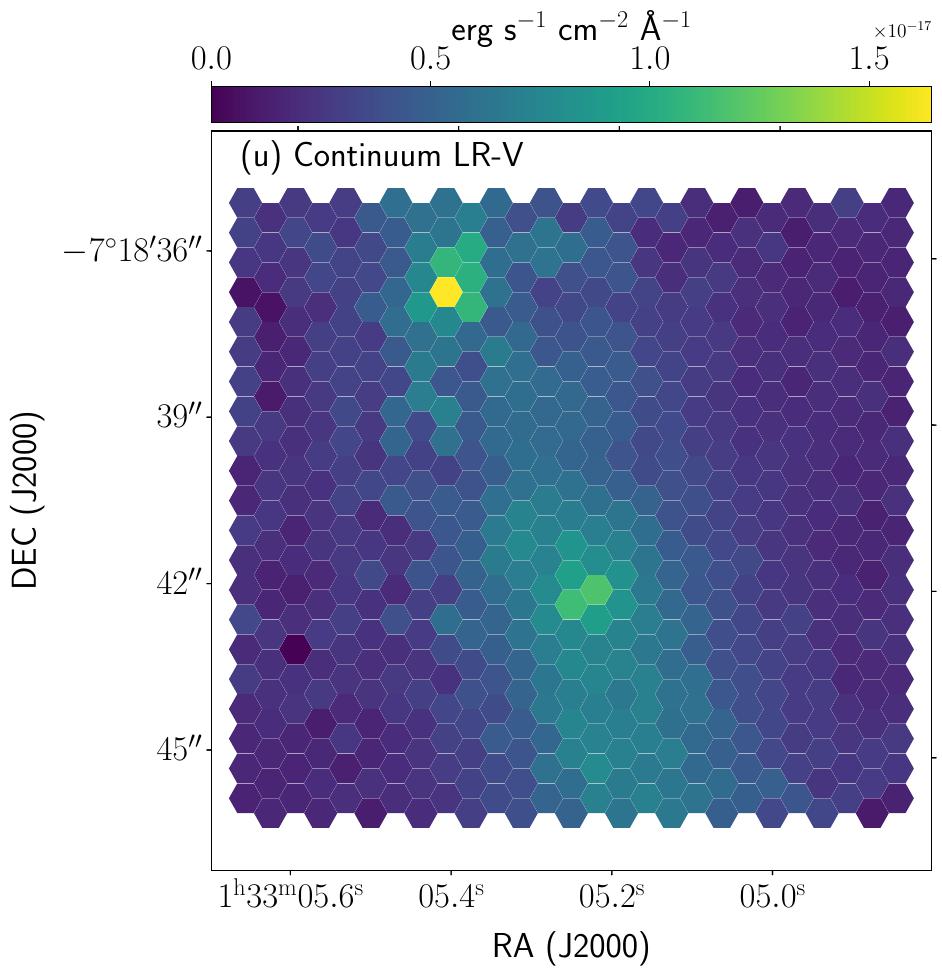}
	\includegraphics[clip, width=0.24\linewidth]{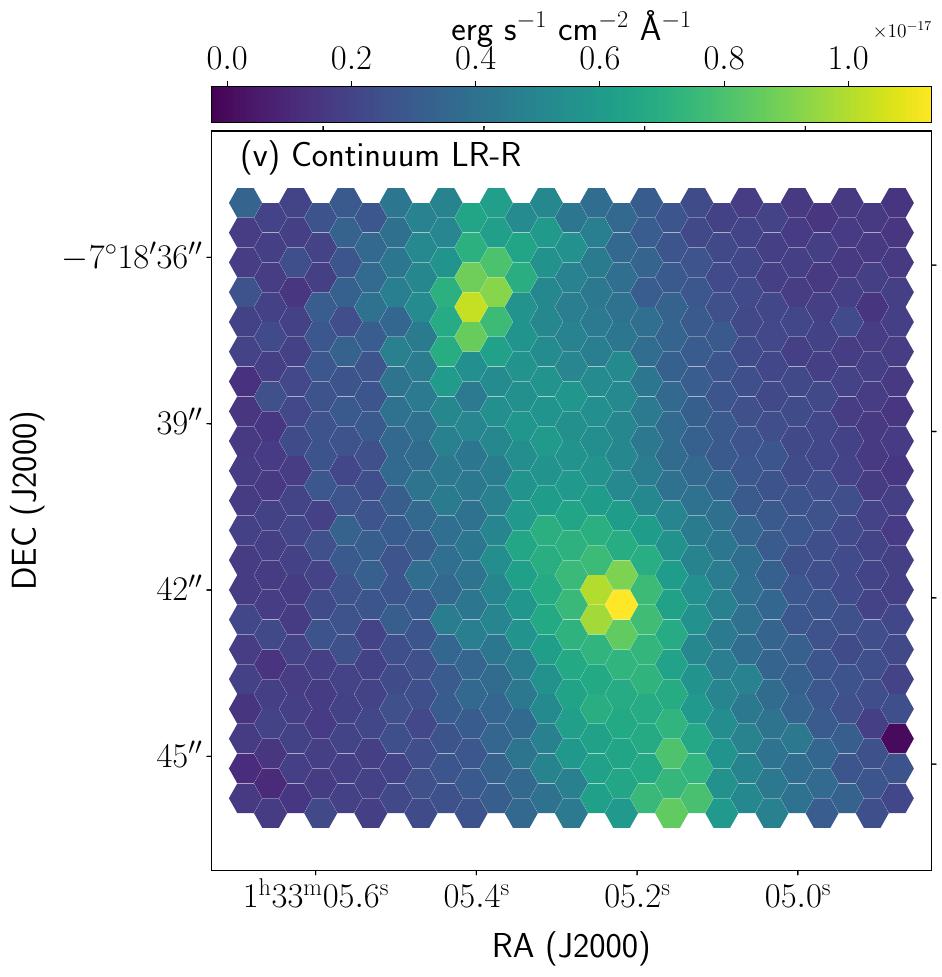}
	\includegraphics[clip, width=0.24\linewidth]{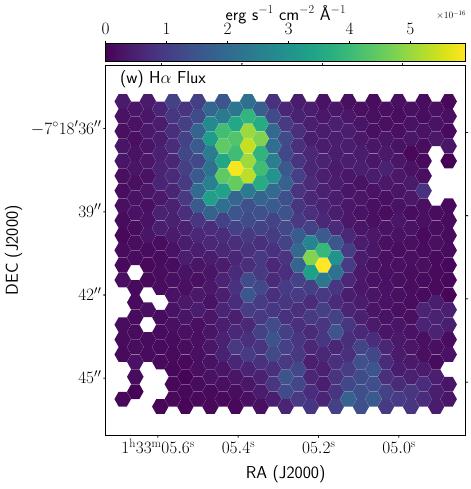}
	\includegraphics[clip, width=0.24\linewidth]{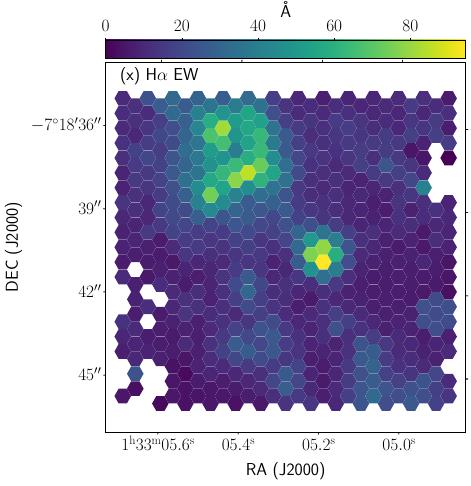}
	\includegraphics[clip, width=0.24\linewidth]{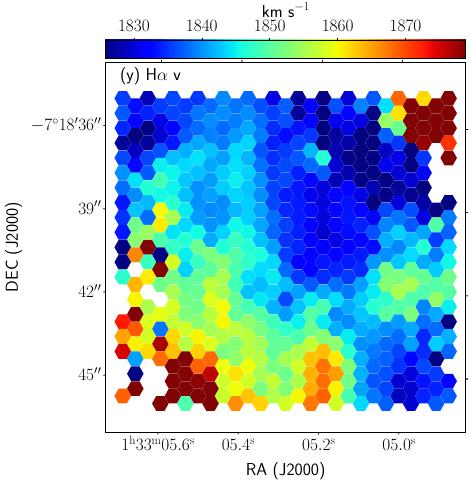}
	\includegraphics[clip, width=0.24\linewidth]{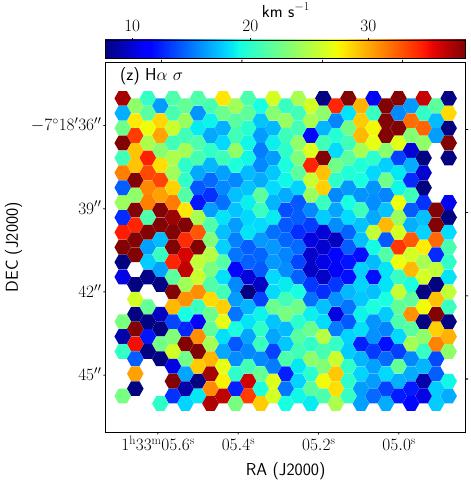}
	\includegraphics[clip, width=0.24\linewidth]{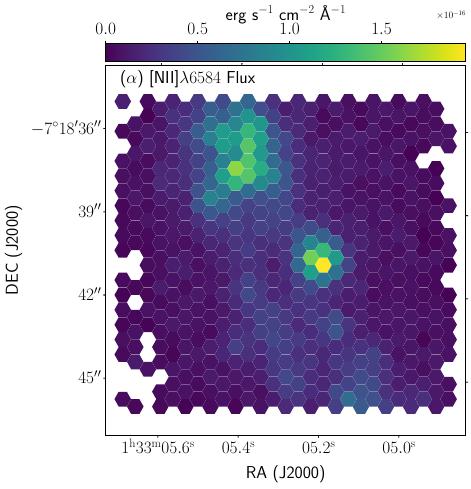}
	\includegraphics[clip, width=0.24\linewidth]{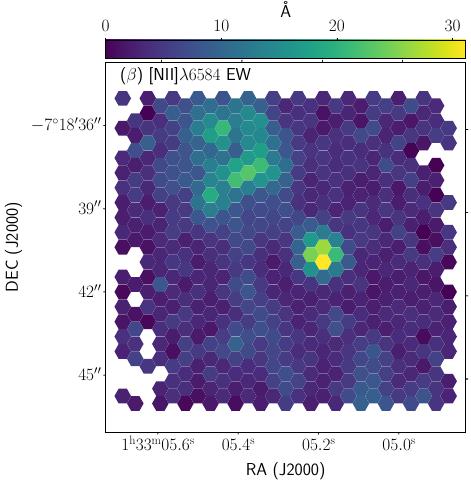}
	\includegraphics[clip, width=0.24\linewidth]{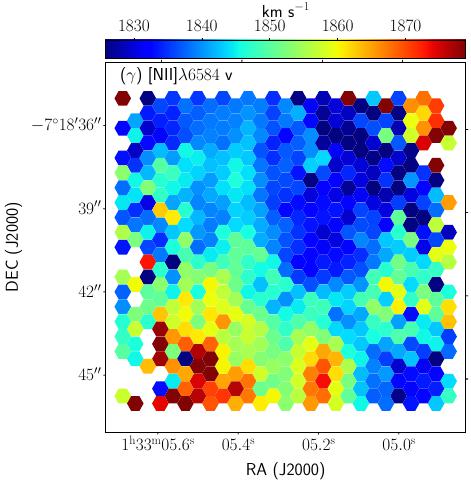}
	\includegraphics[clip, width=0.24\linewidth]{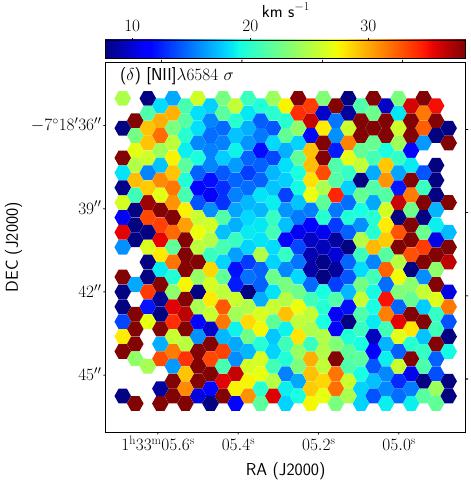}
	\includegraphics[clip, width=0.24\linewidth]{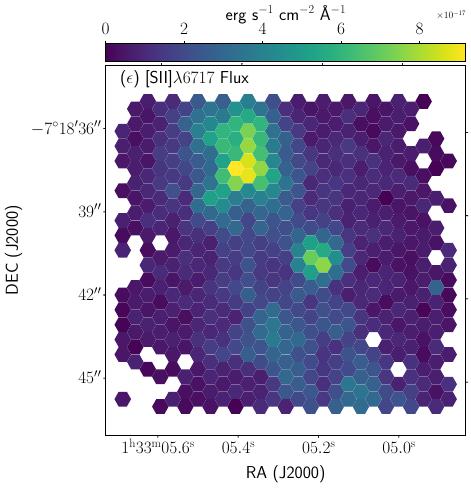}
	\includegraphics[clip, width=0.24\linewidth]{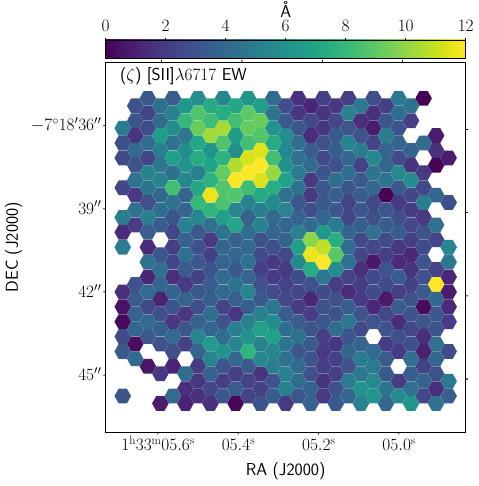}
	\includegraphics[clip, width=0.24\linewidth]{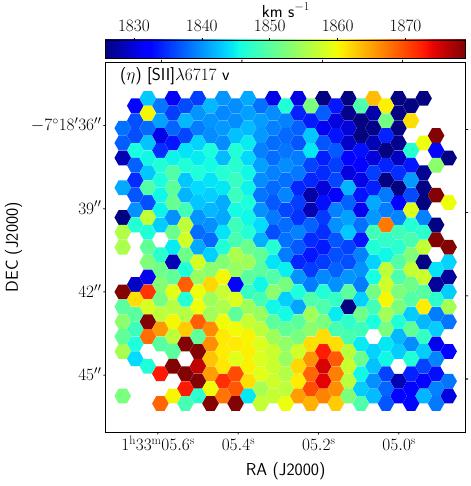}
	\includegraphics[clip, width=0.24\linewidth]{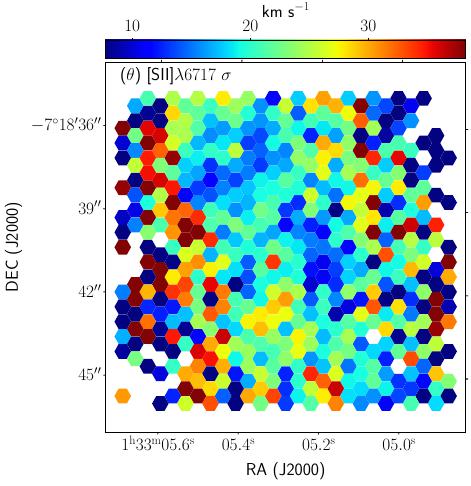}
	\includegraphics[clip, width=0.24\linewidth]{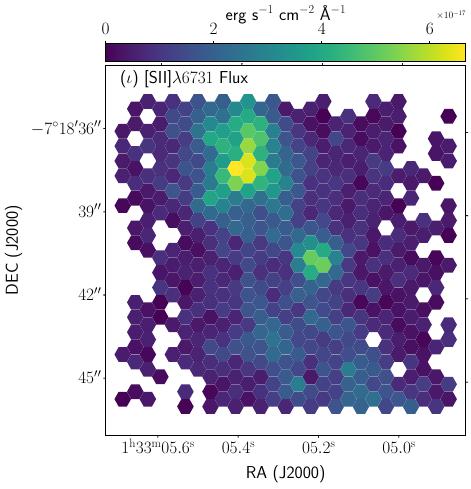}
	\includegraphics[clip, width=0.24\linewidth]{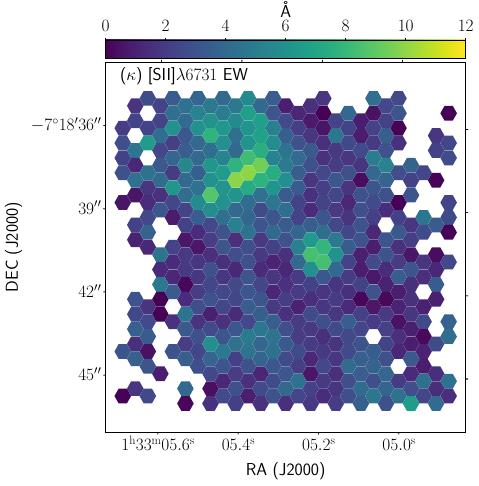}
	\includegraphics[clip, width=0.24\linewidth]{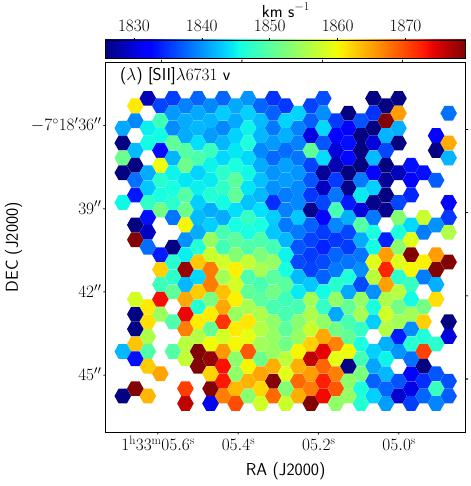}
	\includegraphics[clip, width=0.24\linewidth]{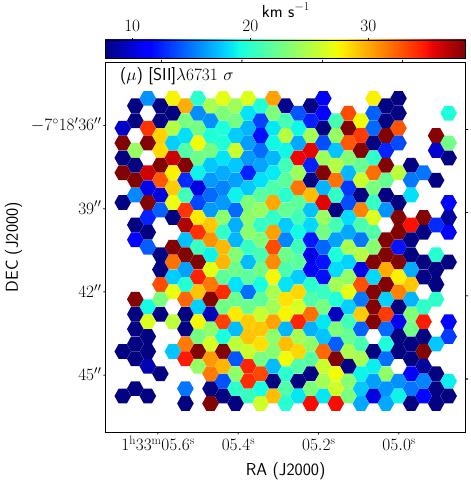}
	\caption{(cont.) NGC~0600 card.}
	\label{fig:NGC0600_card_2}
\end{figure*}

\begin{figure*}[h]
	\centering
	\includegraphics[clip, width=0.35\linewidth]{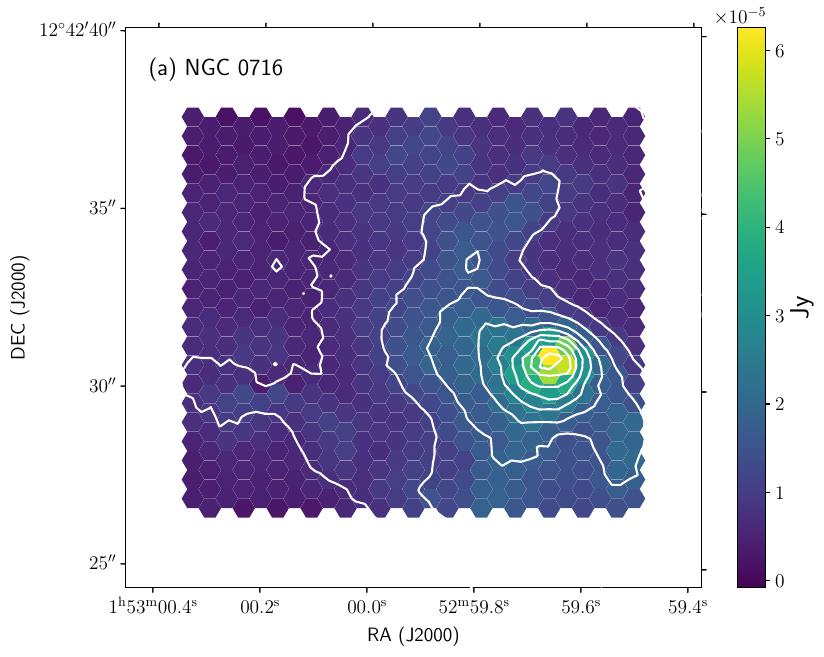}
	\includegraphics[clip, width=0.6\linewidth]{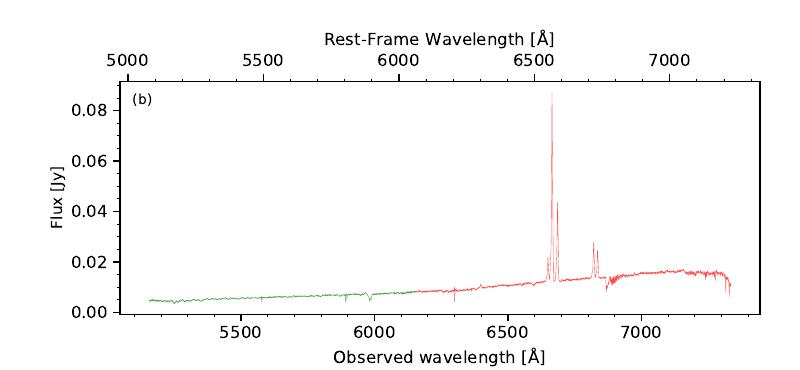}
	\includegraphics[clip, width=0.24\linewidth]{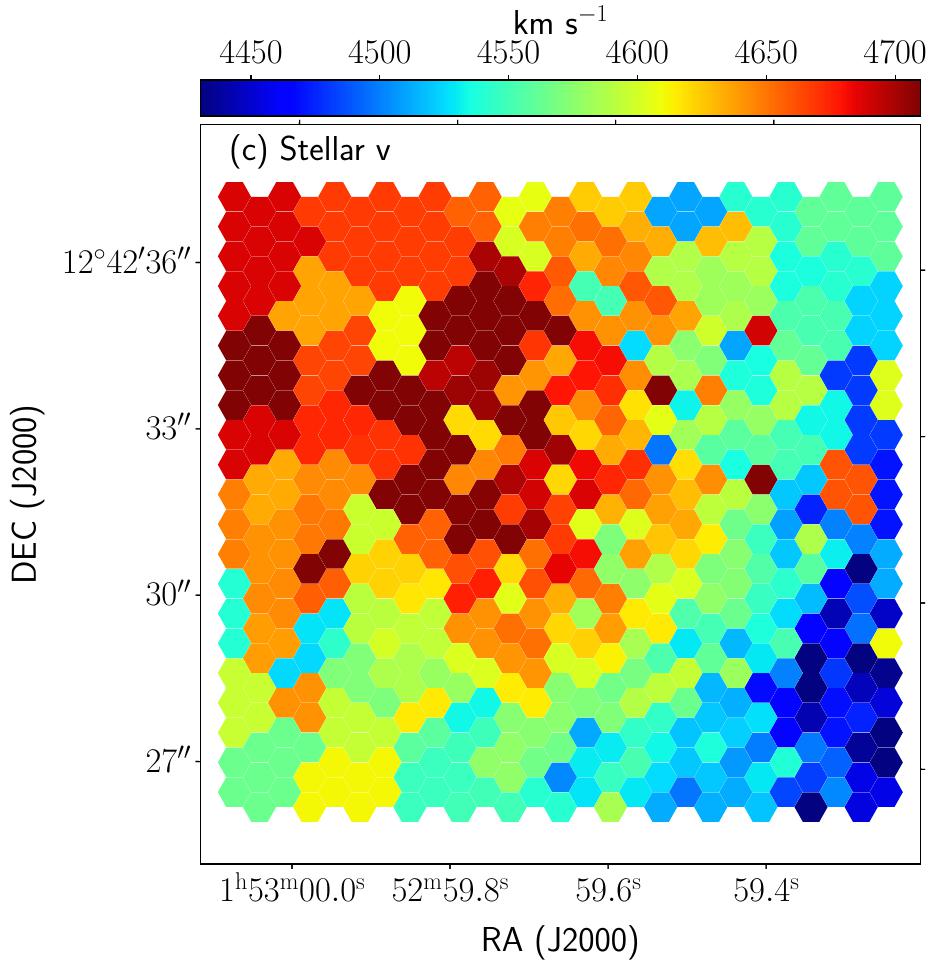}
	\includegraphics[clip, width=0.24\linewidth]{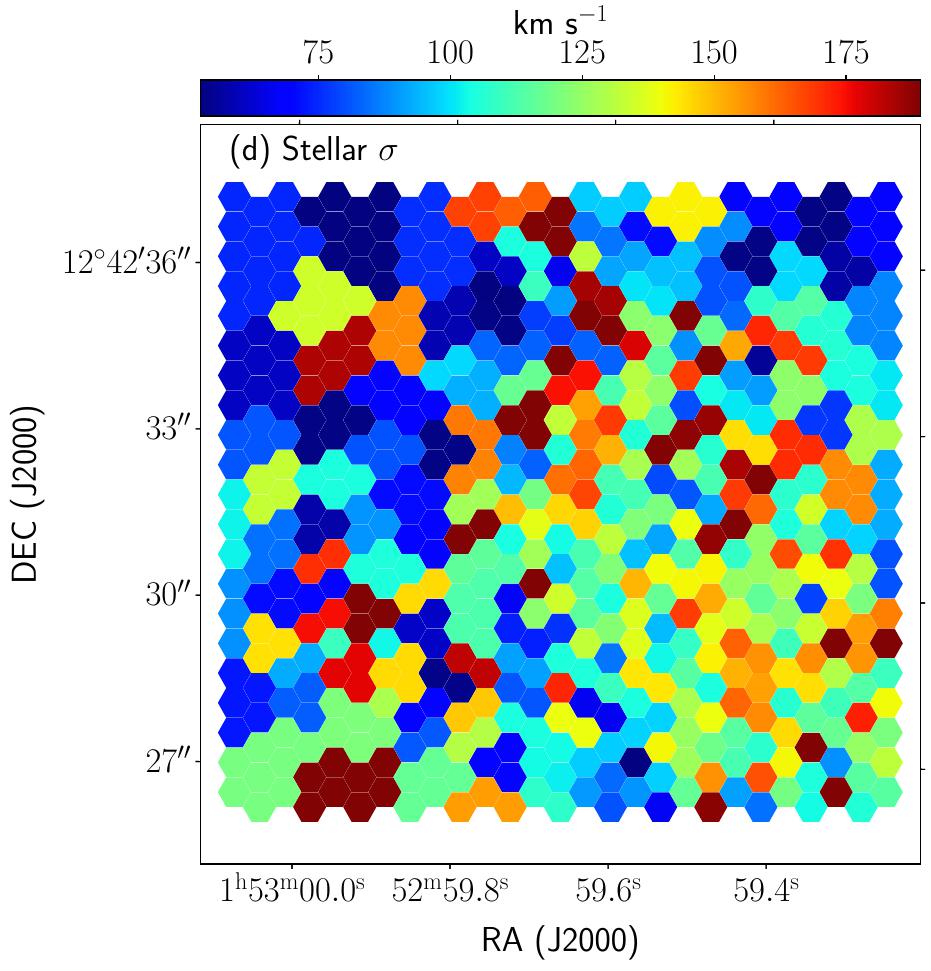}
	\includegraphics[clip, width=0.24\linewidth]{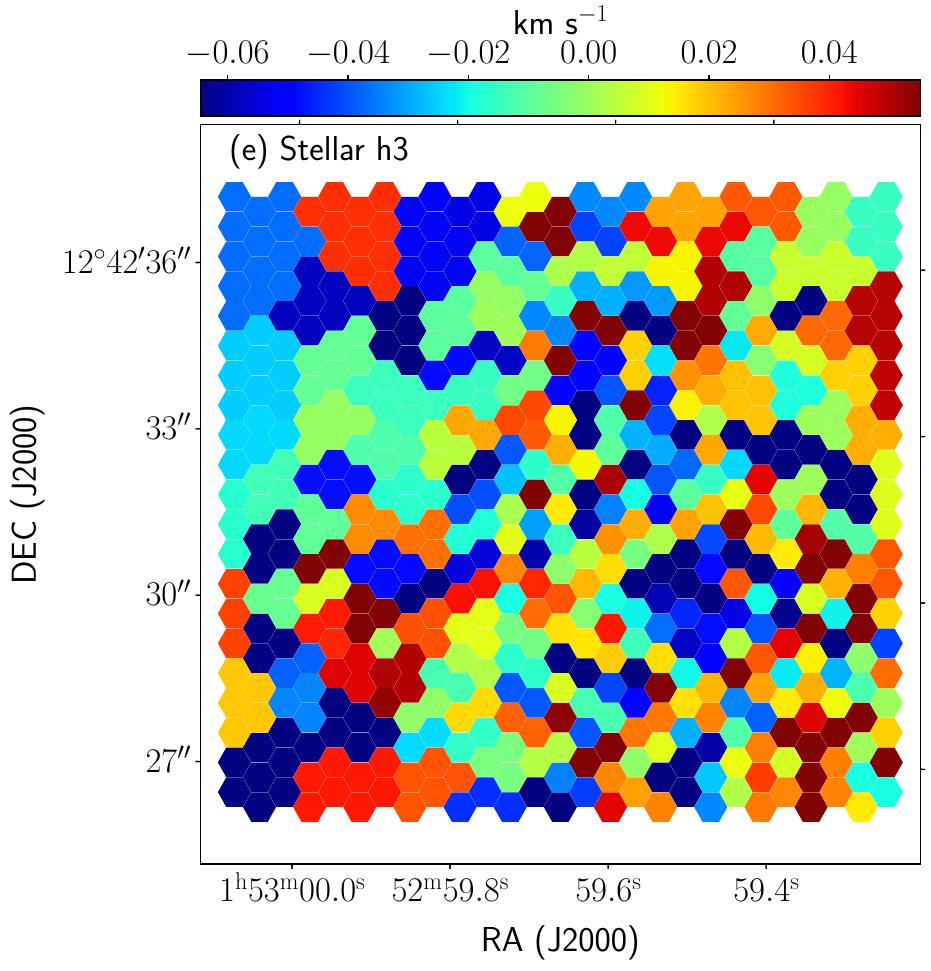}
	\includegraphics[clip, width=0.24\linewidth]{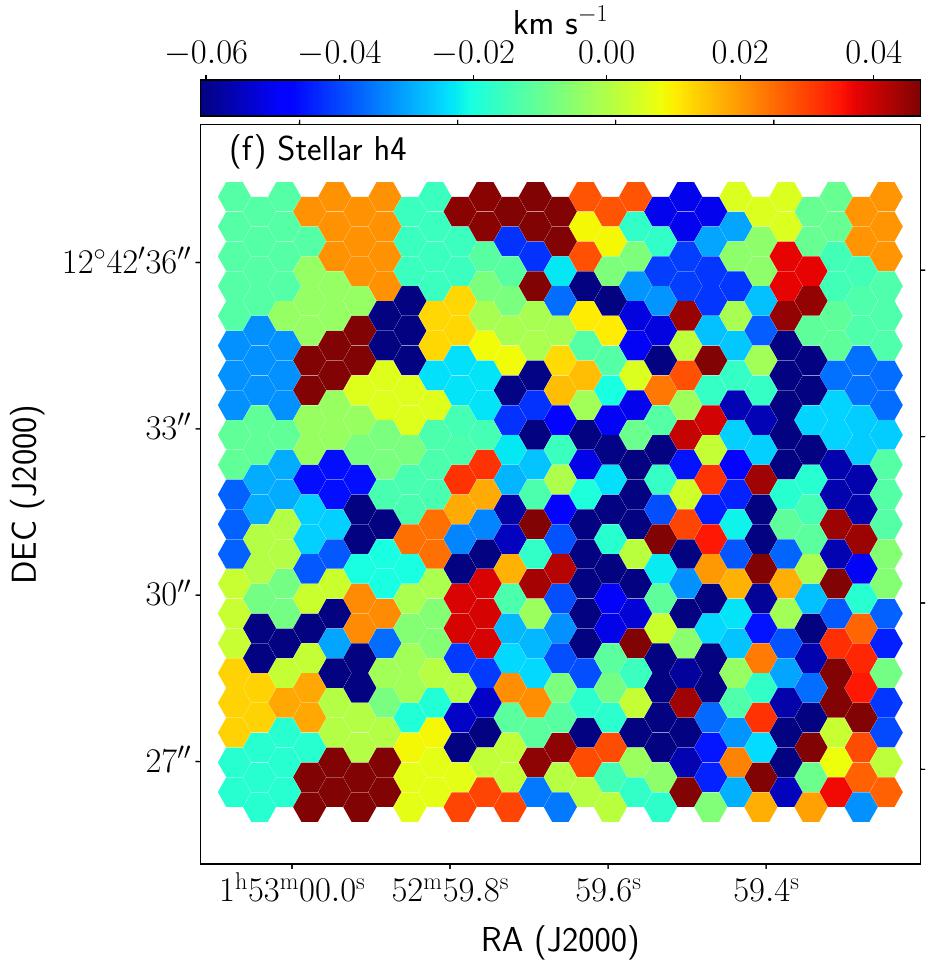}
	
	\vspace{8.8cm}
	
	\includegraphics[clip, width=0.24\linewidth]{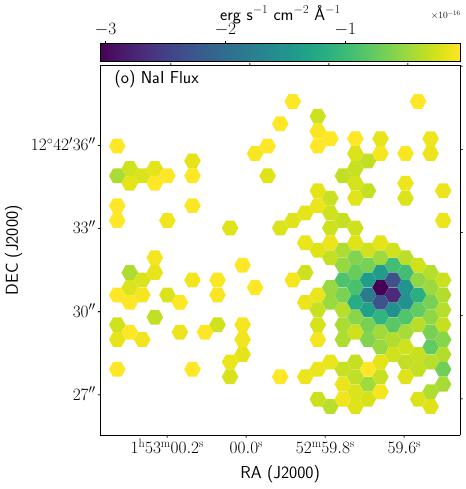}
	\includegraphics[clip, width=0.24\linewidth]{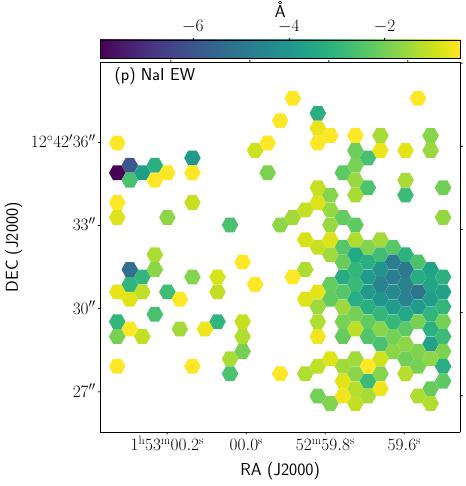}
	\includegraphics[clip, width=0.24\linewidth]{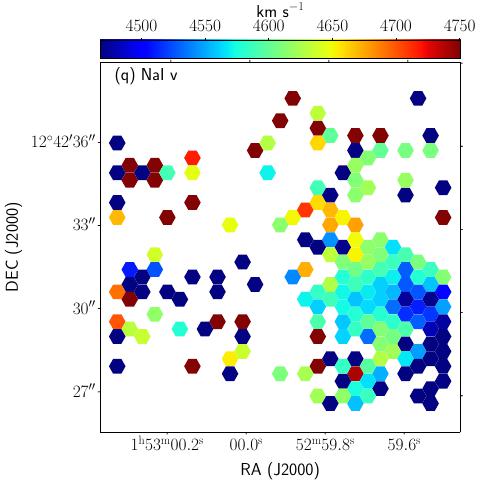}
	\includegraphics[clip, width=0.24\linewidth]{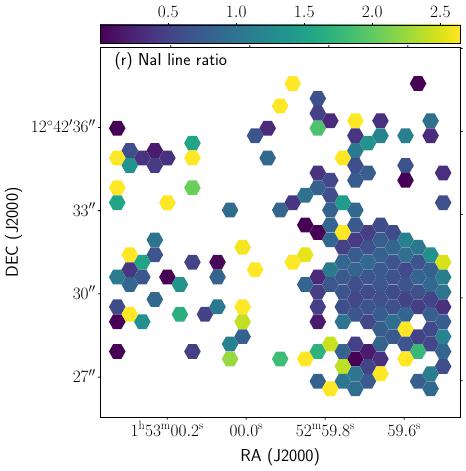}
	\caption{NGC~0716 card.}
	\label{fig:NGC0716_card_1}
\end{figure*}
\addtocounter{figure}{-1}
\begin{figure*}[h]
	\centering
	\includegraphics[clip, width=0.24\linewidth]{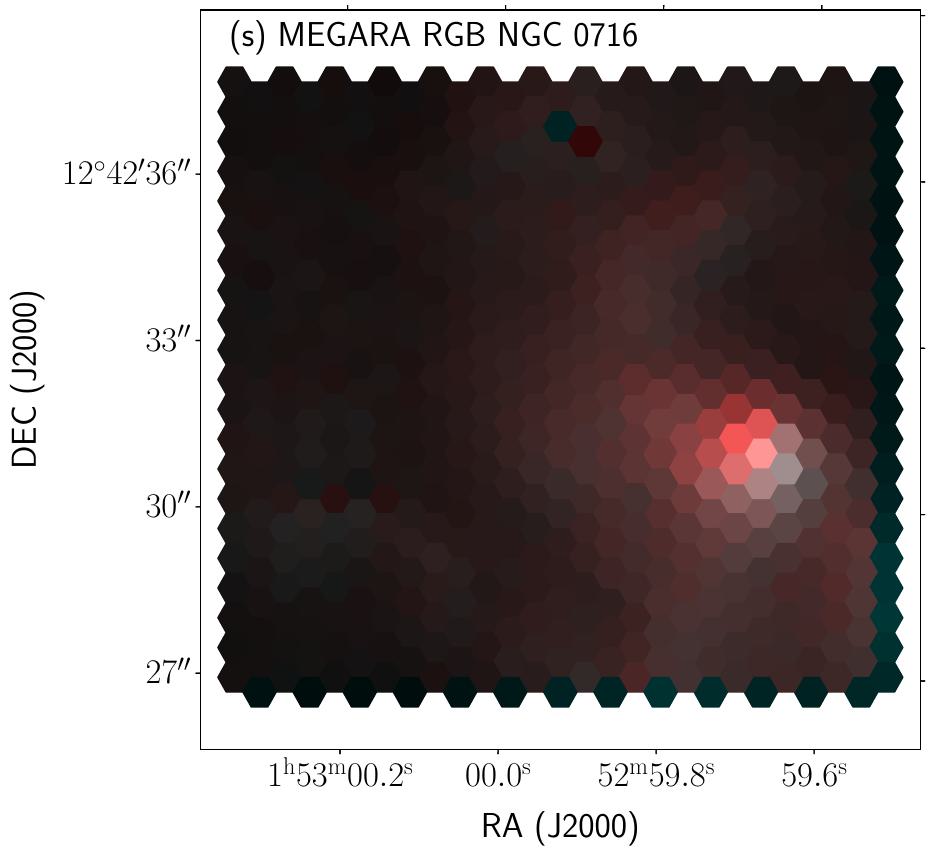}
	\hspace{4,4cm}
	\includegraphics[clip, width=0.24\linewidth]{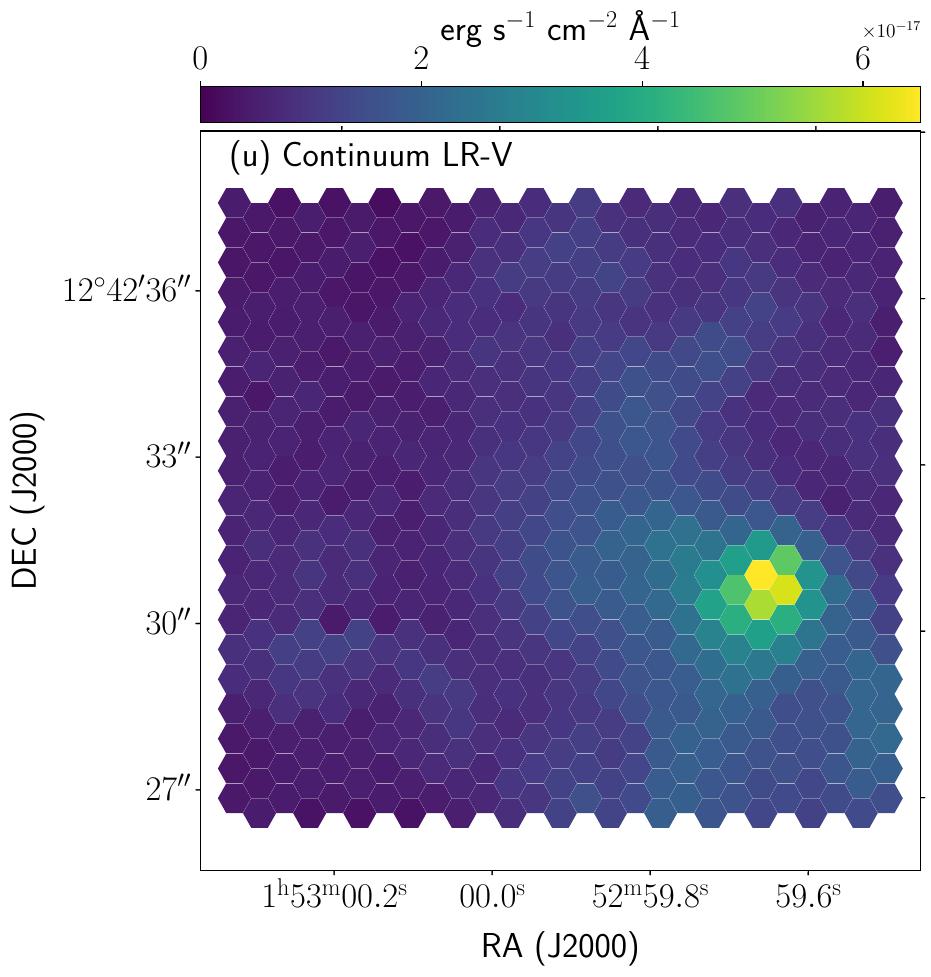}
	\includegraphics[clip, width=0.24\linewidth]{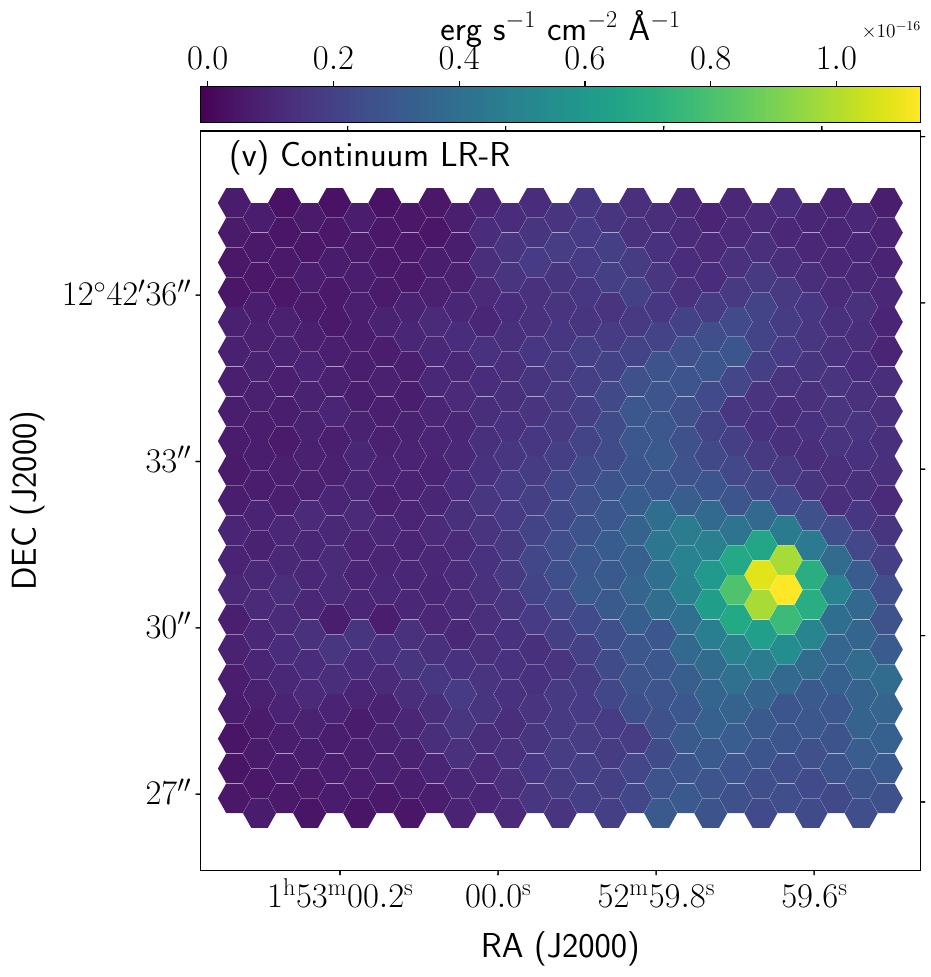}
	\includegraphics[clip, width=0.24\linewidth]{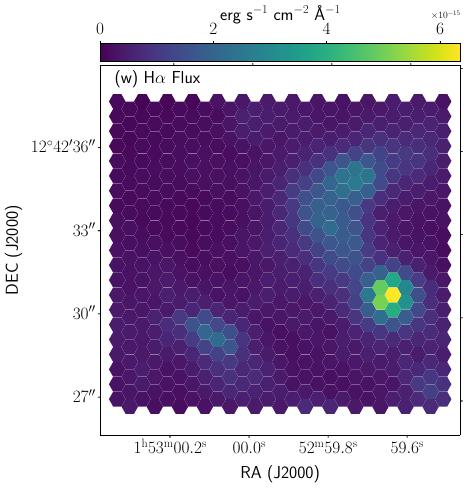}
	\includegraphics[clip, width=0.24\linewidth]{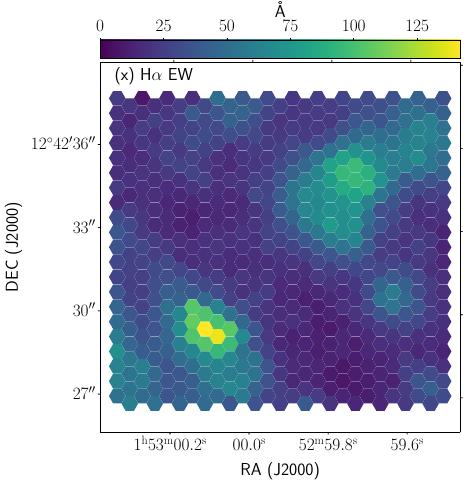}
	\includegraphics[clip, width=0.24\linewidth]{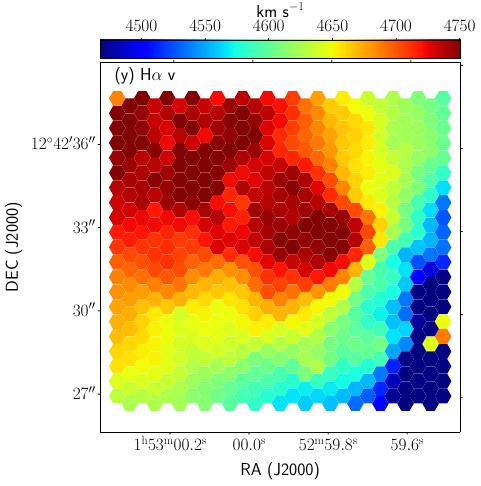}
	\includegraphics[clip, width=0.24\linewidth]{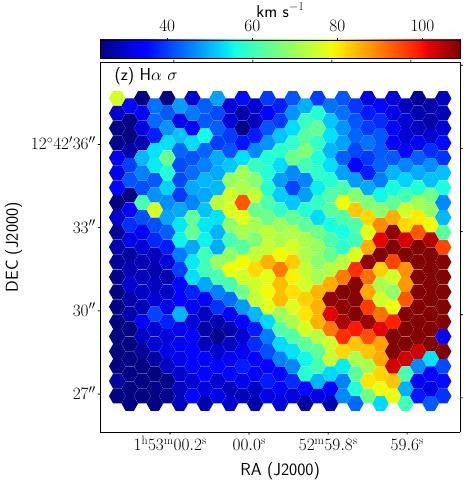}
	\includegraphics[clip, width=0.24\linewidth]{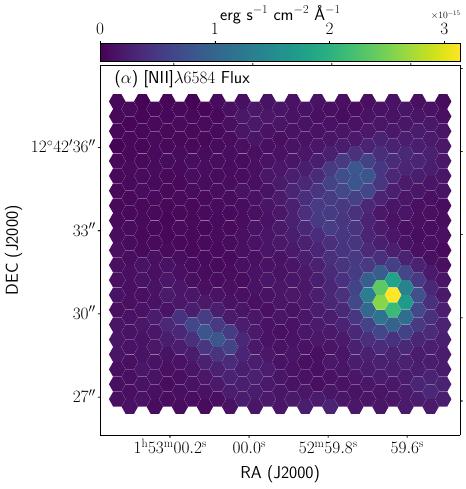}
	\includegraphics[clip, width=0.24\linewidth]{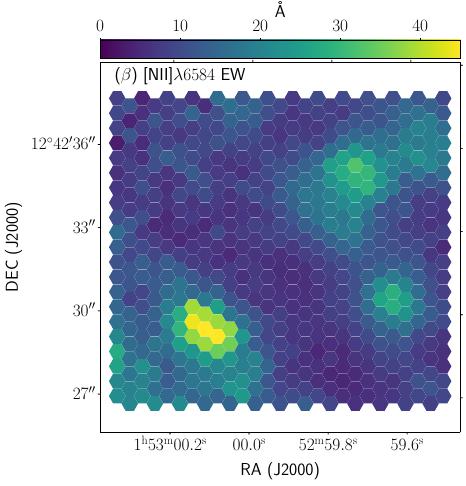}
	\includegraphics[clip, width=0.24\linewidth]{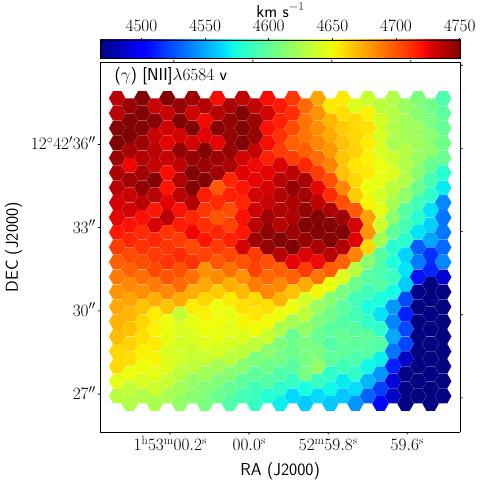}
	\includegraphics[clip, width=0.24\linewidth]{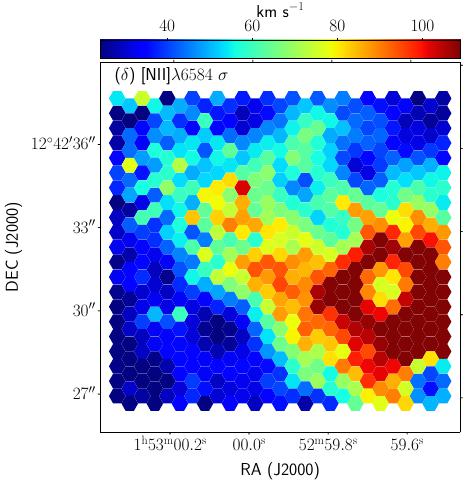}
	\includegraphics[clip, width=0.24\linewidth]{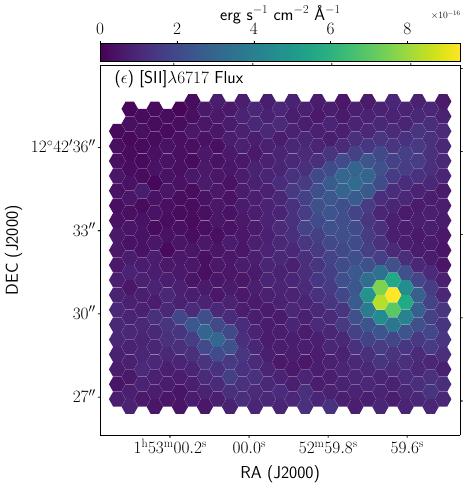}
	\includegraphics[clip, width=0.24\linewidth]{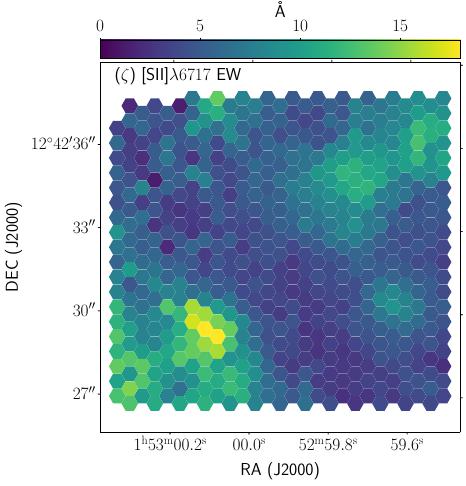}
	\includegraphics[clip, width=0.24\linewidth]{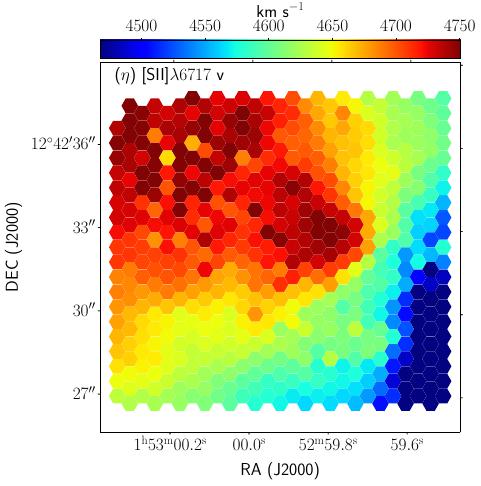}
	\includegraphics[clip, width=0.24\linewidth]{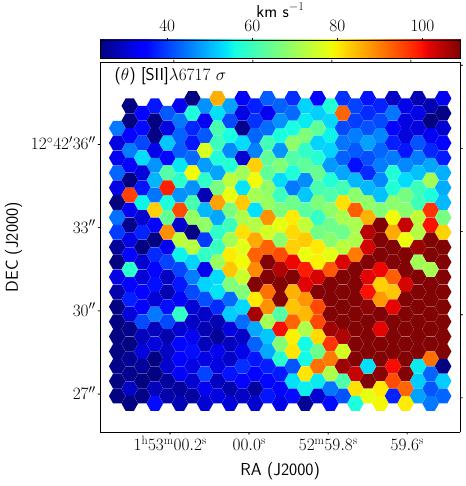}
	\includegraphics[clip, width=0.24\linewidth]{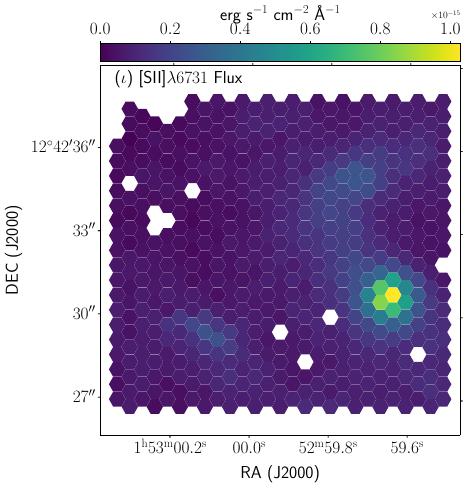}
	\includegraphics[clip, width=0.24\linewidth]{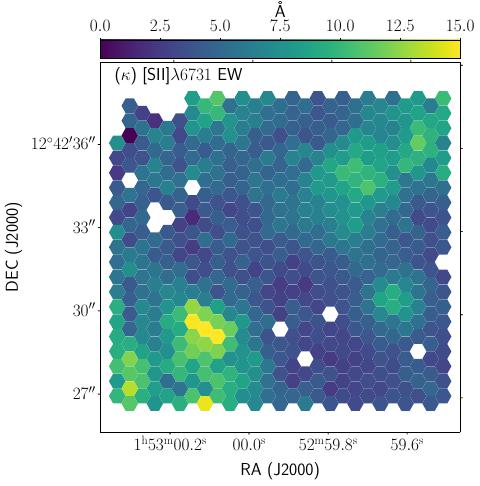}
	\includegraphics[clip, width=0.24\linewidth]{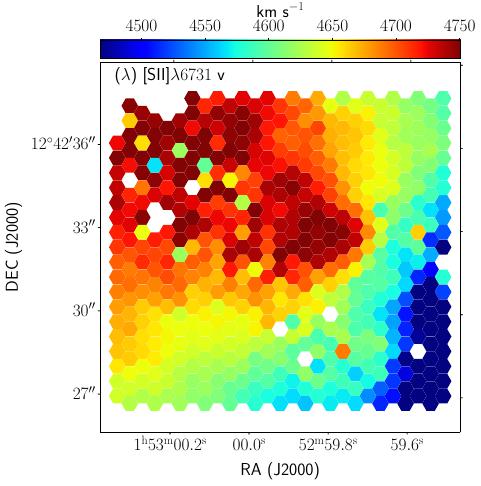}
	\includegraphics[clip, width=0.24\linewidth]{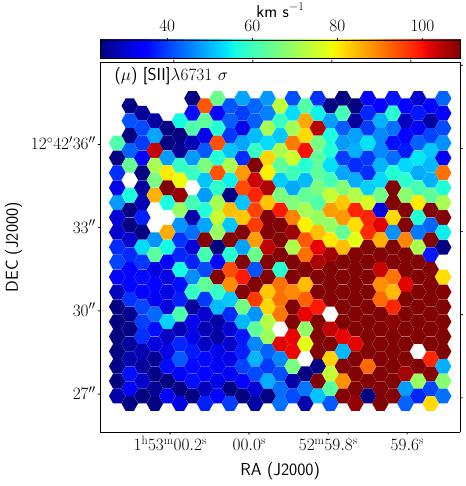}
	\caption{(cont.) NGC~0716 card.}
	\label{fig:NGC0716_card_2}
\end{figure*}

\begin{figure*}[h]
	\centering
	\includegraphics[clip, width=0.35\linewidth]{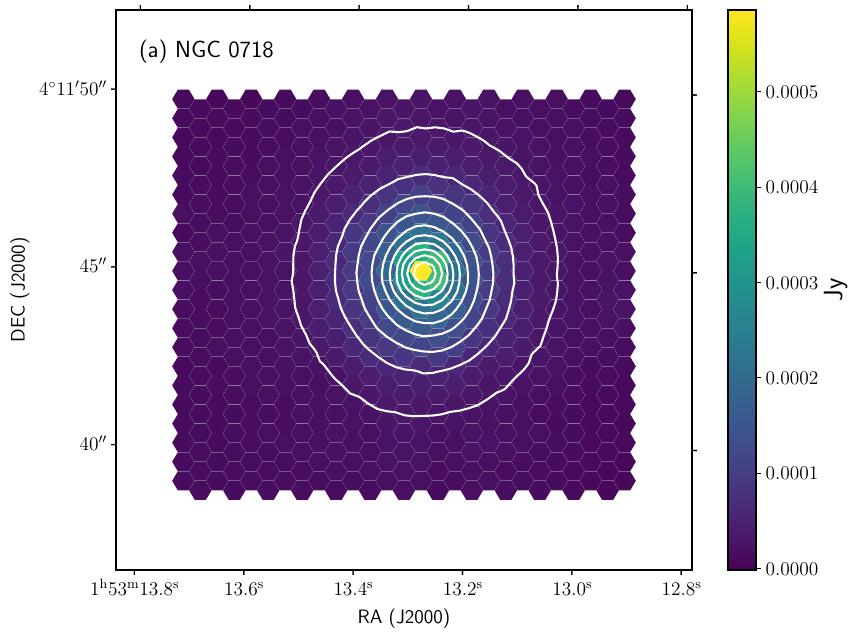}
	\includegraphics[clip, width=0.6\linewidth]{Figuras/Concatenated_Spectra/final_rss_combinado_NGC0718.jpg}
	\includegraphics[clip, width=0.24\linewidth]{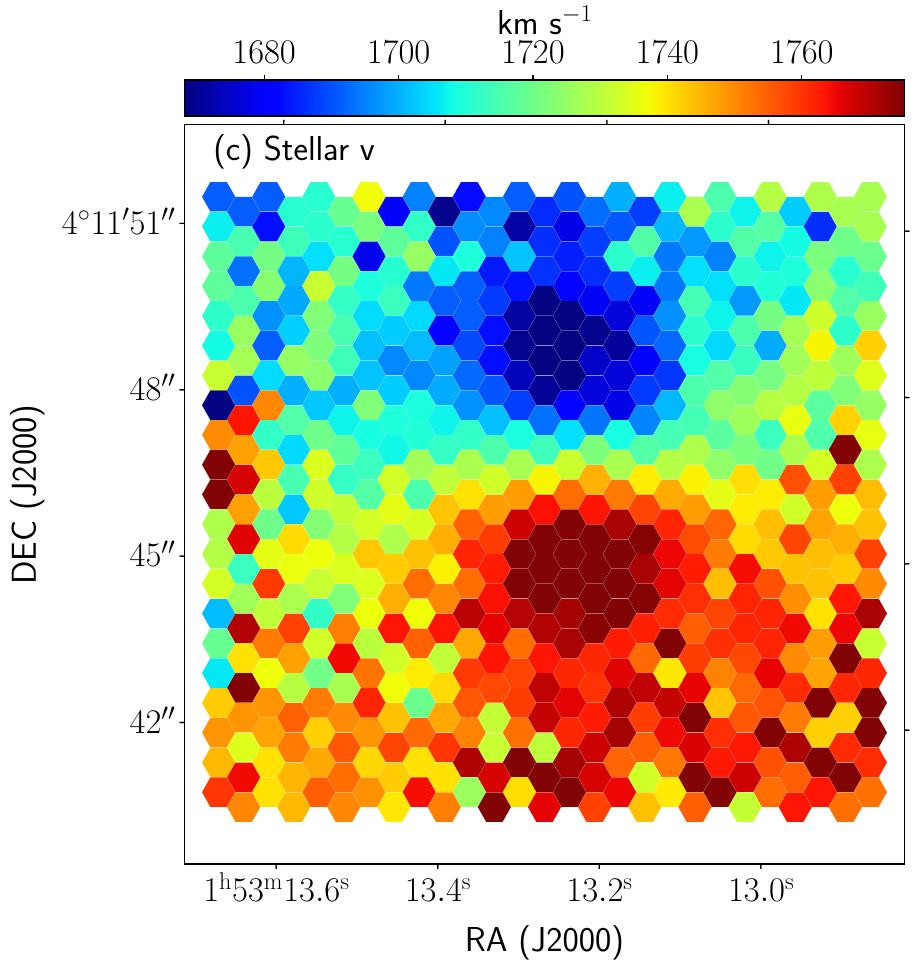}
	\includegraphics[clip, width=0.24\linewidth]{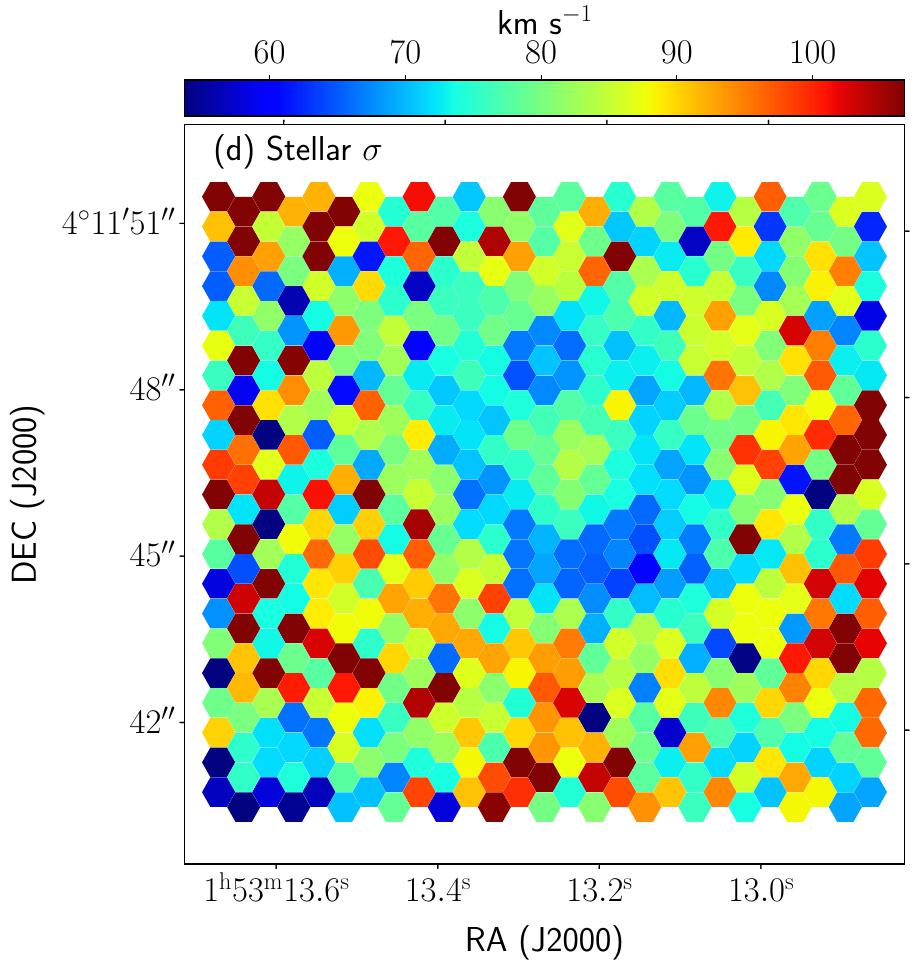}
	\includegraphics[clip, width=0.24\linewidth]{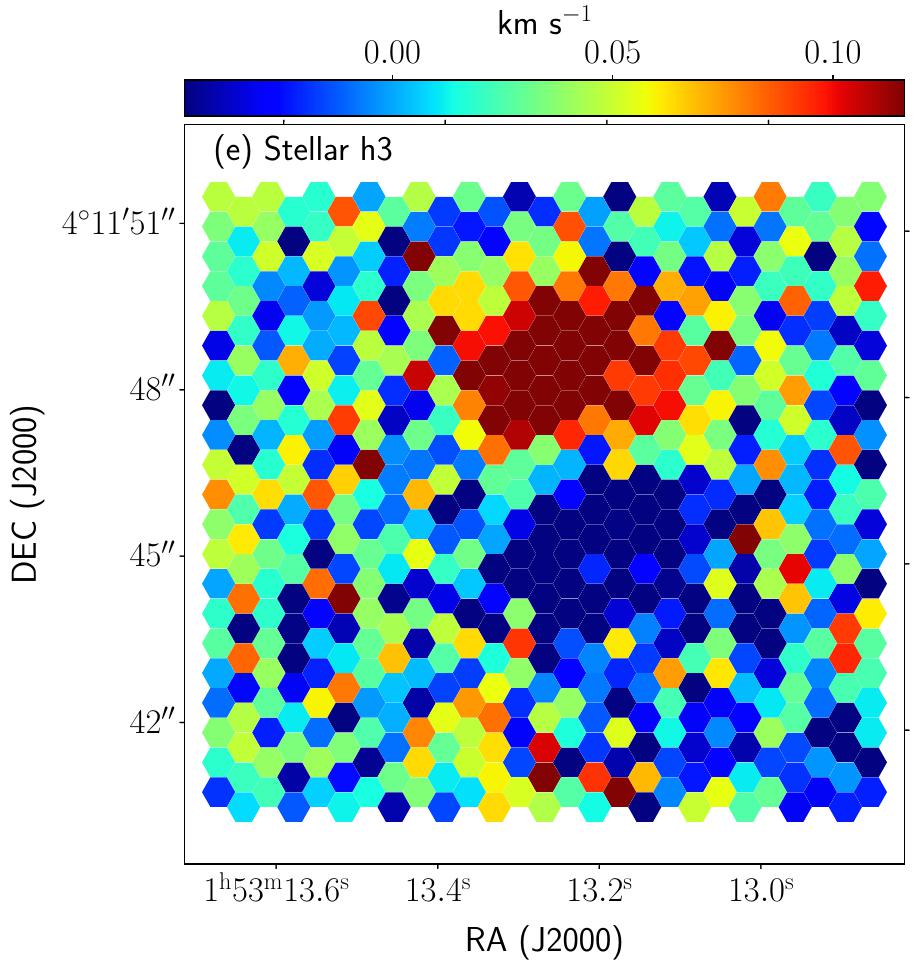}
	\includegraphics[clip, width=0.24\linewidth]{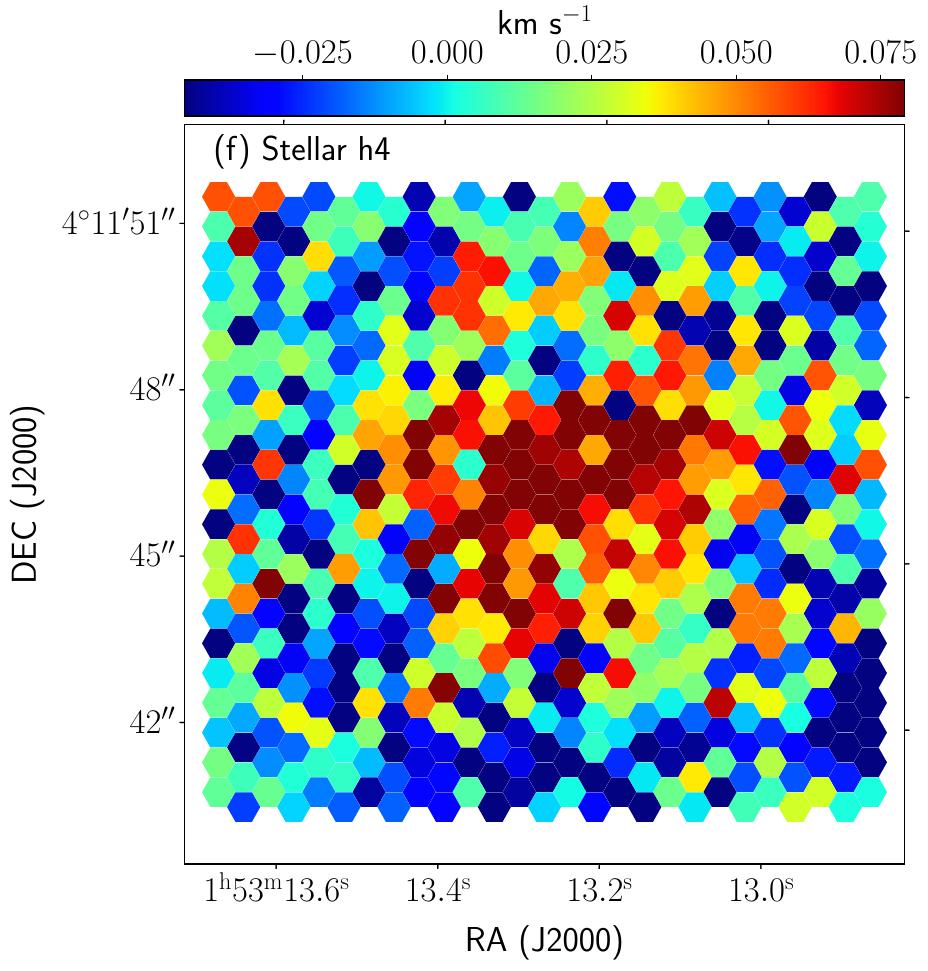}
	\includegraphics[clip, width=0.24\linewidth]{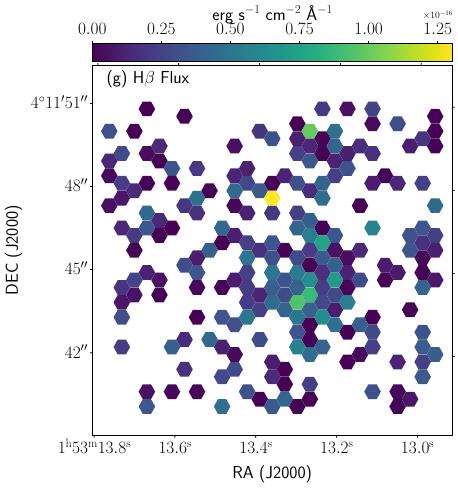}
	\includegraphics[clip, width=0.24\linewidth]{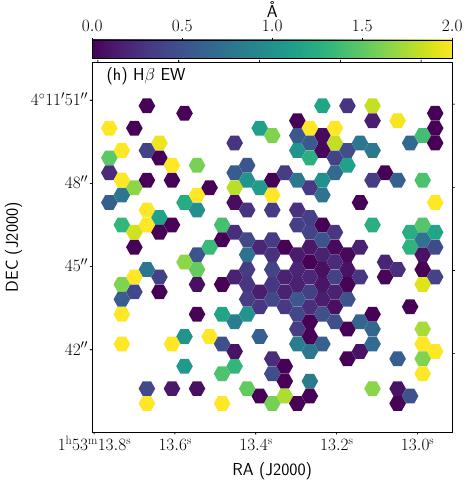}
	\includegraphics[clip, width=0.24\linewidth]{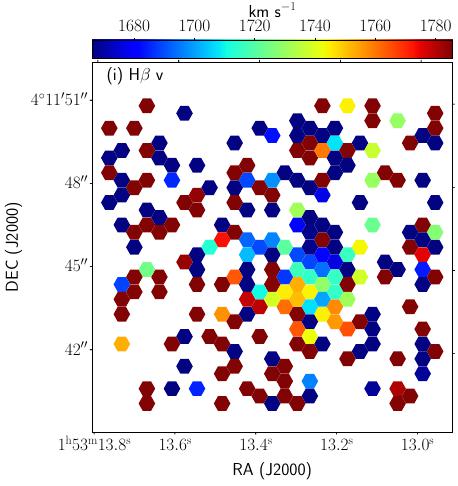}
	\includegraphics[clip, width=0.24\linewidth]{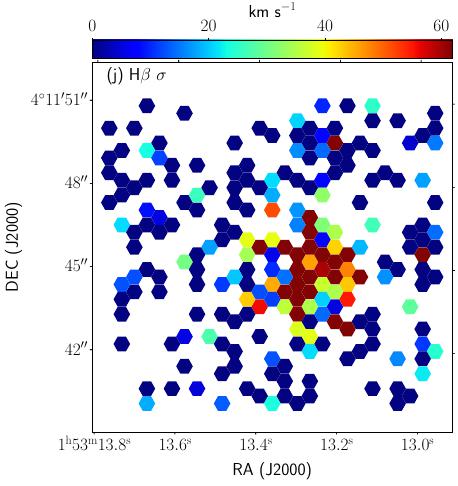}
	\includegraphics[clip, width=0.24\linewidth]{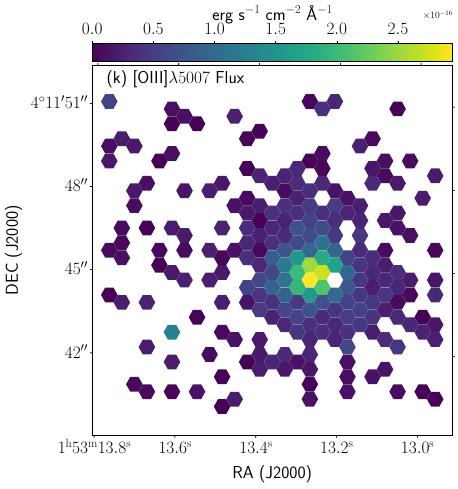}
	\includegraphics[clip, width=0.24\linewidth]{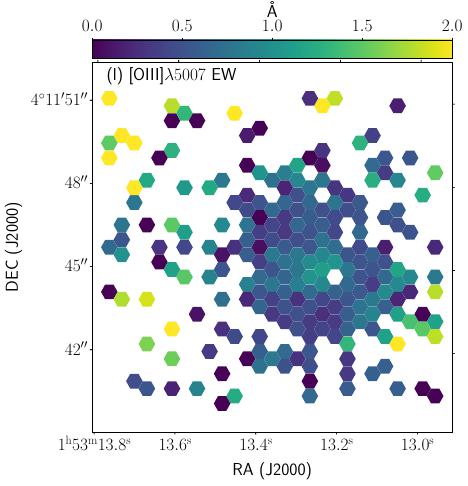}
	\includegraphics[clip, width=0.24\linewidth]{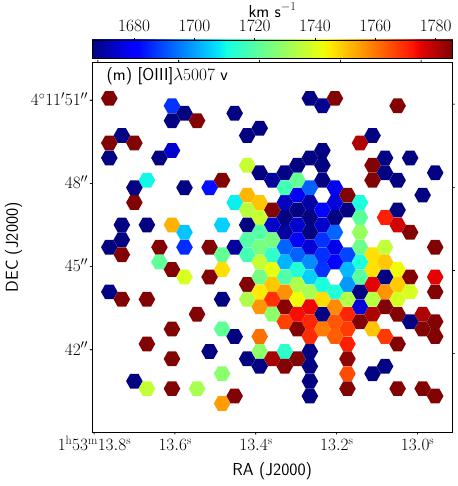}
	\includegraphics[clip, width=0.24\linewidth]{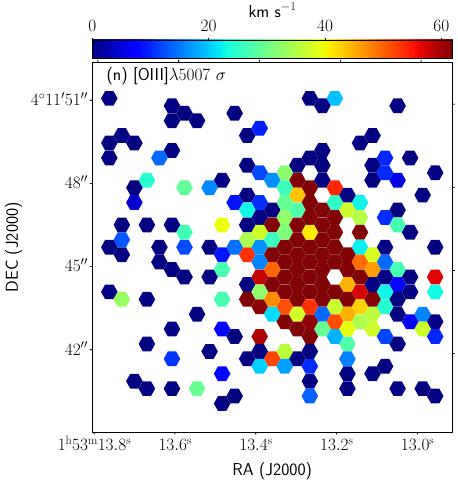}
	\vspace{5cm}
	\caption{NGC~0718 card.}
	\label{fig:NGC0718_card_1}
\end{figure*}
\addtocounter{figure}{-1}
\begin{figure*}[h]
	\centering
	\includegraphics[clip, width=0.24\linewidth]{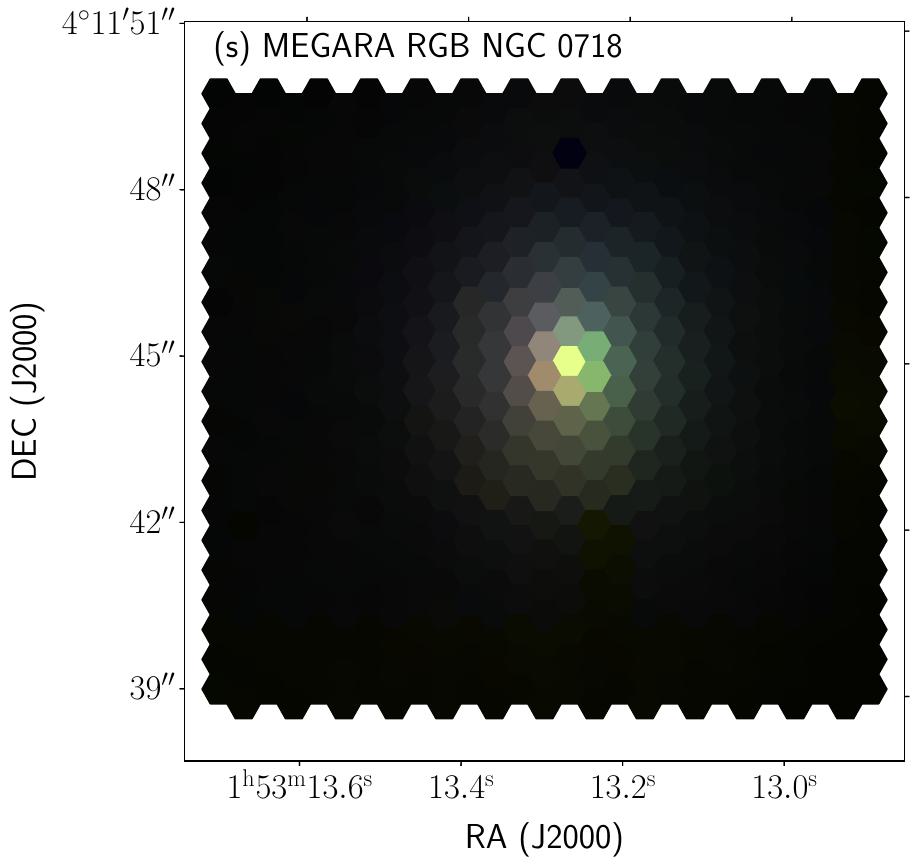}
	\includegraphics[clip, width=0.24\linewidth]{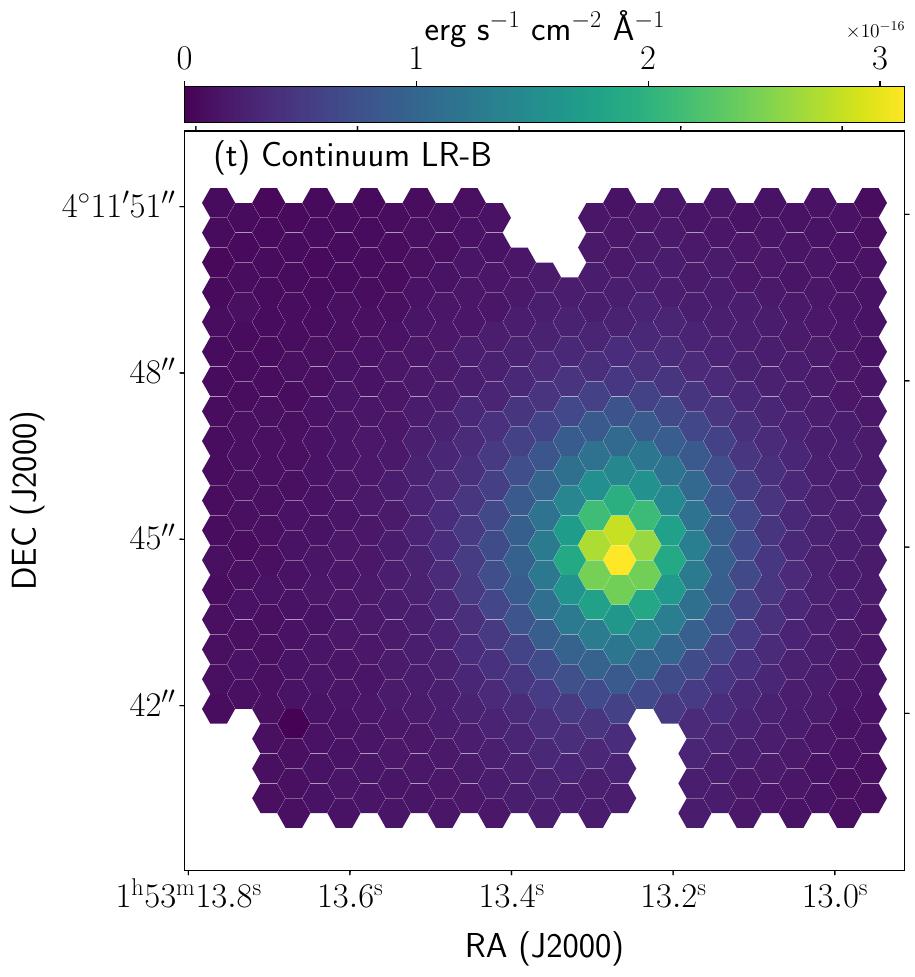}
	\includegraphics[clip, width=0.24\linewidth]{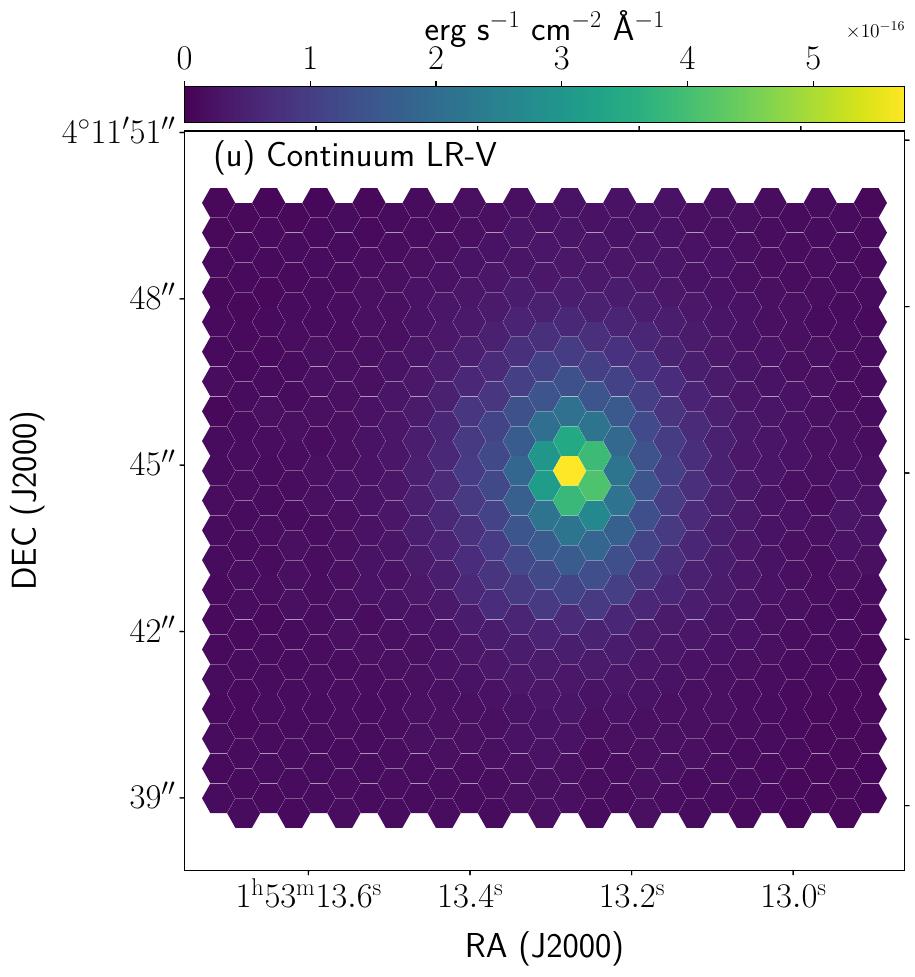}
	\includegraphics[clip, width=0.24\linewidth]{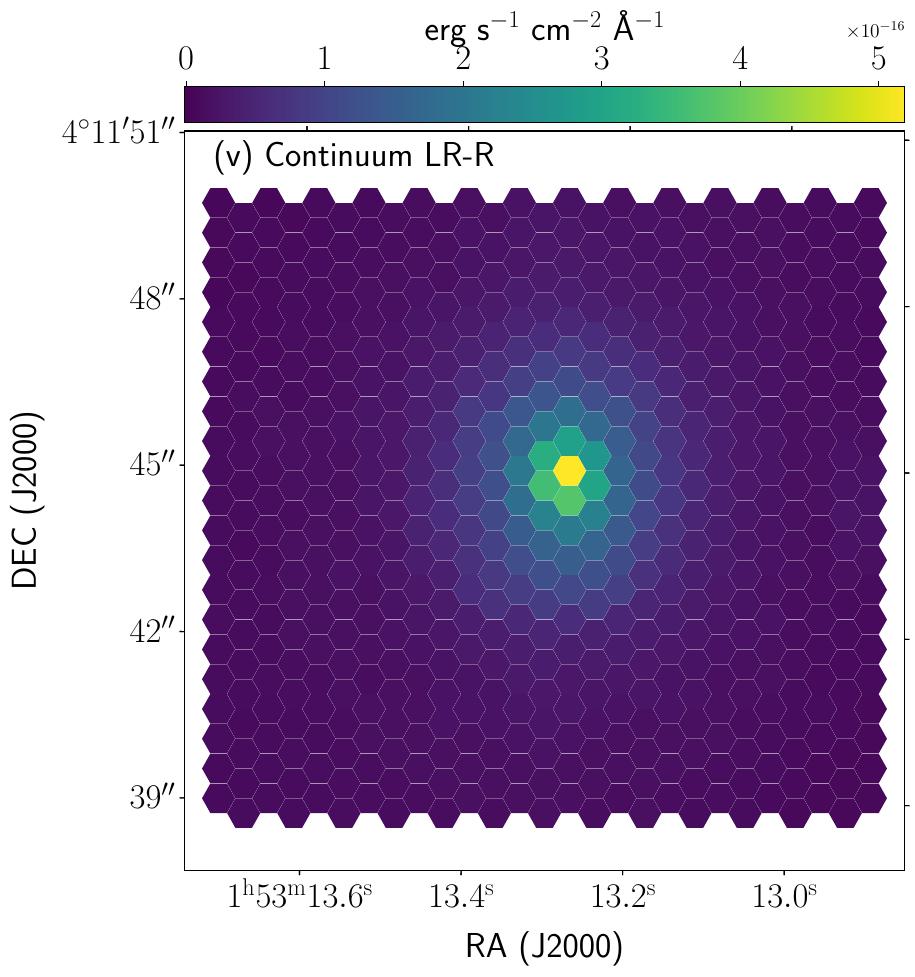}
	\includegraphics[clip, width=0.24\linewidth]{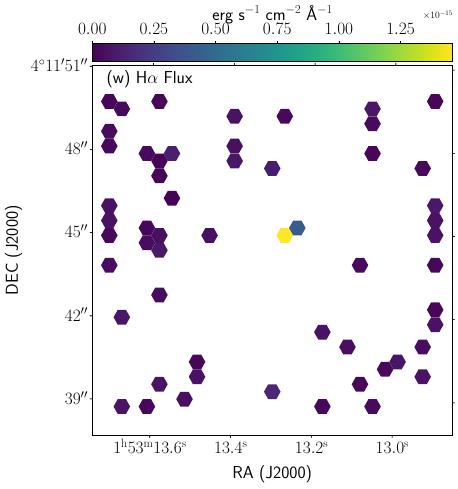}
	\includegraphics[clip, width=0.24\linewidth]{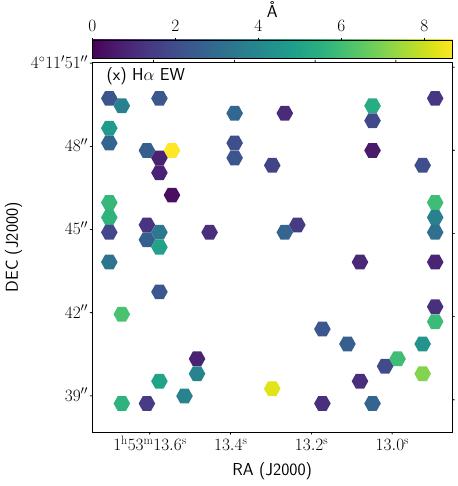}
	\includegraphics[clip, width=0.24\linewidth]{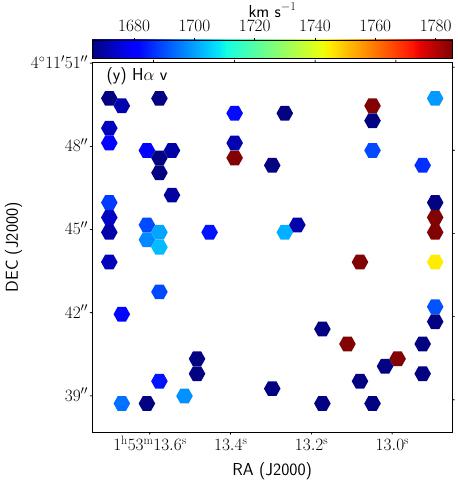}
	\includegraphics[clip, width=0.24\linewidth]{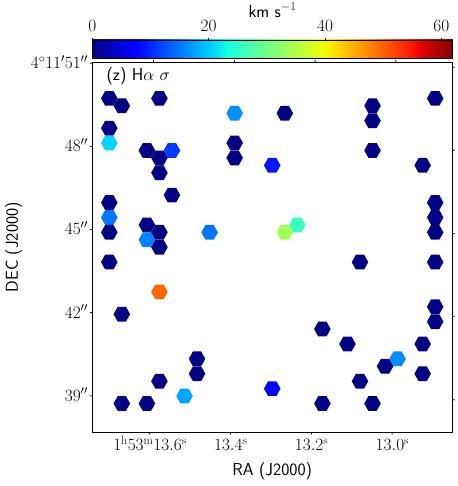}
	\includegraphics[clip, width=0.24\linewidth]{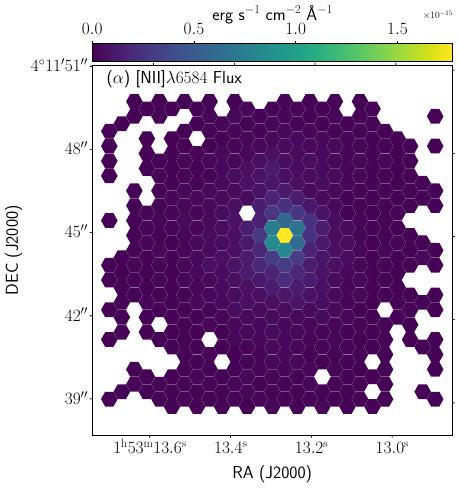}
	\includegraphics[clip, width=0.24\linewidth]{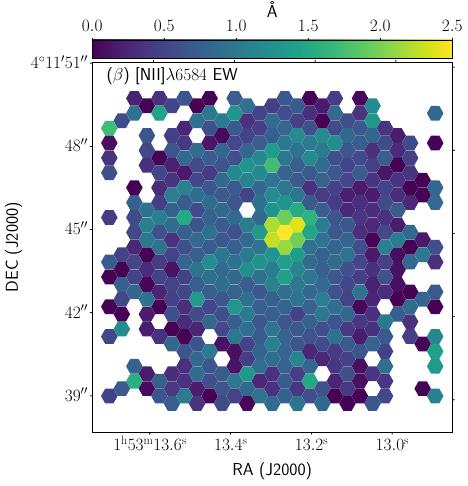}
	\includegraphics[clip, width=0.24\linewidth]{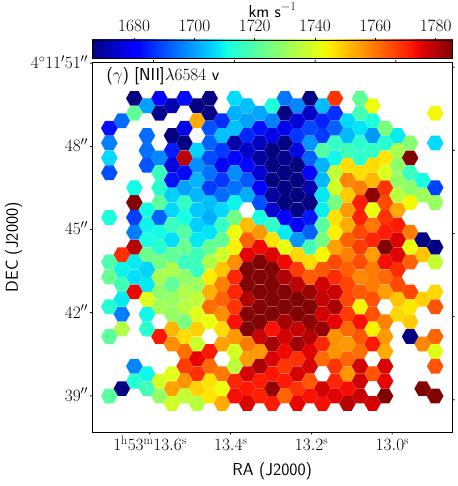}
	\includegraphics[clip, width=0.24\linewidth]{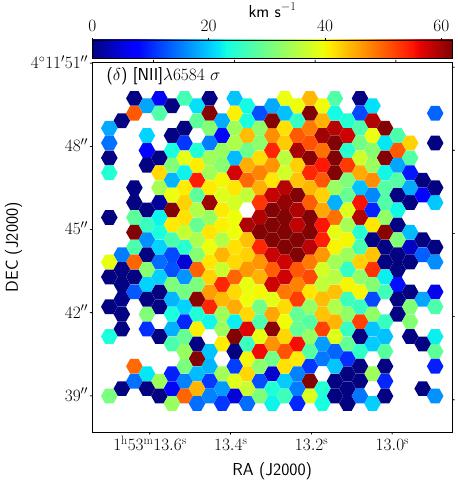}
	\includegraphics[clip, width=0.24\linewidth]{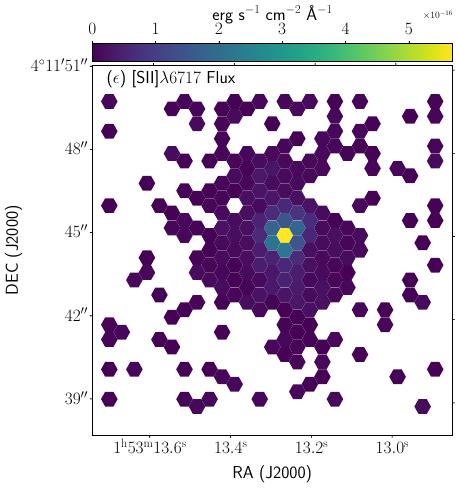}
	\includegraphics[clip, width=0.24\linewidth]{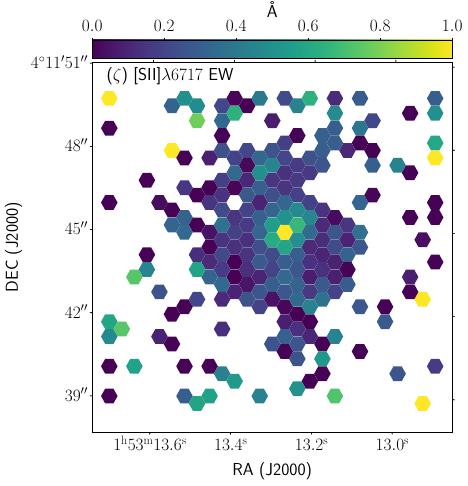}
	\includegraphics[clip, width=0.24\linewidth]{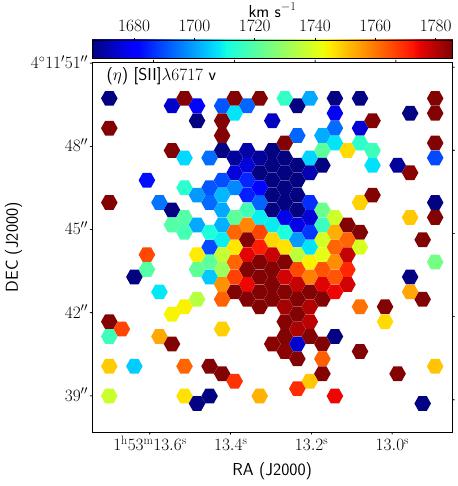}
	\includegraphics[clip, width=0.24\linewidth]{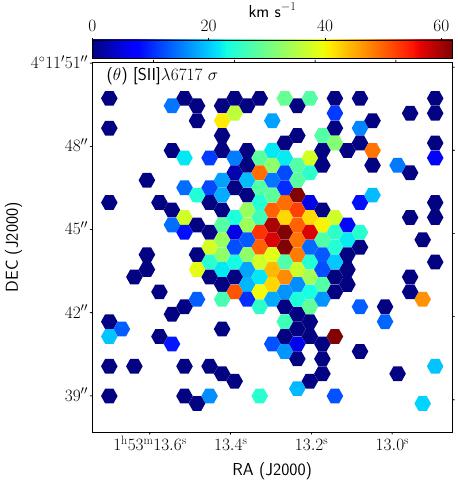}
	\includegraphics[clip, width=0.24\linewidth]{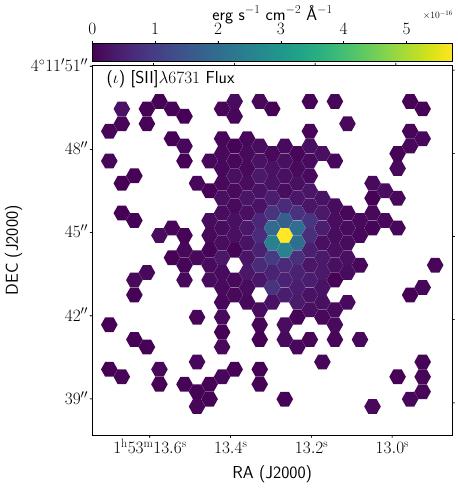}
	\includegraphics[clip, width=0.24\linewidth]{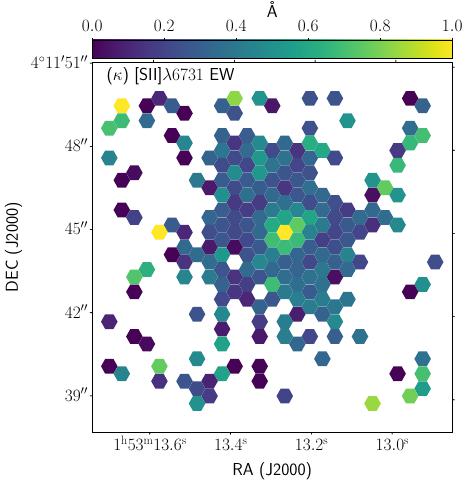}
	\includegraphics[clip, width=0.24\linewidth]{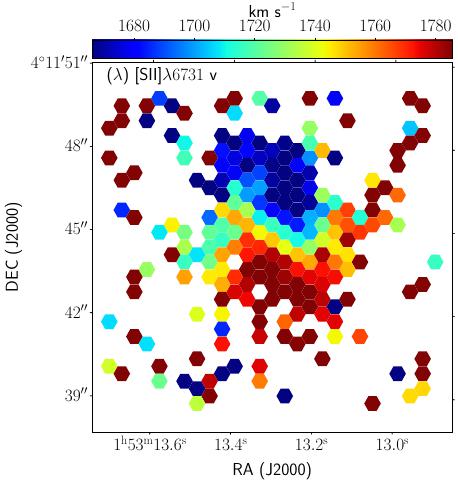}
	\includegraphics[clip, width=0.24\linewidth]{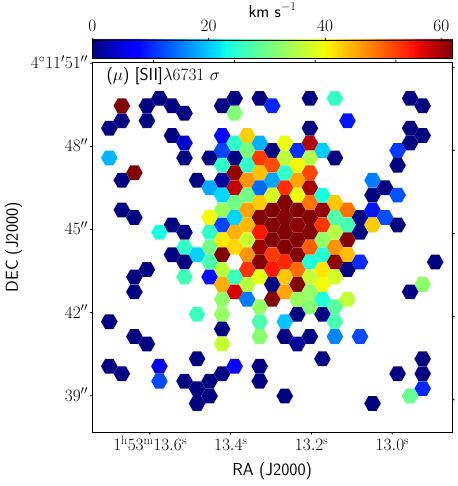}
	\caption{(cont.) NGC~0718 card.}
	\label{fig:NGC0718_card_2}
\end{figure*}

\begin{figure*}[h]
	\centering
	\includegraphics[clip, width=0.35\linewidth]{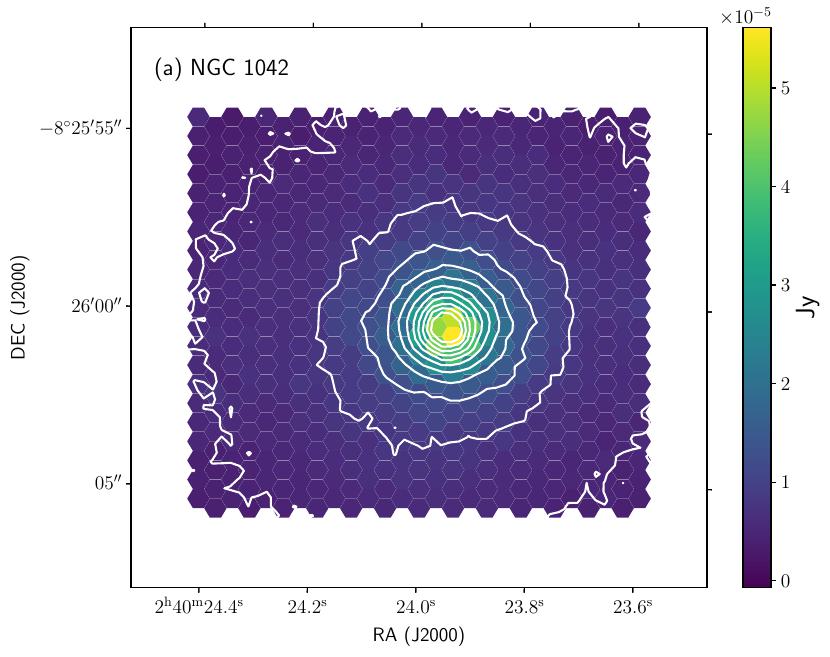}
	\includegraphics[clip, width=0.6\linewidth]{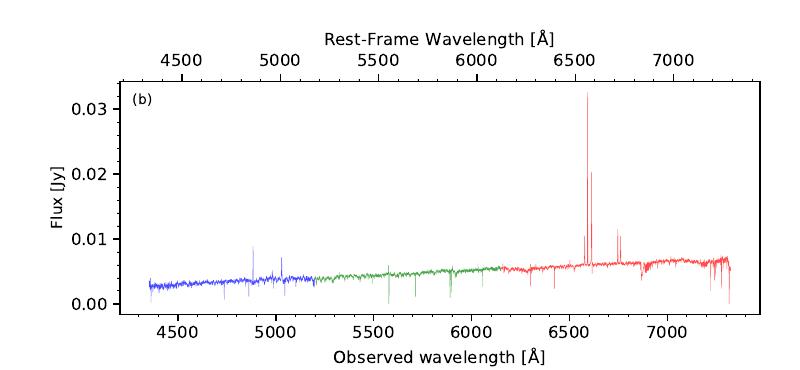}
	\includegraphics[clip, width=0.24\linewidth]{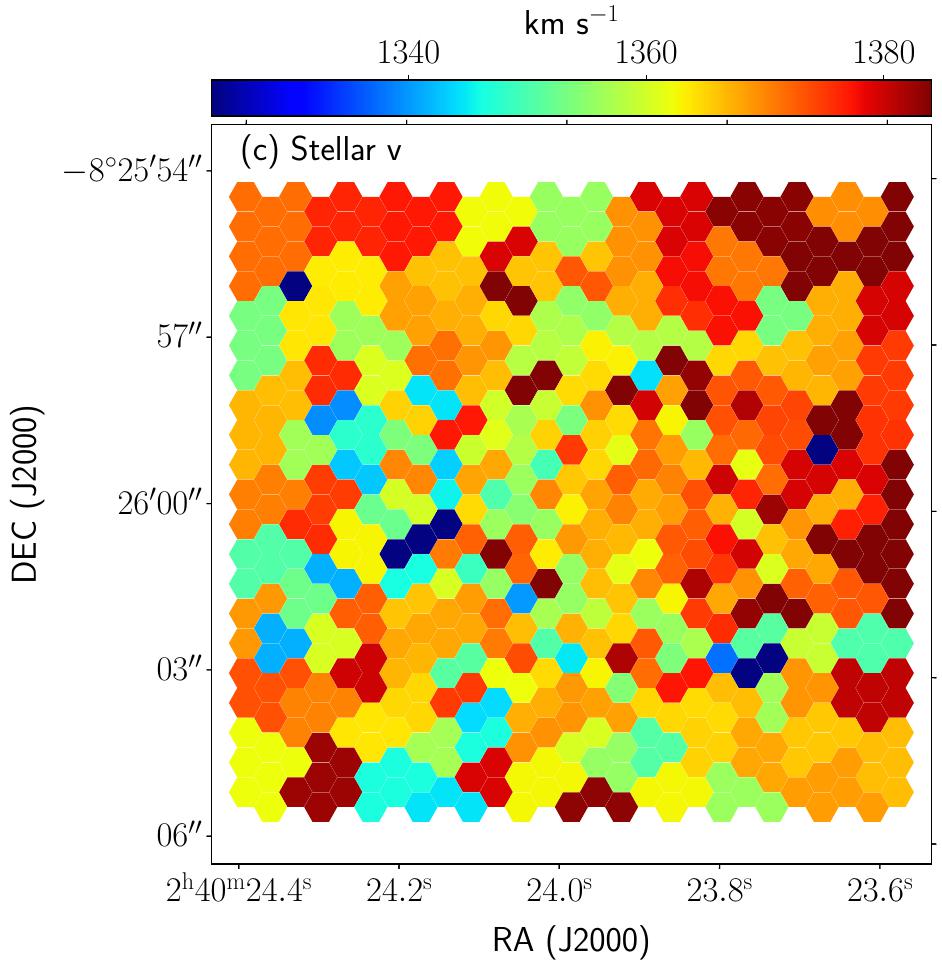}
	\includegraphics[clip, width=0.24\linewidth]{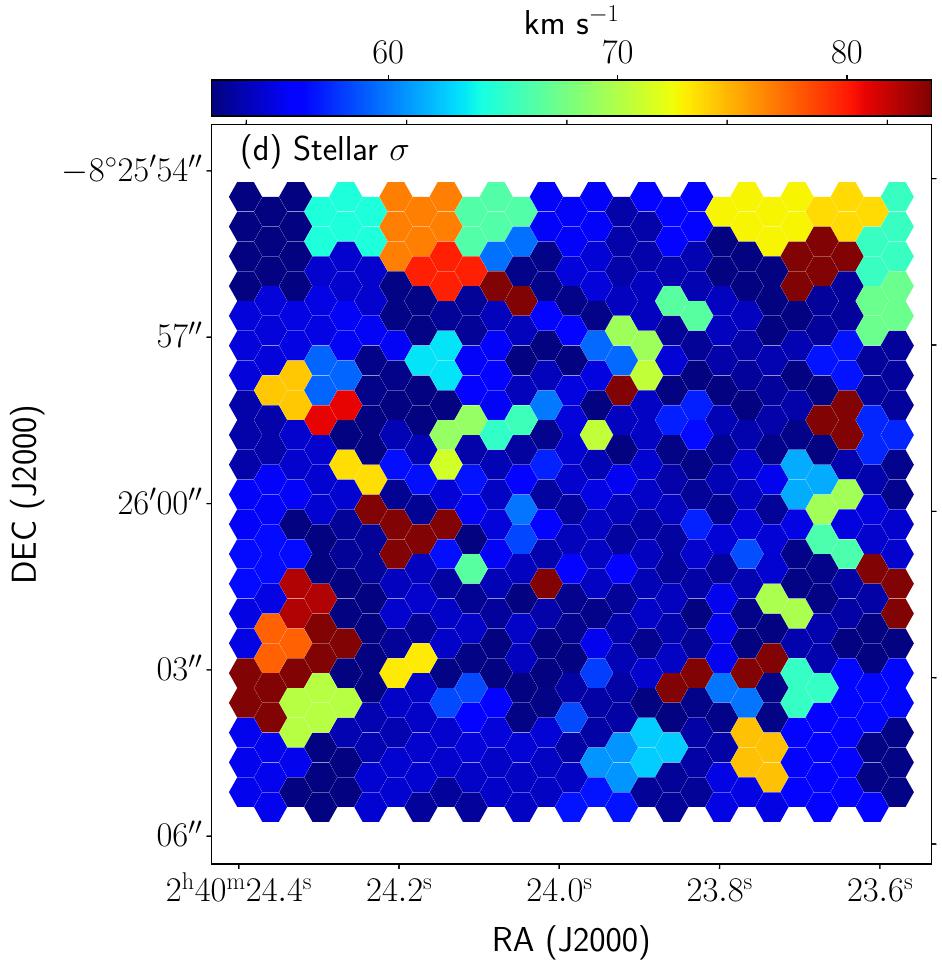}
	\includegraphics[clip, width=0.24\linewidth]{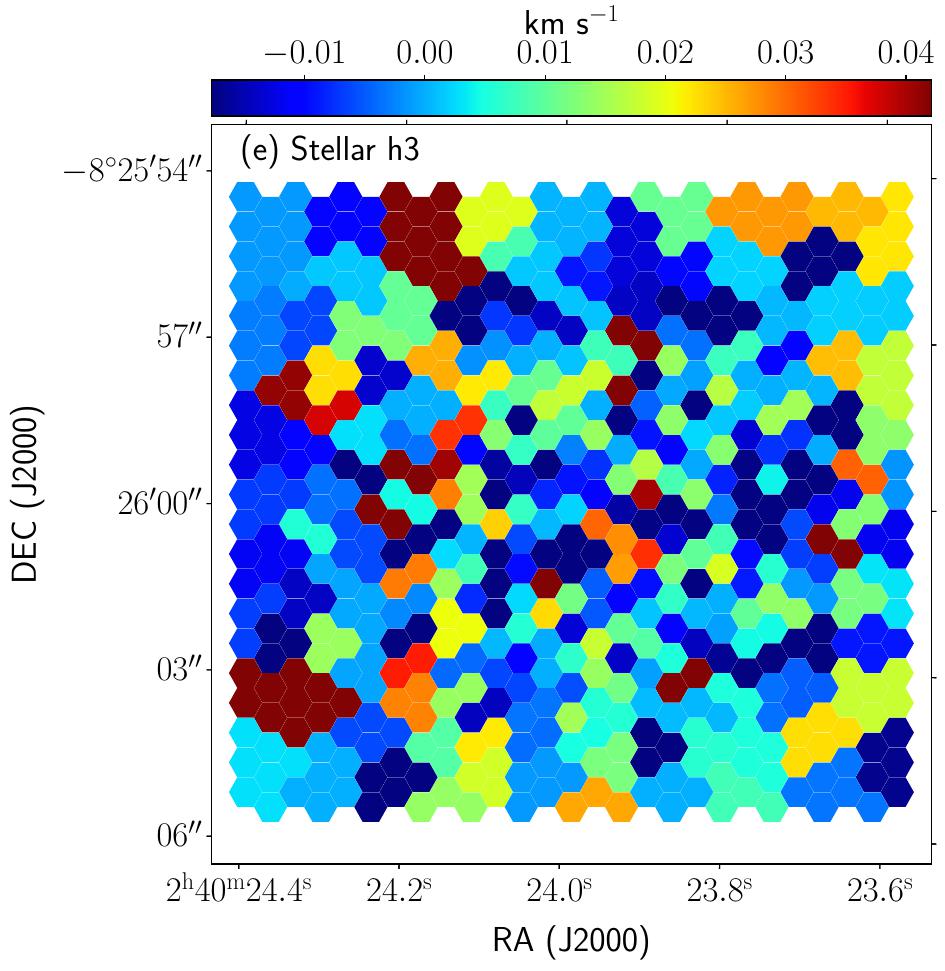}
	\includegraphics[clip, width=0.24\linewidth]{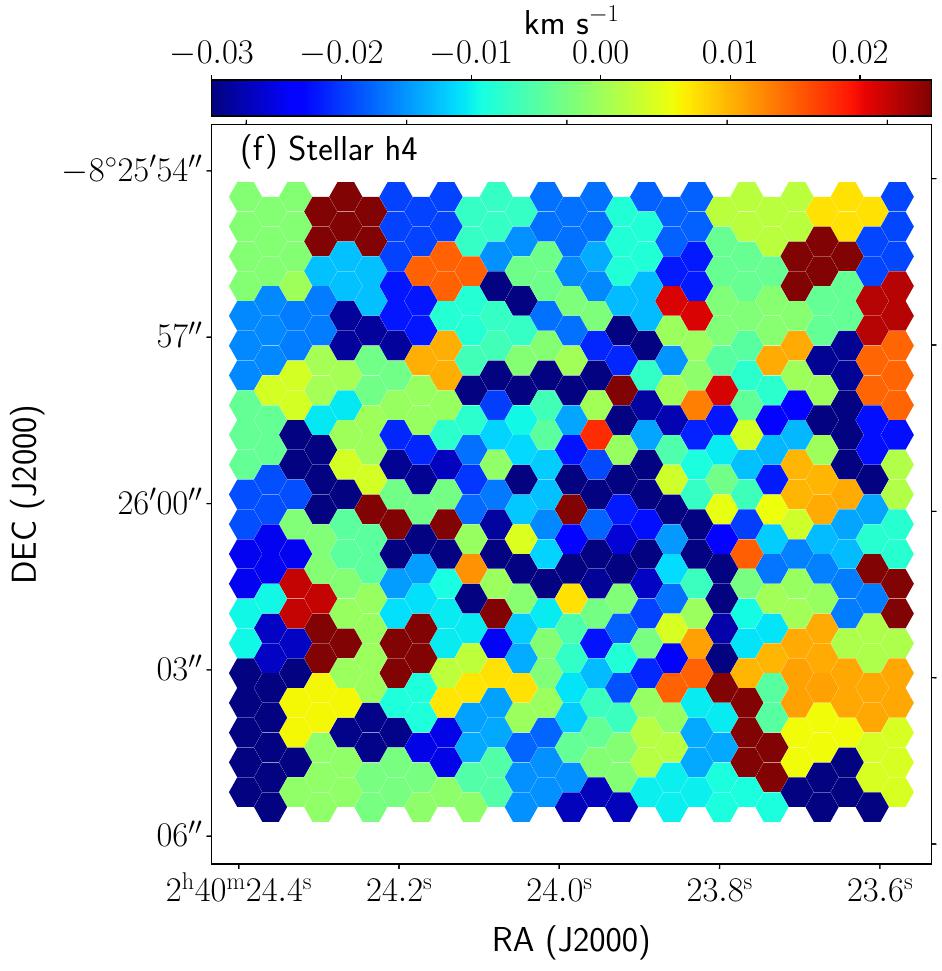}
	\includegraphics[clip, width=0.24\linewidth]{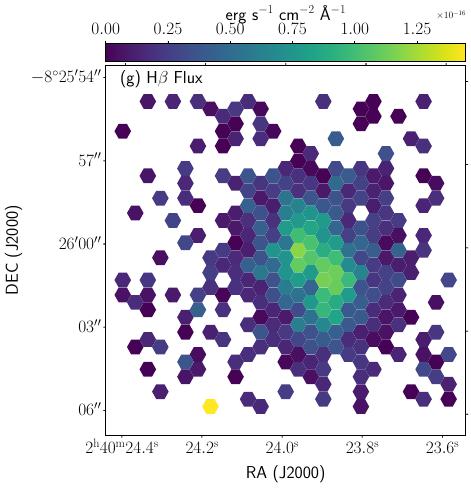}
	\includegraphics[clip, width=0.24\linewidth]{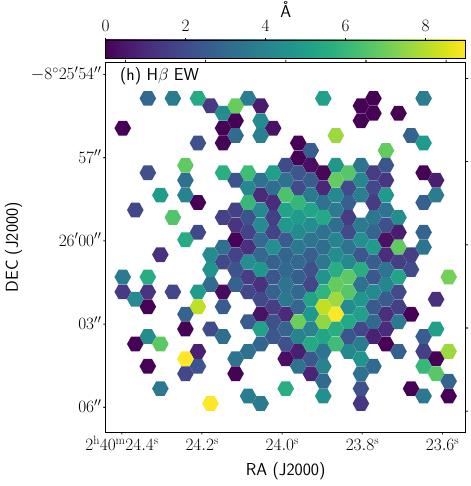}
	\includegraphics[clip, width=0.24\linewidth]{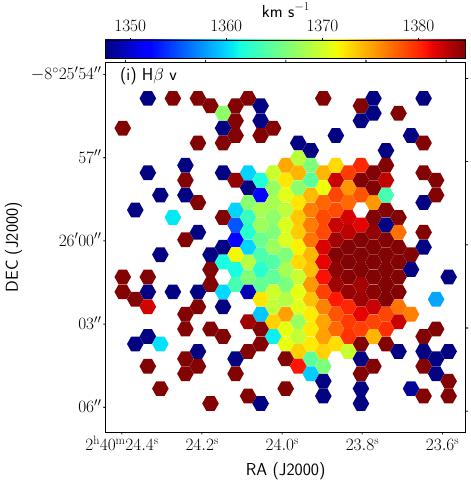}
	\includegraphics[clip, width=0.24\linewidth]{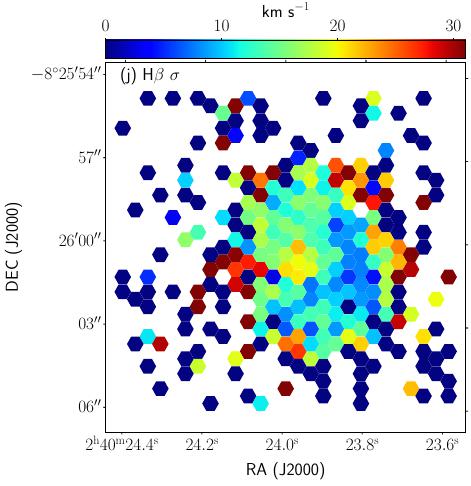}
	\includegraphics[clip, width=0.24\linewidth]{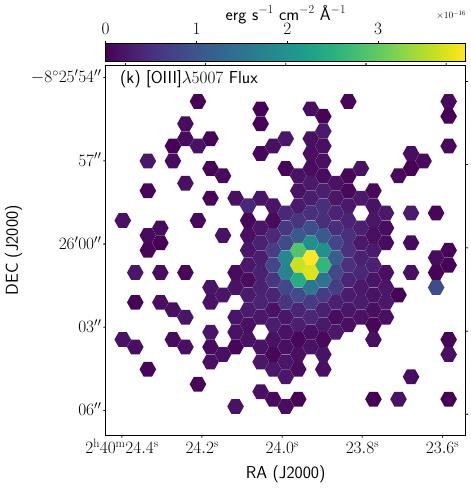}
	\includegraphics[clip, width=0.24\linewidth]{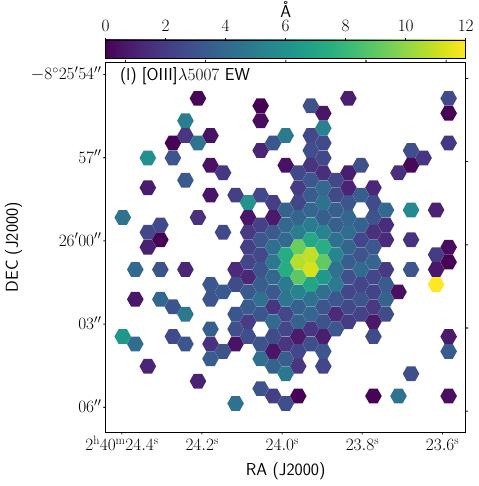}
	\includegraphics[clip, width=0.24\linewidth]{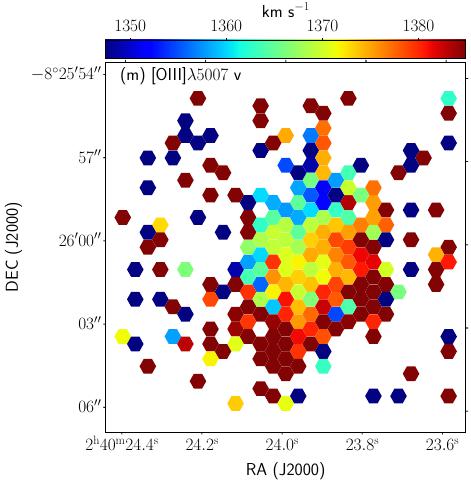}
	\includegraphics[clip, width=0.24\linewidth]{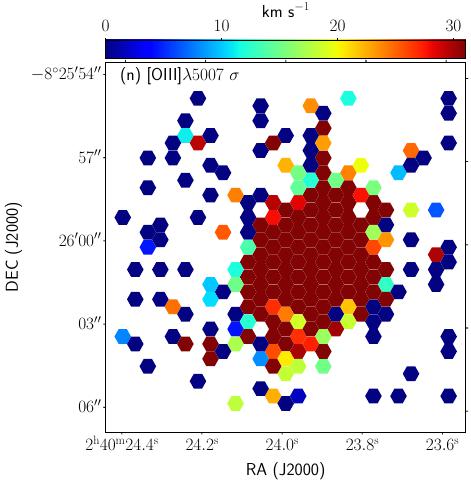}
	\vspace{5cm}
	\caption{NGC~1042 card.}
	\label{fig:NGC1042_card_1}
\end{figure*}
\addtocounter{figure}{-1}
\begin{figure*}[h]
	\centering
	\includegraphics[clip, width=0.24\linewidth]{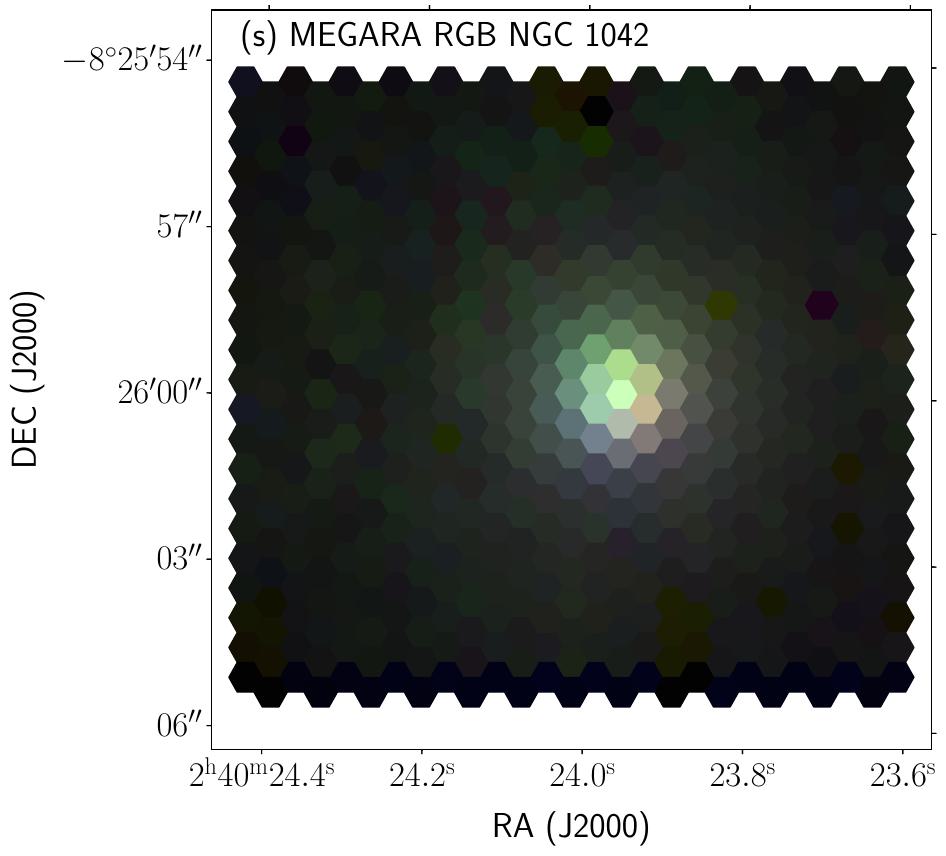}
	\includegraphics[clip, width=0.24\linewidth]{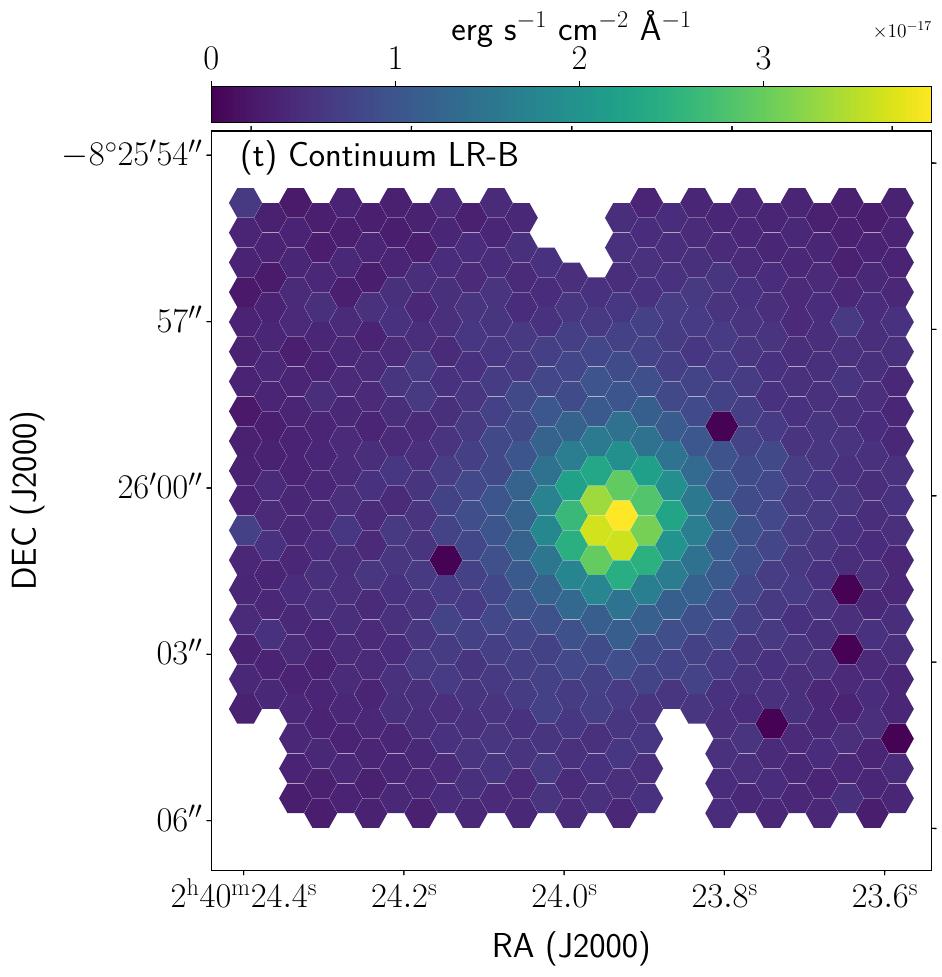}
	\includegraphics[clip, width=0.24\linewidth]{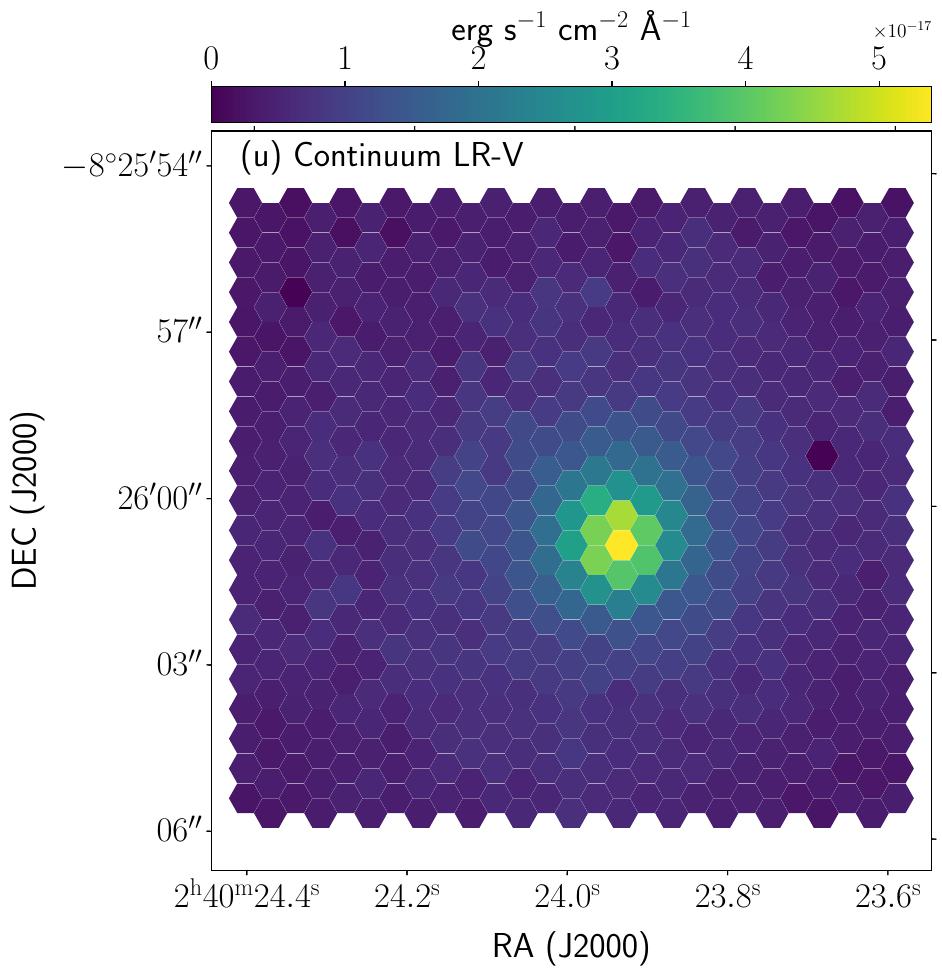}
	\includegraphics[clip, width=0.24\linewidth]{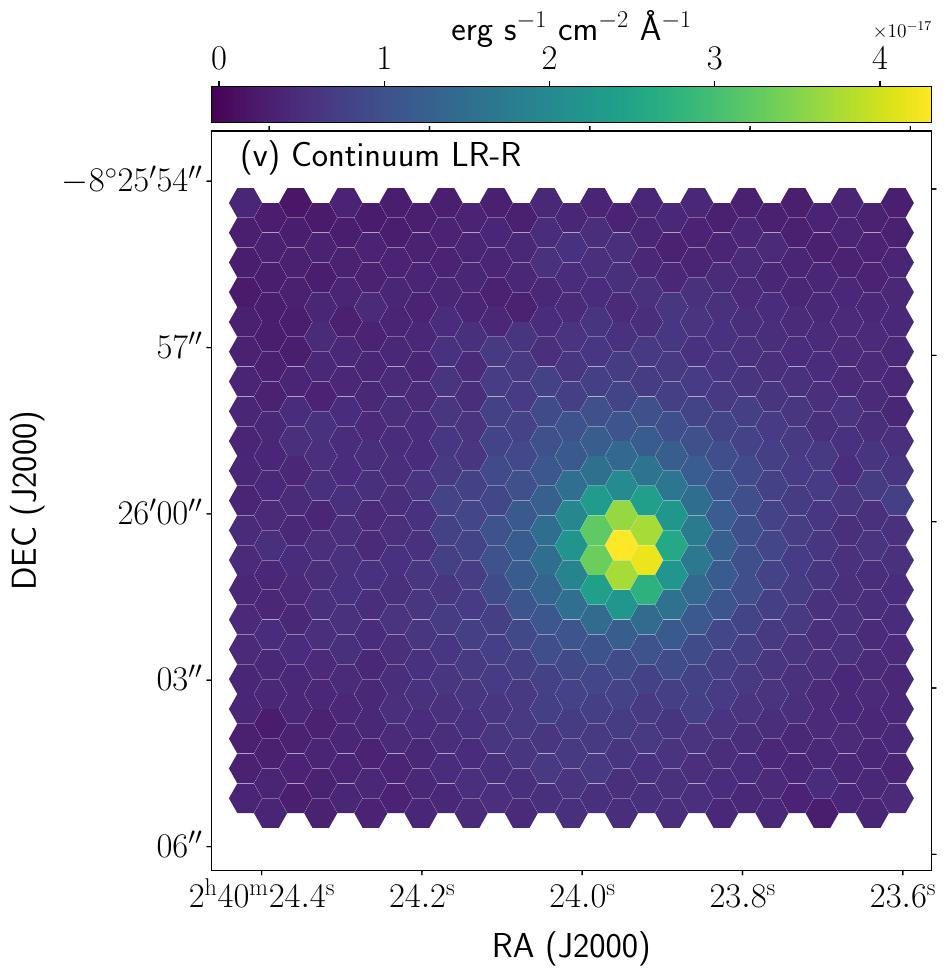}
	\includegraphics[clip, width=0.24\linewidth]{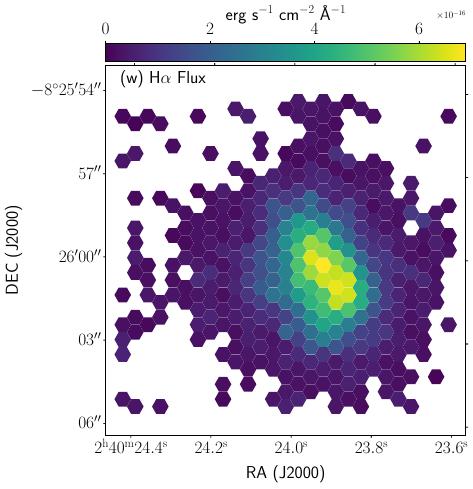}
	\includegraphics[clip, width=0.24\linewidth]{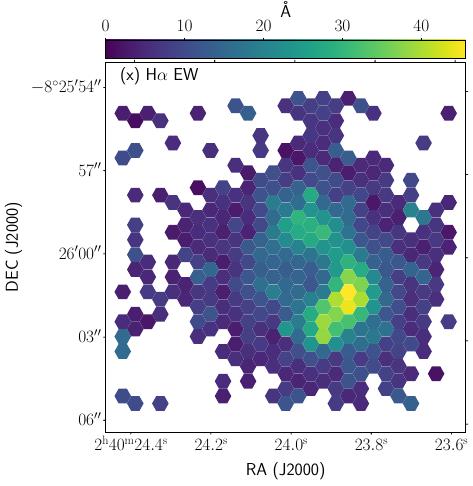}
	\includegraphics[clip, width=0.24\linewidth]{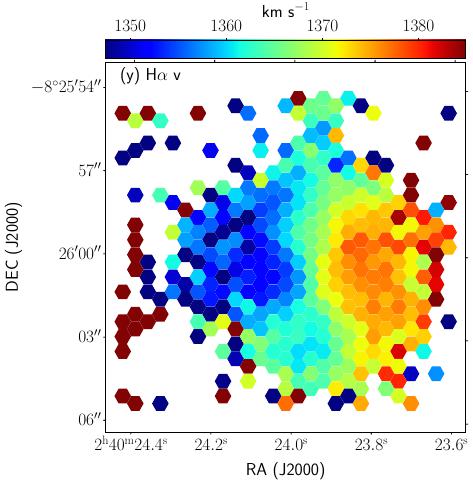}
	\includegraphics[clip, width=0.24\linewidth]{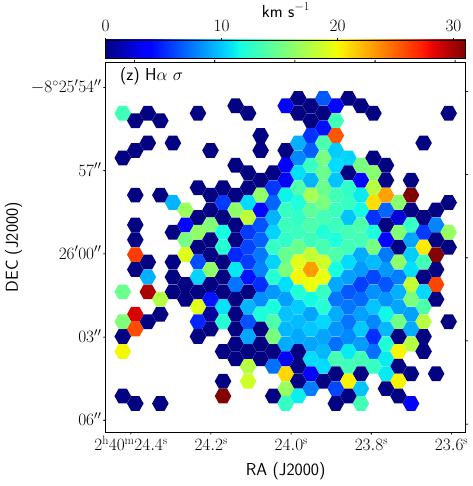}
	\includegraphics[clip, width=0.24\linewidth]{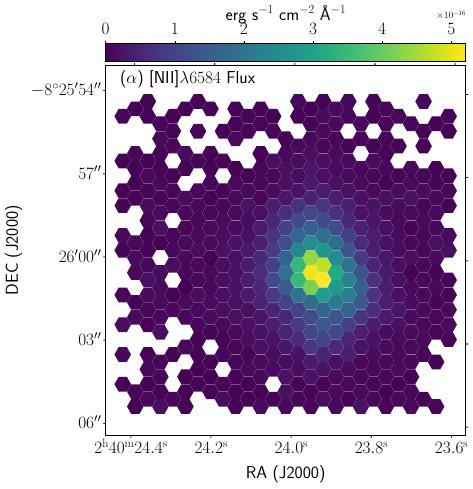}
	\includegraphics[clip, width=0.24\linewidth]{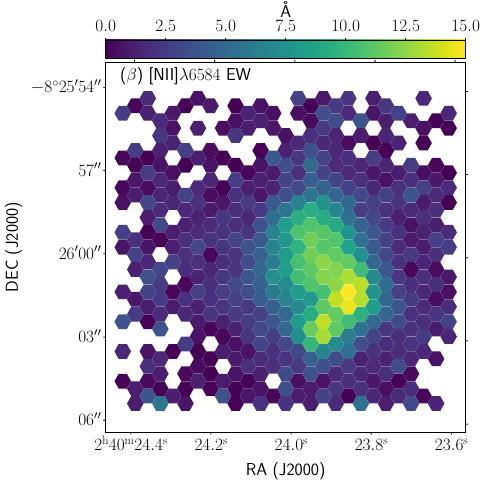}
	\includegraphics[clip, width=0.24\linewidth]{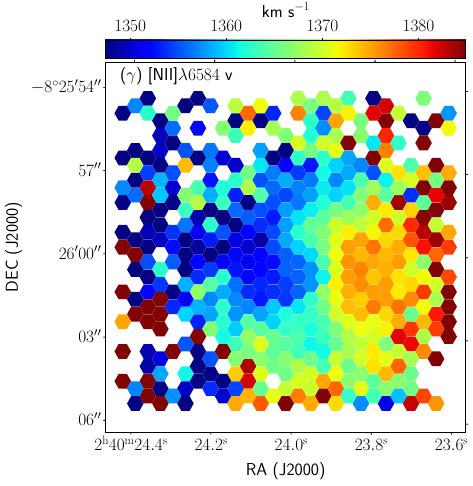}
	\includegraphics[clip, width=0.24\linewidth]{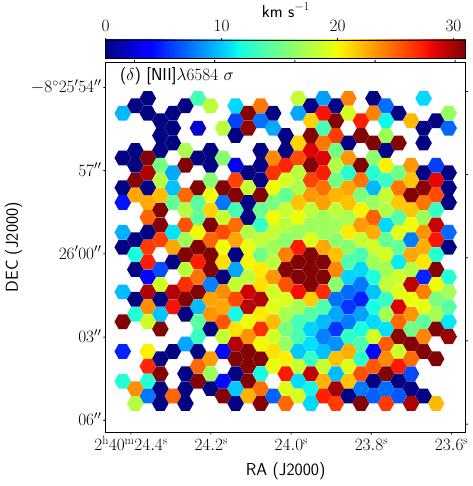}
	\includegraphics[clip, width=0.24\linewidth]{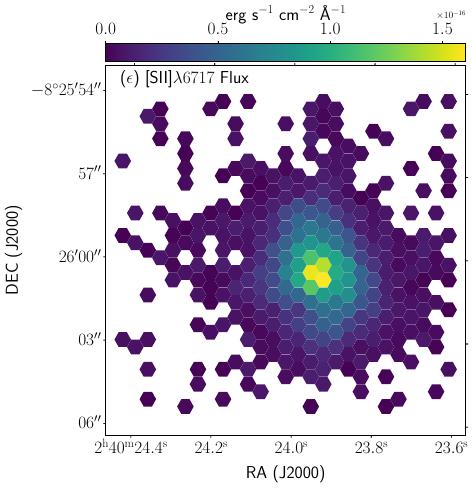}
	\includegraphics[clip, width=0.24\linewidth]{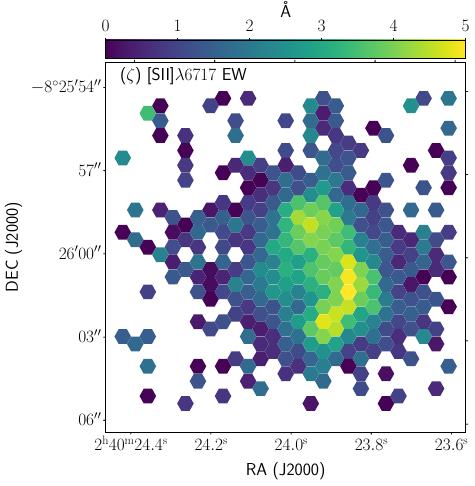}
	\includegraphics[clip, width=0.24\linewidth]{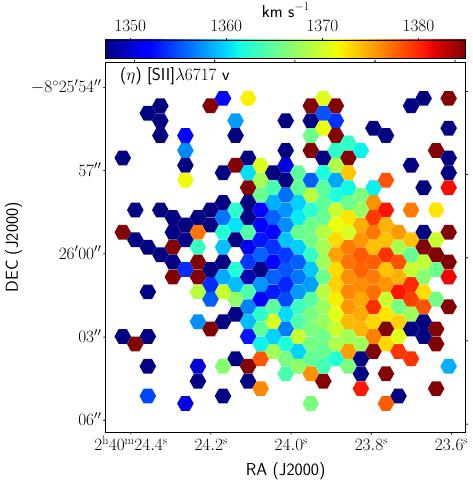}
	\includegraphics[clip, width=0.24\linewidth]{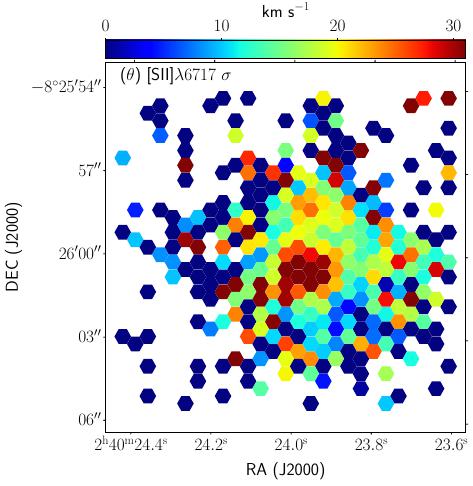}
	\includegraphics[clip, width=0.24\linewidth]{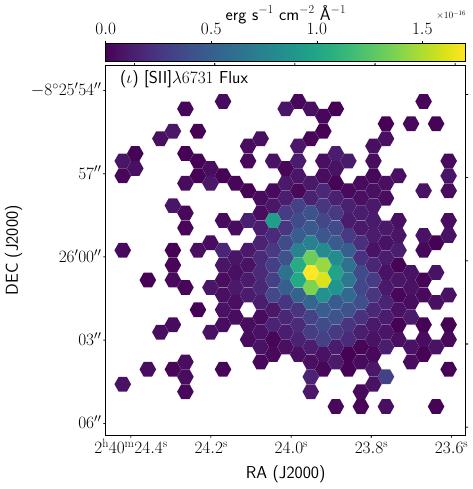}
	\includegraphics[clip, width=0.24\linewidth]{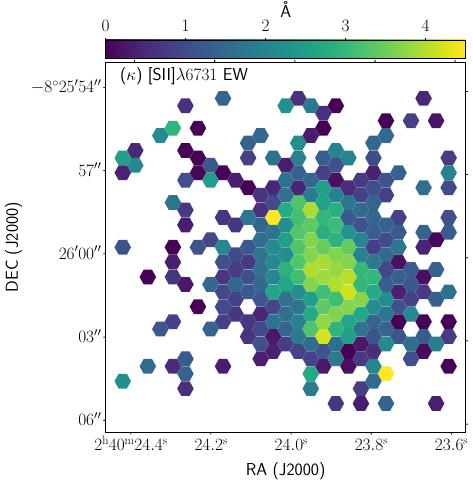}
	\includegraphics[clip, width=0.24\linewidth]{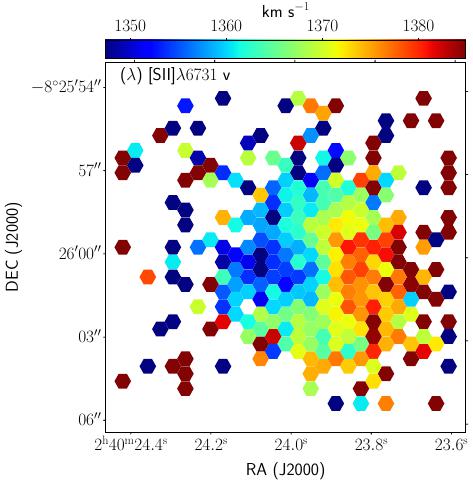}
	\includegraphics[clip, width=0.24\linewidth]{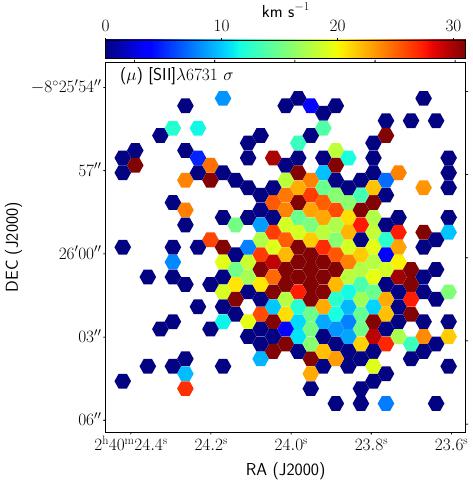}
	\caption{(cont.) NGC~1042 card.}
	\label{fig:NGC1042_card_2}
\end{figure*}

\begin{figure*}[h]
	\centering
	\includegraphics[clip, width=0.35\linewidth]{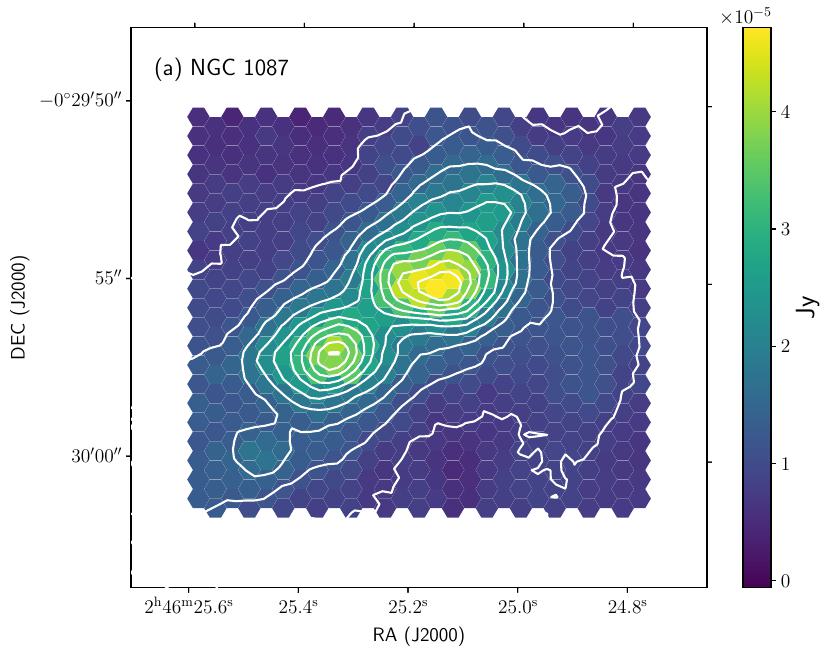}
	\includegraphics[clip, width=0.6\linewidth]{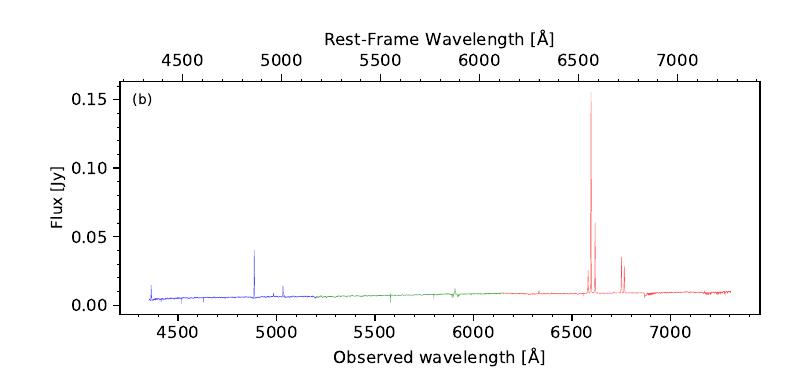}
	\includegraphics[clip, width=0.24\linewidth]{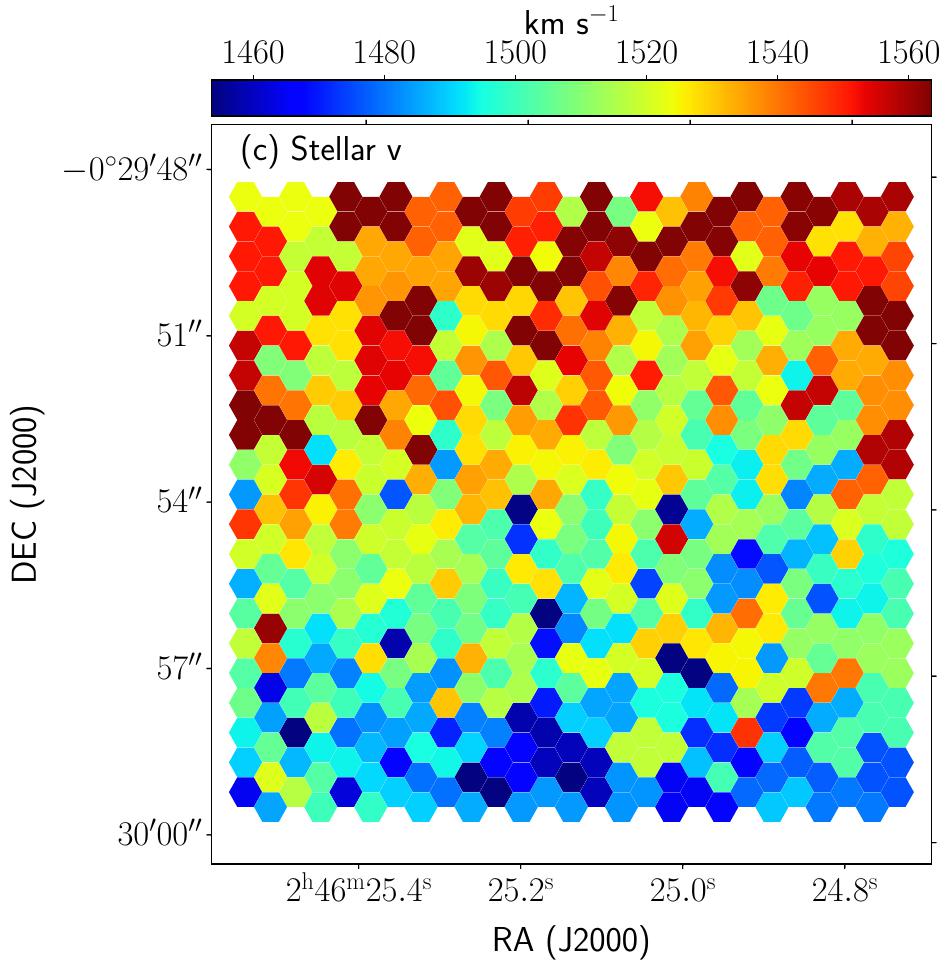}
	\includegraphics[clip, width=0.24\linewidth]{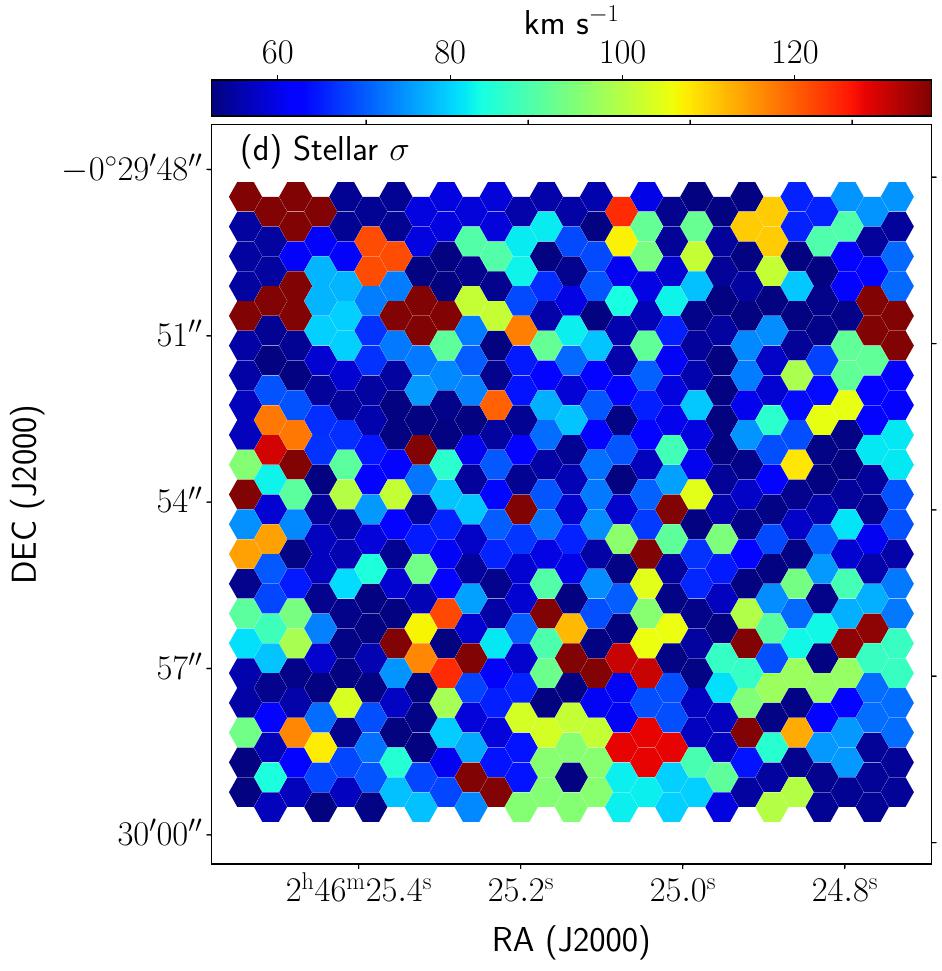}
	\includegraphics[clip, width=0.24\linewidth]{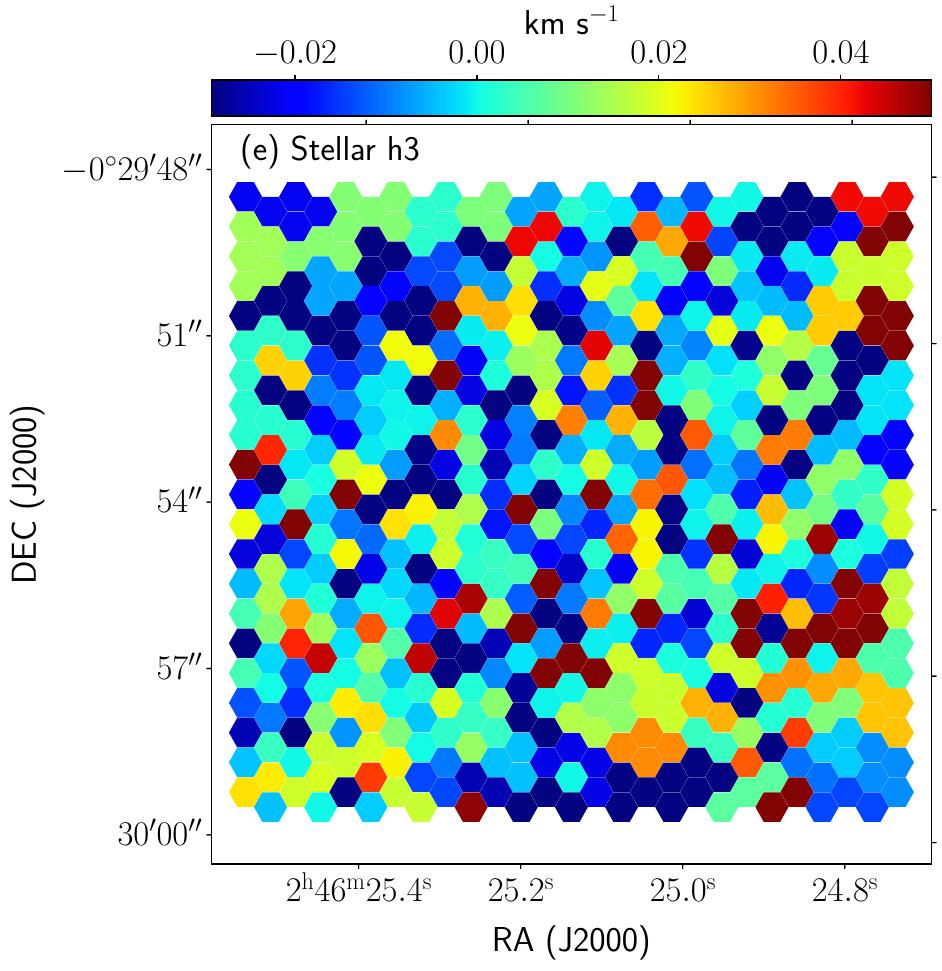}
	\includegraphics[clip, width=0.24\linewidth]{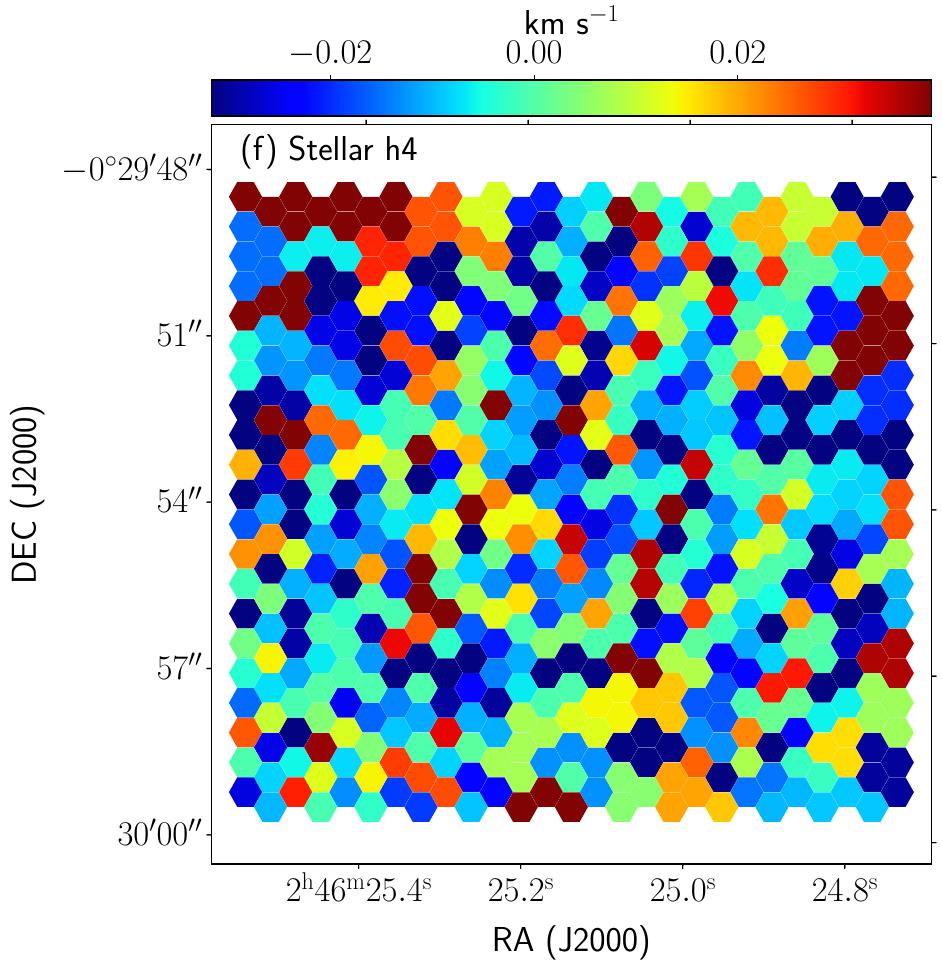}
	\includegraphics[clip, width=0.24\linewidth]{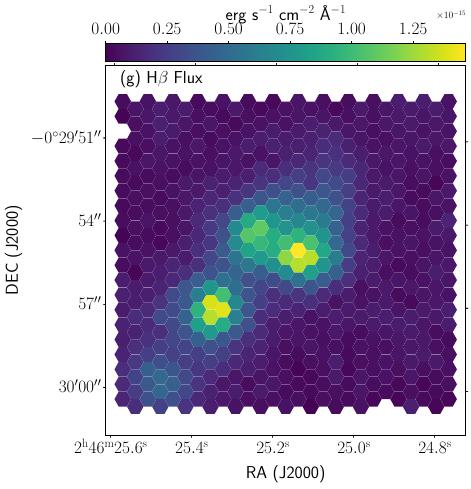}
	\includegraphics[clip, width=0.24\linewidth]{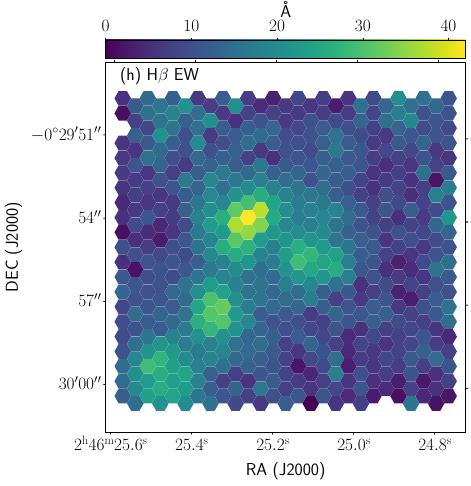}
	\includegraphics[clip, width=0.24\linewidth]{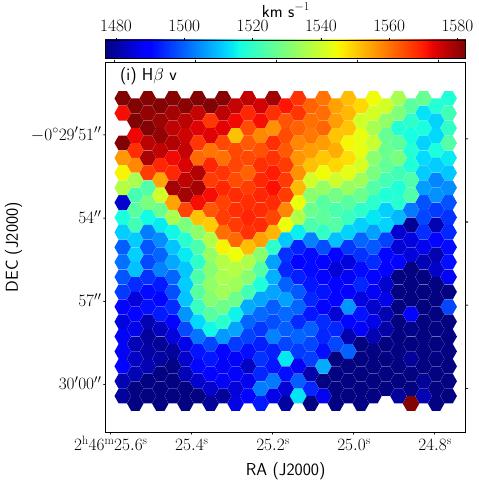}
	\includegraphics[clip, width=0.24\linewidth]{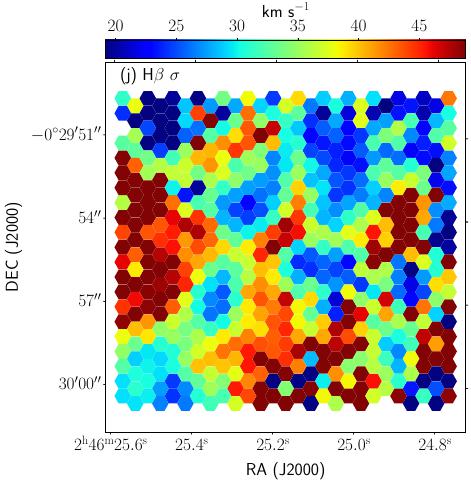}
	\includegraphics[clip, width=0.24\linewidth]{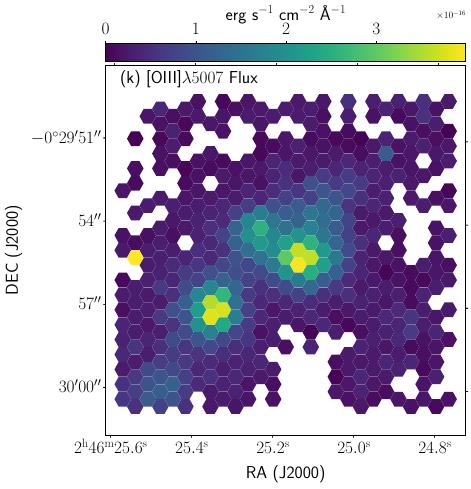}
	\includegraphics[clip, width=0.24\linewidth]{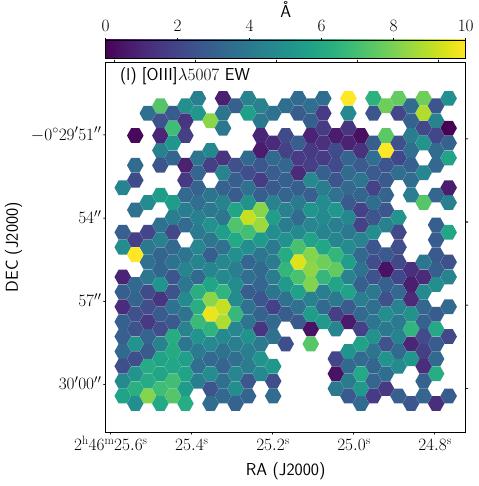}
	\includegraphics[clip, width=0.24\linewidth]{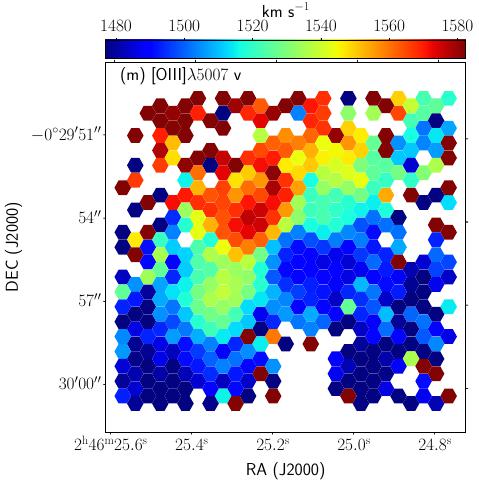}
	\includegraphics[clip, width=0.24\linewidth]{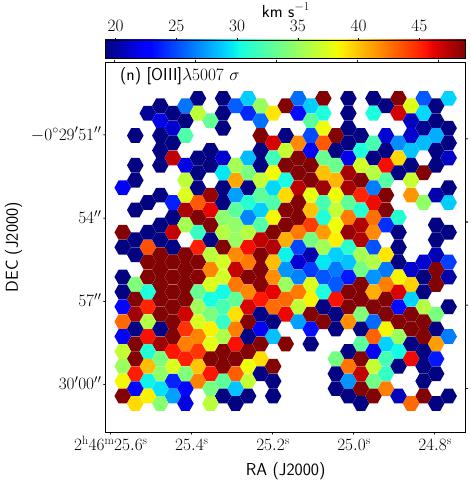}
	\vspace{5cm}
	\caption{NGC~1087 card.}
	\label{fig:NGC1087_card_1}
\end{figure*}
\addtocounter{figure}{-1}
\begin{figure*}[h]
	\centering
	\includegraphics[clip, width=0.24\linewidth]{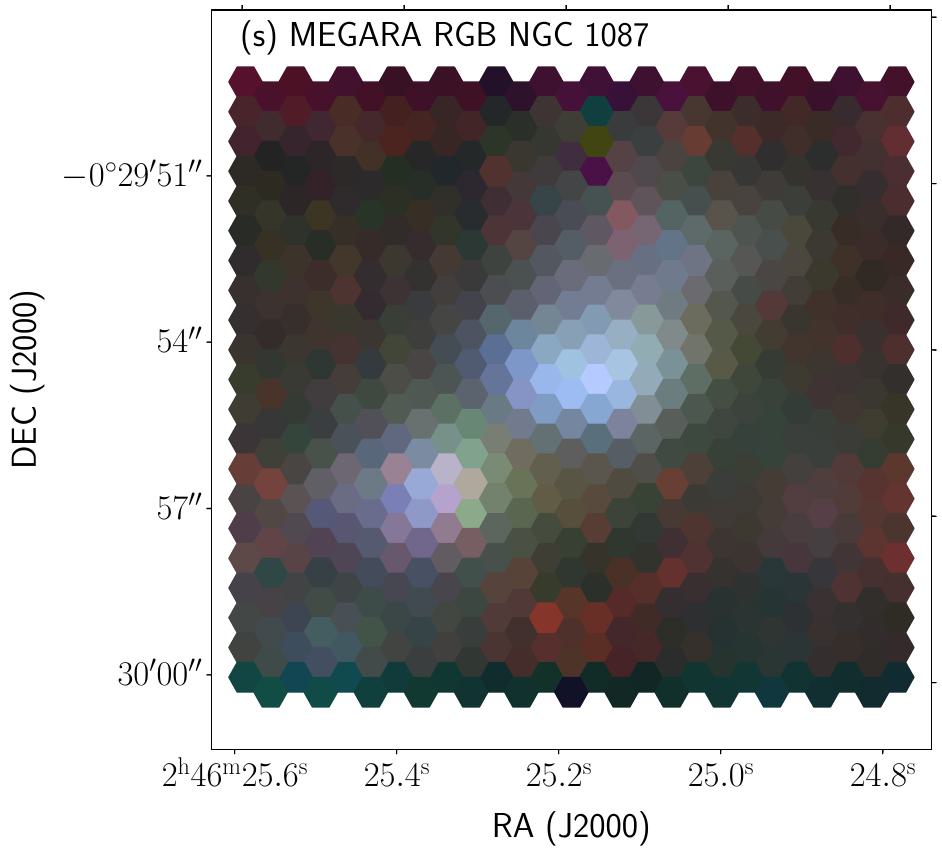}
	\includegraphics[clip, width=0.24\linewidth]{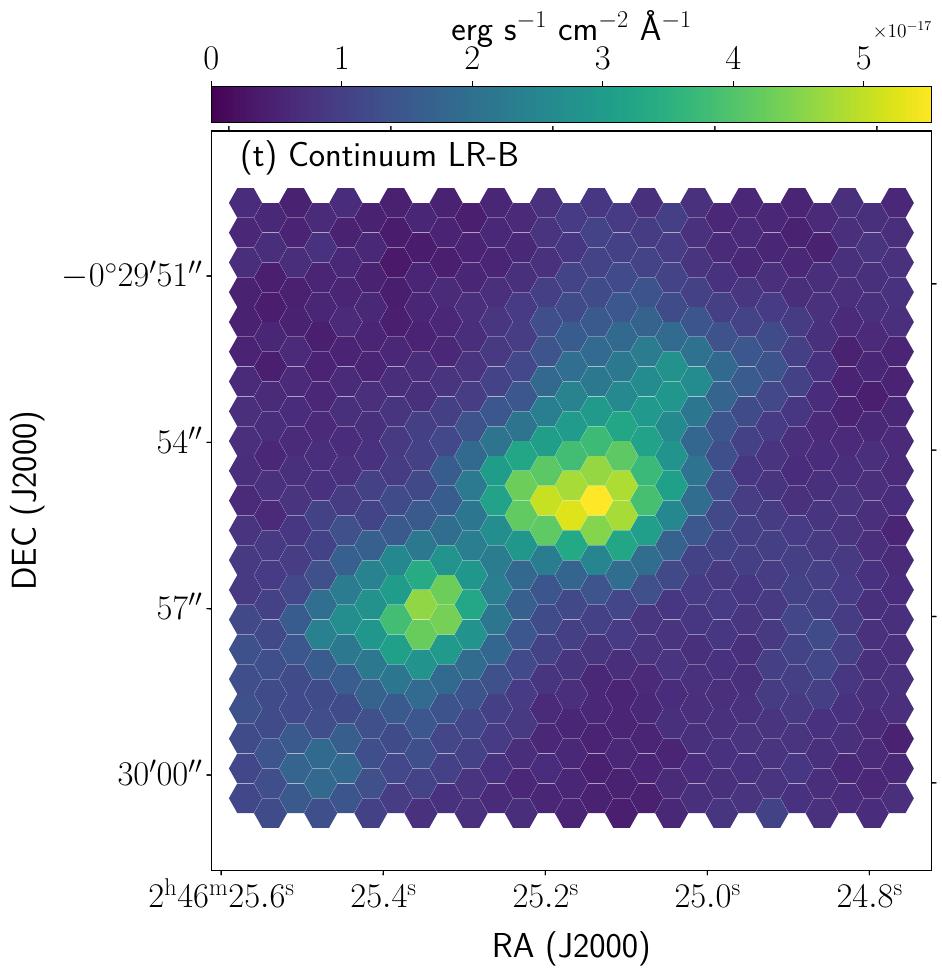}
	\includegraphics[clip, width=0.24\linewidth]{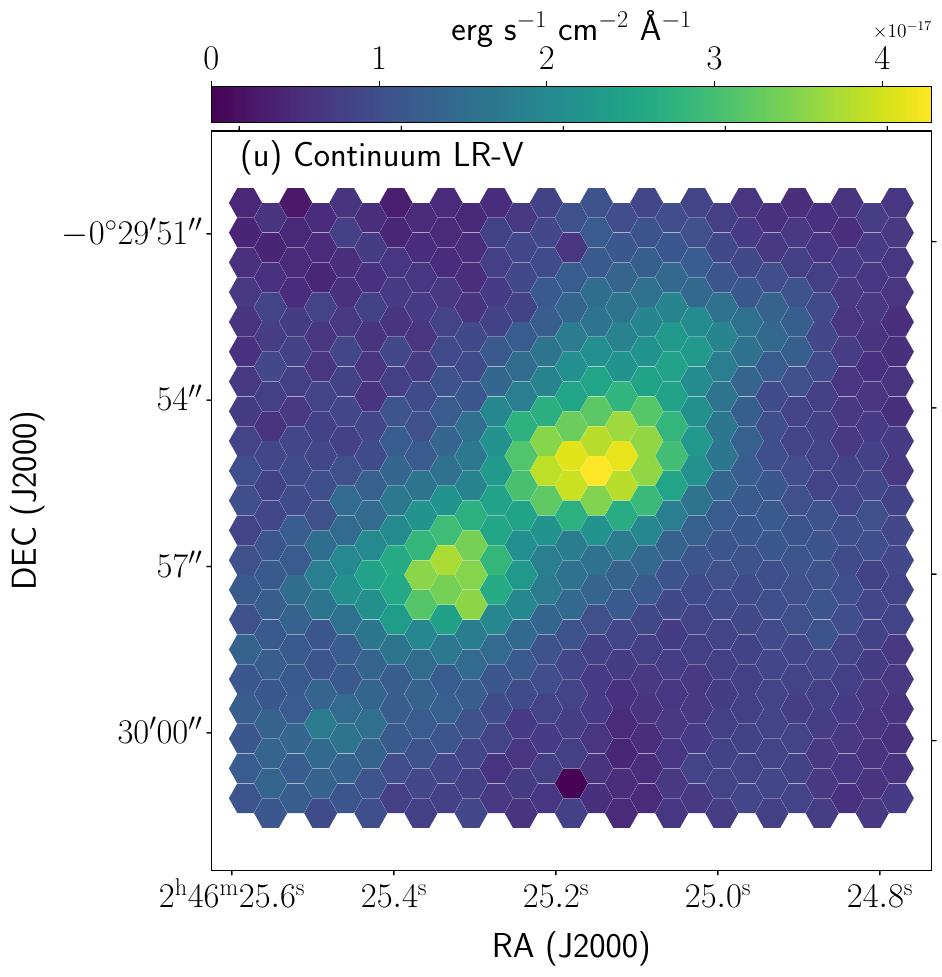}
	\includegraphics[clip, width=0.24\linewidth]{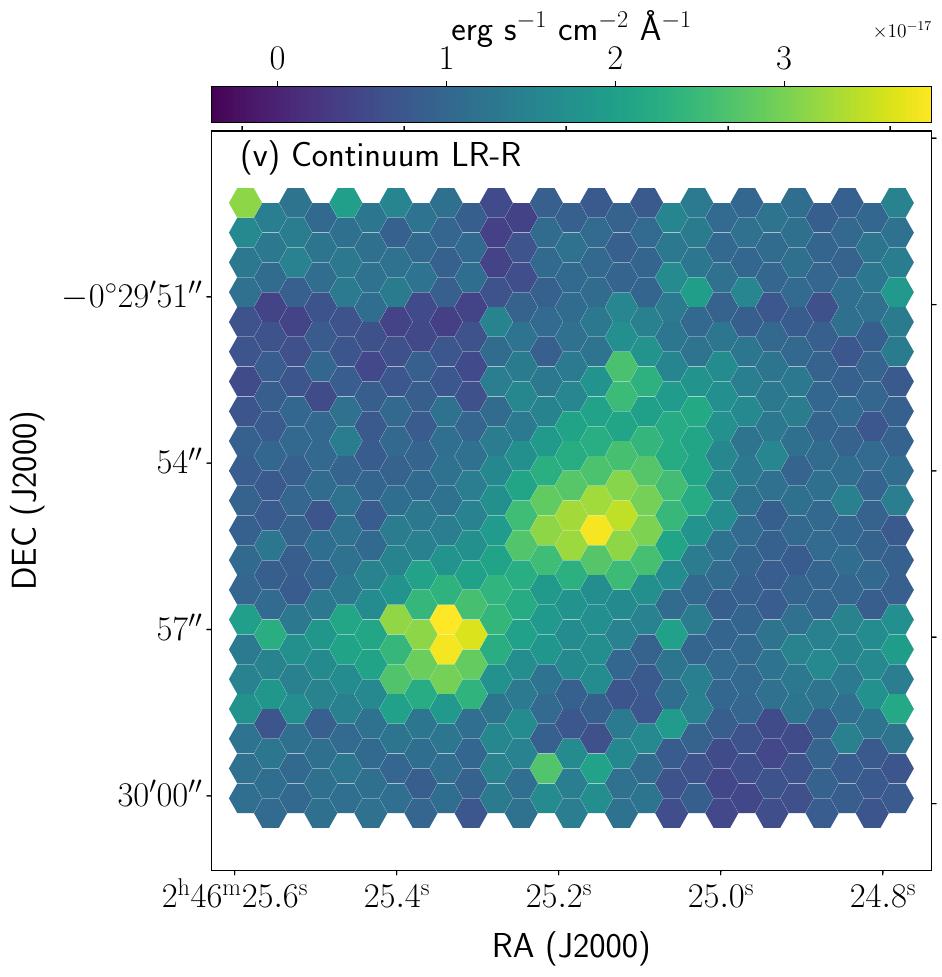}
	\includegraphics[clip, width=0.24\linewidth]{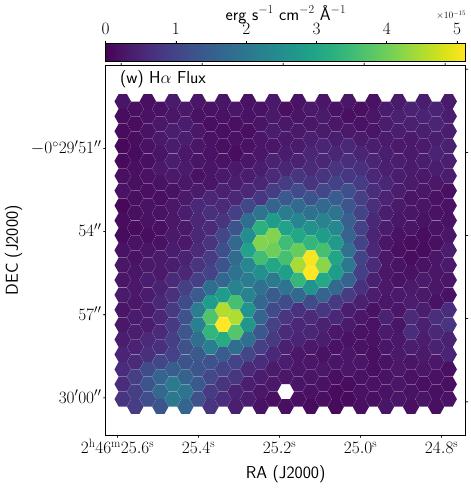}
	\includegraphics[clip, width=0.24\linewidth]{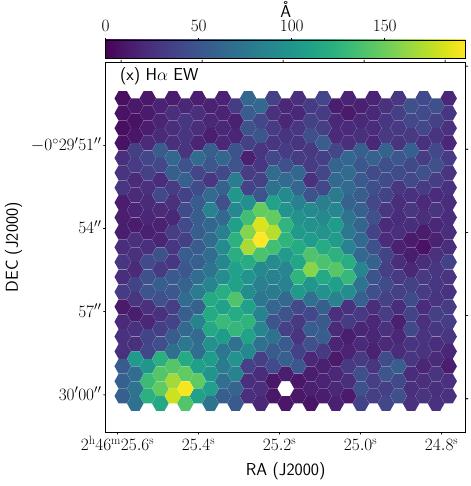}
	\includegraphics[clip, width=0.24\linewidth]{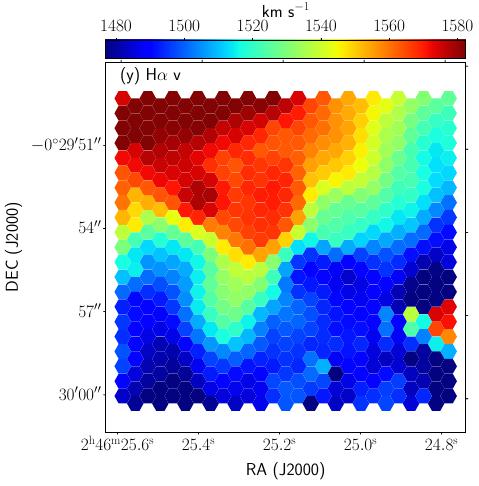}
	\includegraphics[clip, width=0.24\linewidth]{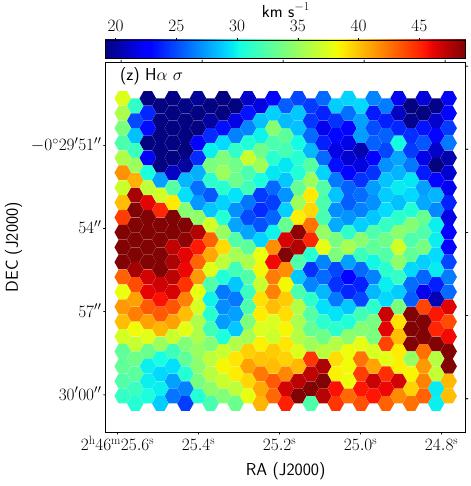}
	\includegraphics[clip, width=0.24\linewidth]{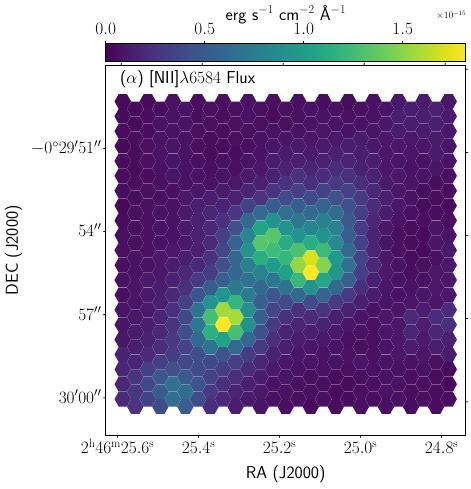}
	\includegraphics[clip, width=0.24\linewidth]{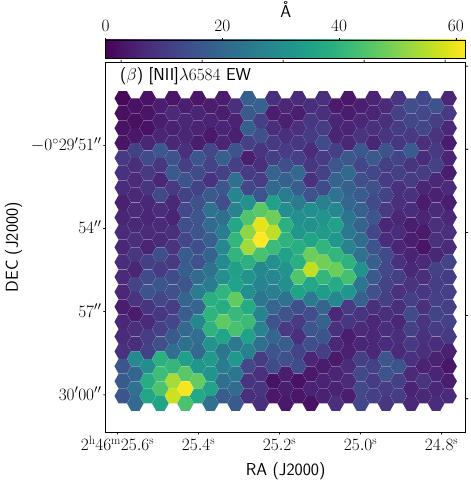}
	\includegraphics[clip, width=0.24\linewidth]{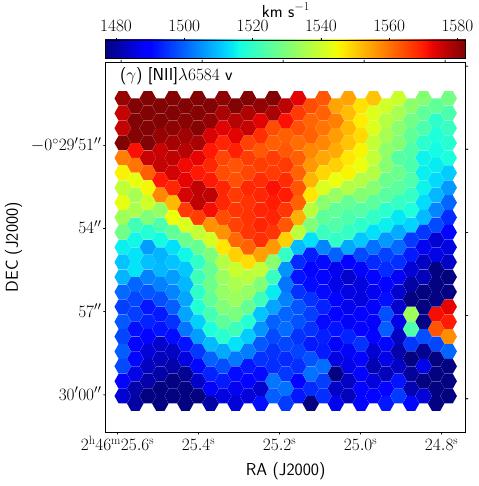}
	\includegraphics[clip, width=0.24\linewidth]{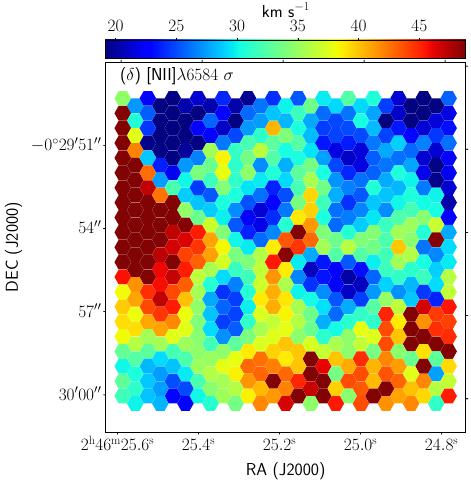}
	\includegraphics[clip, width=0.24\linewidth]{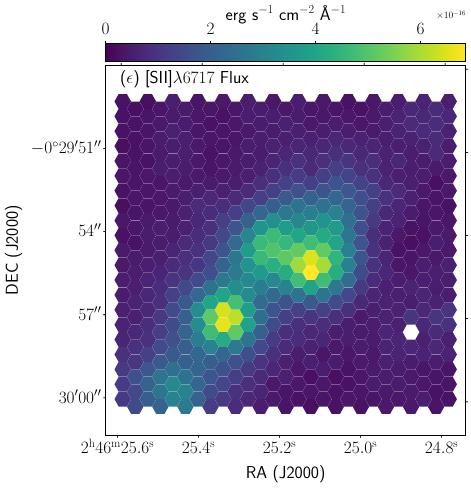}
	\includegraphics[clip, width=0.24\linewidth]{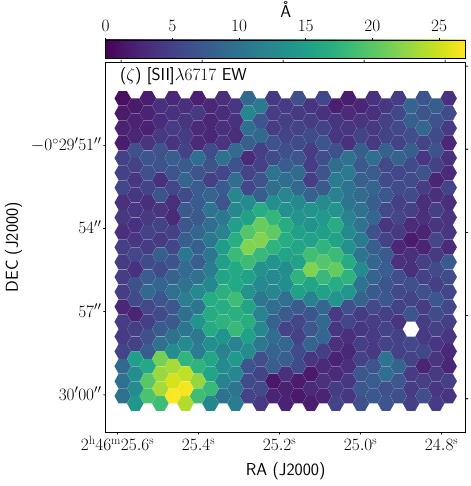}
	\includegraphics[clip, width=0.24\linewidth]{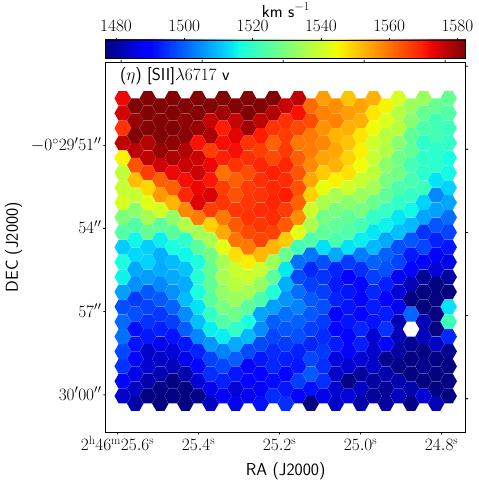}
	\includegraphics[clip, width=0.24\linewidth]{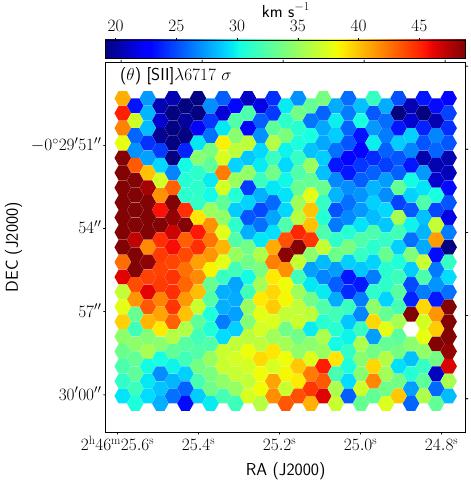}
	\includegraphics[clip, width=0.24\linewidth]{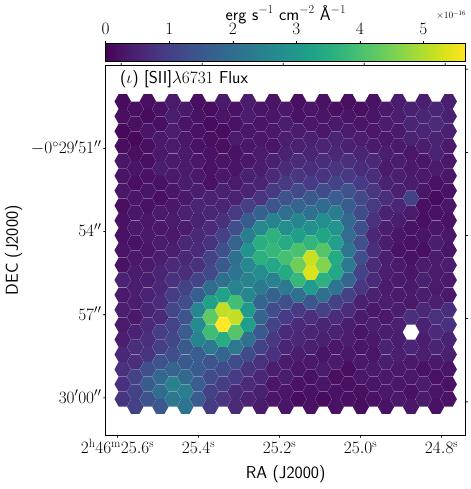}
	\includegraphics[clip, width=0.24\linewidth]{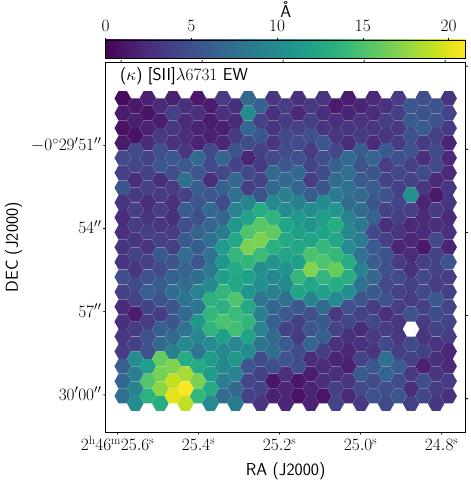}
	\includegraphics[clip, width=0.24\linewidth]{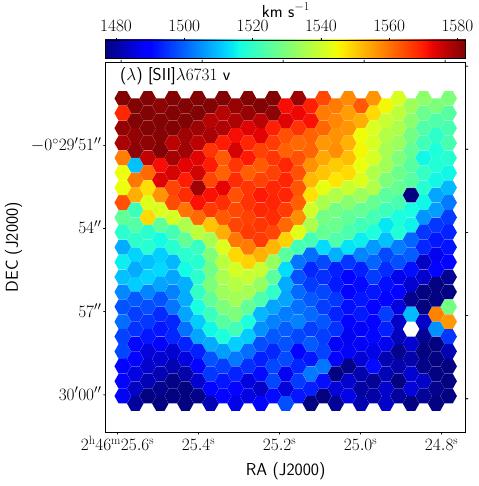}
	\includegraphics[clip, width=0.24\linewidth]{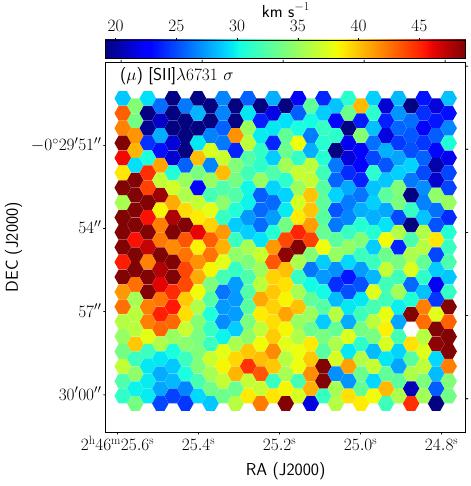}
	\caption{(cont.) NGC~1087 card.}
	\label{fig:NGC1087_card_2}
\end{figure*}

\begin{figure*}[h]
	\centering
	\includegraphics[clip, width=0.35\linewidth]{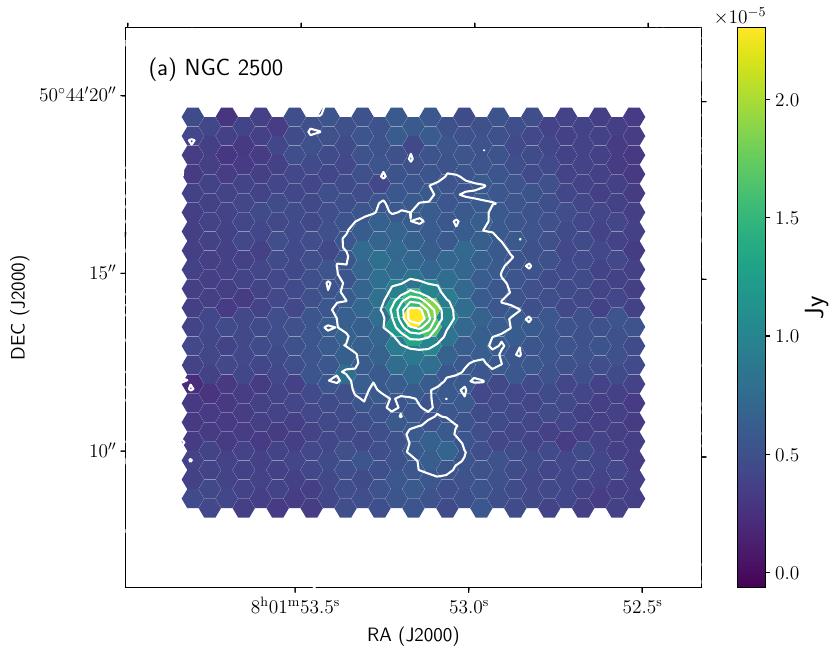}
	\includegraphics[clip, width=0.6\linewidth]{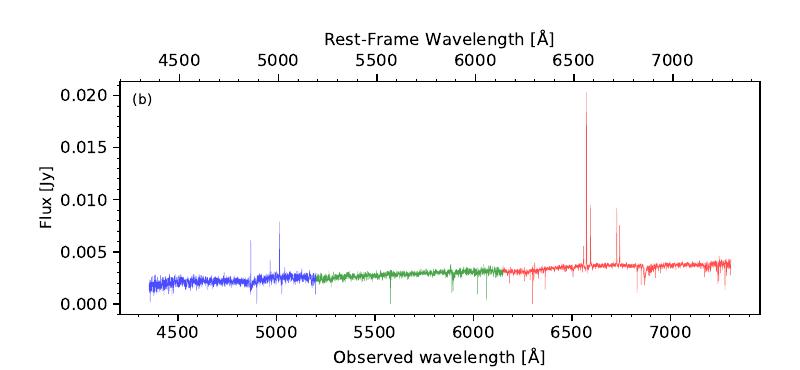}
	\includegraphics[clip, width=0.24\linewidth]{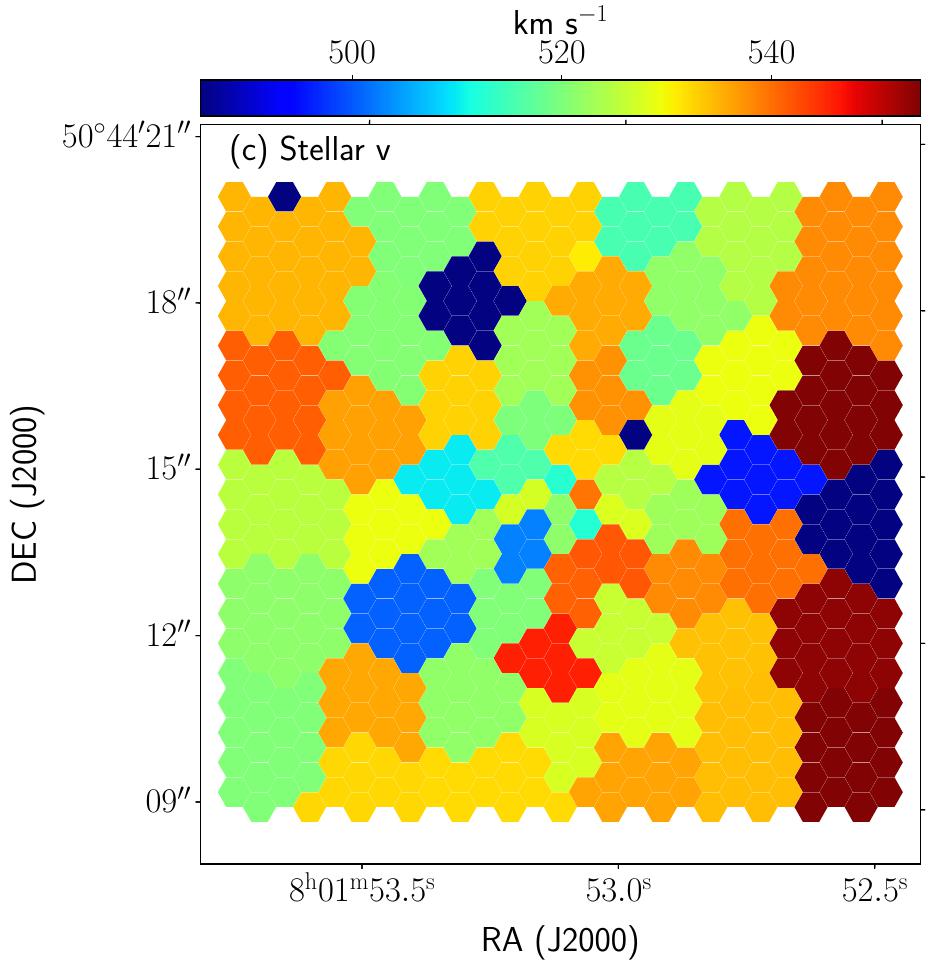}
	\includegraphics[clip, width=0.24\linewidth]{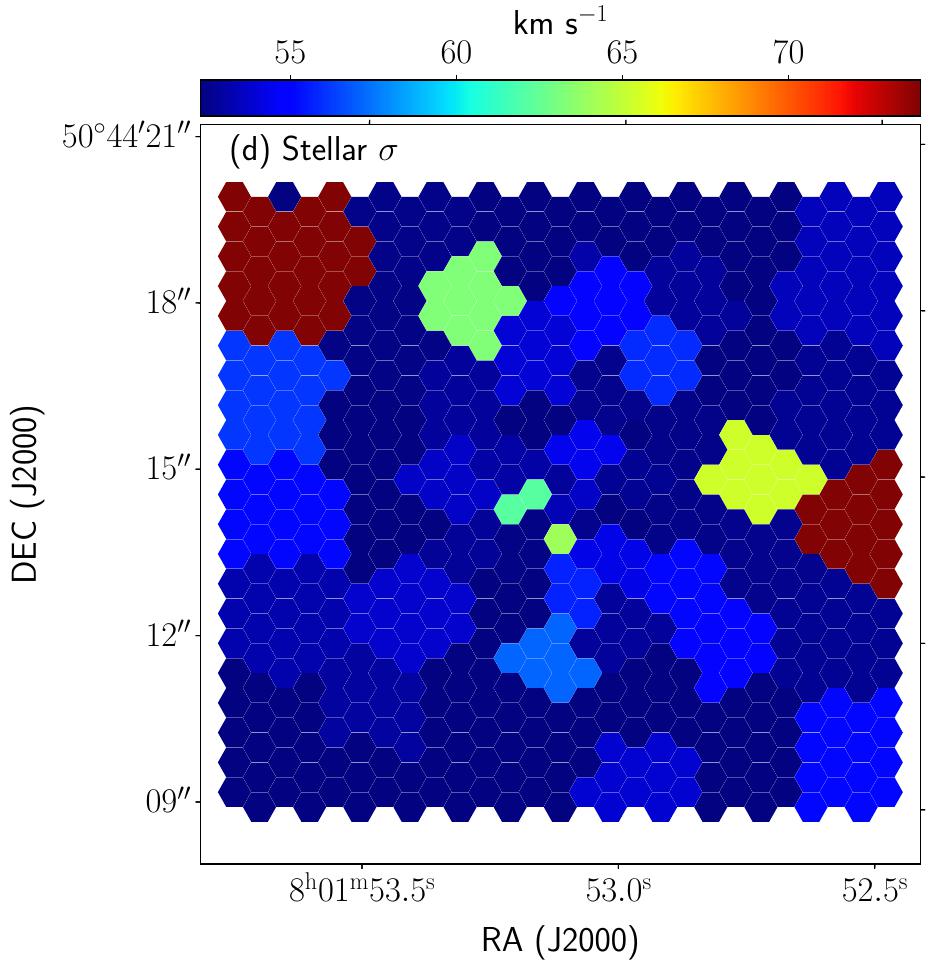}
	\includegraphics[clip, width=0.24\linewidth]{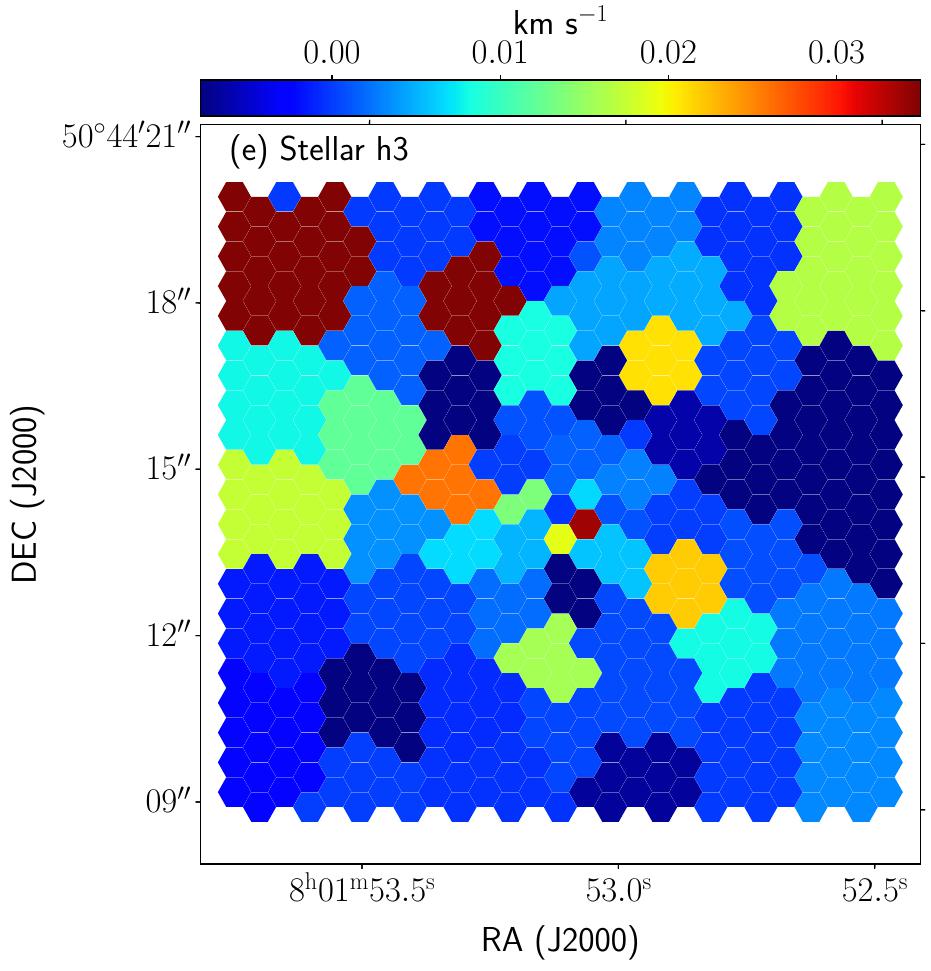}
	\includegraphics[clip, width=0.24\linewidth]{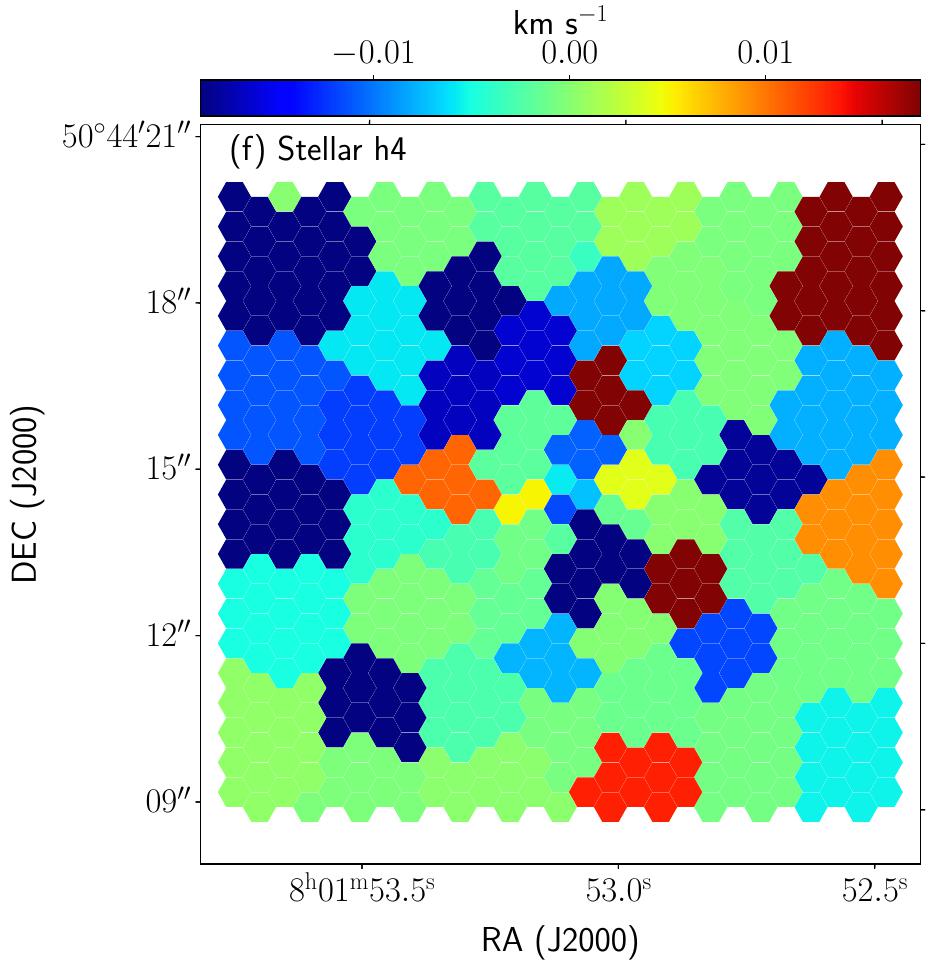}
	\includegraphics[clip, width=0.24\linewidth]{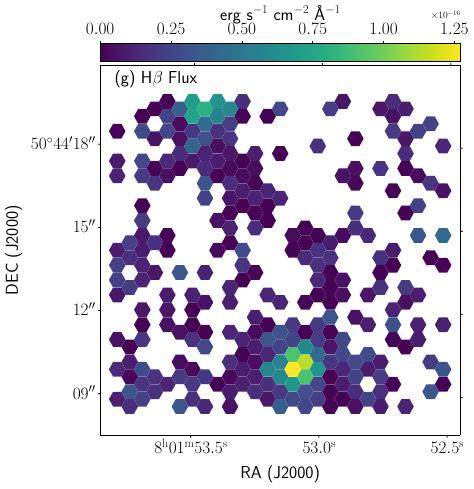}
	\includegraphics[clip, width=0.24\linewidth]{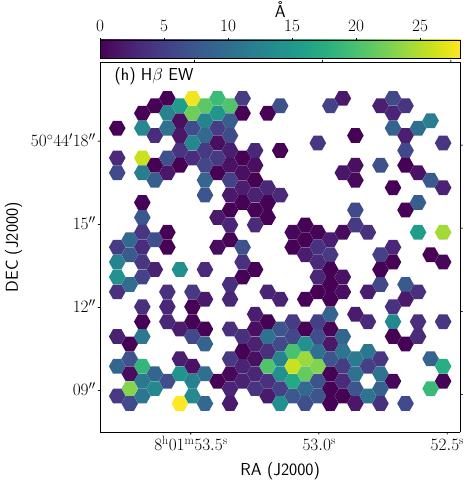}
	\includegraphics[clip, width=0.24\linewidth]{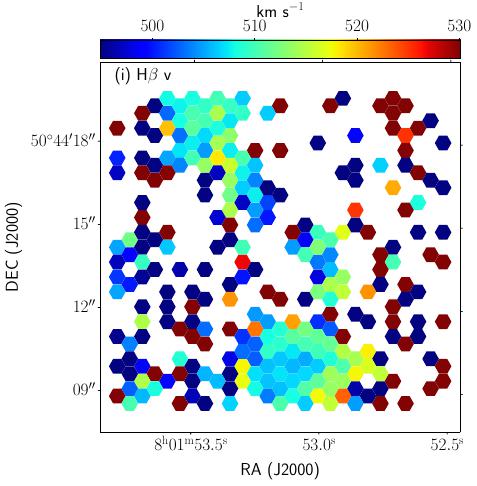}
	\includegraphics[clip, width=0.24\linewidth]{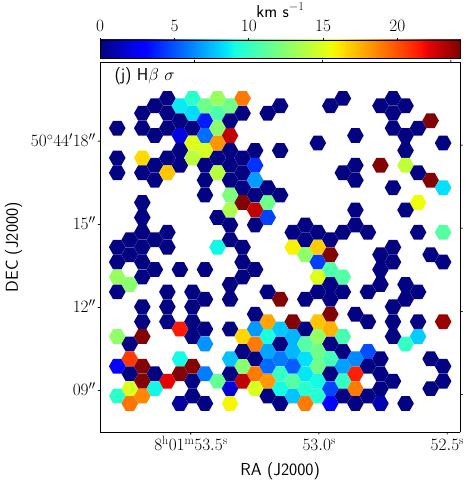}
	\includegraphics[clip, width=0.24\linewidth]{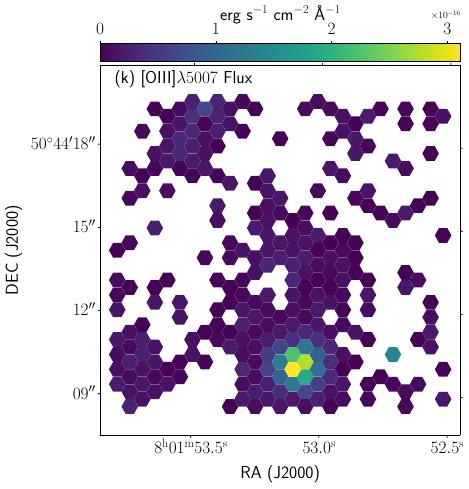}
	\includegraphics[clip, width=0.24\linewidth]{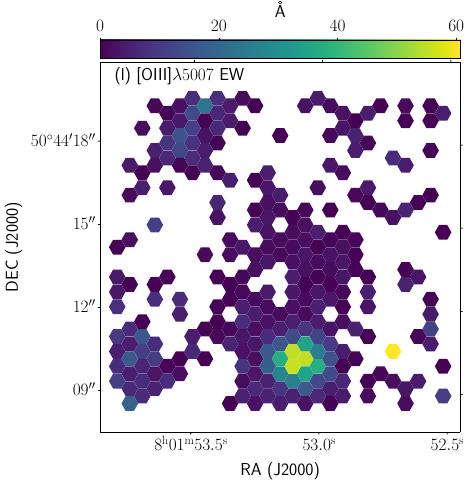}
	\includegraphics[clip, width=0.24\linewidth]{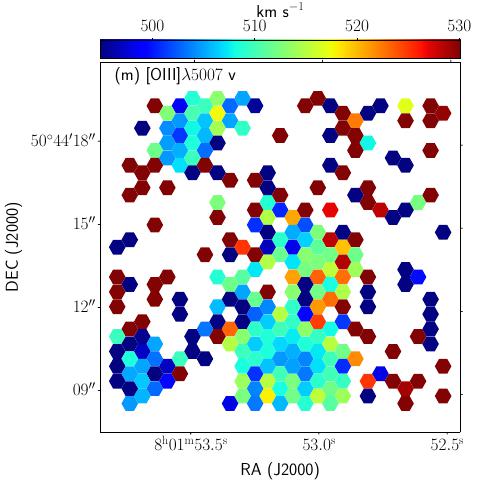}
	\includegraphics[clip, width=0.24\linewidth]{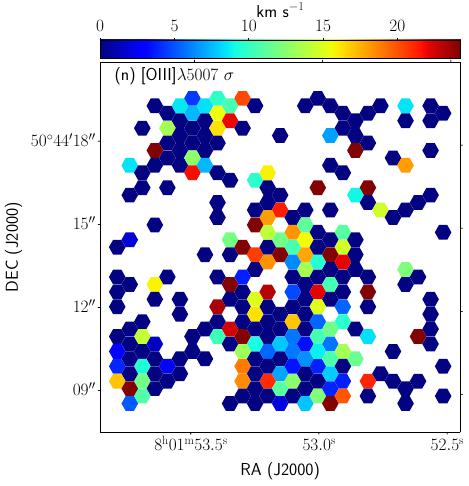}
	\includegraphics[clip, width=0.24\linewidth]{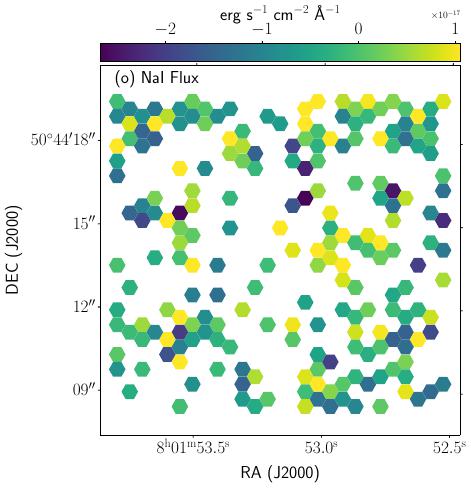}
	\includegraphics[clip, width=0.24\linewidth]{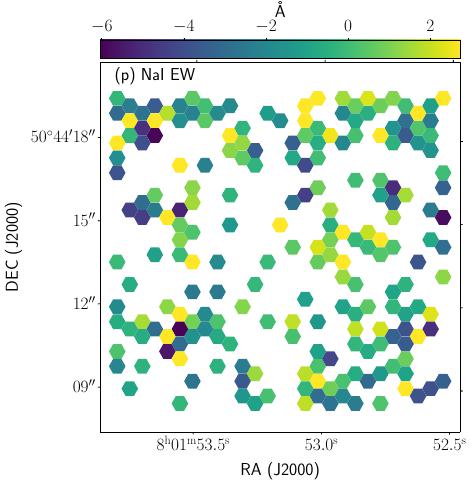}
	\includegraphics[clip, width=0.24\linewidth]{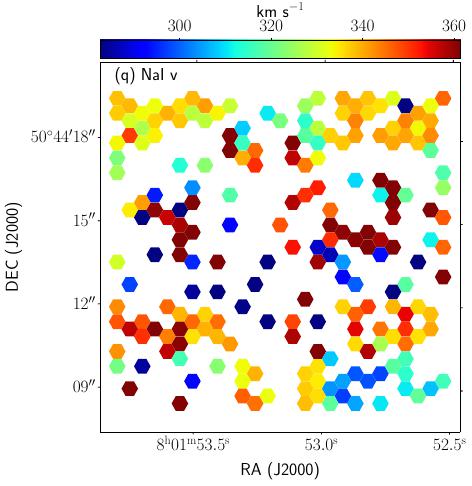}
	\includegraphics[clip, width=0.24\linewidth]{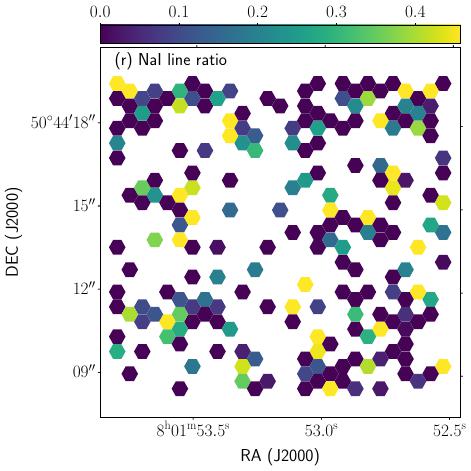}
	\caption{NGC~2500 card.}
	\label{fig:NGC2500_card_1}
\end{figure*}
\addtocounter{figure}{-1}
\begin{figure*}[h]
	\centering
	\includegraphics[clip, width=0.24\linewidth]{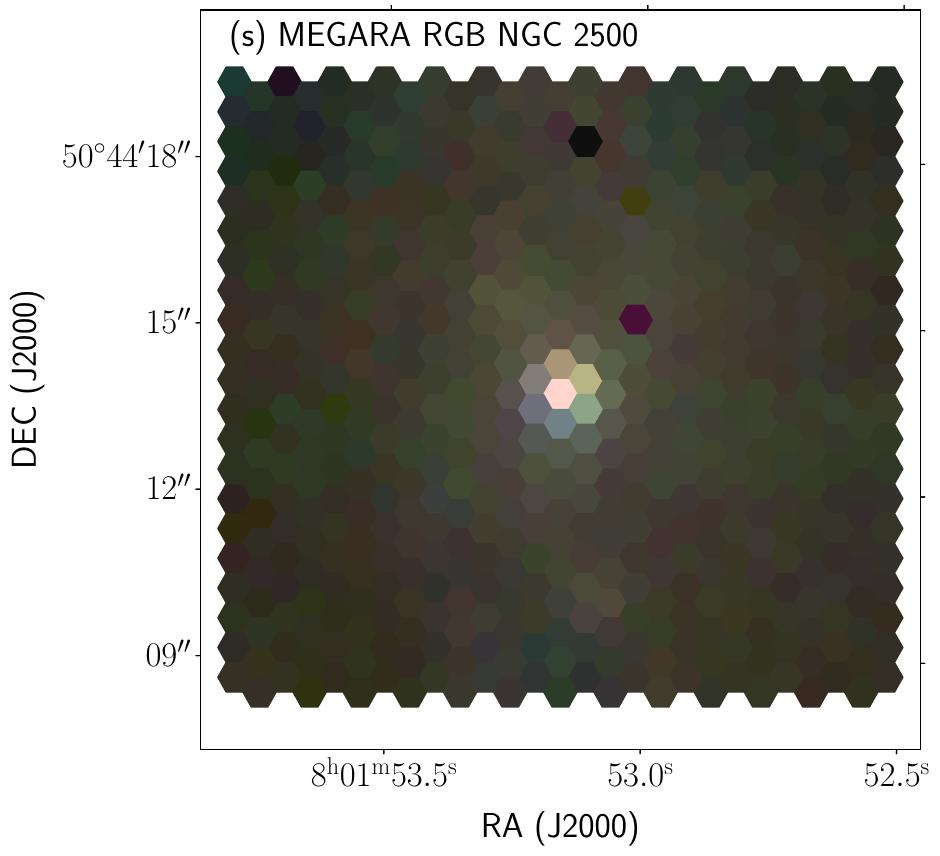}
	\includegraphics[clip, width=0.24\linewidth]{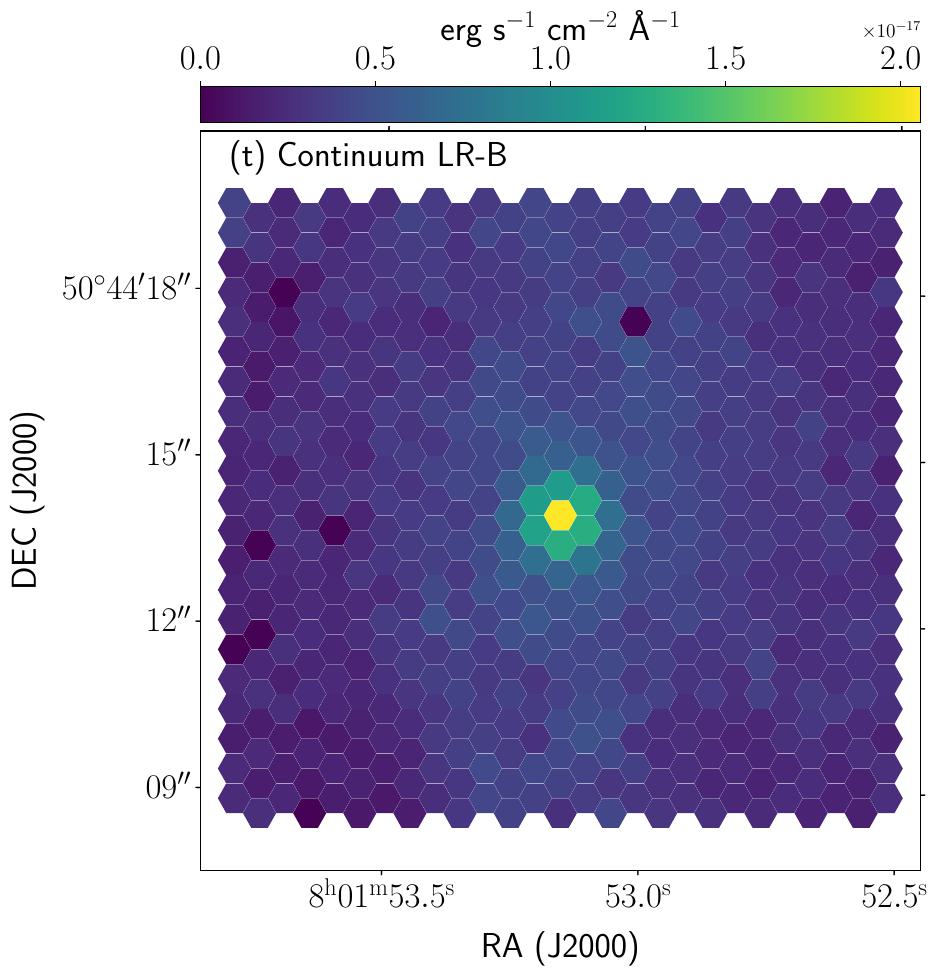}
	\includegraphics[clip, width=0.24\linewidth]{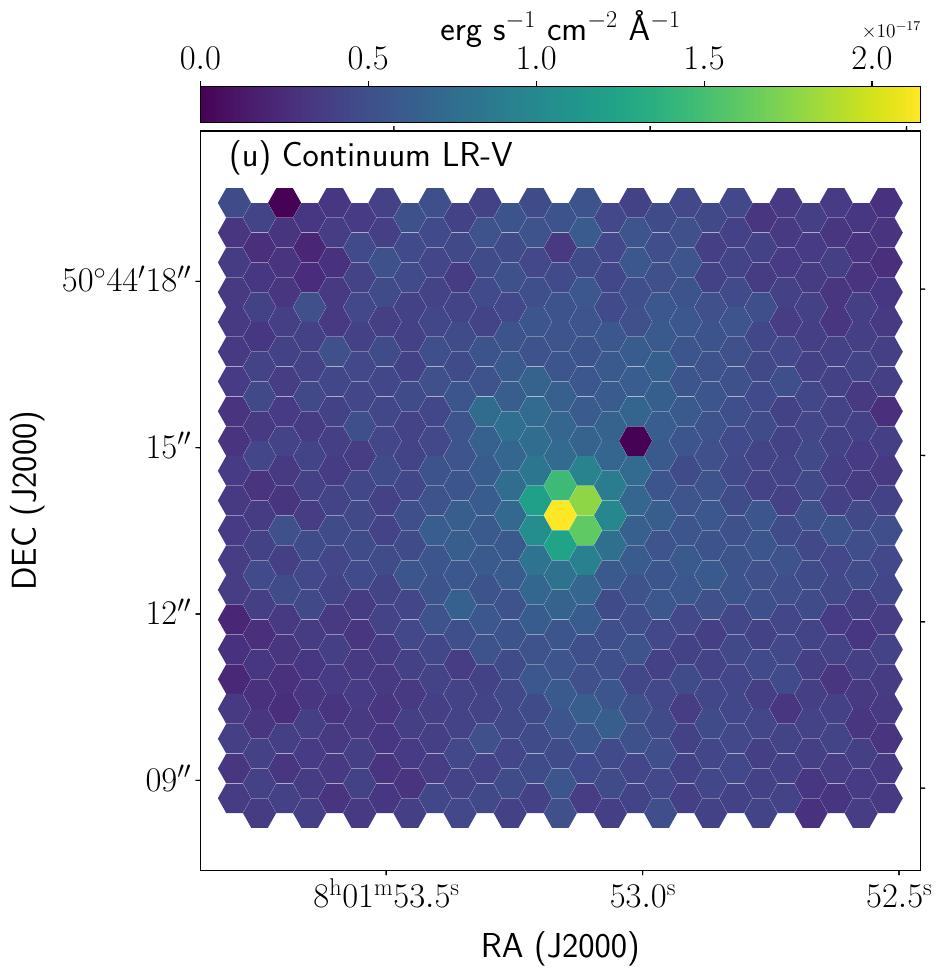}
	\includegraphics[clip, width=0.24\linewidth]{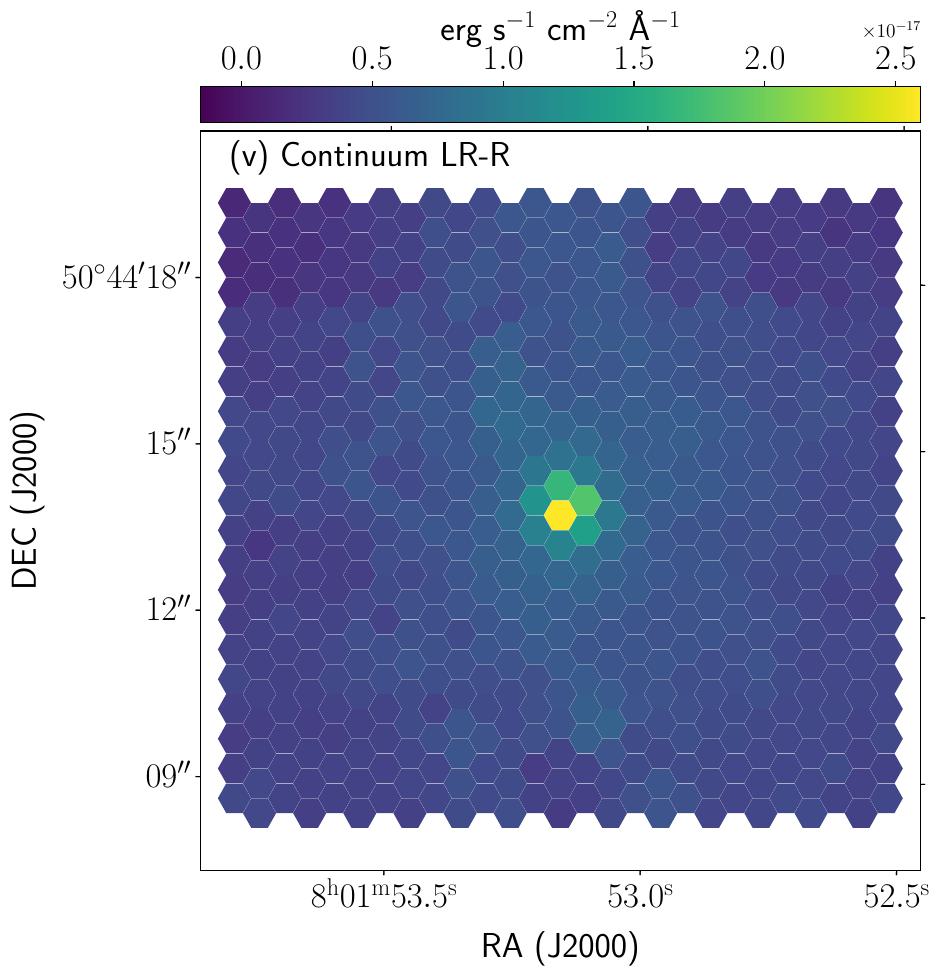}
	\includegraphics[clip, width=0.24\linewidth]{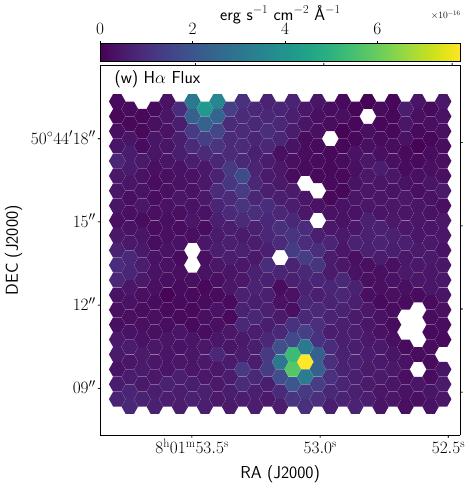}
	\includegraphics[clip, width=0.24\linewidth]{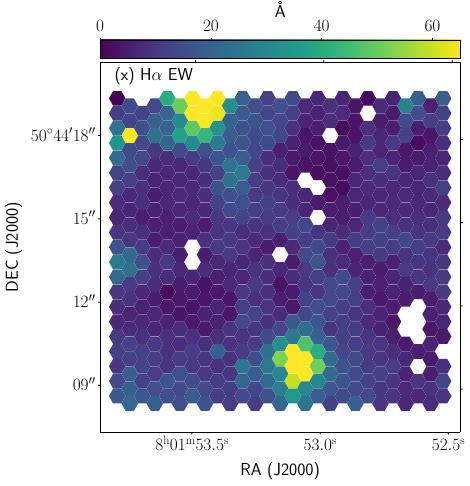}
	\includegraphics[clip, width=0.24\linewidth]{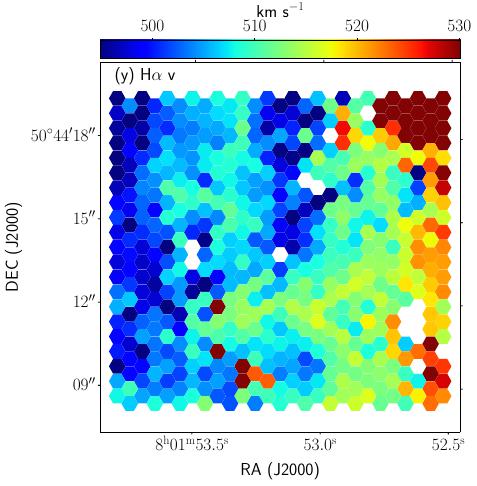}
	\includegraphics[clip, width=0.24\linewidth]{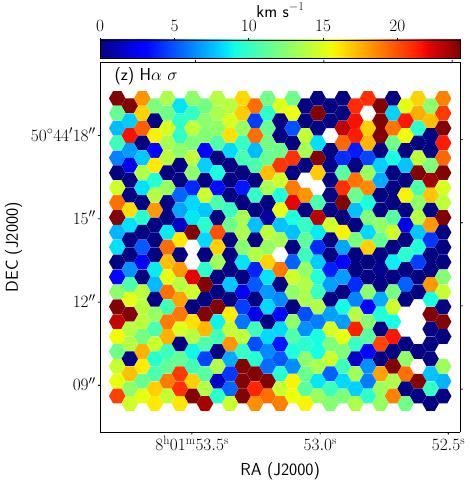}
	\includegraphics[clip, width=0.24\linewidth]{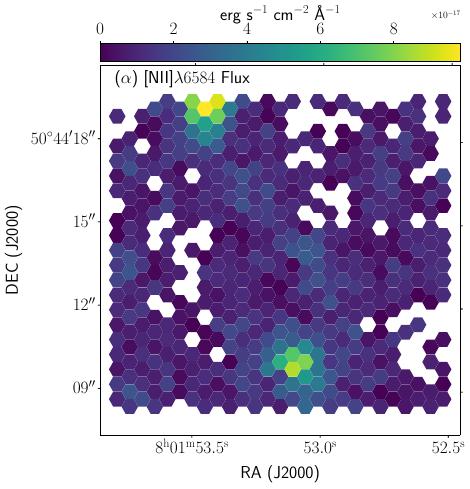}
	\includegraphics[clip, width=0.24\linewidth]{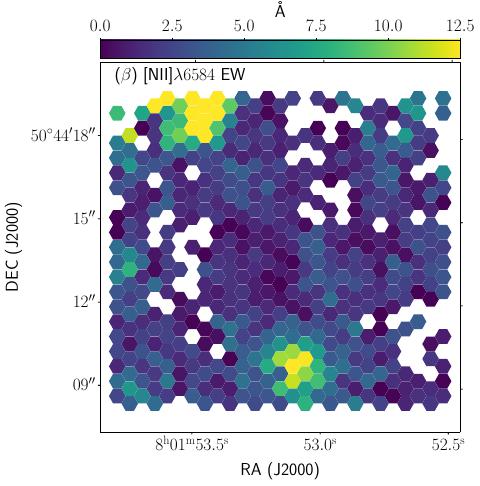}
	\includegraphics[clip, width=0.24\linewidth]{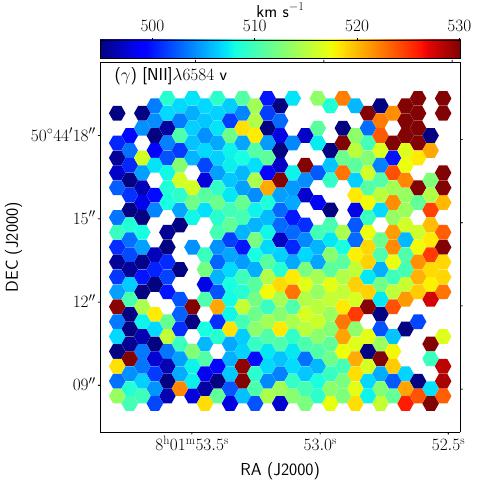}
	\includegraphics[clip, width=0.24\linewidth]{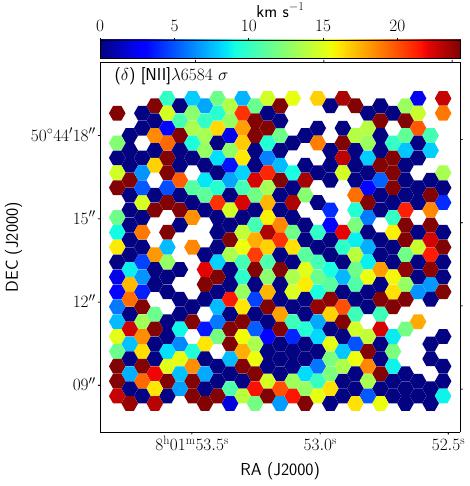}
	\includegraphics[clip, width=0.24\linewidth]{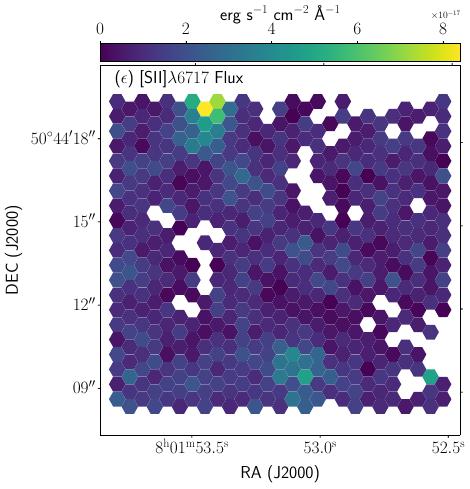}
	\includegraphics[clip, width=0.24\linewidth]{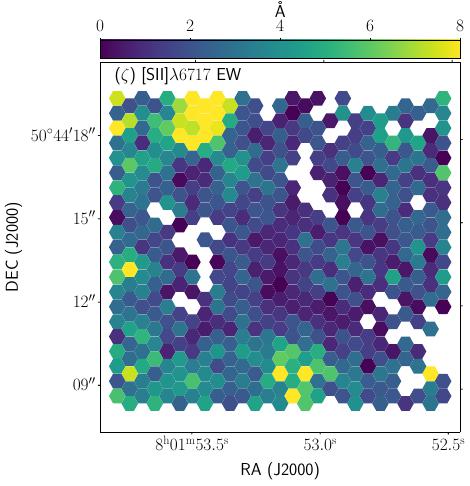}
	\includegraphics[clip, width=0.24\linewidth]{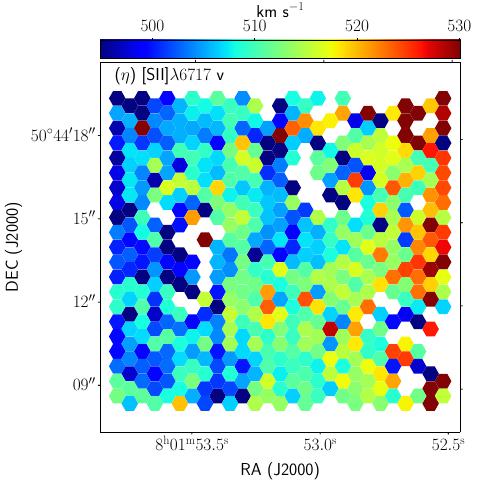}
	\includegraphics[clip, width=0.24\linewidth]{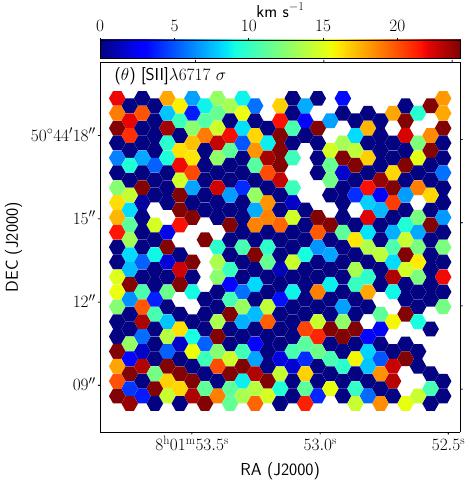}
	\includegraphics[clip, width=0.24\linewidth]{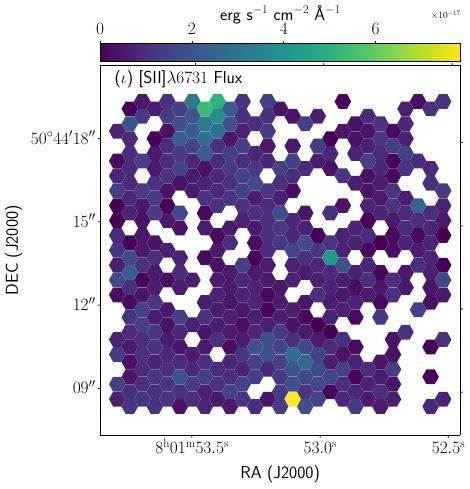}
	\includegraphics[clip, width=0.24\linewidth]{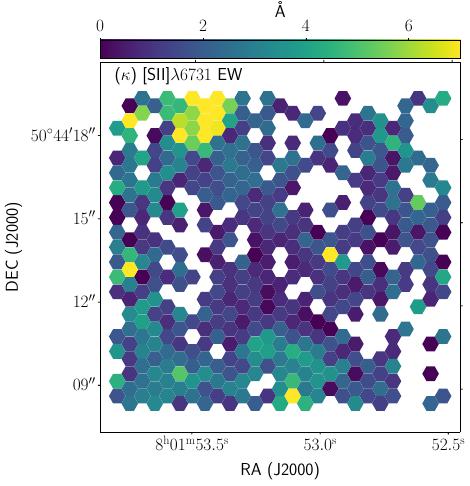}
	\includegraphics[clip, width=0.24\linewidth]{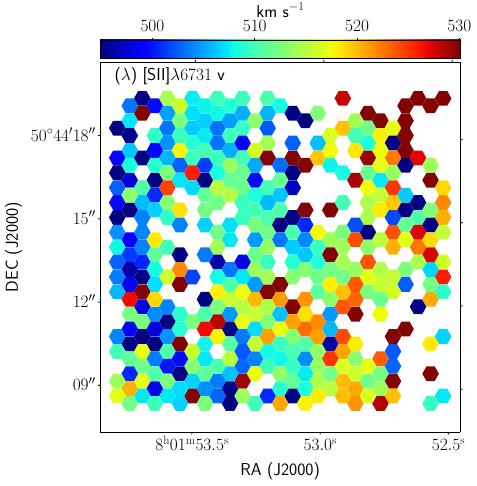}
	\includegraphics[clip, width=0.24\linewidth]{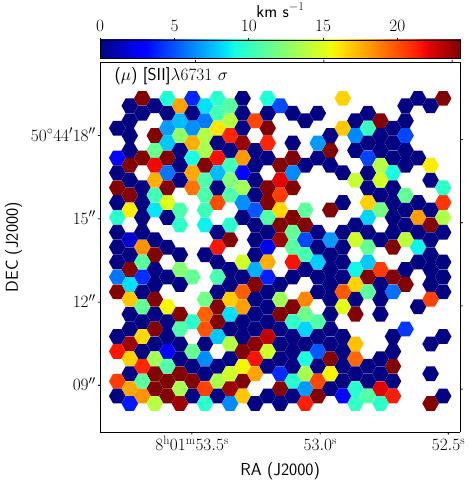}
	\caption{(cont.) NGC~2500 card.}
	\label{fig:NGC2500_card_2}
\end{figure*}

\begin{figure*}[h]
	\centering
	\includegraphics[clip, width=0.35\linewidth]{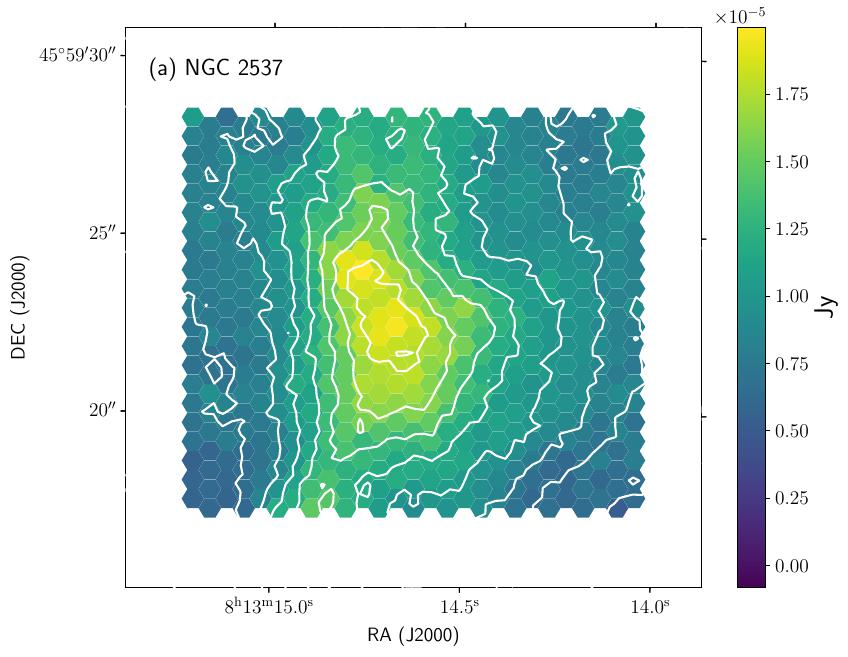}
	\includegraphics[clip, width=0.6\linewidth]{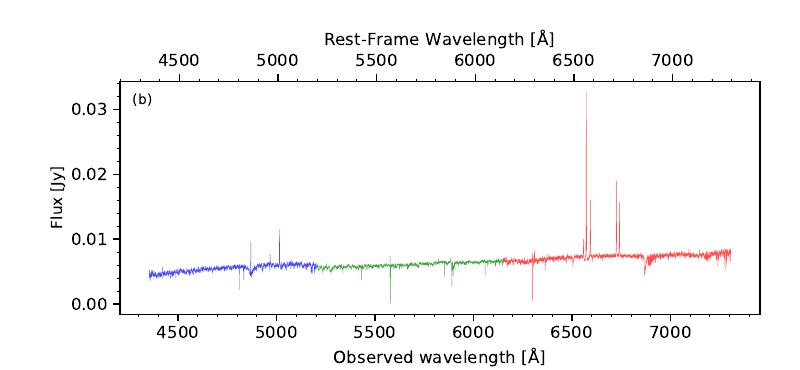}
	\includegraphics[clip, width=0.24\linewidth]{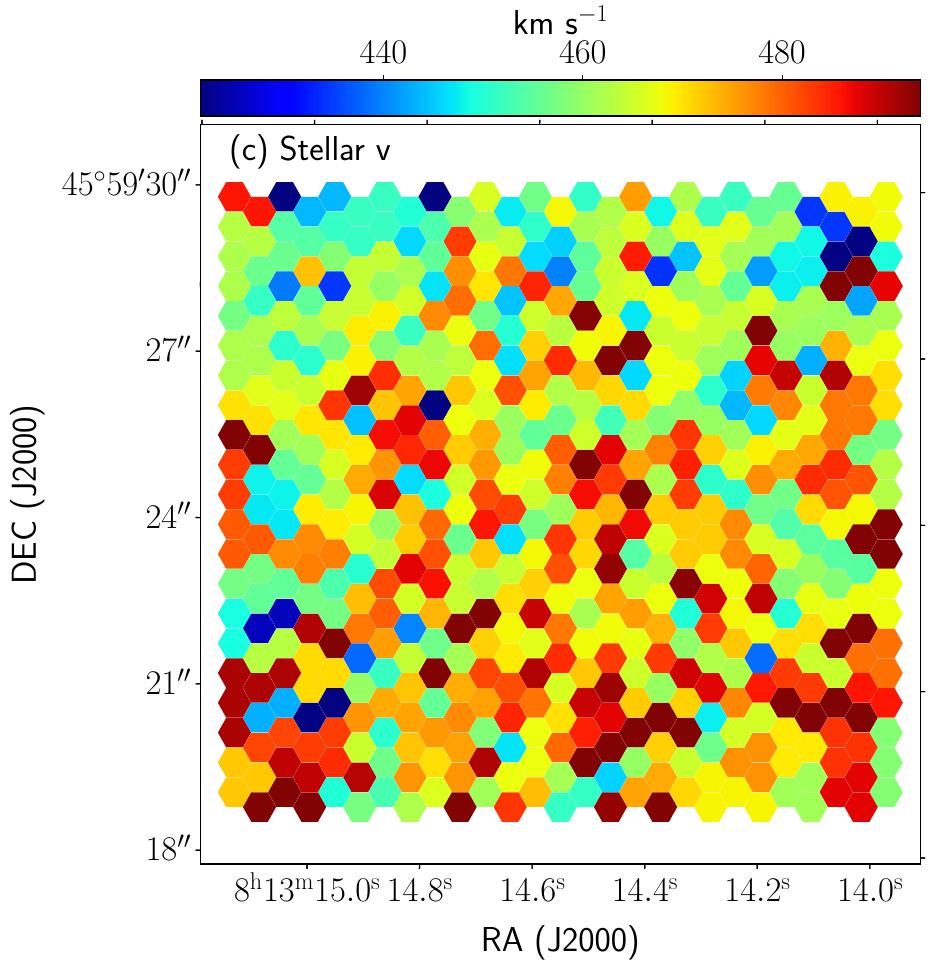}
	\includegraphics[clip, width=0.24\linewidth]{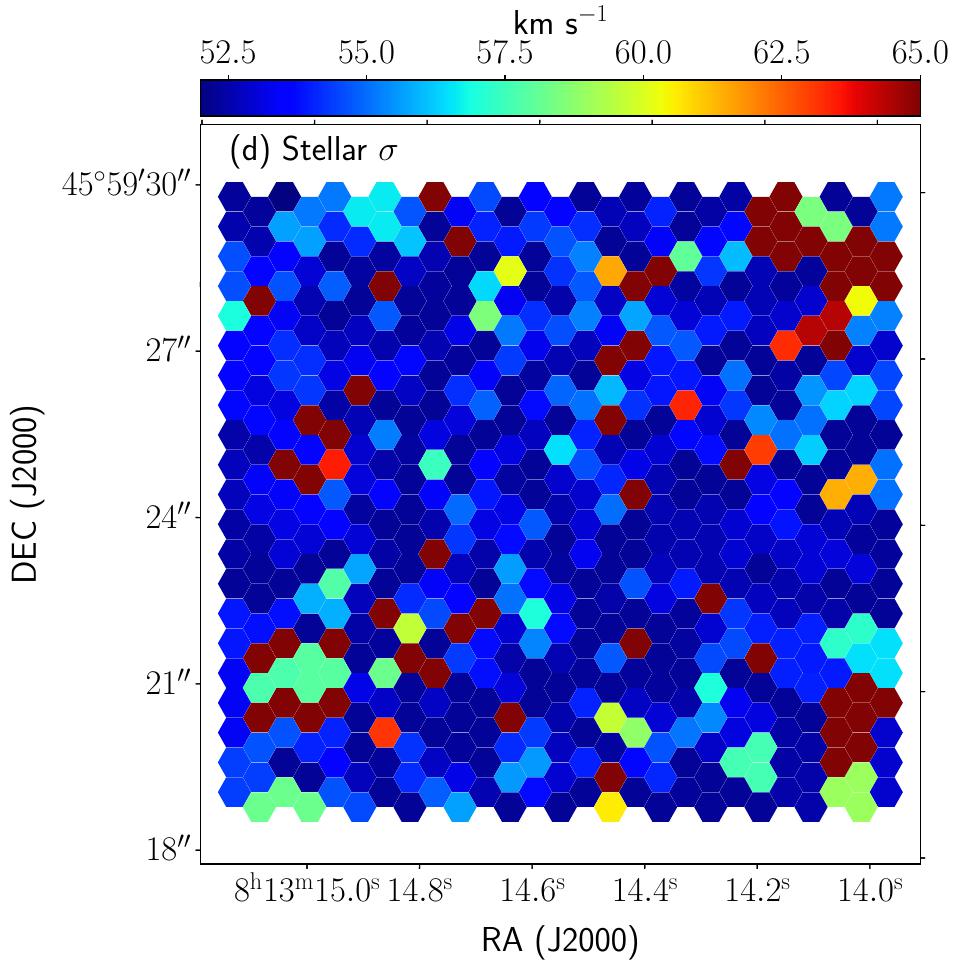}
	\includegraphics[clip, width=0.24\linewidth]{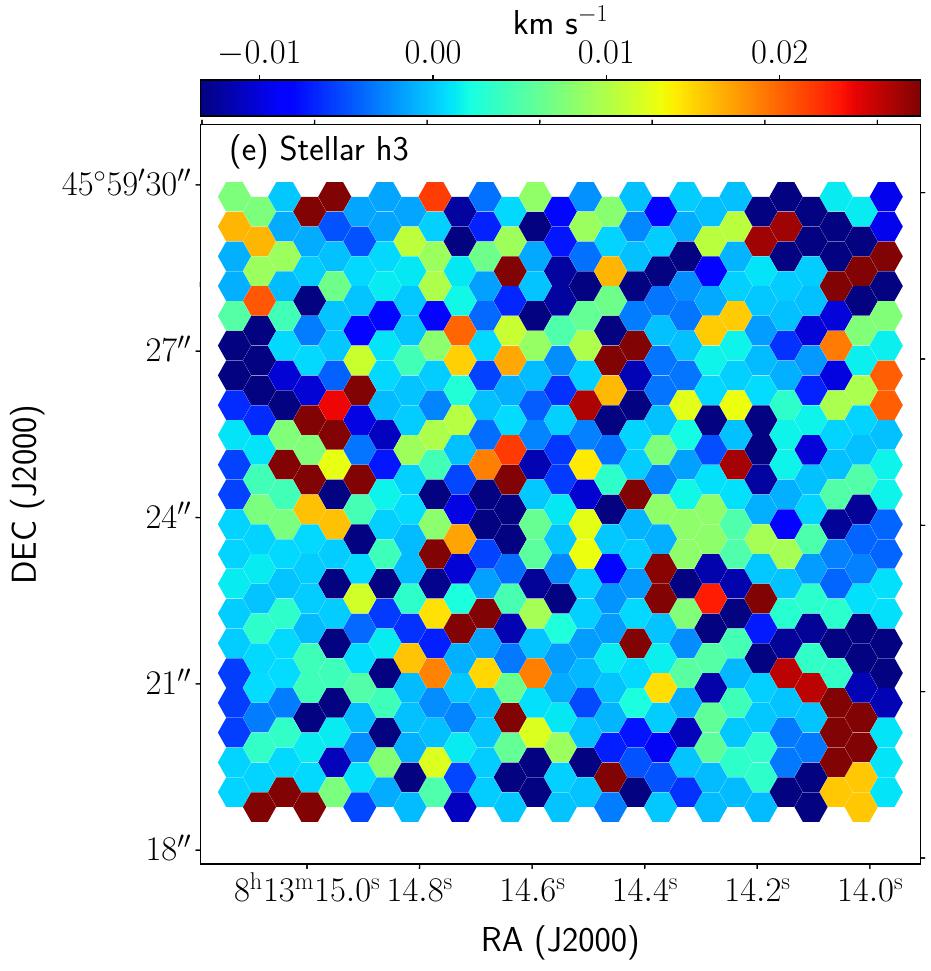}
	\includegraphics[clip, width=0.24\linewidth]{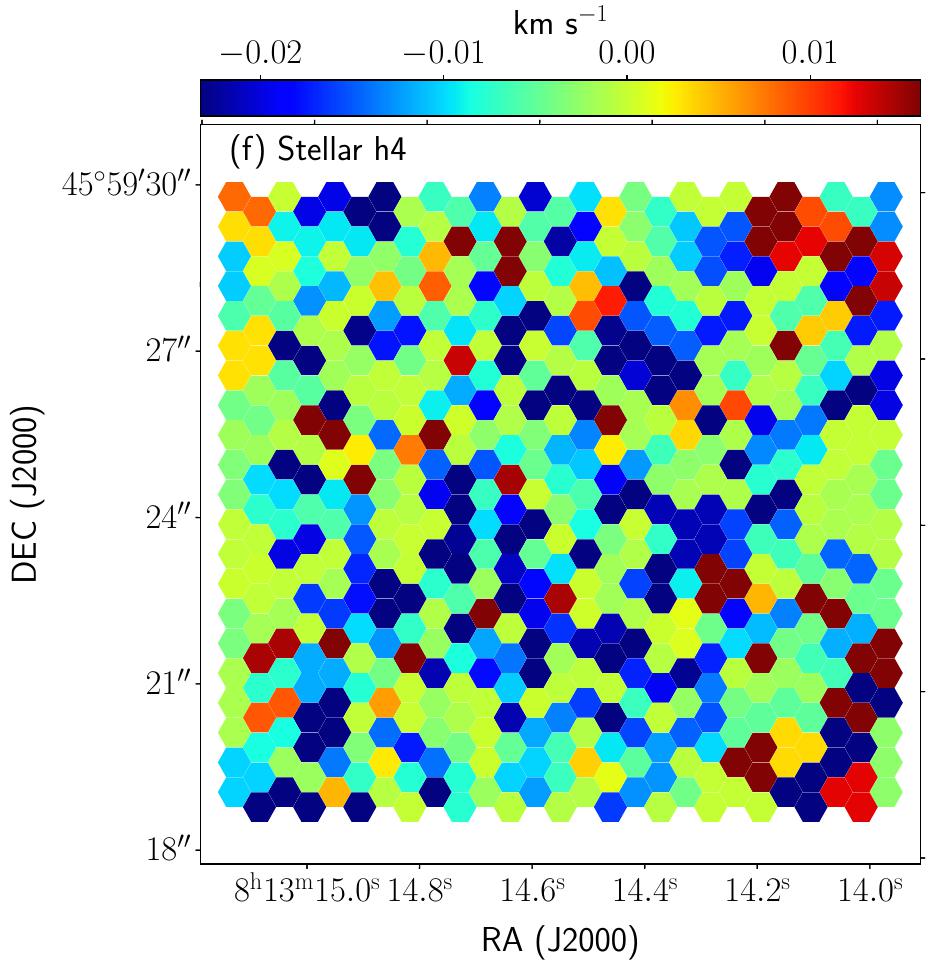}
	\includegraphics[clip, width=0.24\linewidth]{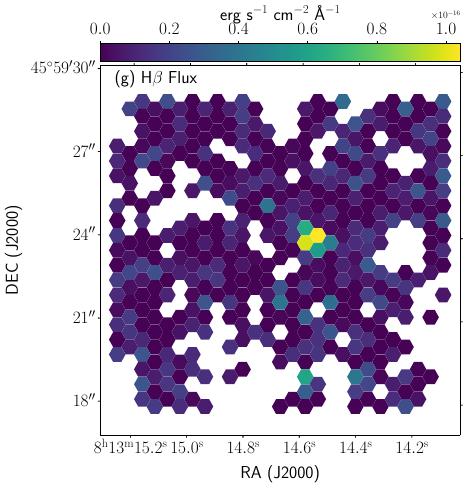}
	\includegraphics[clip, width=0.24\linewidth]{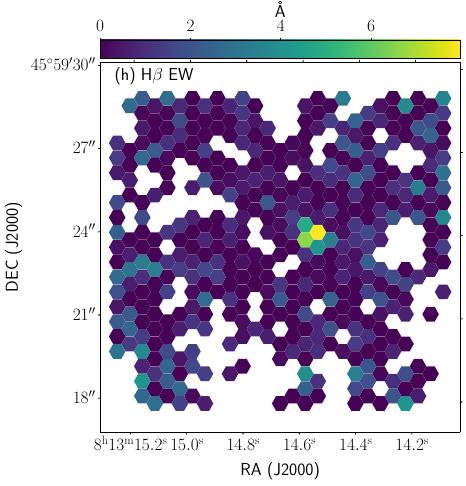}
	\includegraphics[clip, width=0.24\linewidth]{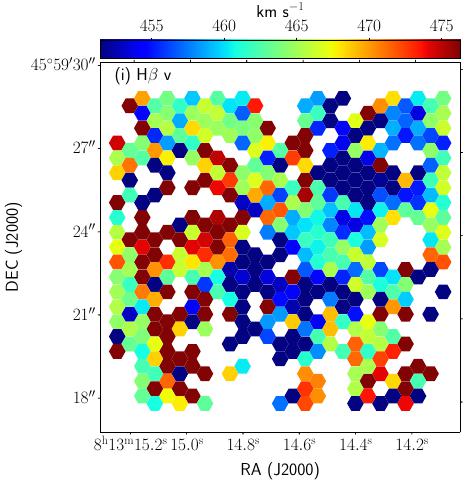}
	\includegraphics[clip, width=0.24\linewidth]{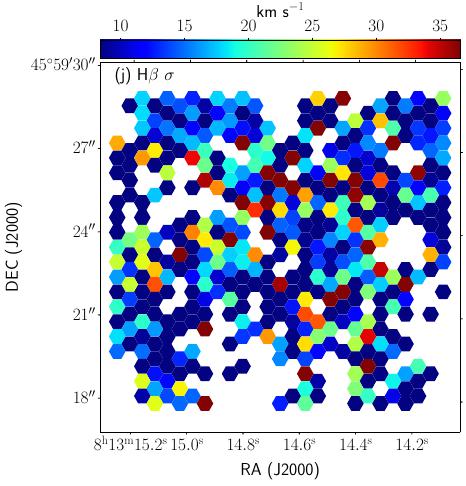}
	\includegraphics[clip, width=0.24\linewidth]{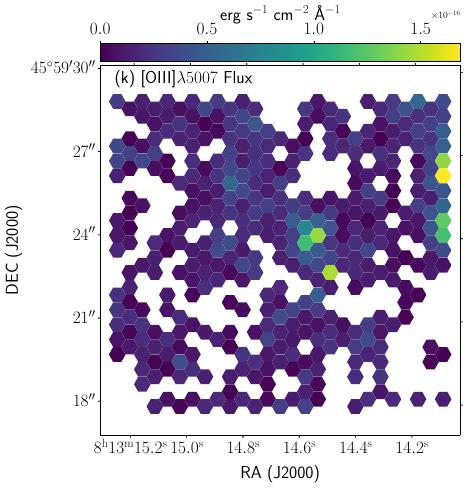}
	\includegraphics[clip, width=0.24\linewidth]{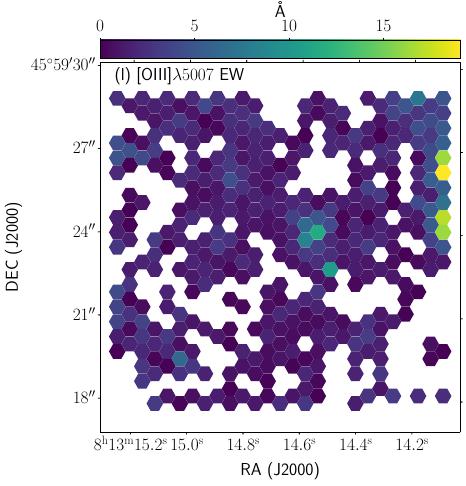}
	\includegraphics[clip, width=0.24\linewidth]{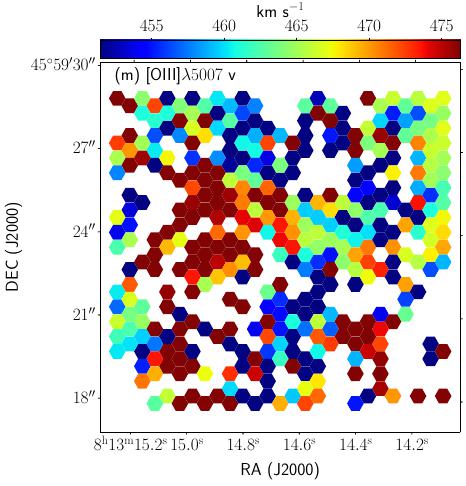}
	\includegraphics[clip, width=0.24\linewidth]{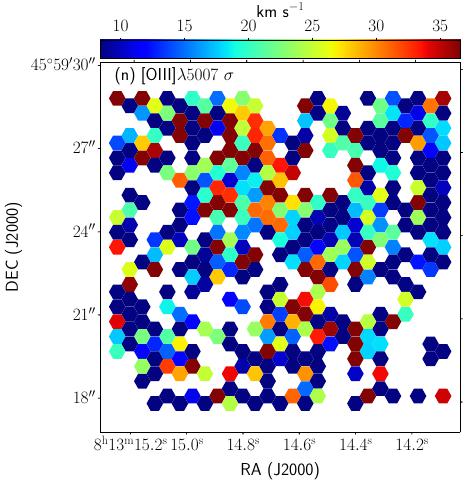}
	\includegraphics[clip, width=0.24\linewidth]{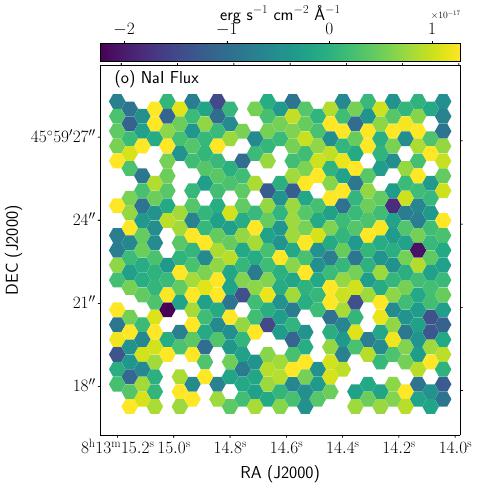}
	\includegraphics[clip, width=0.24\linewidth]{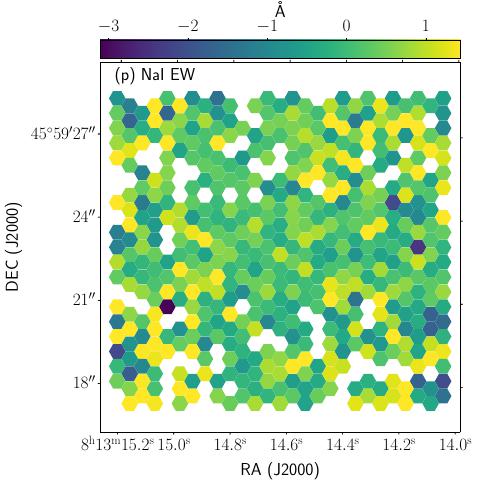}
	\includegraphics[clip, width=0.24\linewidth]{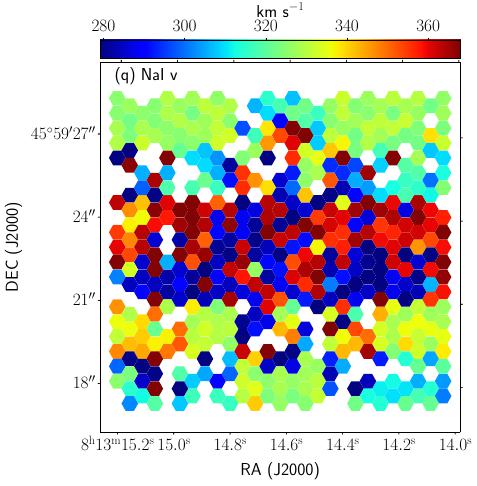}
	\includegraphics[clip, width=0.24\linewidth]{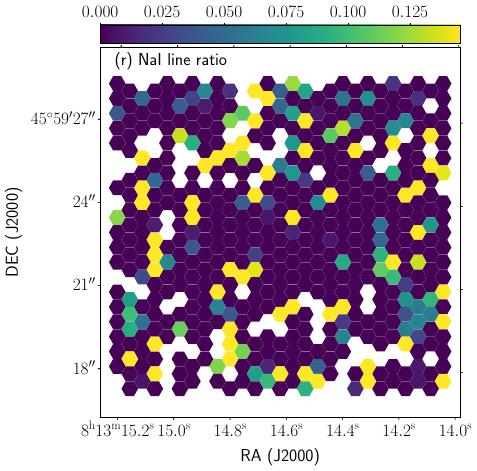}
	\caption{NGC~2537 card.}
	\label{fig:NGC2537_card_1}
\end{figure*}
\addtocounter{figure}{-1}
\begin{figure*}[h]
	\centering
	\includegraphics[clip, width=0.24\linewidth]{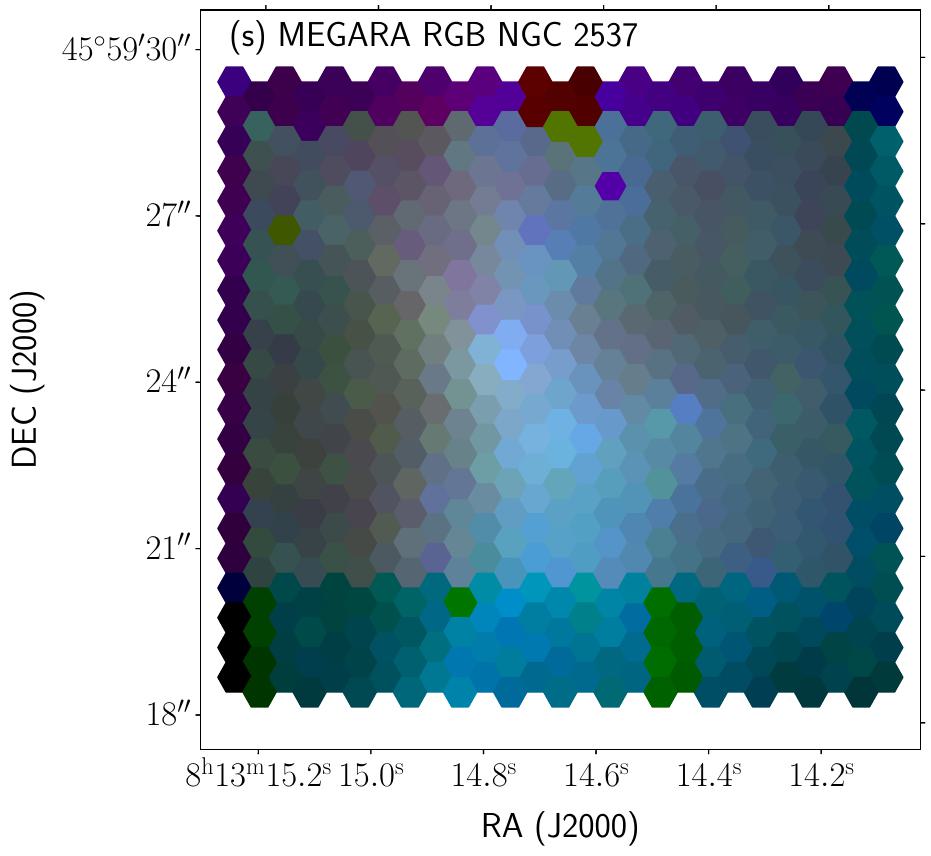}
	\includegraphics[clip, width=0.24\linewidth]{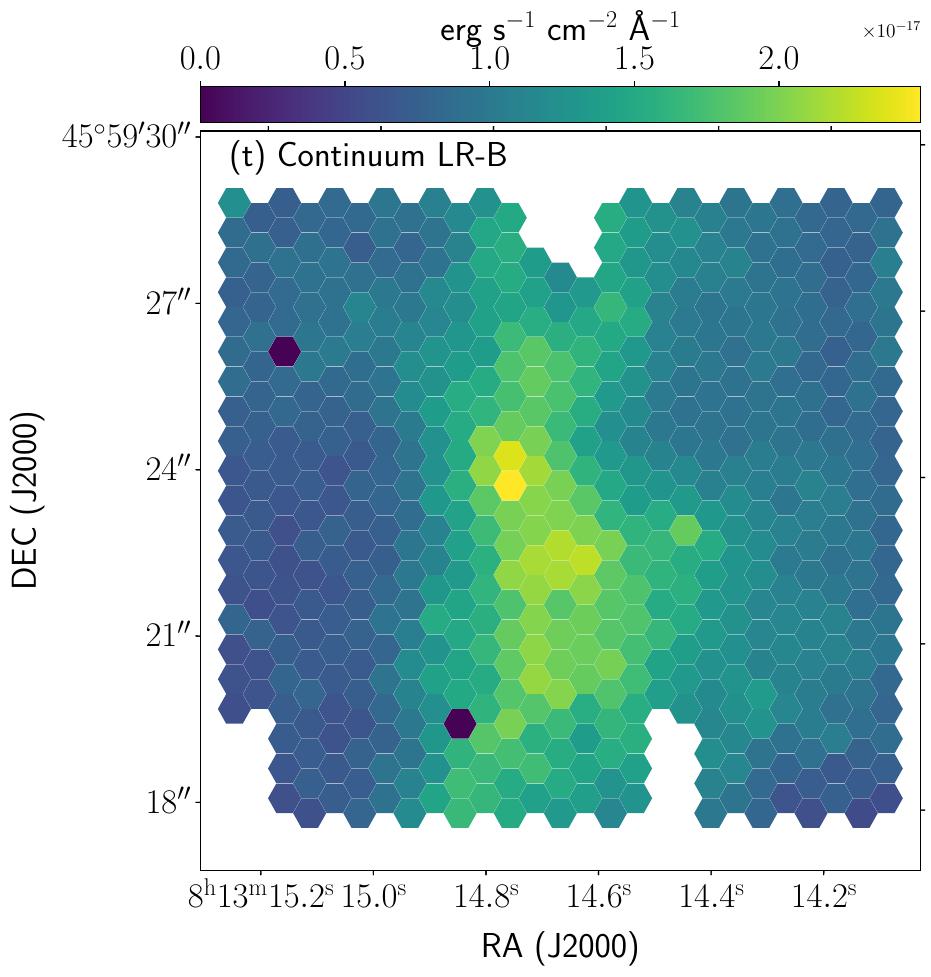}
	\includegraphics[clip, width=0.24\linewidth]{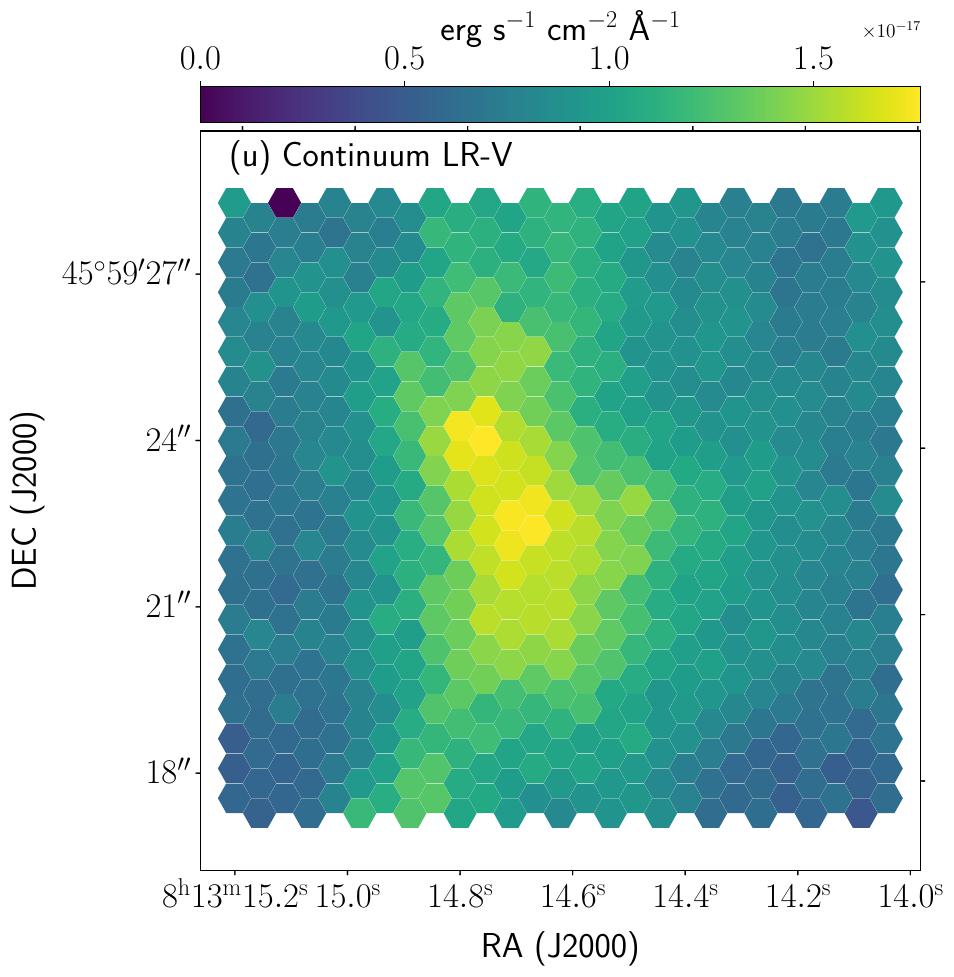}
	\includegraphics[clip, width=0.24\linewidth]{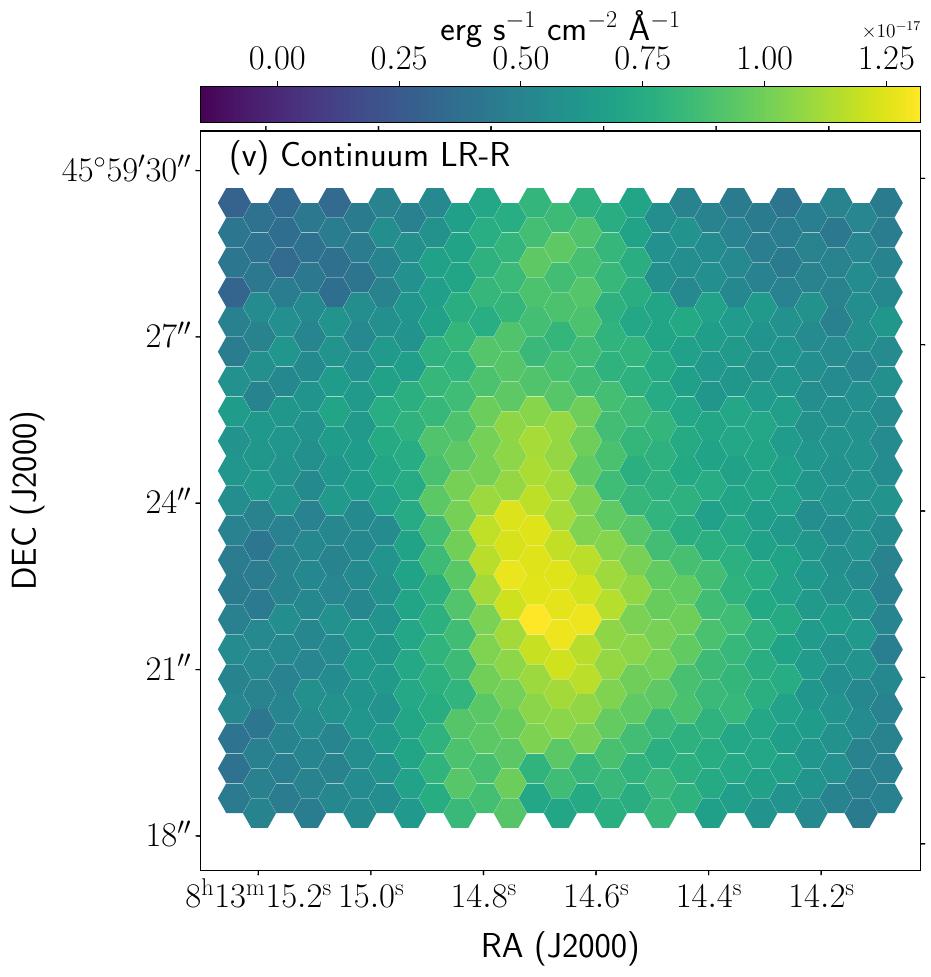}
	\includegraphics[clip, width=0.24\linewidth]{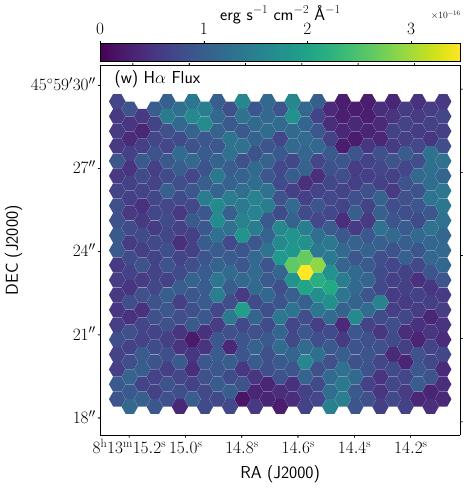}
	\includegraphics[clip, width=0.24\linewidth]{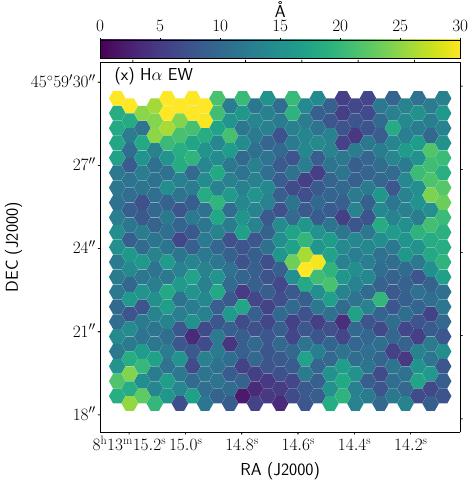}
	\includegraphics[clip, width=0.24\linewidth]{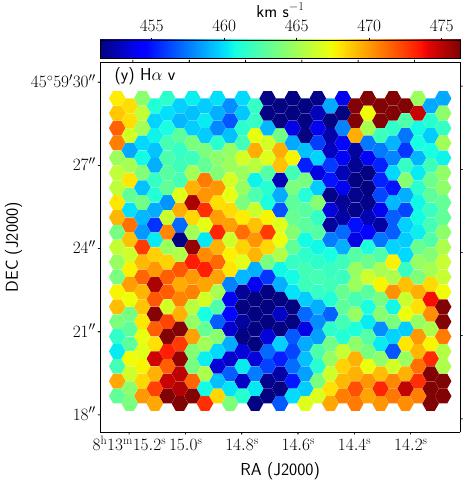}
	\includegraphics[clip, width=0.24\linewidth]{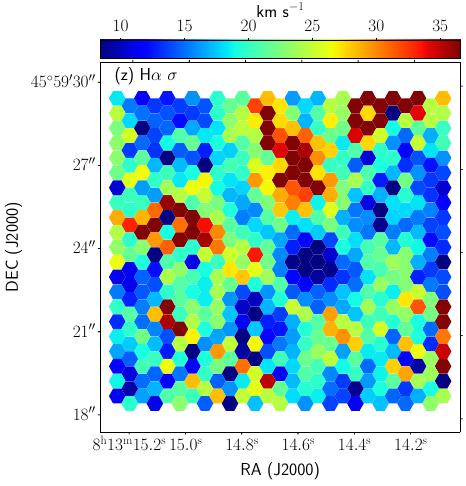}
	\includegraphics[clip, width=0.24\linewidth]{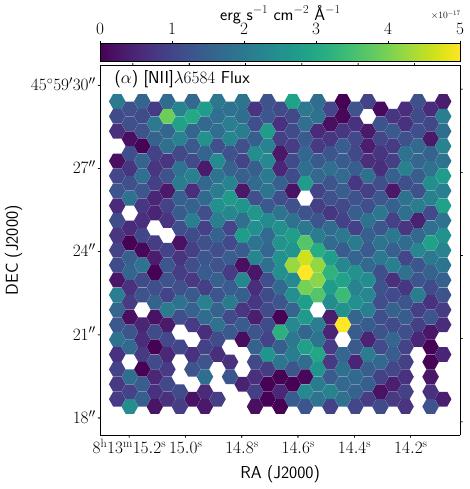}
	\includegraphics[clip, width=0.24\linewidth]{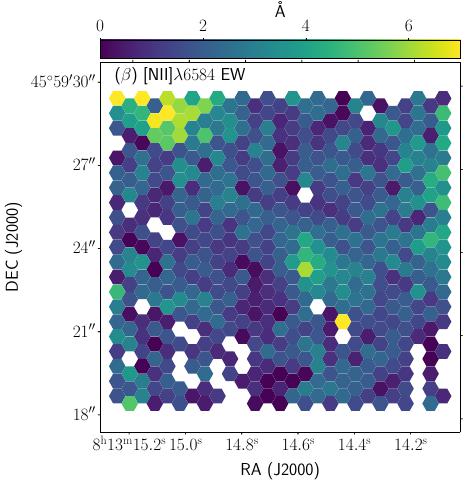}
	\includegraphics[clip, width=0.24\linewidth]{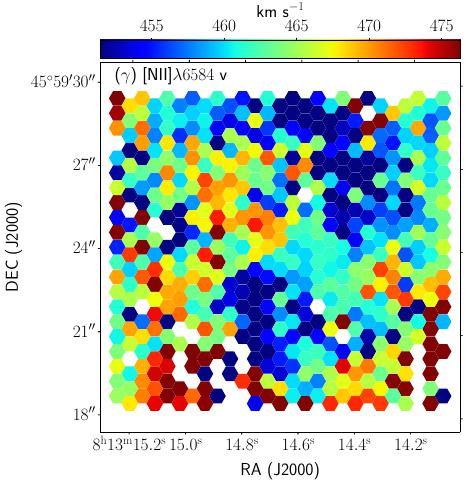}
	\includegraphics[clip, width=0.24\linewidth]{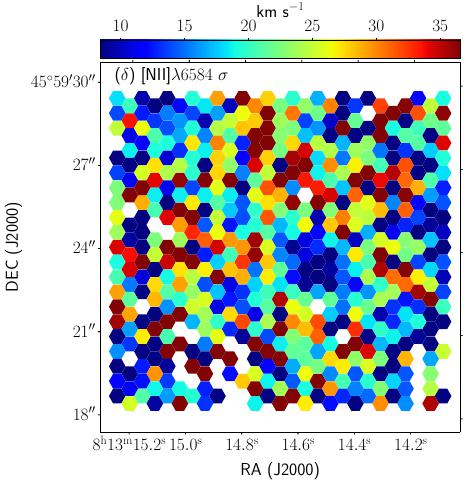}
	\includegraphics[clip, width=0.24\linewidth]{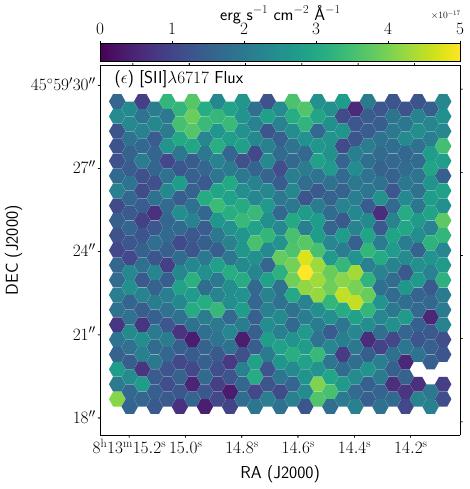}
	\includegraphics[clip, width=0.24\linewidth]{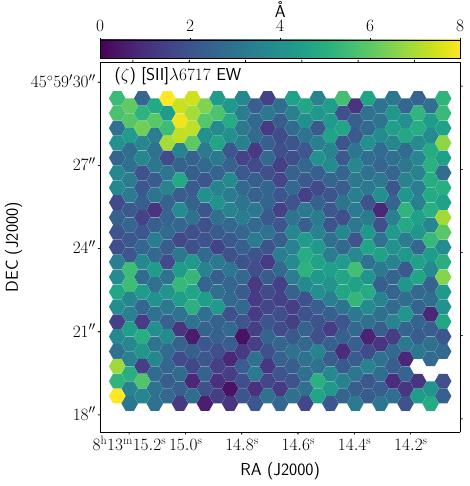}
	\includegraphics[clip, width=0.24\linewidth]{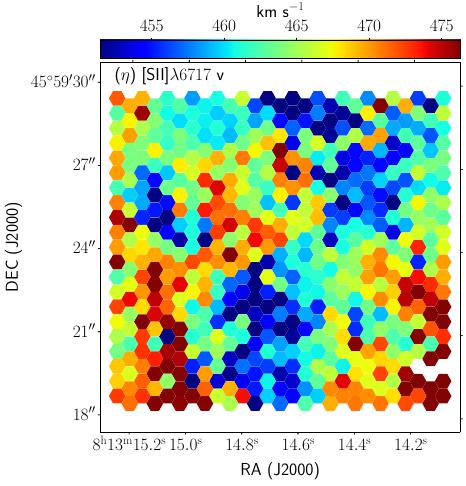}
	\includegraphics[clip, width=0.24\linewidth]{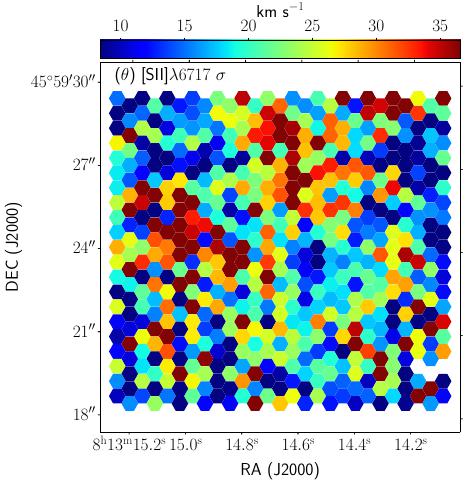}
	\includegraphics[clip, width=0.24\linewidth]{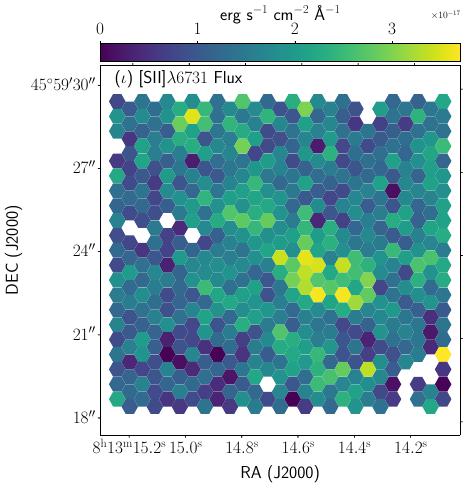}
	\includegraphics[clip, width=0.24\linewidth]{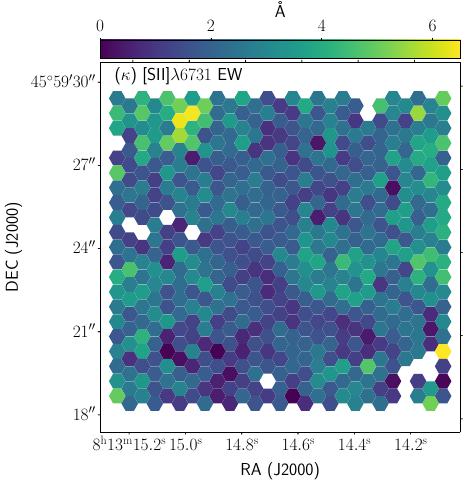}
	\includegraphics[clip, width=0.24\linewidth]{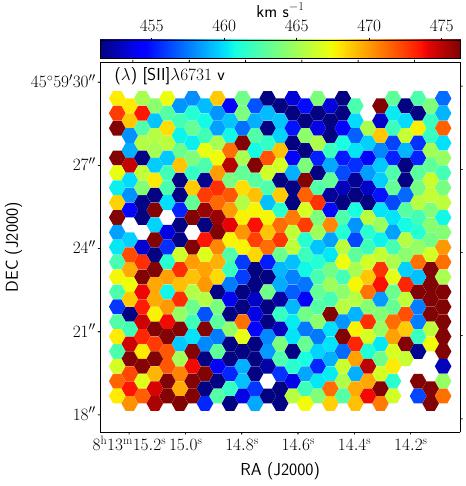}
	\includegraphics[clip, width=0.24\linewidth]{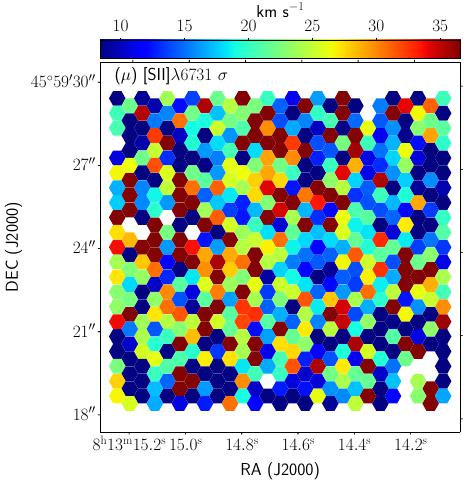}
	\caption{(cont.) NGC~2537 card.}
	\label{fig:NGC2537_card_2}
\end{figure*}

\begin{figure*}[h]
	\centering
	\includegraphics[clip, width=0.35\linewidth]{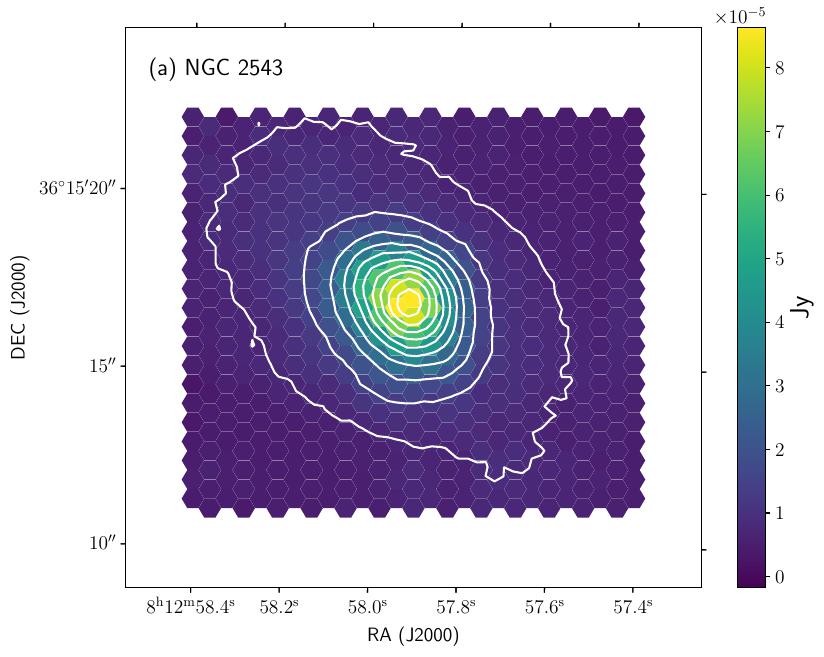}
	\includegraphics[clip, width=0.6\linewidth]{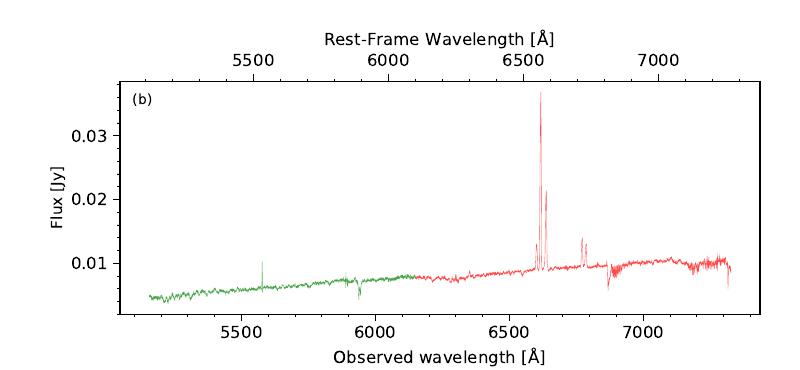}
	\includegraphics[clip, width=0.24\linewidth]{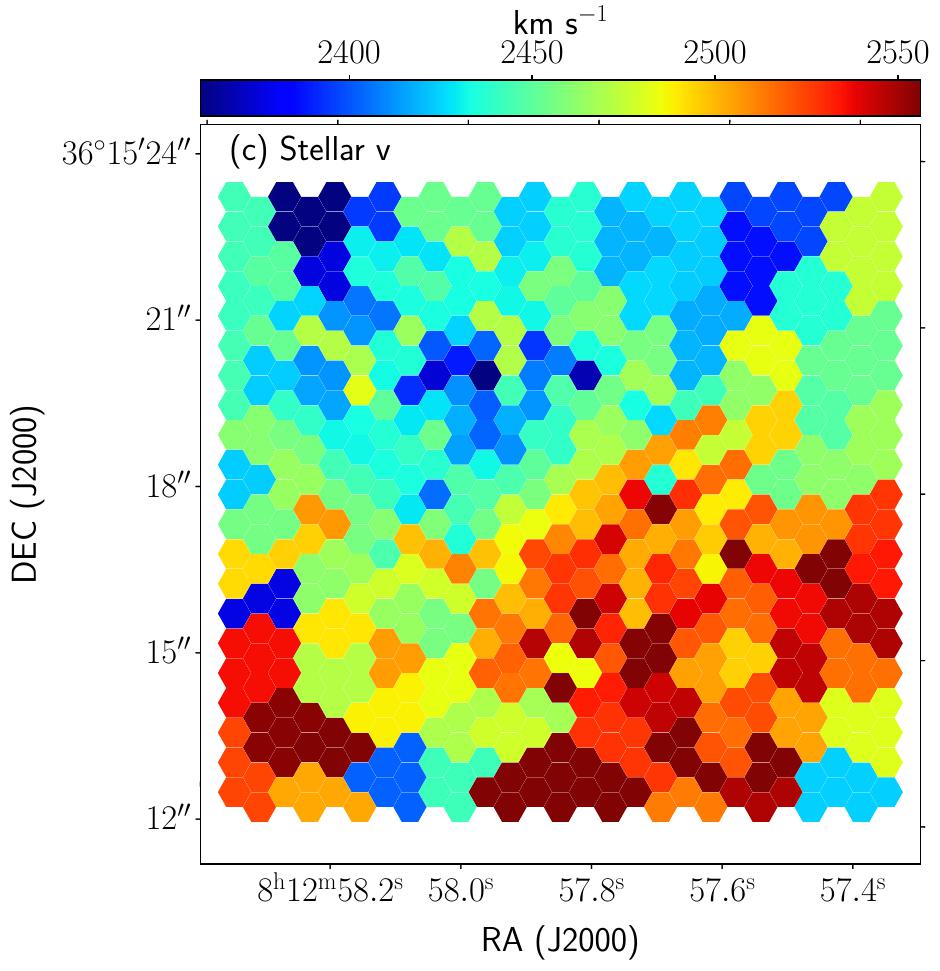}
	\includegraphics[clip, width=0.24\linewidth]{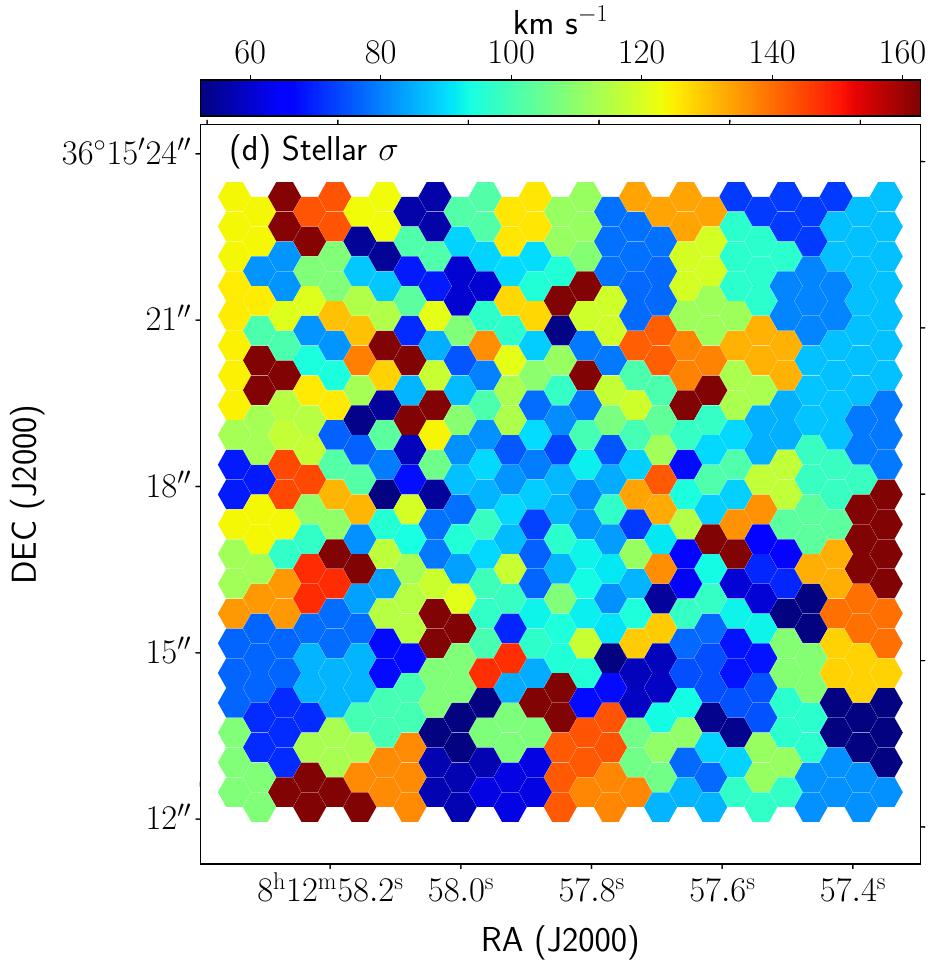}
	\includegraphics[clip, width=0.24\linewidth]{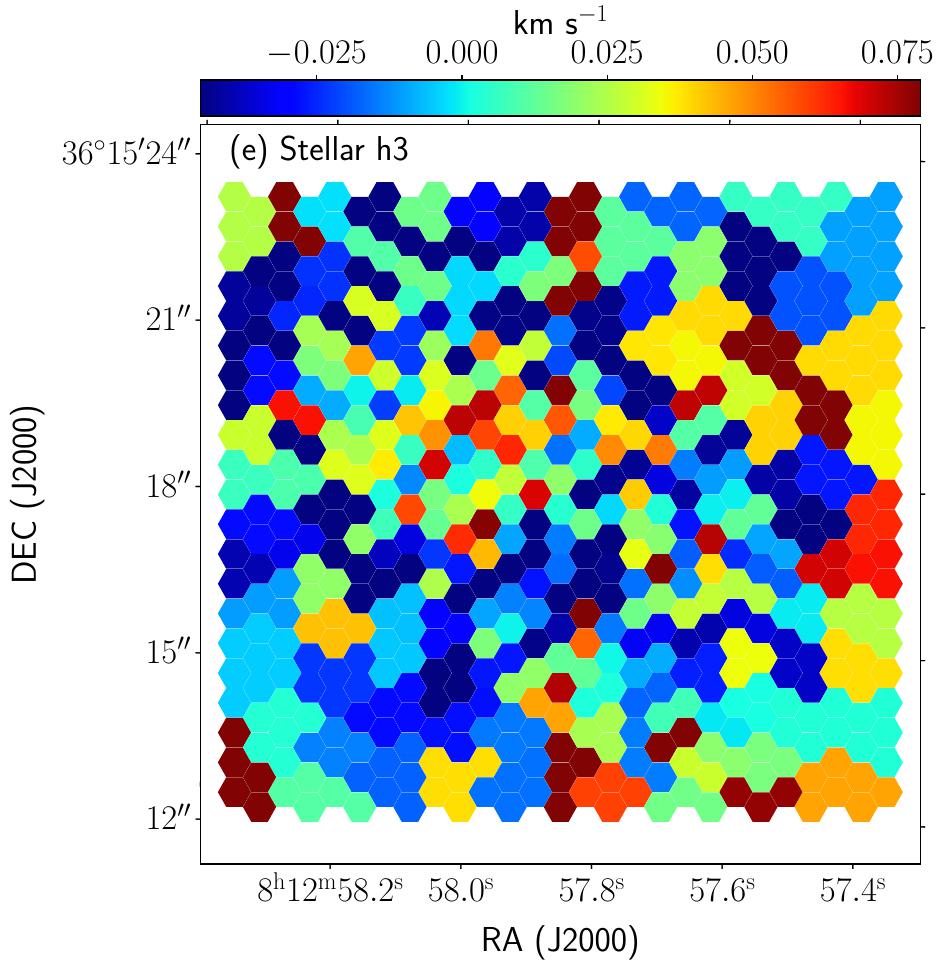}
	\includegraphics[clip, width=0.24\linewidth]{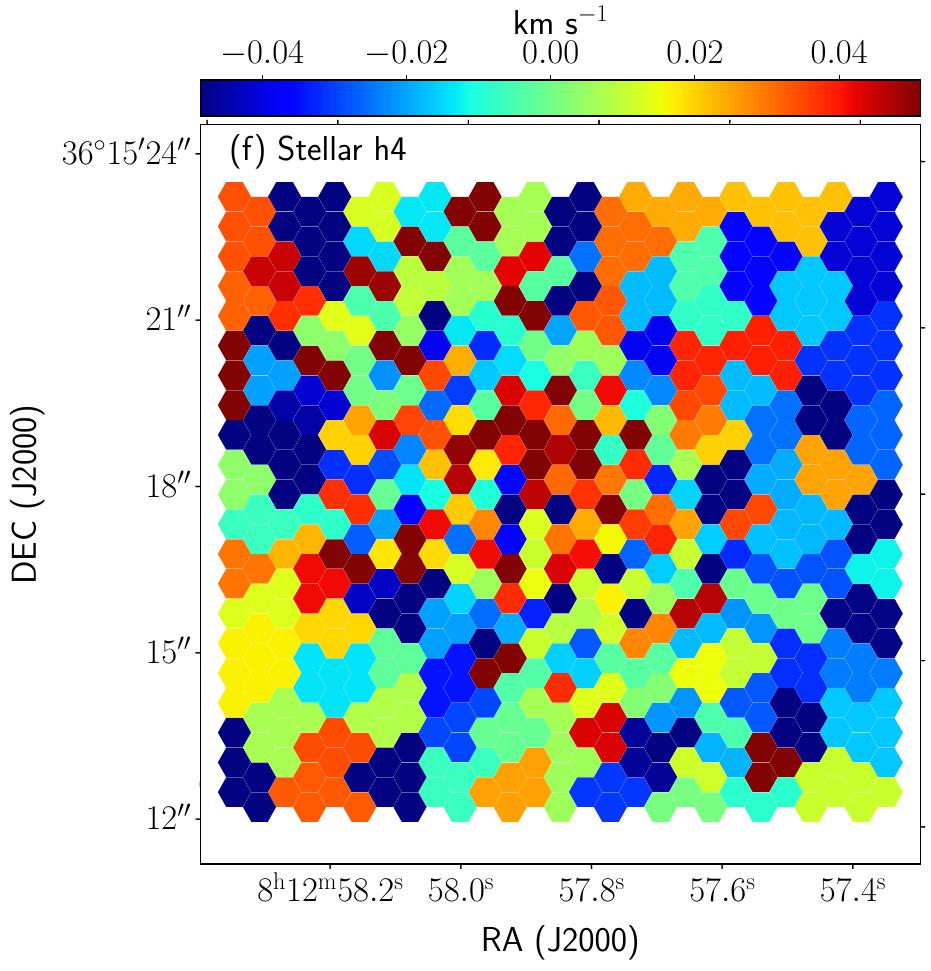}
	
	\vspace{8.8cm}
	
	\includegraphics[clip, width=0.24\linewidth]{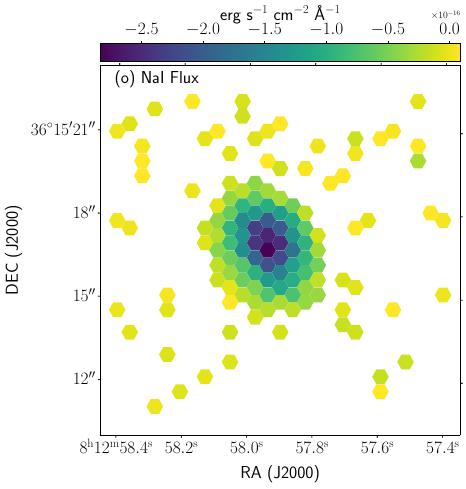}
	\includegraphics[clip, width=0.24\linewidth]{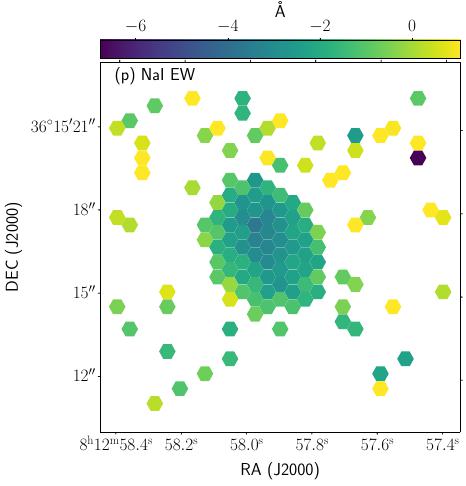}
	\includegraphics[clip, width=0.24\linewidth]{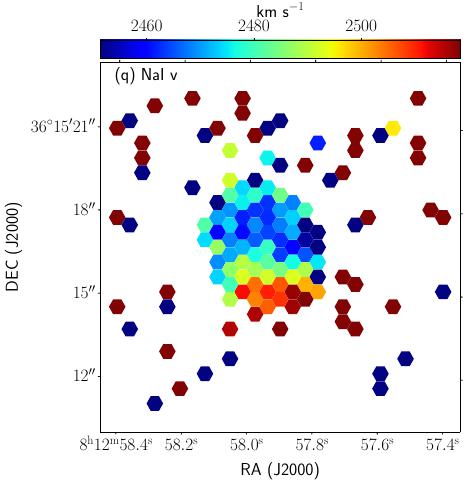}
	\includegraphics[clip, width=0.24\linewidth]{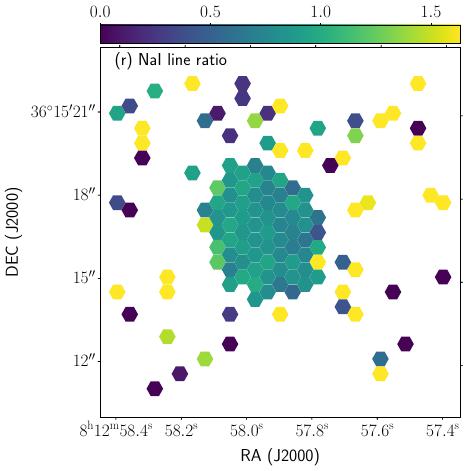}
	\caption{NGC~2543 card.}
	\label{fig:NGC2543_card_1}
\end{figure*}
\addtocounter{figure}{-1}
\begin{figure*}[h]
	\centering
	\includegraphics[clip, width=0.24\linewidth]{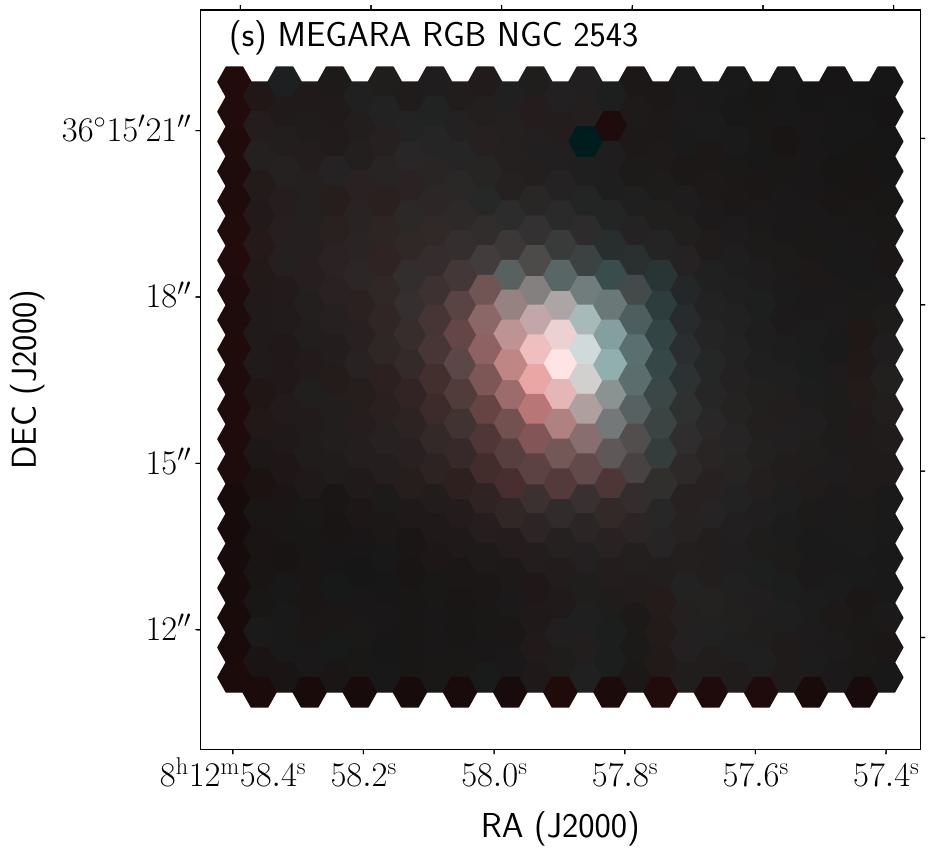}
	\hspace{4,4cm}
	\includegraphics[clip, width=0.24\linewidth]{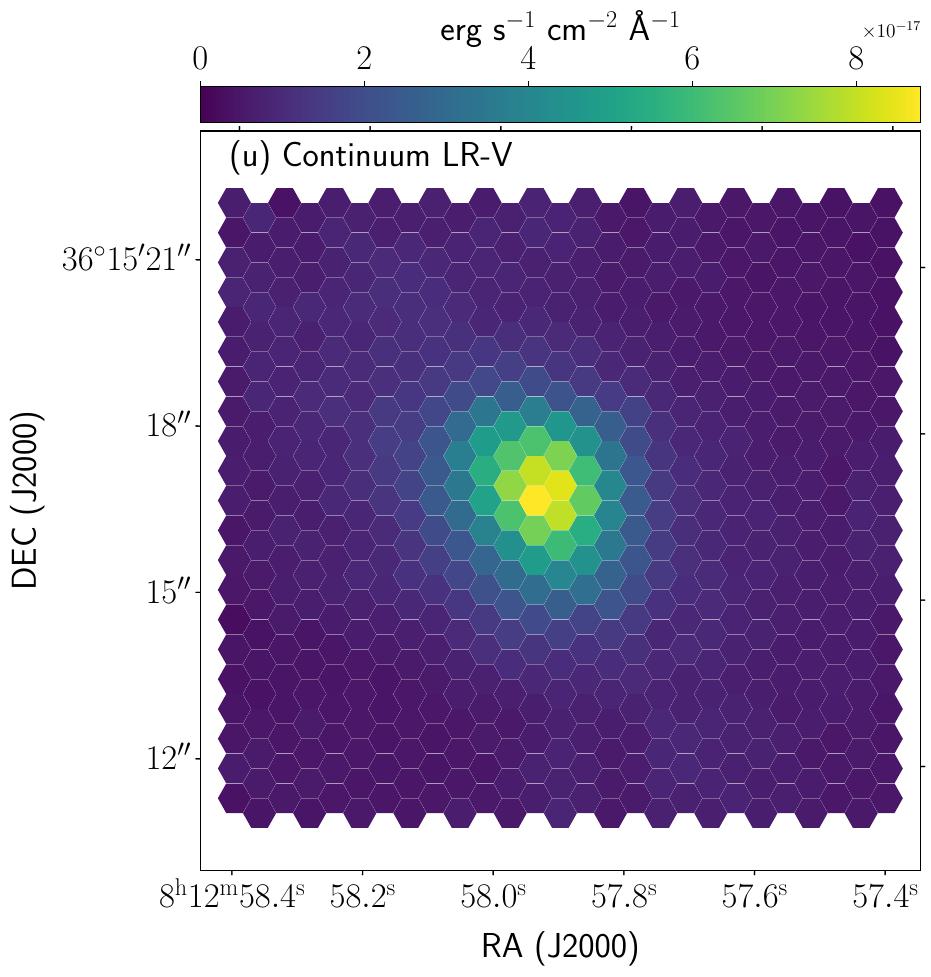}
	\includegraphics[clip, width=0.24\linewidth]{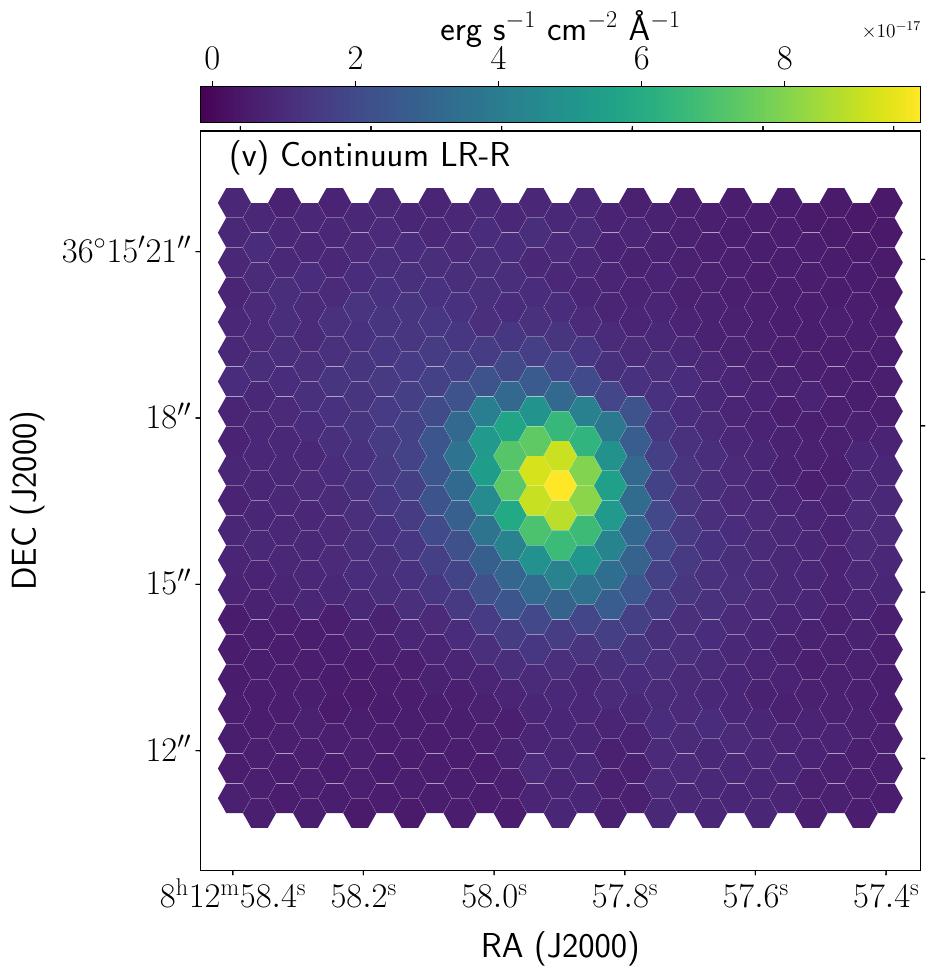}
	\includegraphics[clip, width=0.24\linewidth]{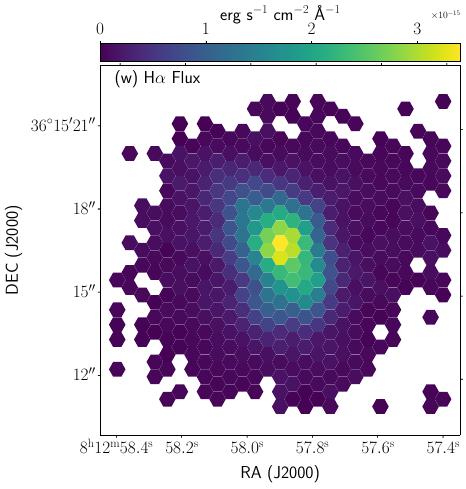}
	\includegraphics[clip, width=0.24\linewidth]{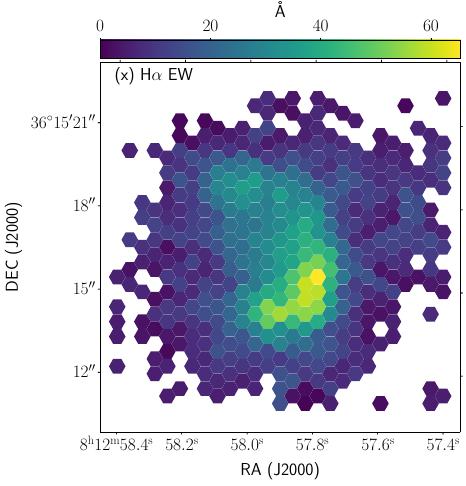}
	\includegraphics[clip, width=0.24\linewidth]{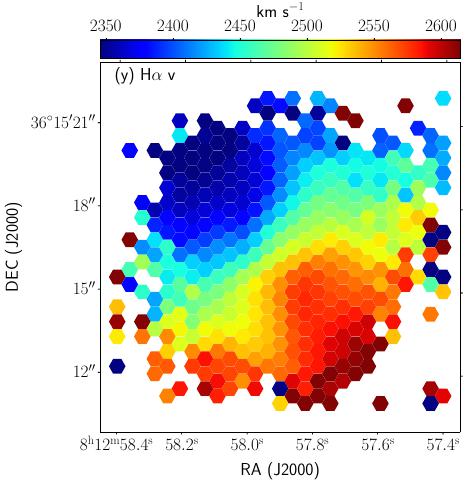}
	\includegraphics[clip, width=0.24\linewidth]{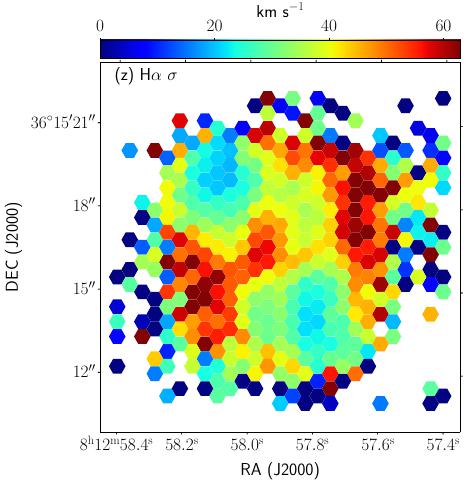}
	\includegraphics[clip, width=0.24\linewidth]{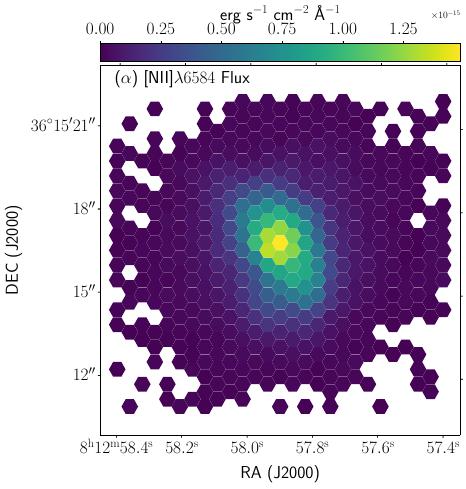}
	\includegraphics[clip, width=0.24\linewidth]{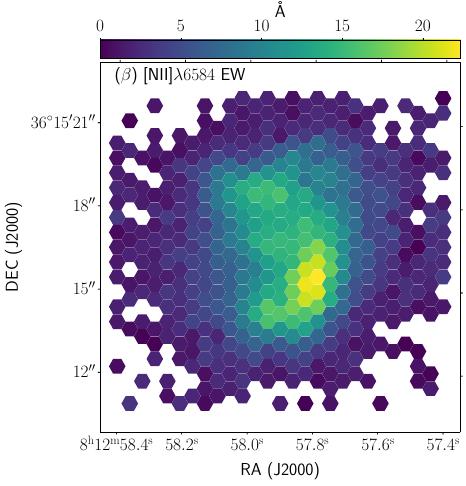}
	\includegraphics[clip, width=0.24\linewidth]{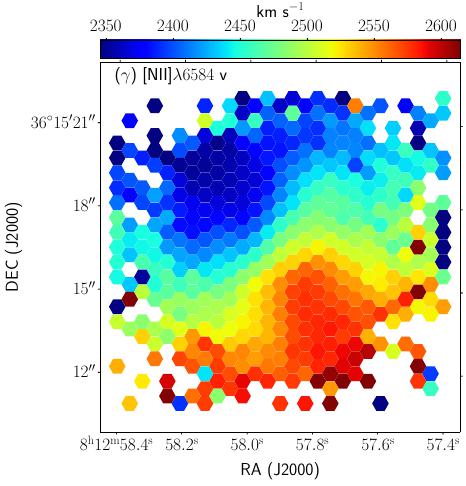}
	\includegraphics[clip, width=0.24\linewidth]{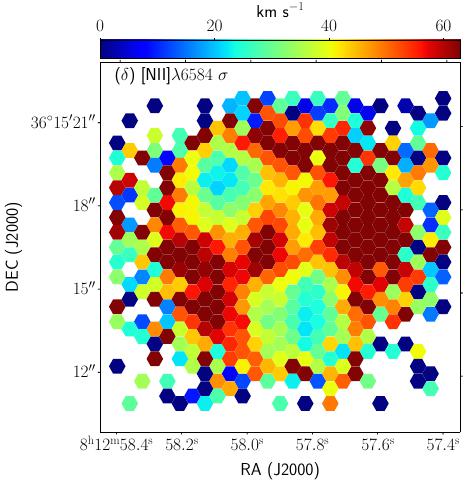}
	\includegraphics[clip, width=0.24\linewidth]{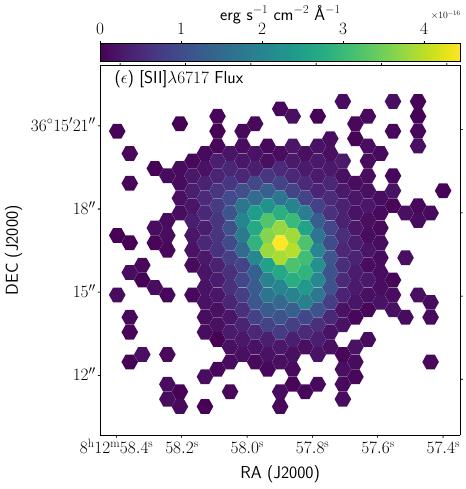}
	\includegraphics[clip, width=0.24\linewidth]{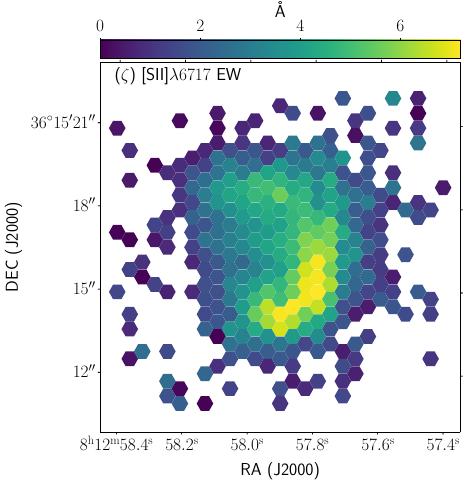}
	\includegraphics[clip, width=0.24\linewidth]{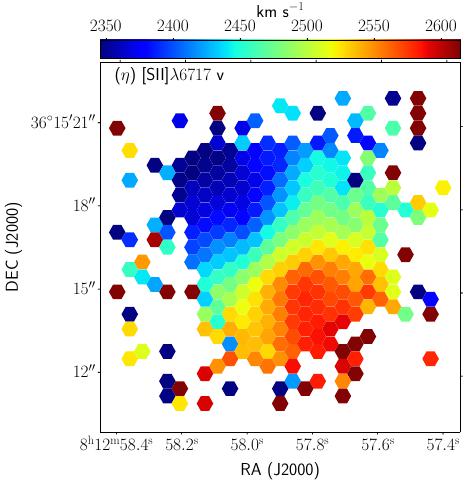}
	\includegraphics[clip, width=0.24\linewidth]{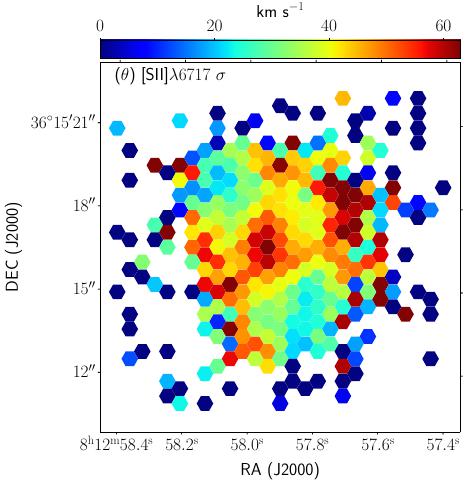}
	\includegraphics[clip, width=0.24\linewidth]{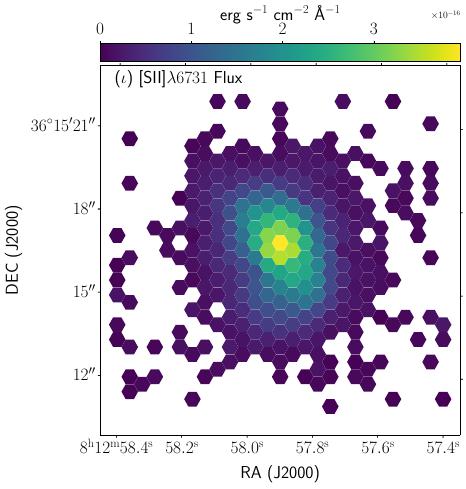}
	\includegraphics[clip, width=0.24\linewidth]{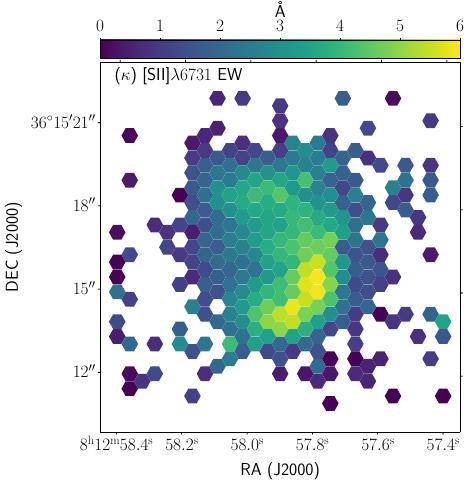}
	\includegraphics[clip, width=0.24\linewidth]{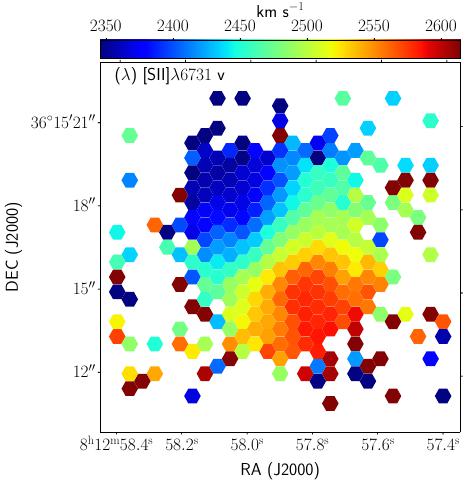}
	\includegraphics[clip, width=0.24\linewidth]{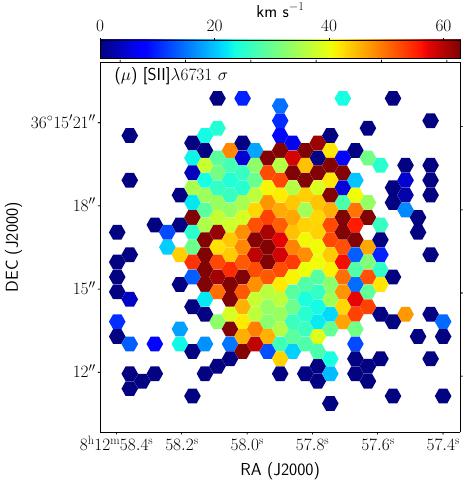}
	\caption{(cont.) NGC~2543 card.}
	\label{fig:NGC2543_card_2}
\end{figure*}

\begin{figure*}[h]
	\centering
	\includegraphics[clip, width=0.35\linewidth]{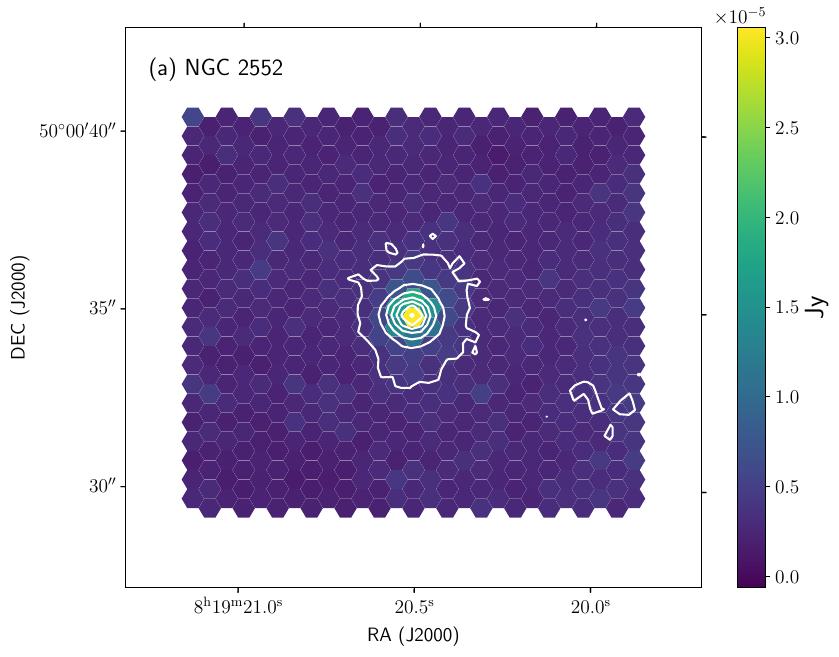}
	\includegraphics[clip, width=0.6\linewidth]{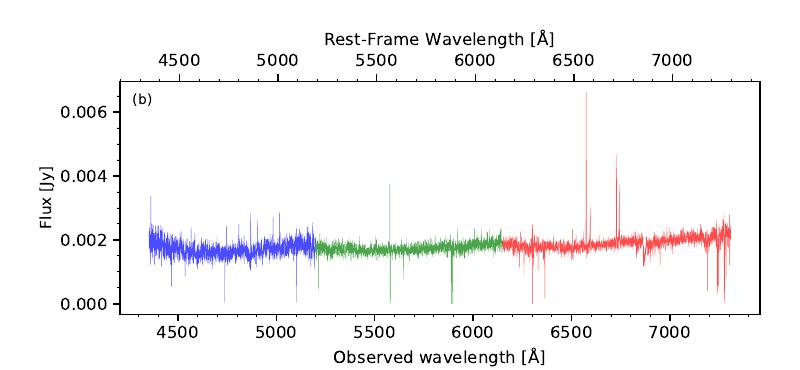}
	\includegraphics[clip, width=0.24\linewidth]{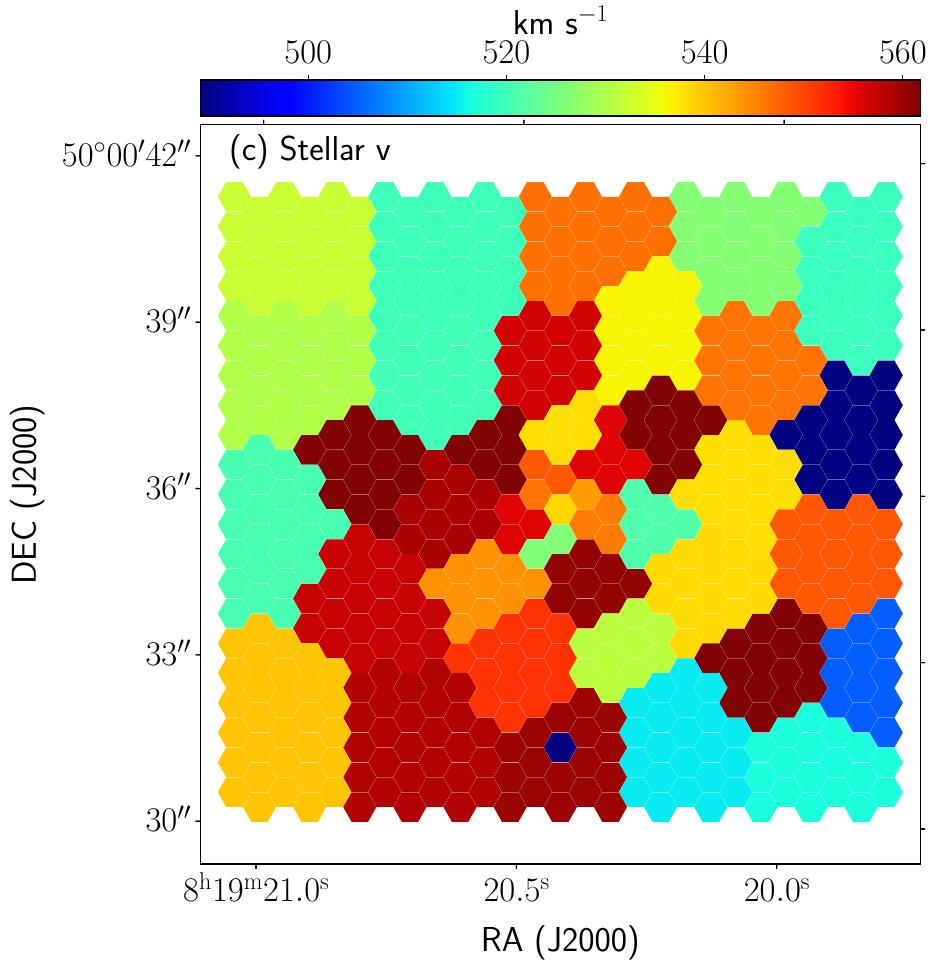}
	\includegraphics[clip, width=0.24\linewidth]{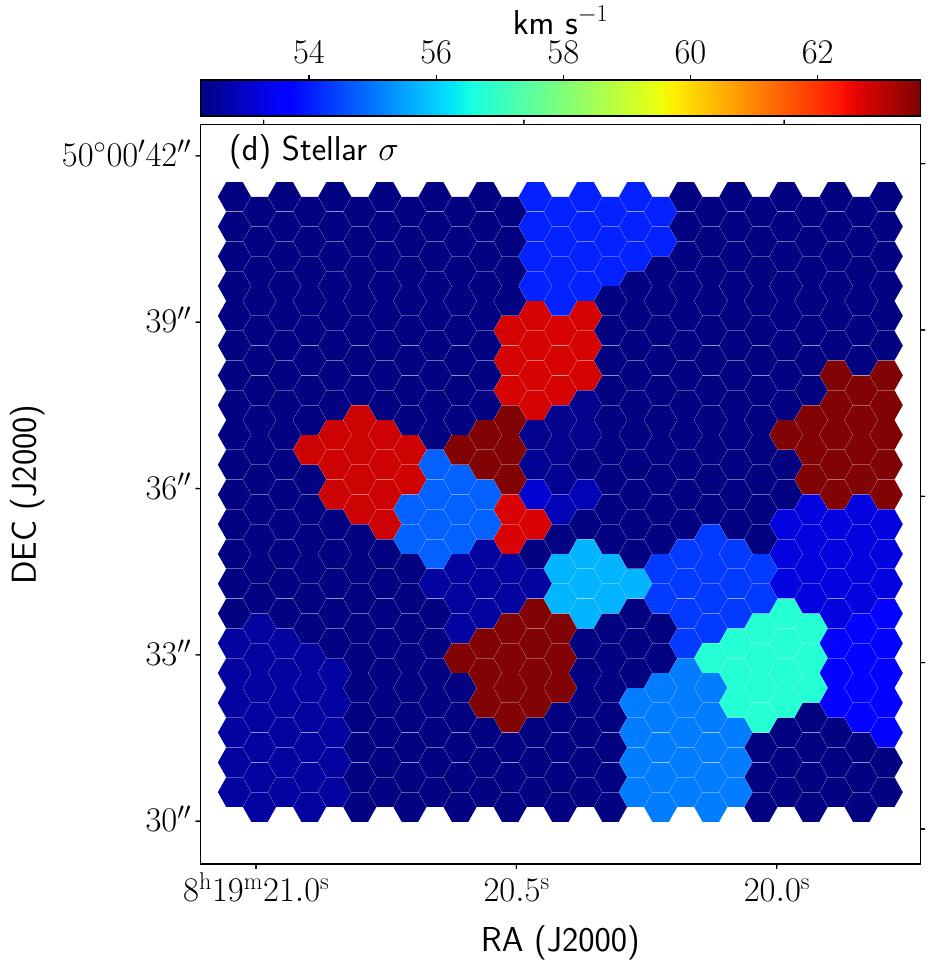}
	\includegraphics[clip, width=0.24\linewidth]{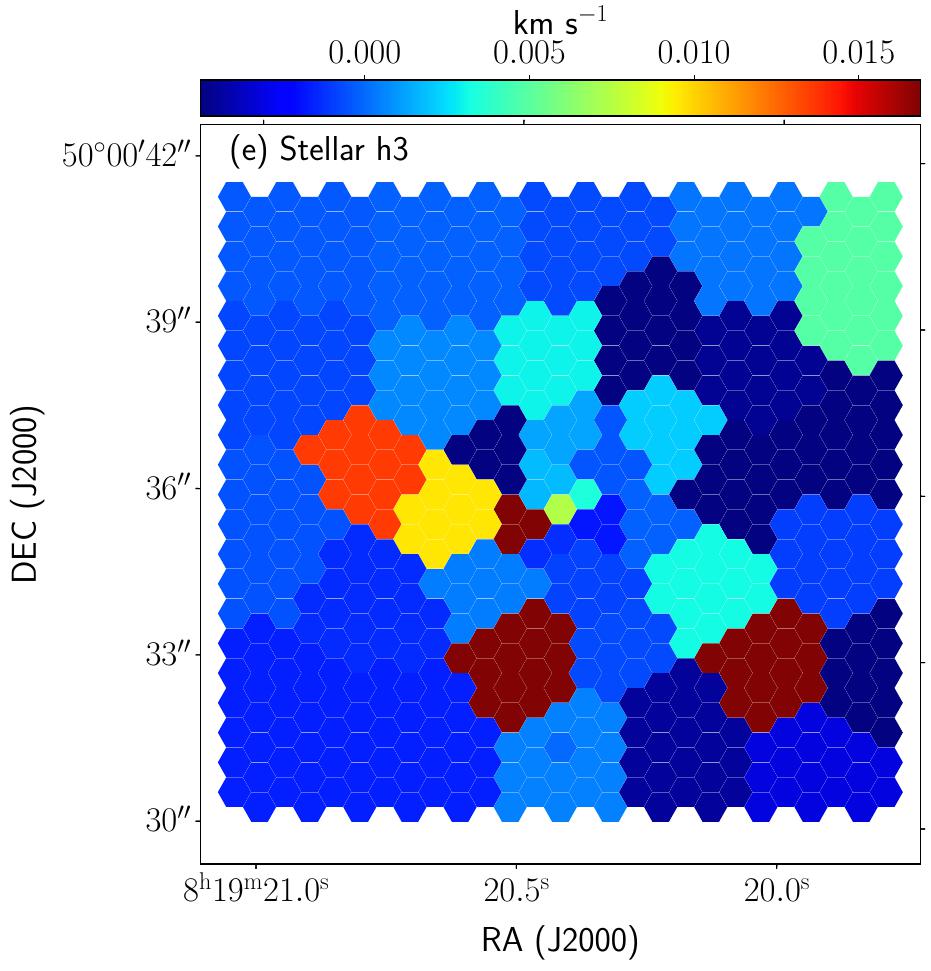}
	\includegraphics[clip, width=0.24\linewidth]{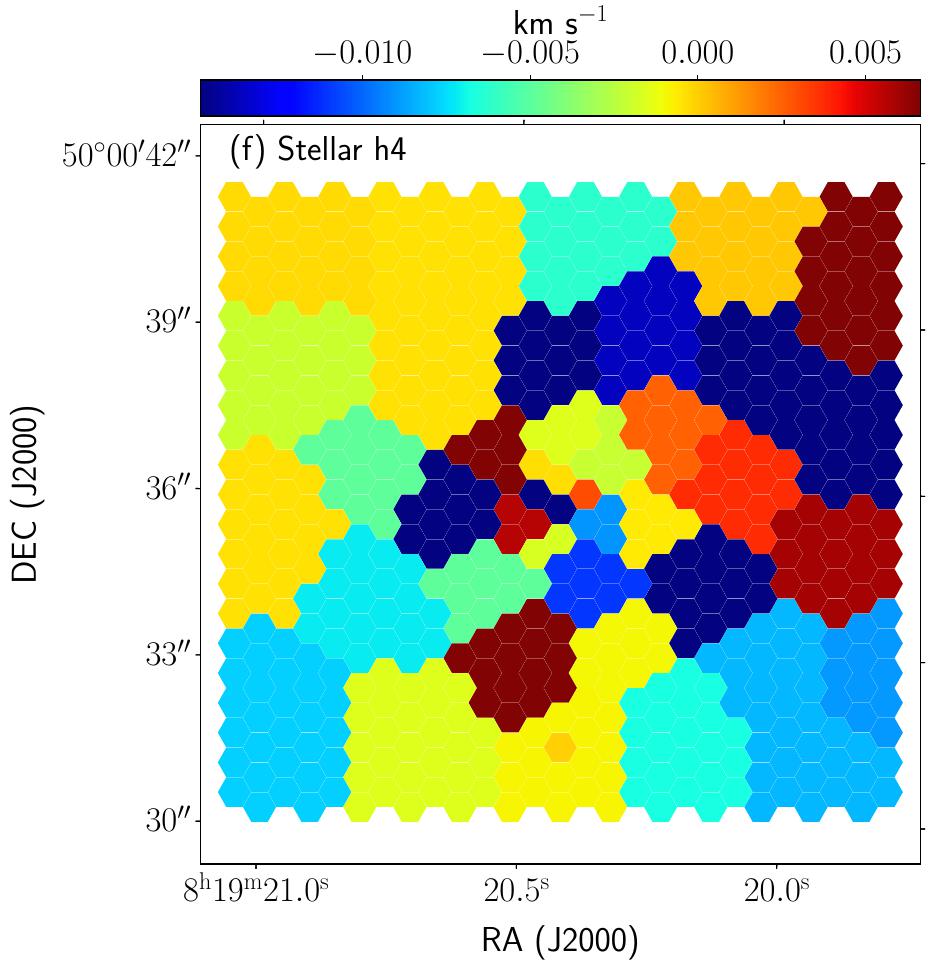}
	\includegraphics[clip, width=0.24\linewidth]{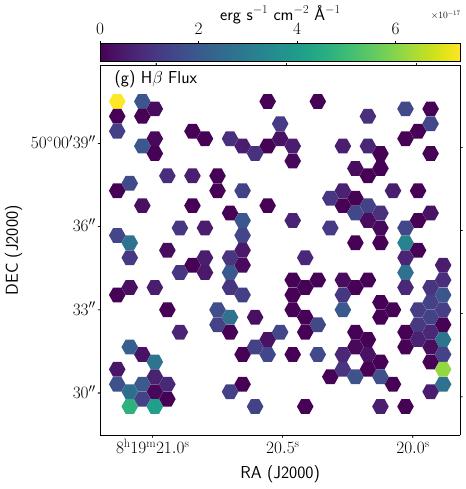}
	\includegraphics[clip, width=0.24\linewidth]{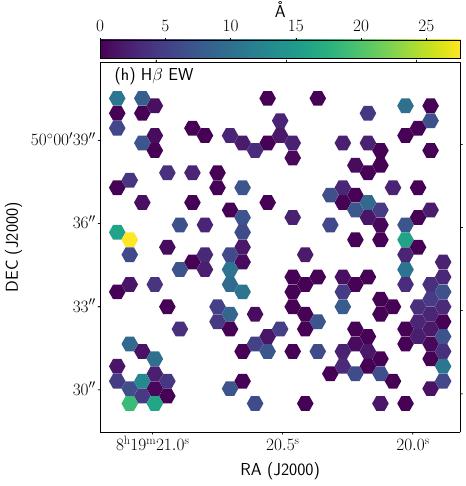}
	\includegraphics[clip, width=0.24\linewidth]{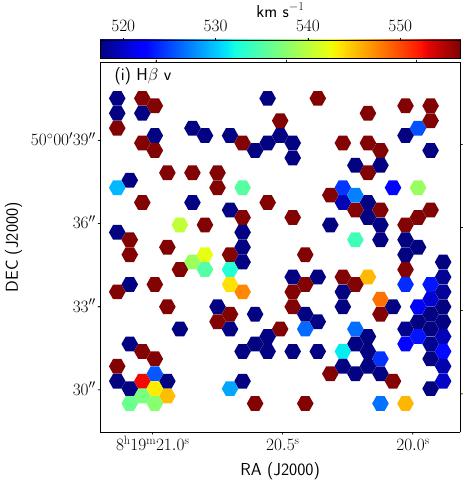}
	\includegraphics[clip, width=0.24\linewidth]{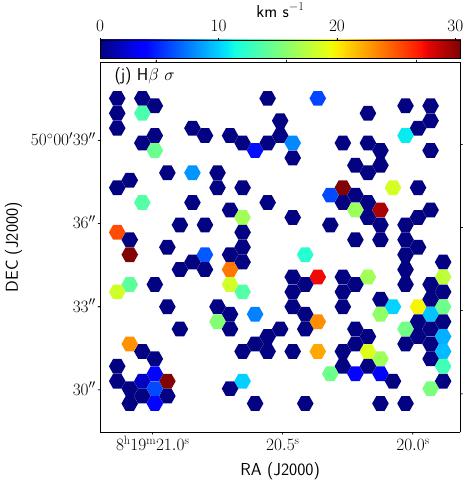}
	\includegraphics[clip, width=0.24\linewidth]{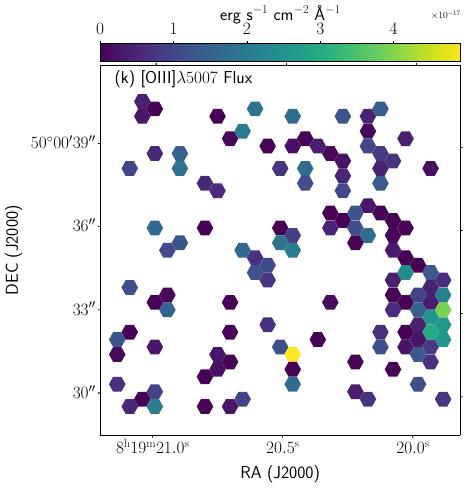}
	\includegraphics[clip, width=0.24\linewidth]{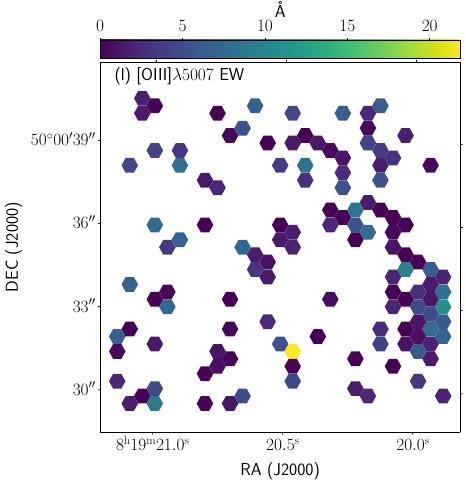}
	\includegraphics[clip, width=0.24\linewidth]{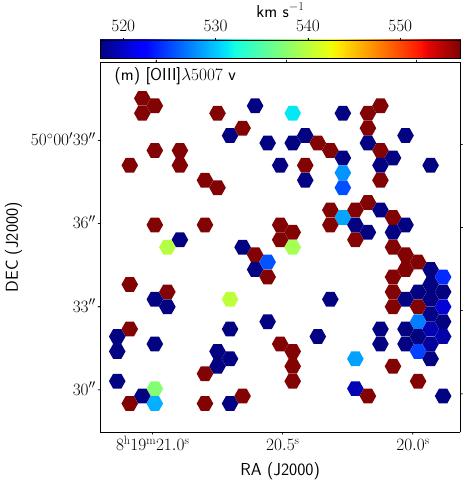}
	\includegraphics[clip, width=0.24\linewidth]{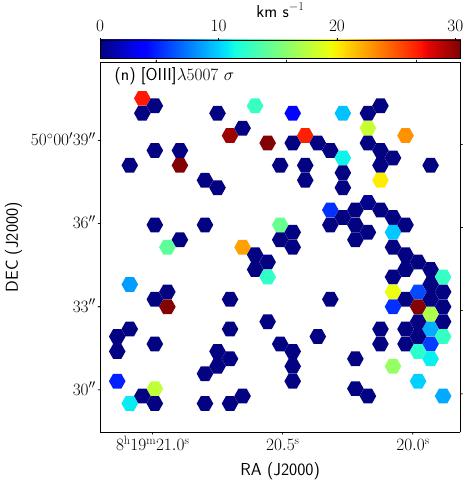}
	\vspace{5cm}
	\caption{NGC~2552 card.}
	\label{fig:NGC2552_card_1}
\end{figure*}
\addtocounter{figure}{-1}
\begin{figure*}[h]
	\centering
	\includegraphics[clip, width=0.24\linewidth]{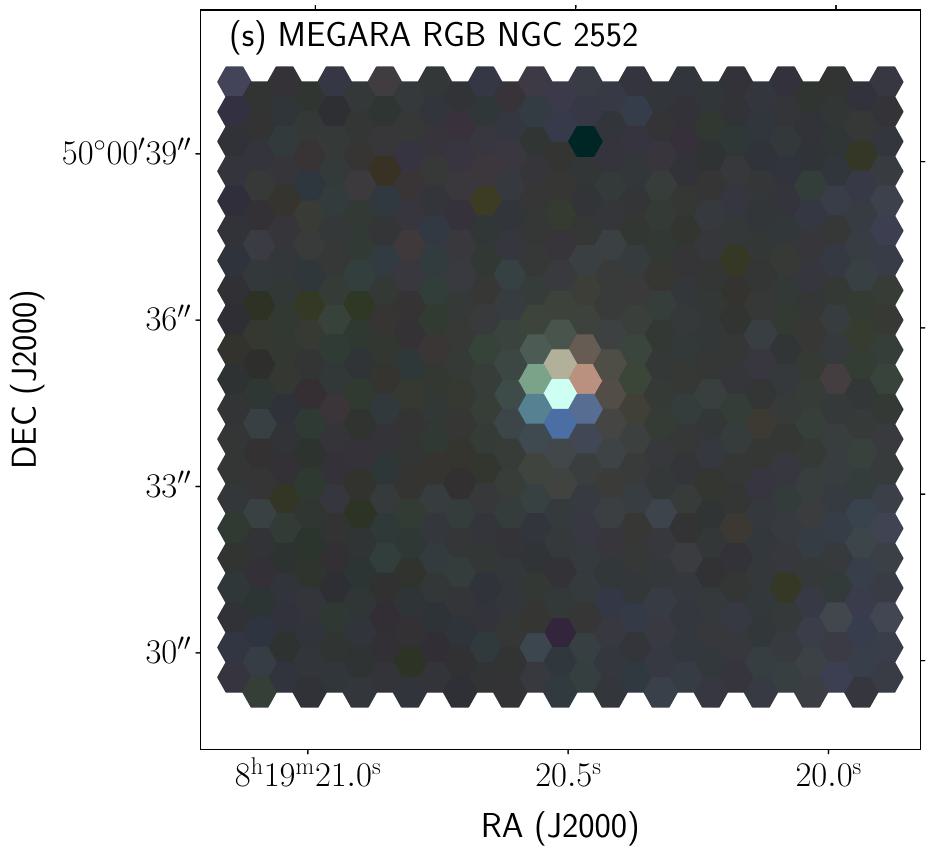}
	\includegraphics[clip, width=0.24\linewidth]{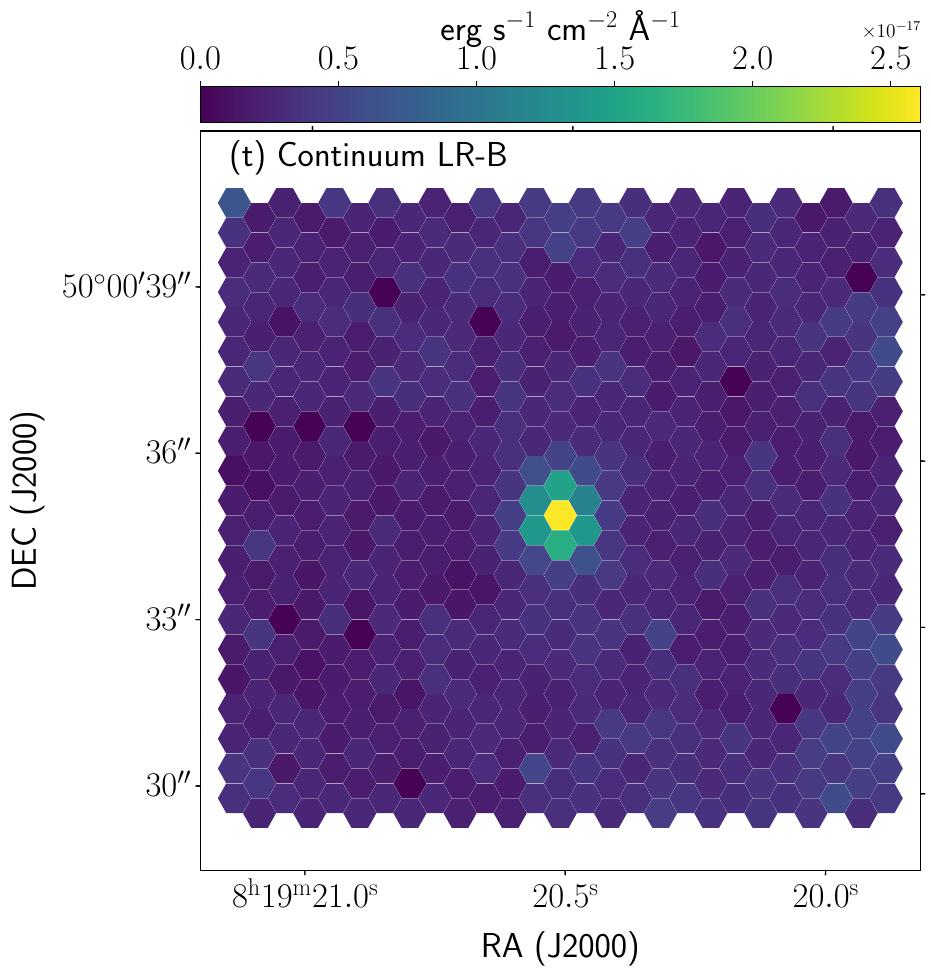}
	\includegraphics[clip, width=0.24\linewidth]{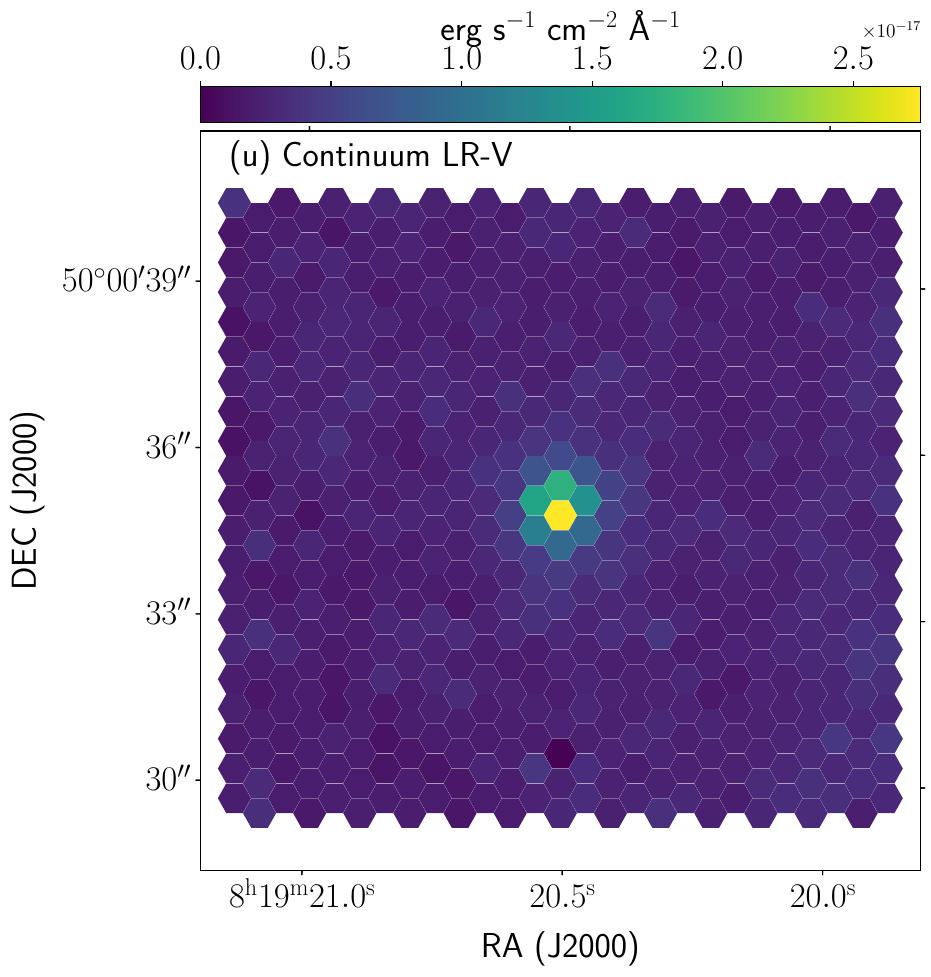}
	\includegraphics[clip, width=0.24\linewidth]{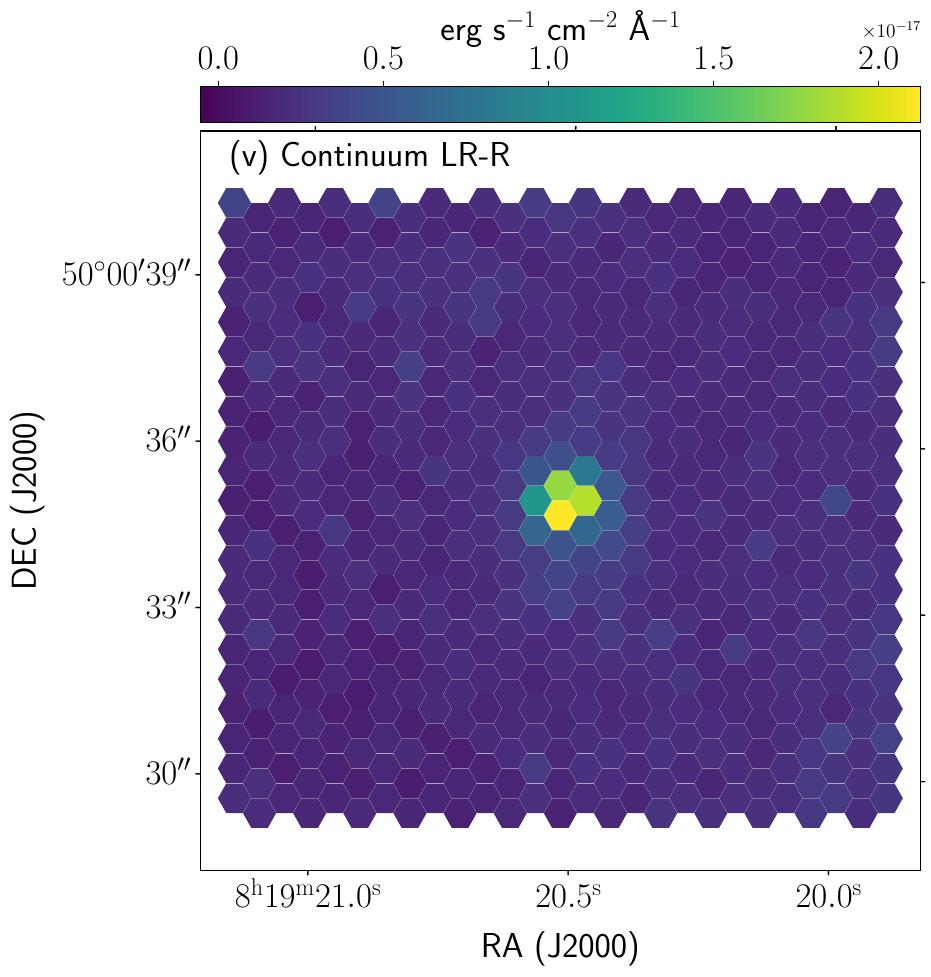}
	\includegraphics[clip, width=0.24\linewidth]{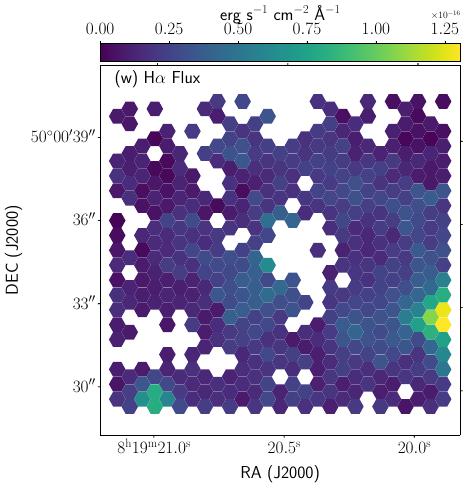}
	\includegraphics[clip, width=0.24\linewidth]{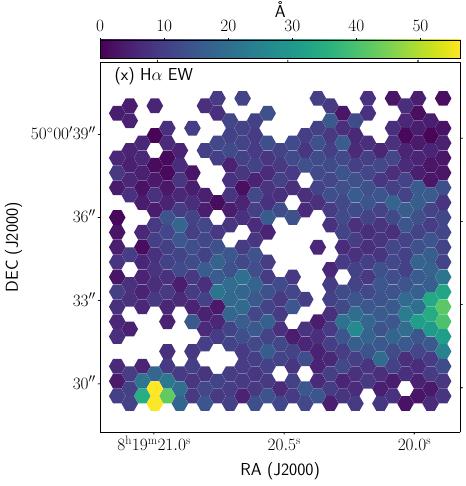}
	\includegraphics[clip, width=0.24\linewidth]{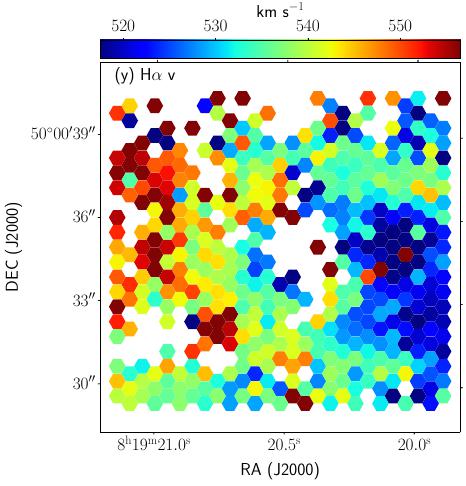}
	\includegraphics[clip, width=0.24\linewidth]{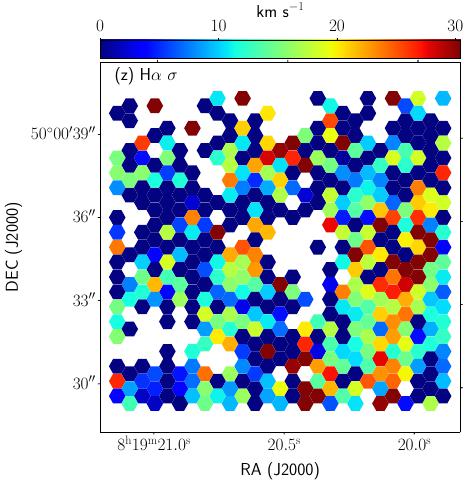}
	\includegraphics[clip, width=0.24\linewidth]{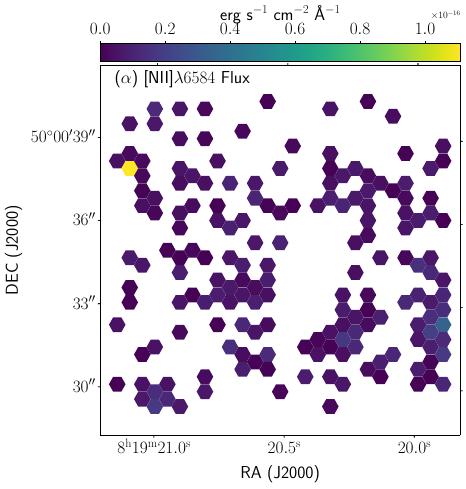}
	\includegraphics[clip, width=0.24\linewidth]{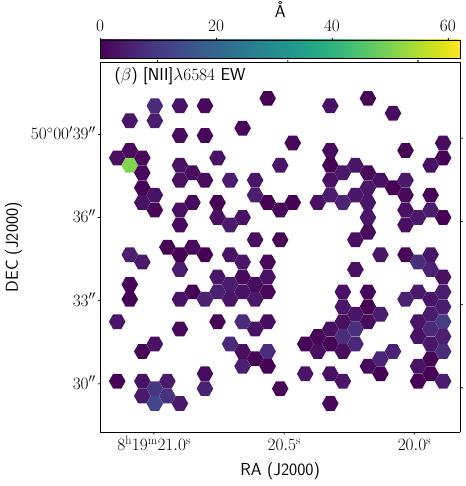}
	\includegraphics[clip, width=0.24\linewidth]{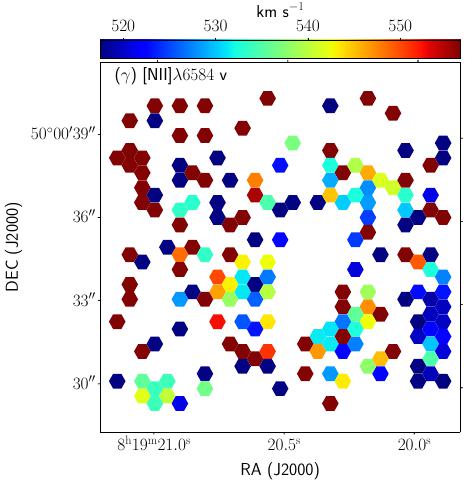}
	\includegraphics[clip, width=0.24\linewidth]{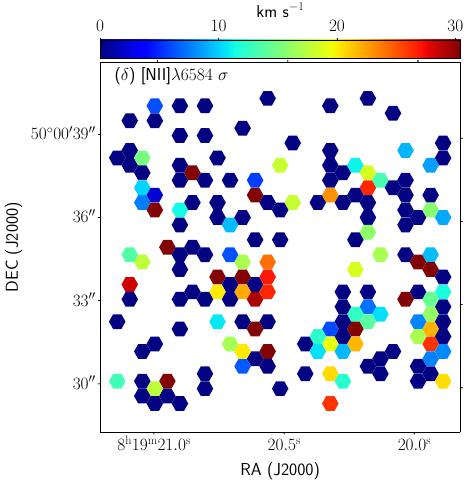}
	\includegraphics[clip, width=0.24\linewidth]{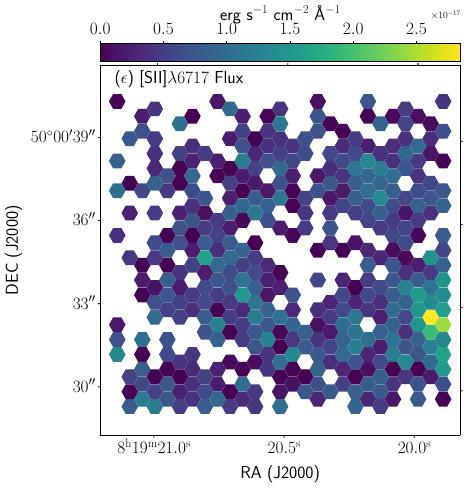}
	\includegraphics[clip, width=0.24\linewidth]{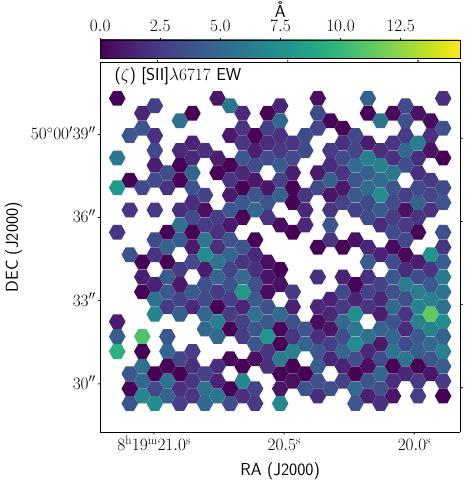}
	\includegraphics[clip, width=0.24\linewidth]{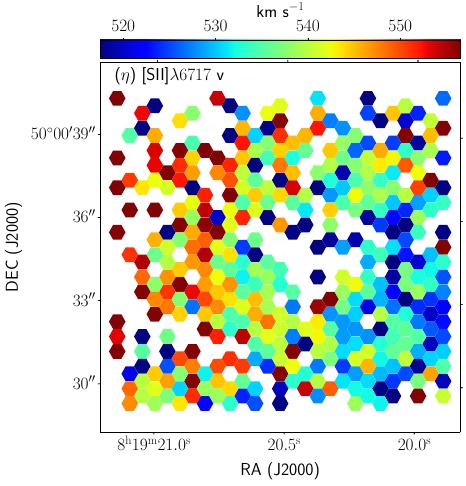}
	\includegraphics[clip, width=0.24\linewidth]{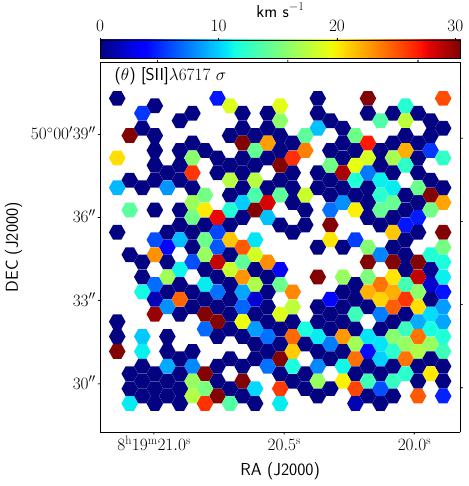}
	\includegraphics[clip, width=0.24\linewidth]{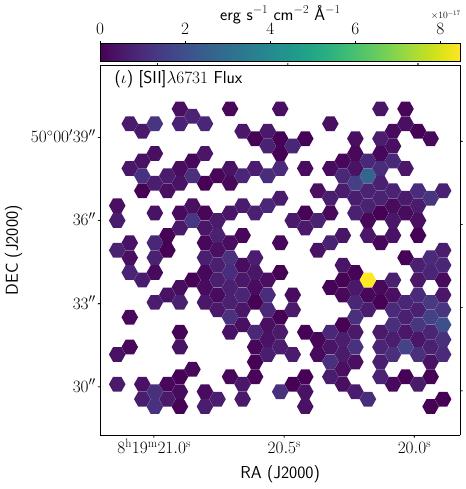}
	\includegraphics[clip, width=0.24\linewidth]{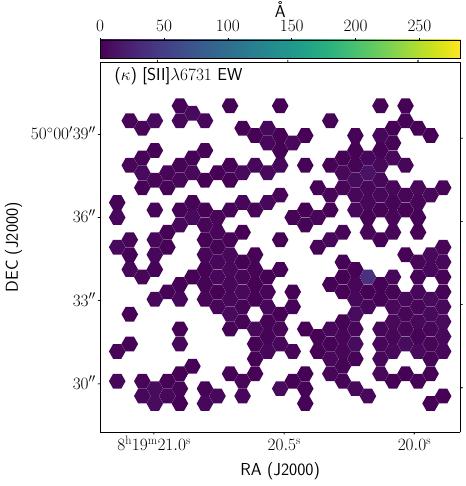}
	\includegraphics[clip, width=0.24\linewidth]{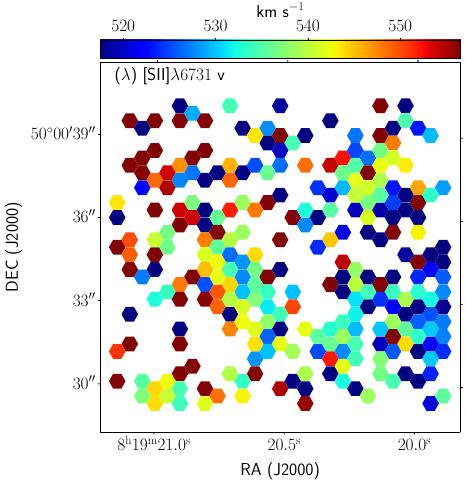}
	\includegraphics[clip, width=0.24\linewidth]{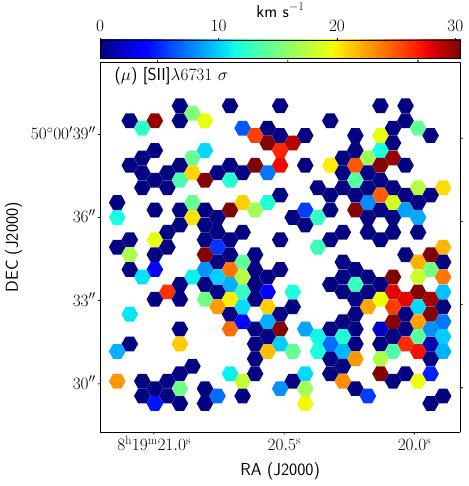}
	\caption{(cont.) NGC~2552 card.}
	\label{fig:NGC2552_card_2}
\end{figure*}

\begin{figure*}[h]
	\centering
	\includegraphics[clip, width=0.35\linewidth]{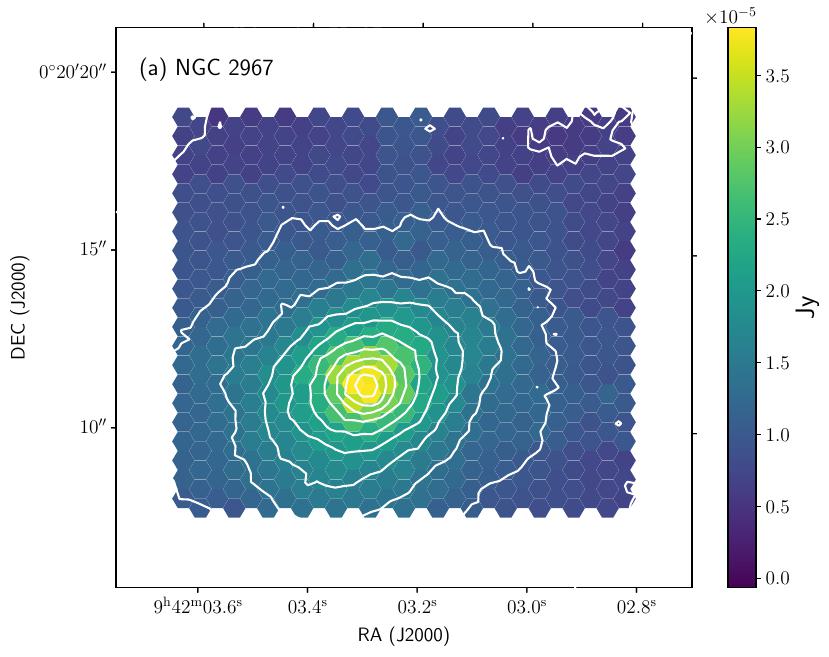}
	\includegraphics[clip, width=0.6\linewidth]{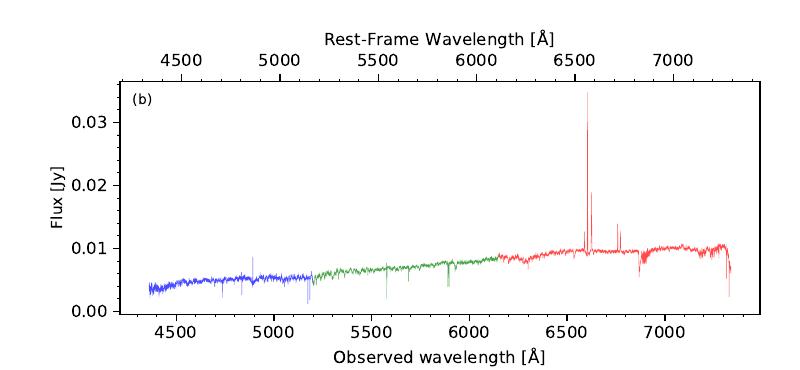}
	\includegraphics[clip, width=0.24\linewidth]{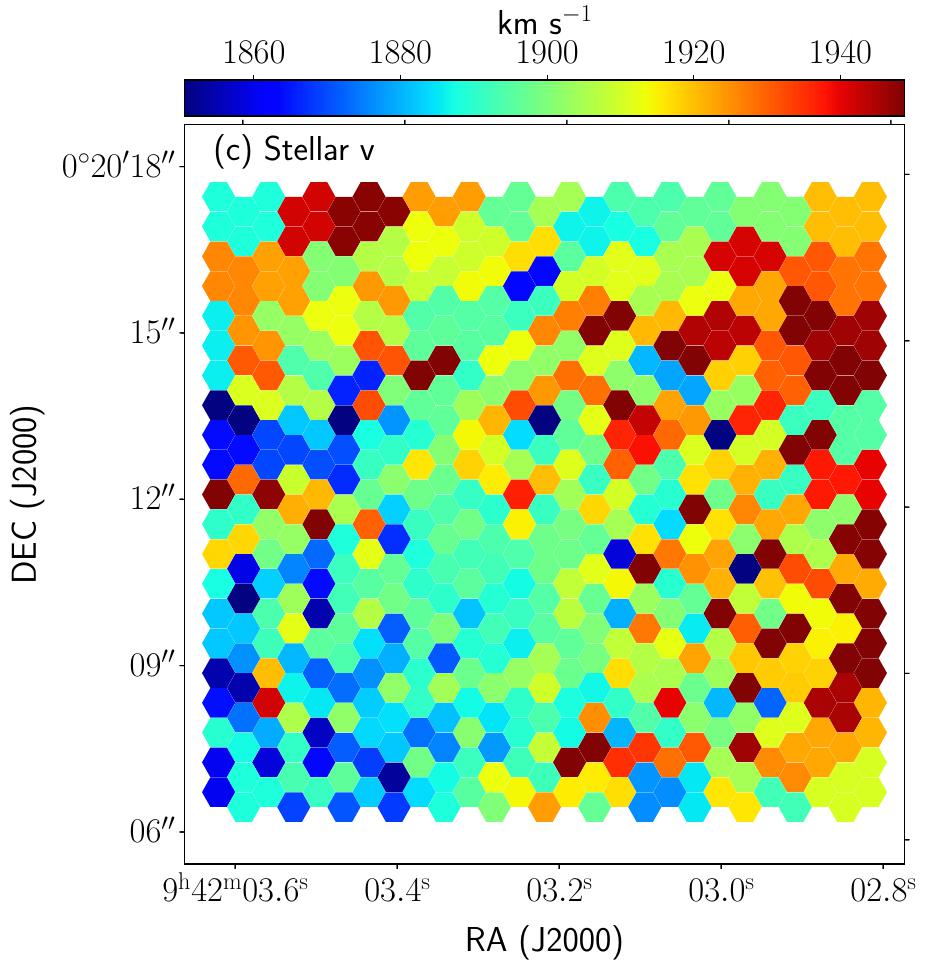}
	\includegraphics[clip, width=0.24\linewidth]{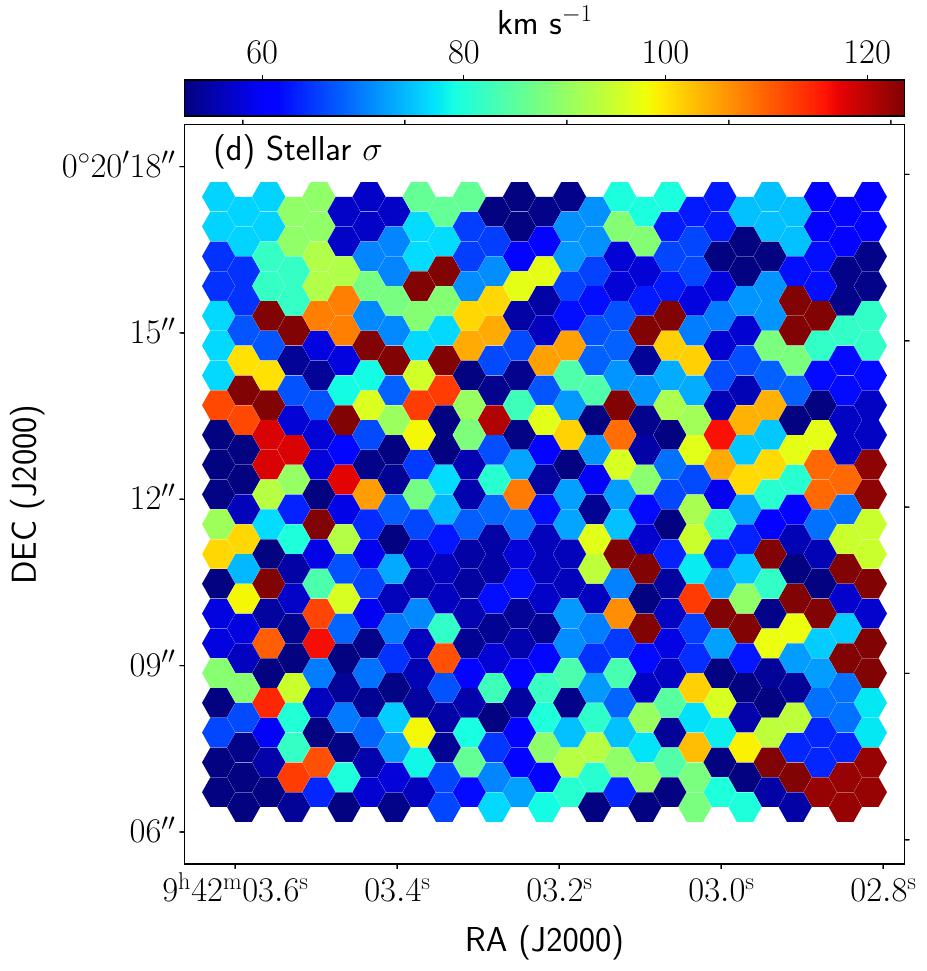}
	\includegraphics[clip, width=0.24\linewidth]{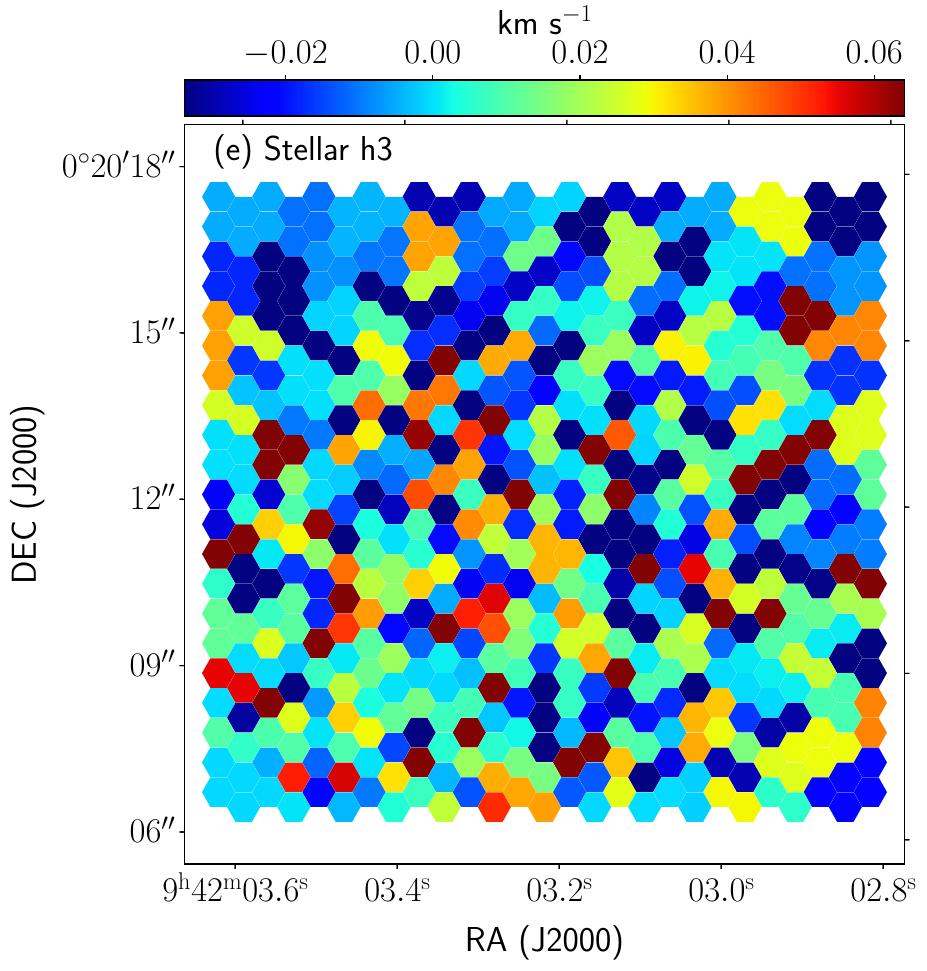}
	\includegraphics[clip, width=0.24\linewidth]{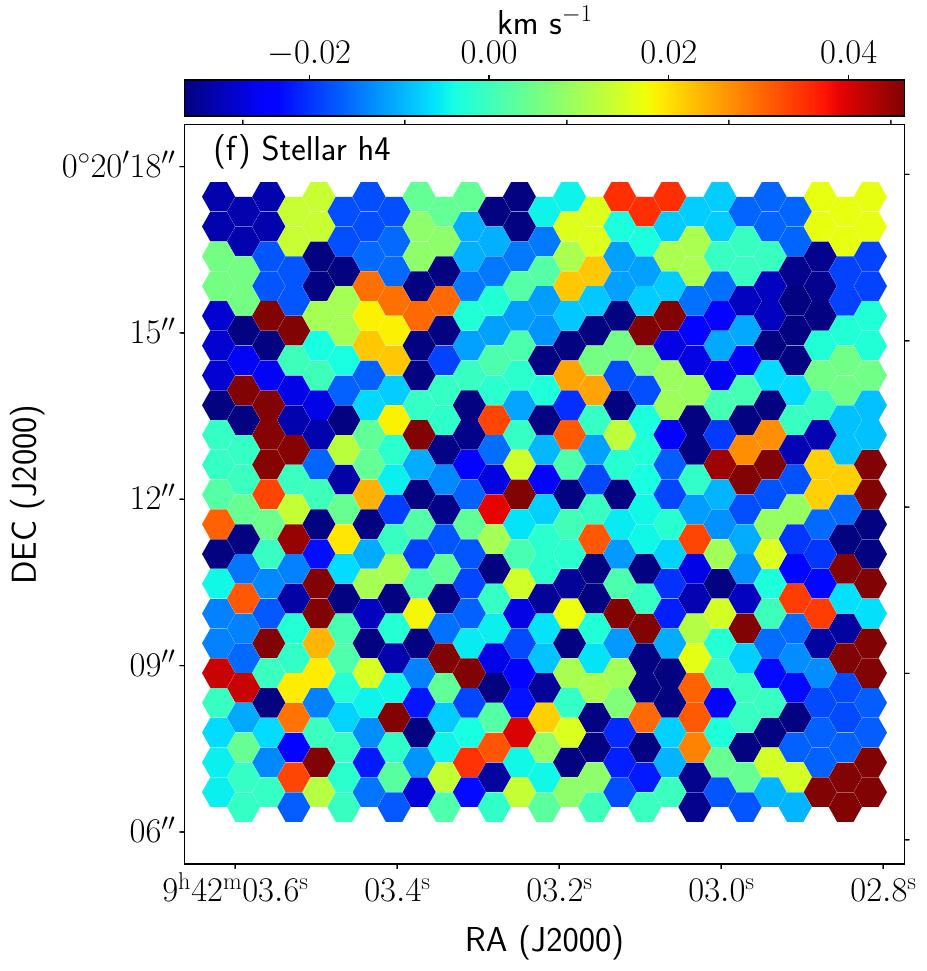}
	\includegraphics[clip, width=0.24\linewidth]{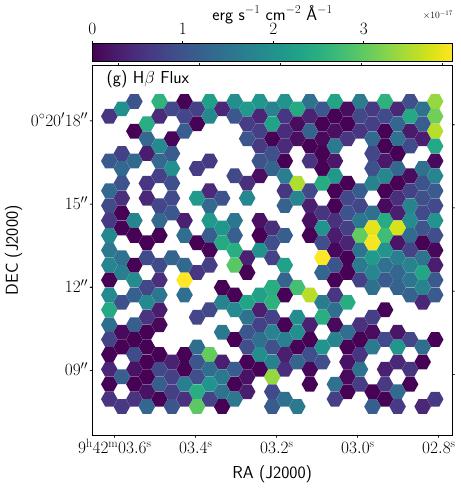}
	\includegraphics[clip, width=0.24\linewidth]{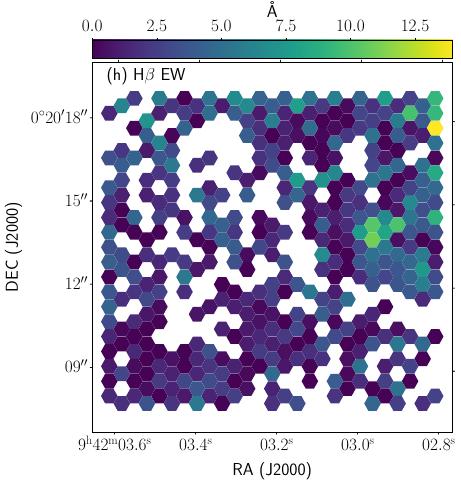}
	\includegraphics[clip, width=0.24\linewidth]{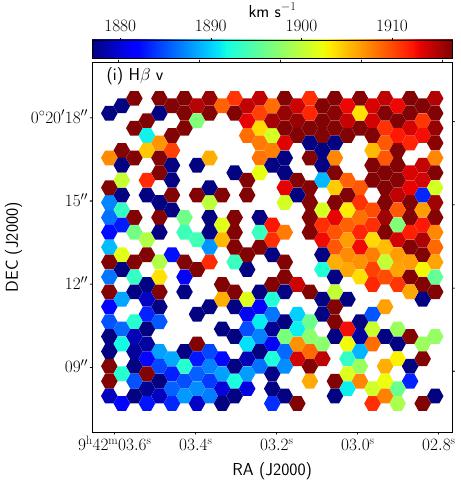}
	\includegraphics[clip, width=0.24\linewidth]{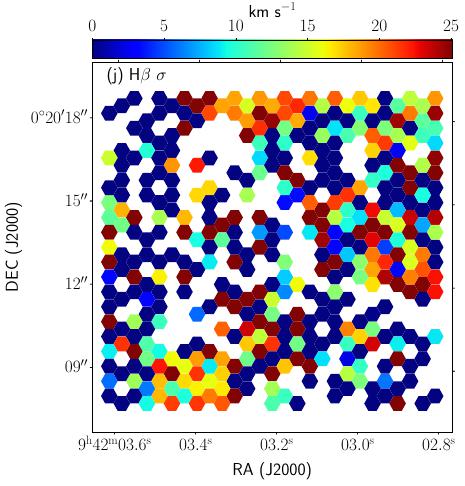}
	\includegraphics[clip, width=0.24\linewidth]{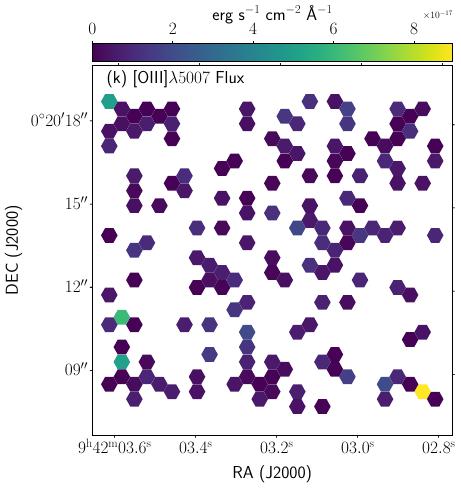}
	\includegraphics[clip, width=0.24\linewidth]{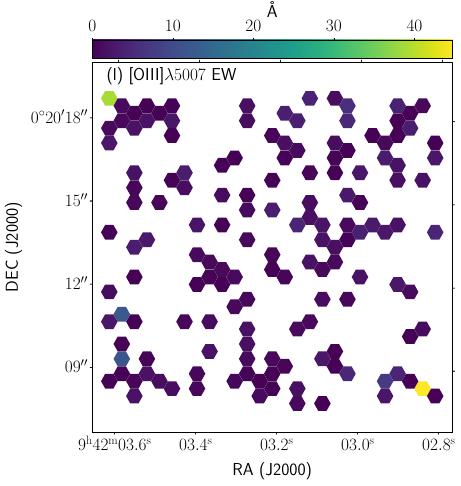}
	\includegraphics[clip, width=0.24\linewidth]{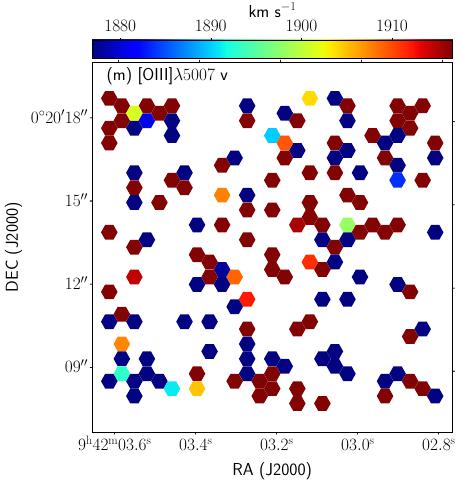}
	\includegraphics[clip, width=0.24\linewidth]{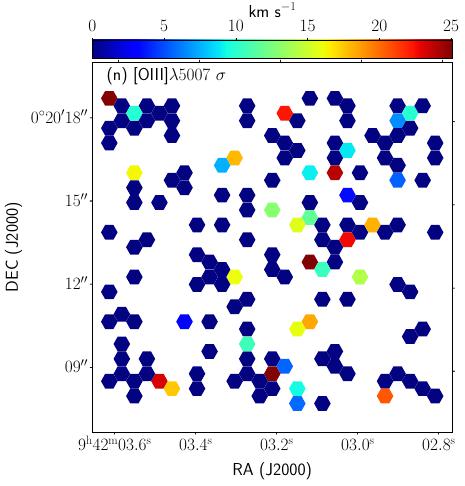}
	\vspace{5cm}
	\caption{NGC~2967 card.}
	\label{fig:NGC2967_card_1}
\end{figure*}
\addtocounter{figure}{-1}
\begin{figure*}[h]
	\centering
	\includegraphics[clip, width=0.24\linewidth]{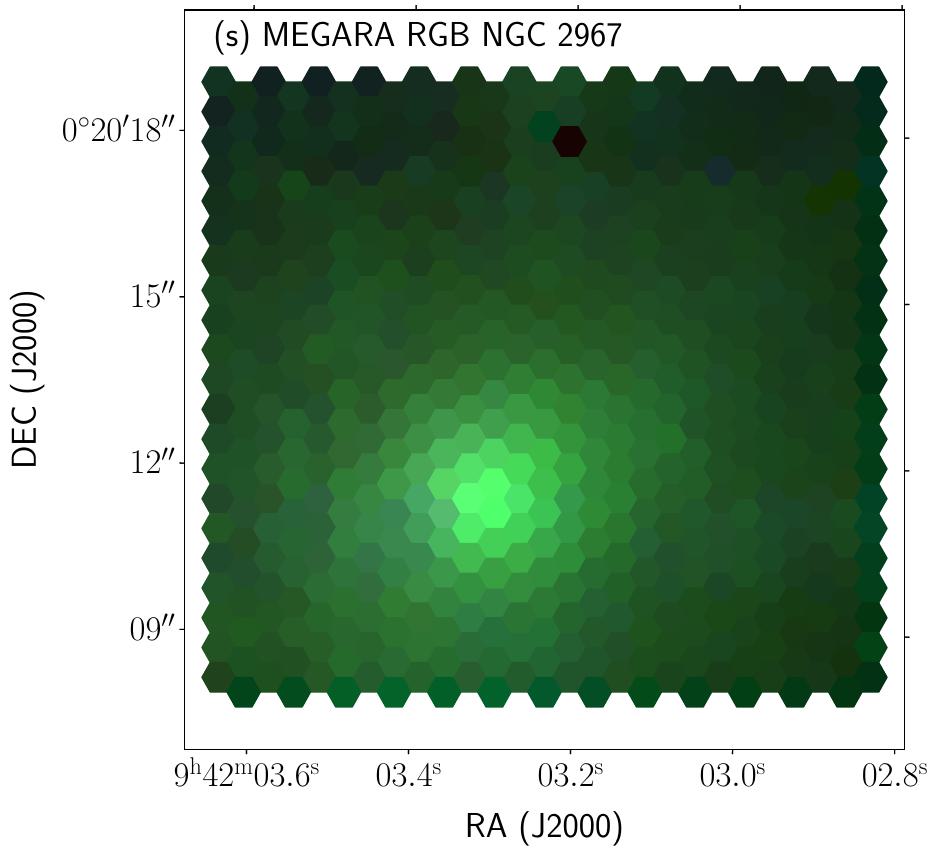}
	\includegraphics[clip, width=0.24\linewidth]{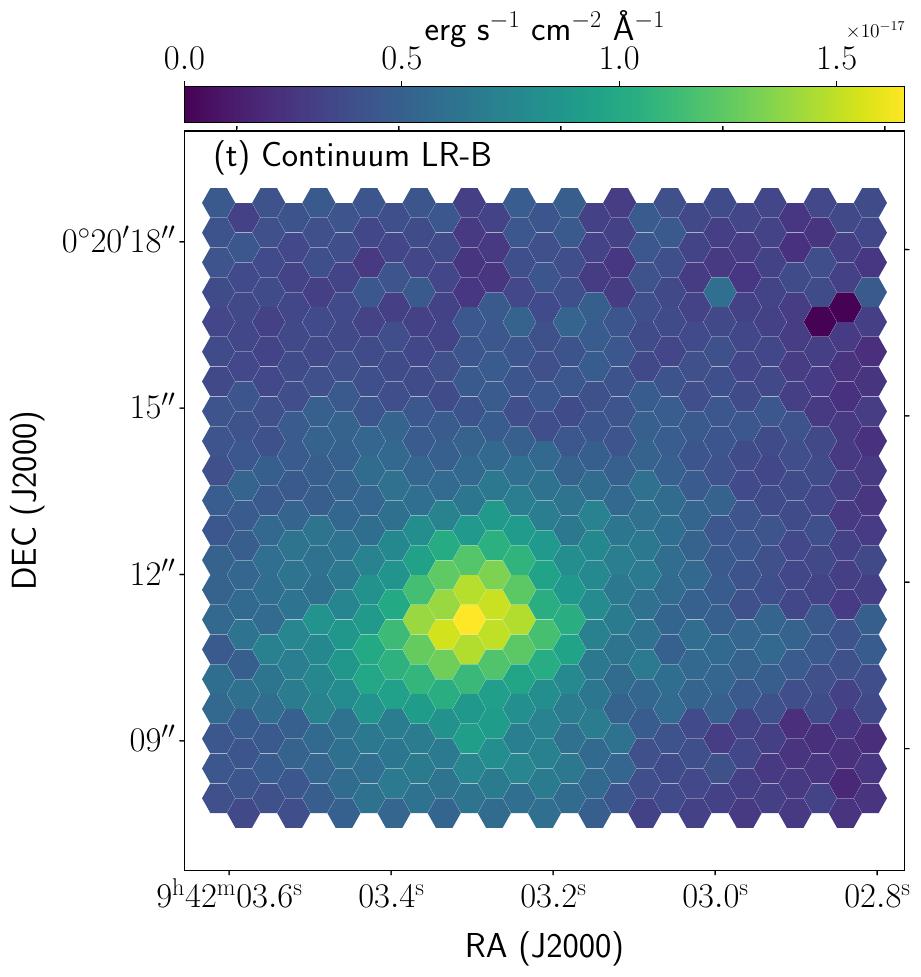}
	\includegraphics[clip, width=0.24\linewidth]{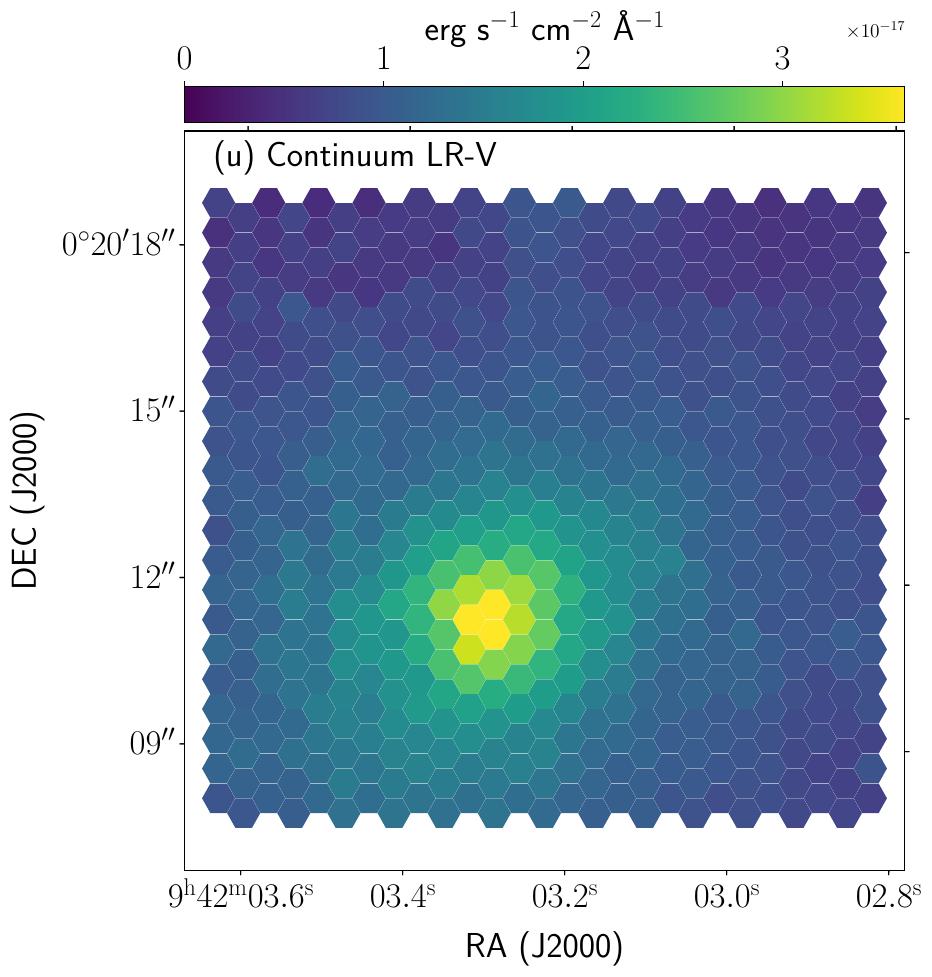}
	\includegraphics[clip, width=0.24\linewidth]{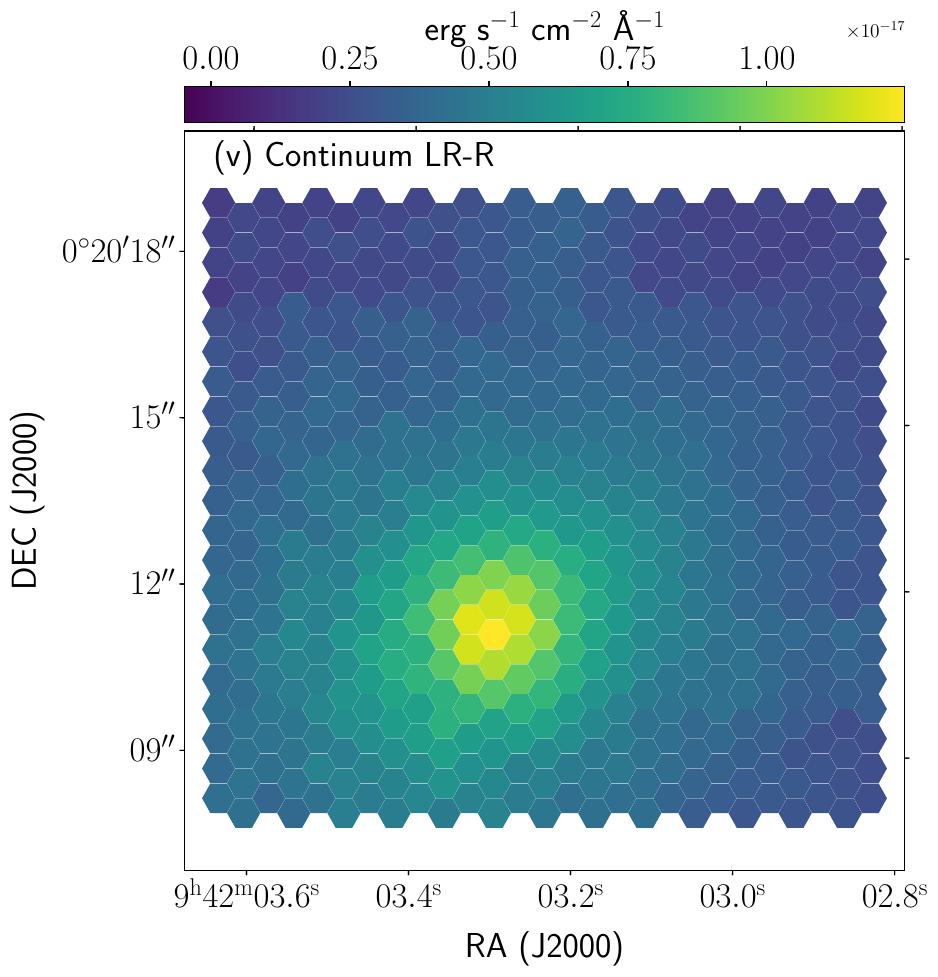}
	\includegraphics[clip, width=0.24\linewidth]{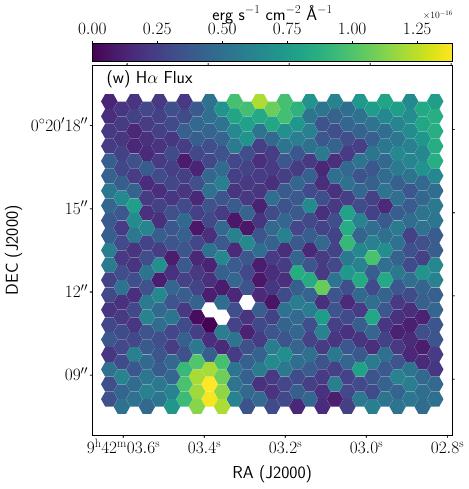}
	\includegraphics[clip, width=0.24\linewidth]{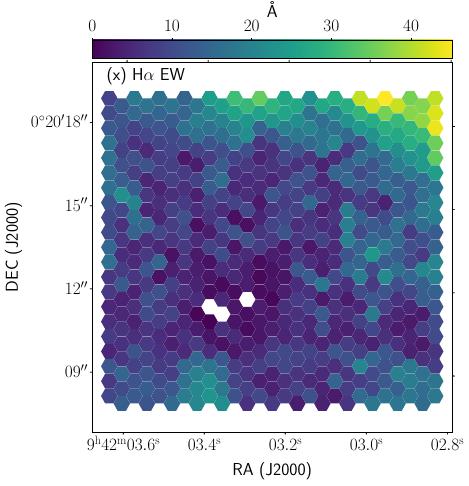}
	\includegraphics[clip, width=0.24\linewidth]{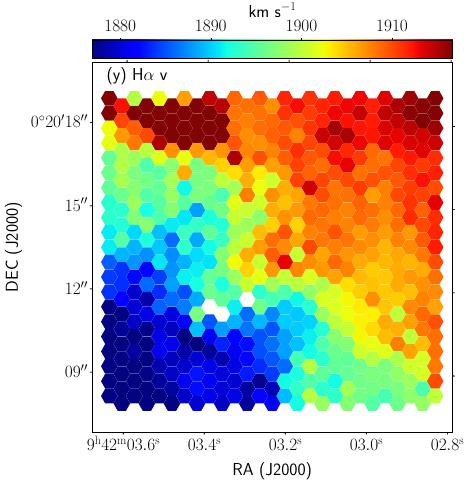}
	\includegraphics[clip, width=0.24\linewidth]{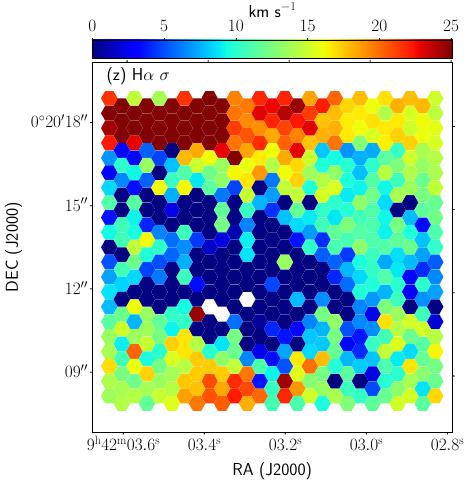}
	\includegraphics[clip, width=0.24\linewidth]{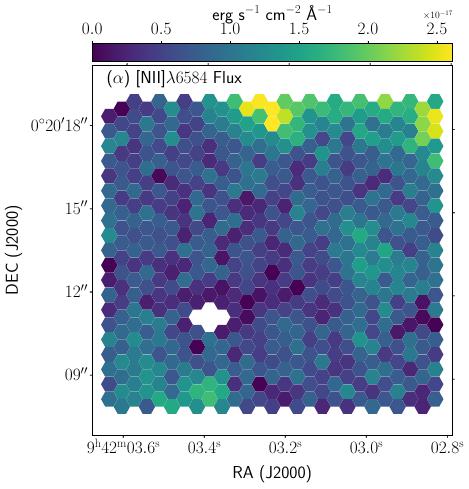}
	\includegraphics[clip, width=0.24\linewidth]{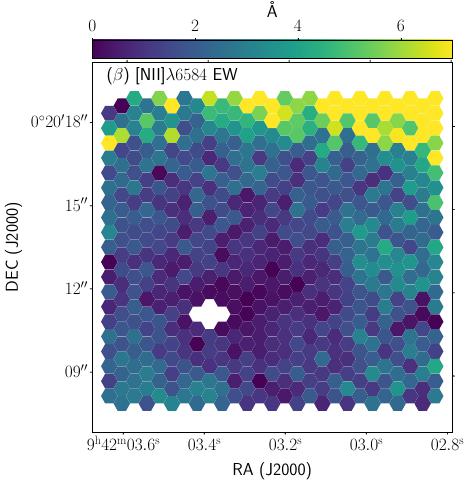}
	\includegraphics[clip, width=0.24\linewidth]{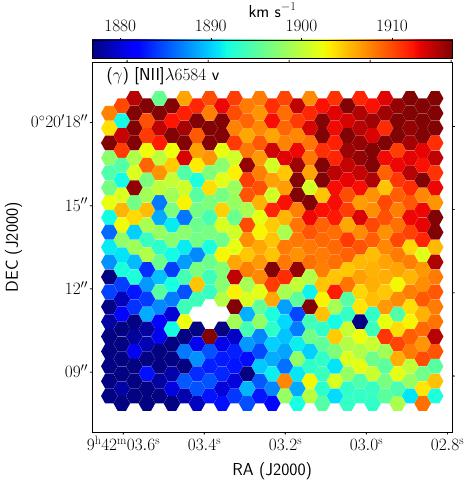}
	\includegraphics[clip, width=0.24\linewidth]{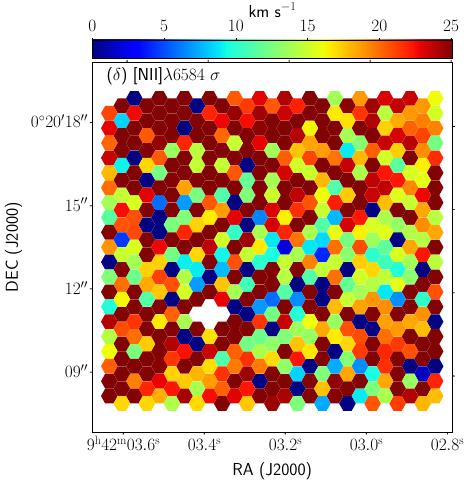}
	\includegraphics[clip, width=0.24\linewidth]{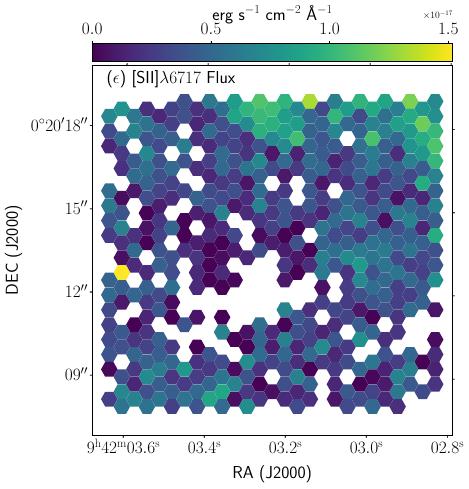}
	\includegraphics[clip, width=0.24\linewidth]{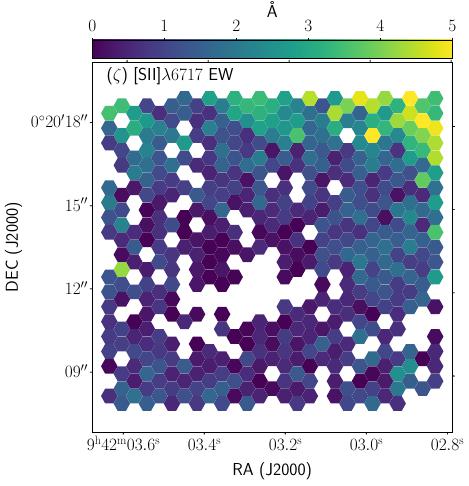}
	\includegraphics[clip, width=0.24\linewidth]{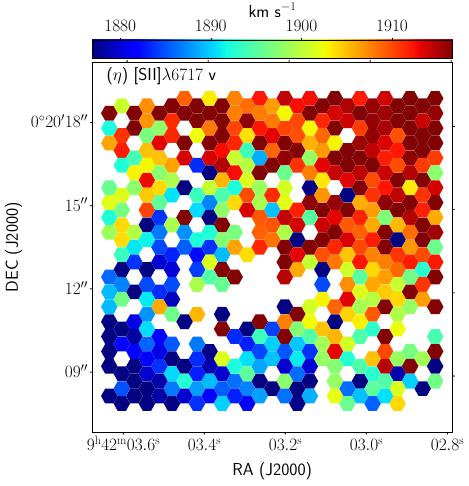}
	\includegraphics[clip, width=0.24\linewidth]{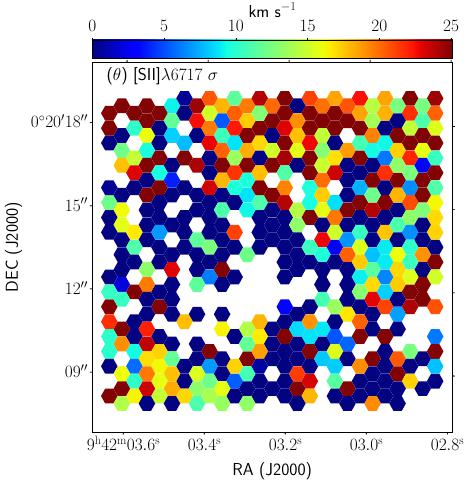}
	\includegraphics[clip, width=0.24\linewidth]{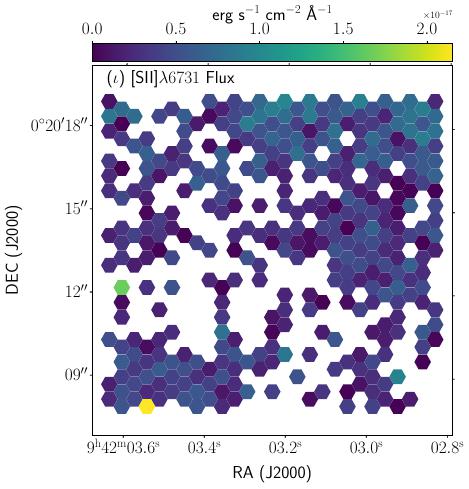}
	\includegraphics[clip, width=0.24\linewidth]{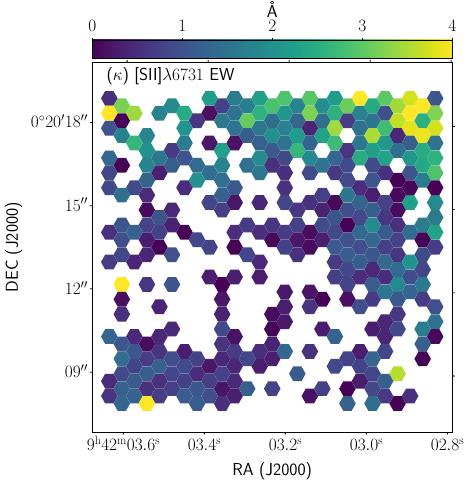}
	\includegraphics[clip, width=0.24\linewidth]{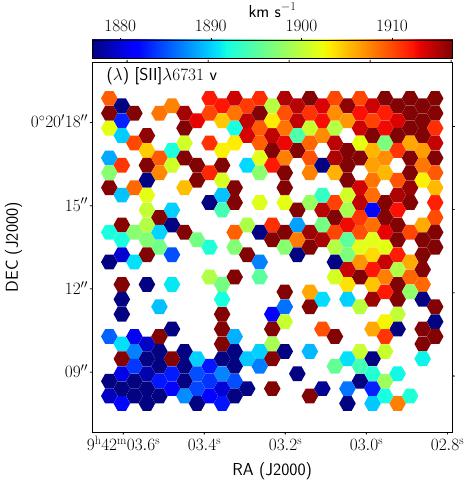}
	\includegraphics[clip, width=0.24\linewidth]{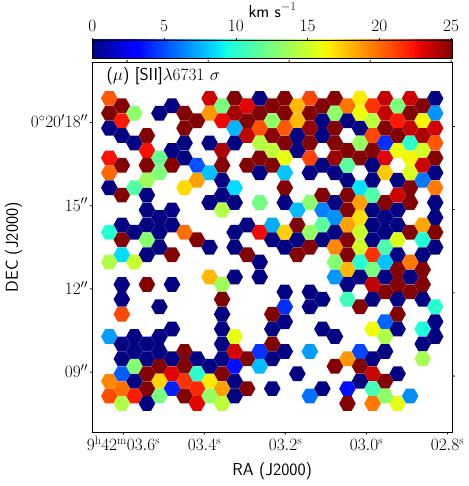}
	\caption{(cont.) NGC~2967 card.}
	\label{fig:NGC2967_card_2}
\end{figure*}

\begin{figure*}[h]
	\centering
	\includegraphics[clip, width=0.35\linewidth]{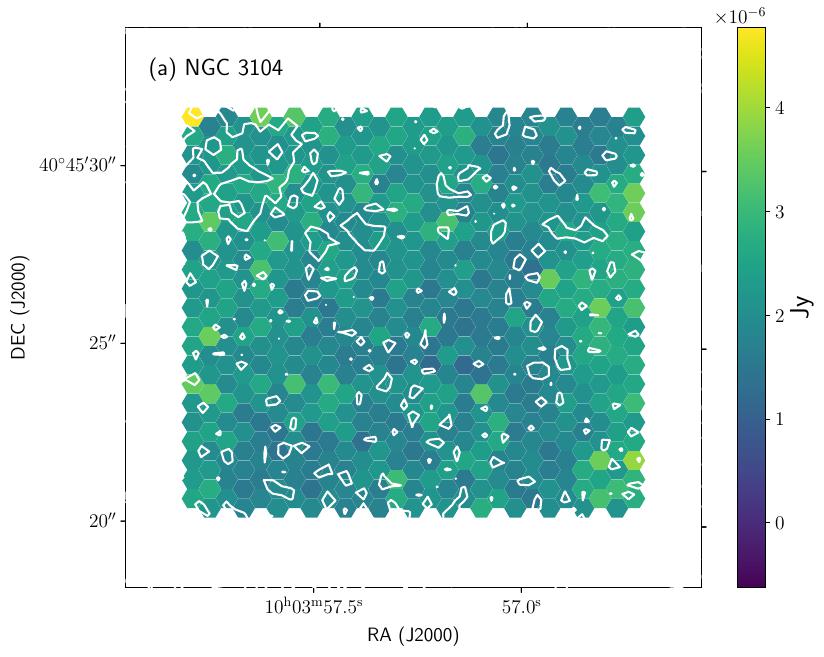}
	\includegraphics[clip, width=0.6\linewidth]{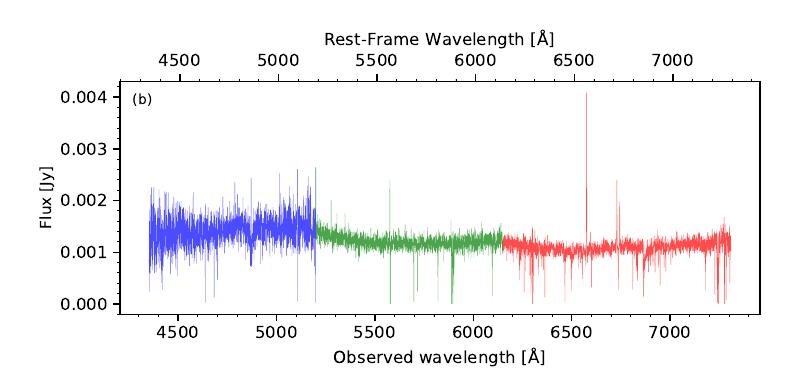}
	\includegraphics[clip, width=0.24\linewidth]{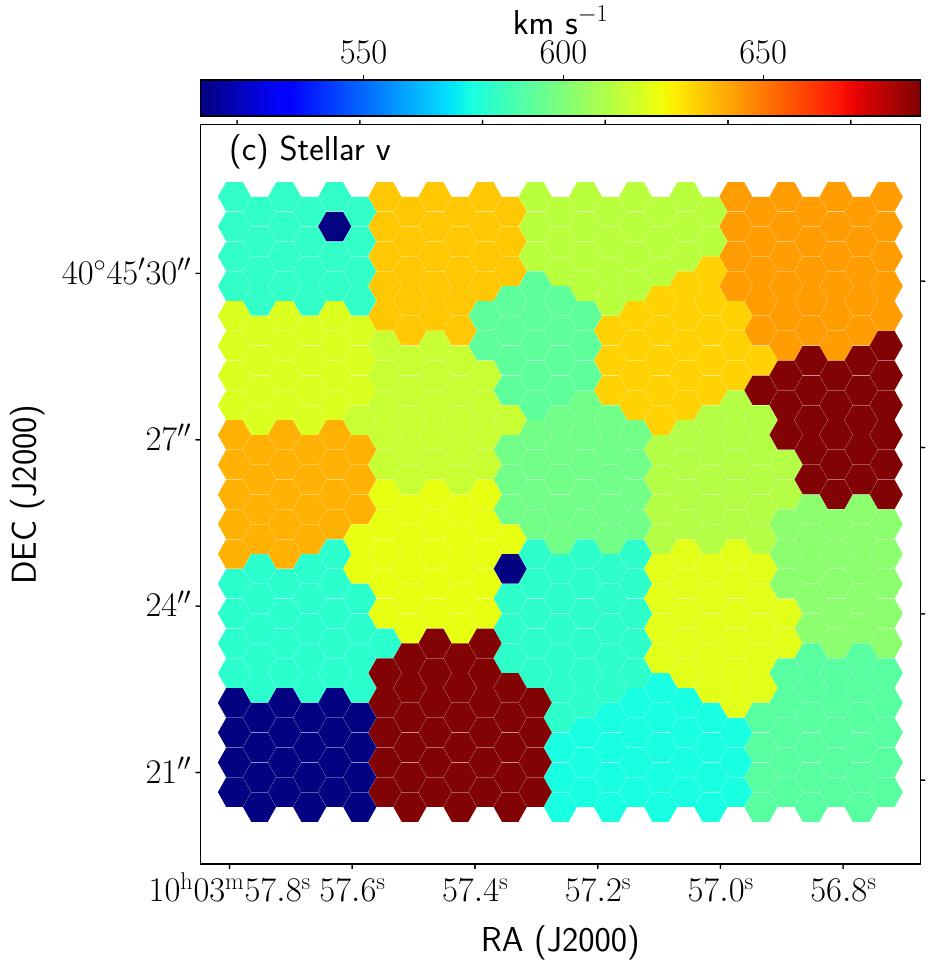}
	\includegraphics[clip, width=0.24\linewidth]{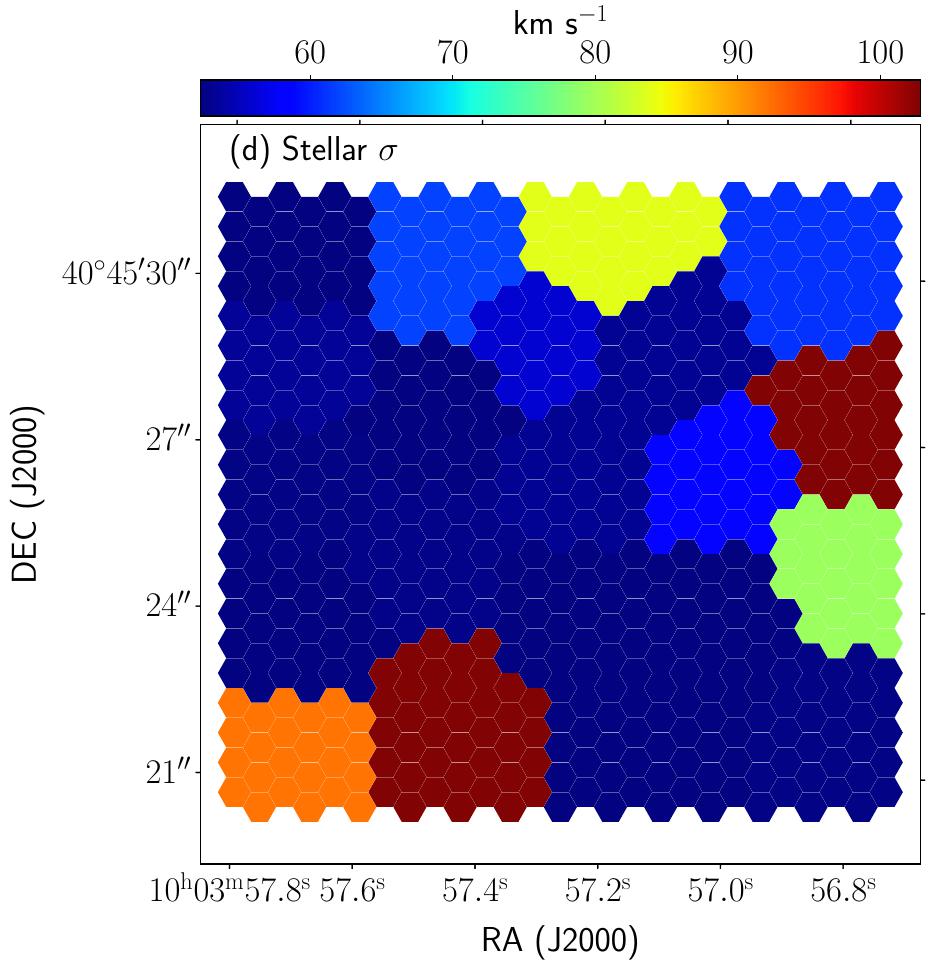}
	\includegraphics[clip, width=0.24\linewidth]{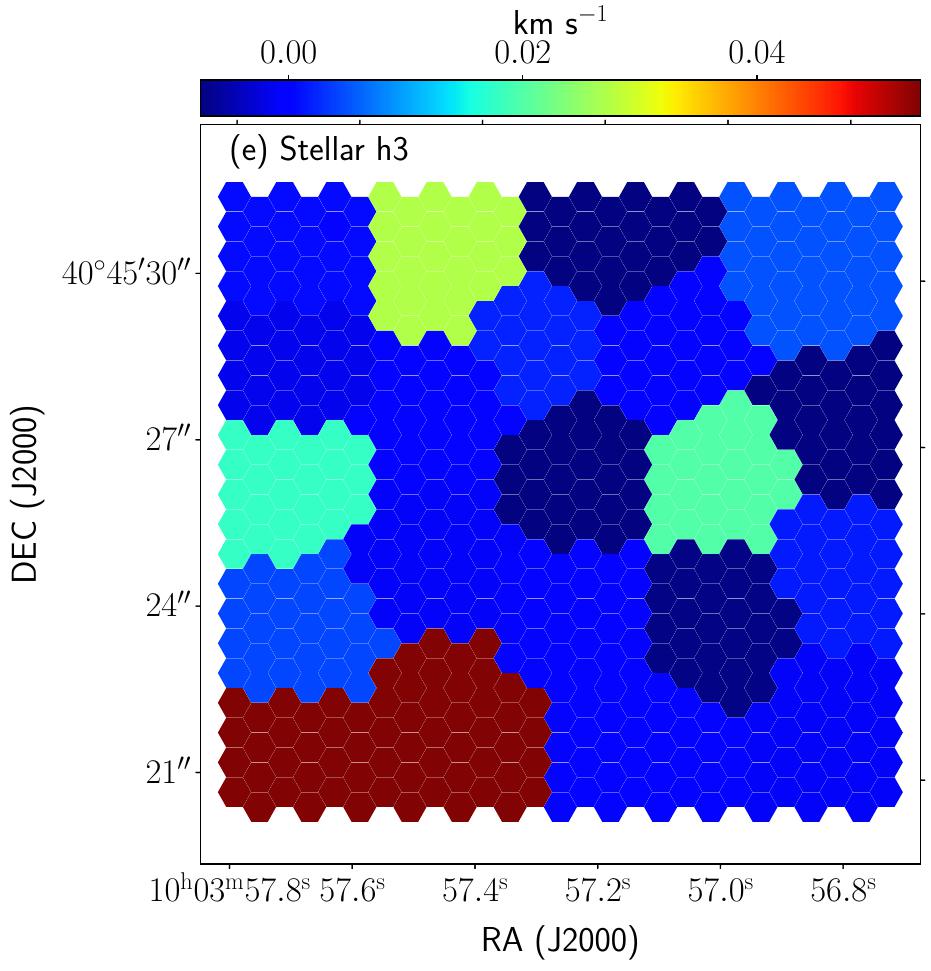}
	\includegraphics[clip, width=0.24\linewidth]{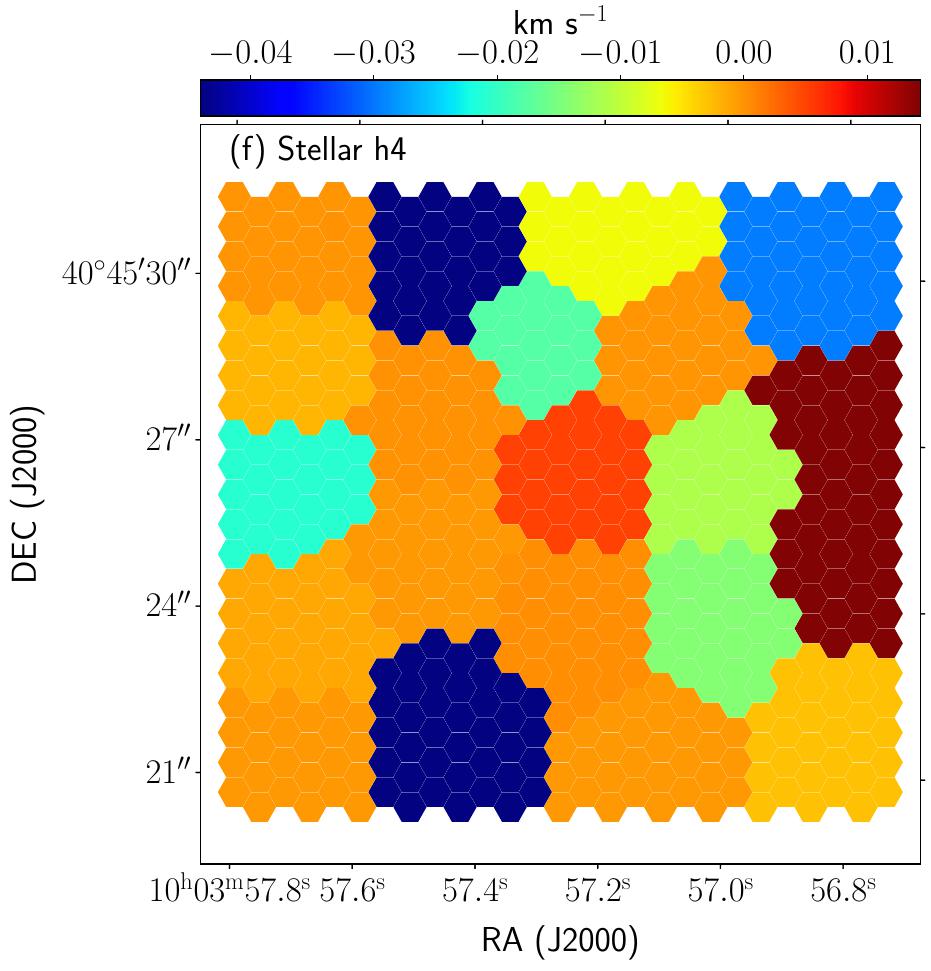}
	\includegraphics[clip, width=0.24\linewidth]{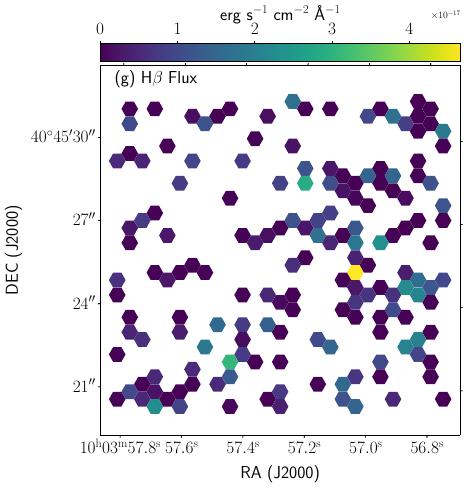}
	\includegraphics[clip, width=0.24\linewidth]{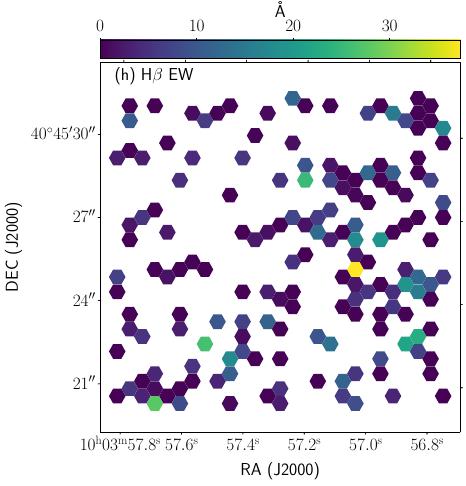}
	\includegraphics[clip, width=0.24\linewidth]{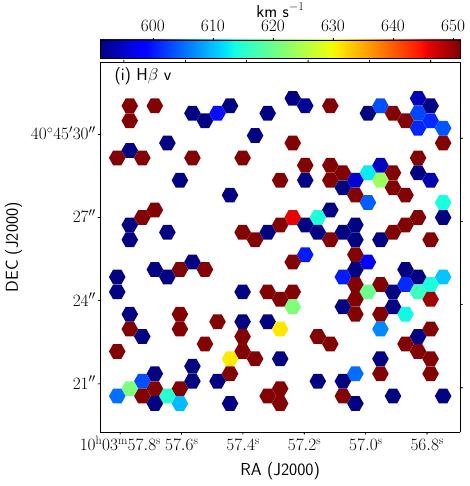}
	\includegraphics[clip, width=0.24\linewidth]{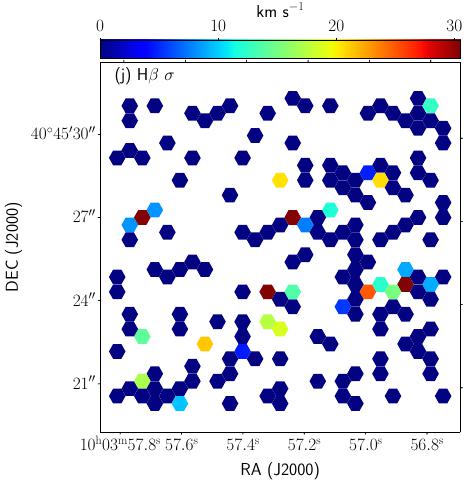}
	\includegraphics[clip, width=0.24\linewidth]{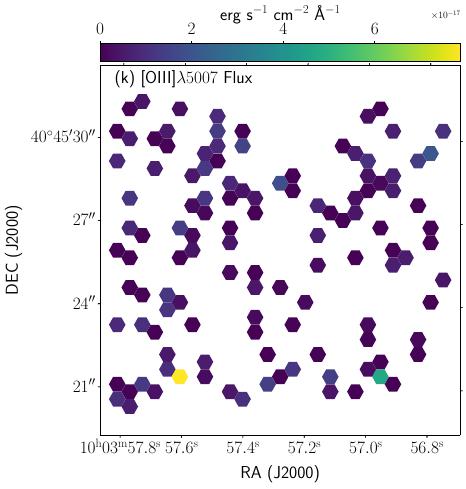}
	\includegraphics[clip, width=0.24\linewidth]{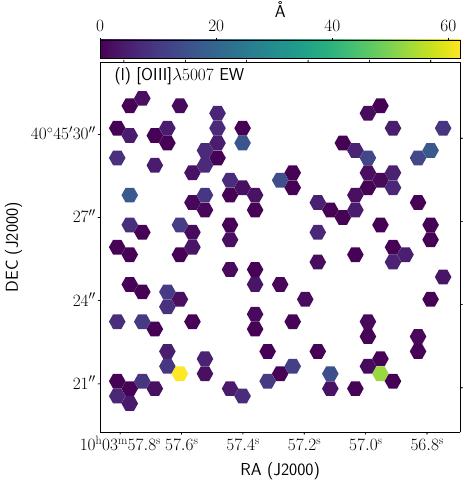}
	\includegraphics[clip, width=0.24\linewidth]{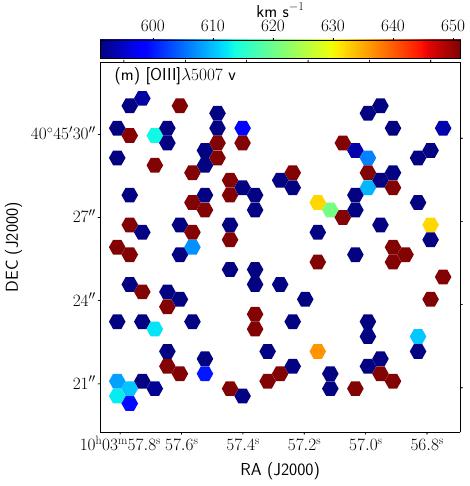}
	\includegraphics[clip, width=0.24\linewidth]{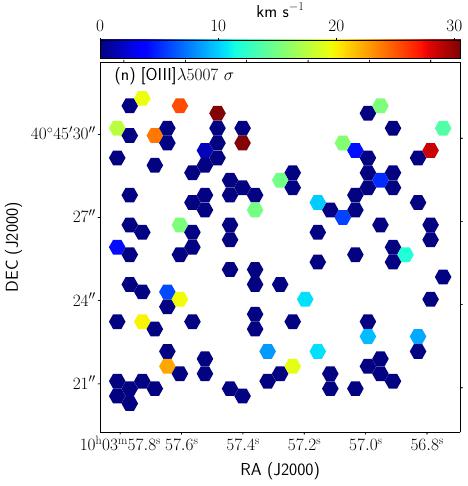}
	\vspace{5cm}
	\caption{NGC~3104 card.}
	\label{fig:NGC3104_card_1}
\end{figure*}
\addtocounter{figure}{-1}
\begin{figure*}[h]
	\centering
	\includegraphics[clip, width=0.24\linewidth]{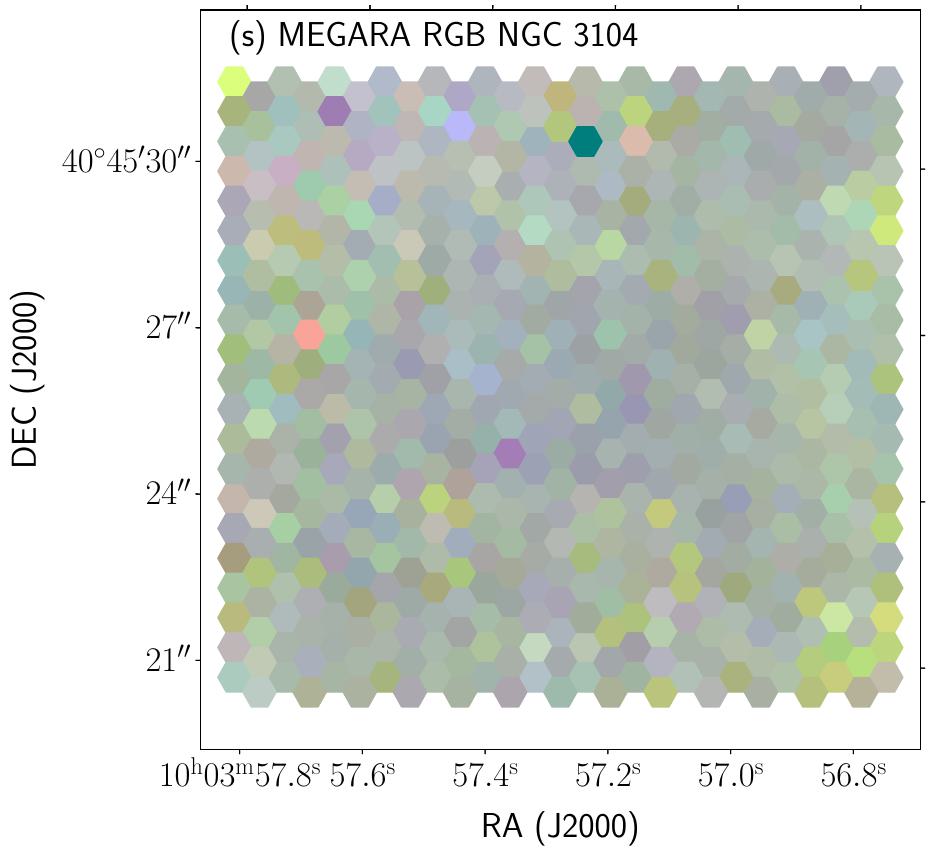}
	\includegraphics[clip, width=0.24\linewidth]{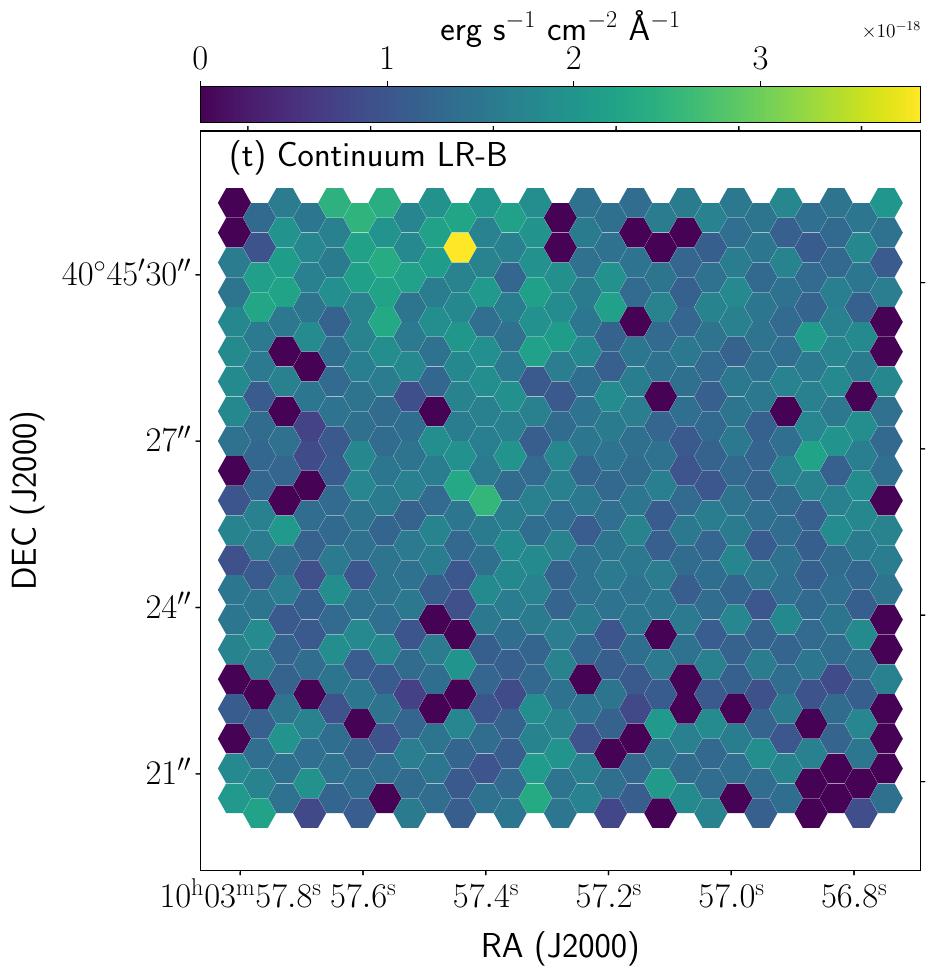}
	\includegraphics[clip, width=0.24\linewidth]{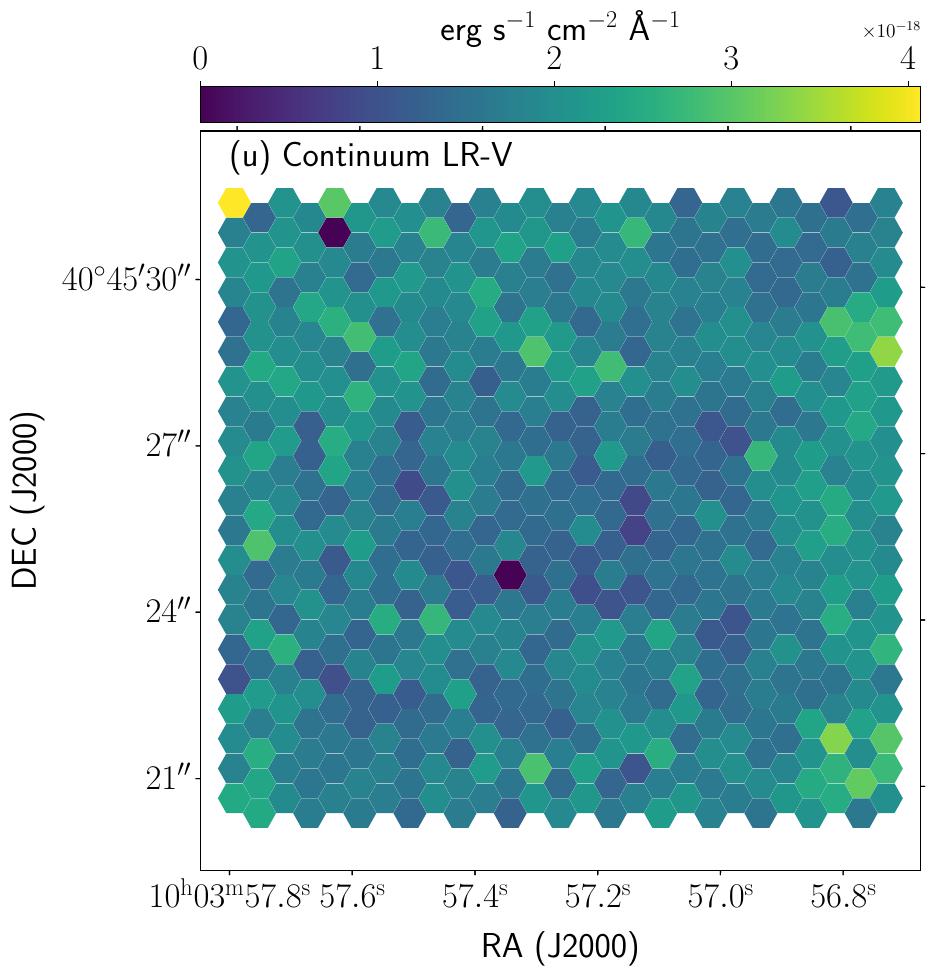}
	\includegraphics[clip, width=0.24\linewidth]{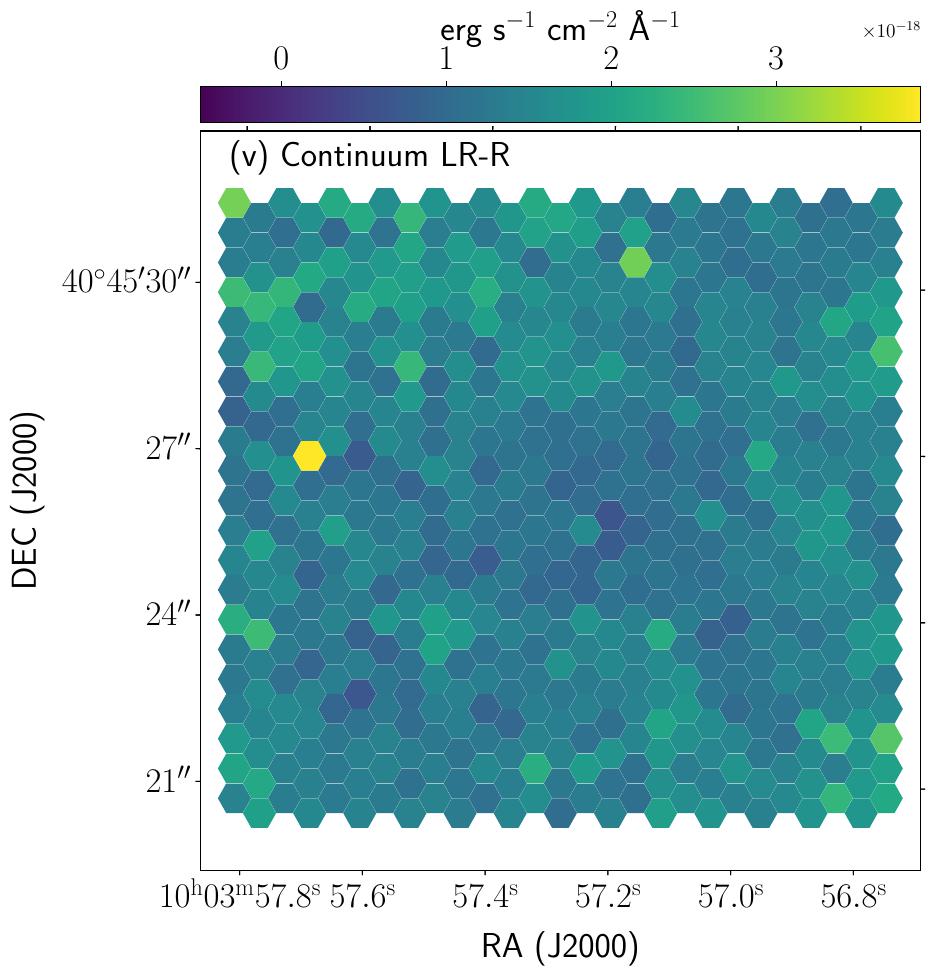}
	\includegraphics[clip, width=0.24\linewidth]{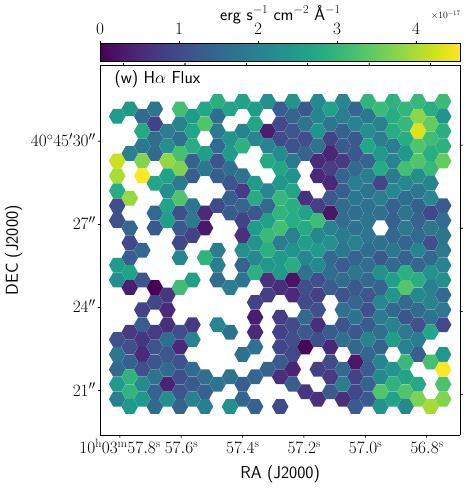}
	\includegraphics[clip, width=0.24\linewidth]{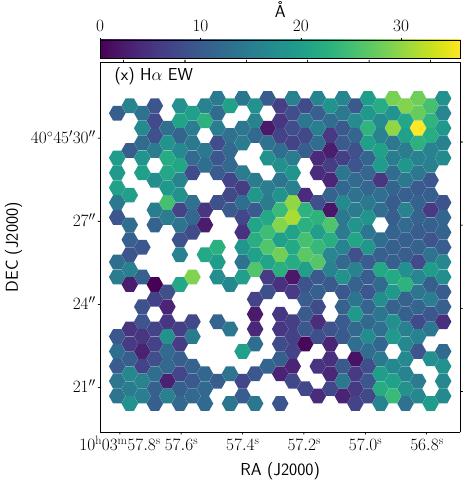}
	\includegraphics[clip, width=0.24\linewidth]{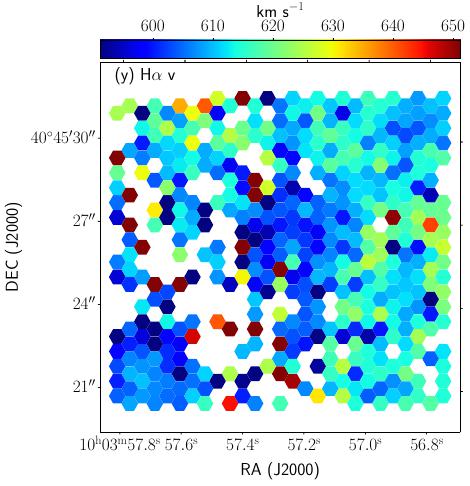}
	\includegraphics[clip, width=0.24\linewidth]{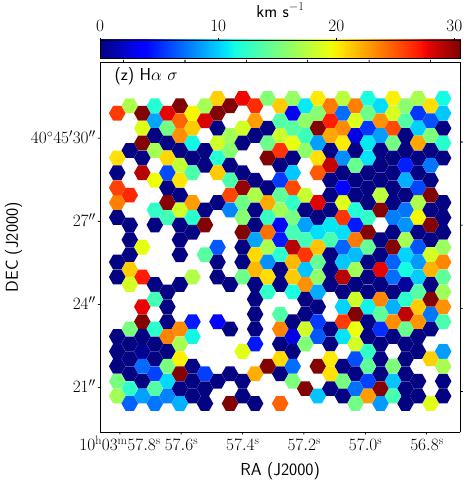}
	\includegraphics[clip, width=0.24\linewidth]{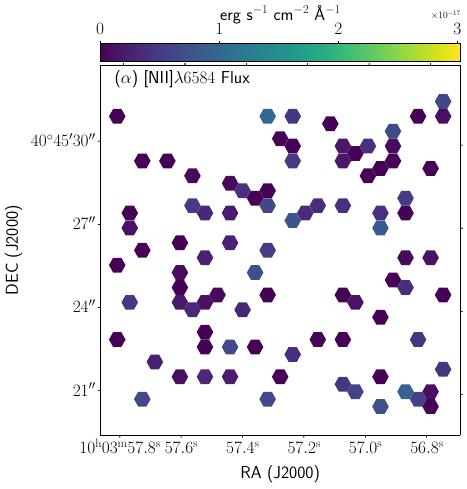}
	\includegraphics[clip, width=0.24\linewidth]{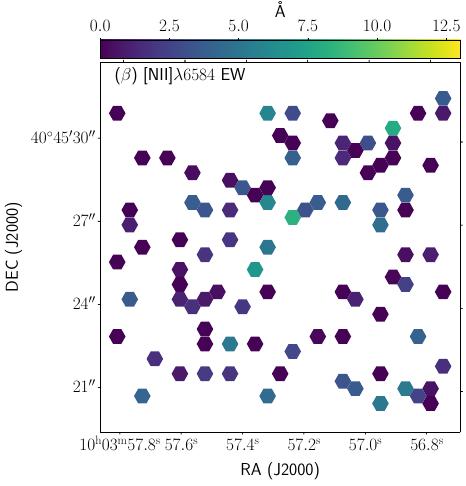}
	\includegraphics[clip, width=0.24\linewidth]{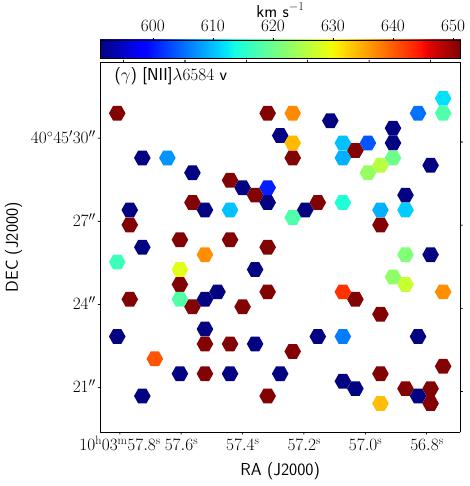}
	\includegraphics[clip, width=0.24\linewidth]{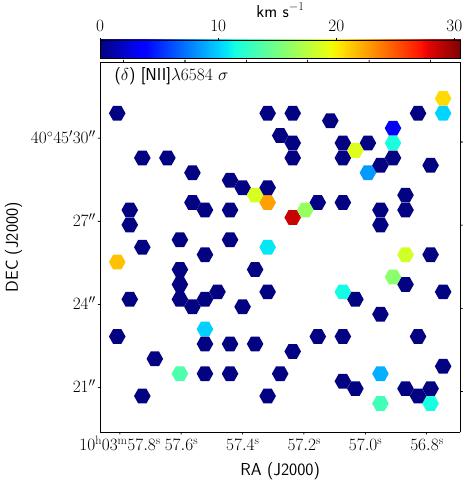}
	\includegraphics[clip, width=0.24\linewidth]{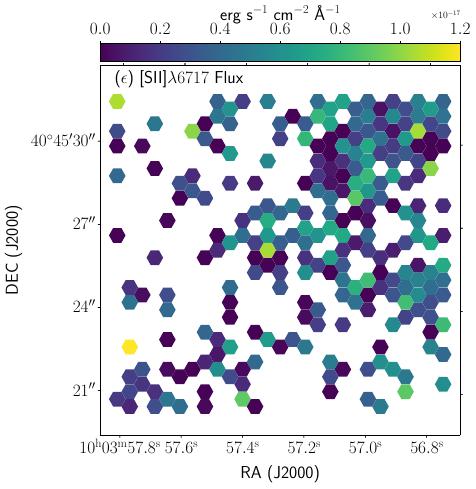}
	\includegraphics[clip, width=0.24\linewidth]{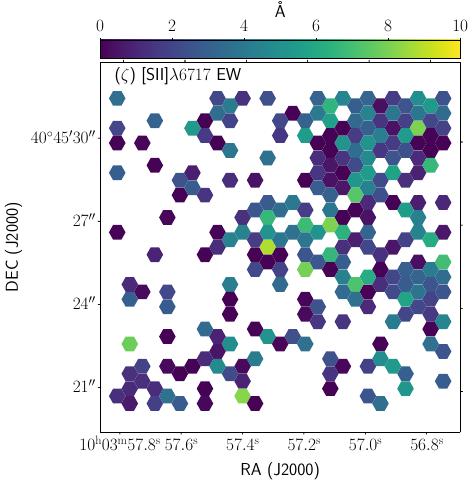}
	\includegraphics[clip, width=0.24\linewidth]{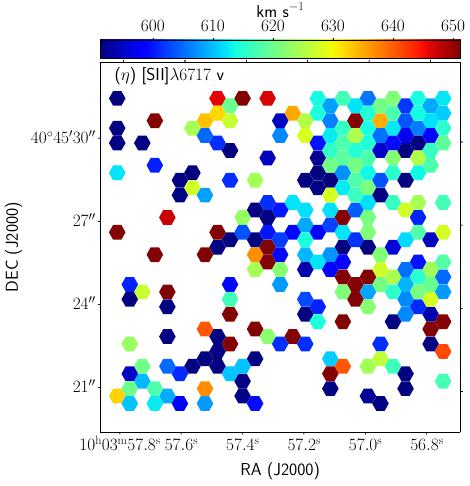}
	\includegraphics[clip, width=0.24\linewidth]{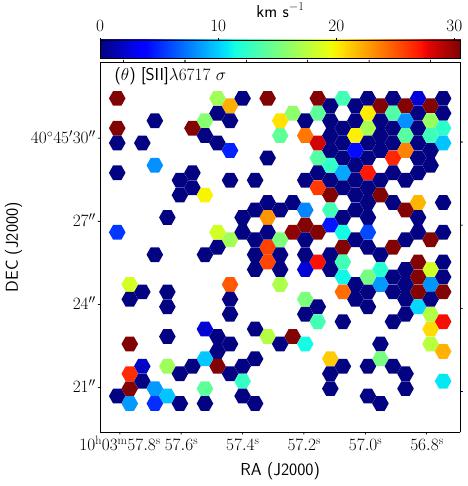}
	\includegraphics[clip, width=0.24\linewidth]{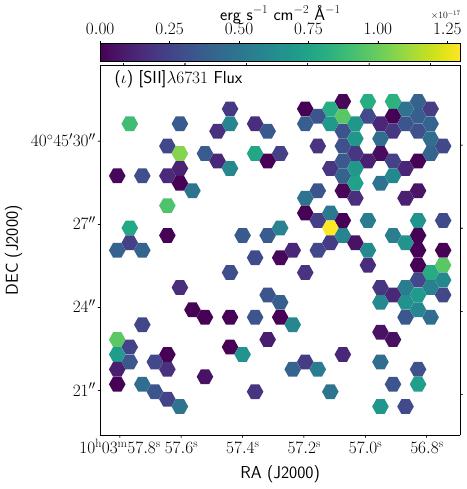}
	\includegraphics[clip, width=0.24\linewidth]{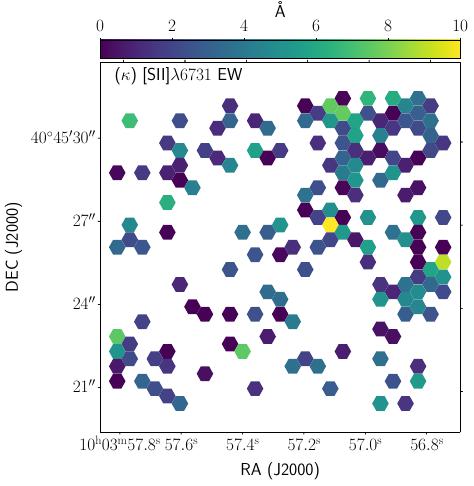}
	\includegraphics[clip, width=0.24\linewidth]{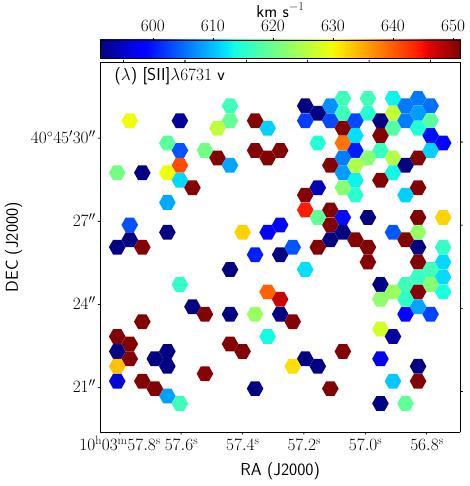}
	\includegraphics[clip, width=0.24\linewidth]{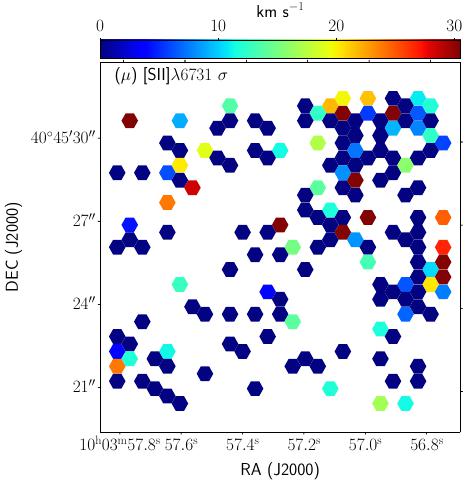}
	\caption{(cont.) NGC~3104 card.}
	\label{fig:NGC3104_card_2}
\end{figure*}

\begin{figure*}[h]
	\centering
	\includegraphics[clip, width=0.35\linewidth]{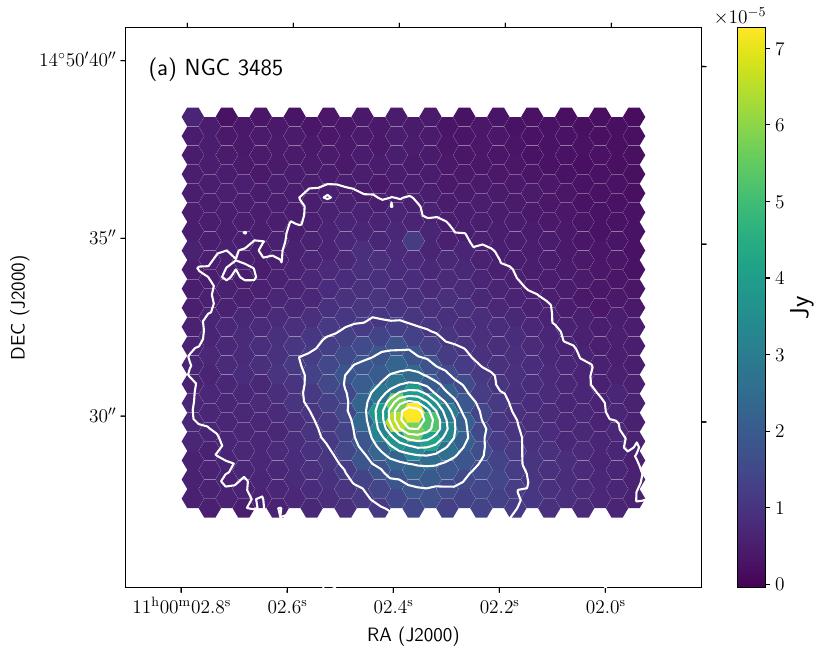}
	\includegraphics[clip, width=0.6\linewidth]{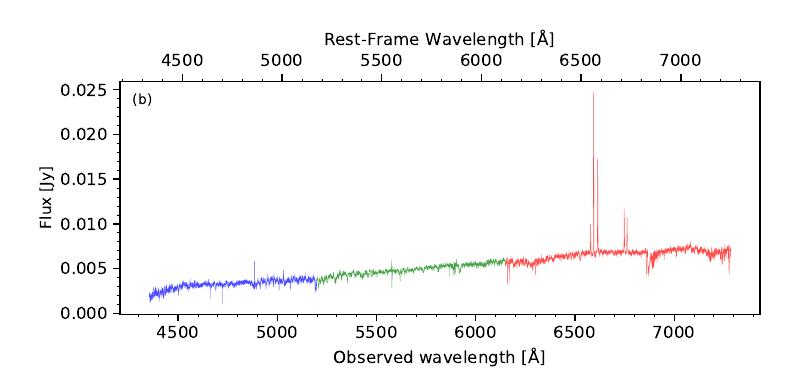}
	\includegraphics[clip, width=0.24\linewidth]{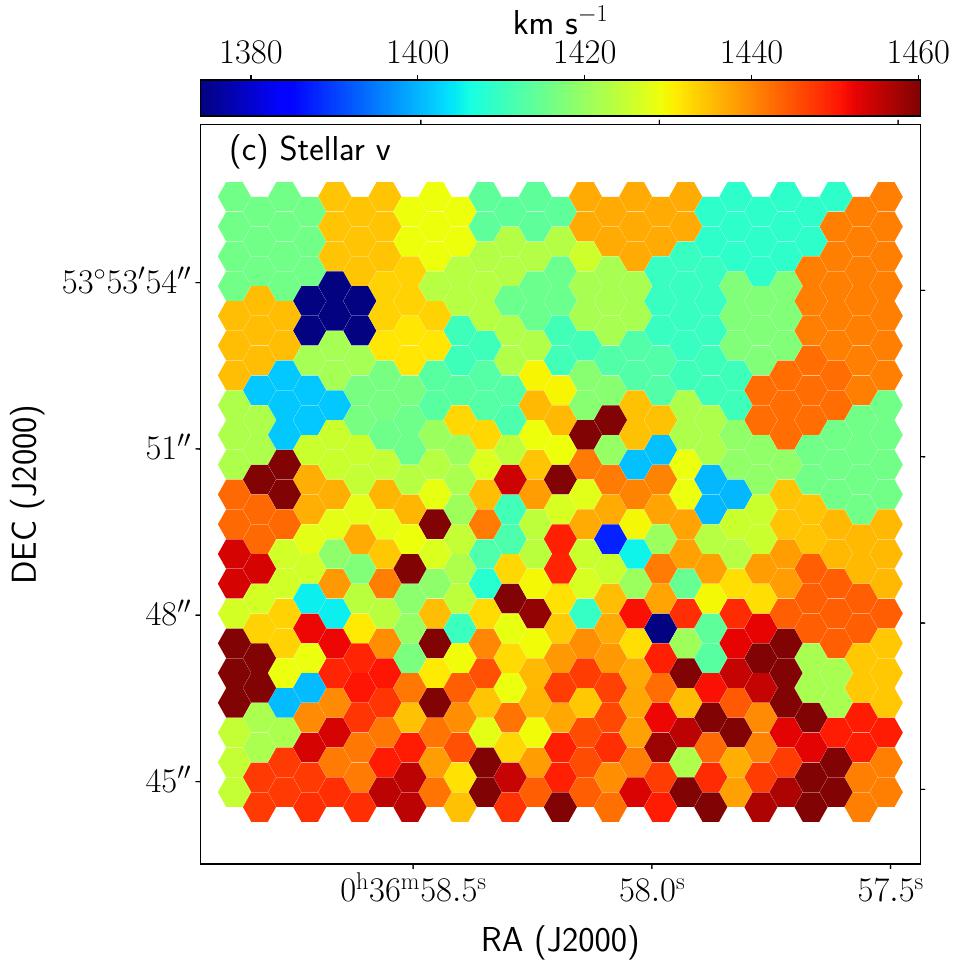}
	\includegraphics[clip, width=0.24\linewidth]{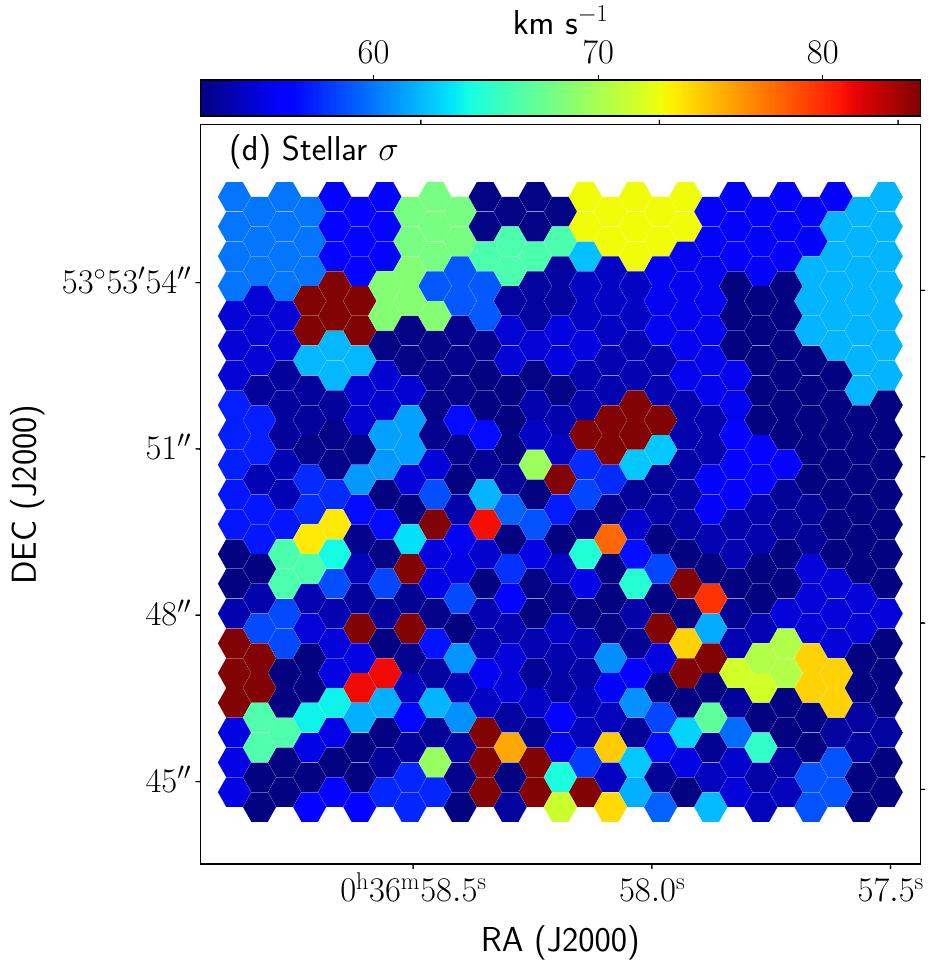}
	\includegraphics[clip, width=0.24\linewidth]{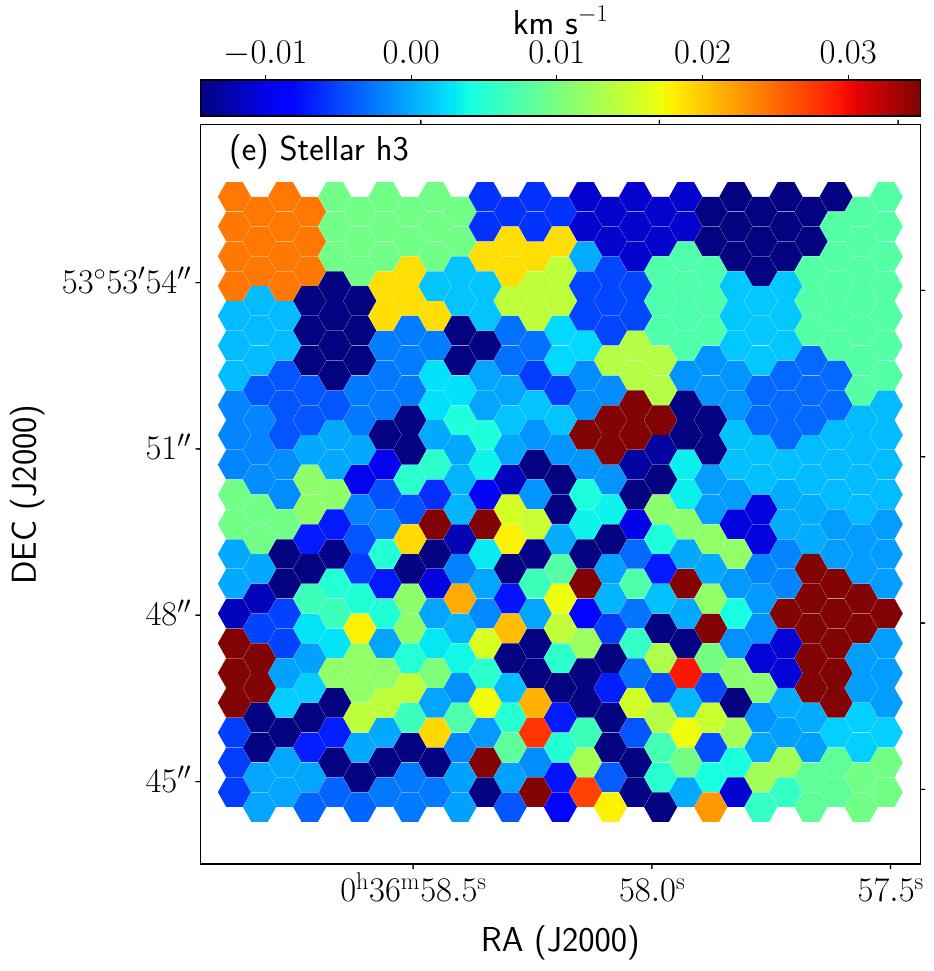}
	\includegraphics[clip, width=0.24\linewidth]{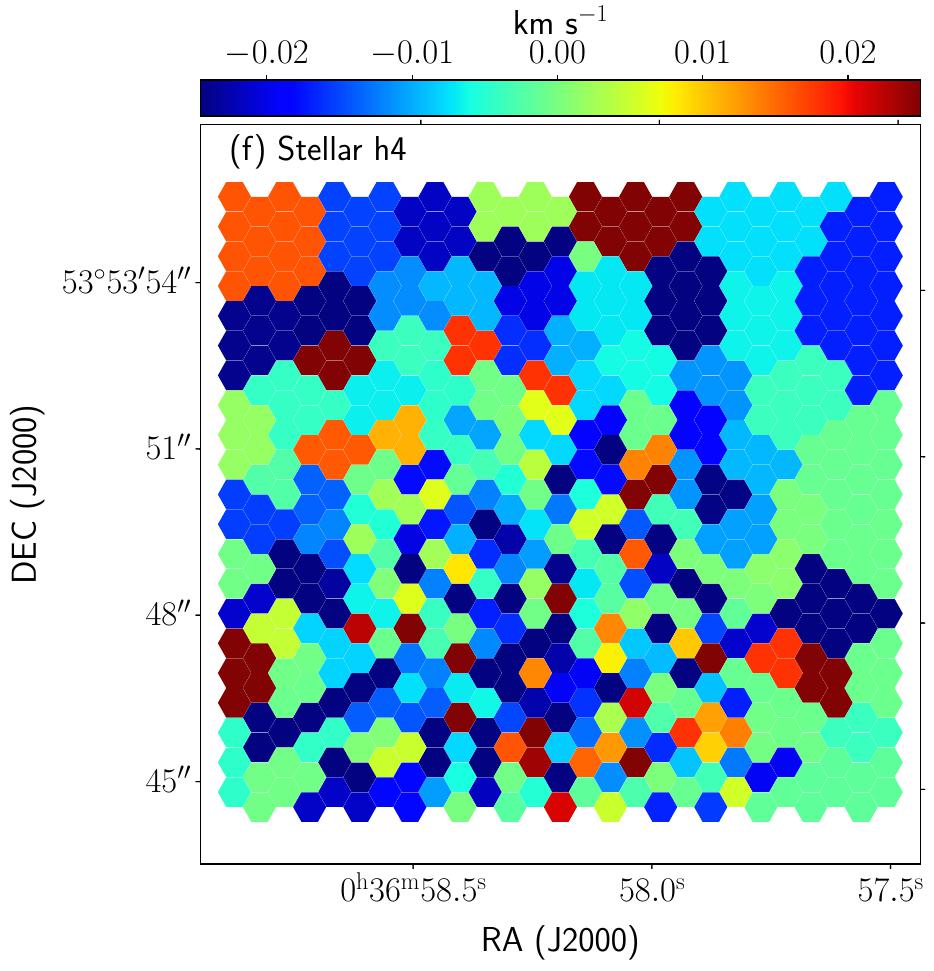}
	\includegraphics[clip, width=0.24\linewidth]{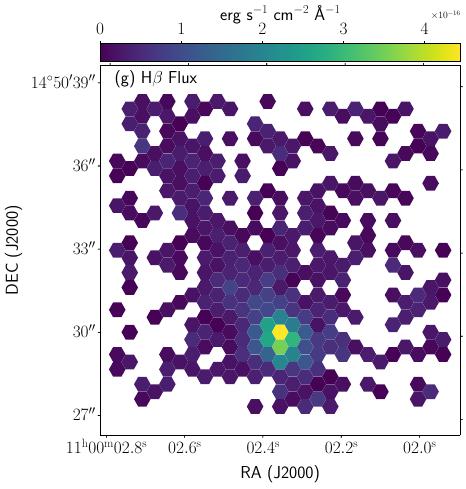}
	\includegraphics[clip, width=0.24\linewidth]{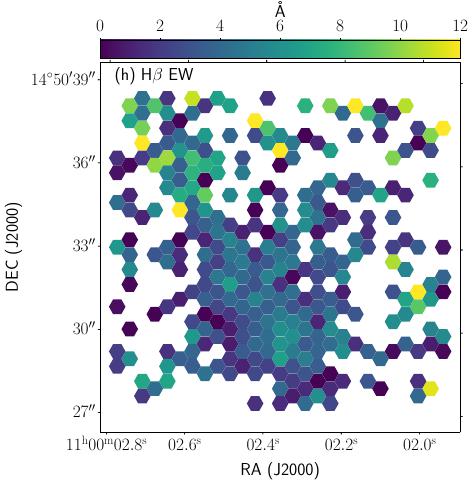}
	\includegraphics[clip, width=0.24\linewidth]{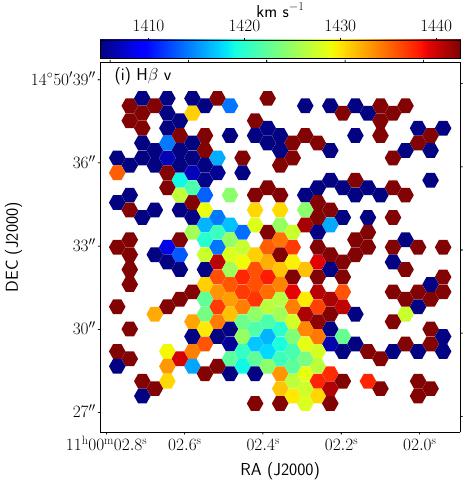}
	\includegraphics[clip, width=0.24\linewidth]{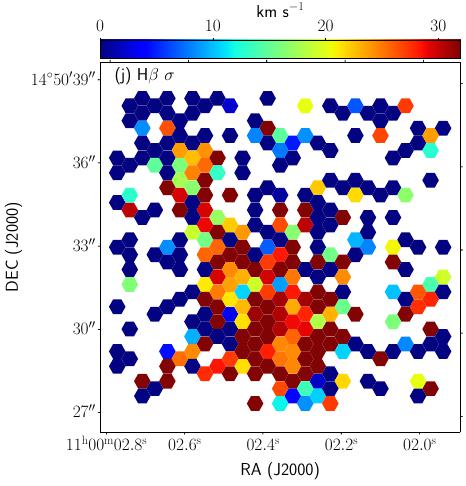}
	\includegraphics[clip, width=0.24\linewidth]{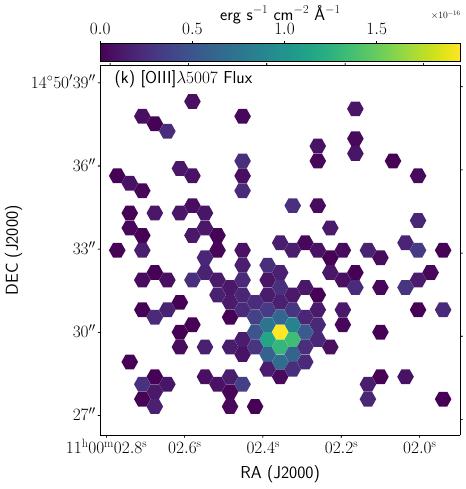}
	\includegraphics[clip, width=0.24\linewidth]{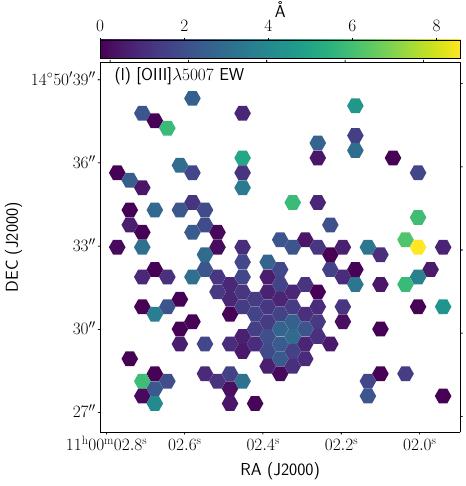}
	\includegraphics[clip, width=0.24\linewidth]{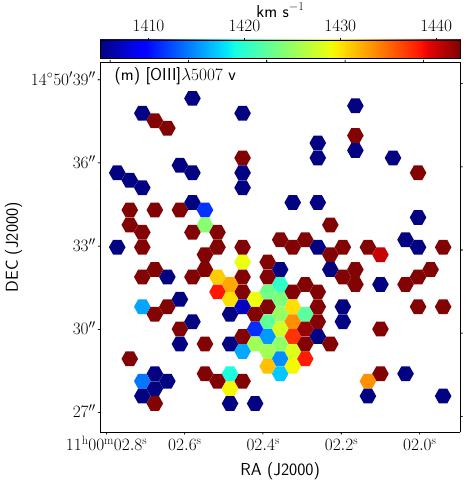}
	\includegraphics[clip, width=0.24\linewidth]{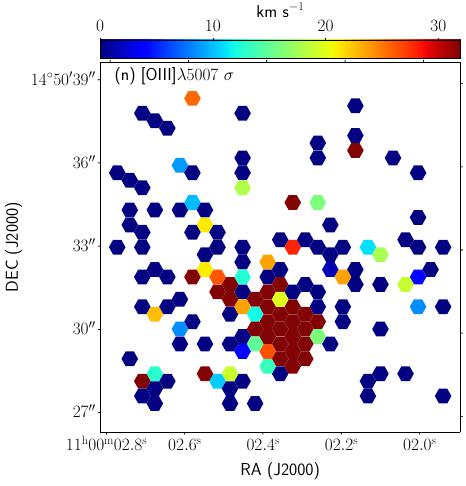}
	\vspace{5cm}
	\caption{NGC~3485 card.}
	\label{fig:NGC3485_card_1}
\end{figure*}
\addtocounter{figure}{-1}
\begin{figure*}[h]
	\centering
	\includegraphics[clip, width=0.24\linewidth]{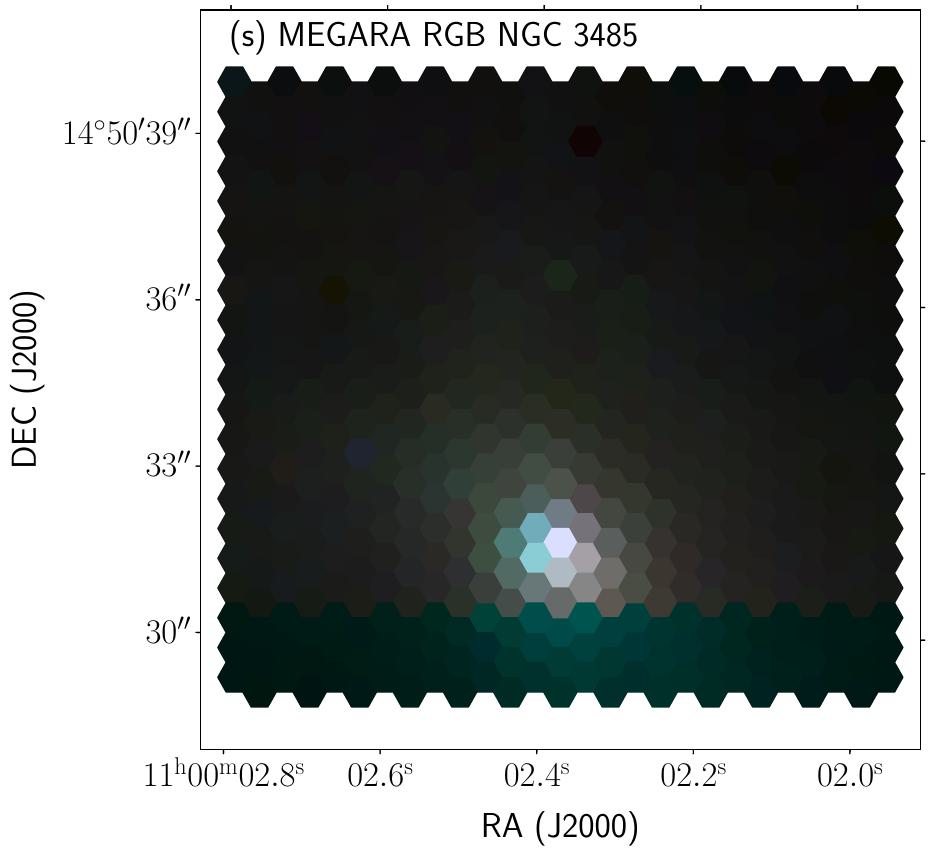}
	\includegraphics[clip, width=0.24\linewidth]{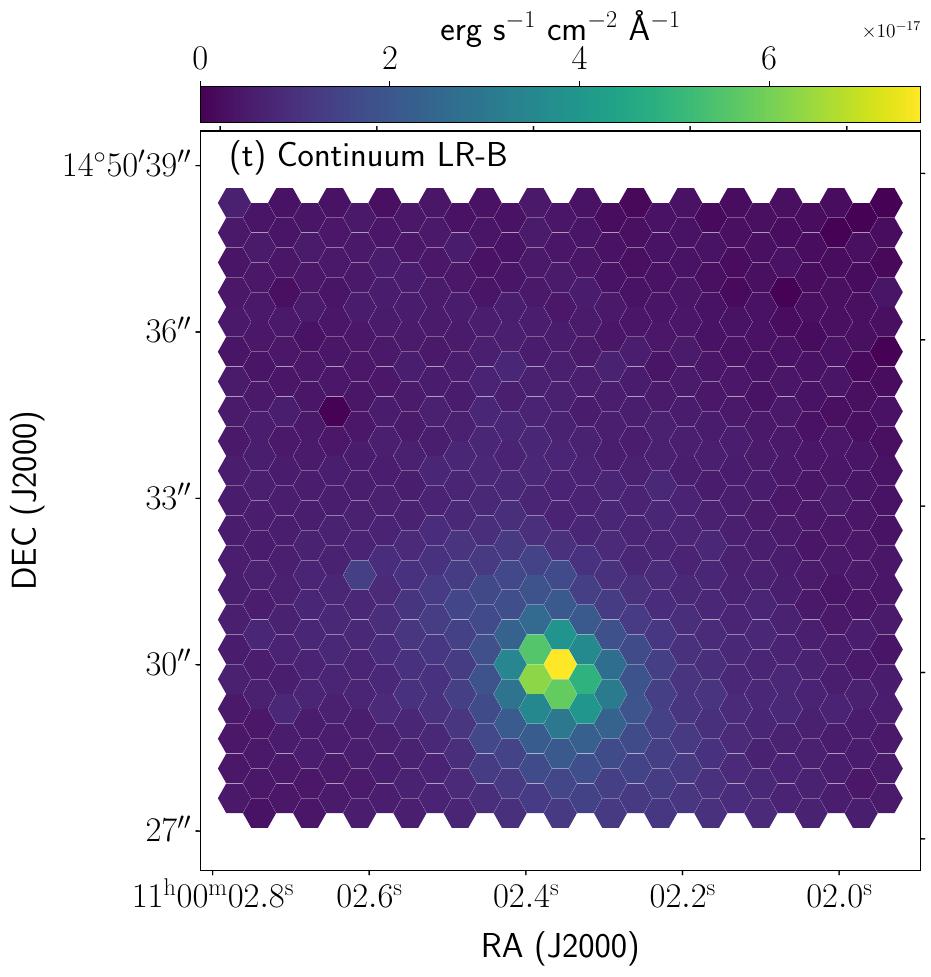}
	\includegraphics[clip, width=0.24\linewidth]{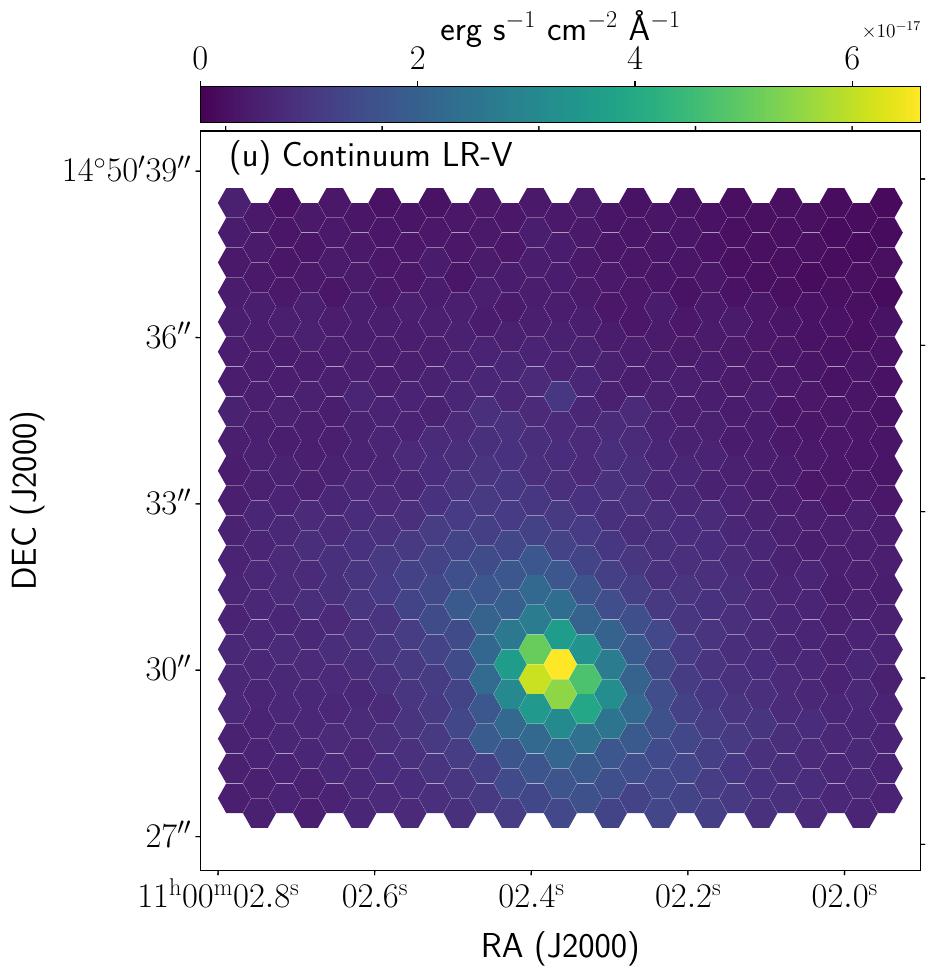}
	\includegraphics[clip, width=0.24\linewidth]{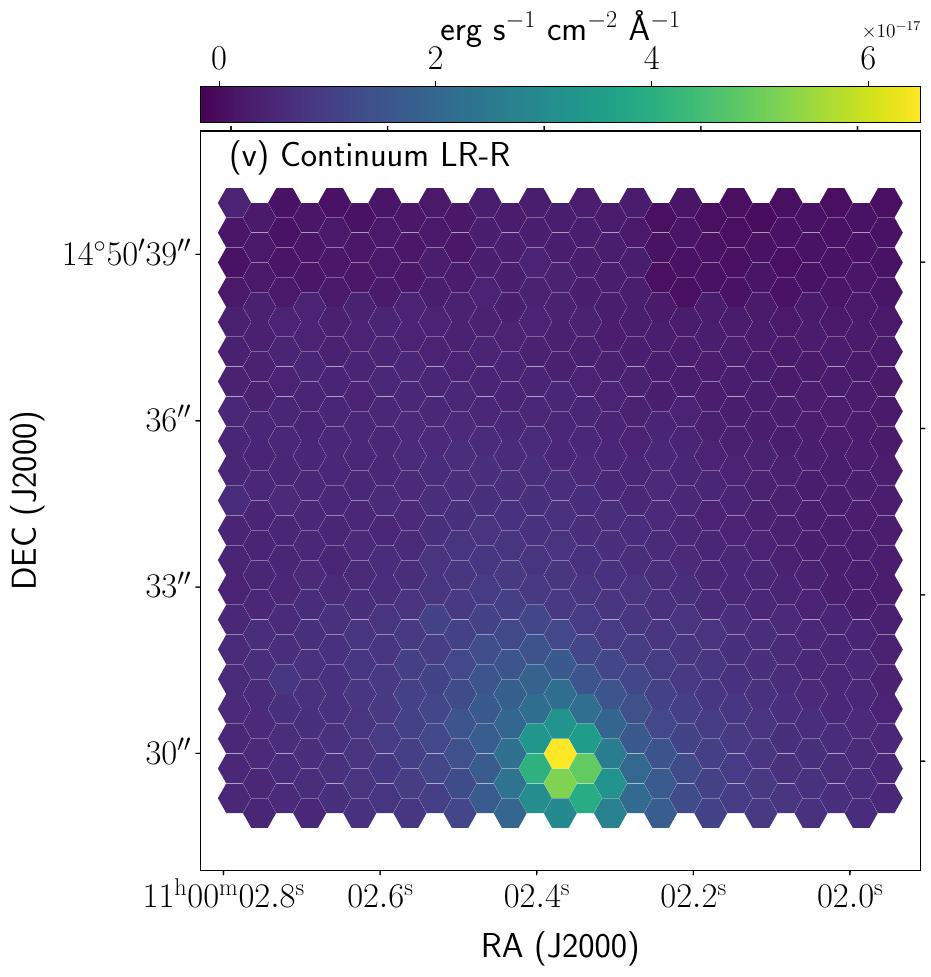}
	\includegraphics[clip, width=0.24\linewidth]{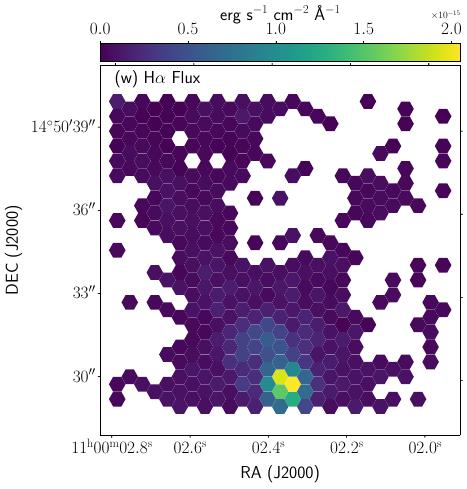}
	\includegraphics[clip, width=0.24\linewidth]{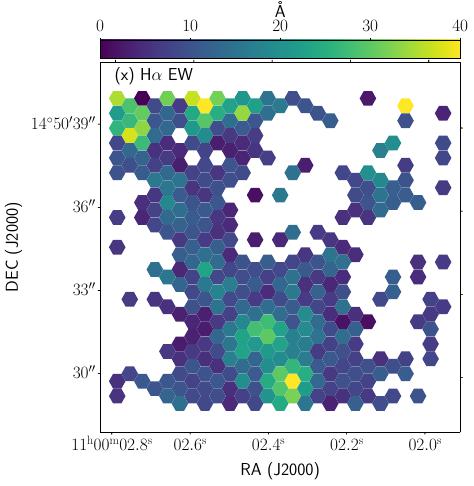}
	\includegraphics[clip, width=0.24\linewidth]{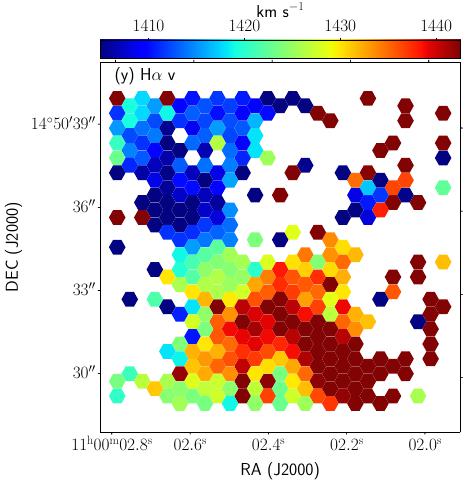}
	\includegraphics[clip, width=0.24\linewidth]{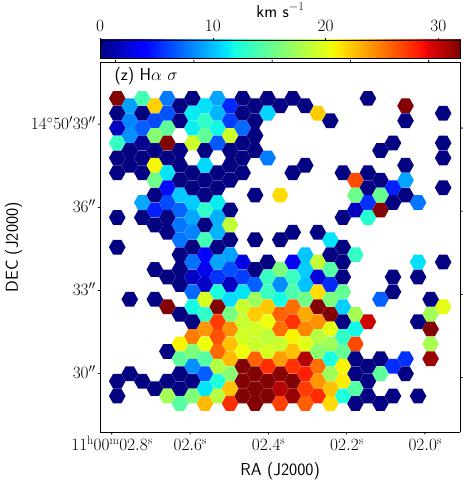}
	\includegraphics[clip, width=0.24\linewidth]{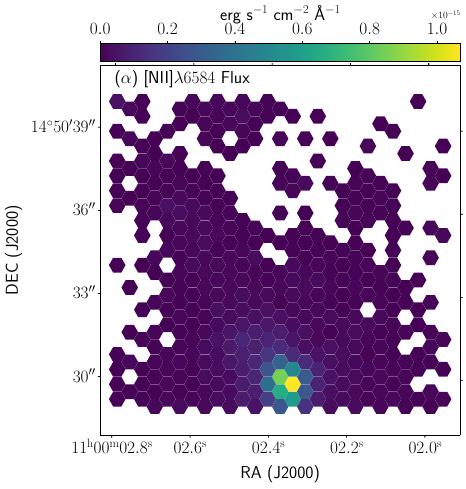}
	\includegraphics[clip, width=0.24\linewidth]{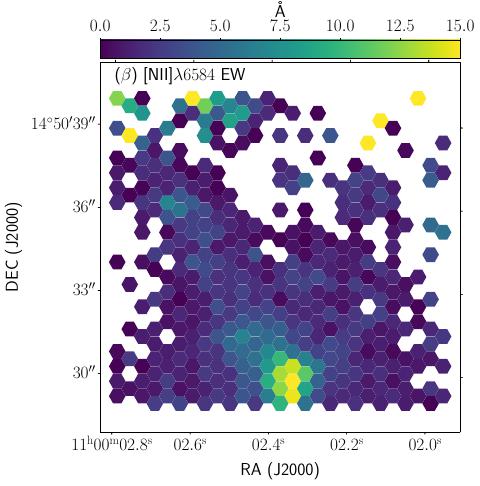}
	\includegraphics[clip, width=0.24\linewidth]{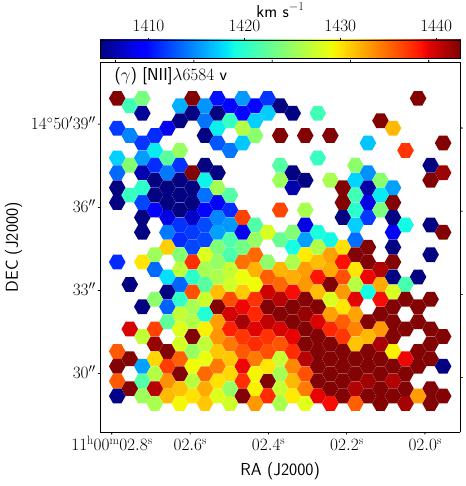}
	\includegraphics[clip, width=0.24\linewidth]{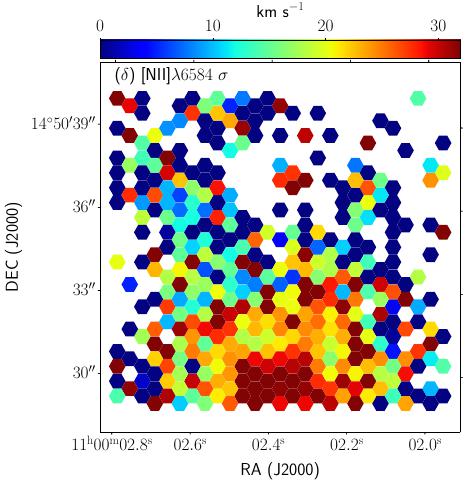}
	\includegraphics[clip, width=0.24\linewidth]{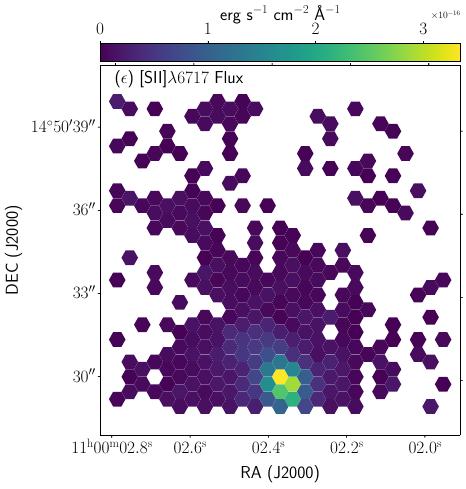}
	\includegraphics[clip, width=0.24\linewidth]{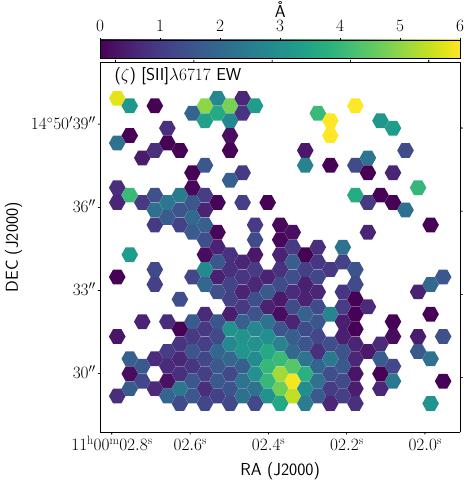}
	\includegraphics[clip, width=0.24\linewidth]{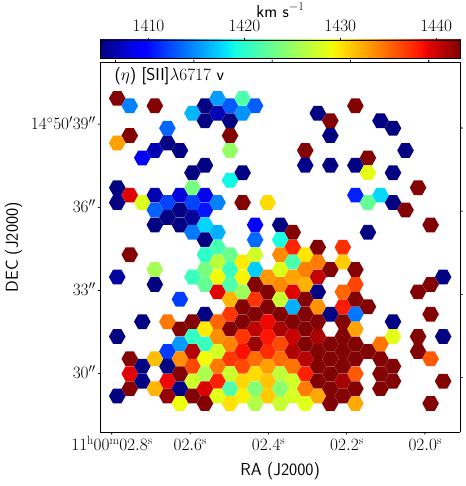}
	\includegraphics[clip, width=0.24\linewidth]{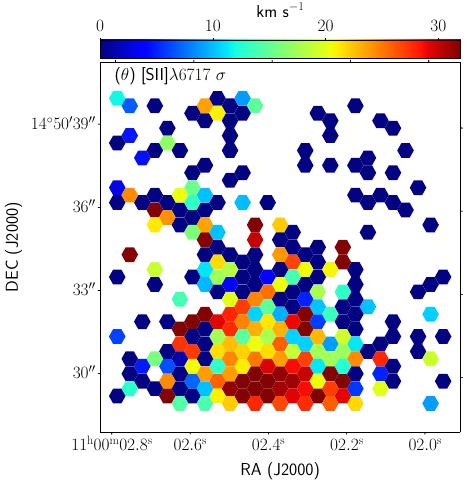}
	\includegraphics[clip, width=0.24\linewidth]{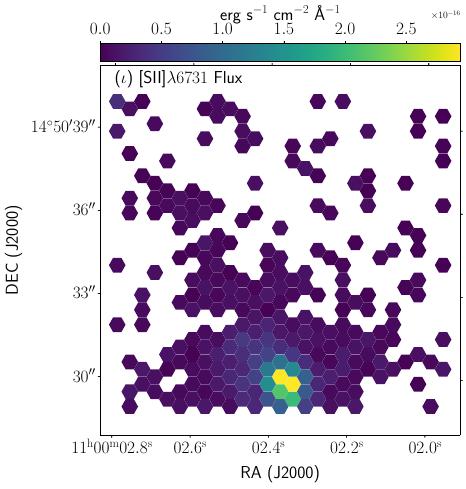}
	\includegraphics[clip, width=0.24\linewidth]{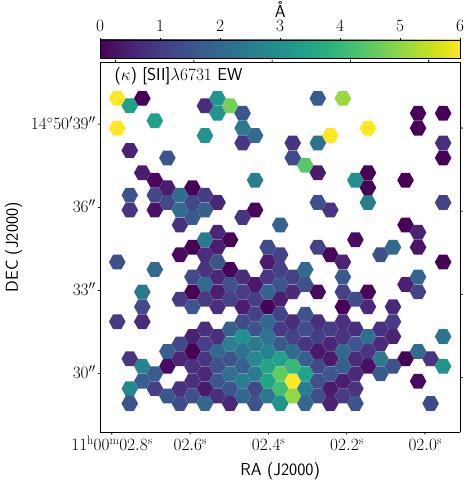}
	\includegraphics[clip, width=0.24\linewidth]{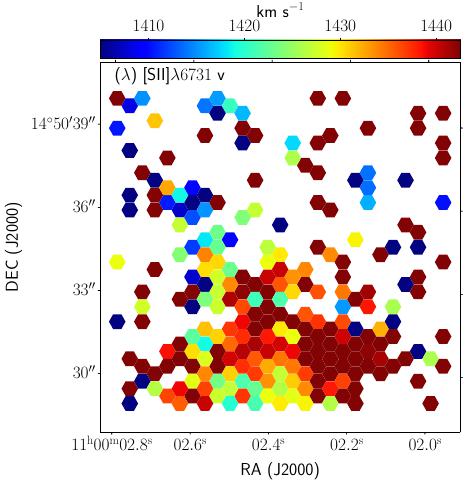}
	\includegraphics[clip, width=0.24\linewidth]{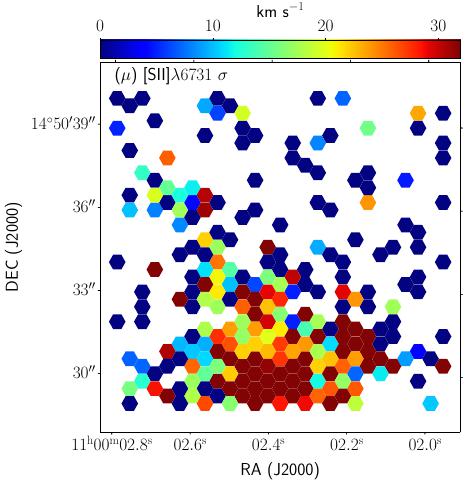}
	\caption{(cont.) NGC~3485 card.}
	\label{fig:NGC3485_card_2}
\end{figure*}

\begin{figure*}[h]
	\centering
	\includegraphics[clip, width=0.35\linewidth]{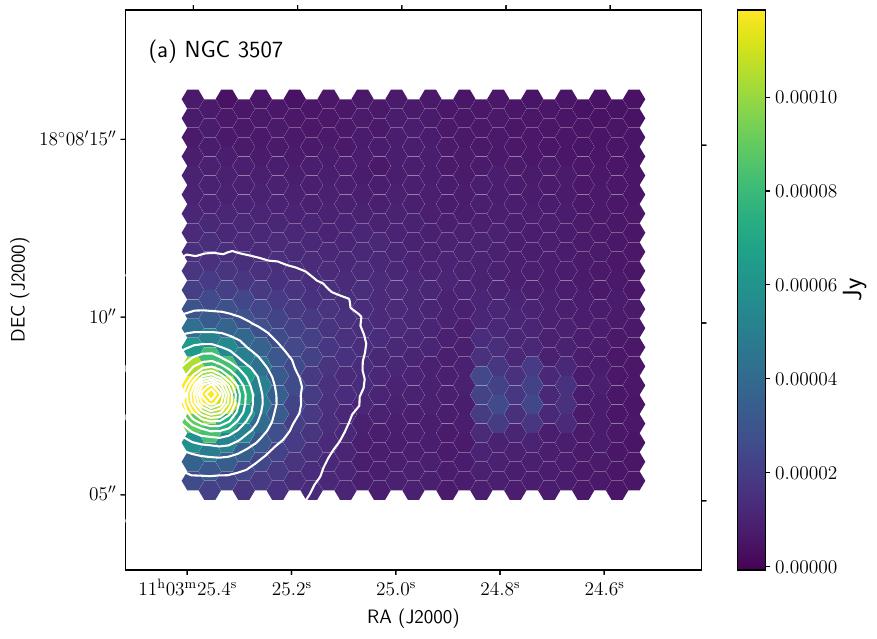}
	\includegraphics[clip, width=0.6\linewidth]{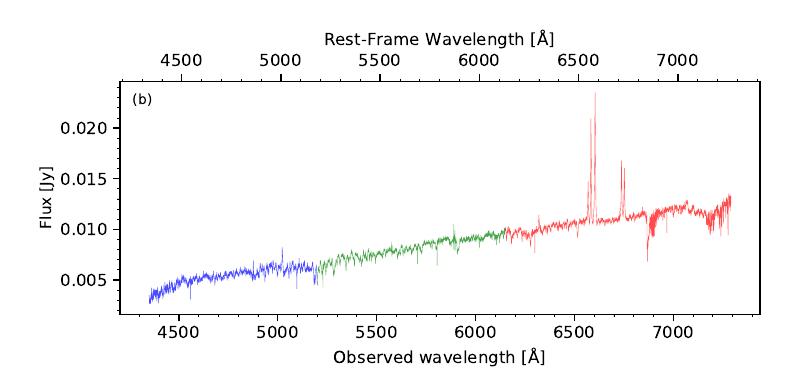}
	\includegraphics[clip, width=0.24\linewidth]{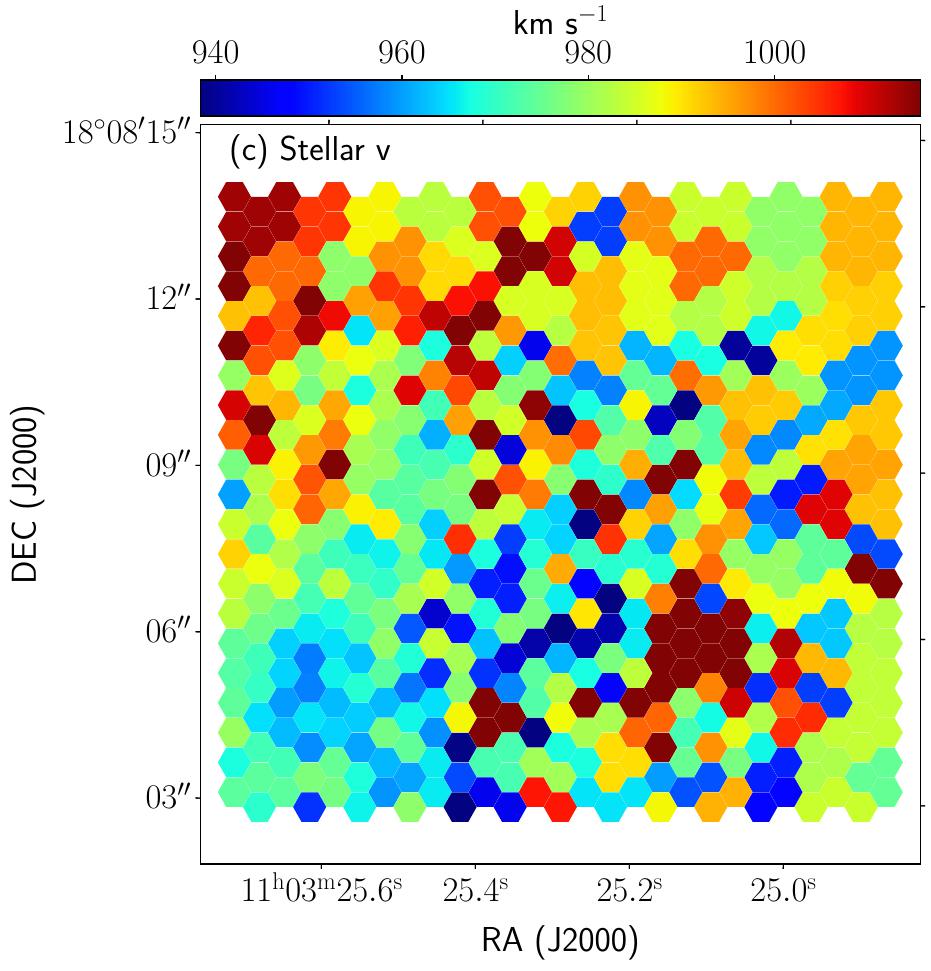}
	\includegraphics[clip, width=0.24\linewidth]{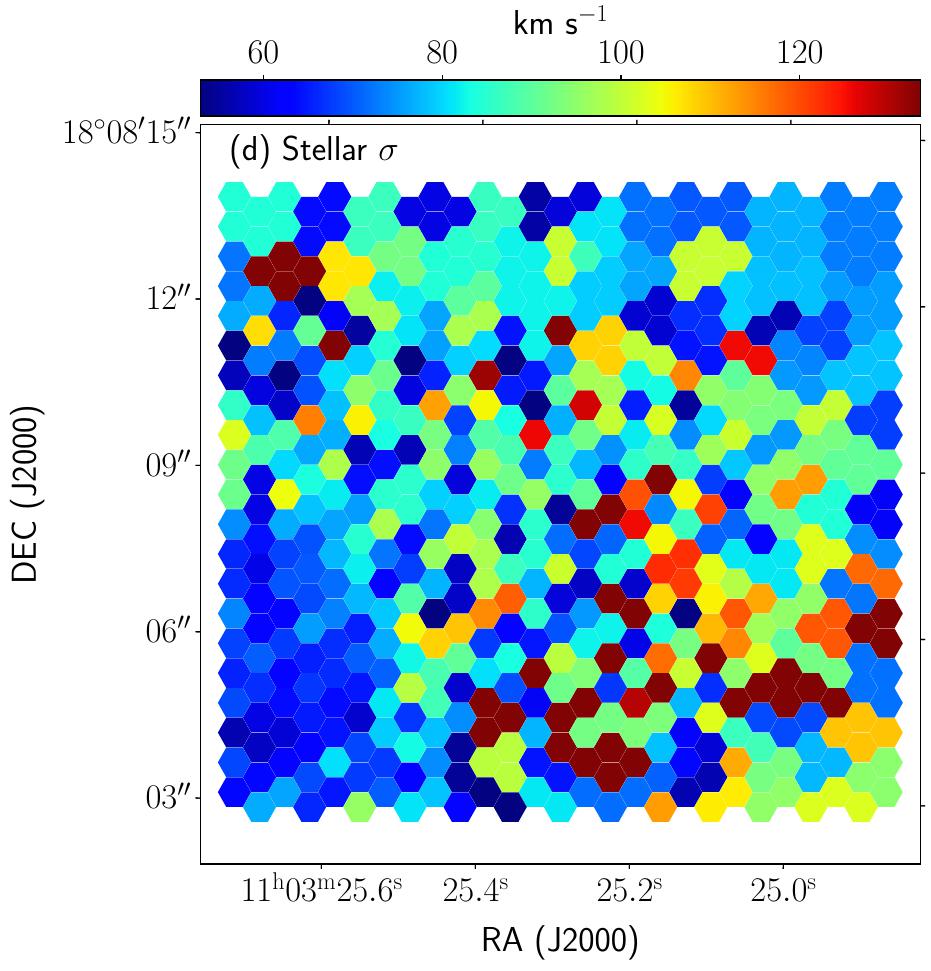}
	\includegraphics[clip, width=0.24\linewidth]{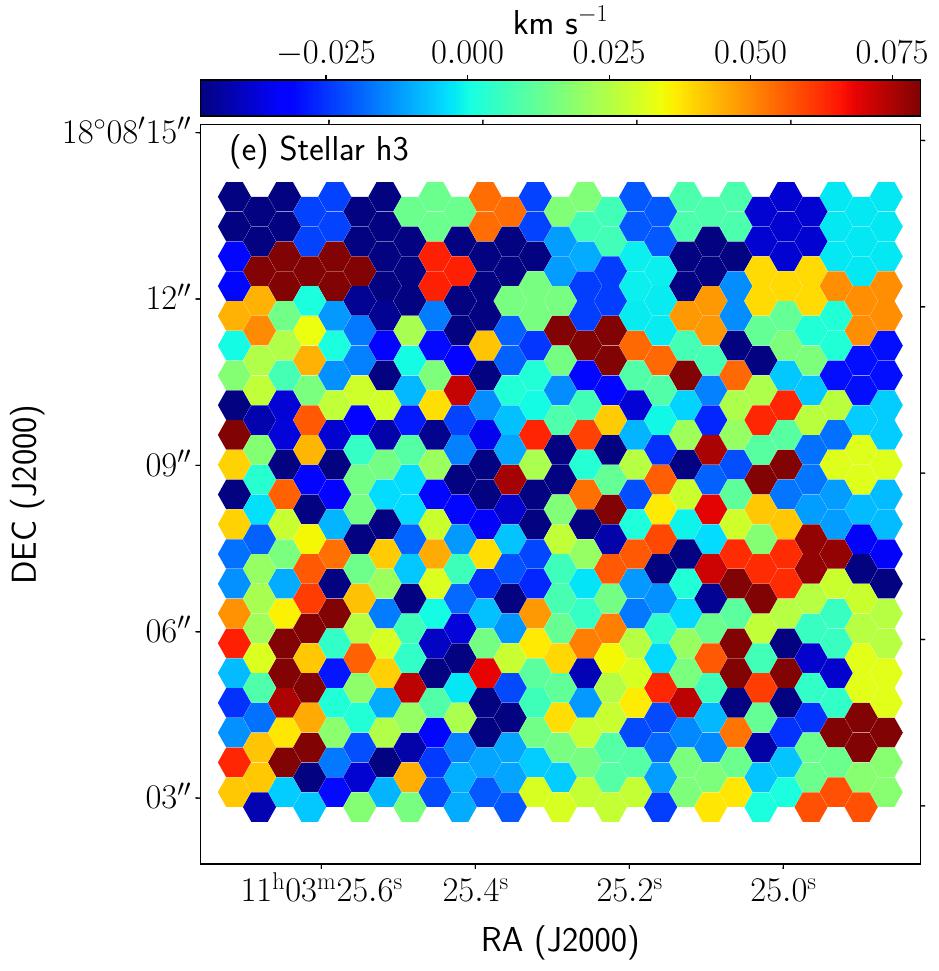}
	\includegraphics[clip, width=0.24\linewidth]{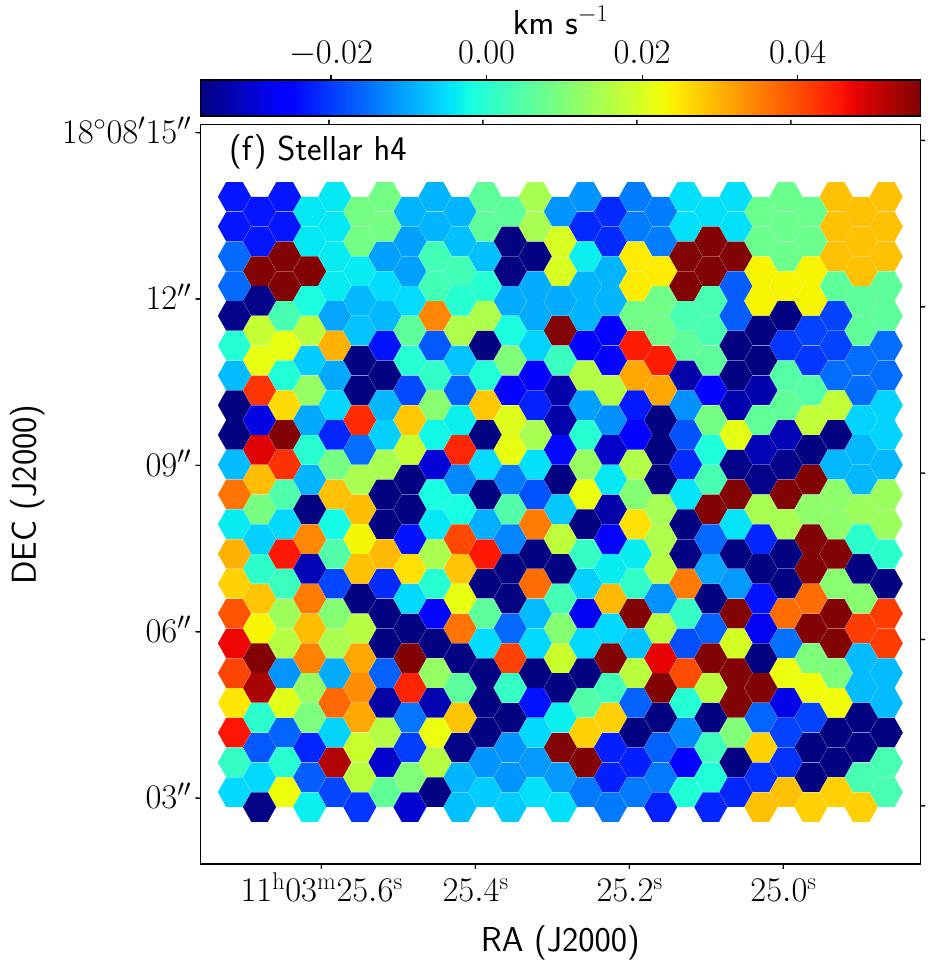}
	\includegraphics[clip, width=0.24\linewidth]{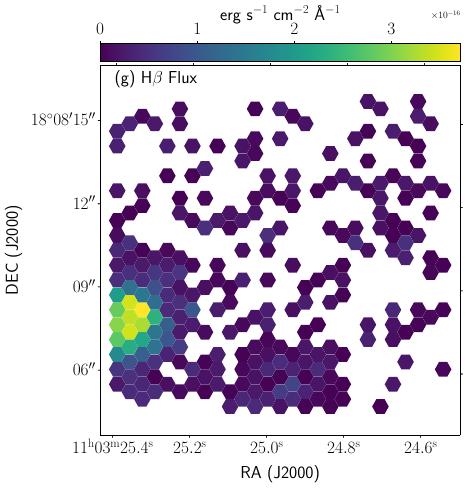}
	\includegraphics[clip, width=0.24\linewidth]{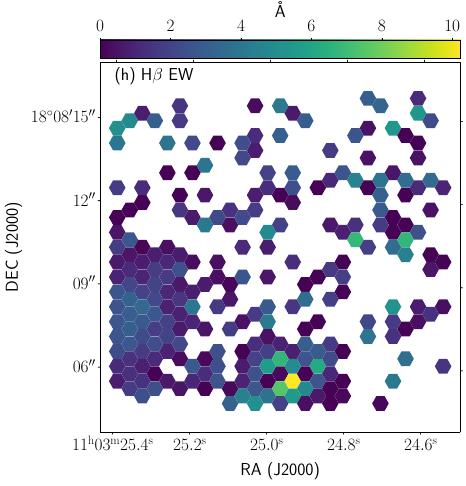}
	\includegraphics[clip, width=0.24\linewidth]{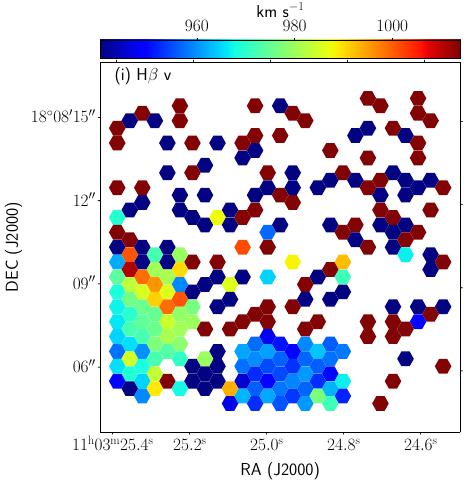}
	\includegraphics[clip, width=0.24\linewidth]{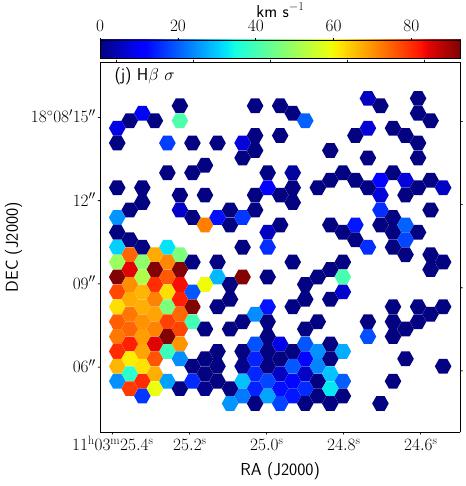}
	\includegraphics[clip, width=0.24\linewidth]{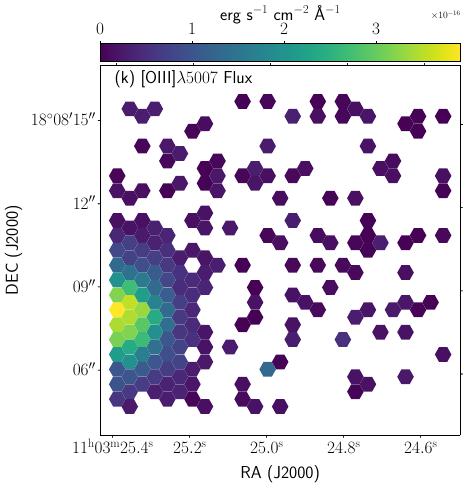}
	\includegraphics[clip, width=0.24\linewidth]{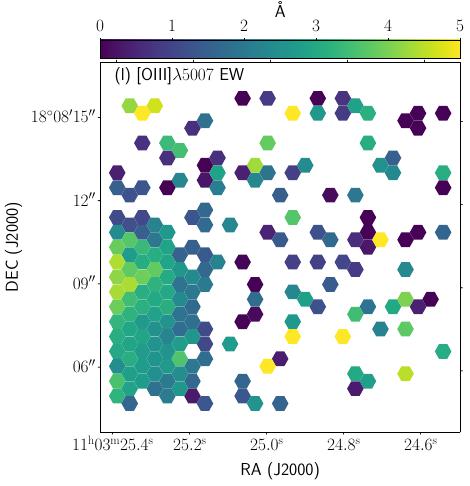}
	\includegraphics[clip, width=0.24\linewidth]{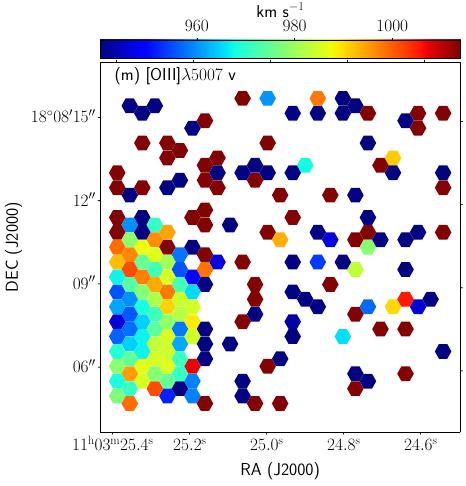}
	\includegraphics[clip, width=0.24\linewidth]{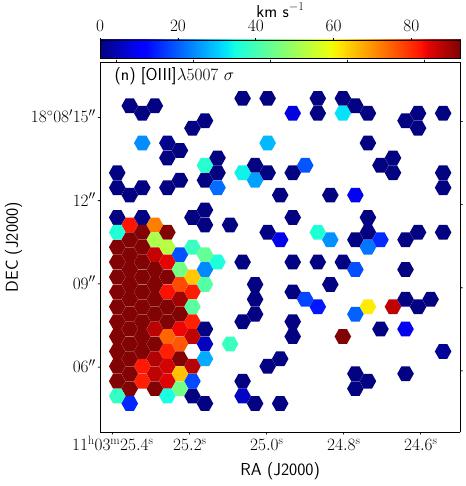}
	\vspace{5cm}
	\caption{NGC~3507 card.}
	\label{fig:NGC3507_card_1}
\end{figure*}
\addtocounter{figure}{-1}
\begin{figure*}[h]
	\centering
	\includegraphics[clip, width=0.24\linewidth]{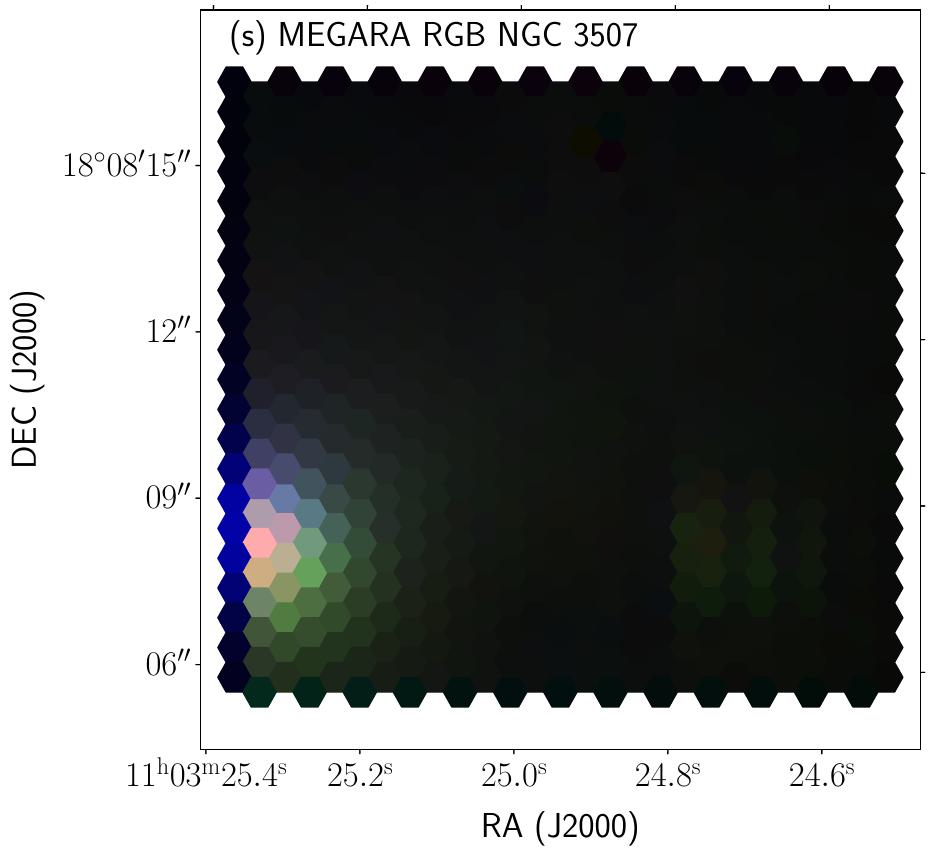}
	\includegraphics[clip, width=0.24\linewidth]{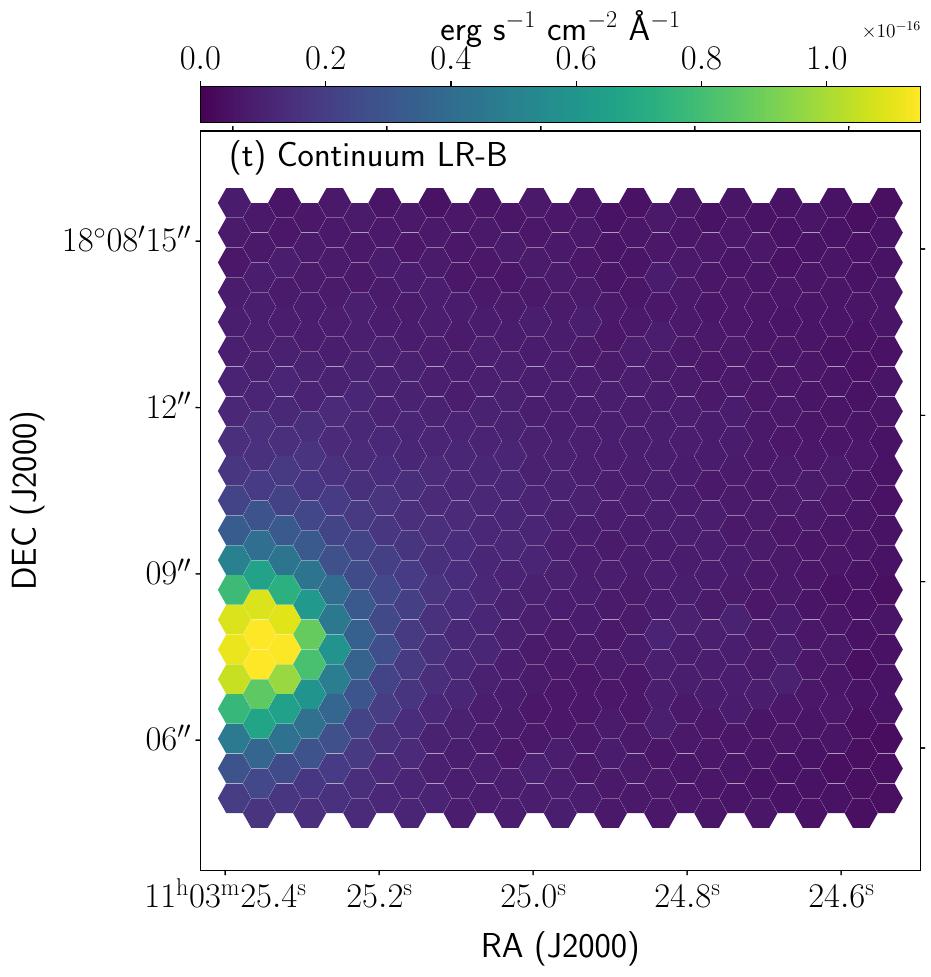}
	\includegraphics[clip, width=0.24\linewidth]{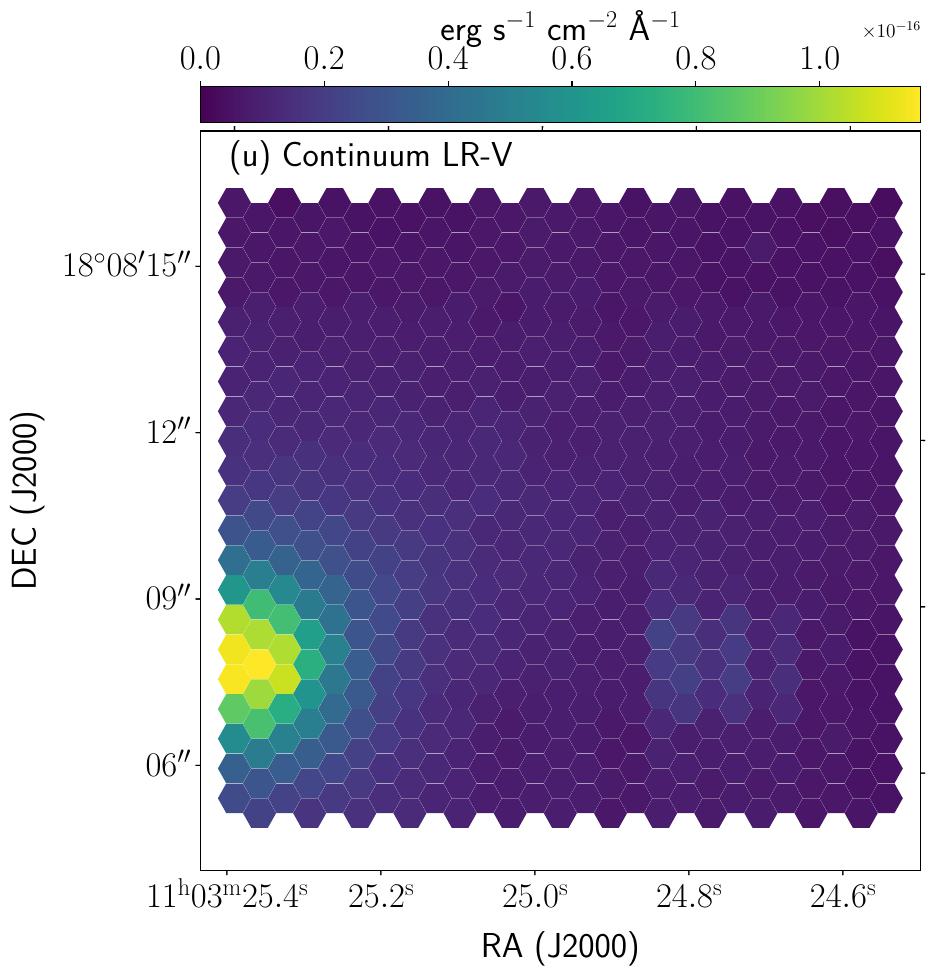}
	\includegraphics[clip, width=0.24\linewidth]{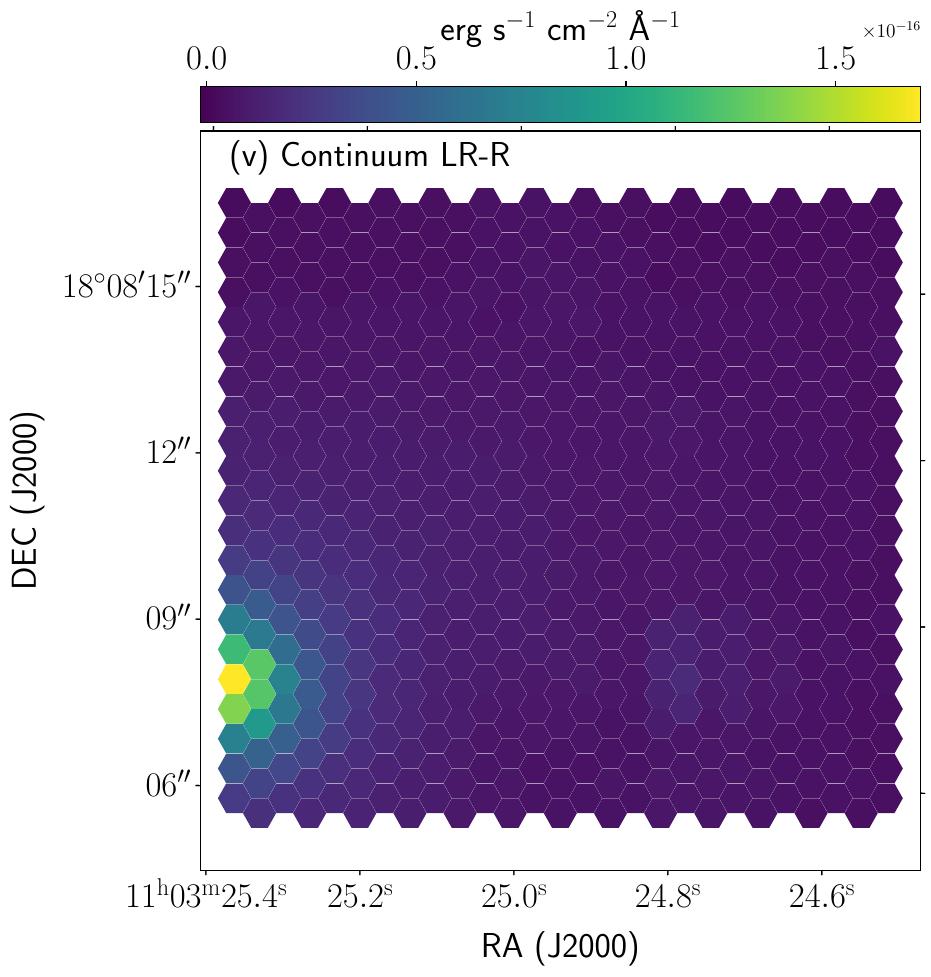}
	\includegraphics[clip, width=0.24\linewidth]{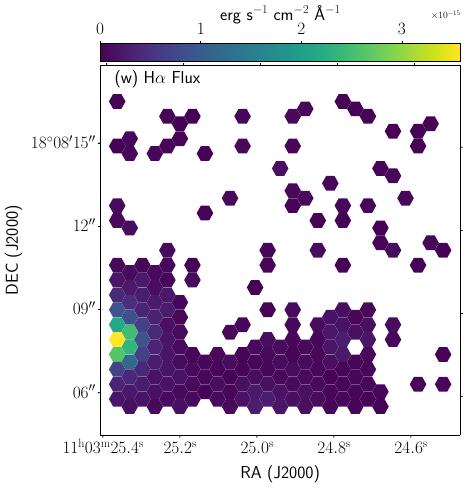}
	\includegraphics[clip, width=0.24\linewidth]{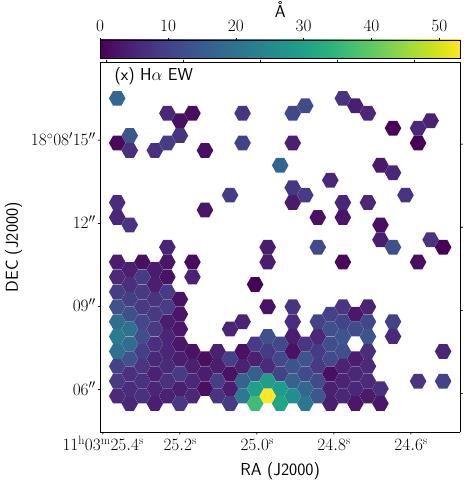}
	\includegraphics[clip, width=0.24\linewidth]{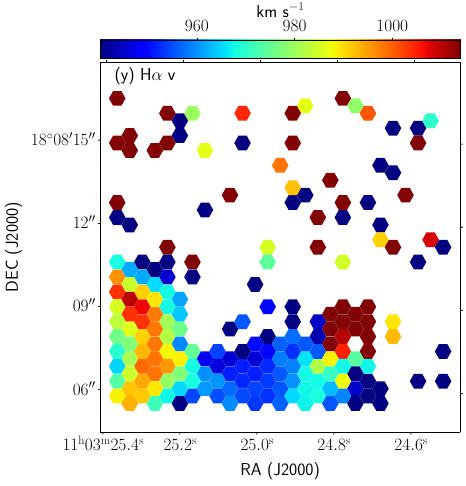}
	\includegraphics[clip, width=0.24\linewidth]{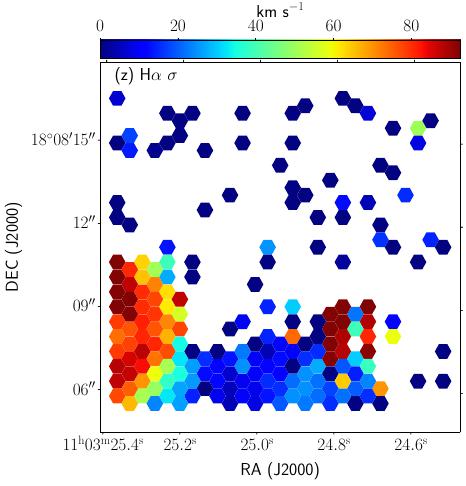}
	\includegraphics[clip, width=0.24\linewidth]{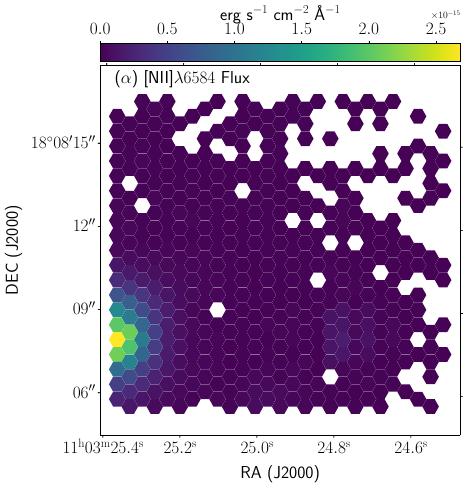}
	\includegraphics[clip, width=0.24\linewidth]{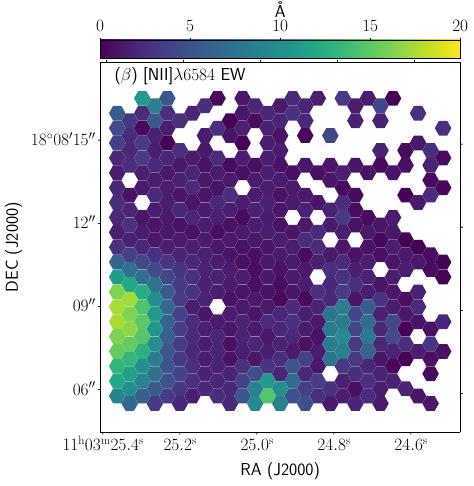}
	\includegraphics[clip, width=0.24\linewidth]{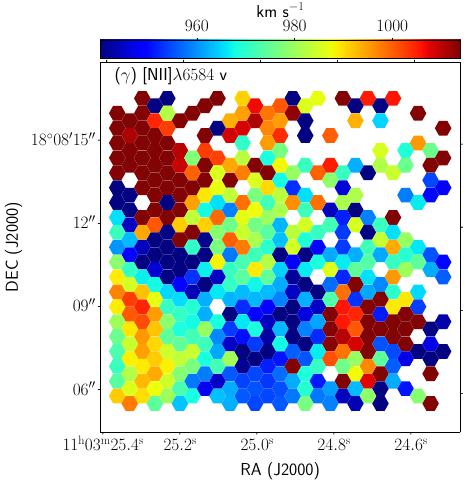}
	\includegraphics[clip, width=0.24\linewidth]{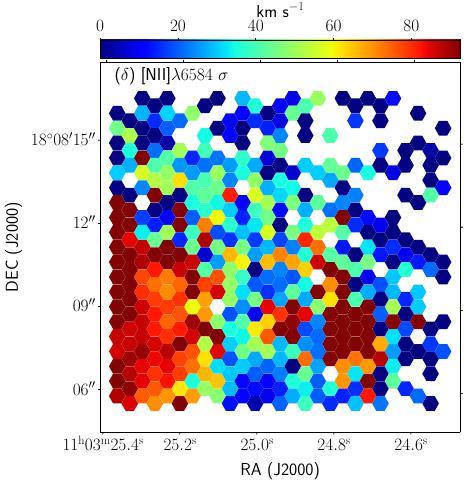}
	\includegraphics[clip, width=0.24\linewidth]{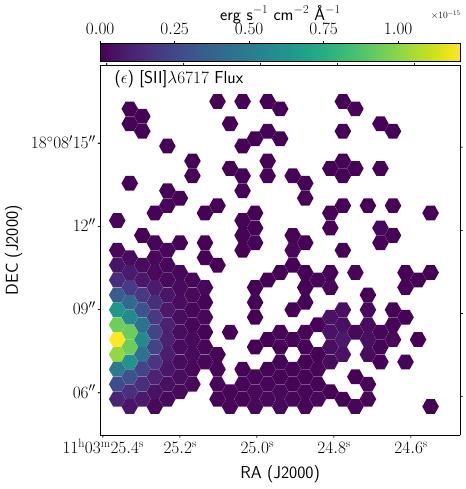}
	\includegraphics[clip, width=0.24\linewidth]{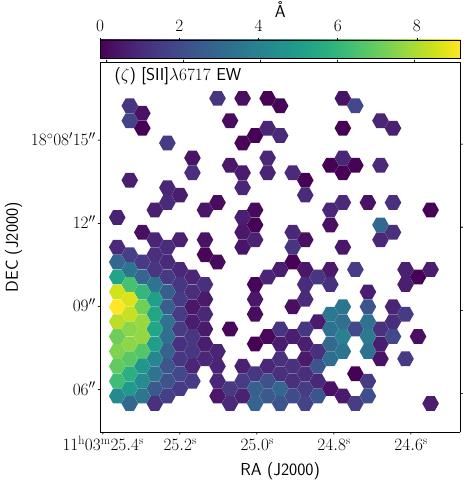}
	\includegraphics[clip, width=0.24\linewidth]{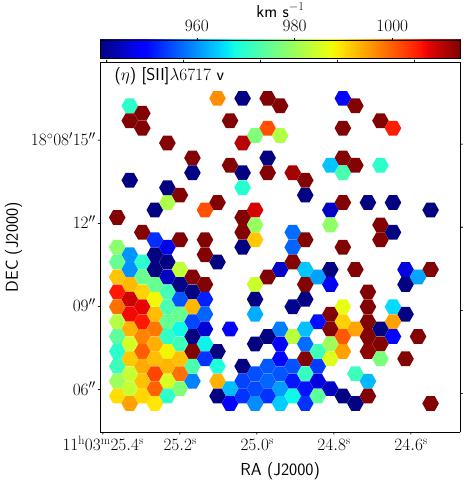}
	\includegraphics[clip, width=0.24\linewidth]{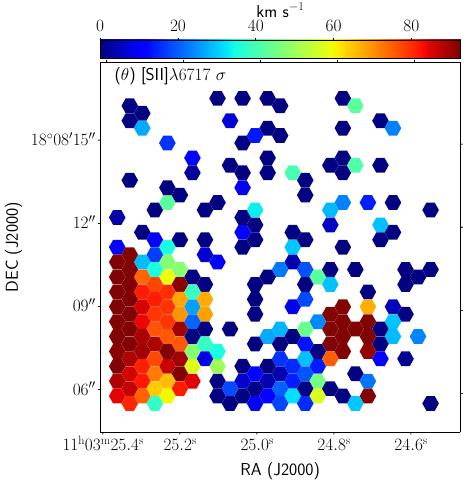}
	\includegraphics[clip, width=0.24\linewidth]{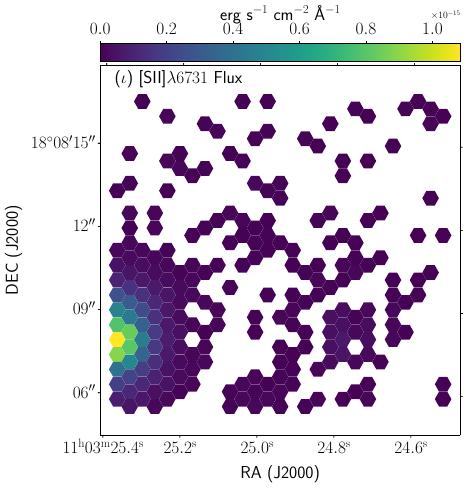}
	\includegraphics[clip, width=0.24\linewidth]{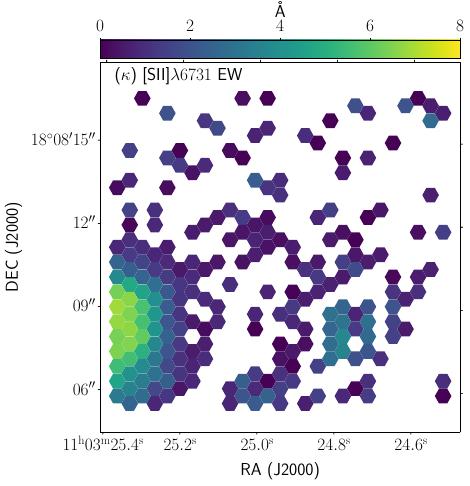}
	\includegraphics[clip, width=0.24\linewidth]{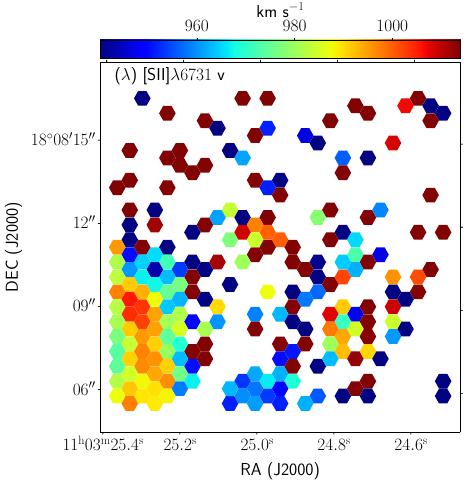}
	\includegraphics[clip, width=0.24\linewidth]{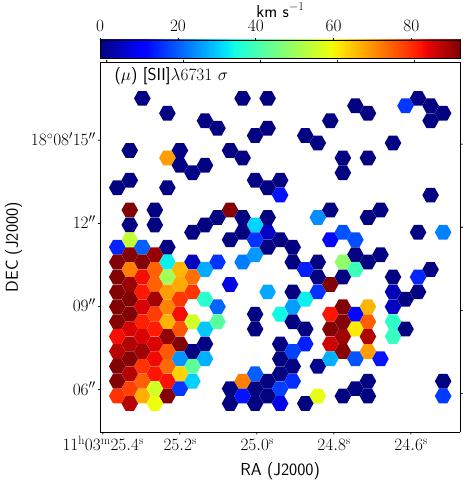}
	\caption{(cont.) NGC~3507 card.}
	\label{fig:NGC3507_card_2}
\end{figure*}
\clearpage
\begin{figure*}[h]
	\centering
	\includegraphics[clip, width=0.35\linewidth]{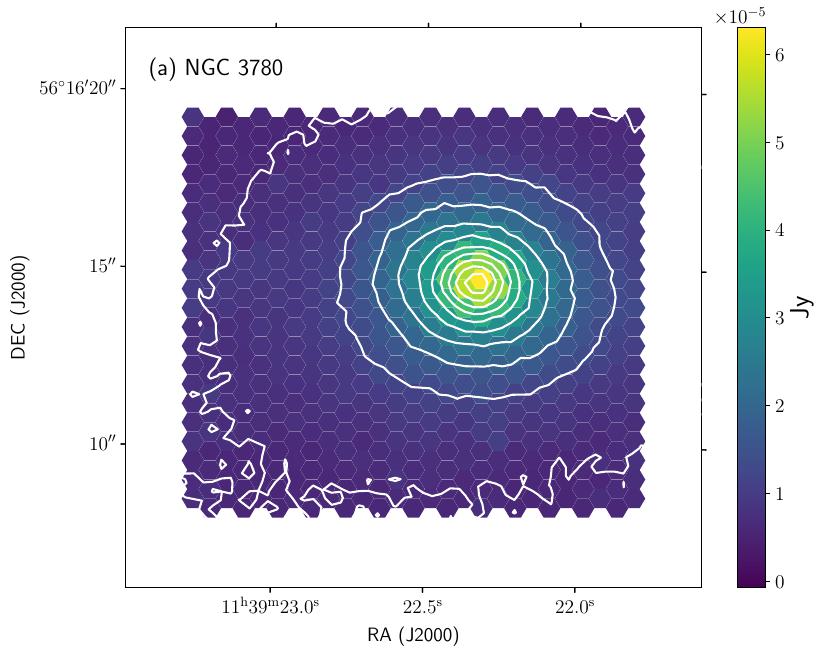}
	\includegraphics[clip, width=0.6\linewidth]{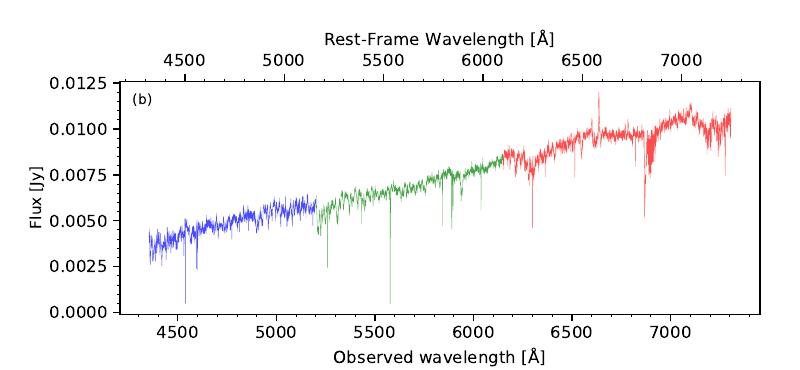}
	\includegraphics[clip, width=0.24\linewidth]{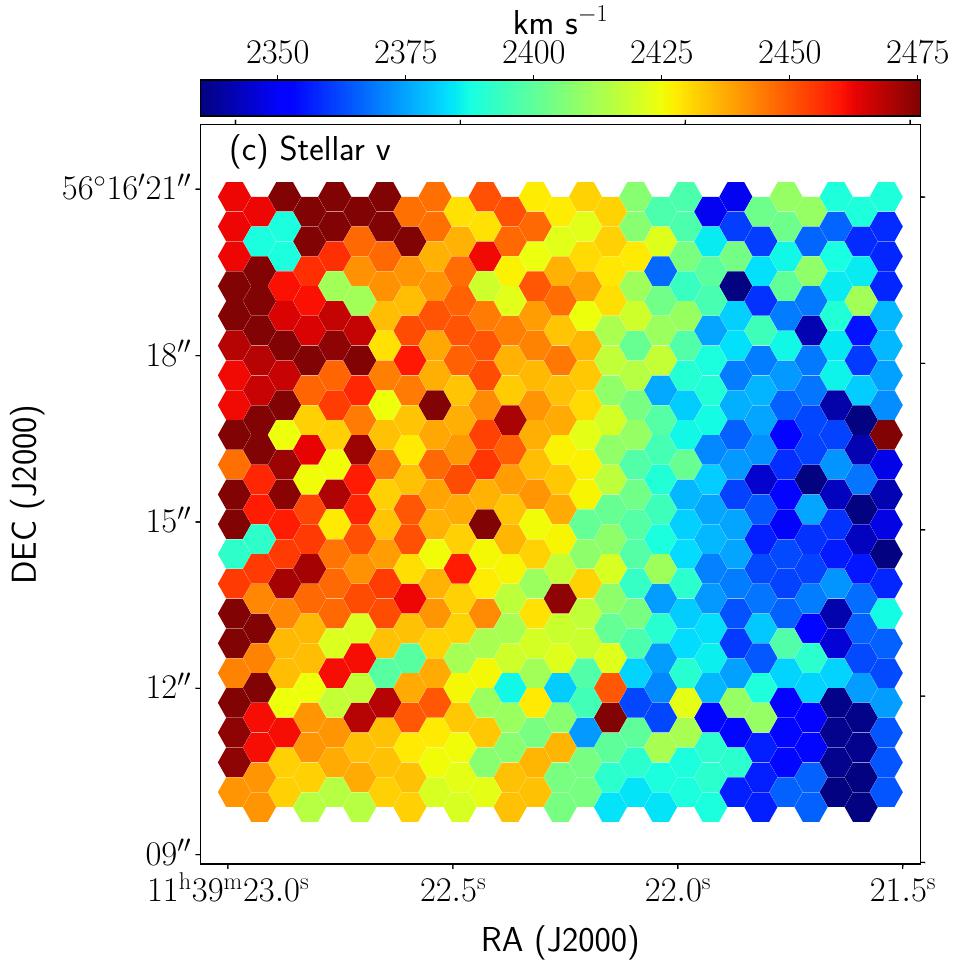}
	\includegraphics[clip, width=0.24\linewidth]{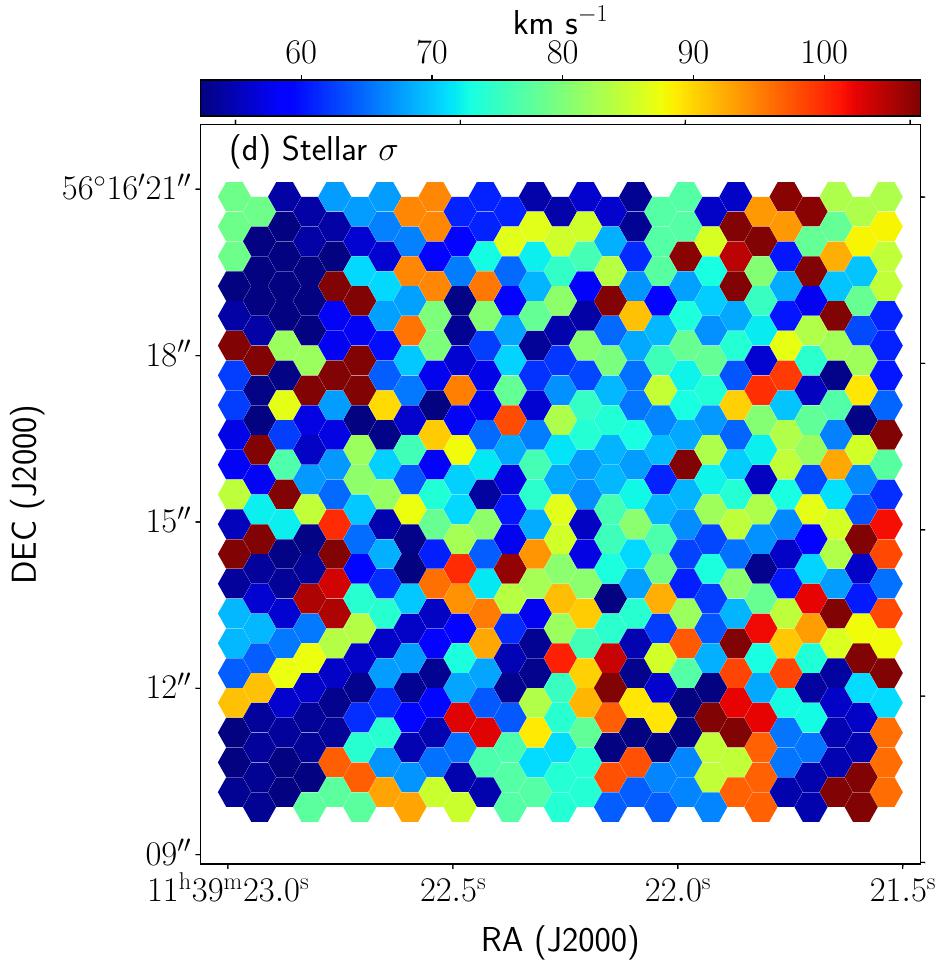}
	\includegraphics[clip, width=0.24\linewidth]{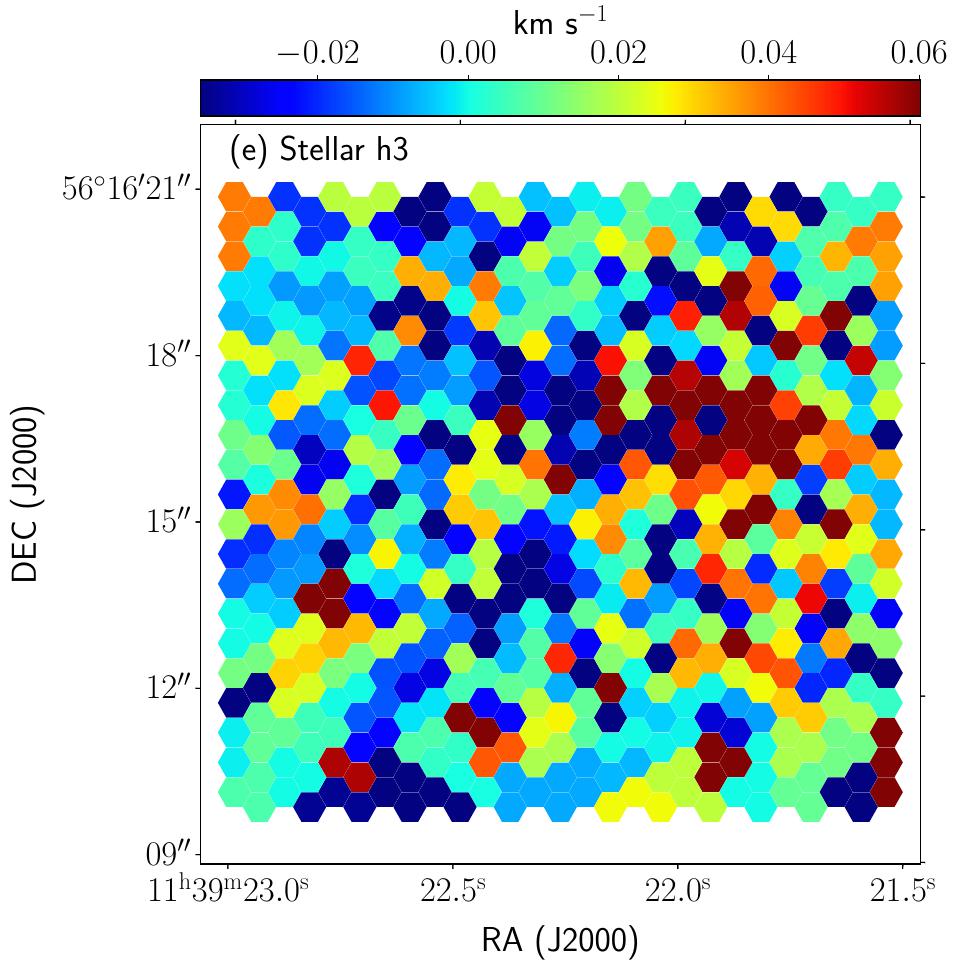}
	\includegraphics[clip, width=0.24\linewidth]{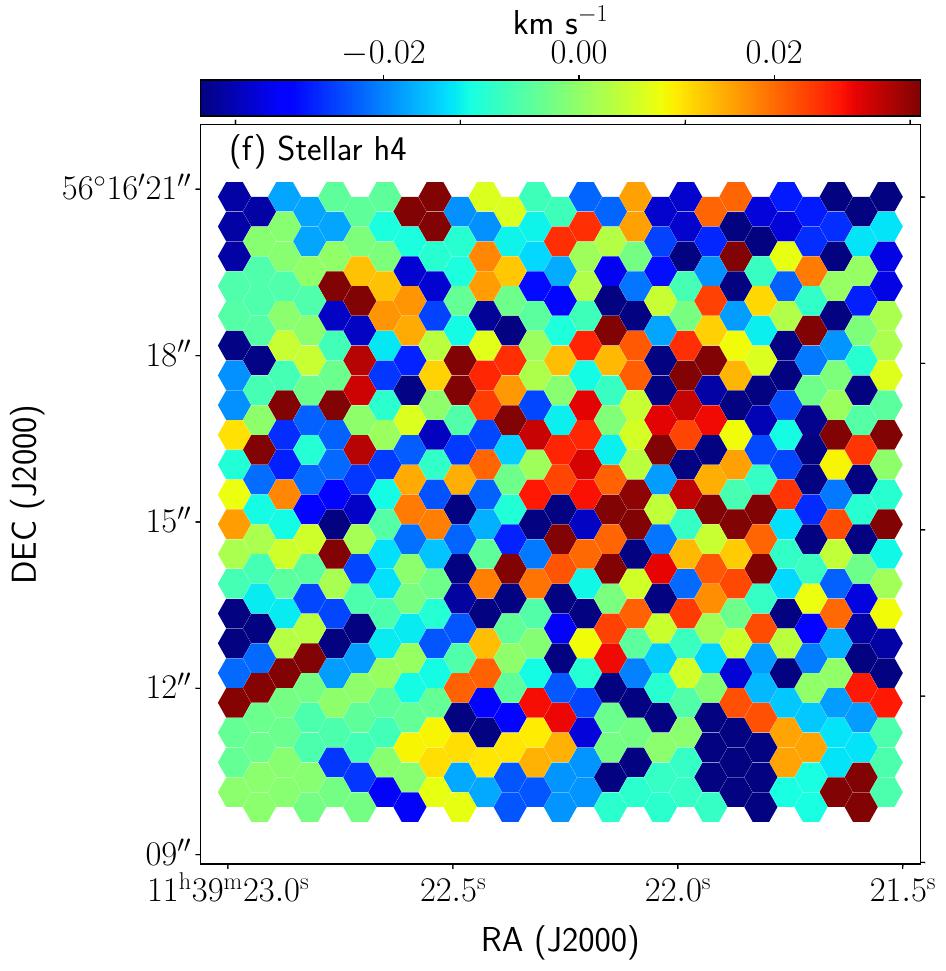}
	\includegraphics[clip, width=0.24\linewidth]{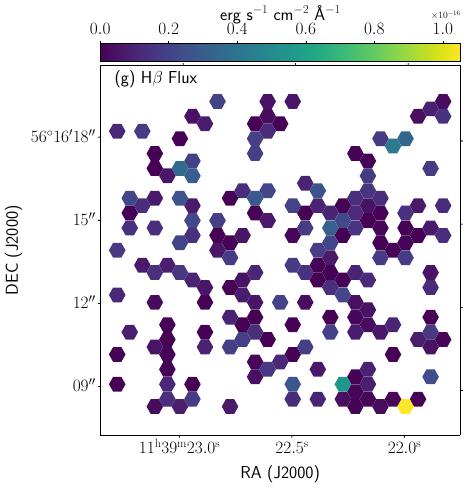}
	\includegraphics[clip, width=0.24\linewidth]{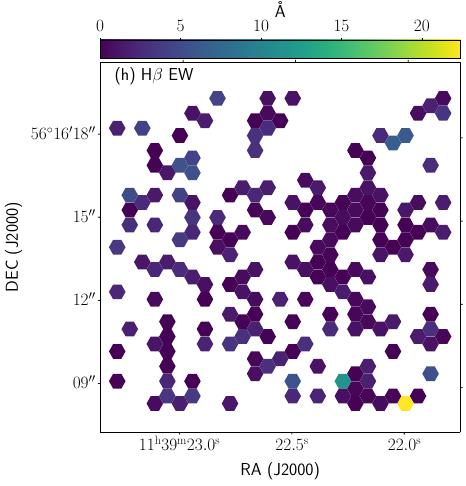}
	\includegraphics[clip, width=0.24\linewidth]{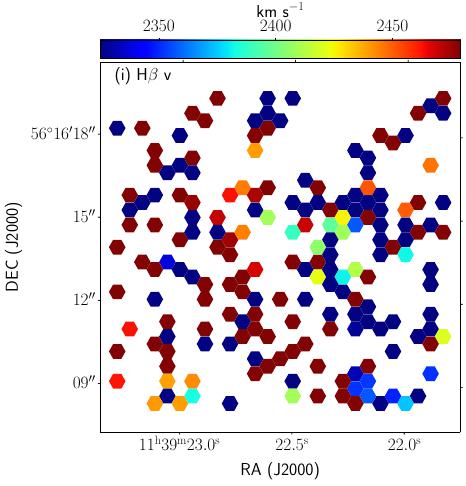}
	\includegraphics[clip, width=0.24\linewidth]{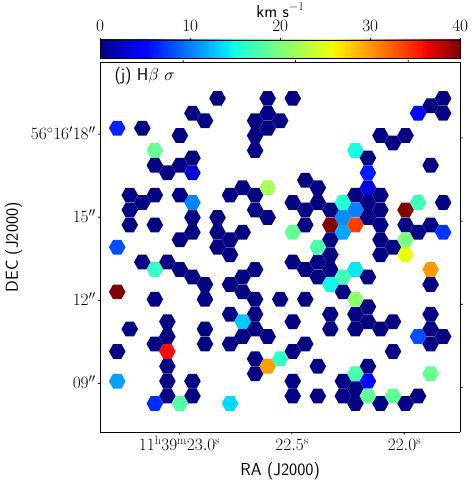}
	\includegraphics[clip, width=0.24\linewidth]{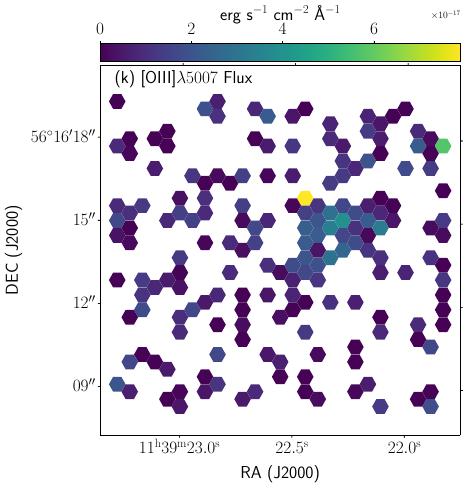}
	\includegraphics[clip, width=0.24\linewidth]{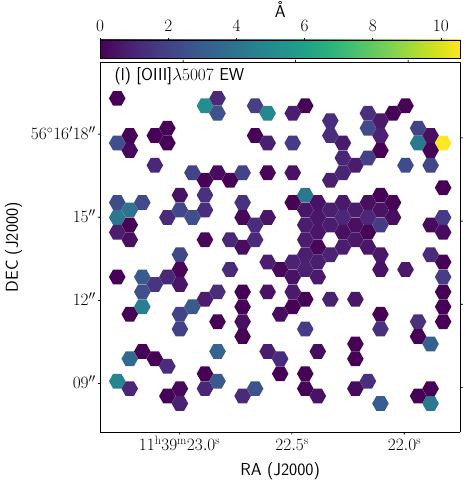}
	\includegraphics[clip, width=0.24\linewidth]{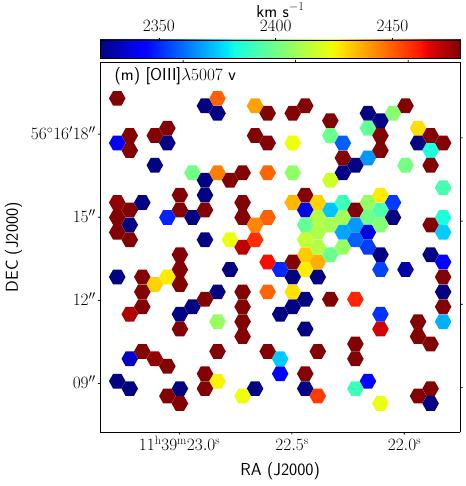}
	\includegraphics[clip, width=0.24\linewidth]{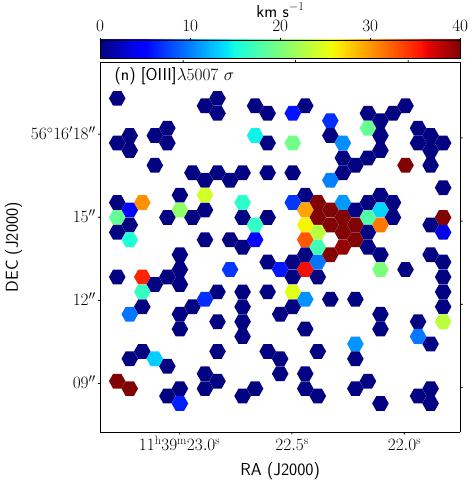}
	\vspace{5cm}
	\caption{NGC~3780 card.}
	\label{fig:NGC3780_card_1}
\end{figure*}
\addtocounter{figure}{-1}
\begin{figure*}[h]
	\centering
	\includegraphics[clip, width=0.24\linewidth]{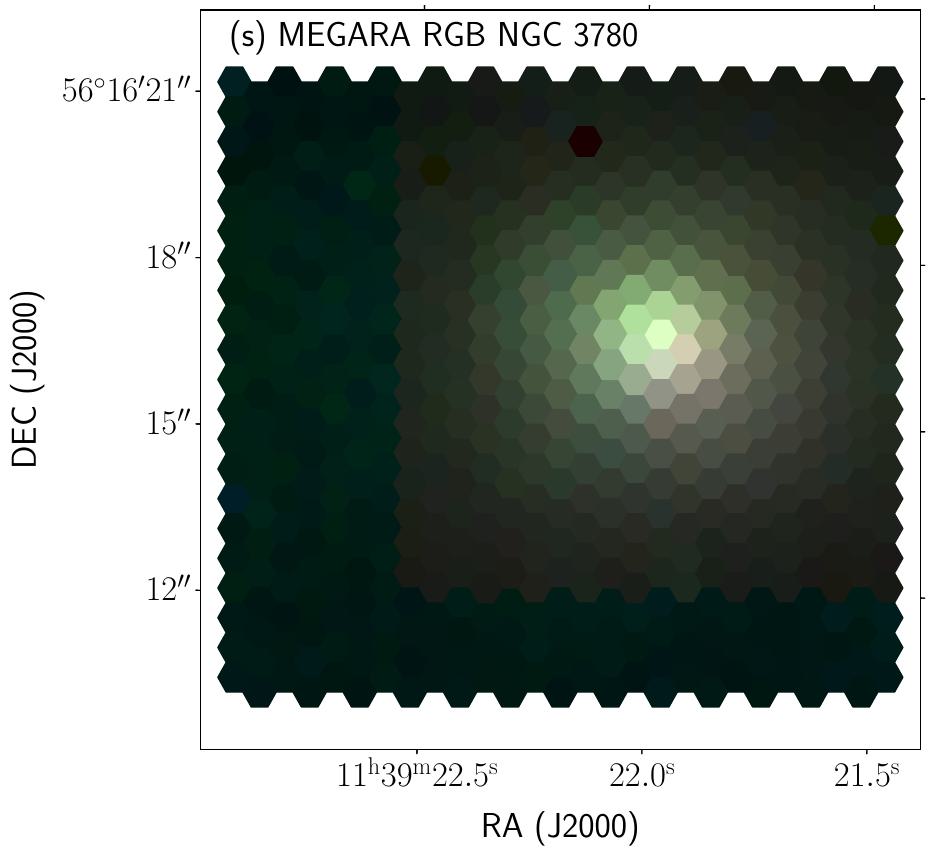}
	\includegraphics[clip, width=0.24\linewidth]{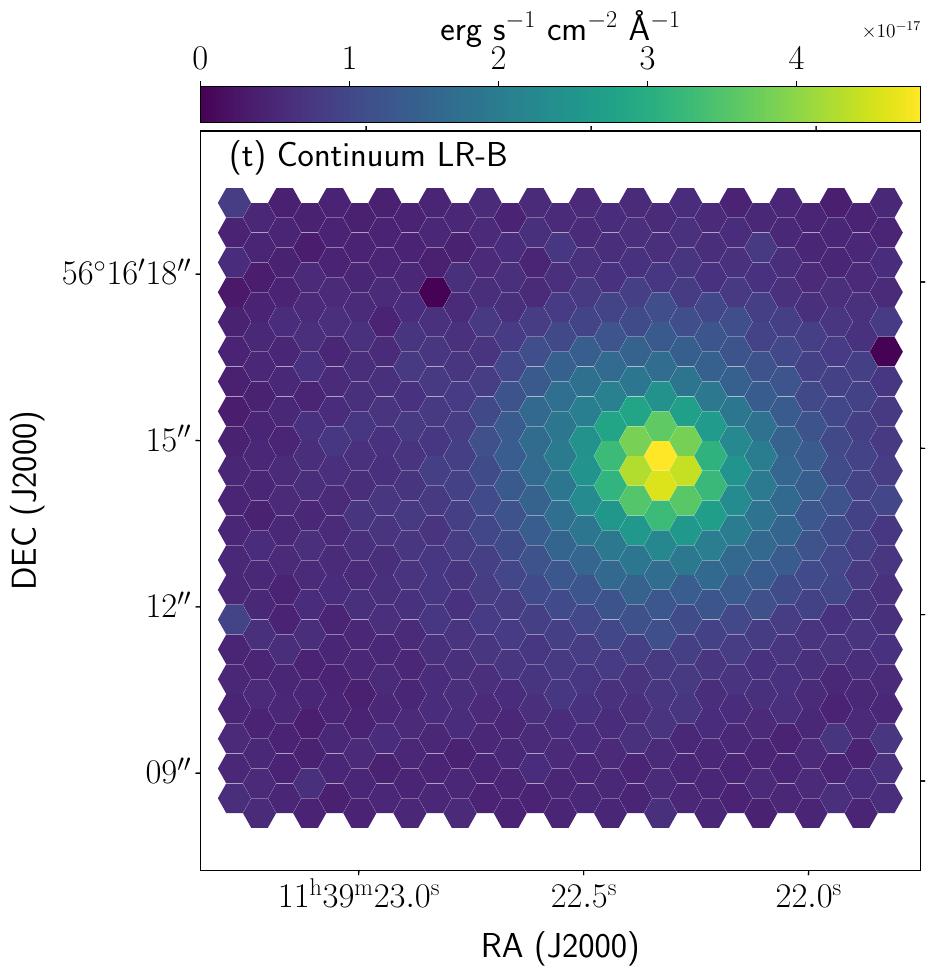}
	\includegraphics[clip, width=0.24\linewidth]{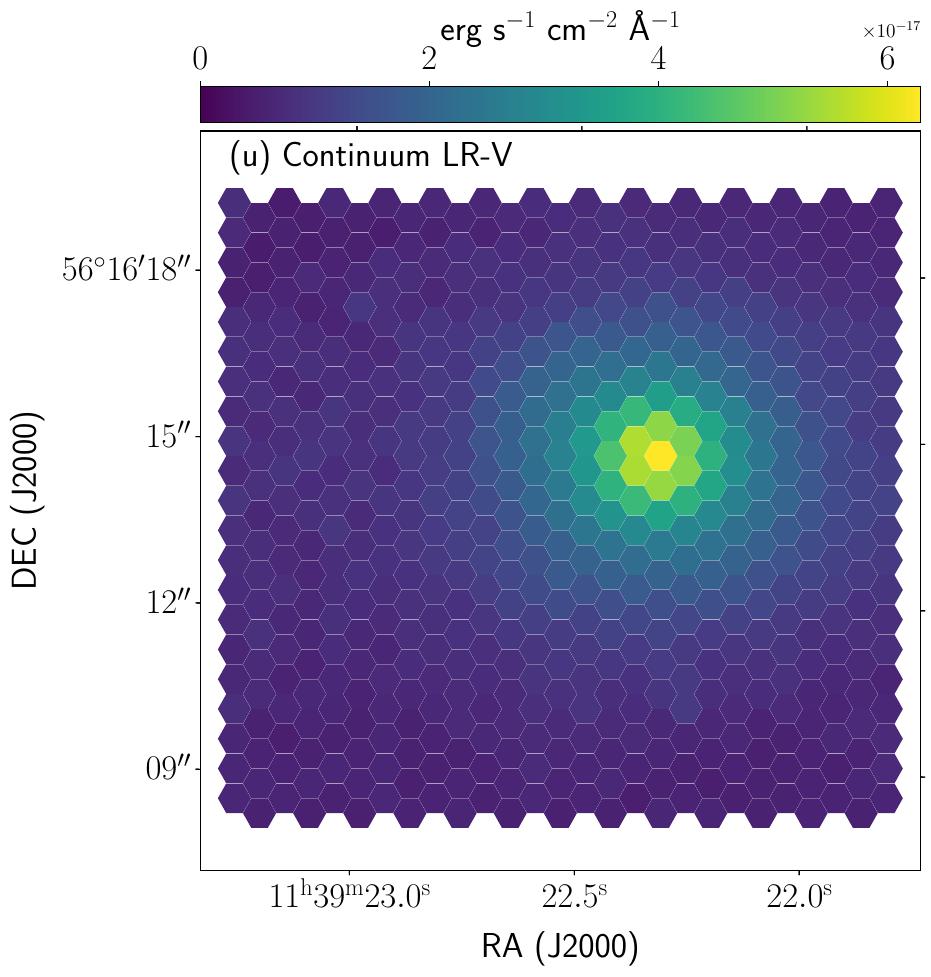}
	\includegraphics[clip, width=0.24\linewidth]{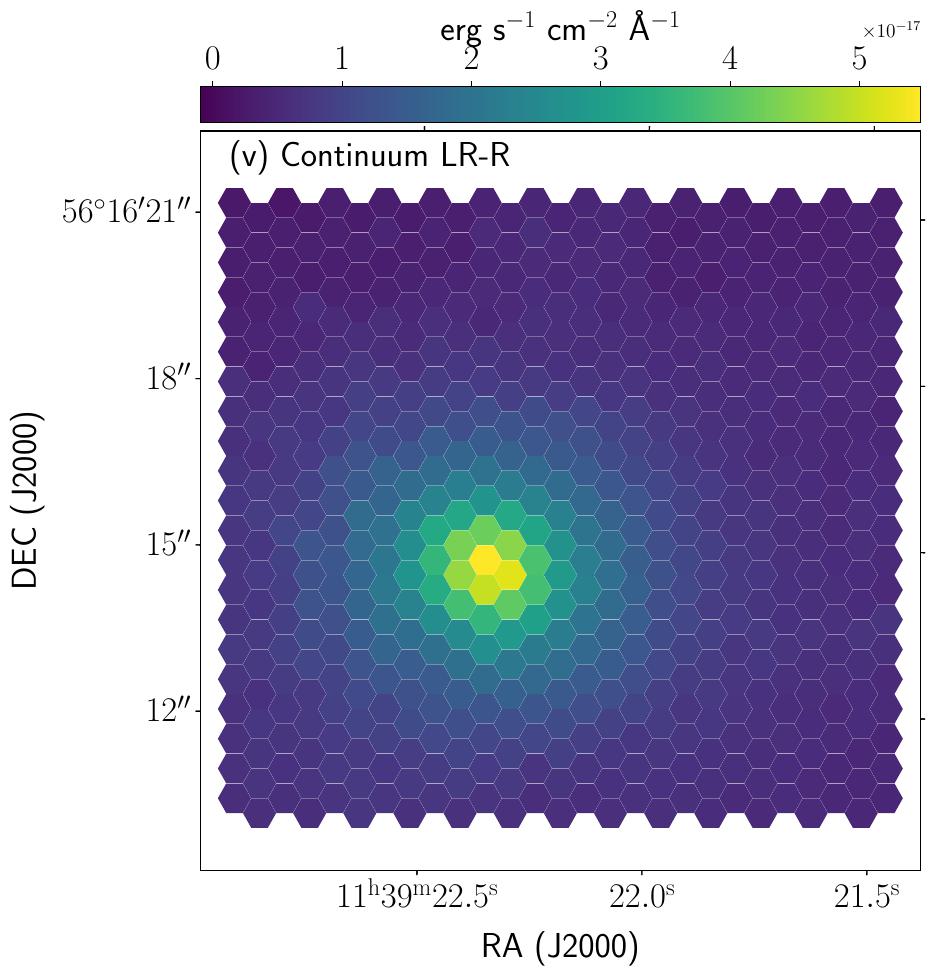}
	\includegraphics[clip, width=0.24\linewidth]{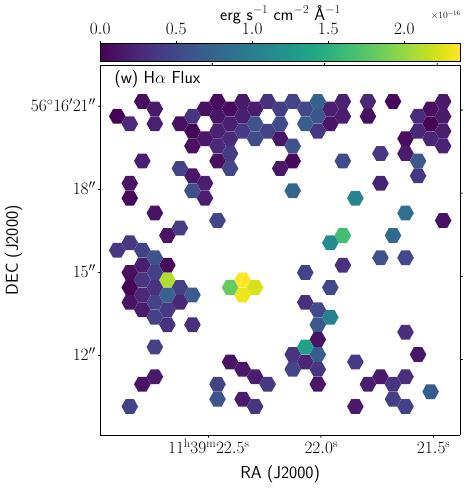}
	\includegraphics[clip, width=0.24\linewidth]{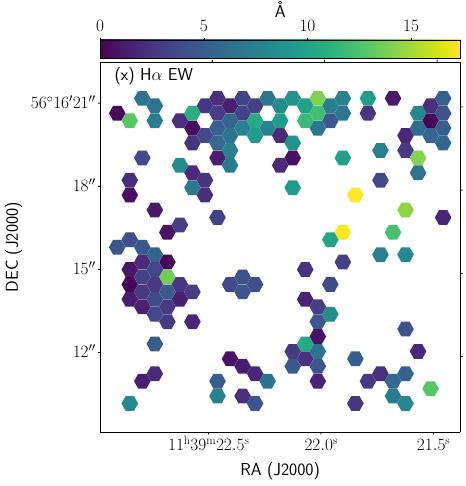}
	\includegraphics[clip, width=0.24\linewidth]{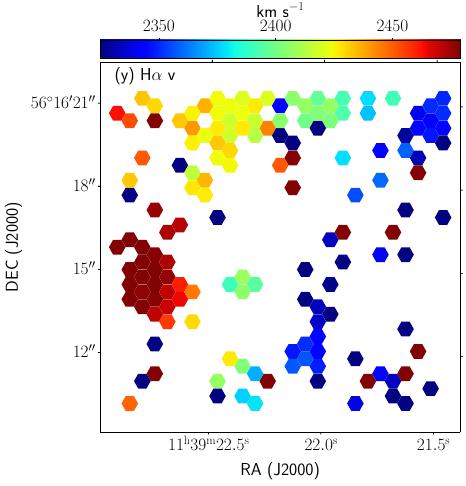}
	\includegraphics[clip, width=0.24\linewidth]{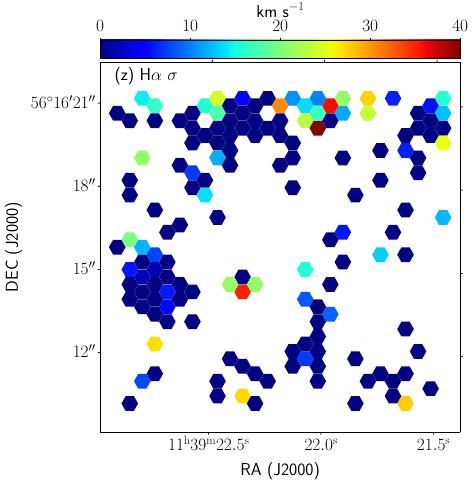}
	\includegraphics[clip, width=0.24\linewidth]{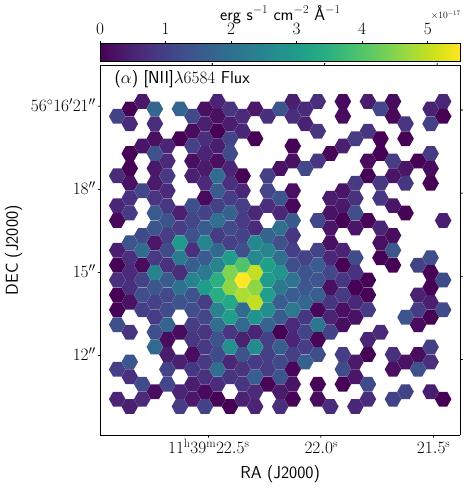}
	\includegraphics[clip, width=0.24\linewidth]{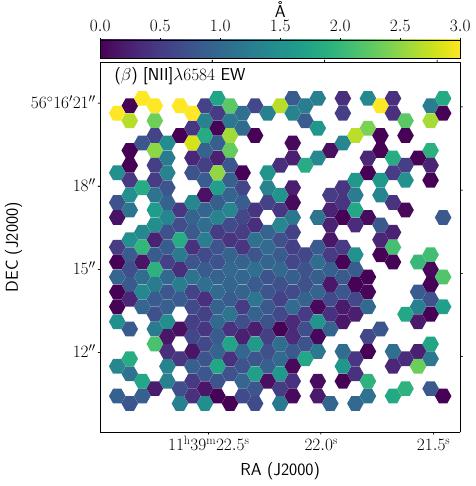}
	\includegraphics[clip, width=0.24\linewidth]{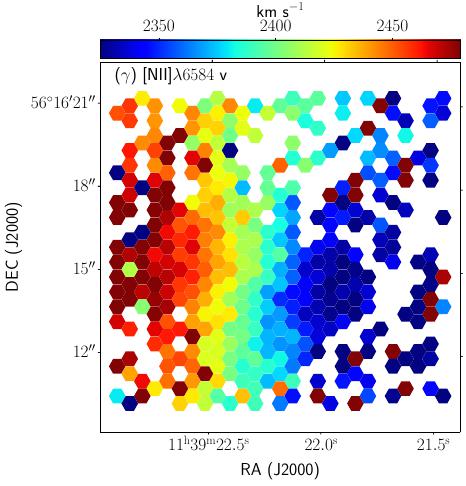}
	\includegraphics[clip, width=0.24\linewidth]{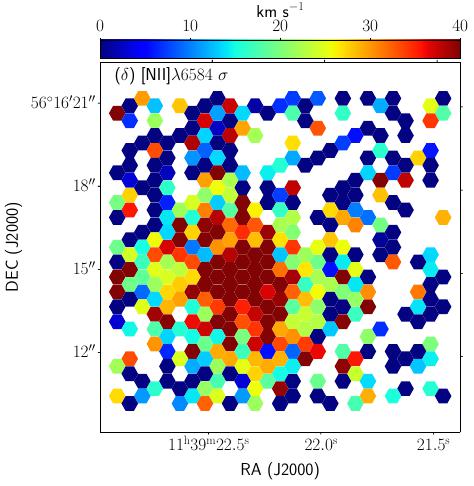}
	\includegraphics[clip, width=0.24\linewidth]{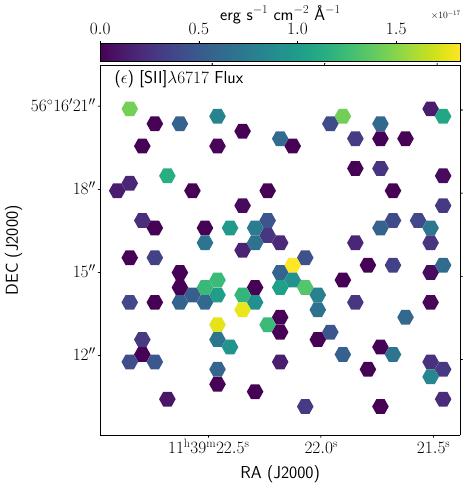}
	\includegraphics[clip, width=0.24\linewidth]{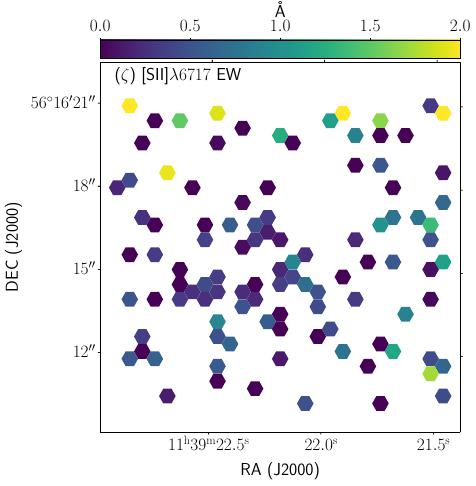}
	\includegraphics[clip, width=0.24\linewidth]{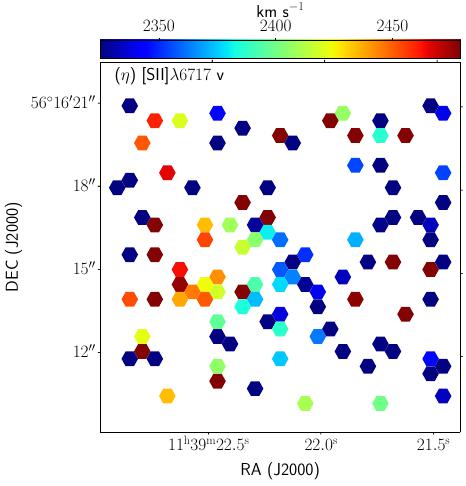}
	\includegraphics[clip, width=0.24\linewidth]{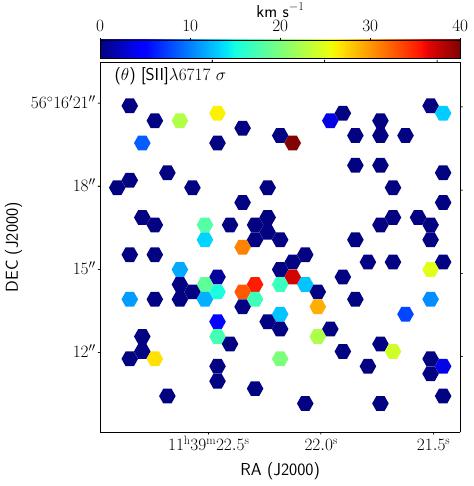}
	\includegraphics[clip, width=0.24\linewidth]{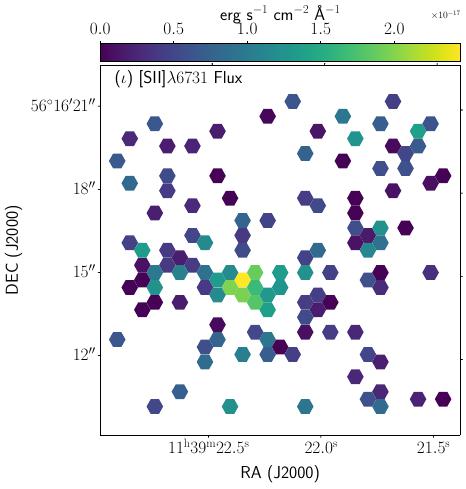}
	\includegraphics[clip, width=0.24\linewidth]{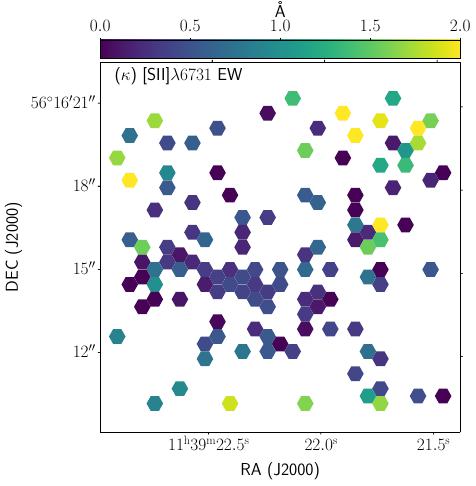}
	\includegraphics[clip, width=0.24\linewidth]{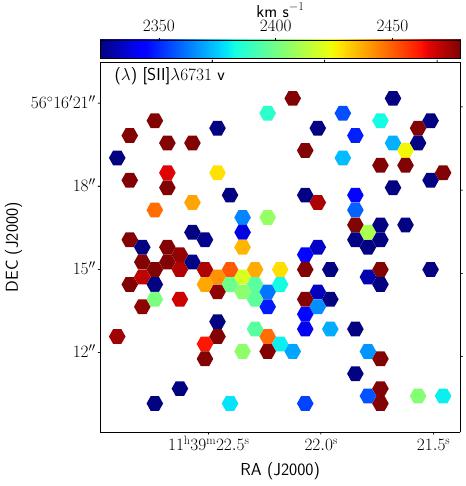}
	\includegraphics[clip, width=0.24\linewidth]{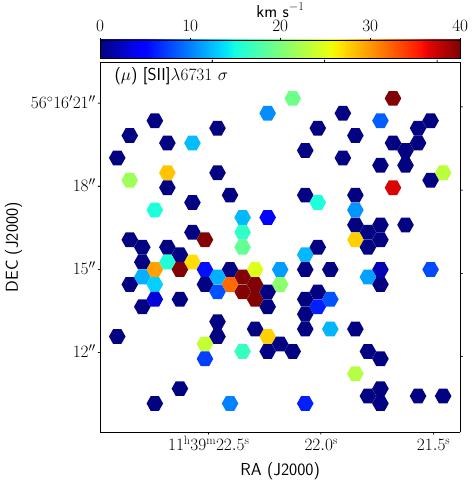}
	\caption{(cont.) NGC~3780 card.}
	\label{fig:NGC3780_card_2}
\end{figure*}

\begin{figure*}[h]
	\centering
	\includegraphics[clip, width=0.35\linewidth]{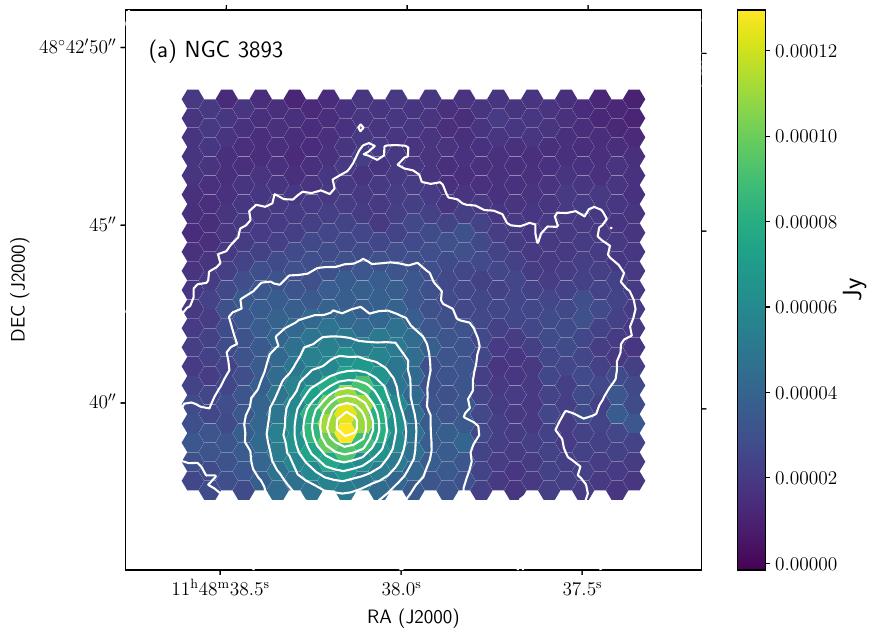}
	\includegraphics[clip, width=0.6\linewidth]{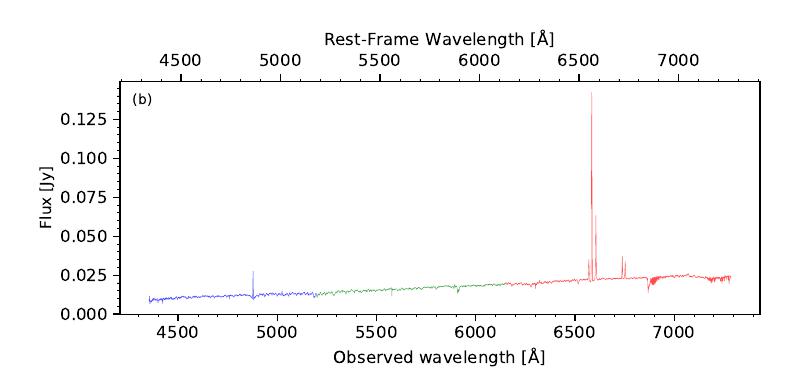}
	\includegraphics[clip, width=0.24\linewidth]{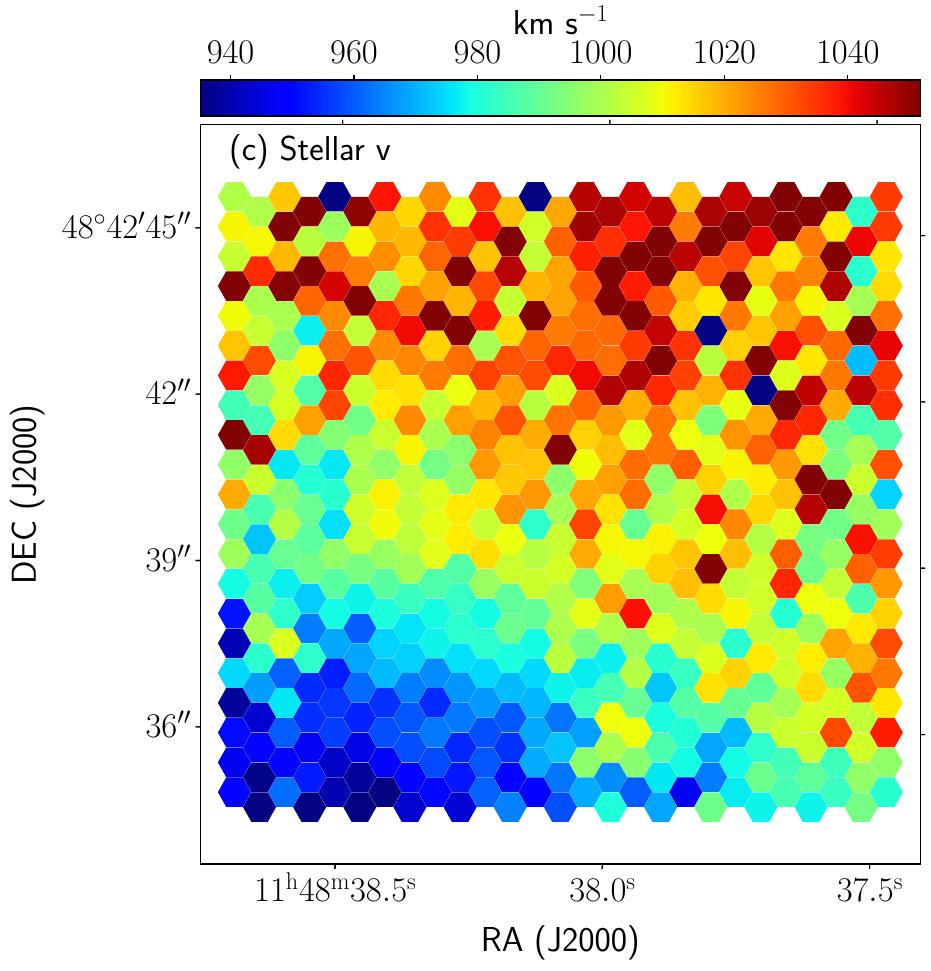}
	\includegraphics[clip, width=0.24\linewidth]{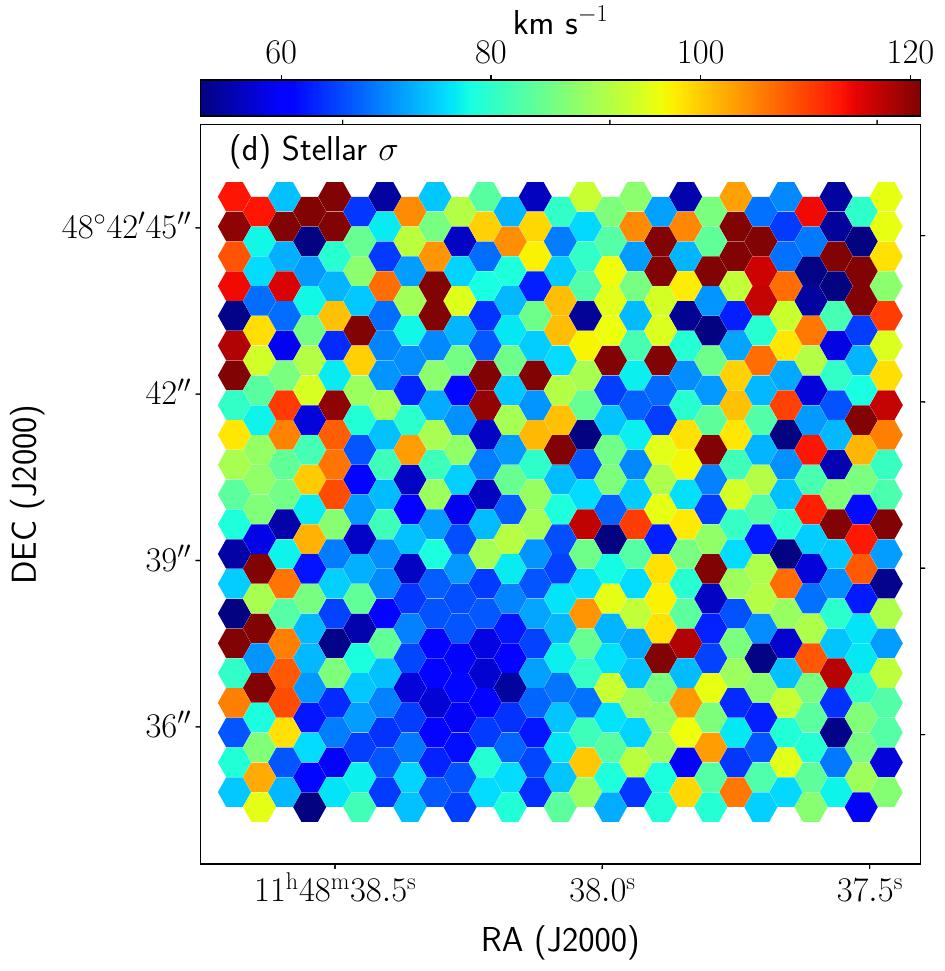}
	\includegraphics[clip, width=0.24\linewidth]{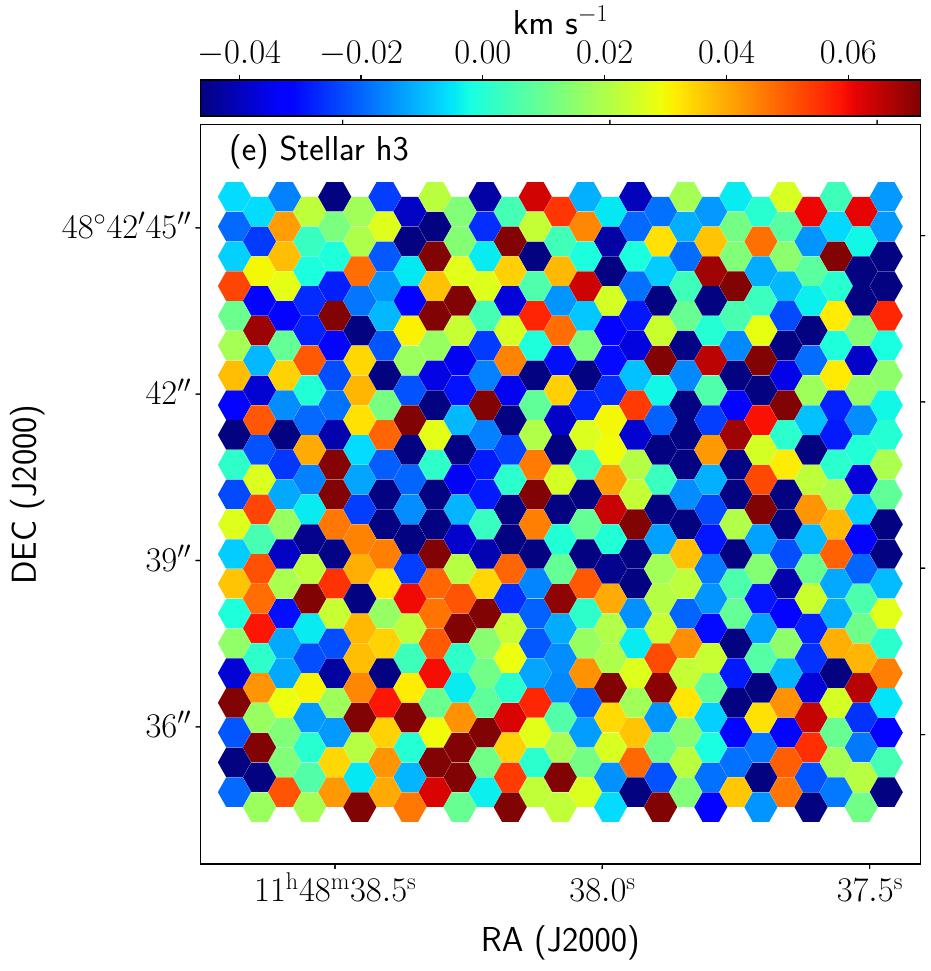}
	\includegraphics[clip, width=0.24\linewidth]{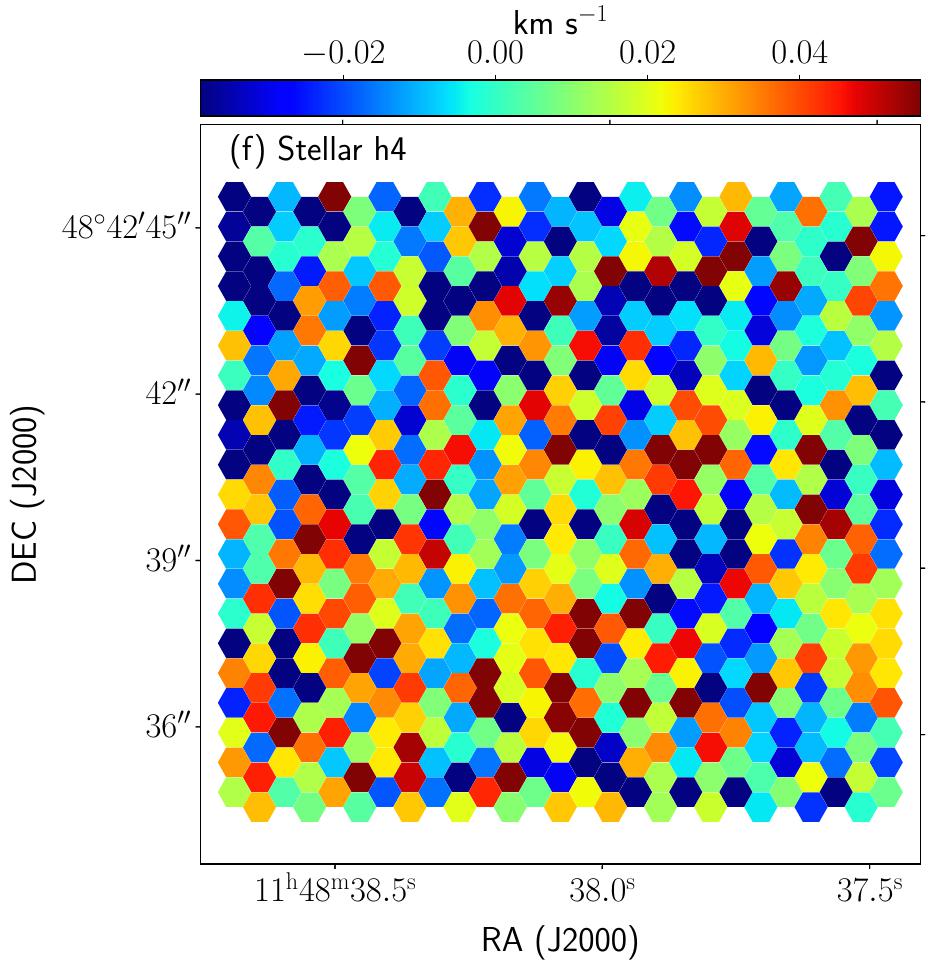}
	\includegraphics[clip, width=0.24\linewidth]{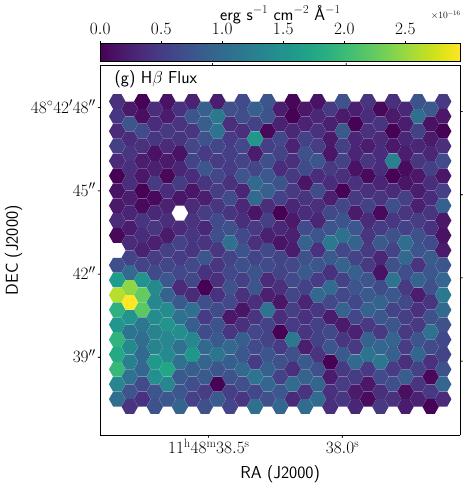}
	\includegraphics[clip, width=0.24\linewidth]{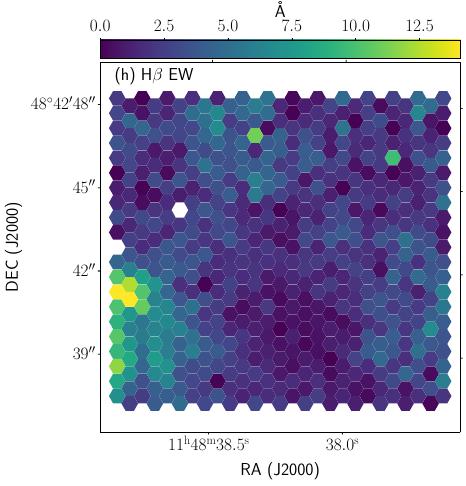}
	\includegraphics[clip, width=0.24\linewidth]{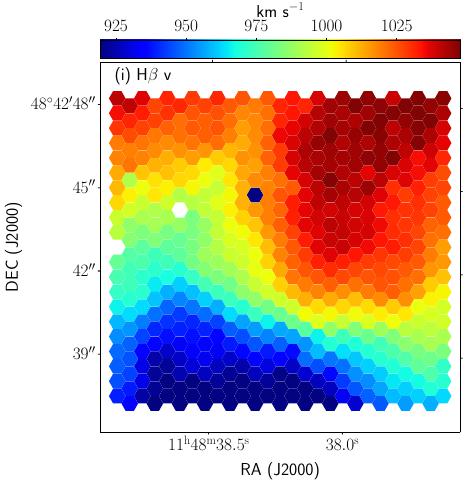}
	\includegraphics[clip, width=0.24\linewidth]{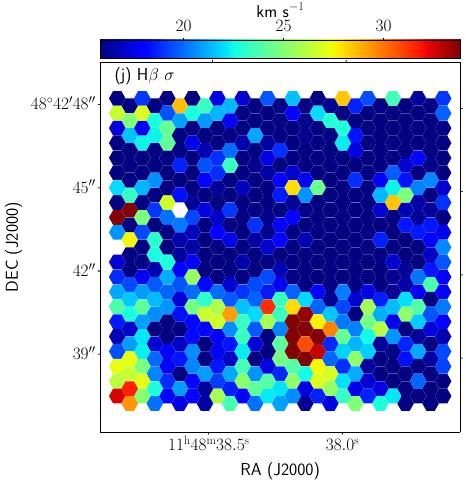}
	\includegraphics[clip, width=0.24\linewidth]{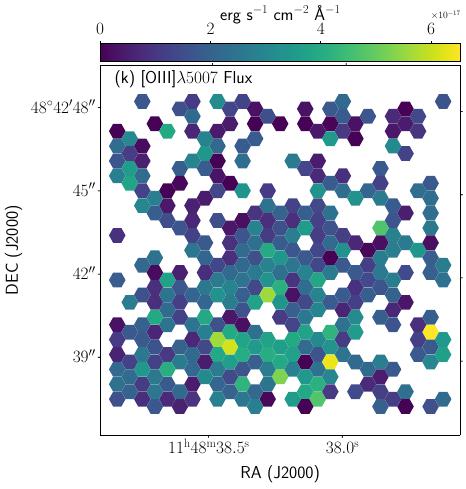}
	\includegraphics[clip, width=0.24\linewidth]{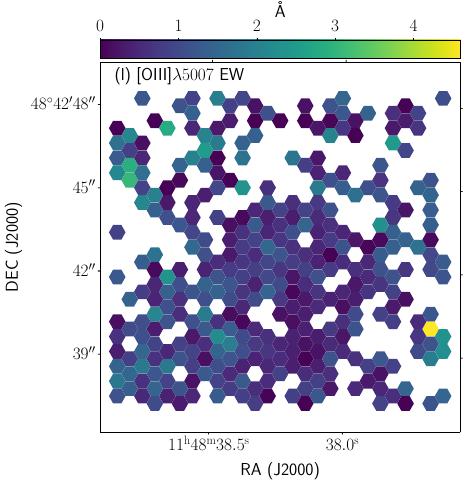}
	\includegraphics[clip, width=0.24\linewidth]{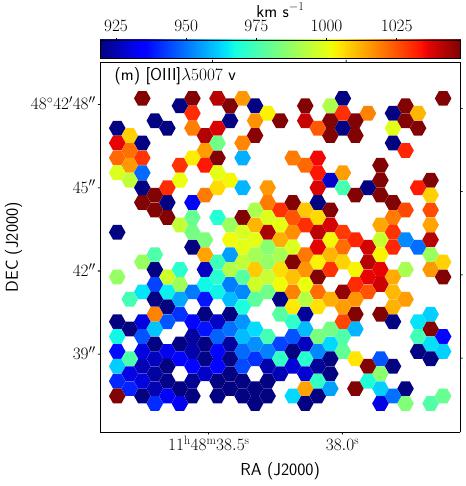}
	\includegraphics[clip, width=0.24\linewidth]{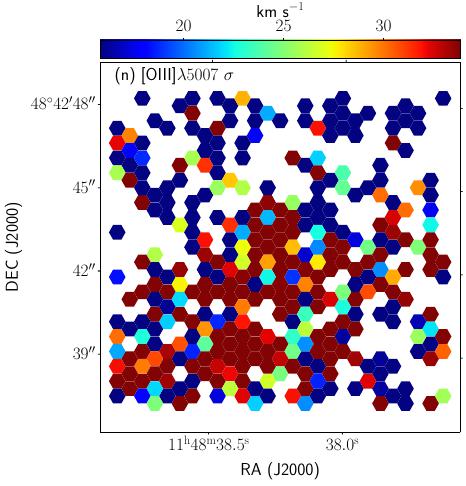}
	\includegraphics[clip, width=0.24\linewidth]{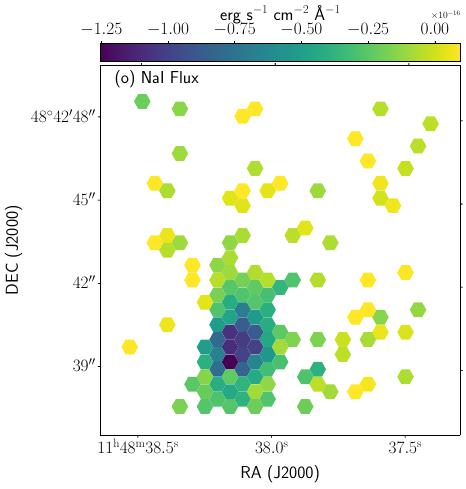}
	\includegraphics[clip, width=0.24\linewidth]{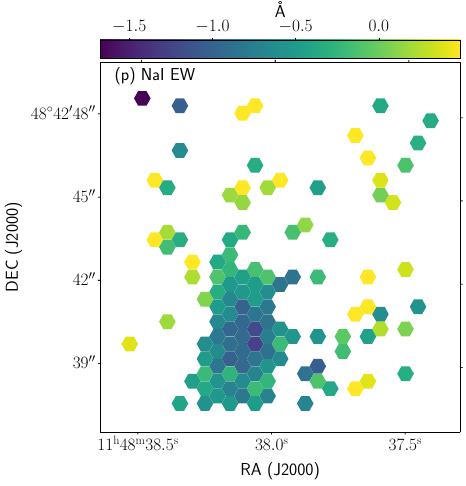}
	\includegraphics[clip, width=0.24\linewidth]{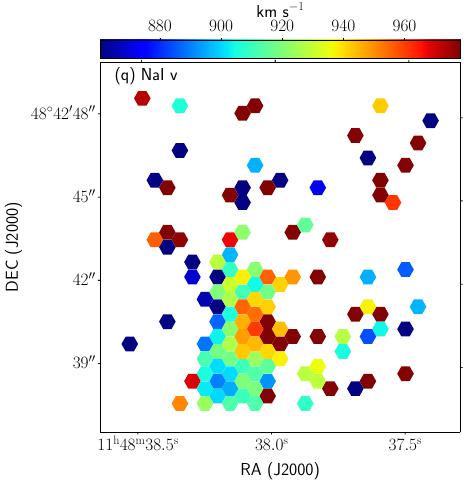}
	\includegraphics[clip, width=0.24\linewidth]{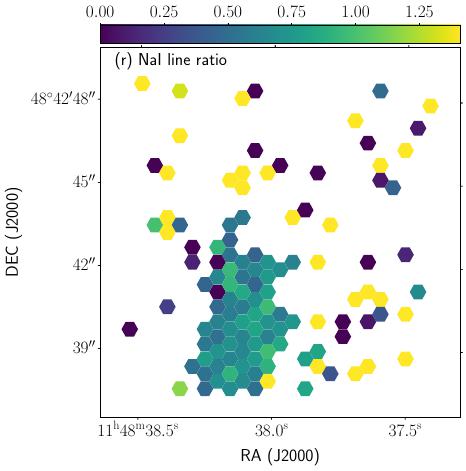}
	\caption{NGC~3893 card.}
	\label{fig:NGC3893_card_1}
\end{figure*}
\addtocounter{figure}{-1}
\begin{figure*}[h]
	\centering
	\includegraphics[clip, width=0.24\linewidth]{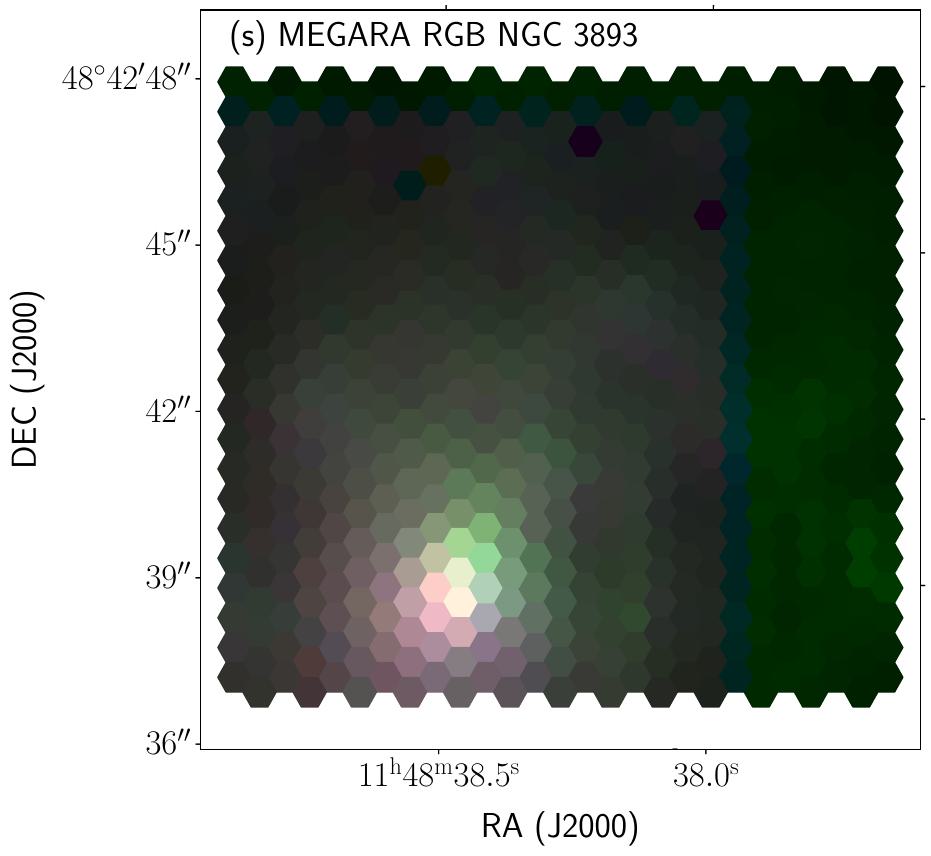}
	\includegraphics[clip, width=0.24\linewidth]{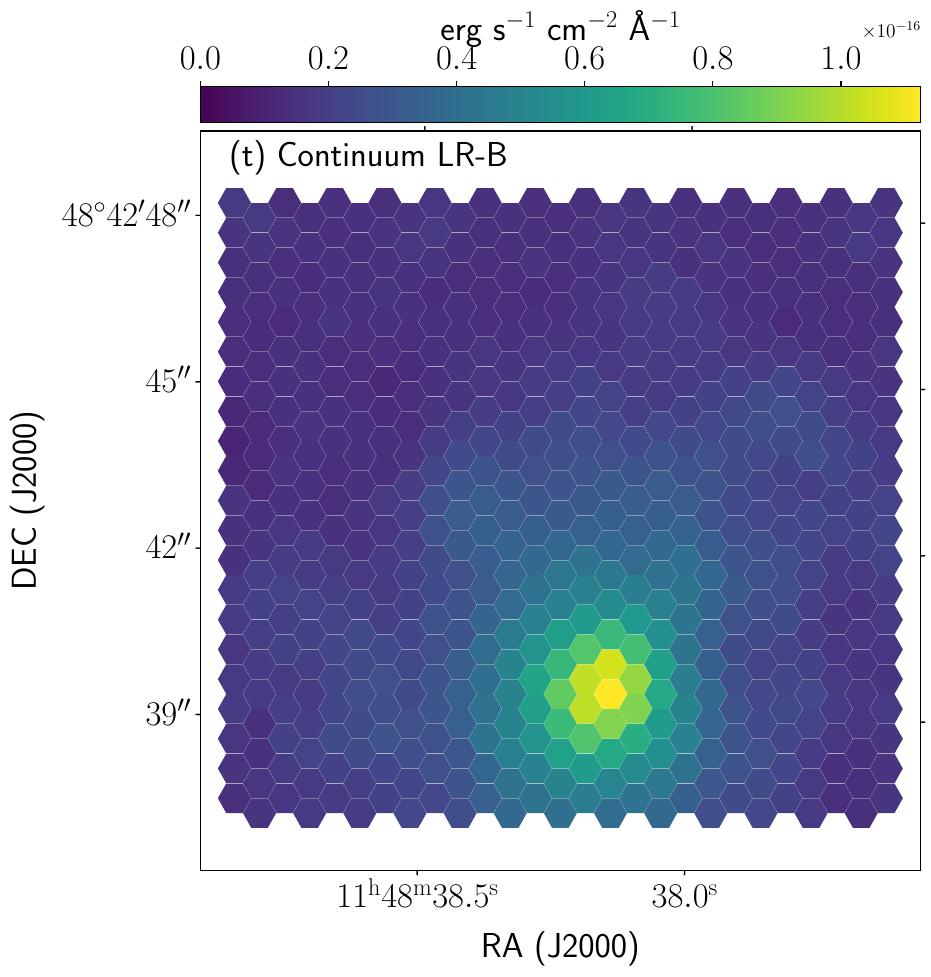}
	\includegraphics[clip, width=0.24\linewidth]{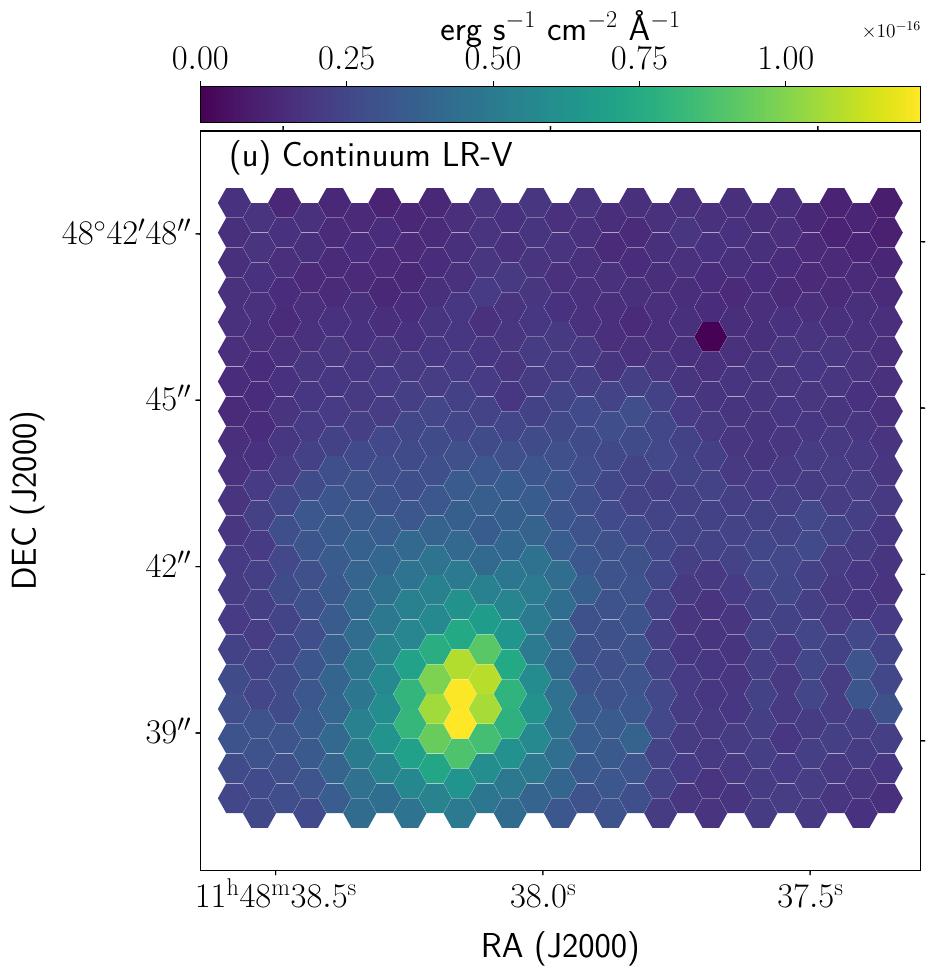}
	\includegraphics[clip, width=0.24\linewidth]{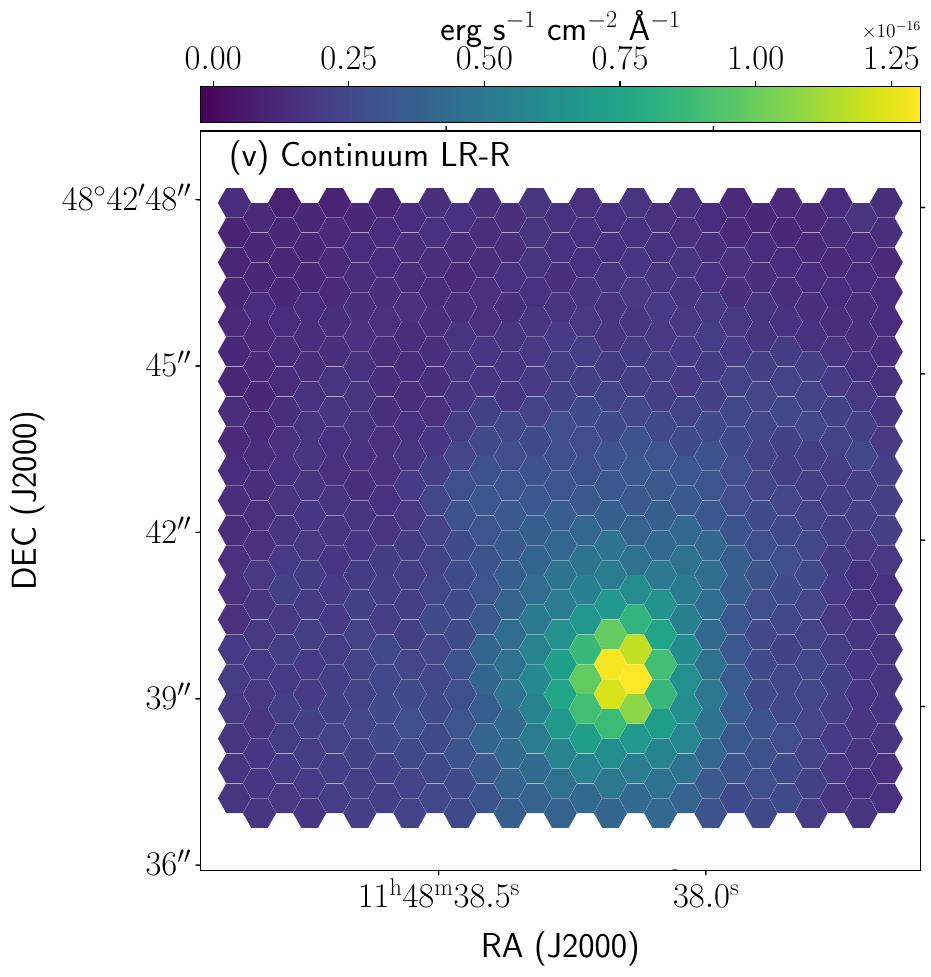}
	\includegraphics[clip, width=0.24\linewidth]{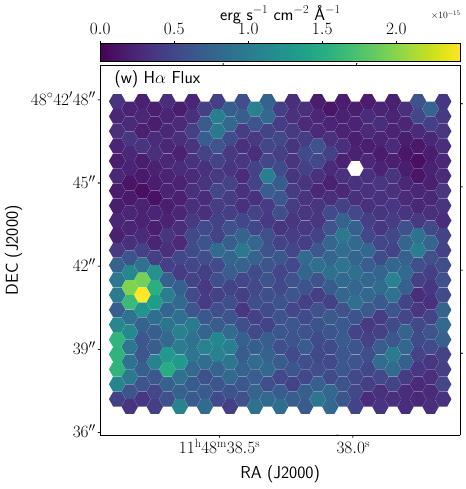}
	\includegraphics[clip, width=0.24\linewidth]{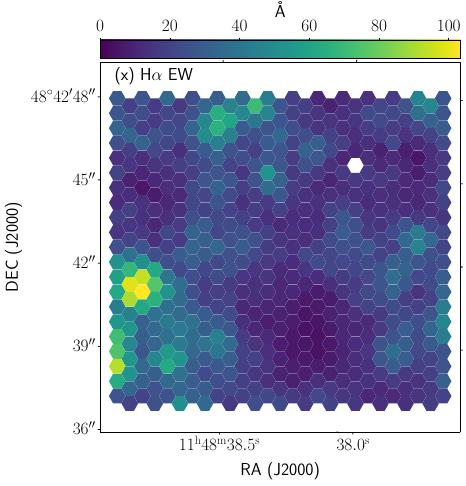}
	\includegraphics[clip, width=0.24\linewidth]{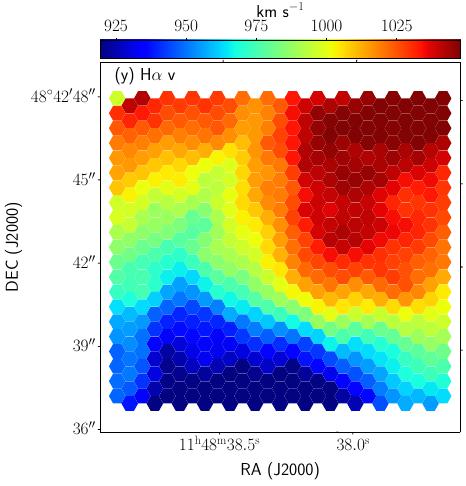}
	\includegraphics[clip, width=0.24\linewidth]{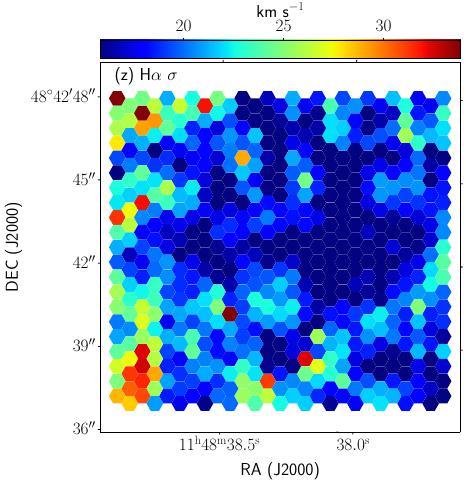}
	\includegraphics[clip, width=0.24\linewidth]{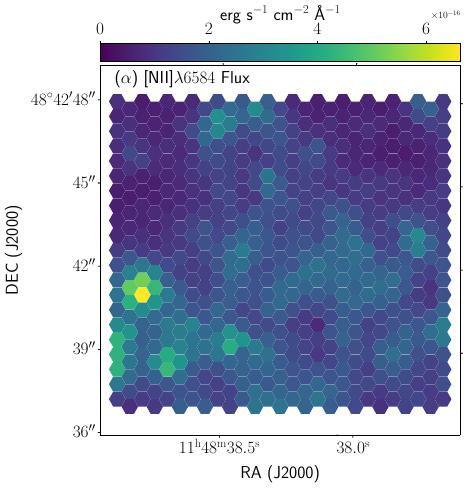}
	\includegraphics[clip, width=0.24\linewidth]{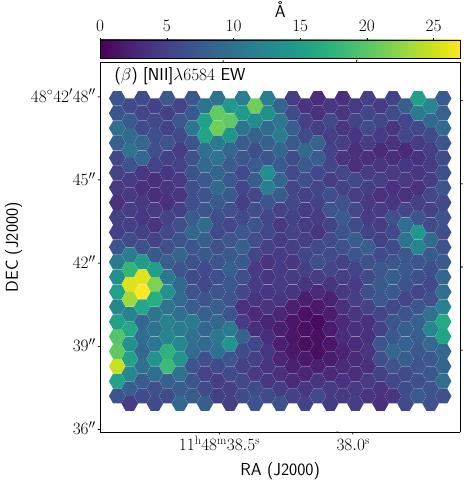}
	\includegraphics[clip, width=0.24\linewidth]{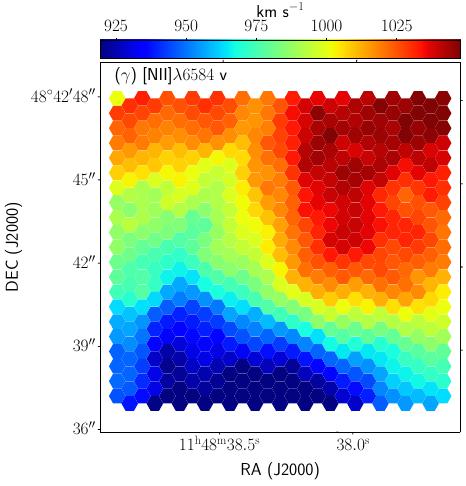}
	\includegraphics[clip, width=0.24\linewidth]{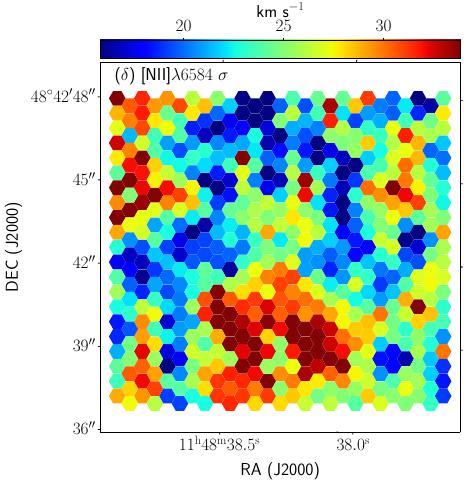}
	\includegraphics[clip, width=0.24\linewidth]{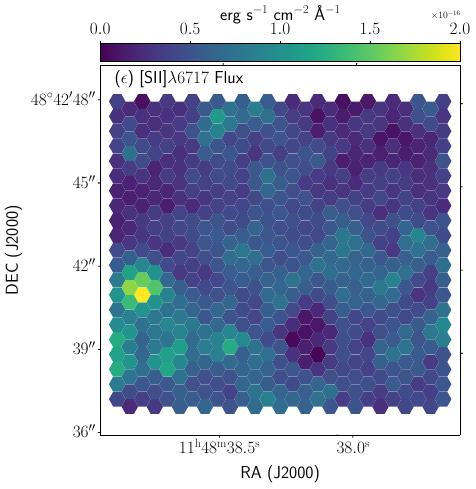}
	\includegraphics[clip, width=0.24\linewidth]{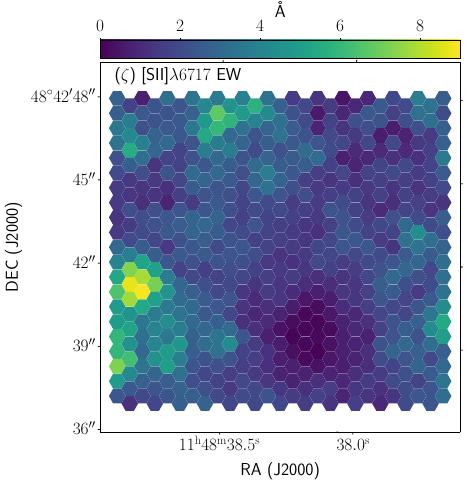}
	\includegraphics[clip, width=0.24\linewidth]{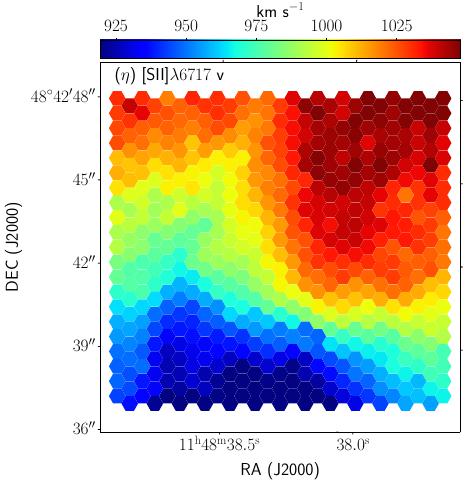}
	\includegraphics[clip, width=0.24\linewidth]{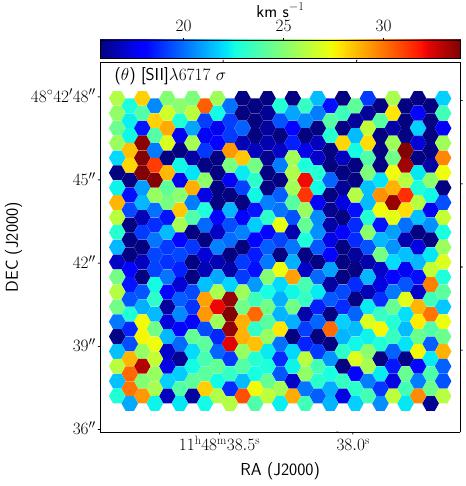}
	\includegraphics[clip, width=0.24\linewidth]{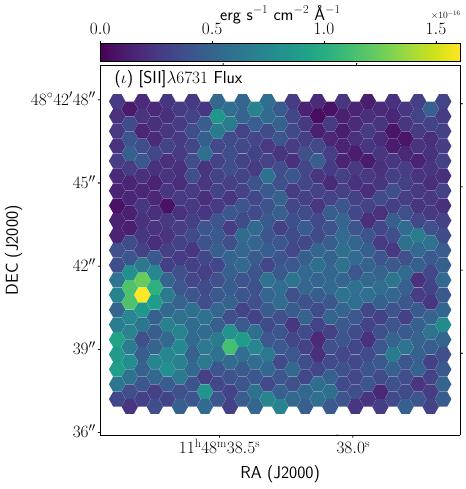}
	\includegraphics[clip, width=0.24\linewidth]{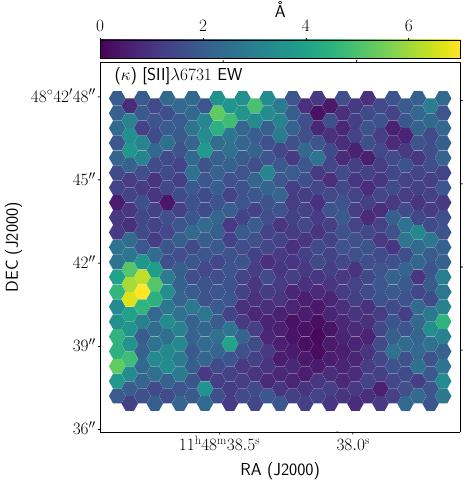}
	\includegraphics[clip, width=0.24\linewidth]{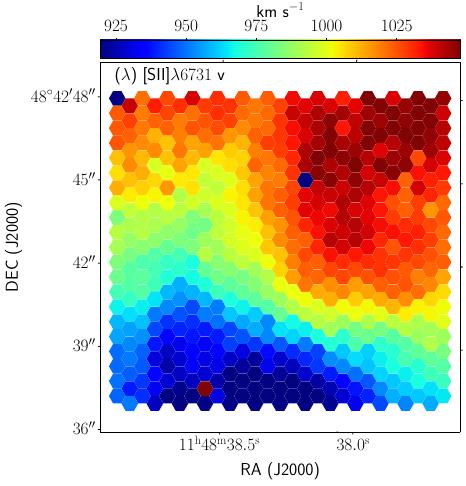}
	\includegraphics[clip, width=0.24\linewidth]{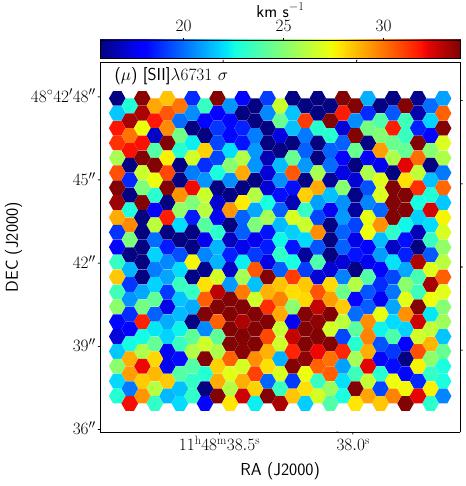}
	\caption{(cont.) NGC~3893 card.}
	\label{fig:NGC3893_card_2}
\end{figure*}

\begin{figure*}[h]
	\centering
	\includegraphics[clip, width=0.35\linewidth]{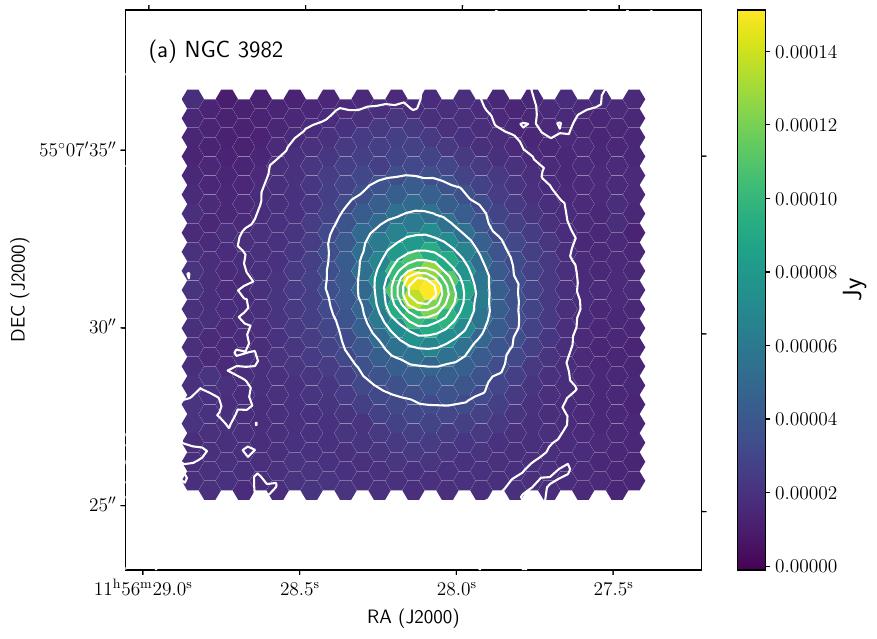}
	\includegraphics[clip, width=0.6\linewidth]{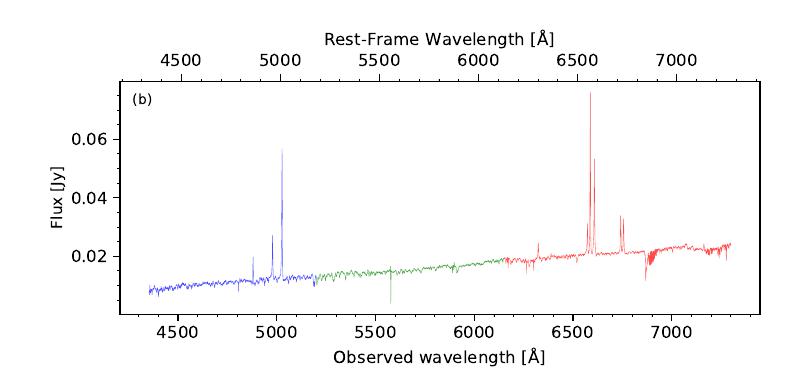}
	\includegraphics[clip, width=0.24\linewidth]{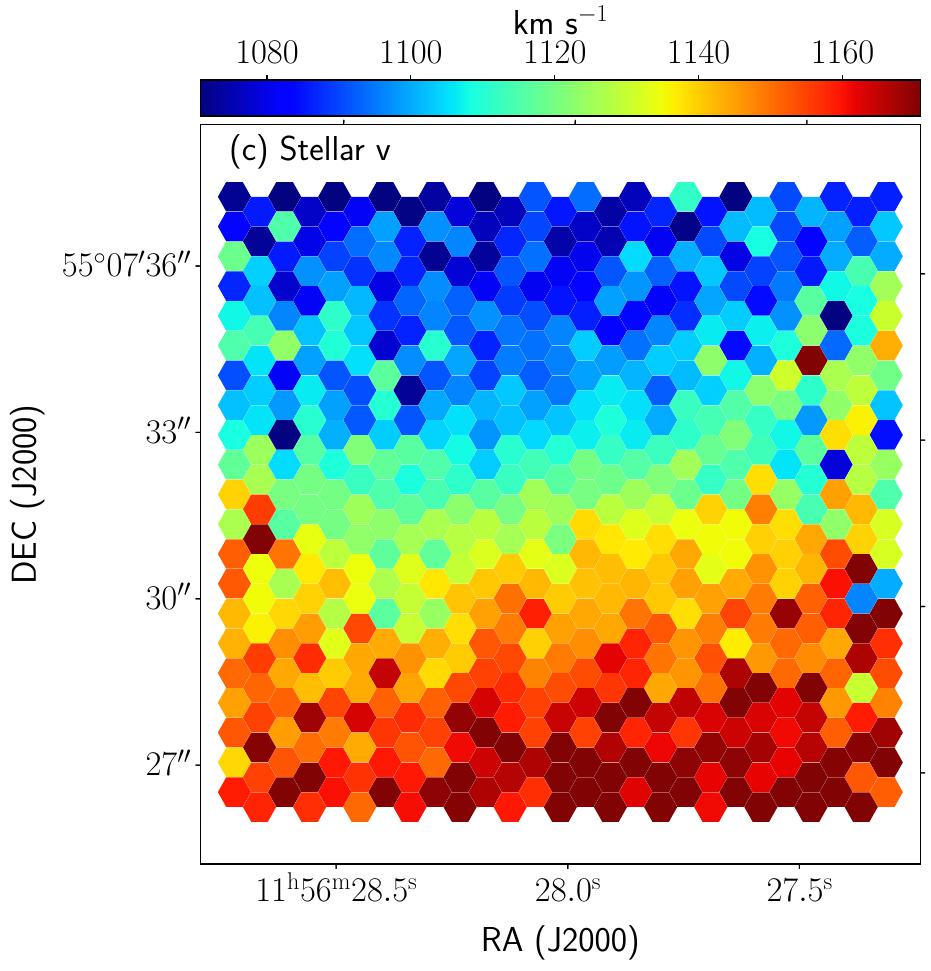}
	\includegraphics[clip, width=0.24\linewidth]{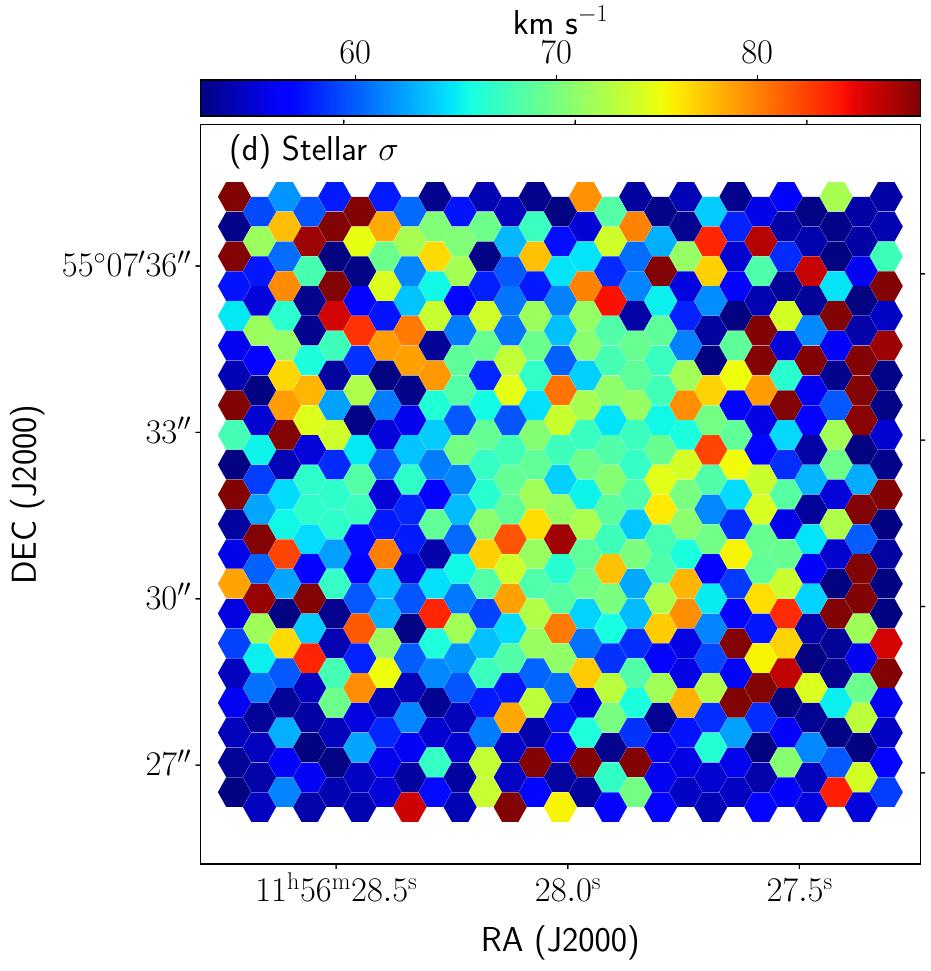}
	\includegraphics[clip, width=0.24\linewidth]{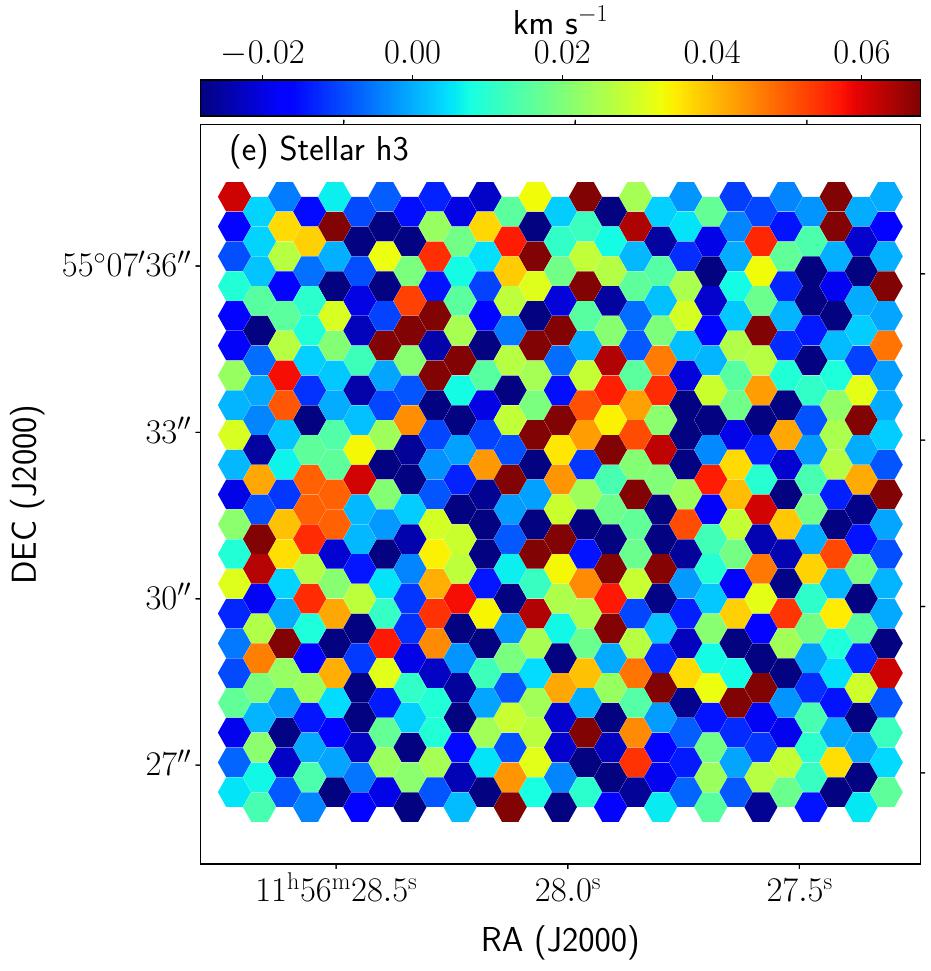}
	\includegraphics[clip, width=0.24\linewidth]{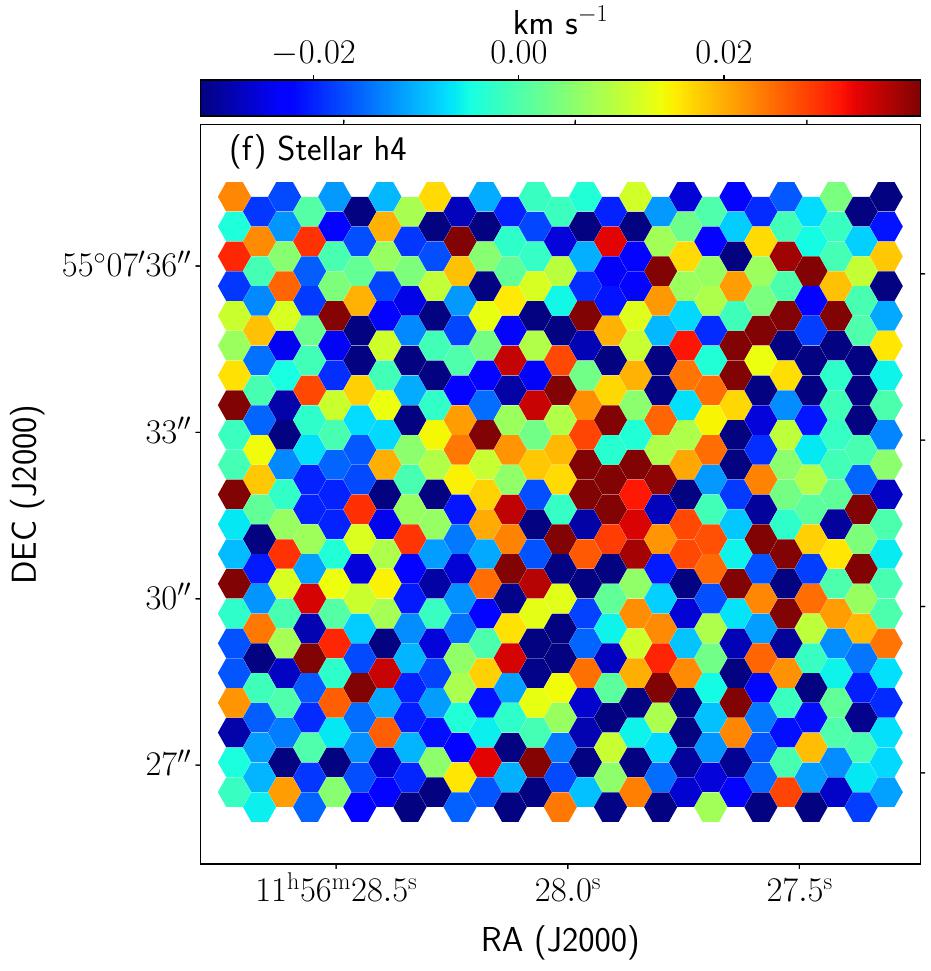}
	\includegraphics[clip, width=0.24\linewidth]{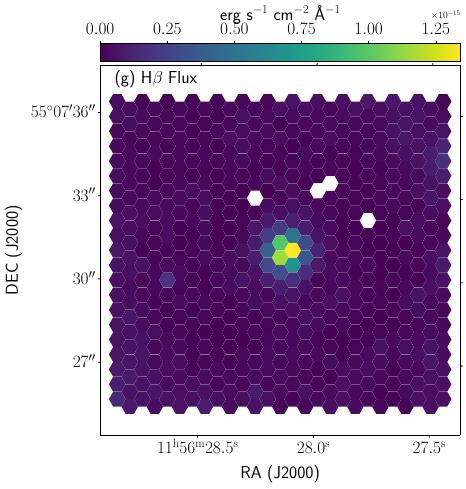}
	\includegraphics[clip, width=0.24\linewidth]{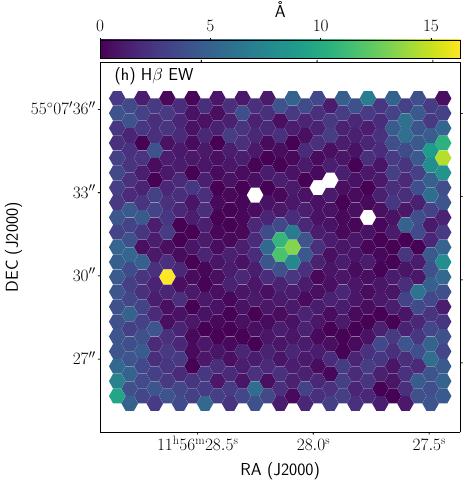}
	\includegraphics[clip, width=0.24\linewidth]{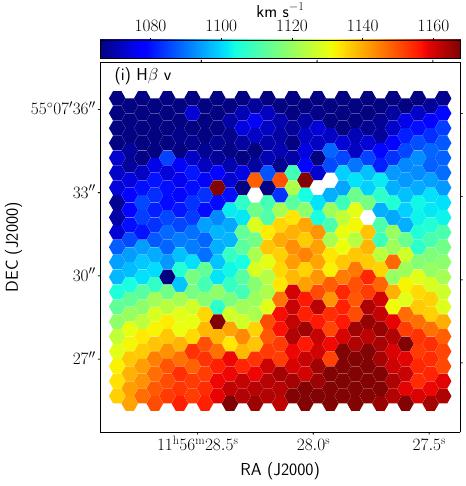}
	\includegraphics[clip, width=0.24\linewidth]{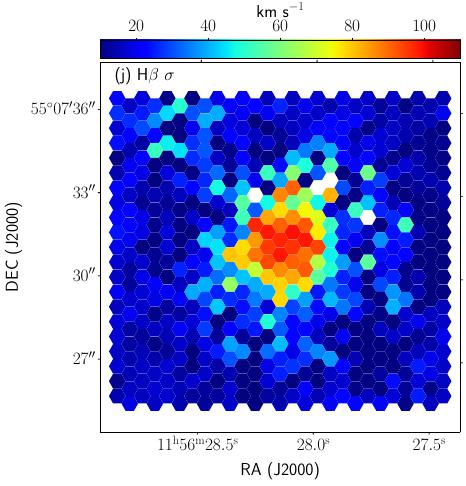}
	\includegraphics[clip, width=0.24\linewidth]{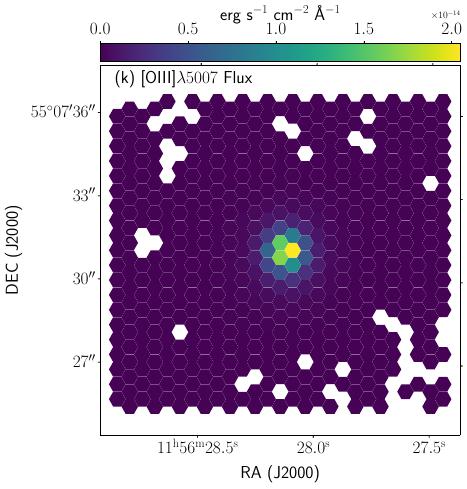}
	\includegraphics[clip, width=0.24\linewidth]{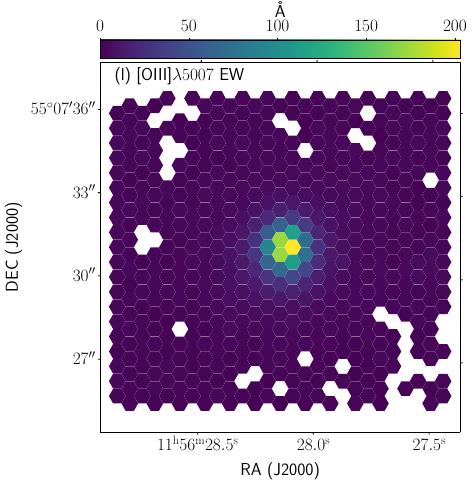}
	\includegraphics[clip, width=0.24\linewidth]{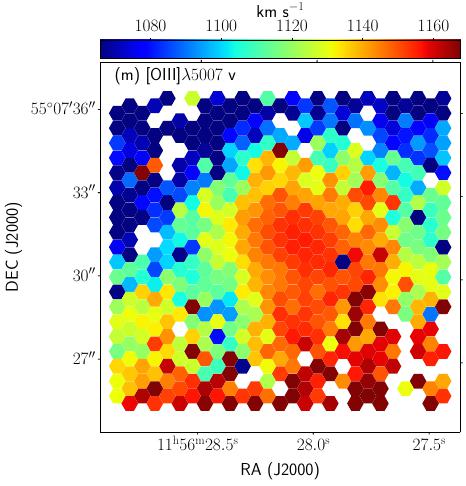}
	\includegraphics[clip, width=0.24\linewidth]{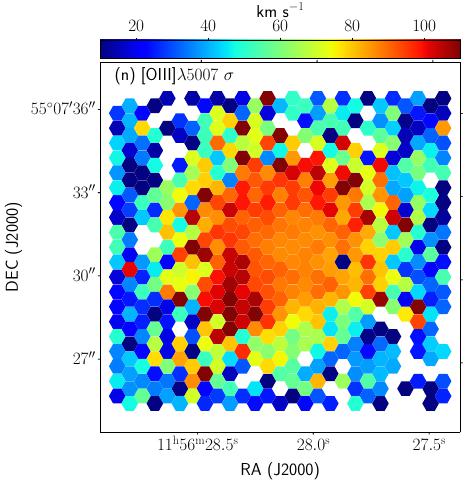}
	\vspace{5cm}
	\caption{NGC~3982 card.}
	\label{fig:NGC3982_card_1}
\end{figure*}
\addtocounter{figure}{-1}
\begin{figure*}[h]
	\centering
	\includegraphics[clip, width=0.24\linewidth]{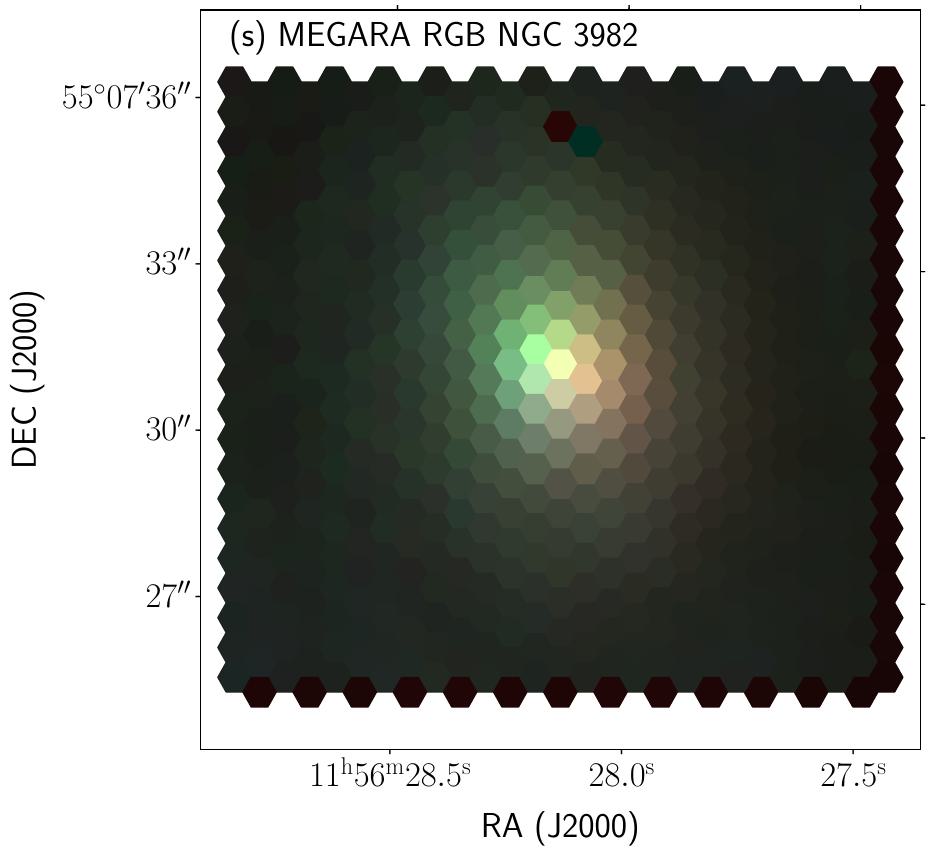}
	\includegraphics[clip, width=0.24\linewidth]{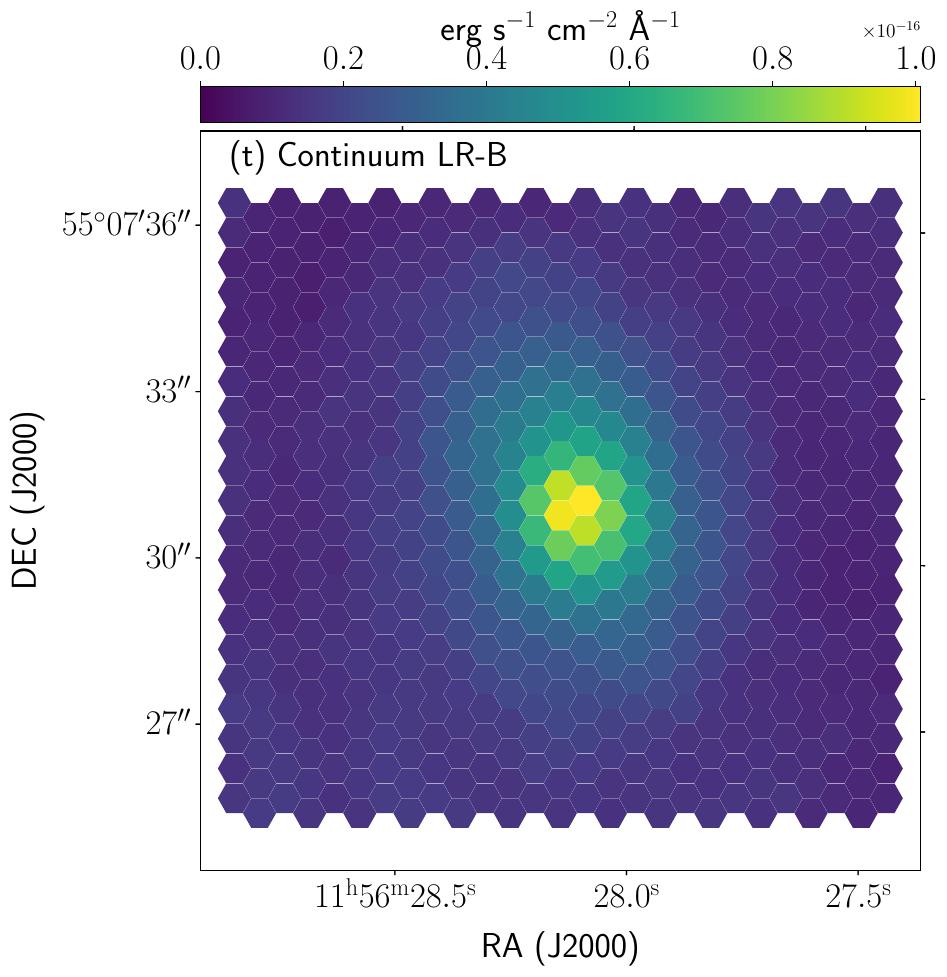}
	\includegraphics[clip, width=0.24\linewidth]{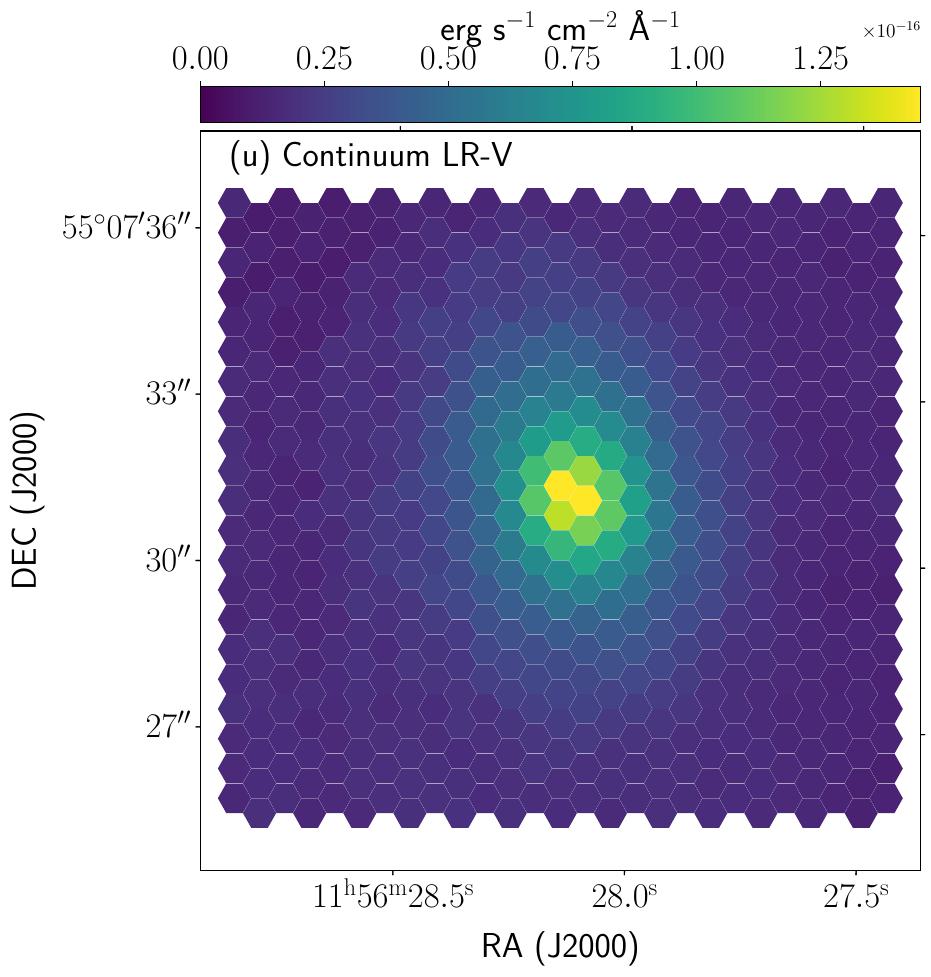}
	\includegraphics[clip, width=0.24\linewidth]{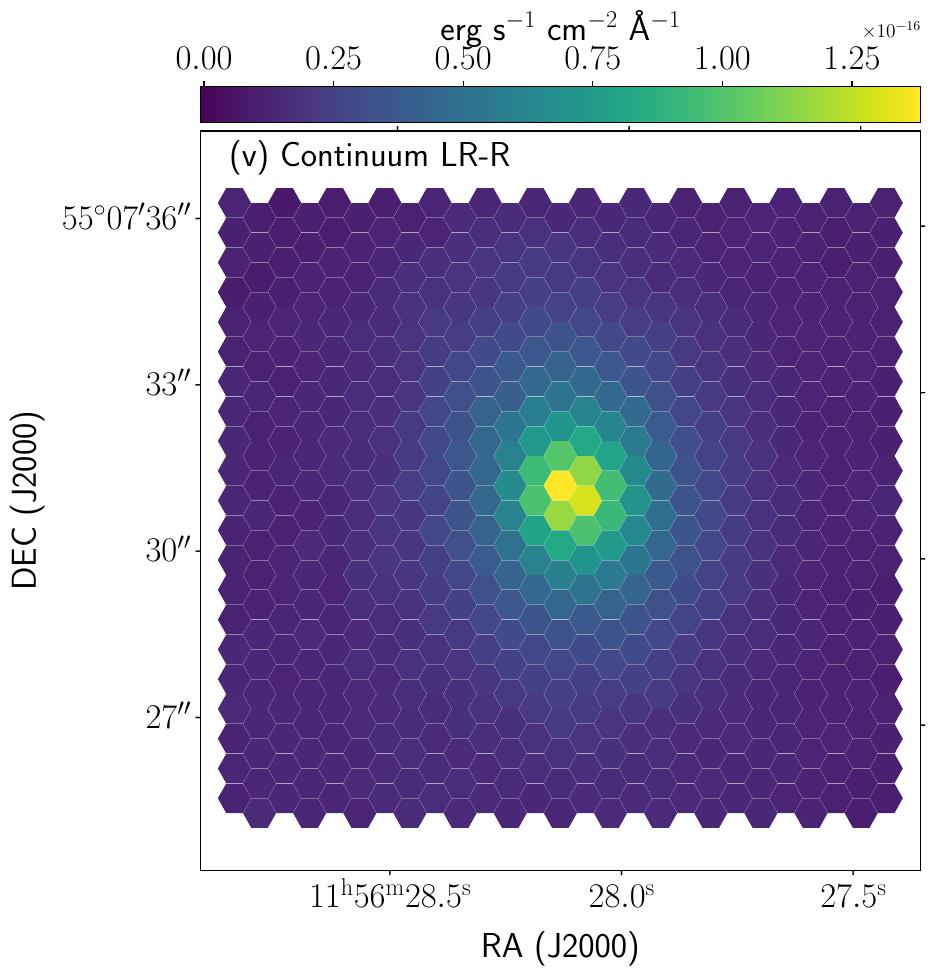}
	\includegraphics[clip, width=0.24\linewidth]{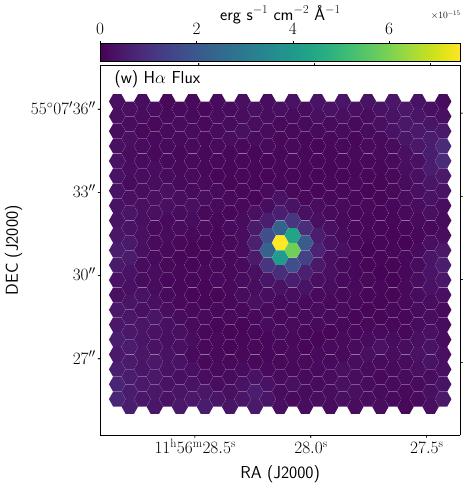}
	\includegraphics[clip, width=0.24\linewidth]{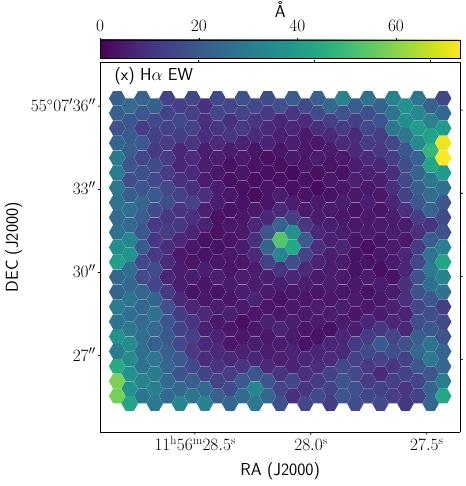}
	\includegraphics[clip, width=0.24\linewidth]{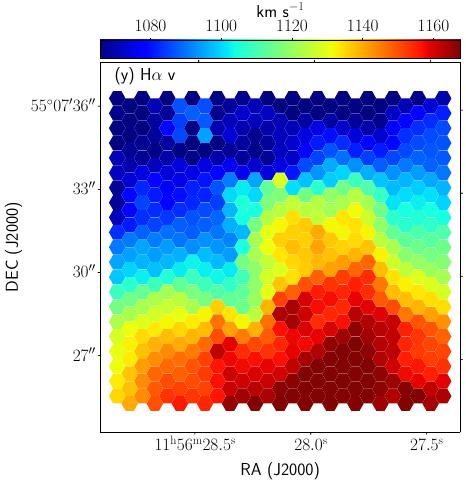}
	\includegraphics[clip, width=0.24\linewidth]{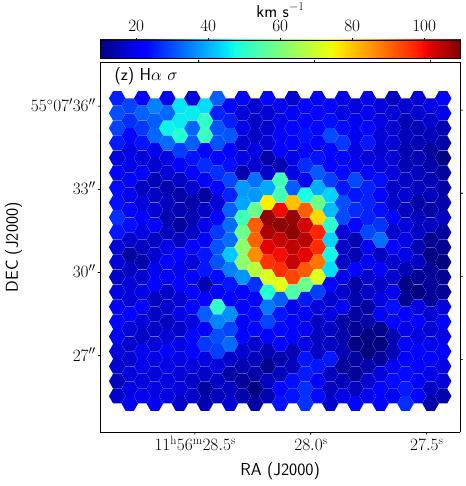}
	\includegraphics[clip, width=0.24\linewidth]{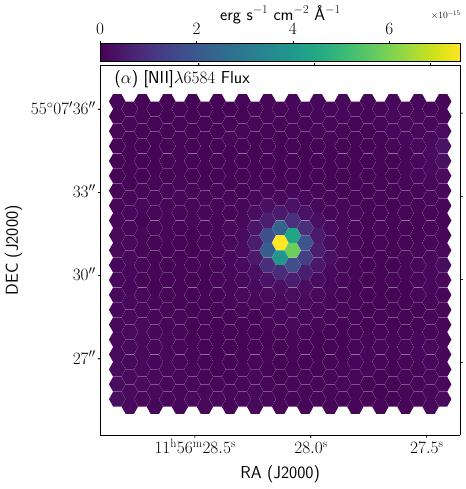}
	\includegraphics[clip, width=0.24\linewidth]{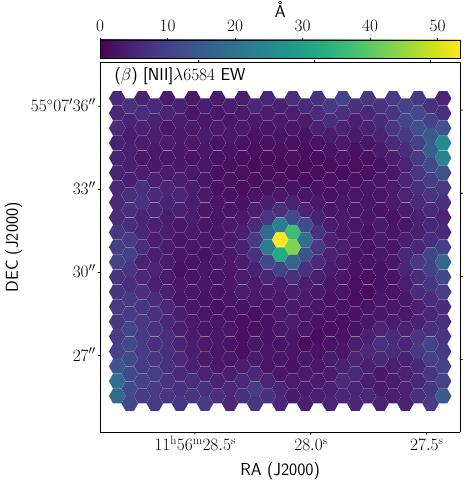}
	\includegraphics[clip, width=0.24\linewidth]{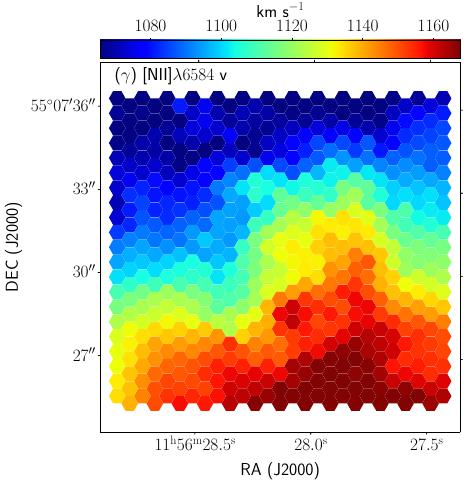}
	\includegraphics[clip, width=0.24\linewidth]{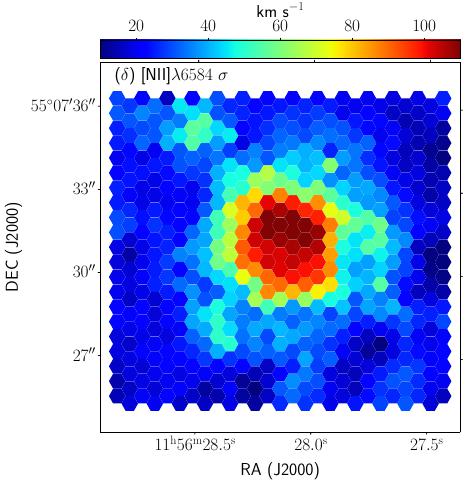}
	\includegraphics[clip, width=0.24\linewidth]{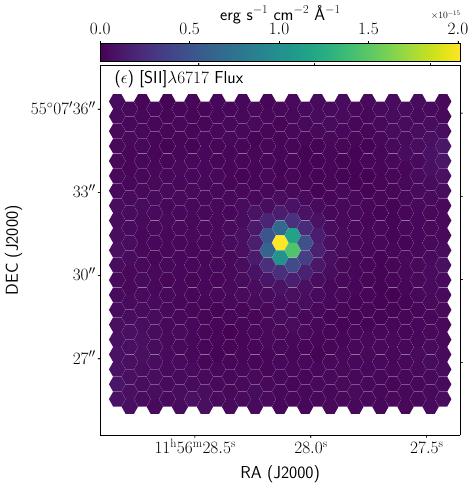}
	\includegraphics[clip, width=0.24\linewidth]{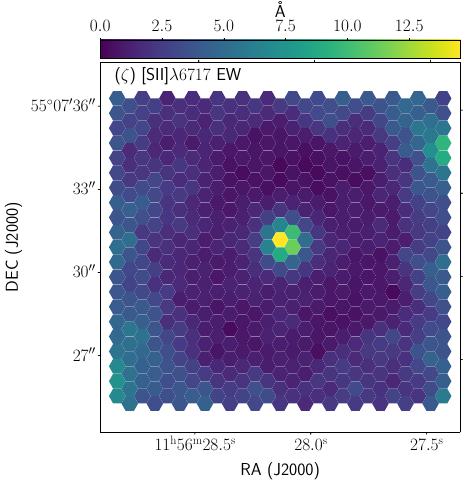}
	\includegraphics[clip, width=0.24\linewidth]{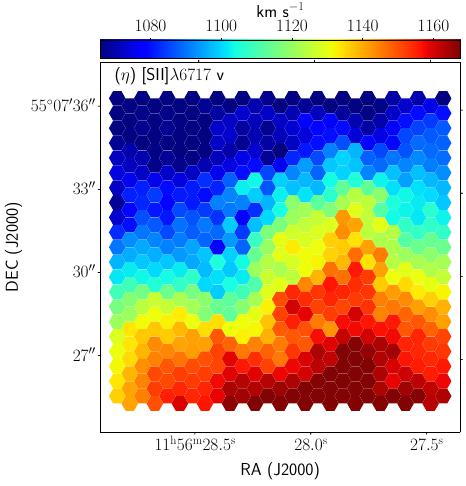}
	\includegraphics[clip, width=0.24\linewidth]{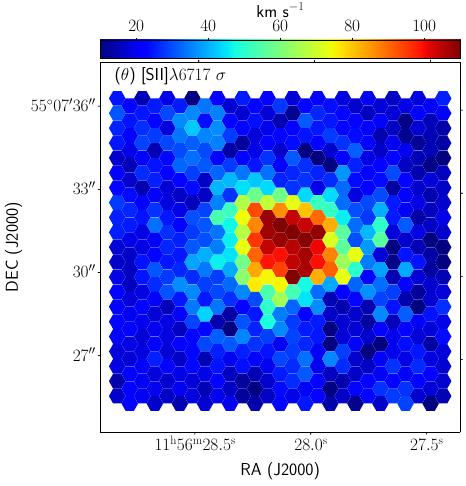}
	\includegraphics[clip, width=0.24\linewidth]{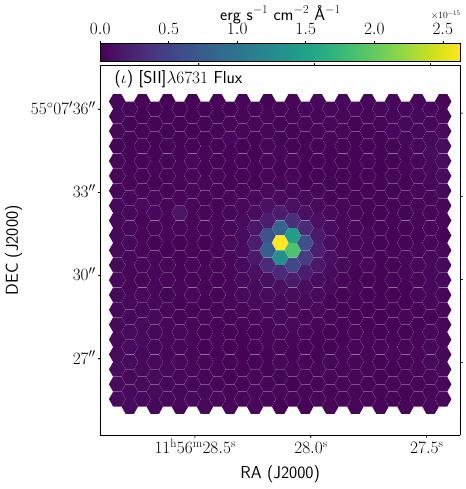}
	\includegraphics[clip, width=0.24\linewidth]{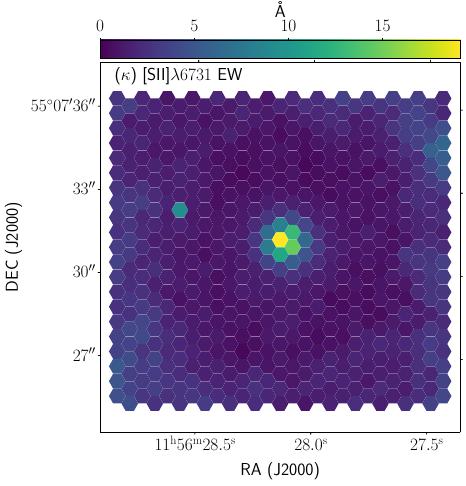}
	\includegraphics[clip, width=0.24\linewidth]{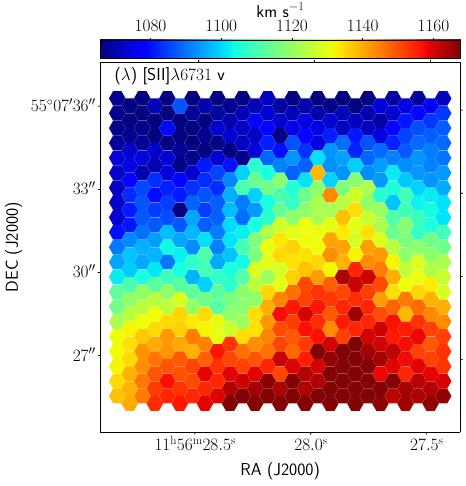}
	\includegraphics[clip, width=0.24\linewidth]{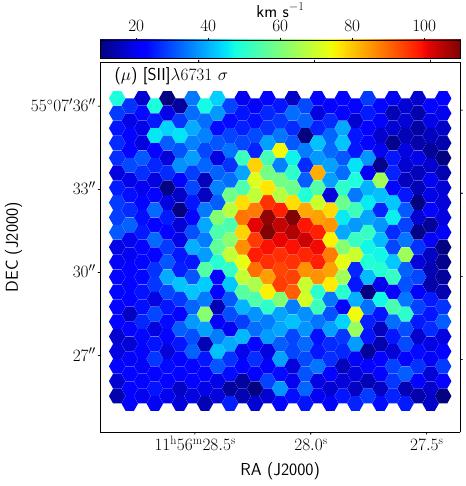}
	\caption{(cont.) NGC~3982 card.}
	\label{fig:NGC3982_card_2}
\end{figure*}

\begin{figure*}[h]
	\centering
	\includegraphics[clip, width=0.35\linewidth]{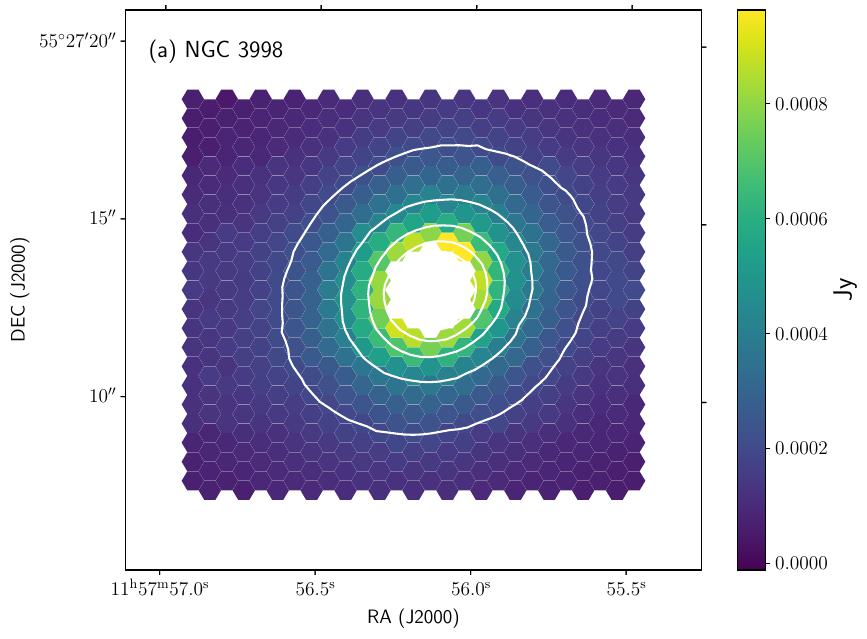}
	\includegraphics[clip, width=0.6\linewidth]{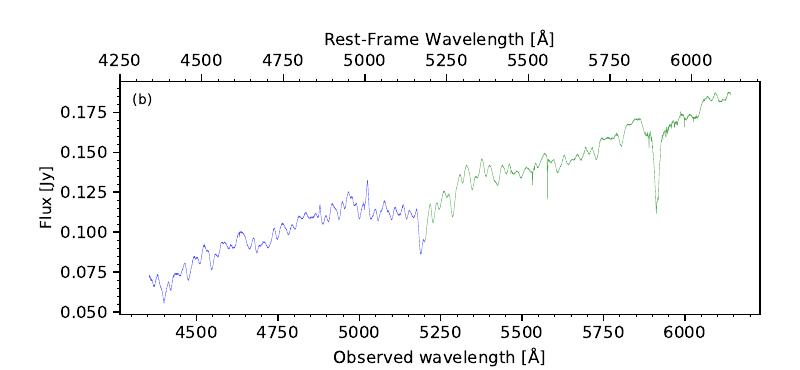}
	\includegraphics[clip, width=0.24\linewidth]{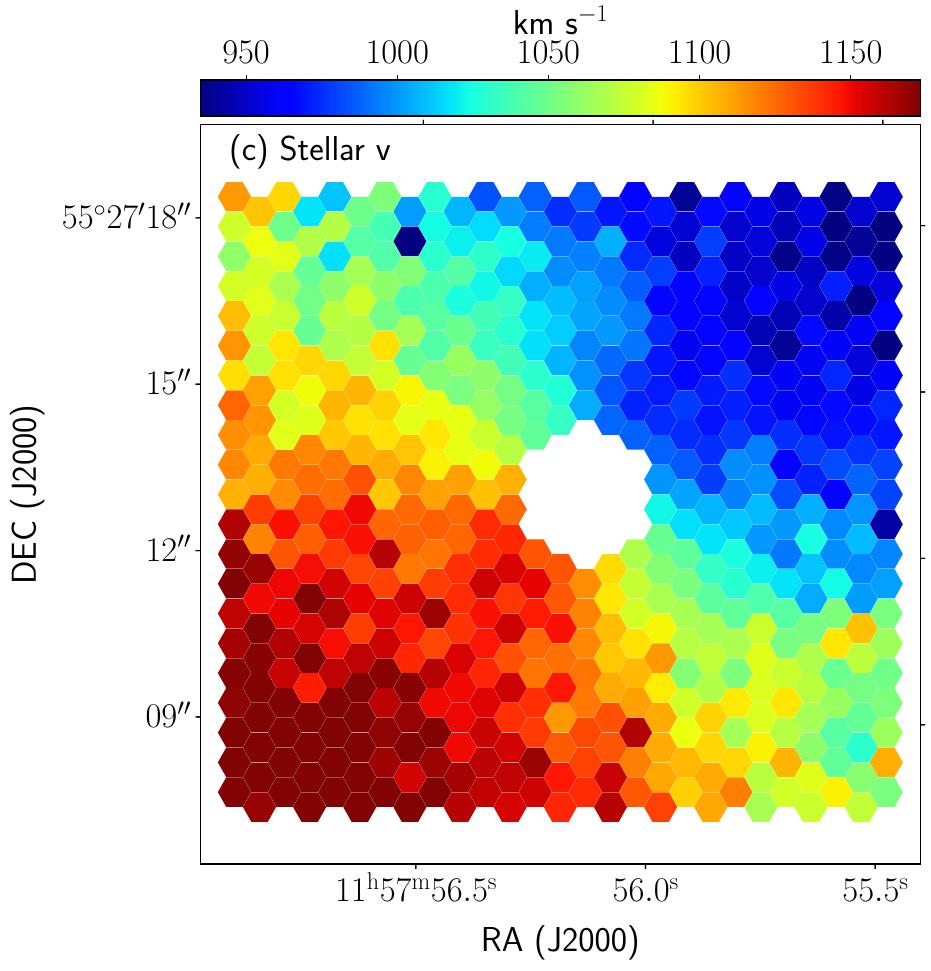}
	\includegraphics[clip, width=0.24\linewidth]{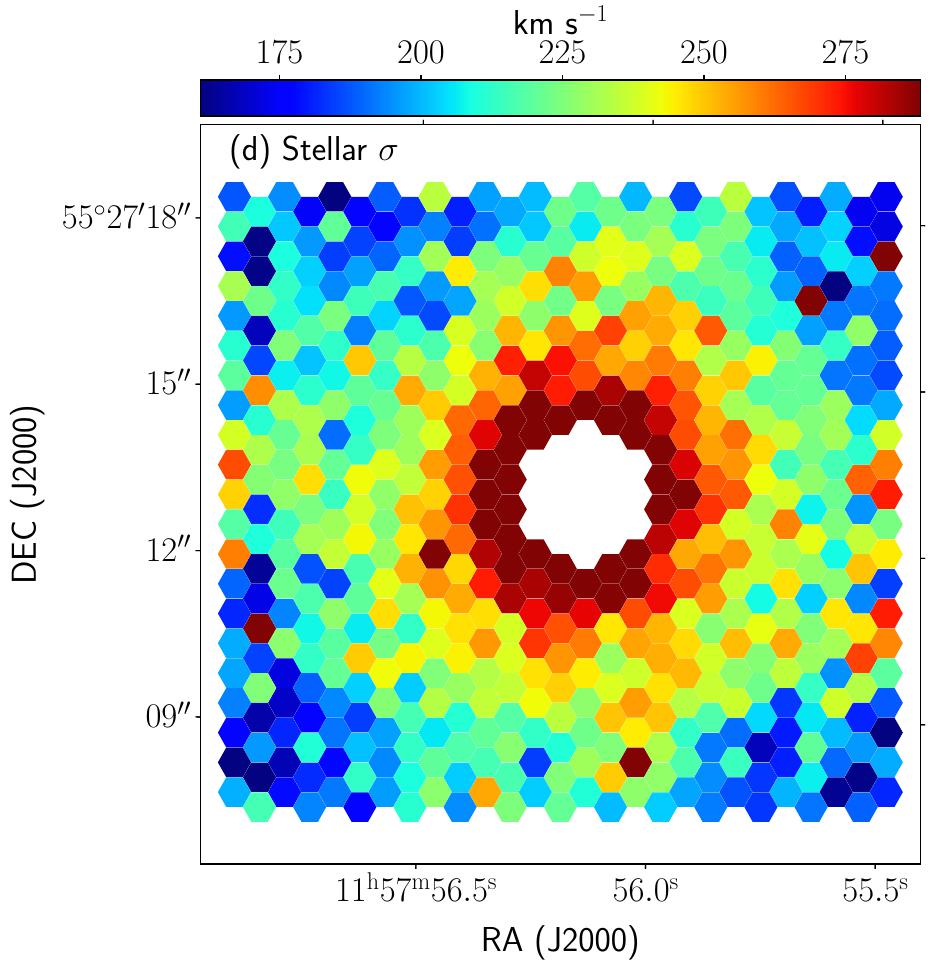}
	\includegraphics[clip, width=0.24\linewidth]{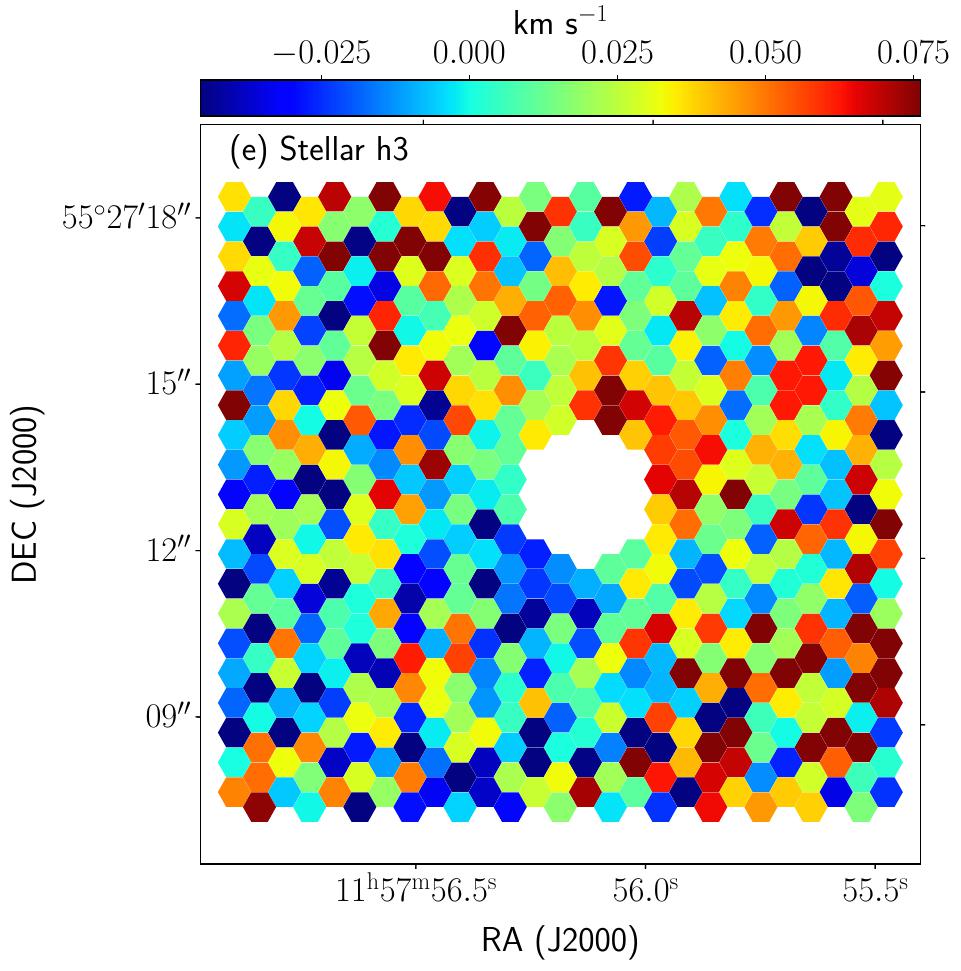}
	\includegraphics[clip, width=0.24\linewidth]{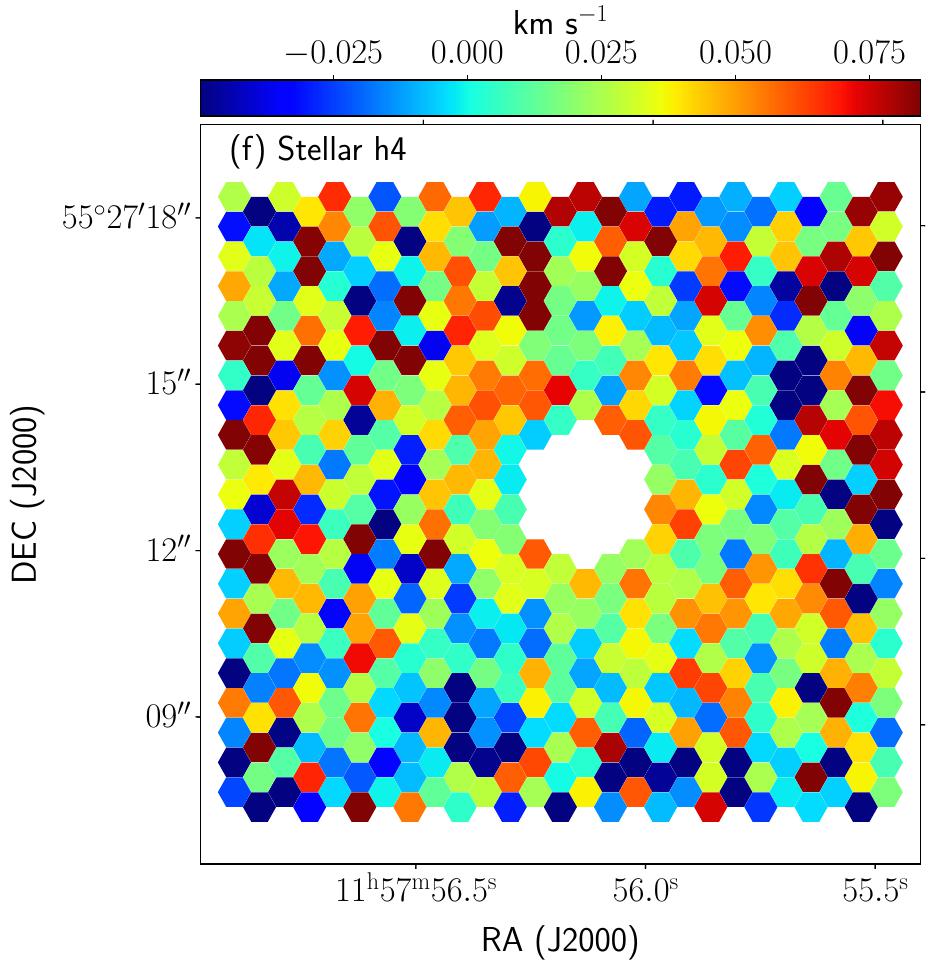}
	\includegraphics[clip, width=0.24\linewidth]{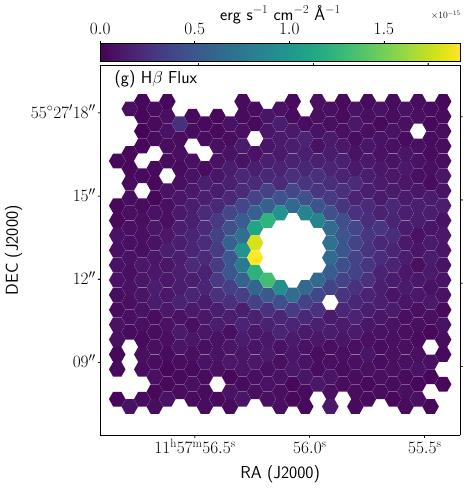}
	\includegraphics[clip, width=0.24\linewidth]{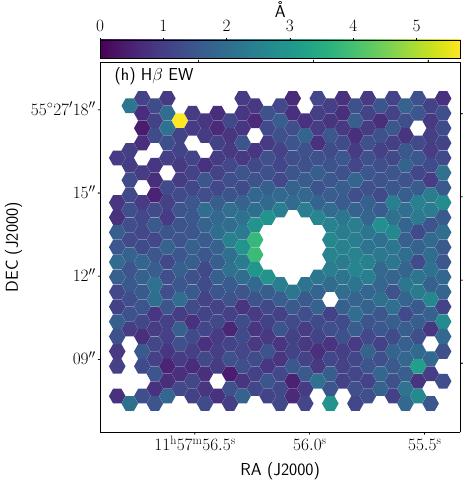}
	\includegraphics[clip, width=0.24\linewidth]{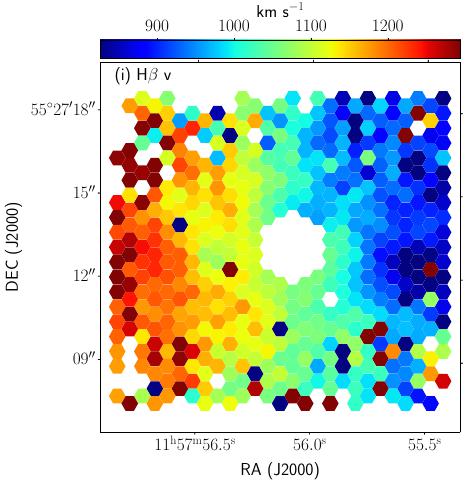}
	\includegraphics[clip, width=0.24\linewidth]{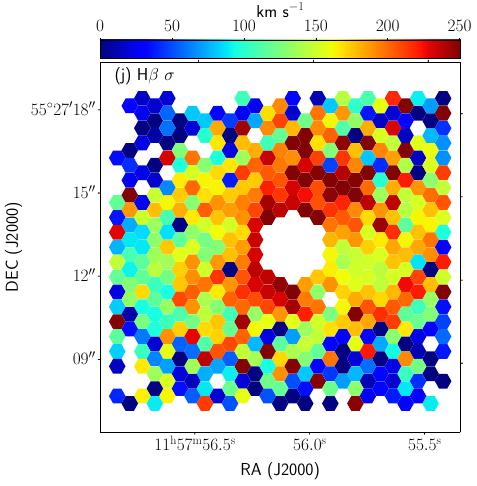}
	\includegraphics[clip, width=0.24\linewidth]{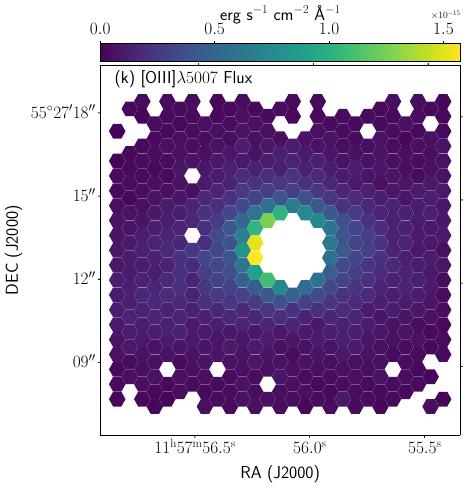}
	\includegraphics[clip, width=0.24\linewidth]{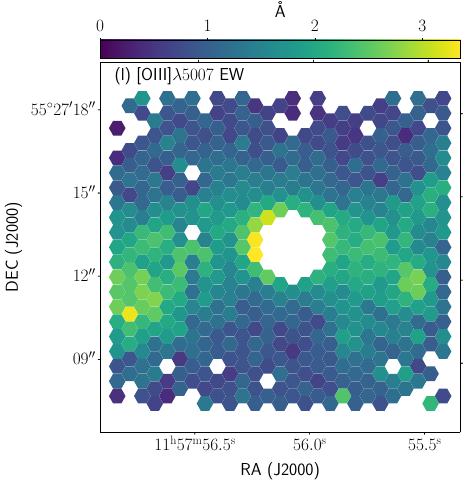}
	\includegraphics[clip, width=0.24\linewidth]{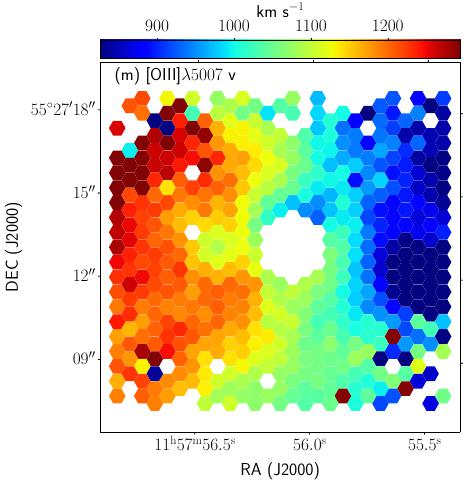}
	\includegraphics[clip, width=0.24\linewidth]{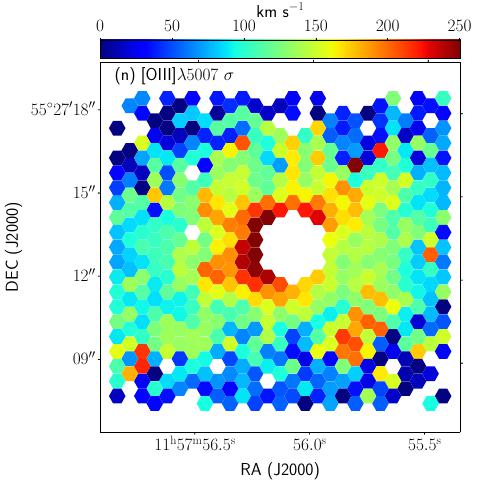}
	\includegraphics[clip, width=0.24\linewidth]{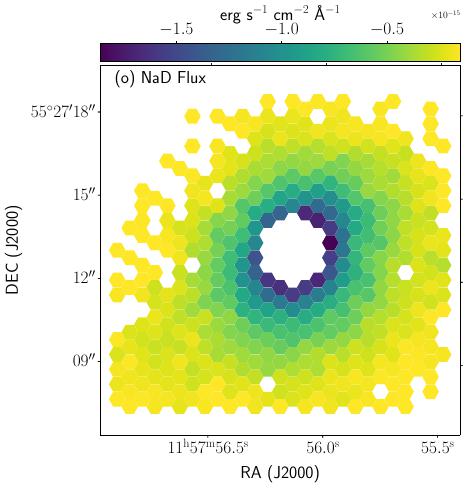}
	\includegraphics[clip, width=0.24\linewidth]{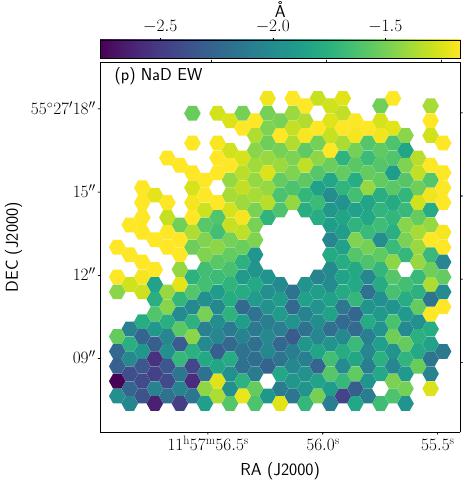}
	\includegraphics[clip, width=0.24\linewidth]{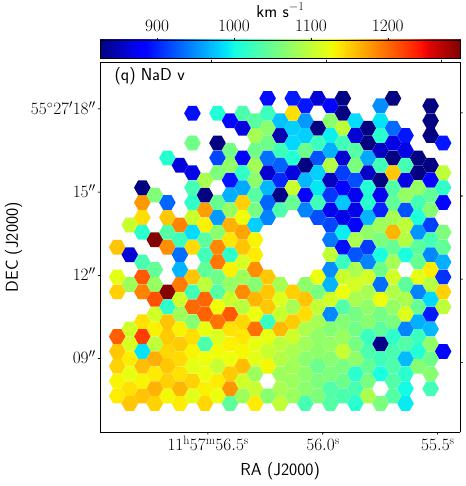}
	\includegraphics[clip, width=0.24\linewidth]{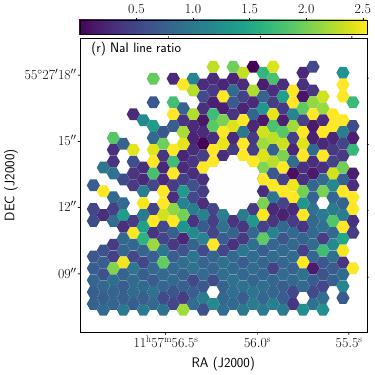}
	\caption{NGC~3998 card.}
	\label{fig:NGC3998_card_1}
\end{figure*}
\addtocounter{figure}{-1}
\begin{figure*}[h]
	\centering
	\includegraphics[clip, width=0.24\linewidth]{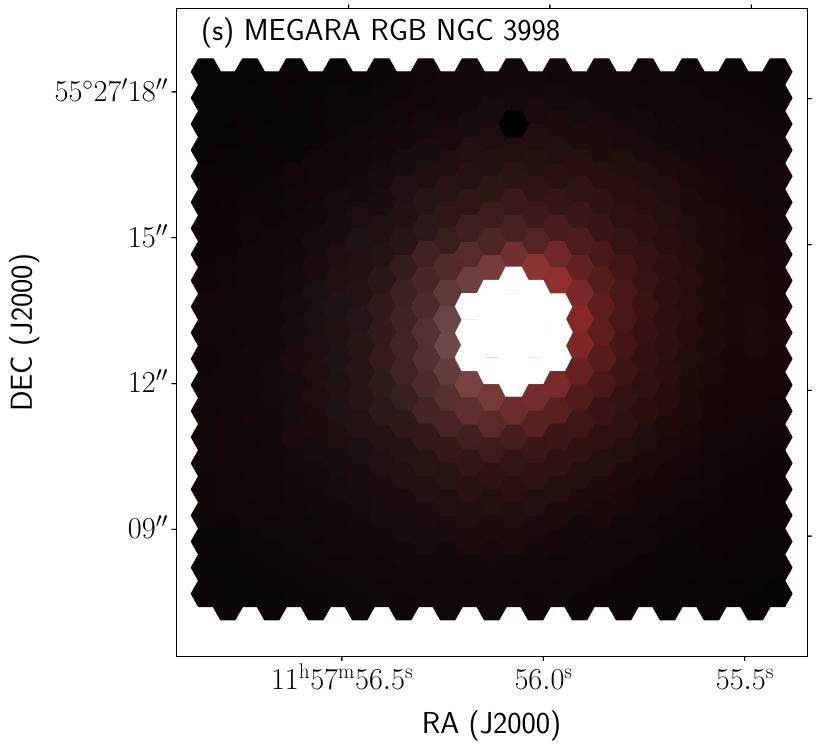}
	\includegraphics[clip, width=0.24\linewidth]{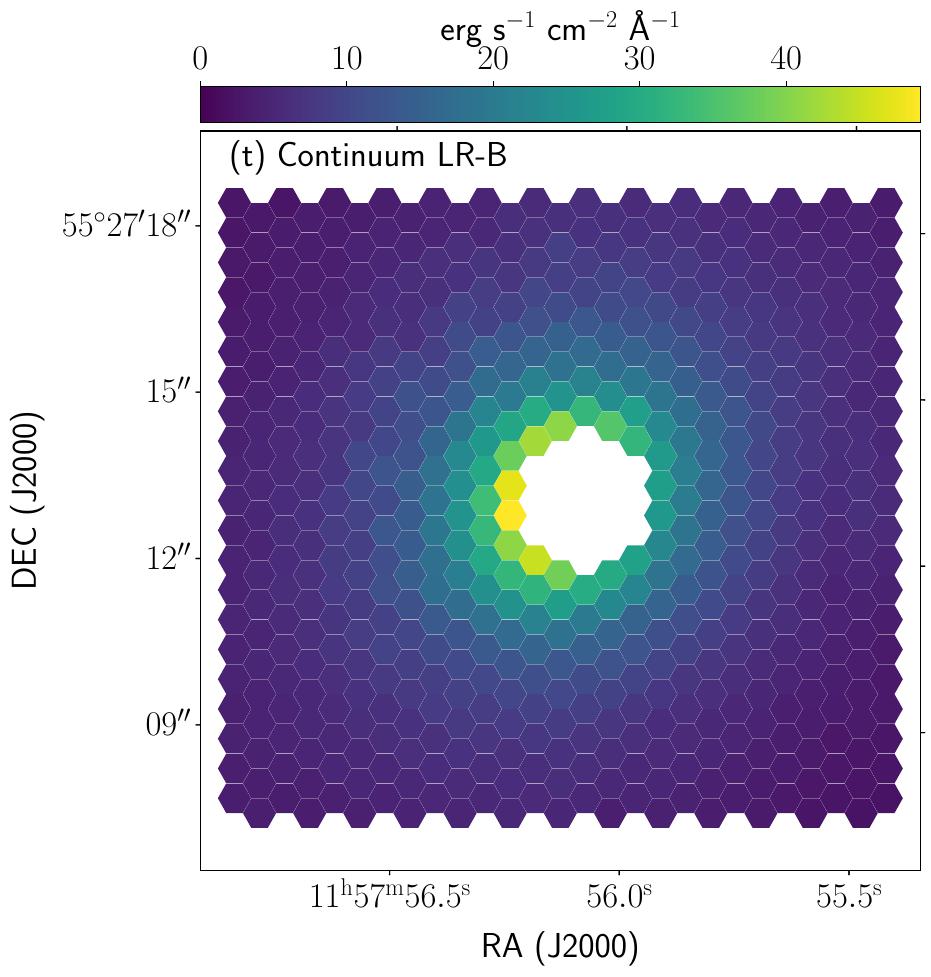}
	\includegraphics[clip, width=0.24\linewidth]{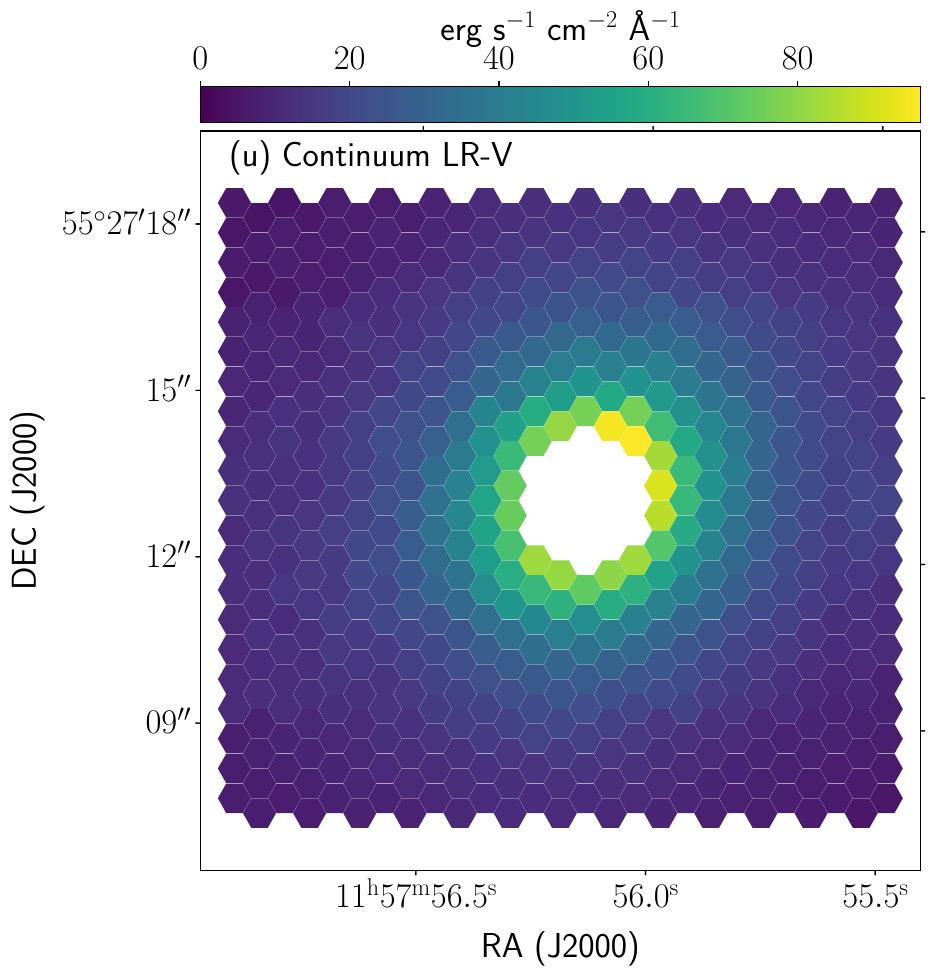}
	\hspace{4cm}
	\caption{(cont.) NGC~3998 card.}
	\label{fig:NGC3998_card_2}
\end{figure*}

\begin{figure*}[h]
	\centering
	\includegraphics[clip, width=0.35\linewidth]{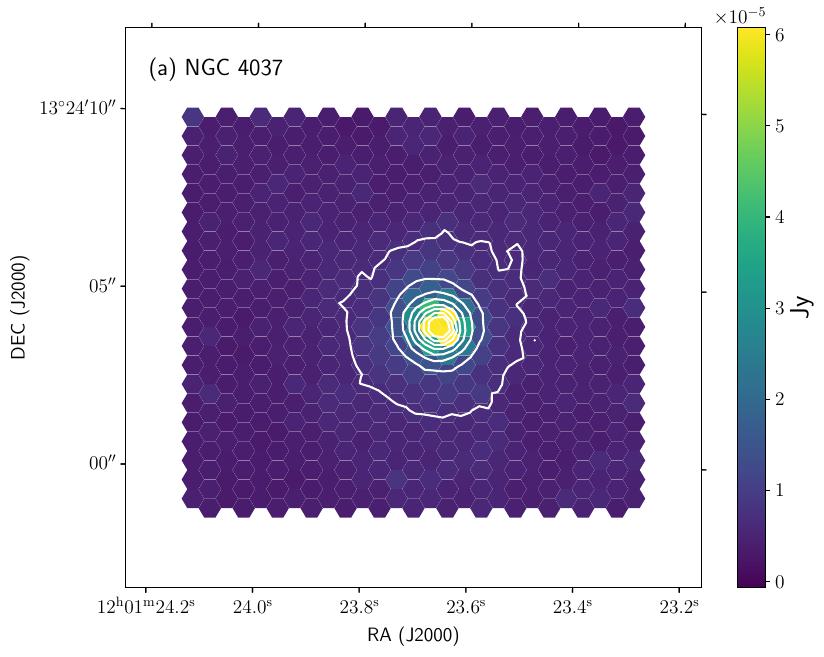}
	\includegraphics[clip, width=0.6\linewidth]{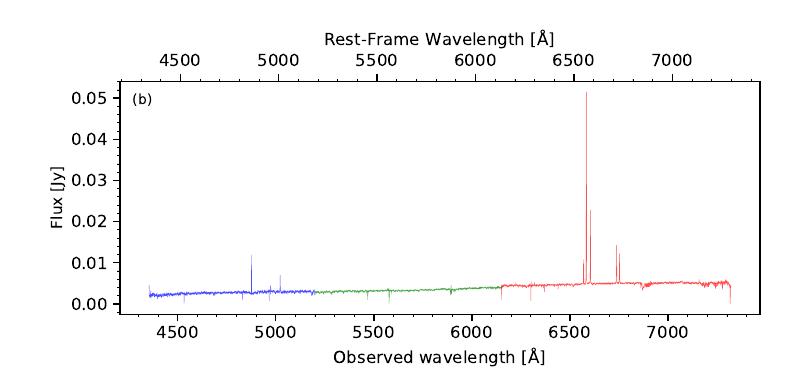}
	\includegraphics[clip, width=0.24\linewidth]{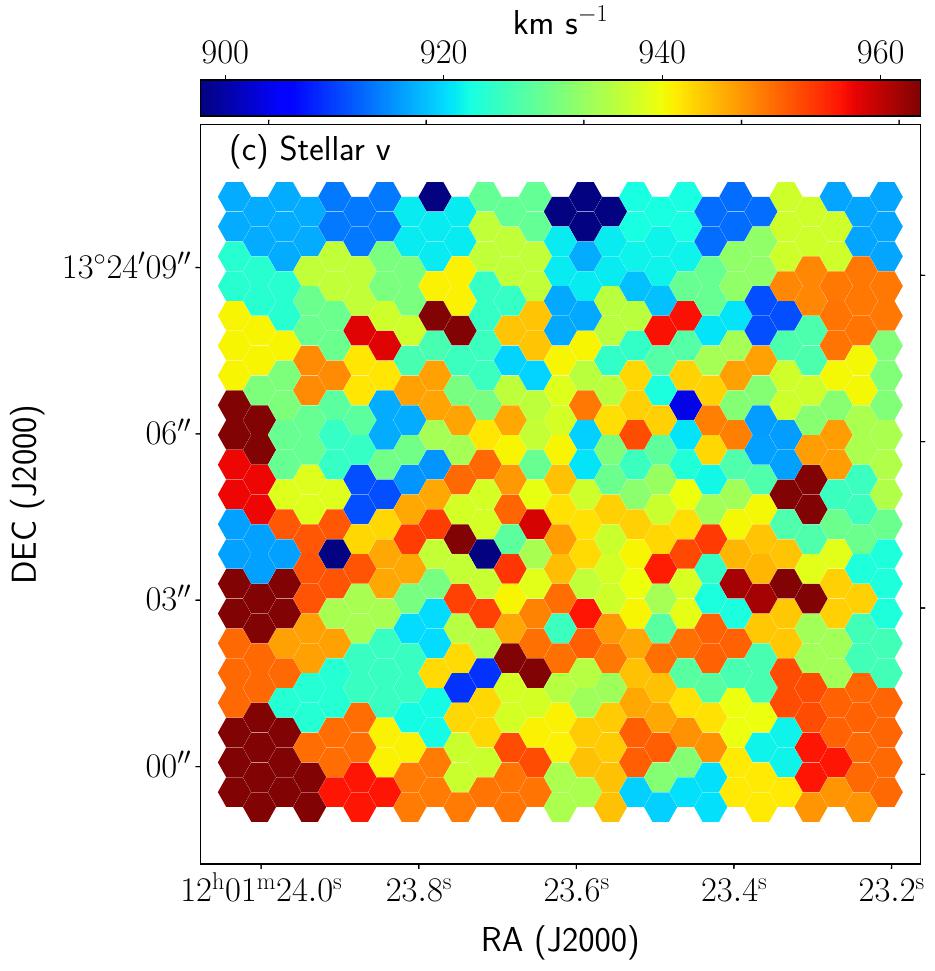}
	\includegraphics[clip, width=0.24\linewidth]{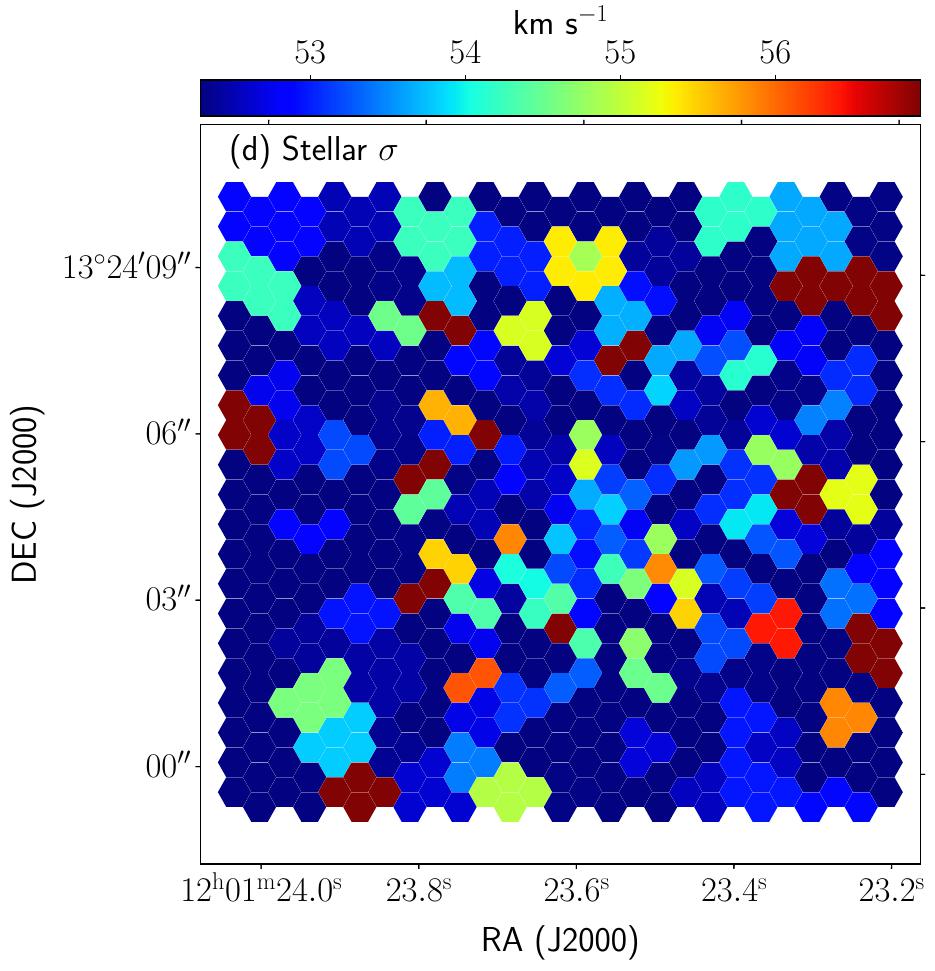}
	\includegraphics[clip, width=0.24\linewidth]{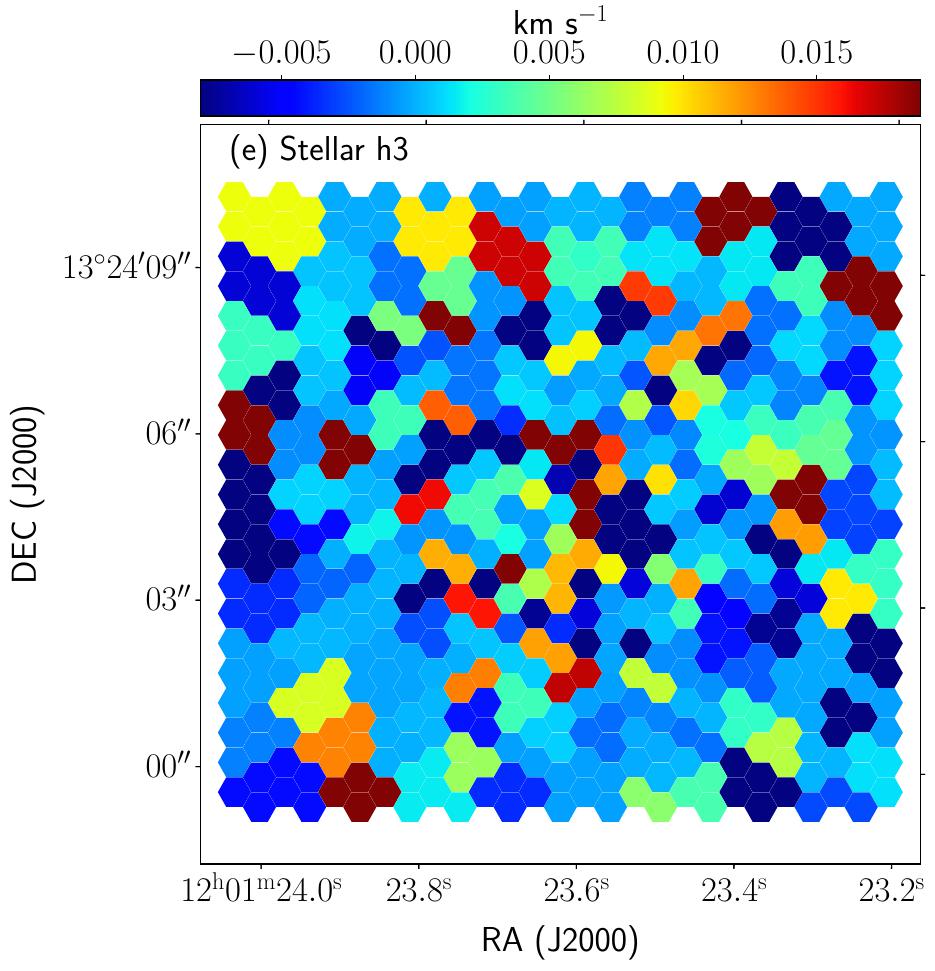}
	\includegraphics[clip, width=0.24\linewidth]{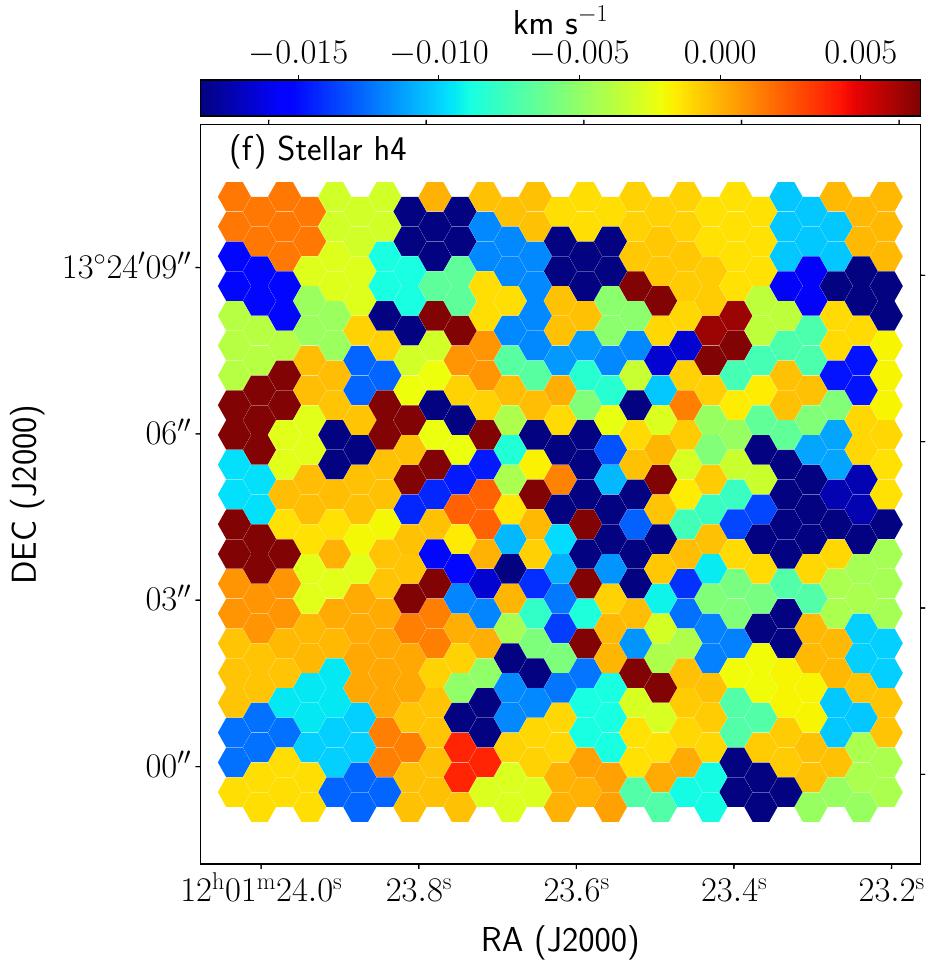}
	\includegraphics[clip, width=0.24\linewidth]{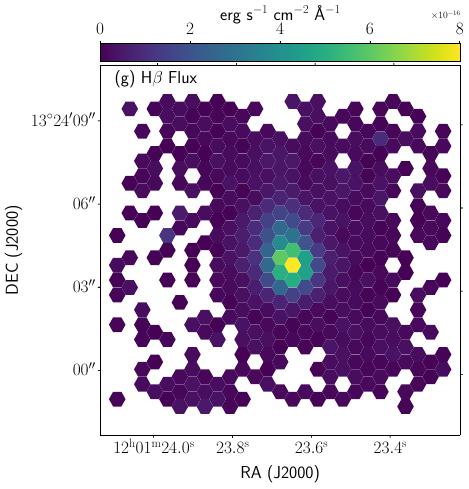}
	\includegraphics[clip, width=0.24\linewidth]{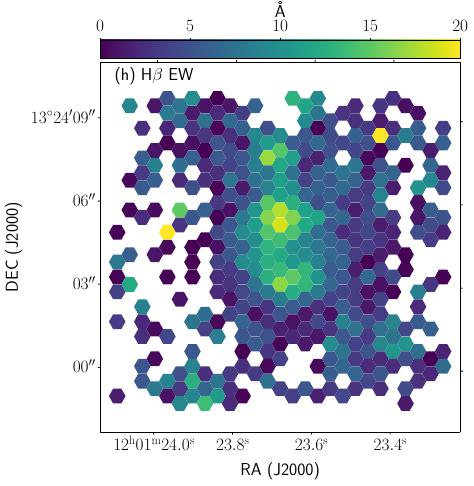}
	\includegraphics[clip, width=0.24\linewidth]{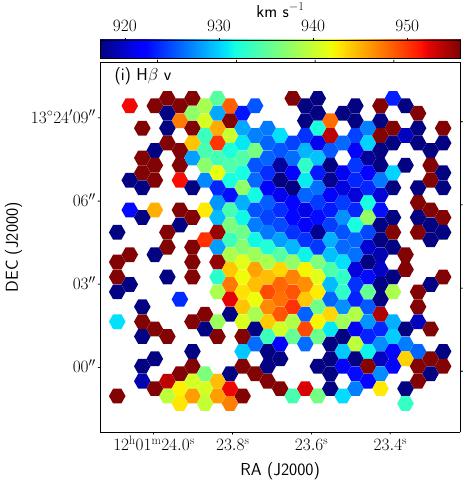}
	\includegraphics[clip, width=0.24\linewidth]{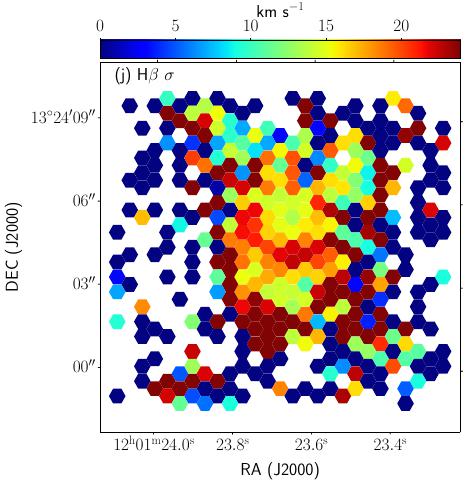}
	\includegraphics[clip, width=0.24\linewidth]{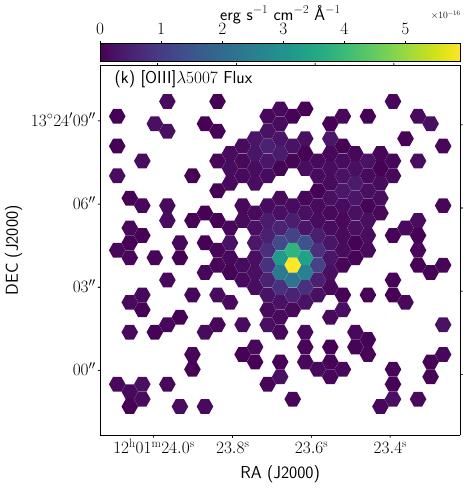}
	\includegraphics[clip, width=0.24\linewidth]{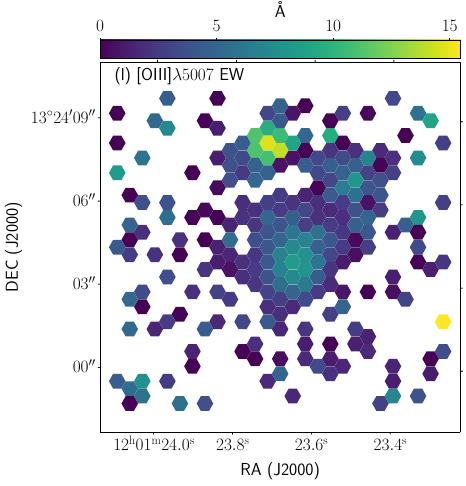}
	\includegraphics[clip, width=0.24\linewidth]{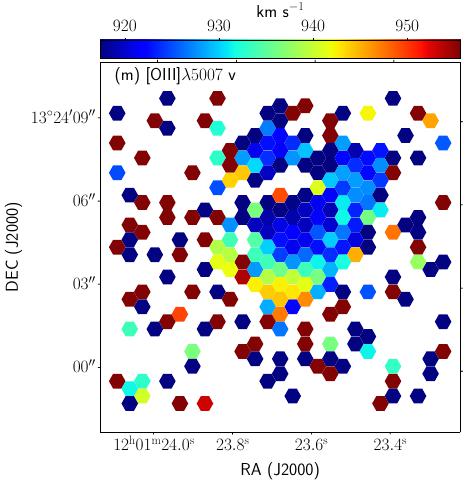}
	\includegraphics[clip, width=0.24\linewidth]{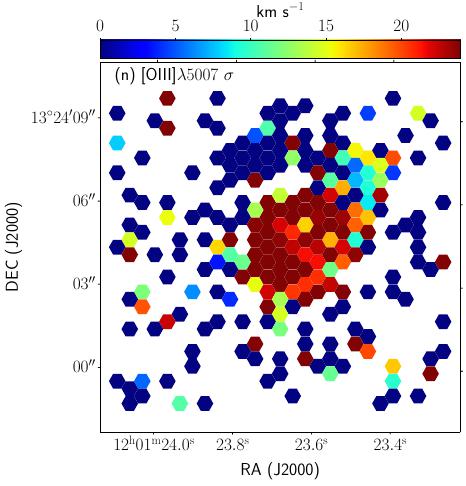}
	\vspace{5cm}
	\caption{NGC~4037 card.}
	\label{fig:NGC4037_card_1}
\end{figure*}
\addtocounter{figure}{-1}
\begin{figure*}[h]
	\centering
	\includegraphics[clip, width=0.24\linewidth]{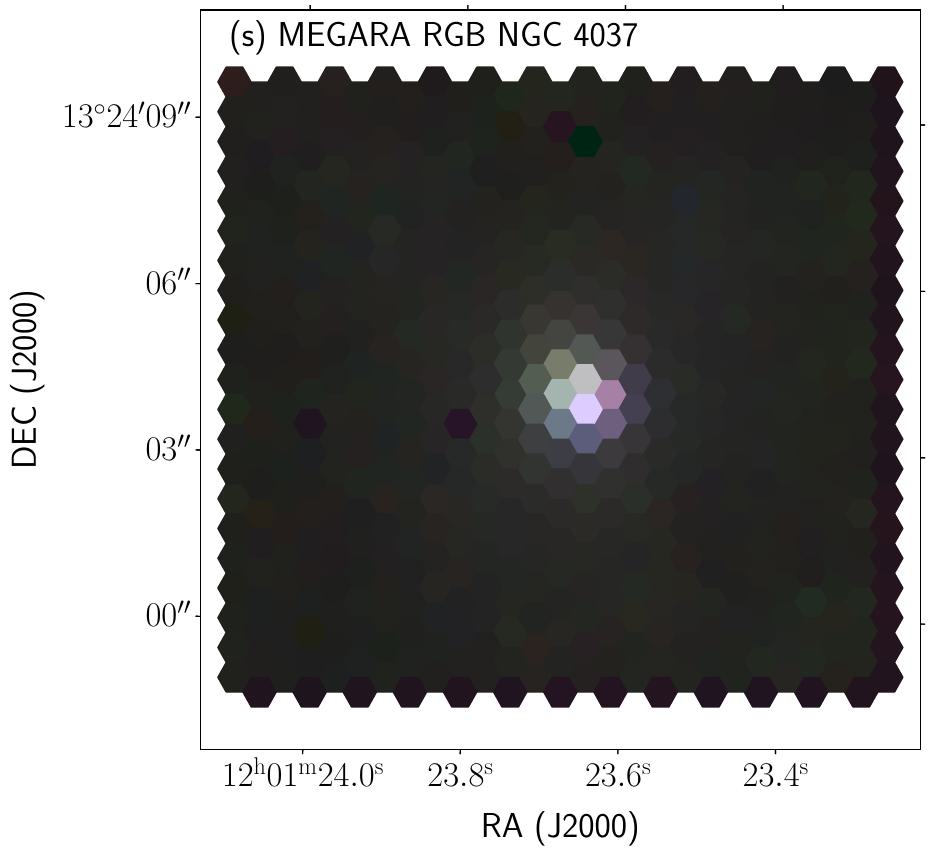}
	\includegraphics[clip, width=0.24\linewidth]{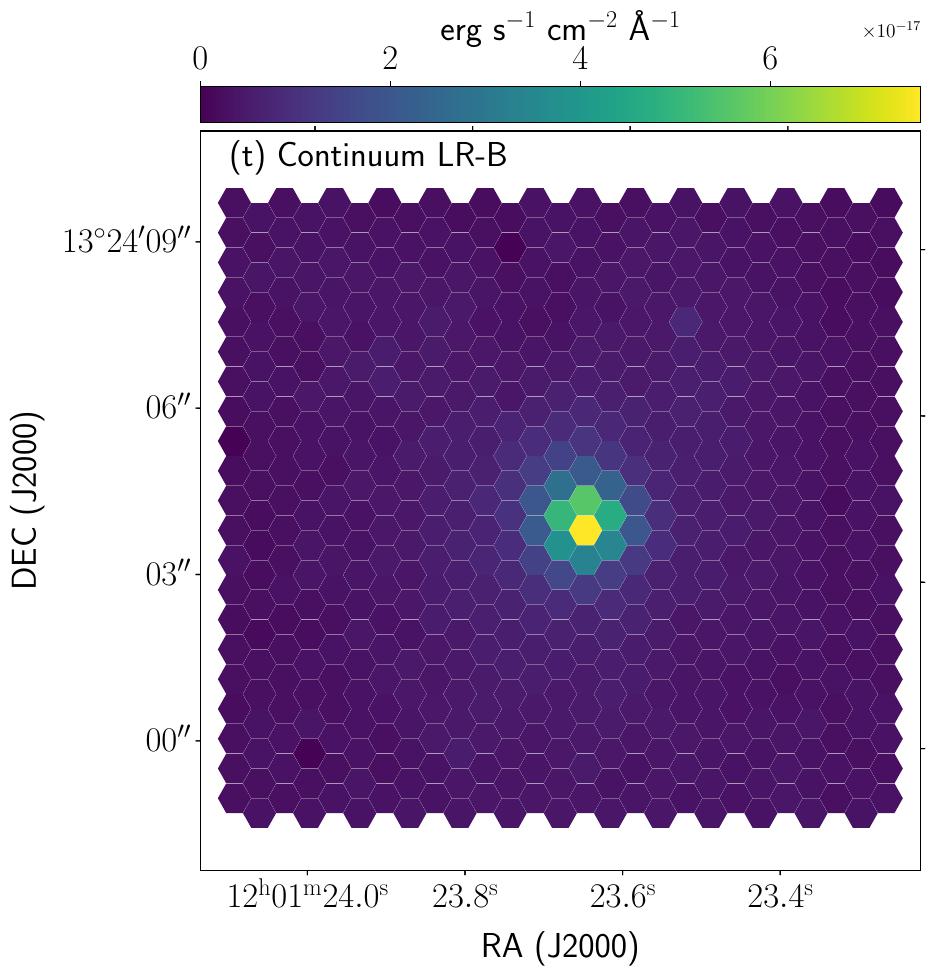}
	\includegraphics[clip, width=0.24\linewidth]{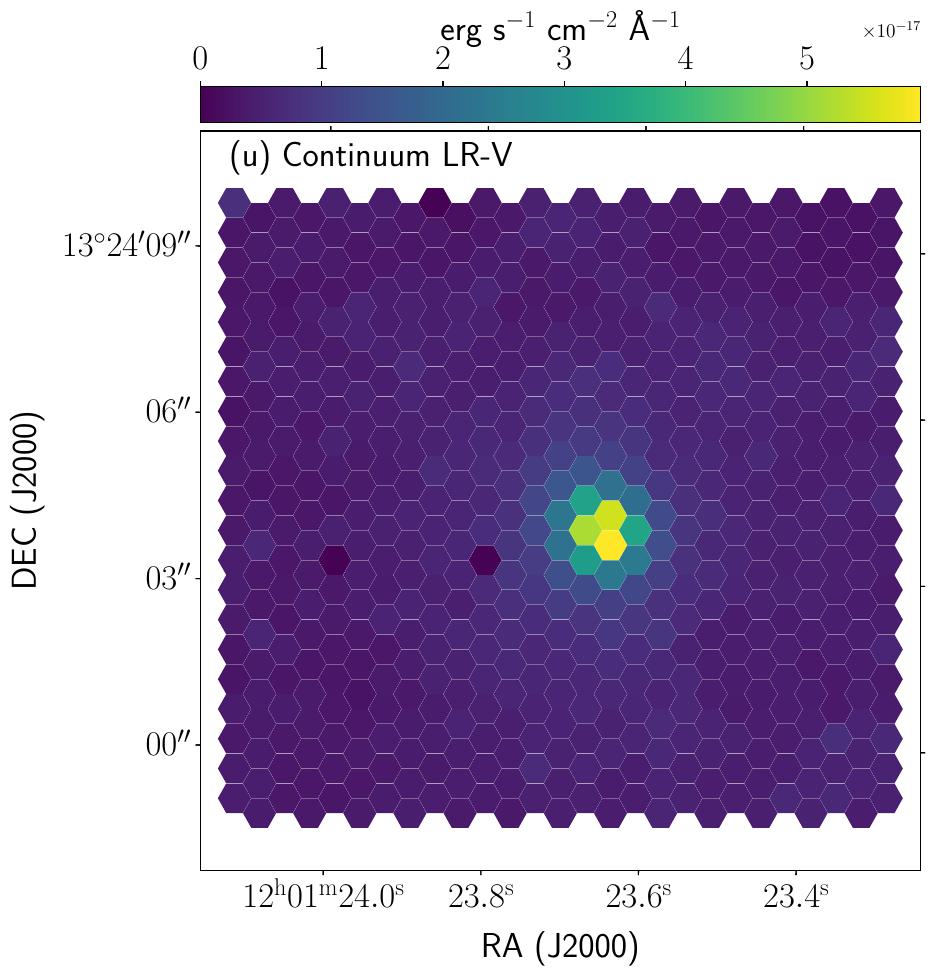}
	\includegraphics[clip, width=0.24\linewidth]{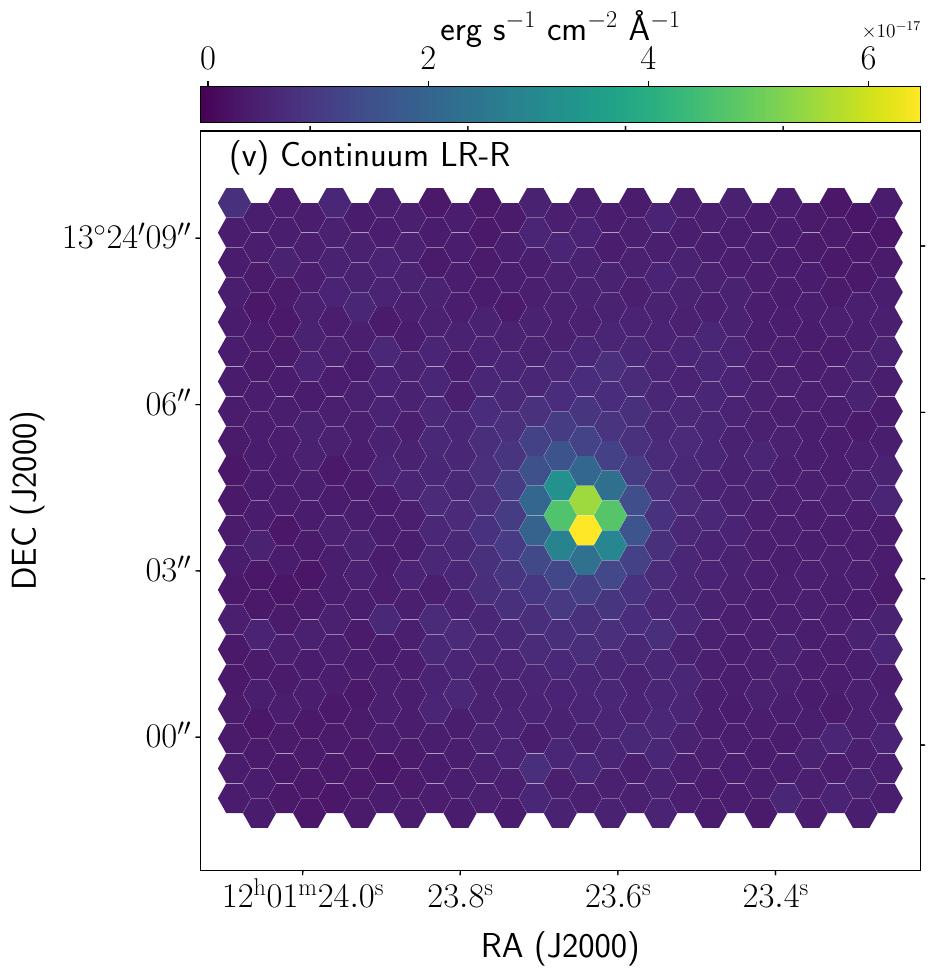}
	\includegraphics[clip, width=0.24\linewidth]{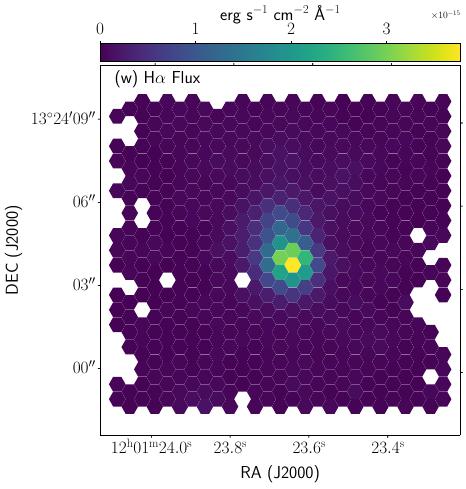}
	\includegraphics[clip, width=0.24\linewidth]{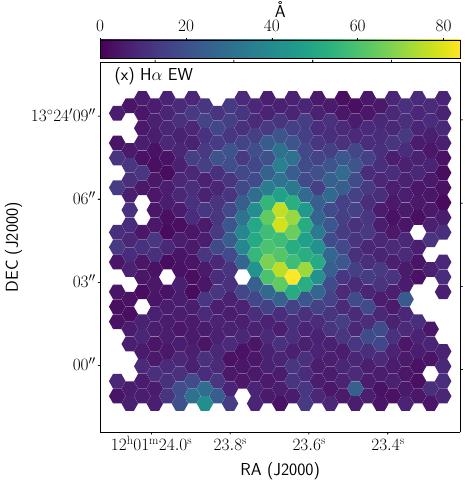}
	\includegraphics[clip, width=0.24\linewidth]{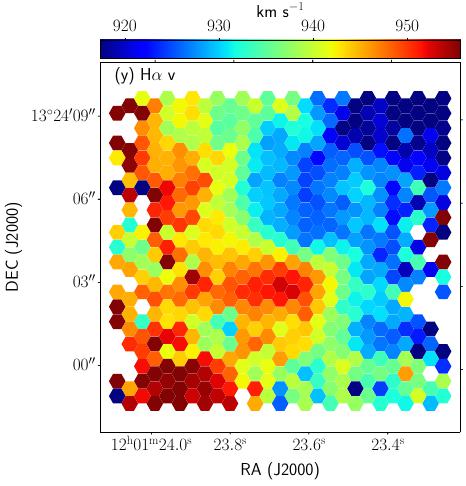}
	\includegraphics[clip, width=0.24\linewidth]{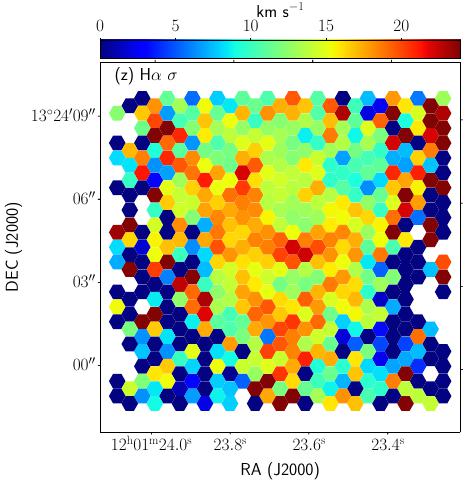}
	\includegraphics[clip, width=0.24\linewidth]{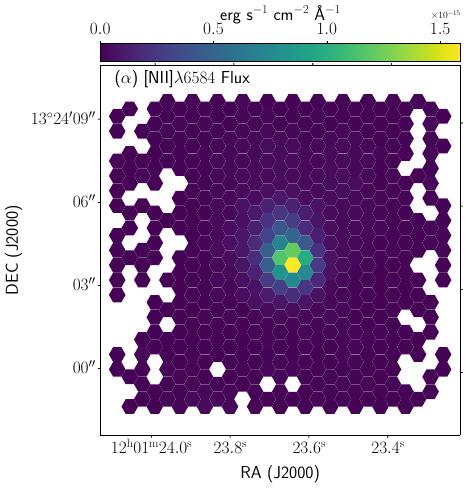}
	\includegraphics[clip, width=0.24\linewidth]{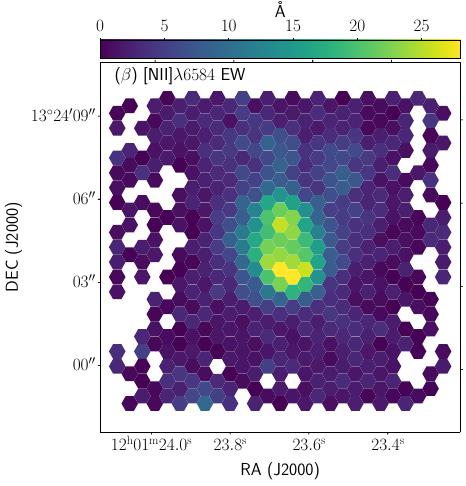}
	\includegraphics[clip, width=0.24\linewidth]{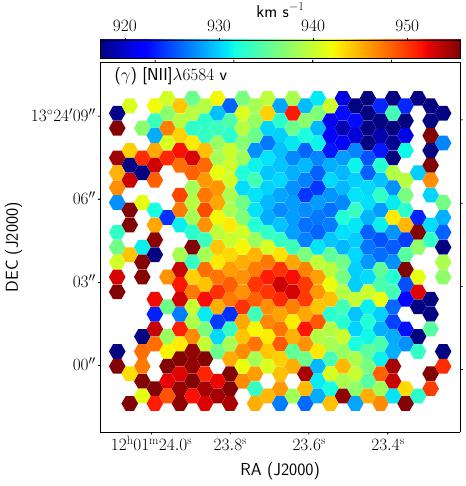}
	\includegraphics[clip, width=0.24\linewidth]{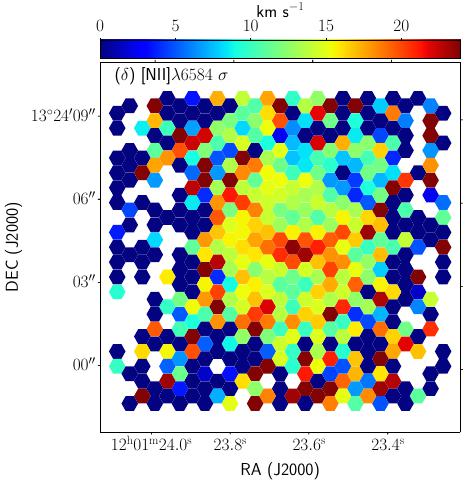}
	\includegraphics[clip, width=0.24\linewidth]{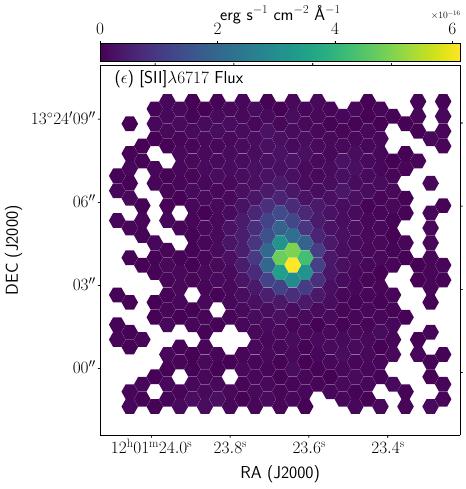}
	\includegraphics[clip, width=0.24\linewidth]{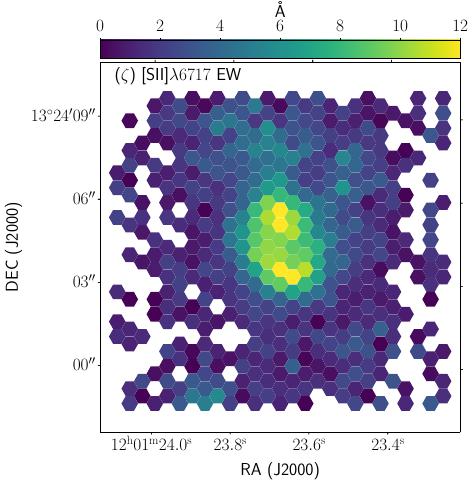}
	\includegraphics[clip, width=0.24\linewidth]{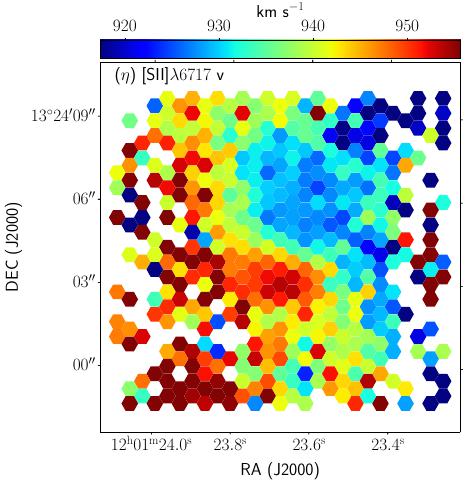}
	\includegraphics[clip, width=0.24\linewidth]{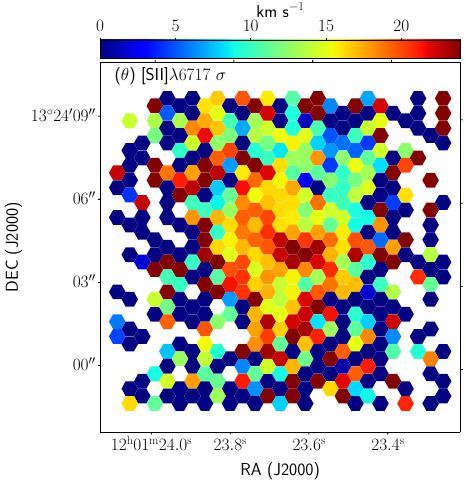}
	\includegraphics[clip, width=0.24\linewidth]{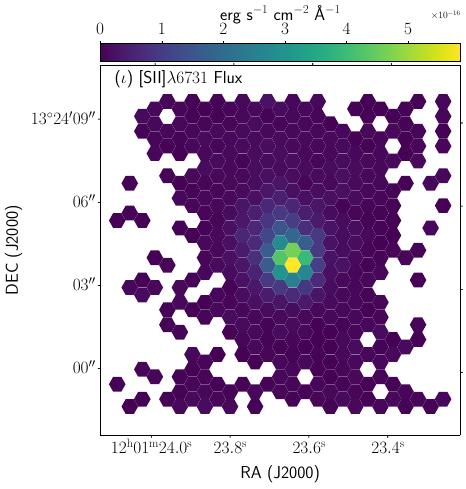}
	\includegraphics[clip, width=0.24\linewidth]{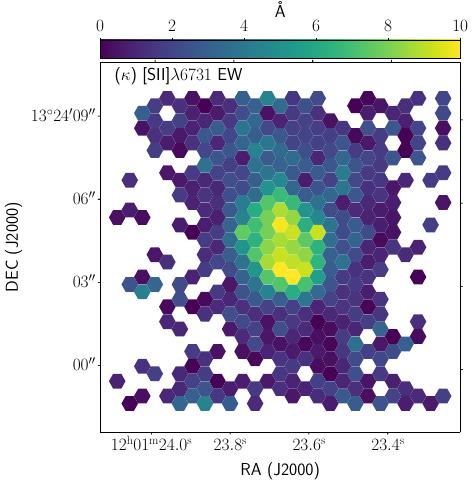}
	\includegraphics[clip, width=0.24\linewidth]{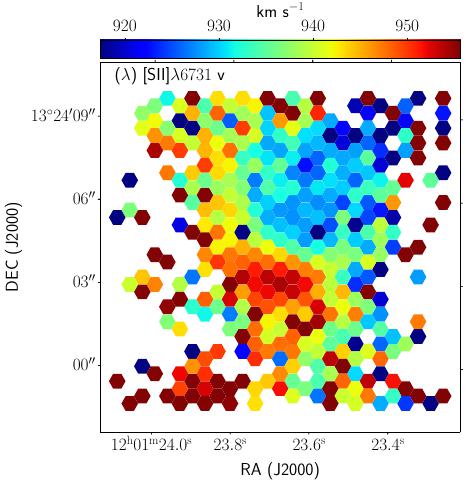}
	\includegraphics[clip, width=0.24\linewidth]{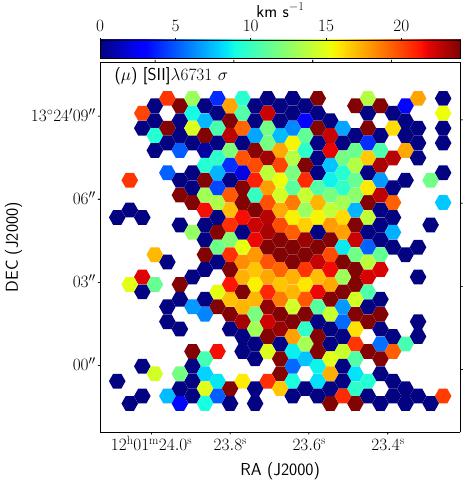}
	\caption{(cont.) NGC~4037 card.}
	\label{fig:NGC4037_card_2}
\end{figure*}

\begin{figure*}[h]
	\centering
	\includegraphics[clip, width=0.35\linewidth]{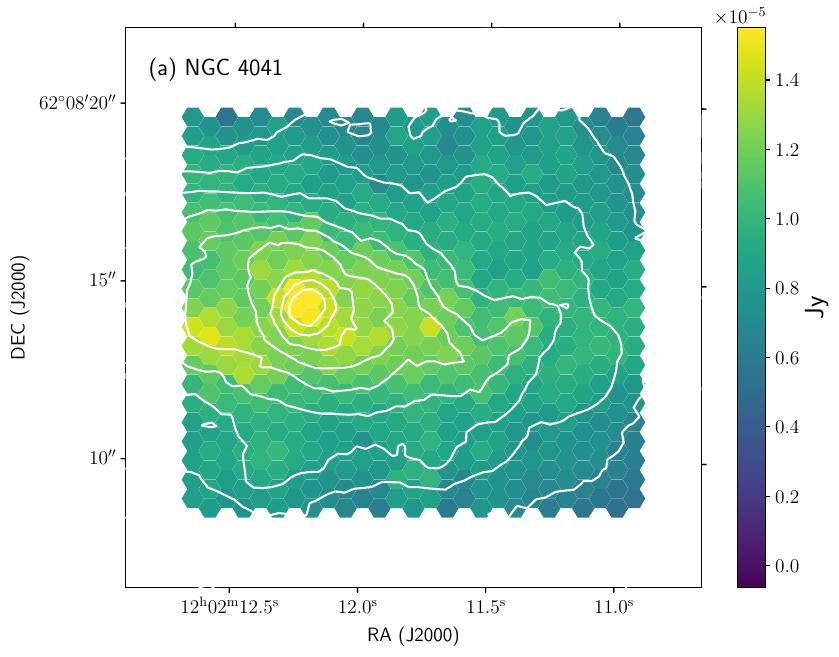}
	\includegraphics[clip, width=0.6\linewidth]{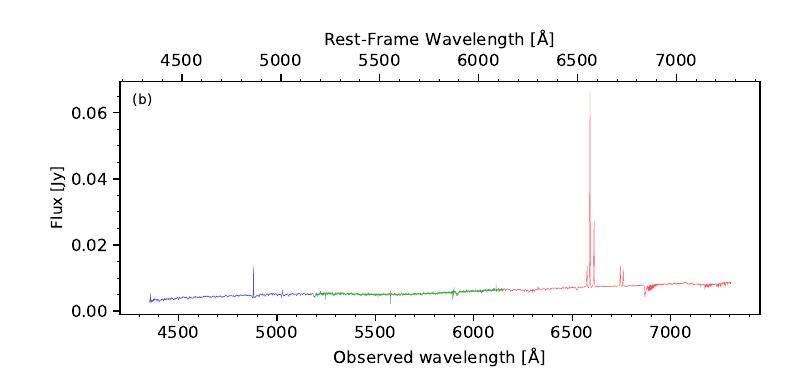}
	\includegraphics[clip, width=0.24\linewidth]{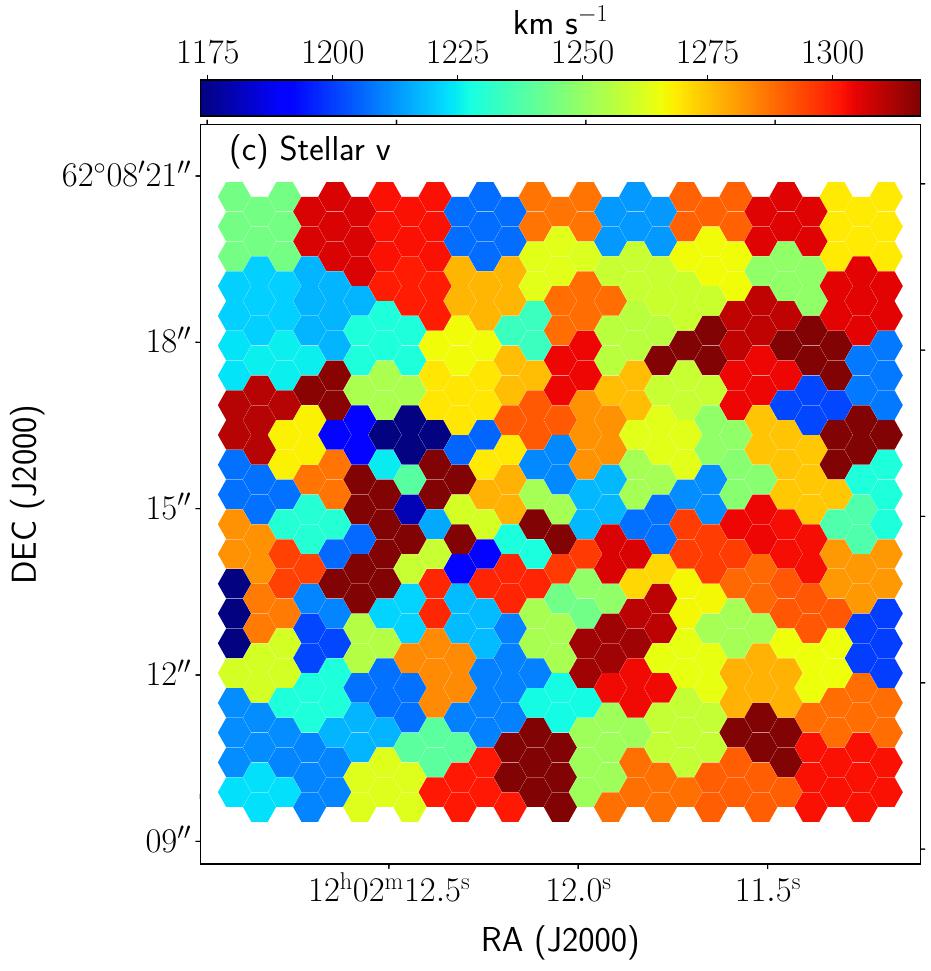}
	\includegraphics[clip, width=0.24\linewidth]{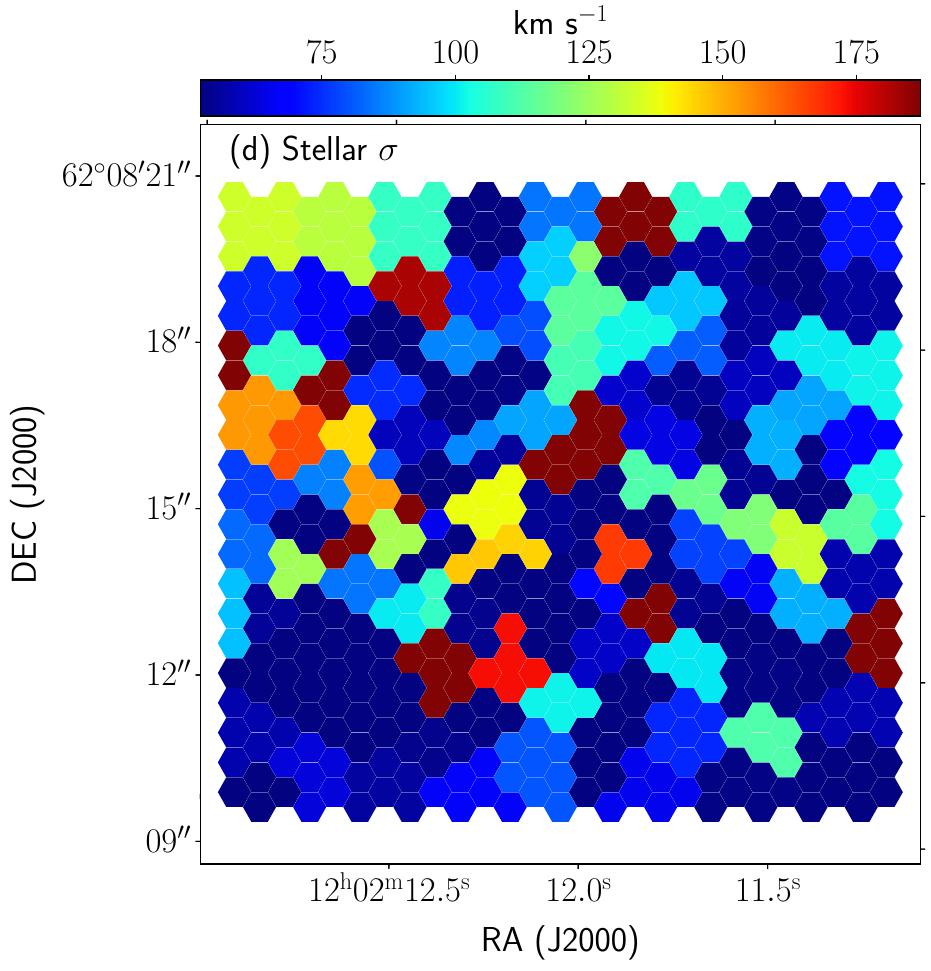}
	\includegraphics[clip, width=0.24\linewidth]{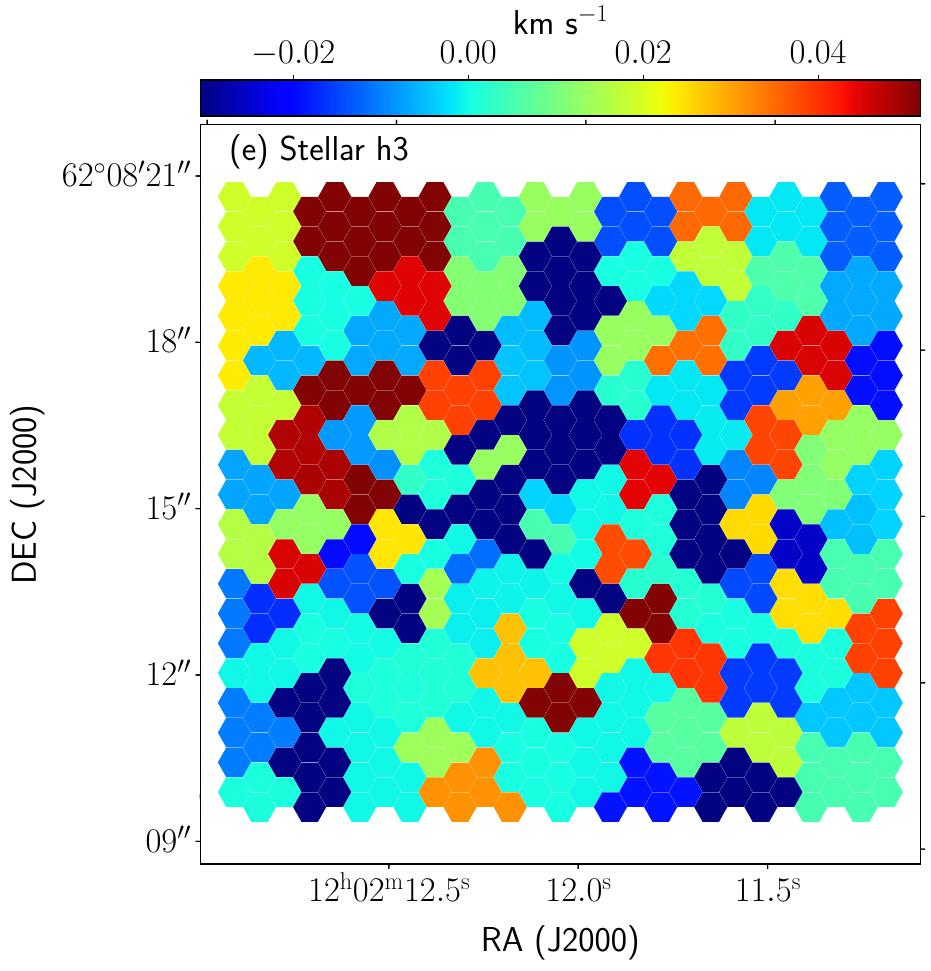}
	\includegraphics[clip, width=0.24\linewidth]{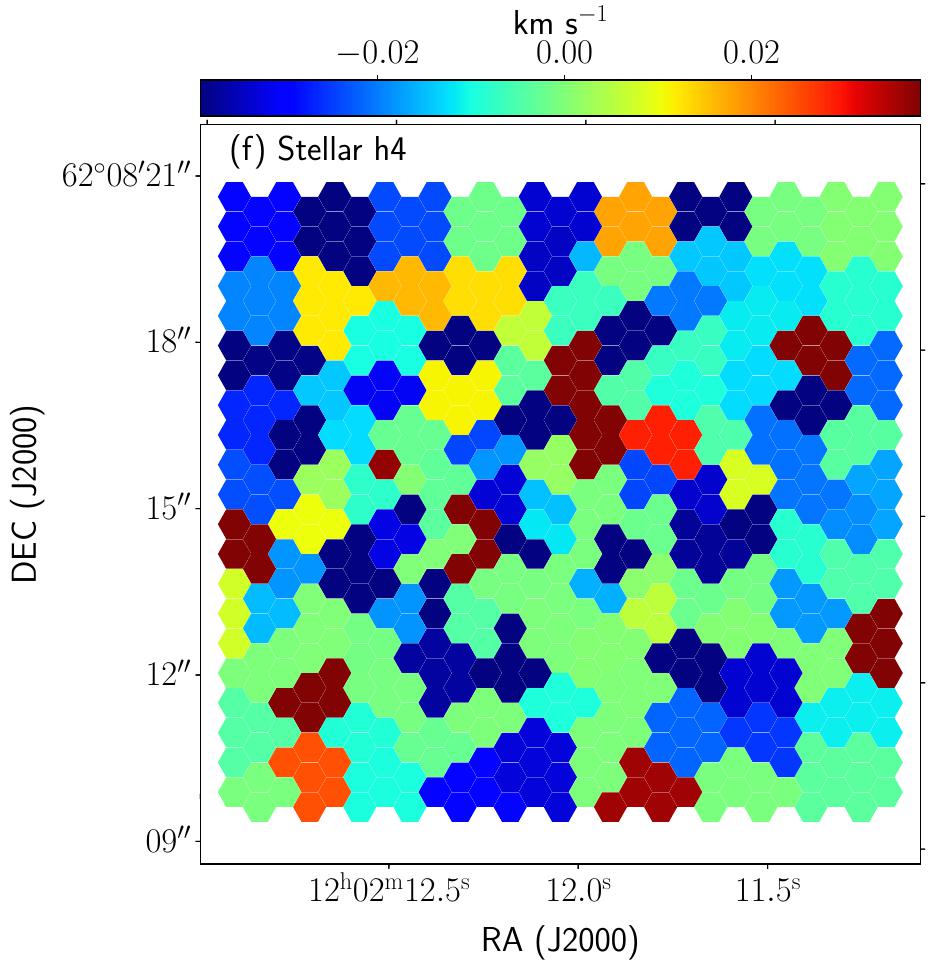}
	\includegraphics[clip, width=0.24\linewidth]{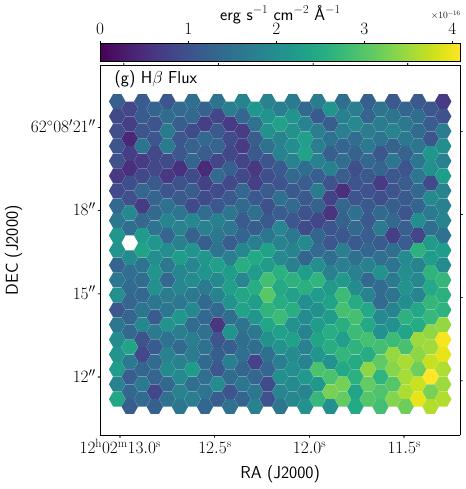}
	\includegraphics[clip, width=0.24\linewidth]{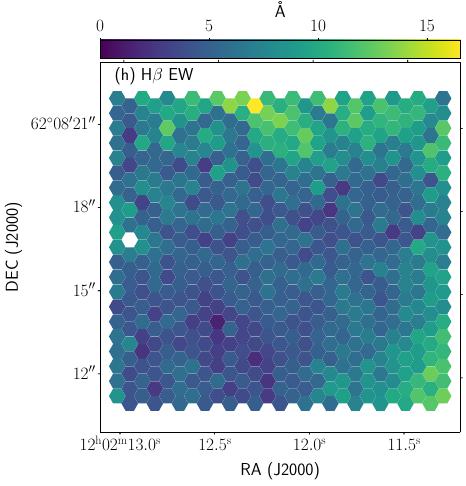}
	\includegraphics[clip, width=0.24\linewidth]{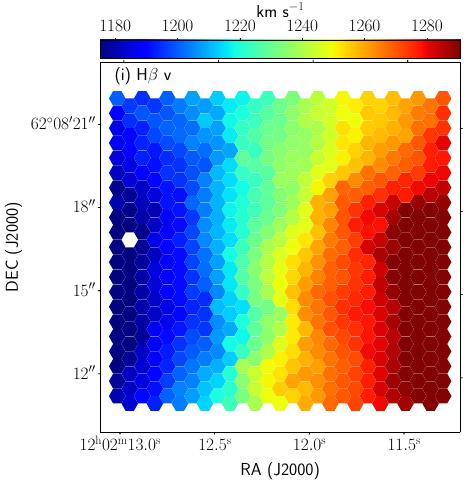}
	\includegraphics[clip, width=0.24\linewidth]{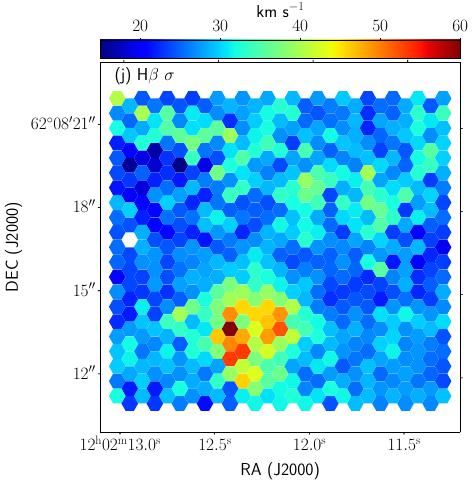}
	\includegraphics[clip, width=0.24\linewidth]{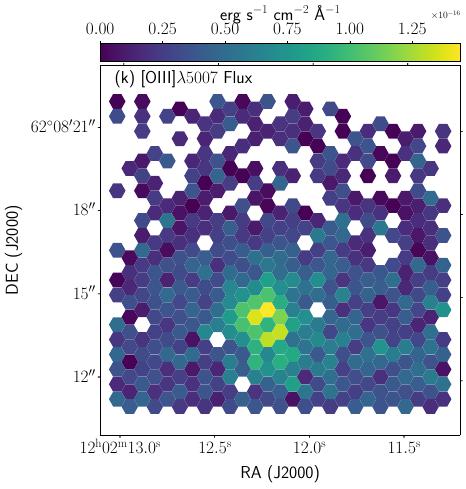}
	\includegraphics[clip, width=0.24\linewidth]{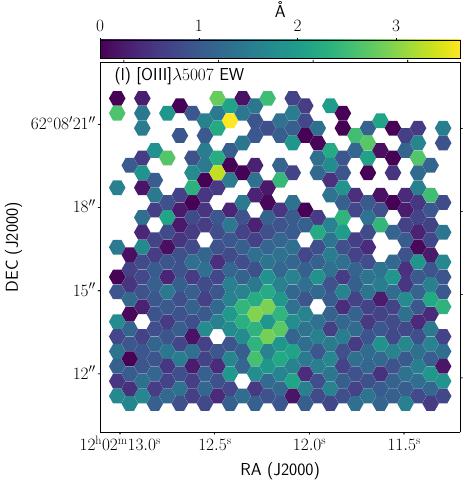}
	\includegraphics[clip, width=0.24\linewidth]{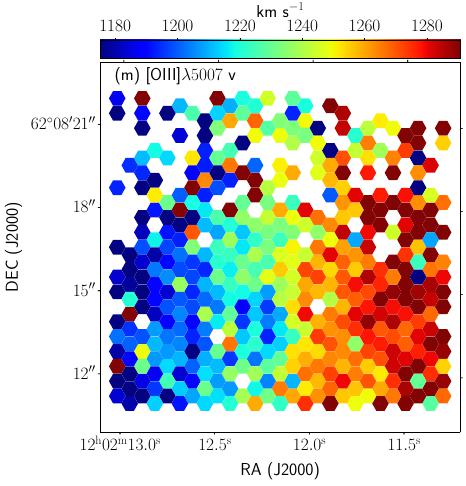}
	\includegraphics[clip, width=0.24\linewidth]{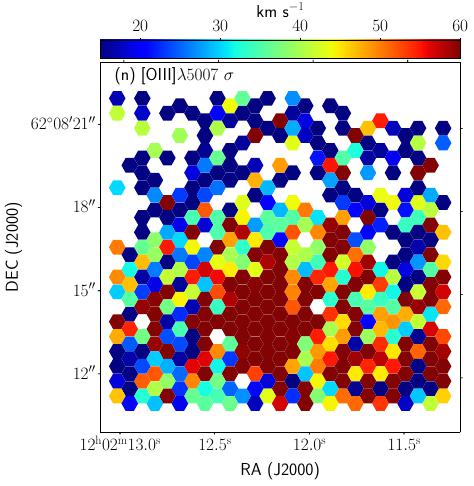}
	\vspace{5cm}
	\caption{NGC~4041 card.}
	\label{fig:NGC4041_card_1}
\end{figure*}
\addtocounter{figure}{-1}
\begin{figure*}[h]
	\centering
	\includegraphics[clip, width=0.24\linewidth]{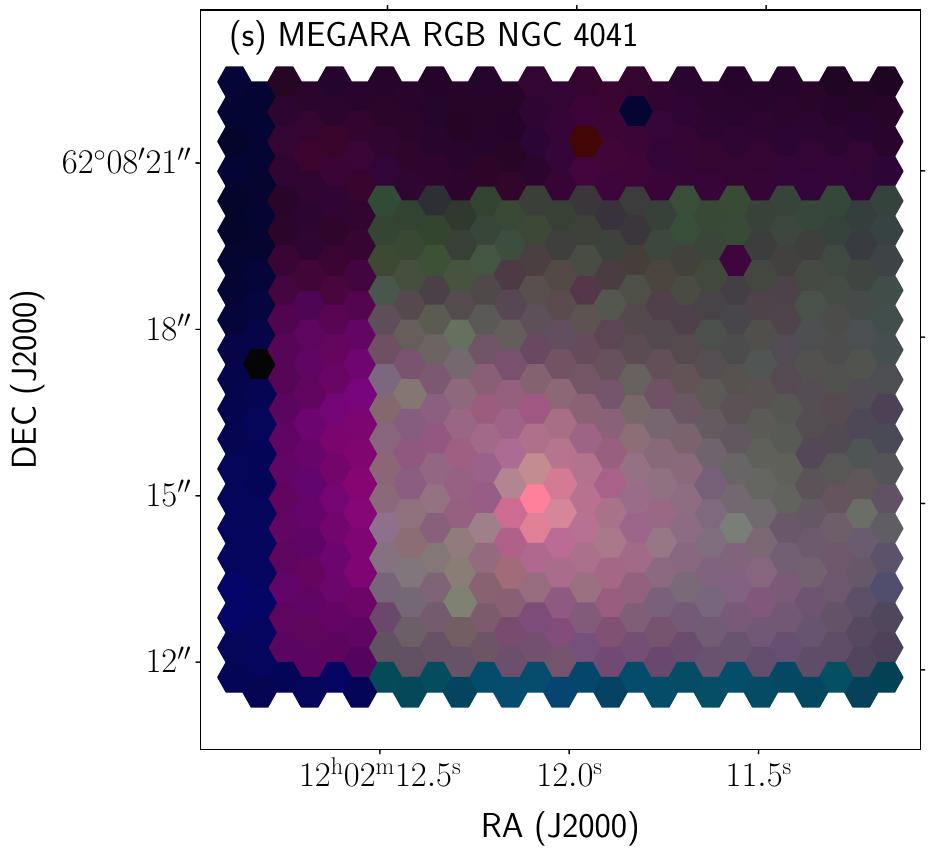}
	\includegraphics[clip, width=0.24\linewidth]{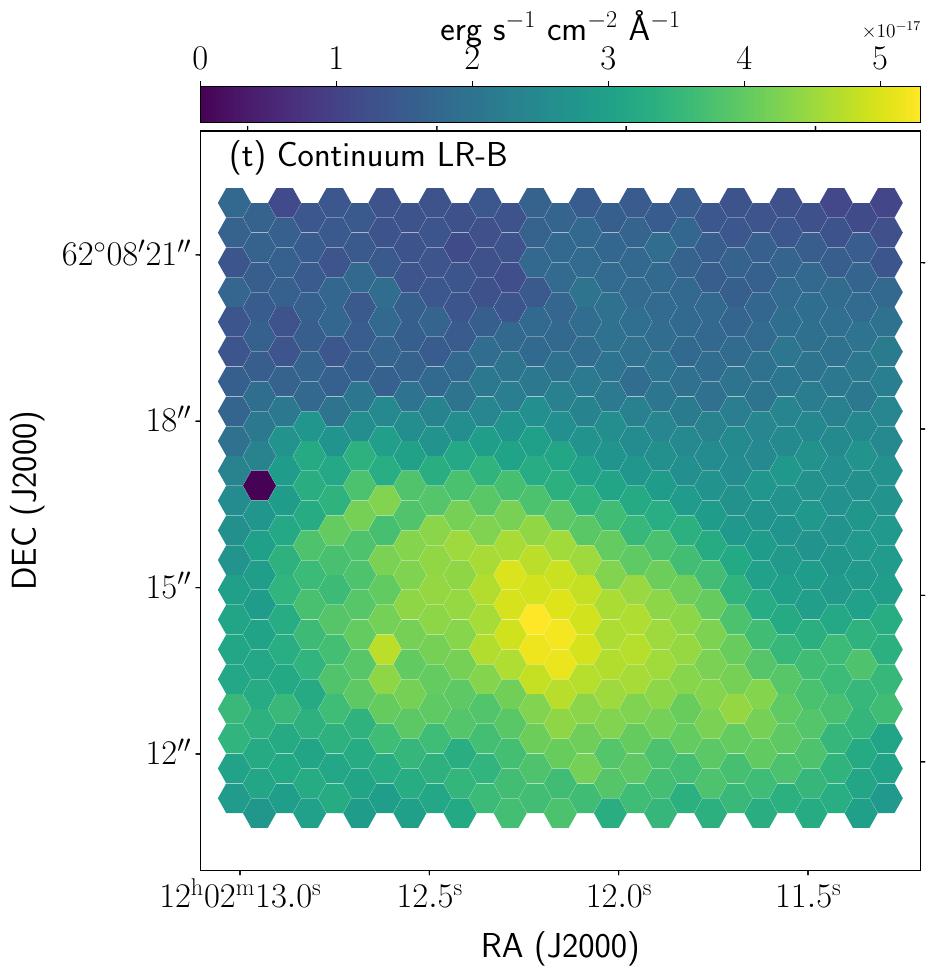}
	\includegraphics[clip, width=0.24\linewidth]{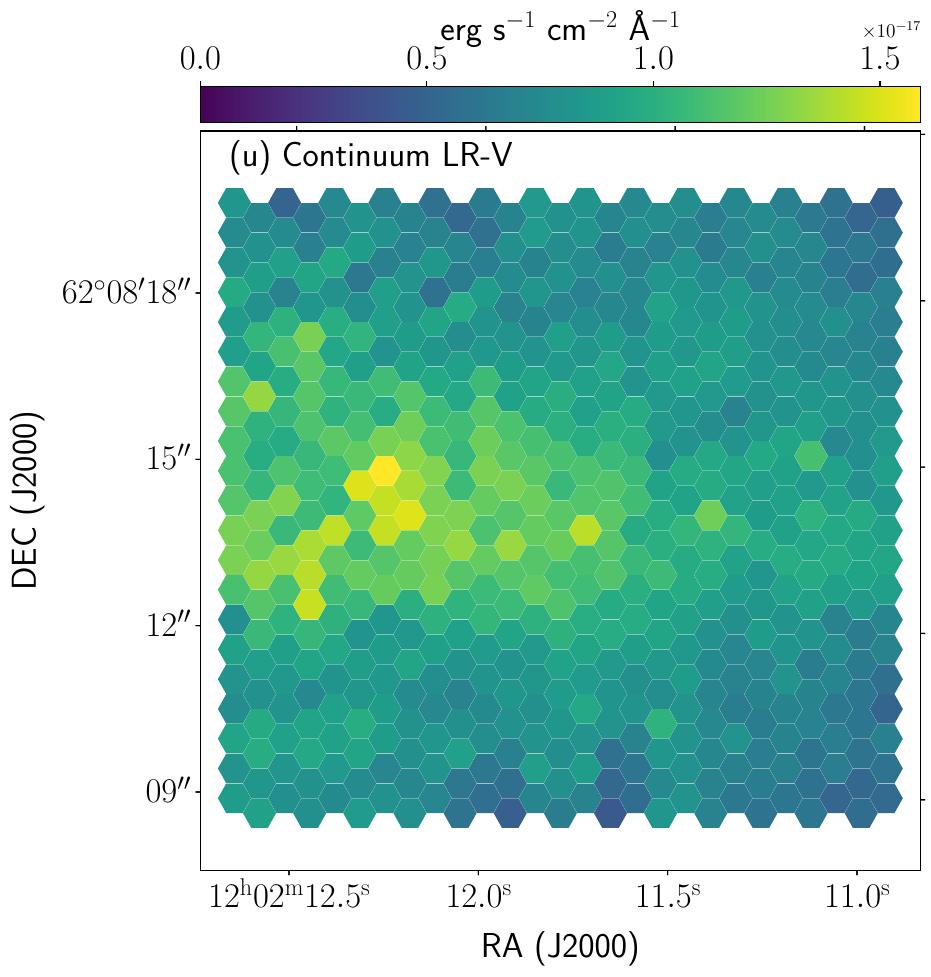}
	\includegraphics[clip, width=0.24\linewidth]{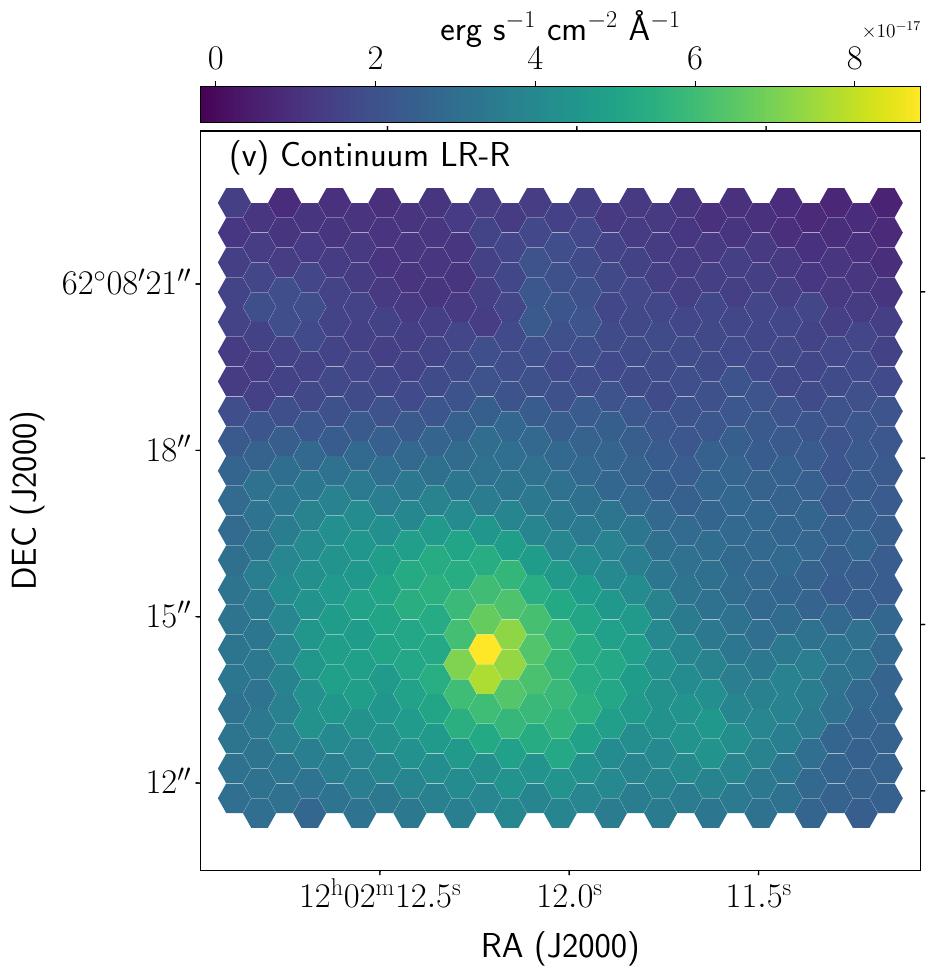}
	\includegraphics[clip, width=0.24\linewidth]{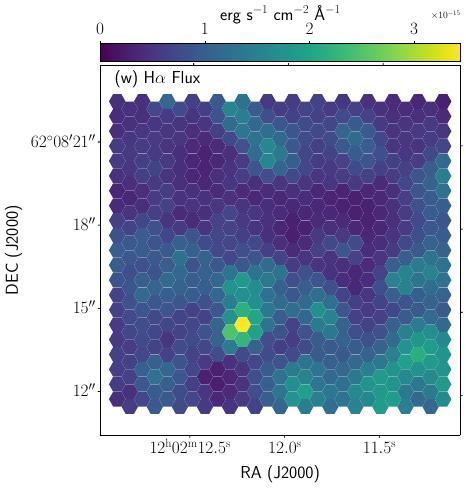}
	\includegraphics[clip, width=0.24\linewidth]{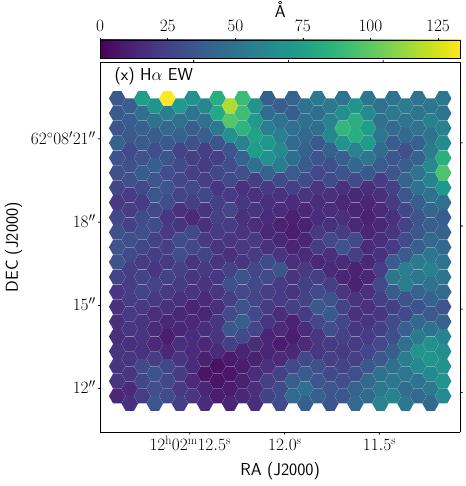}
	\includegraphics[clip, width=0.24\linewidth]{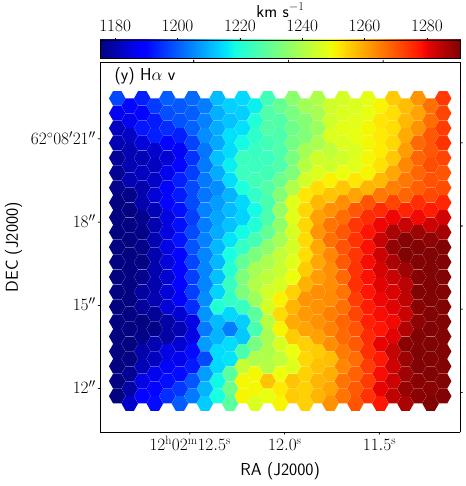}
	\includegraphics[clip, width=0.24\linewidth]{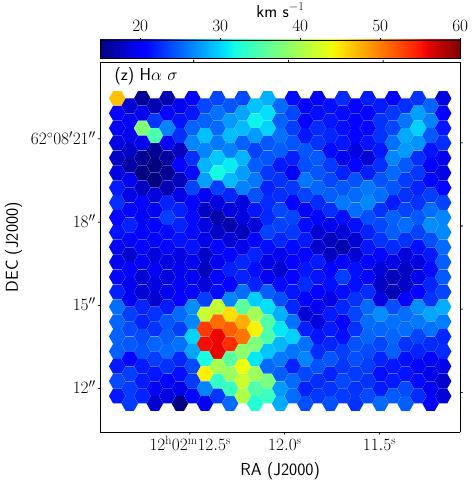}
	\includegraphics[clip, width=0.24\linewidth]{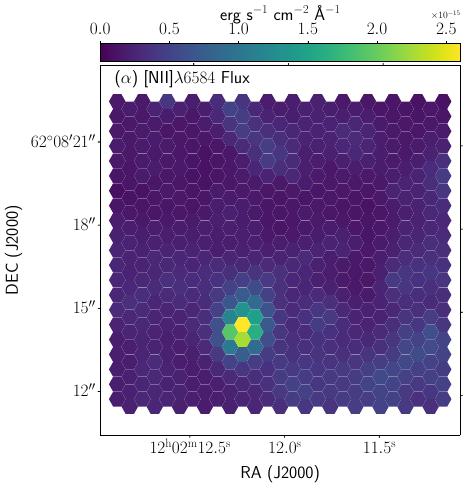}
	\includegraphics[clip, width=0.24\linewidth]{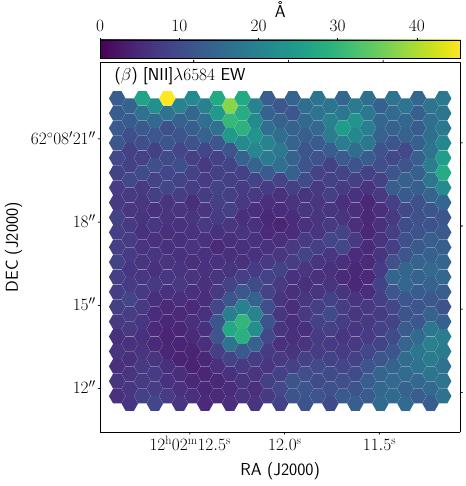}
	\includegraphics[clip, width=0.24\linewidth]{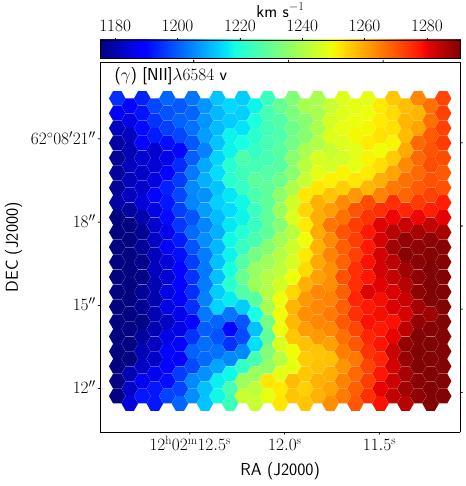}
	\includegraphics[clip, width=0.24\linewidth]{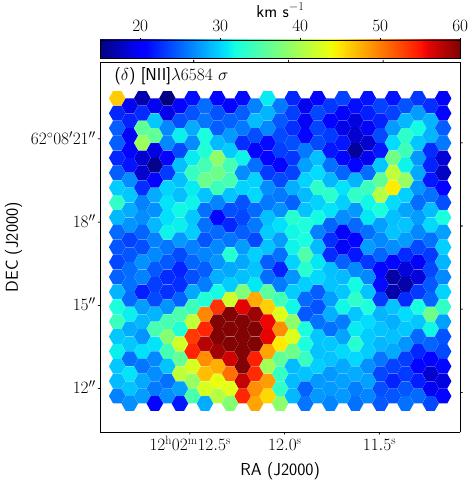}
	\includegraphics[clip, width=0.24\linewidth]{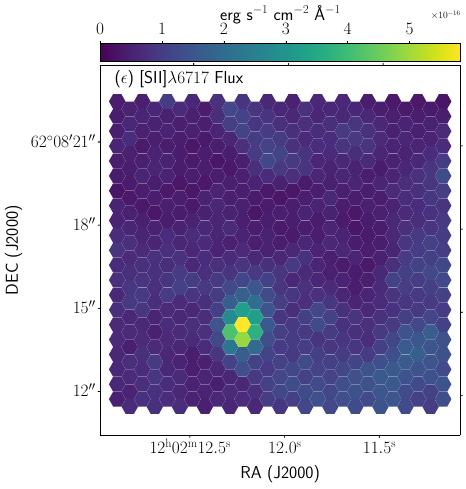}
	\includegraphics[clip, width=0.24\linewidth]{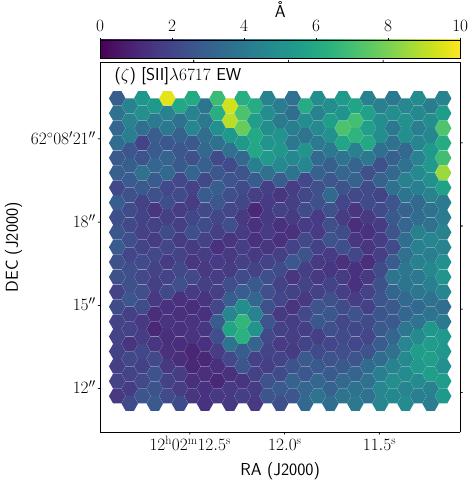}
	\includegraphics[clip, width=0.24\linewidth]{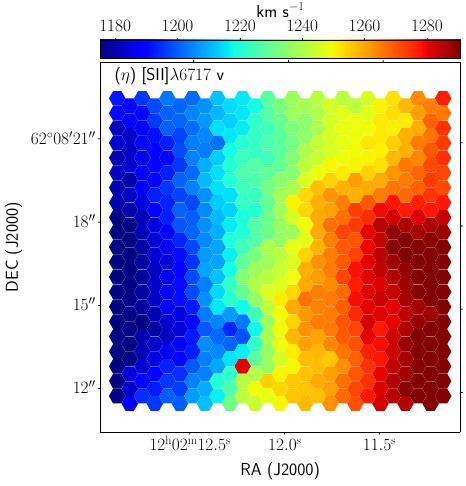}
	\includegraphics[clip, width=0.24\linewidth]{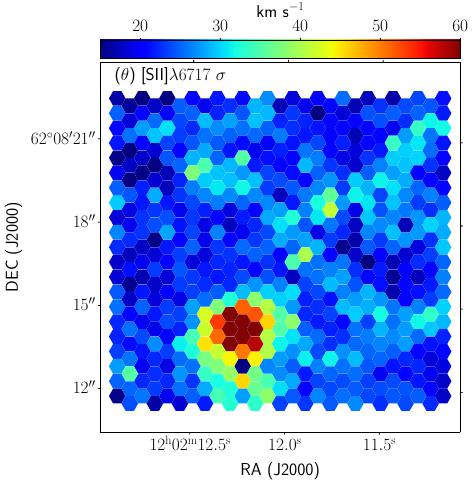}
	\includegraphics[clip, width=0.24\linewidth]{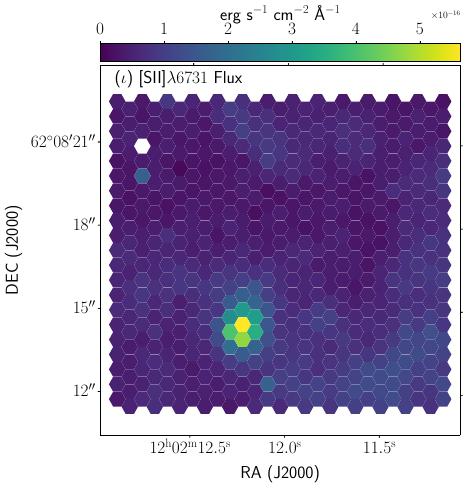}
	\includegraphics[clip, width=0.24\linewidth]{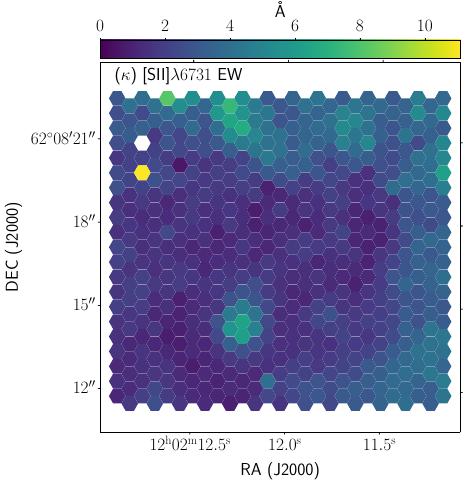}
	\includegraphics[clip, width=0.24\linewidth]{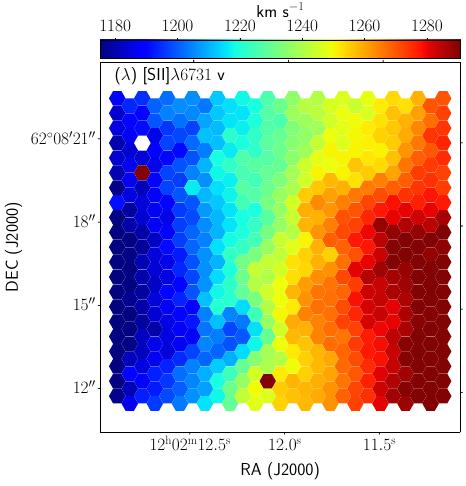}
	\includegraphics[clip, width=0.24\linewidth]{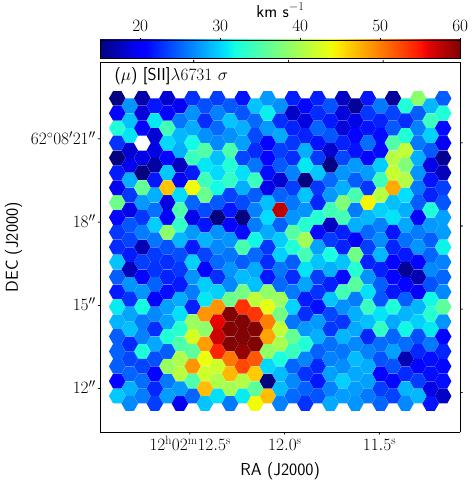}
	\caption{(cont.) NGC~4041 card.}
	\label{fig:NGC4041_card_2}
\end{figure*}

\begin{figure*}[h]
	\centering
	\includegraphics[clip, width=0.35\linewidth]{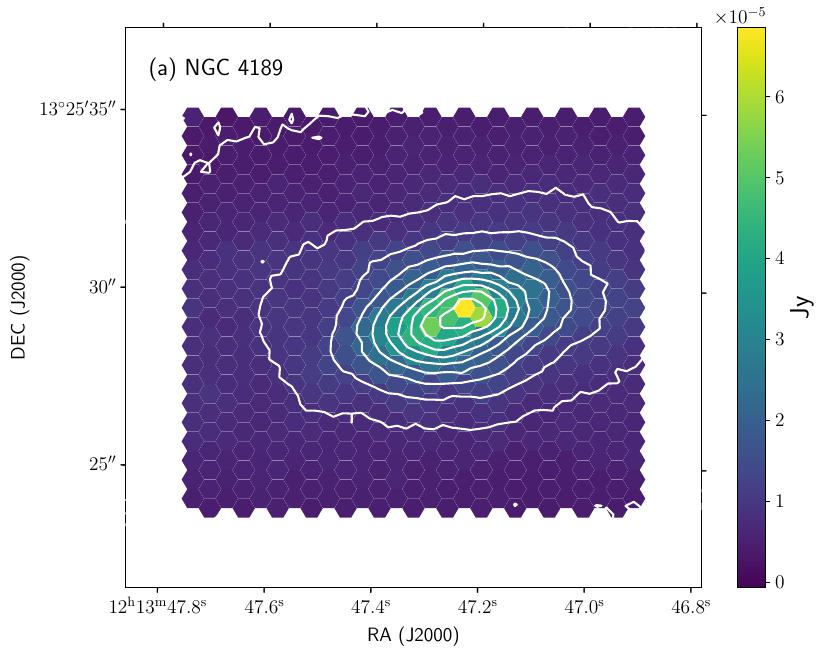}
	\includegraphics[clip, width=0.6\linewidth]{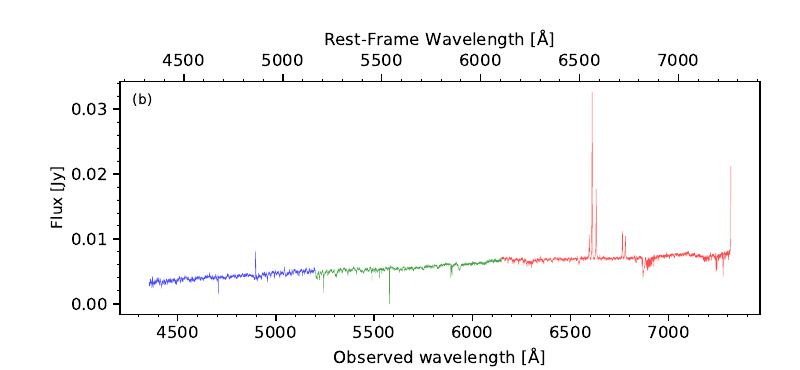}
	\includegraphics[clip, width=0.24\linewidth]{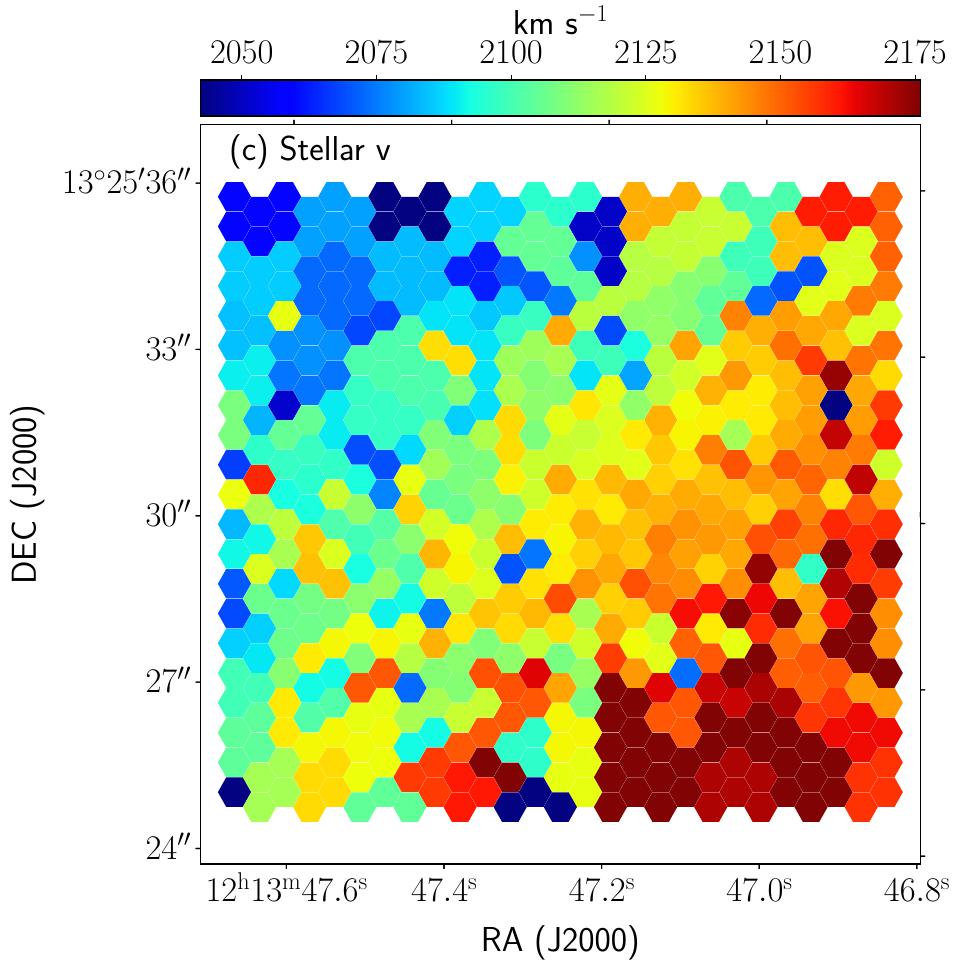}
	\includegraphics[clip, width=0.24\linewidth]{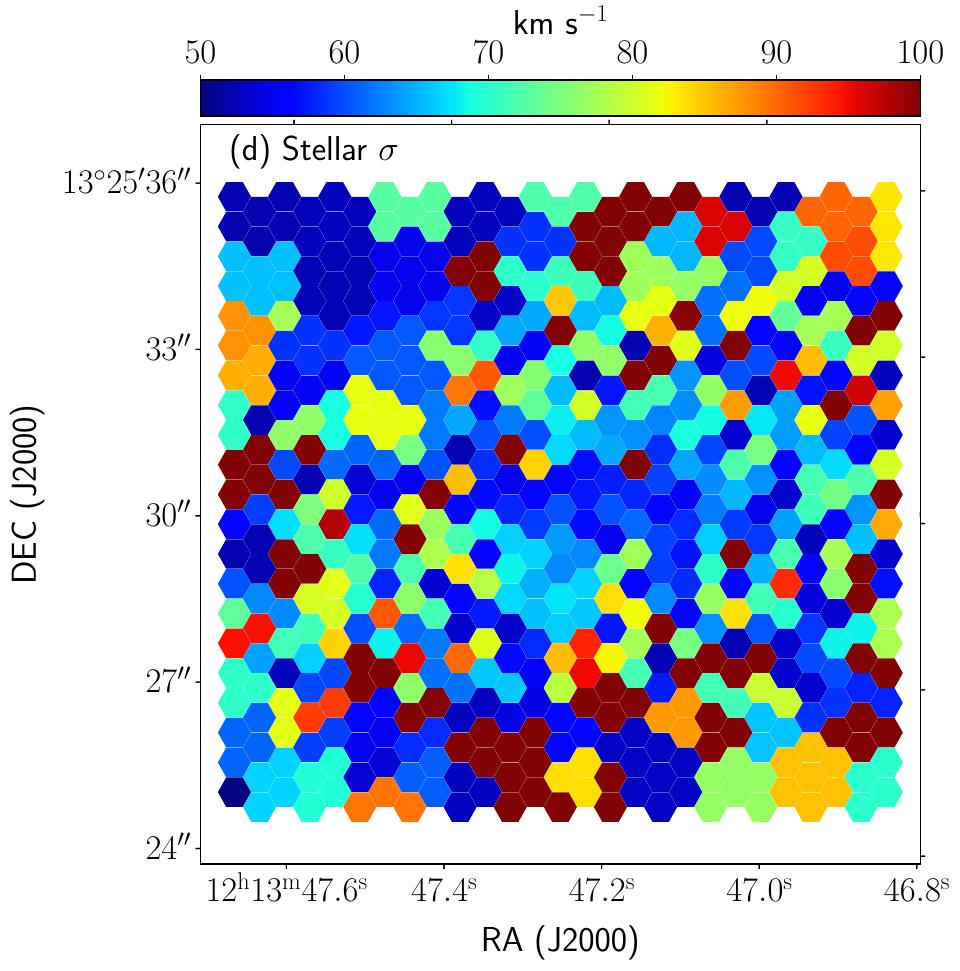}
	\includegraphics[clip, width=0.24\linewidth]{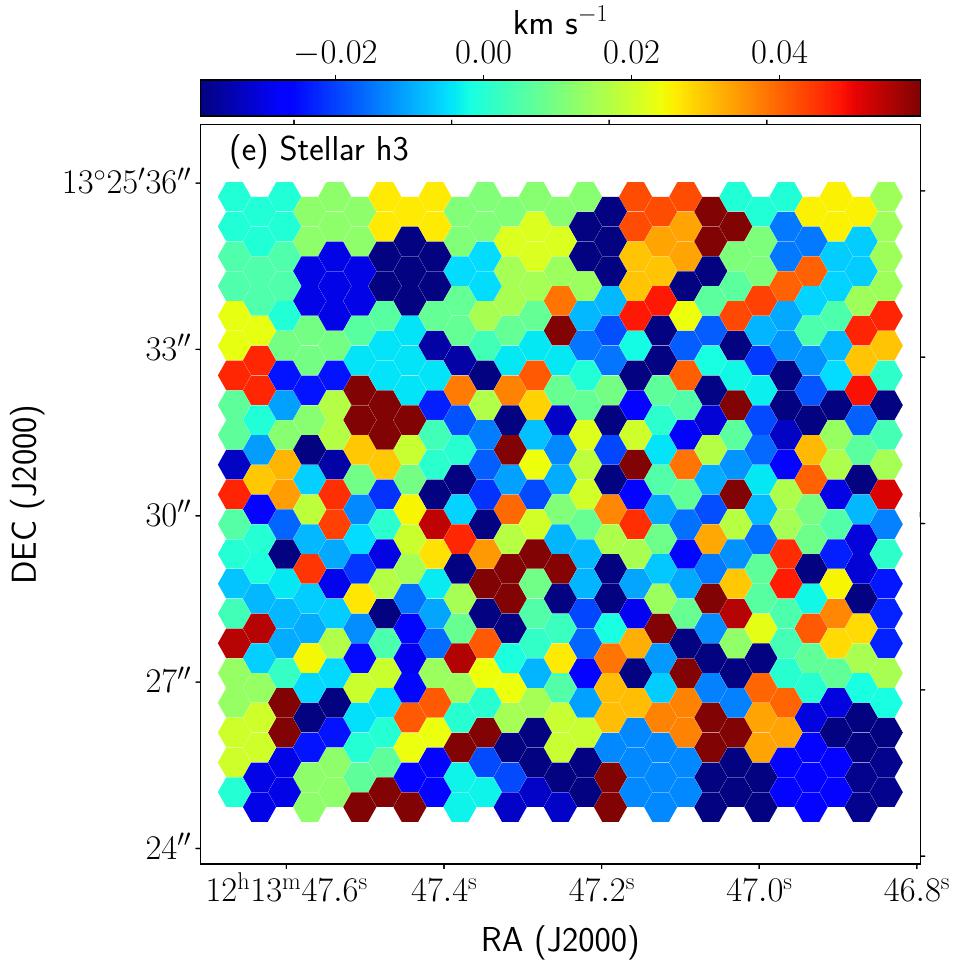}
	\includegraphics[clip, width=0.24\linewidth]{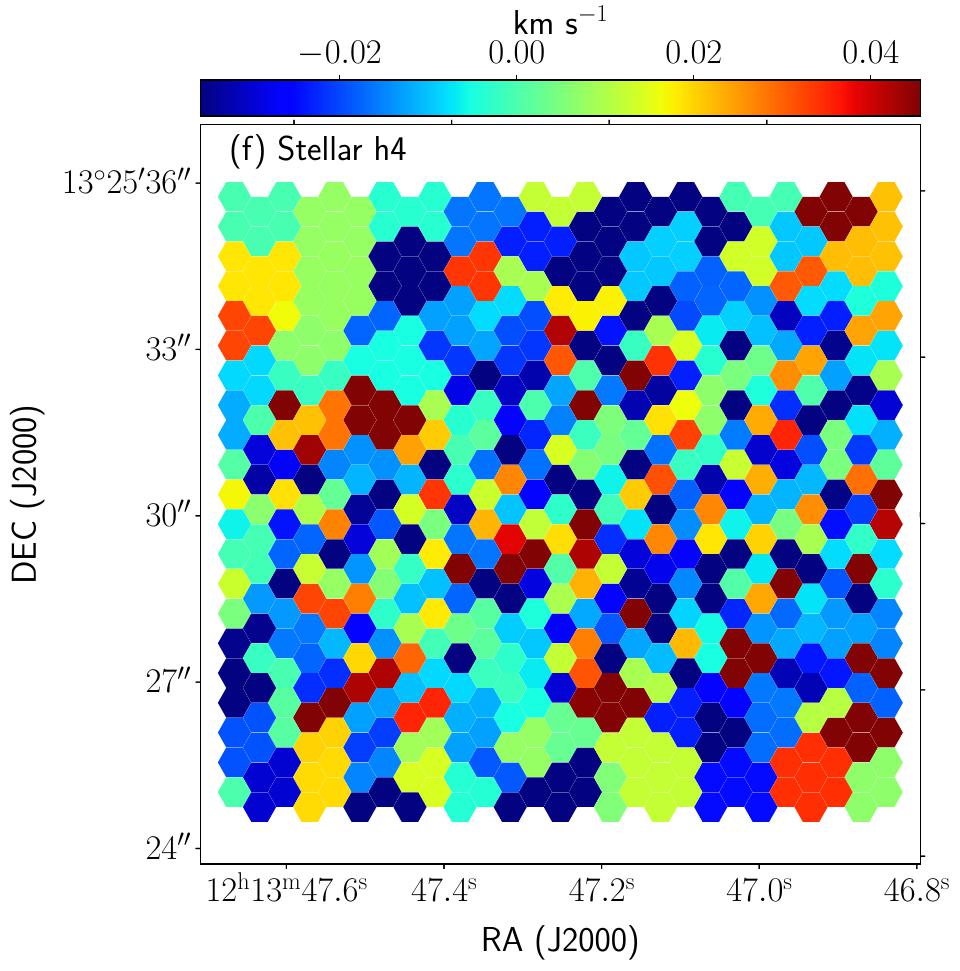}
	\includegraphics[clip, width=0.24\linewidth]{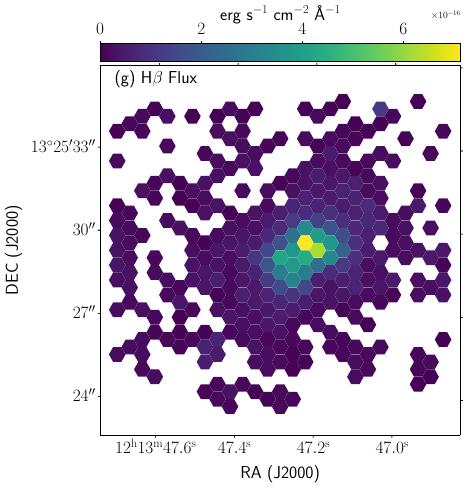}
	\includegraphics[clip, width=0.24\linewidth]{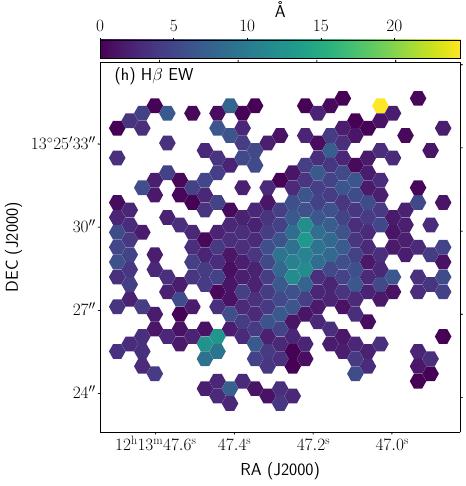}
	\includegraphics[clip, width=0.24\linewidth]{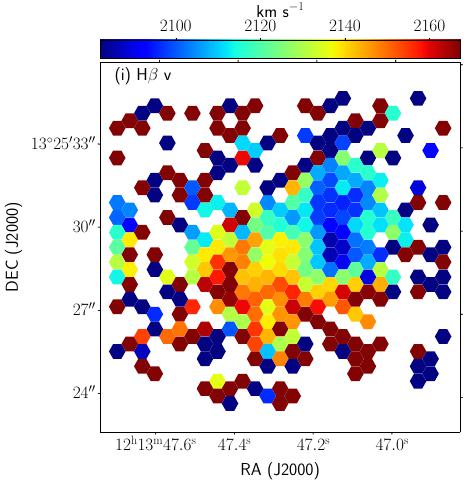}
	\includegraphics[clip, width=0.24\linewidth]{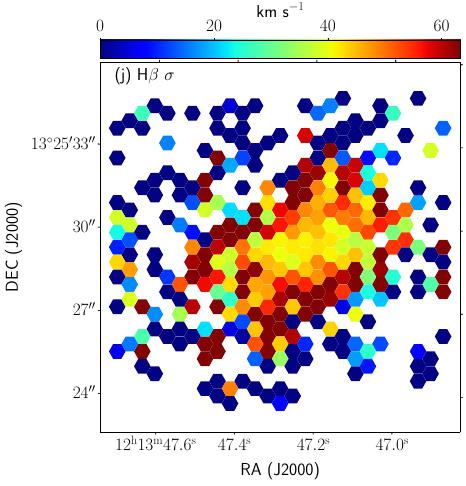}
	\includegraphics[clip, width=0.24\linewidth]{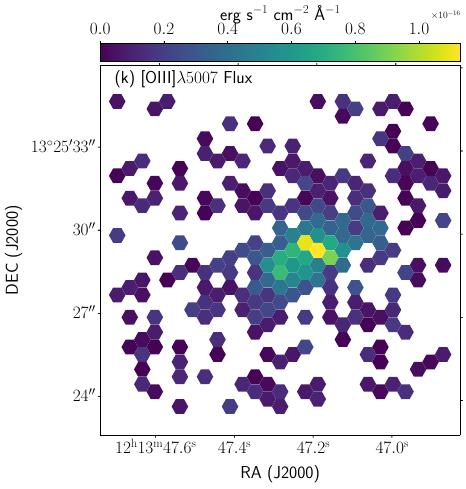}
	\includegraphics[clip, width=0.24\linewidth]{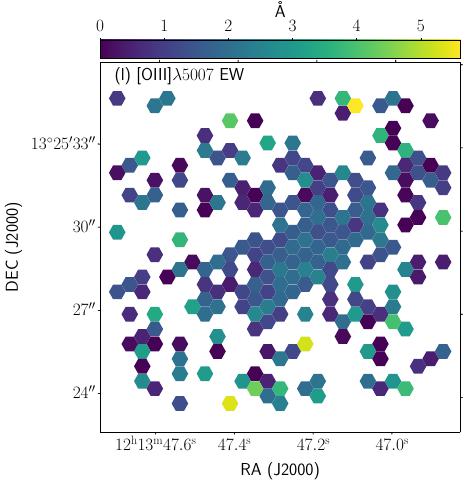}
	\includegraphics[clip, width=0.24\linewidth]{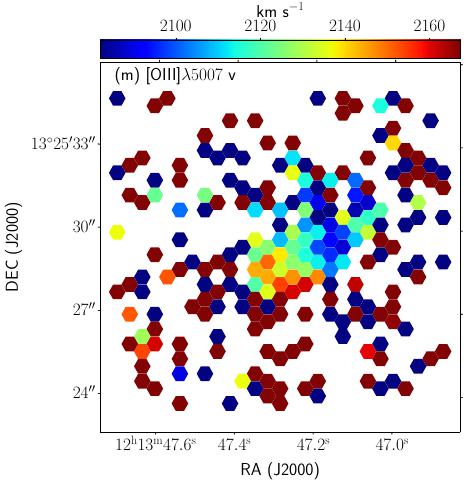}
	\includegraphics[clip, width=0.24\linewidth]{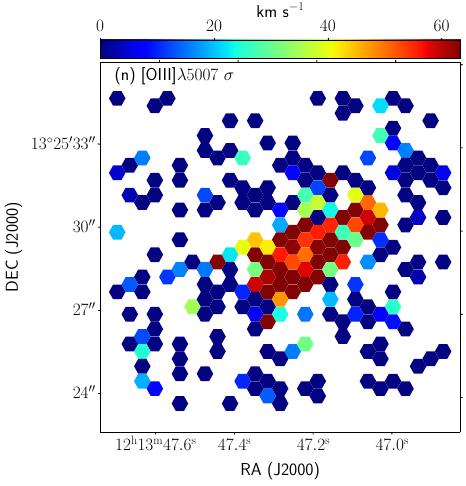}
	\includegraphics[clip, width=0.24\linewidth]{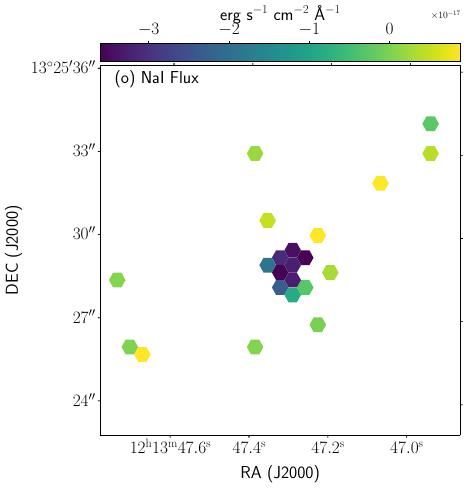}
	\includegraphics[clip, width=0.24\linewidth]{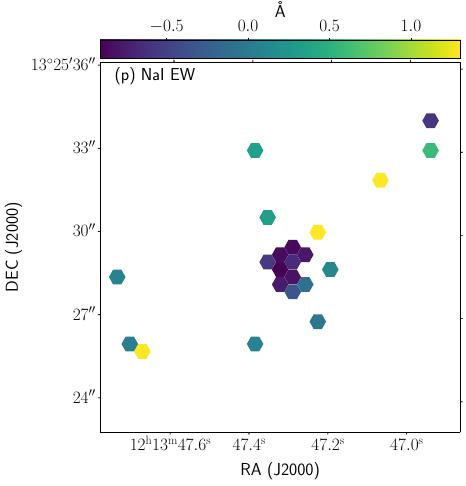}
	\includegraphics[clip, width=0.24\linewidth]{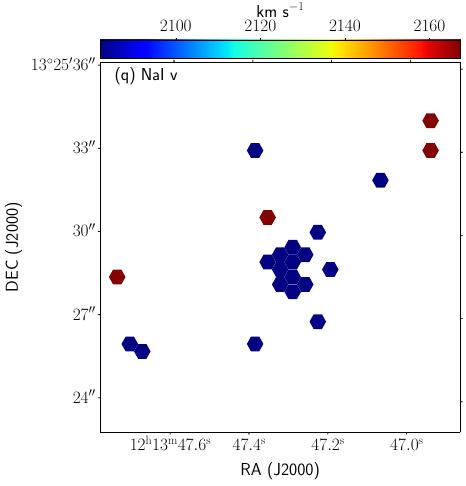}
	\includegraphics[clip, width=0.24\linewidth]{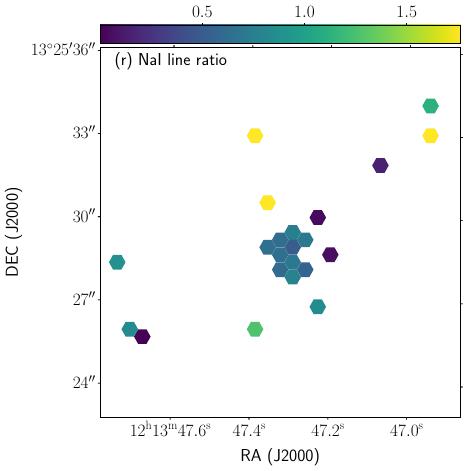}
	\caption{NGC~4189 card.}
	\label{fig:NGC4189_card_1}
\end{figure*}
\addtocounter{figure}{-1}
\begin{figure*}[h]
	\centering
	\includegraphics[clip, width=0.24\linewidth]{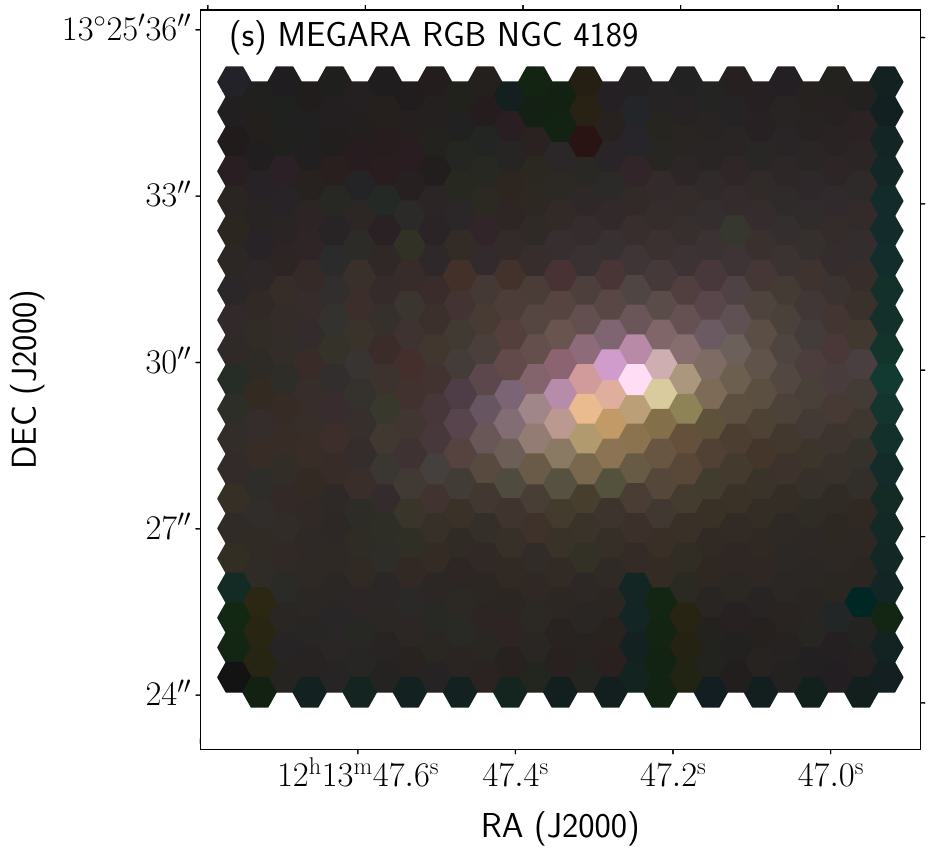}
	\includegraphics[clip, width=0.24\linewidth]{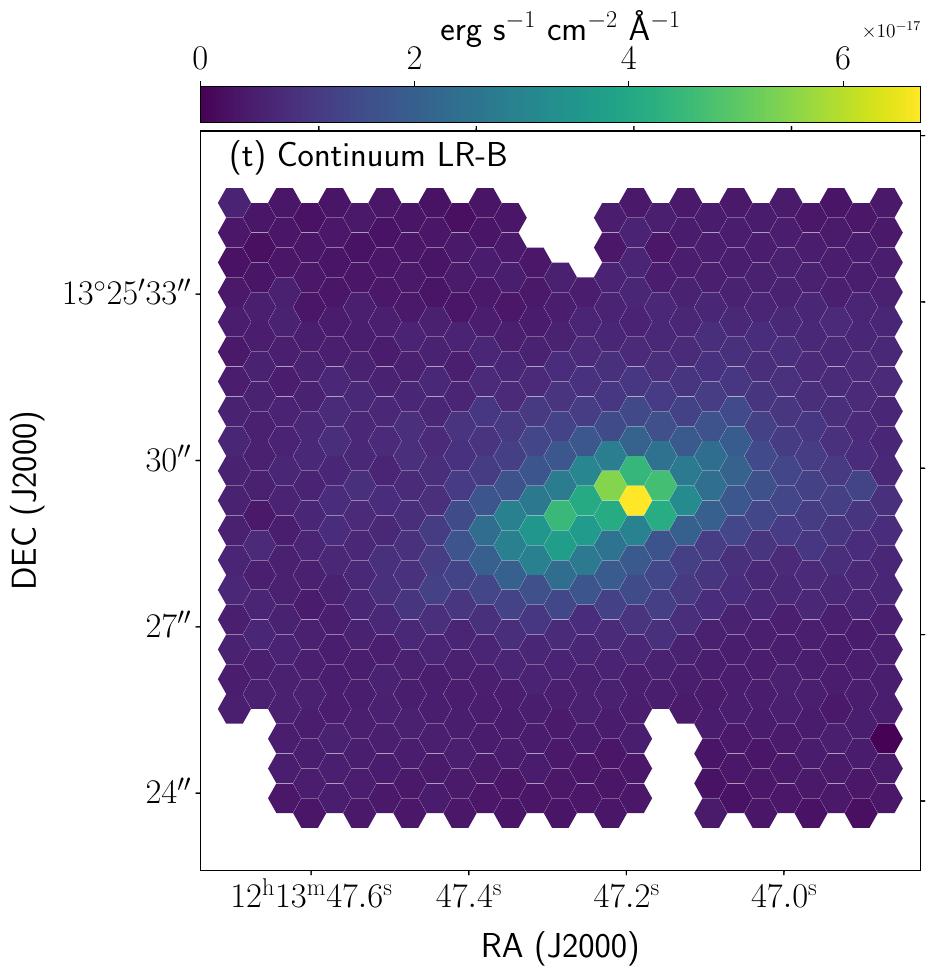}
	\includegraphics[clip, width=0.24\linewidth]{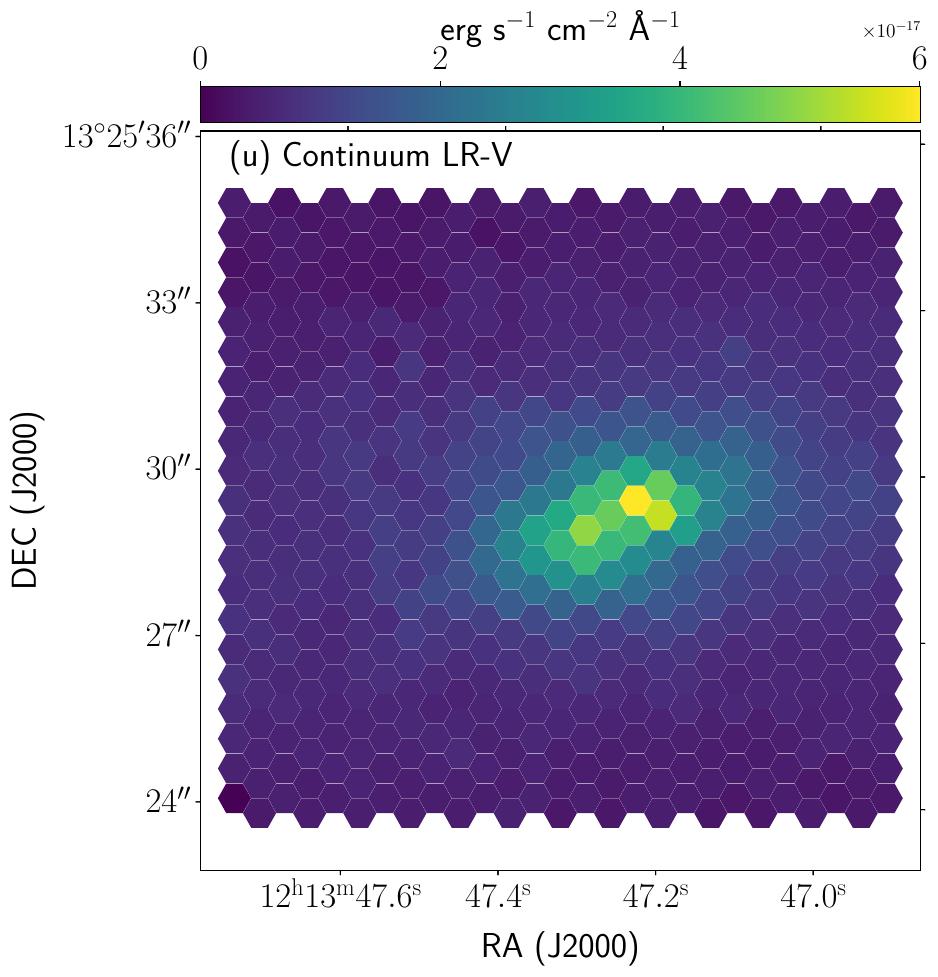}
	\includegraphics[clip, width=0.24\linewidth]{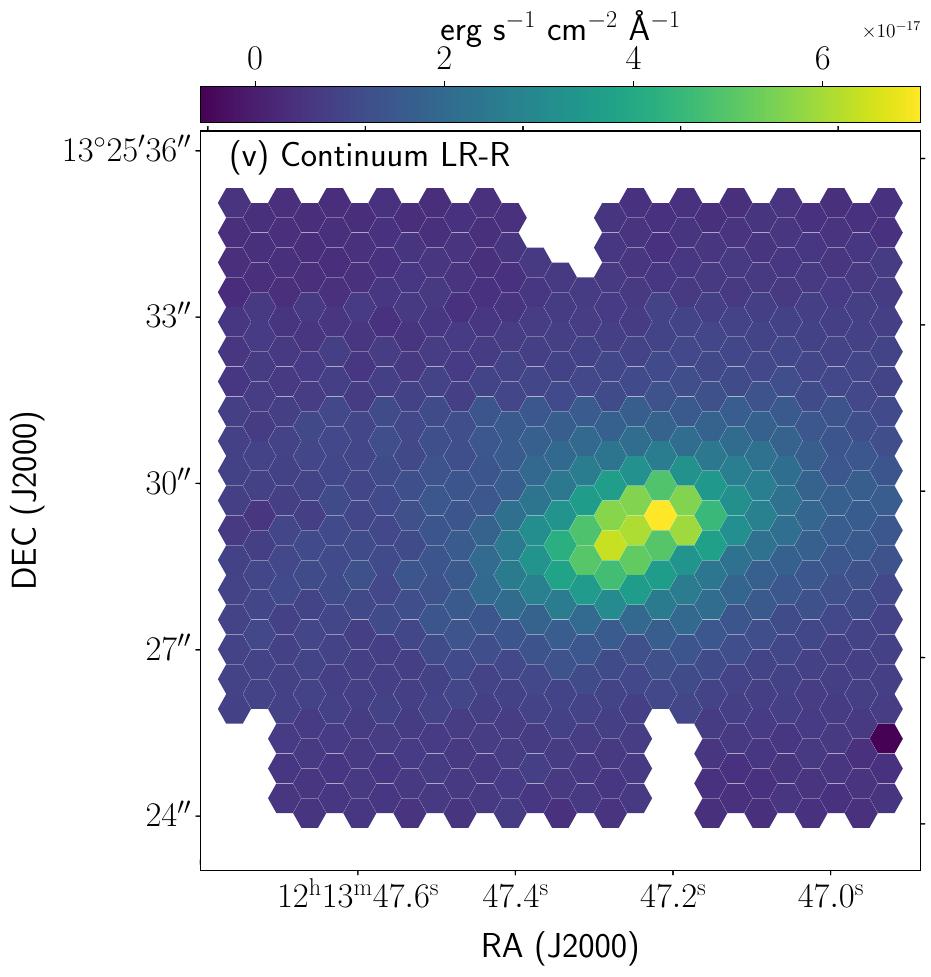}
	\includegraphics[clip, width=0.24\linewidth]{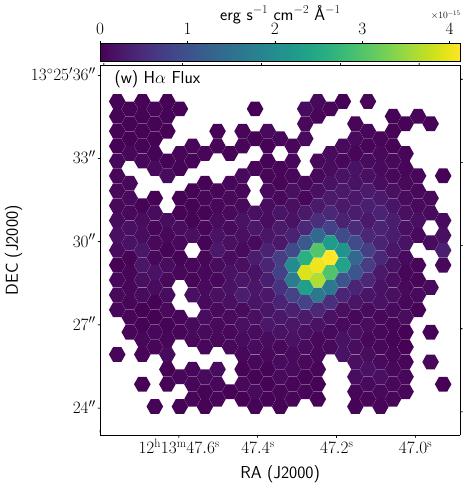}
	\includegraphics[clip, width=0.24\linewidth]{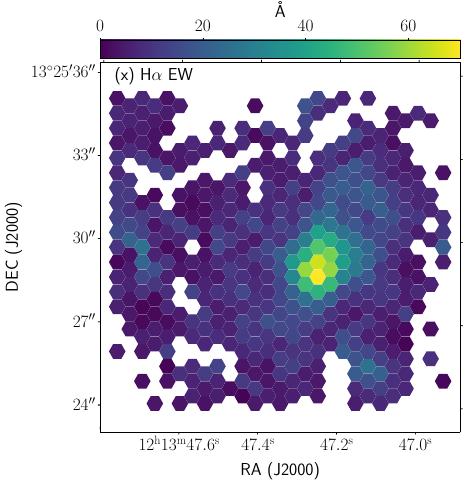}
	\includegraphics[clip, width=0.24\linewidth]{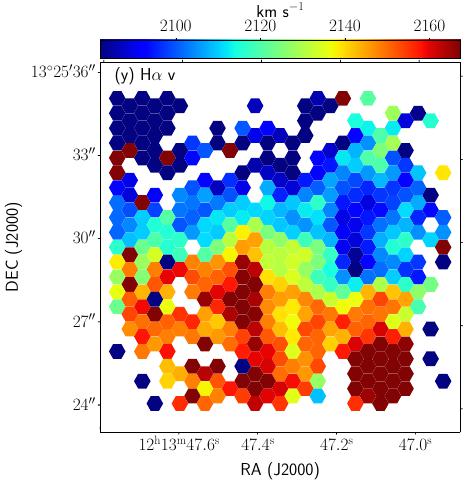}
	\includegraphics[clip, width=0.24\linewidth]{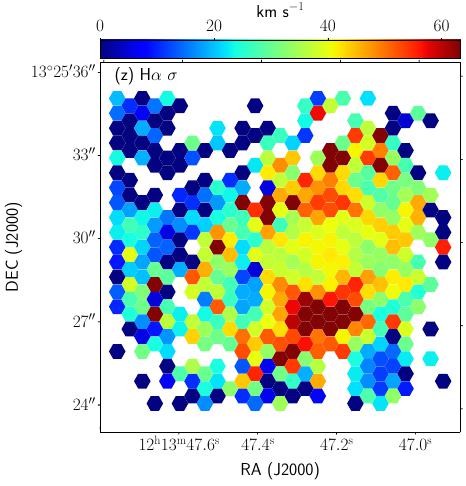}
	\includegraphics[clip, width=0.24\linewidth]{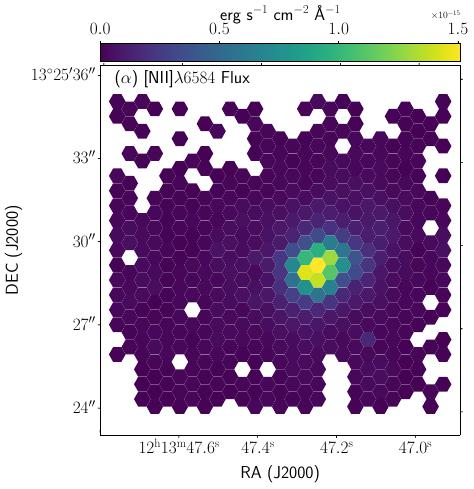}
	\includegraphics[clip, width=0.24\linewidth]{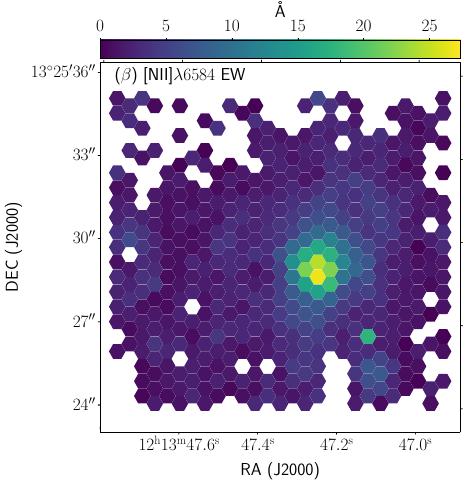}
	\includegraphics[clip, width=0.24\linewidth]{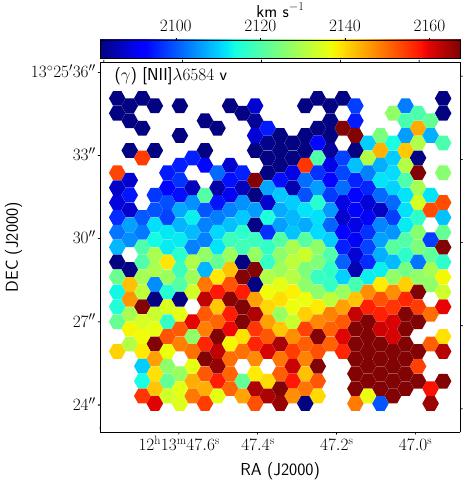}
	\includegraphics[clip, width=0.24\linewidth]{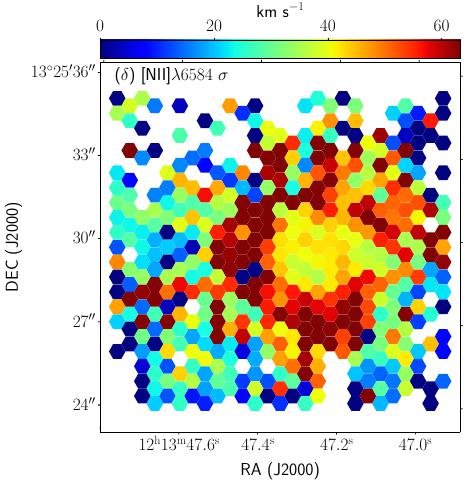}
	\includegraphics[clip, width=0.24\linewidth]{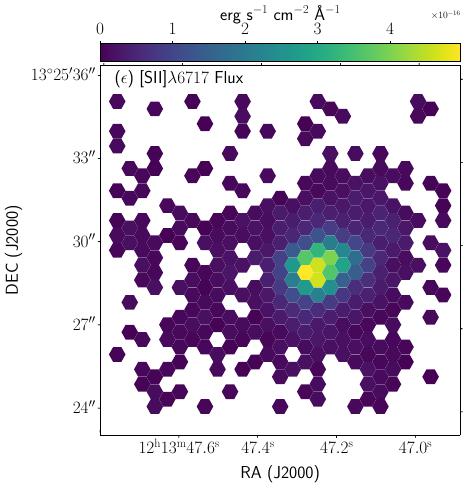}
	\includegraphics[clip, width=0.24\linewidth]{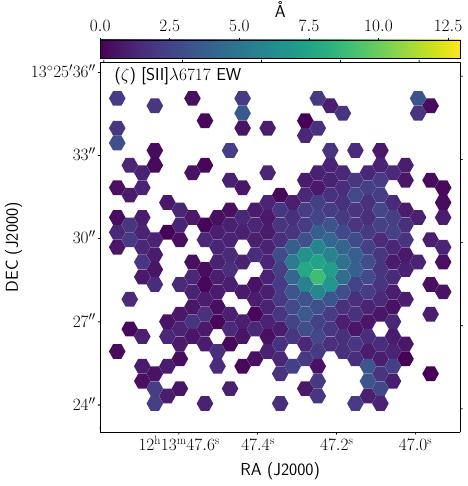}
	\includegraphics[clip, width=0.24\linewidth]{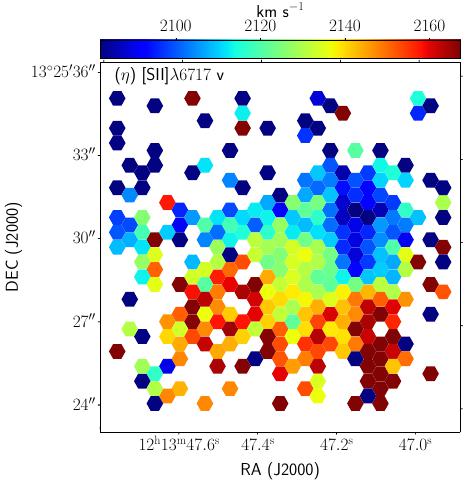}
	\includegraphics[clip, width=0.24\linewidth]{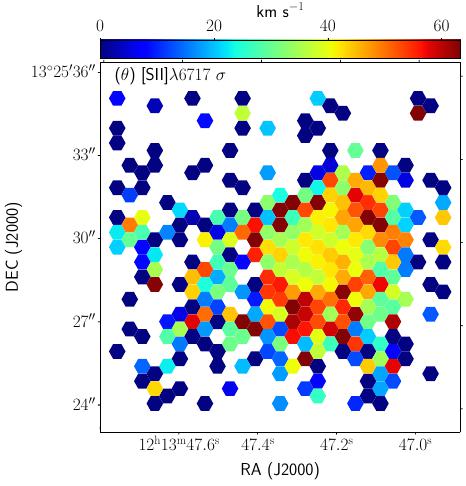}
	\includegraphics[clip, width=0.24\linewidth]{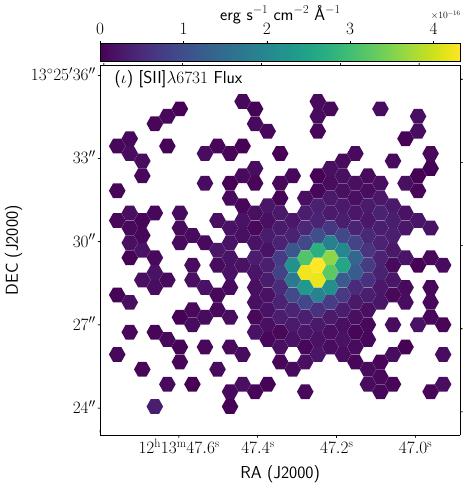}
	\includegraphics[clip, width=0.24\linewidth]{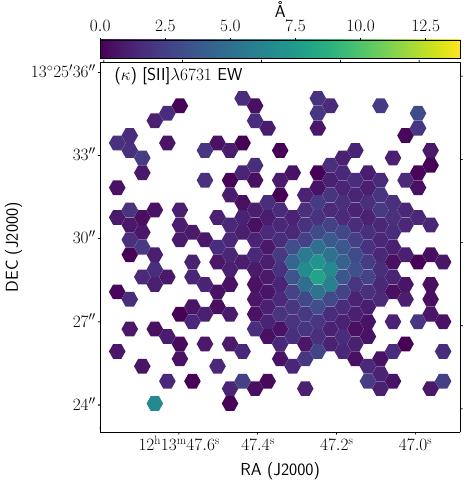}
	\includegraphics[clip, width=0.24\linewidth]{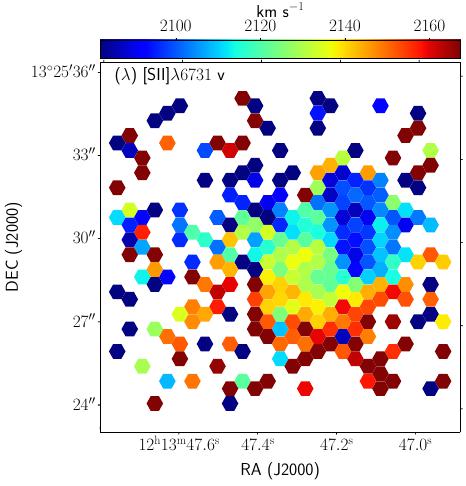}
	\includegraphics[clip, width=0.24\linewidth]{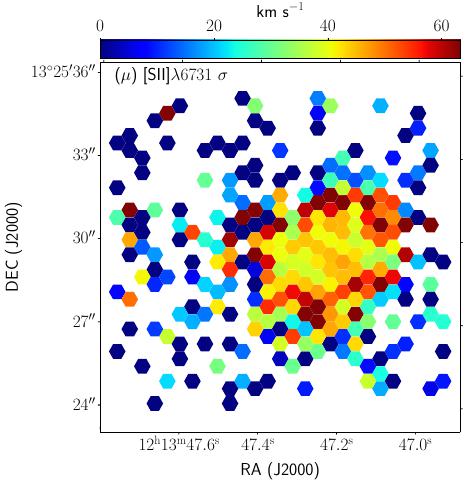}
	\caption{(cont.) NGC~4189 card.}
	\label{fig:NGC4189_card_2}
\end{figure*}

\begin{figure*}[h]
	\centering
	\includegraphics[clip, width=0.35\linewidth]{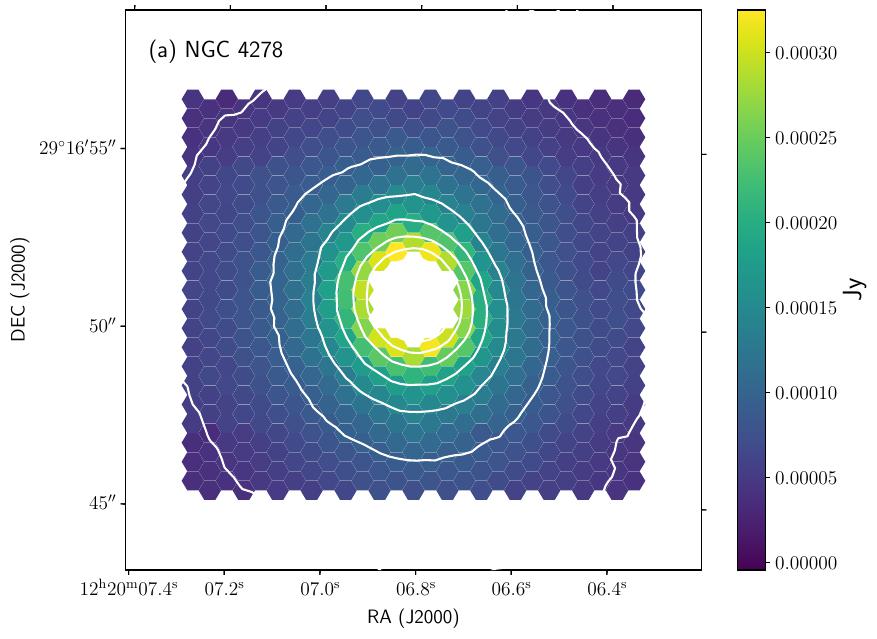}
	\includegraphics[clip, width=0.6\linewidth]{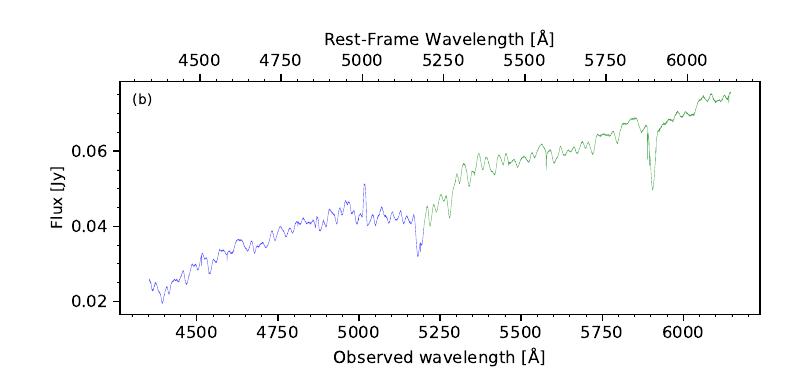}
	\includegraphics[clip, width=0.24\linewidth]{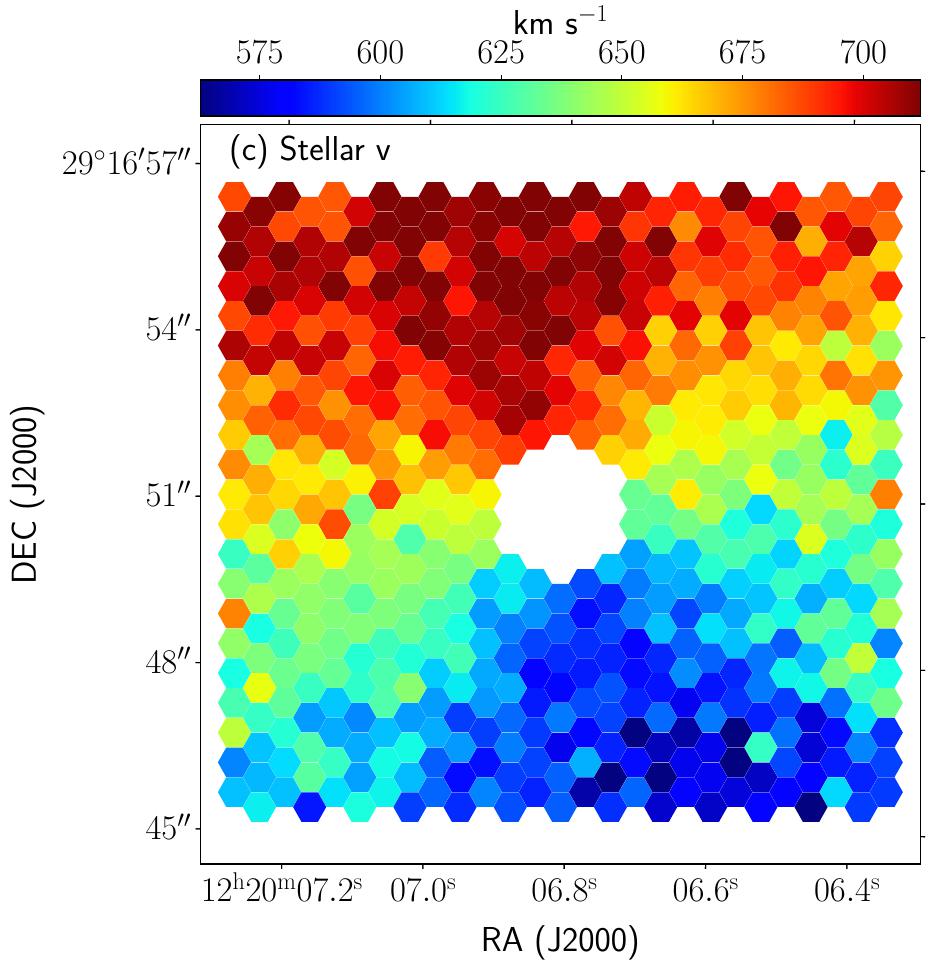}
	\includegraphics[clip, width=0.24\linewidth]{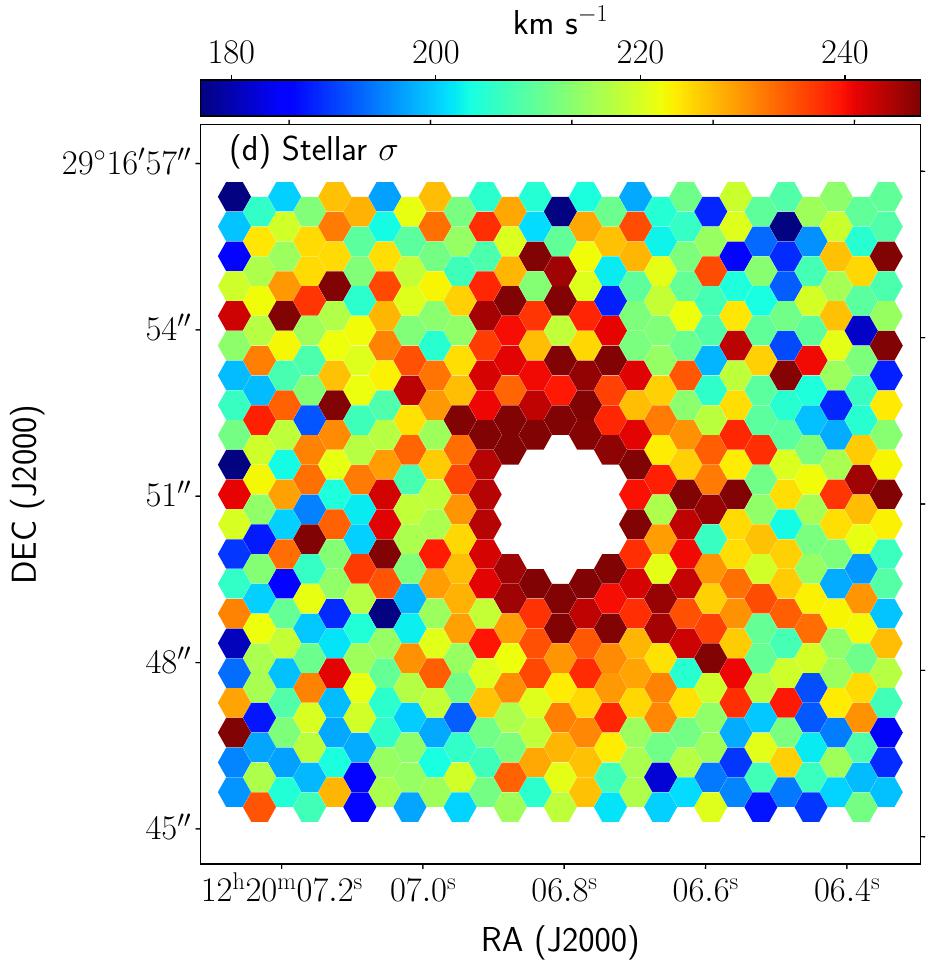}
	\includegraphics[clip, width=0.24\linewidth]{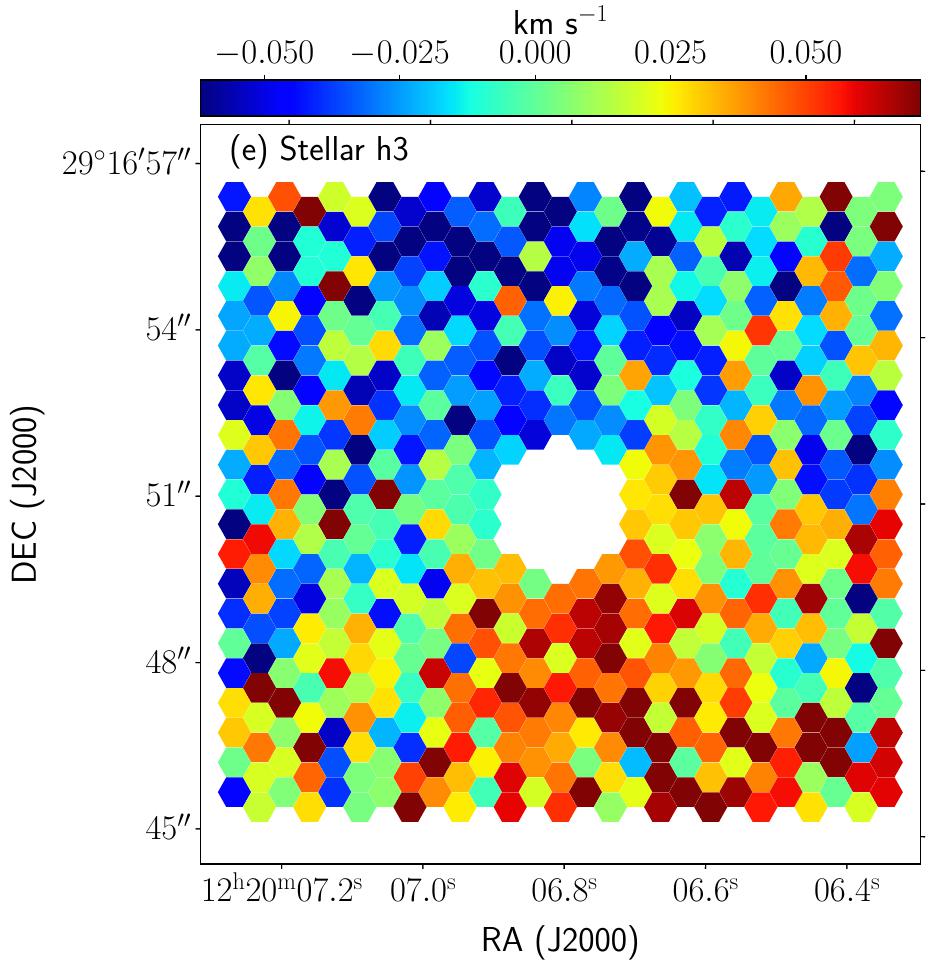}
	\includegraphics[clip, width=0.24\linewidth]{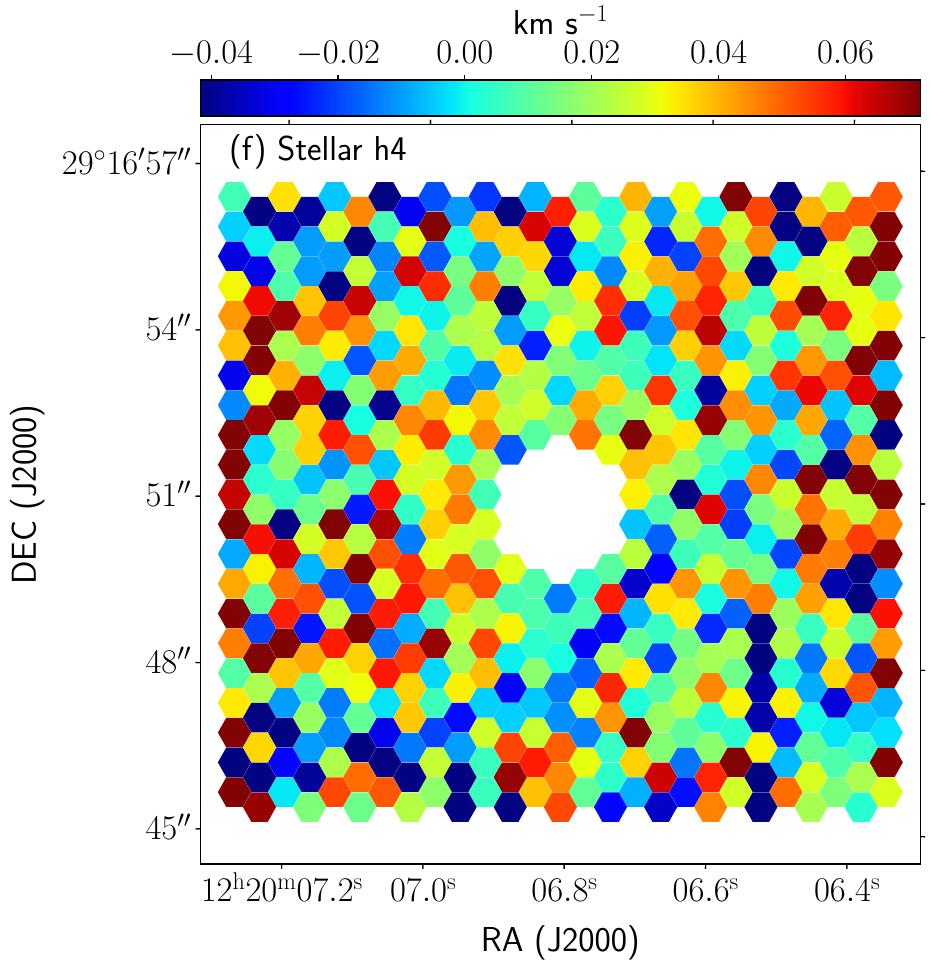}
	\includegraphics[clip, width=0.24\linewidth]{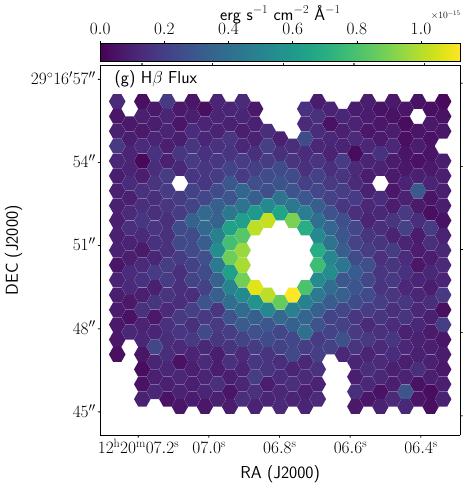}
	\includegraphics[clip, width=0.24\linewidth]{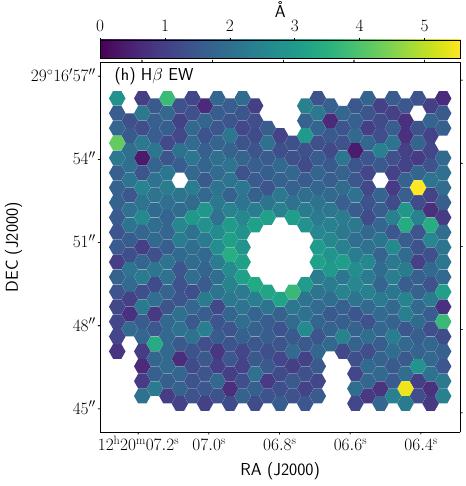}
	\includegraphics[clip, width=0.24\linewidth]{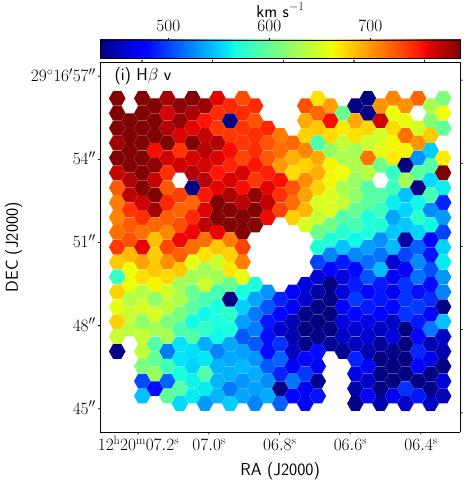}
	\includegraphics[clip, width=0.24\linewidth]{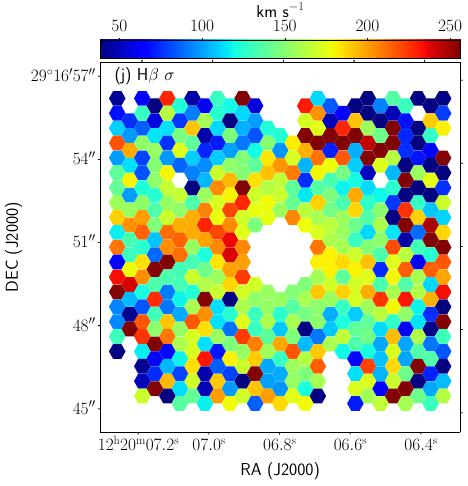}
	\includegraphics[clip, width=0.24\linewidth]{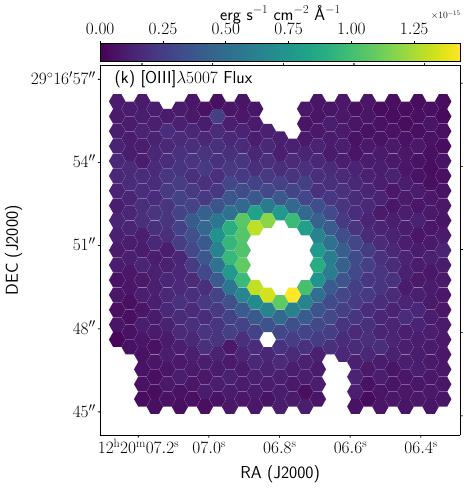}
	\includegraphics[clip, width=0.24\linewidth]{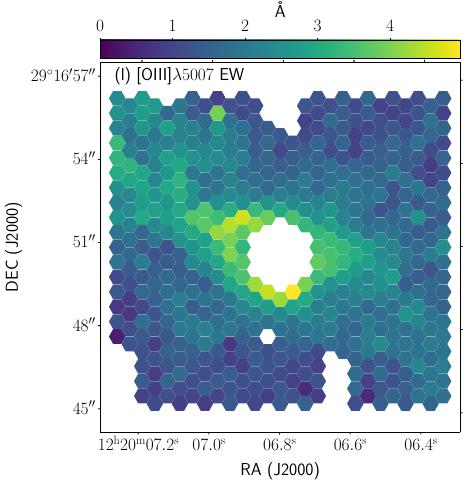}
	\includegraphics[clip, width=0.24\linewidth]{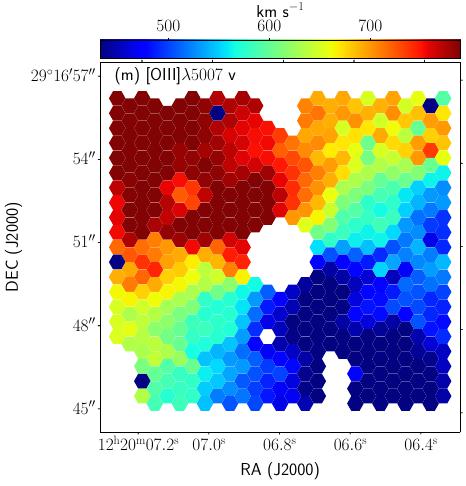}
	\includegraphics[clip, width=0.24\linewidth]{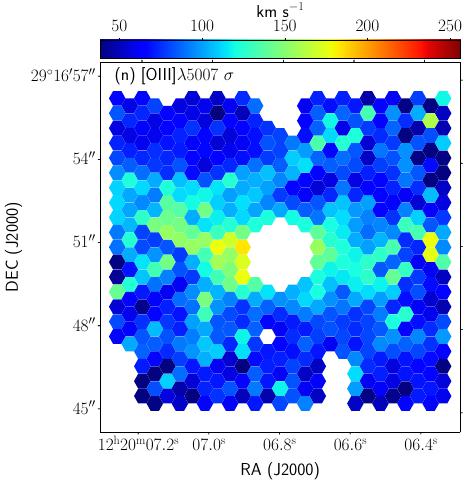}
	\includegraphics[clip, width=0.24\linewidth]{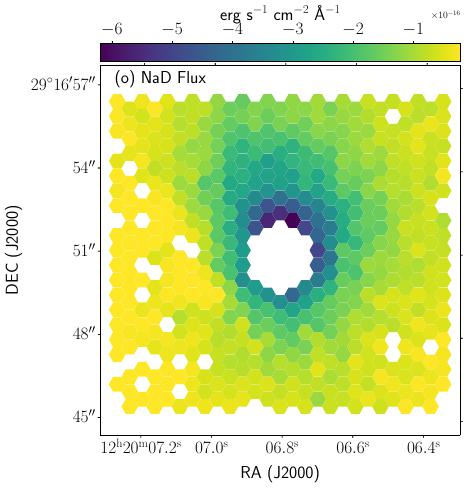}
	\includegraphics[clip, width=0.24\linewidth]{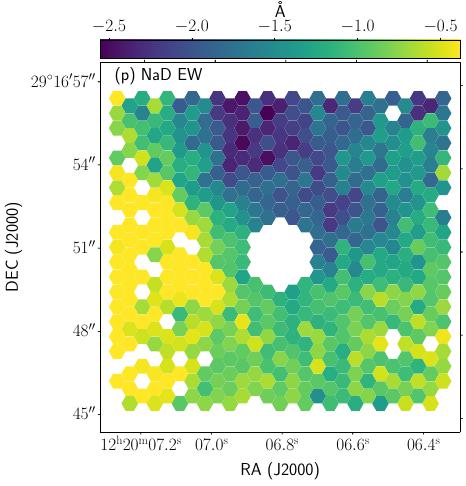}
	\includegraphics[clip, width=0.24\linewidth]{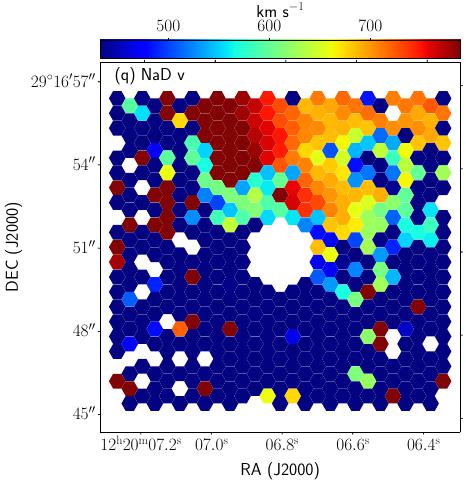}
	\includegraphics[clip, width=0.24\linewidth]{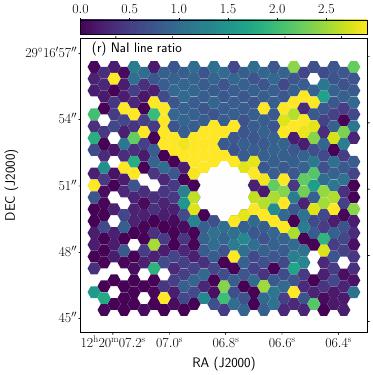}
	\caption{NGC~4278 card.}
	\label{fig:NGC4278_card_1}
\end{figure*}
\addtocounter{figure}{-1}
\begin{figure*}[h]
	\centering
	\includegraphics[clip, width=0.24\linewidth]{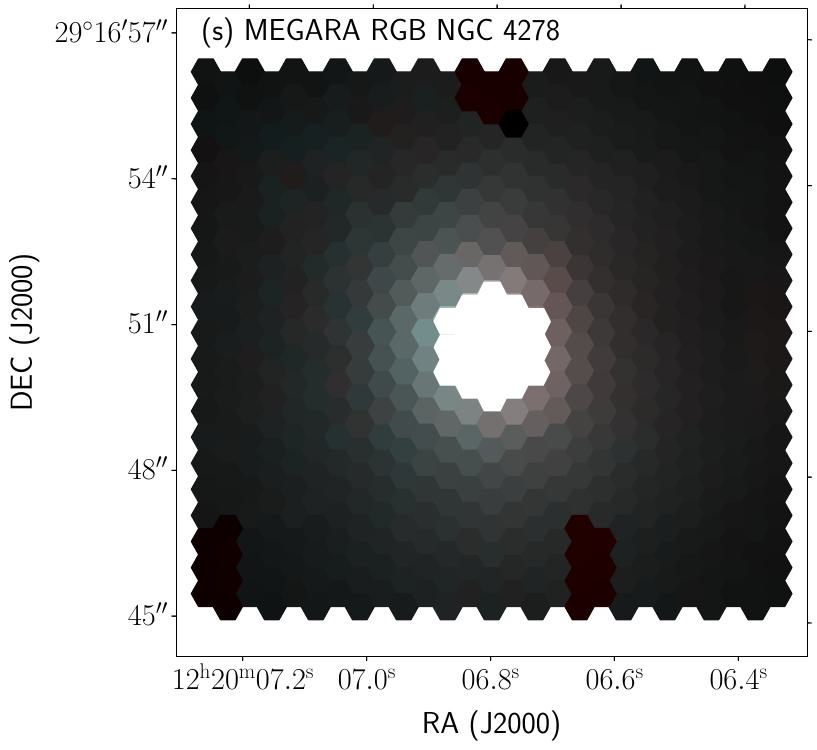}
	\includegraphics[clip, width=0.24\linewidth]{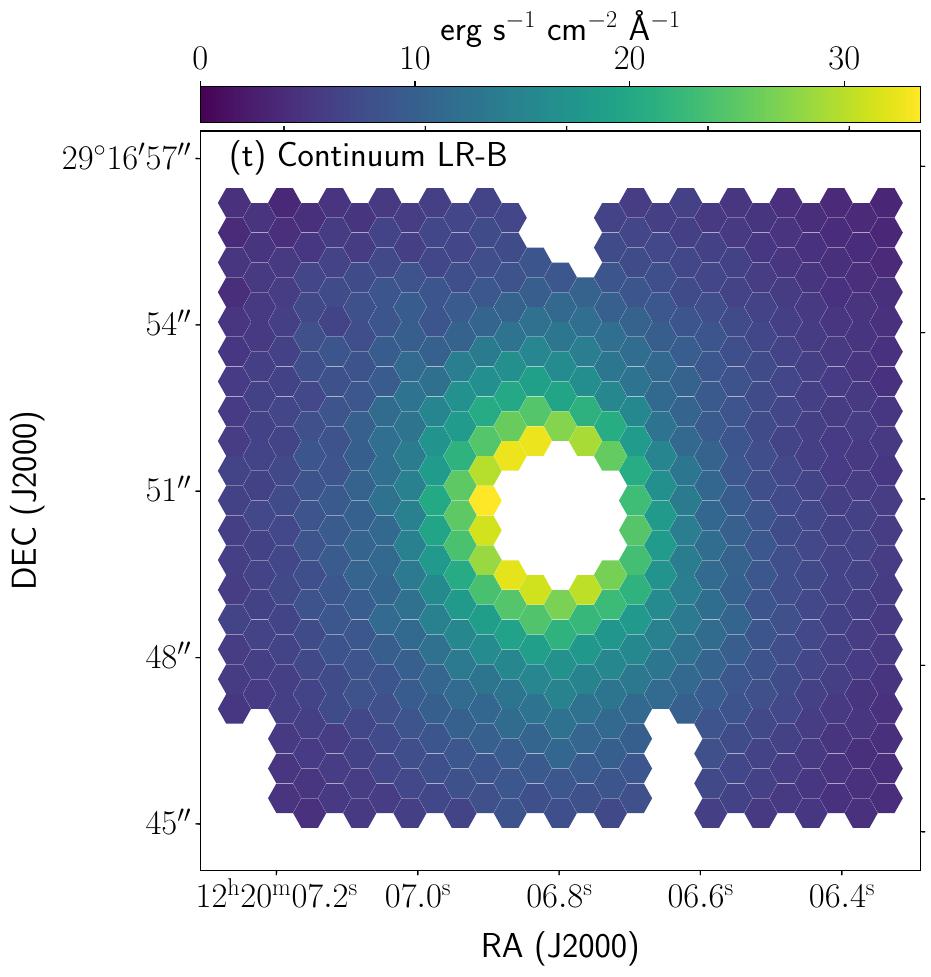}
	\includegraphics[clip, width=0.24\linewidth]{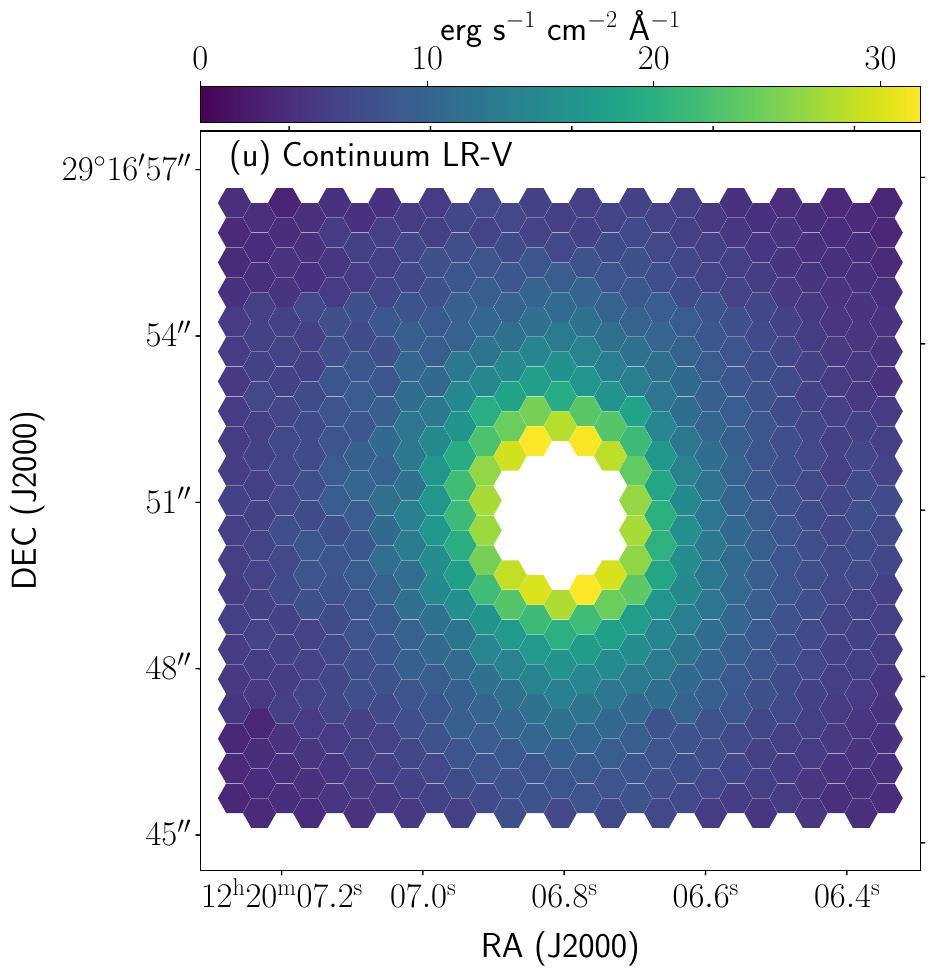}
	\hspace{4cm}
	\caption{(cont.) NGC~4278 card.}
	\label{fig:NGC4278_card_2}
\end{figure*}

\begin{figure*}[h]
	\centering
	\includegraphics[clip, width=0.35\linewidth]{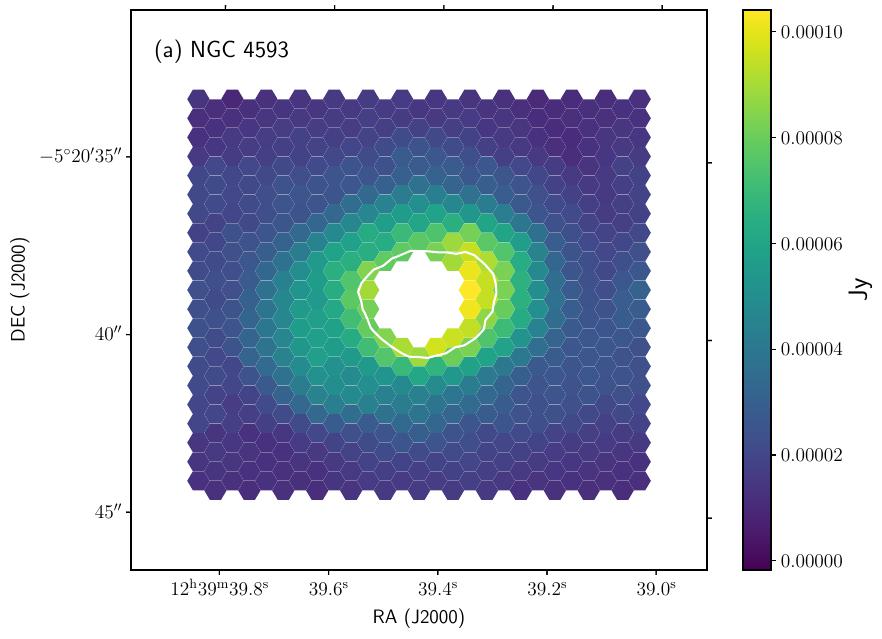}
	\includegraphics[clip, width=0.6\linewidth]{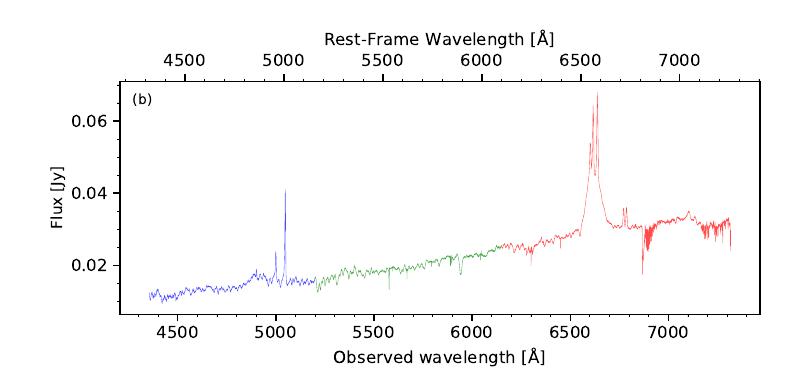}
	\includegraphics[clip, width=0.24\linewidth]{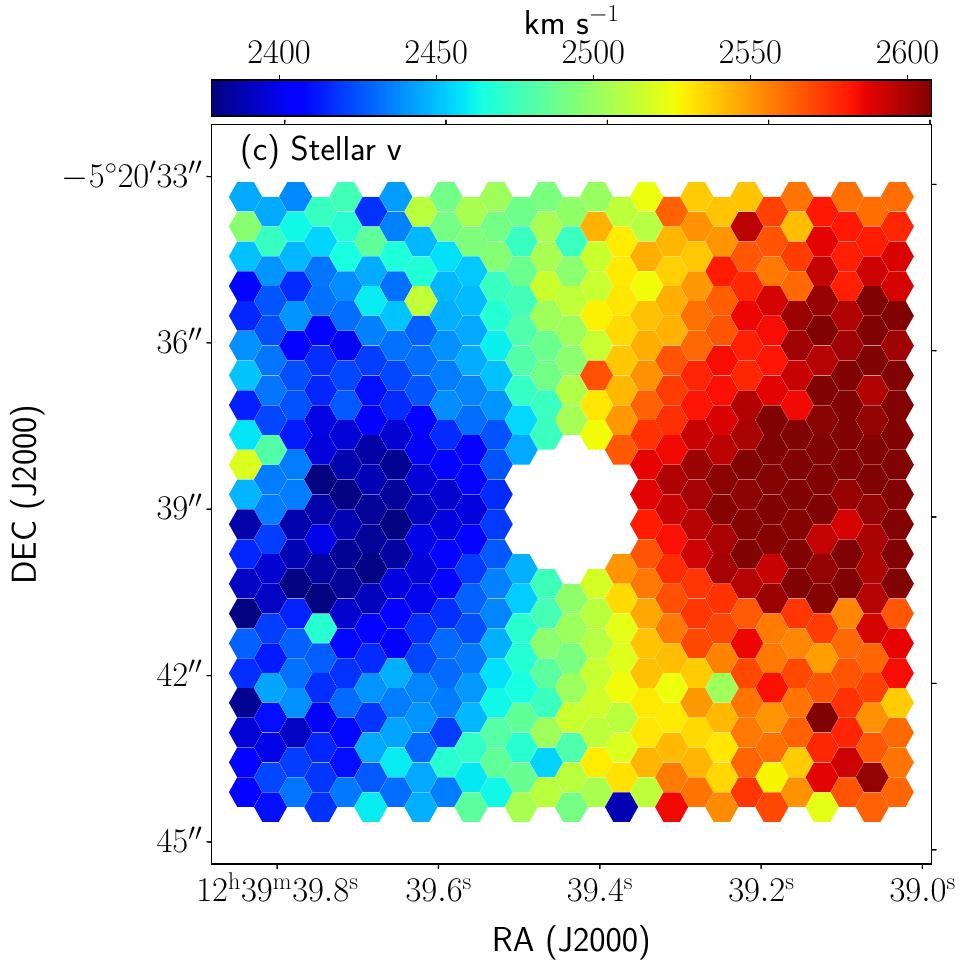}
	\includegraphics[clip, width=0.24\linewidth]{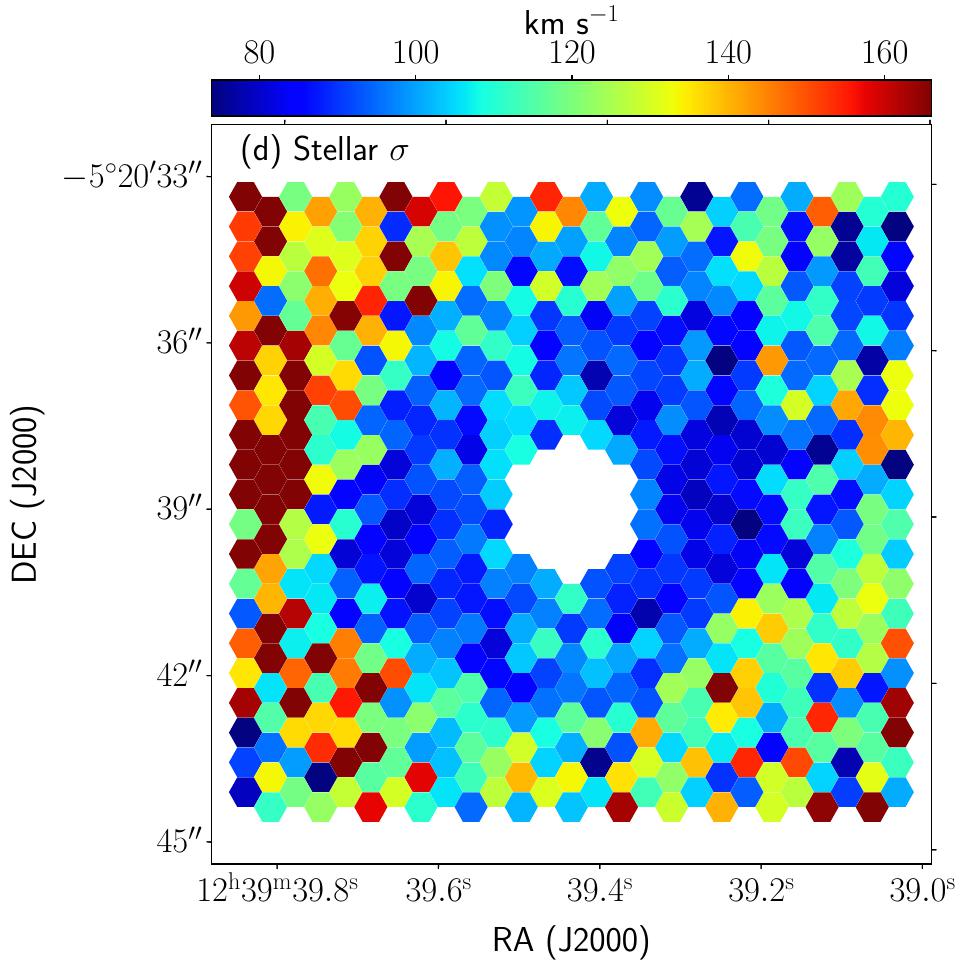}
	\includegraphics[clip, width=0.24\linewidth]{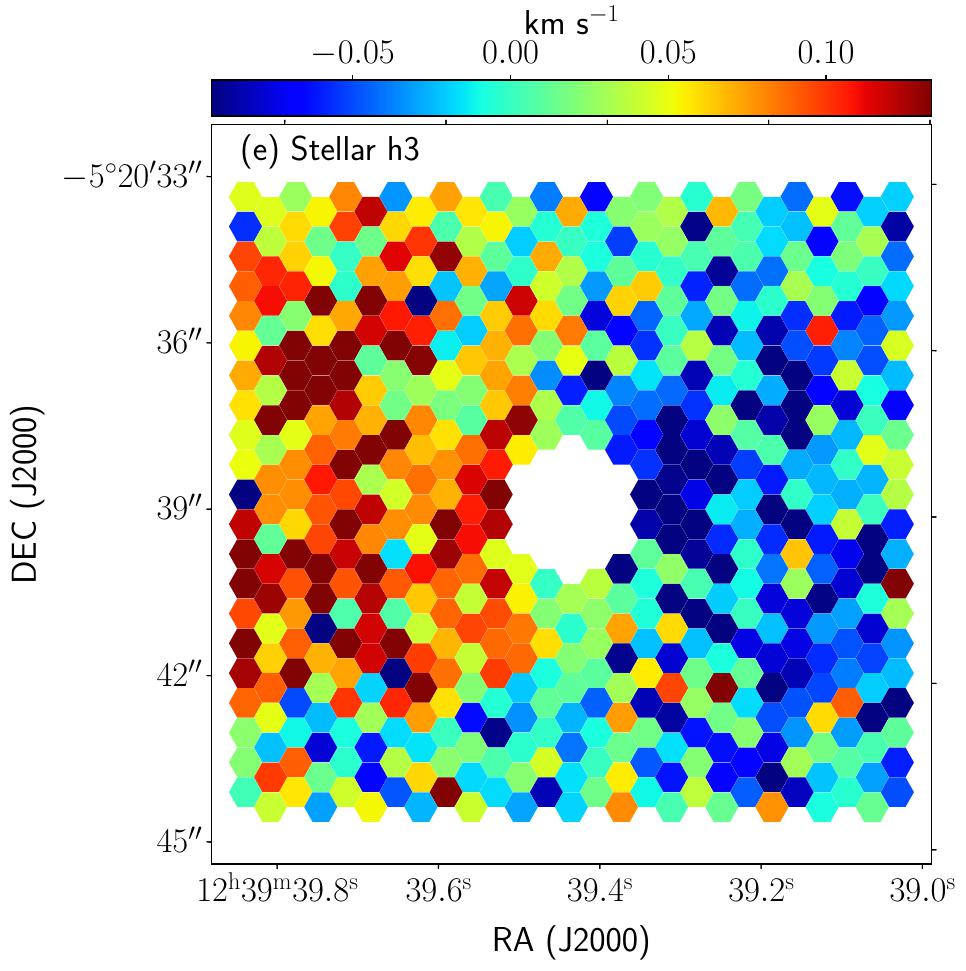}
	\includegraphics[clip, width=0.24\linewidth]{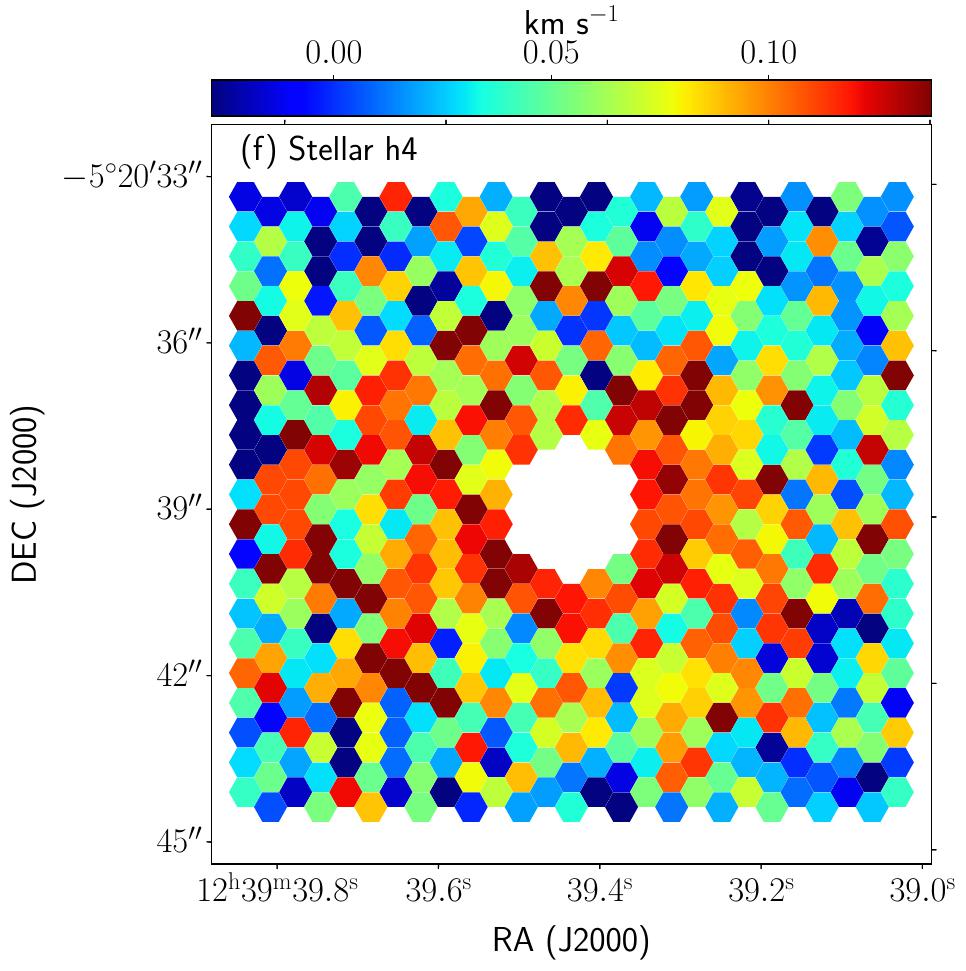}
	\includegraphics[clip, width=0.24\linewidth]{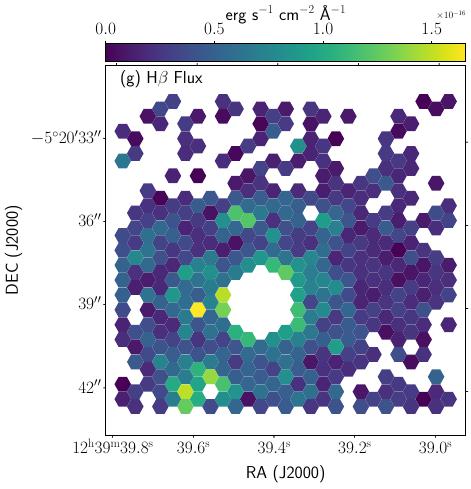}
	\includegraphics[clip, width=0.24\linewidth]{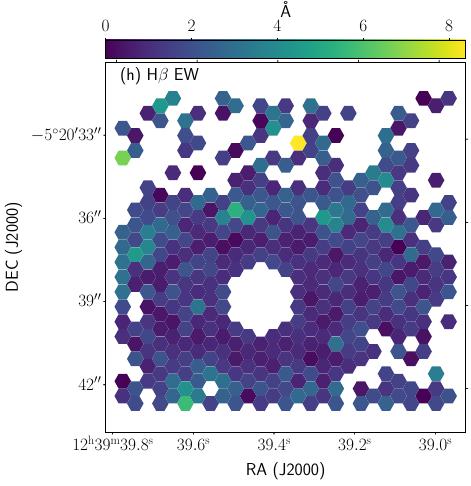}
	\includegraphics[clip, width=0.24\linewidth]{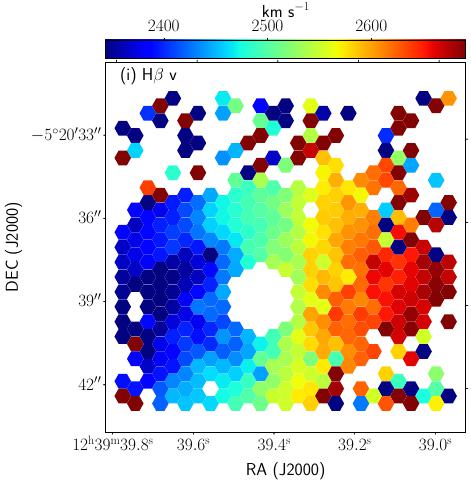}
	\includegraphics[clip, width=0.24\linewidth]{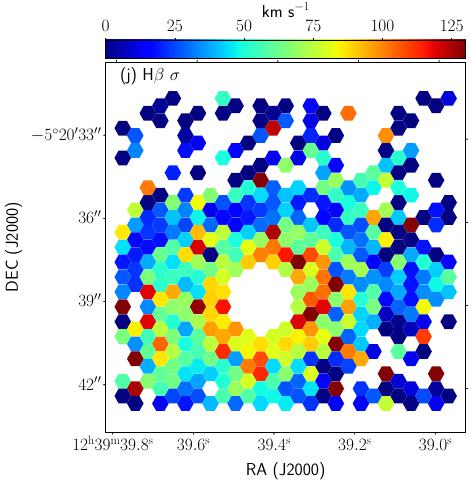}
	\includegraphics[clip, width=0.24\linewidth]{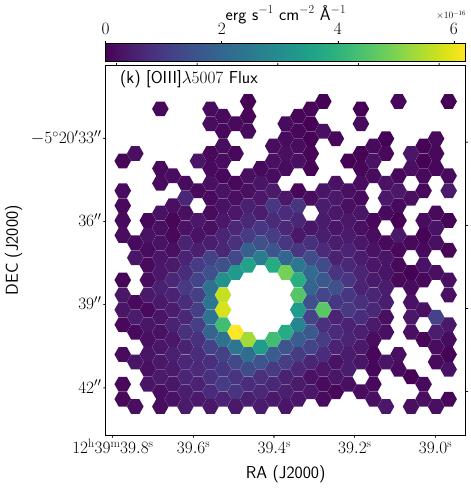}
	\includegraphics[clip, width=0.24\linewidth]{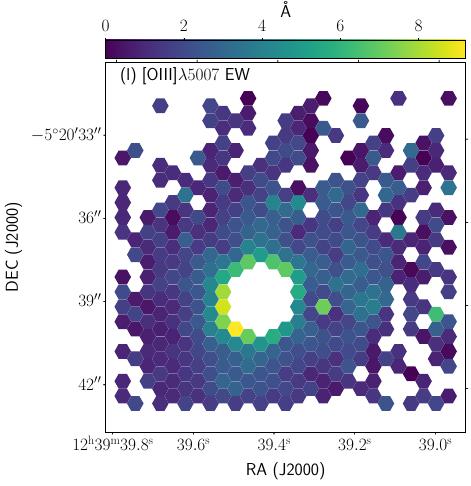}
	\includegraphics[clip, width=0.24\linewidth]{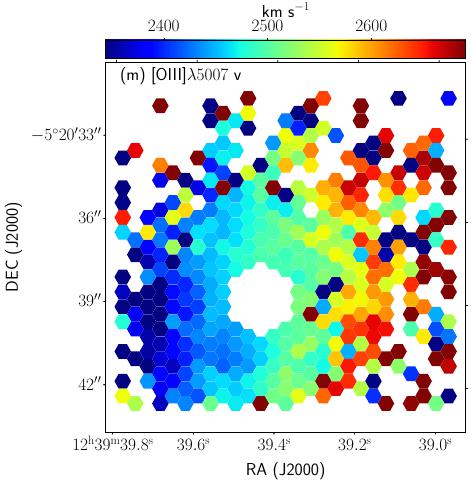}
	\includegraphics[clip, width=0.24\linewidth]{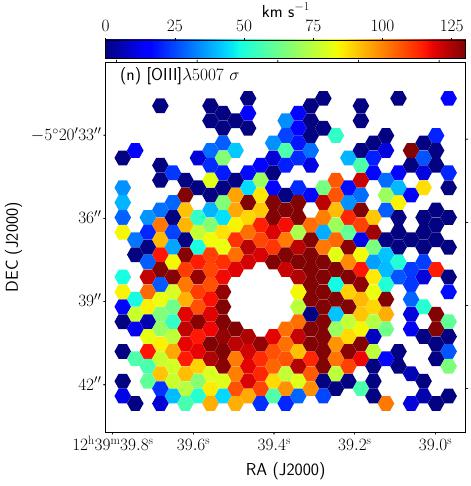}
	\includegraphics[clip, width=0.24\linewidth]{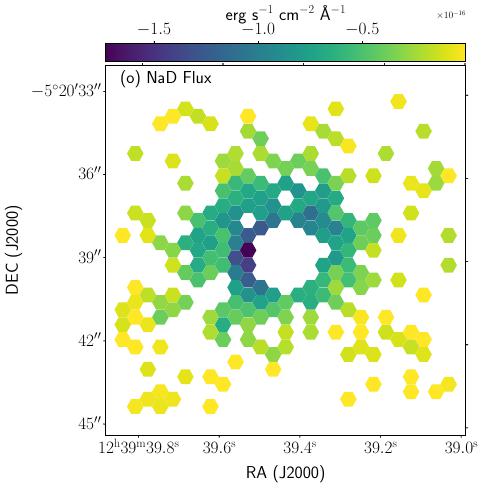}
	\includegraphics[clip, width=0.24\linewidth]{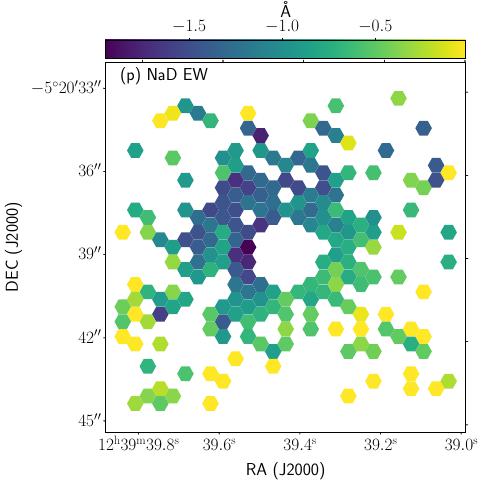}
	\includegraphics[clip, width=0.24\linewidth]{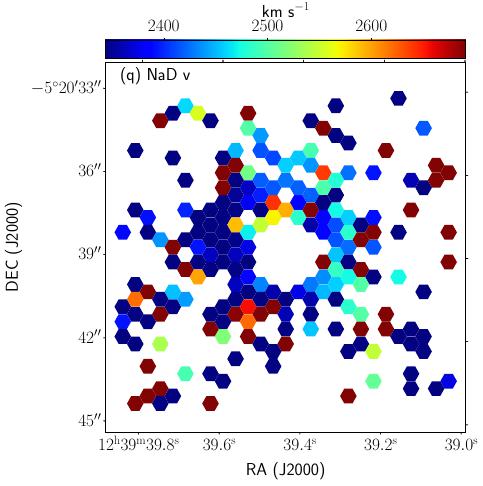}
	\includegraphics[clip, width=0.24\linewidth]{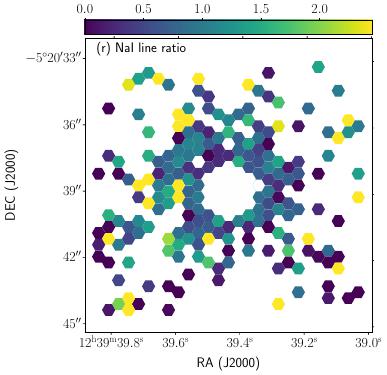}
	\caption{NGC~4593 card.}
	\label{fig:NGC4593_card_1}
\end{figure*}
\addtocounter{figure}{-1}
\begin{figure*}[h]
	\centering
	\includegraphics[clip, width=0.24\linewidth]{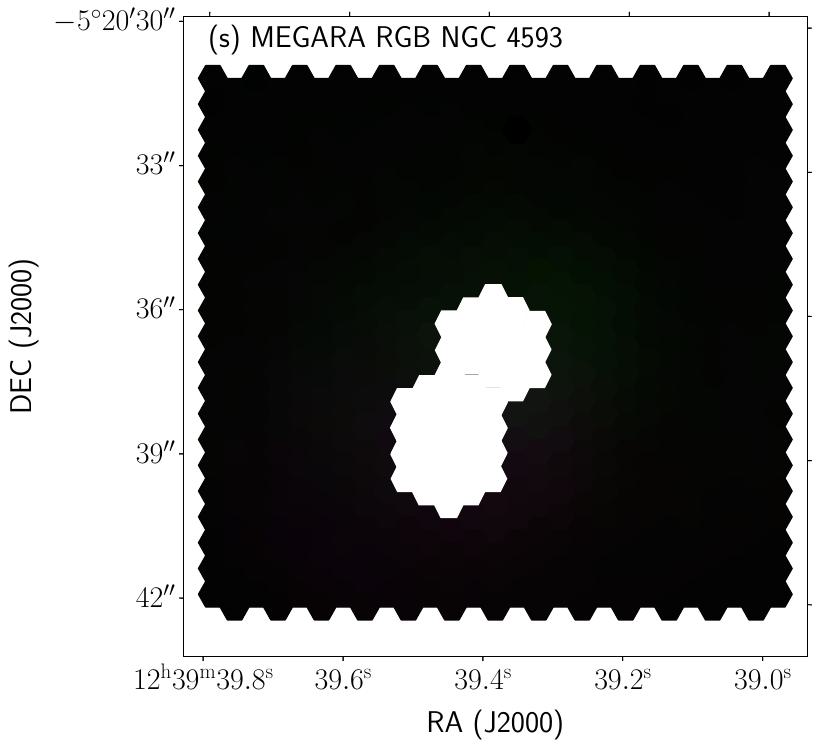}
	\includegraphics[clip, width=0.24\linewidth]{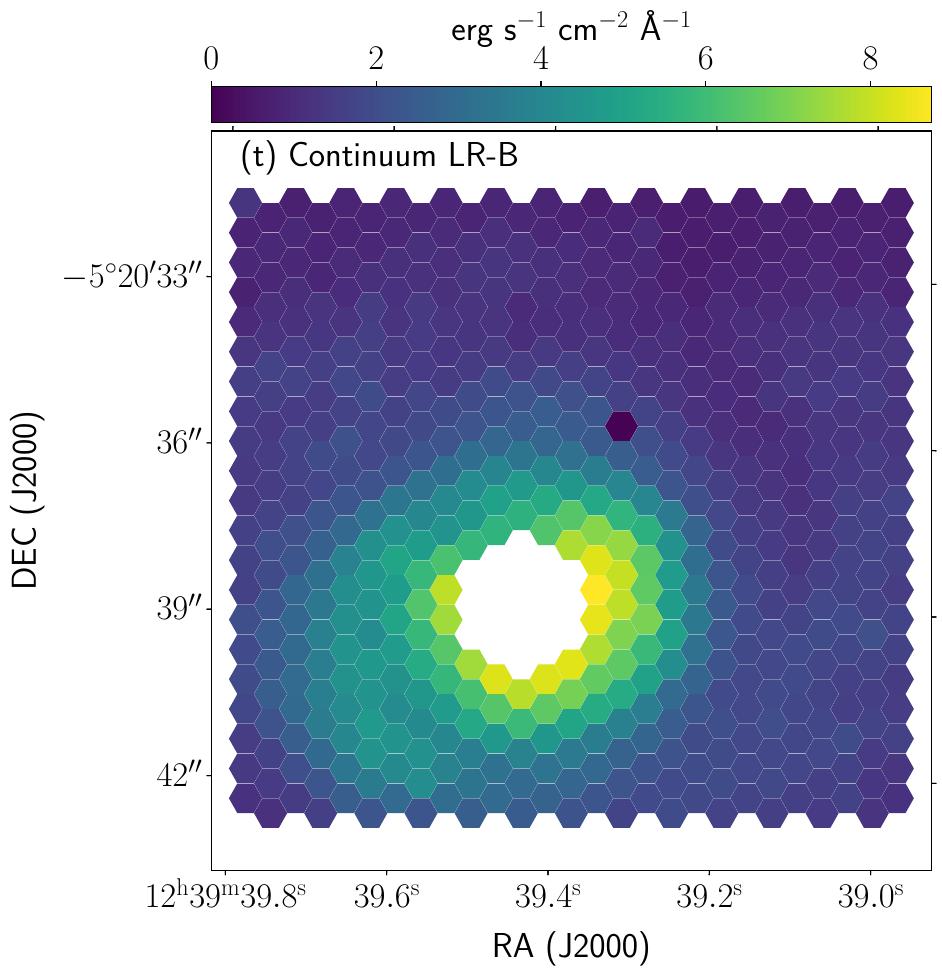}
	\includegraphics[clip, width=0.24\linewidth]{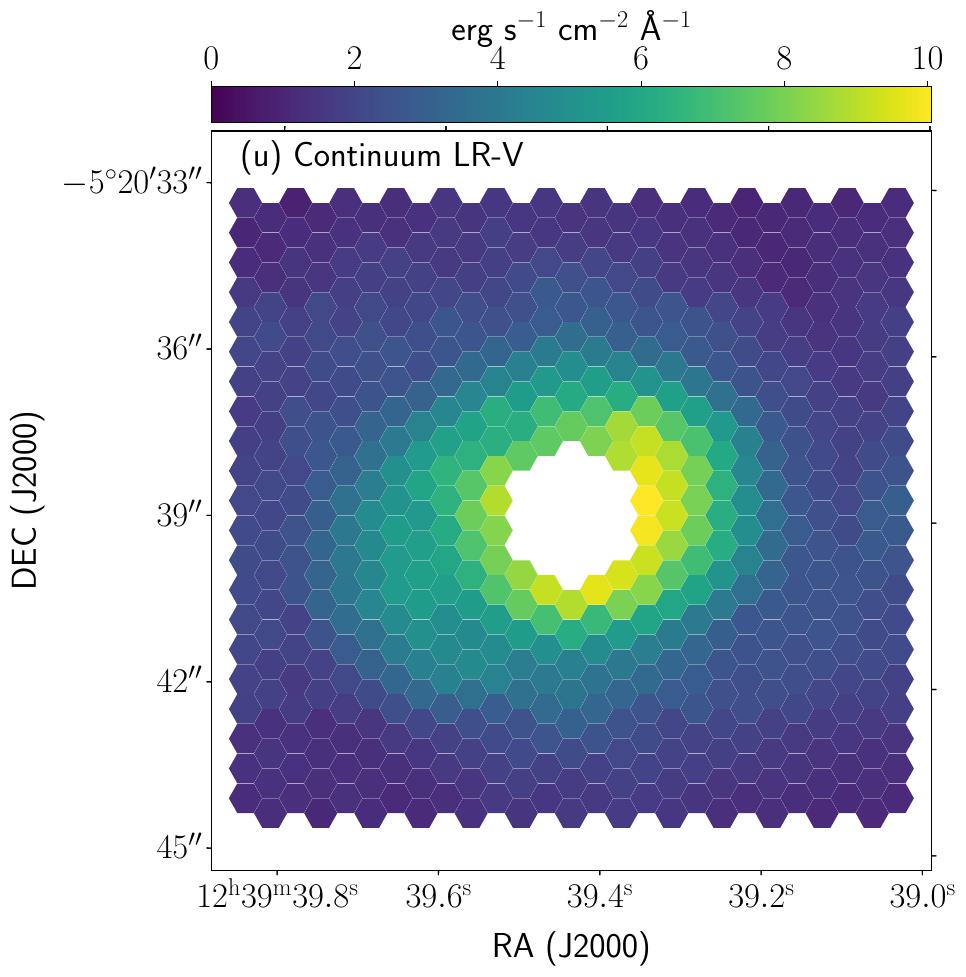}
	\includegraphics[clip, width=0.24\linewidth]{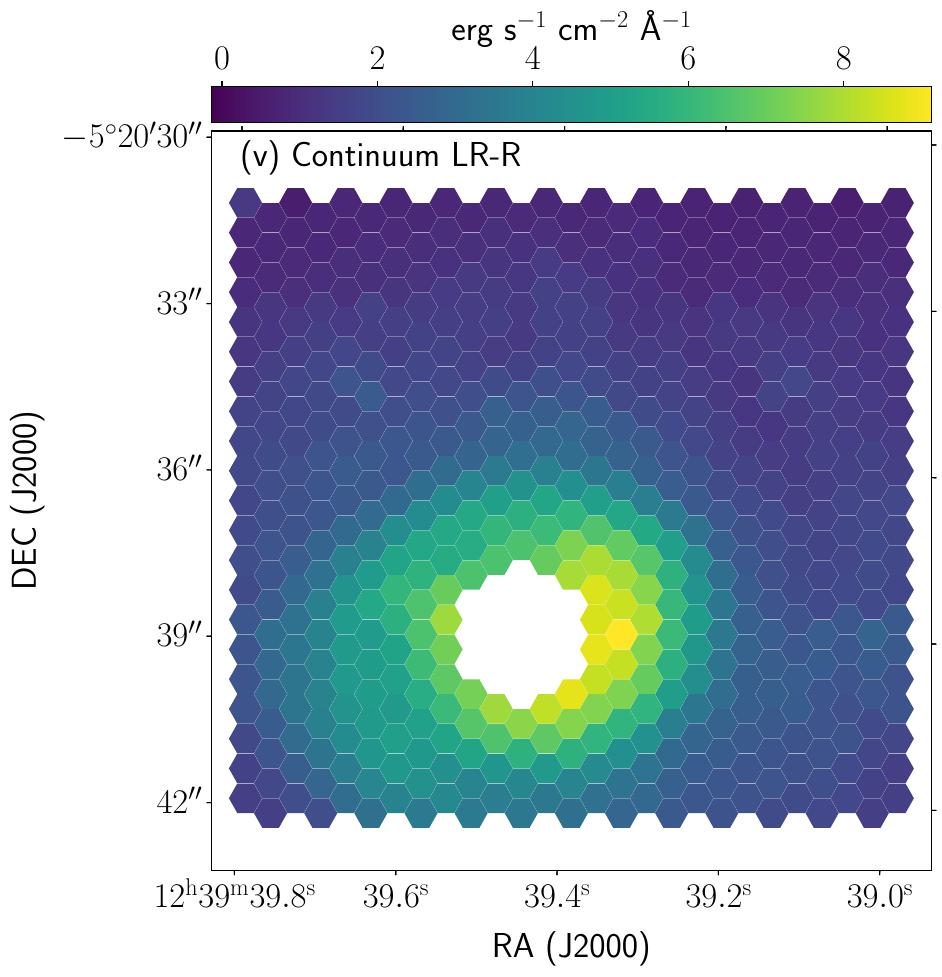}
	\includegraphics[clip, width=0.24\linewidth]{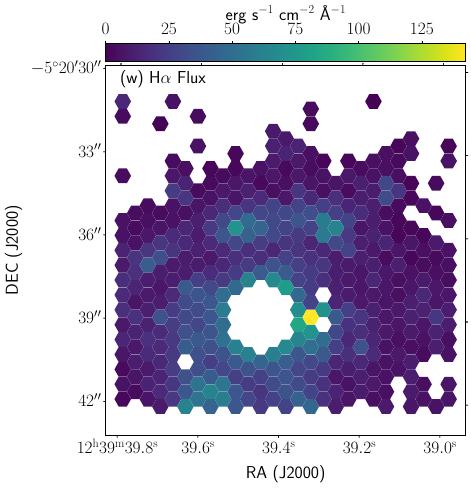}
	\includegraphics[clip, width=0.24\linewidth]{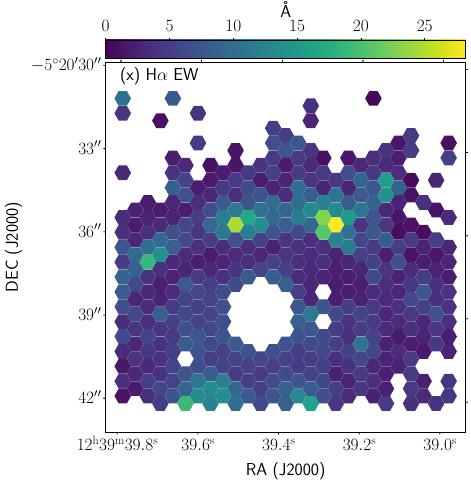}
	\includegraphics[clip, width=0.24\linewidth]{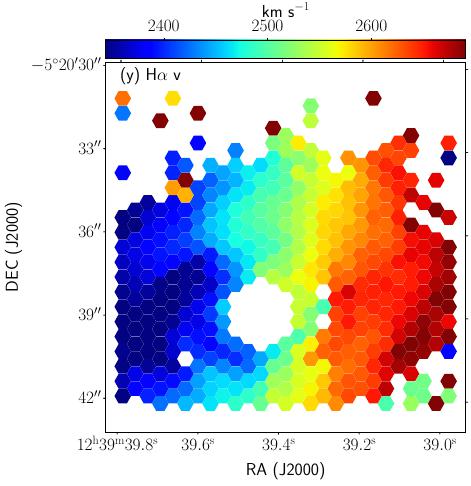}
	\includegraphics[clip, width=0.24\linewidth]{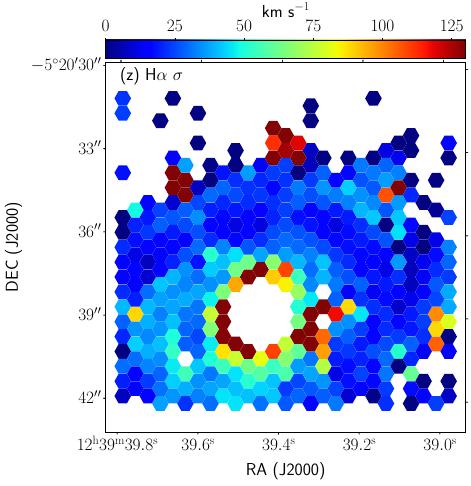}
	\includegraphics[clip, width=0.24\linewidth]{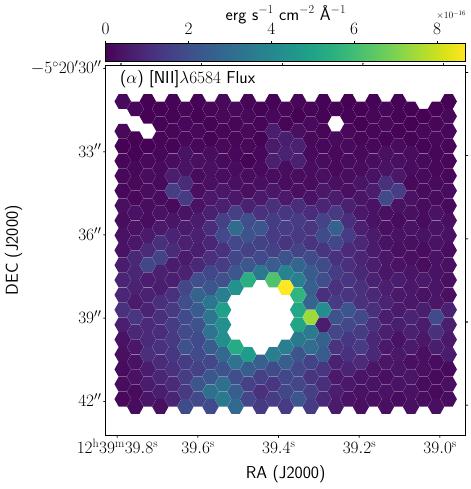}
	\includegraphics[clip, width=0.24\linewidth]{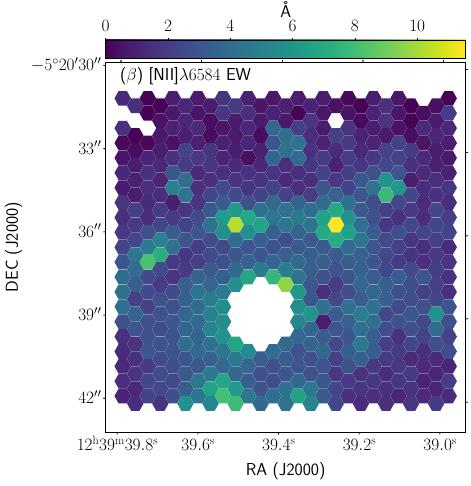}
	\includegraphics[clip, width=0.24\linewidth]{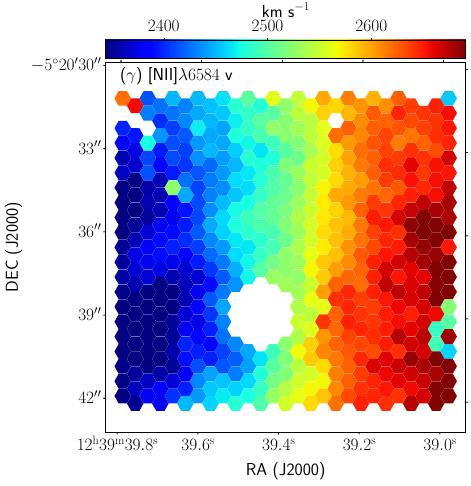}
	\includegraphics[clip, width=0.24\linewidth]{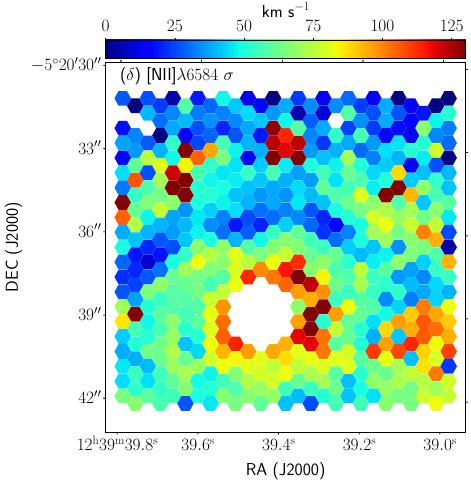}
	\includegraphics[clip, width=0.24\linewidth]{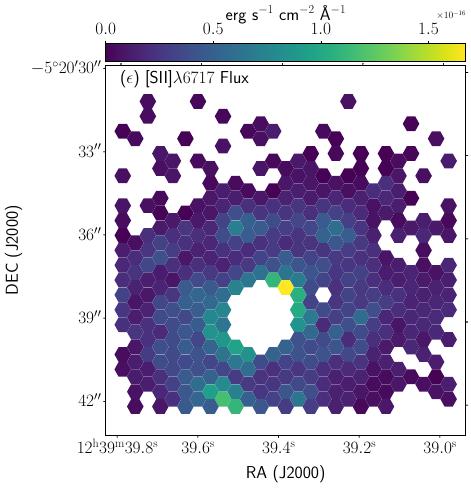}
	\includegraphics[clip, width=0.24\linewidth]{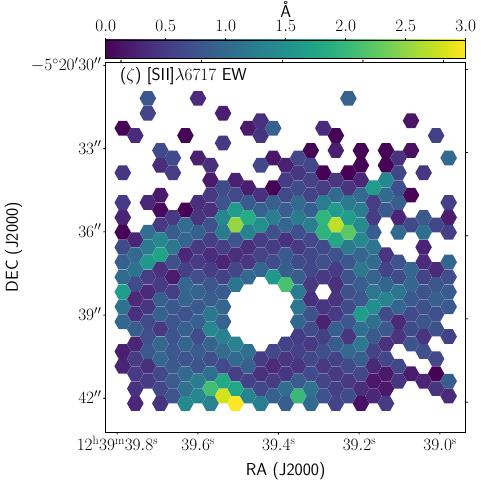}
	\includegraphics[clip, width=0.24\linewidth]{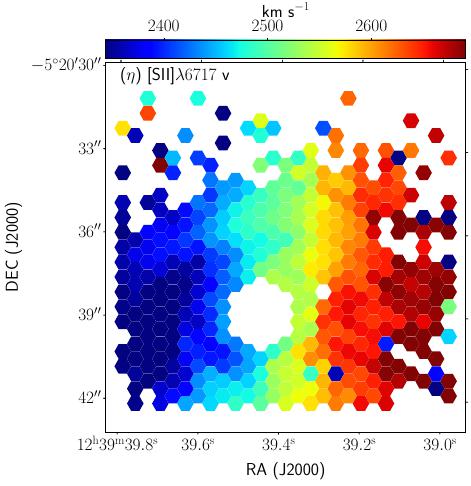}
	\includegraphics[clip, width=0.24\linewidth]{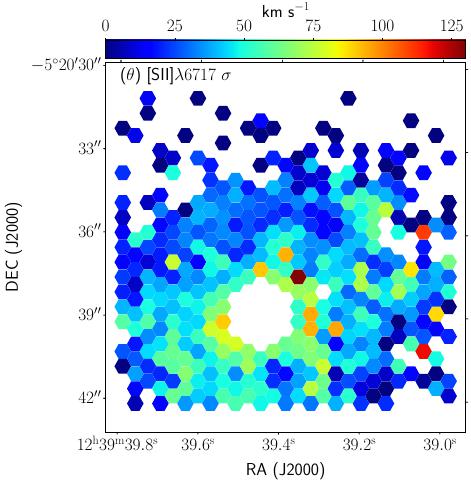}
	\includegraphics[clip, width=0.24\linewidth]{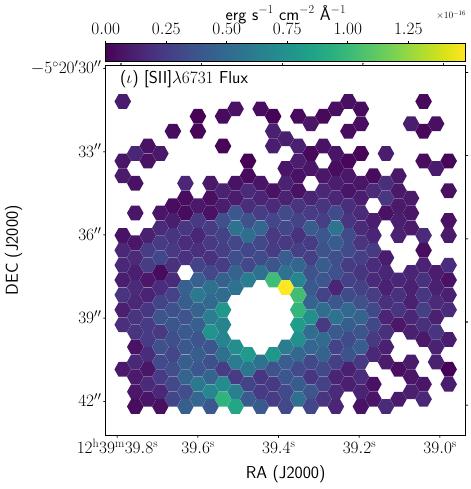}
	\includegraphics[clip, width=0.24\linewidth]{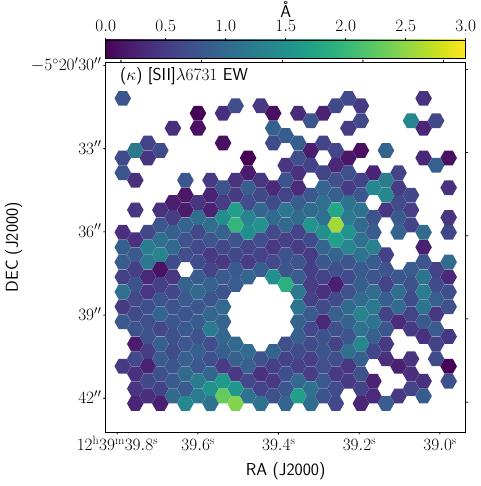}
	\includegraphics[clip, width=0.24\linewidth]{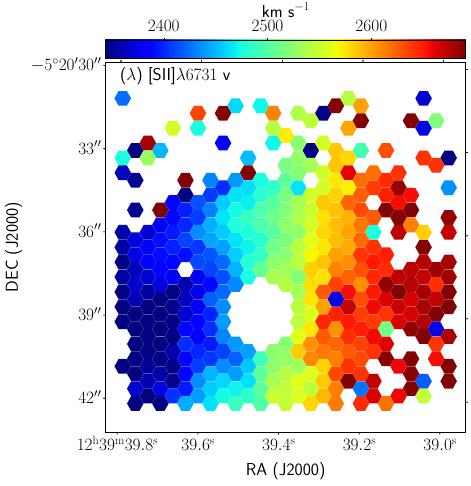}
	\includegraphics[clip, width=0.24\linewidth]{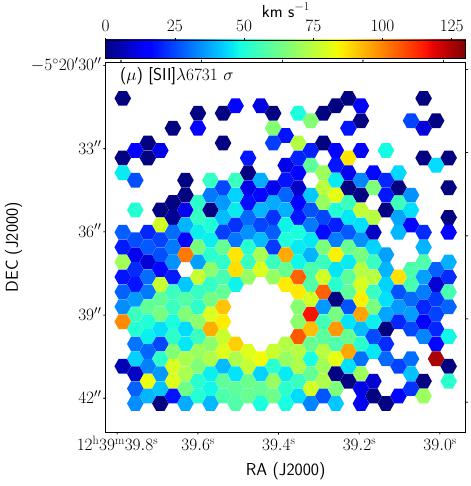}
	\caption{(cont.) NGC~4593 card.}
	\label{fig:NGC4593_card_2}
\end{figure*}

\begin{figure*}[h]
	\centering
	\includegraphics[clip, width=0.35\linewidth]{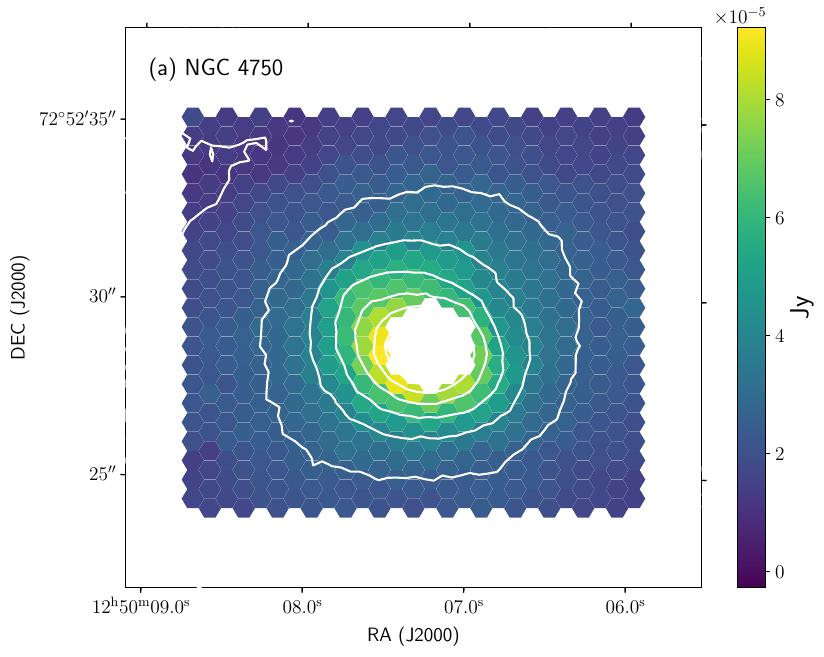}
	\includegraphics[clip, width=0.6\linewidth]{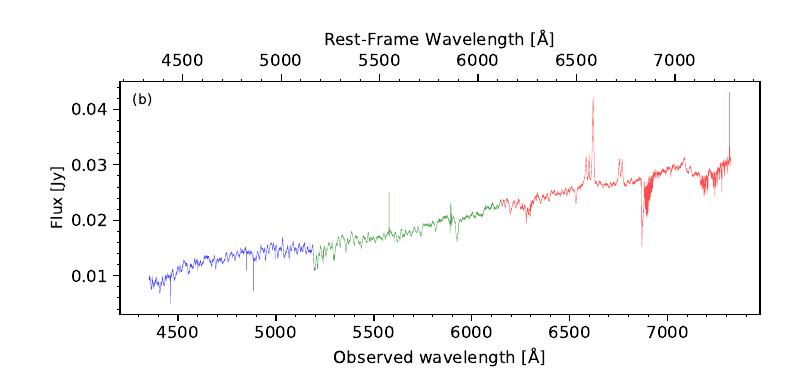}
	\includegraphics[clip, width=0.24\linewidth]{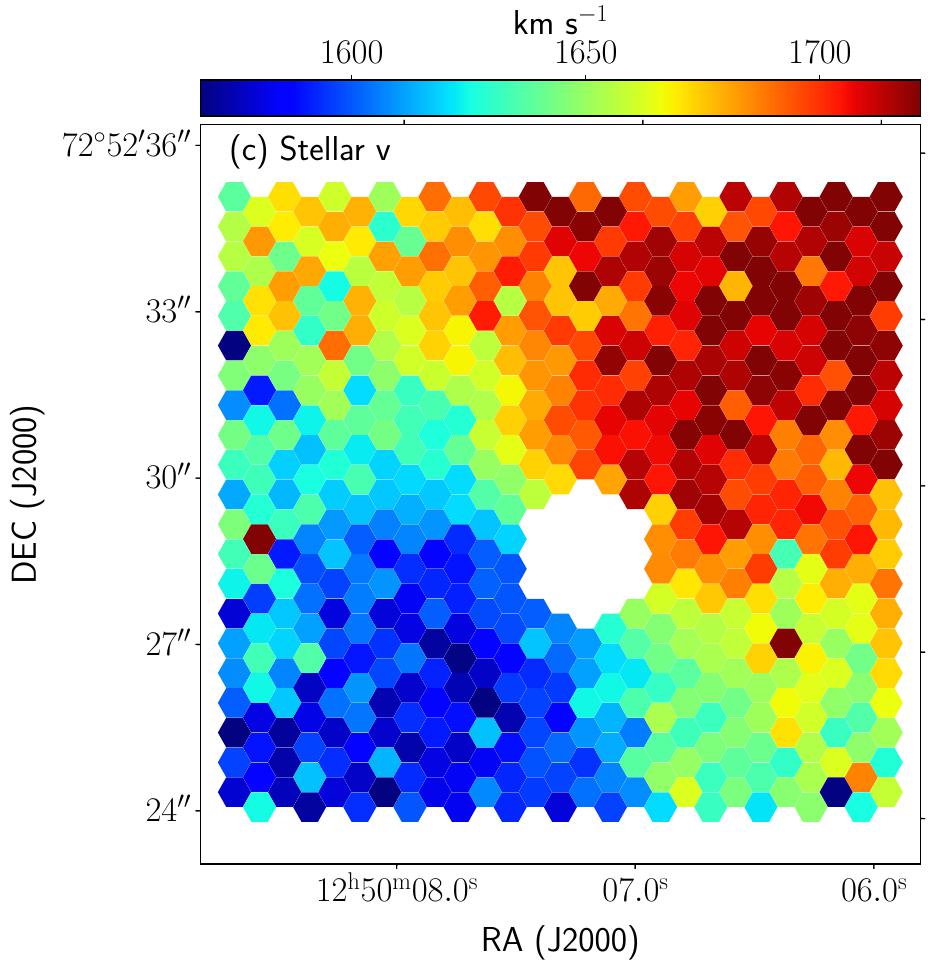}
	\includegraphics[clip, width=0.24\linewidth]{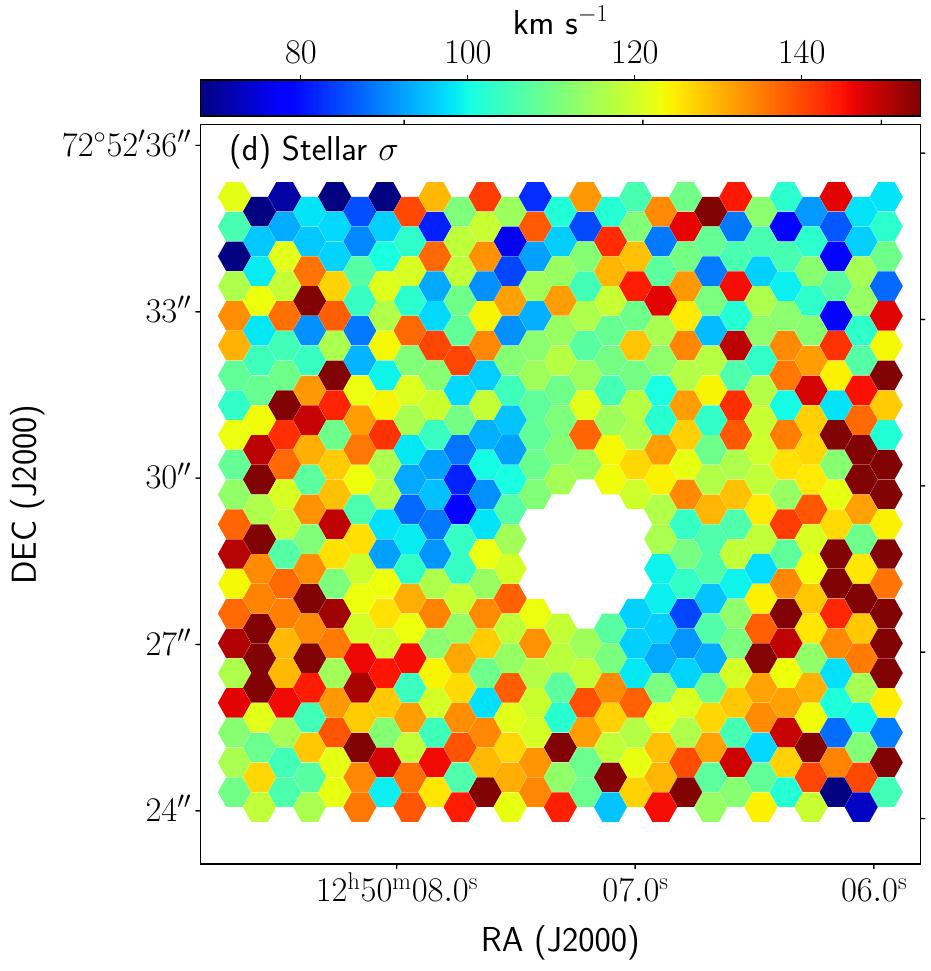}
	\includegraphics[clip, width=0.24\linewidth]{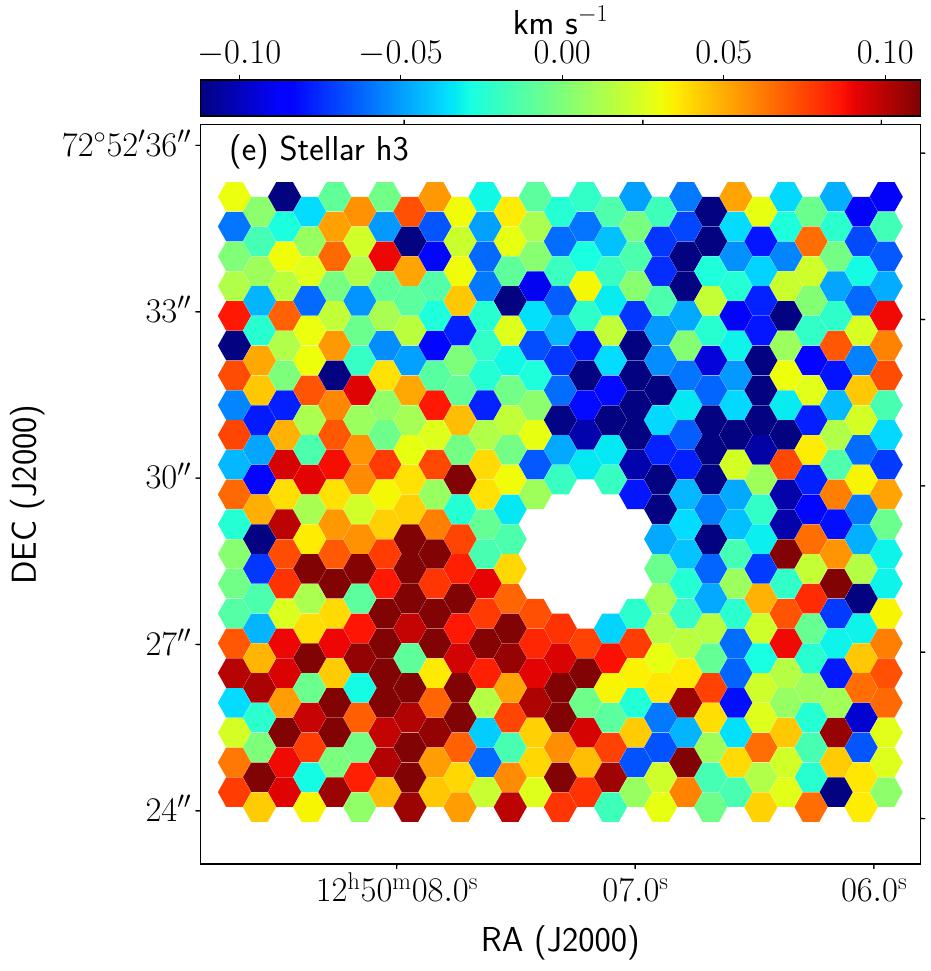}
	\includegraphics[clip, width=0.24\linewidth]{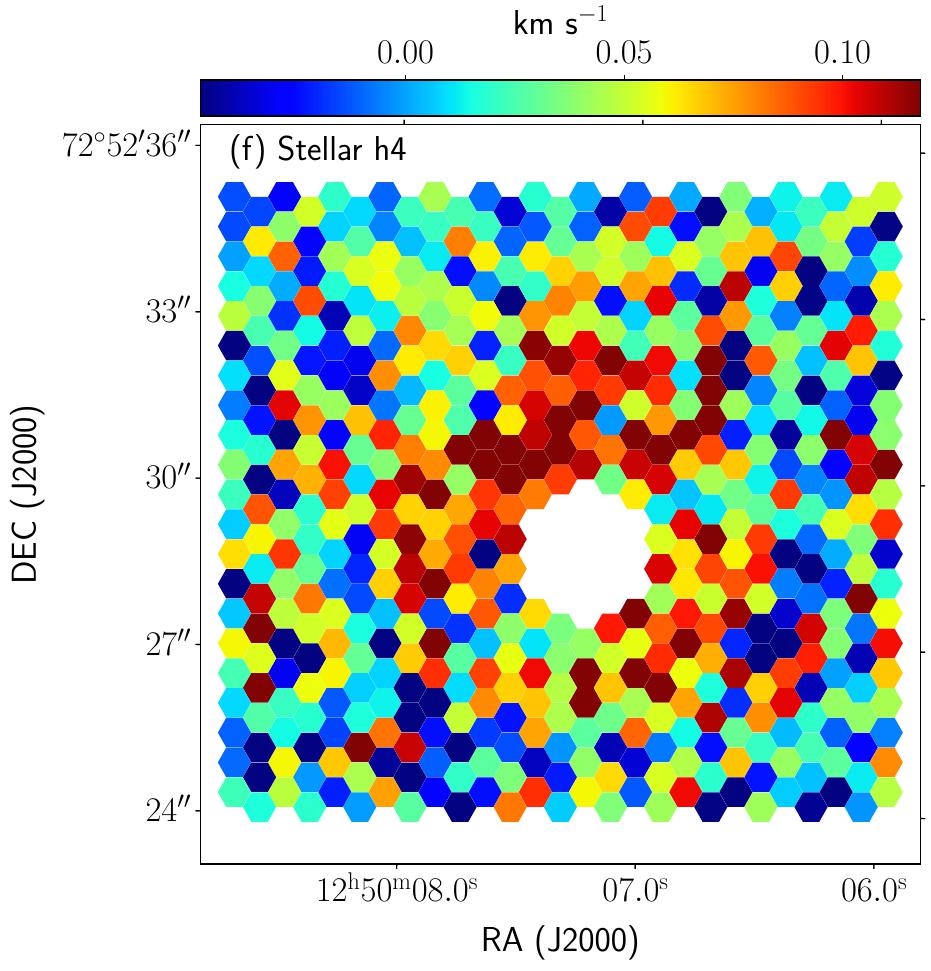}
	\includegraphics[clip, width=0.24\linewidth]{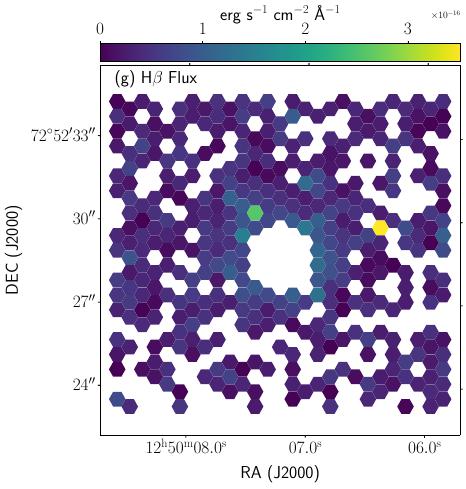}
	\includegraphics[clip, width=0.24\linewidth]{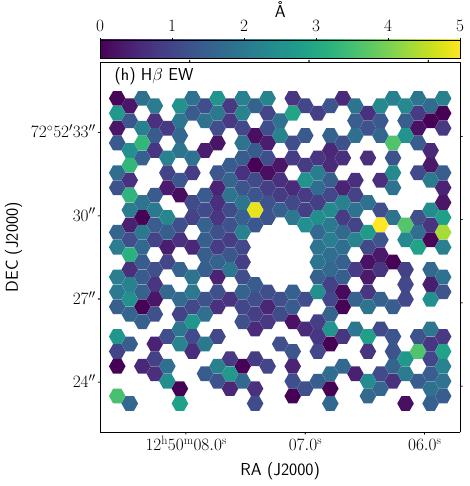}
	\includegraphics[clip, width=0.24\linewidth]{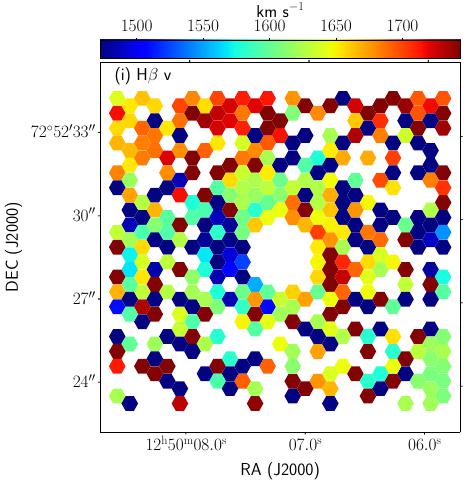}
	\includegraphics[clip, width=0.24\linewidth]{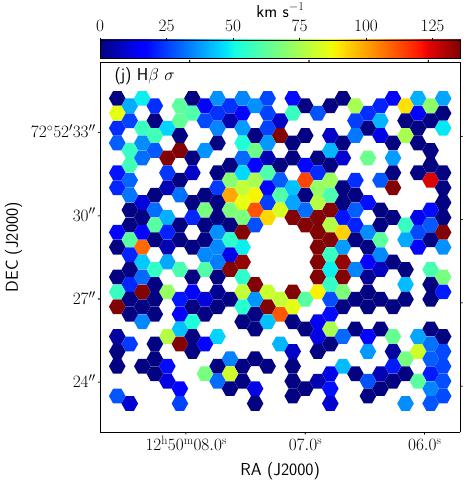}
	\includegraphics[clip, width=0.24\linewidth]{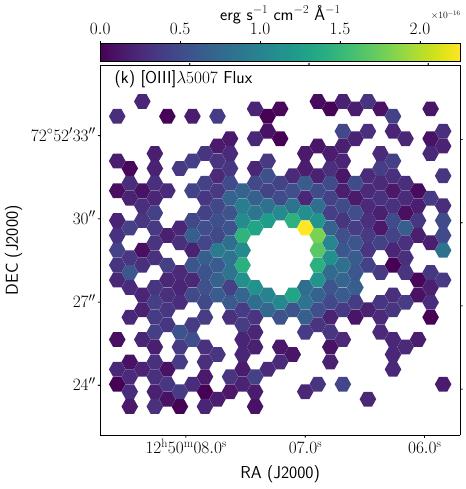}
	\includegraphics[clip, width=0.24\linewidth]{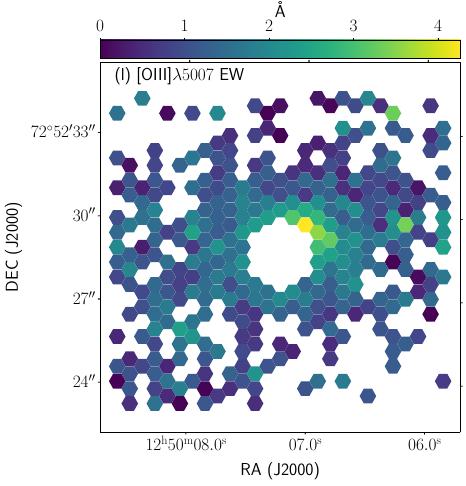}
	\includegraphics[clip, width=0.24\linewidth]{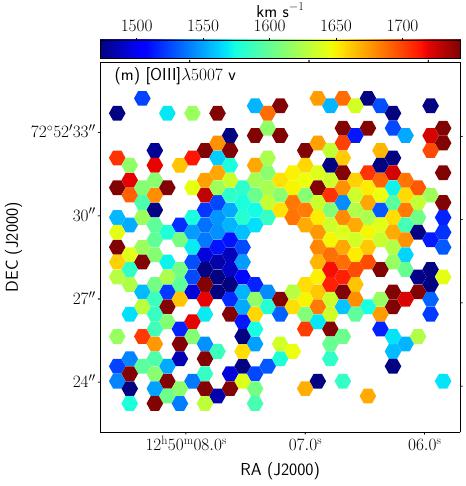}
	\includegraphics[clip, width=0.24\linewidth]{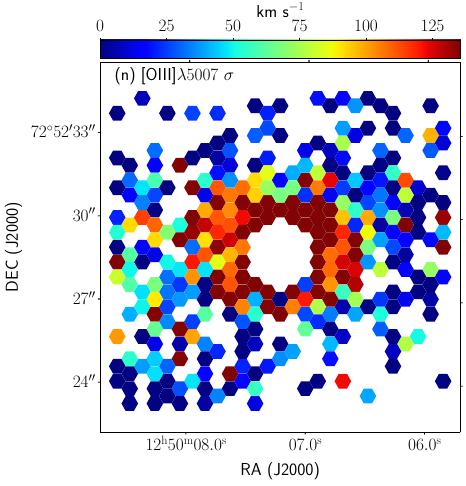}
	\vspace{5cm}
	\caption{NGC~4750 card.}
	\label{fig:NGC4750_card_1}
\end{figure*}
\addtocounter{figure}{-1}
\begin{figure*}[h]
	\centering
	\includegraphics[clip, width=0.24\linewidth]{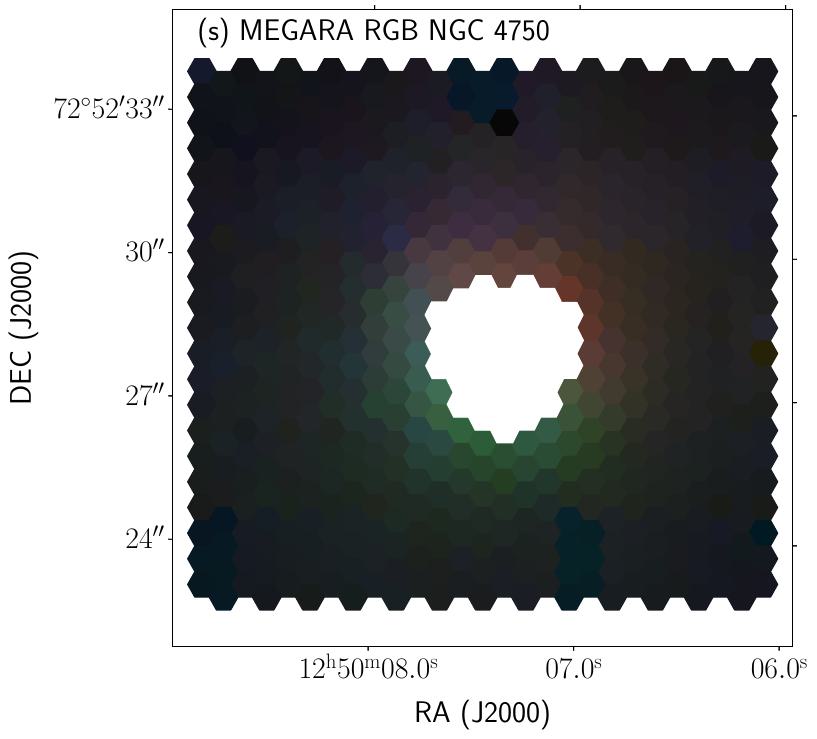}
	\includegraphics[clip, width=0.24\linewidth]{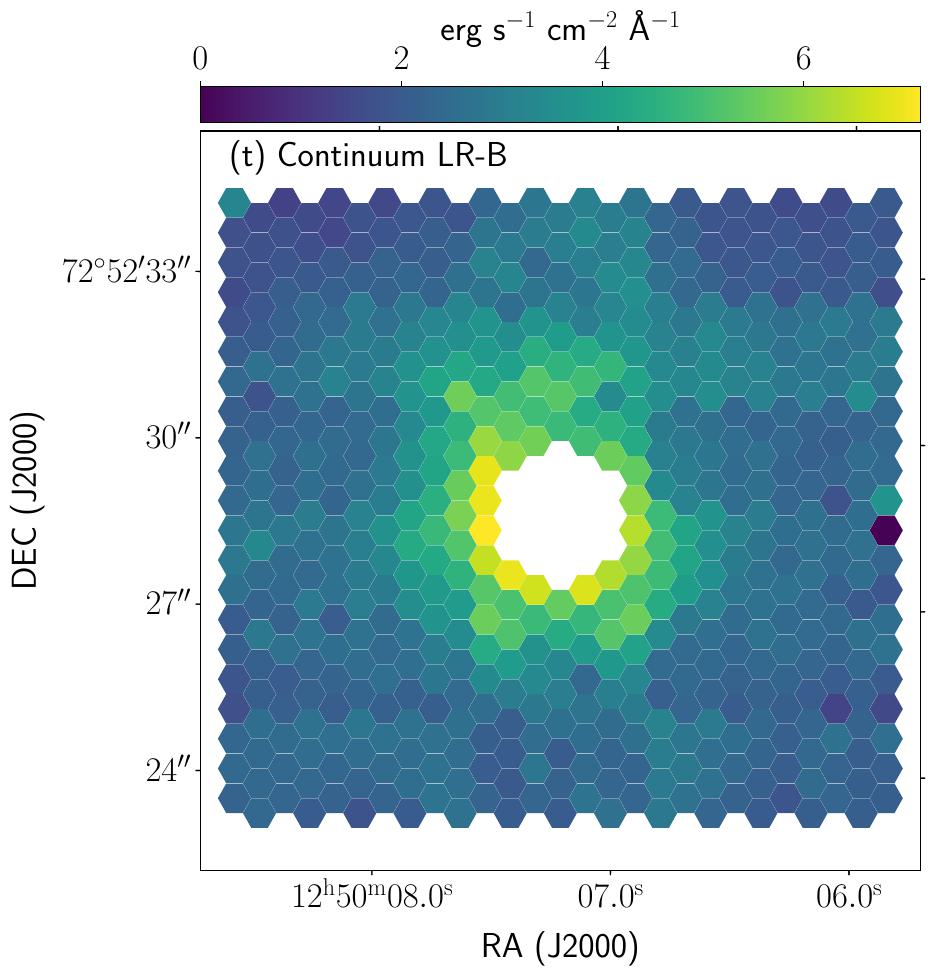}
	\includegraphics[clip, width=0.24\linewidth]{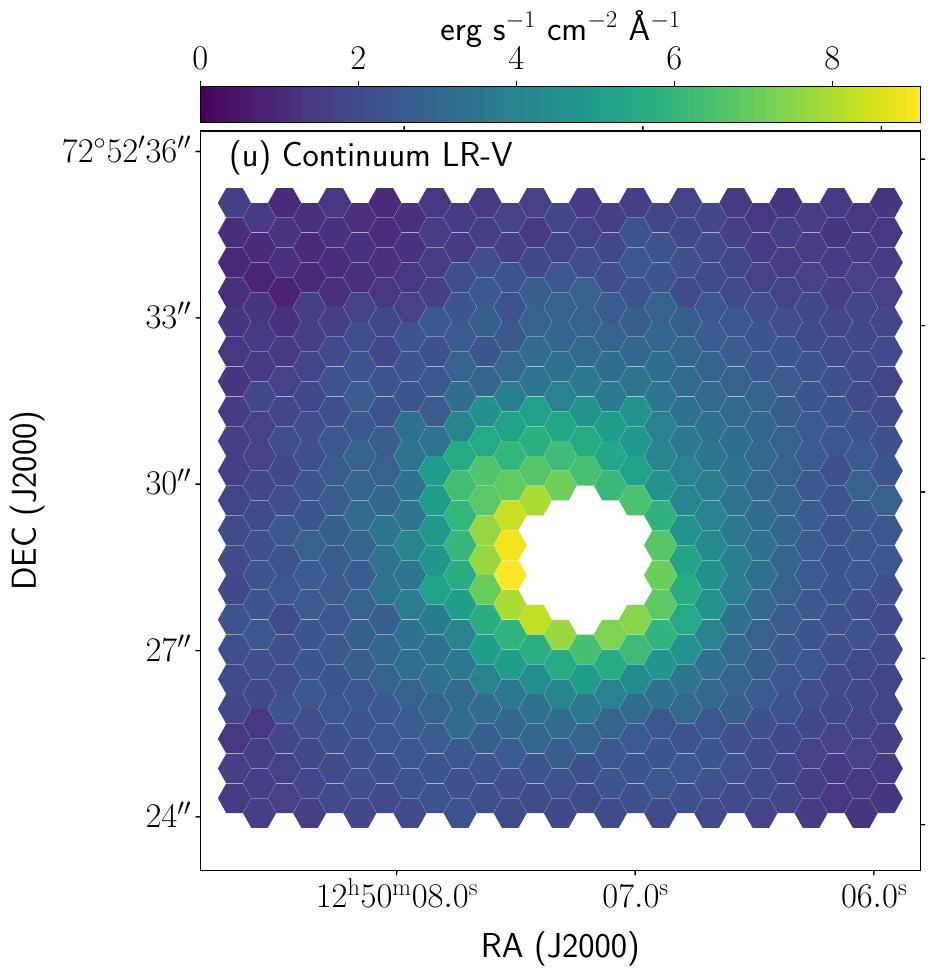}
	\includegraphics[clip, width=0.24\linewidth]{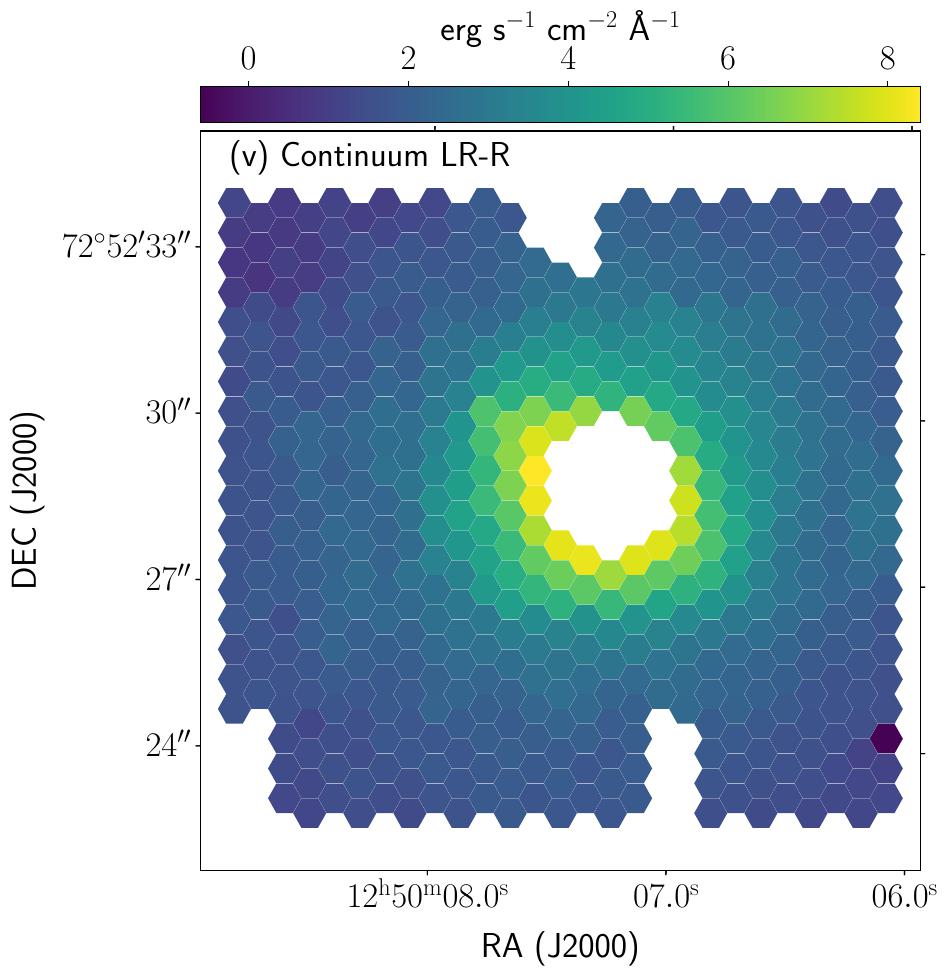}
	\includegraphics[clip, width=0.24\linewidth]{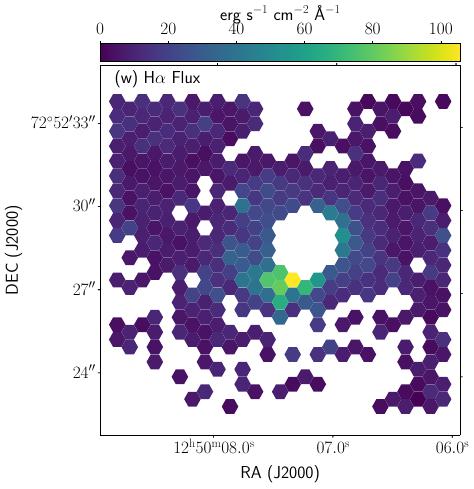}
	\includegraphics[clip, width=0.24\linewidth]{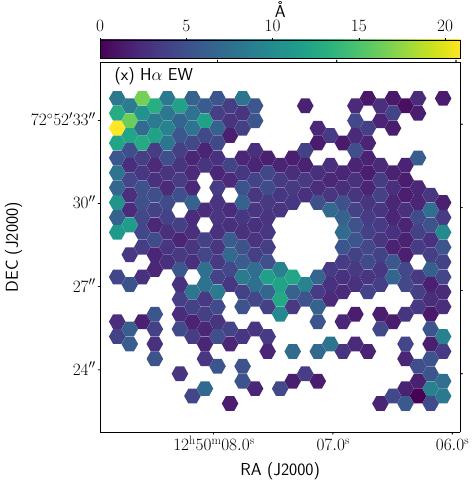}
	\includegraphics[clip, width=0.24\linewidth]{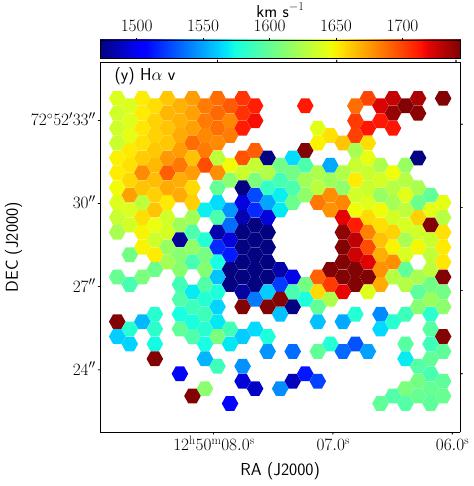}
	\includegraphics[clip, width=0.24\linewidth]{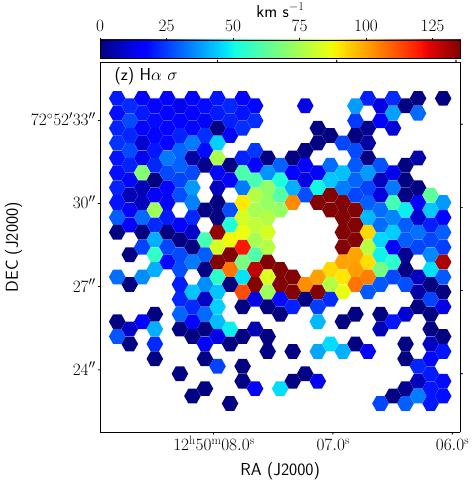}
	\includegraphics[clip, width=0.24\linewidth]{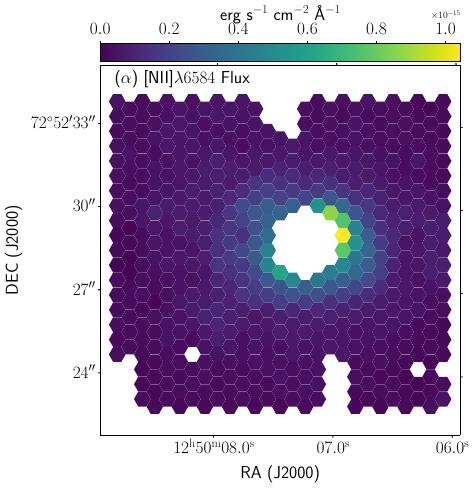}
	\includegraphics[clip, width=0.24\linewidth]{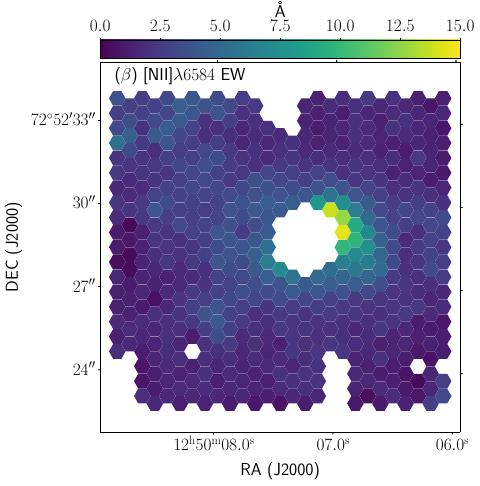}
	\includegraphics[clip, width=0.24\linewidth]{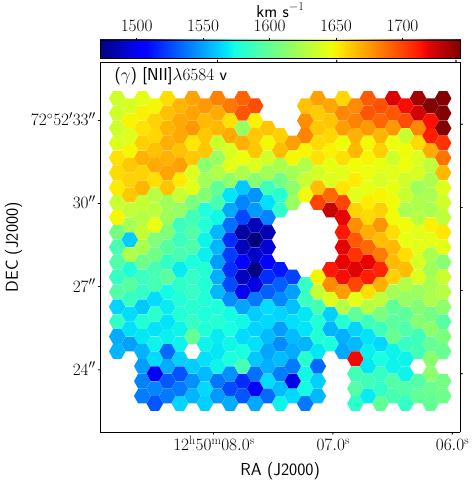}
	\includegraphics[clip, width=0.24\linewidth]{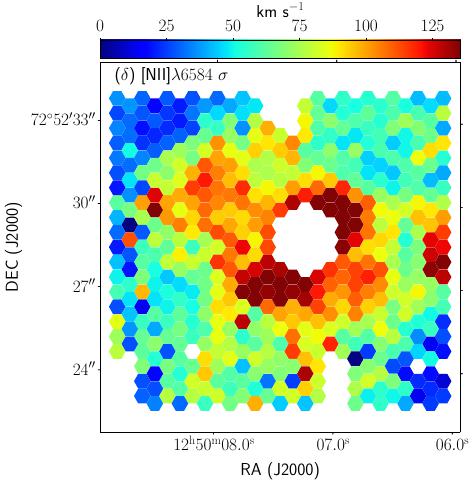}
	\includegraphics[clip, width=0.24\linewidth]{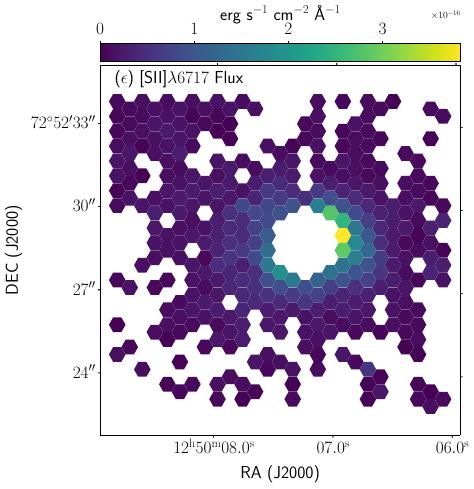}
	\includegraphics[clip, width=0.24\linewidth]{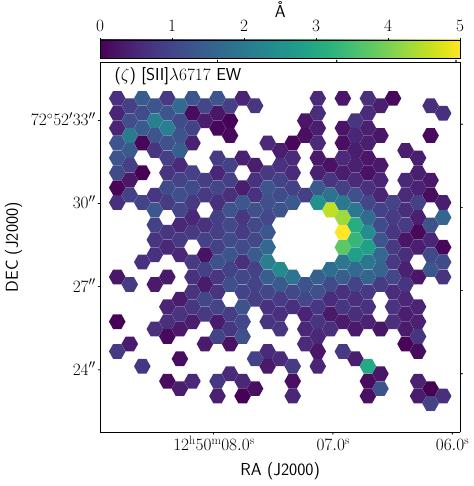}
	\includegraphics[clip, width=0.24\linewidth]{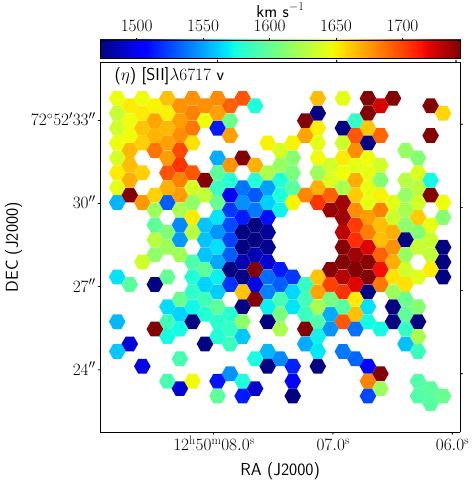}
	\includegraphics[clip, width=0.24\linewidth]{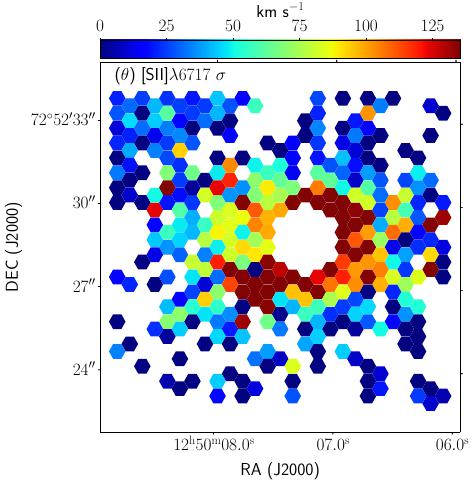}
	\includegraphics[clip, width=0.24\linewidth]{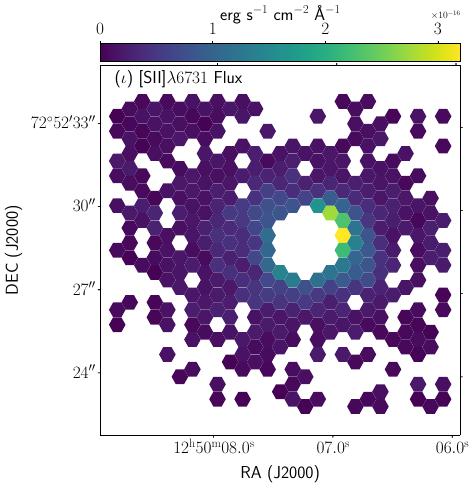}
	\includegraphics[clip, width=0.24\linewidth]{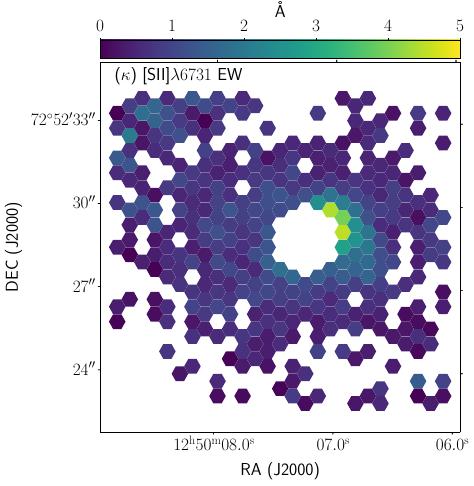}
	\includegraphics[clip, width=0.24\linewidth]{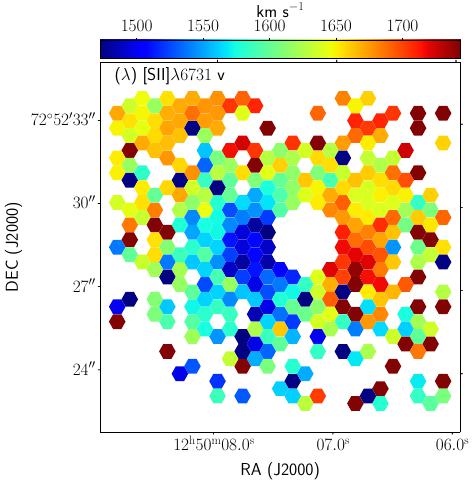}
	\includegraphics[clip, width=0.24\linewidth]{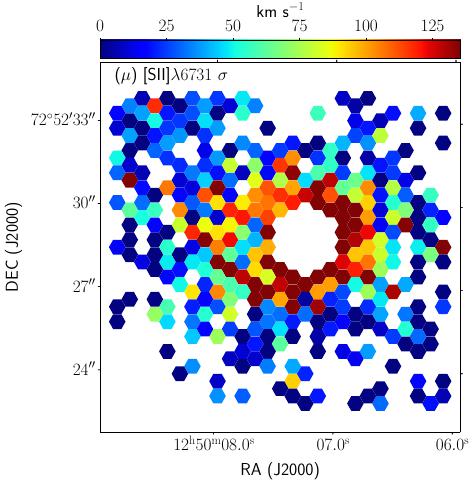}
	\caption{(cont.) NGC~4750 card.}
	\label{fig:NGC4750_card_2}
\end{figure*}

\begin{figure*}[h]
	\centering
	\includegraphics[clip, width=0.35\linewidth]{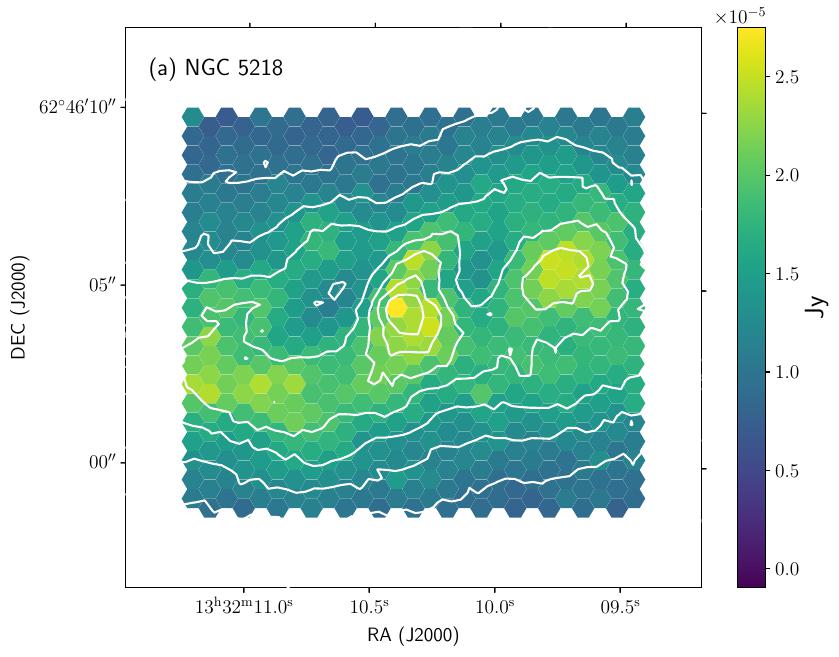}
	\includegraphics[clip, width=0.6\linewidth]{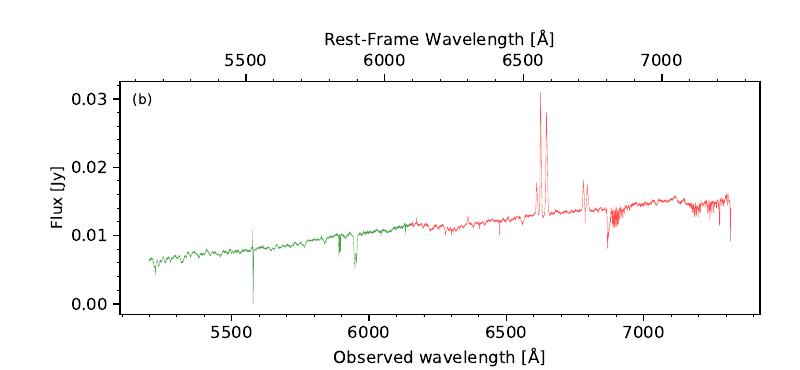}
	\includegraphics[clip, width=0.24\linewidth]{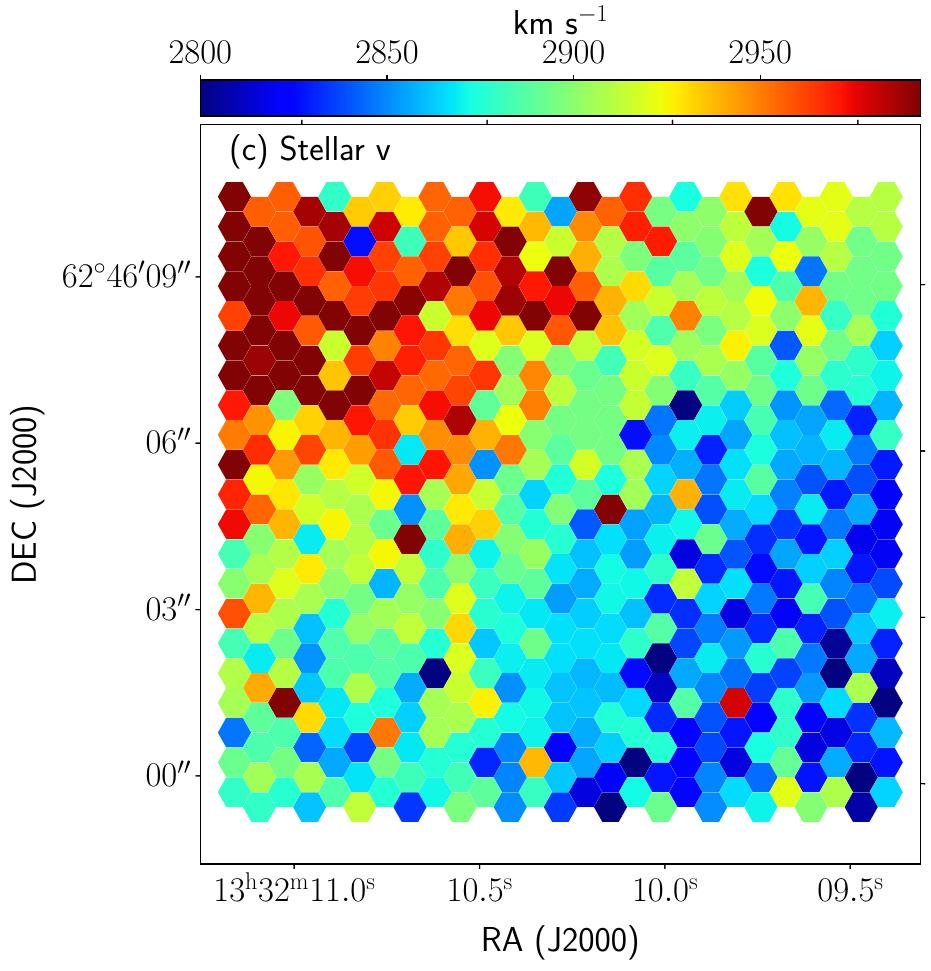}
	\includegraphics[clip, width=0.24\linewidth]{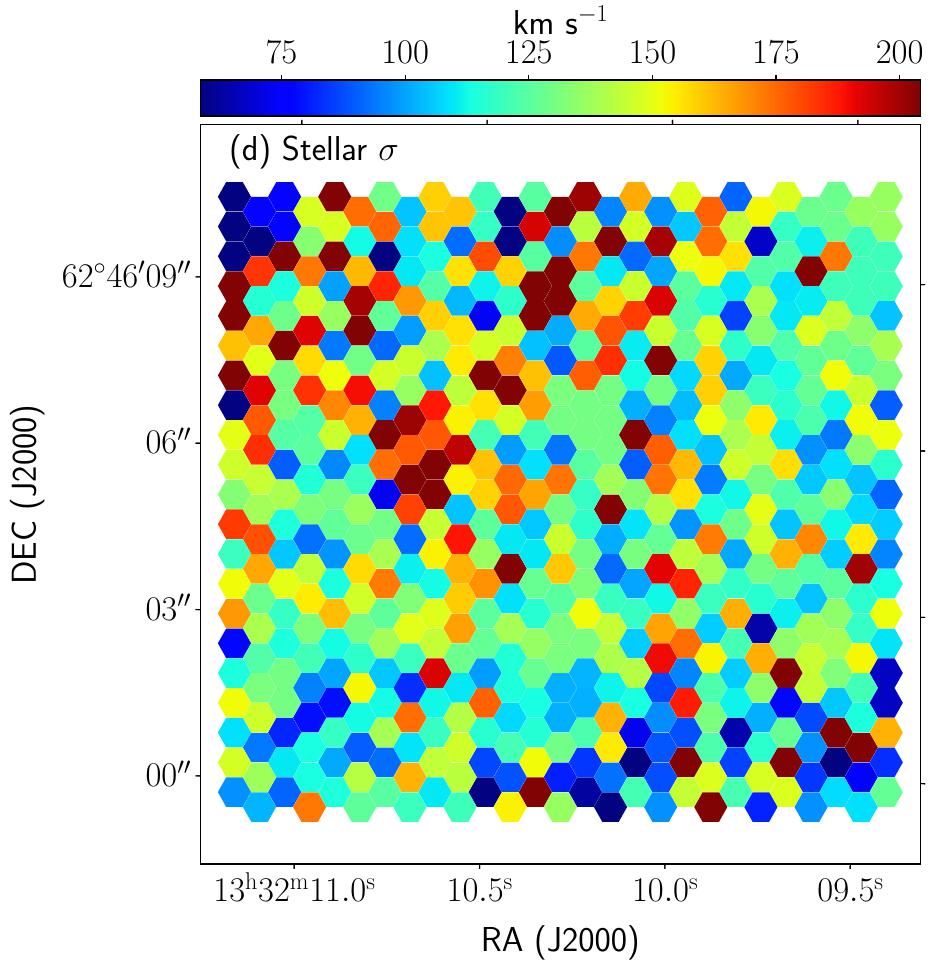}
	\includegraphics[clip, width=0.24\linewidth]{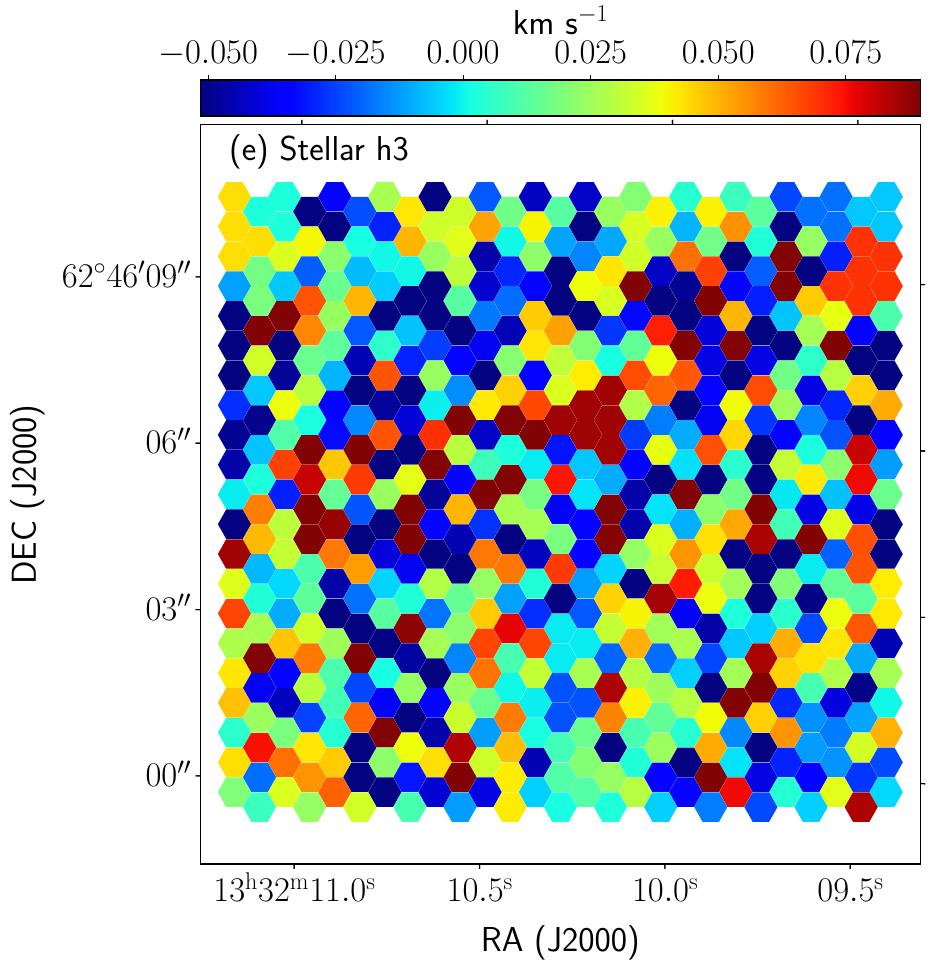}
	\includegraphics[clip, width=0.24\linewidth]{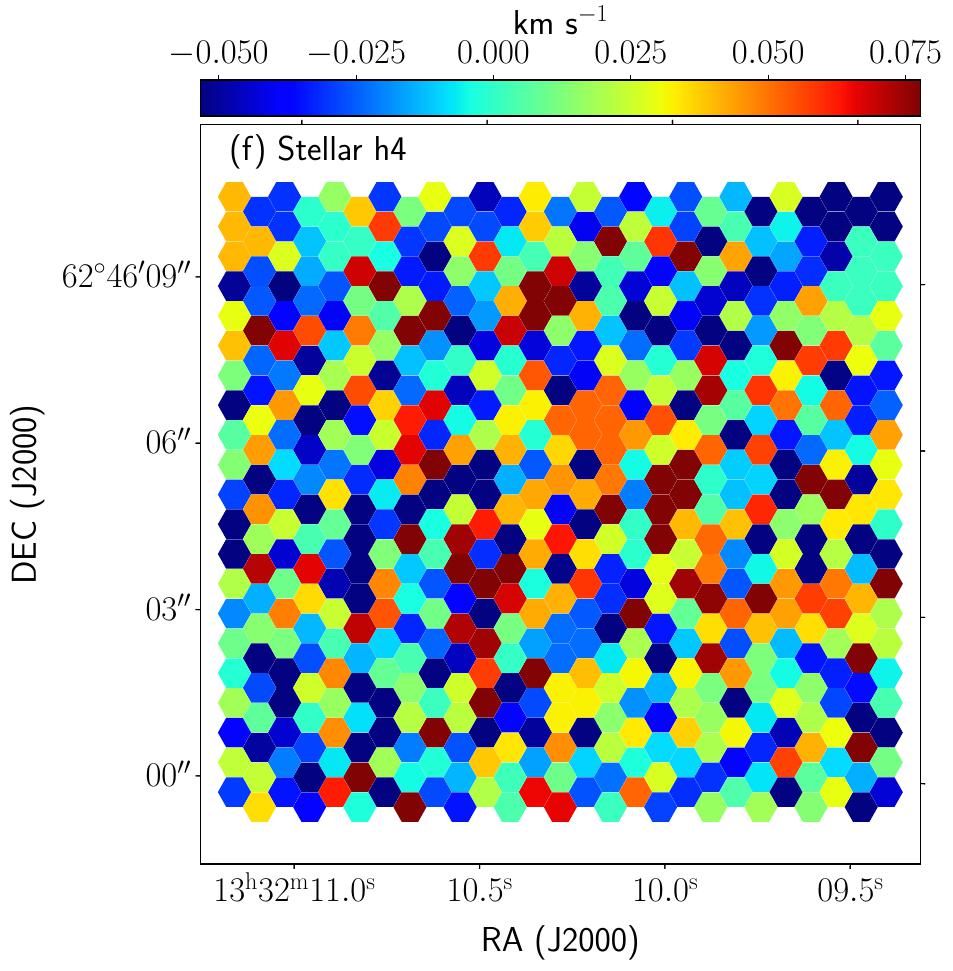}
	
	\vspace{8.8cm}
	
	\includegraphics[clip, width=0.24\linewidth]{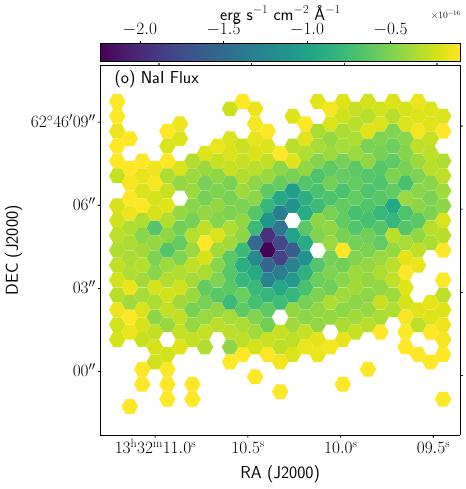}
	\includegraphics[clip, width=0.24\linewidth]{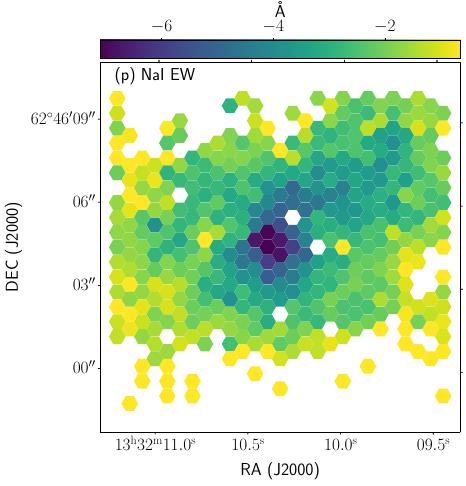}
	\includegraphics[clip, width=0.24\linewidth]{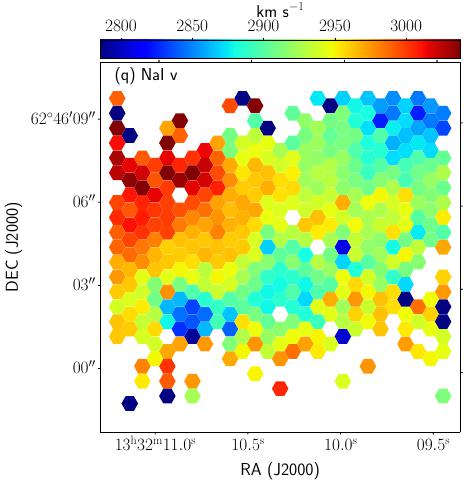}
	\includegraphics[clip, width=0.24\linewidth]{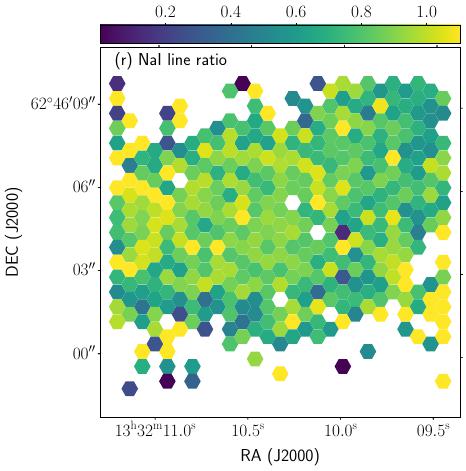}
	\caption{NGC~5218 card.}
	\label{fig:NGC5218_card_1}
\end{figure*}
\addtocounter{figure}{-1}
\begin{figure*}[h]
	\centering
	\includegraphics[clip, width=0.24\linewidth]{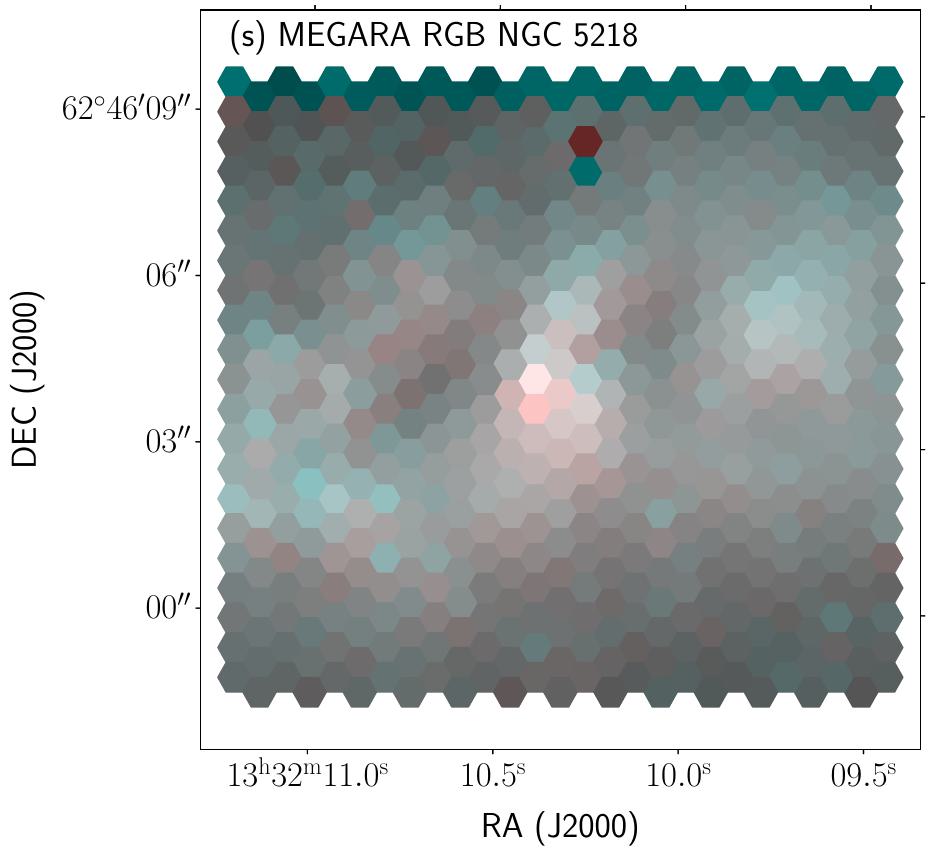}
	\hspace{4,4cm}
	\includegraphics[clip, width=0.24\linewidth]{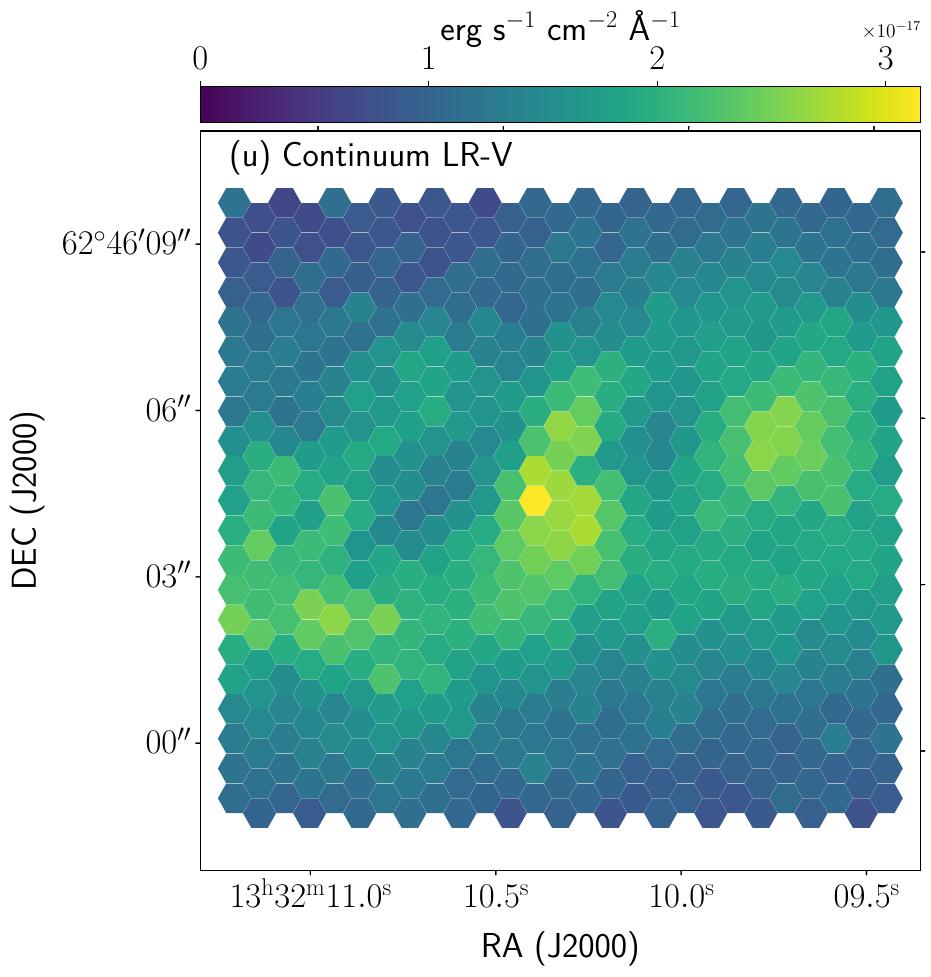}
	\includegraphics[clip, width=0.24\linewidth]{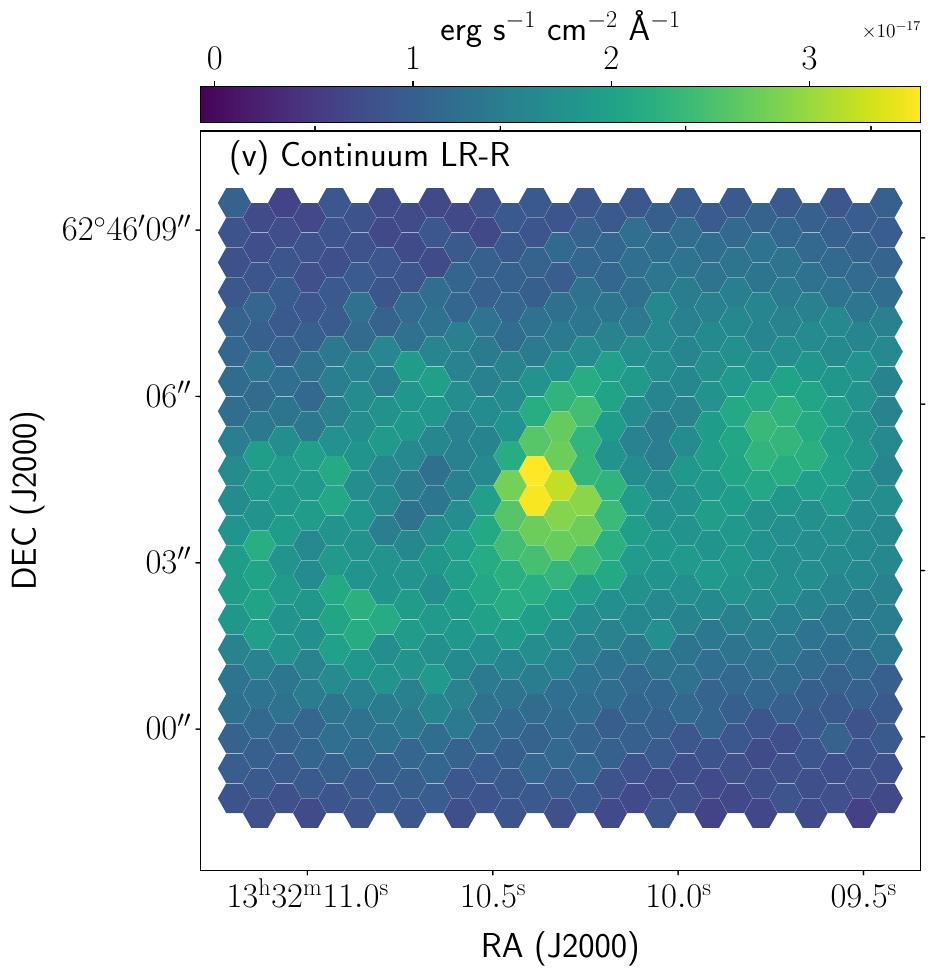}
	\includegraphics[clip, width=0.24\linewidth]{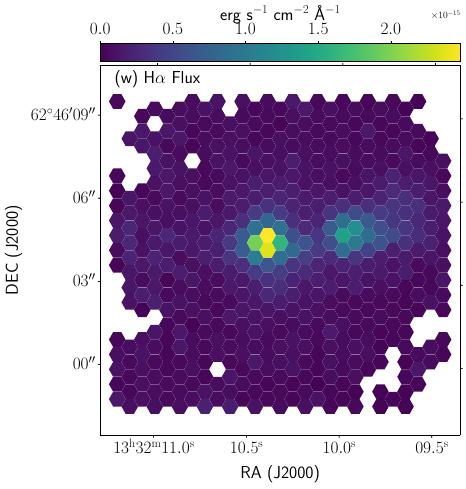}
	\includegraphics[clip, width=0.24\linewidth]{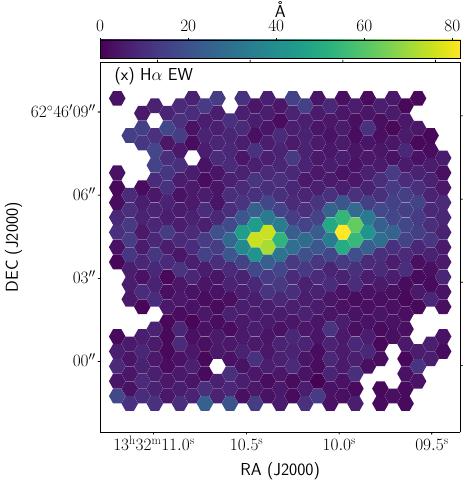}
	\includegraphics[clip, width=0.24\linewidth]{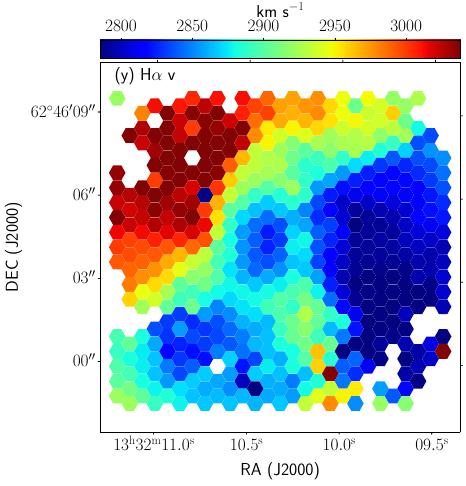}
	\includegraphics[clip, width=0.24\linewidth]{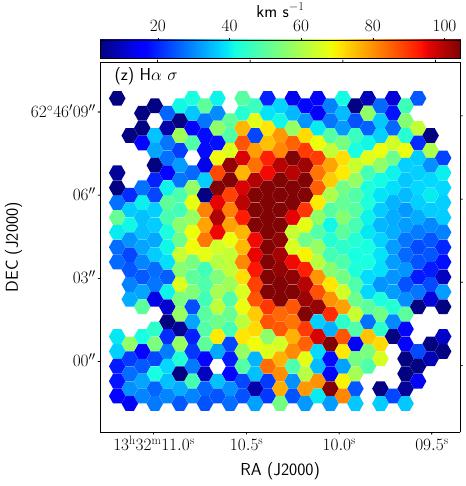}
	\includegraphics[clip, width=0.24\linewidth]{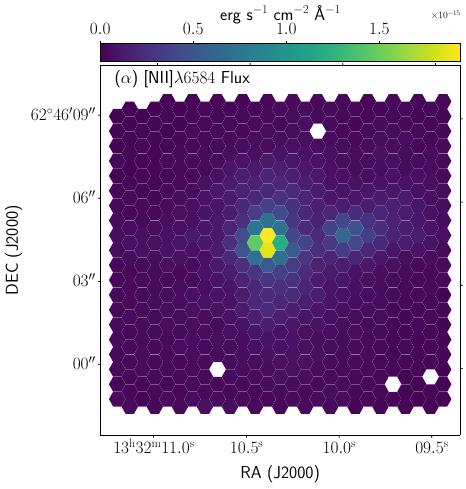}
	\includegraphics[clip, width=0.24\linewidth]{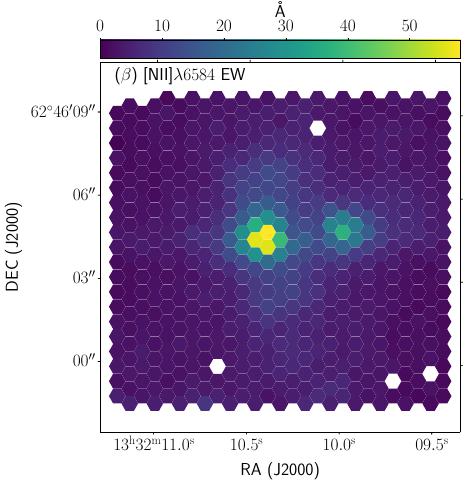}
	\includegraphics[clip, width=0.24\linewidth]{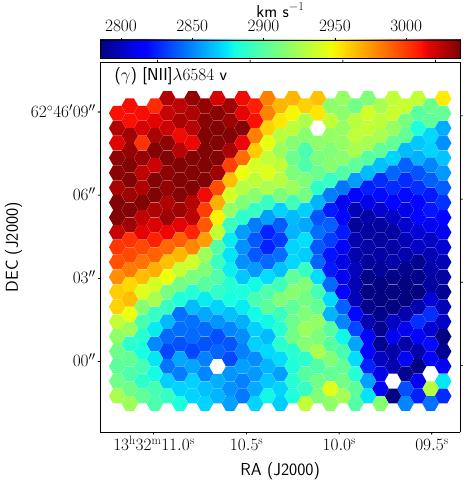}
	\includegraphics[clip, width=0.24\linewidth]{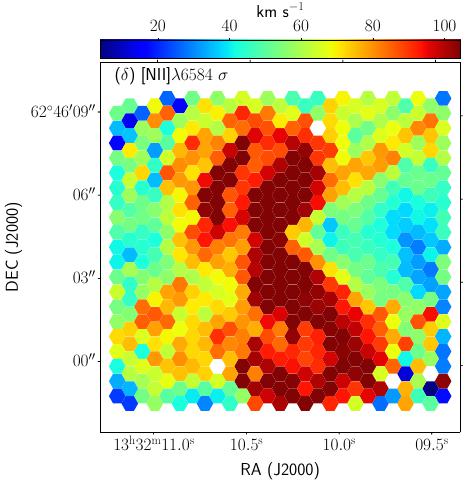}
	\includegraphics[clip, width=0.24\linewidth]{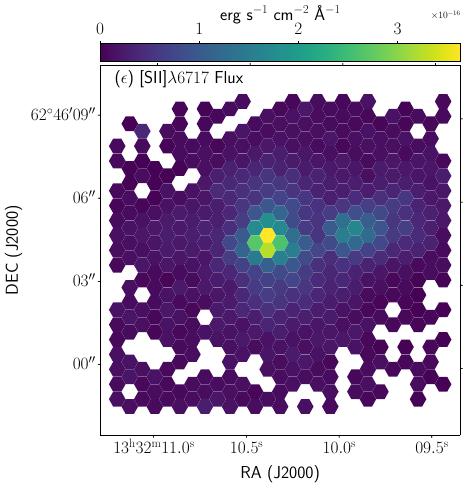}
	\includegraphics[clip, width=0.24\linewidth]{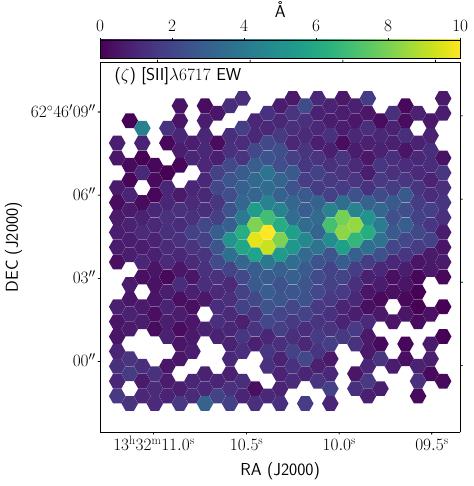}
	\includegraphics[clip, width=0.24\linewidth]{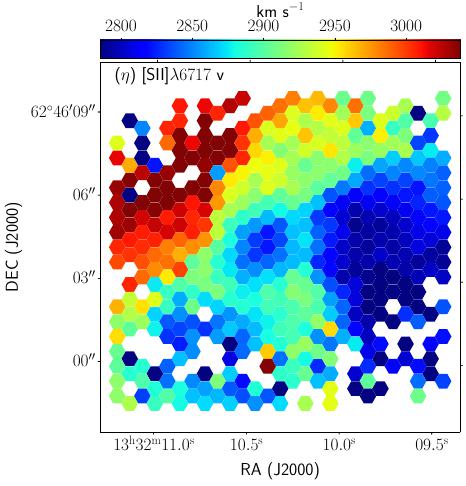}
	\includegraphics[clip, width=0.24\linewidth]{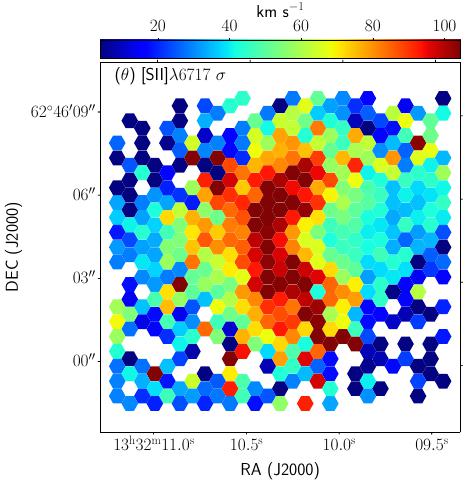}
	\includegraphics[clip, width=0.24\linewidth]{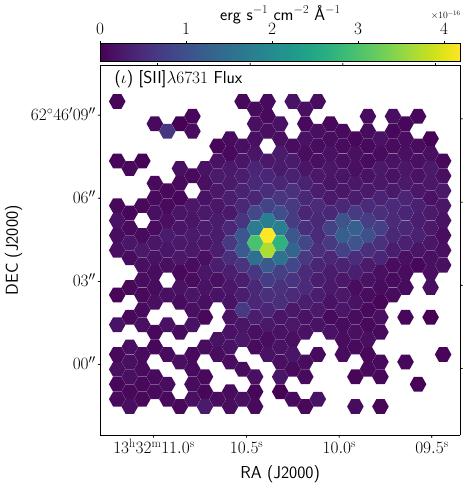}
	\includegraphics[clip, width=0.24\linewidth]{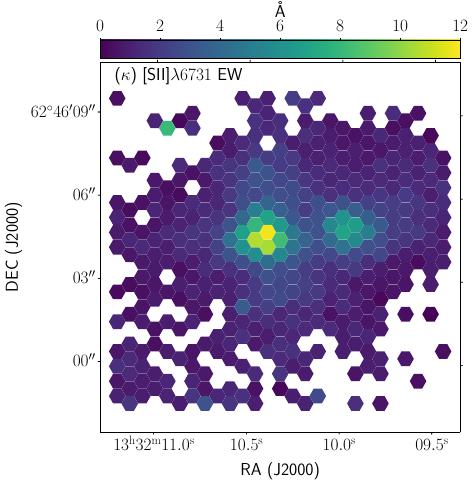}
	\includegraphics[clip, width=0.24\linewidth]{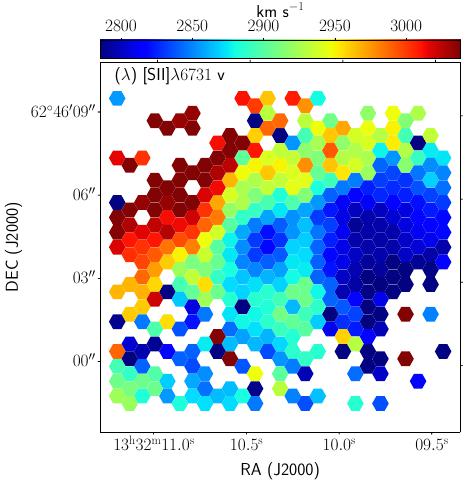}
	\includegraphics[clip, width=0.24\linewidth]{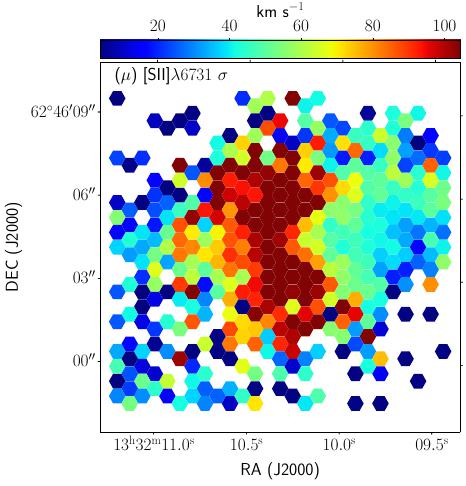}
	\caption{(cont.) NGC~5218 card.}
	\label{fig:NGC5218_card_2}
\end{figure*}

\begin{figure*}[h]
	\centering
	\includegraphics[clip, width=0.35\linewidth]{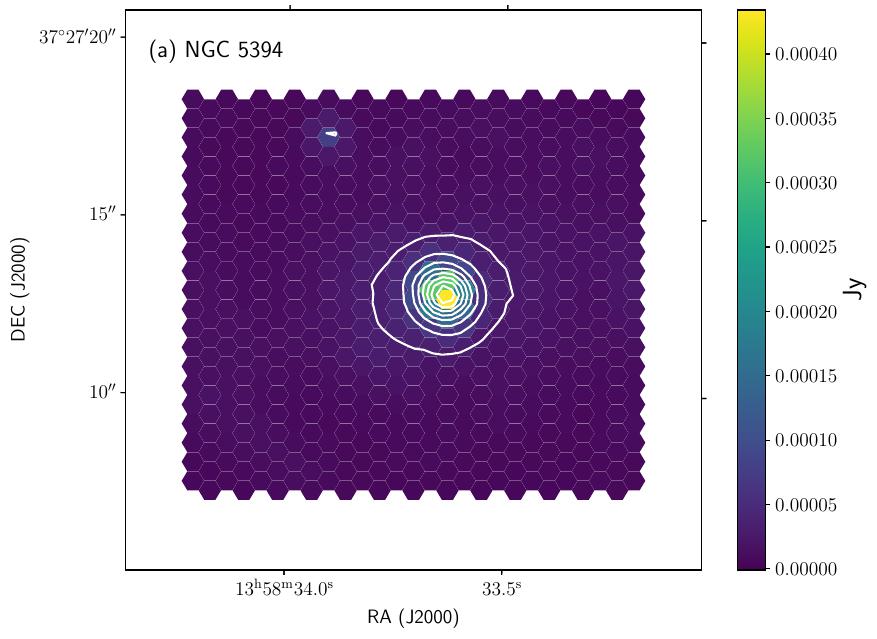}
	\includegraphics[clip, width=0.6\linewidth]{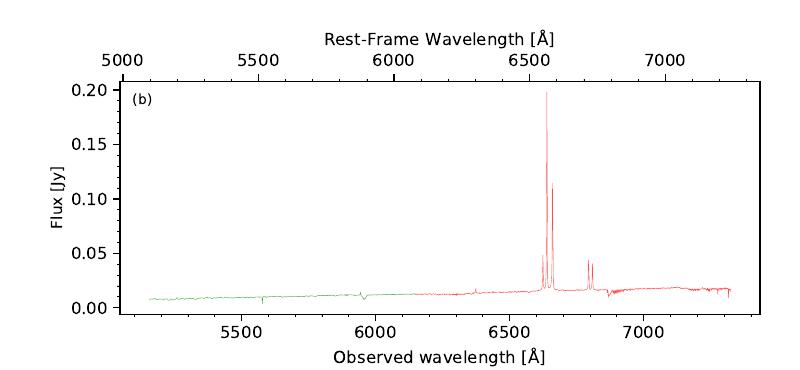}
	\includegraphics[clip, width=0.24\linewidth]{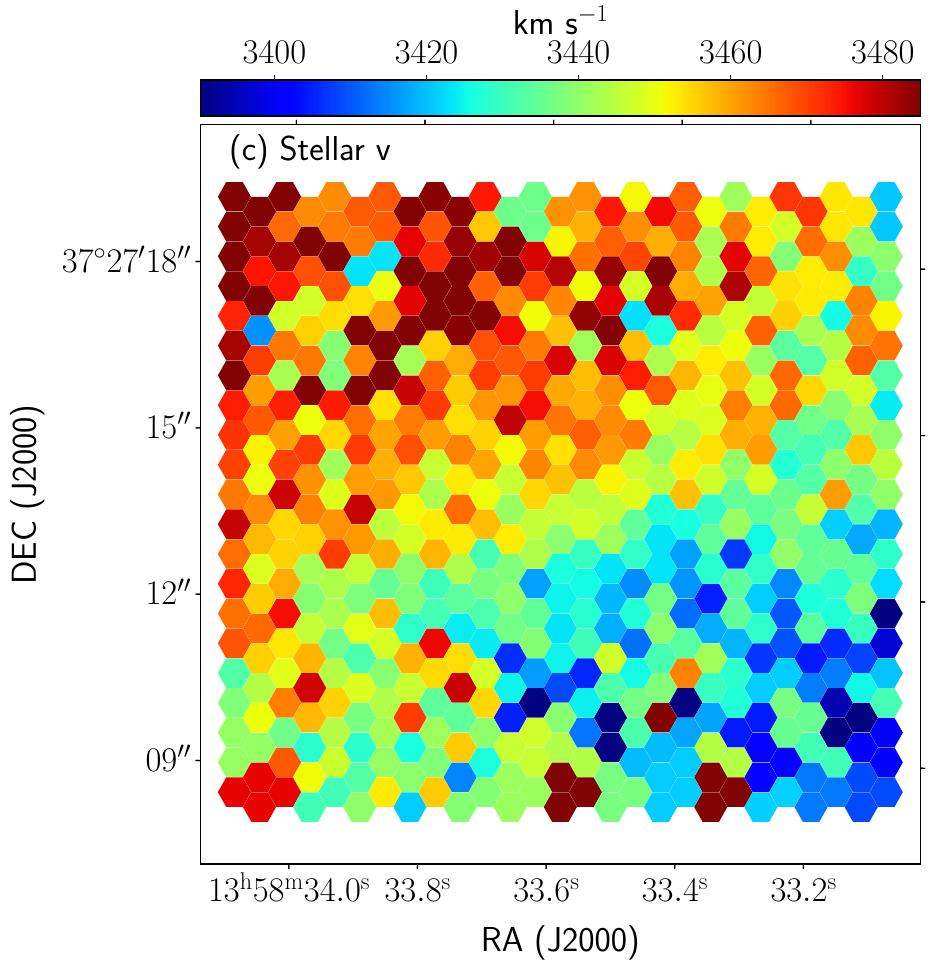}
	\includegraphics[clip, width=0.24\linewidth]{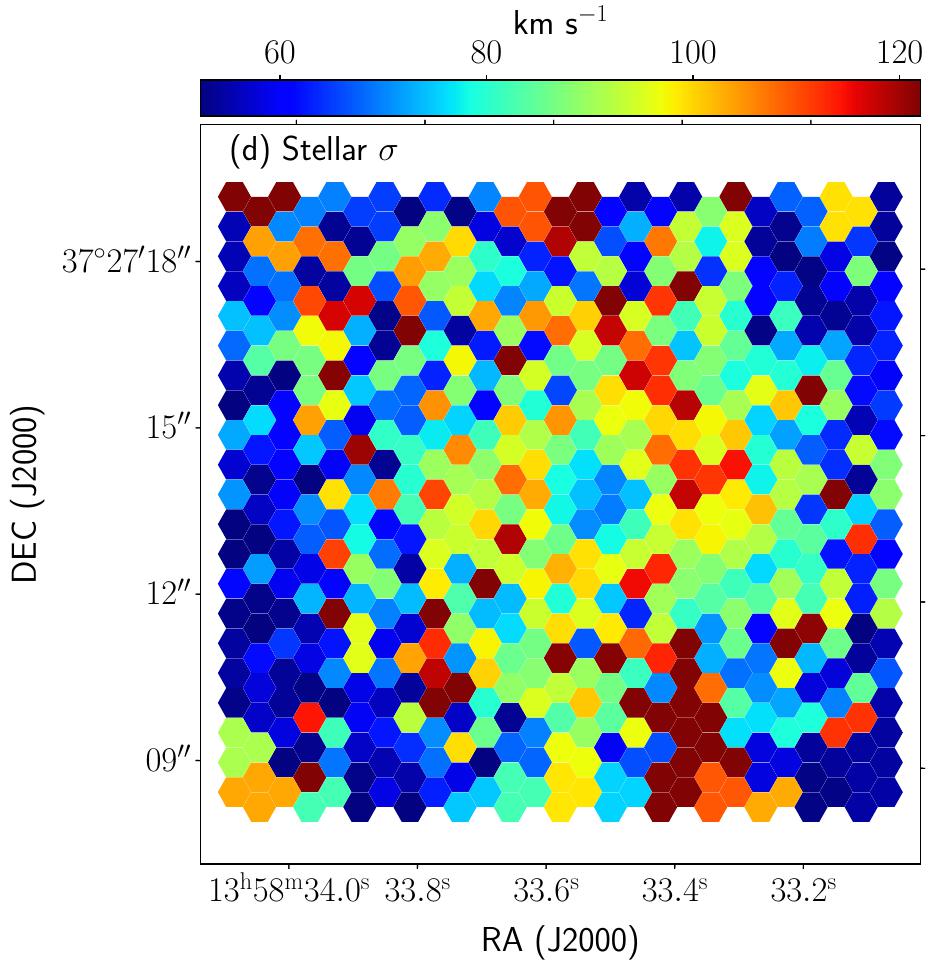}
	\includegraphics[clip, width=0.24\linewidth]{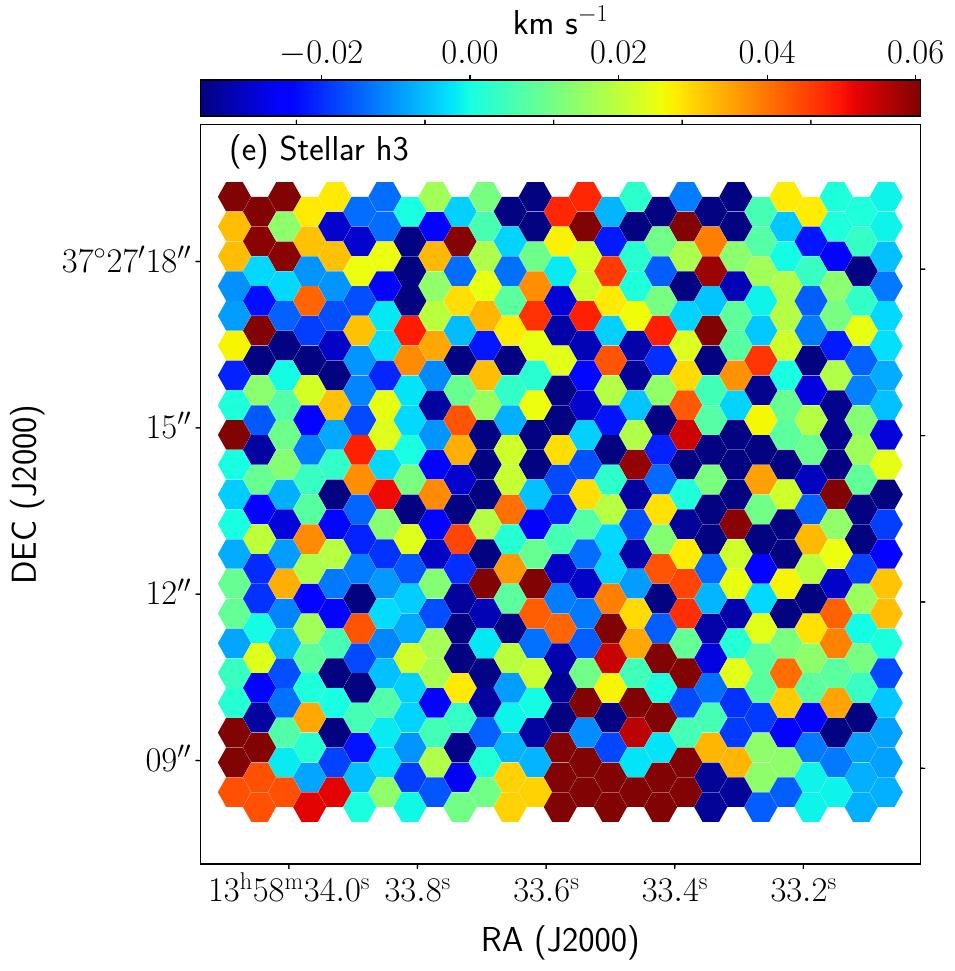}
	\includegraphics[clip, width=0.24\linewidth]{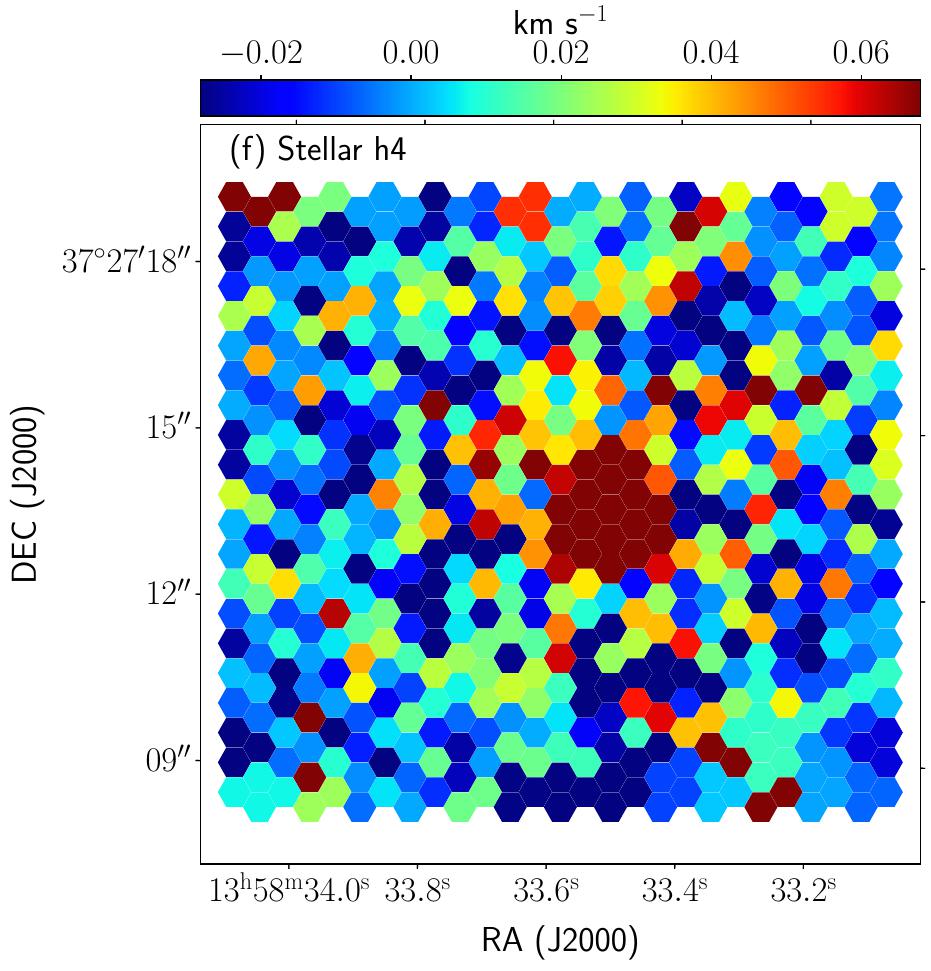}
	
	\vspace{8.8cm}
	
	\includegraphics[clip, width=0.24\linewidth]{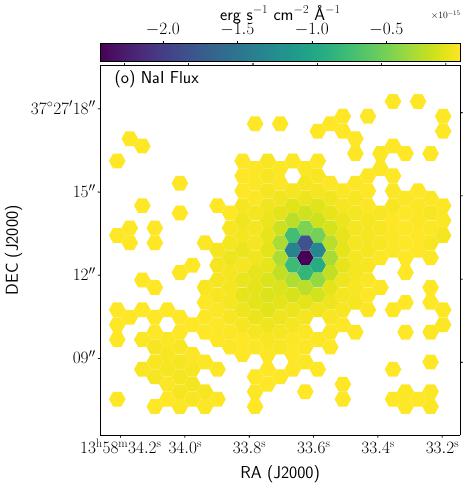}
	\includegraphics[clip, width=0.24\linewidth]{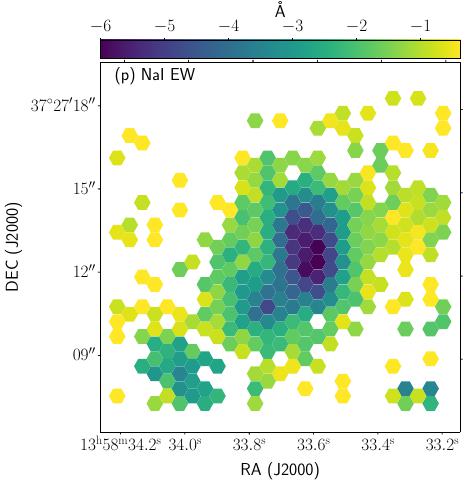}
	\includegraphics[clip, width=0.24\linewidth]{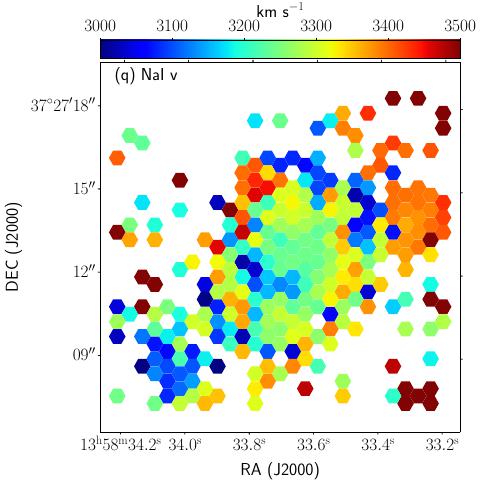}
	\includegraphics[clip, width=0.24\linewidth]{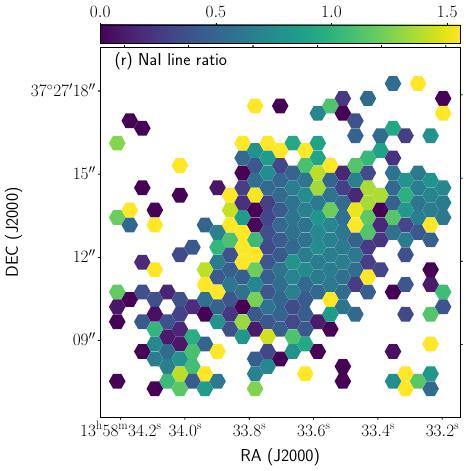}
	\caption{NGC~5394 card.}
	\label{fig:NGC5394_card_1}
\end{figure*}
\addtocounter{figure}{-1}
\begin{figure*}[h]
	\centering
	\includegraphics[clip, width=0.24\linewidth]{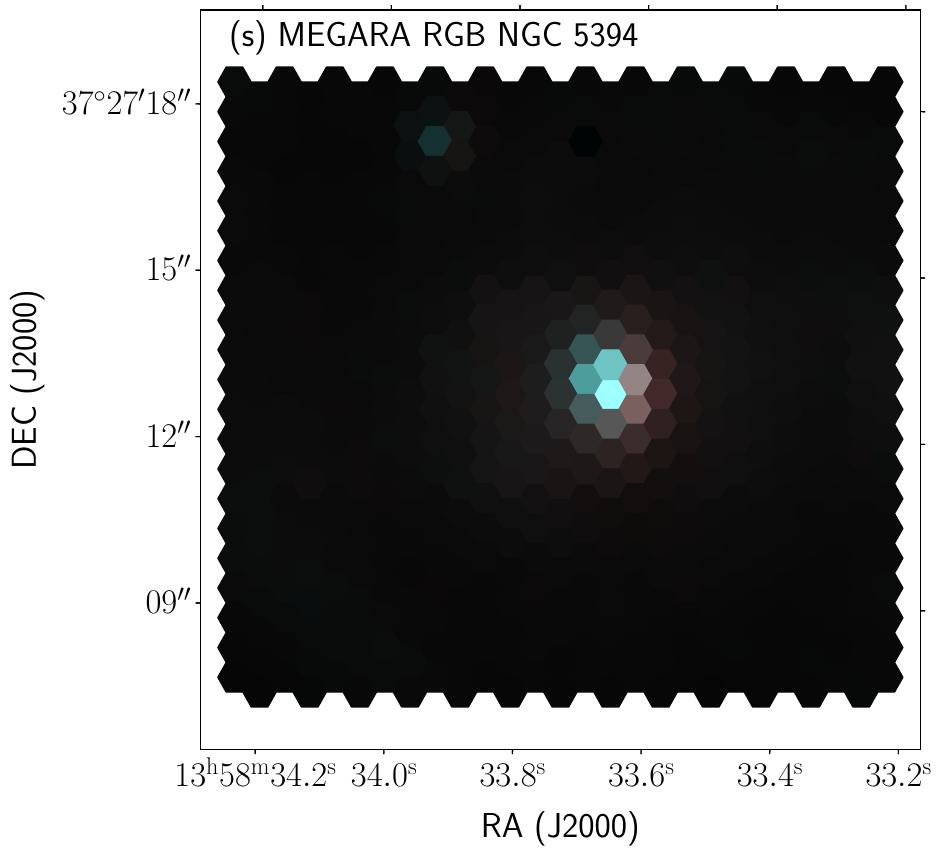}
	\hspace{4,4cm}
	\includegraphics[clip, width=0.24\linewidth]{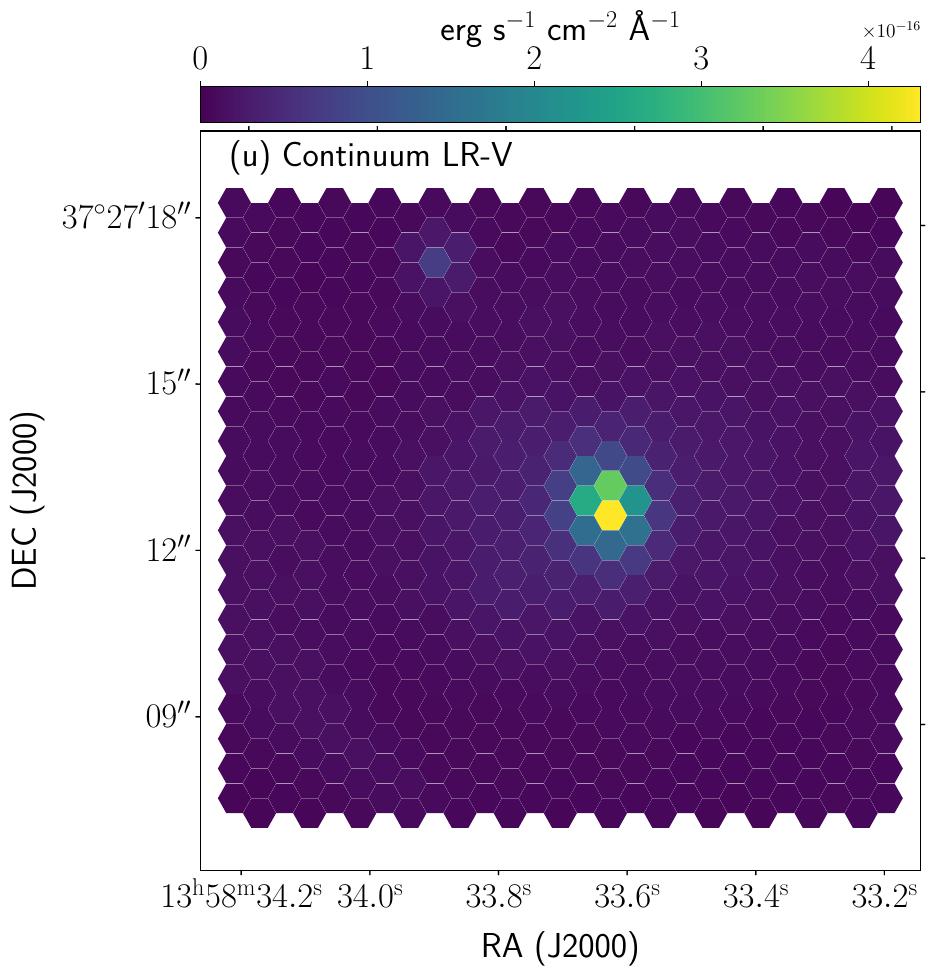}
	\includegraphics[clip, width=0.24\linewidth]{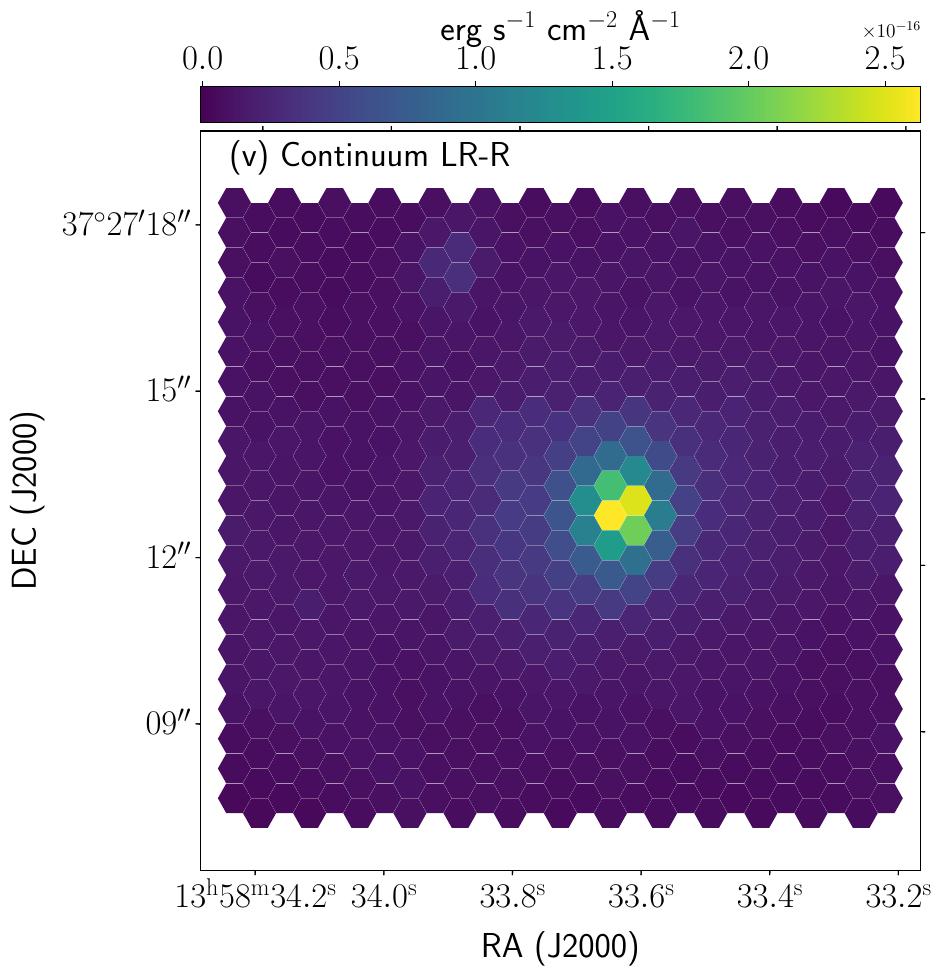}
	\includegraphics[clip, width=0.24\linewidth]{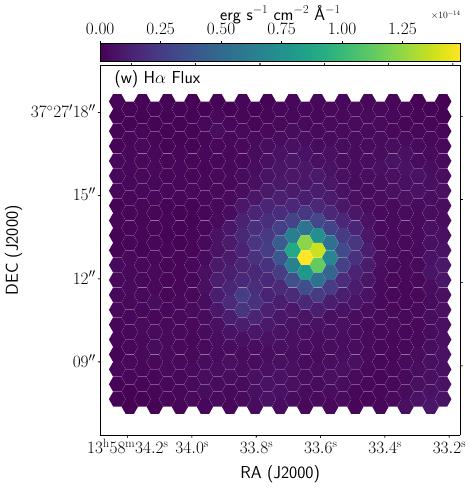}
	\includegraphics[clip, width=0.24\linewidth]{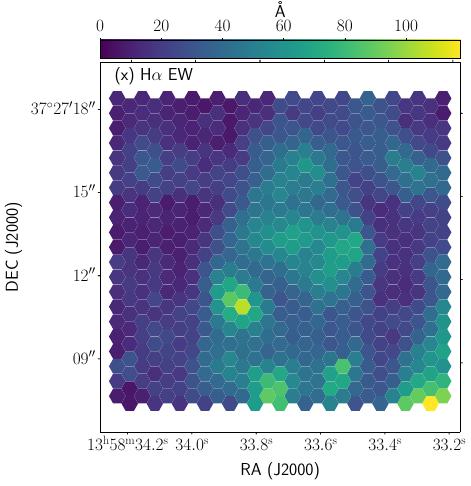}
	\includegraphics[clip, width=0.24\linewidth]{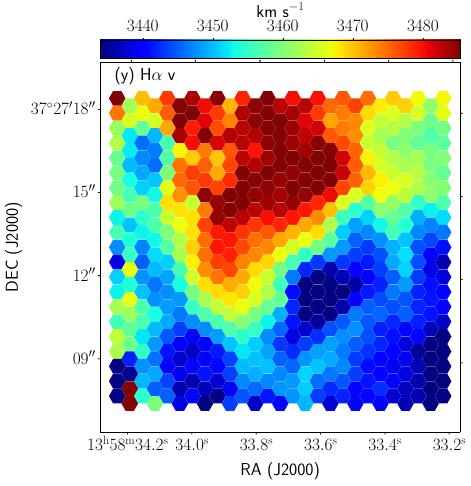}
	\includegraphics[clip, width=0.24\linewidth]{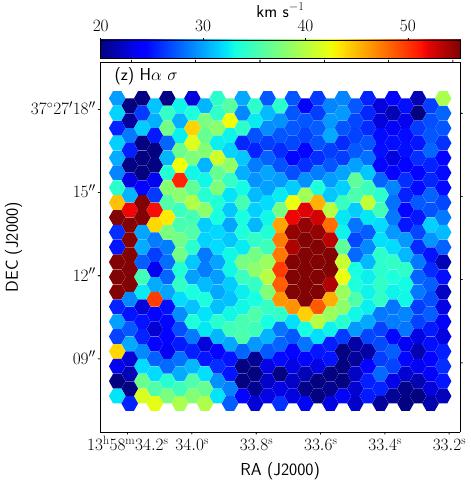}
	\includegraphics[clip, width=0.24\linewidth]{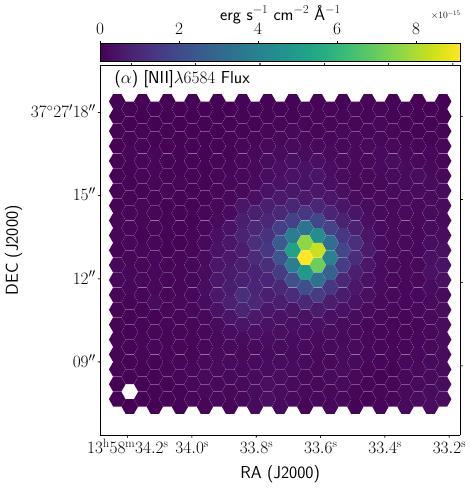}
	\includegraphics[clip, width=0.24\linewidth]{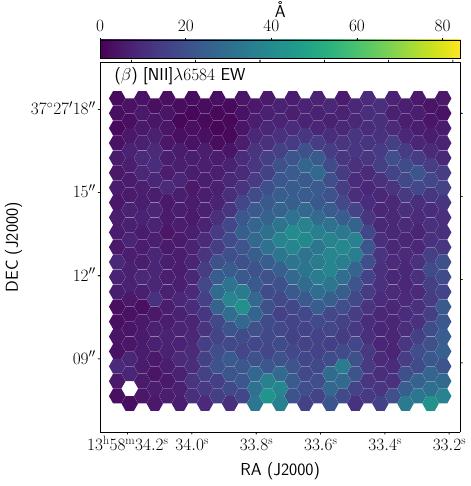}
	\includegraphics[clip, width=0.24\linewidth]{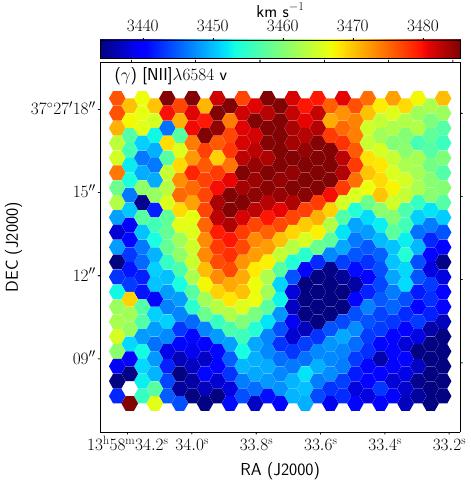}
	\includegraphics[clip, width=0.24\linewidth]{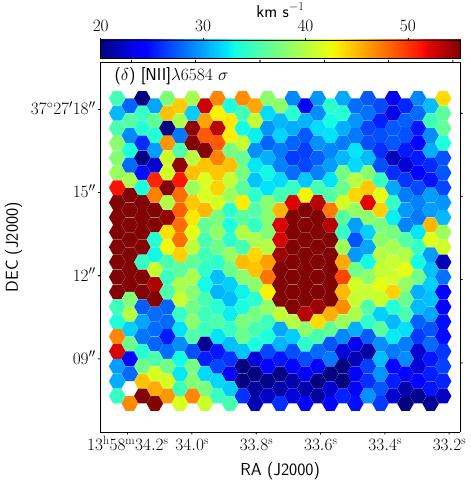}
	\includegraphics[clip, width=0.24\linewidth]{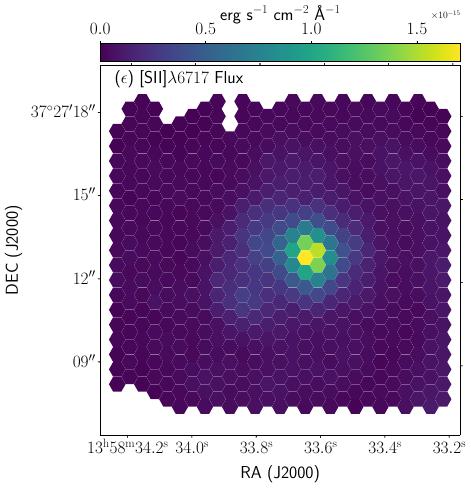}
	\includegraphics[clip, width=0.24\linewidth]{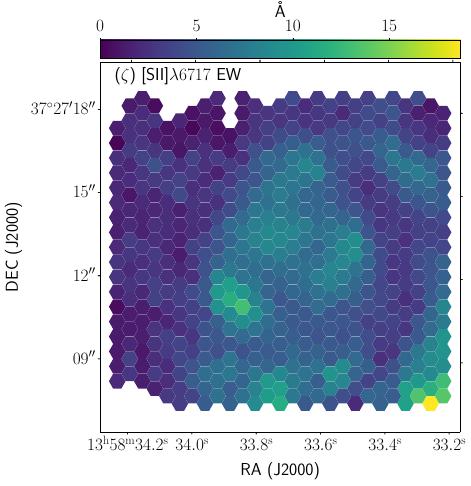}
	\includegraphics[clip, width=0.24\linewidth]{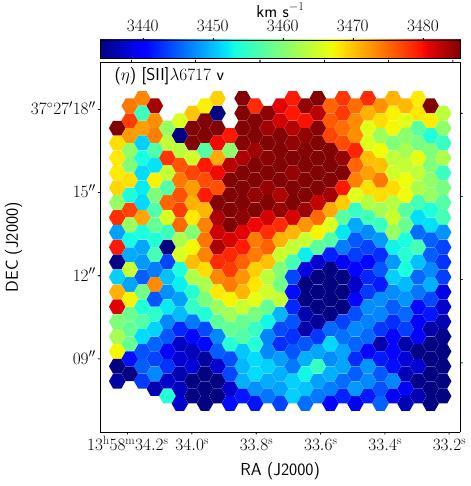}
	\includegraphics[clip, width=0.24\linewidth]{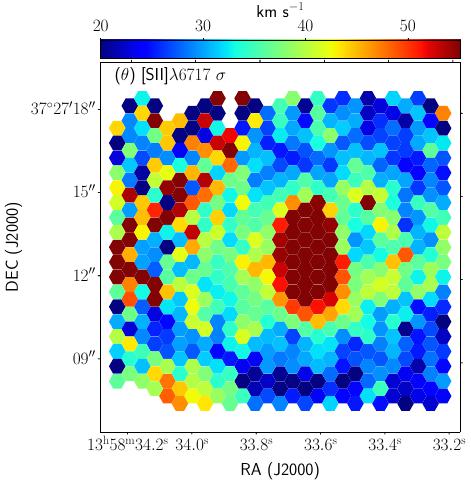}
	\includegraphics[clip, width=0.24\linewidth]{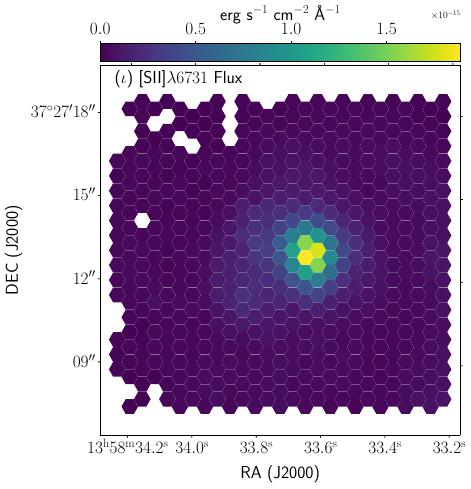}
	\includegraphics[clip, width=0.24\linewidth]{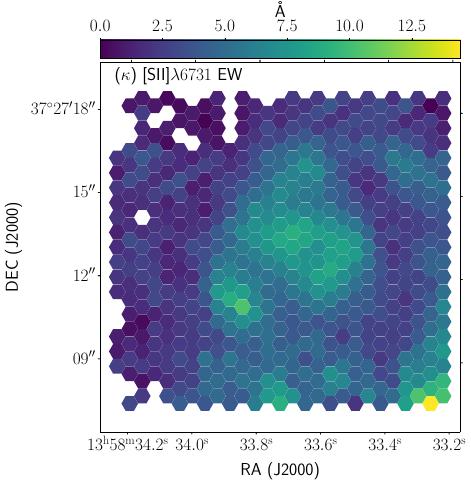}
	\includegraphics[clip, width=0.24\linewidth]{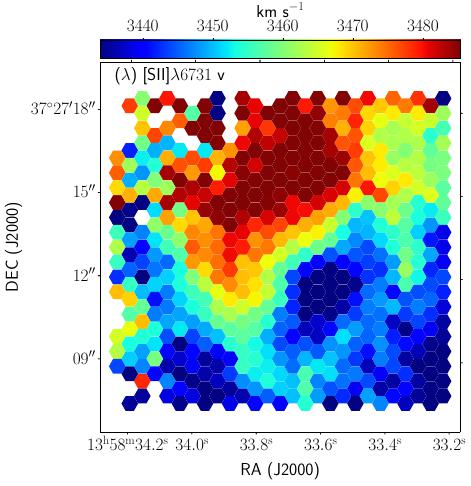}
	\includegraphics[clip, width=0.24\linewidth]{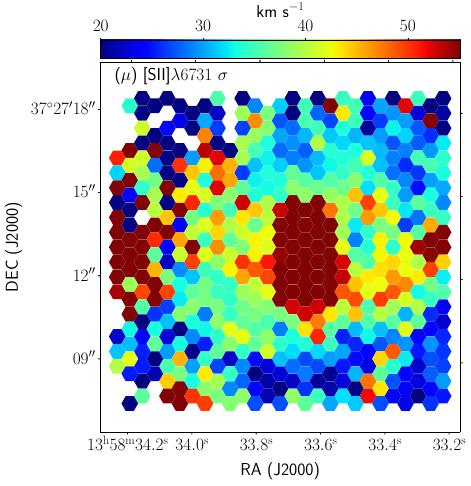}
	\caption{(cont.) NGC~5394 card.}
	\label{fig:NGC5394_card_2}
\end{figure*}

\begin{figure*}[h]
	\centering
	\includegraphics[clip, width=0.35\linewidth]{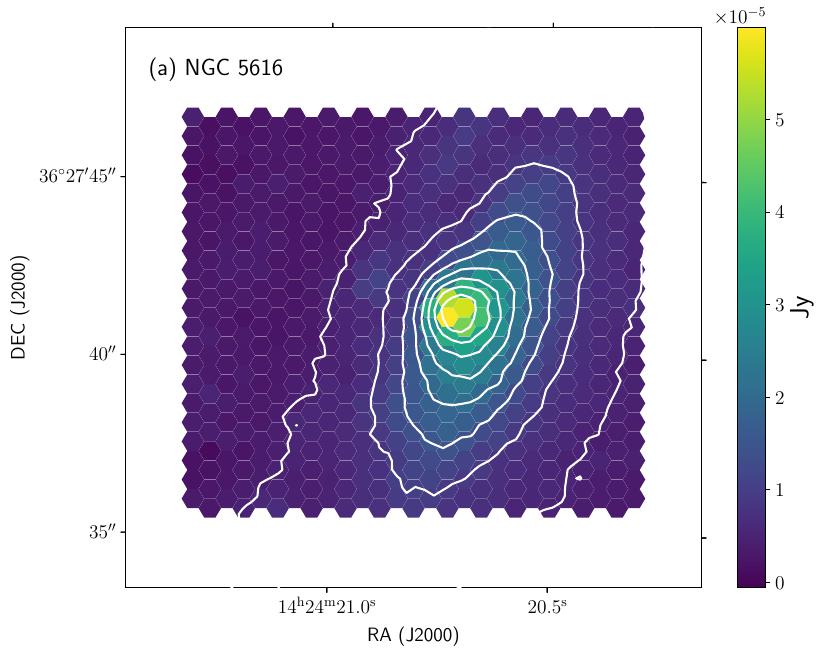}
	\includegraphics[clip, width=0.6\linewidth]{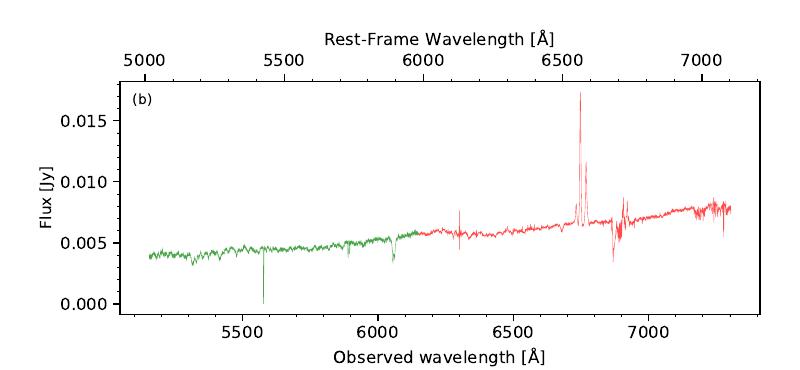}
	\includegraphics[clip, width=0.24\linewidth]{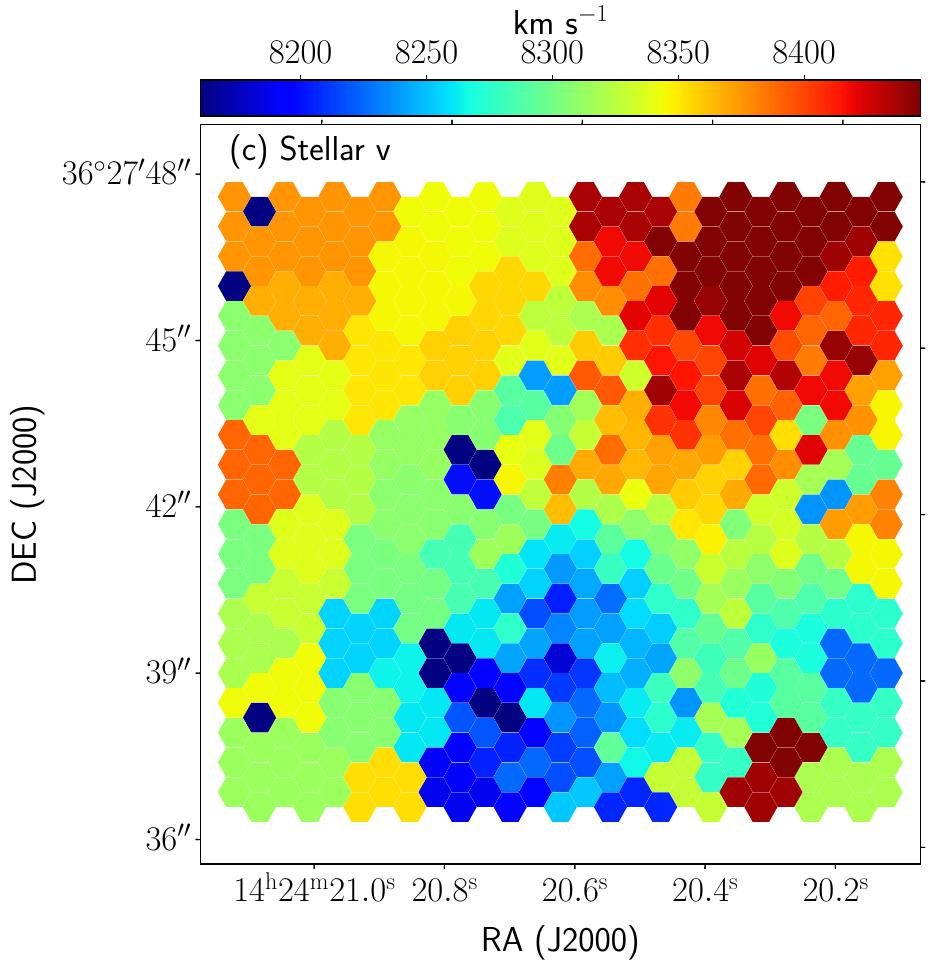}
	\includegraphics[clip, width=0.24\linewidth]{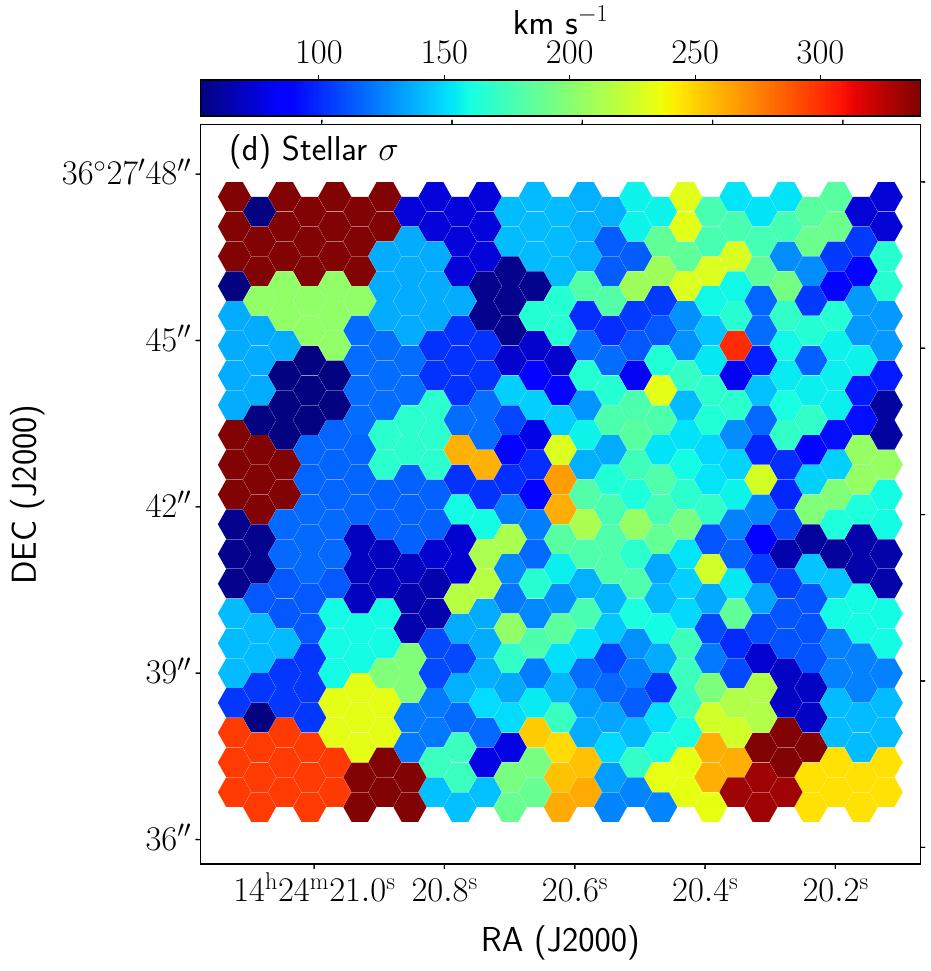}
	\includegraphics[clip, width=0.24\linewidth]{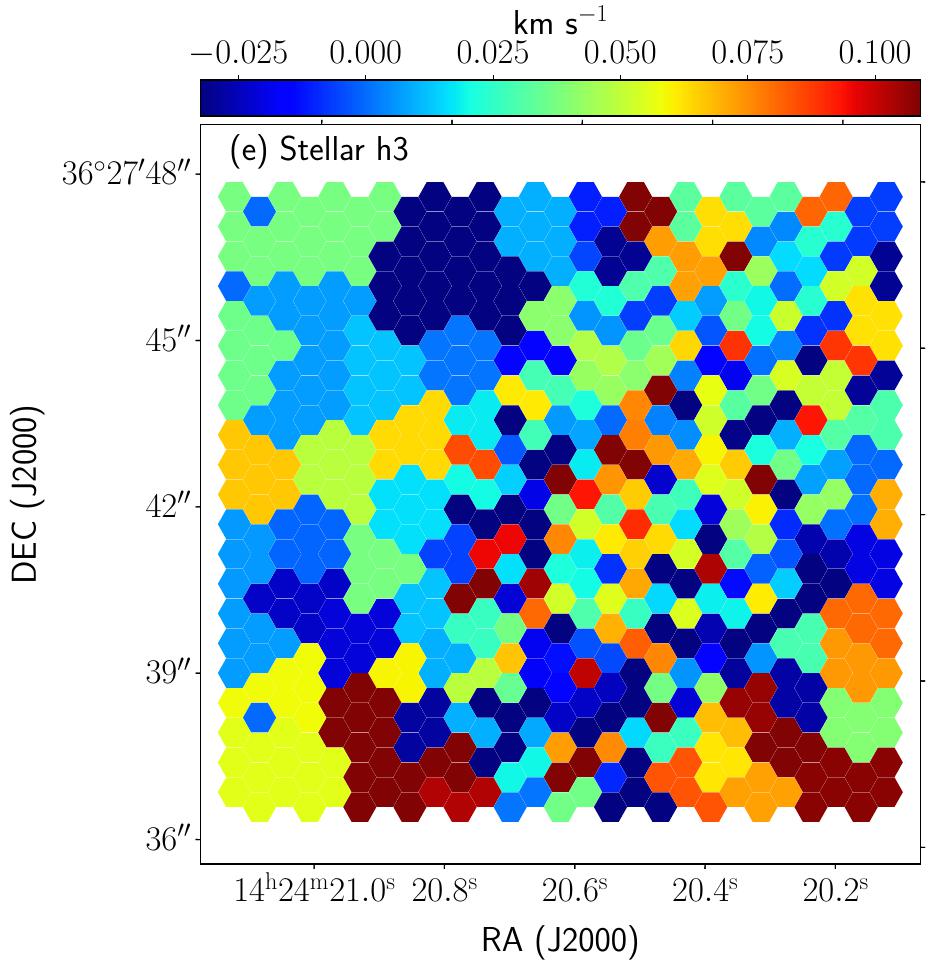}
	\includegraphics[clip, width=0.24\linewidth]{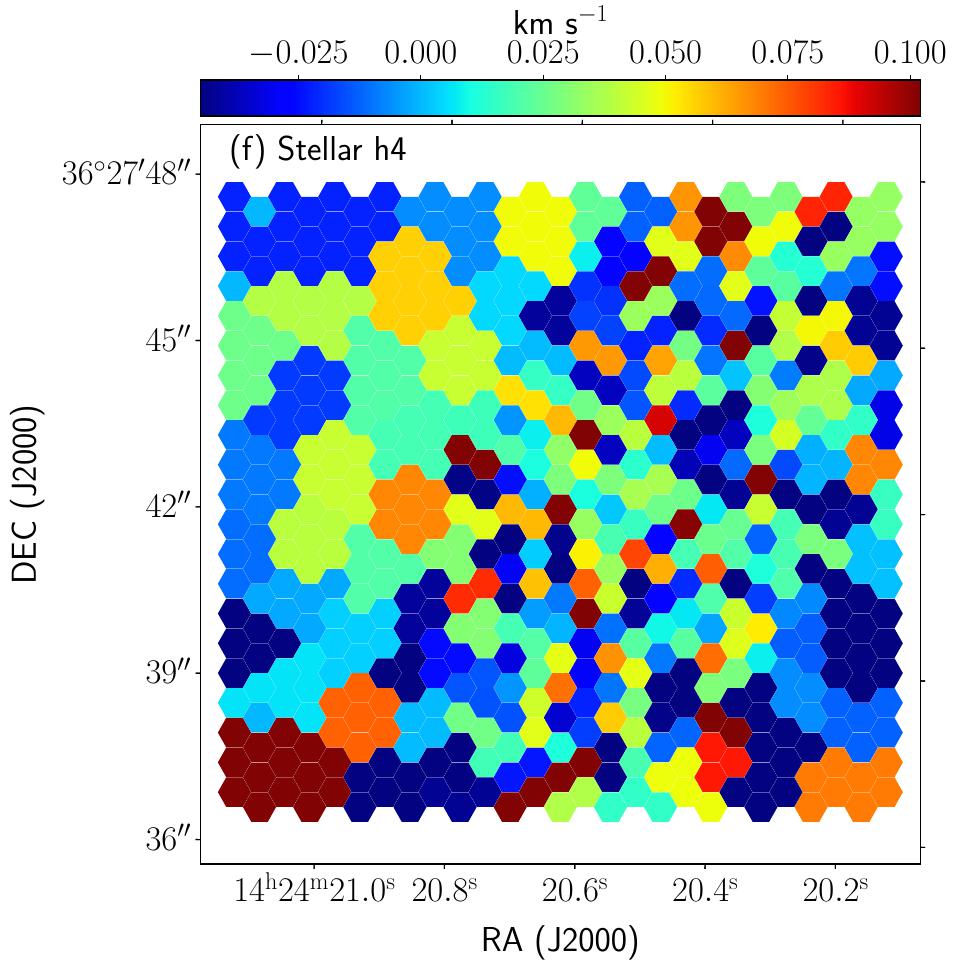}
	
	\vspace{8.8cm}
	
	\includegraphics[clip, width=0.24\linewidth]{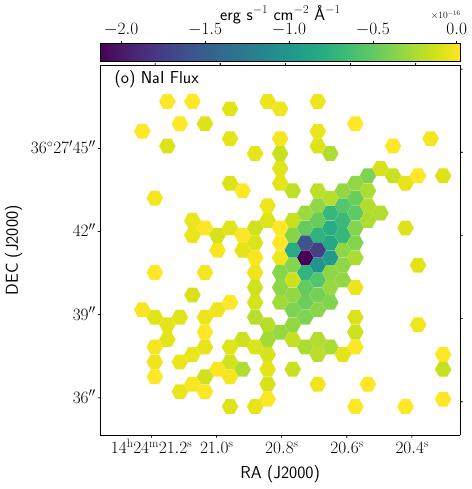}
	\includegraphics[clip, width=0.24\linewidth]{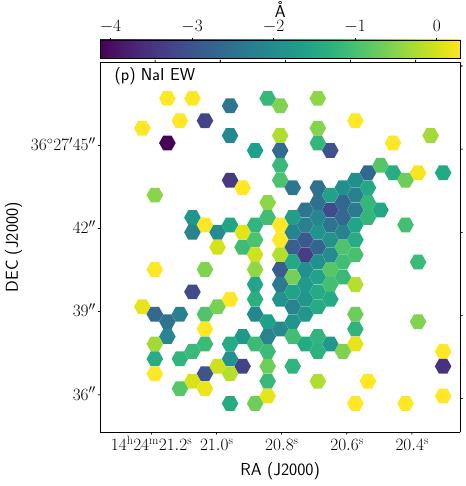}
	\includegraphics[clip, width=0.24\linewidth]{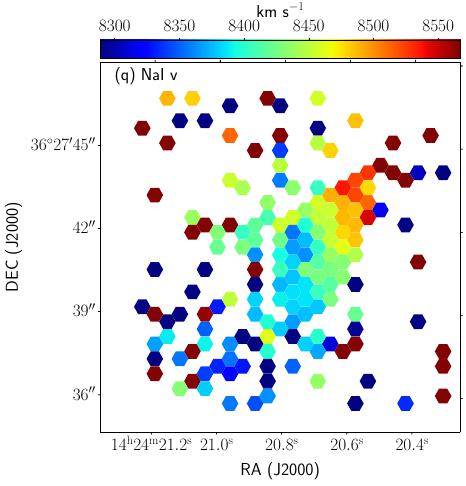}
	\includegraphics[clip, width=0.24\linewidth]{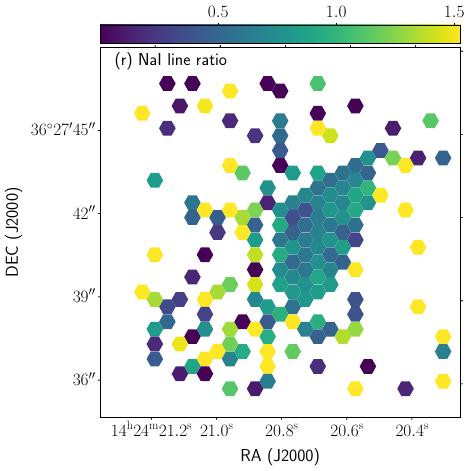}
	\caption{NGC~5616 card.}
	\label{fig:NGC5616_card_1}
\end{figure*}
\addtocounter{figure}{-1}
\begin{figure*}[h]
	\centering
	\includegraphics[clip, width=0.24\linewidth]{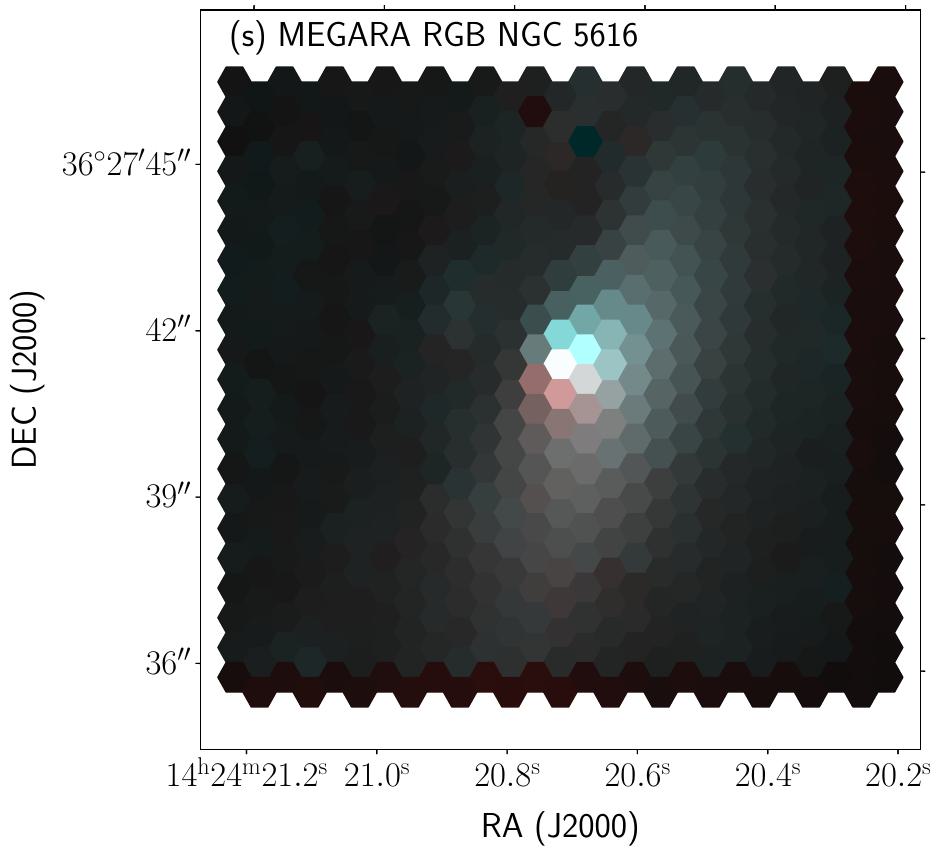}
	\hspace{4,4cm}
	\includegraphics[clip, width=0.24\linewidth]{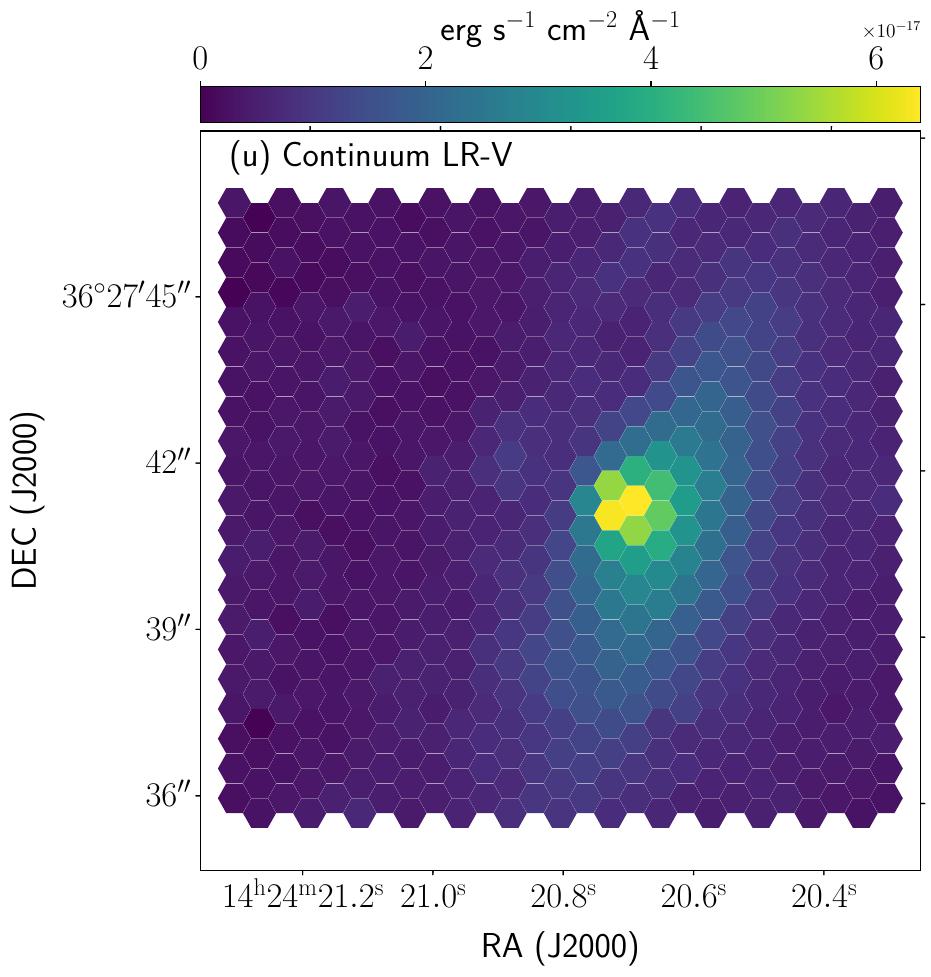}
	\includegraphics[clip, width=0.24\linewidth]{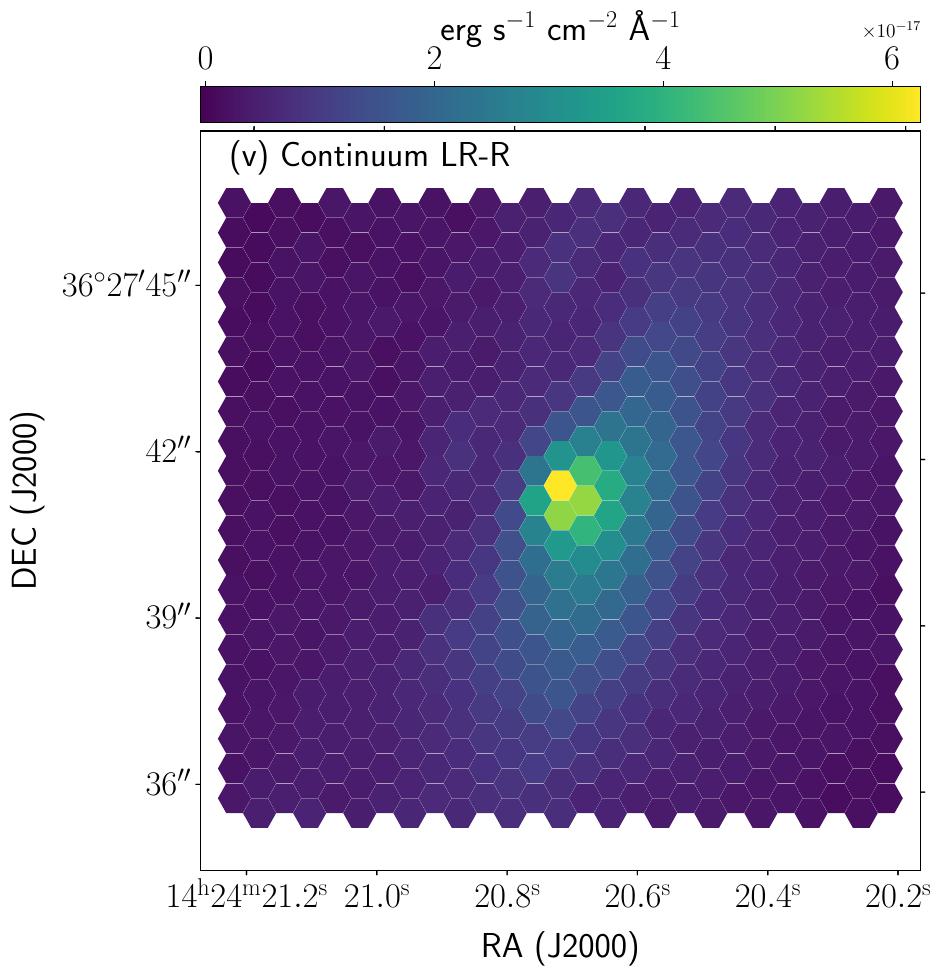}
	\includegraphics[clip, width=0.24\linewidth]{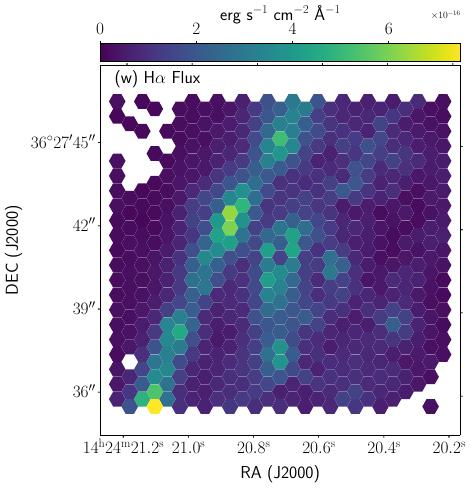}
	\includegraphics[clip, width=0.24\linewidth]{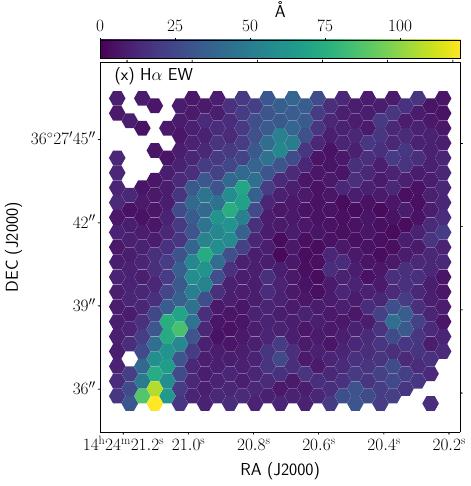}
	\includegraphics[clip, width=0.24\linewidth]{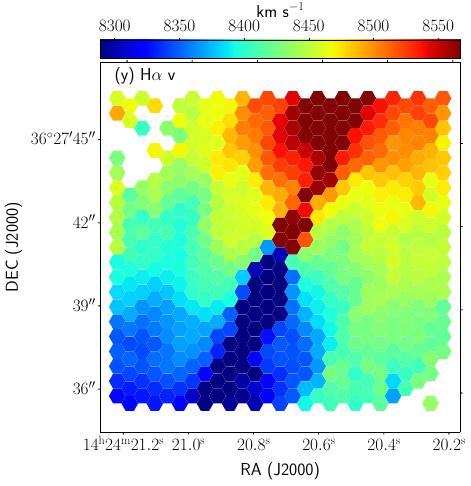}
	\includegraphics[clip, width=0.24\linewidth]{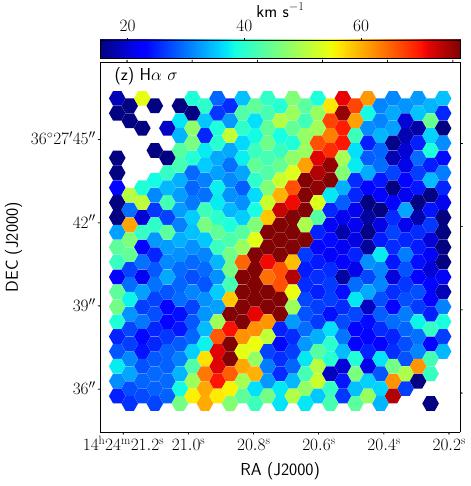}
	\includegraphics[clip, width=0.24\linewidth]{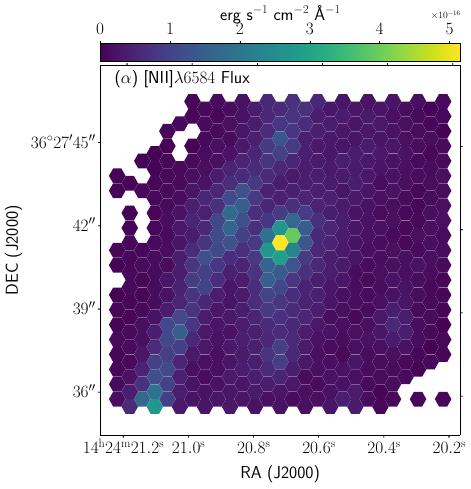}
	\includegraphics[clip, width=0.24\linewidth]{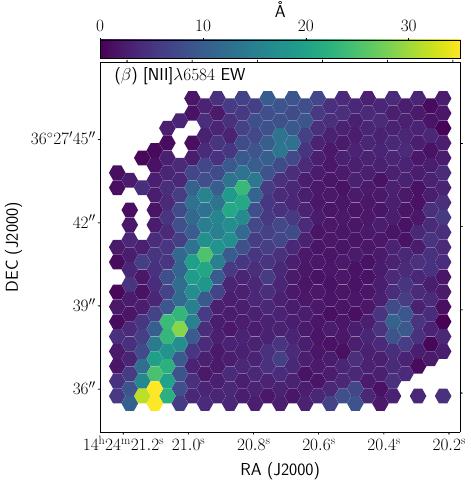}
	\includegraphics[clip, width=0.24\linewidth]{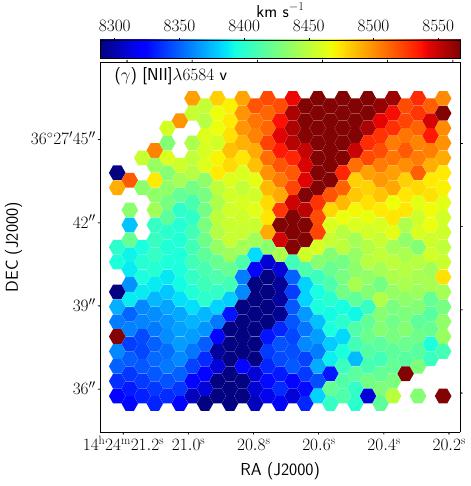}
	\includegraphics[clip, width=0.24\linewidth]{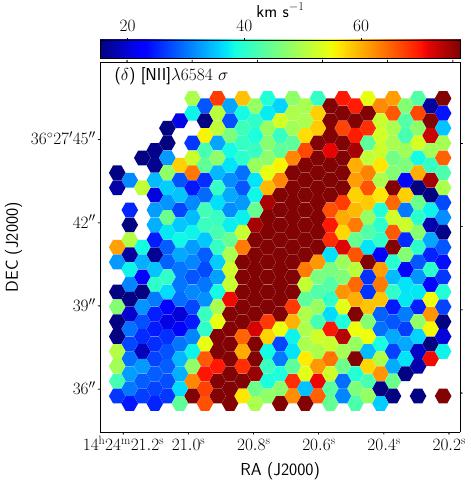}
	\includegraphics[clip, width=0.24\linewidth]{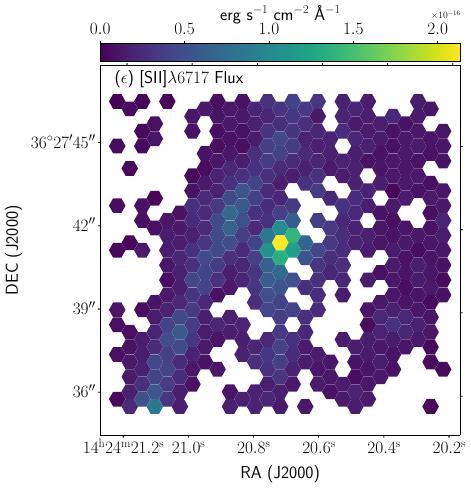}
	\includegraphics[clip, width=0.24\linewidth]{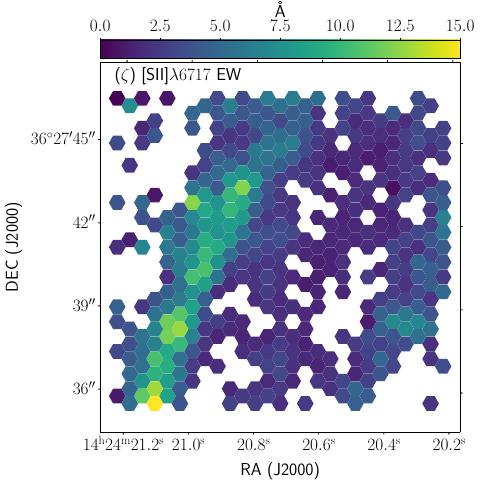}
	\includegraphics[clip, width=0.24\linewidth]{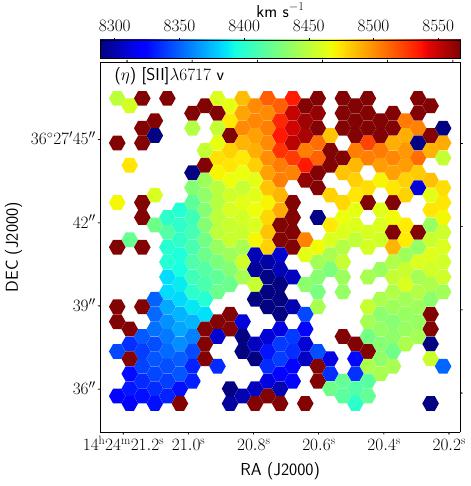}
	\includegraphics[clip, width=0.24\linewidth]{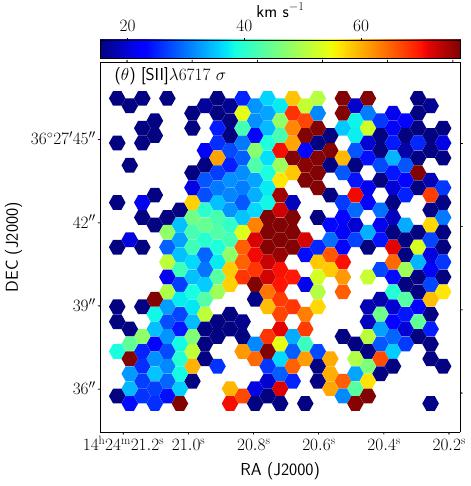}
	\includegraphics[clip, width=0.24\linewidth]{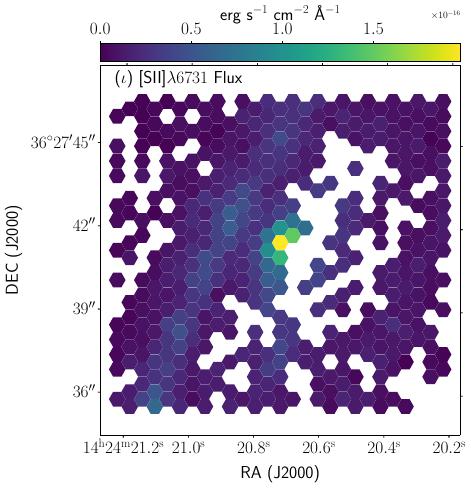}
	\includegraphics[clip, width=0.24\linewidth]{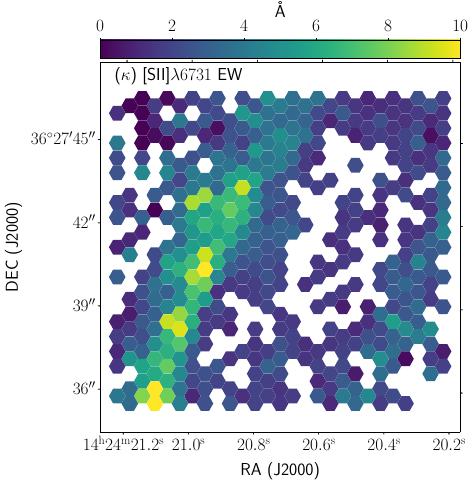}
	\includegraphics[clip, width=0.24\linewidth]{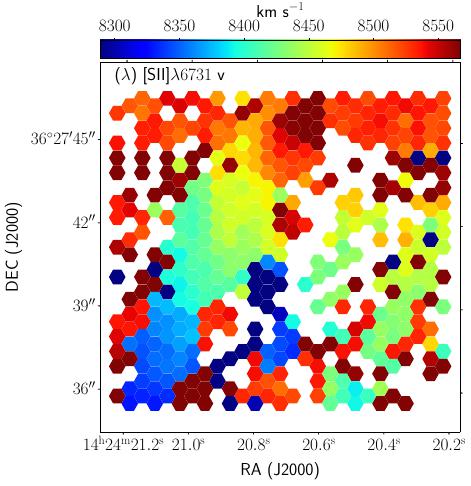}
	\includegraphics[clip, width=0.24\linewidth]{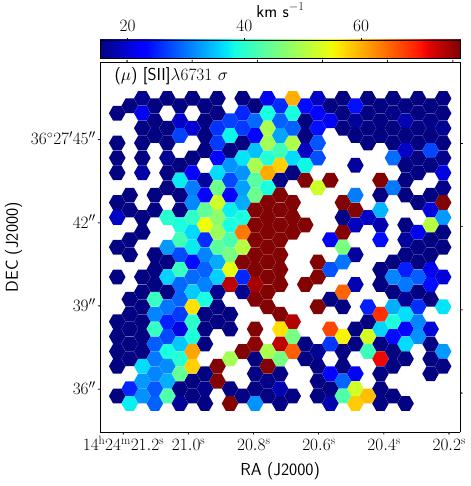}
	\caption{(cont.) NGC~5616 card.}
	\label{fig:NGC5616_card_2}
\end{figure*}

\begin{figure*}[h]
	\centering
	\includegraphics[clip, width=0.35\linewidth]{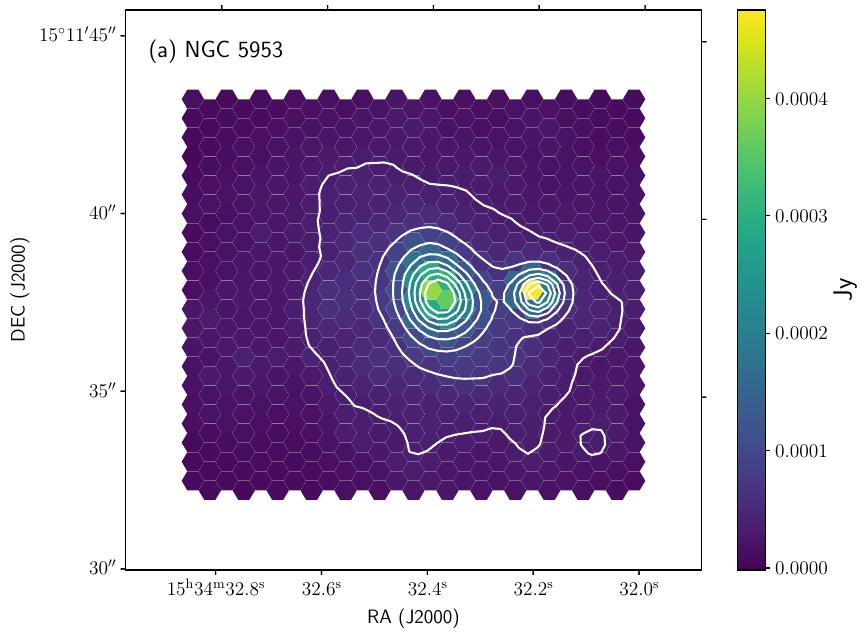}
	\includegraphics[clip, width=0.6\linewidth]{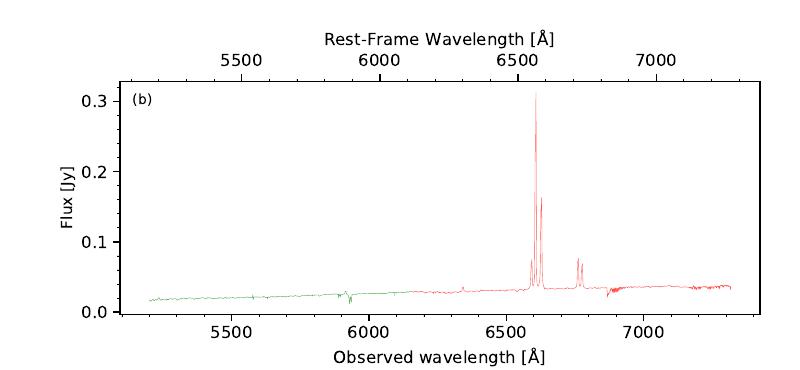}
	\includegraphics[clip, width=0.24\linewidth]{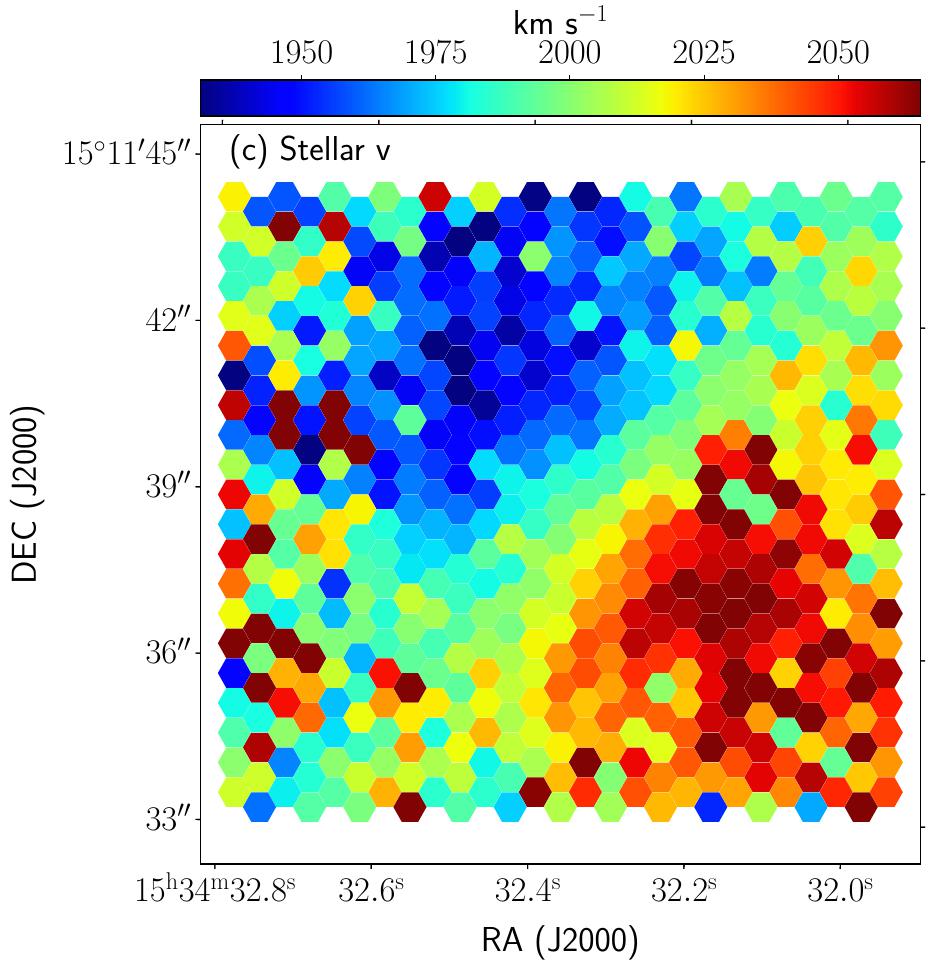}
	\includegraphics[clip, width=0.24\linewidth]{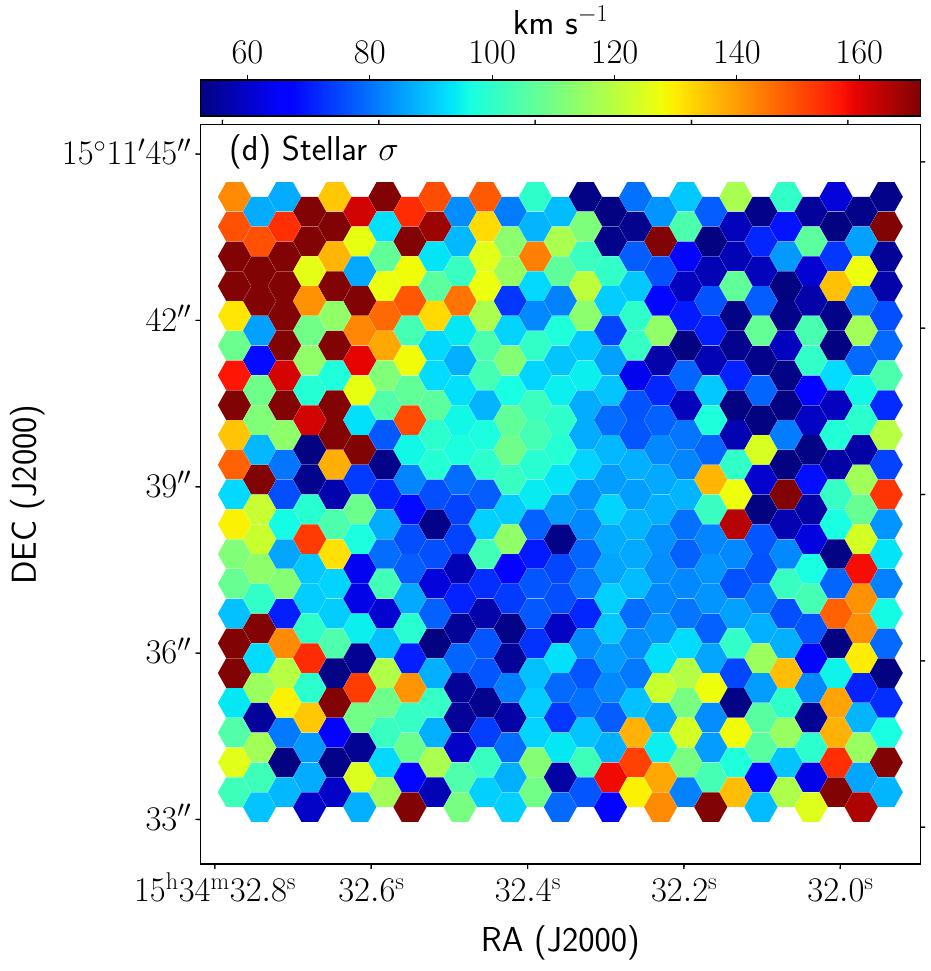}
	\includegraphics[clip, width=0.24\linewidth]{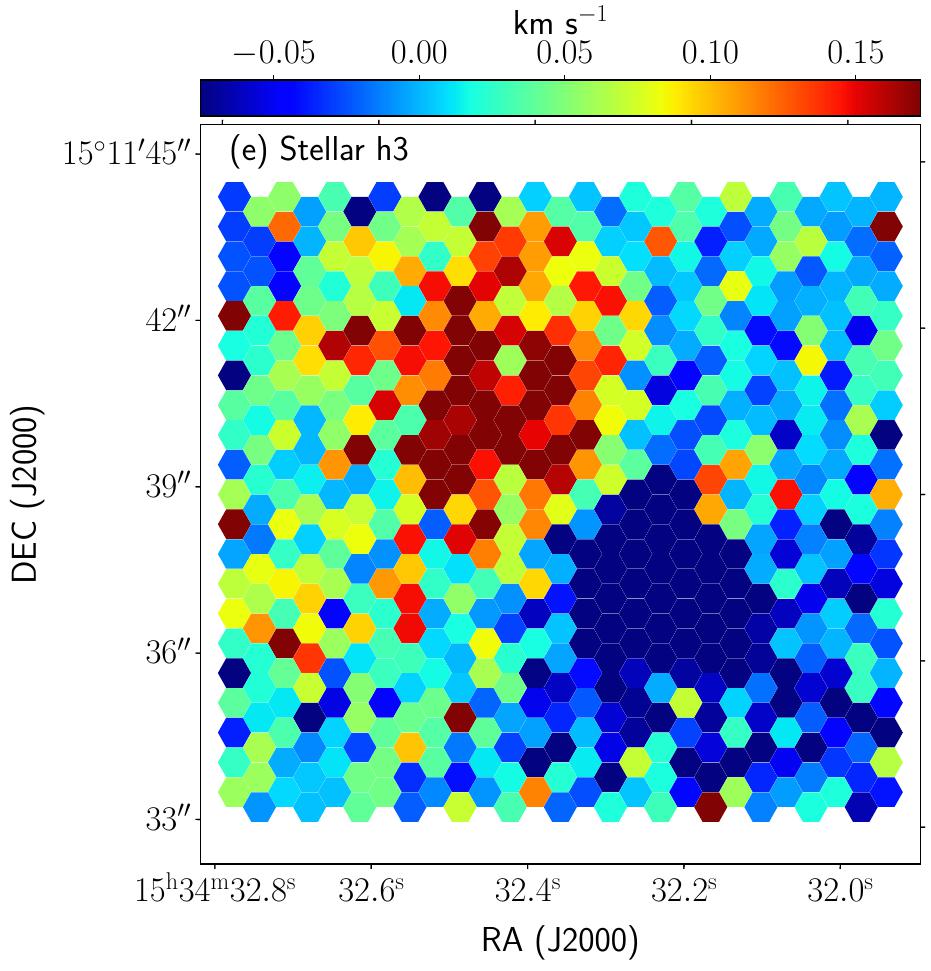}
	\includegraphics[clip, width=0.24\linewidth]{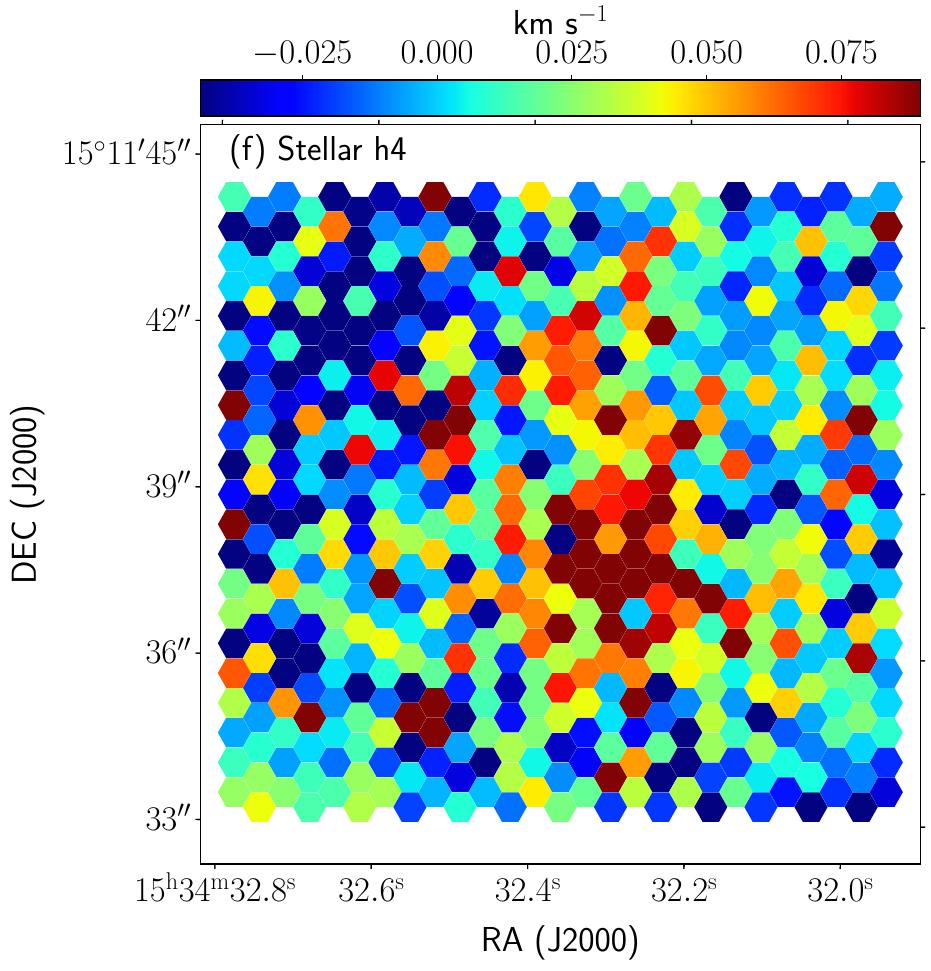}
	
	\vspace{8.8cm}
	
	\includegraphics[clip, width=0.24\linewidth]{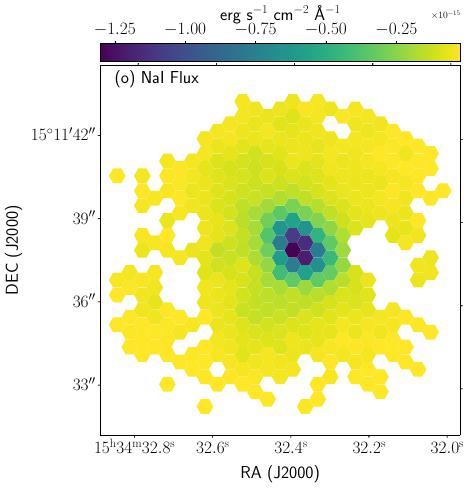}
	\includegraphics[clip, width=0.24\linewidth]{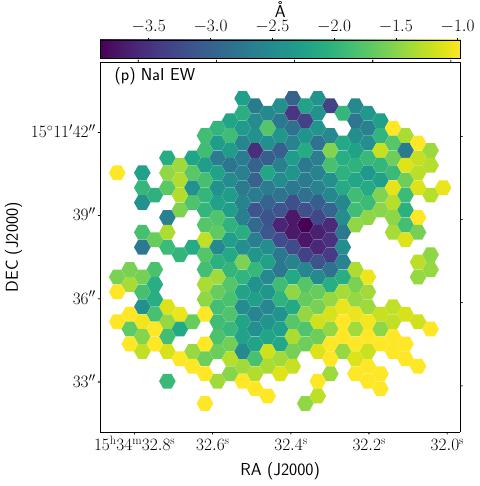}
	\includegraphics[clip, width=0.24\linewidth]{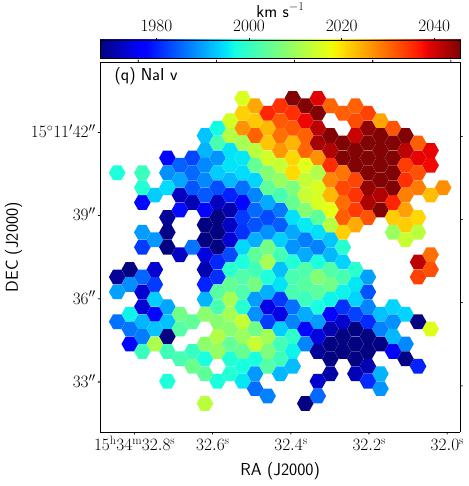}
	\includegraphics[clip, width=0.24\linewidth]{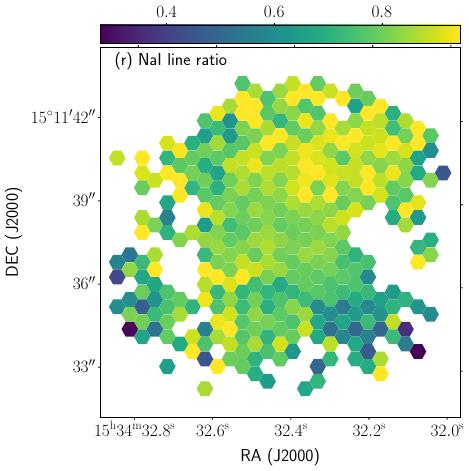}
	\caption{NGC~5953 card.}
	\label{fig:NGC5953_card_1}
\end{figure*}
\addtocounter{figure}{-1}
\begin{figure*}[h]
	\centering
	\includegraphics[clip, width=0.24\linewidth]{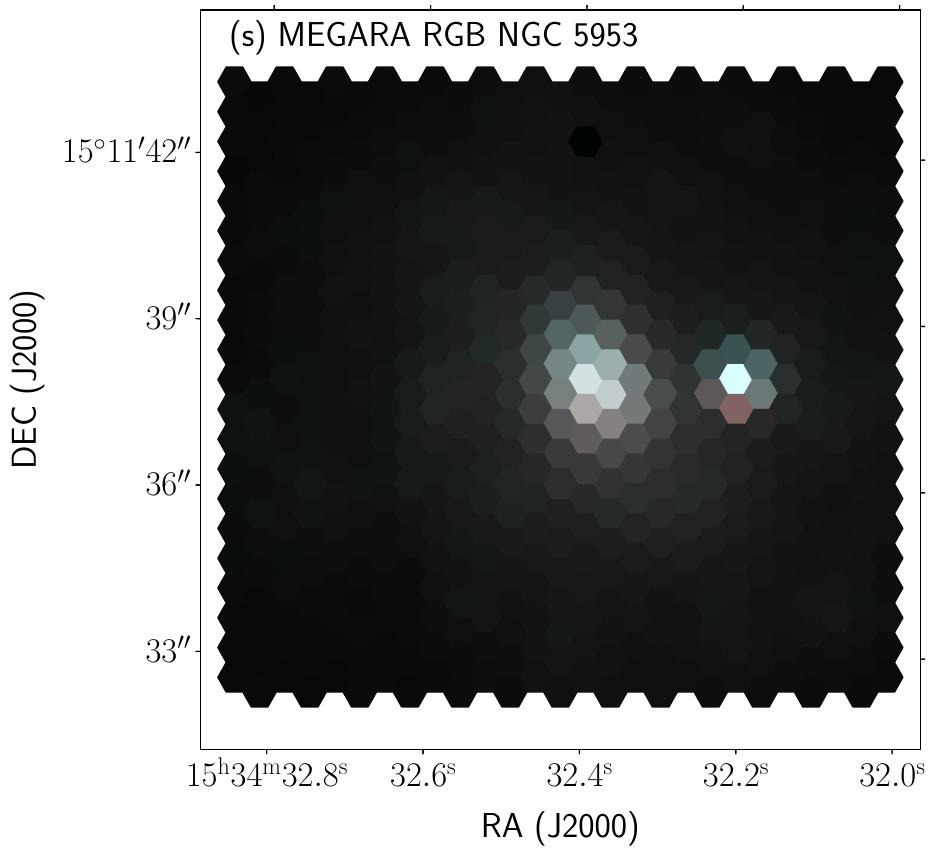}
	\hspace{4,4cm}
	\includegraphics[clip, width=0.24\linewidth]{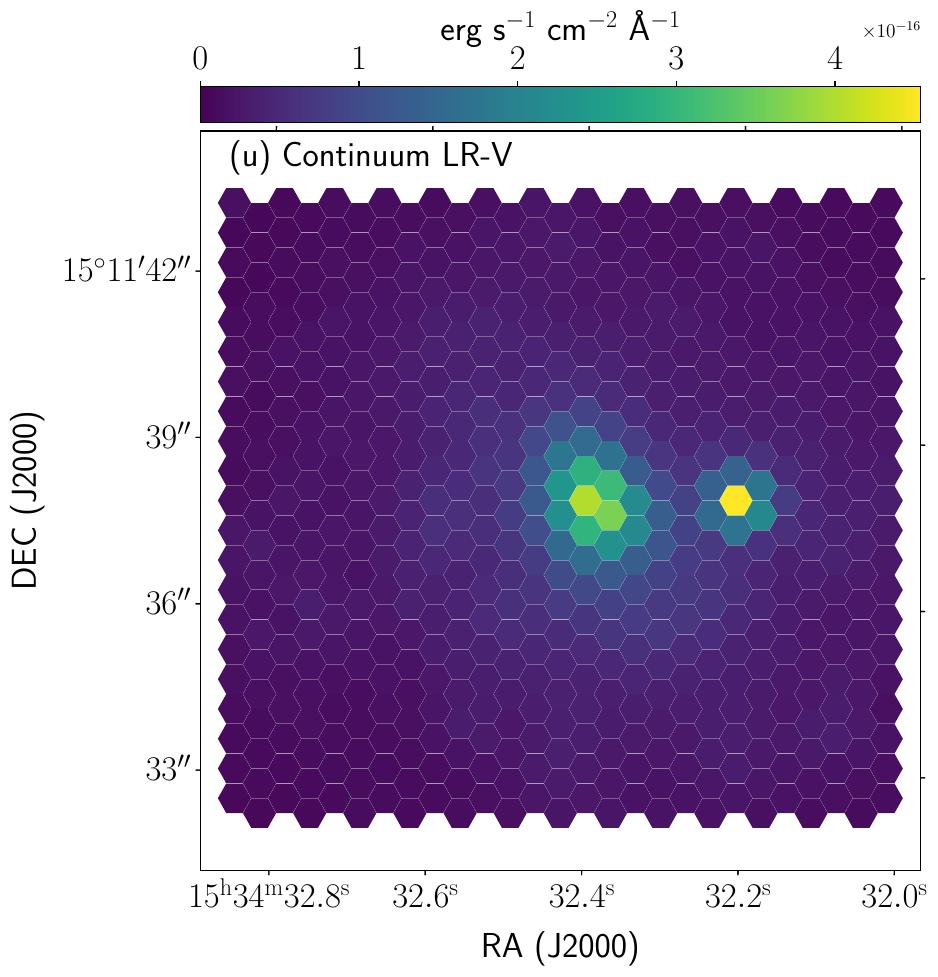}
	\includegraphics[clip, width=0.24\linewidth]{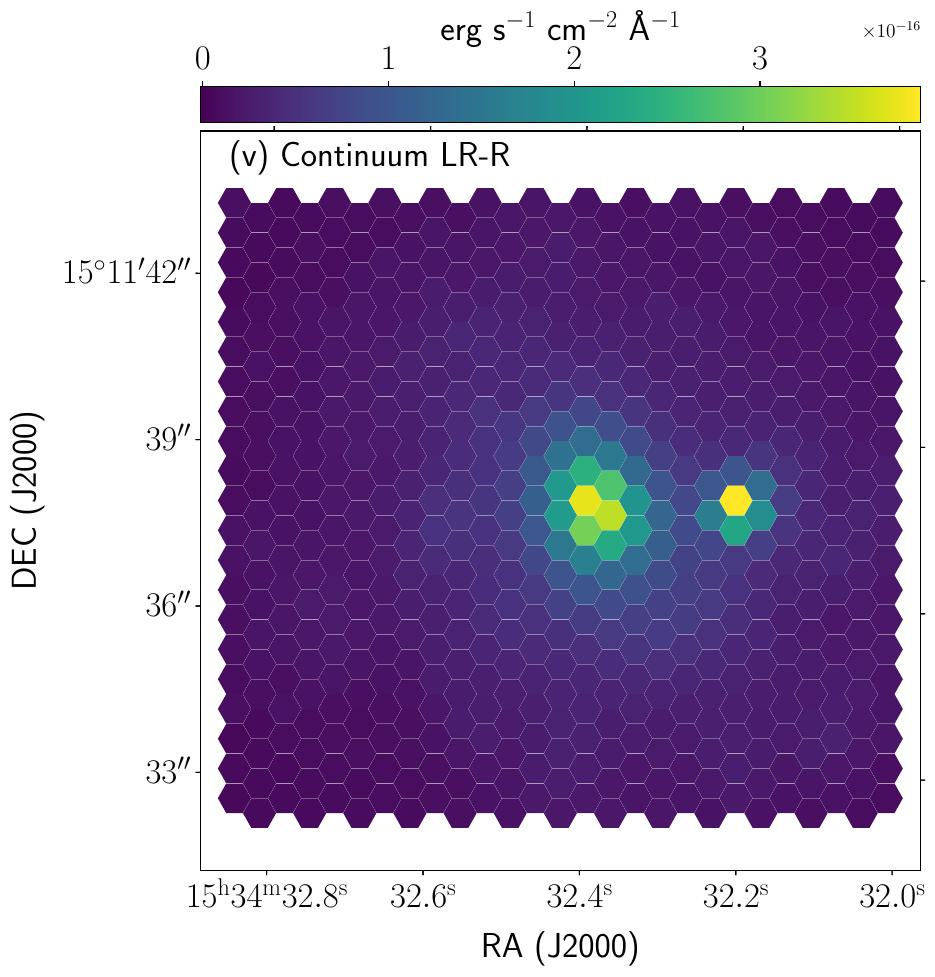}
	\includegraphics[clip, width=0.24\linewidth]{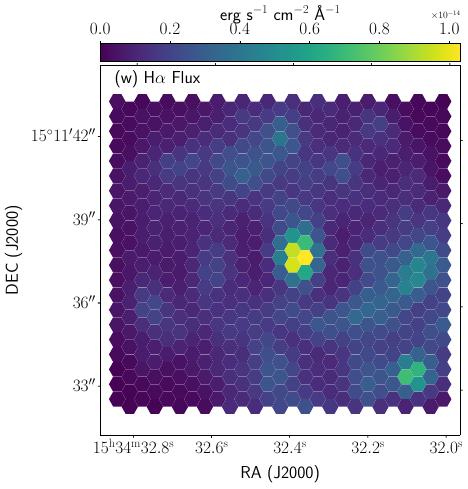}
	\includegraphics[clip, width=0.24\linewidth]{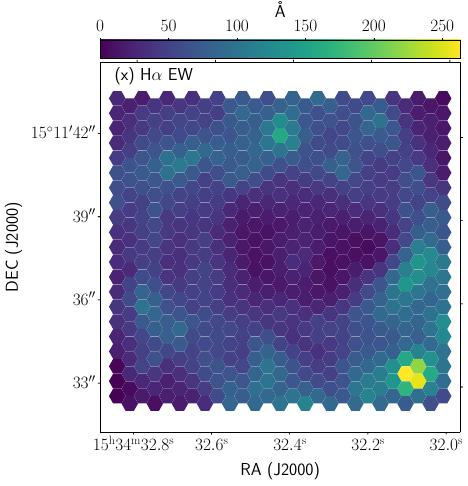}
	\includegraphics[clip, width=0.24\linewidth]{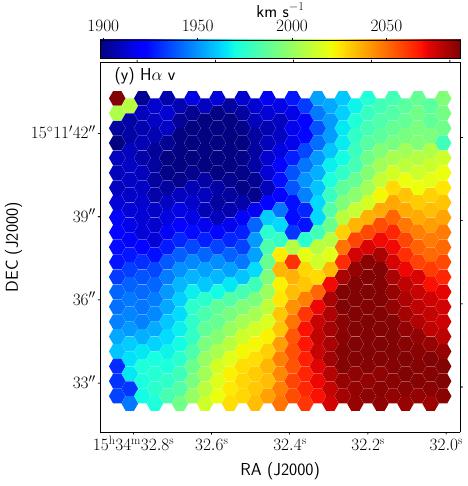}
	\includegraphics[clip, width=0.24\linewidth]{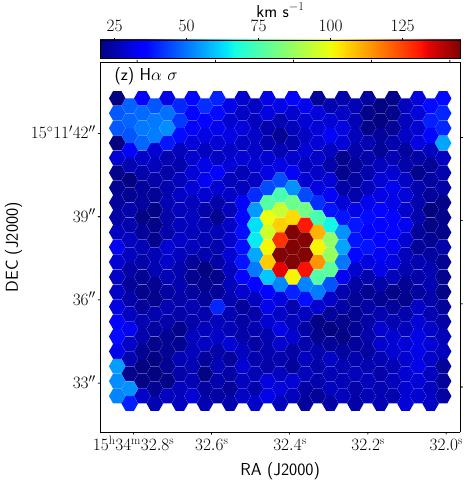}
	\includegraphics[clip, width=0.24\linewidth]{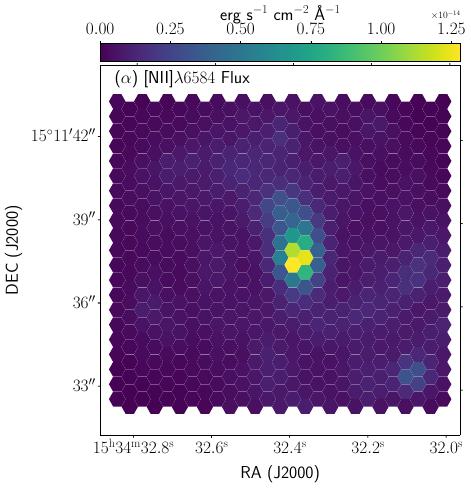}
	\includegraphics[clip, width=0.24\linewidth]{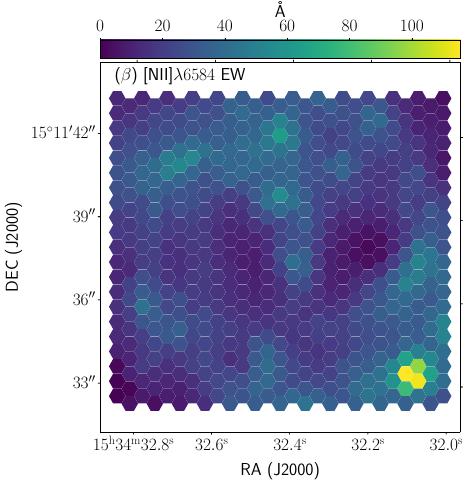}
	\includegraphics[clip, width=0.24\linewidth]{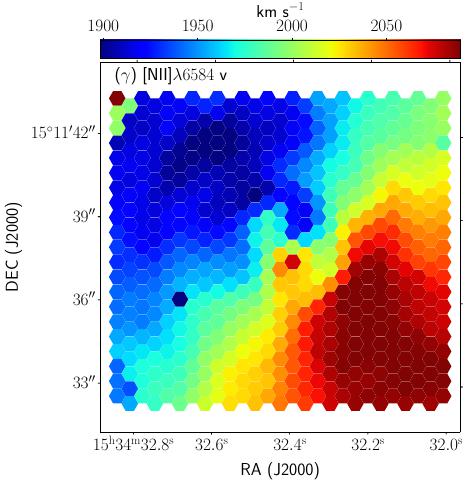}
	\includegraphics[clip, width=0.24\linewidth]{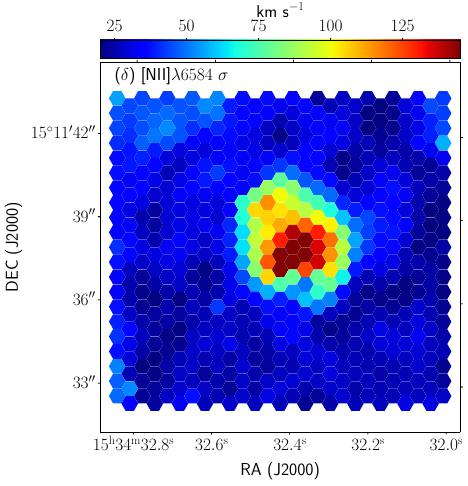}
	\includegraphics[clip, width=0.24\linewidth]{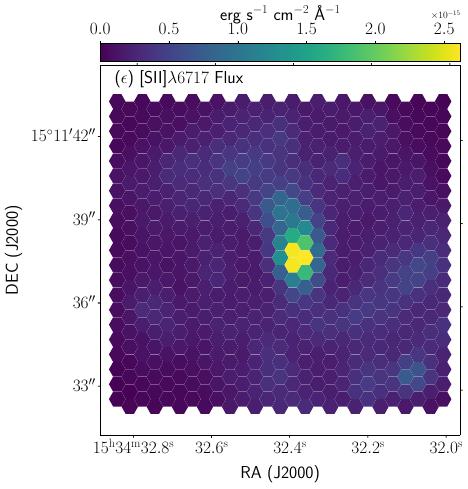}
	\includegraphics[clip, width=0.24\linewidth]{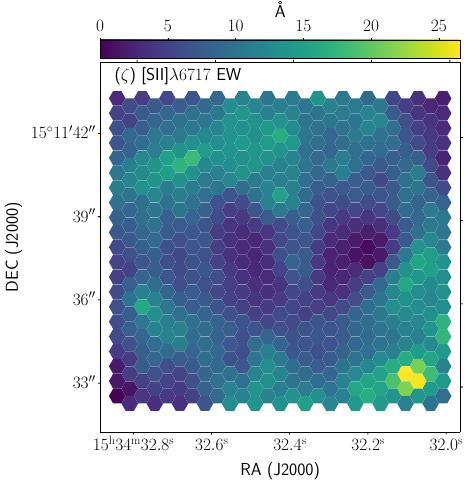}
	\includegraphics[clip, width=0.24\linewidth]{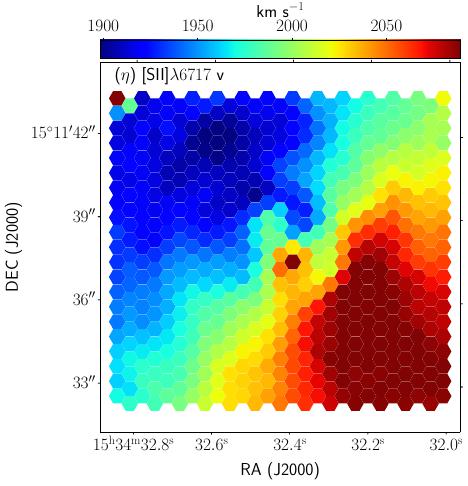}
	\includegraphics[clip, width=0.24\linewidth]{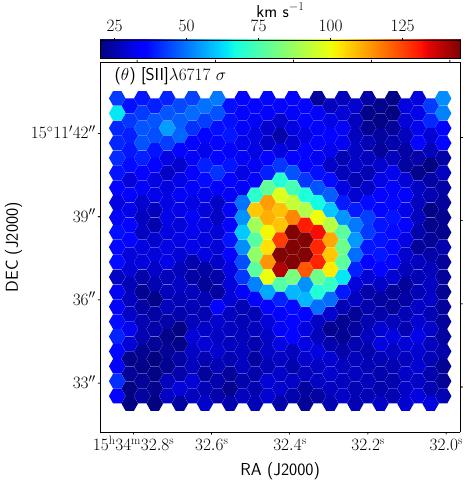}
	\includegraphics[clip, width=0.24\linewidth]{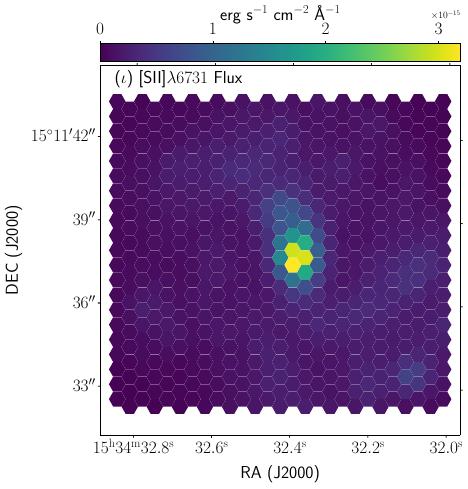}
	\includegraphics[clip, width=0.24\linewidth]{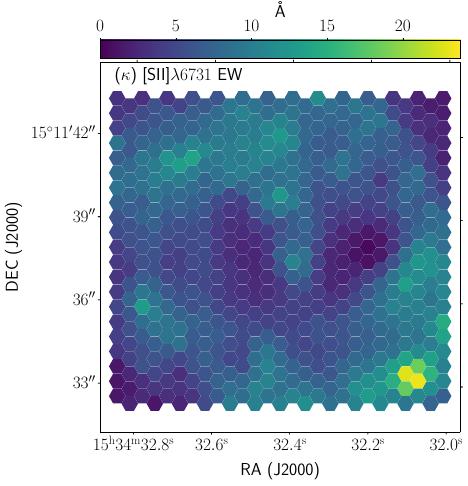}
	\includegraphics[clip, width=0.24\linewidth]{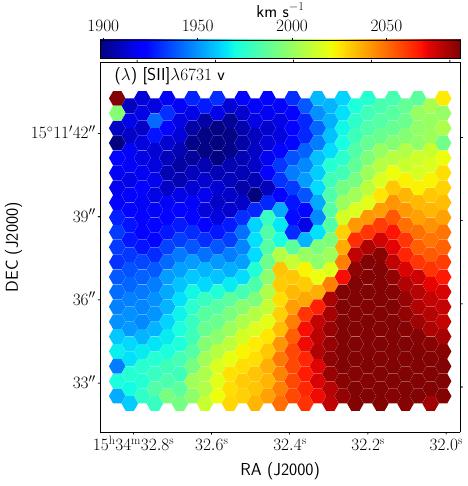}
	\includegraphics[clip, width=0.24\linewidth]{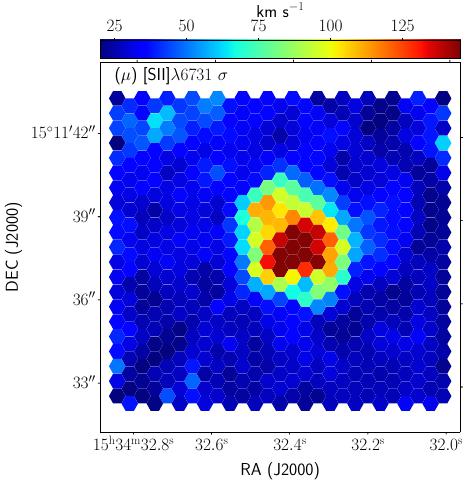}
	\caption{(cont.) NGC~5953 card.}
	\label{fig:NGC5953_card_2}
\end{figure*}

\begin{figure*}[h]
	\centering
	\includegraphics[clip, width=0.35\linewidth]{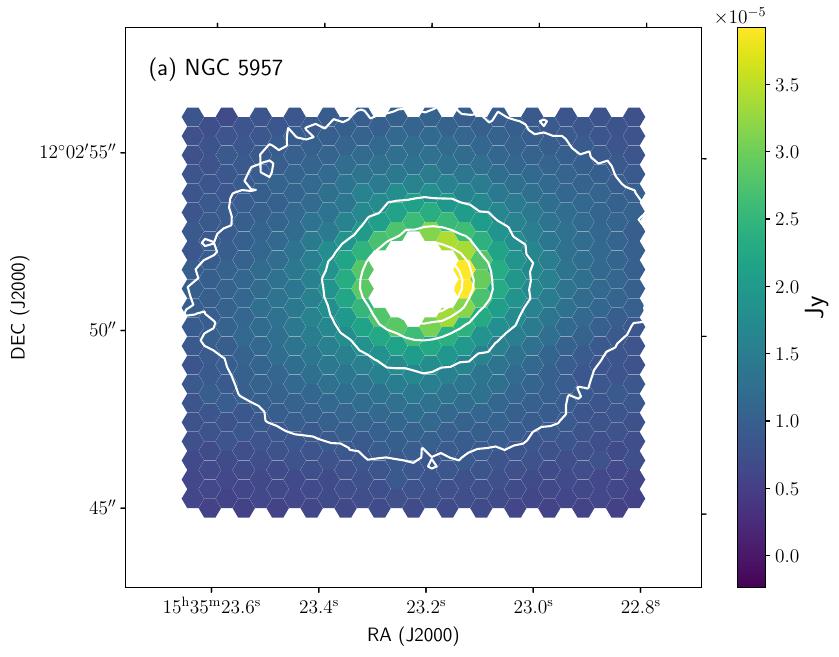}
	\includegraphics[clip, width=0.6\linewidth]{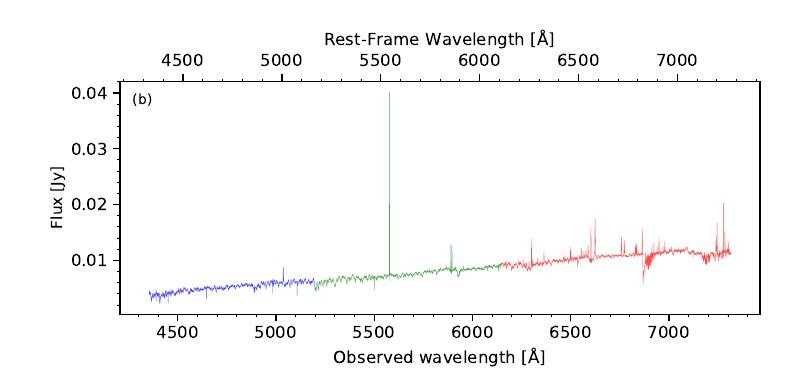}
	\includegraphics[clip, width=0.24\linewidth]{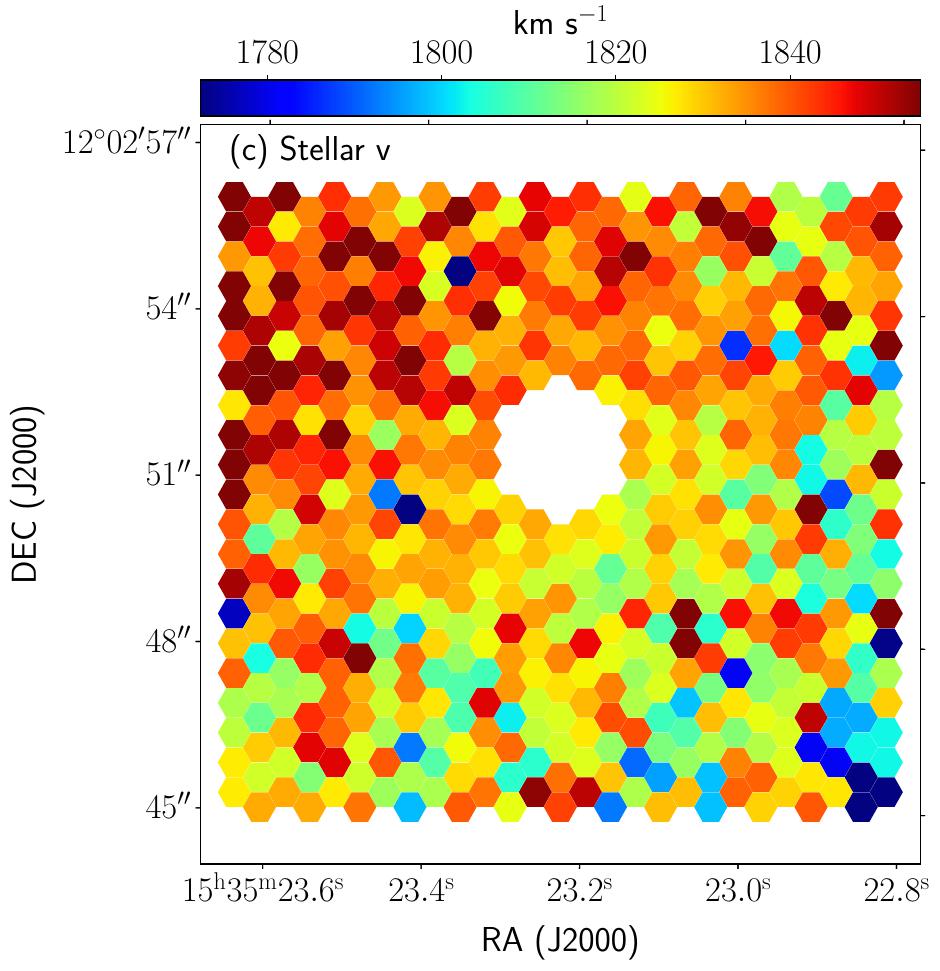}
	\includegraphics[clip, width=0.24\linewidth]{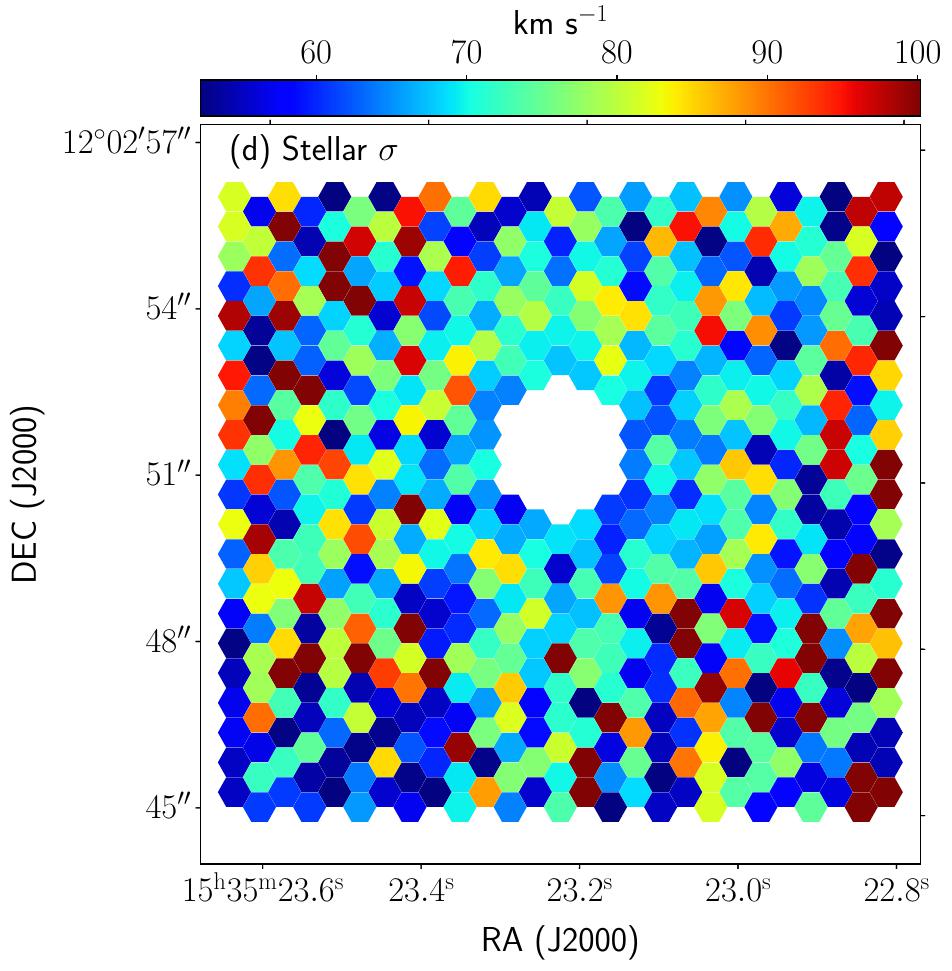}
	\includegraphics[clip, width=0.24\linewidth]{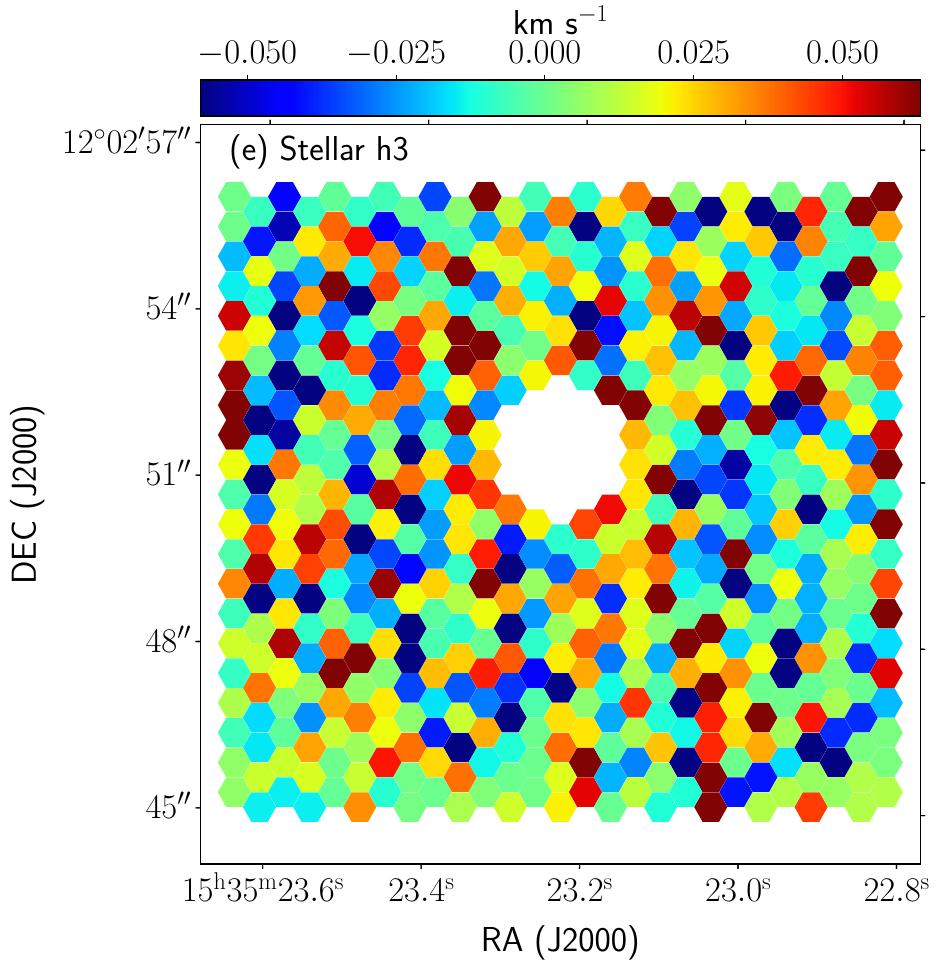}
	\includegraphics[clip, width=0.24\linewidth]{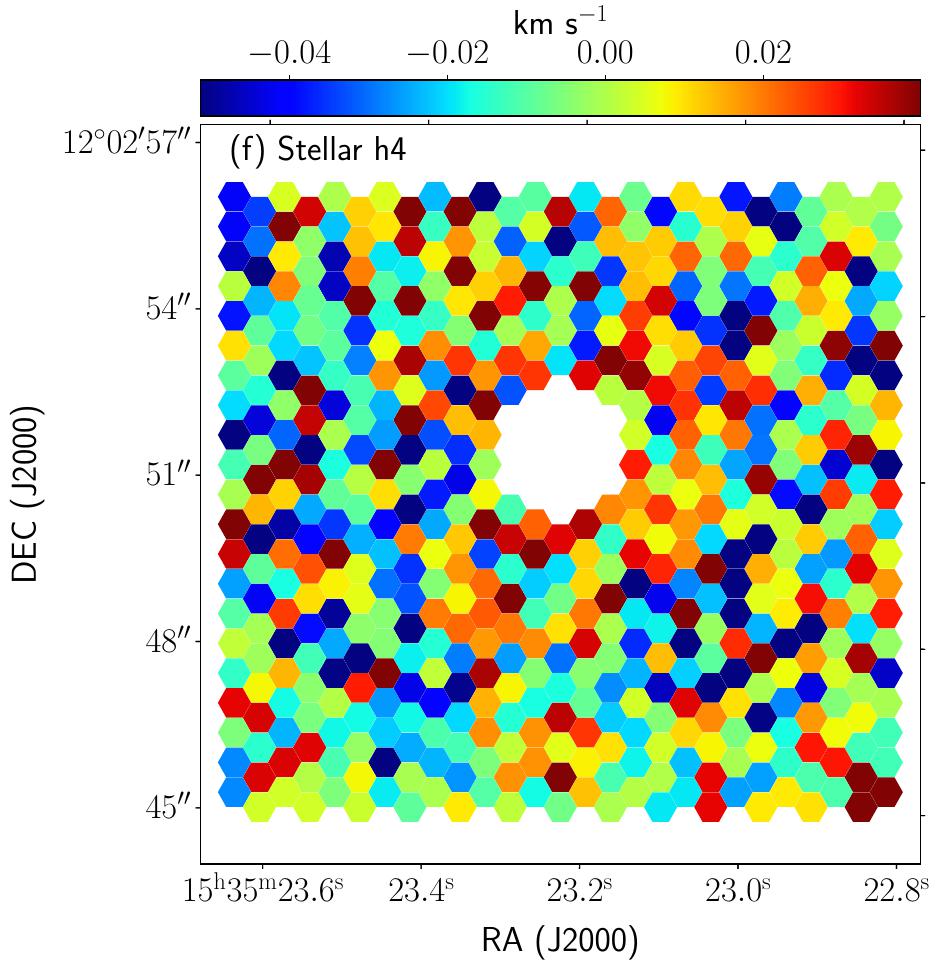}
	\includegraphics[clip, width=0.24\linewidth]{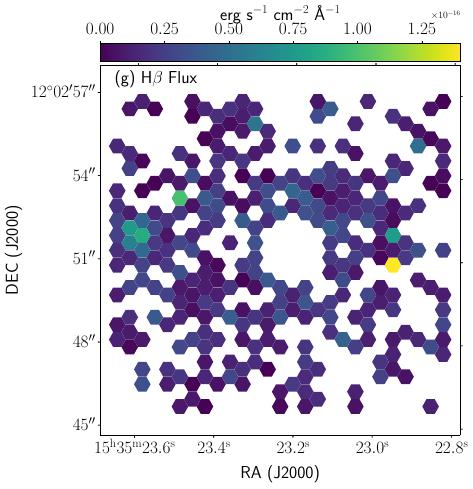}
	\includegraphics[clip, width=0.24\linewidth]{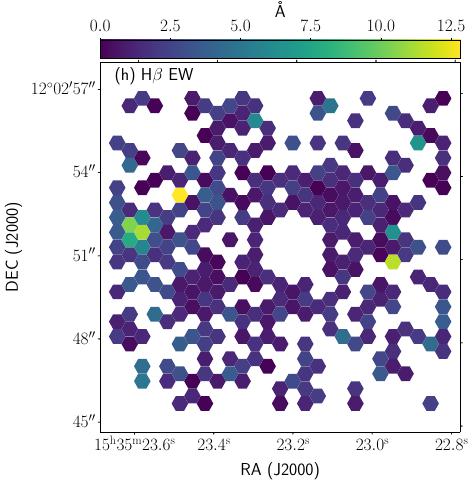}
	\includegraphics[clip, width=0.24\linewidth]{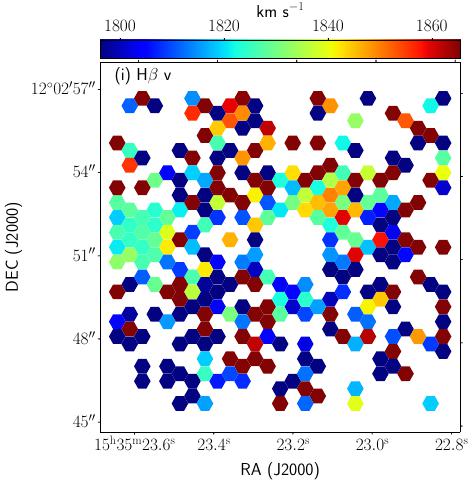}
	\includegraphics[clip, width=0.24\linewidth]{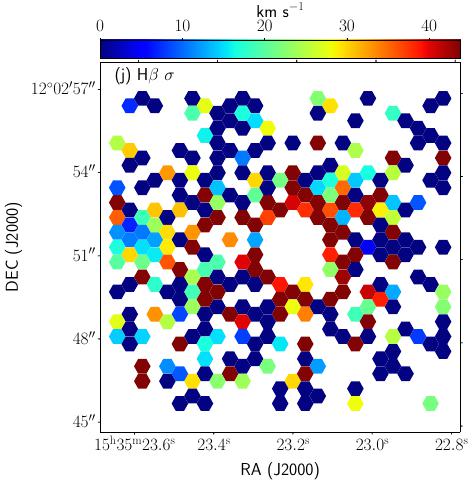}
	\includegraphics[clip, width=0.24\linewidth]{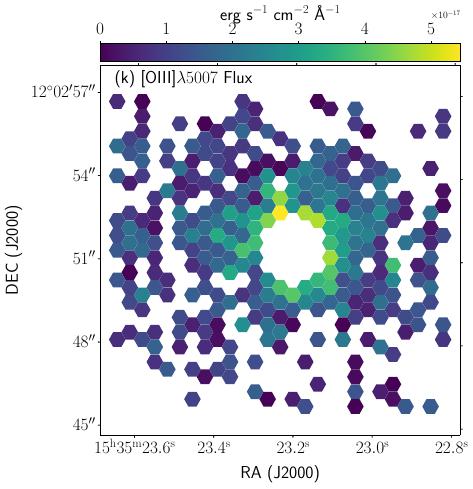}
	\includegraphics[clip, width=0.24\linewidth]{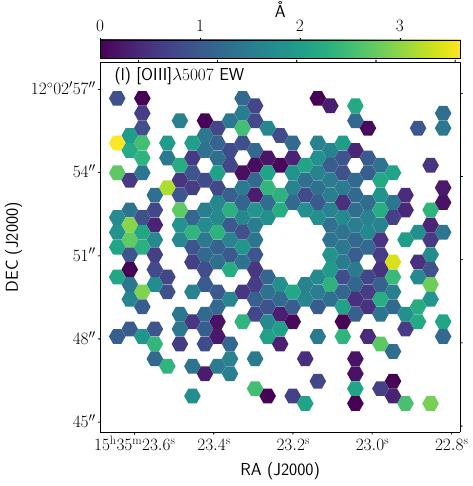}
	\includegraphics[clip, width=0.24\linewidth]{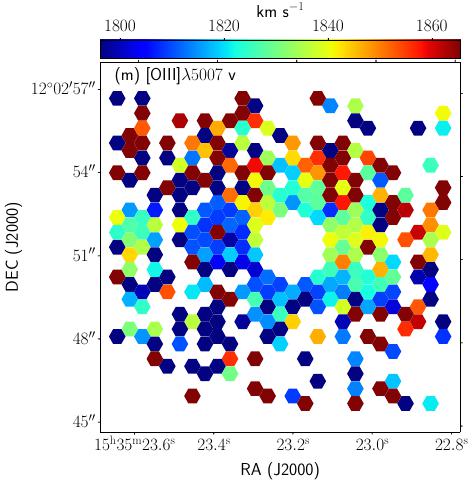}
	\includegraphics[clip, width=0.24\linewidth]{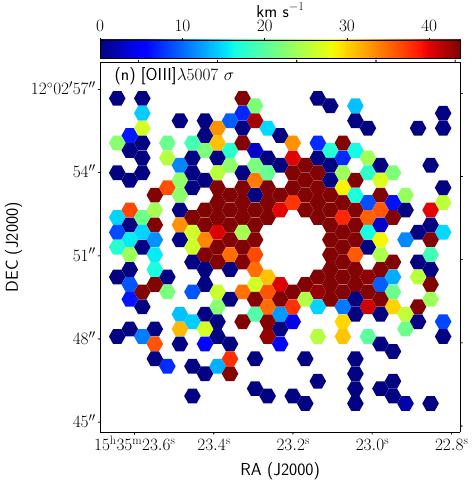}
	\vspace{5cm}
	\caption{NGC~5957 card.}
	\label{fig:NGC5957_card_1}
\end{figure*}
\addtocounter{figure}{-1}
\begin{figure*}[h]
	\centering
	\includegraphics[clip, width=0.24\linewidth]{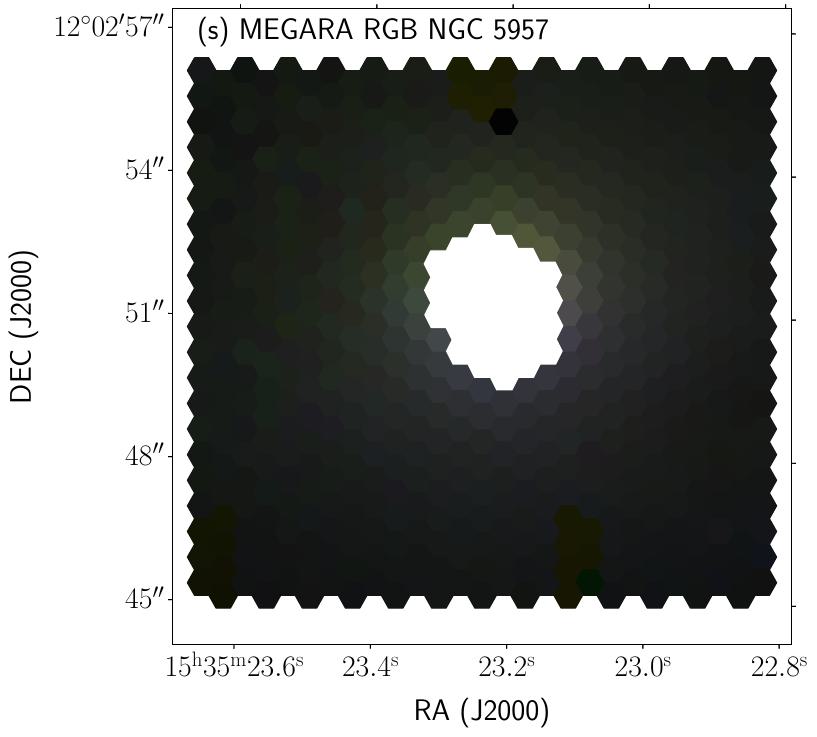}
	\includegraphics[clip, width=0.24\linewidth]{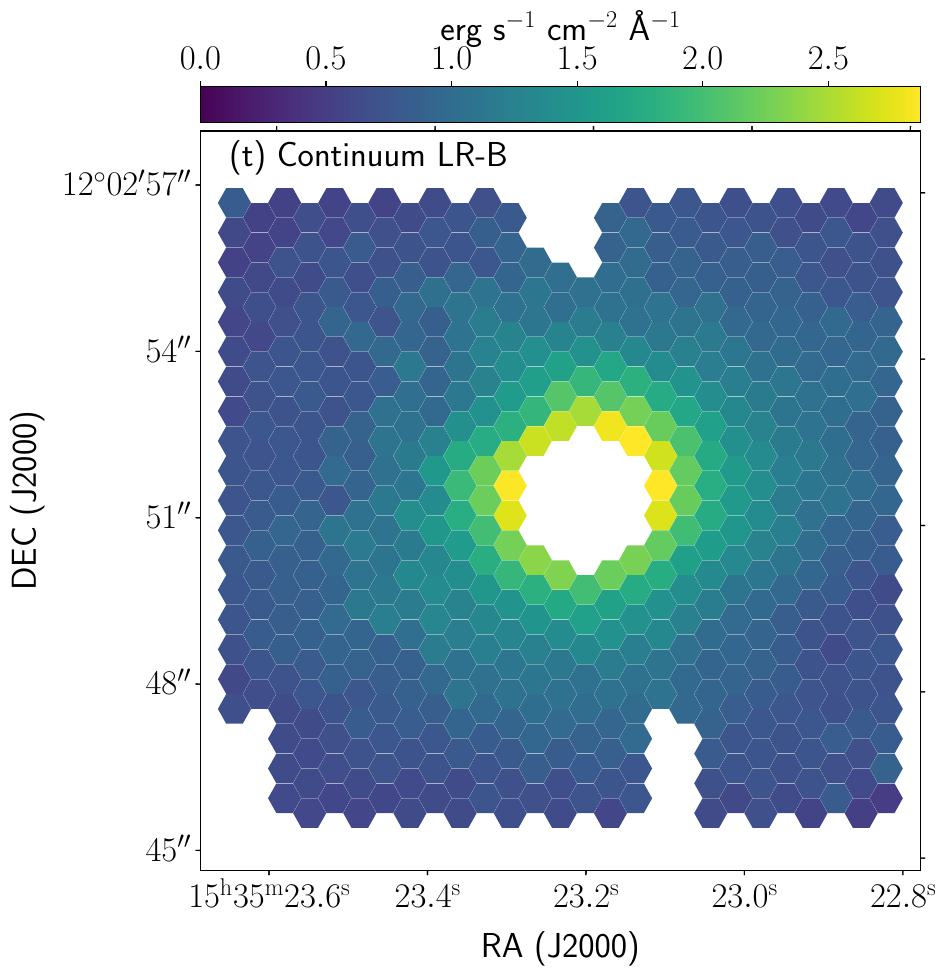}
	\includegraphics[clip, width=0.24\linewidth]{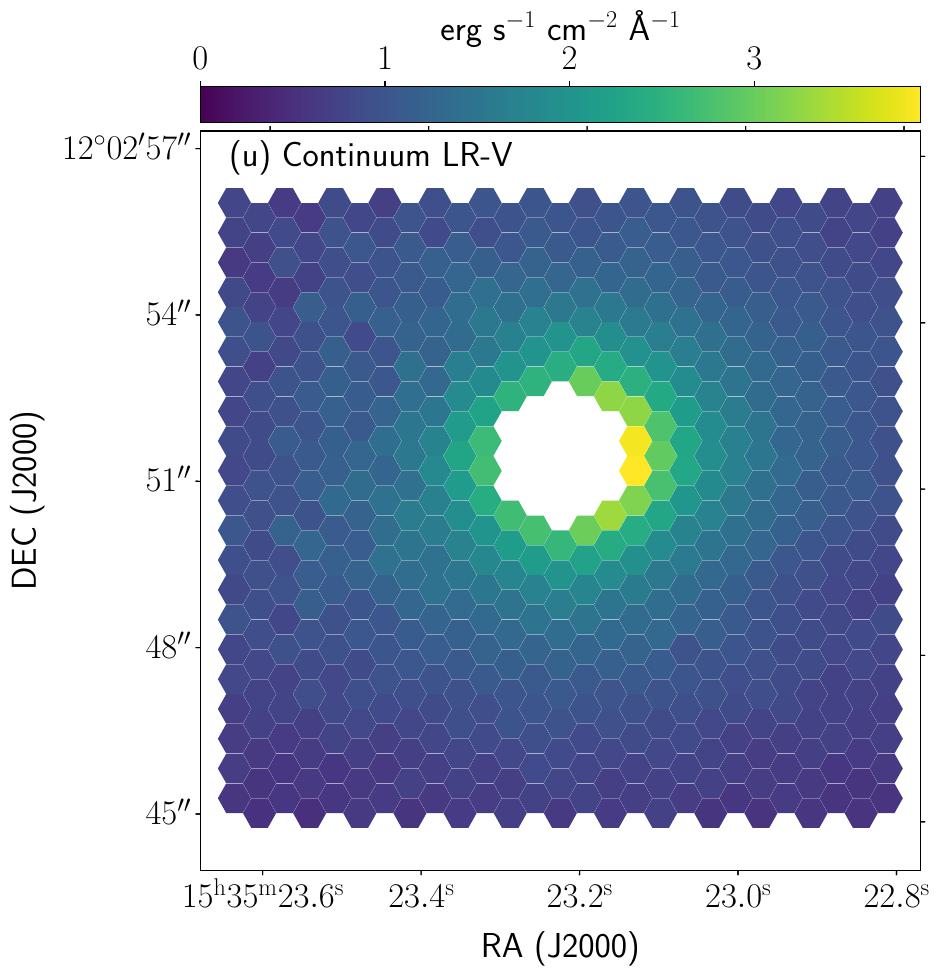}
	\includegraphics[clip, width=0.24\linewidth]{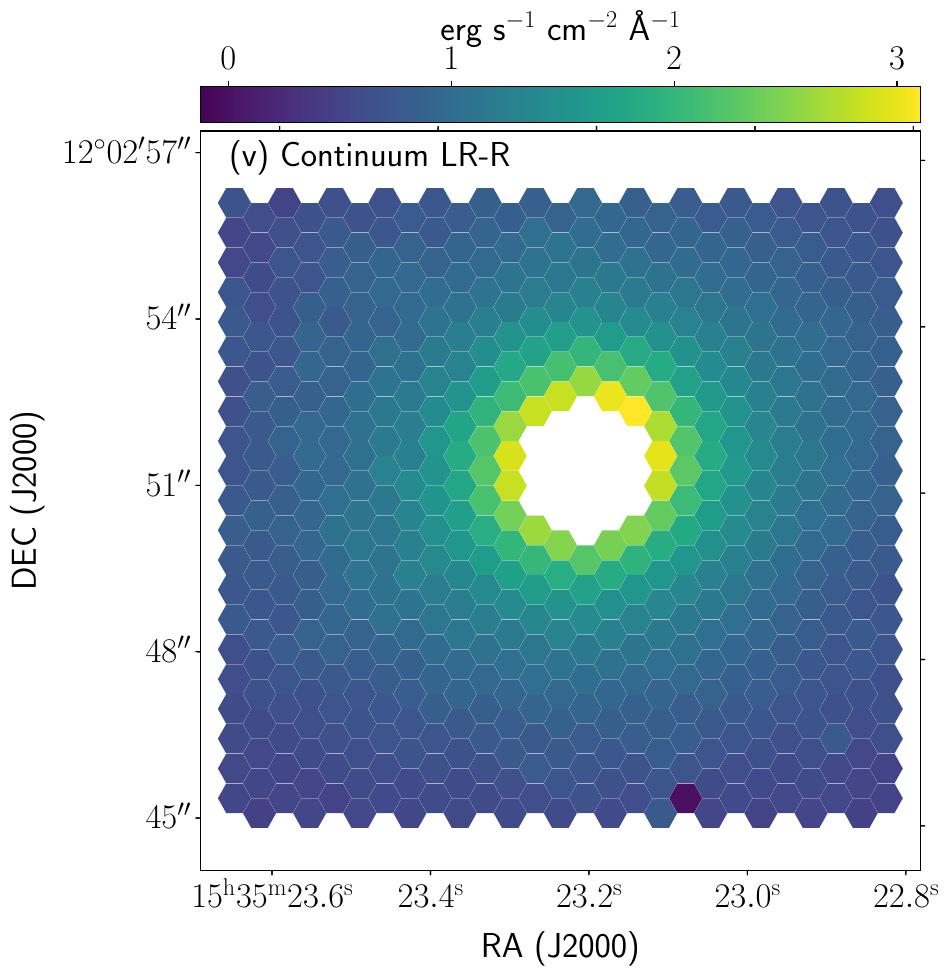}
	\includegraphics[clip, width=0.24\linewidth]{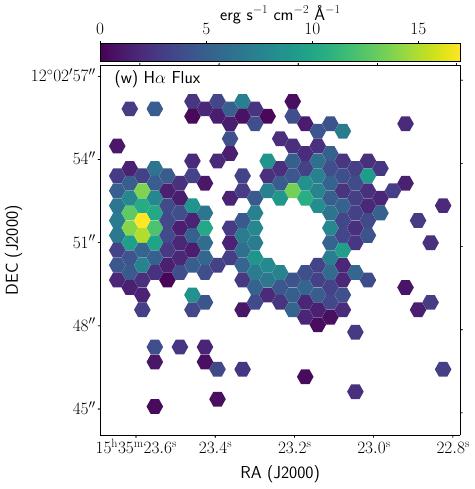}
	\includegraphics[clip, width=0.24\linewidth]{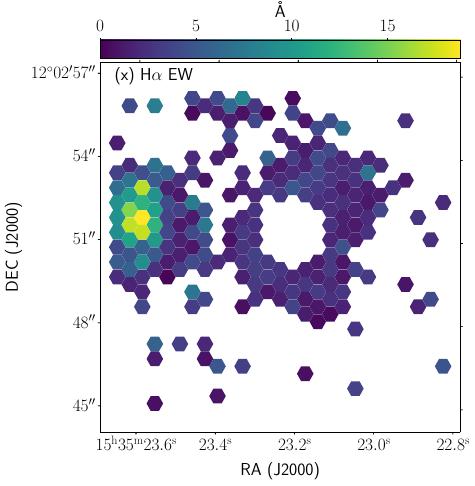}
	\includegraphics[clip, width=0.24\linewidth]{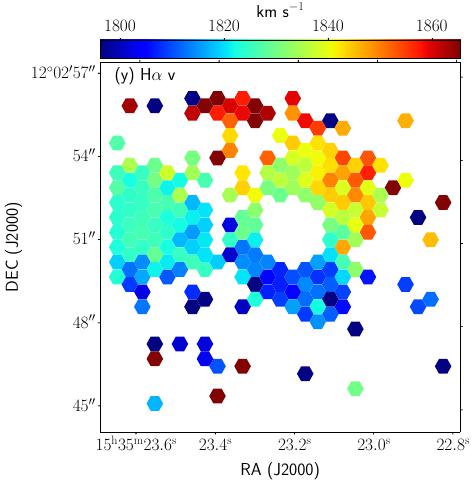}
	\includegraphics[clip, width=0.24\linewidth]{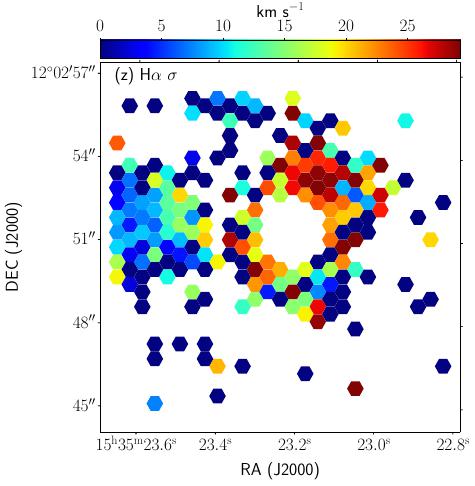}
	\includegraphics[clip, width=0.24\linewidth]{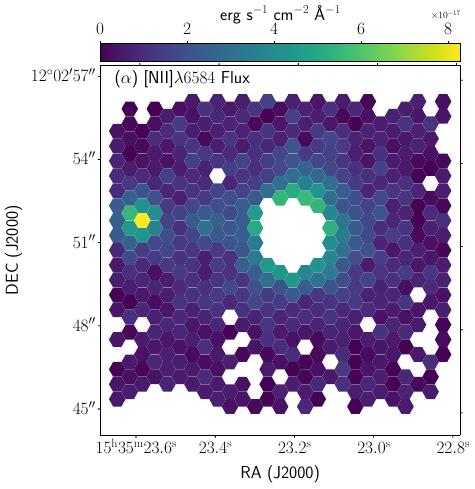}
	\includegraphics[clip, width=0.24\linewidth]{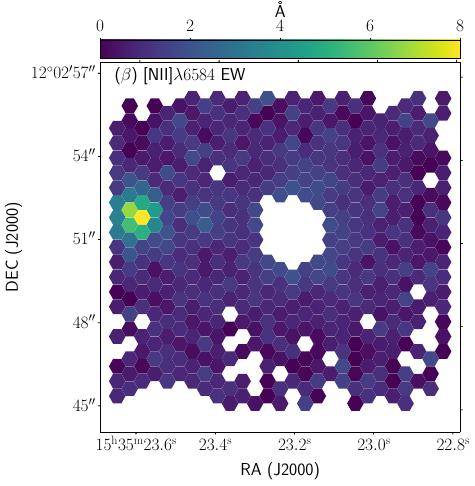}
	\includegraphics[clip, width=0.24\linewidth]{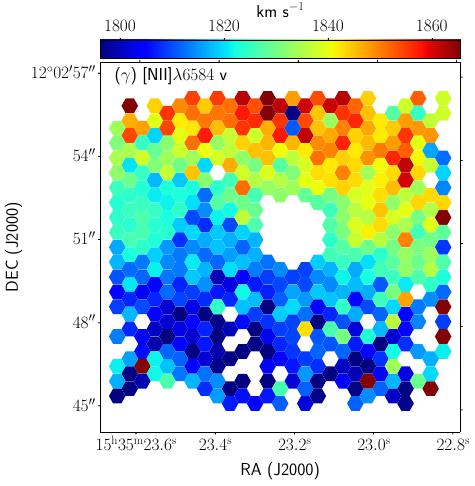}
	\includegraphics[clip, width=0.24\linewidth]{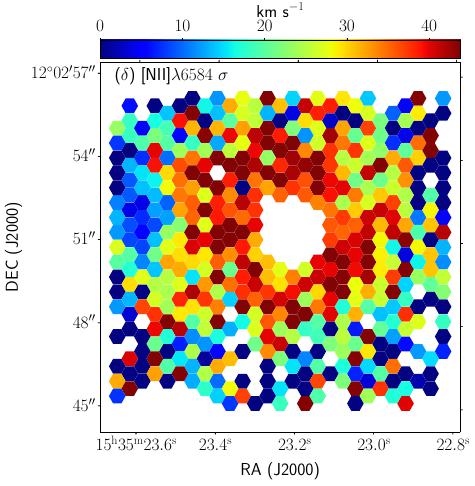}
	\includegraphics[clip, width=0.24\linewidth]{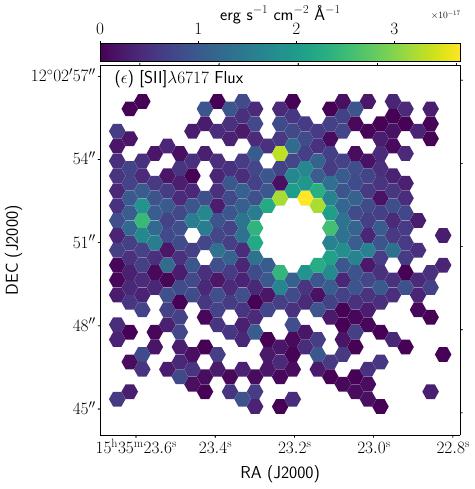}
	\includegraphics[clip, width=0.24\linewidth]{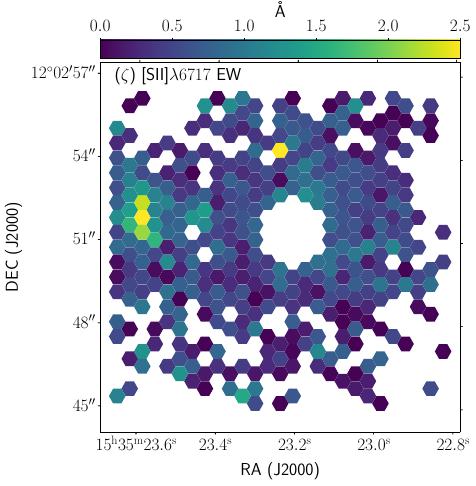}
	\includegraphics[clip, width=0.24\linewidth]{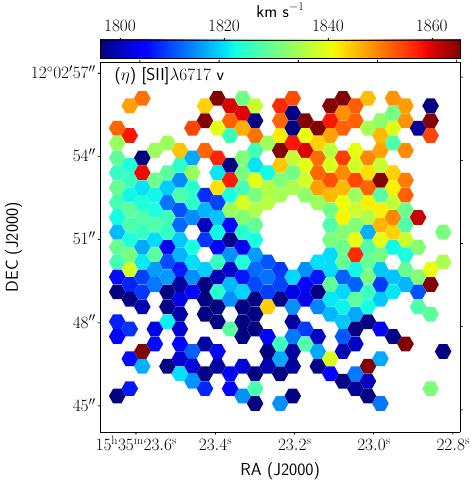}
	\includegraphics[clip, width=0.24\linewidth]{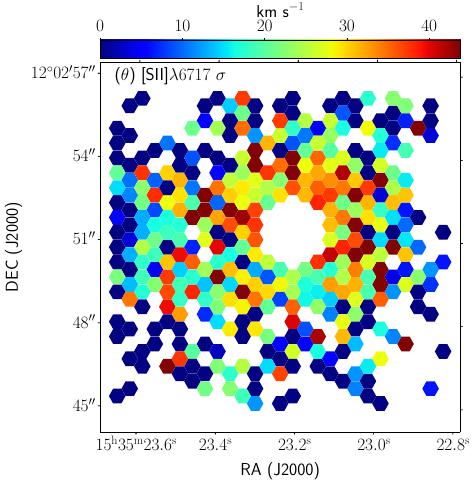}
	\includegraphics[clip, width=0.24\linewidth]{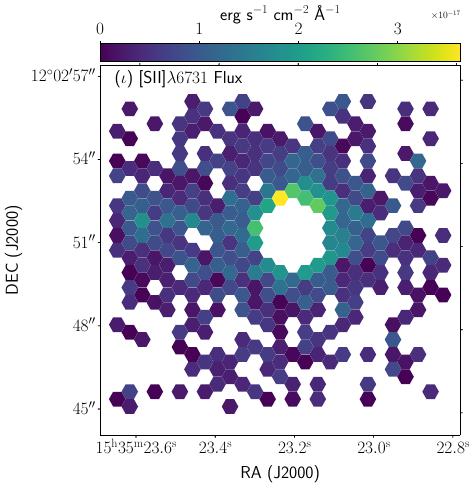}
	\includegraphics[clip, width=0.24\linewidth]{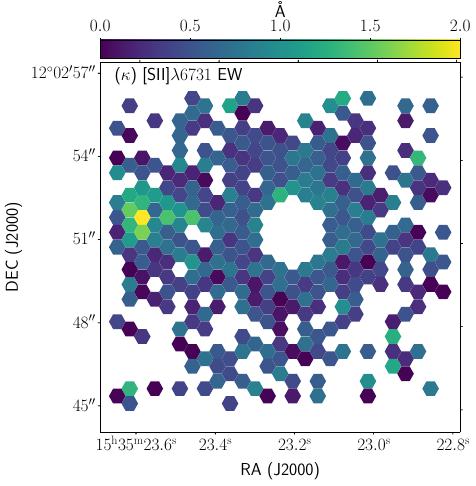}
	\includegraphics[clip, width=0.24\linewidth]{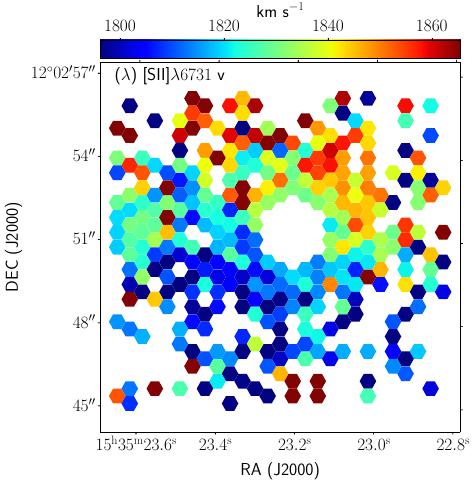}
	\includegraphics[clip, width=0.24\linewidth]{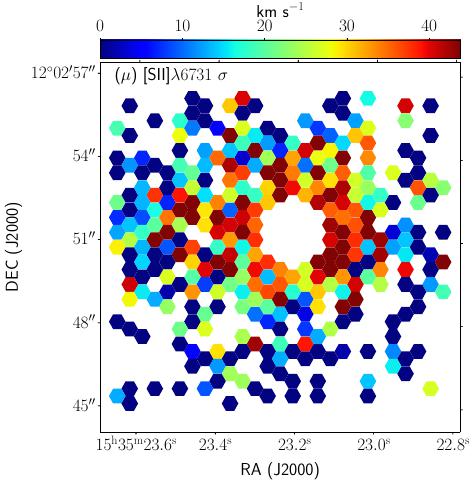}
	\caption{(cont.) NGC~5957 card.}
	\label{fig:NGC5957_card_2}
\end{figure*}

\begin{figure*}[h]
	\centering
	\includegraphics[clip, width=0.35\linewidth]{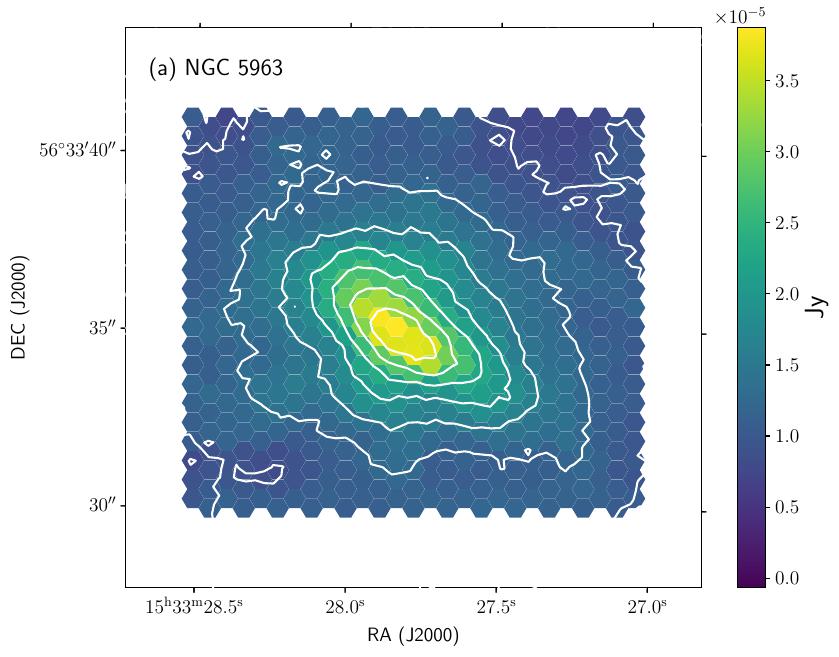}
	\includegraphics[clip, width=0.6\linewidth]{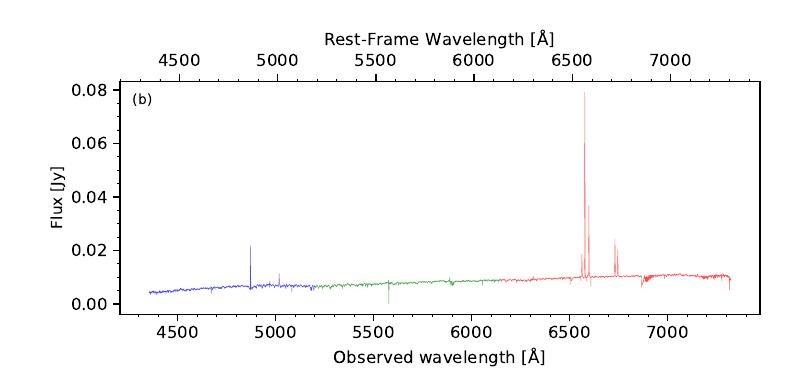}
	\includegraphics[clip, width=0.24\linewidth]{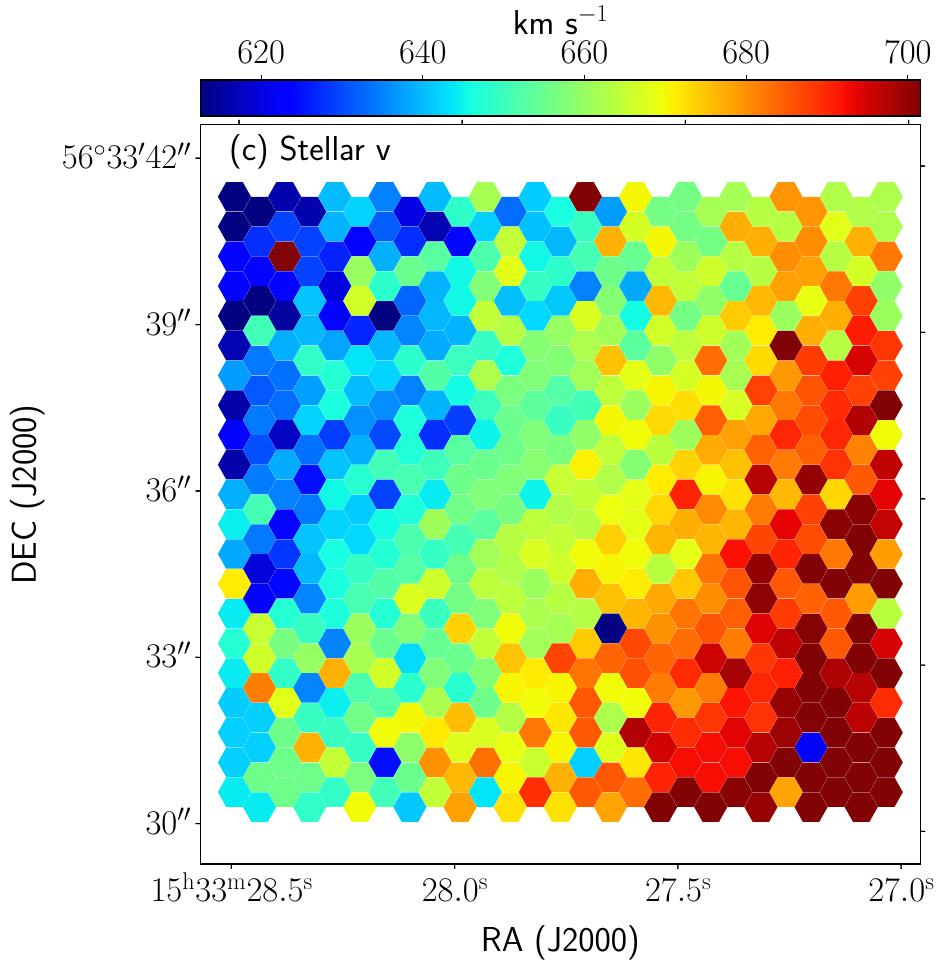}
	\includegraphics[clip, width=0.24\linewidth]{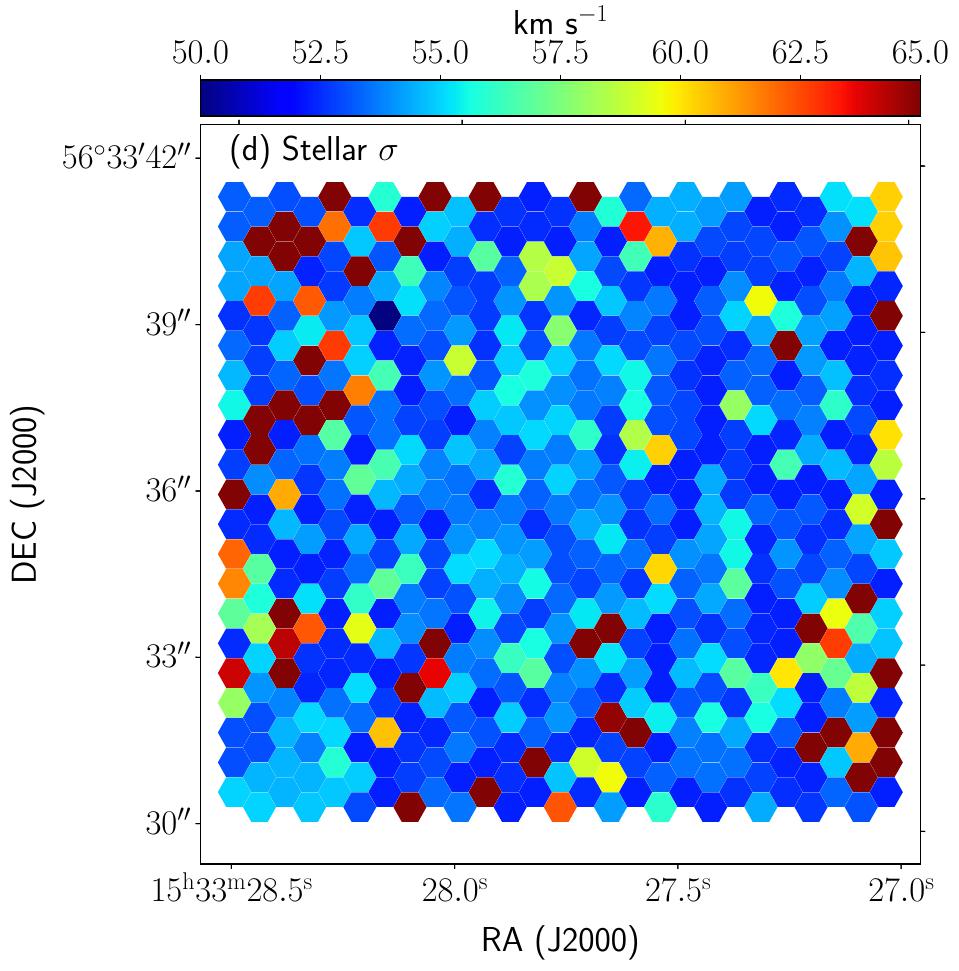}
	\includegraphics[clip, width=0.24\linewidth]{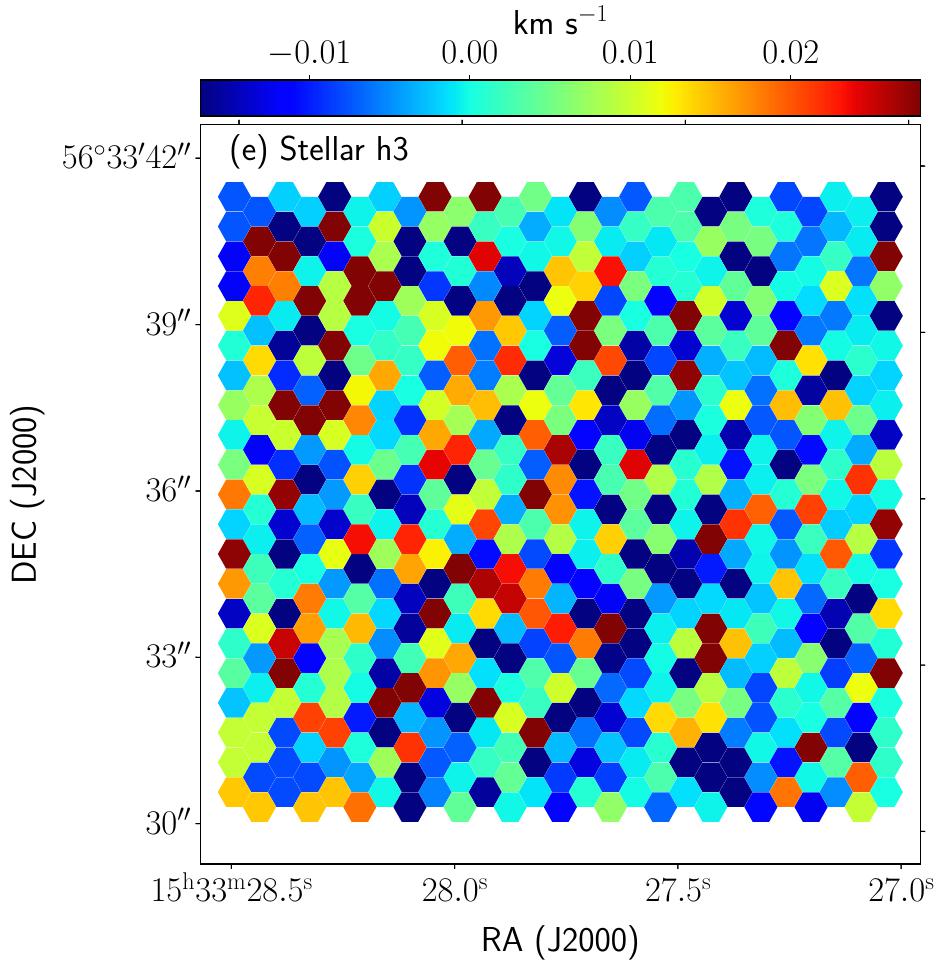}
	\includegraphics[clip, width=0.24\linewidth]{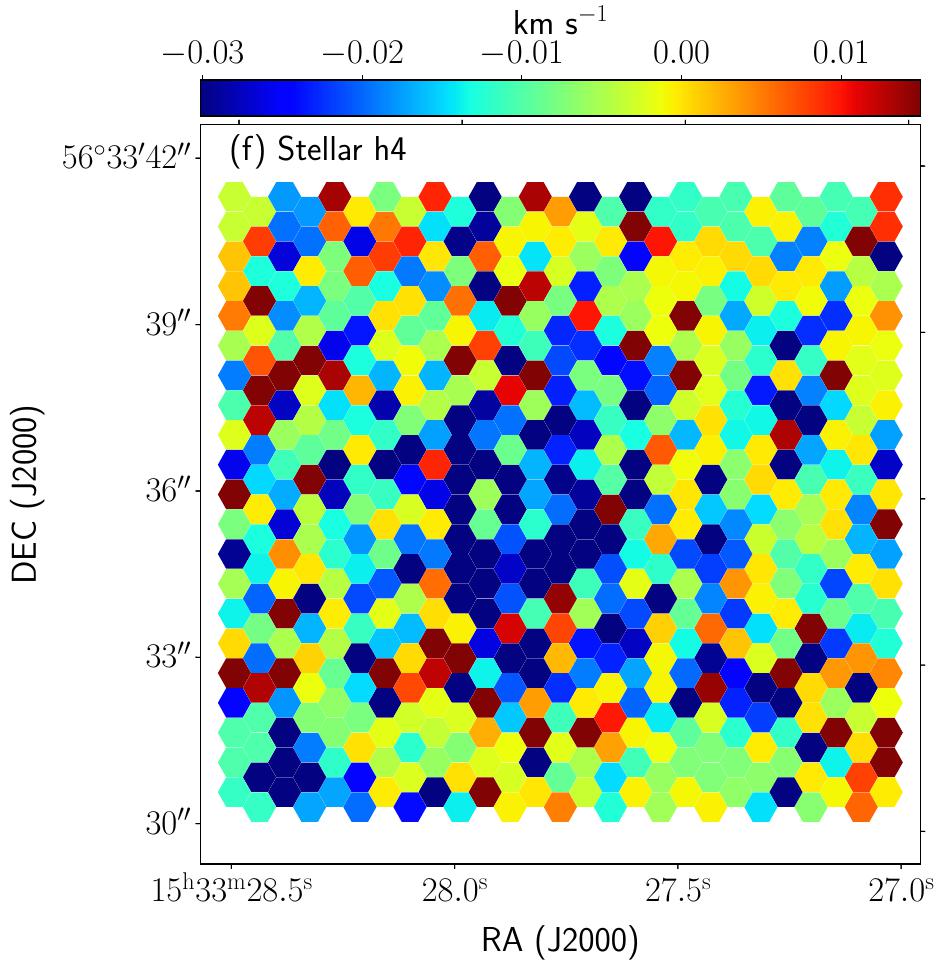}
	\includegraphics[clip, width=0.24\linewidth]{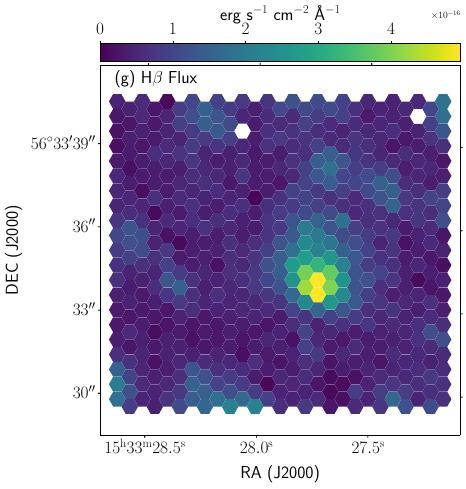}
	\includegraphics[clip, width=0.24\linewidth]{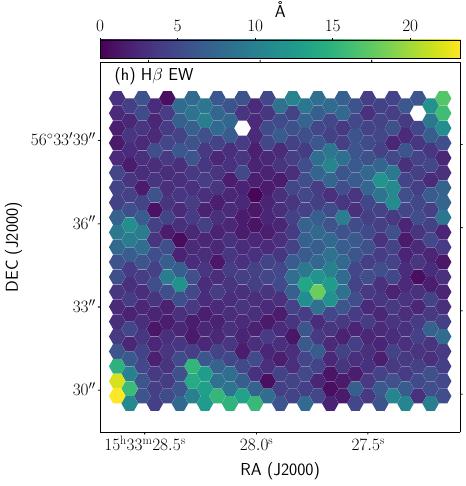}
	\includegraphics[clip, width=0.24\linewidth]{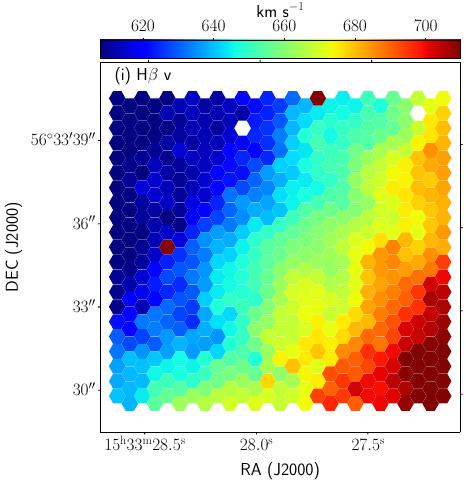}
	\includegraphics[clip, width=0.24\linewidth]{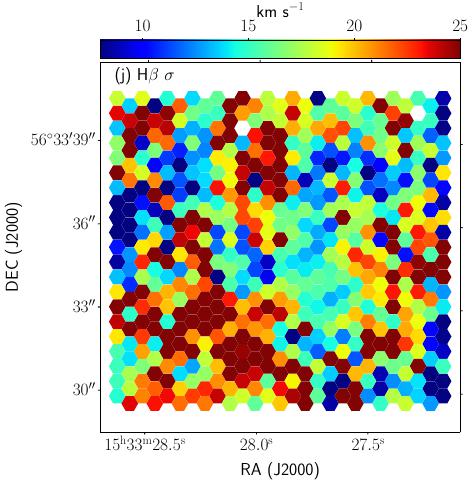}
	\includegraphics[clip, width=0.24\linewidth]{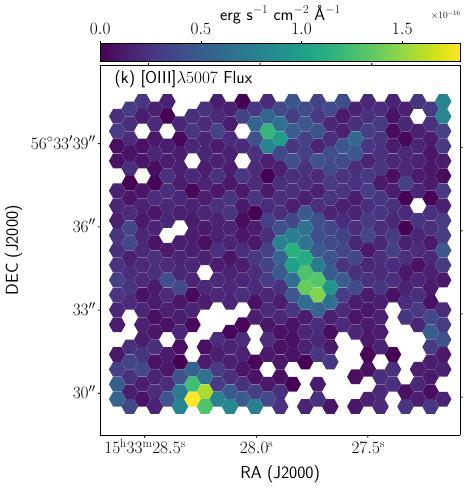}
	\includegraphics[clip, width=0.24\linewidth]{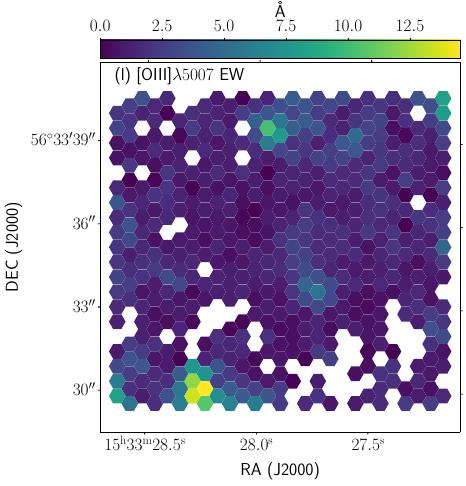}
	\includegraphics[clip, width=0.24\linewidth]{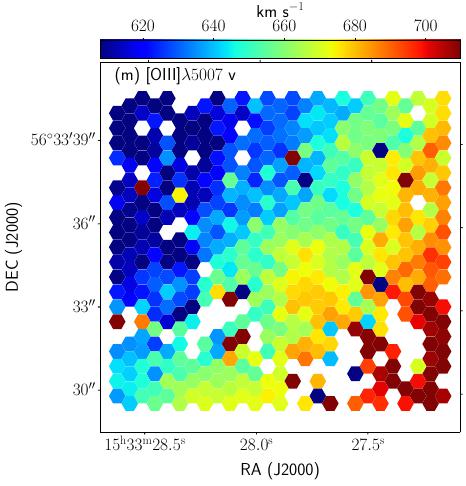}
	\includegraphics[clip, width=0.24\linewidth]{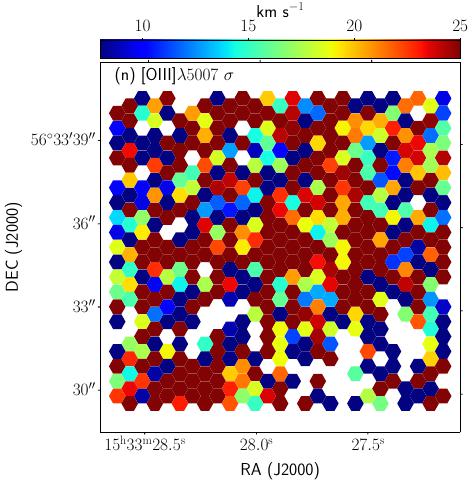}
	\includegraphics[clip, width=0.24\linewidth]{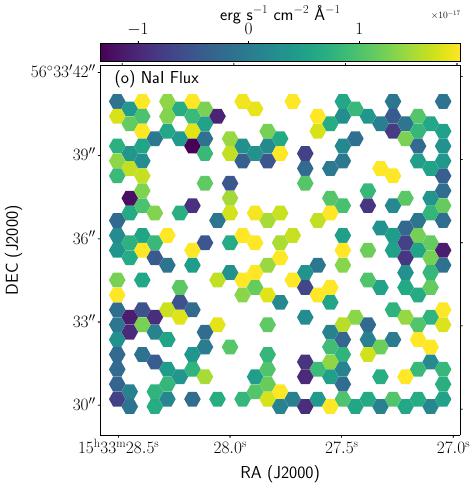}
	\includegraphics[clip, width=0.24\linewidth]{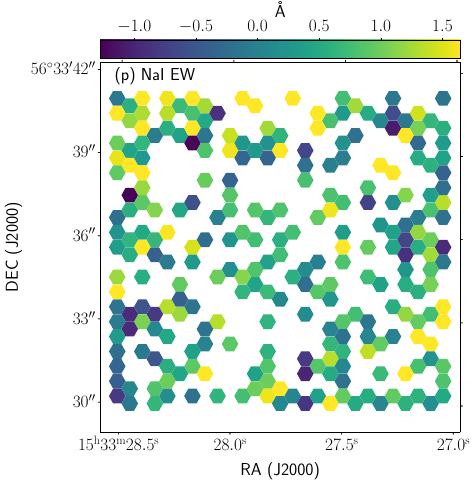}
	\includegraphics[clip, width=0.24\linewidth]{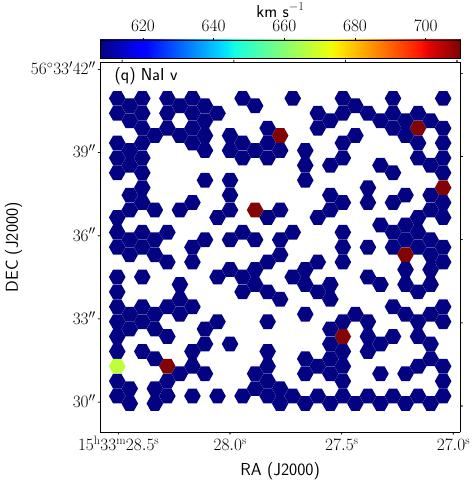}
	\includegraphics[clip, width=0.24\linewidth]{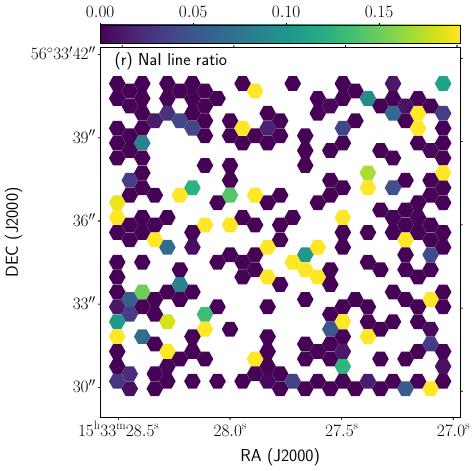}
	\caption{NGC~5963 card.}
	\label{fig:NGC5963_card_1}
\end{figure*}
\clearpage
\addtocounter{figure}{-1}
\begin{figure*}[h]
	\centering
	\includegraphics[clip, width=0.24\linewidth]{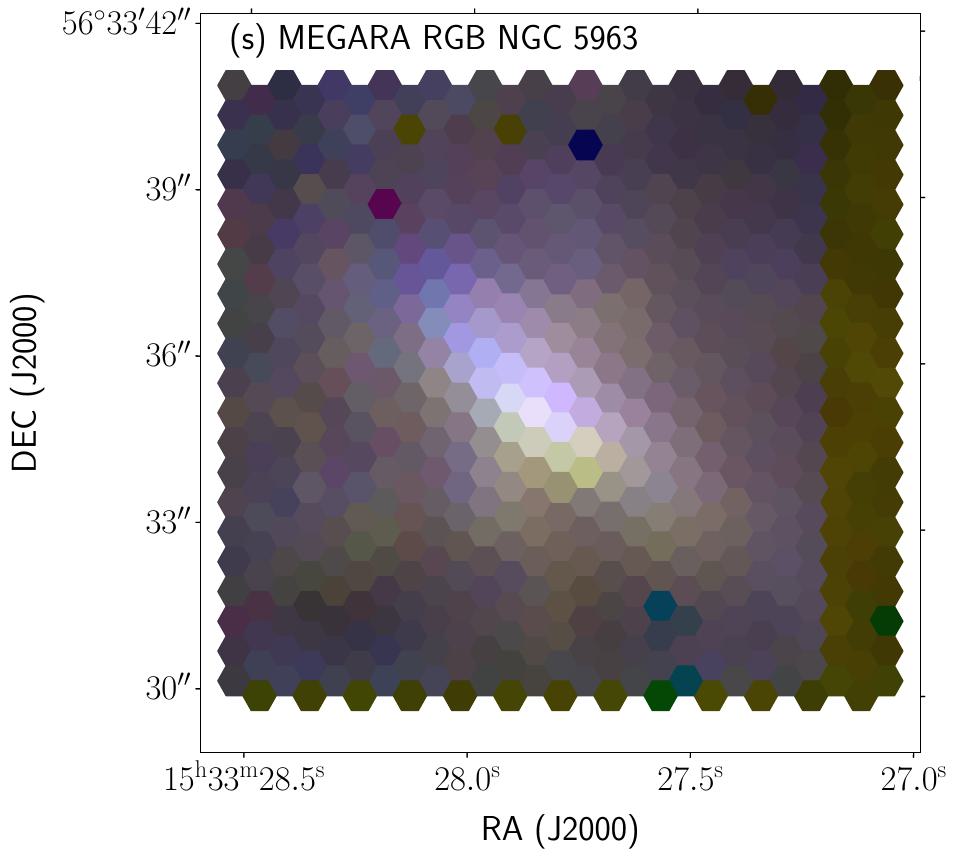}
	\includegraphics[clip, width=0.24\linewidth]{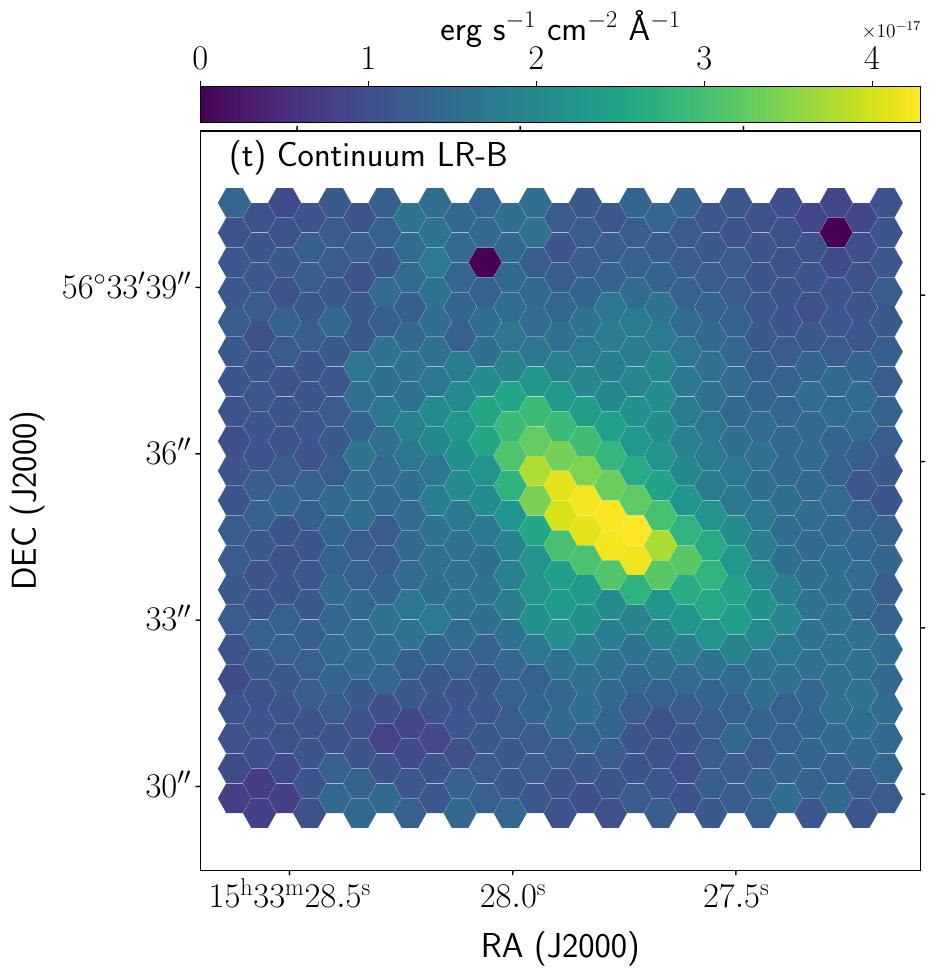}
	\includegraphics[clip, width=0.24\linewidth]{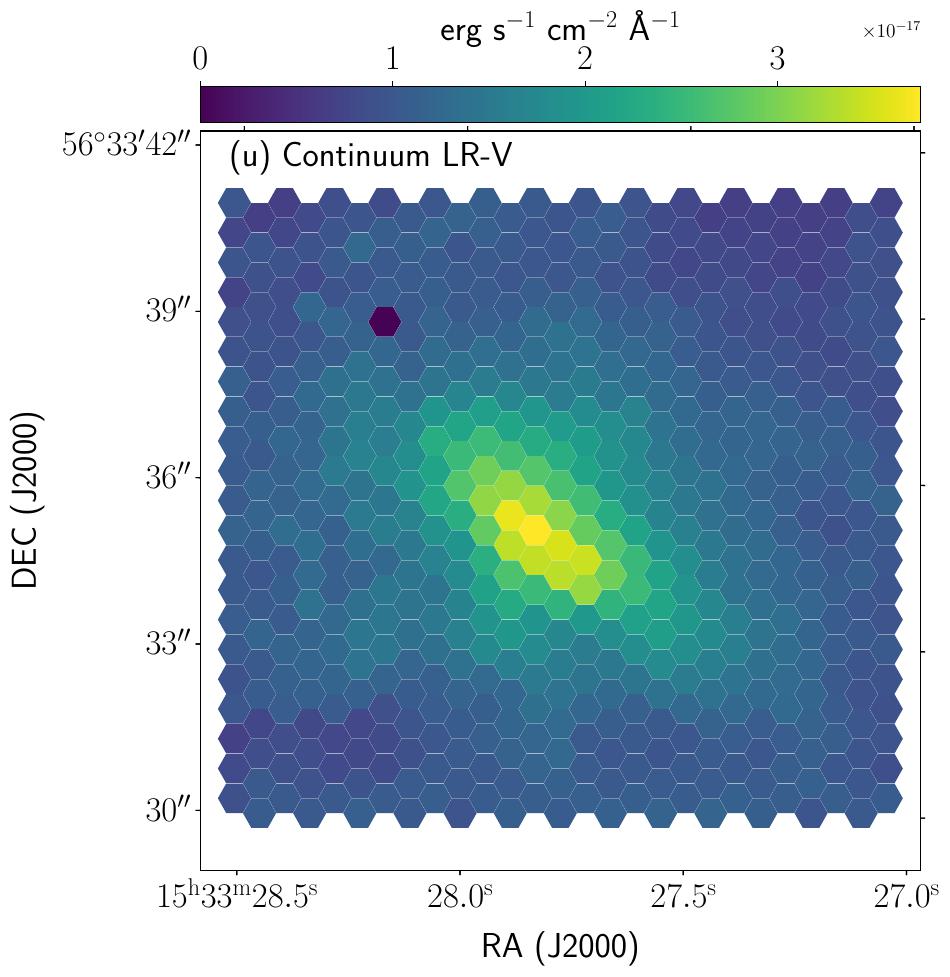}
	\includegraphics[clip, width=0.24\linewidth]{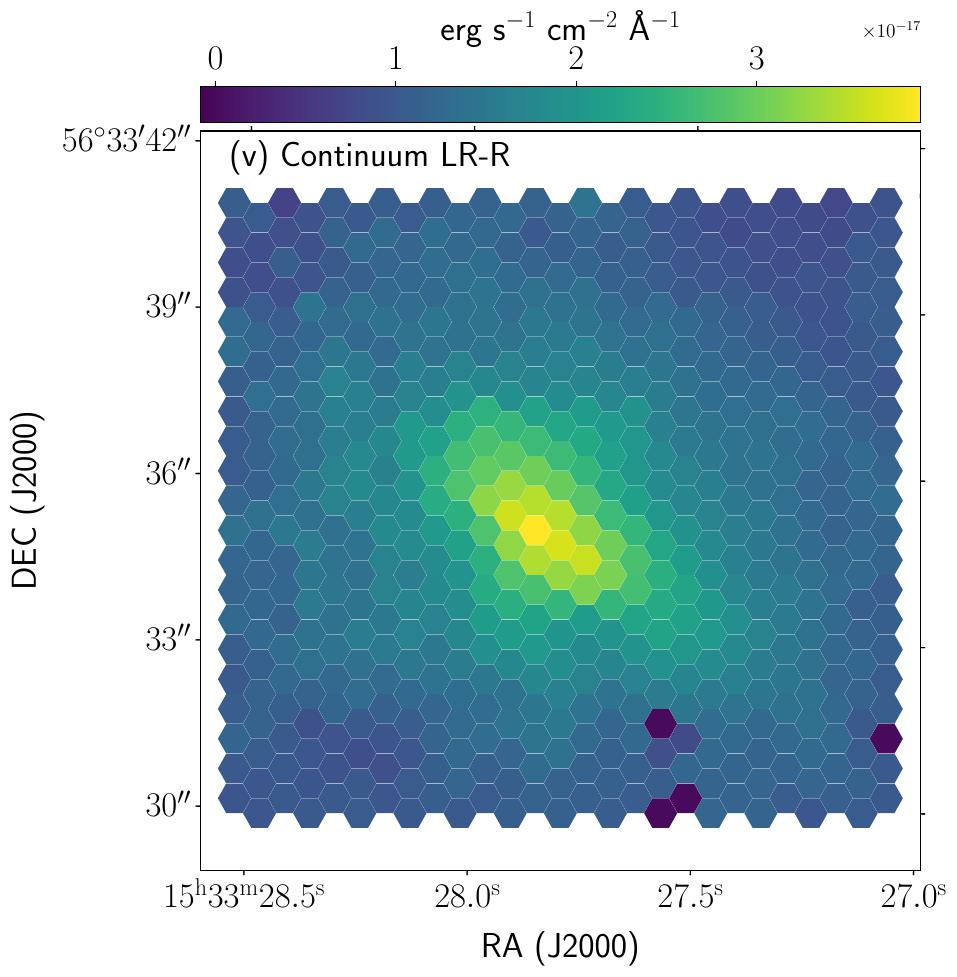}
	\includegraphics[clip, width=0.24\linewidth]{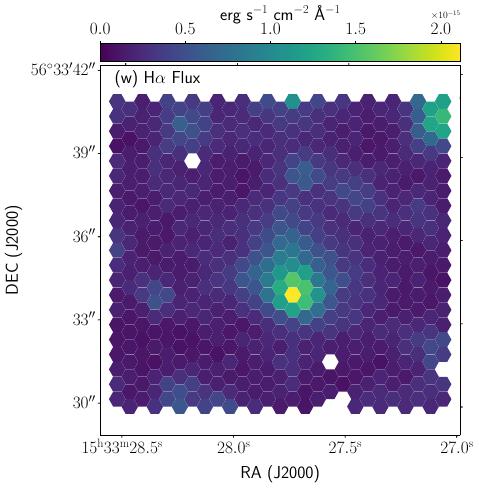}
	\includegraphics[clip, width=0.24\linewidth]{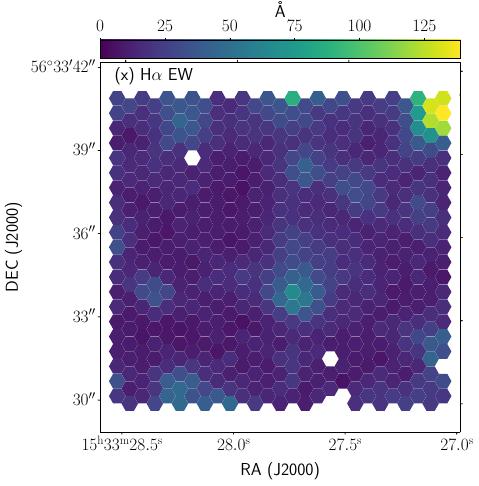}
	\includegraphics[clip, width=0.24\linewidth]{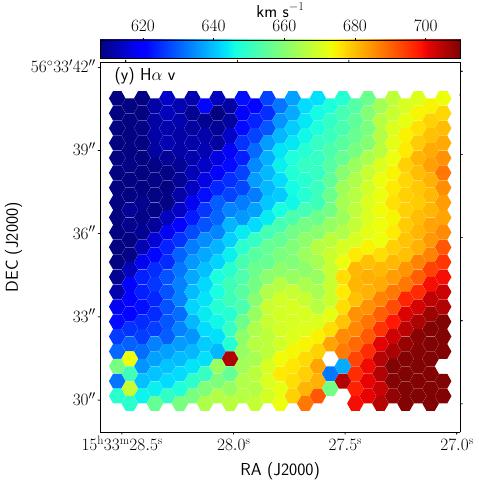}
	\includegraphics[clip, width=0.24\linewidth]{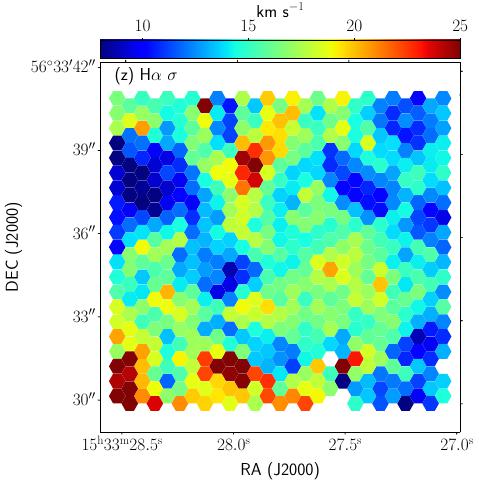}
	\includegraphics[clip, width=0.24\linewidth]{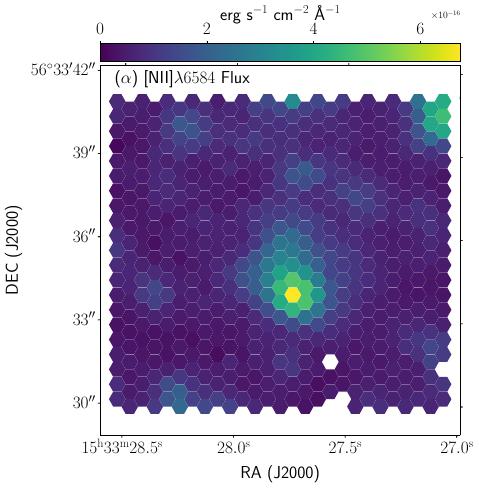}
	\includegraphics[clip, width=0.24\linewidth]{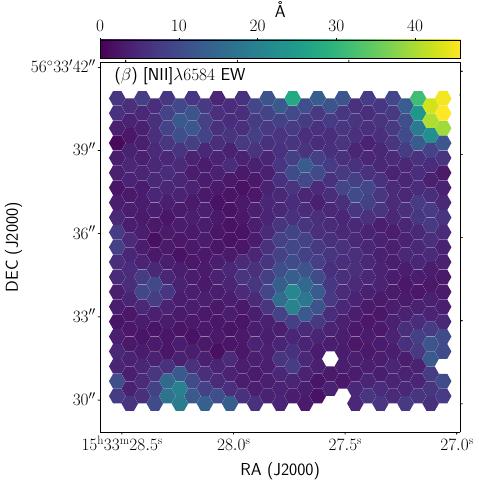}
	\includegraphics[clip, width=0.24\linewidth]{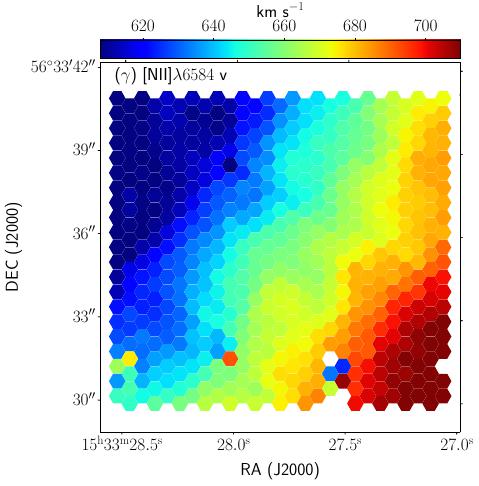}
	\includegraphics[clip, width=0.24\linewidth]{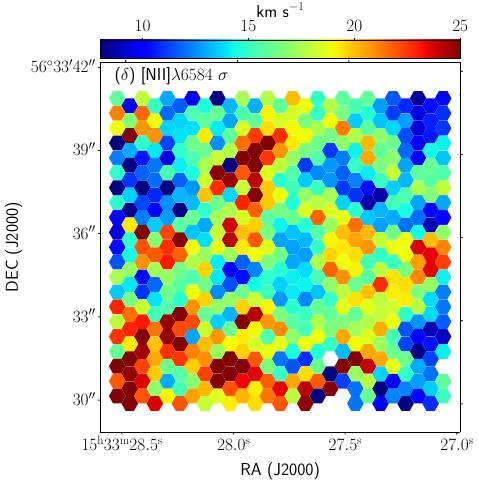}
	\includegraphics[clip, width=0.24\linewidth]{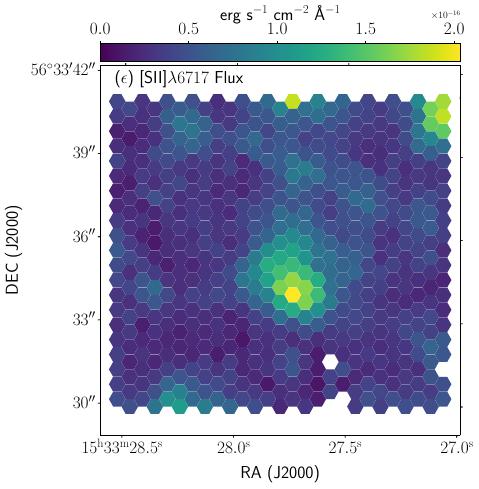}
	\includegraphics[clip, width=0.24\linewidth]{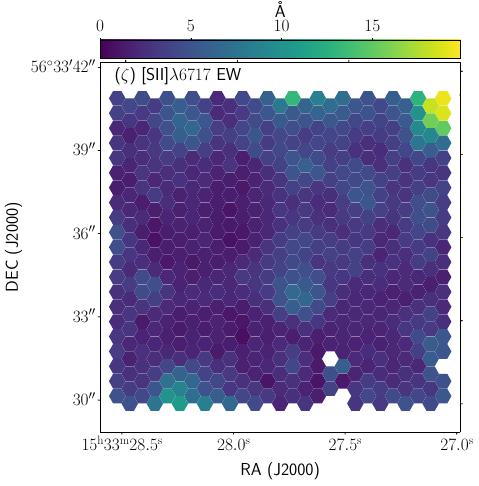}
	\includegraphics[clip, width=0.24\linewidth]{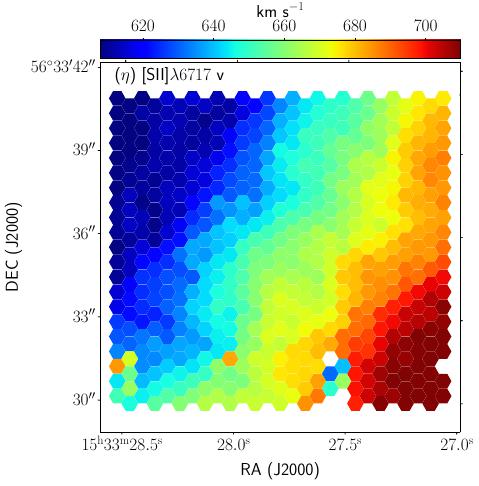}
	\includegraphics[clip, width=0.24\linewidth]{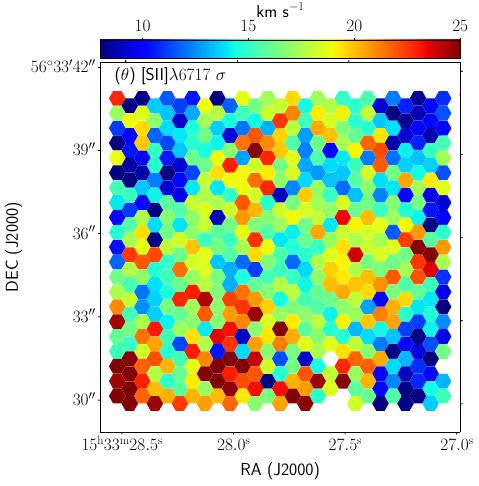}
	\includegraphics[clip, width=0.24\linewidth]{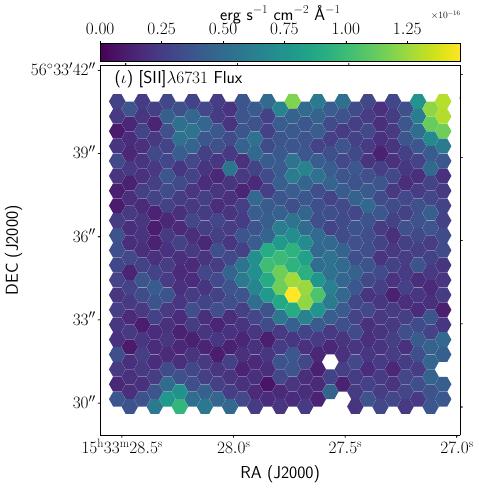}
	\includegraphics[clip, width=0.24\linewidth]{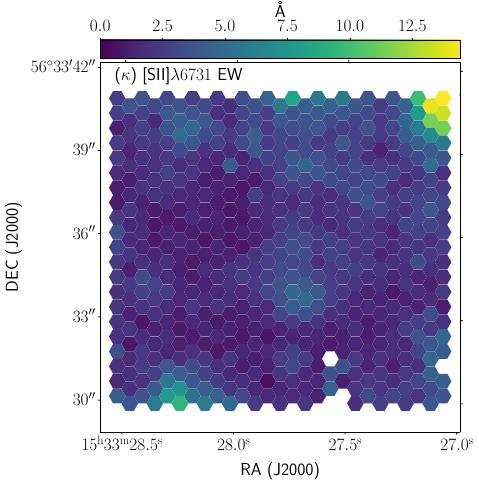}
	\includegraphics[clip, width=0.24\linewidth]{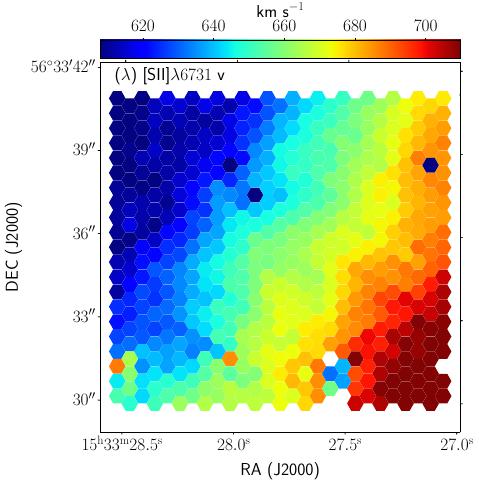}
	\includegraphics[clip, width=0.24\linewidth]{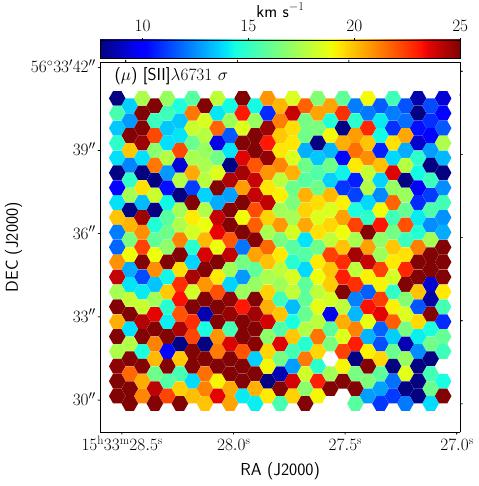}
	\caption{(cont.) NGC~5963 card.}
	\label{fig:NGC5963_card_2}
\end{figure*}

\begin{figure*}[h]
	\centering
	\includegraphics[clip, width=0.35\linewidth]{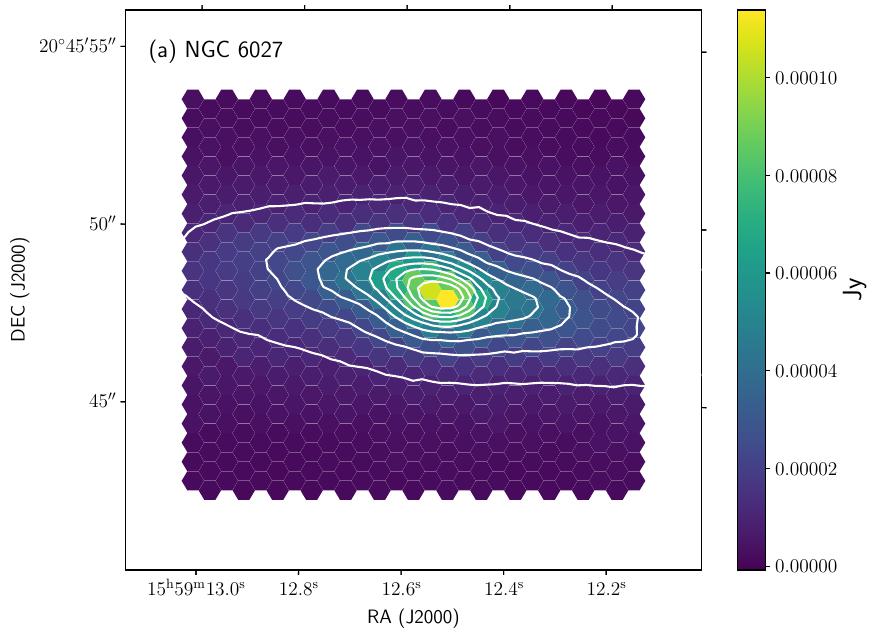}
	\includegraphics[clip, width=0.6\linewidth]{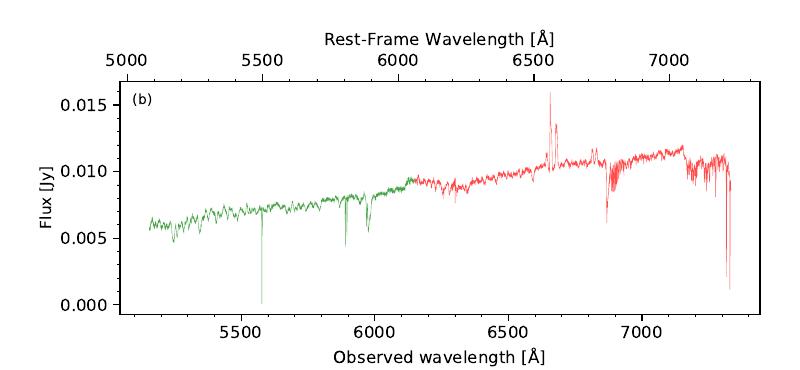}
	\includegraphics[clip, width=0.24\linewidth]{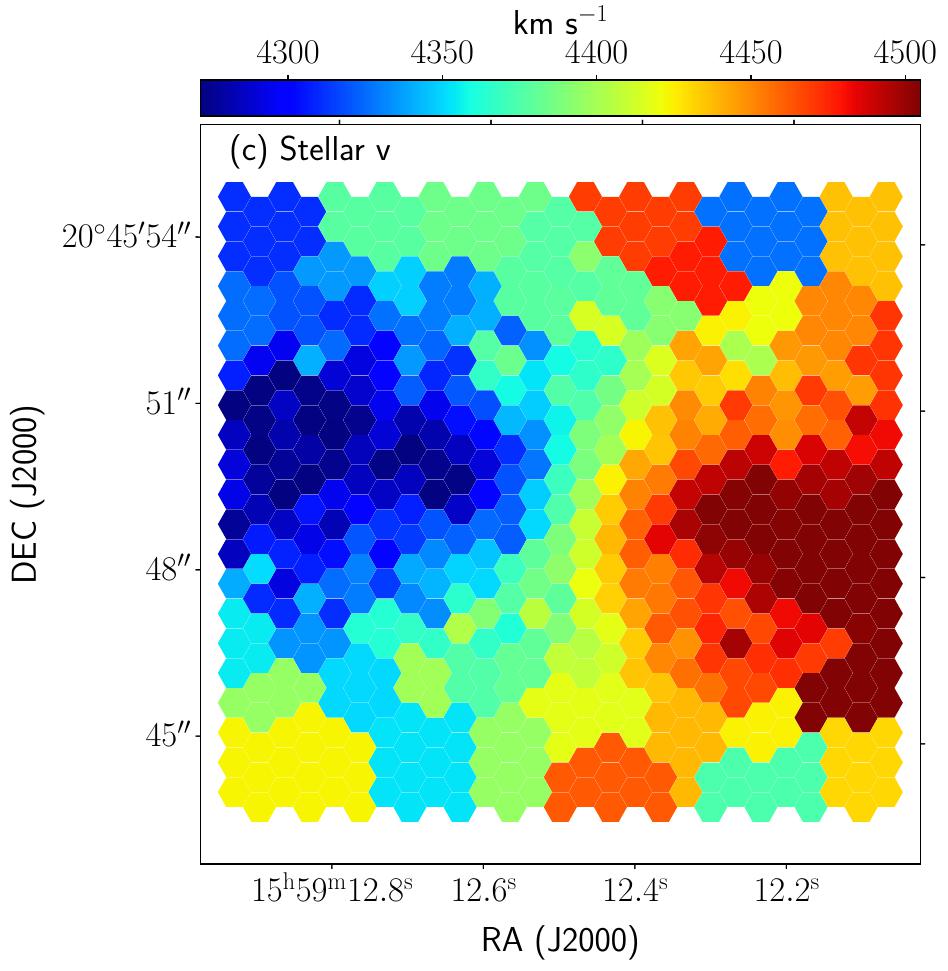}
	\includegraphics[clip, width=0.24\linewidth]{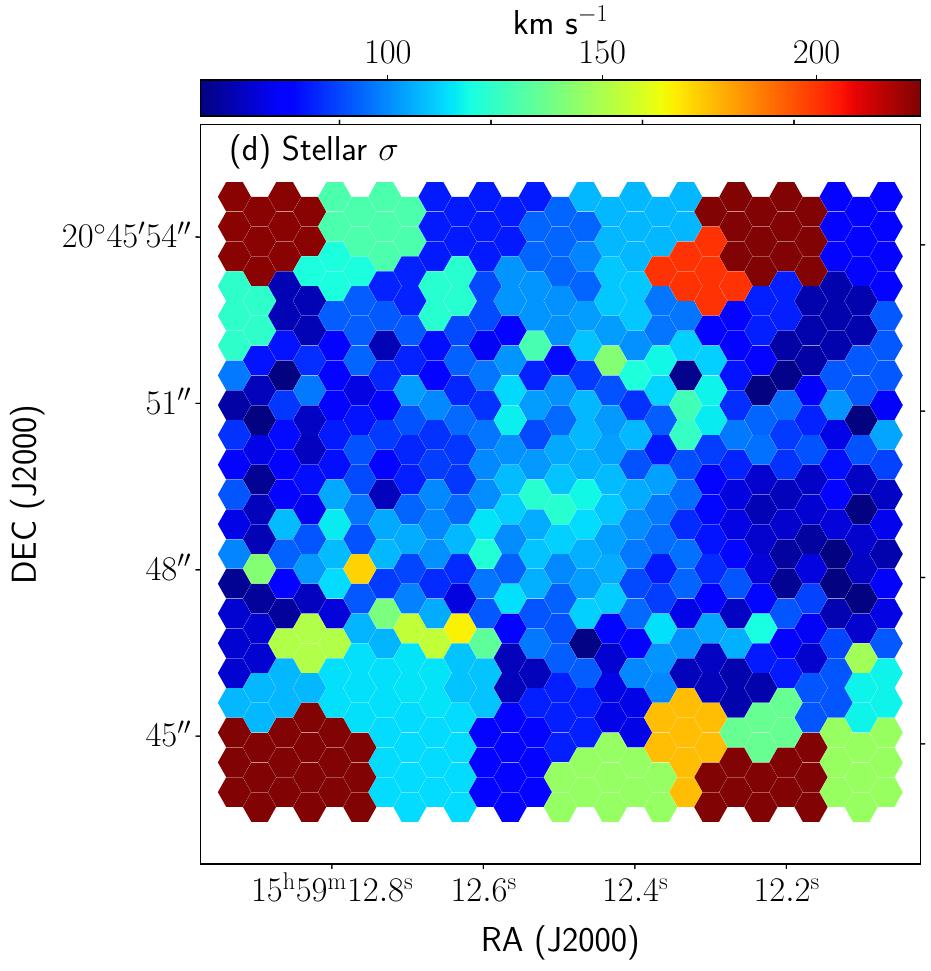}
	\includegraphics[clip, width=0.24\linewidth]{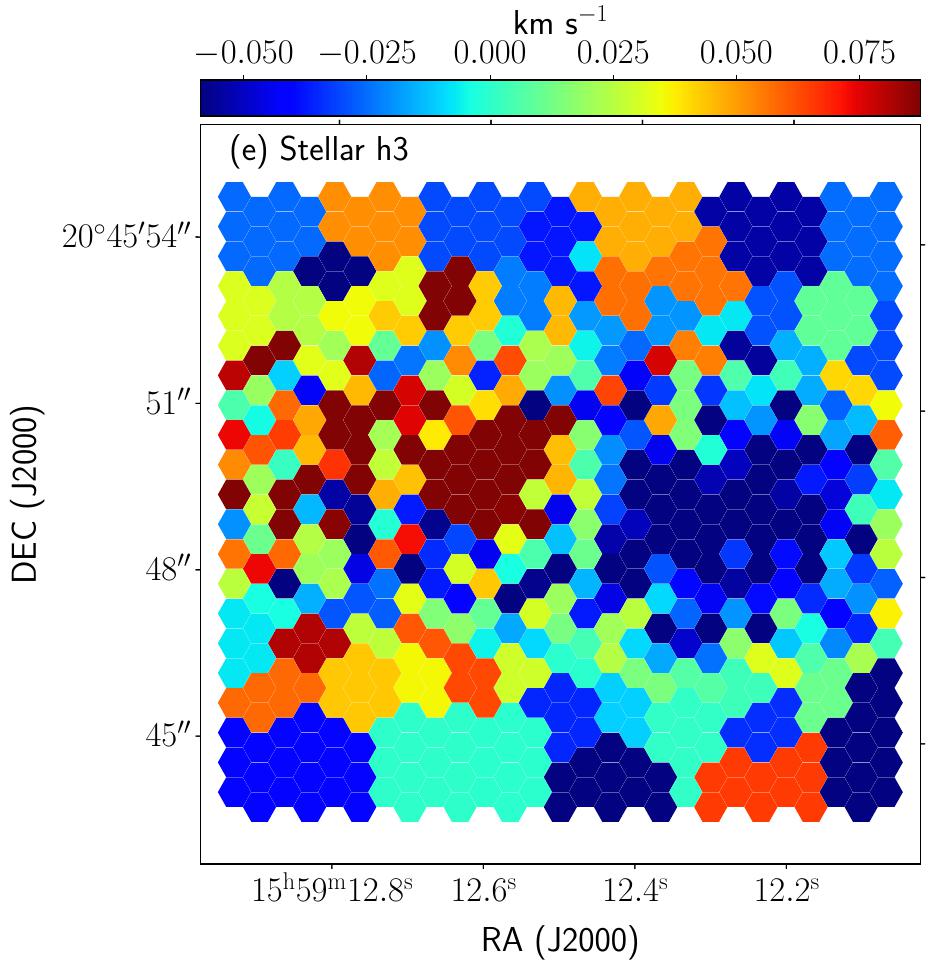}
	\includegraphics[clip, width=0.24\linewidth]{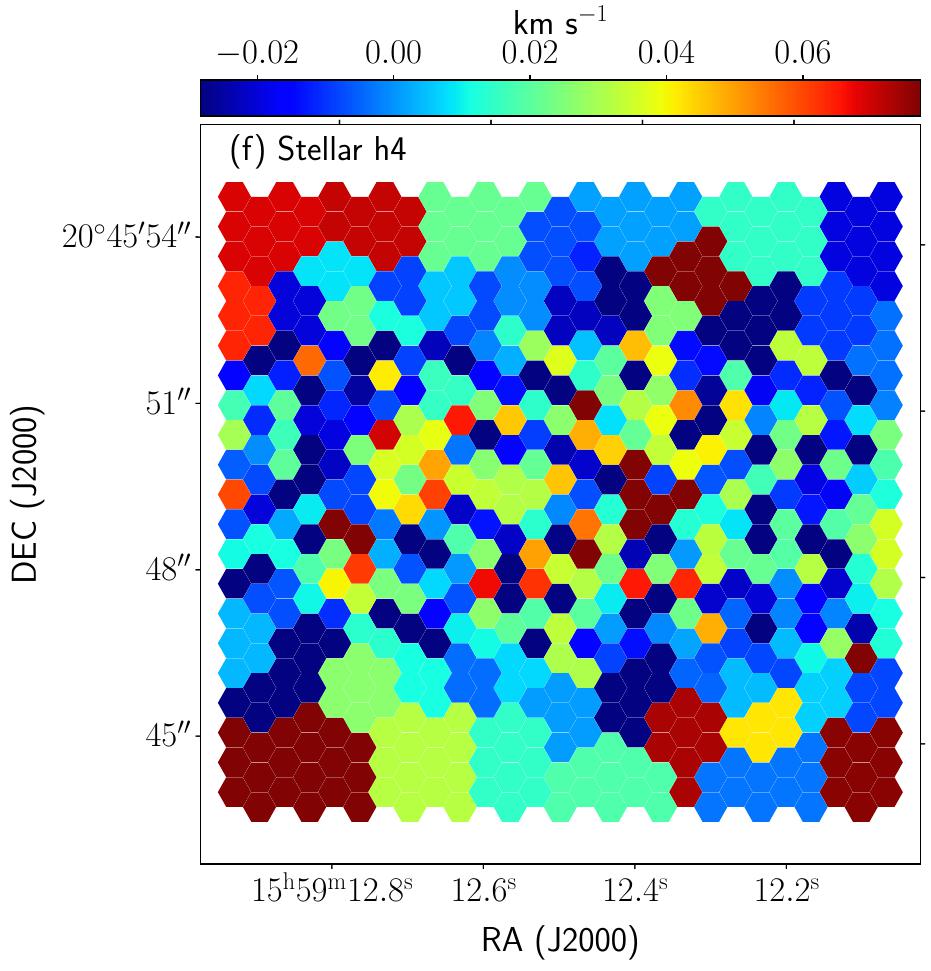}
	
	\vspace{8.8cm}
	
	\includegraphics[clip, width=0.24\linewidth]{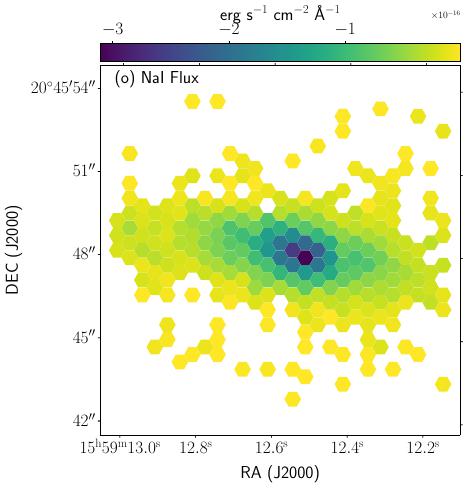}
	\includegraphics[clip, width=0.24\linewidth]{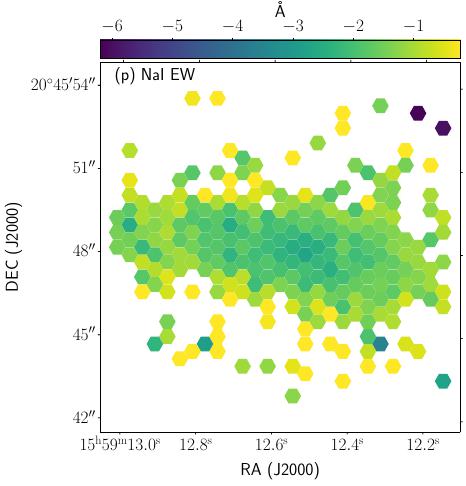}
	\includegraphics[clip, width=0.24\linewidth]{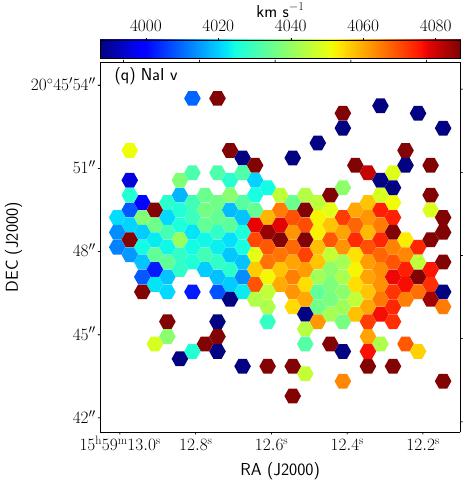}
	\includegraphics[clip, width=0.24\linewidth]{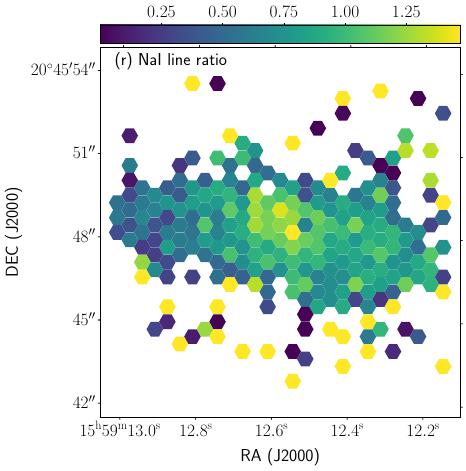}
	\caption{NGC~6027 card.}
	\label{fig:NGC6027_card_1}
\end{figure*}
\addtocounter{figure}{-1}
\begin{figure*}[h]
	\centering
	\includegraphics[clip, width=0.24\linewidth]{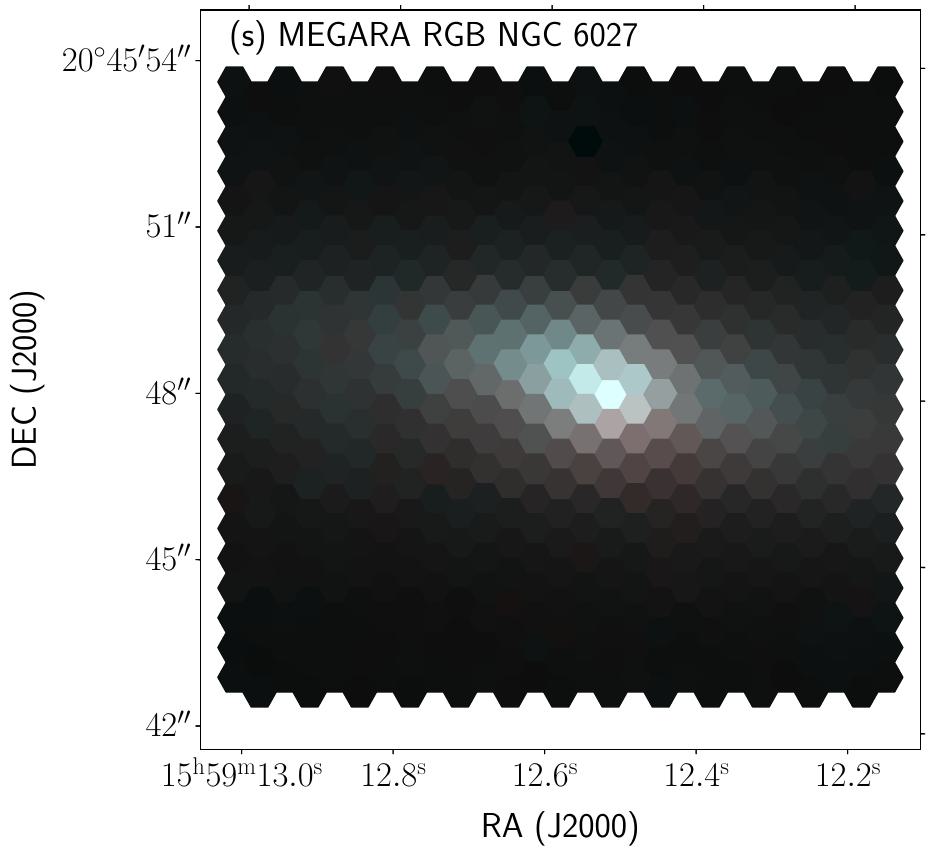}
	\hspace{4,4cm}
	\includegraphics[clip, width=0.24\linewidth]{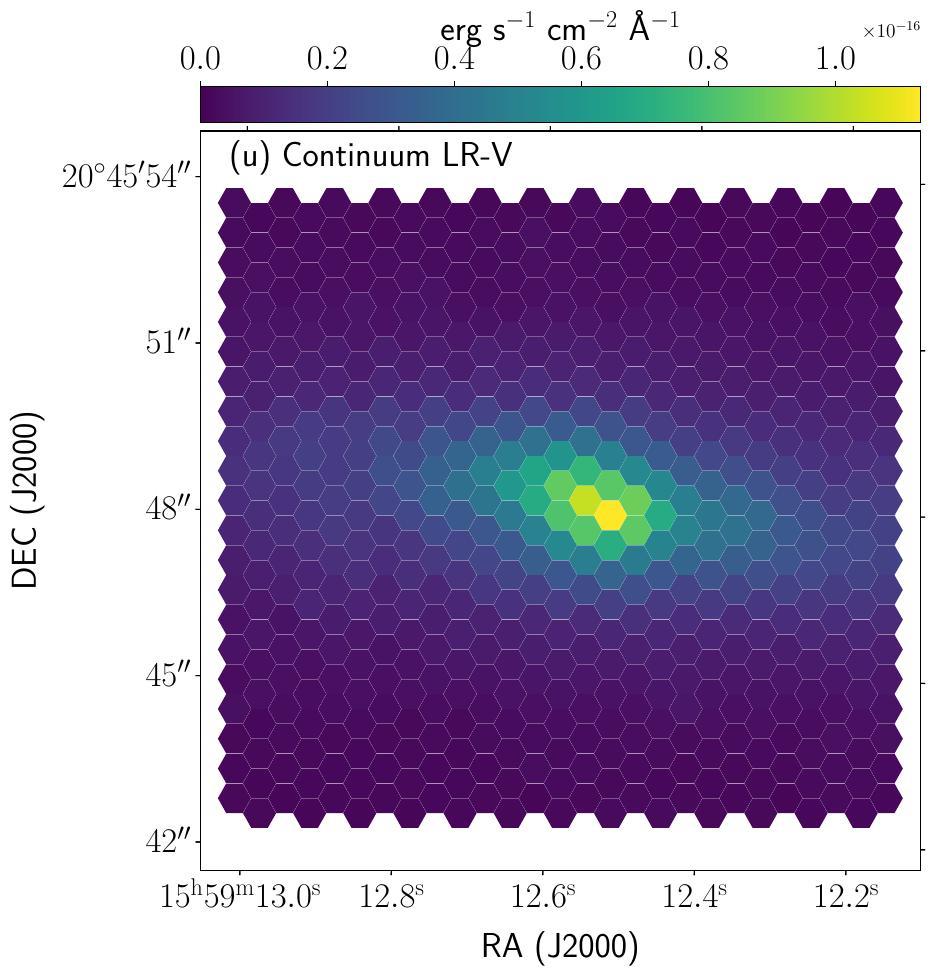}
	\includegraphics[clip, width=0.24\linewidth]{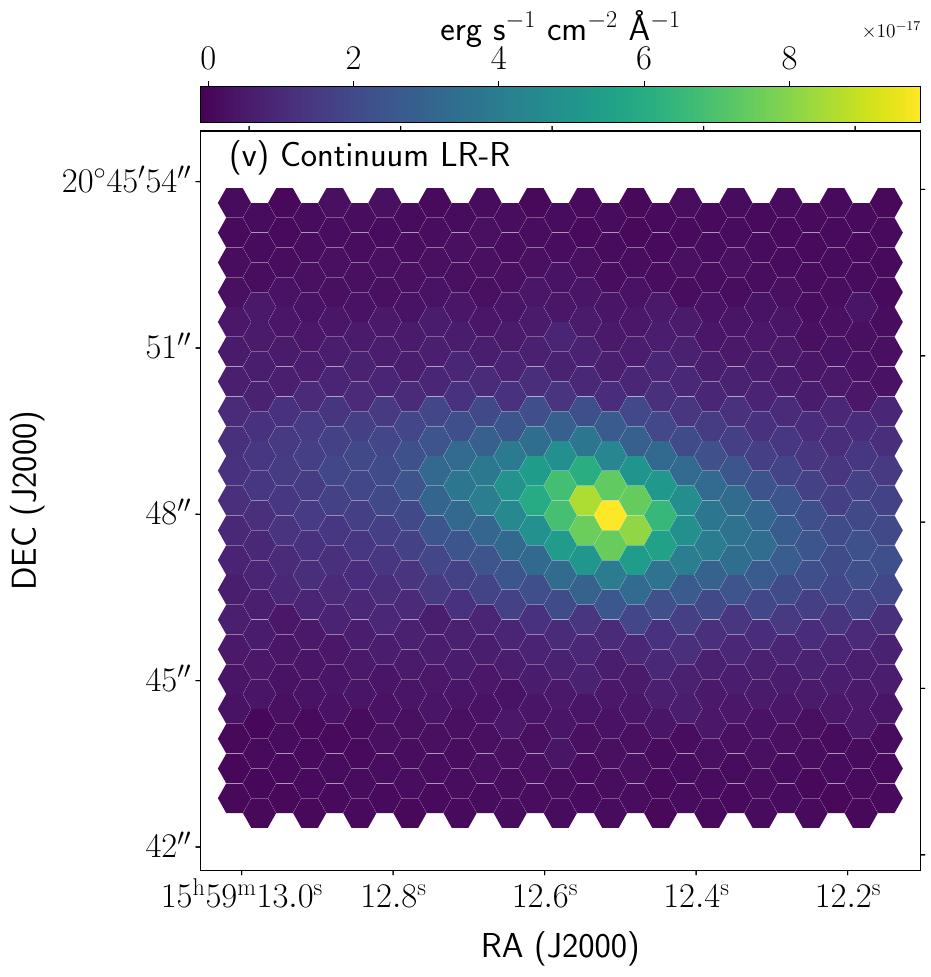}
	\includegraphics[clip, width=0.24\linewidth]{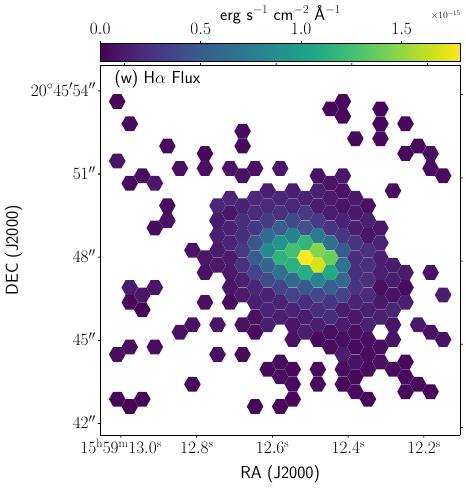}
	\includegraphics[clip, width=0.24\linewidth]{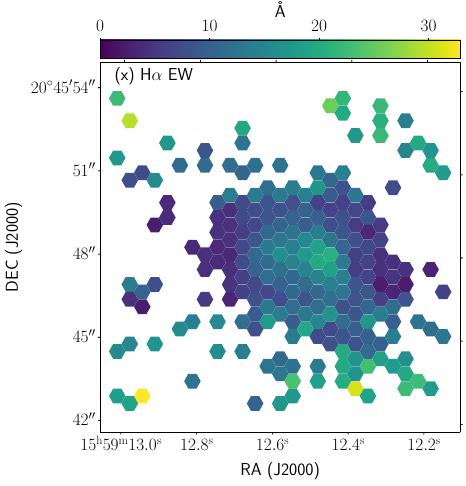}
	\includegraphics[clip, width=0.24\linewidth]{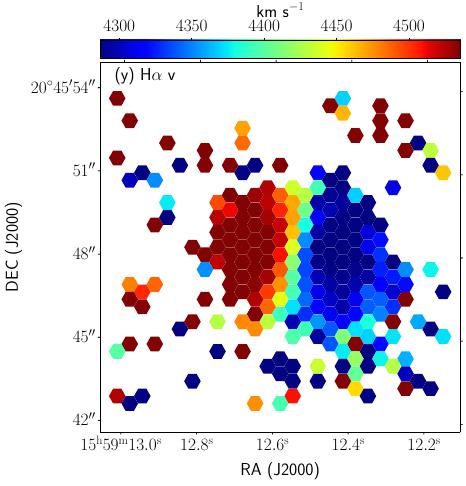}
	\includegraphics[clip, width=0.24\linewidth]{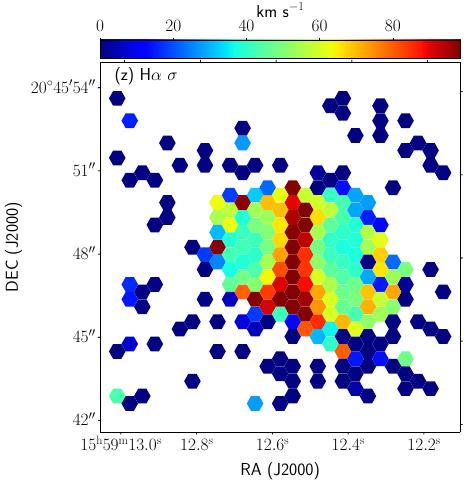}
	\includegraphics[clip, width=0.24\linewidth]{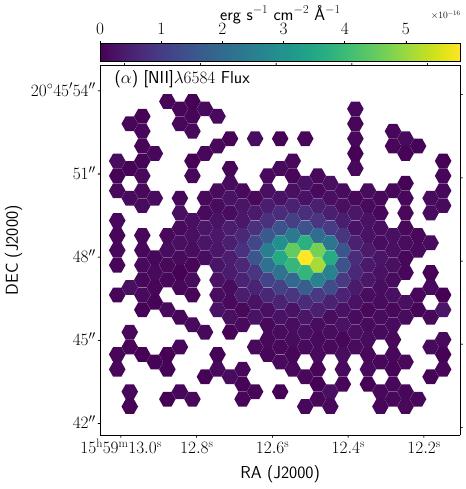}
	\includegraphics[clip, width=0.24\linewidth]{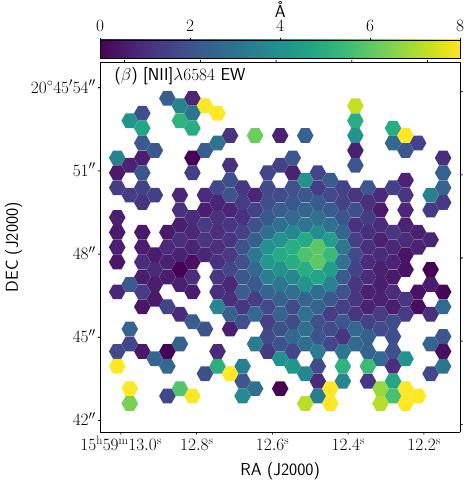}
	\includegraphics[clip, width=0.24\linewidth]{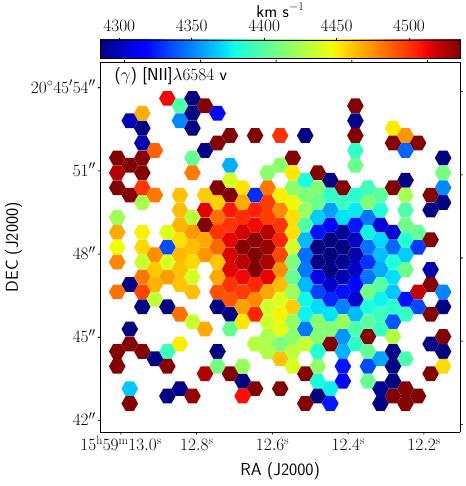}
	\includegraphics[clip, width=0.24\linewidth]{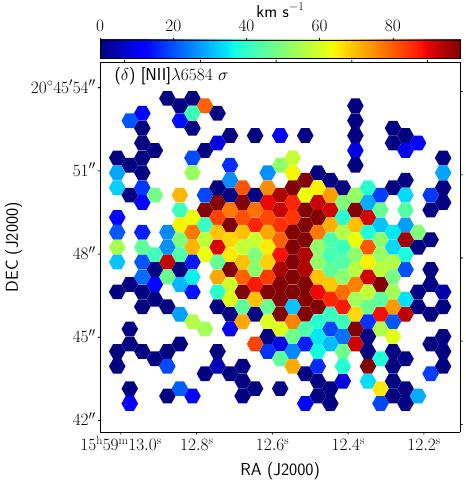}
	\includegraphics[clip, width=0.24\linewidth]{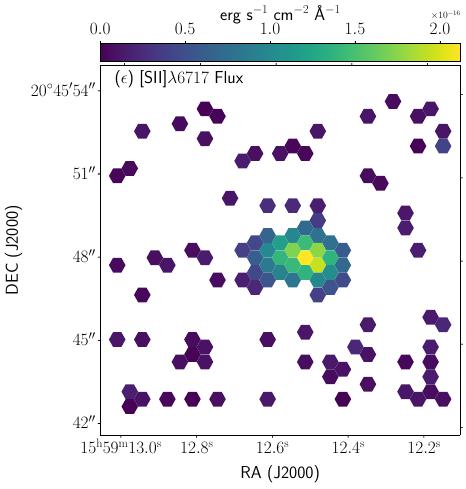}
	\includegraphics[clip, width=0.24\linewidth]{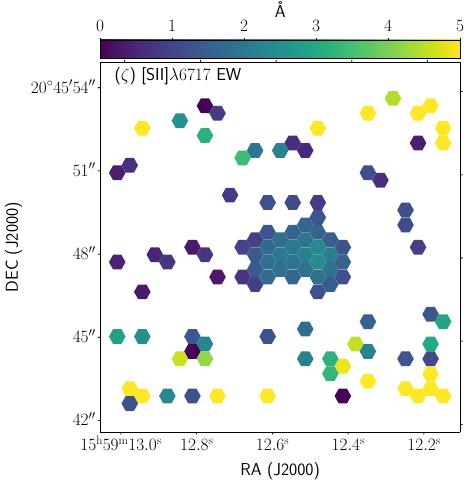}
	\includegraphics[clip, width=0.24\linewidth]{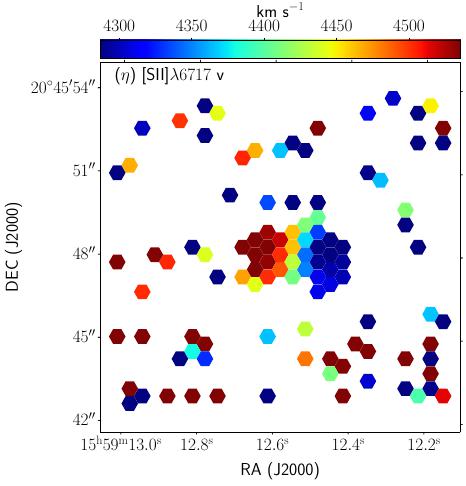}
	\includegraphics[clip, width=0.24\linewidth]{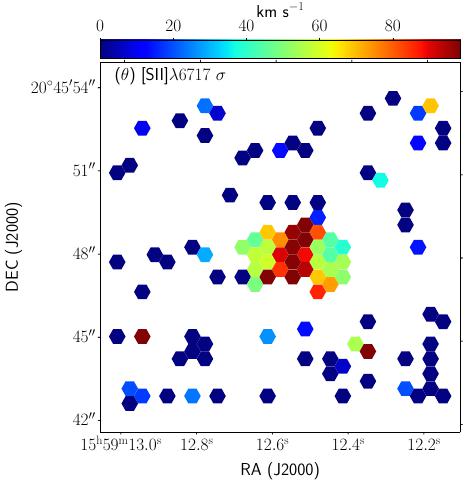}
	\includegraphics[clip, width=0.24\linewidth]{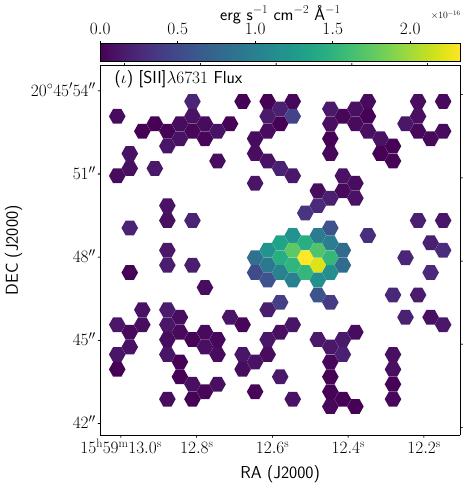}
	\includegraphics[clip, width=0.24\linewidth]{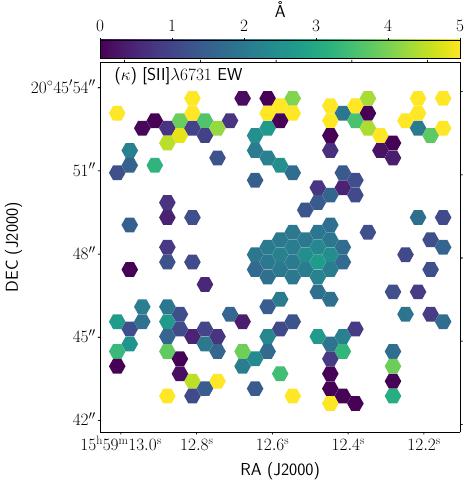}
	\includegraphics[clip, width=0.24\linewidth]{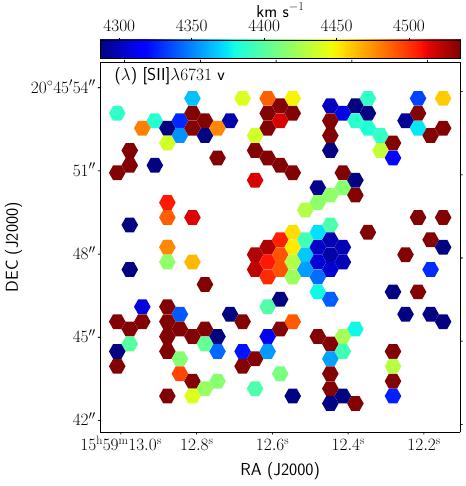}
	\includegraphics[clip, width=0.24\linewidth]{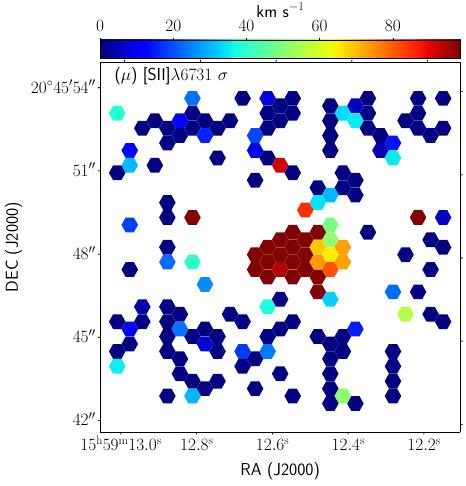}
	\caption{(cont.) NGC~6027 card.}
	\label{fig:NGC6027_card_2}
\end{figure*}

\begin{figure*}[h]
	\centering
	\includegraphics[clip, width=0.35\linewidth]{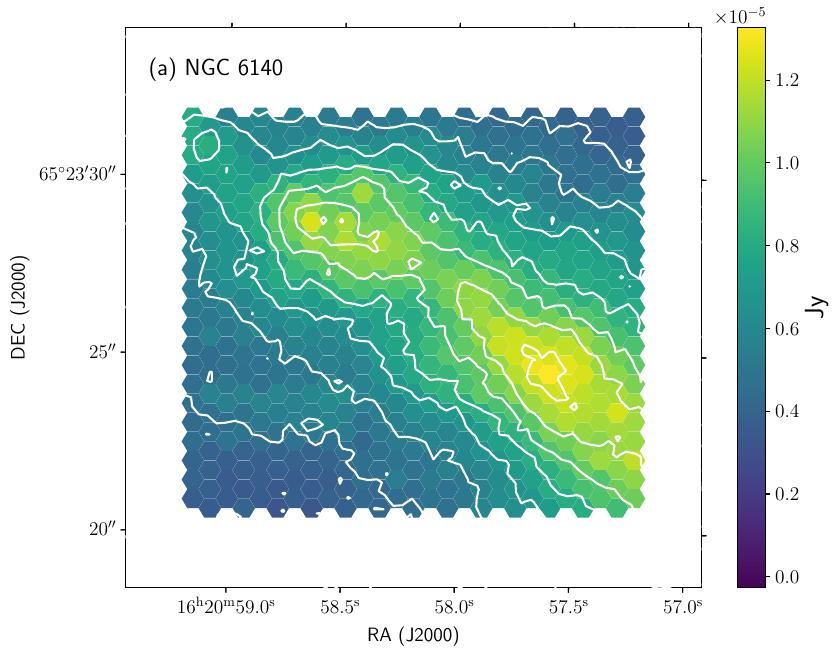}
	\includegraphics[clip, width=0.6\linewidth]{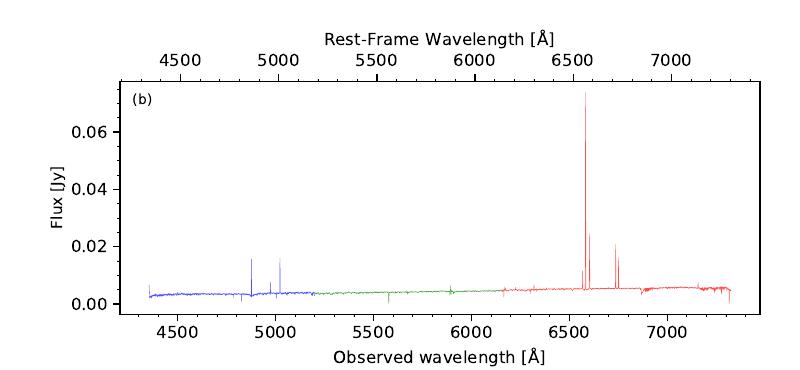}
	\includegraphics[clip, width=0.24\linewidth]{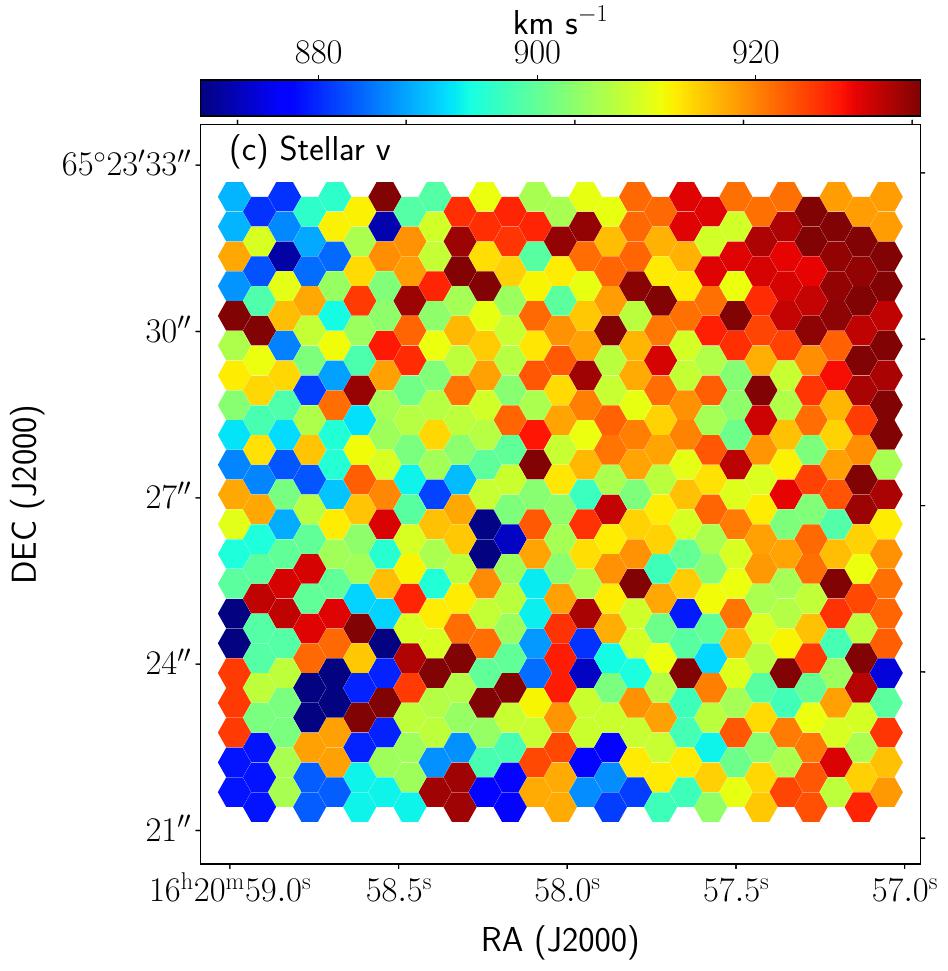}
	\includegraphics[clip, width=0.24\linewidth]{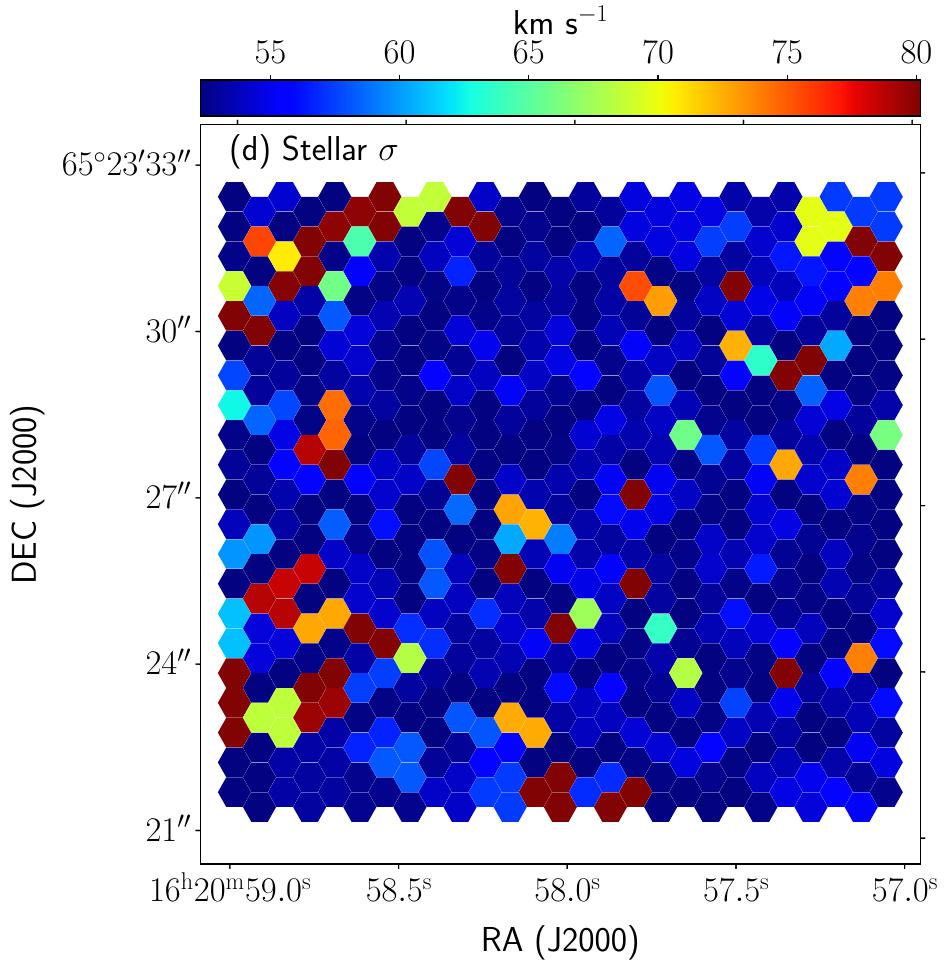}
	\includegraphics[clip, width=0.24\linewidth]{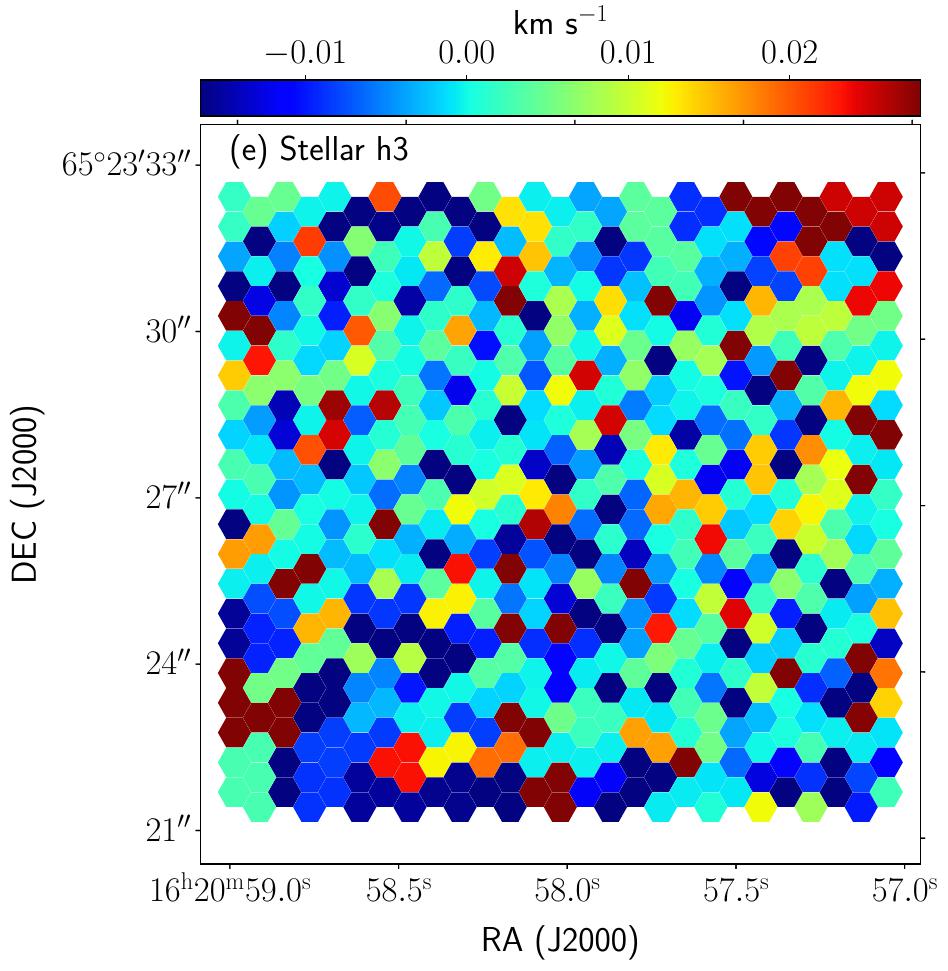}
	\includegraphics[clip, width=0.24\linewidth]{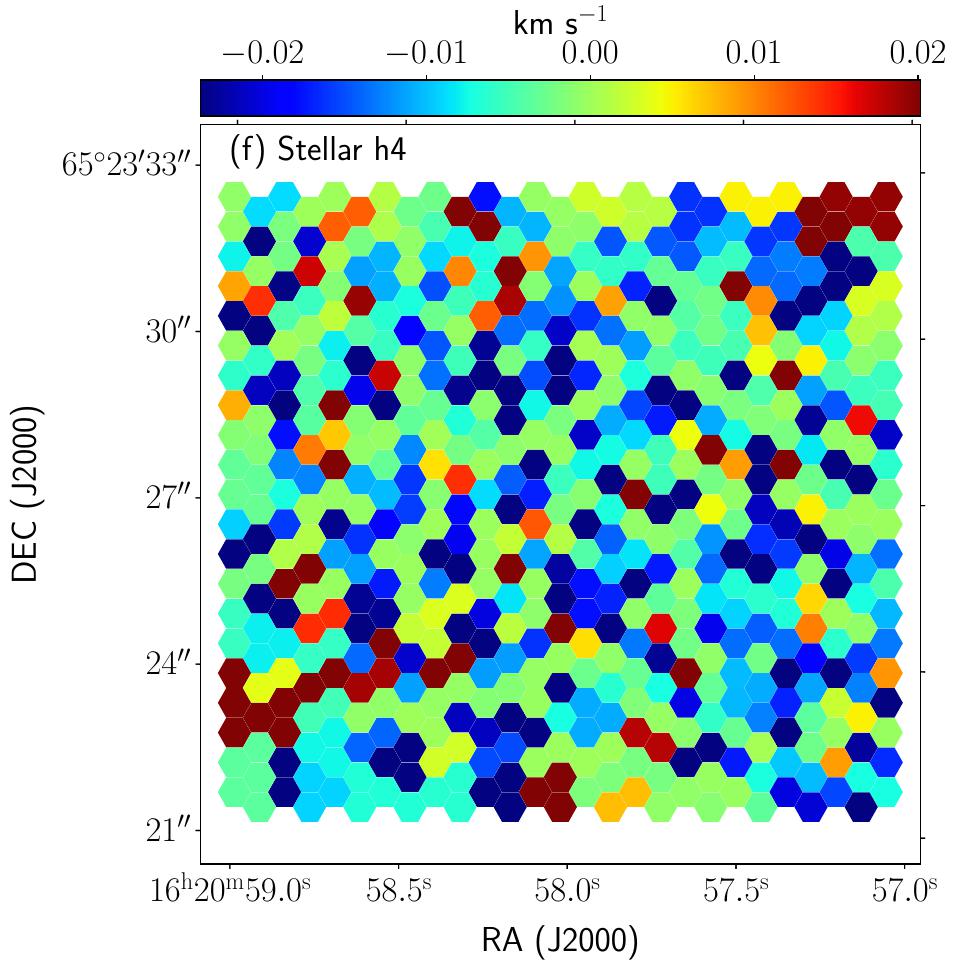}
	\includegraphics[clip, width=0.24\linewidth]{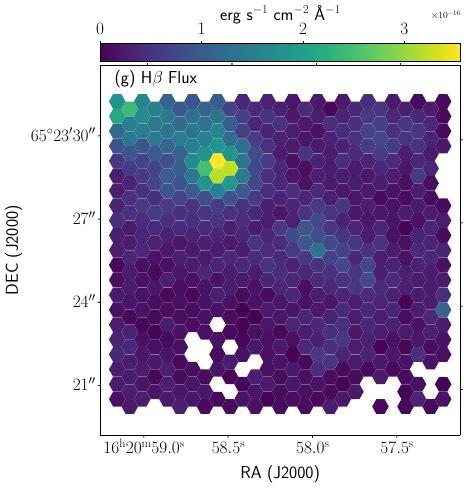}
	\includegraphics[clip, width=0.24\linewidth]{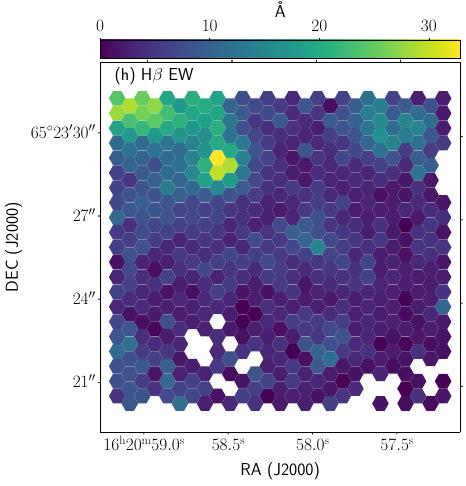}
	\includegraphics[clip, width=0.24\linewidth]{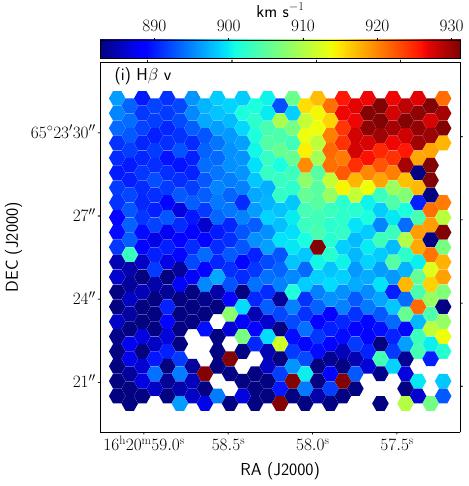}
	\includegraphics[clip, width=0.24\linewidth]{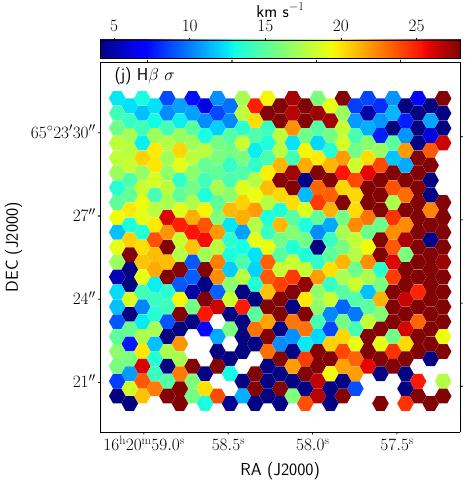}
	\includegraphics[clip, width=0.24\linewidth]{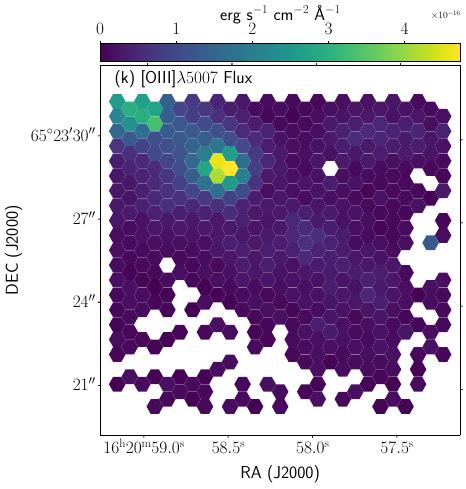}
	\includegraphics[clip, width=0.24\linewidth]{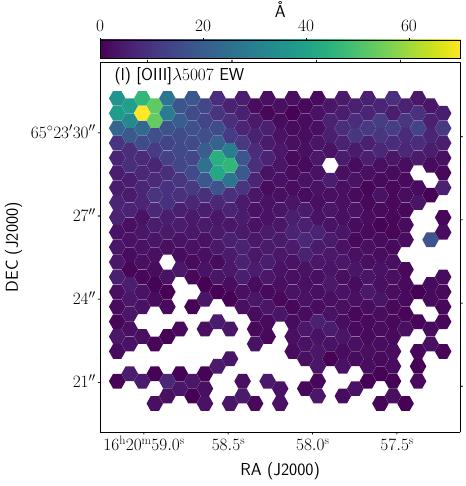}
	\includegraphics[clip, width=0.24\linewidth]{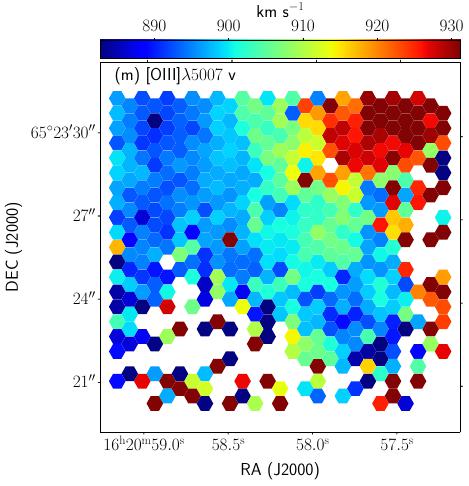}
	\includegraphics[clip, width=0.24\linewidth]{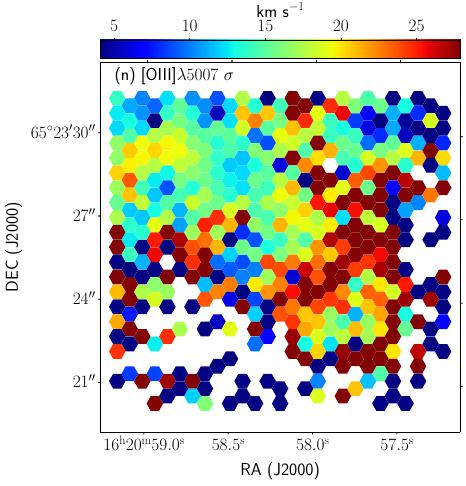}
	\vspace{5cm}
	\caption{NGC~6140 card.}
	\label{fig:NGC6140_card_1}
\end{figure*}
\addtocounter{figure}{-1}
\begin{figure*}[h]
	\centering
	\includegraphics[clip, width=0.24\linewidth]{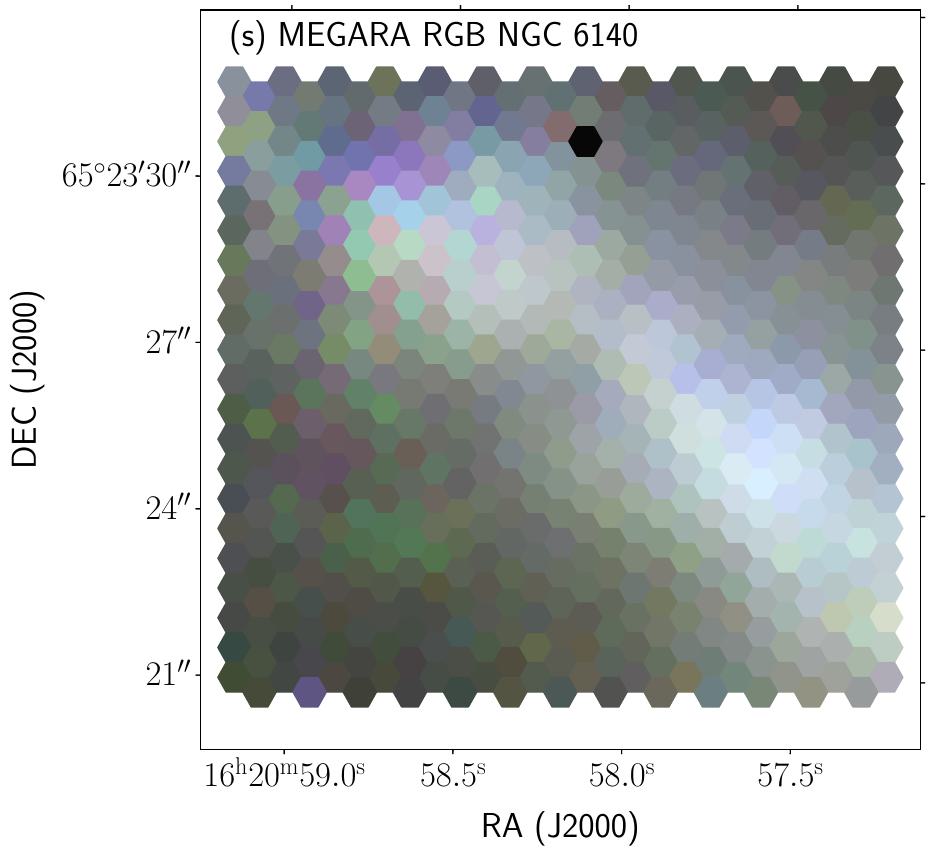}
	\includegraphics[clip, width=0.24\linewidth]{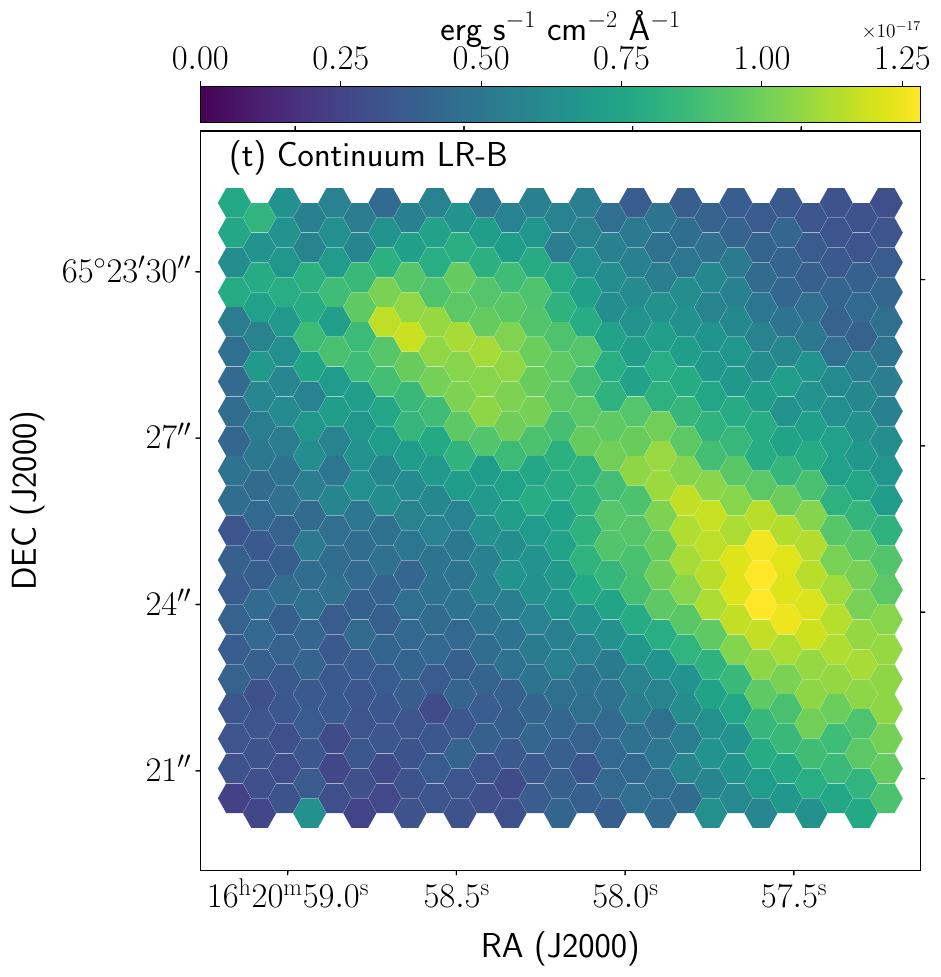}
	\includegraphics[clip, width=0.24\linewidth]{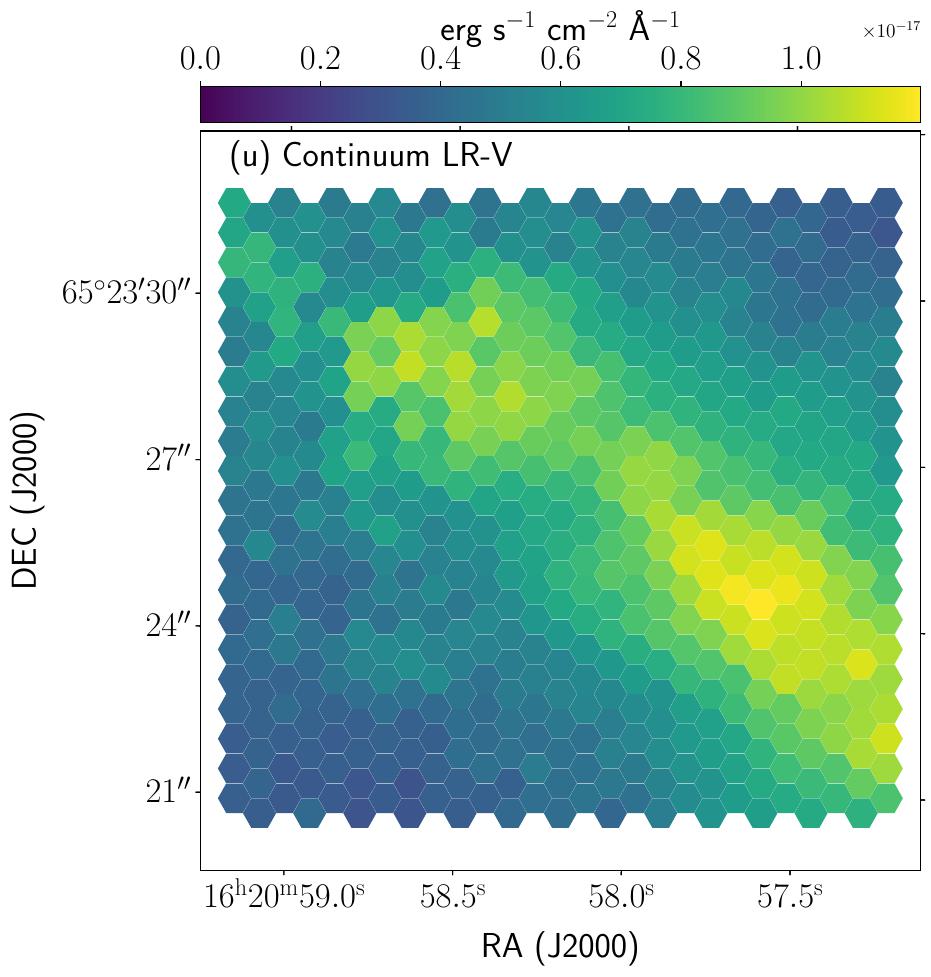}
	\includegraphics[clip, width=0.24\linewidth]{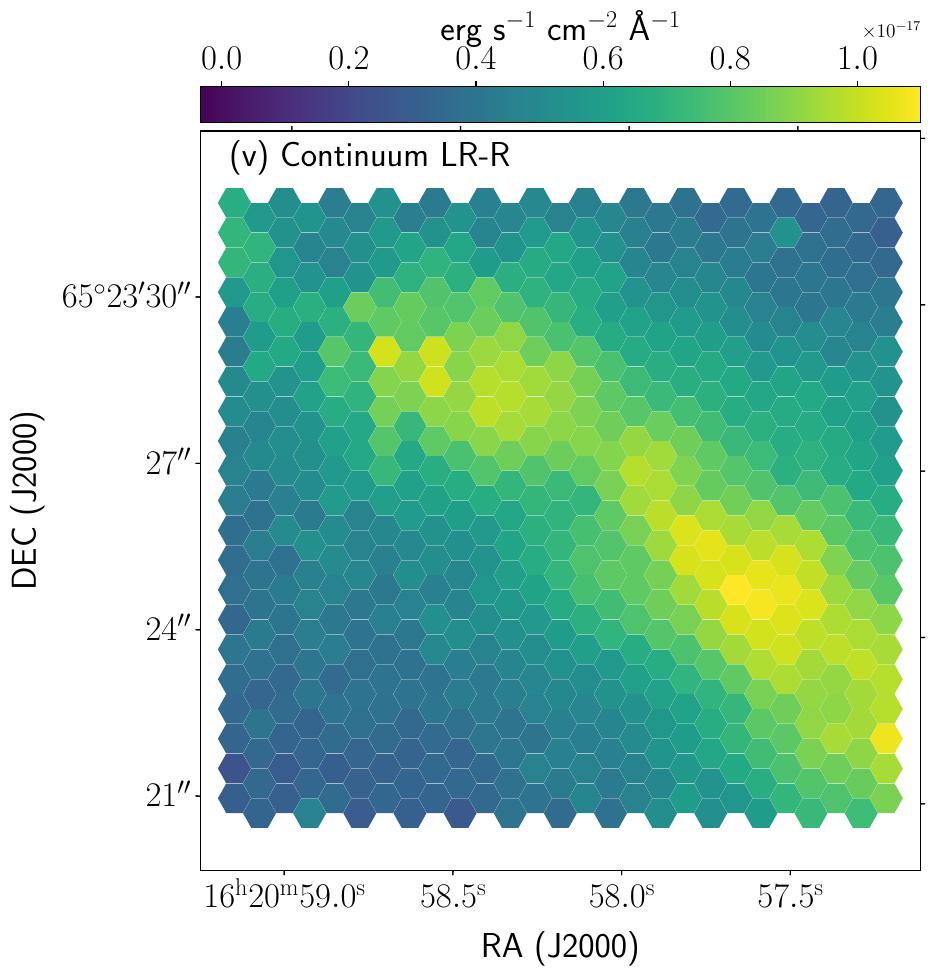}
	\includegraphics[clip, width=0.24\linewidth]{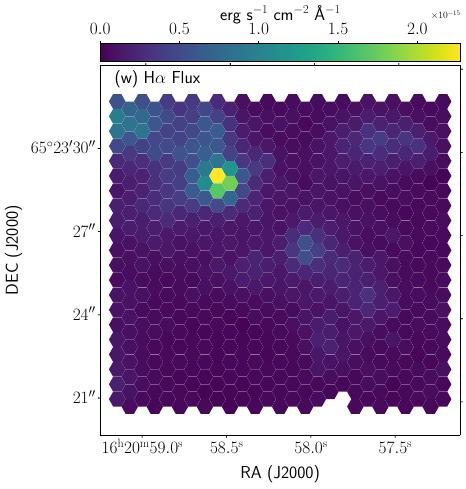}
	\includegraphics[clip, width=0.24\linewidth]{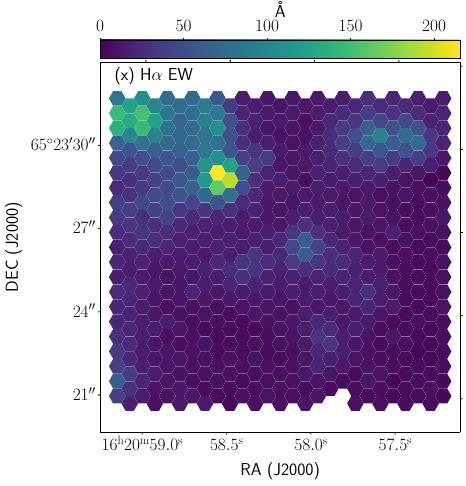}
	\includegraphics[clip, width=0.24\linewidth]{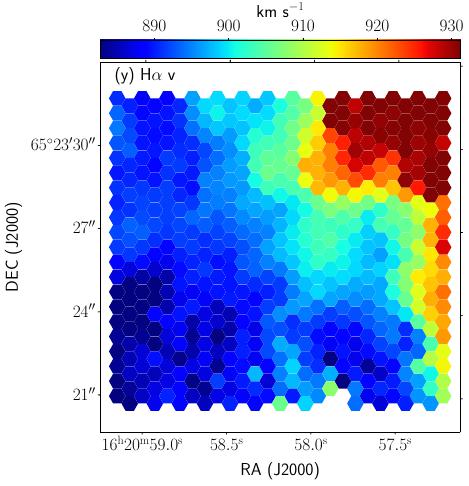}
	\includegraphics[clip, width=0.24\linewidth]{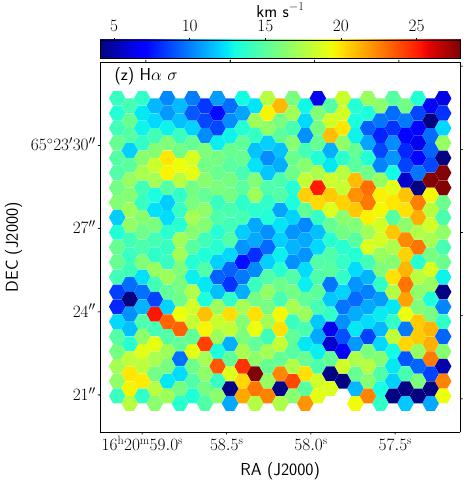}
	\includegraphics[clip, width=0.24\linewidth]{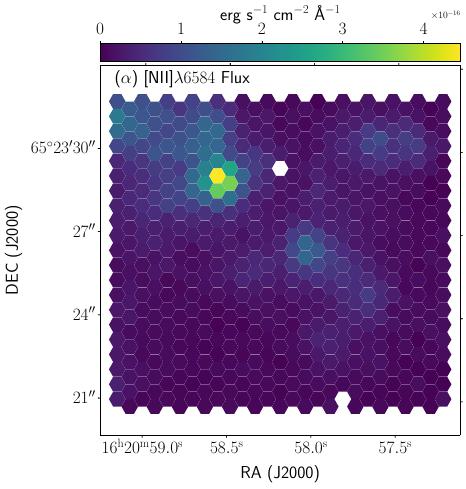}
	\includegraphics[clip, width=0.24\linewidth]{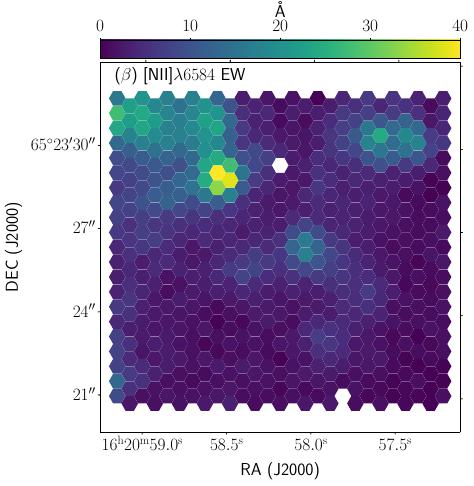}
	\includegraphics[clip, width=0.24\linewidth]{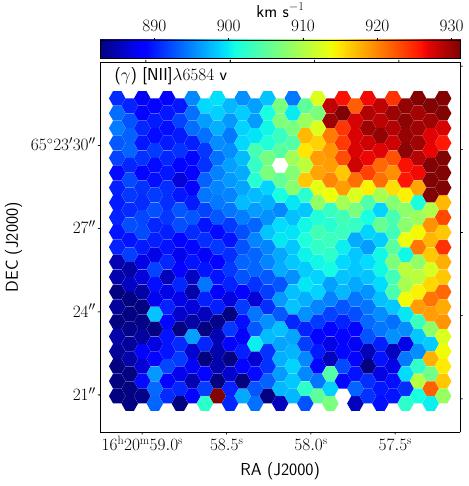}
	\includegraphics[clip, width=0.24\linewidth]{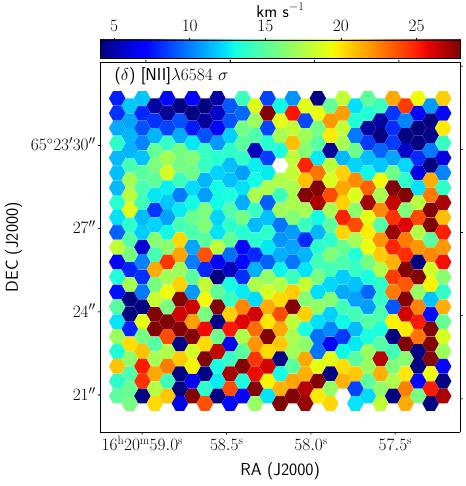}
	\includegraphics[clip, width=0.24\linewidth]{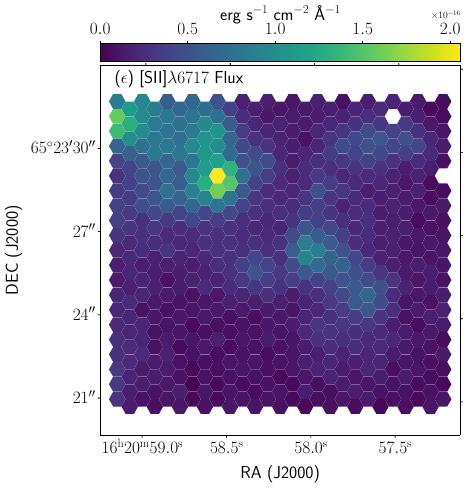}
	\includegraphics[clip, width=0.24\linewidth]{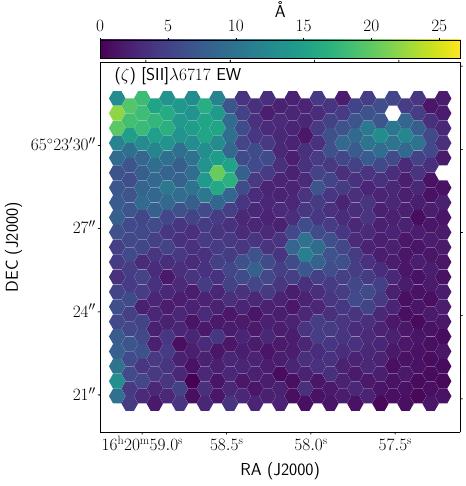}
	\includegraphics[clip, width=0.24\linewidth]{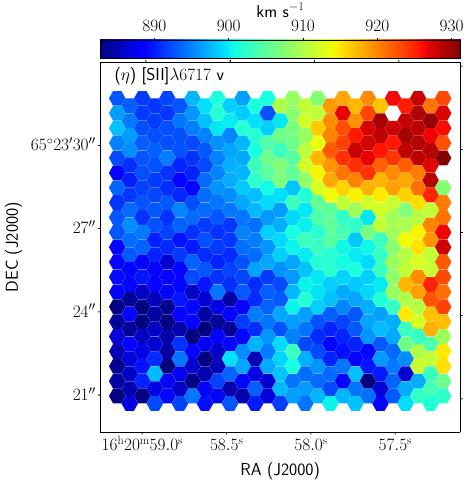}
	\includegraphics[clip, width=0.24\linewidth]{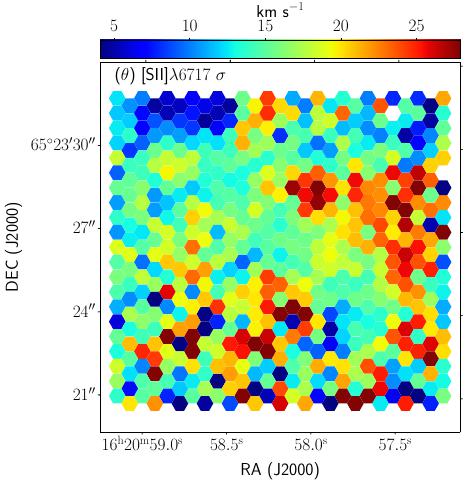}
	\includegraphics[clip, width=0.24\linewidth]{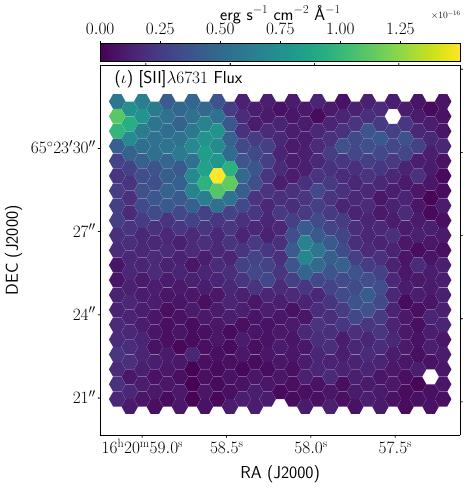}
	\includegraphics[clip, width=0.24\linewidth]{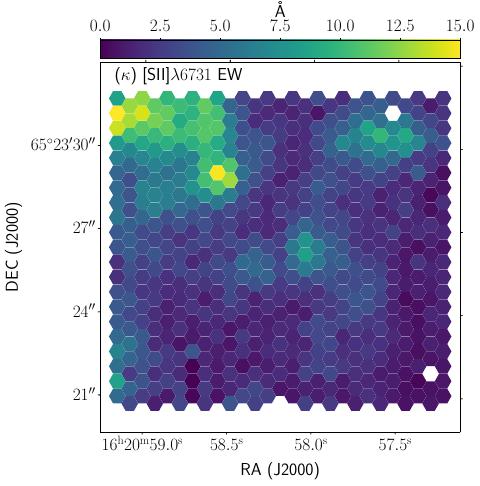}
	\includegraphics[clip, width=0.24\linewidth]{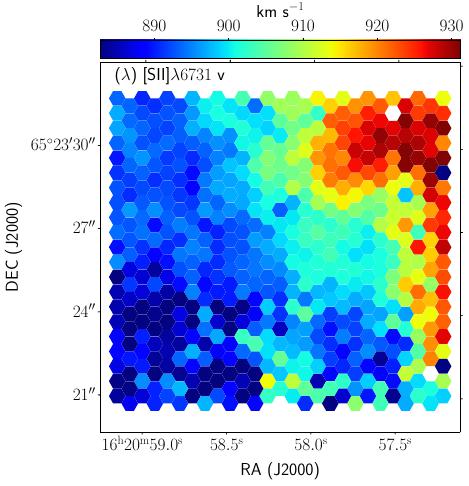}
	\includegraphics[clip, width=0.24\linewidth]{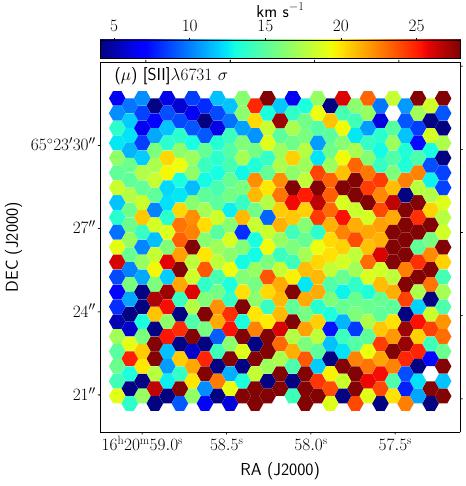}
	\caption{(cont.) NGC~6140 card.}
	\label{fig:NGC6140_card_2}
\end{figure*}

\begin{figure*}[h]
	\centering
	\includegraphics[clip, width=0.35\linewidth]{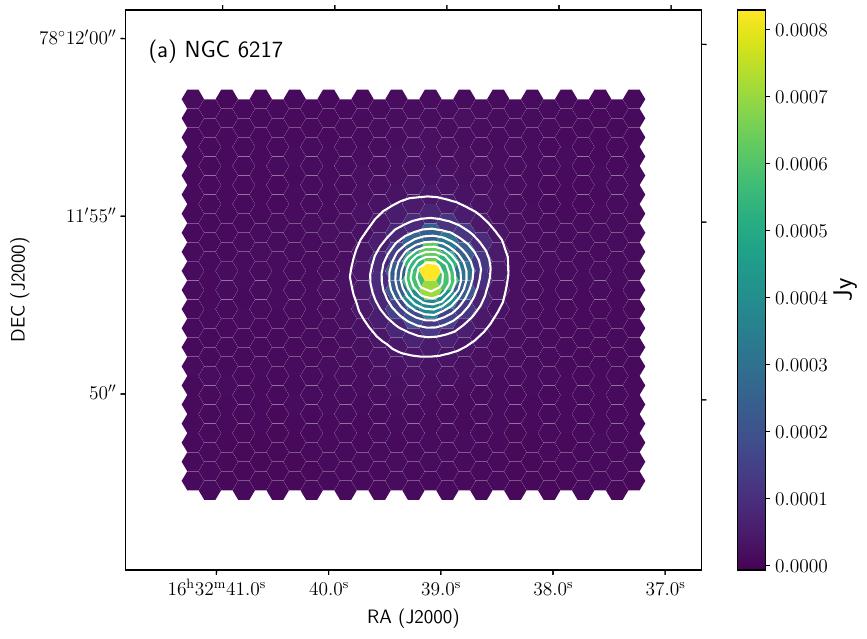}
	\includegraphics[clip, width=0.6\linewidth]{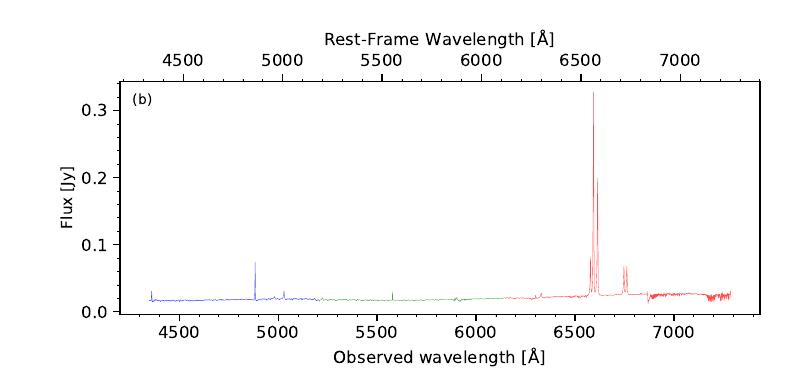}
	\includegraphics[clip, width=0.24\linewidth]{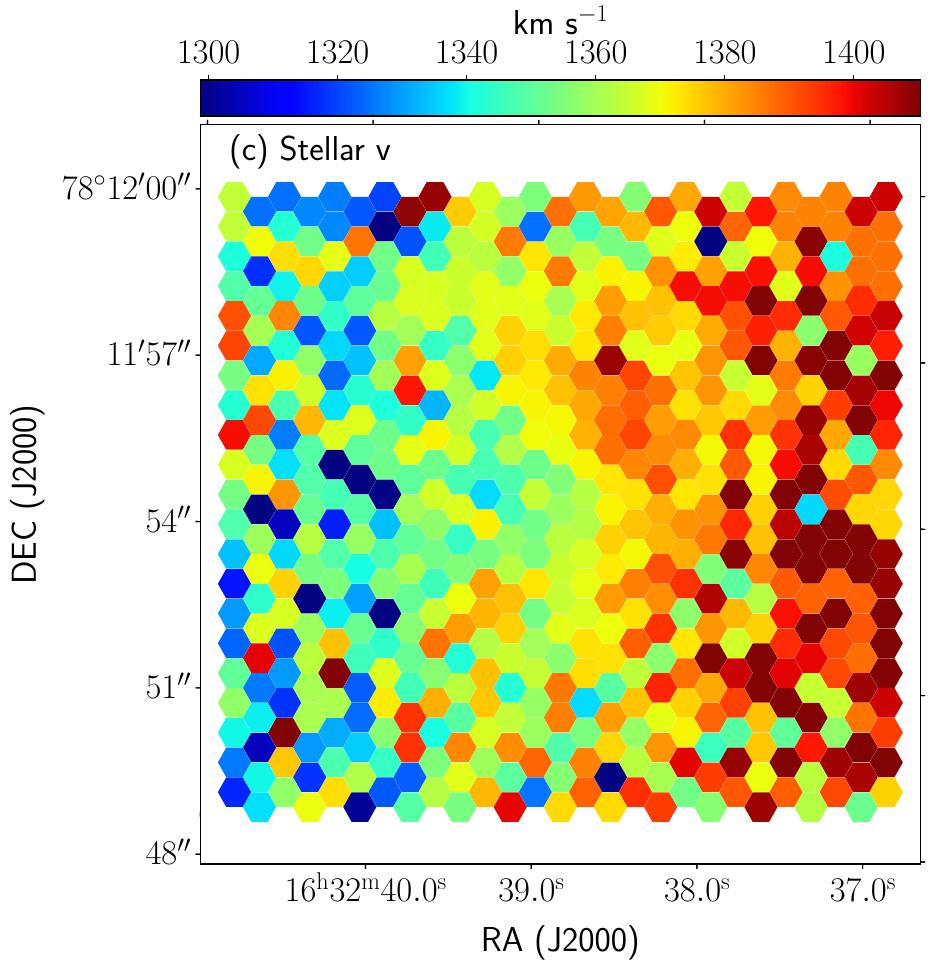}
	\includegraphics[clip, width=0.24\linewidth]{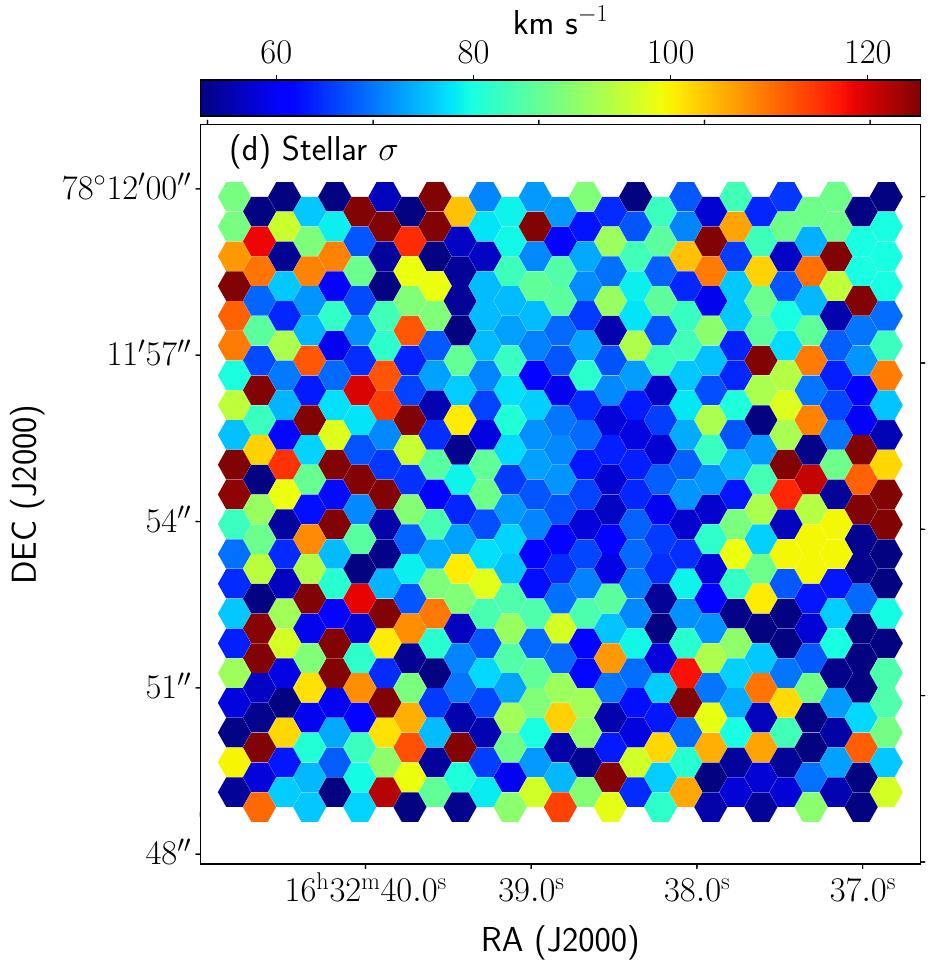}
	\includegraphics[clip, width=0.24\linewidth]{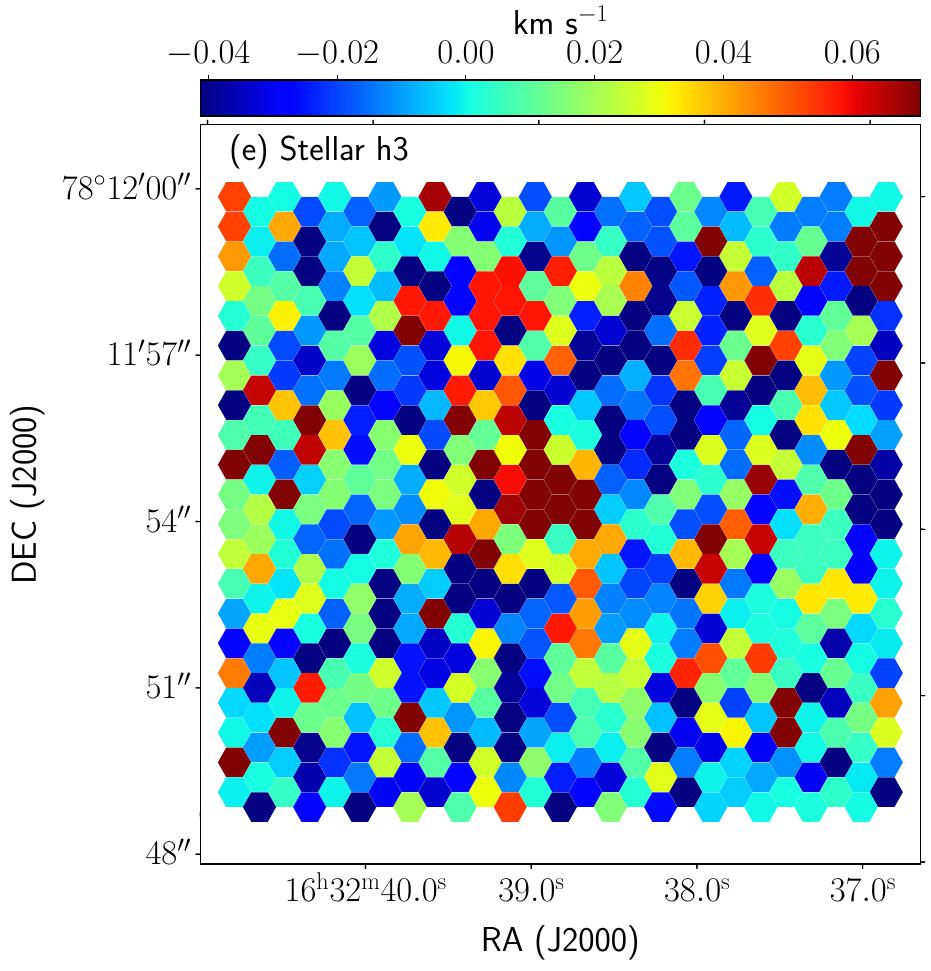}
	\includegraphics[clip, width=0.24\linewidth]{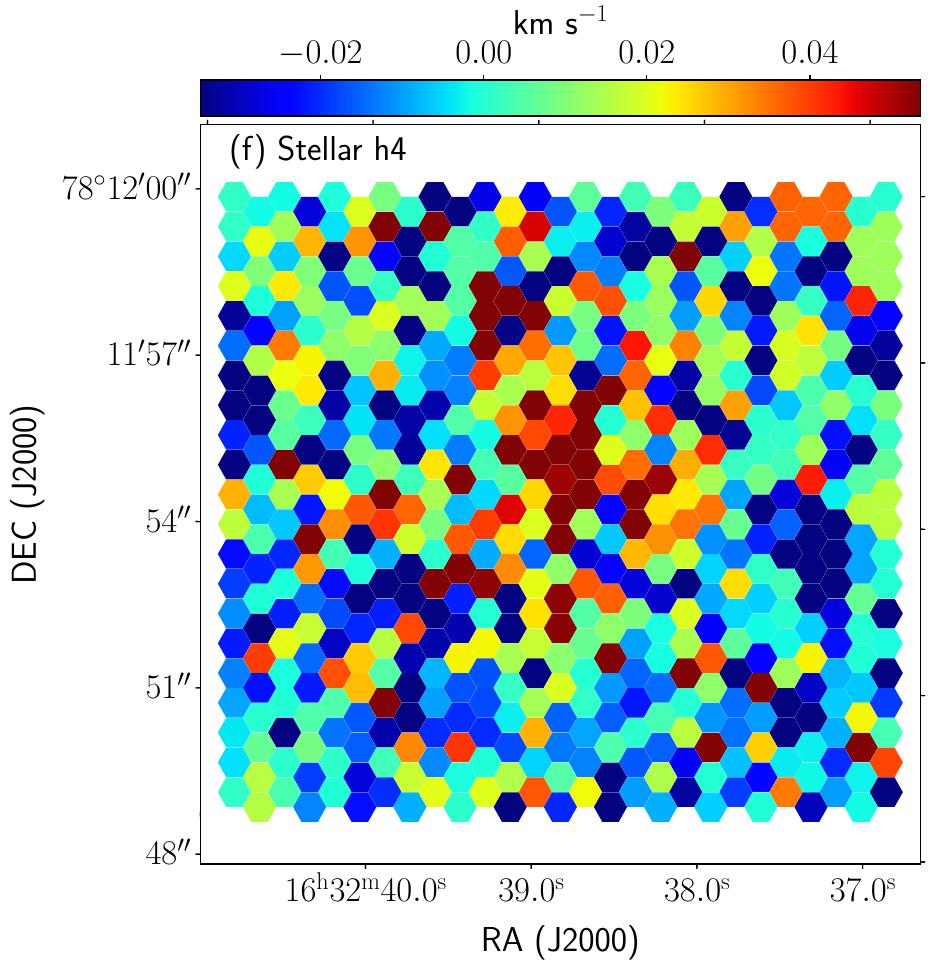}
	\includegraphics[clip, width=0.24\linewidth]{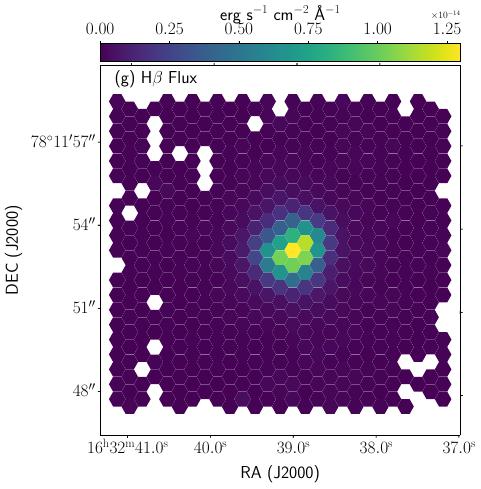}
	\includegraphics[clip, width=0.24\linewidth]{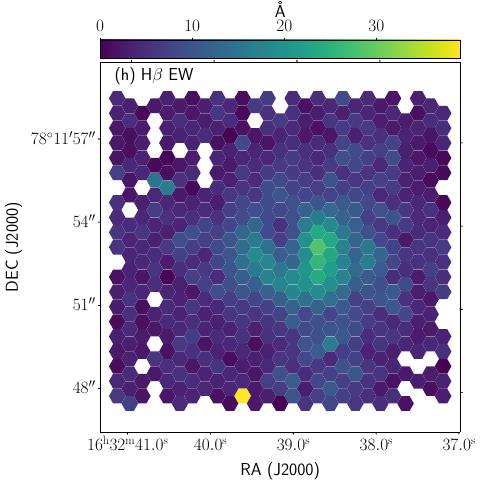}
	\includegraphics[clip, width=0.24\linewidth]{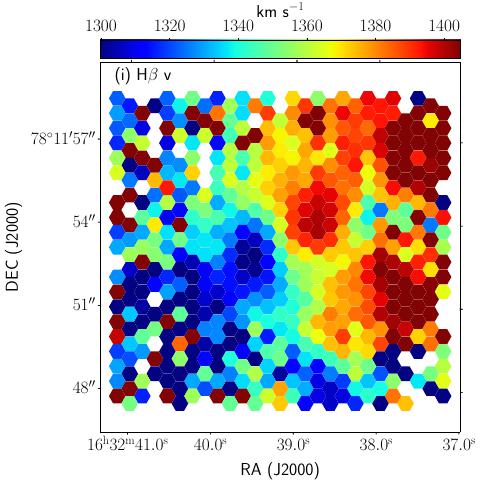}
	\includegraphics[clip, width=0.24\linewidth]{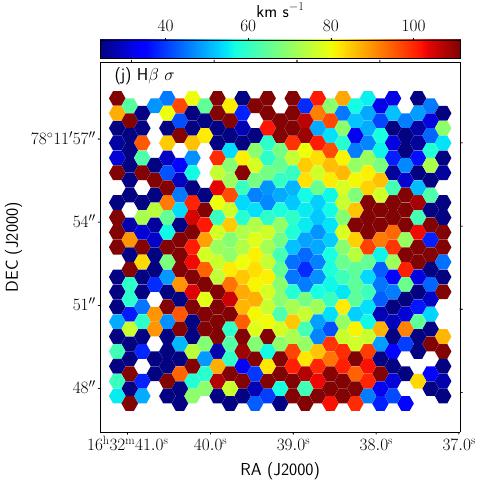}
	\includegraphics[clip, width=0.24\linewidth]{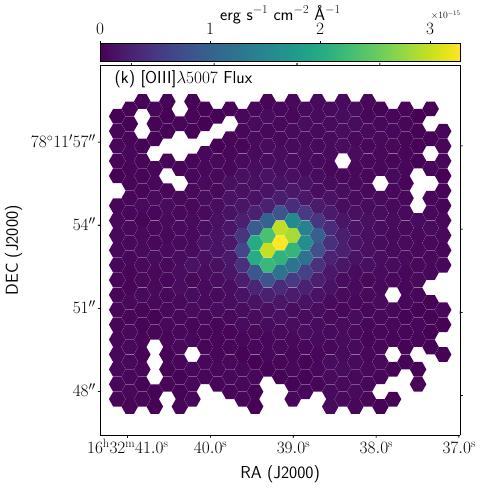}
	\includegraphics[clip, width=0.24\linewidth]{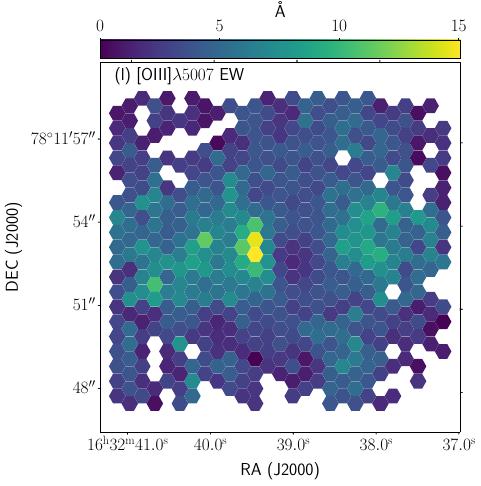}
	\includegraphics[clip, width=0.24\linewidth]{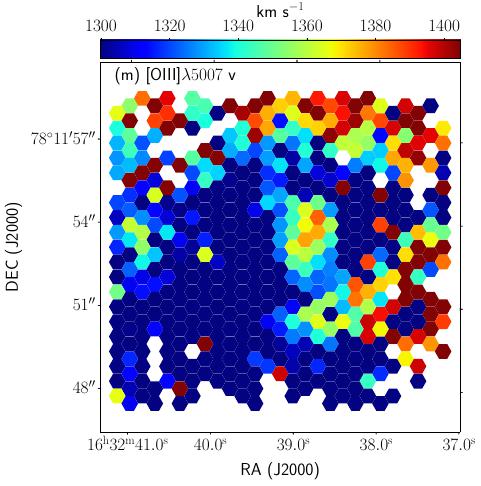}
	\includegraphics[clip, width=0.24\linewidth]{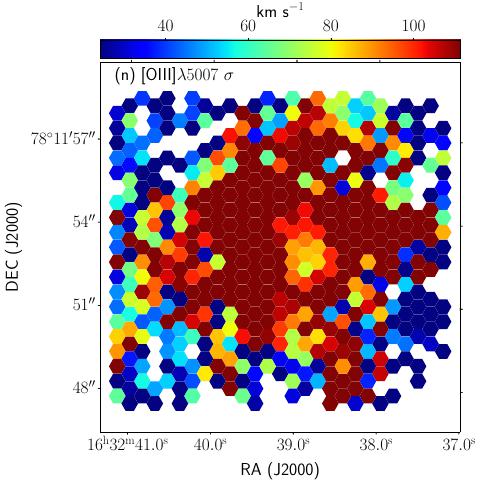}
	\includegraphics[clip, width=0.24\linewidth]{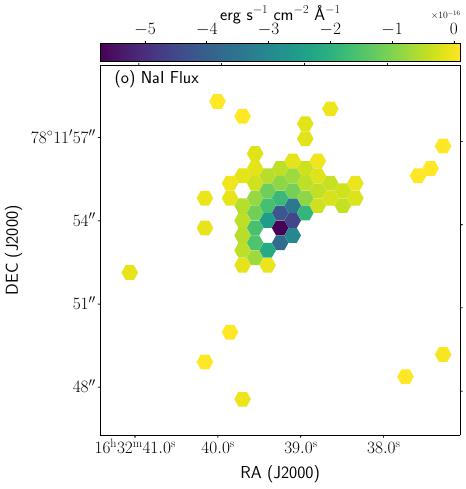}
	\includegraphics[clip, width=0.24\linewidth]{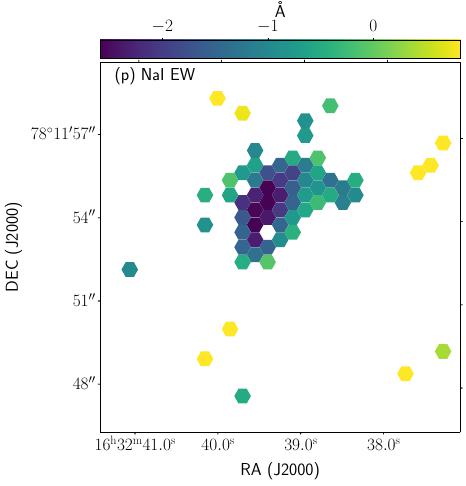}
	\includegraphics[clip, width=0.24\linewidth]{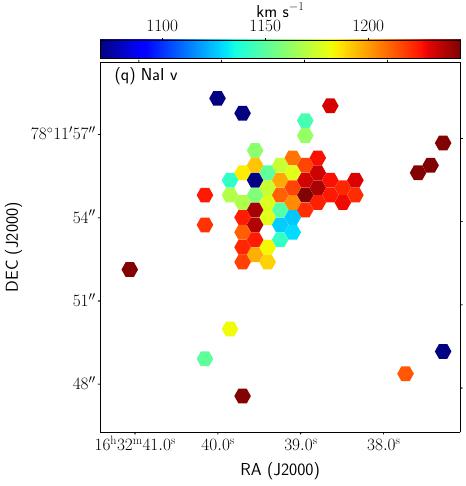}
	\includegraphics[clip, width=0.24\linewidth]{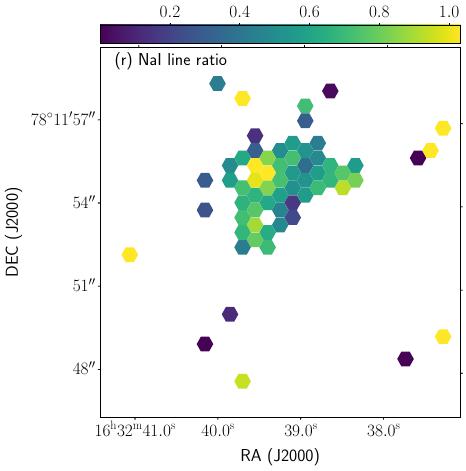}
	\caption{NGC~6217 card.}
	\label{fig:NGC6217_card_1}
\end{figure*}
\addtocounter{figure}{-1}
\begin{figure*}[h]
	\centering
	\includegraphics[clip, width=0.24\linewidth]{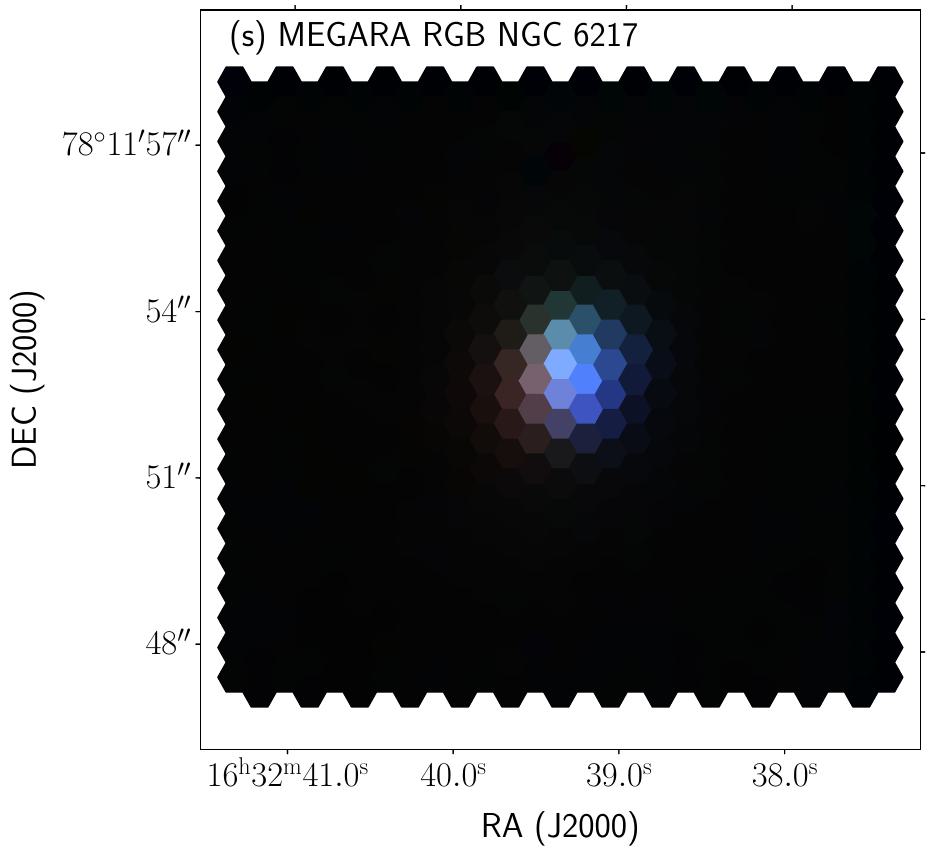}
	\includegraphics[clip, width=0.24\linewidth]{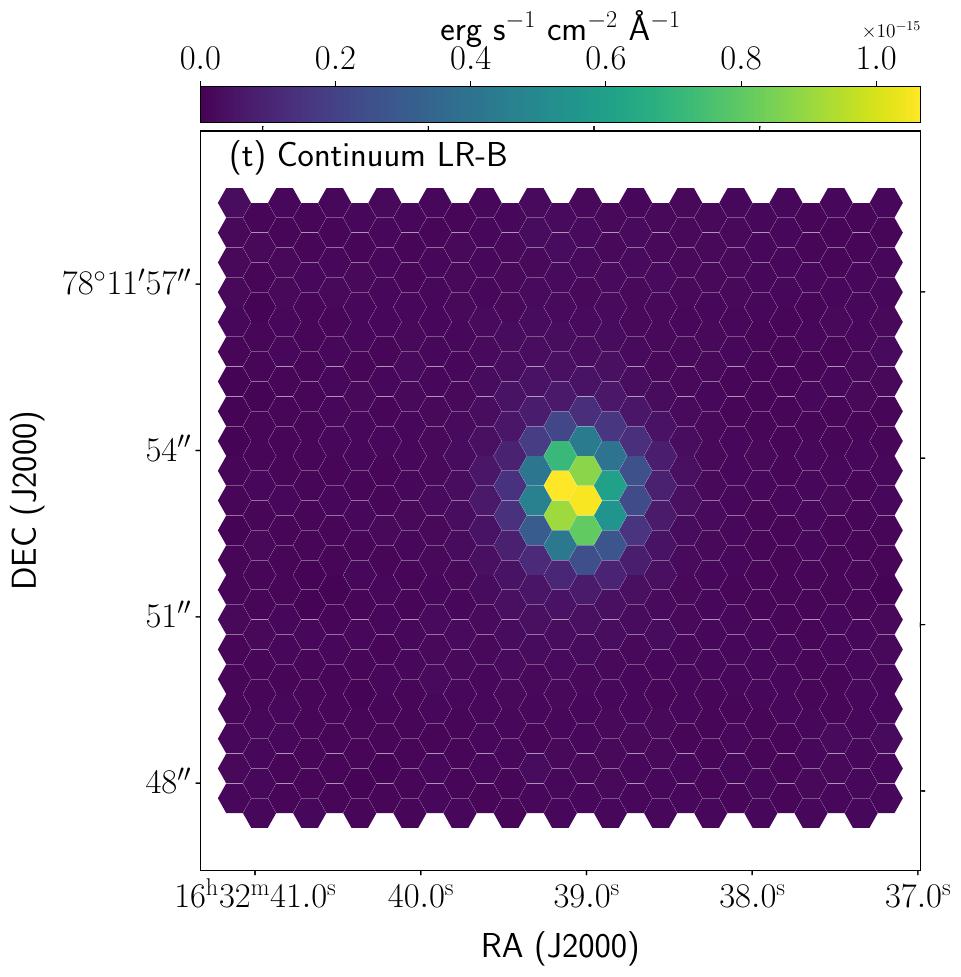}
	\includegraphics[clip, width=0.24\linewidth]{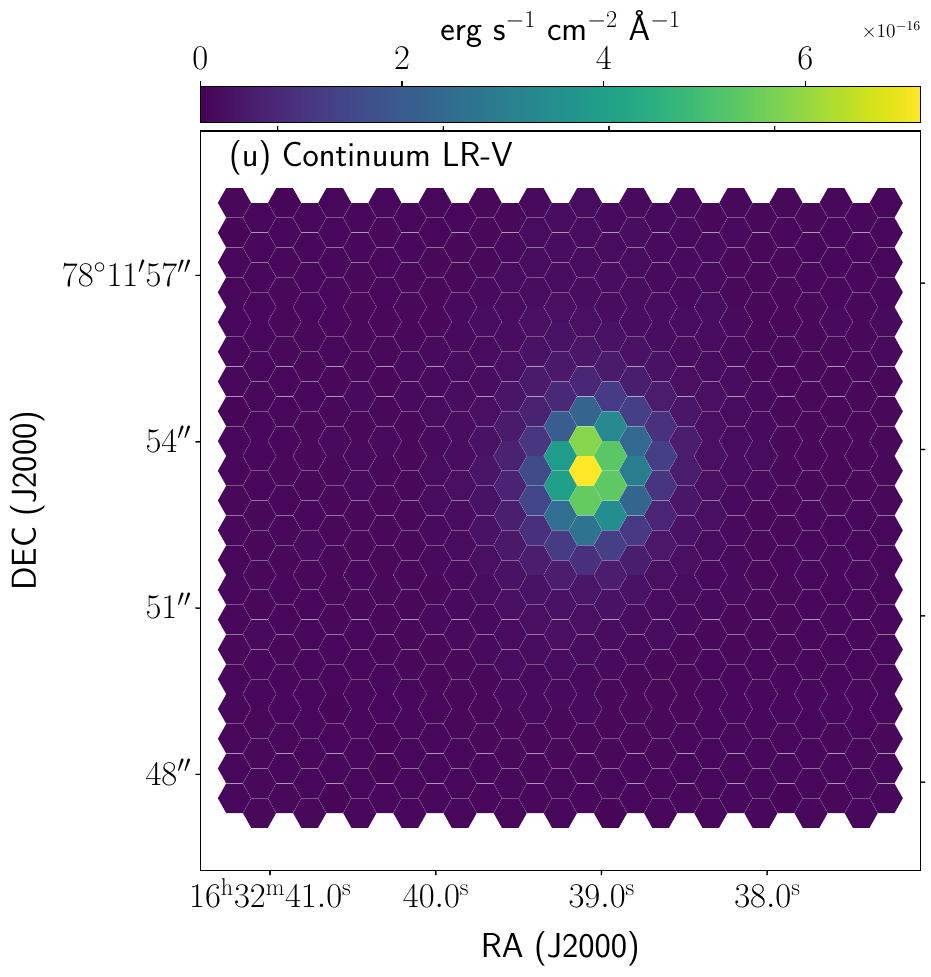}
	\includegraphics[clip, width=0.24\linewidth]{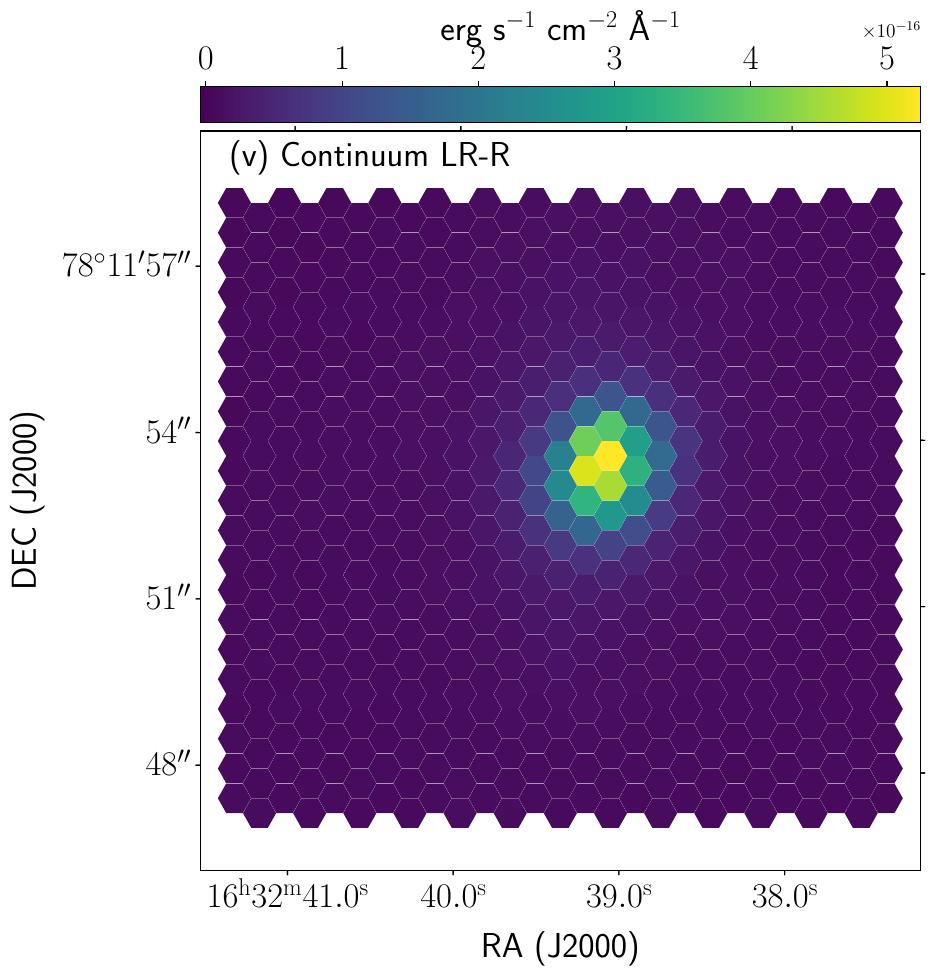}
	\includegraphics[clip, width=0.24\linewidth]{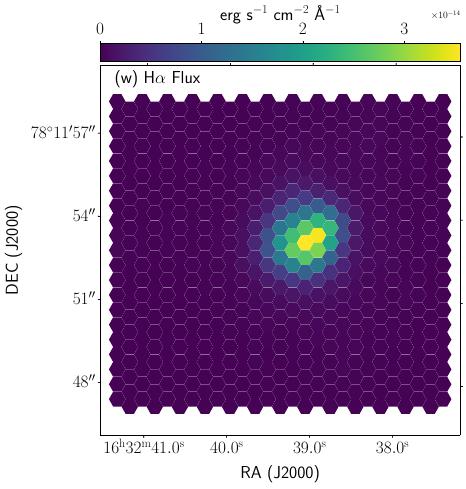}
	\includegraphics[clip, width=0.24\linewidth]{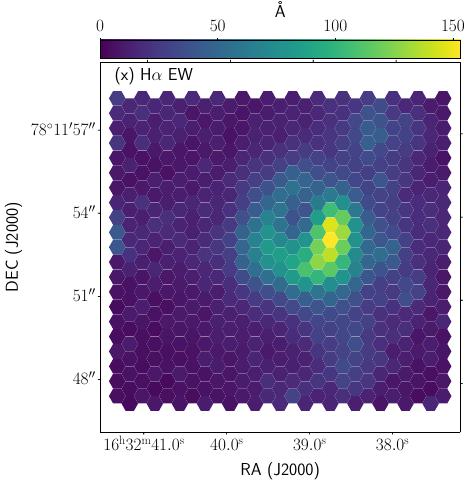}
	\includegraphics[clip, width=0.24\linewidth]{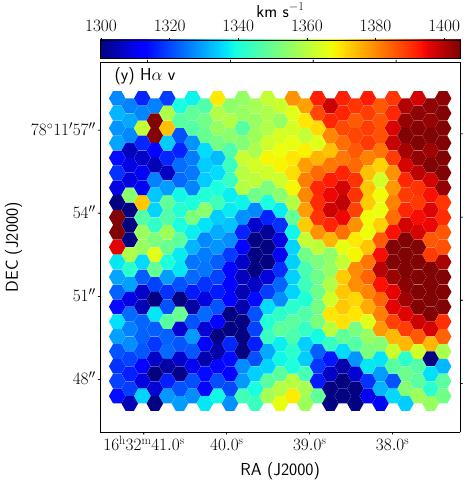}
	\includegraphics[clip, width=0.24\linewidth]{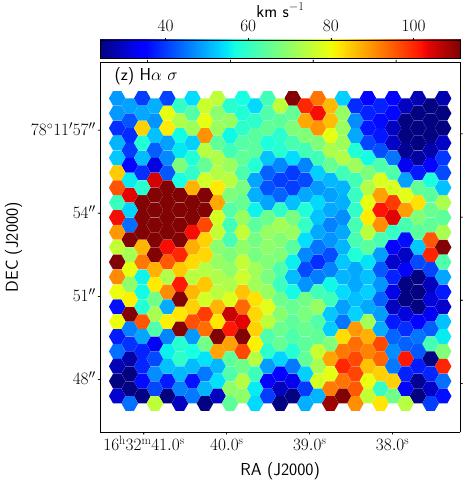}
	\includegraphics[clip, width=0.24\linewidth]{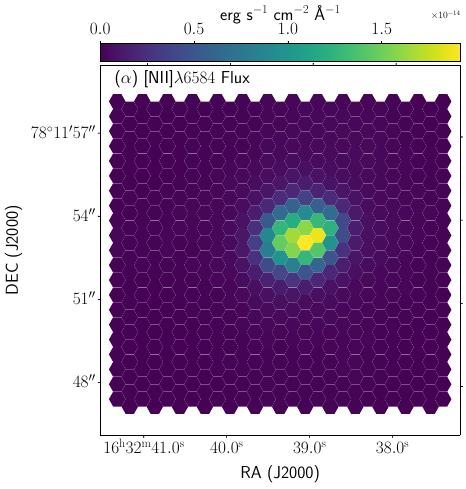}
	\includegraphics[clip, width=0.24\linewidth]{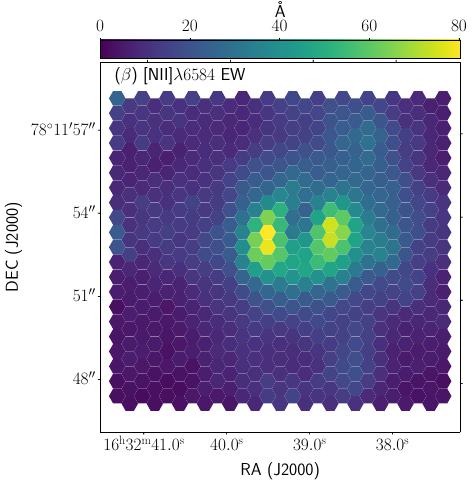}
	\includegraphics[clip, width=0.24\linewidth]{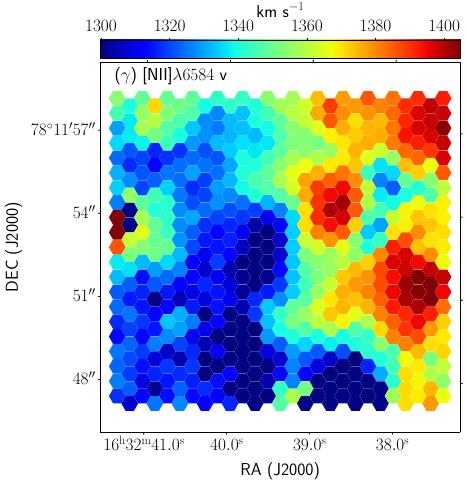}
	\includegraphics[clip, width=0.24\linewidth]{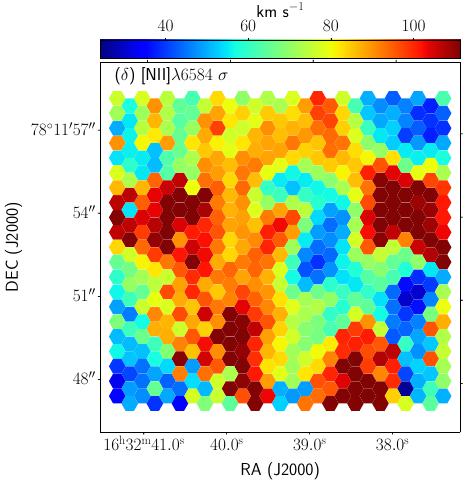}
	\includegraphics[clip, width=0.24\linewidth]{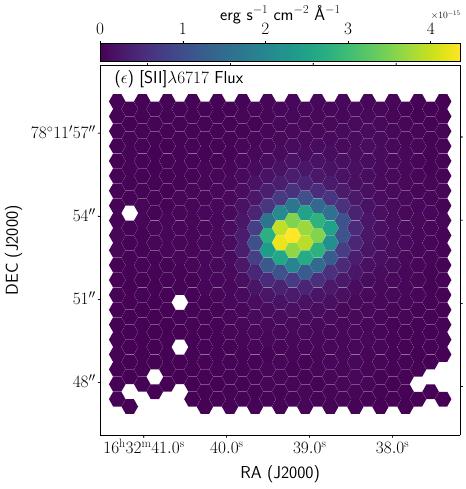}
	\includegraphics[clip, width=0.24\linewidth]{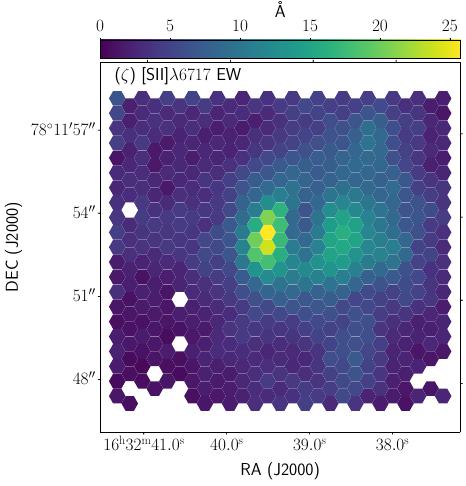}
	\includegraphics[clip, width=0.24\linewidth]{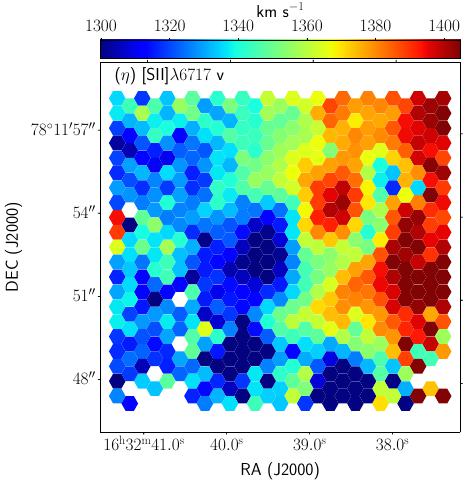}
	\includegraphics[clip, width=0.24\linewidth]{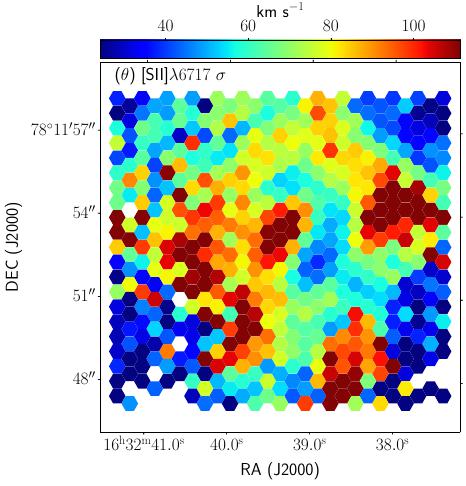}
	\includegraphics[clip, width=0.24\linewidth]{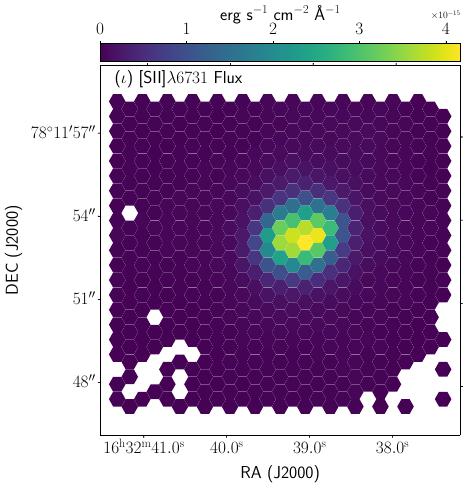}
	\includegraphics[clip, width=0.24\linewidth]{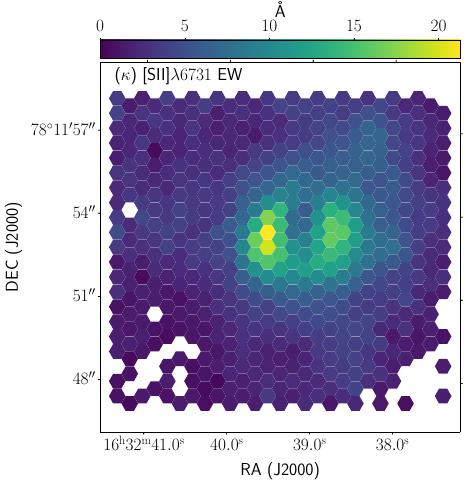}
	\includegraphics[clip, width=0.24\linewidth]{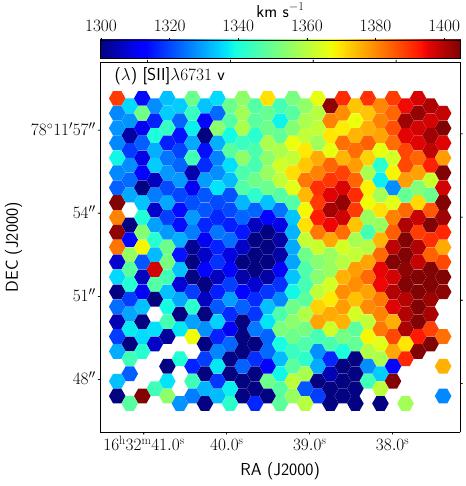}
	\includegraphics[clip, width=0.24\linewidth]{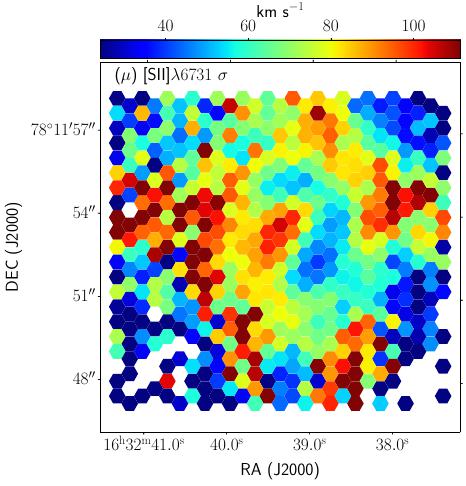}
	\caption{(cont.) NGC~6217 card.}
	\label{fig:NGC6217_card_2}
\end{figure*}

\begin{figure*}[h]
	\centering
	\includegraphics[clip, width=0.35\linewidth]{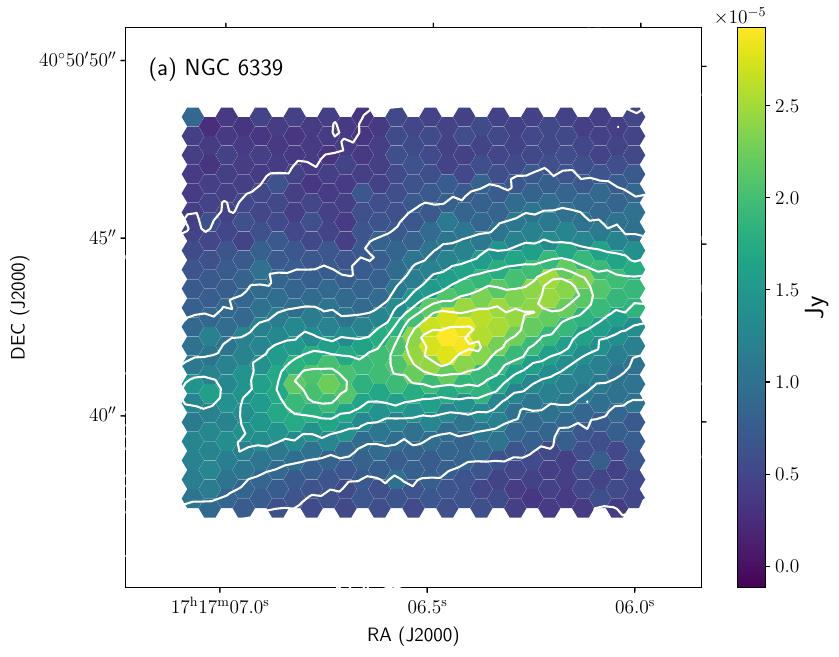}
	\includegraphics[clip, width=0.6\linewidth]{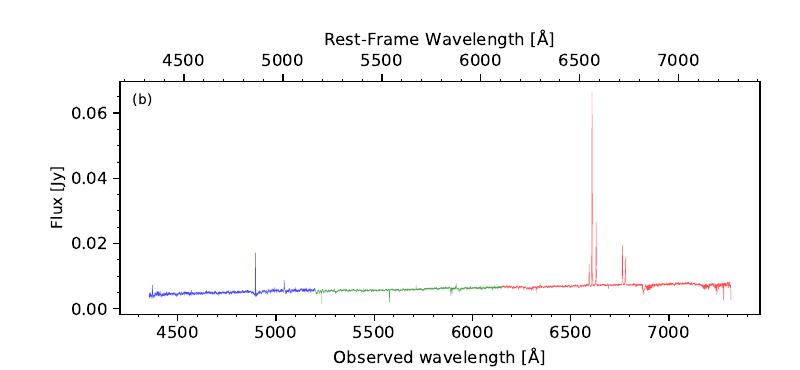}
	\includegraphics[clip, width=0.24\linewidth]{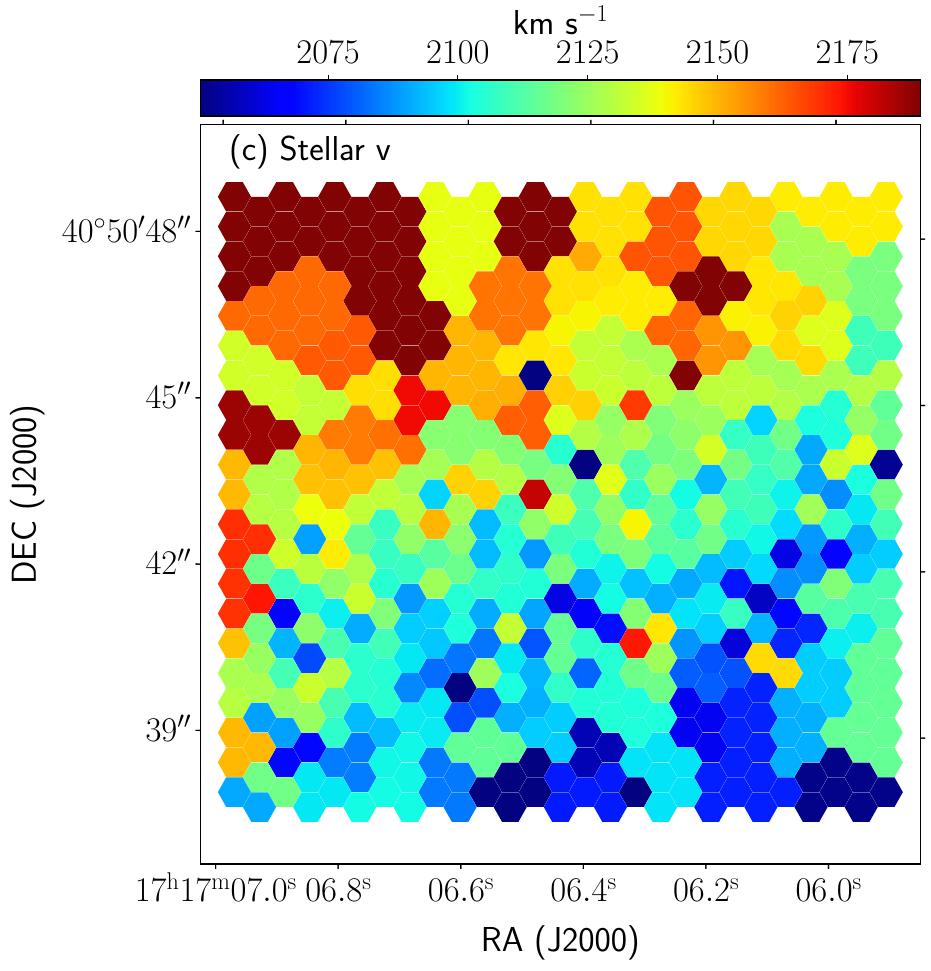}
	\includegraphics[clip, width=0.24\linewidth]{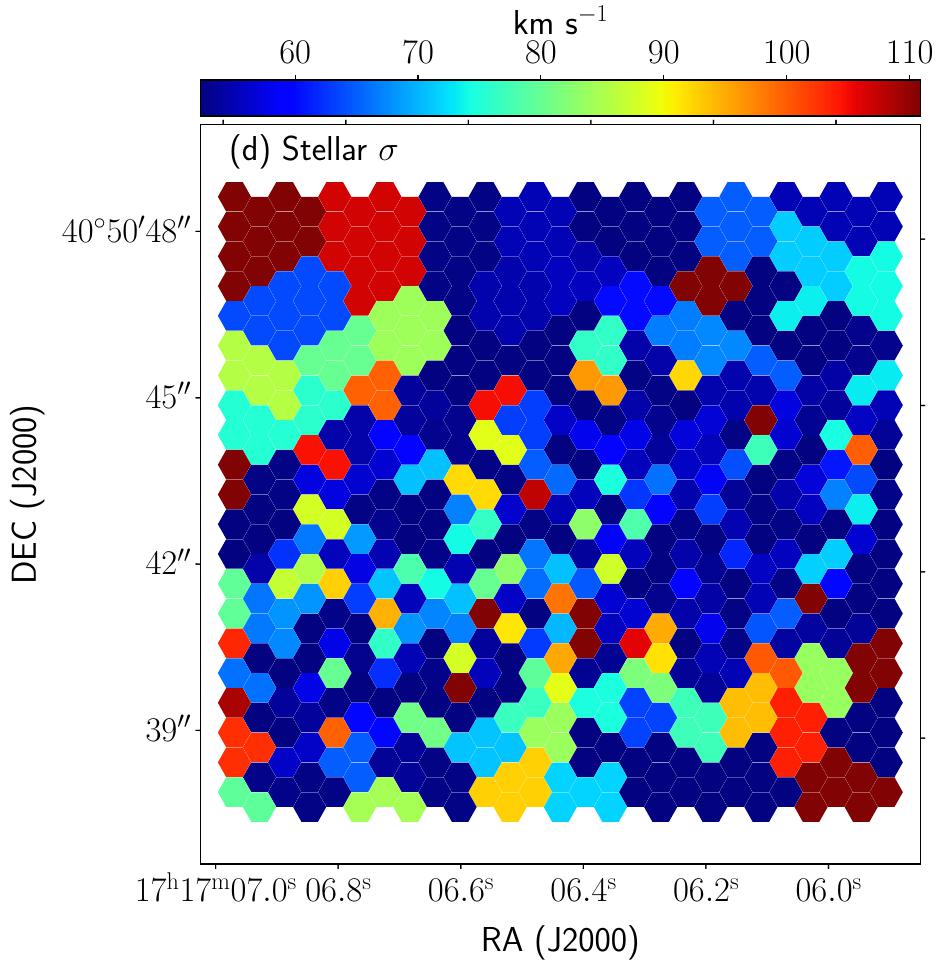}
	\includegraphics[clip, width=0.24\linewidth]{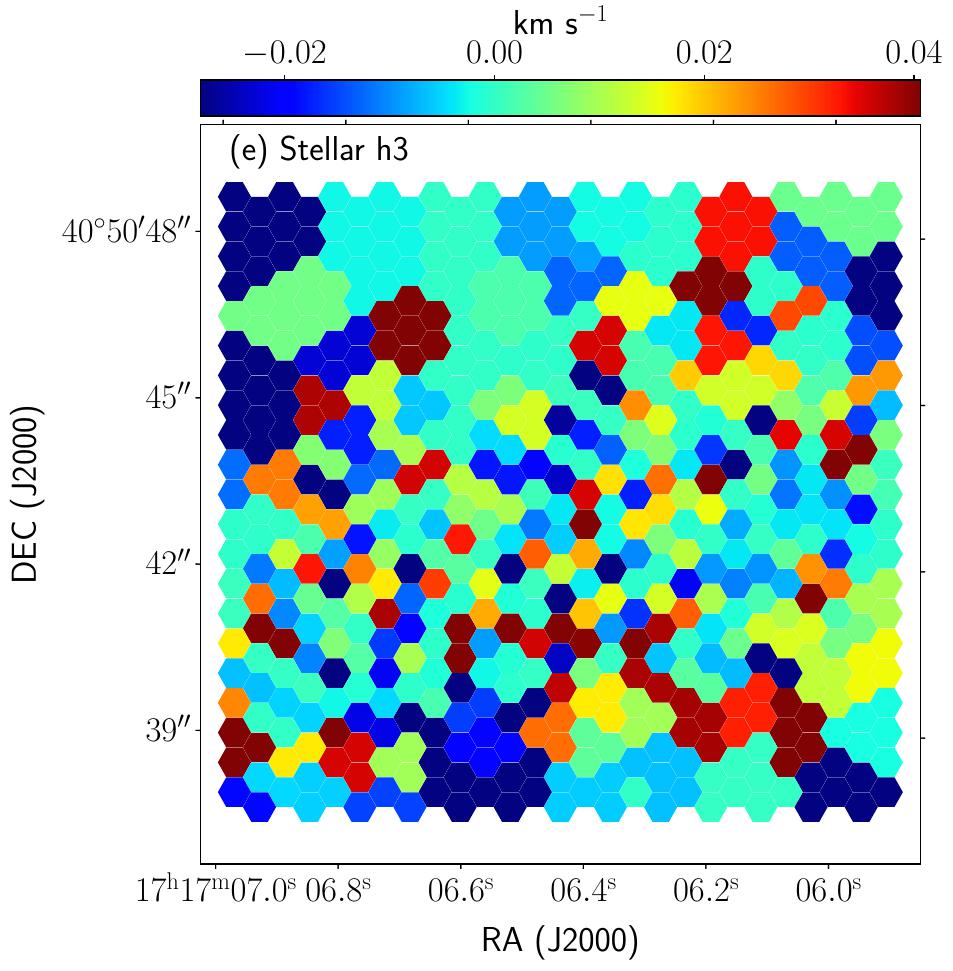}
	\includegraphics[clip, width=0.24\linewidth]{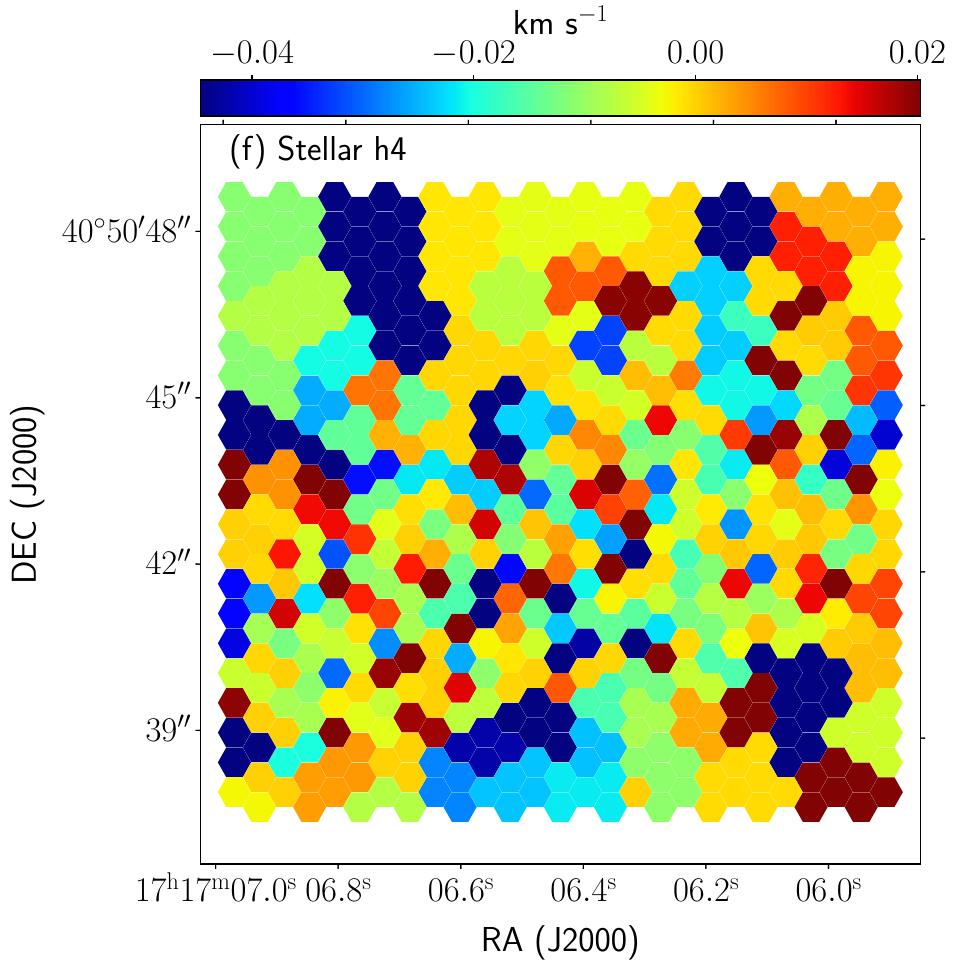}
	\includegraphics[clip, width=0.24\linewidth]{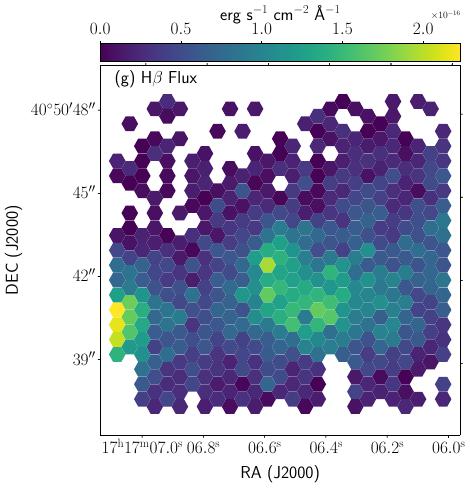}
	\includegraphics[clip, width=0.24\linewidth]{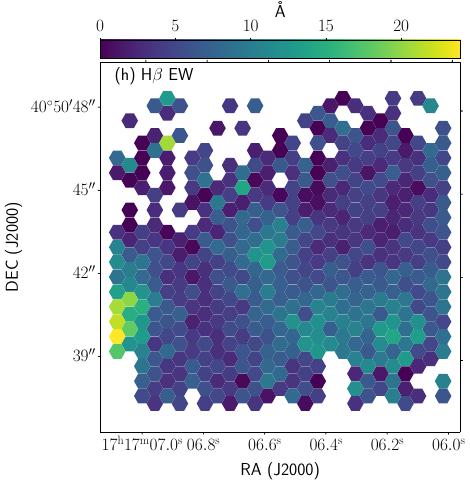}
	\includegraphics[clip, width=0.24\linewidth]{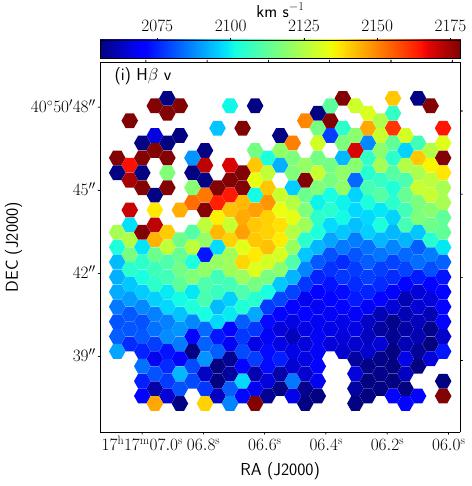}
	\includegraphics[clip, width=0.24\linewidth]{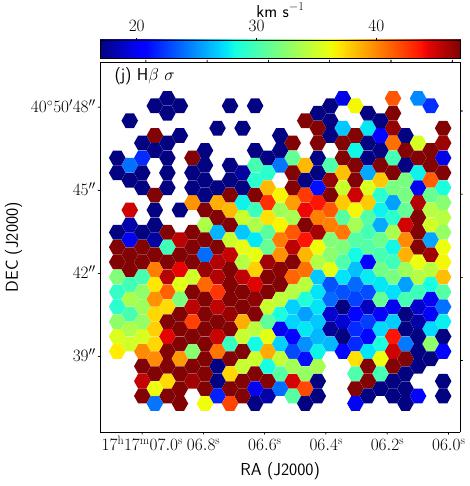}
	\includegraphics[clip, width=0.24\linewidth]{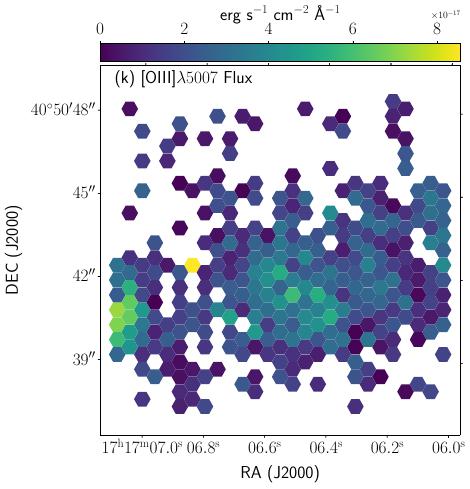}
	\includegraphics[clip, width=0.24\linewidth]{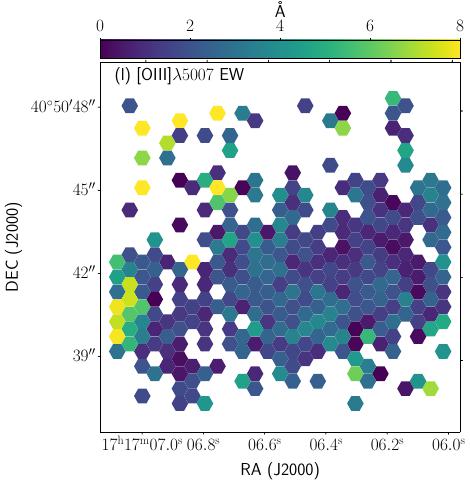}
	\includegraphics[clip, width=0.24\linewidth]{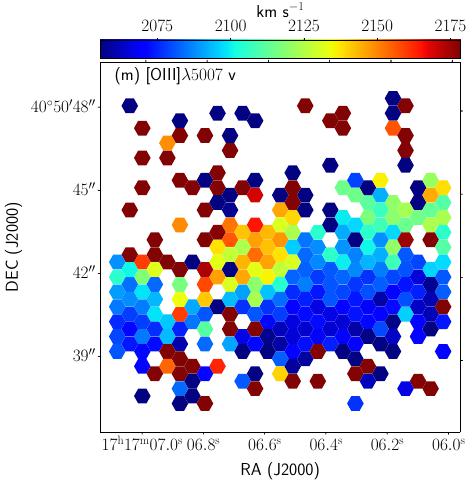}
	\includegraphics[clip, width=0.24\linewidth]{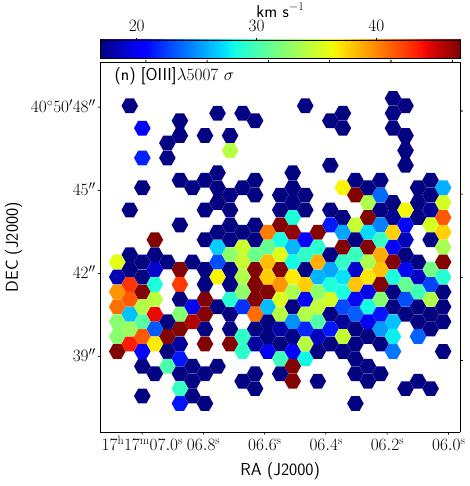}
	\vspace{5cm}
	\caption{NGC~6339 card.}
	\label{fig:NGC6339_card_1}
\end{figure*}
\addtocounter{figure}{-1}
\begin{figure*}[h]
	\centering
	\includegraphics[clip, width=0.24\linewidth]{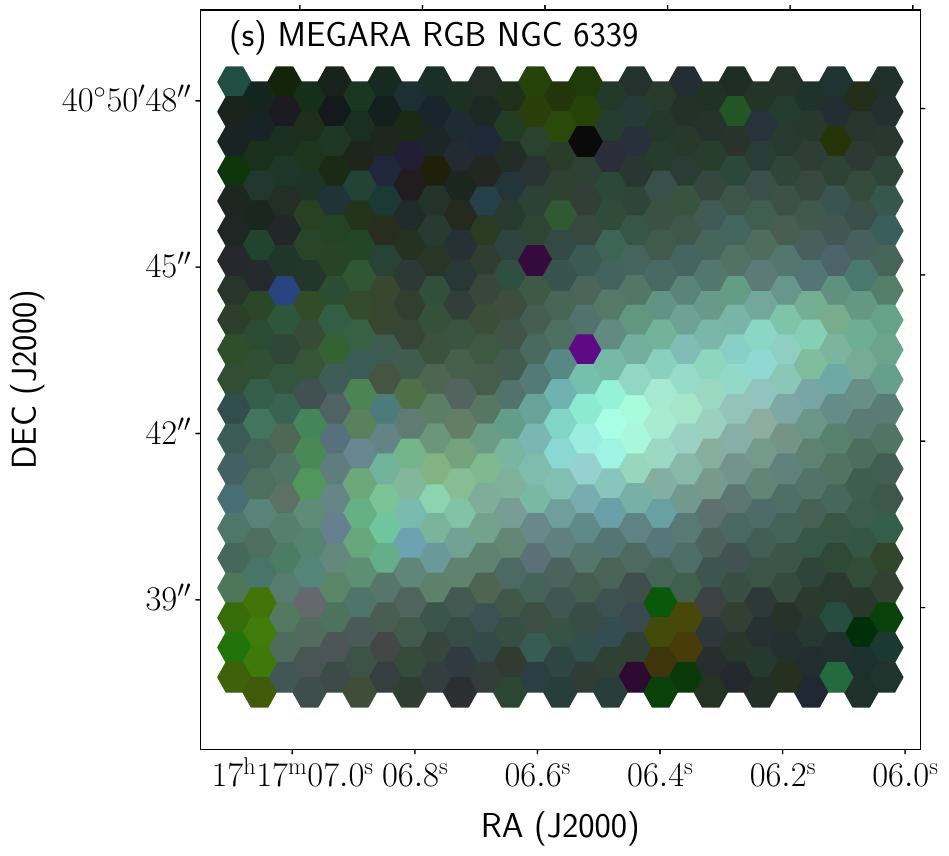}
	\includegraphics[clip, width=0.24\linewidth]{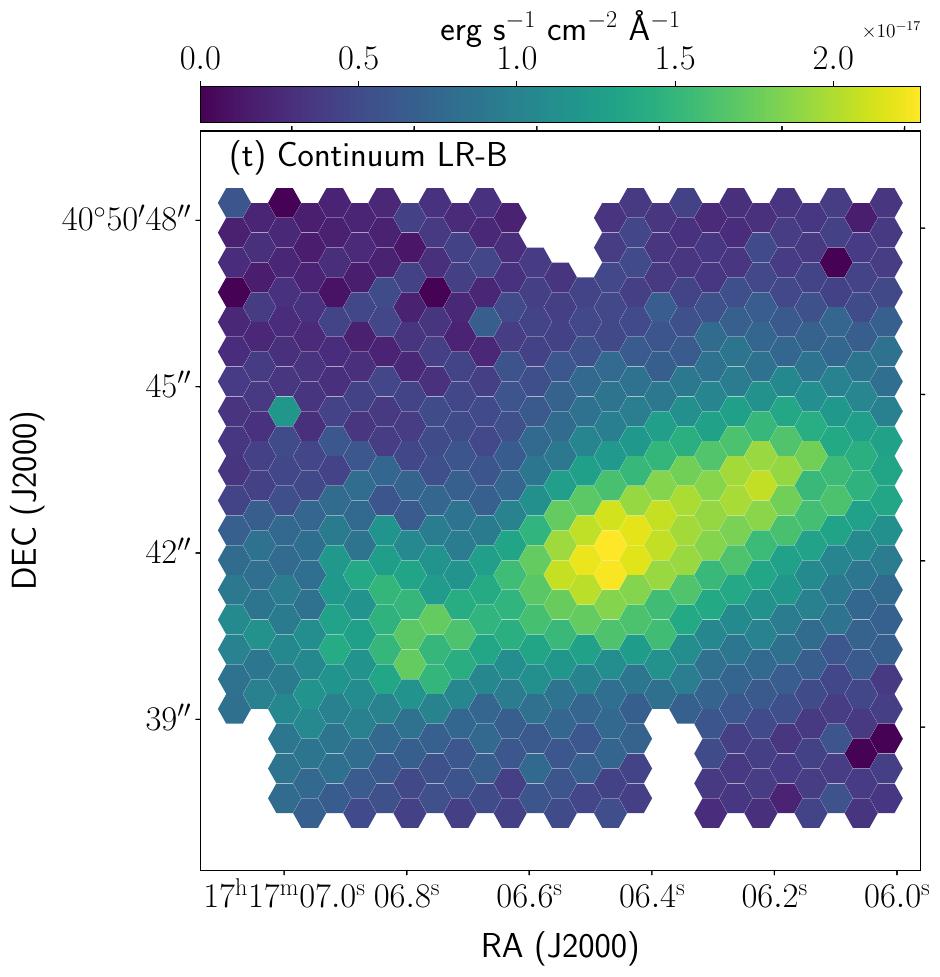}
	\includegraphics[clip, width=0.24\linewidth]{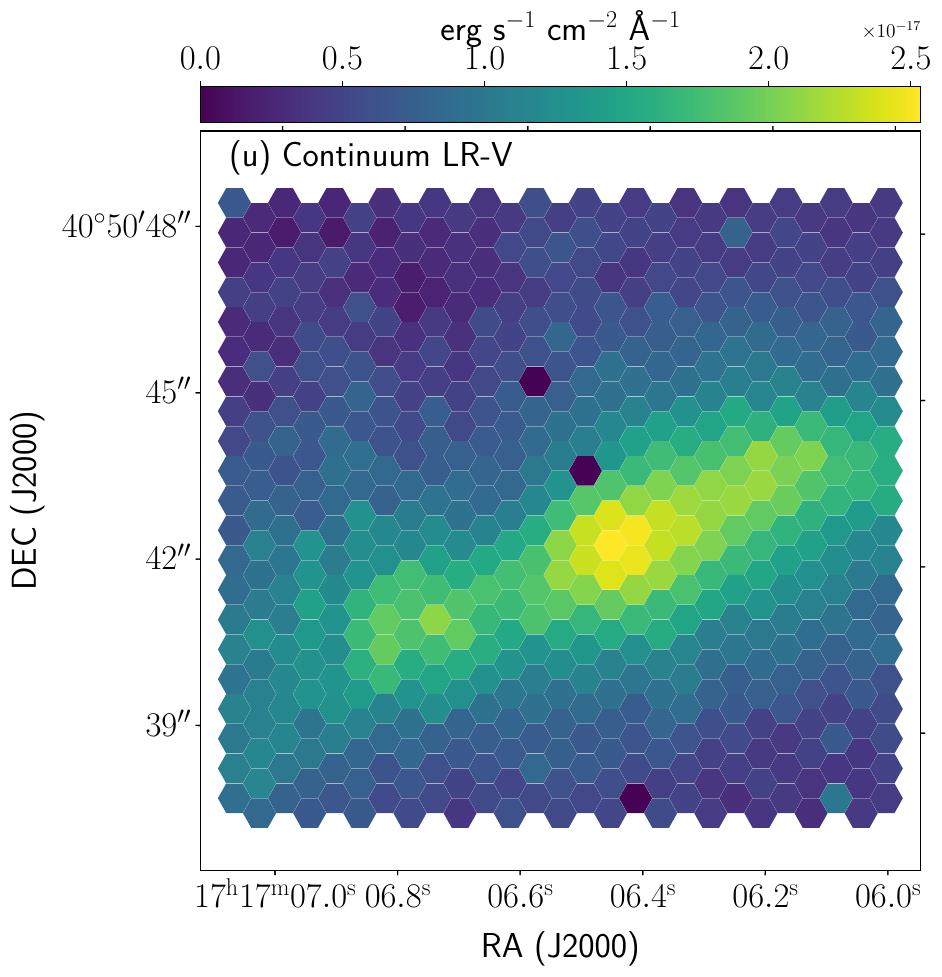}
	\includegraphics[clip, width=0.24\linewidth]{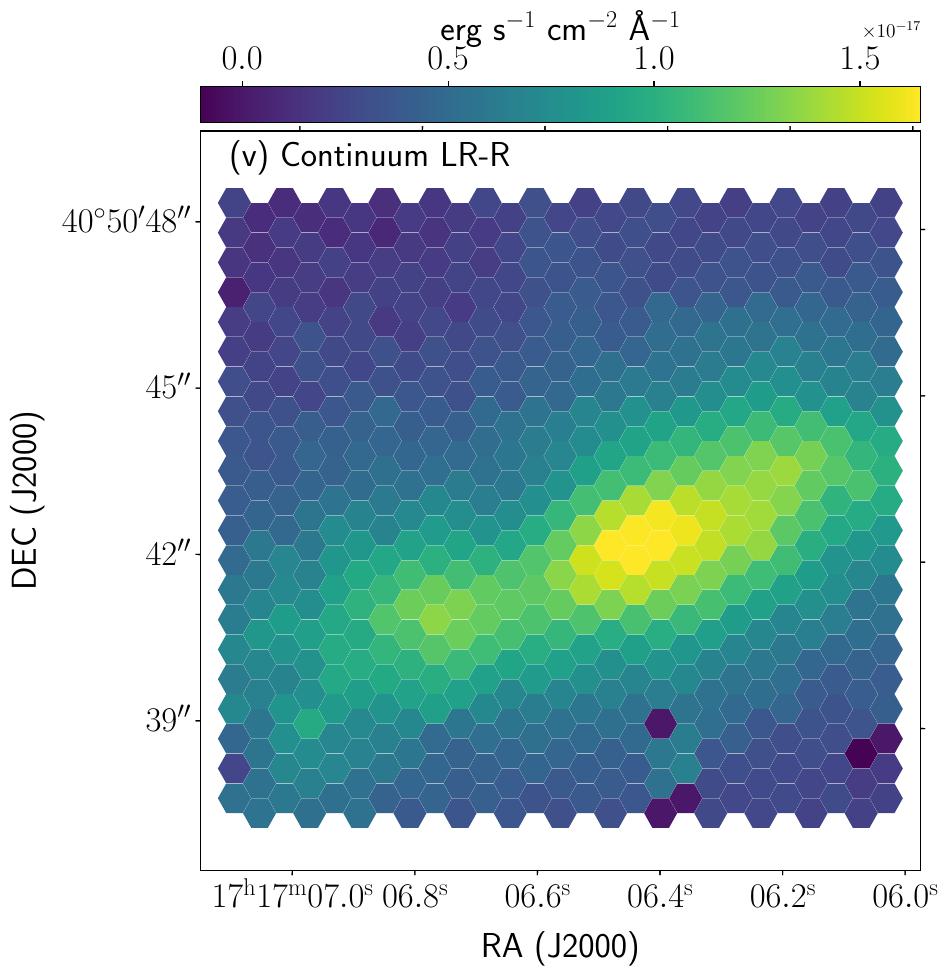}
	\includegraphics[clip, width=0.24\linewidth]{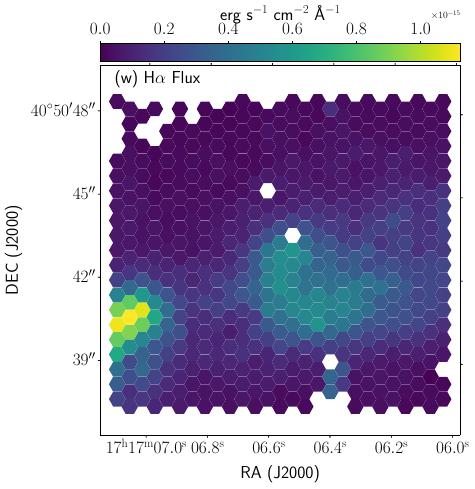}
	\includegraphics[clip, width=0.24\linewidth]{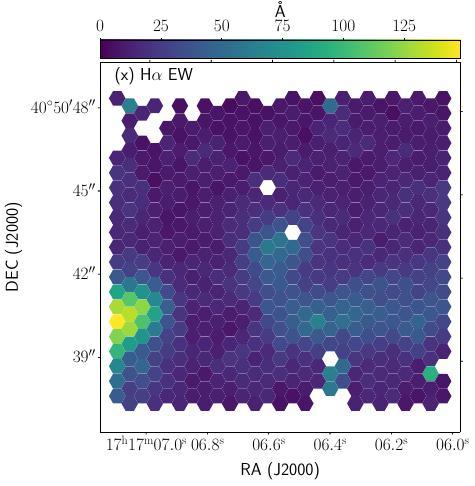}
	\includegraphics[clip, width=0.24\linewidth]{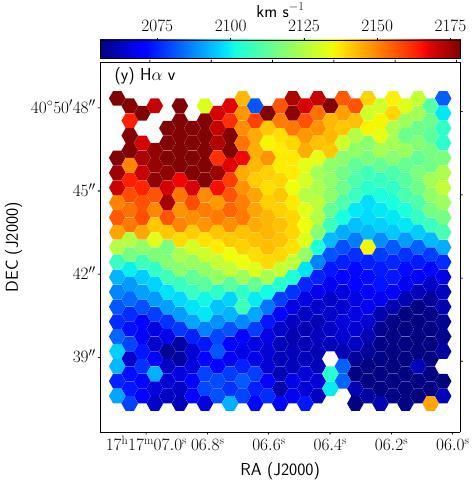}
	\includegraphics[clip, width=0.24\linewidth]{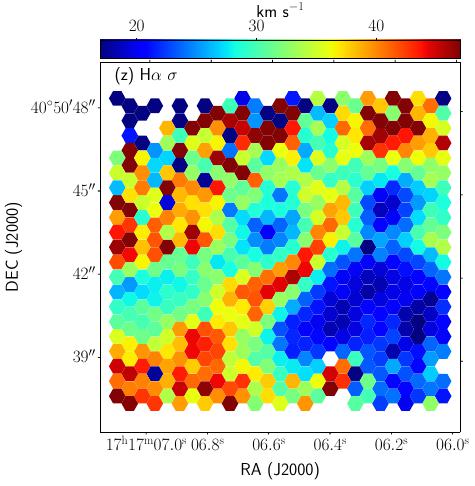}
	\includegraphics[clip, width=0.24\linewidth]{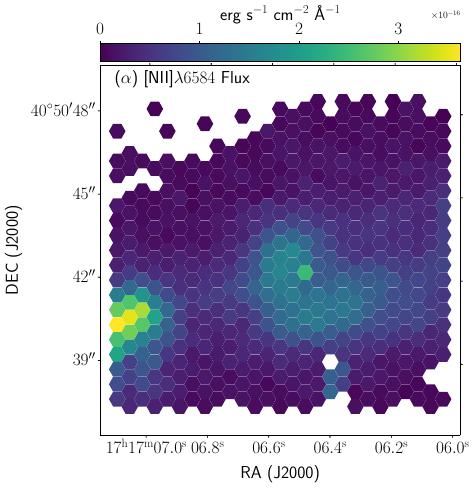}
	\includegraphics[clip, width=0.24\linewidth]{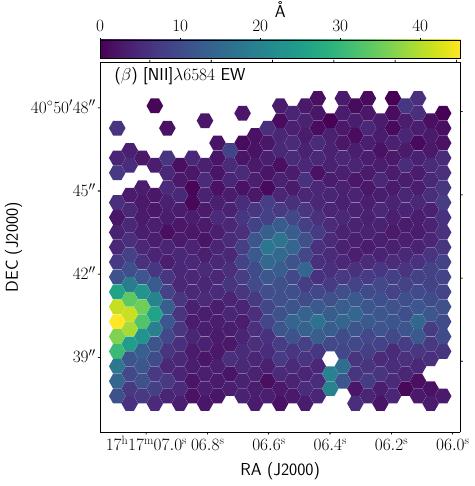}
	\includegraphics[clip, width=0.24\linewidth]{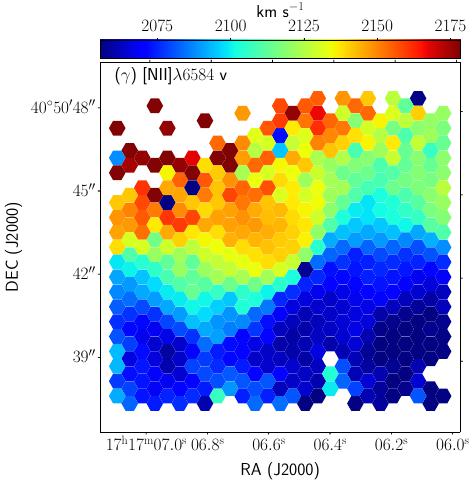}
	\includegraphics[clip, width=0.24\linewidth]{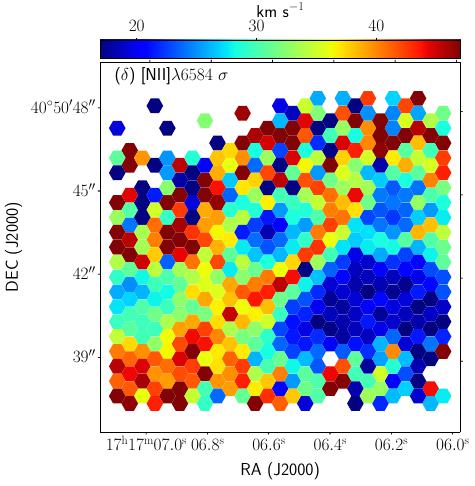}
	\includegraphics[clip, width=0.24\linewidth]{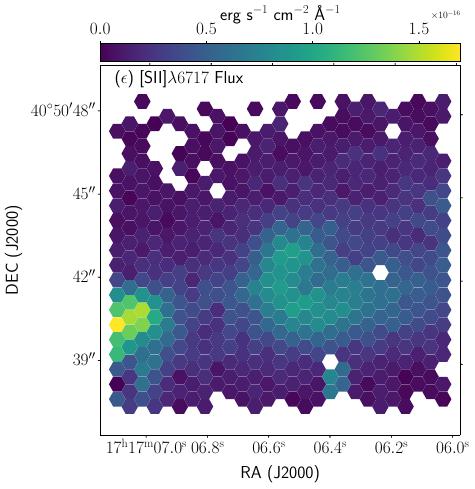}
	\includegraphics[clip, width=0.24\linewidth]{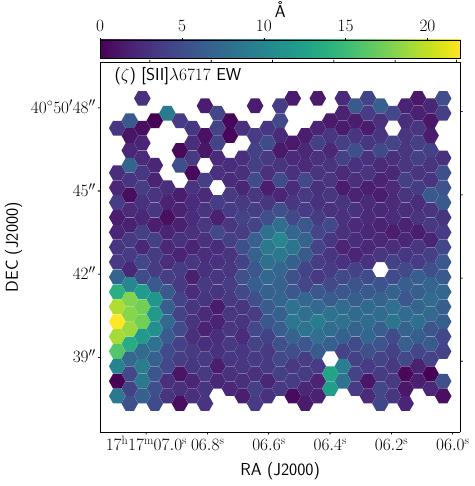}
	\includegraphics[clip, width=0.24\linewidth]{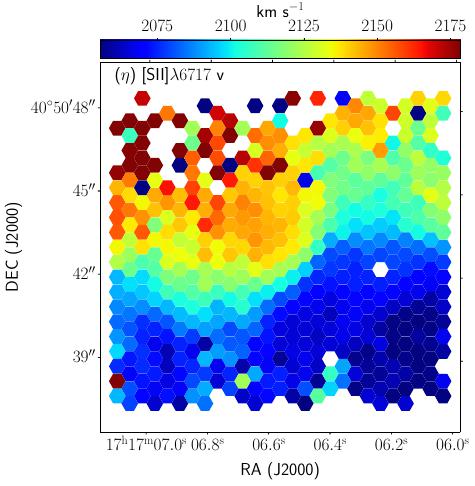}
	\includegraphics[clip, width=0.24\linewidth]{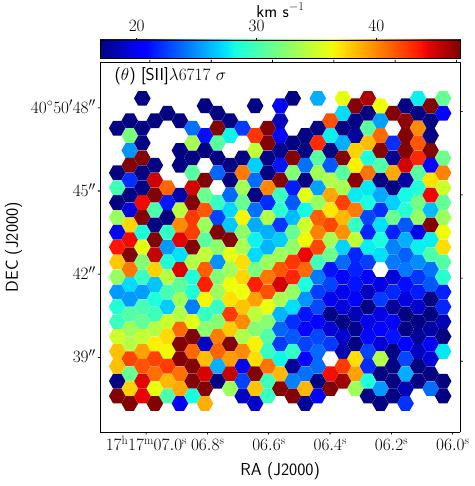}
	\includegraphics[clip, width=0.24\linewidth]{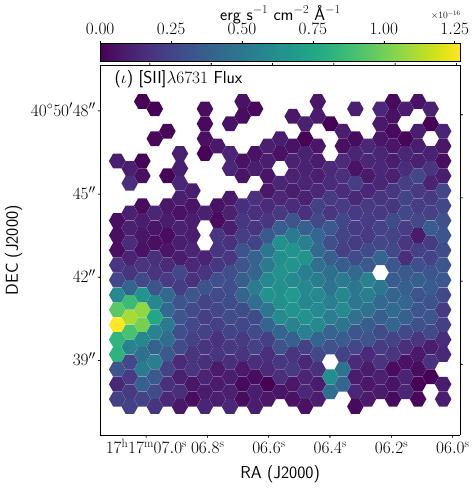}
	\includegraphics[clip, width=0.24\linewidth]{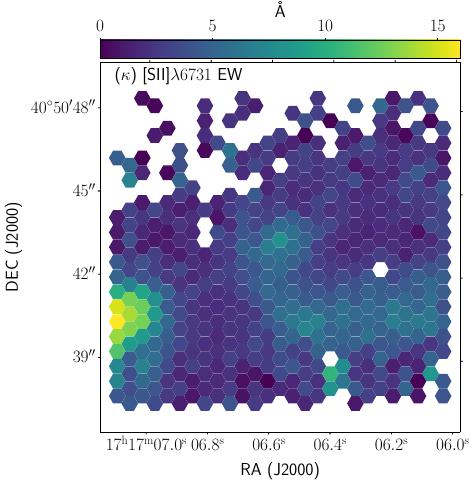}
	\includegraphics[clip, width=0.24\linewidth]{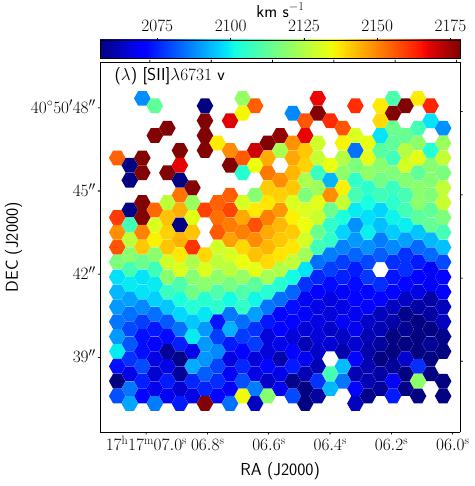}
	\includegraphics[clip, width=0.24\linewidth]{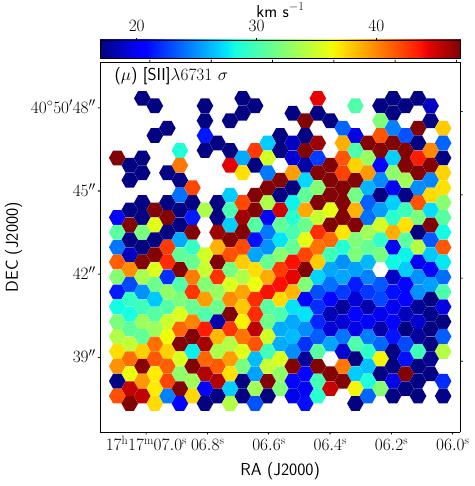}
	\caption{(cont.) NGC~6339 card.}
	\label{fig:NGC6339_card_2}
\end{figure*}

\begin{figure*}[h]
	\centering
	\includegraphics[clip, width=0.35\linewidth]{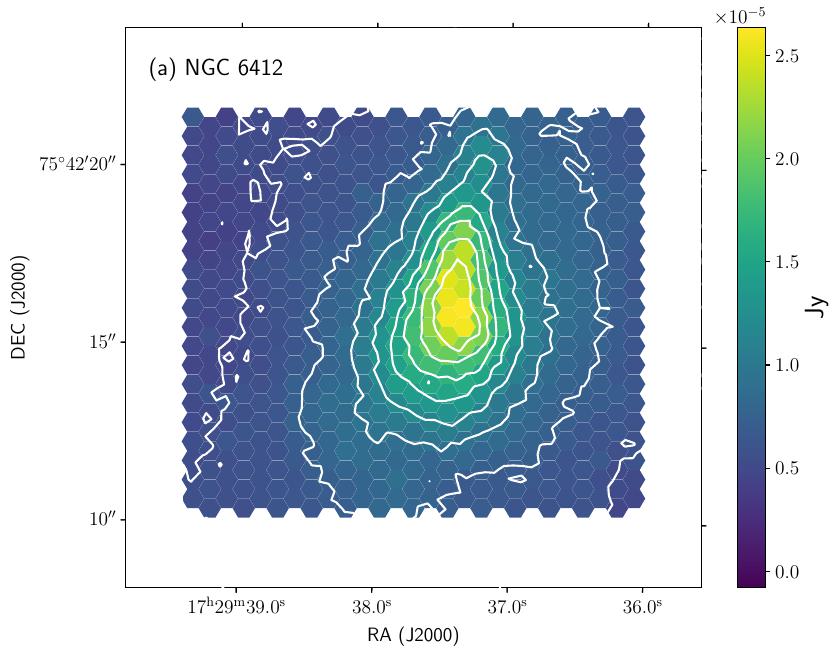}
	\includegraphics[clip, width=0.6\linewidth]{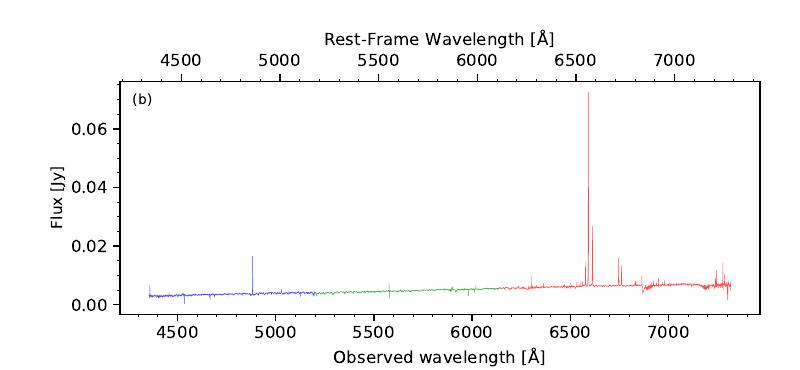}
	\includegraphics[clip, width=0.24\linewidth]{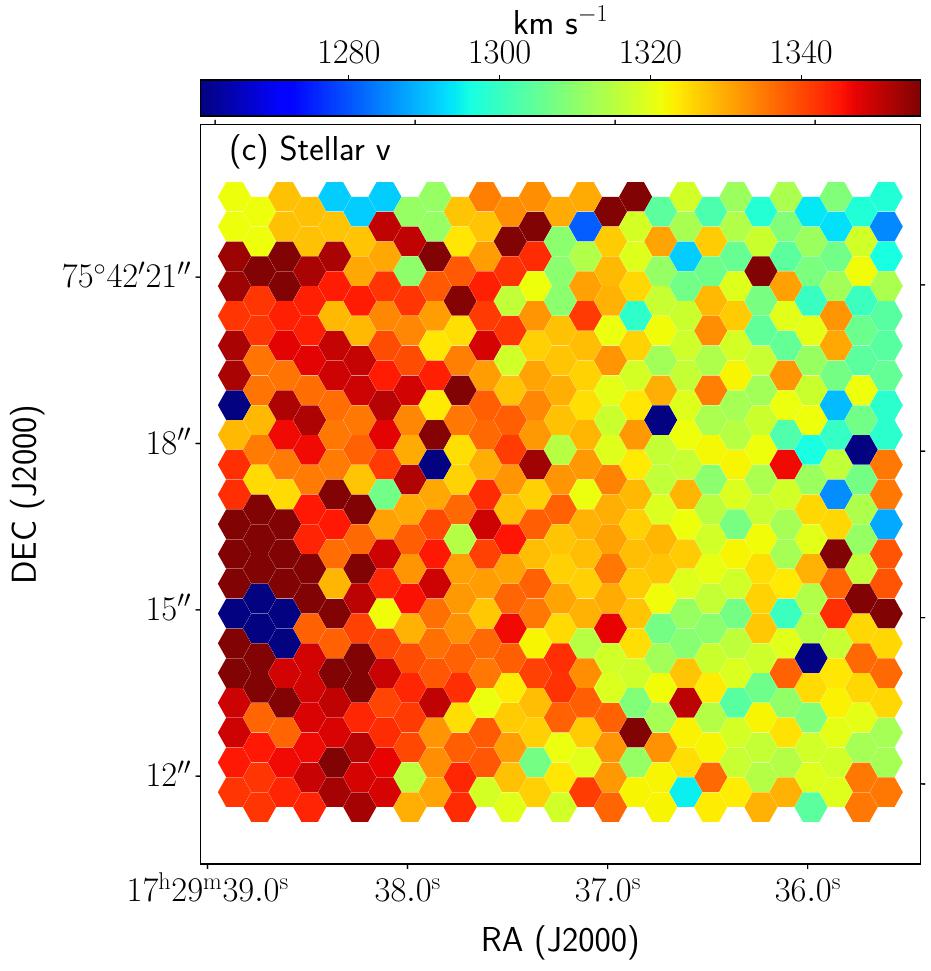}
	\includegraphics[clip, width=0.24\linewidth]{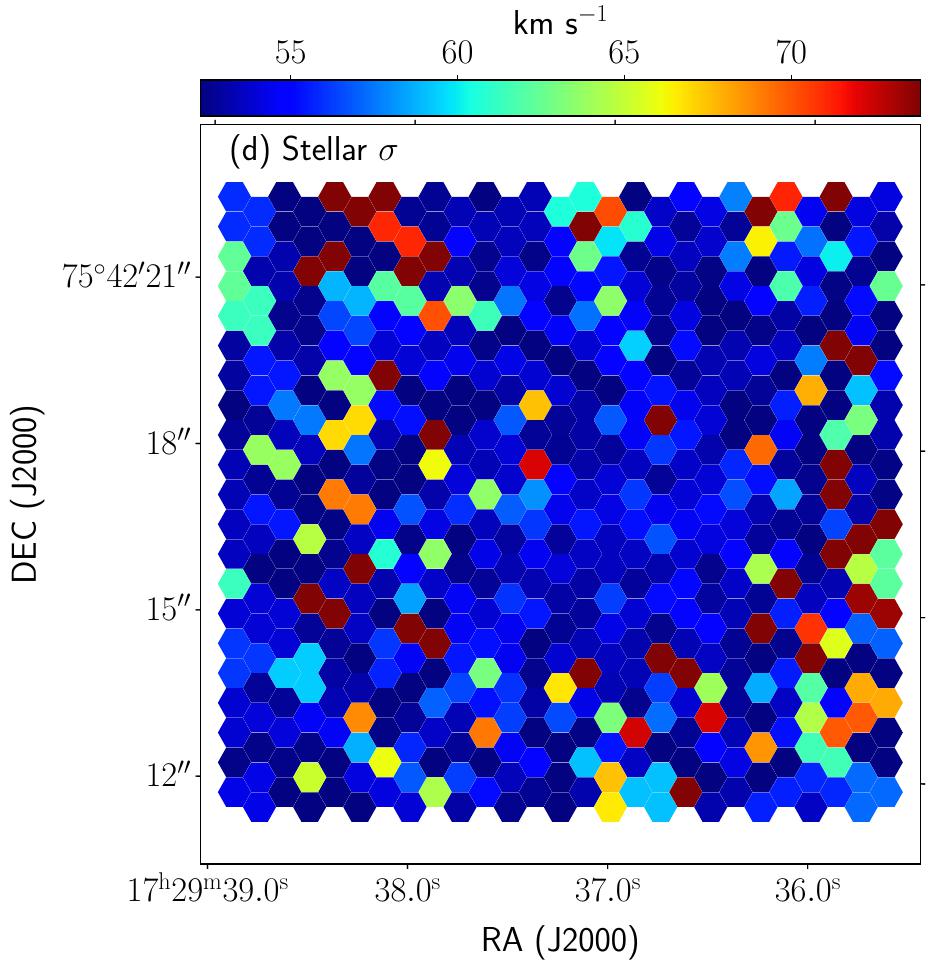}
	\includegraphics[clip, width=0.24\linewidth]{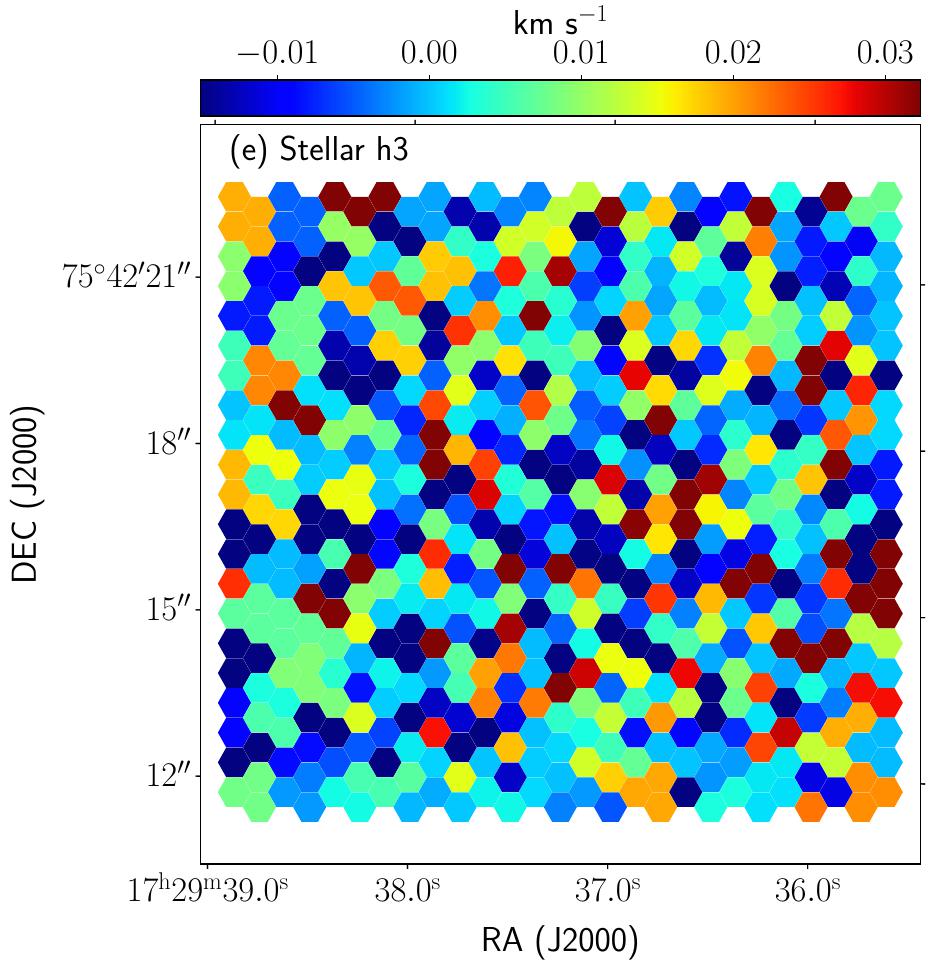}
	\includegraphics[clip, width=0.24\linewidth]{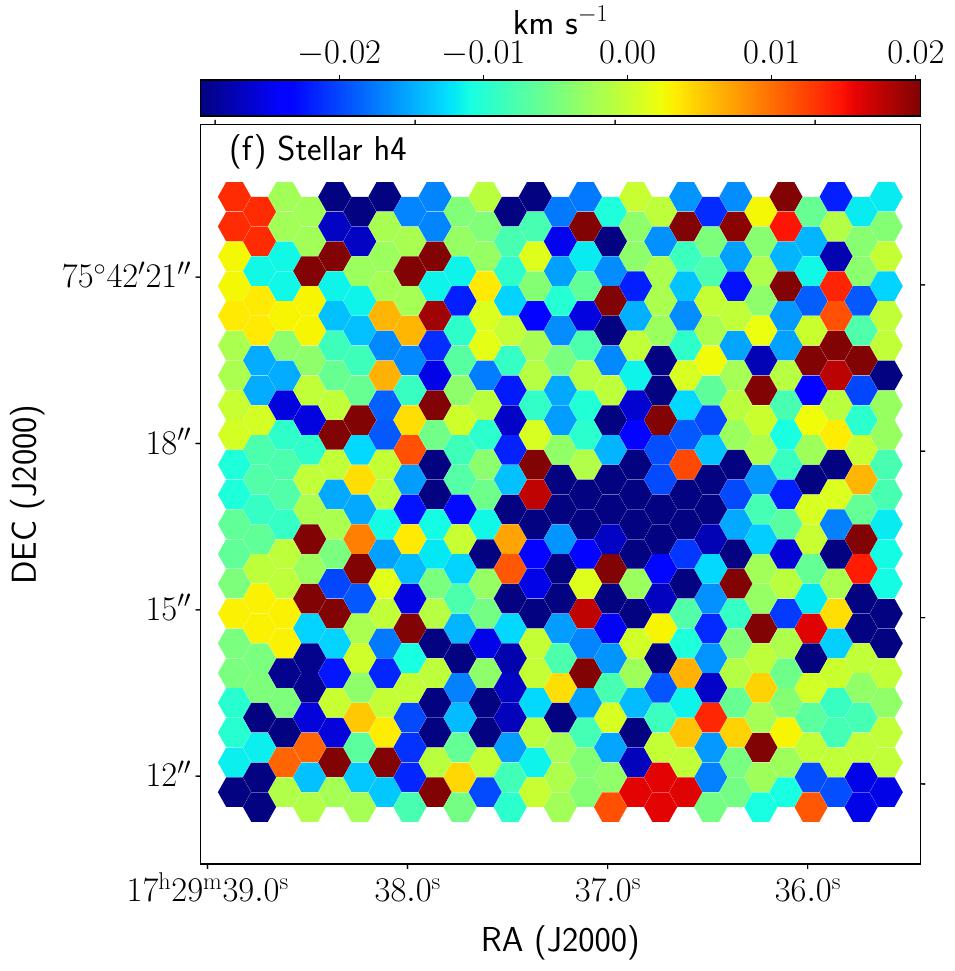}
	\includegraphics[clip, width=0.24\linewidth]{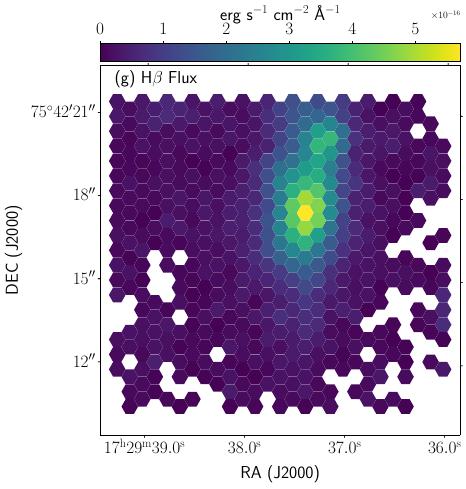}
	\includegraphics[clip, width=0.24\linewidth]{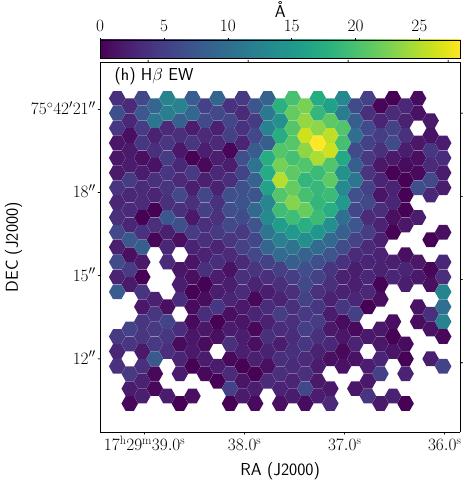}
	\includegraphics[clip, width=0.24\linewidth]{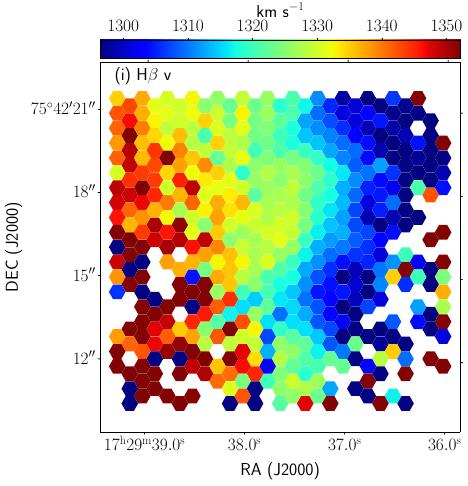}
	\includegraphics[clip, width=0.24\linewidth]{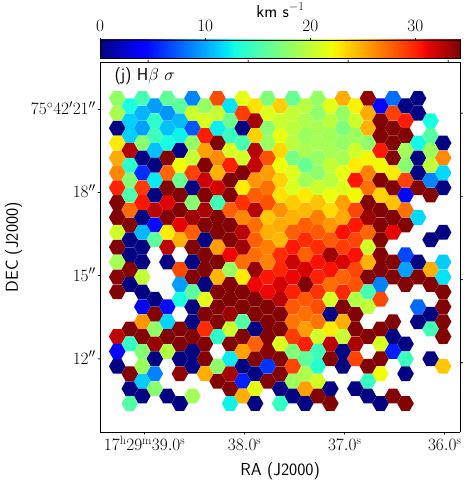}
	\includegraphics[clip, width=0.24\linewidth]{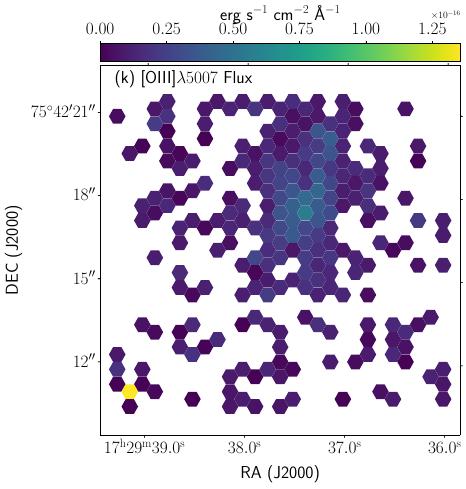}
	\includegraphics[clip, width=0.24\linewidth]{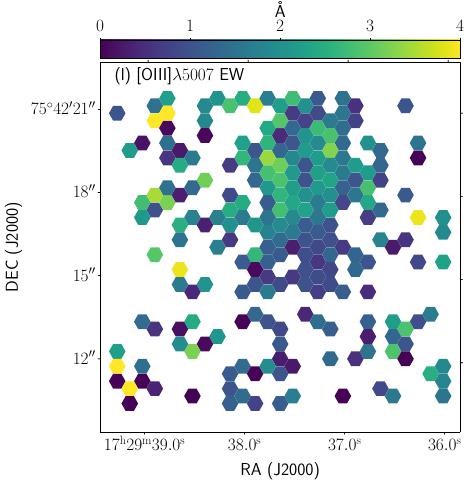}
	\includegraphics[clip, width=0.24\linewidth]{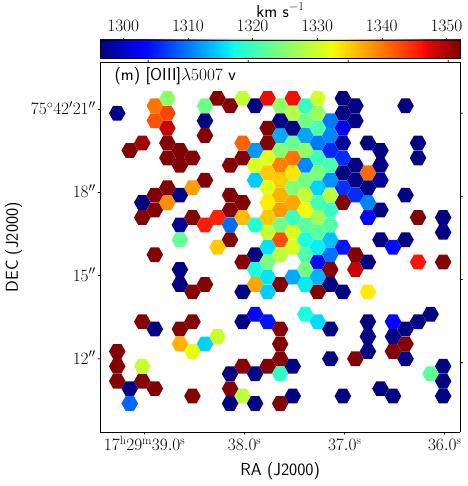}
	\includegraphics[clip, width=0.24\linewidth]{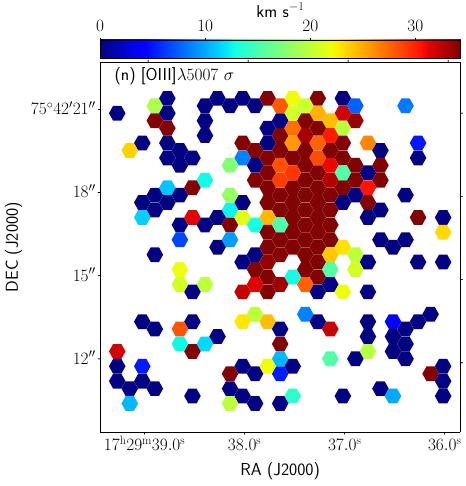}
	\vspace{5cm}
	\caption{NGC~6412 card.}
	\label{fig:NGC6412_card_1}
\end{figure*}
\addtocounter{figure}{-1}
\begin{figure*}[h]
	\centering
	\includegraphics[clip, width=0.24\linewidth]{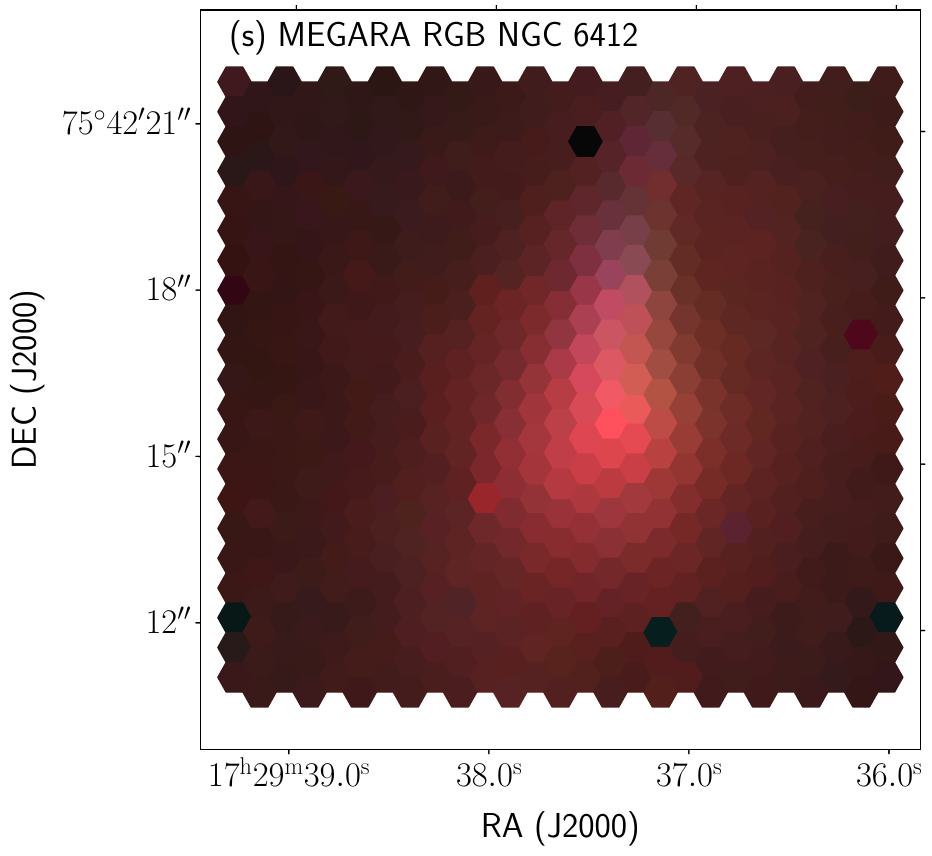}
	\includegraphics[clip, width=0.24\linewidth]{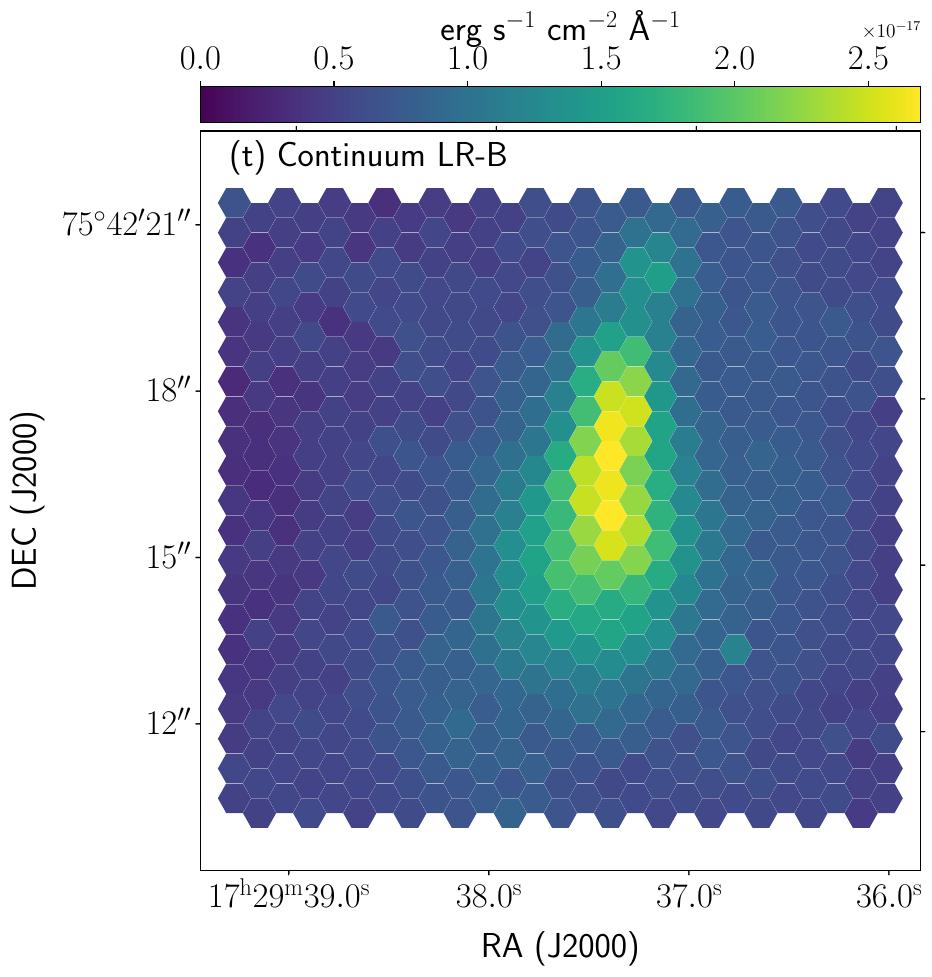}
	\includegraphics[clip, width=0.24\linewidth]{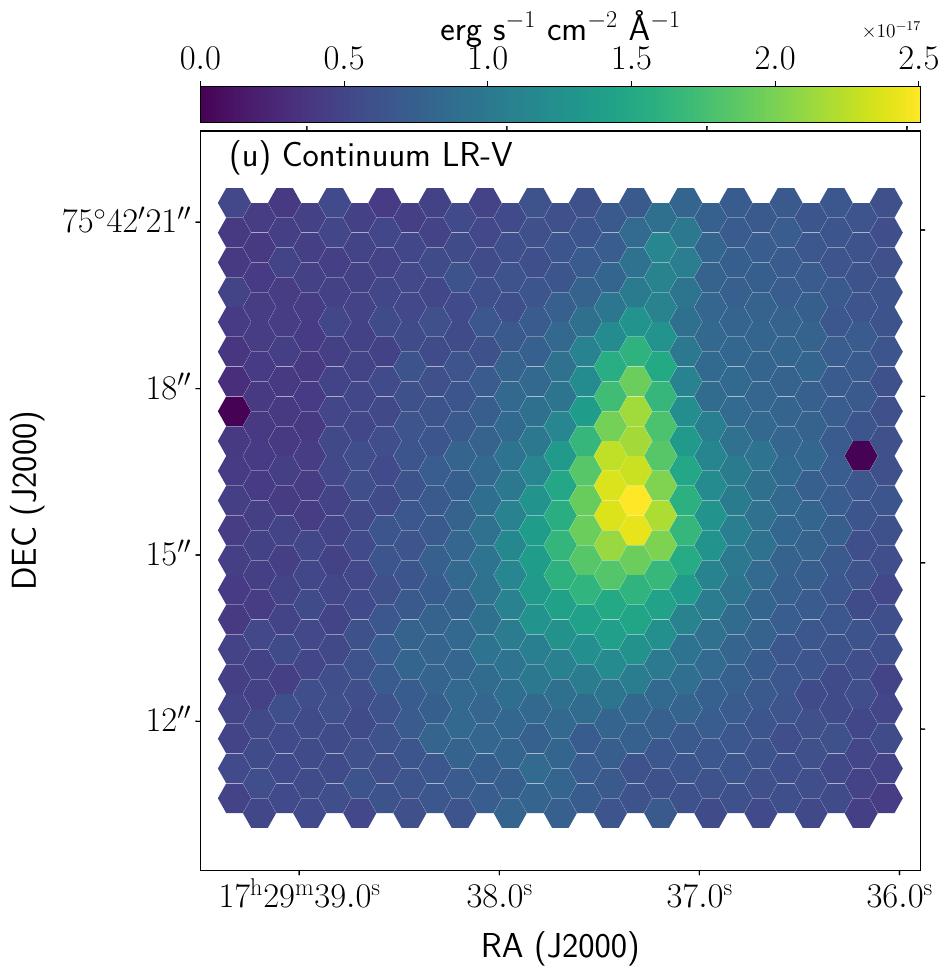}
	\includegraphics[clip, width=0.24\linewidth]{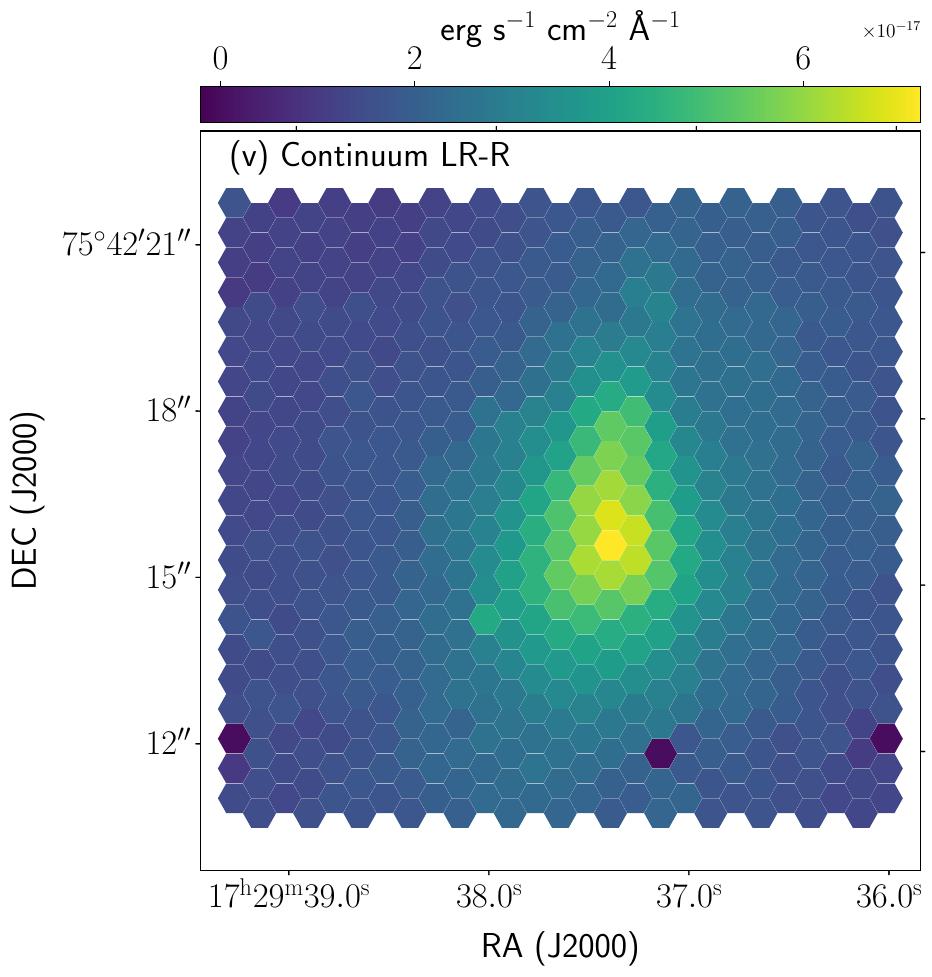}
	\includegraphics[clip, width=0.24\linewidth]{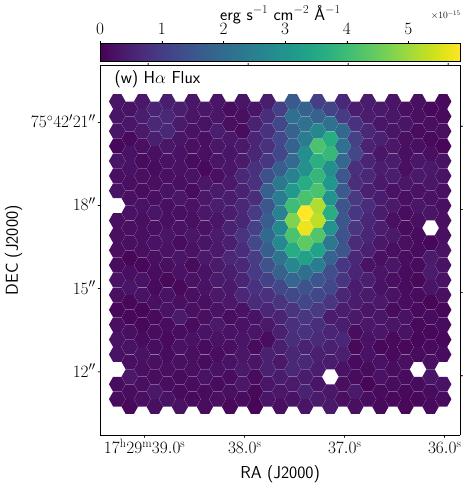}
	\includegraphics[clip, width=0.24\linewidth]{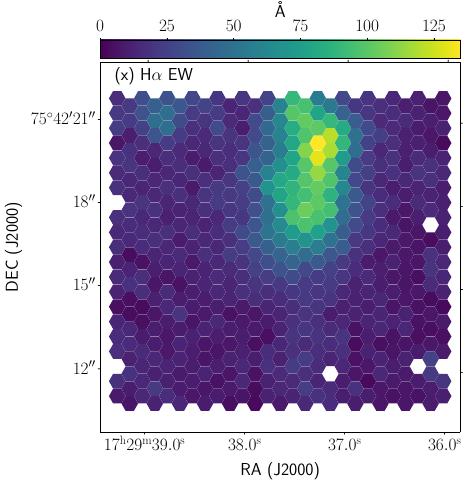}
	\includegraphics[clip, width=0.24\linewidth]{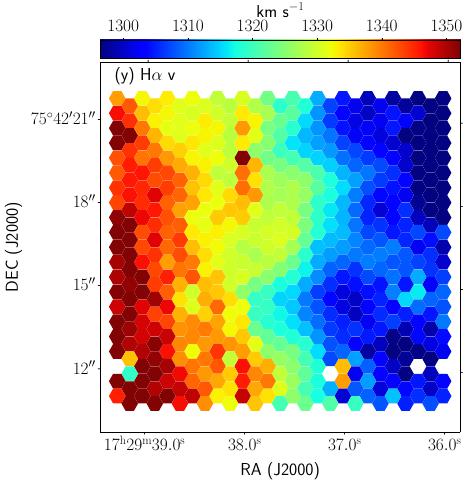}
	\includegraphics[clip, width=0.24\linewidth]{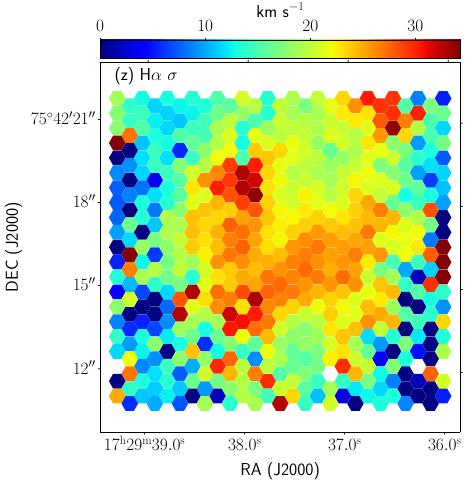}
	\includegraphics[clip, width=0.24\linewidth]{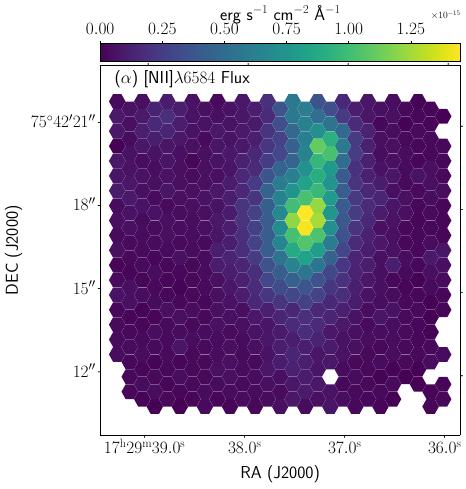}
	\includegraphics[clip, width=0.24\linewidth]{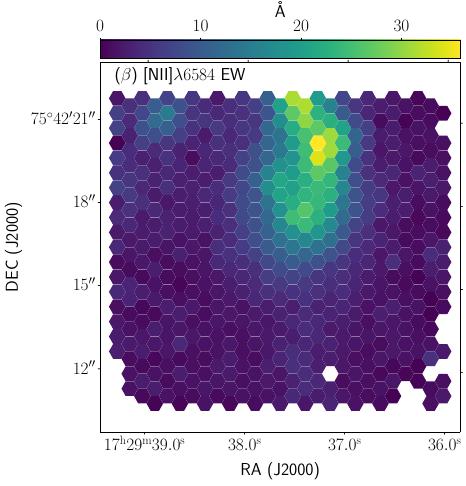}
	\includegraphics[clip, width=0.24\linewidth]{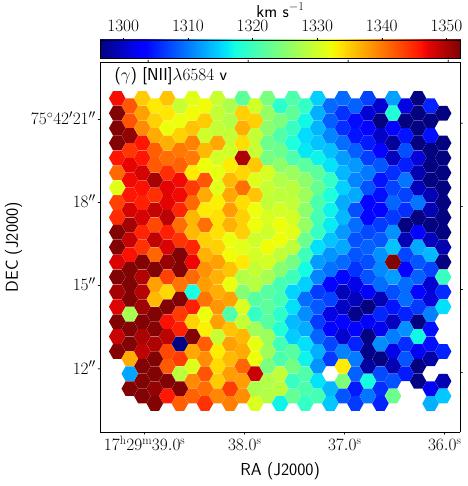}
	\includegraphics[clip, width=0.24\linewidth]{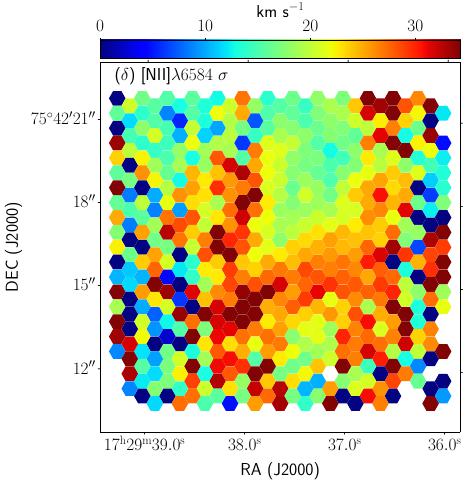}
	\includegraphics[clip, width=0.24\linewidth]{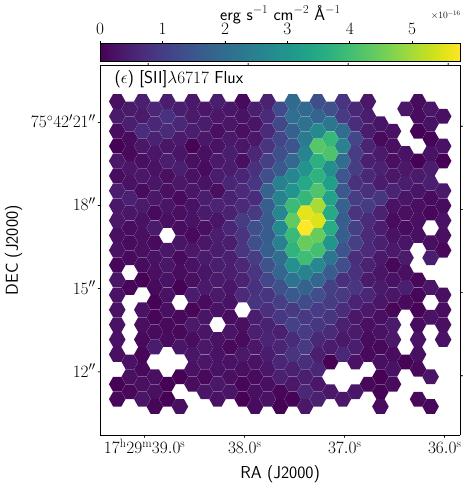}
	\includegraphics[clip, width=0.24\linewidth]{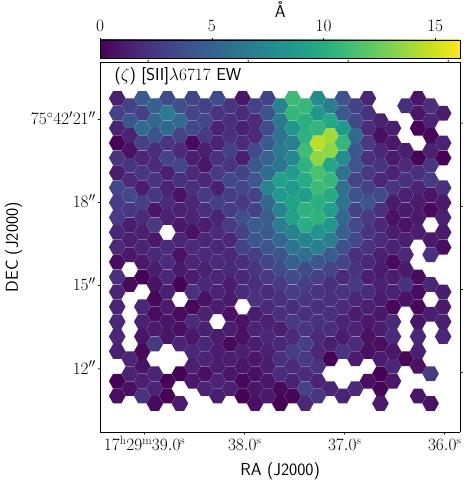}
	\includegraphics[clip, width=0.24\linewidth]{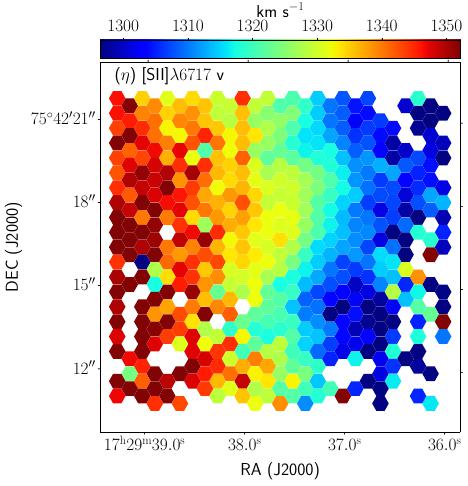}
	\includegraphics[clip, width=0.24\linewidth]{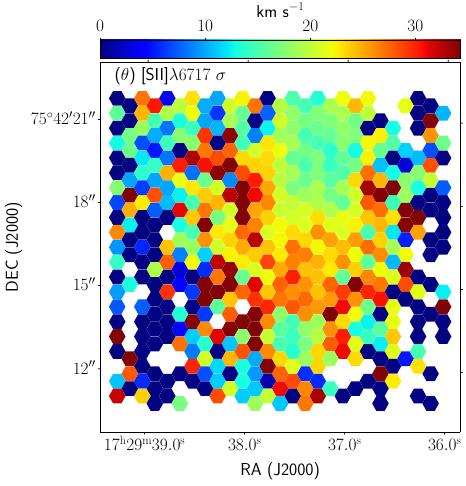}
	\includegraphics[clip, width=0.24\linewidth]{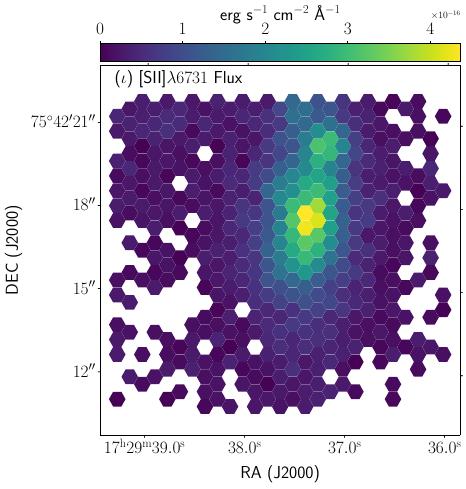}
	\includegraphics[clip, width=0.24\linewidth]{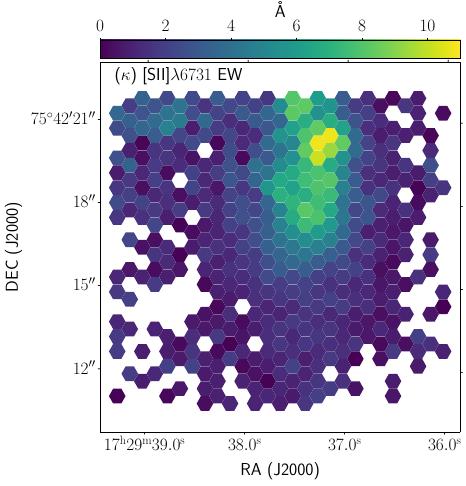}
	\includegraphics[clip, width=0.24\linewidth]{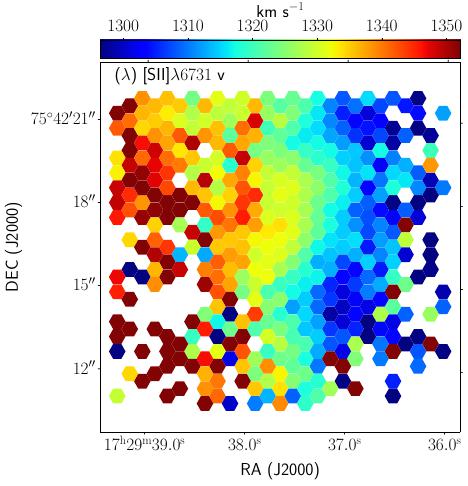}
	\includegraphics[clip, width=0.24\linewidth]{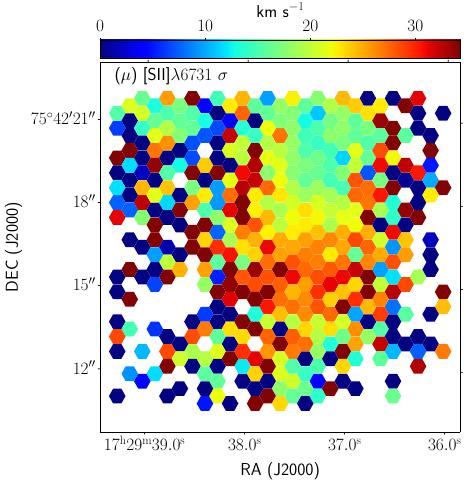}
	\caption{(cont.) NGC~6412 card.}
	\label{fig:NGC6412_card_2}
\end{figure*}

\begin{figure*}[h]
	\centering
	\includegraphics[clip, width=0.35\linewidth]{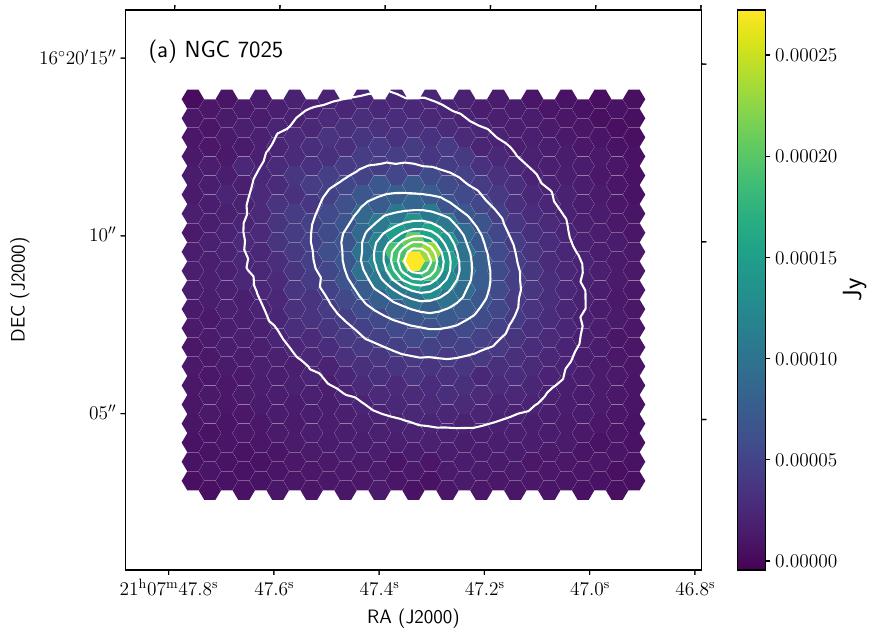}
	\includegraphics[clip, width=0.6\linewidth]{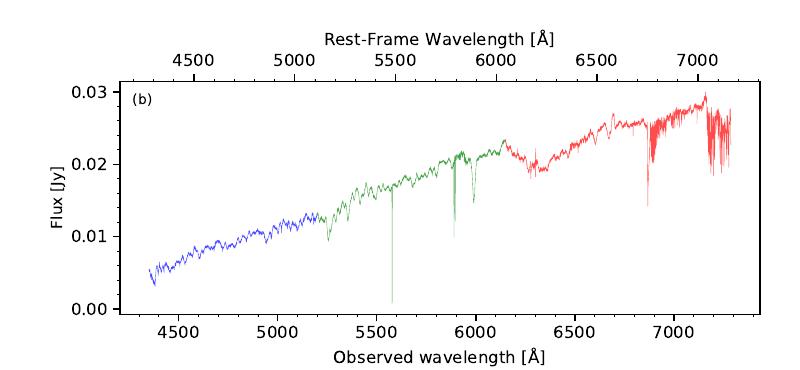}
	\includegraphics[clip, width=0.24\linewidth]{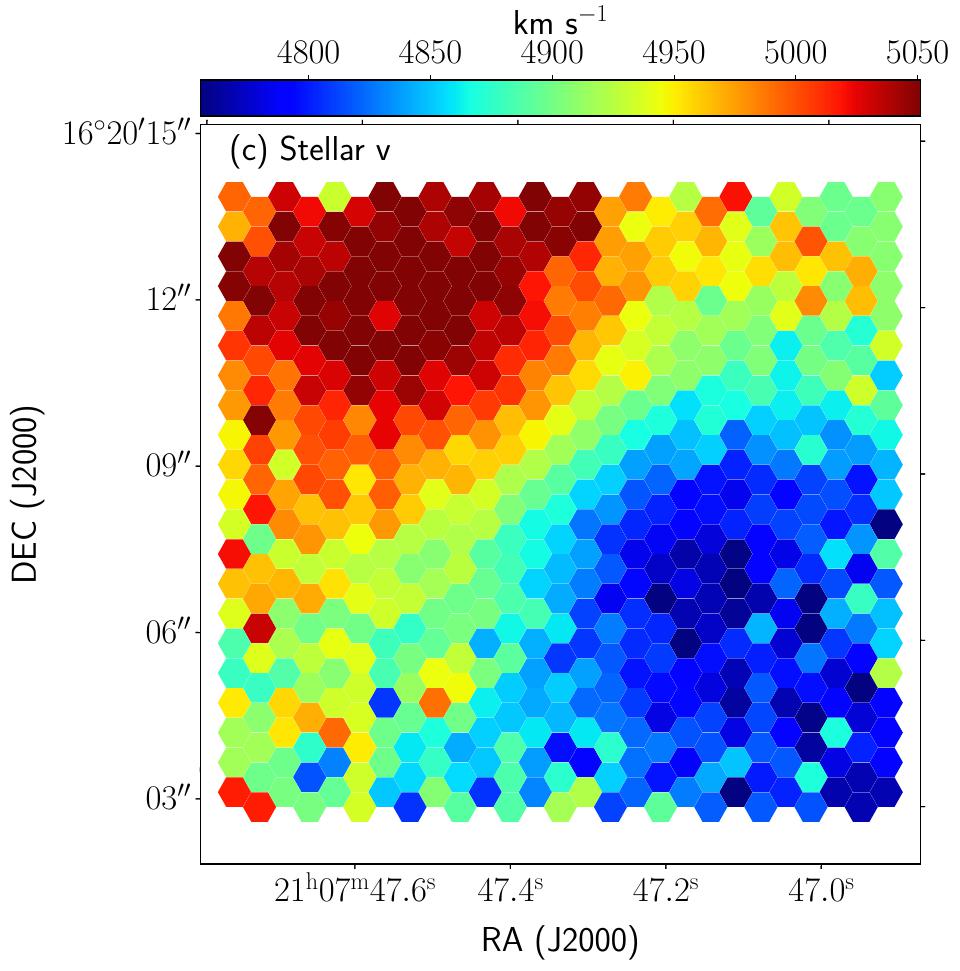}
	\includegraphics[clip, width=0.24\linewidth]{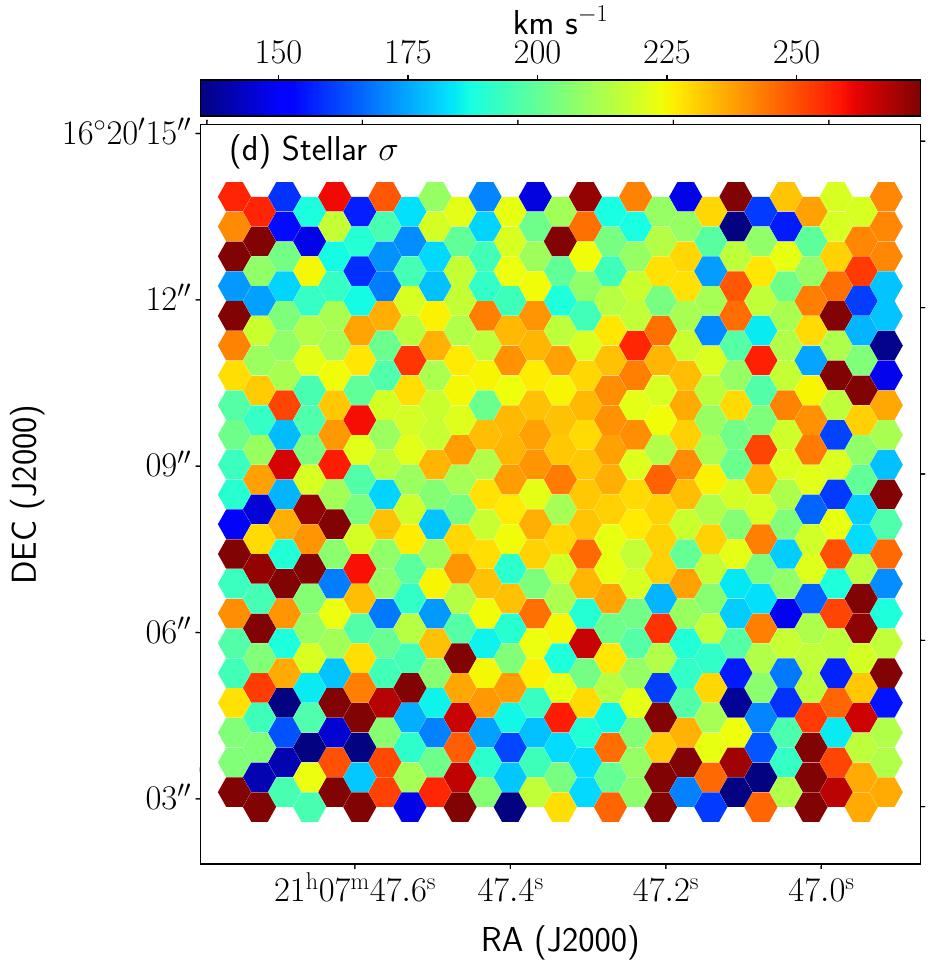}
	\includegraphics[clip, width=0.24\linewidth]{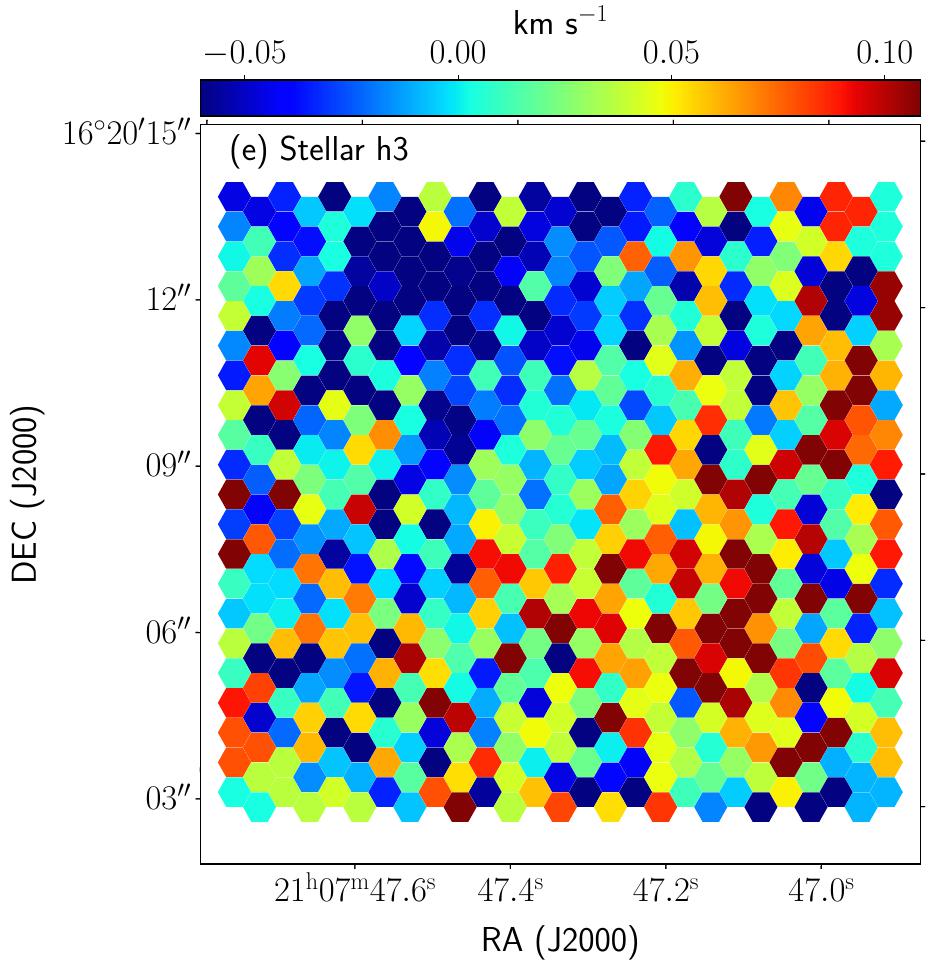}
	\includegraphics[clip, width=0.24\linewidth]{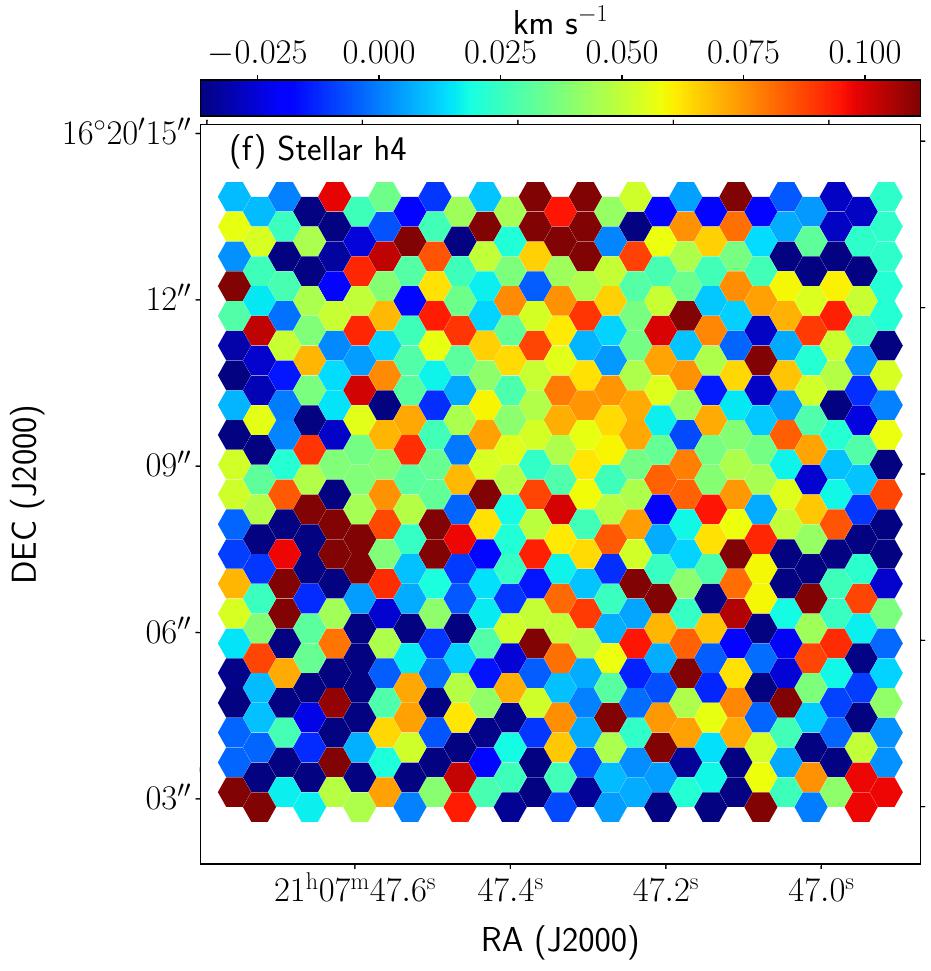}
	\includegraphics[clip, width=0.24\linewidth]{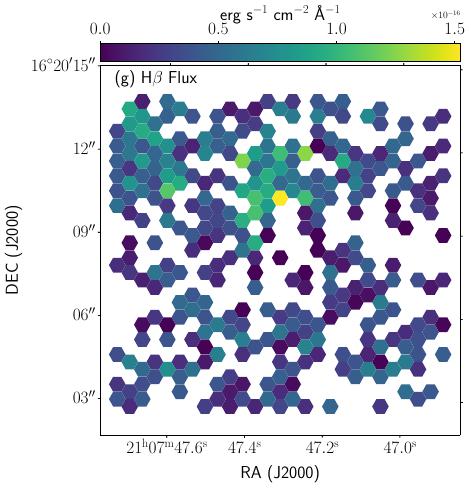}
	\includegraphics[clip, width=0.24\linewidth]{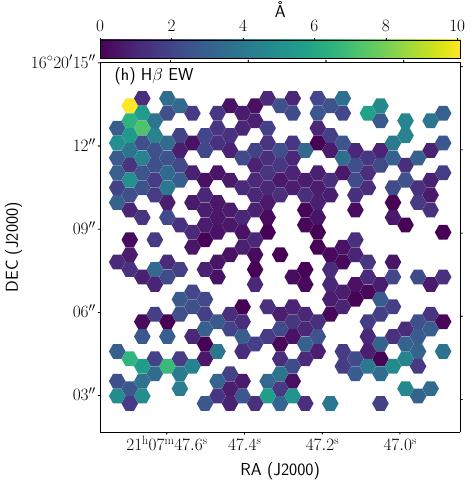}
	\includegraphics[clip, width=0.24\linewidth]{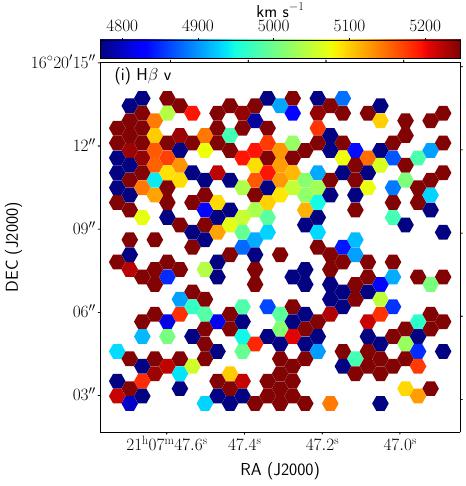}
	\includegraphics[clip, width=0.24\linewidth]{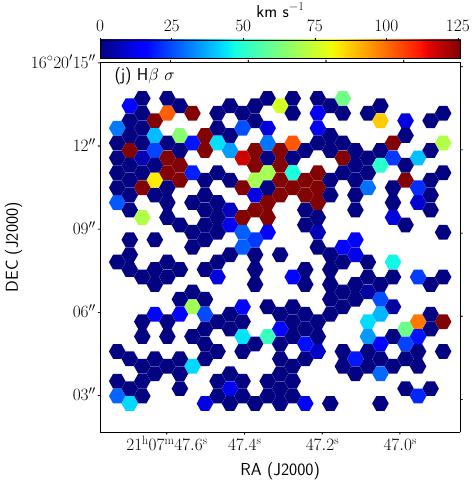}
	\includegraphics[clip, width=0.24\linewidth]{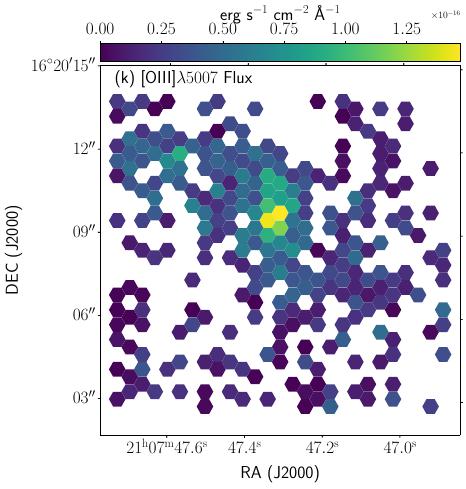}
	\includegraphics[clip, width=0.24\linewidth]{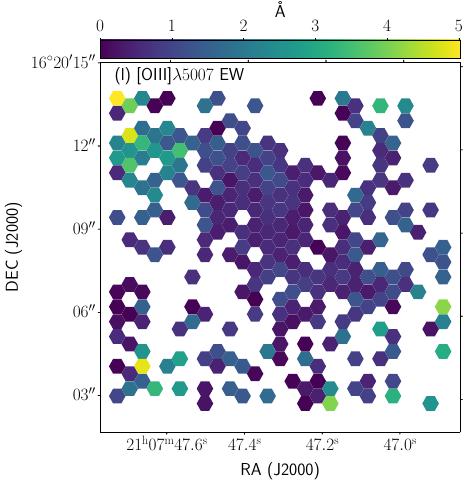}
	\includegraphics[clip, width=0.24\linewidth]{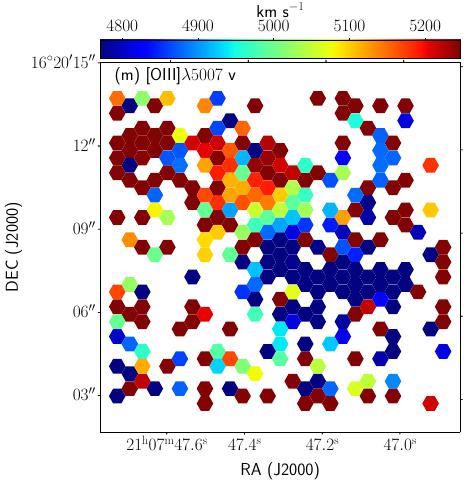}
	\includegraphics[clip, width=0.24\linewidth]{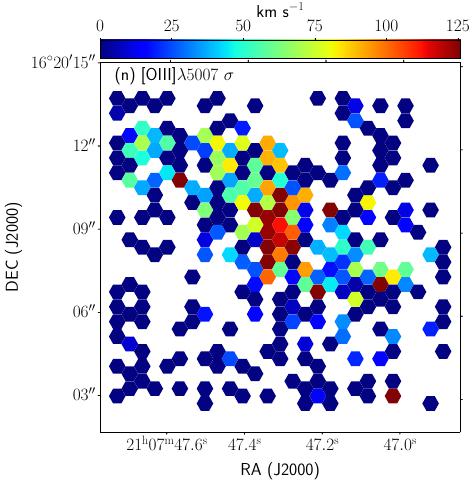}
	\includegraphics[clip, width=0.24\linewidth]{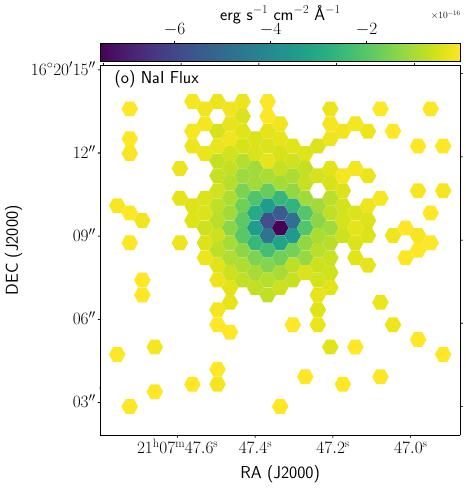}
	\includegraphics[clip, width=0.24\linewidth]{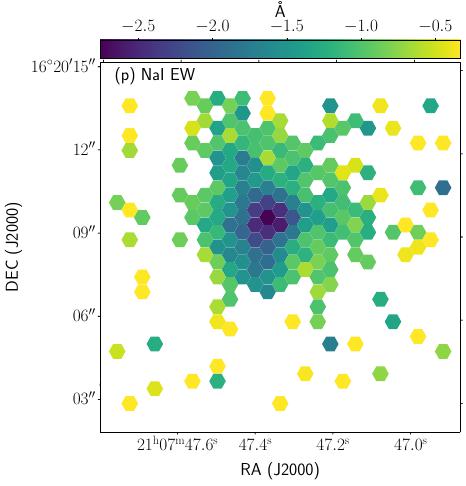}
	\includegraphics[clip, width=0.24\linewidth]{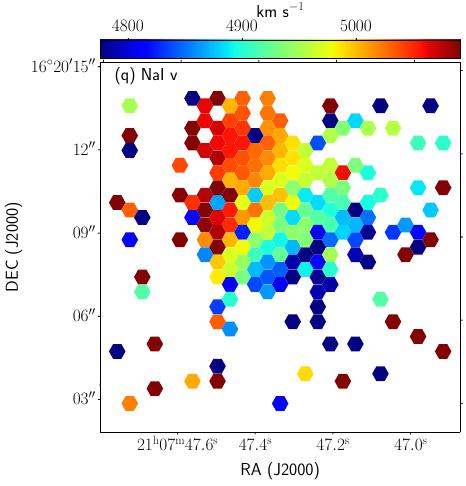}
	\includegraphics[clip, width=0.24\linewidth]{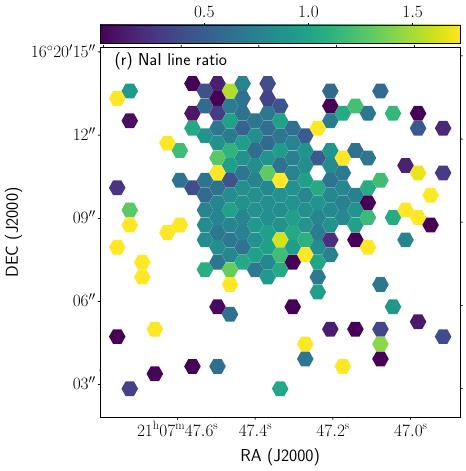}
	\caption{NGC~7025 card.}
	\label{fig:NGC7025_card_1}
\end{figure*}
\addtocounter{figure}{-1}
\begin{figure*}[h]
	\centering
	\includegraphics[clip, width=0.24\linewidth]{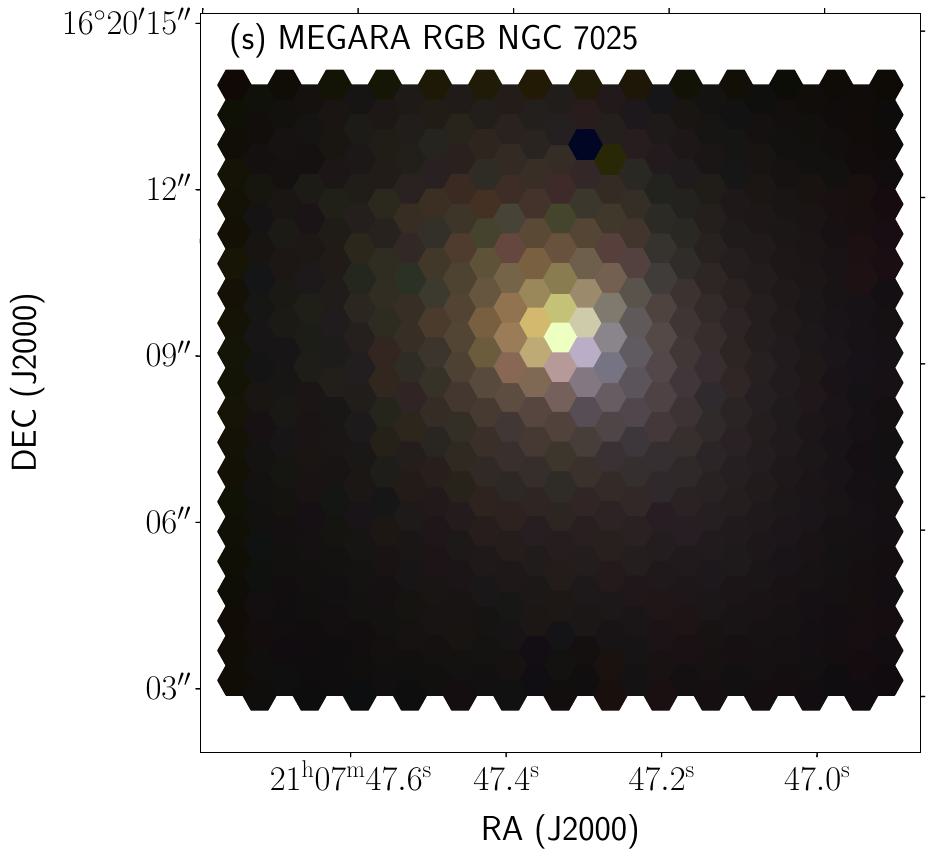}
	\includegraphics[clip, width=0.24\linewidth]{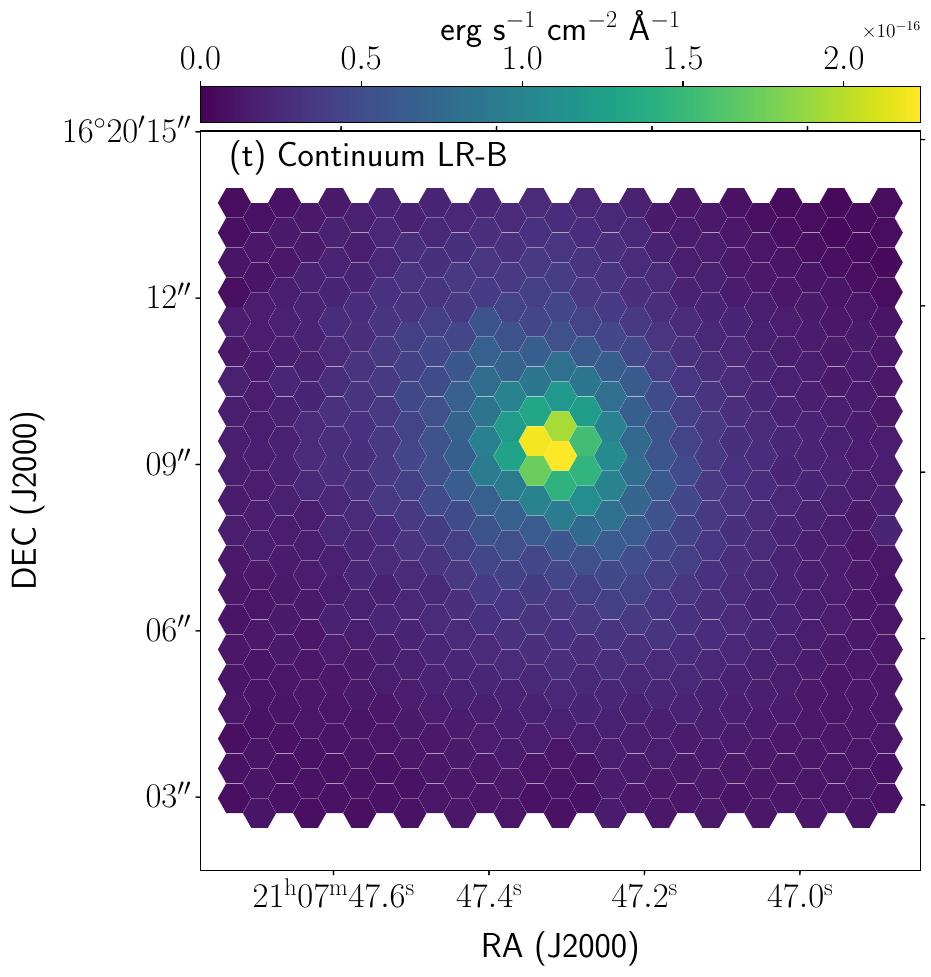}
	\includegraphics[clip, width=0.24\linewidth]{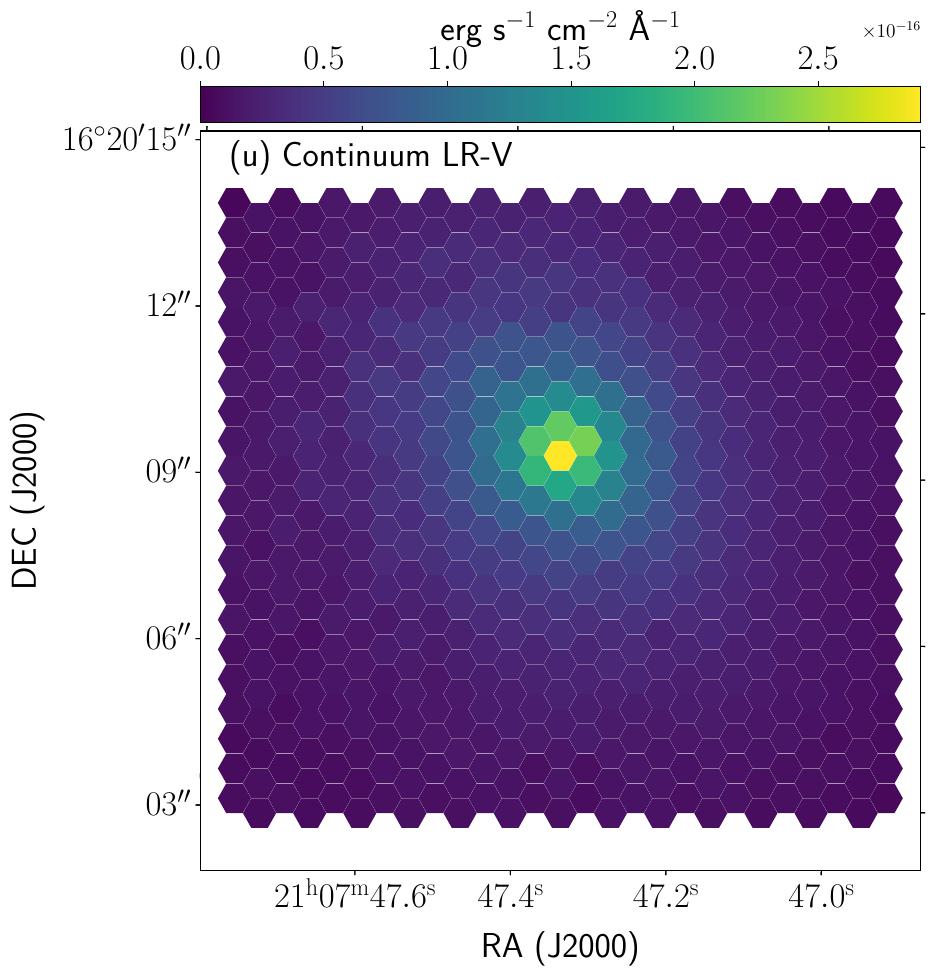}
	\includegraphics[clip, width=0.24\linewidth]{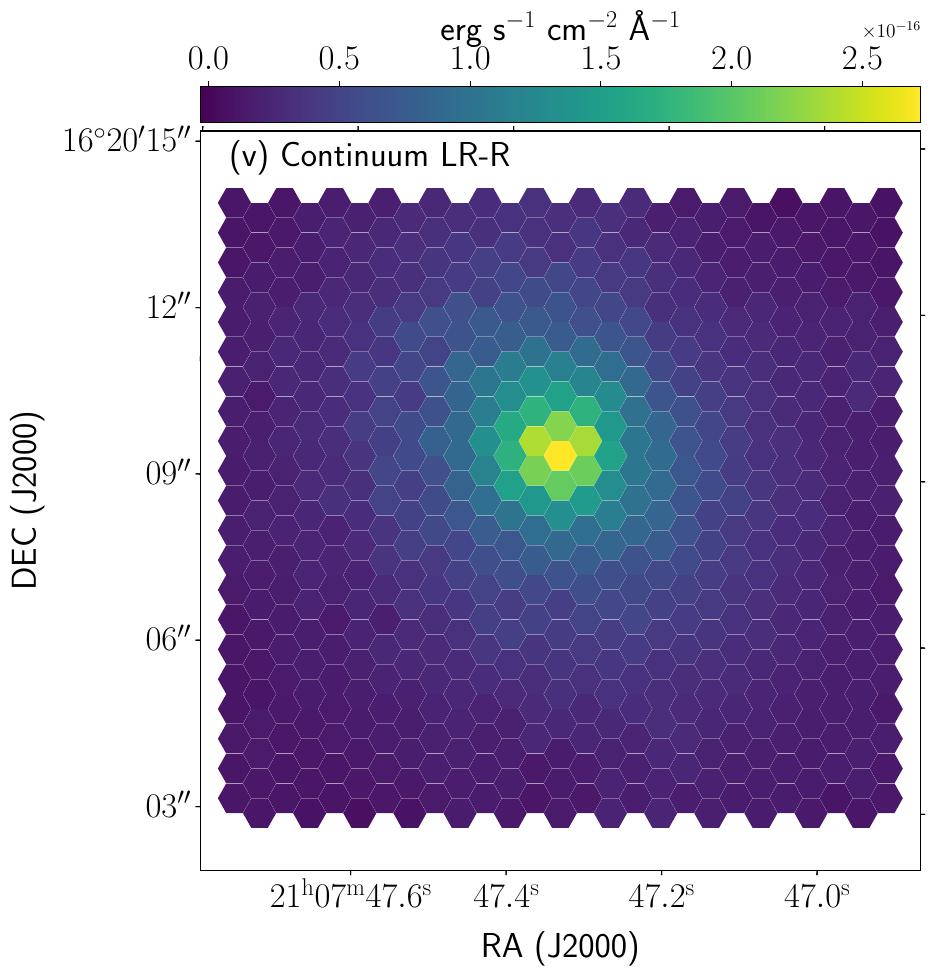}
	\includegraphics[clip, width=0.24\linewidth]{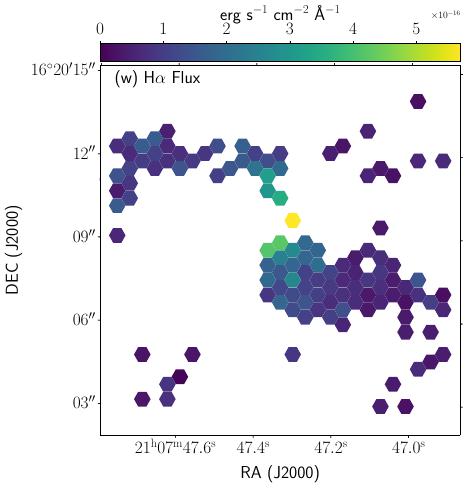}
	\includegraphics[clip, width=0.24\linewidth]{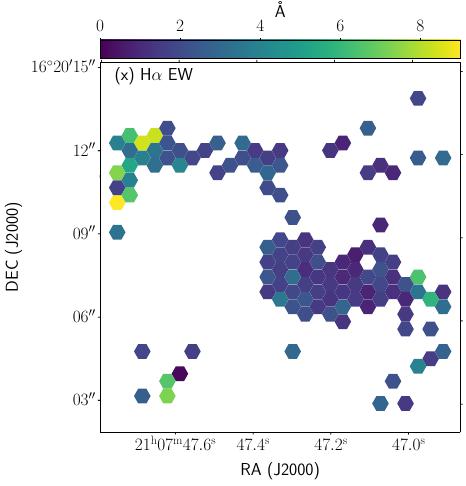}
	\includegraphics[clip, width=0.24\linewidth]{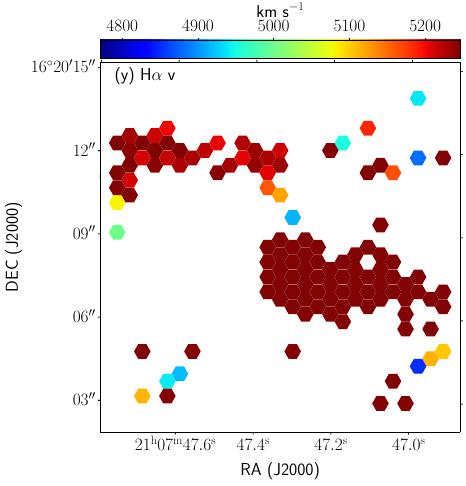}
	\includegraphics[clip, width=0.24\linewidth]{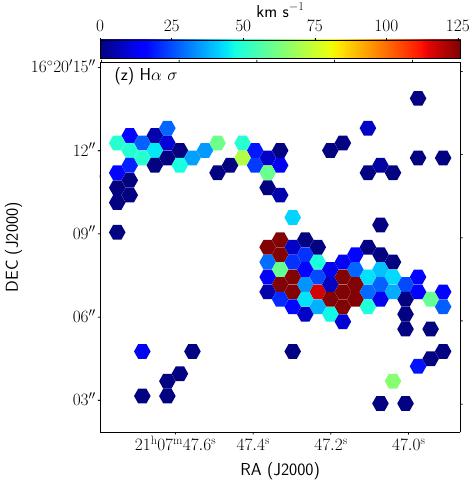}
	\includegraphics[clip, width=0.24\linewidth]{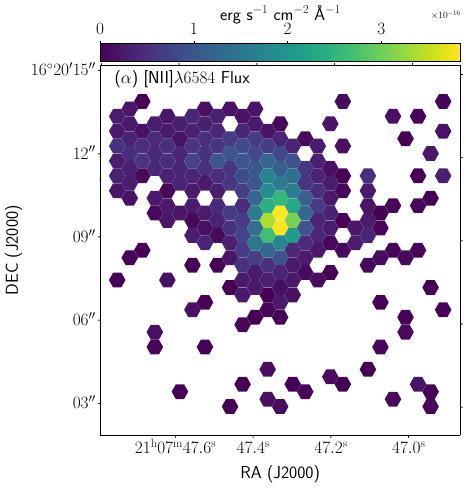}
	\includegraphics[clip, width=0.24\linewidth]{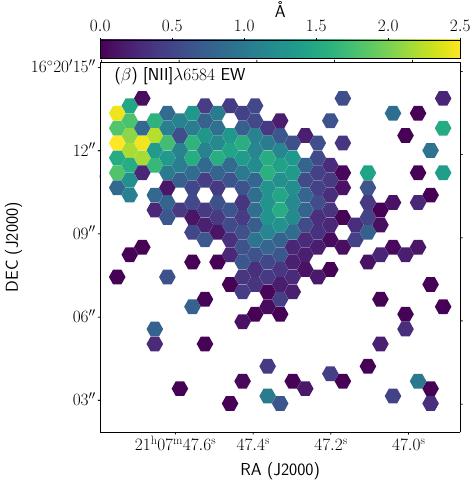}
	\includegraphics[clip, width=0.24\linewidth]{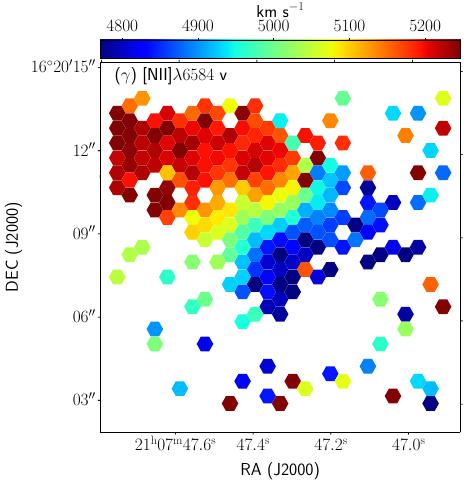}
	\includegraphics[clip, width=0.24\linewidth]{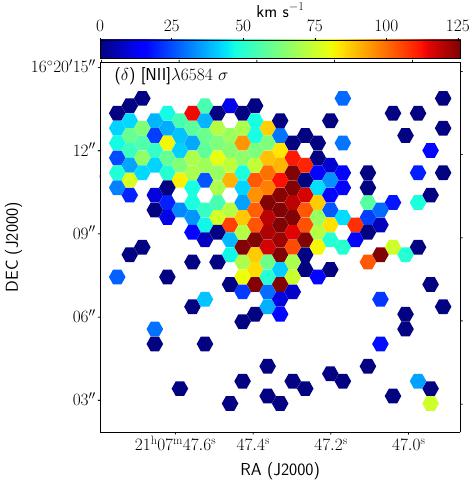}
	\includegraphics[clip, width=0.24\linewidth]{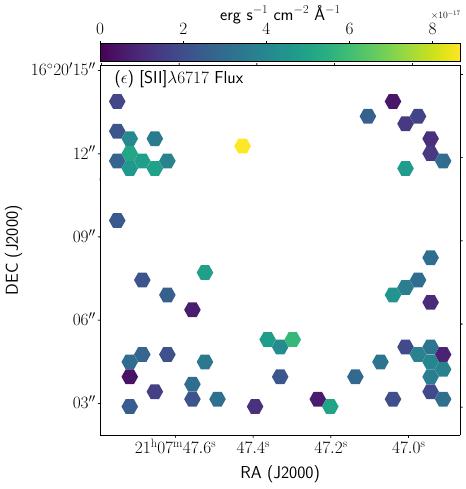}
	\includegraphics[clip, width=0.24\linewidth]{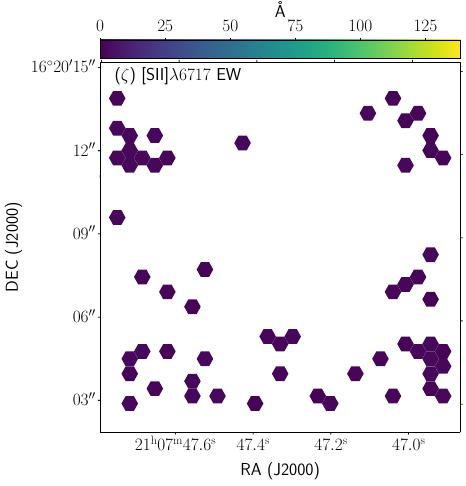}
	\includegraphics[clip, width=0.24\linewidth]{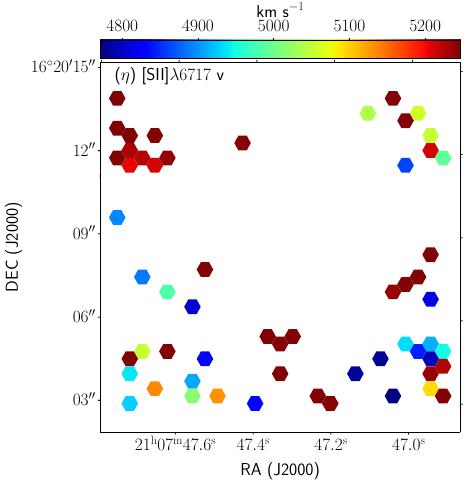}
	\includegraphics[clip, width=0.24\linewidth]{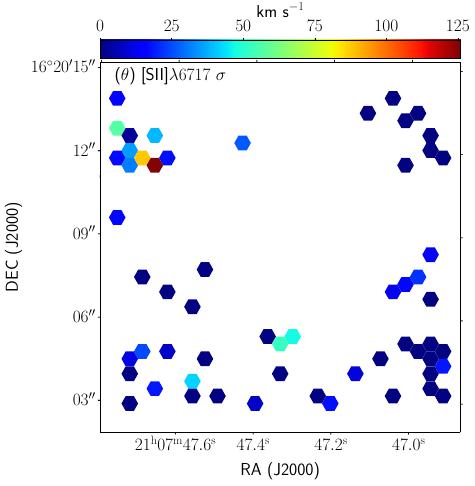}
	\includegraphics[clip, width=0.24\linewidth]{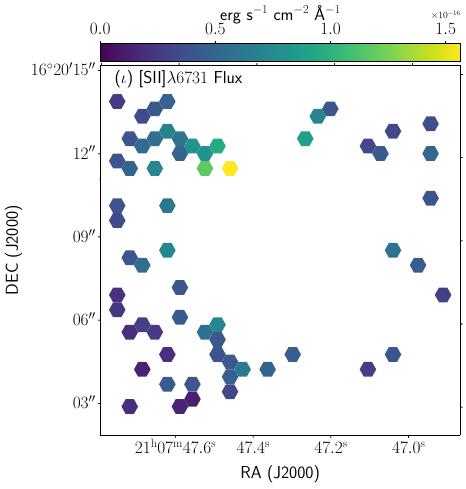}
	\includegraphics[clip, width=0.24\linewidth]{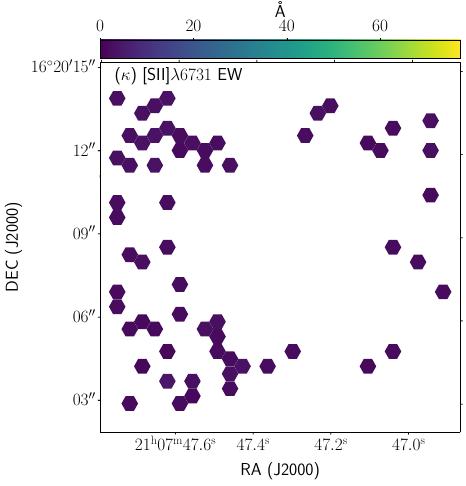}
	\includegraphics[clip, width=0.24\linewidth]{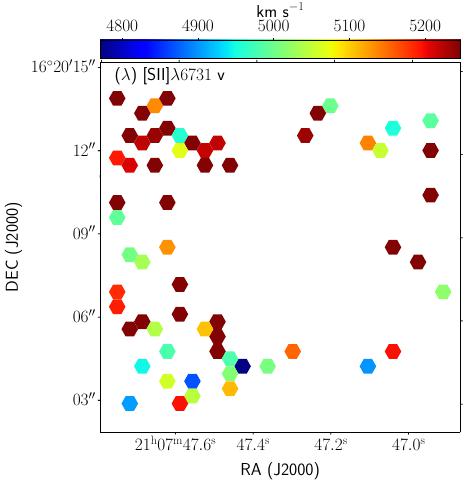}
	\includegraphics[clip, width=0.24\linewidth]{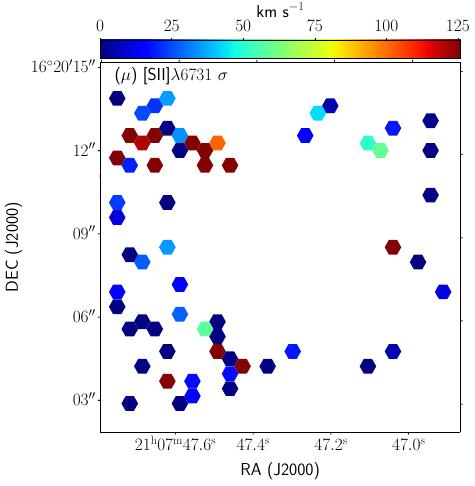}
	\caption{(cont.) NGC~7025 card.}
	\label{fig:NGC7025_card_2}
\end{figure*}

\begin{figure*}[h]
	\centering
	\includegraphics[clip, width=0.35\linewidth]{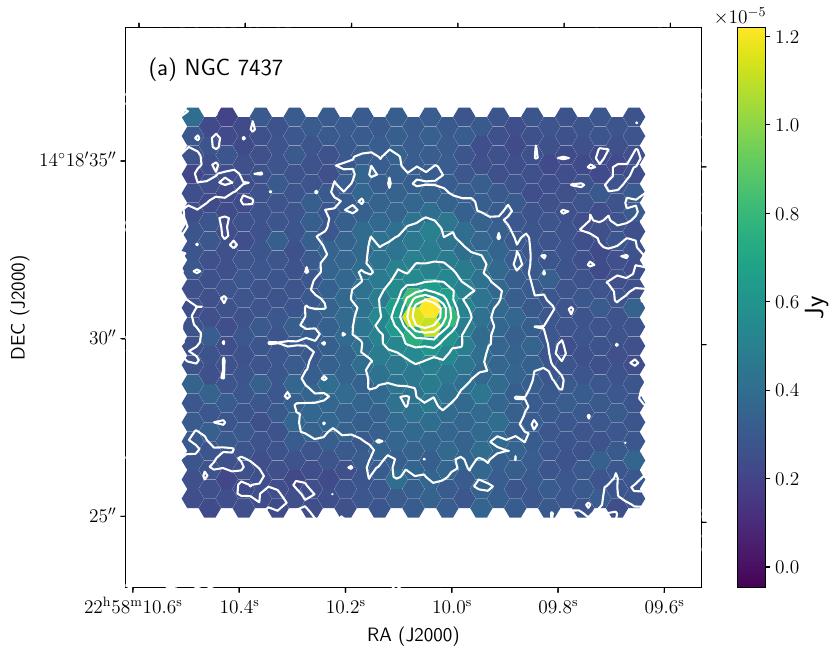}
	\includegraphics[clip, width=0.6\linewidth]{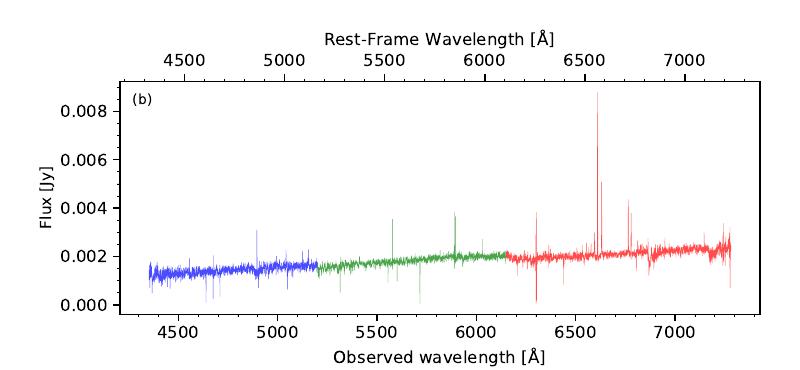}
	\includegraphics[clip, width=0.24\linewidth]{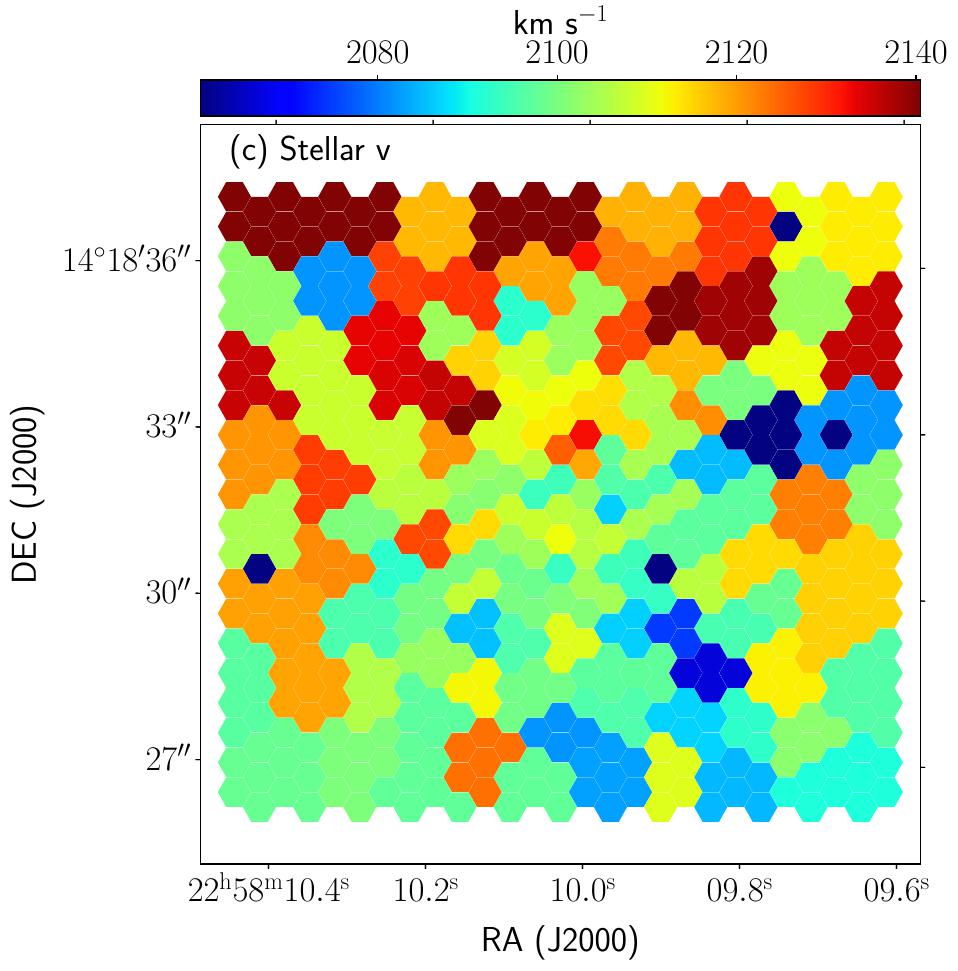}
	\includegraphics[clip, width=0.24\linewidth]{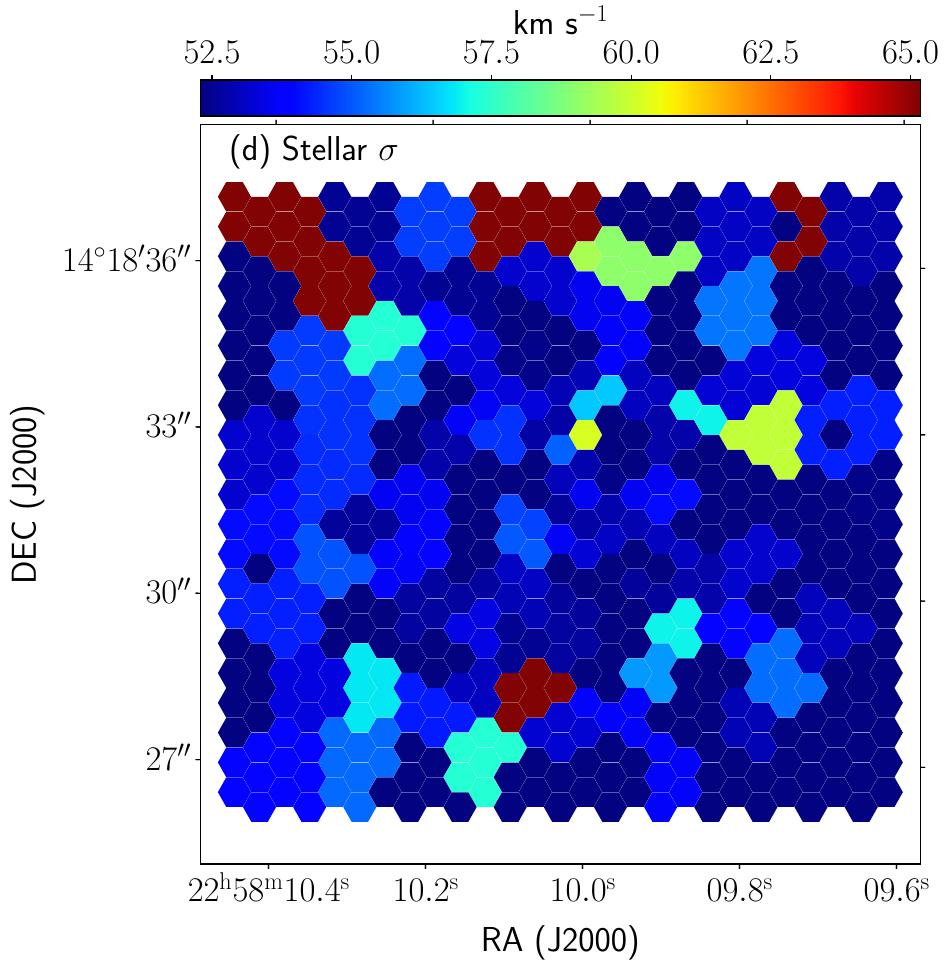}
	\includegraphics[clip, width=0.24\linewidth]{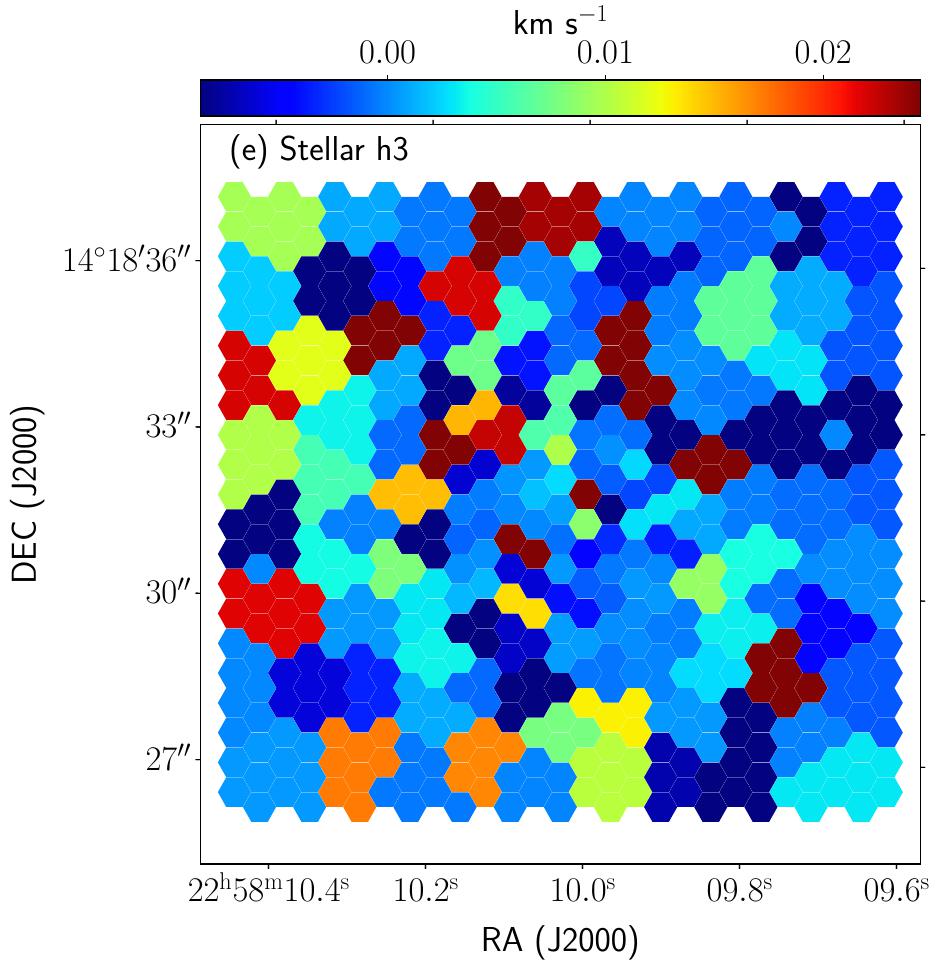}
	\includegraphics[clip, width=0.24\linewidth]{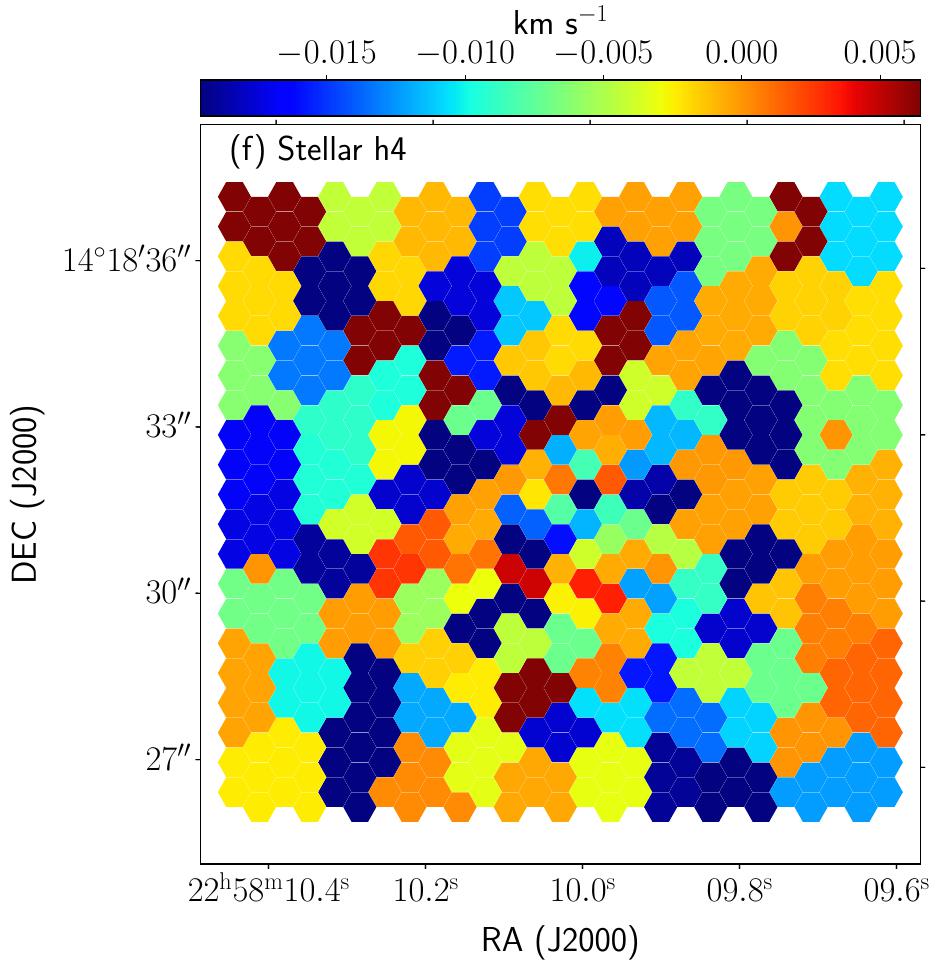}
	\includegraphics[clip, width=0.24\linewidth]{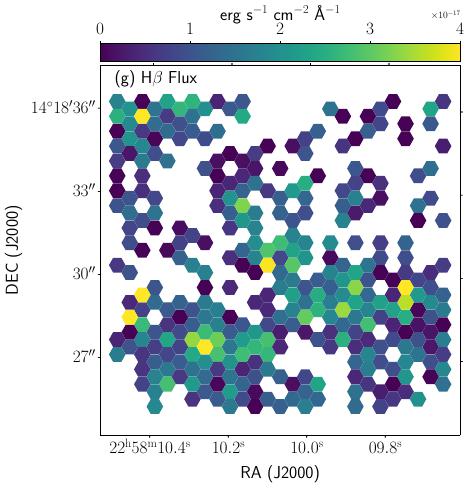}
	\includegraphics[clip, width=0.24\linewidth]{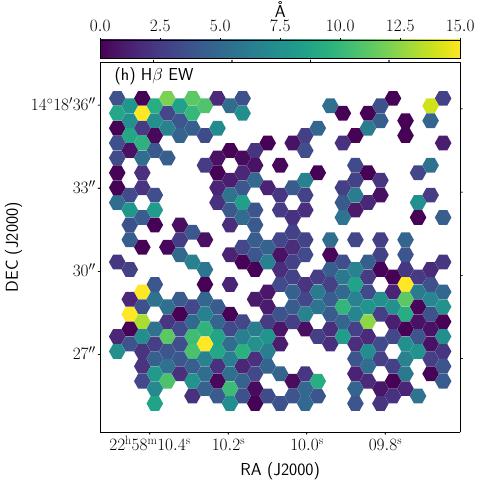}
	\includegraphics[clip, width=0.24\linewidth]{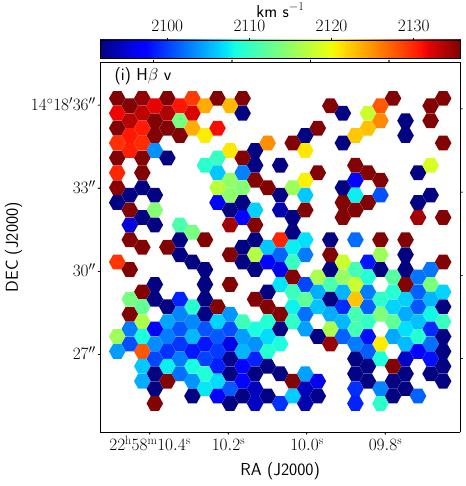}
	\includegraphics[clip, width=0.24\linewidth]{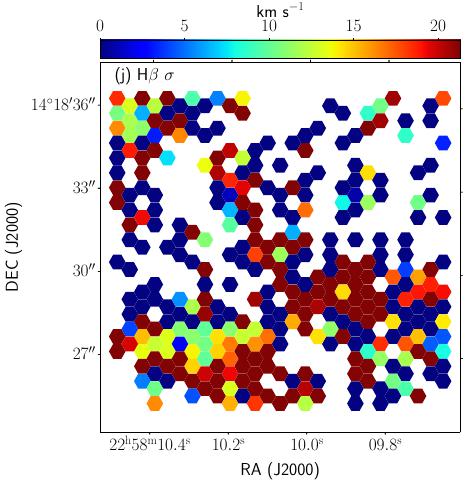}
	\includegraphics[clip, width=0.24\linewidth]{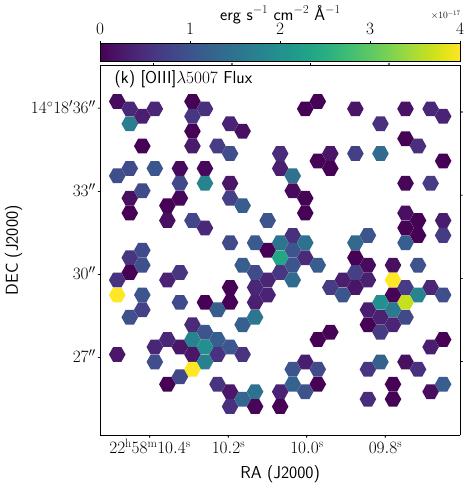}
	\includegraphics[clip, width=0.24\linewidth]{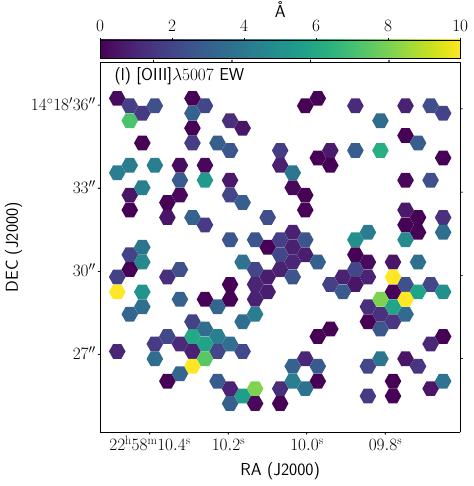}
	\includegraphics[clip, width=0.24\linewidth]{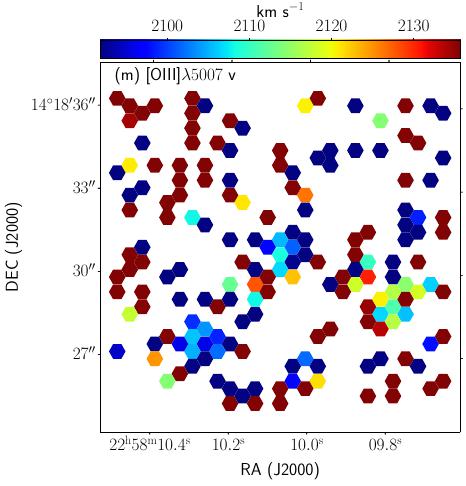}
	\includegraphics[clip, width=0.24\linewidth]{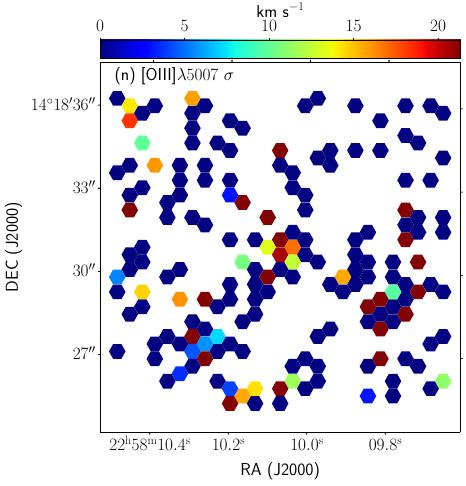}
	\vspace{5cm}
	\caption{NGC~7437 card.}
	\label{fig:NGC7437_card_1}
\end{figure*}
\addtocounter{figure}{-1}
\begin{figure*}[h]
	\centering
	\includegraphics[clip, width=0.24\linewidth]{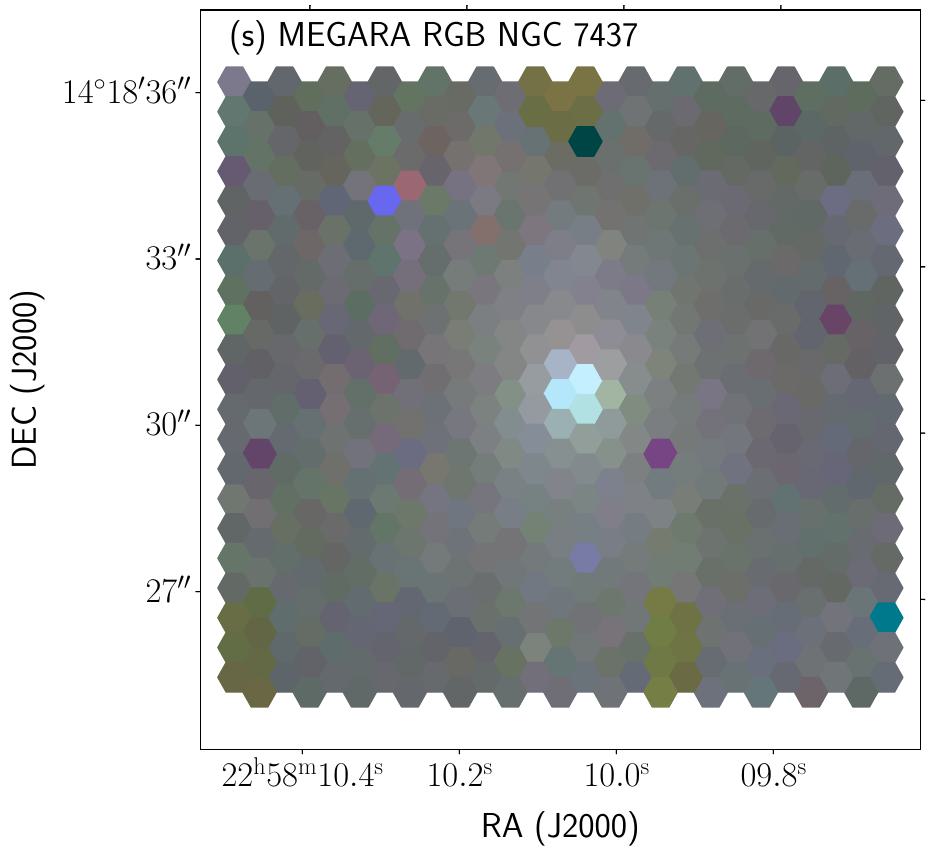}
	\includegraphics[clip, width=0.24\linewidth]{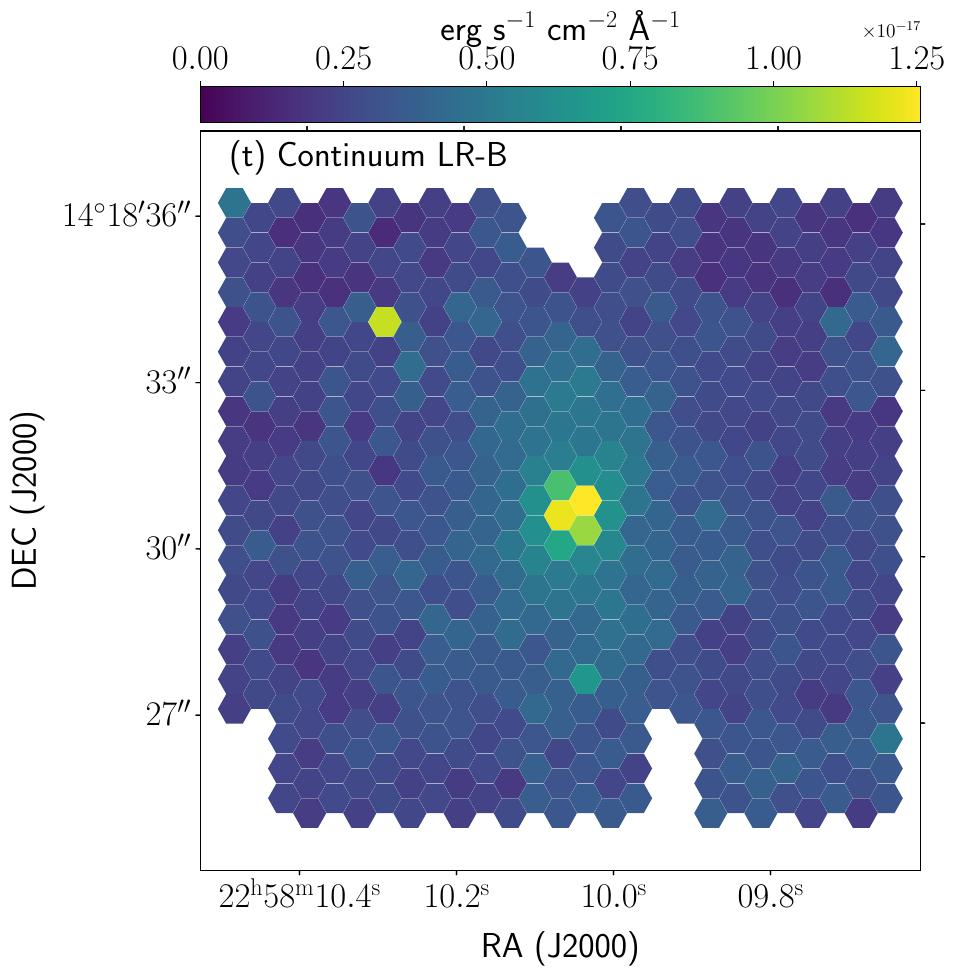}
	\includegraphics[clip, width=0.24\linewidth]{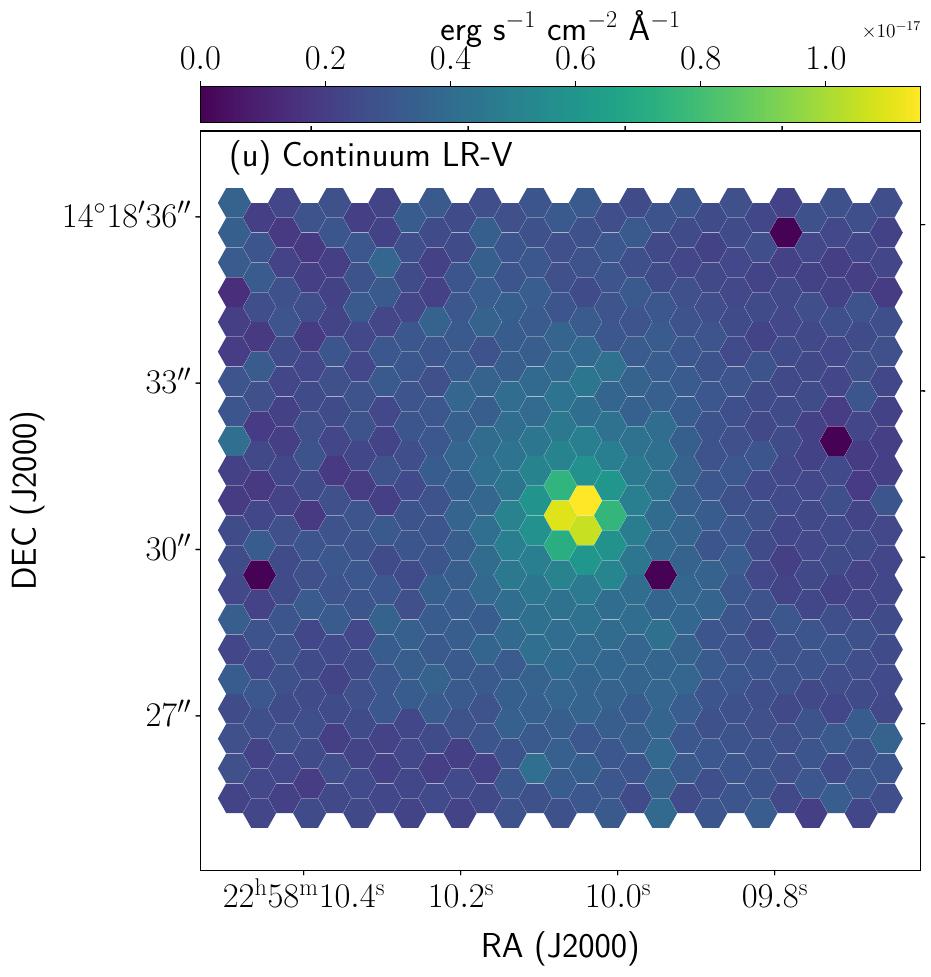}
	\includegraphics[clip, width=0.24\linewidth]{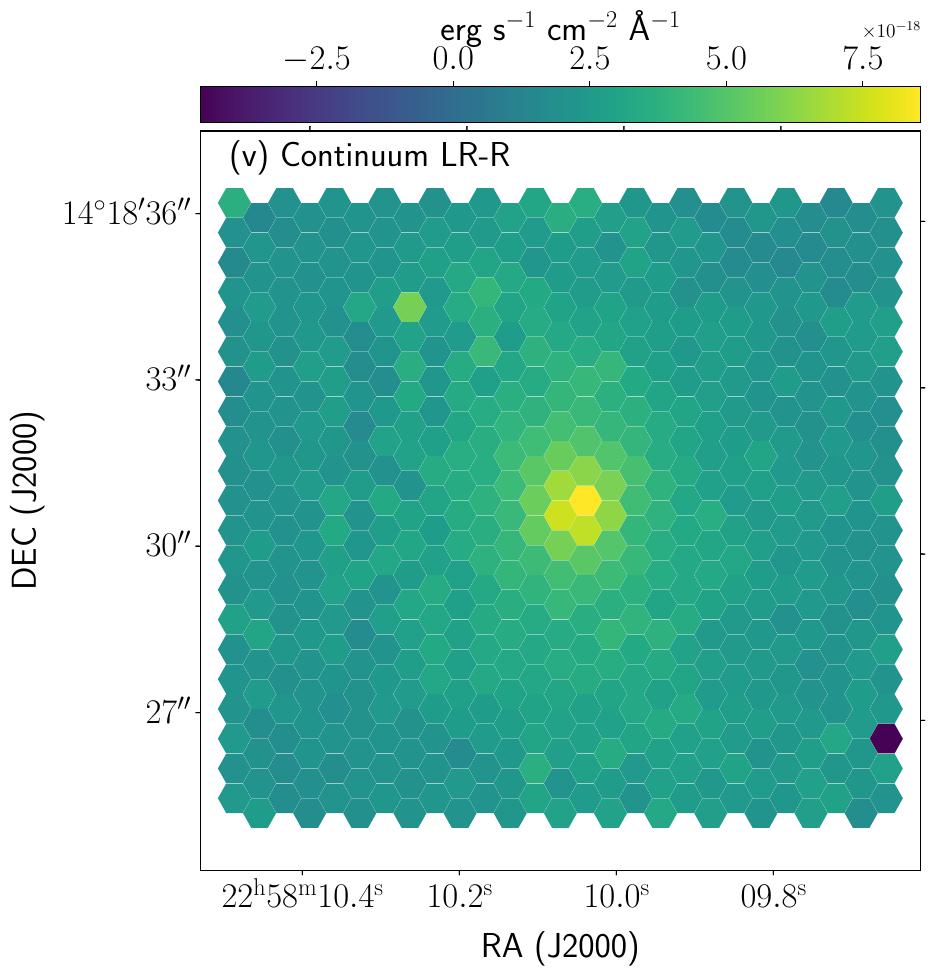}
	\includegraphics[clip, width=0.24\linewidth]{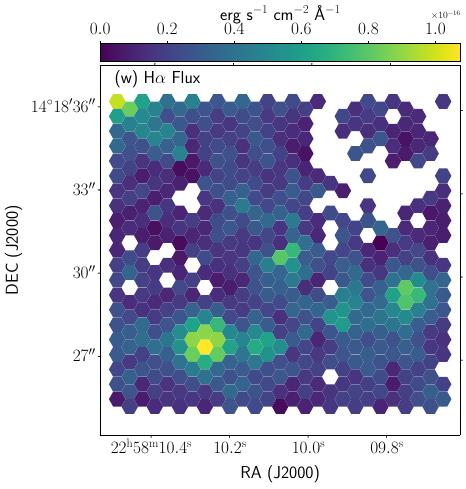}
	\includegraphics[clip, width=0.24\linewidth]{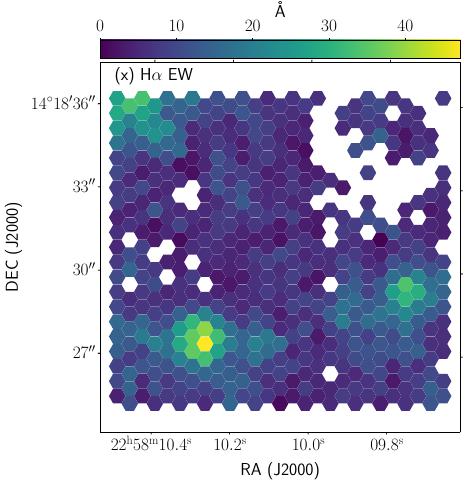}
	\includegraphics[clip, width=0.24\linewidth]{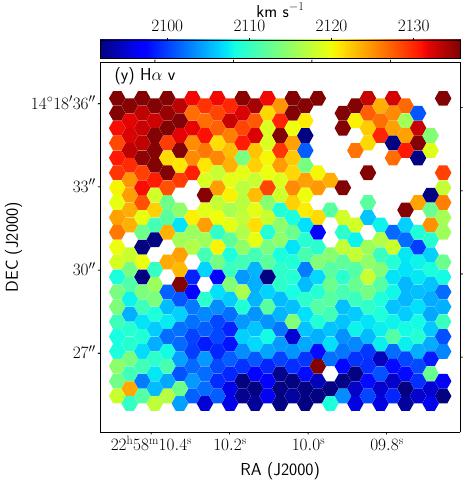}
	\includegraphics[clip, width=0.24\linewidth]{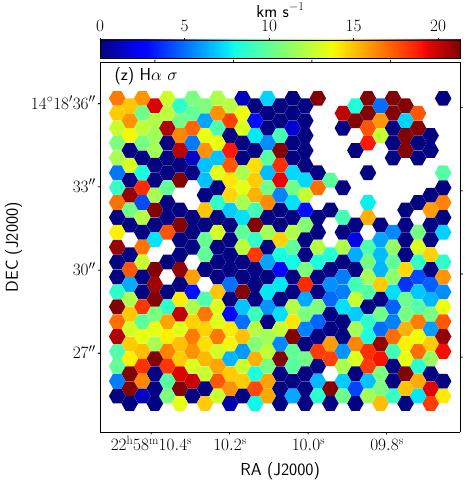}
	\includegraphics[clip, width=0.24\linewidth]{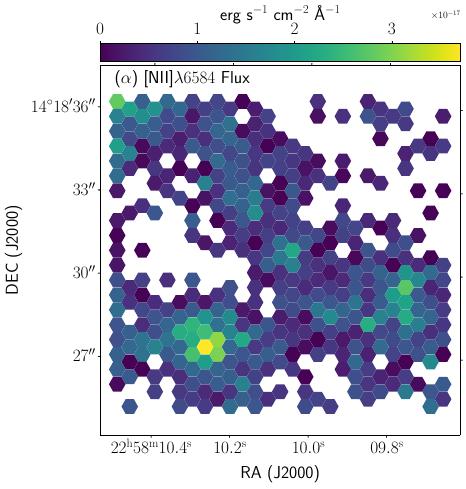}
	\includegraphics[clip, width=0.24\linewidth]{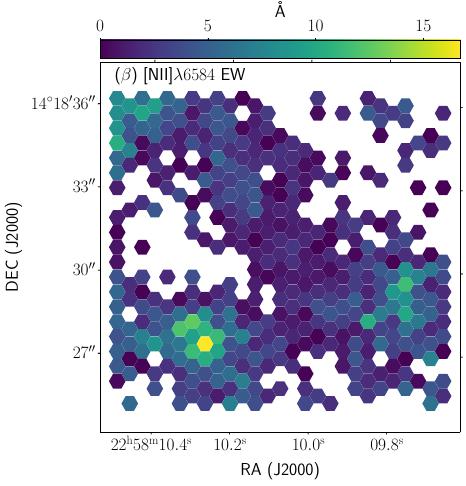}
	\includegraphics[clip, width=0.24\linewidth]{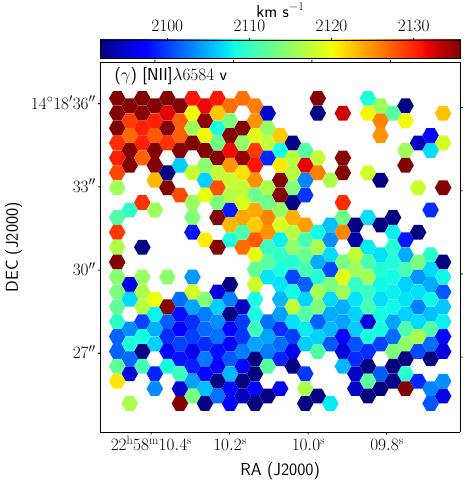}
	\includegraphics[clip, width=0.24\linewidth]{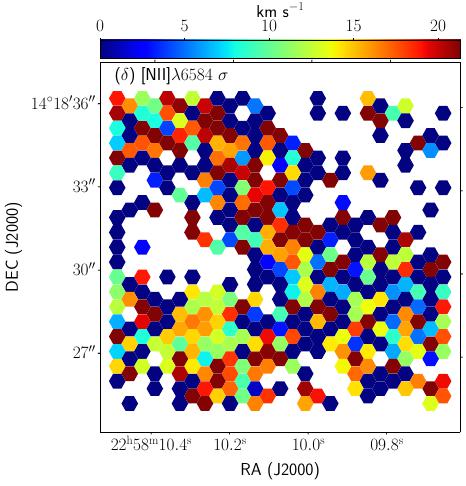}
	\includegraphics[clip, width=0.24\linewidth]{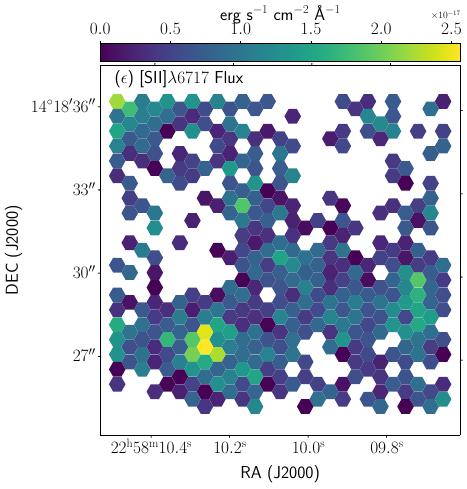}
	\includegraphics[clip, width=0.24\linewidth]{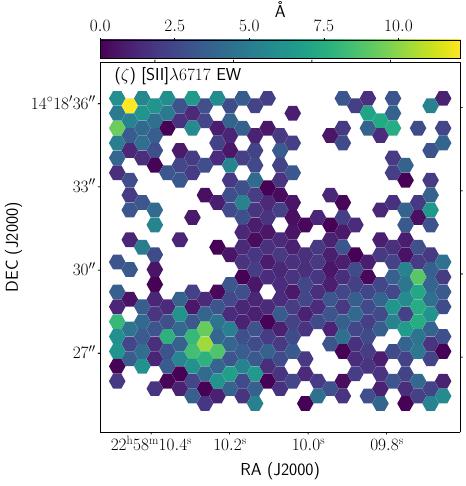}
	\includegraphics[clip, width=0.24\linewidth]{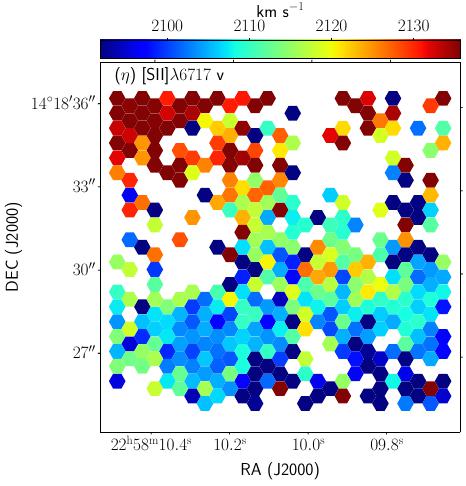}
	\includegraphics[clip, width=0.24\linewidth]{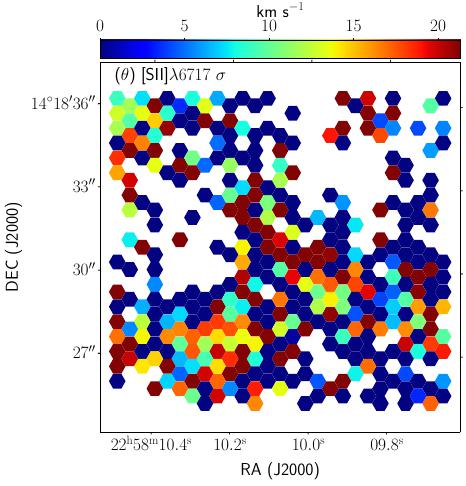}
	\includegraphics[clip, width=0.24\linewidth]{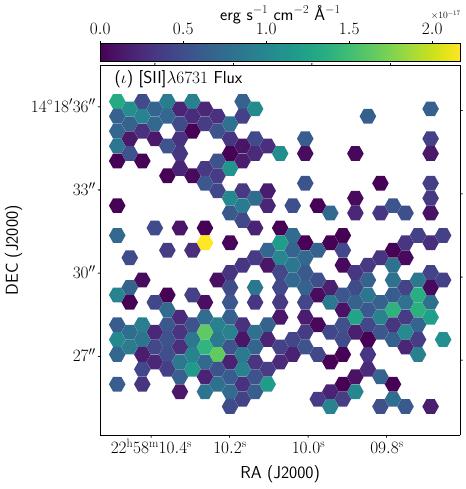}
	\includegraphics[clip, width=0.24\linewidth]{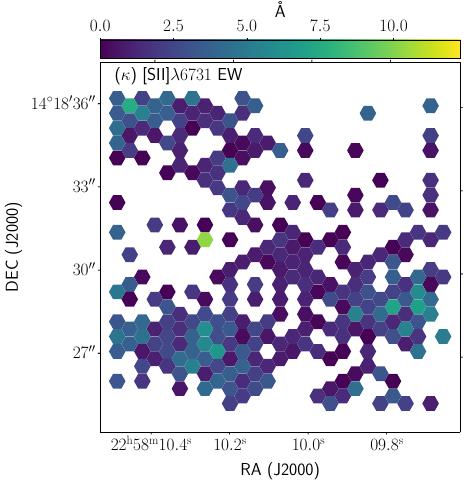}
	\includegraphics[clip, width=0.24\linewidth]{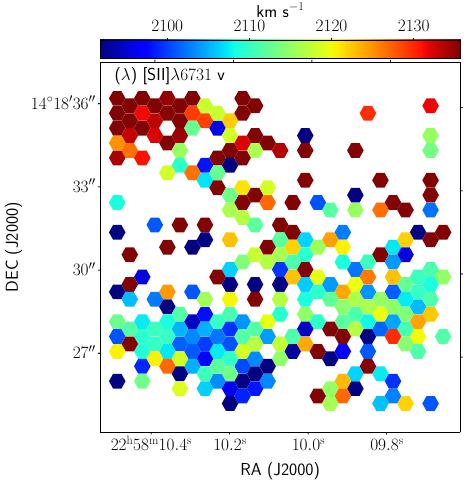}
	\includegraphics[clip, width=0.24\linewidth]{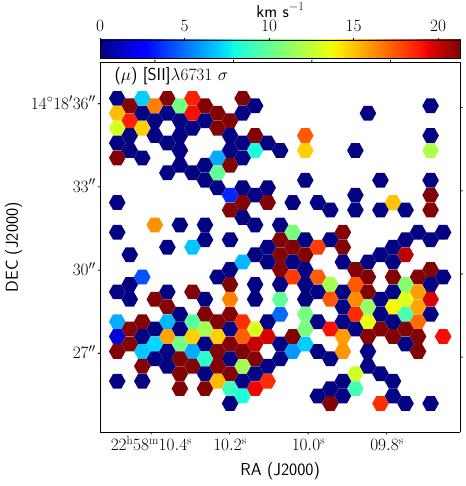}
	\caption{(cont.) NGC~7437 card.}
	\label{fig:NGC7437_card_2}
\end{figure*}

\begin{figure*}[h]
	\centering
	\includegraphics[clip, width=0.35\linewidth]{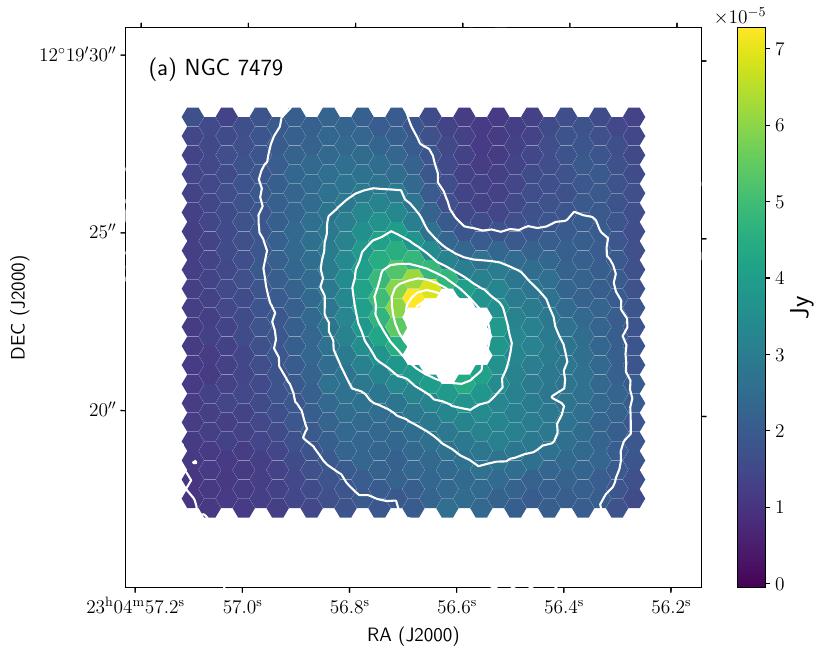}
	\includegraphics[clip, width=0.6\linewidth]{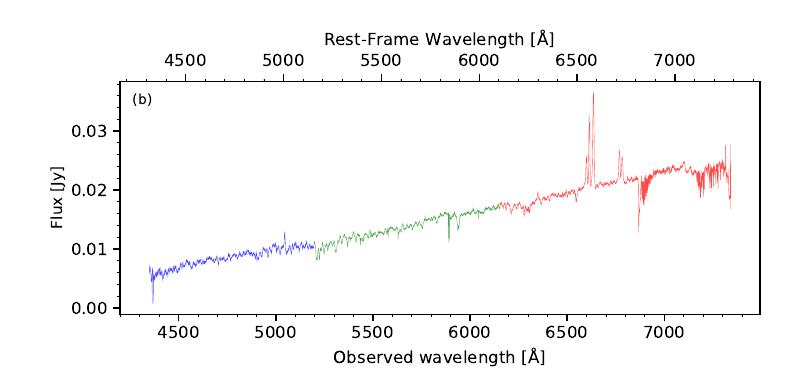}
	\includegraphics[clip, width=0.24\linewidth]{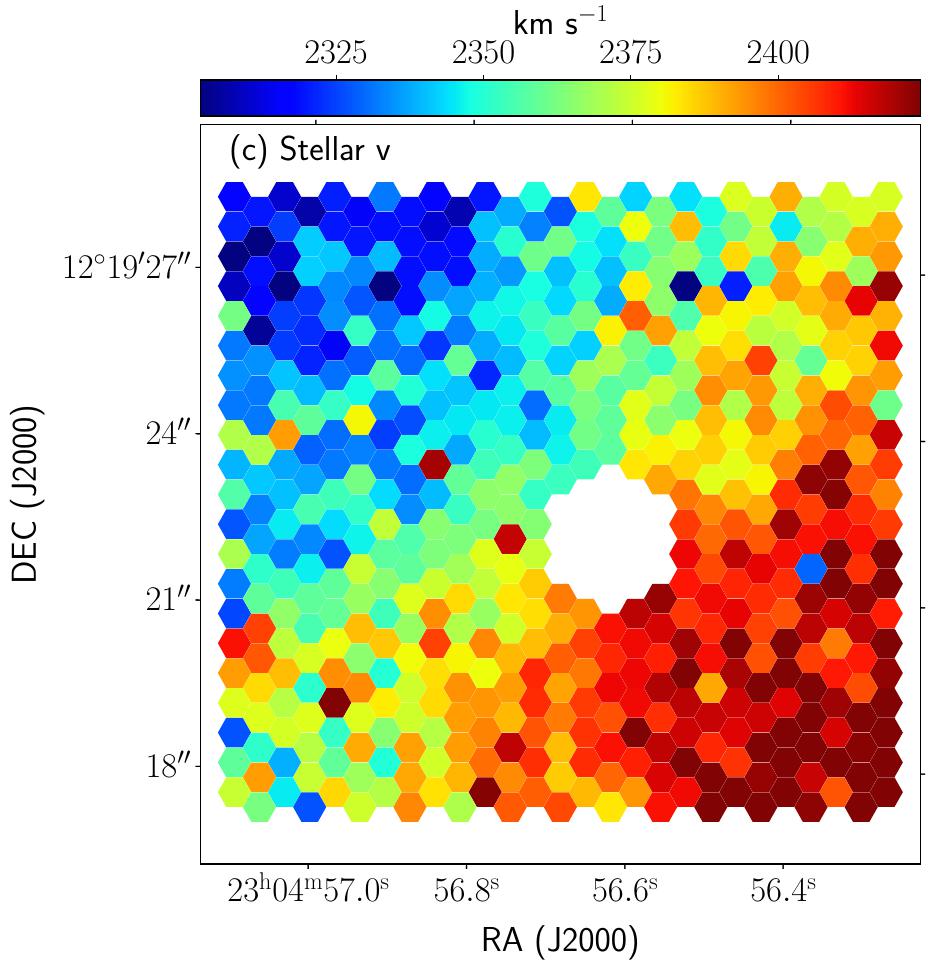}
	\includegraphics[clip, width=0.24\linewidth]{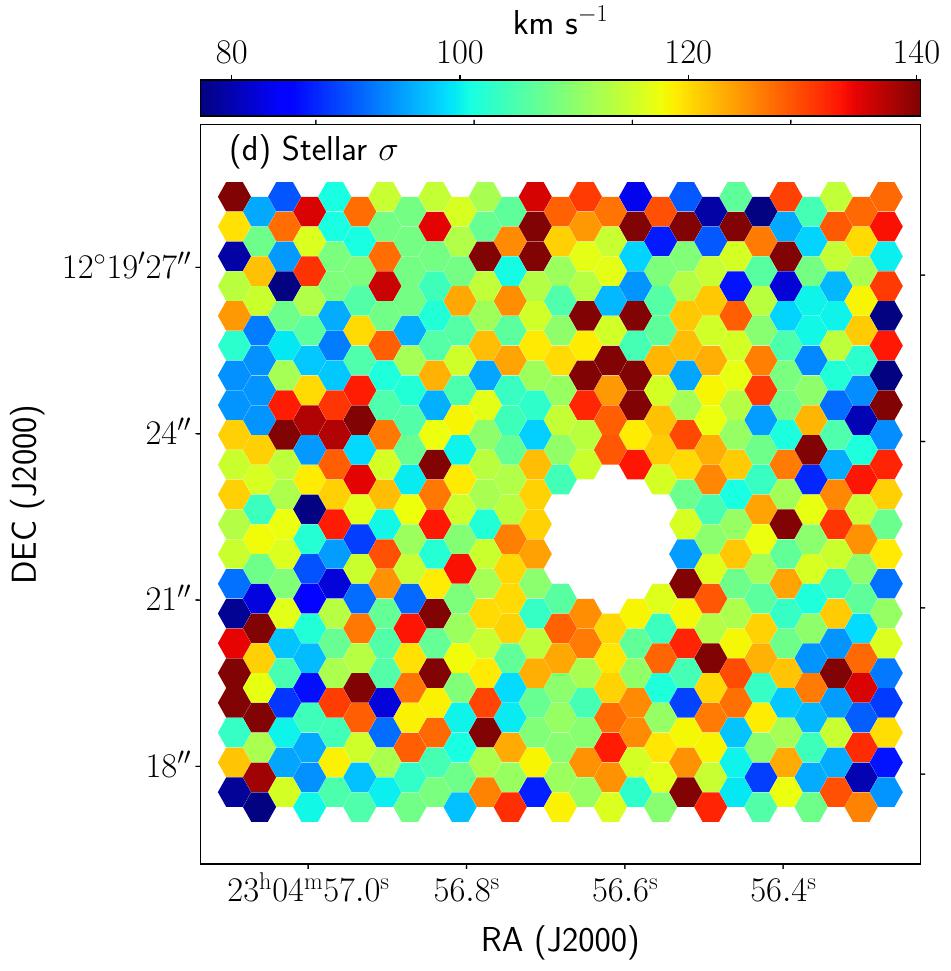}
	\includegraphics[clip, width=0.24\linewidth]{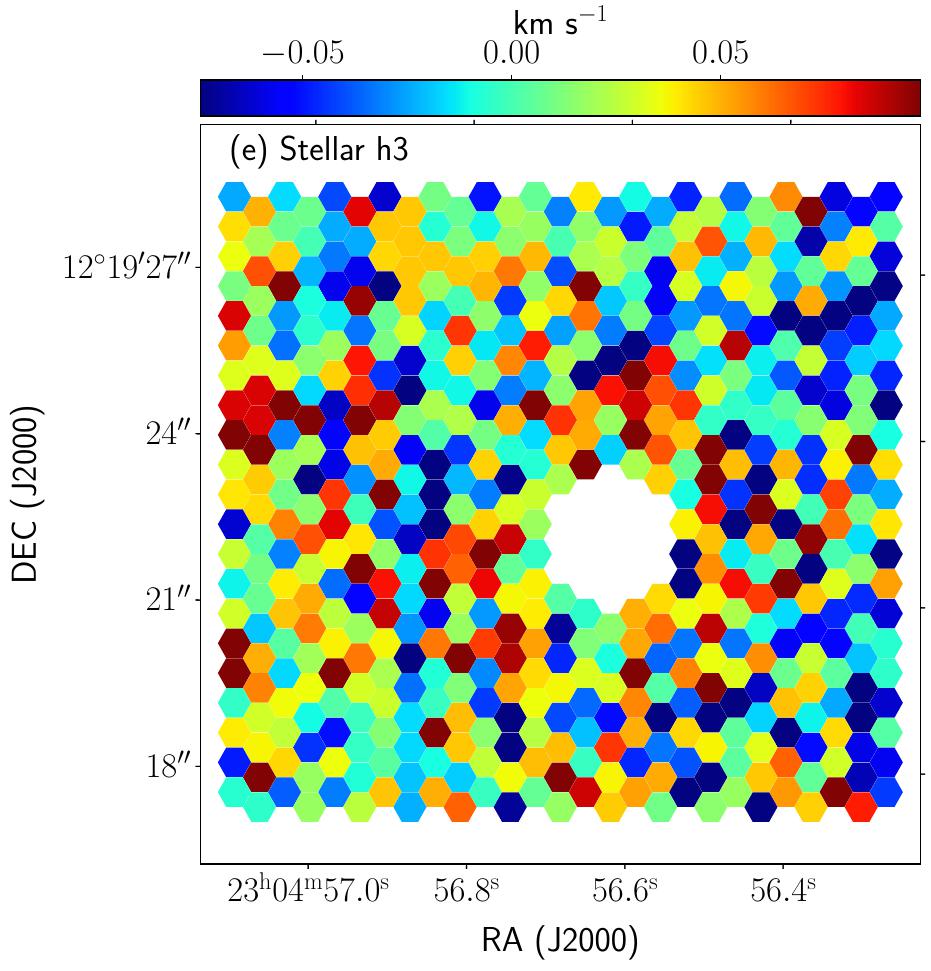}
	\includegraphics[clip, width=0.24\linewidth]{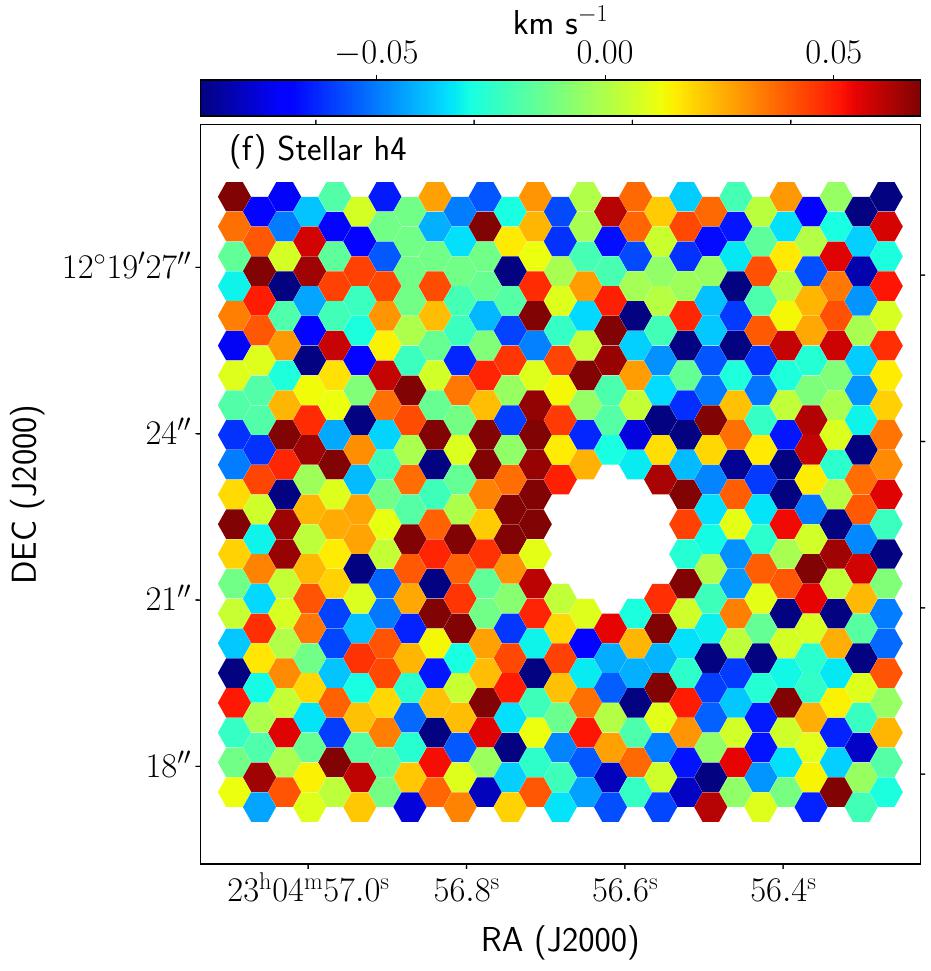}
	\includegraphics[clip, width=0.24\linewidth]{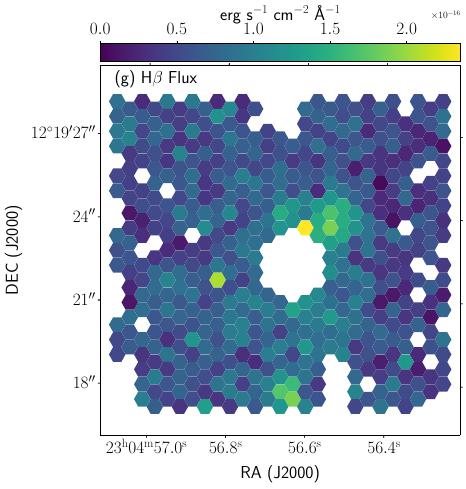}
	\includegraphics[clip, width=0.24\linewidth]{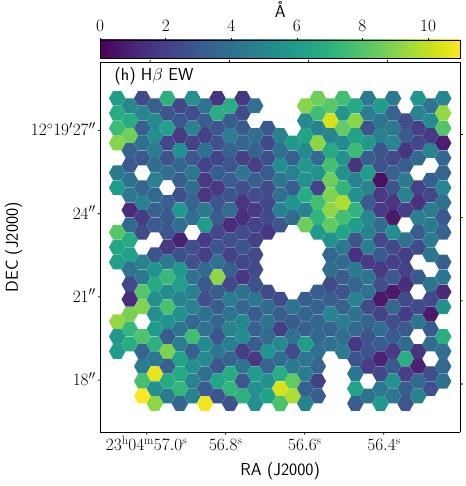}
	\includegraphics[clip, width=0.24\linewidth]{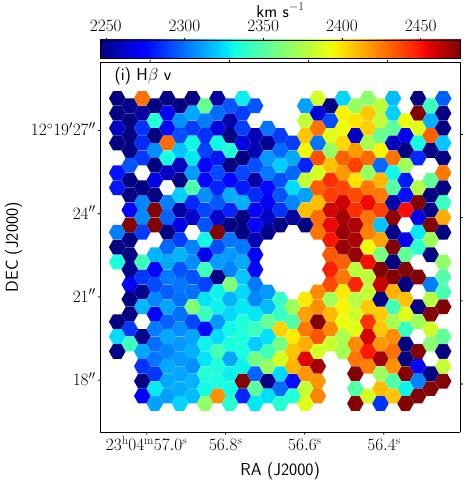}
	\includegraphics[clip, width=0.24\linewidth]{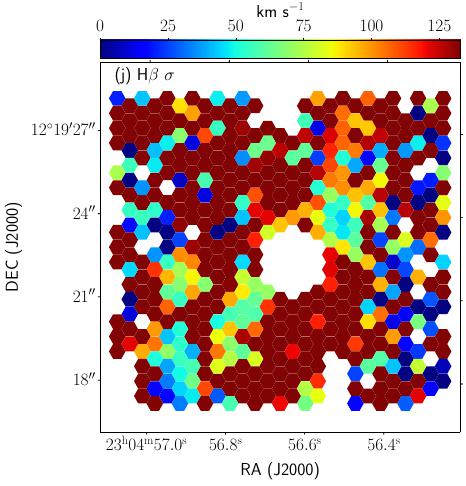}
	\includegraphics[clip, width=0.24\linewidth]{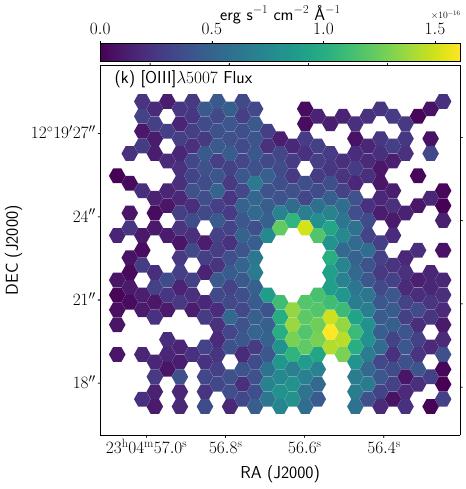}
	\includegraphics[clip, width=0.24\linewidth]{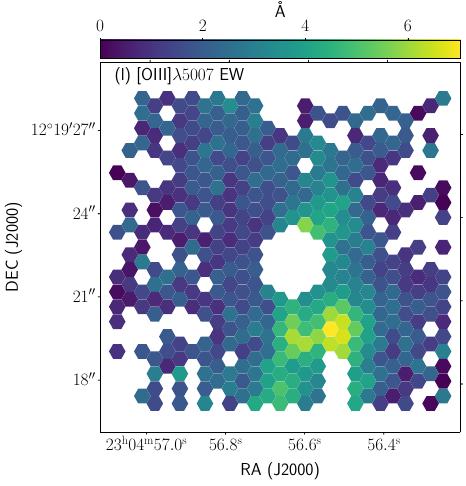}
	\includegraphics[clip, width=0.24\linewidth]{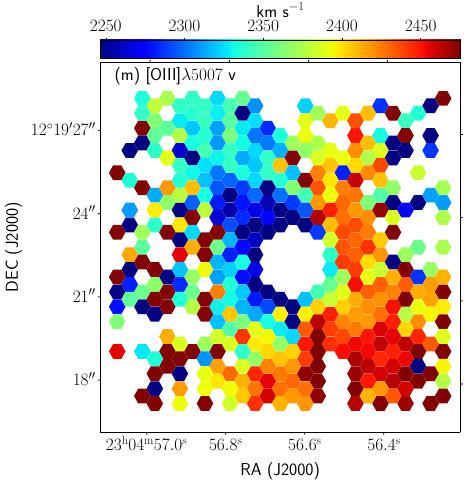}
	\includegraphics[clip, width=0.24\linewidth]{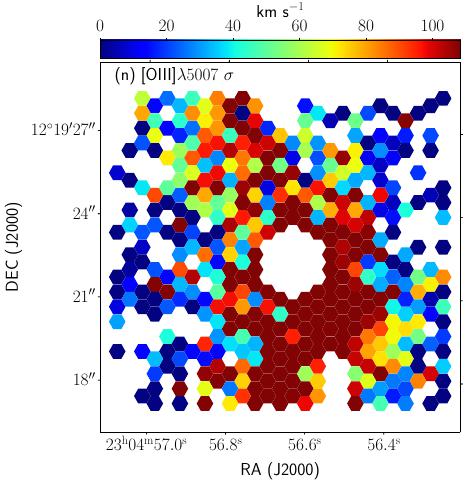}
	\vspace{5cm}
	\caption{NGC~7479 card.}
	\label{fig:NGC7479_card_1}
\end{figure*}
\addtocounter{figure}{-1}
\begin{figure*}[h]
	\centering
	\includegraphics[clip, width=0.24\linewidth]{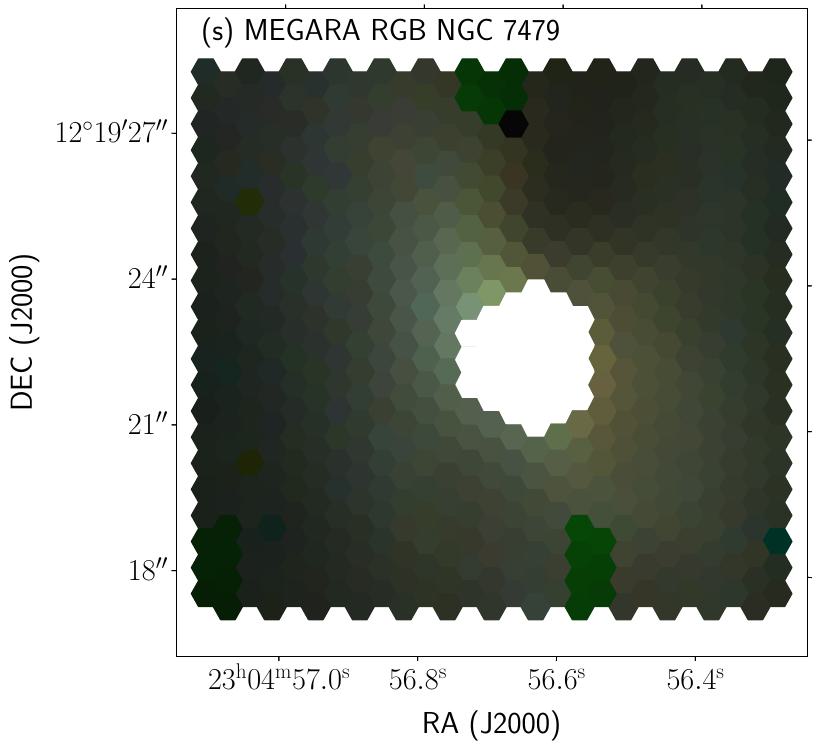}
	\includegraphics[clip, width=0.24\linewidth]{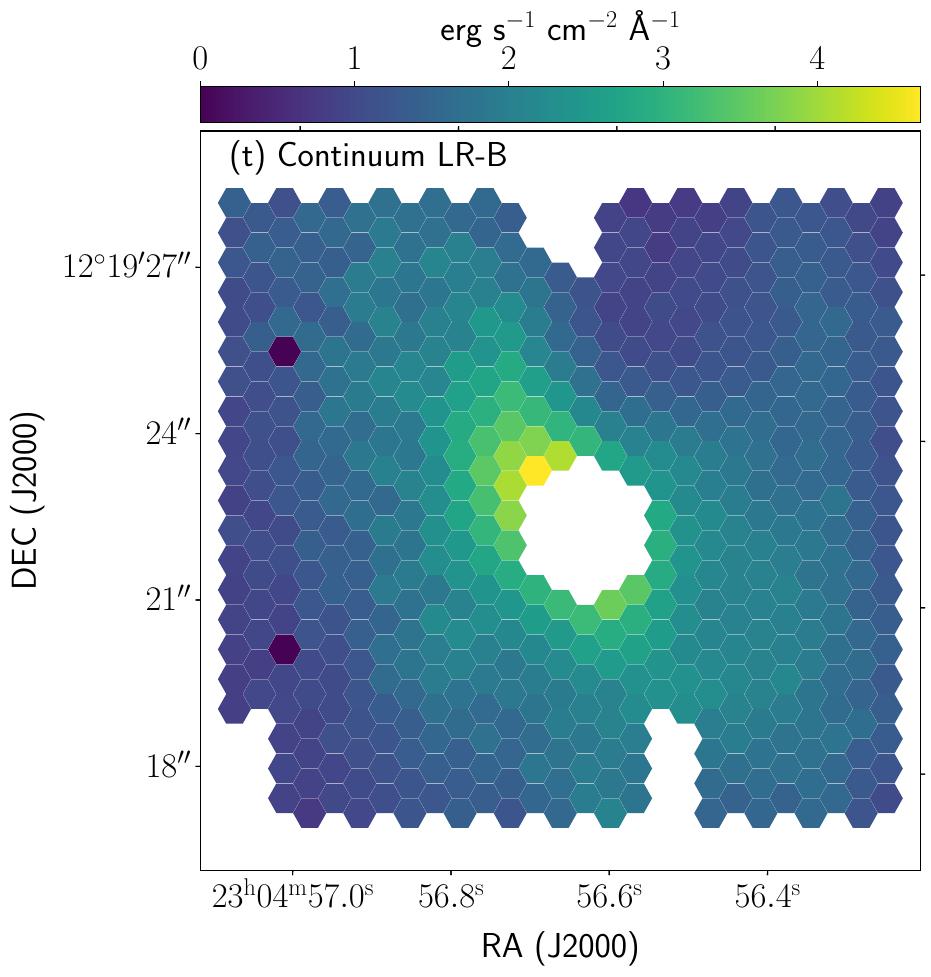}
	\includegraphics[clip, width=0.24\linewidth]{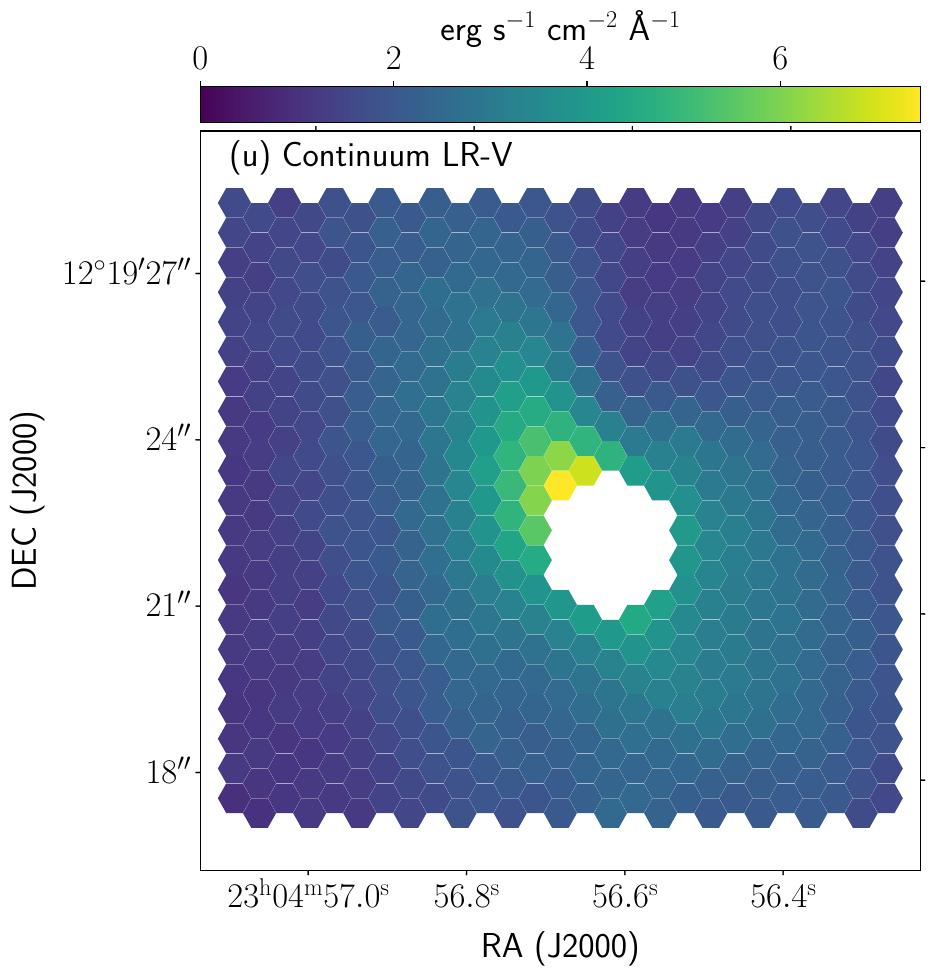}
	\includegraphics[clip, width=0.24\linewidth]{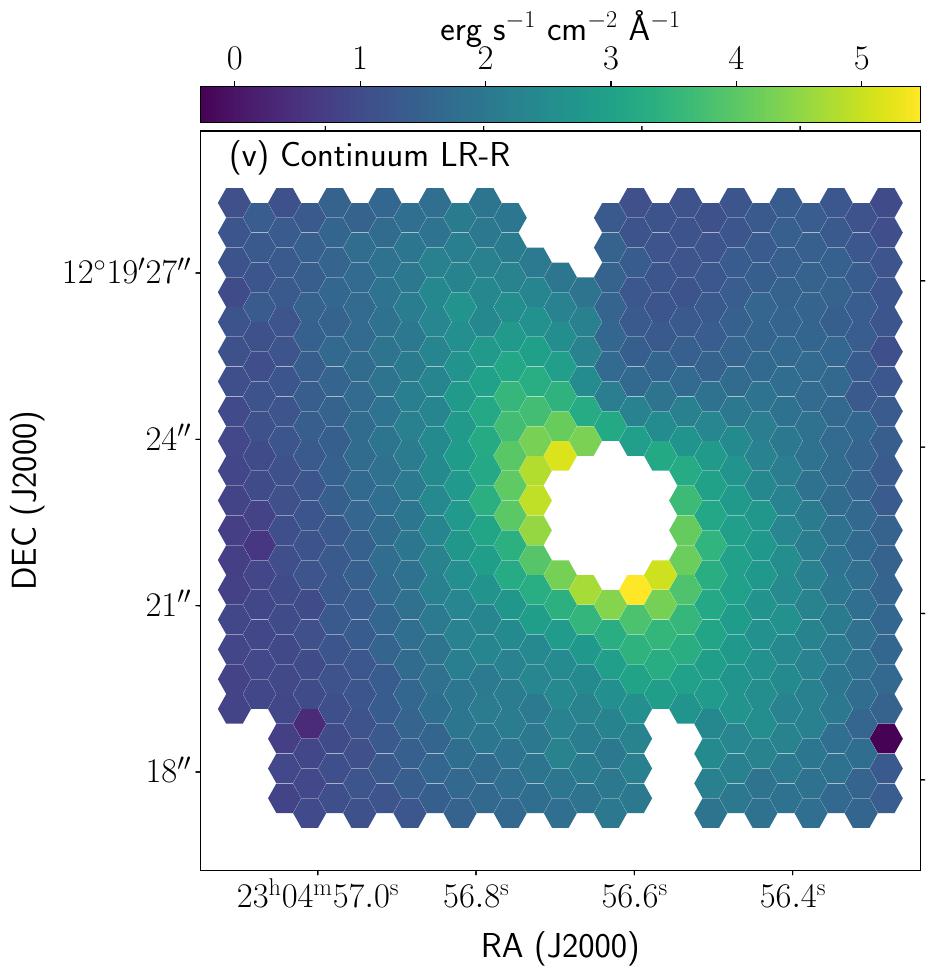}
	\includegraphics[clip, width=0.24\linewidth]{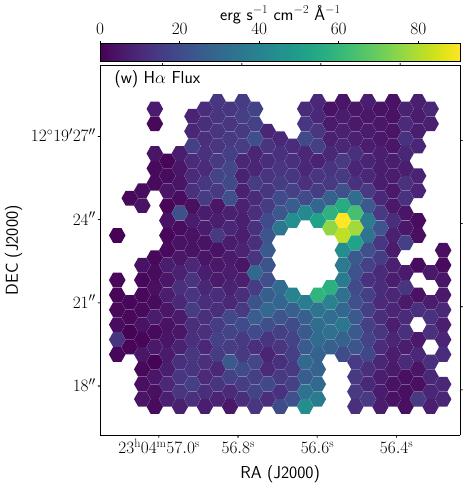}
	\includegraphics[clip, width=0.24\linewidth]{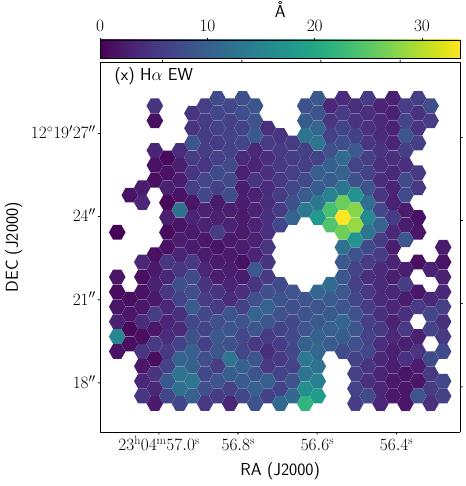}
	\includegraphics[clip, width=0.24\linewidth]{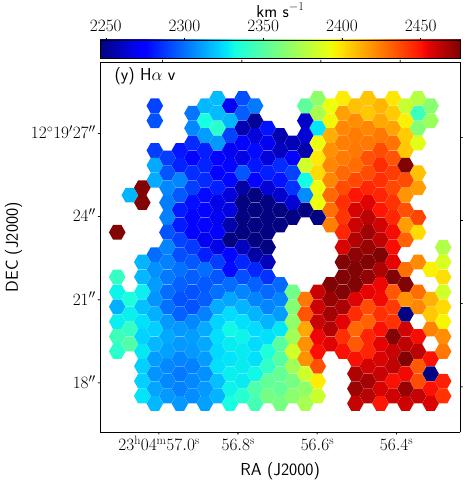}
	\includegraphics[clip, width=0.24\linewidth]{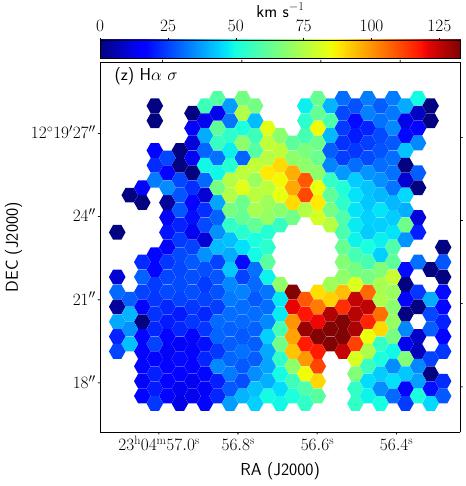}
	\includegraphics[clip, width=0.24\linewidth]{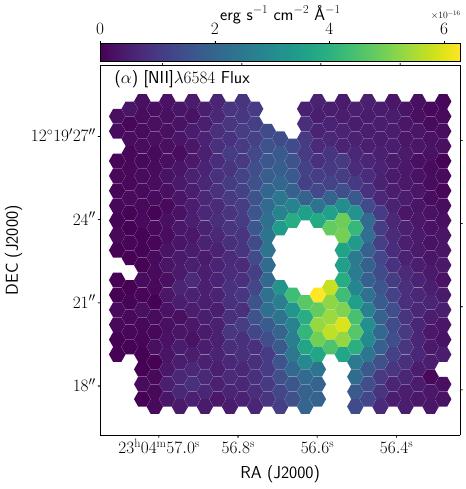}
	\includegraphics[clip, width=0.24\linewidth]{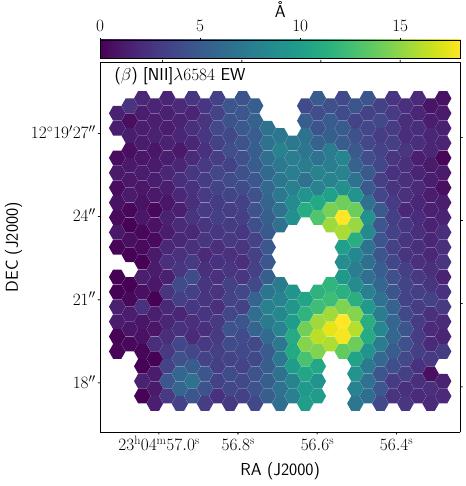}
	\includegraphics[clip, width=0.24\linewidth]{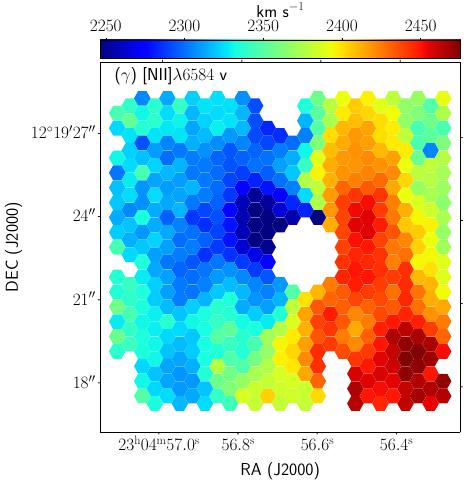}
	\includegraphics[clip, width=0.24\linewidth]{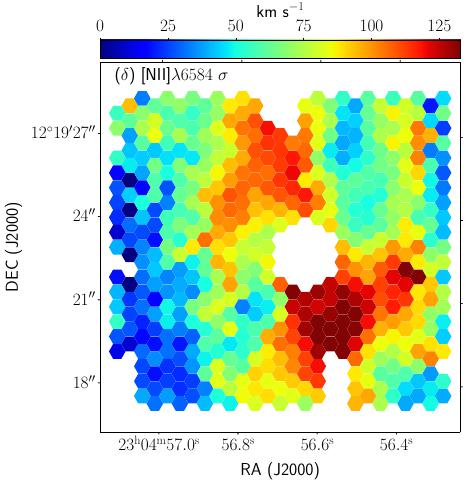}
	\includegraphics[clip, width=0.24\linewidth]{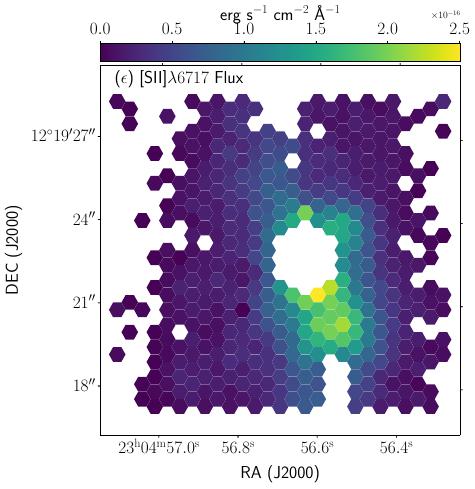}
	\includegraphics[clip, width=0.24\linewidth]{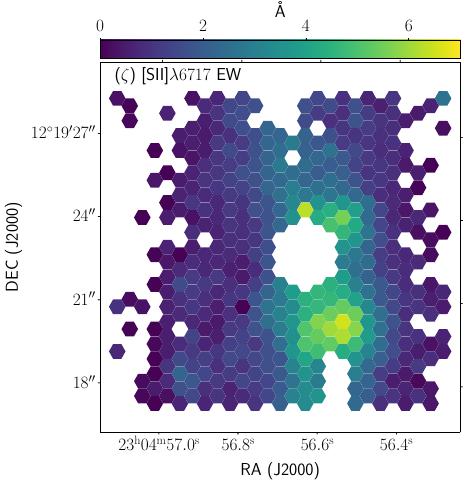}
	\includegraphics[clip, width=0.24\linewidth]{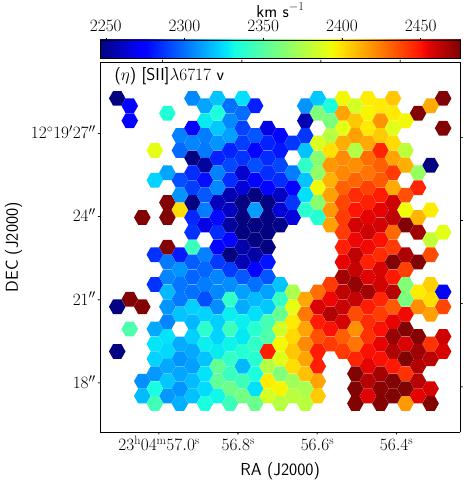}
	\includegraphics[clip, width=0.24\linewidth]{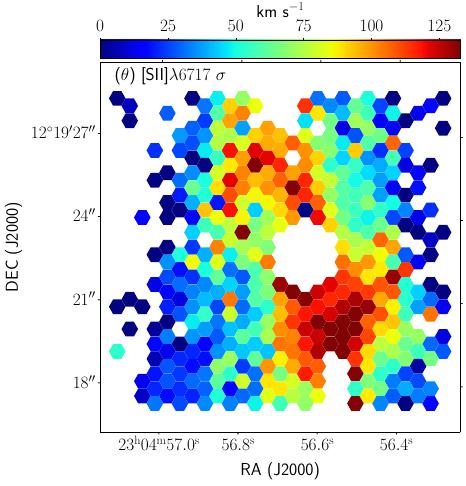}
	\includegraphics[clip, width=0.24\linewidth]{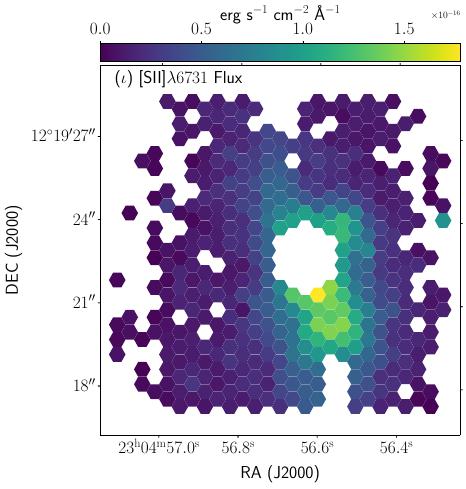}
	\includegraphics[clip, width=0.24\linewidth]{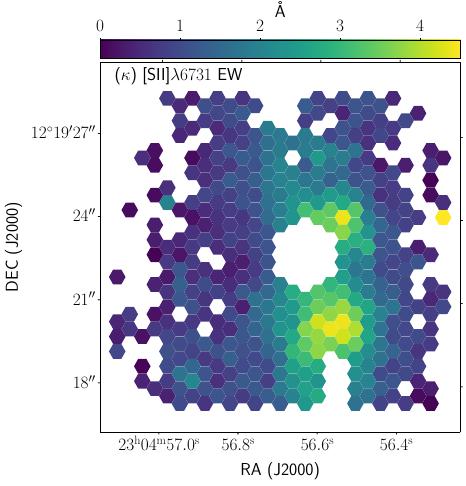}
	\includegraphics[clip, width=0.24\linewidth]{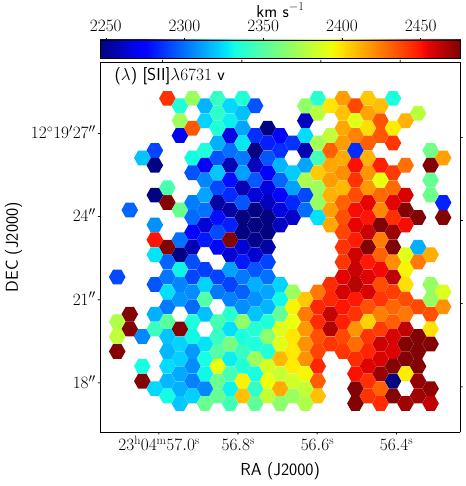}
	\includegraphics[clip, width=0.24\linewidth]{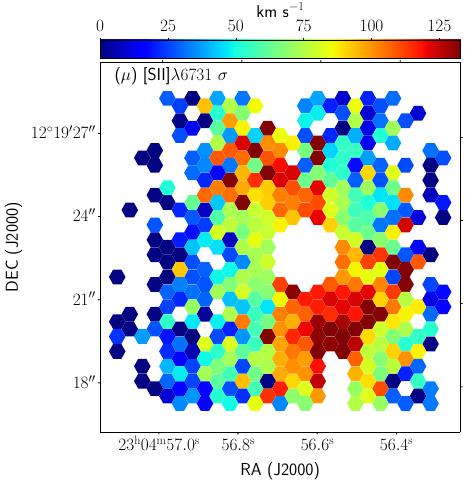}
	\caption{(cont.) NGC~7479 card.}
	\label{fig:NGC7479_card_2}
\end{figure*}

\begin{figure*}[h]
	\centering
	\includegraphics[clip, width=0.35\linewidth]{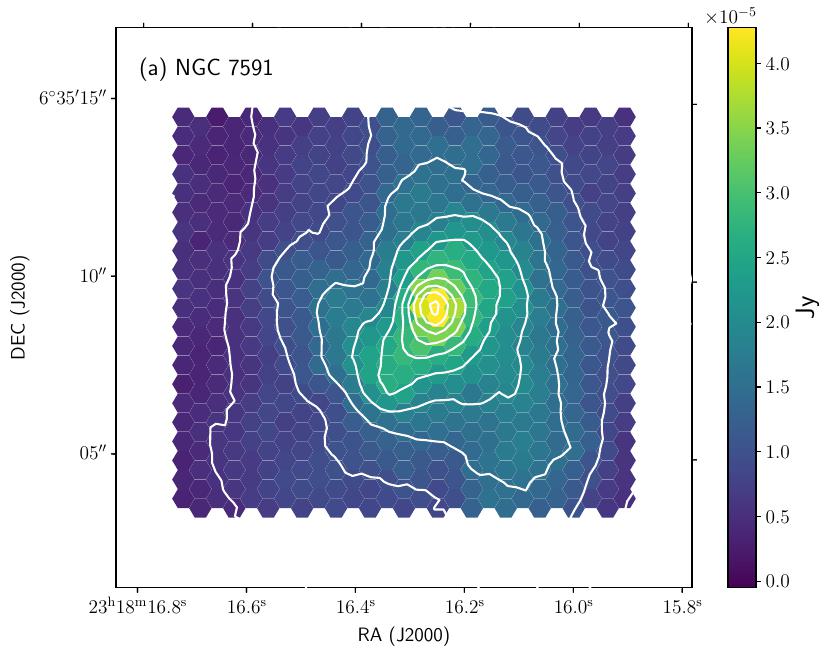}
	\includegraphics[clip, width=0.6\linewidth]{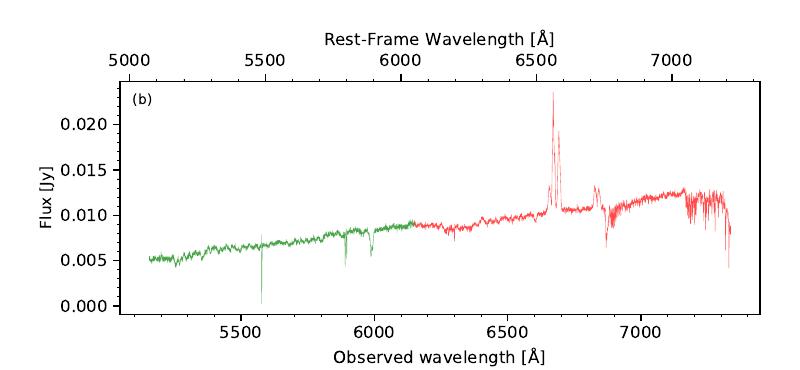}
	\includegraphics[clip, width=0.24\linewidth]{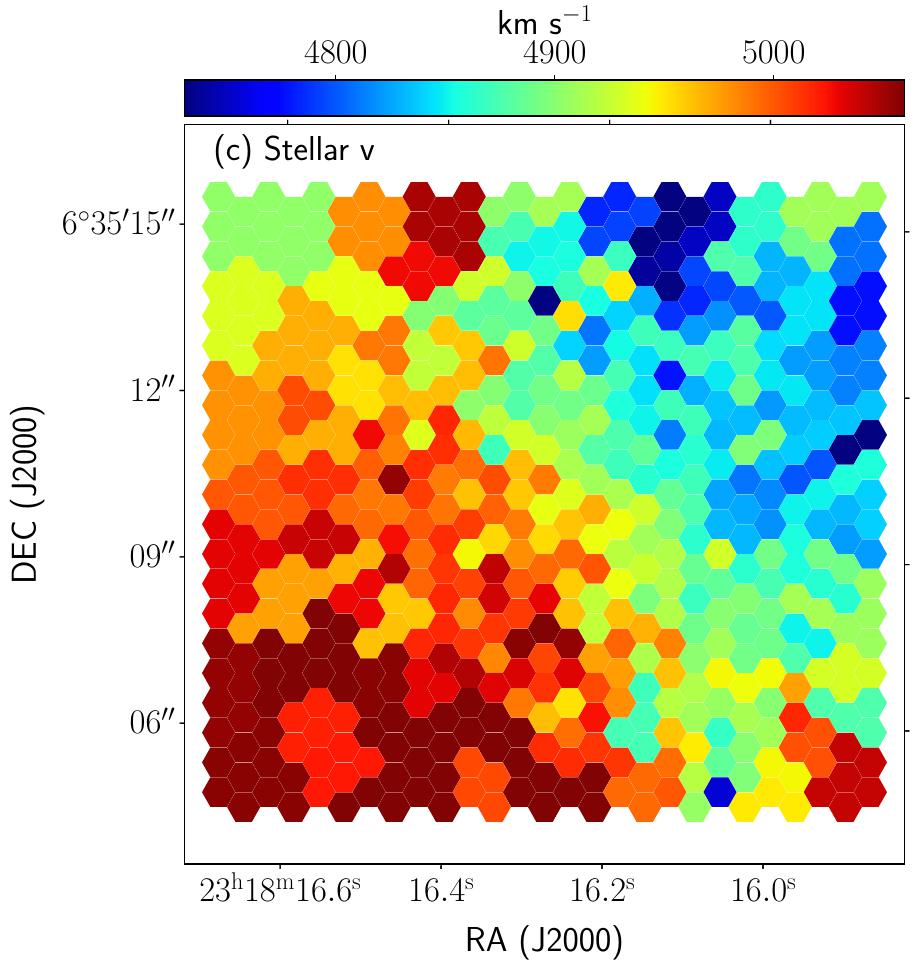}
	\includegraphics[clip, width=0.24\linewidth]{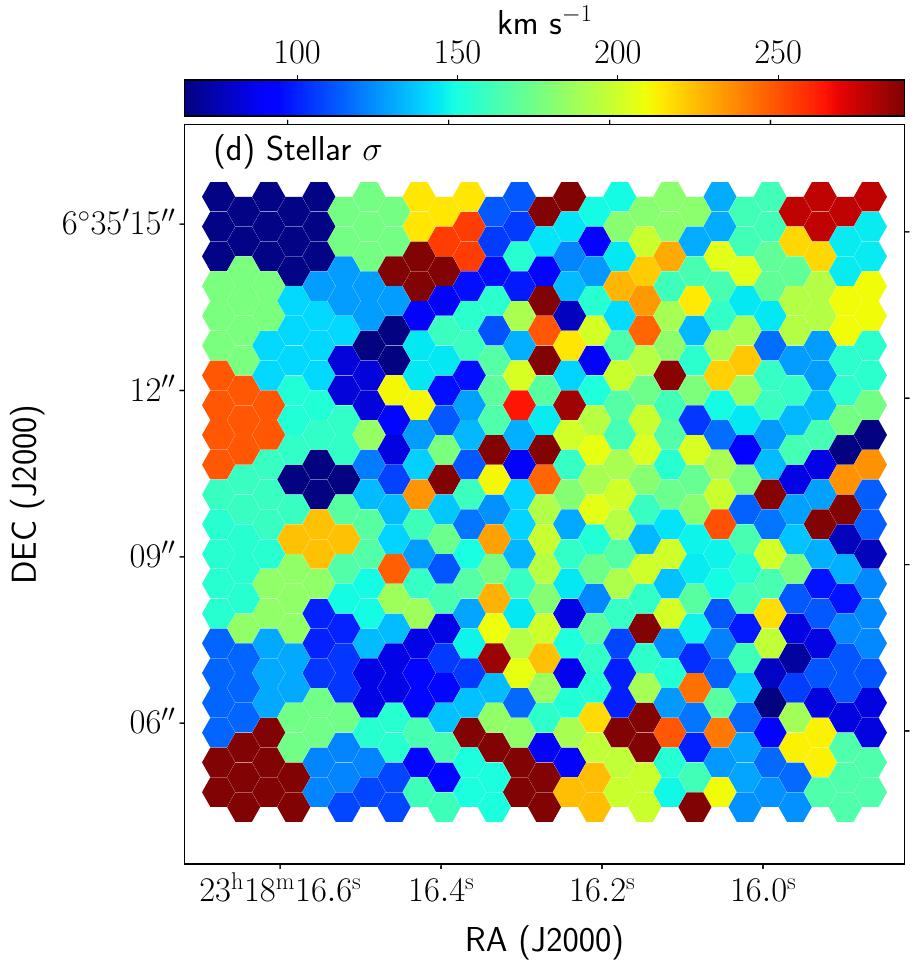}
	\includegraphics[clip, width=0.24\linewidth]{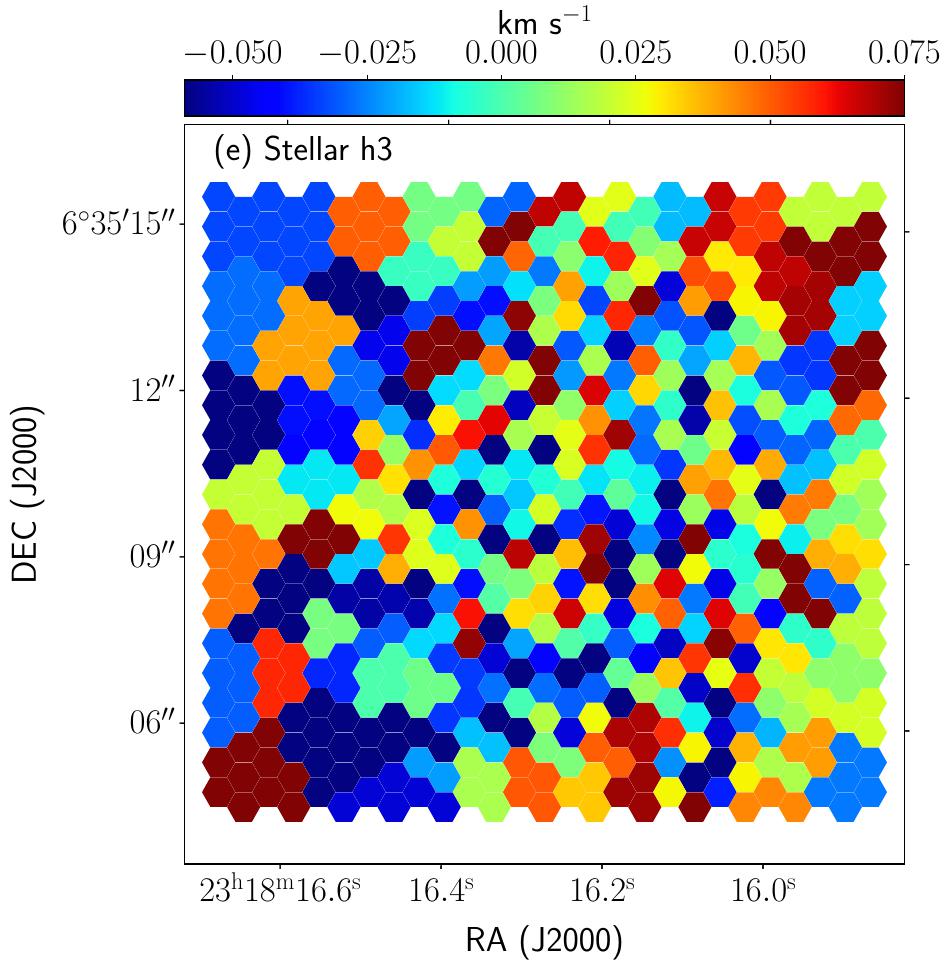}
	\includegraphics[clip, width=0.24\linewidth]{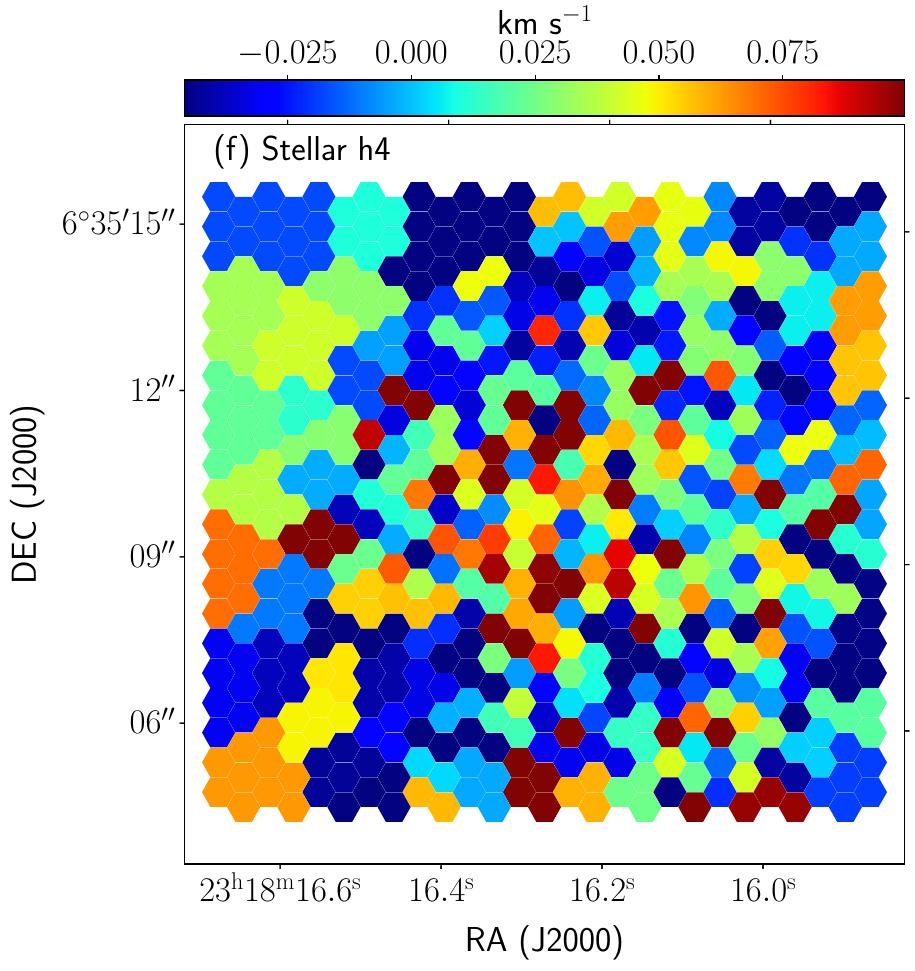}
	
	\vspace{8.8cm}
	
	\includegraphics[clip, width=0.24\linewidth]{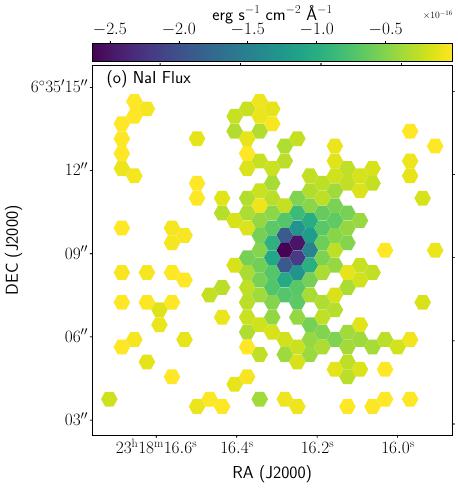}
	\includegraphics[clip, width=0.24\linewidth]{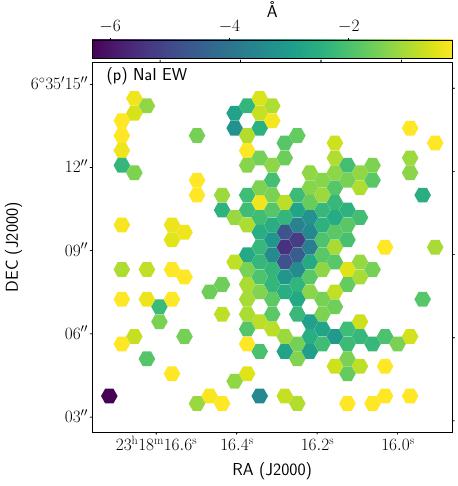}
	\includegraphics[clip, width=0.24\linewidth]{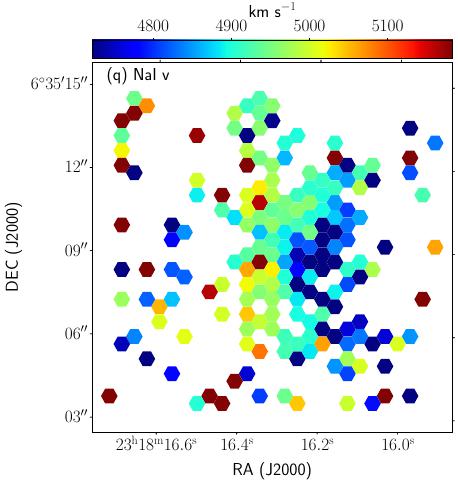}
	\includegraphics[clip, width=0.24\linewidth]{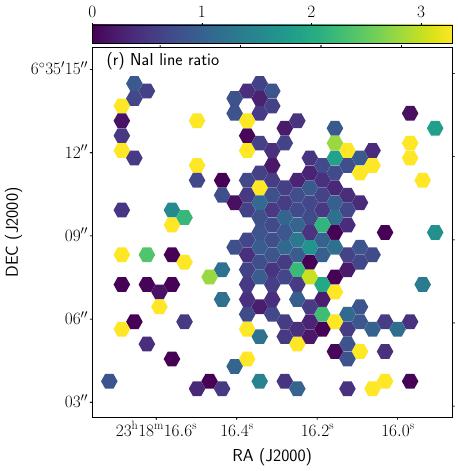}
	\caption{NGC~7591 card.}
	\label{fig:NGC7591_card_1}
\end{figure*}
\addtocounter{figure}{-1}
\begin{figure*}[h]
	\centering
	\includegraphics[clip, width=0.24\linewidth]{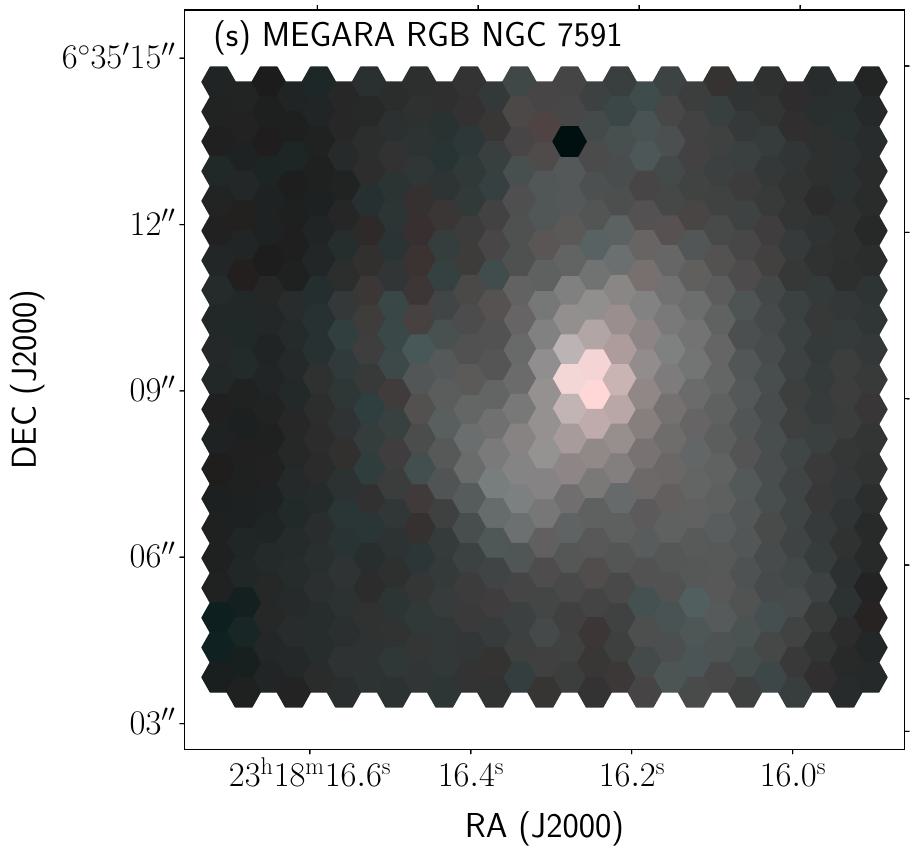}
	\hspace{4,4cm}
	\includegraphics[clip, width=0.24\linewidth]{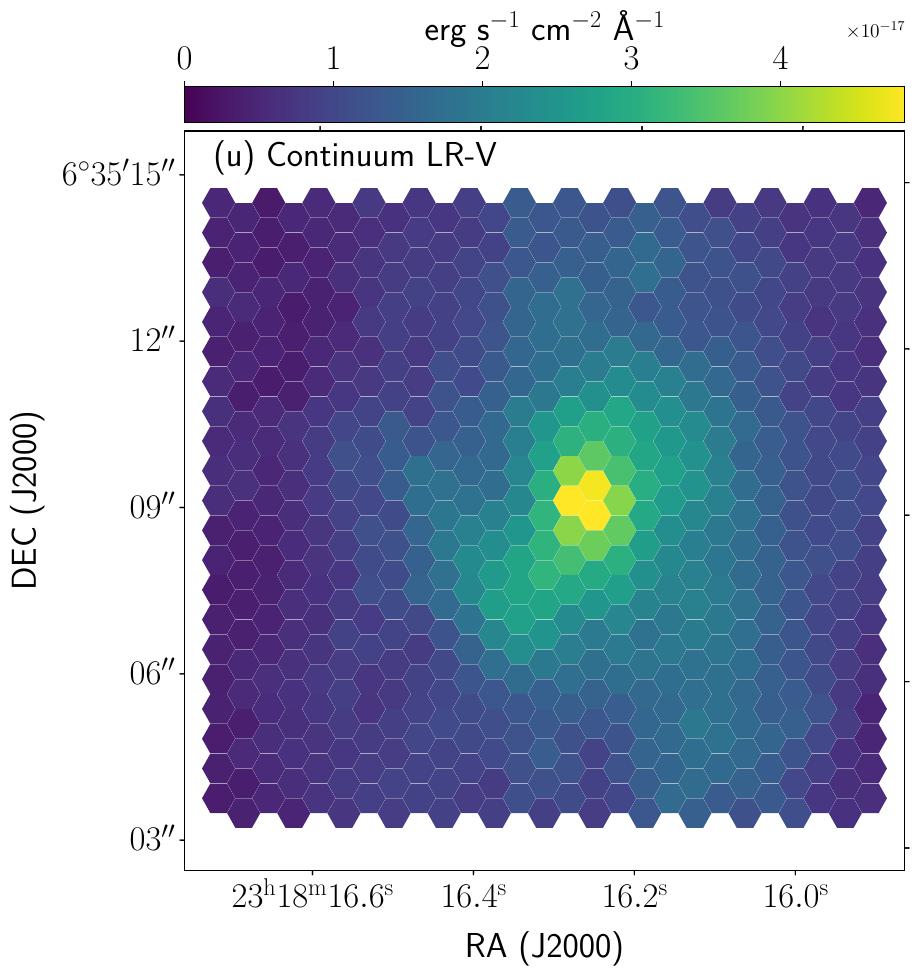}
	\includegraphics[clip, width=0.24\linewidth]{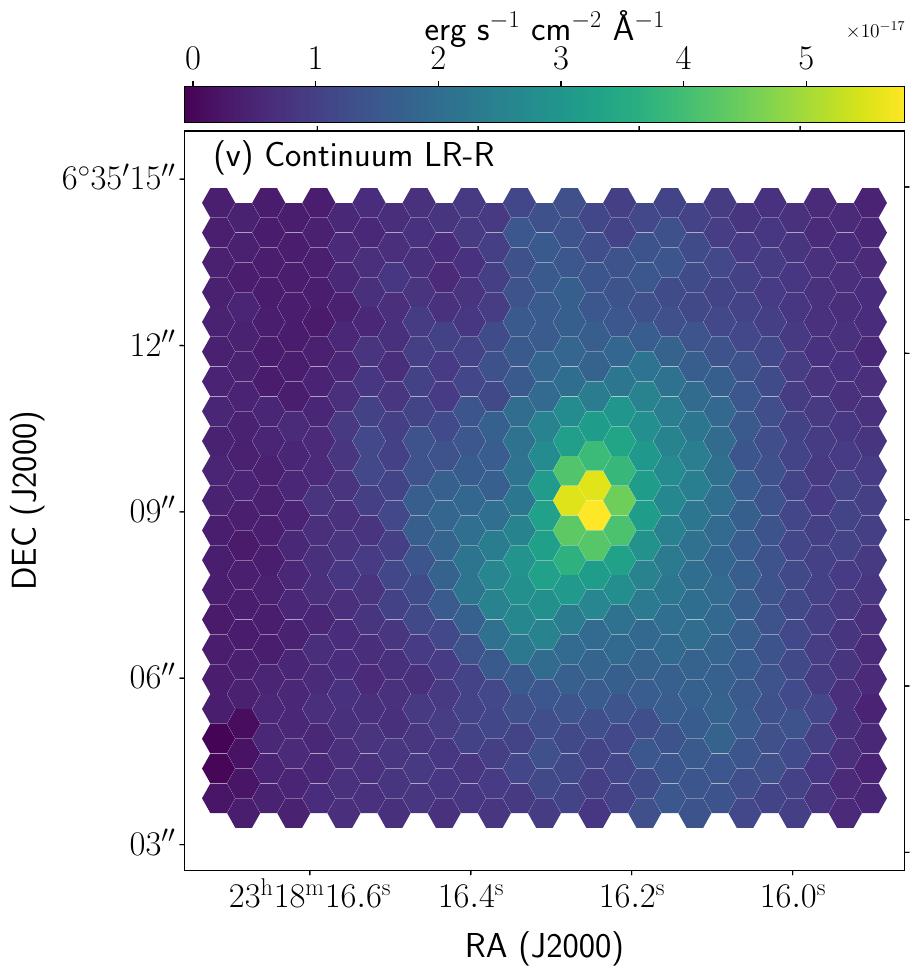}
	\includegraphics[clip, width=0.24\linewidth]{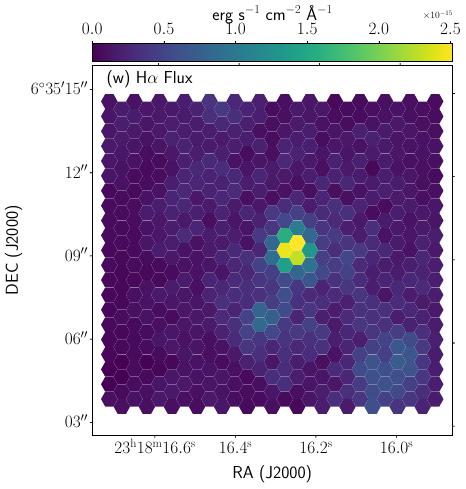}
	\includegraphics[clip, width=0.24\linewidth]{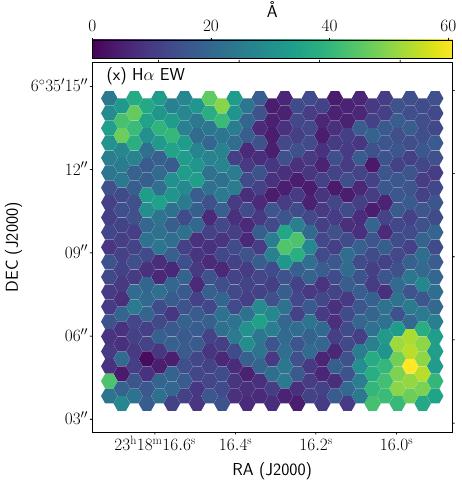}
	\includegraphics[clip, width=0.24\linewidth]{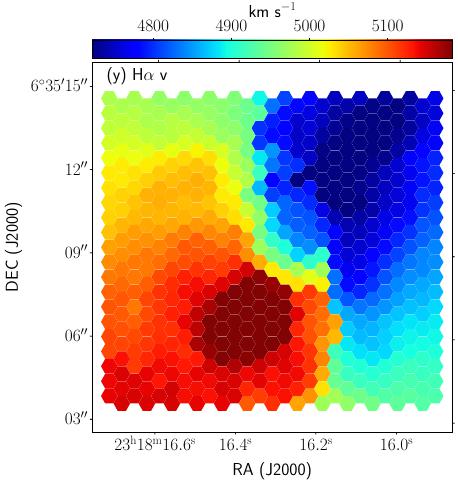}
	\includegraphics[clip, width=0.24\linewidth]{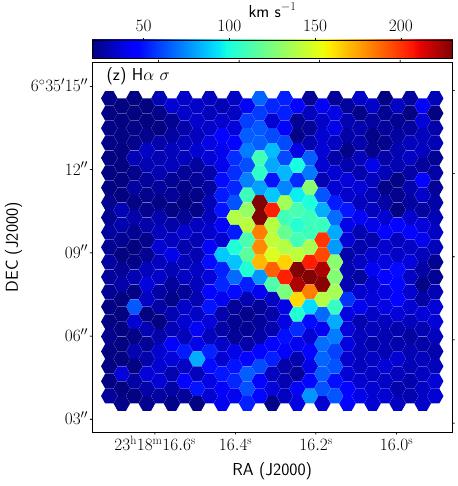}
	\includegraphics[clip, width=0.24\linewidth]{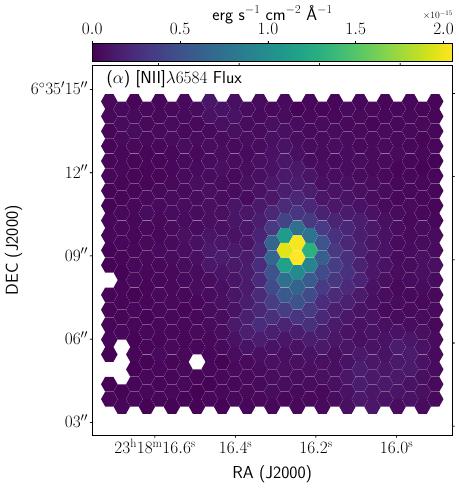}
	\includegraphics[clip, width=0.24\linewidth]{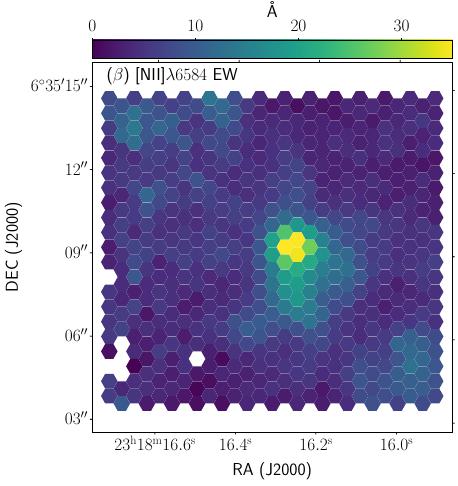}
	\includegraphics[clip, width=0.24\linewidth]{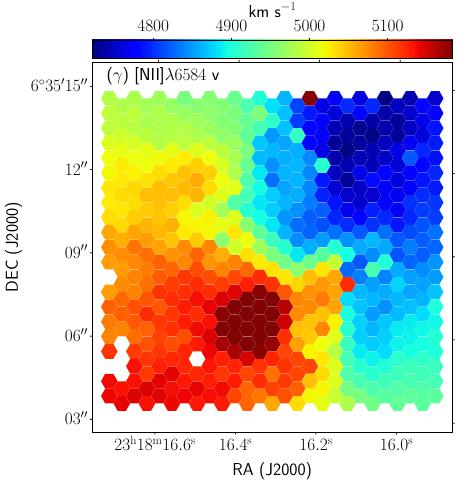}
	\includegraphics[clip, width=0.24\linewidth]{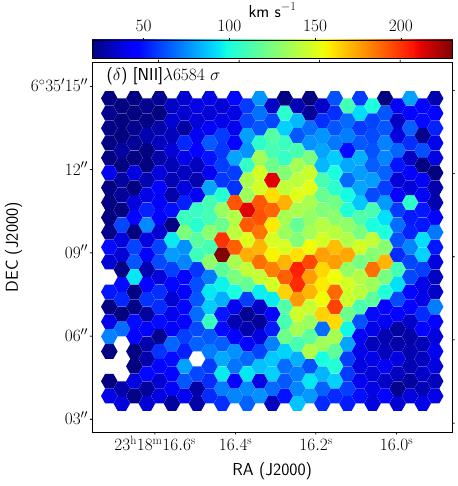}
	\includegraphics[clip, width=0.24\linewidth]{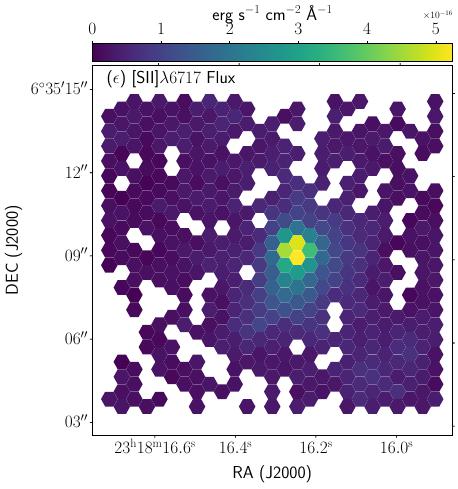}
	\includegraphics[clip, width=0.24\linewidth]{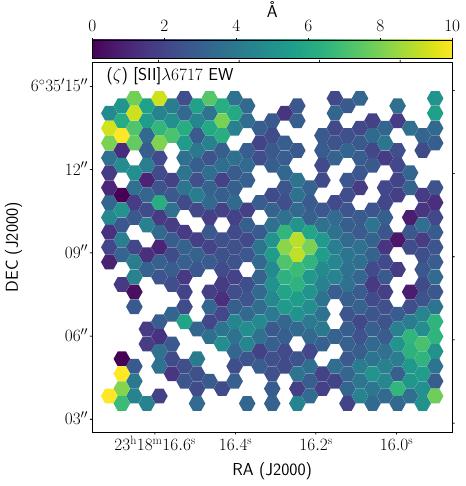}
	\includegraphics[clip, width=0.24\linewidth]{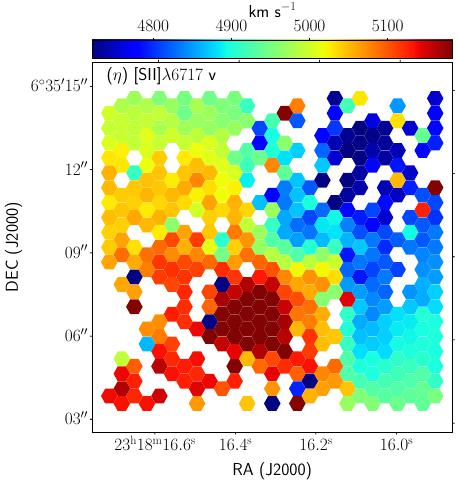}
	\includegraphics[clip, width=0.24\linewidth]{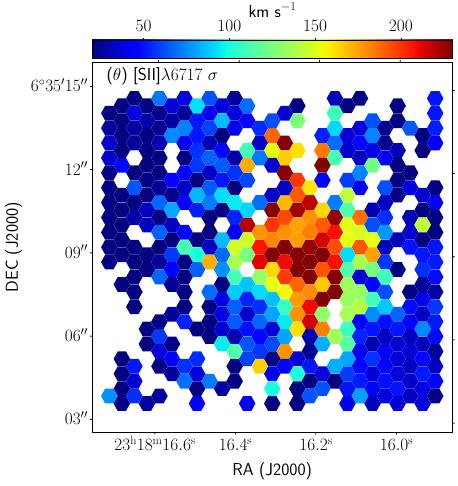}
	\includegraphics[clip, width=0.24\linewidth]{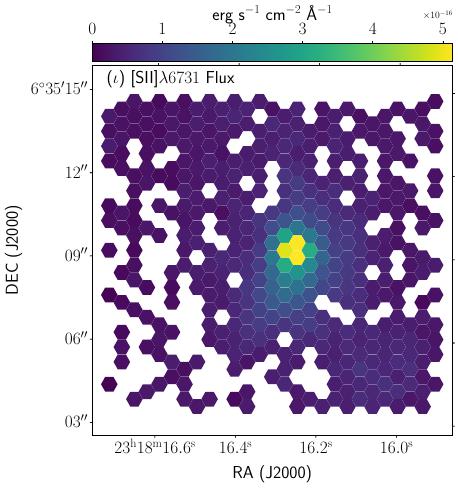}
	\includegraphics[clip, width=0.24\linewidth]{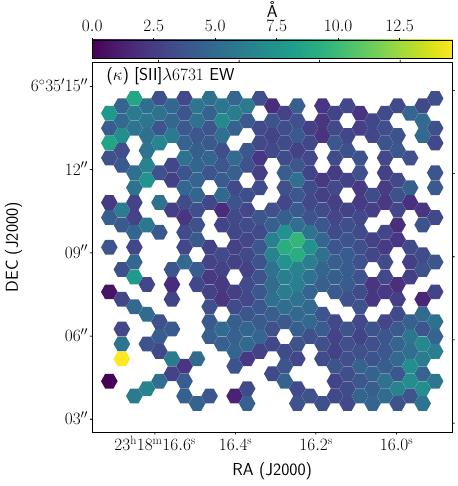}
	\includegraphics[clip, width=0.24\linewidth]{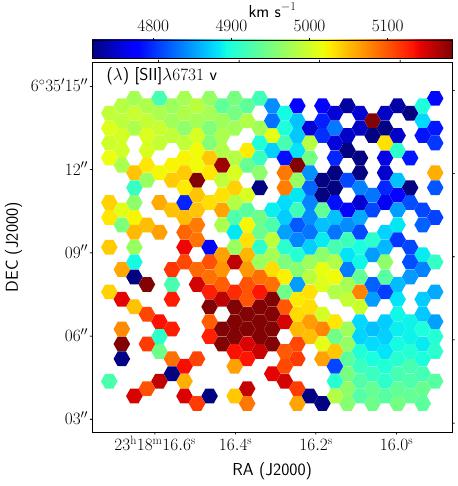}
	\includegraphics[clip, width=0.24\linewidth]{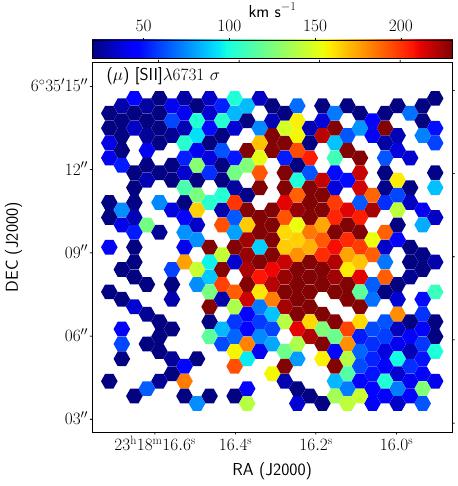}
	\caption{(cont.) NGC~7591 card.}
	\label{fig:NGC7591_card_2}
\end{figure*}

\begin{figure*}[h]
	\centering
	\includegraphics[clip, width=0.35\linewidth]{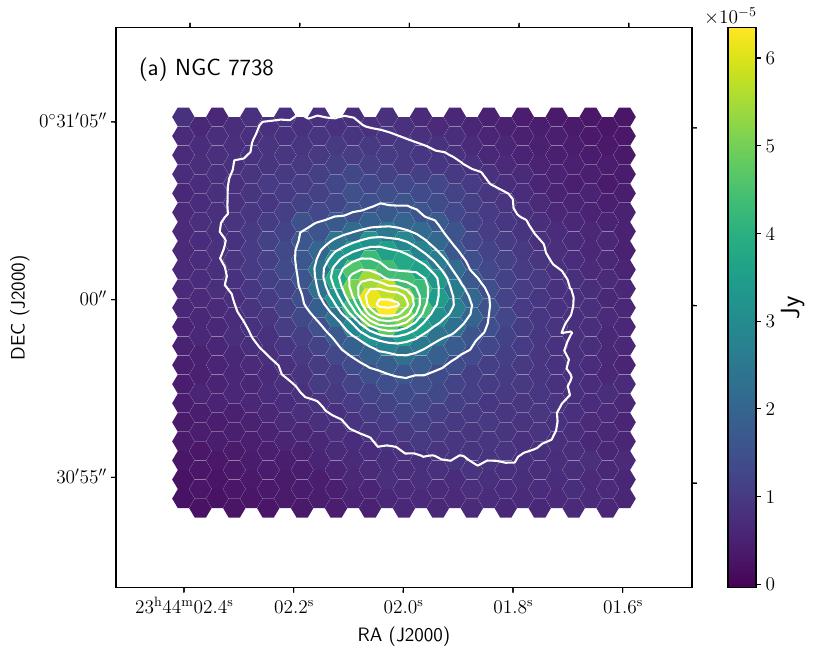}
	\includegraphics[clip, width=0.6\linewidth]{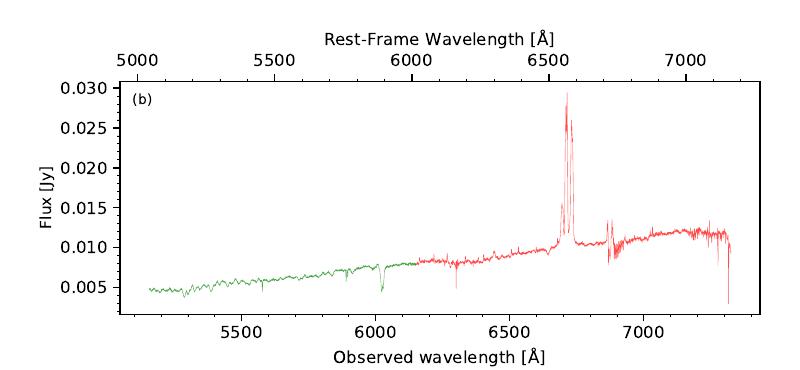}
	\includegraphics[clip, width=0.24\linewidth]{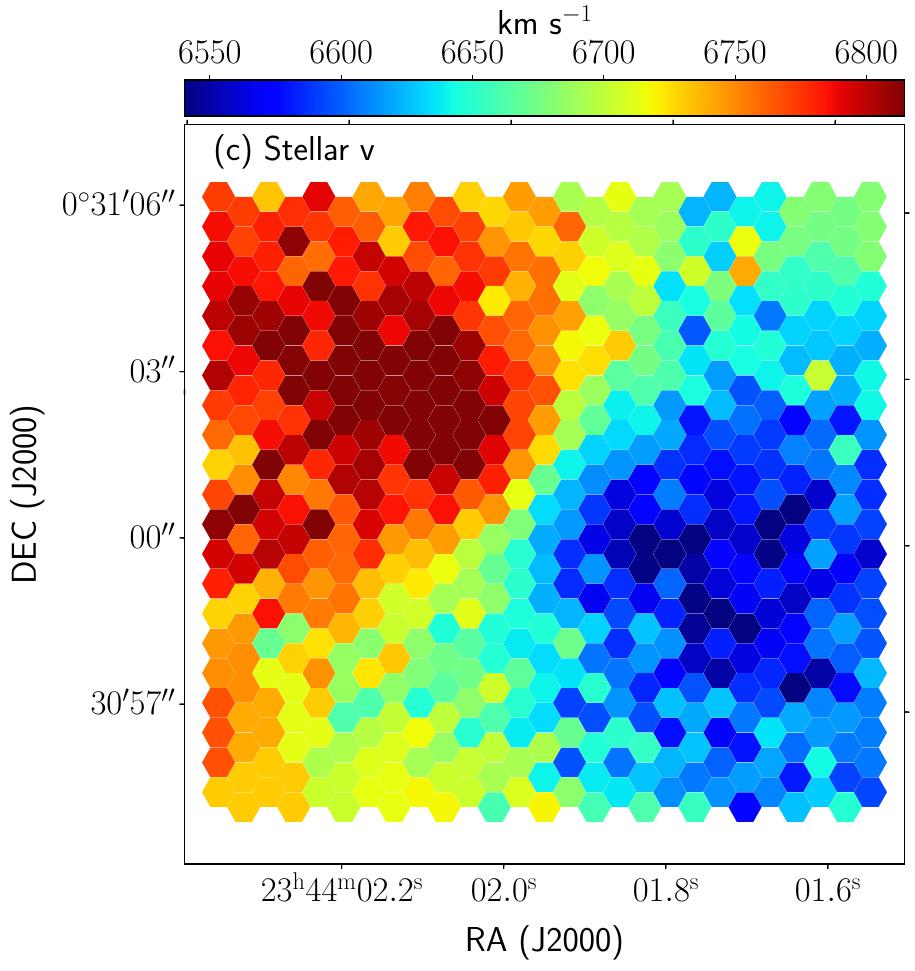}
	\includegraphics[clip, width=0.24\linewidth]{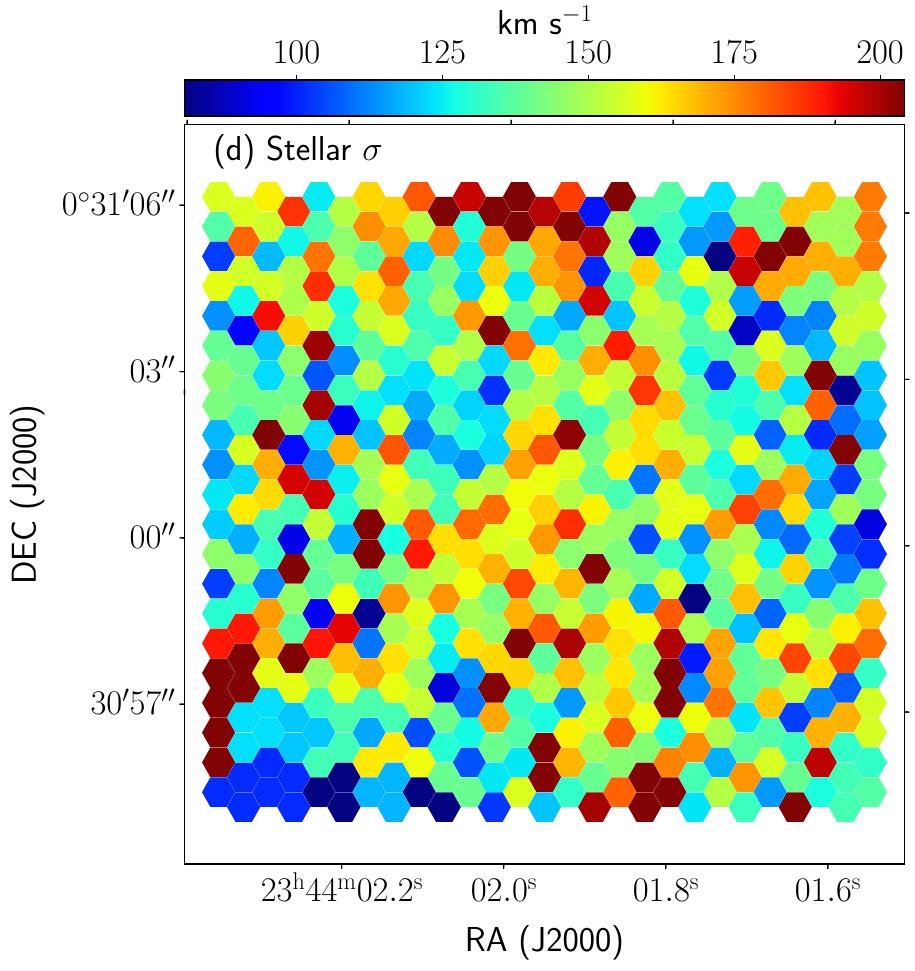}
	\includegraphics[clip, width=0.24\linewidth]{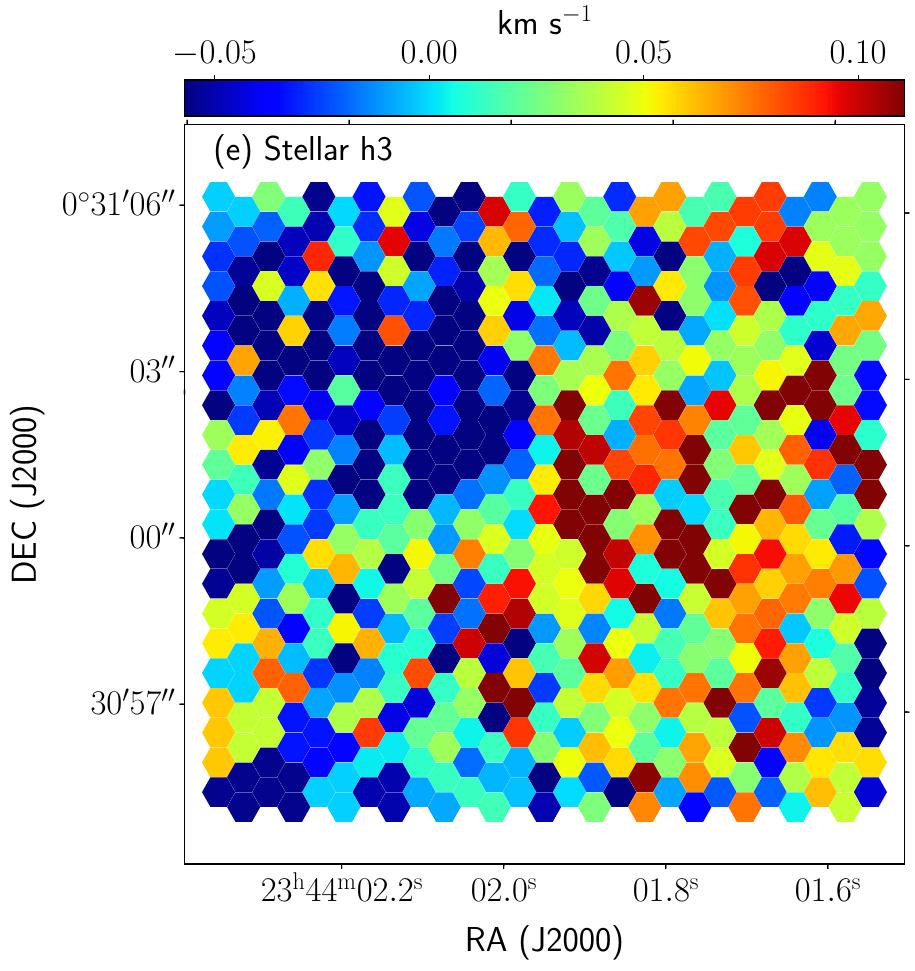}
	\includegraphics[clip, width=0.24\linewidth]{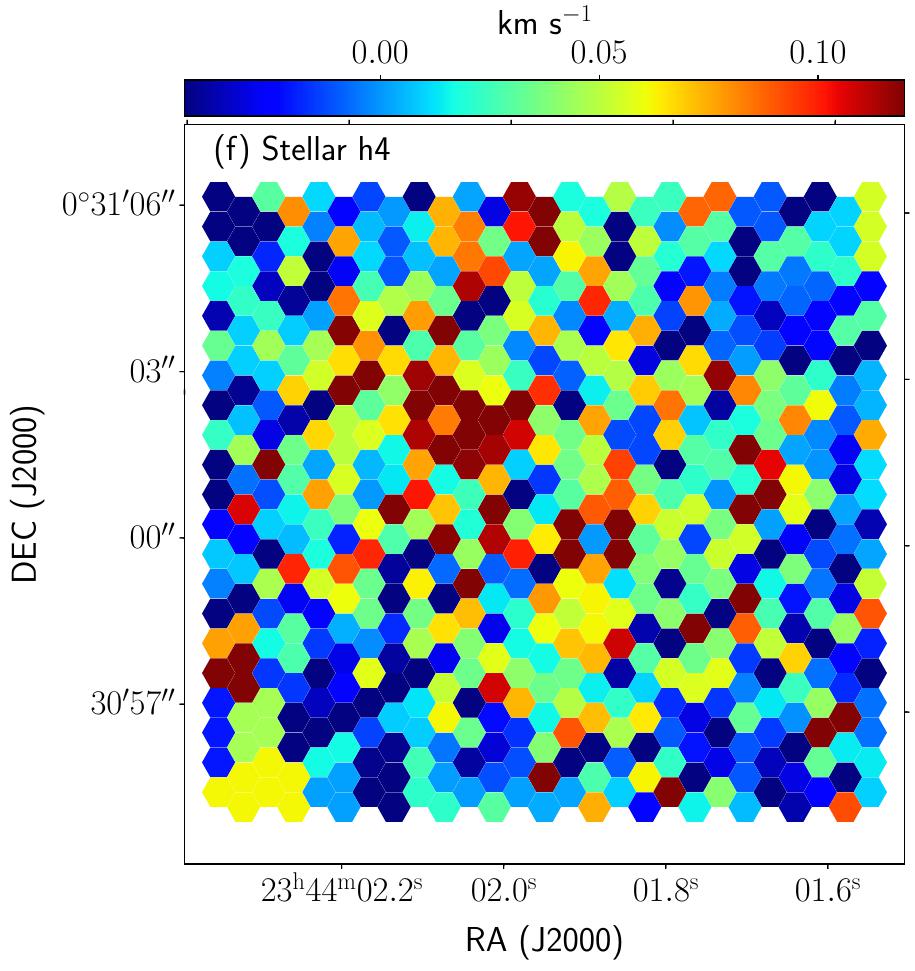}
	
	\vspace{8.8cm}
	
	\includegraphics[clip, width=0.24\linewidth]{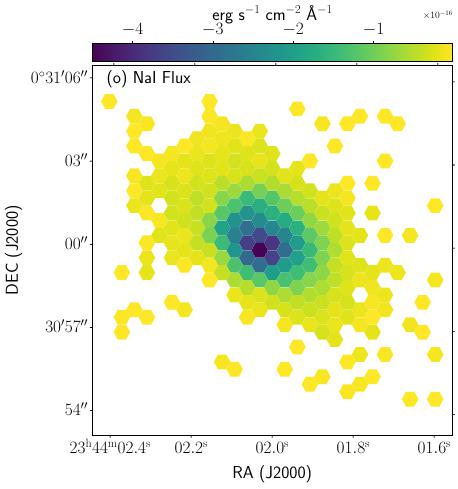}
	\includegraphics[clip, width=0.24\linewidth]{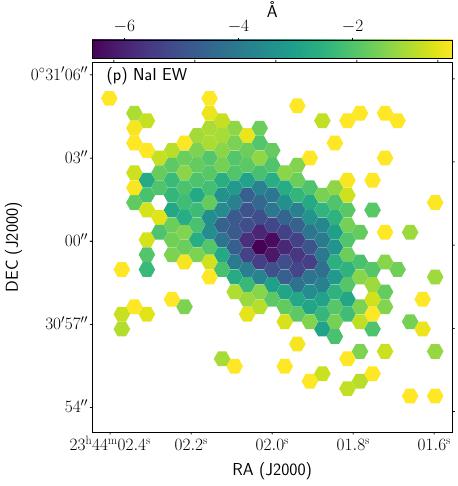}
	\includegraphics[clip, width=0.24\linewidth]{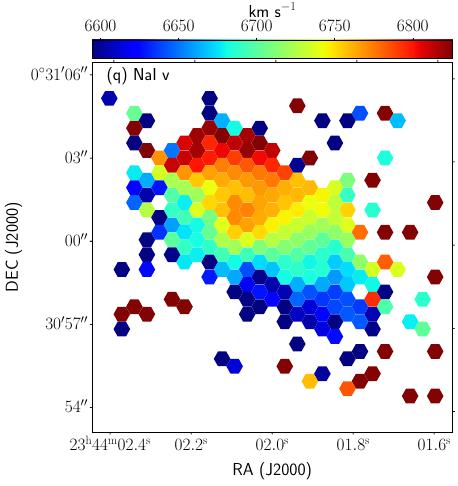}
	\includegraphics[clip, width=0.24\linewidth]{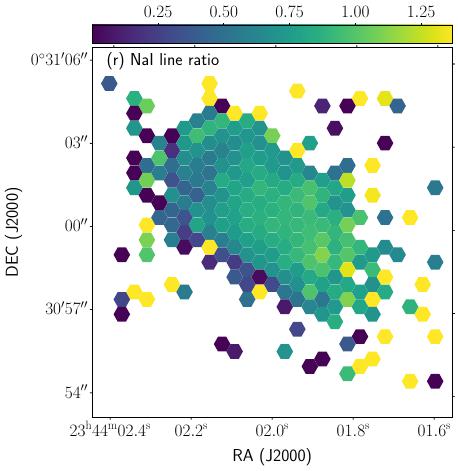}
	\caption{NGC~7738 card.}
	\label{fig:NGC7738_card_1}
\end{figure*}
\addtocounter{figure}{-1}
\begin{figure*}[h]
	\centering
	\includegraphics[clip, width=0.24\linewidth]{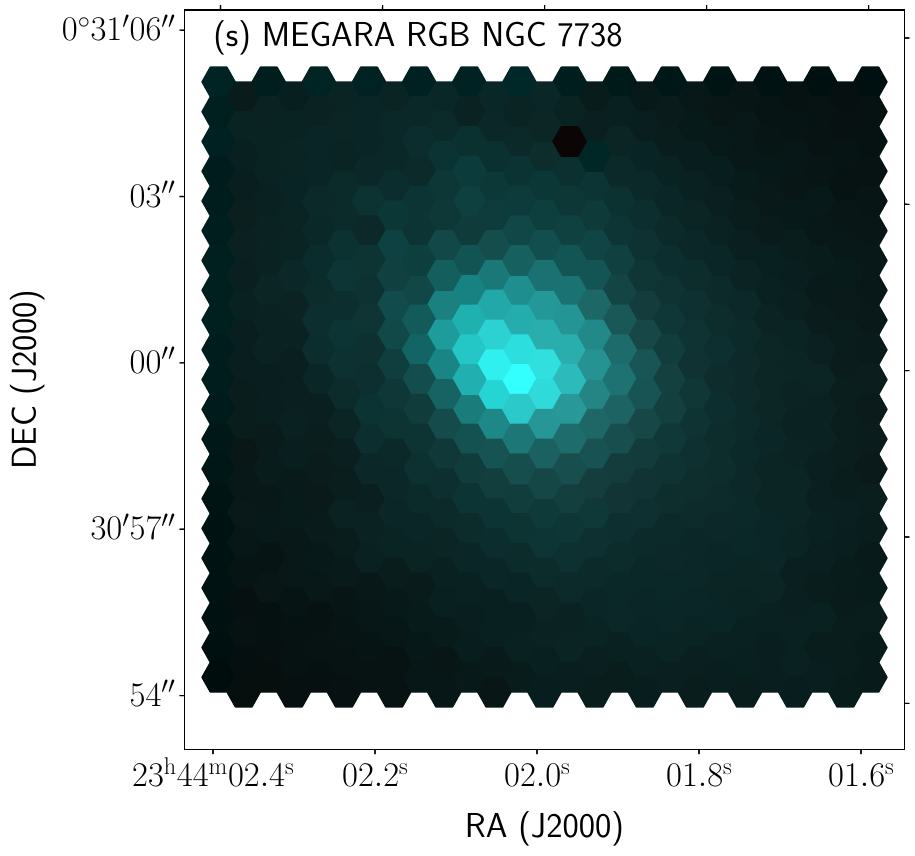}
	\hspace{4,4cm}
	\includegraphics[clip, width=0.24\linewidth]{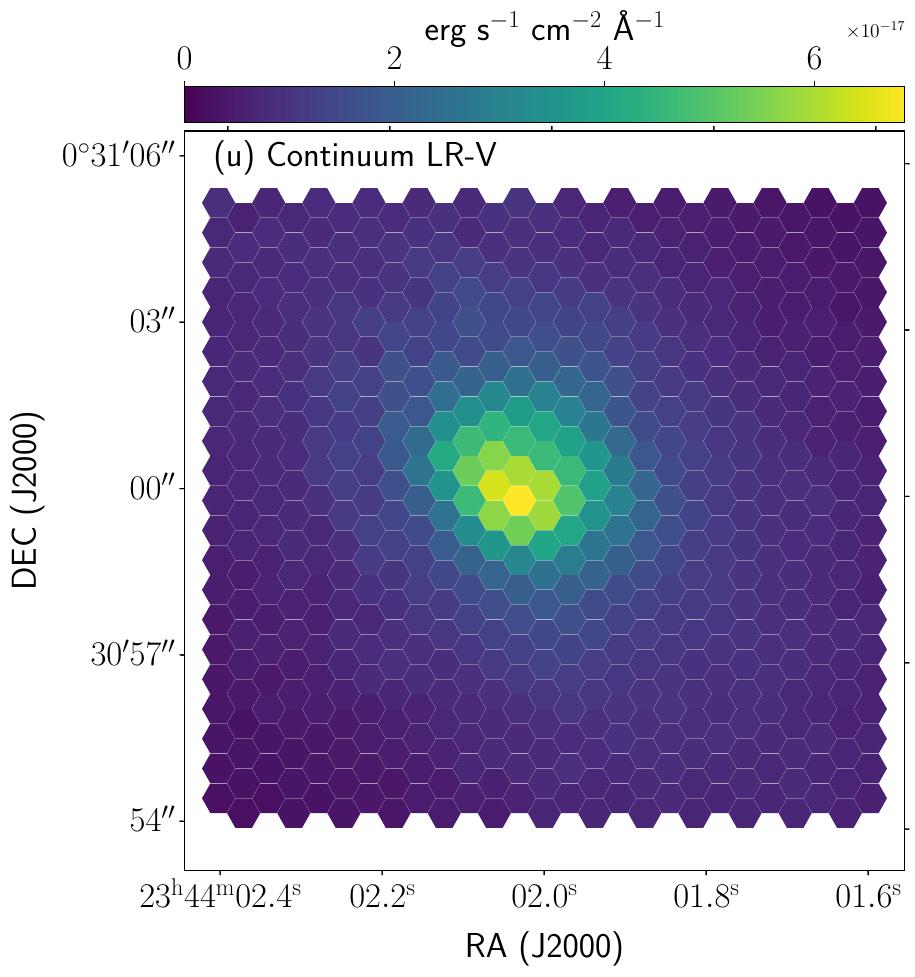}
	\includegraphics[clip, width=0.24\linewidth]{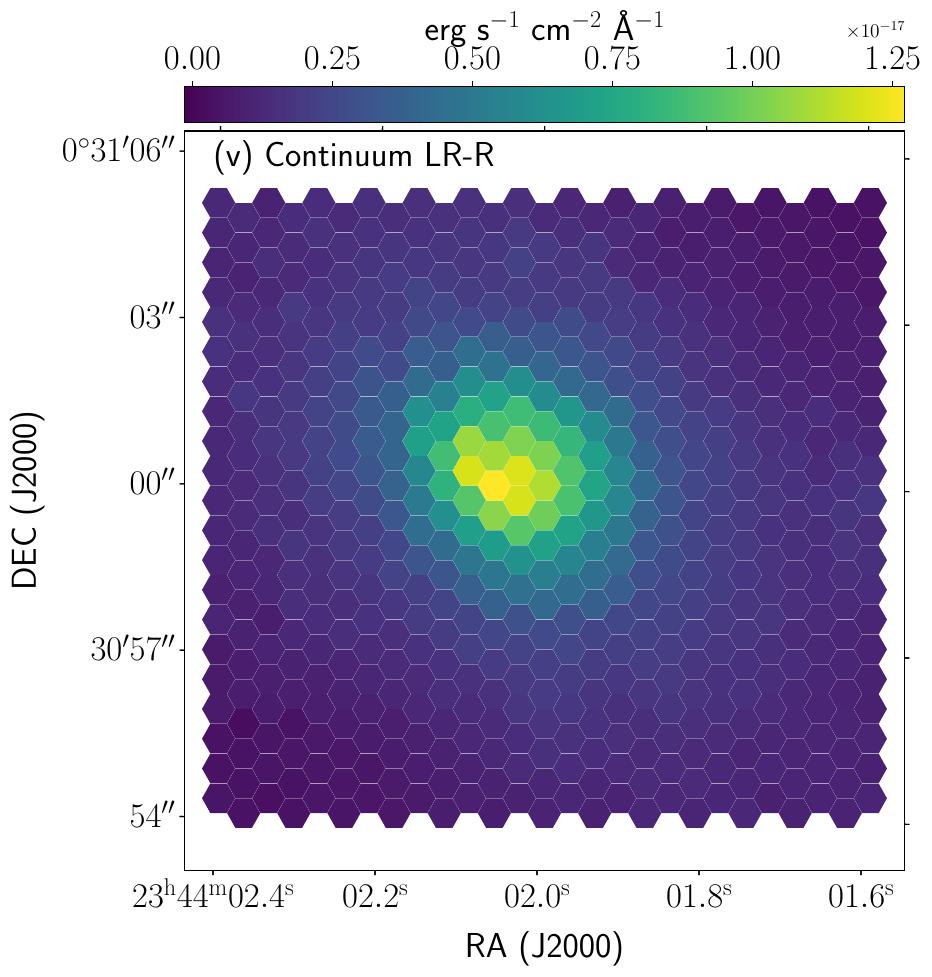}
	\includegraphics[clip, width=0.24\linewidth]{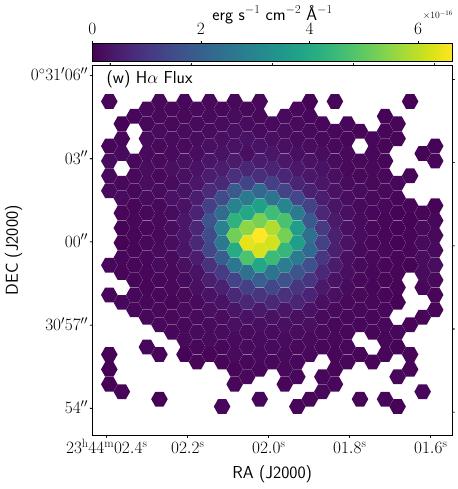}
	\includegraphics[clip, width=0.24\linewidth]{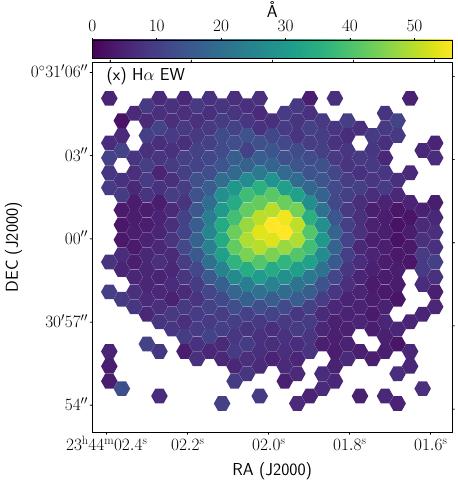}
	\includegraphics[clip, width=0.24\linewidth]{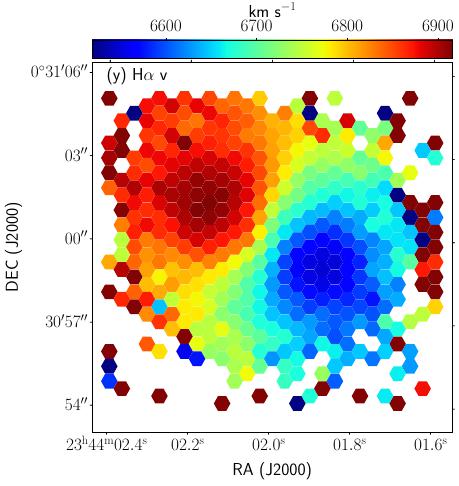}
	\includegraphics[clip, width=0.24\linewidth]{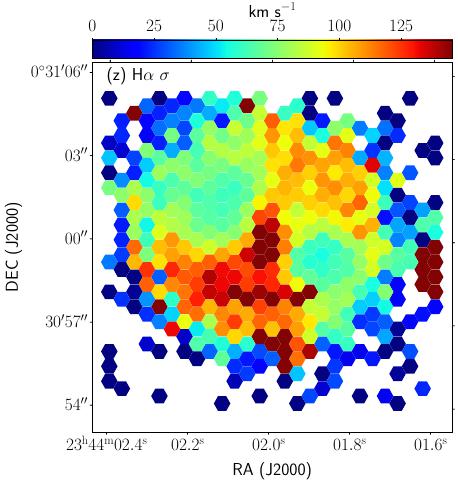}
	\includegraphics[clip, width=0.24\linewidth]{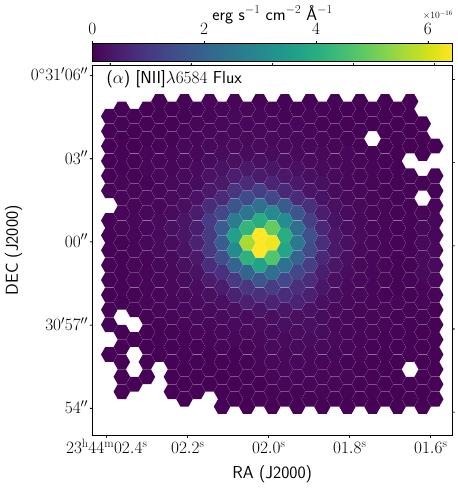}
	\includegraphics[clip, width=0.24\linewidth]{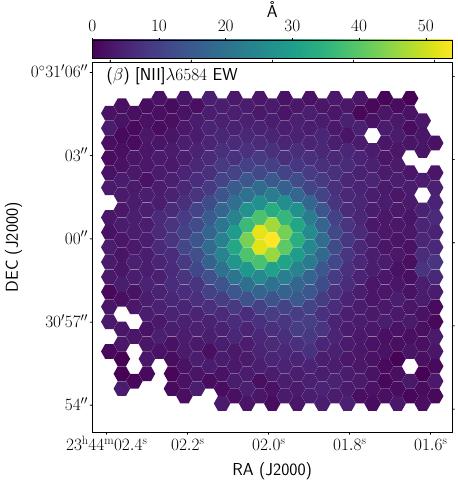}
	\includegraphics[clip, width=0.24\linewidth]{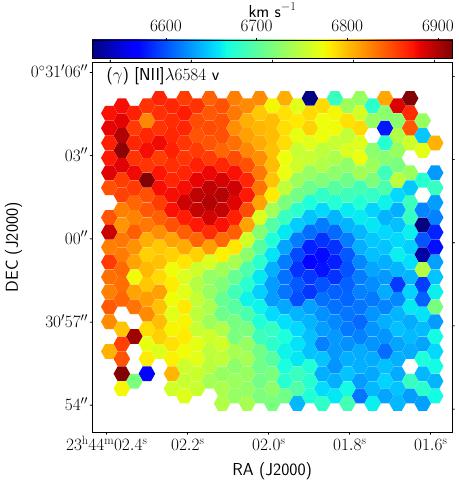}
	\includegraphics[clip, width=0.24\linewidth]{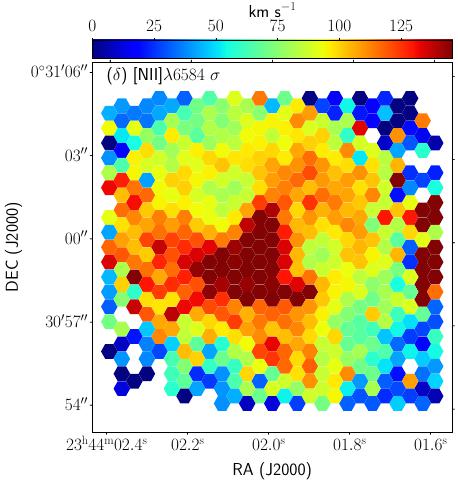}
	\includegraphics[clip, width=0.24\linewidth]{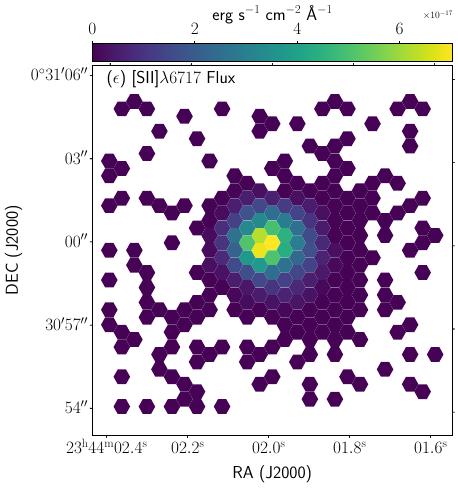}
	\includegraphics[clip, width=0.24\linewidth]{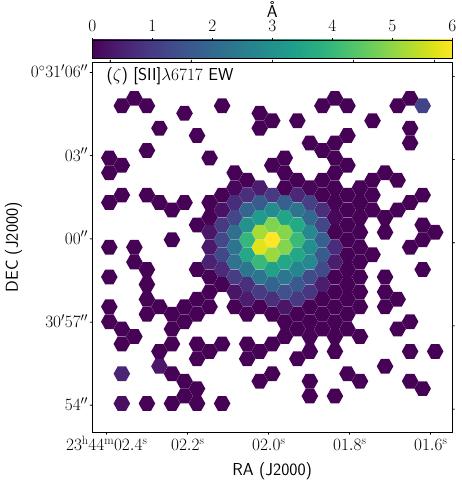}
	\includegraphics[clip, width=0.24\linewidth]{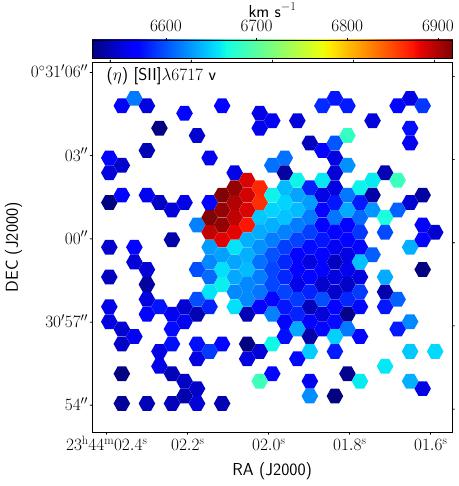}
	\includegraphics[clip, width=0.24\linewidth]{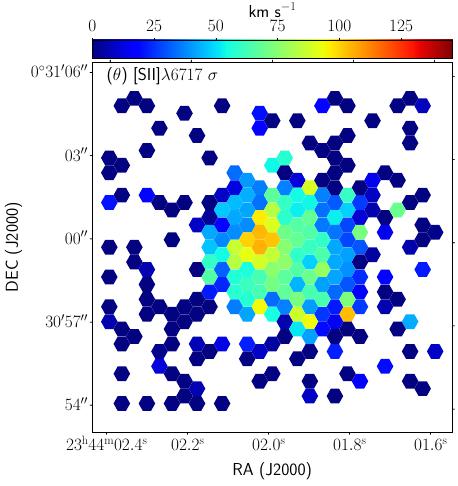}
	\includegraphics[clip, width=0.24\linewidth]{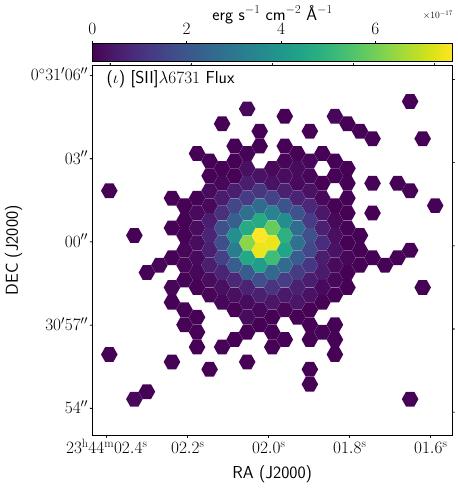}
	\includegraphics[clip, width=0.24\linewidth]{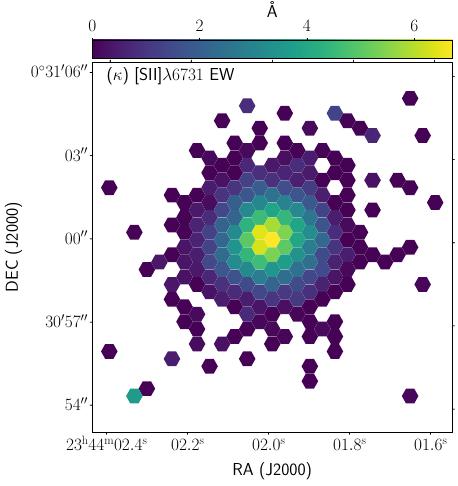}
	\includegraphics[clip, width=0.24\linewidth]{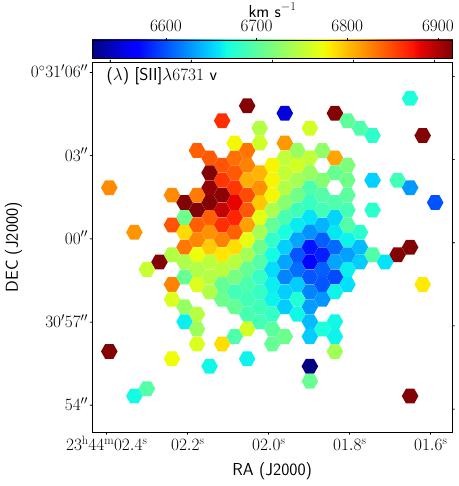}
	\includegraphics[clip, width=0.24\linewidth]{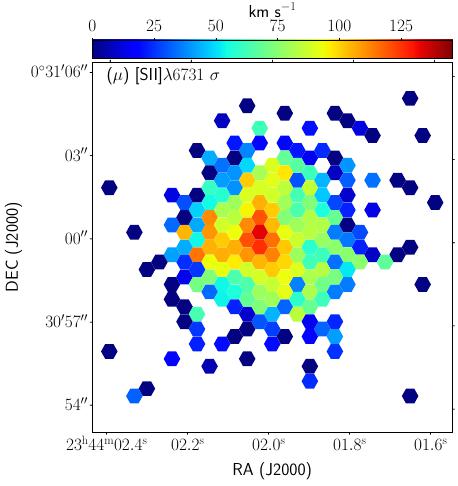}
	\caption{(cont.) NGC~7738 card. The velocity field of the $[\mathrm{S II}]\lambda6717$ line (subfigure* $\eta$) is affected by atmospheric telluric absorptions.}
	\label{fig:NGC7738_card_2}
\end{figure*}

\begin{figure*}[h]
	\centering
	\includegraphics[clip, width=0.35\linewidth]{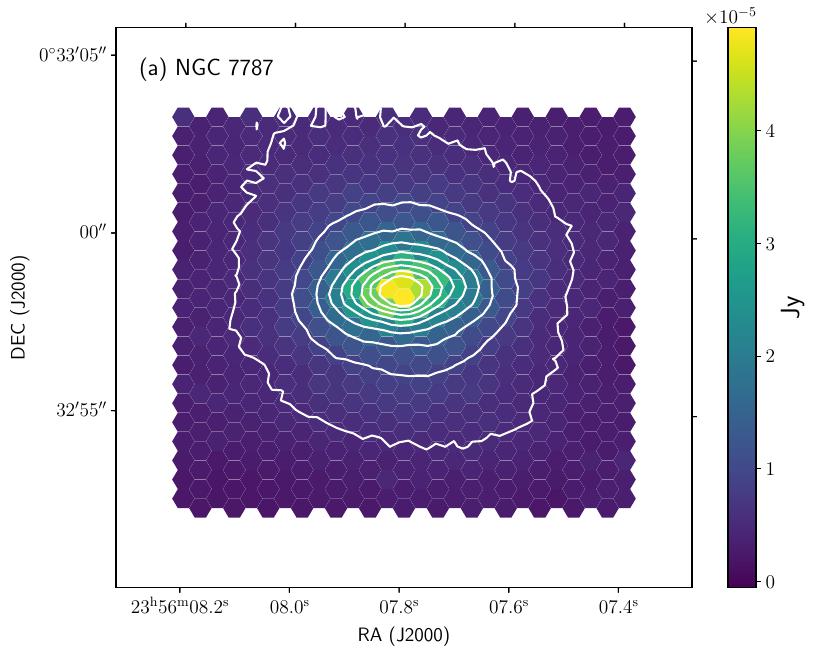}
	\includegraphics[clip, width=0.6\linewidth]{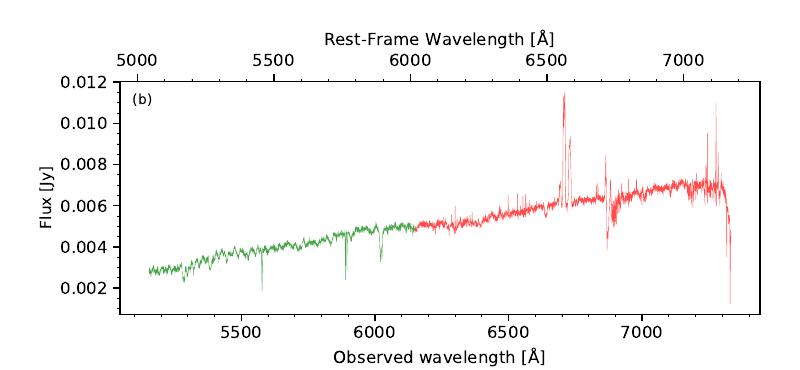}
	\includegraphics[clip, width=0.24\linewidth]{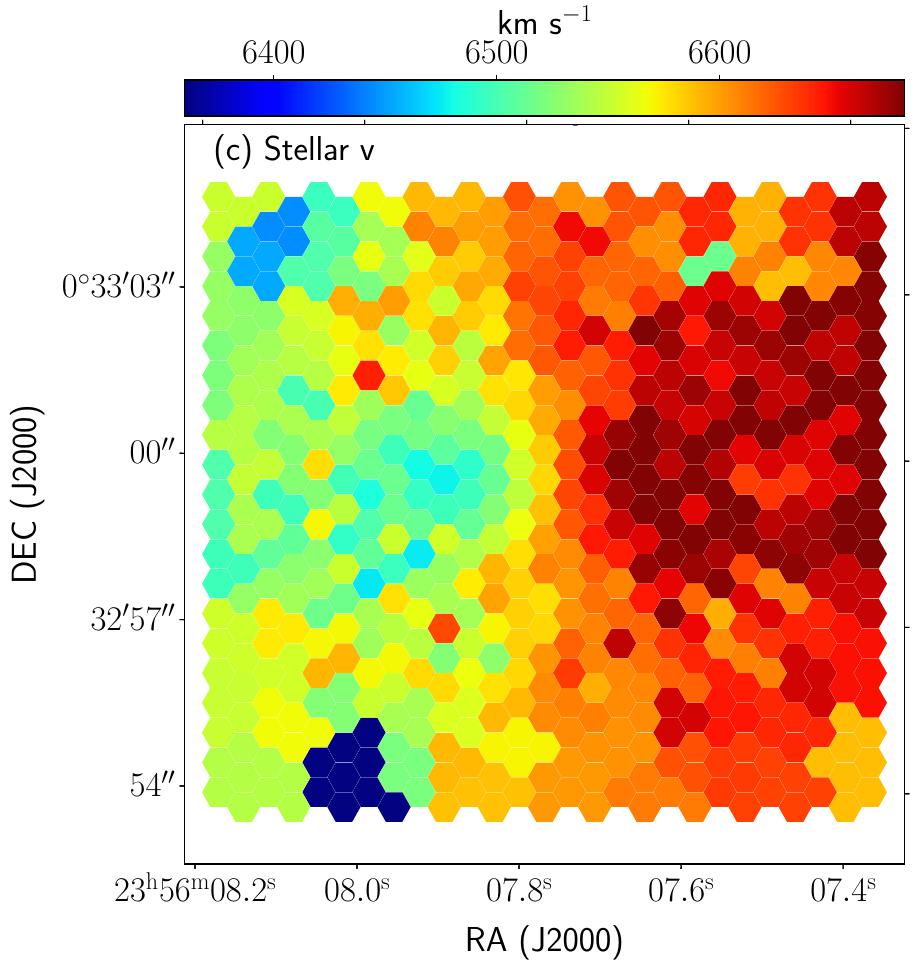}
	\includegraphics[clip, width=0.24\linewidth]{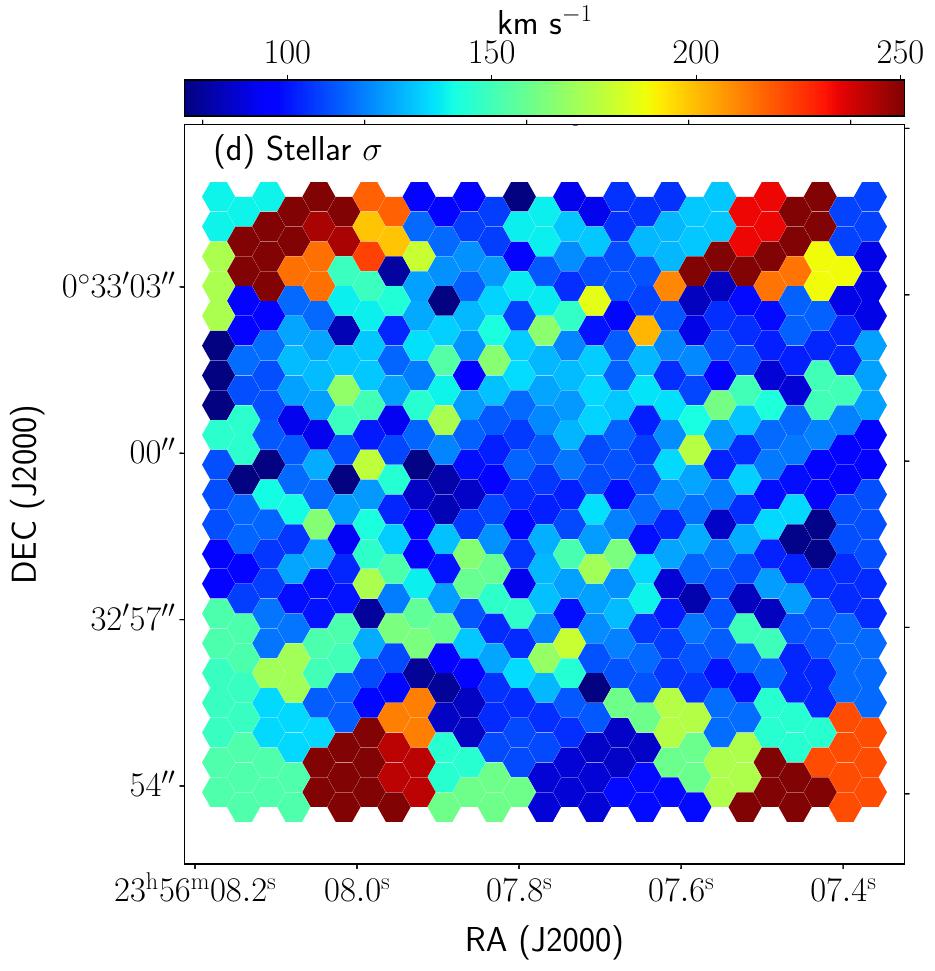}
	\includegraphics[clip, width=0.24\linewidth]{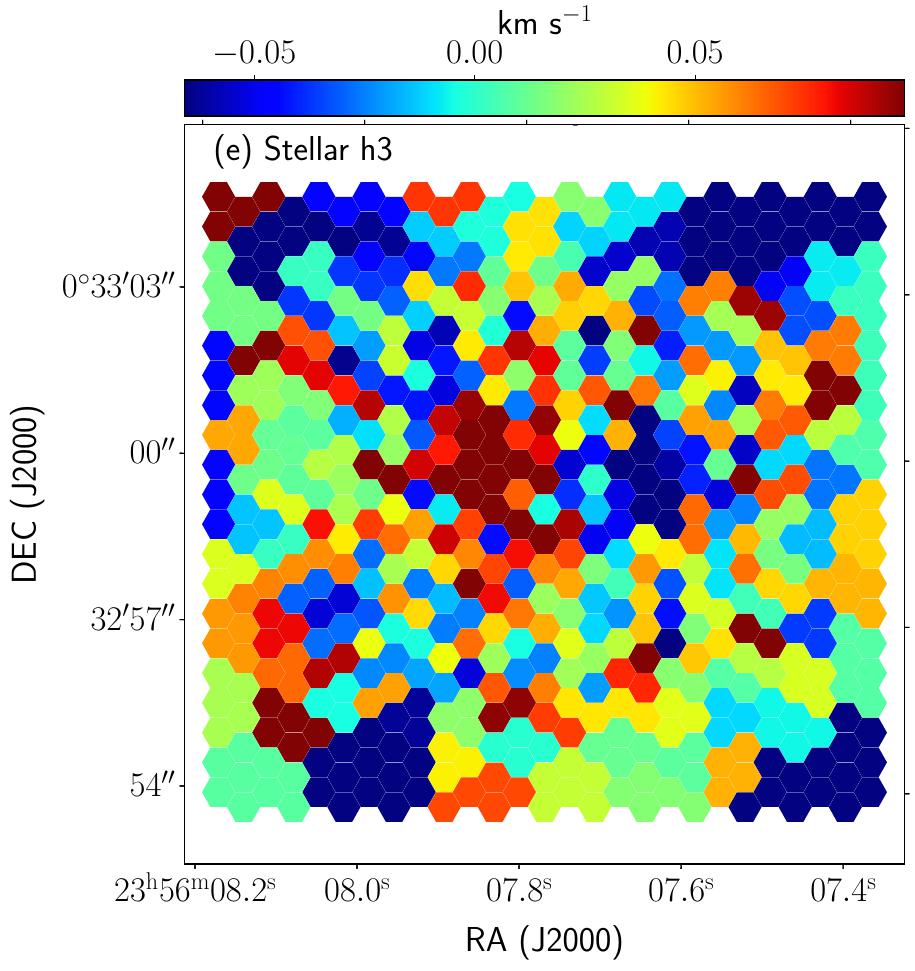}
	\includegraphics[clip, width=0.24\linewidth]{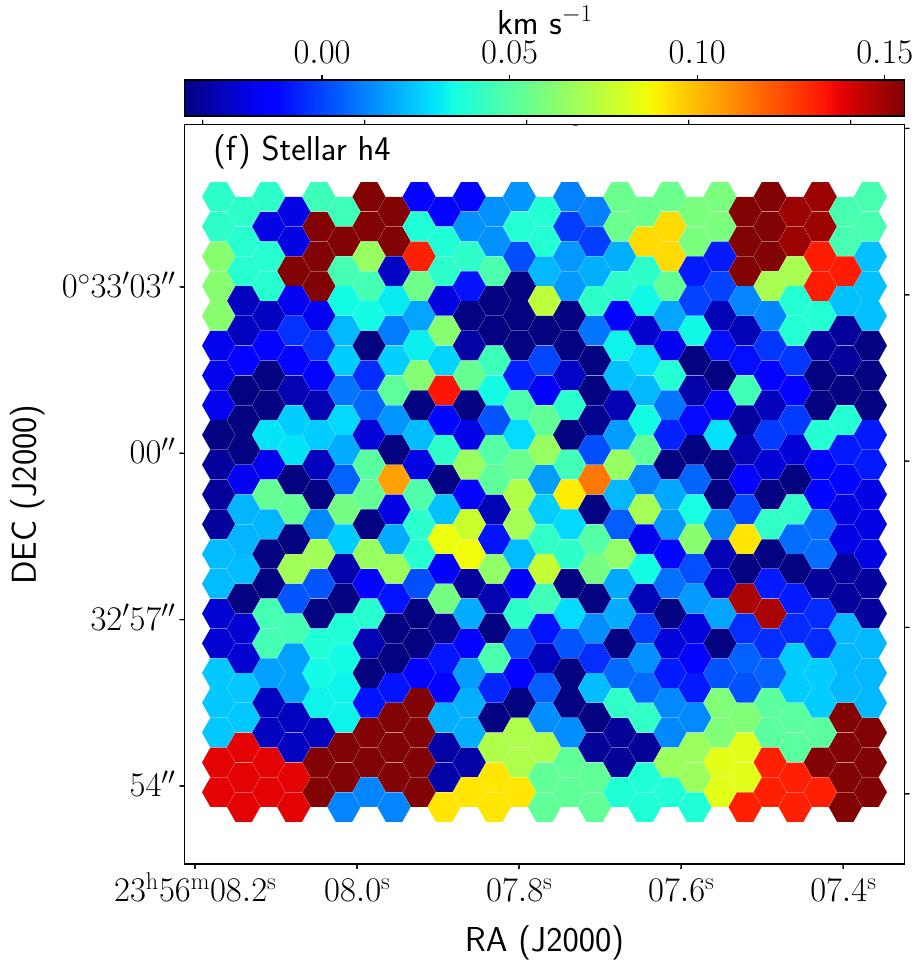}
	
	\vspace{8.8cm}
	
	\includegraphics[clip, width=0.24\linewidth]{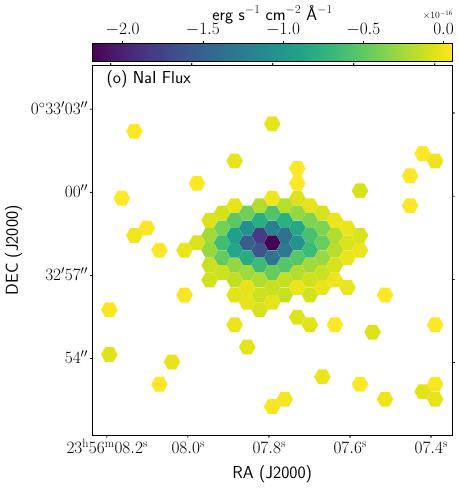}
	\includegraphics[clip, width=0.24\linewidth]{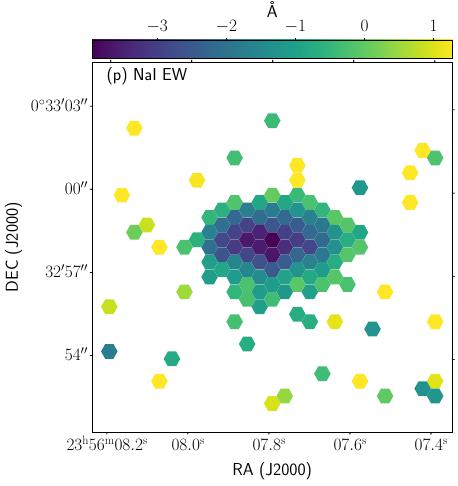}
	\includegraphics[clip, width=0.24\linewidth]{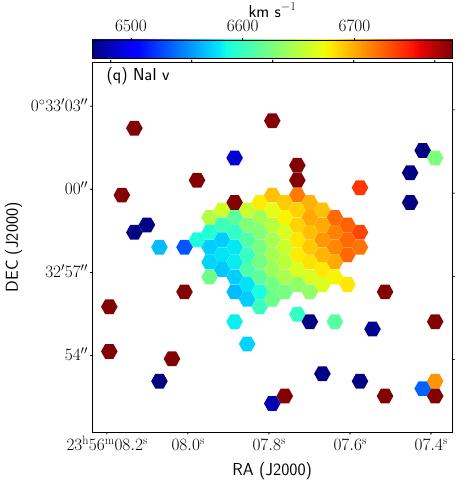}
	\includegraphics[clip, width=0.24\linewidth]{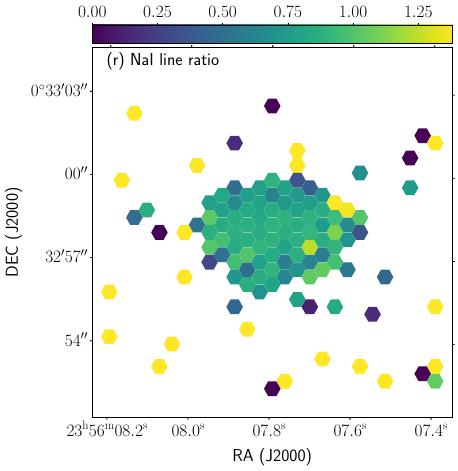}
	\caption{NGC~7787 card.}
	\label{fig:NGC7787_card_1}
\end{figure*}
\addtocounter{figure}{-1}
\begin{figure*}[h]
	\centering
	\includegraphics[clip, width=0.24\linewidth]{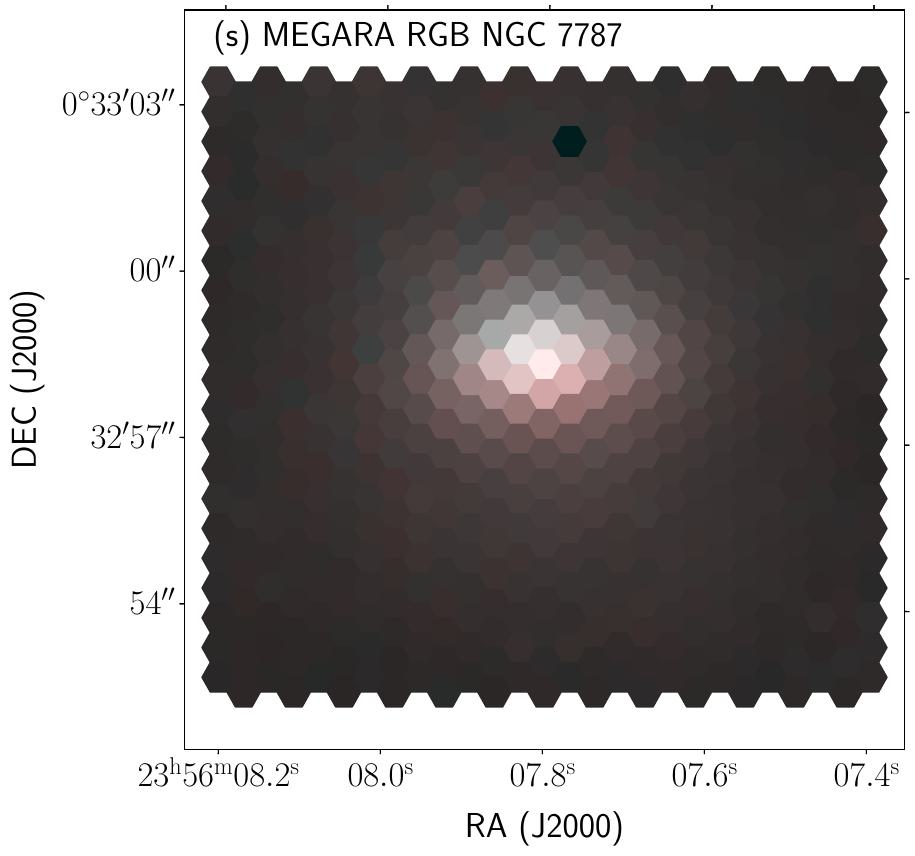}
	\hspace{4,4cm}
	\includegraphics[clip, width=0.24\linewidth]{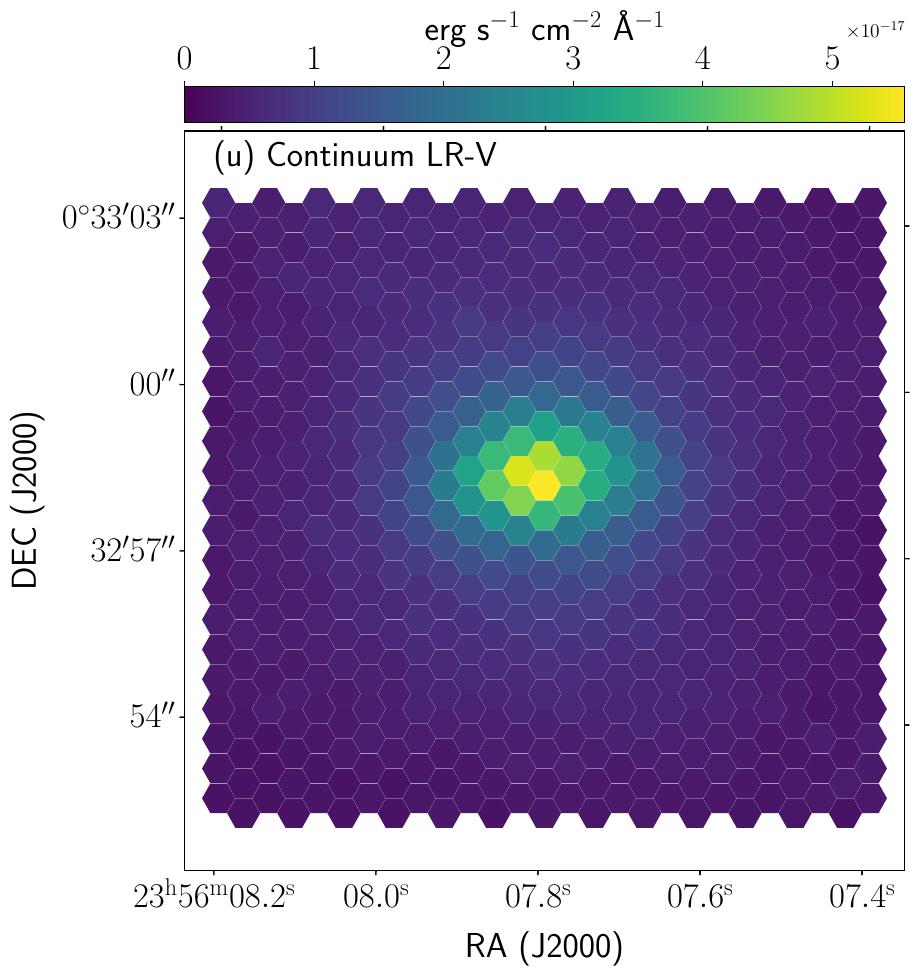}
	\includegraphics[clip, width=0.24\linewidth]{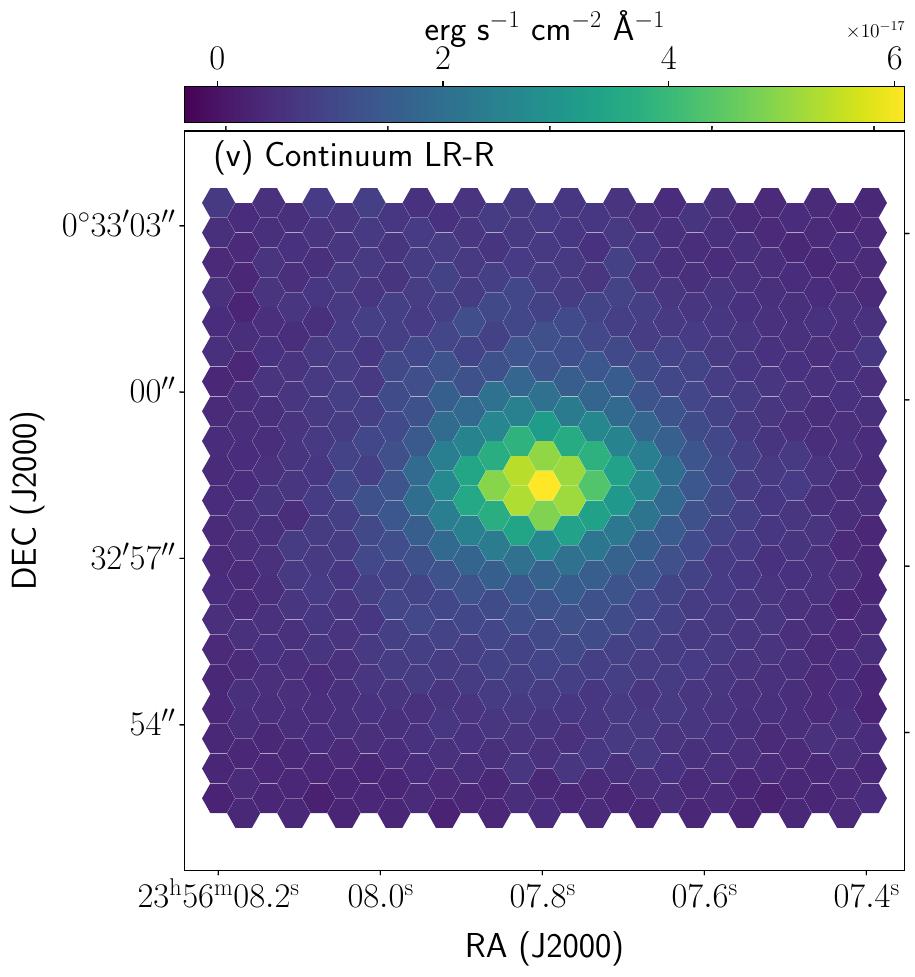}
	\includegraphics[clip, width=0.24\linewidth]{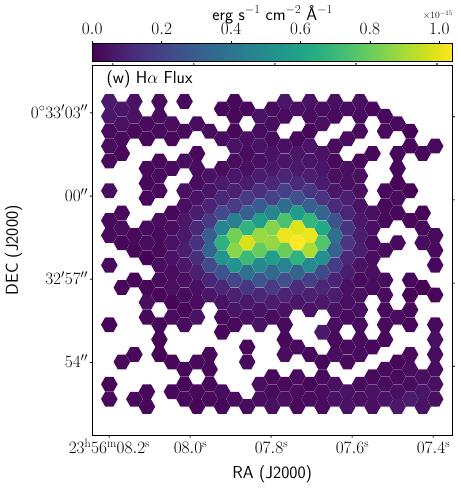}
	\includegraphics[clip, width=0.24\linewidth]{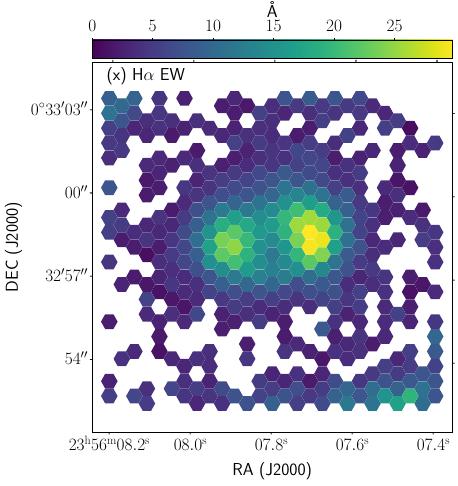}
	\includegraphics[clip, width=0.24\linewidth]{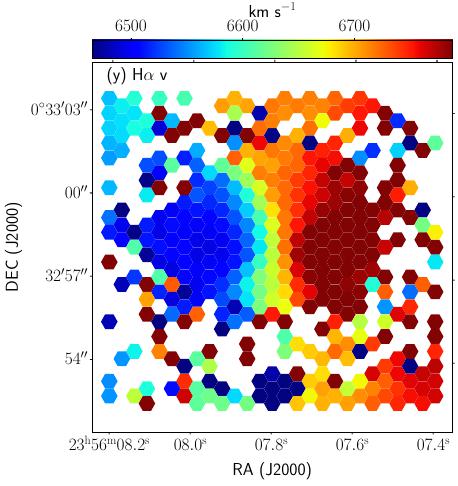}
	\includegraphics[clip, width=0.24\linewidth]{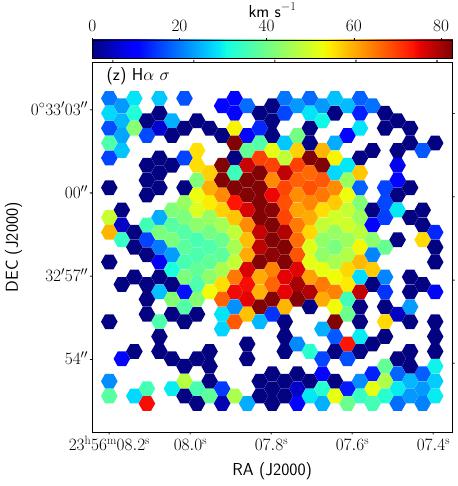}
	\includegraphics[clip, width=0.24\linewidth]{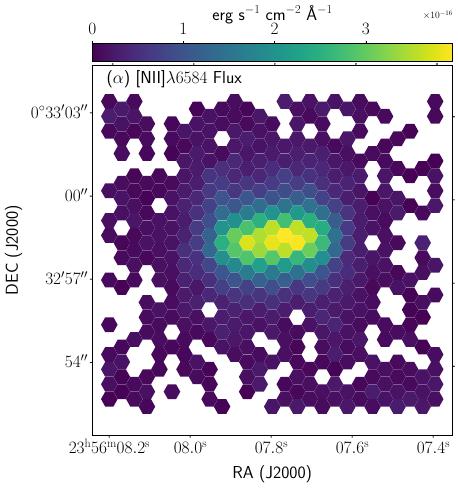}
	\includegraphics[clip, width=0.24\linewidth]{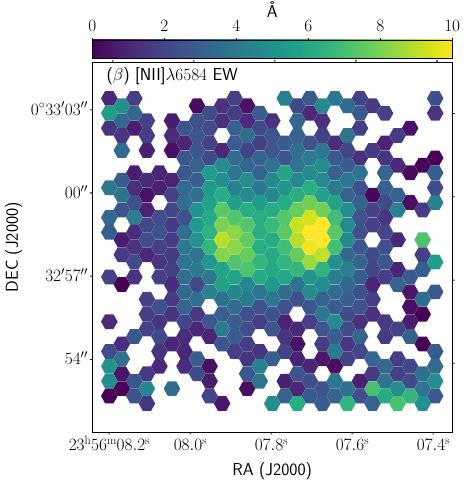}
	\includegraphics[clip, width=0.24\linewidth]{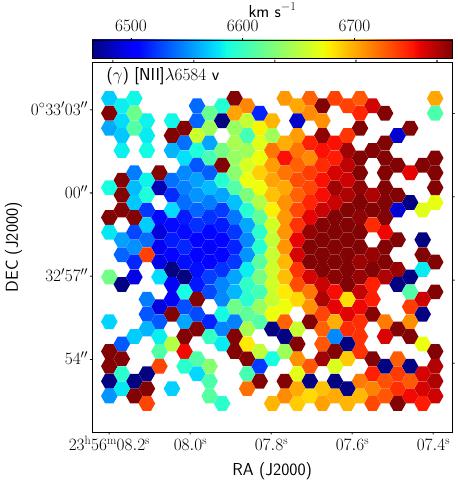}
	\includegraphics[clip, width=0.24\linewidth]{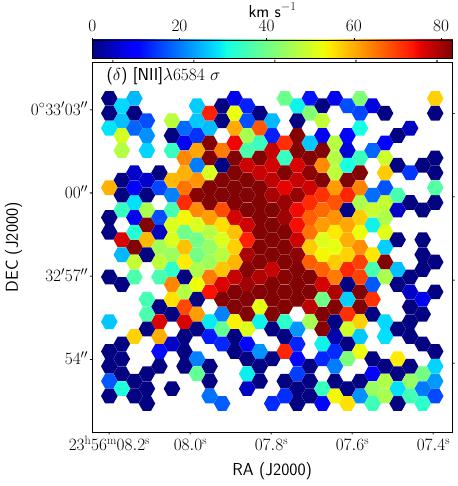}
	\includegraphics[clip, width=0.24\linewidth]{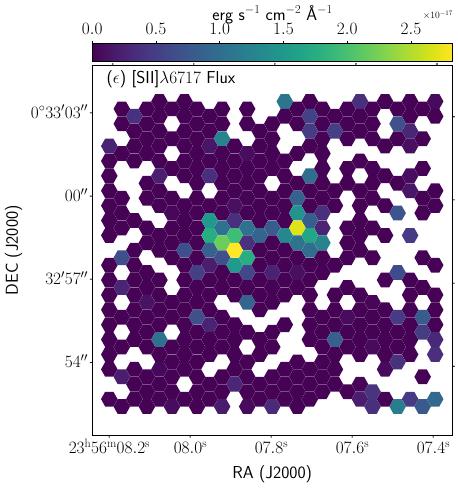}
	\includegraphics[clip, width=0.24\linewidth]{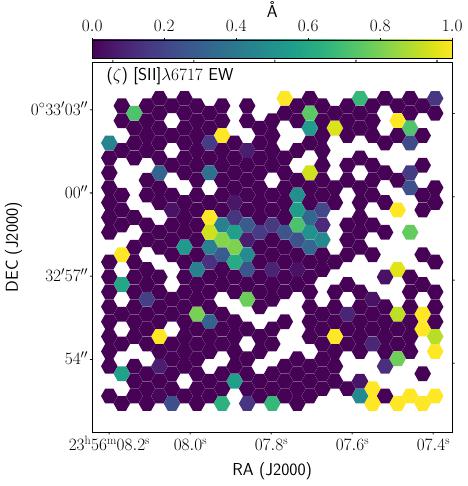}
	\includegraphics[clip, width=0.24\linewidth]{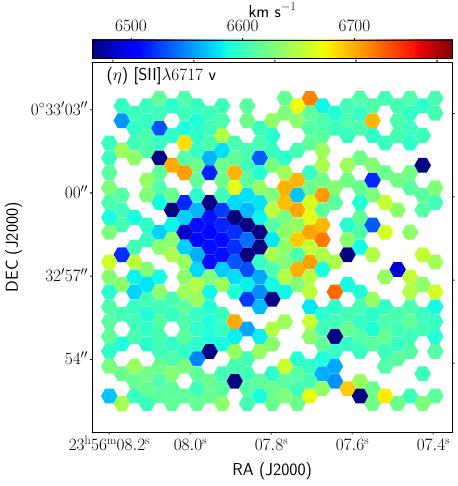}
	\includegraphics[clip, width=0.24\linewidth]{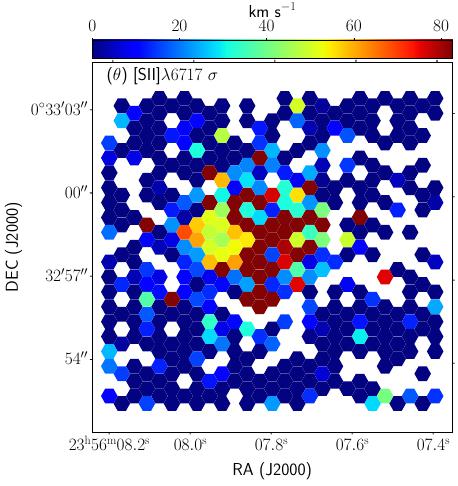}
	\includegraphics[clip, width=0.24\linewidth]{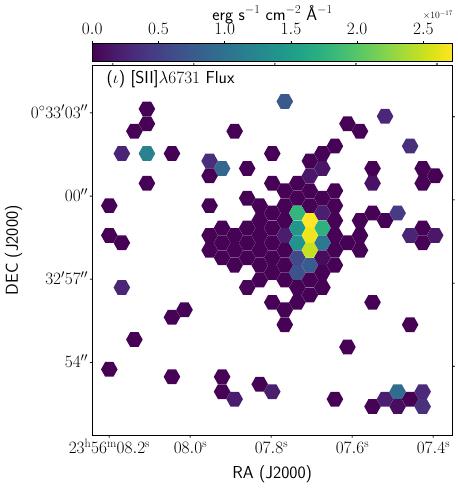}
	\includegraphics[clip, width=0.24\linewidth]{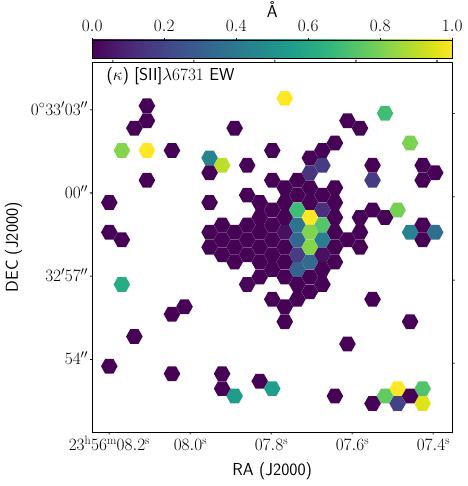}
	\includegraphics[clip, width=0.24\linewidth]{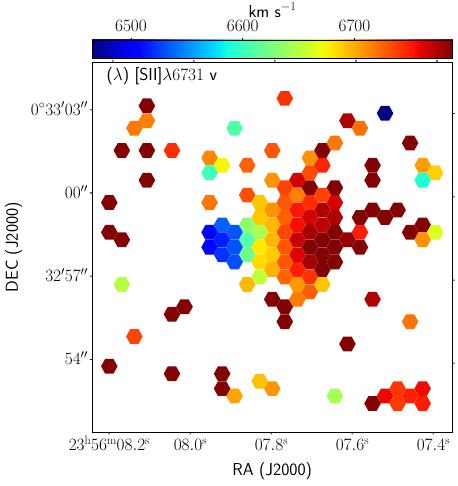}
	\includegraphics[clip, width=0.24\linewidth]{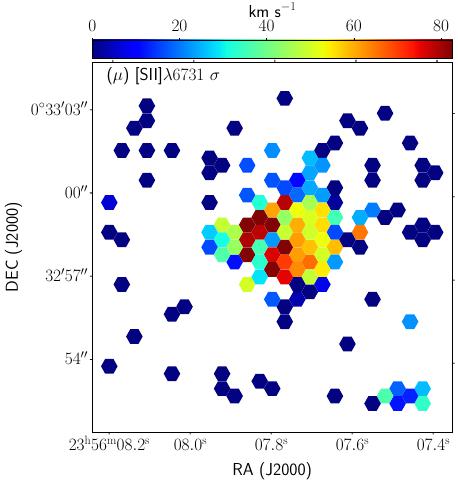}
	\caption{(cont.) NGC~7787 card.The velocity field of the $[\mathrm{S II}]\lambda6717$ line (subfigure* $\eta$) is affected by atmospheric telluric absorptions.}
	\label{fig:NGC7787_card_2}
\end{figure*}

\begin{figure*}[h]
	\centering
	\includegraphics[clip, width=0.35\linewidth]{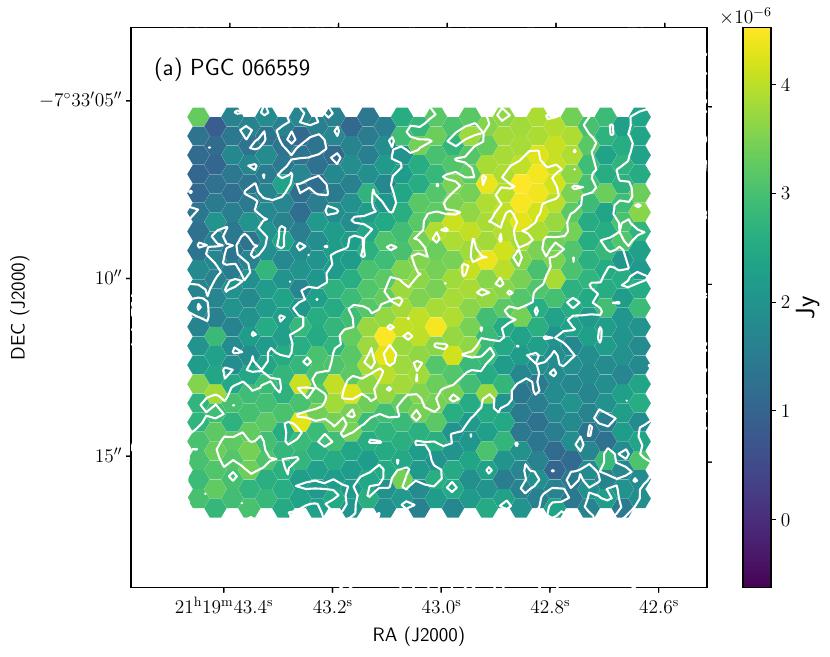}
	\includegraphics[clip, width=0.6\linewidth]{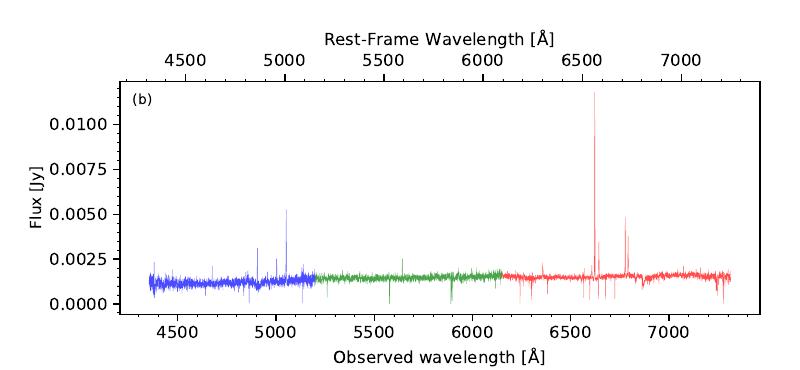}
	\includegraphics[clip, width=0.24\linewidth]{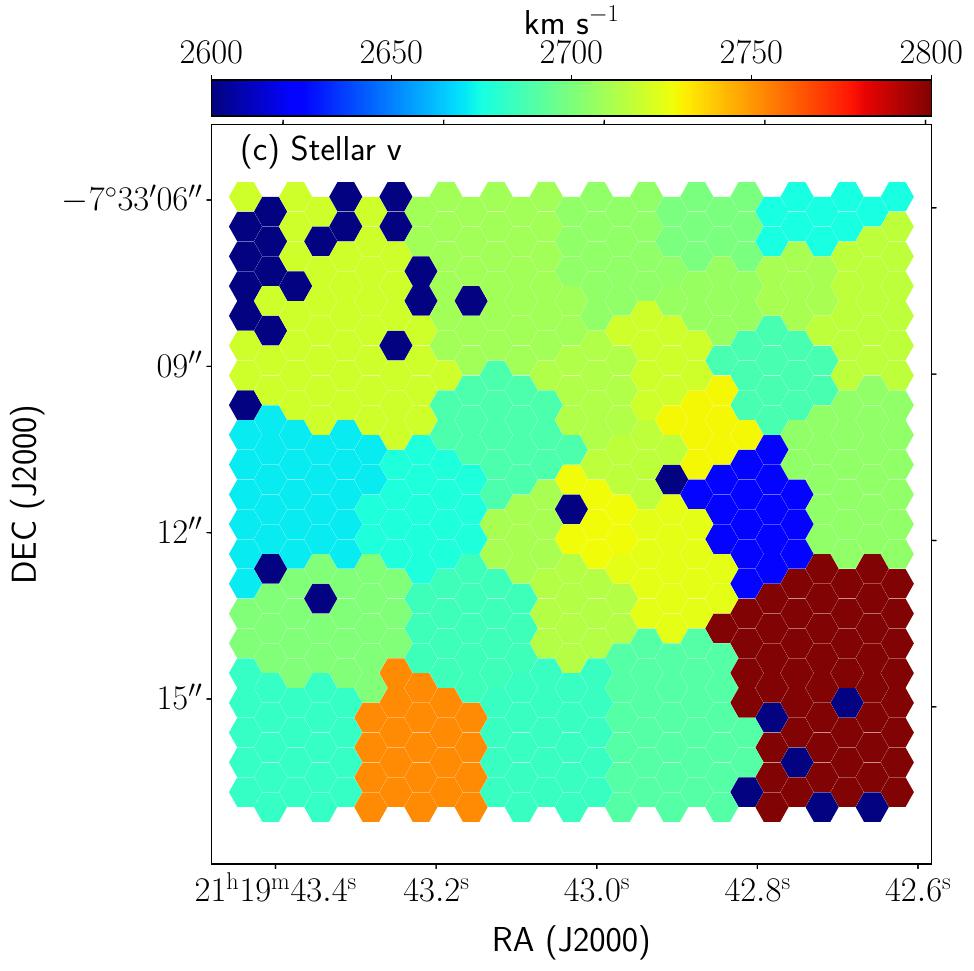}
	\includegraphics[clip, width=0.24\linewidth]{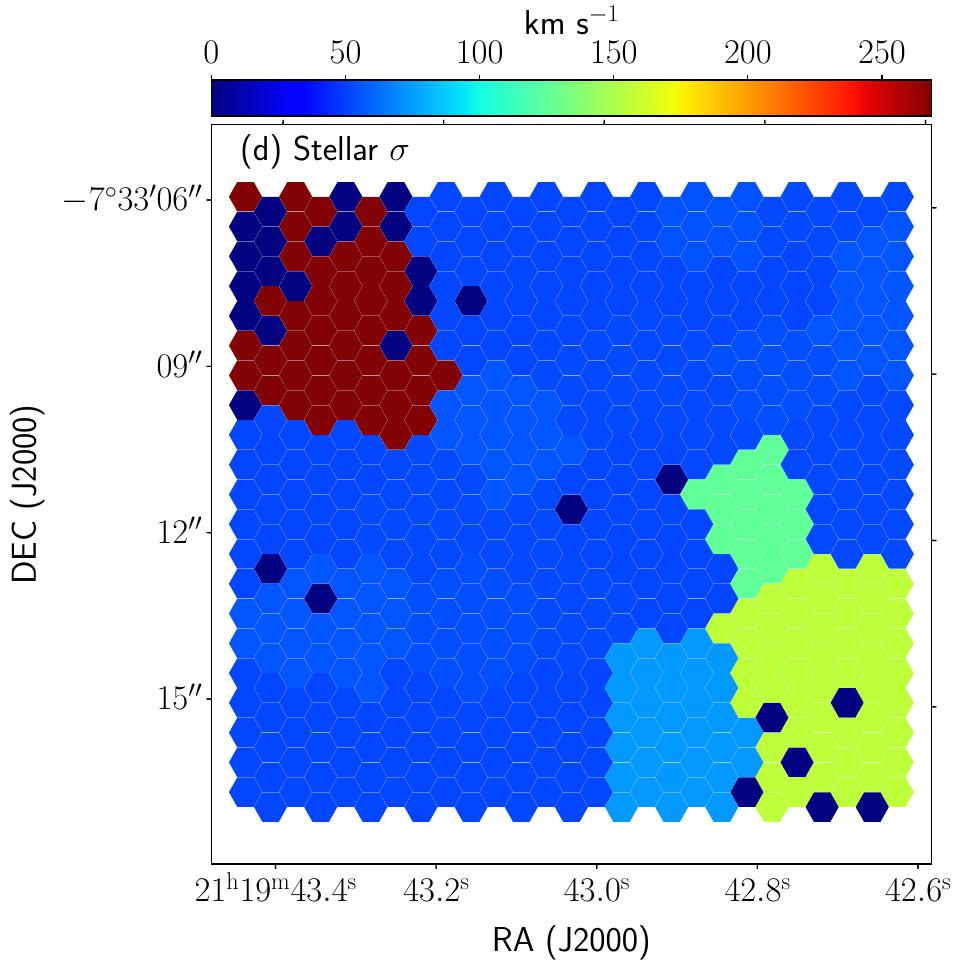}
	\includegraphics[clip, width=0.24\linewidth]{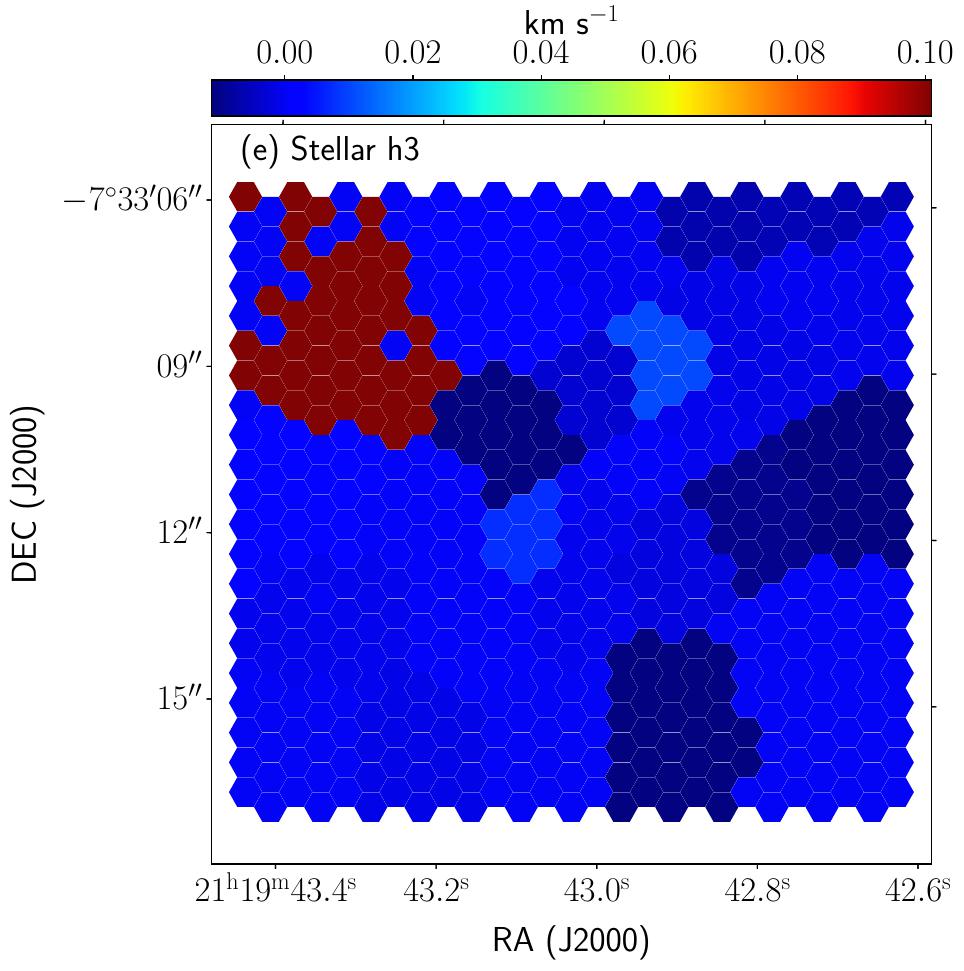}
	\includegraphics[clip, width=0.24\linewidth]{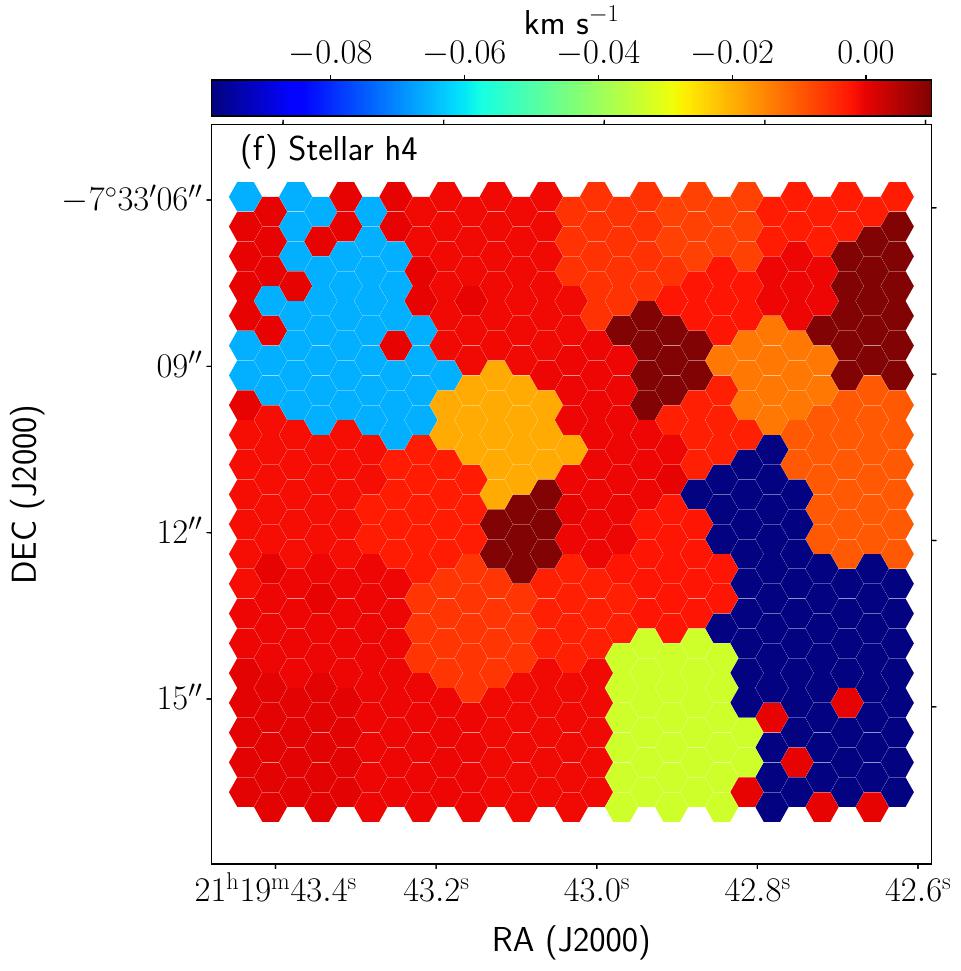}
	\includegraphics[clip, width=0.24\linewidth]{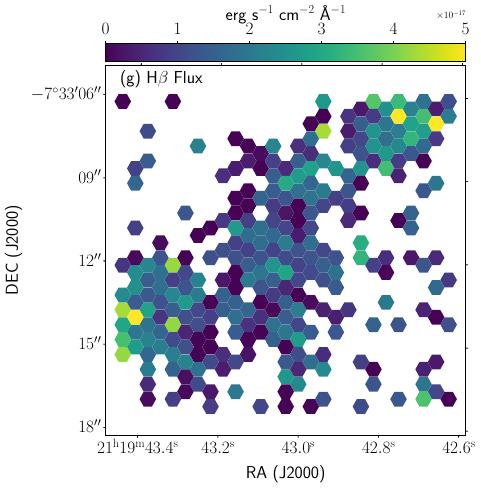}
	\includegraphics[clip, width=0.24\linewidth]{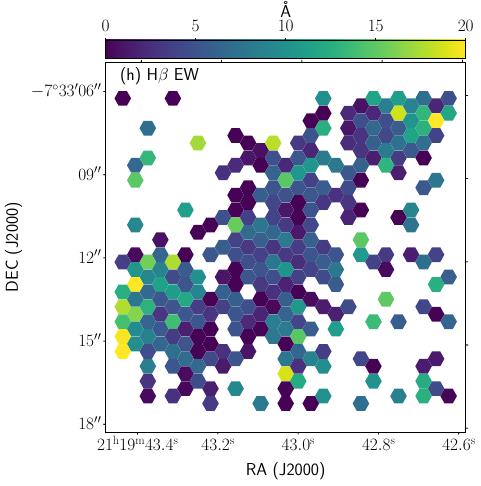}
	\includegraphics[clip, width=0.24\linewidth]{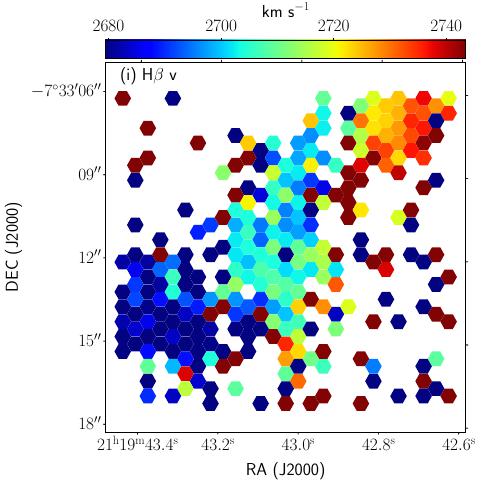}
	\includegraphics[clip, width=0.24\linewidth]{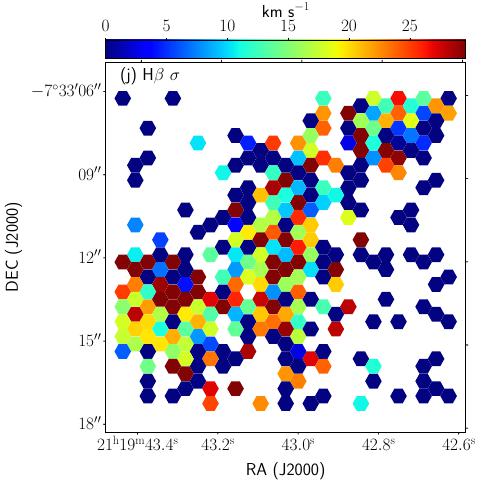}
	\includegraphics[clip, width=0.24\linewidth]{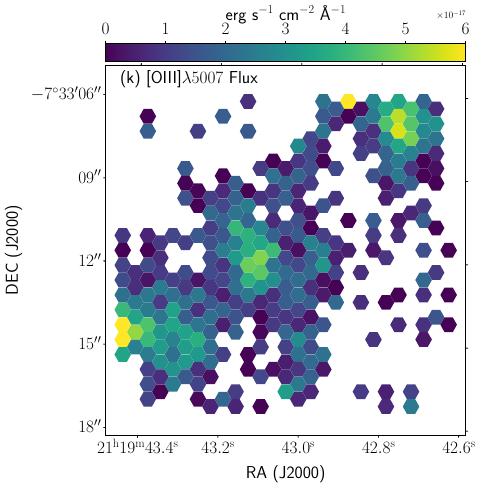}
	\includegraphics[clip, width=0.24\linewidth]{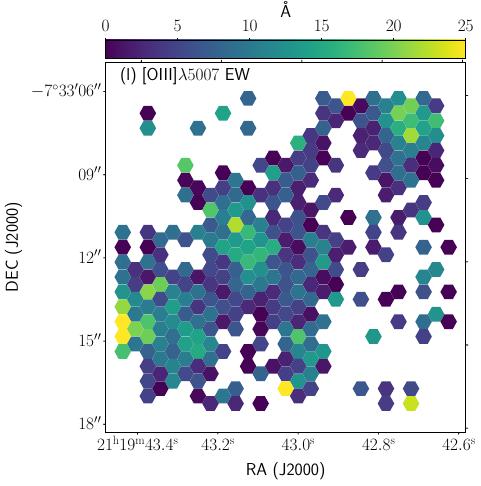}
	\includegraphics[clip, width=0.24\linewidth]{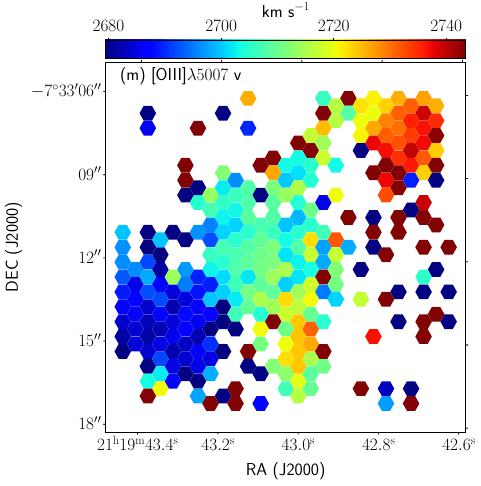}
	\includegraphics[clip, width=0.24\linewidth]{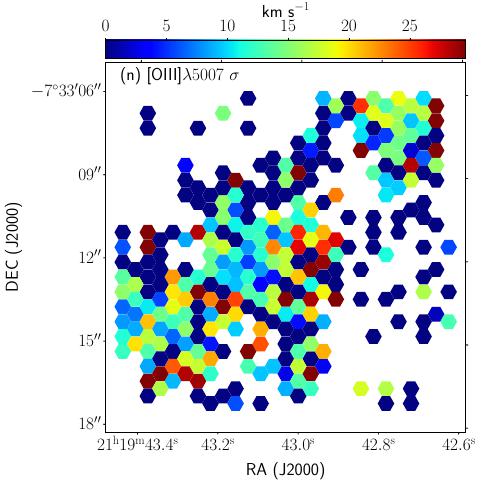}
	\vspace{5cm}
	\caption{PGC~066559 card.}
	\label{fig:PGC066559_card_1}
\end{figure*}
\addtocounter{figure}{-1}
\begin{figure*}[h]
	\centering
	\includegraphics[clip, width=0.24\linewidth]{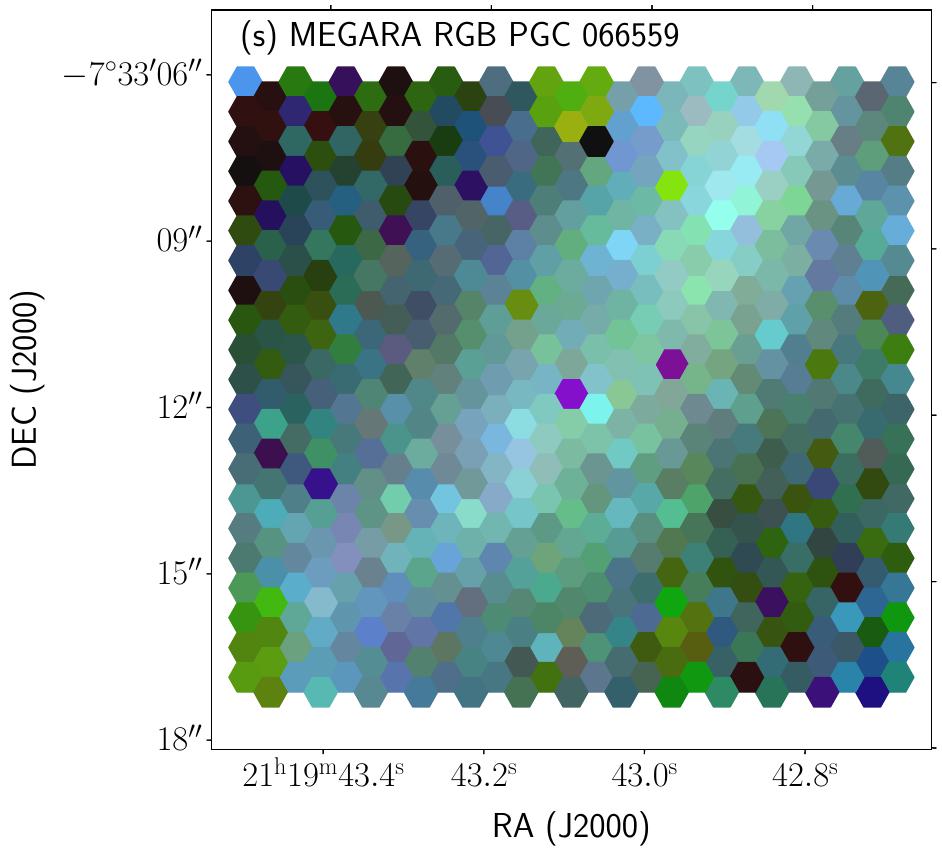}
	\includegraphics[clip, width=0.24\linewidth]{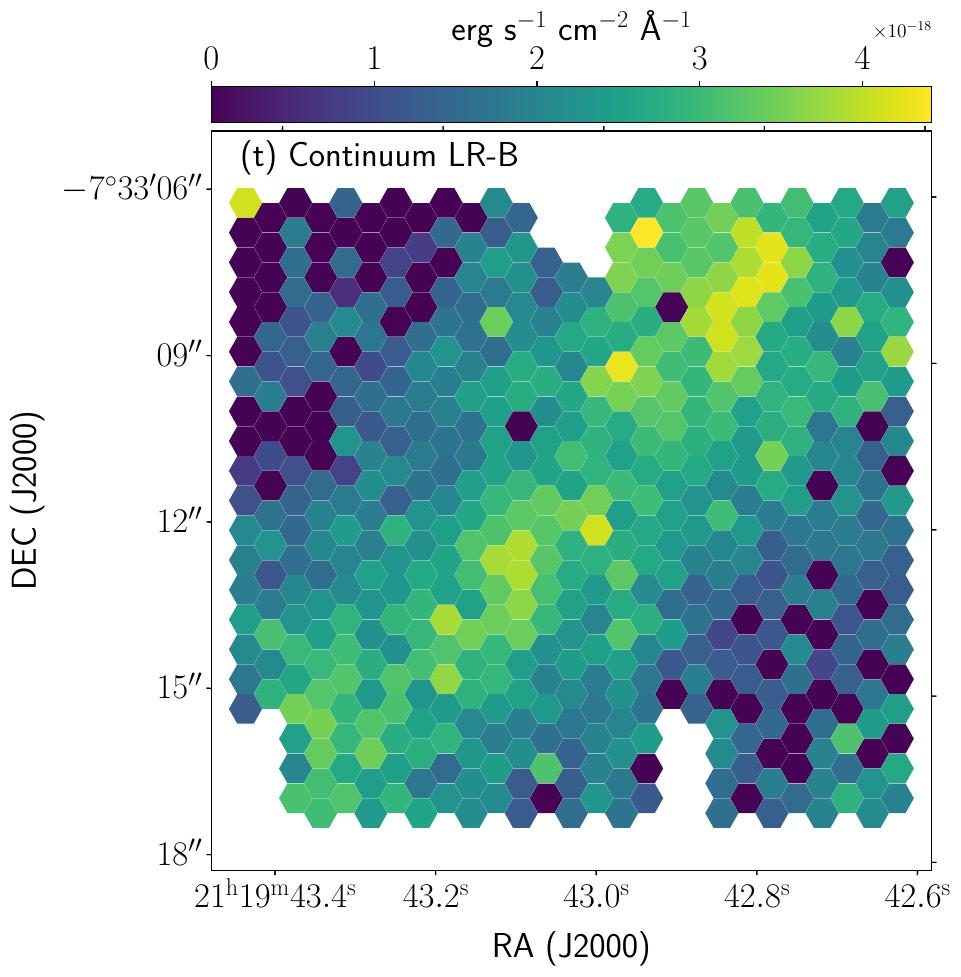}
	\includegraphics[clip, width=0.24\linewidth]{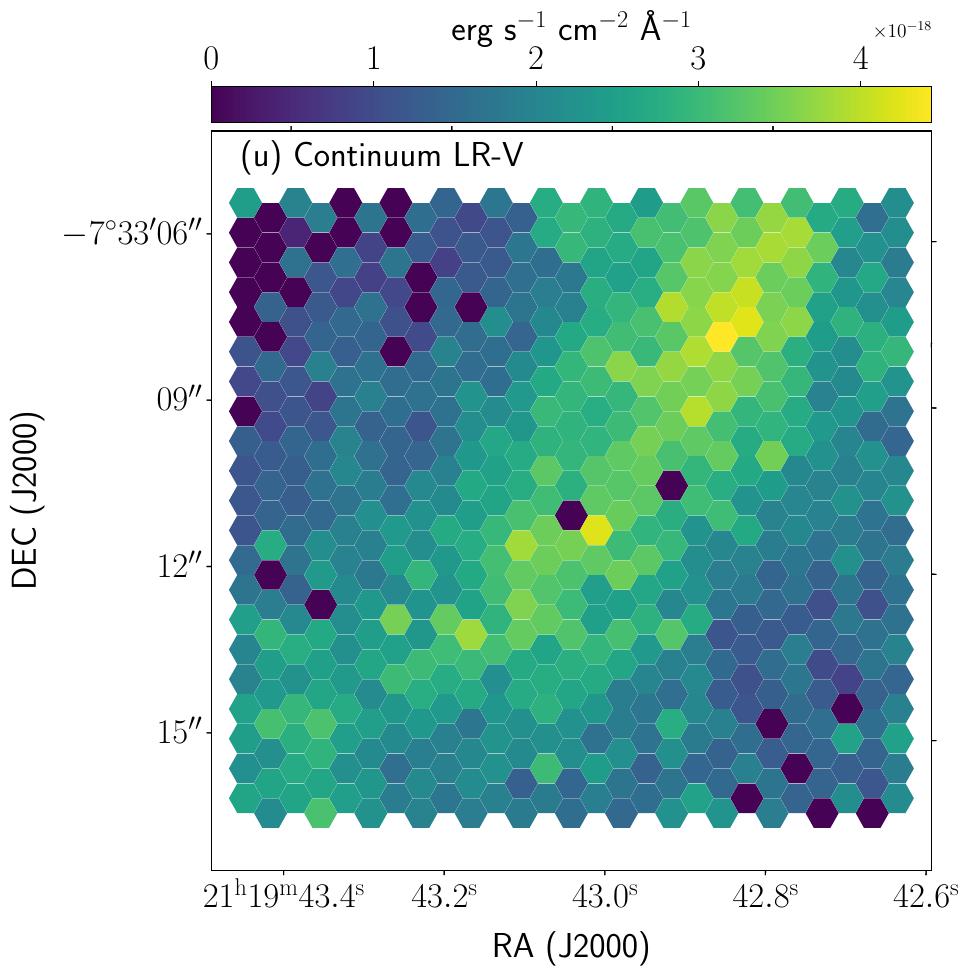}
	\includegraphics[clip, width=0.24\linewidth]{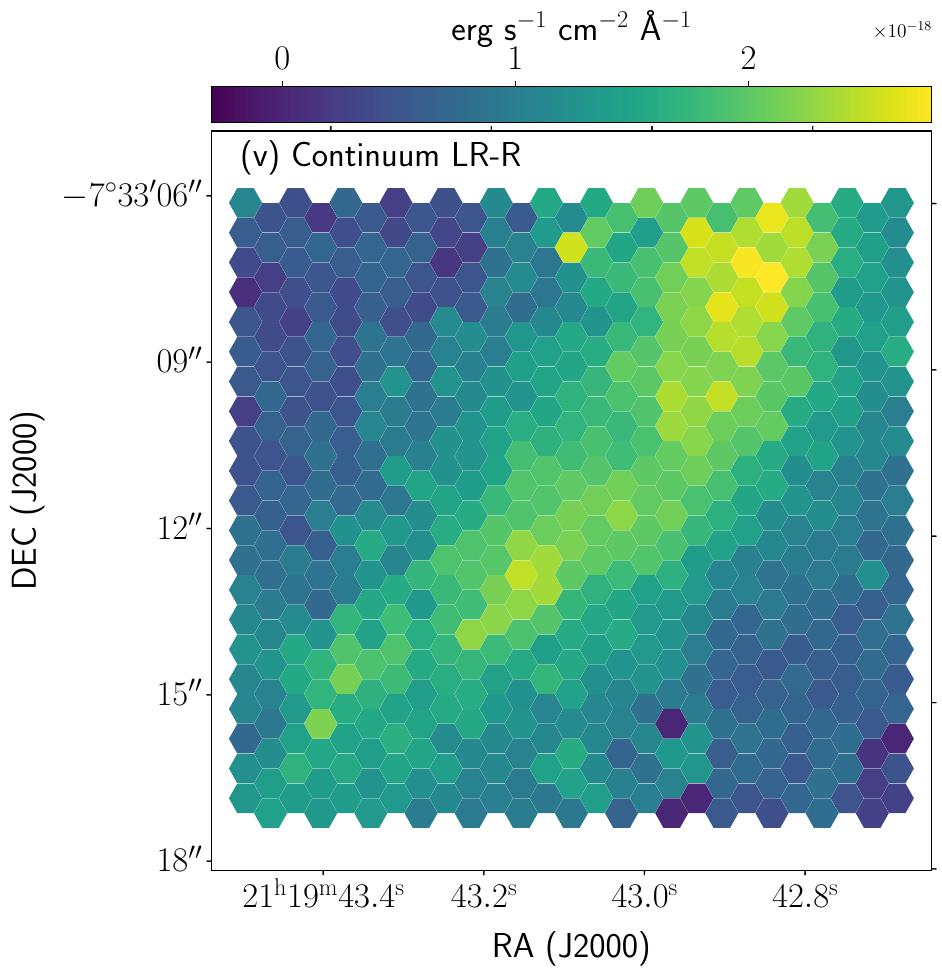}
	\includegraphics[clip, width=0.24\linewidth]{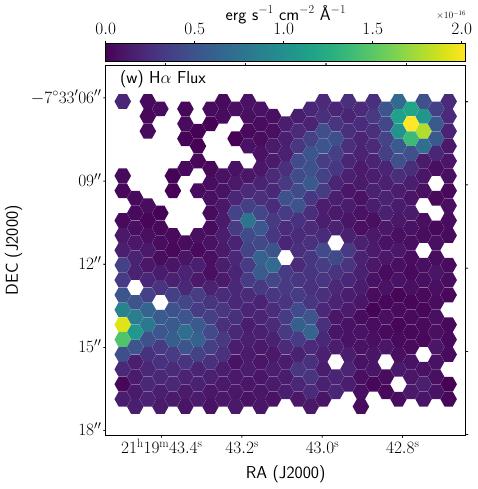}
	\includegraphics[clip, width=0.24\linewidth]{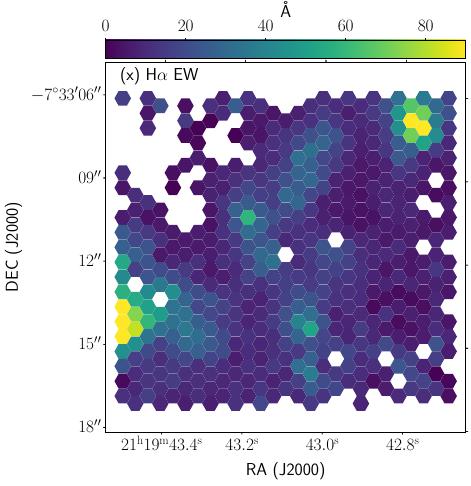}
	\includegraphics[clip, width=0.24\linewidth]{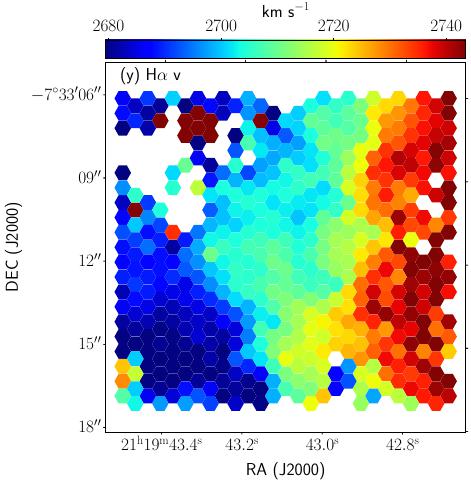}
	\includegraphics[clip, width=0.24\linewidth]{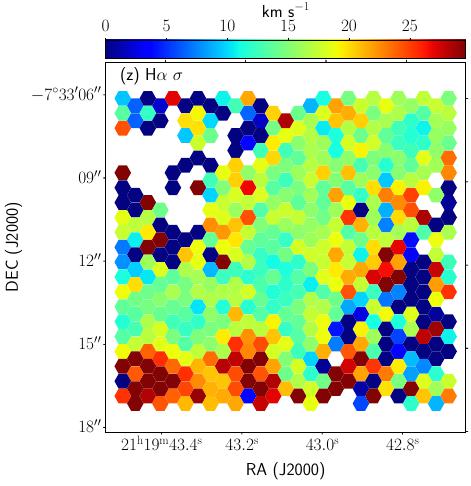}
	\includegraphics[clip, width=0.24\linewidth]{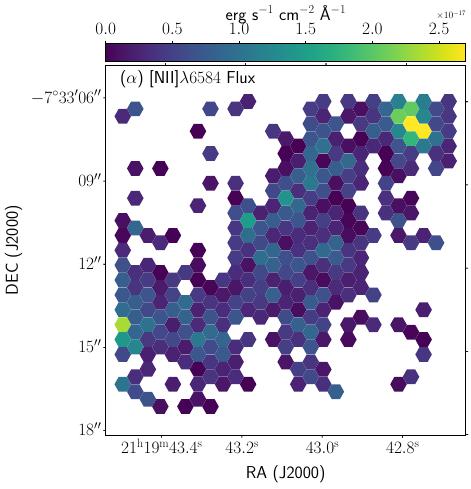}
	\includegraphics[clip, width=0.24\linewidth]{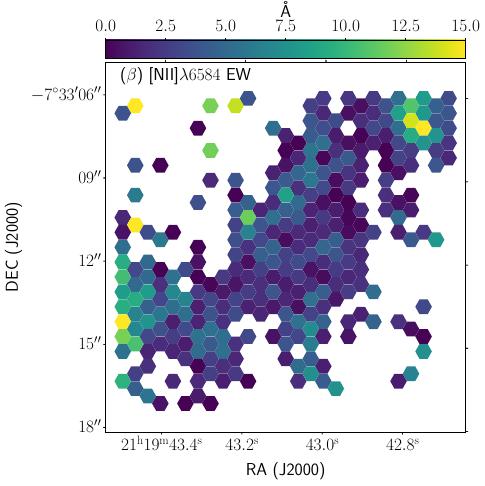}
	\includegraphics[clip, width=0.24\linewidth]{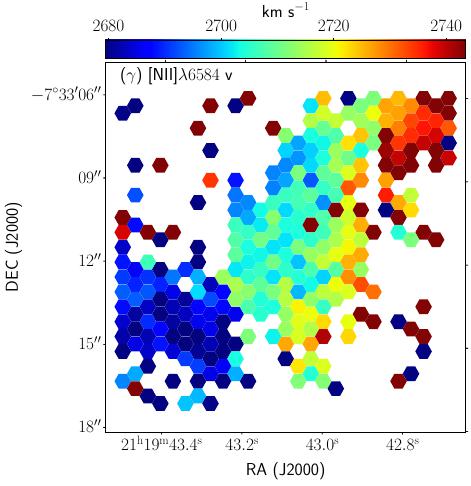}
	\includegraphics[clip, width=0.24\linewidth]{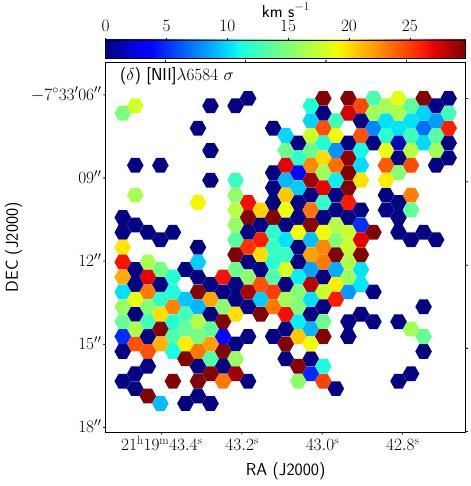}
	\includegraphics[clip, width=0.24\linewidth]{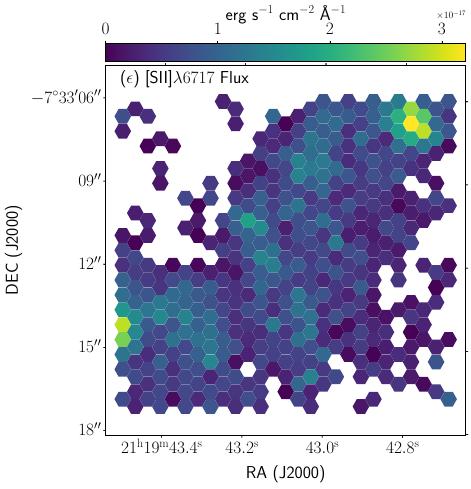}
	\includegraphics[clip, width=0.24\linewidth]{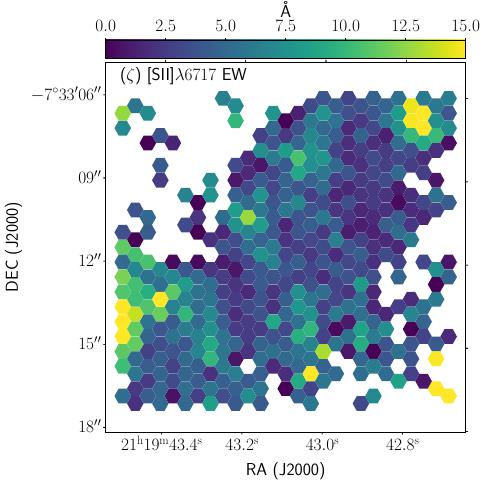}
	\includegraphics[clip, width=0.24\linewidth]{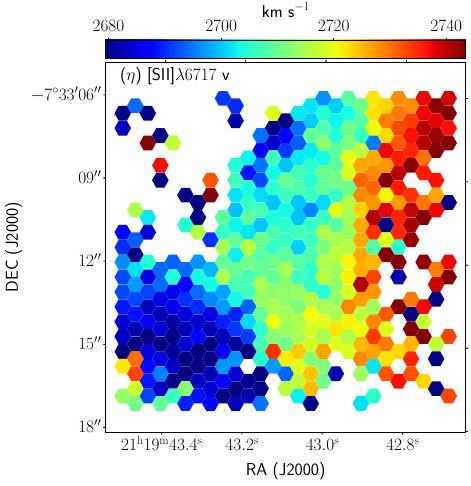}
	\includegraphics[clip, width=0.24\linewidth]{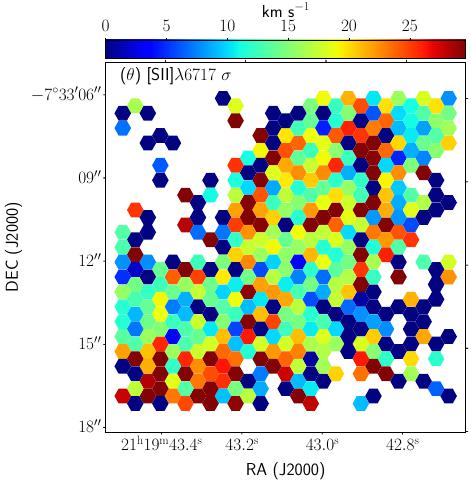}
	\includegraphics[clip, width=0.24\linewidth]{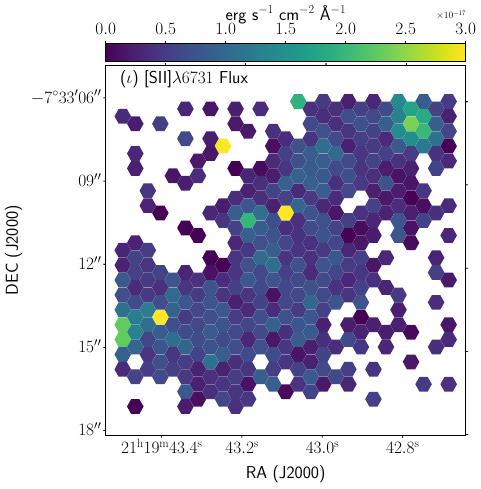}
	\includegraphics[clip, width=0.24\linewidth]{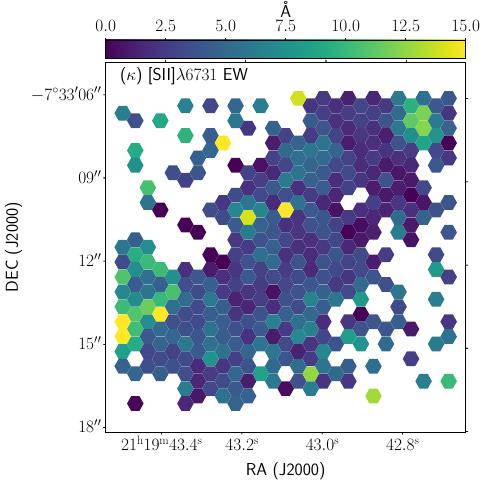}
	\includegraphics[clip, width=0.24\linewidth]{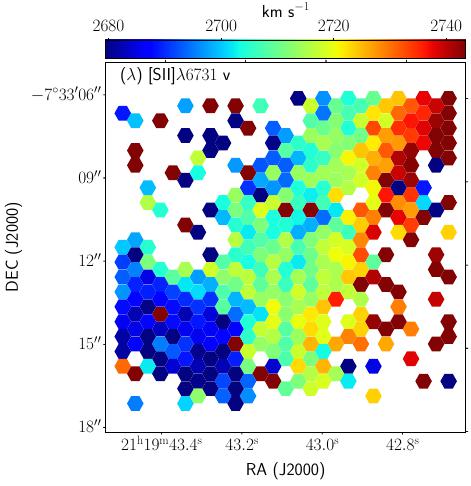}
	\includegraphics[clip, width=0.24\linewidth]{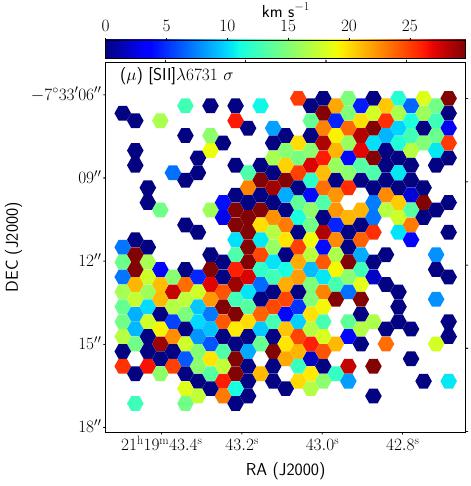}
	\caption{(cont.) PGC~066559 card.}
	\label{fig:PGC066559_card_2}
\end{figure*}

\end{appendix}

\end{document}